# Operationalism, Causality, and Quantum Theory: a mostly time symmetric perspective

Lucien Hardy
*Perimeter Institute,*
*31 Caroline Street North,*
*Waterloo, Ontario N2L 2Y5, Canada*

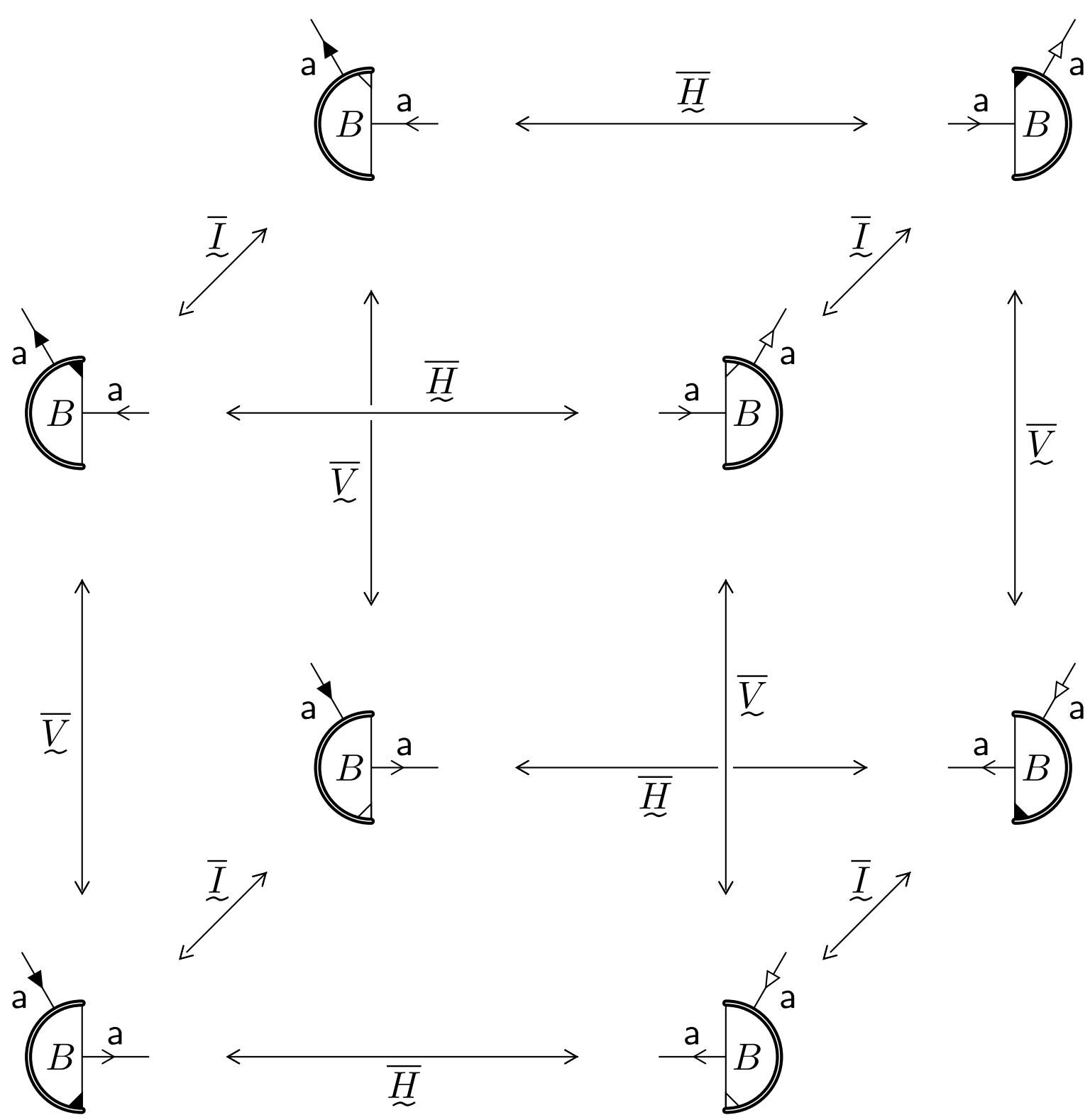

**Abstract**

This is a book about operational probabilistic theories. The standard approach in such theories is from a *time forward* perspective. In this book we mostly take a *time symmetric* perspective. This presents a branding problem. Is this a niche book merely *about* time symmetry? No. This is a comprehensive book about operational probabilistic theories, but from a time symmetric perspective (mostly).

In fact, this book consists of two books: (1) a simple book about simple operations having simple causal structure (where all the inputs are before all the outputs); and (2) a complex book about complex operations that can have complicated causal structure (a complex operation comes equipped with a causal diagram). For the simple case we are able to show that the time symmetric perspective is equivalent to the time forward perspective.

In each book we set up (A) operational probabilistic theories (OPTs) in terms of operations, (B) Operational Quantum Theory (OQT) in terms of operator tensors which correspond to operations, and (C) the theory of Hilbert objects which can be doubled up to give operator tensors.

Operations are required to be *physical*. Physicality guarantees that circuits built out of operations have probabilities between 0 and 1 and that certain causality conditions are met. We prove composition theorems for both simple and complex operations – that when we wire together operations the resulting networks are also physical (these theorems are especially interesting in the case of complex operations).

The theory of complex operations can be used to model physics happening in (discrete) spacetime. We use this to address Sorkin's impossible measurements. It turns out that if the operations are physical then there is no anomalous signalling.

We develop new diagrammatic notation to deal with Hilbert objects, particularly in the complex case. We discuss the *conjuposition* group of transformations on Hilbert objects. This includes mirrors to notate doubling up and some mirror theorems. We use this framework to prove time symmetric causal dilation theorems for a variety of causal diagrams.

END MATTER

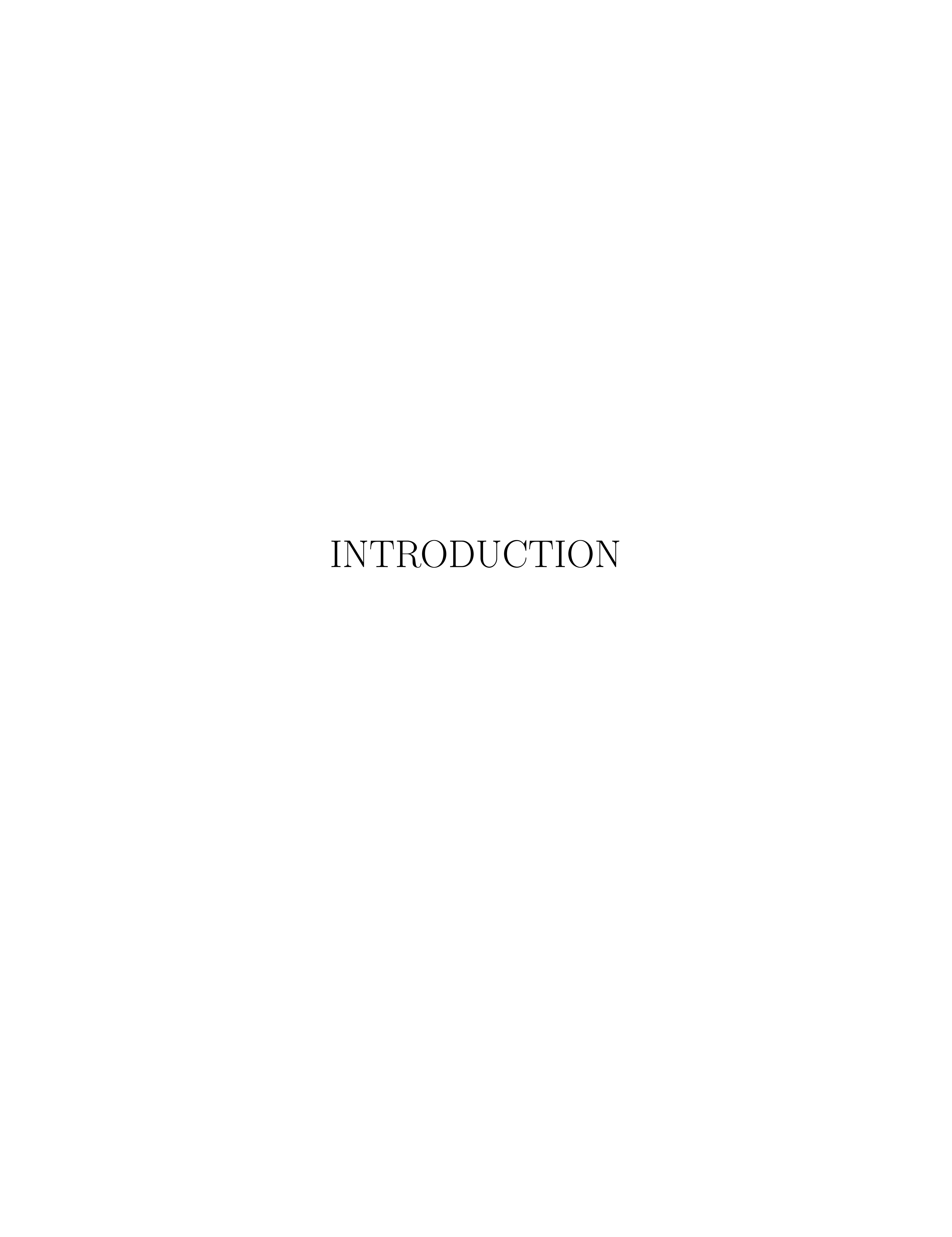

# INTRODUCTION

# 1 Structure of this book

Operational theories are built out of operations. An operation corresponds to one use of an apparatus. This apparatus may have holes in it allowing systems to go into the apparatus (these are inputs) and come out of the apparatus (these are outputs).

This book is, essentially, comprised of two books - the *Simple Book* and the *Complex Book*. The Simple Book concerns operations with simple causal structure (where the outputs are after the inputs). The Complex Book concerns operations with complex causal structure. The word "complex" here is used in the sense of "complicated". This concerns operations where the outputs are not necessarily all after the inputs. Each book consists of three parts.

**Operational Probabilistic Theories.** Here we start with an operational description of experiments then add probabilities to set up a linear framework for calculating probabilities for circuits.

**Operational Quantum Theory.** Here we set up a correspondence from operations to operators which can be used to state the axioms of Operational Quantum Theory.

**Hilbert Objects.** Here we set up Hilbert Objects - these are elements of a Hilbert space defined by the input/output wires. They are essential if we are to prove theorems. In particular, we will prove various Stinespring-style dilation theorems including, in the complex case, causal dilation theorems.

In the Simple Book we have simple operations, simple operators, simple left Hilbert objects, and simple right Hilbert objects, which we represent pictorially as

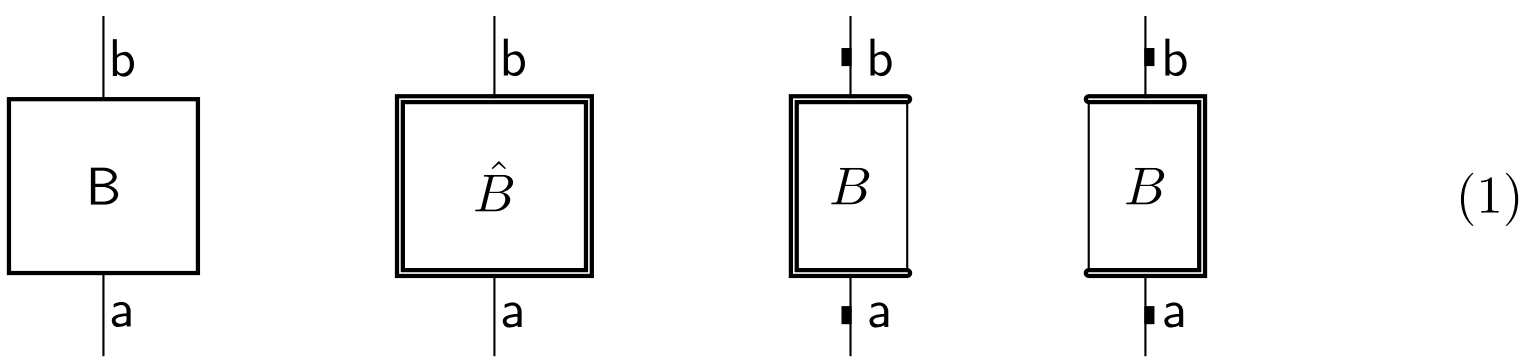

(1)

respectively. In the Complex Book we have complex operations, complex operators, complex left Hilbert objects, and complex right Hilbert objects which we represent pictorially as

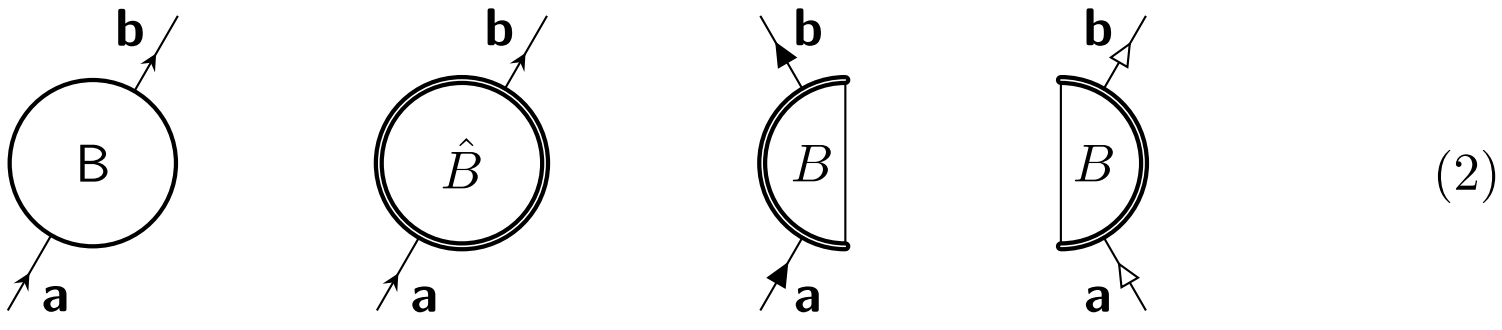

(2)

respectively.

The standard approach in operational theories is from a *time forward* perspective. In this book we mostly take a *time symmetric* perspective. This presents a branding problem. Is this a niche book merely *about* time symmetry? No. This is a comprehensive book about operational probabilistic theories from a time symmetric perspective (mostly). The idea I want to get across here is that we can take a time forward, a time backward, or a time symmetric perspective. Whilst the time forward perspective accords with how we see the world for various contingent reasons, the time symmetric perspective is, in some respects, more fundamental. Furthermore, for simple operational theories we can prove that the time symmetric perspective is empirically equivalent to the time forward perspective (and the time backward perspective). The different perspectives simply "package" the probabilities in different ways. Interestingly, whether this equivalence between these temporal perspectives holds for complex operational theories is, at the time of writing, an open question. It is possible that time symmetric complex operational theories are more constrained than time forward complex operational theories. If this turns out to be the case then it is reasonable to suppose nature would choose the more constrained time symmetric theory.

The simple part of this book has its roots in my earlier work, Hardy [2021], on time symmetry in operational theories.

# 2 Basic ideas and motivations

## 2.1 Picturalism

Throughout we adopt a *pictorial* approach. If you flip through the pages of this book you will notice that almost all mathematical expressions, equivalences, inequalities, and equations are given in pictorial form. The idea that tensor calculations can be represented by diagrams goes back to Penrose [1971] (see also Penrose and Rindler [1984]) and was given a categorical foundation by Joyal and Street [1991]. This was adopted in the categorical approach to Quantum theory pioneered by Abramsky and Coecke [2004]. The pictorial aspect of this categorical approach was emphasised in Coecke [2005]. Building on this work, Selinger [2007, 2011] provided more foundational developments of the diagrammatic framework including application to completely positive maps which are very relevant for the present work. The beautiful book "Picturing Quantum Processes" by Coecke and Kissinger [2017] is the canonical reference. Such approaches are sometimes referred to as *process theories*. Categorical thinking has been pursued by other people thinking about Quantum Theory. For example see Baez [2006].

This pictorial approach of Coecke and others was enthusiastically adopted and adapted by researches thinking from an operational perspective including Chiribella, D'Ariano, and Perinotti [2010, 2011] and Hardy [2013a, 2011b, 2012]. We will discuss operationalism below. Picturalism has, by now, become standard currency in the Quantum Foundations community. In the present book we adapt the diagrammatic notation of earlier works a little for the simple case introducing some new elements. For the complex case we have a more substantial adaptation.

Many similar mathematical structures to those considered in process theories appear in the study of *tensor networks* (see Vidal [2003] Vidal [2008] and Verstraete et al. [2008]) although the philosophy and motivation is different. The field of tensor networks has its origins in the study of efficient representation of quantum states, numerical simulations, and many body physics. Reviews can be found in Schollwöck [2011] and Orús [2014]. Tensor networks and process theories were independent developments though both were influenced by Penrose's earlier work mentioned above.

Pictures can make quite complicated proofs much easier to follow. Understanding pictorial expressions is like eating delicious wholesome cake whilst understanding the corresponding symbolic expressions is like eating broken glass laced with nails and razor blades. Marshall McLuhan's famous dictum "the medium is the message" (see McLuhan [1964]) is clearly apt here. McLuhan's idea is that the form of a medium embeds itself in the message it transmits, shaping how the message is perceived and understood. The choice between symbolic or diagrammatic medium in the context of a mathematical proof certainly impacts on how the latter is perceived and understood.

The idea that mathematical expressions can be expressed by pictures in place of symbolic expressions is quite striking when one first encounters it. Feynman diagrams are probably the best known example in the physics literature. There has been some discussion of diagrammatic proofs in the philosophical literature. For example, see the public lecture by James Brown (at Brown [2004]) and also his academic work on the subject in Brown [1997, 2008]. Major philosophical works on this subject include Peirce [1933] and Shin [1994]. Also see the Stanford Encyclopedia review at Shin et al. [2025].

A challenge for picturalism is to take these pictures out of the notebook into published form. Fortunately, the TikZ drawing package is incredibly versatile. The pictures in this book have many common elements and so it would be laborious to write fresh TikZ code afresh for each new drawing. Almost all the pictures in this book are drawn using the CompoZition package which comprises of commands using TikZ (the "Z" in CompoZition is homage to TikZ). This package is designed so that the code has a similar "look and feel" to LaTeX code for typesetting for mathematical. For example the picture

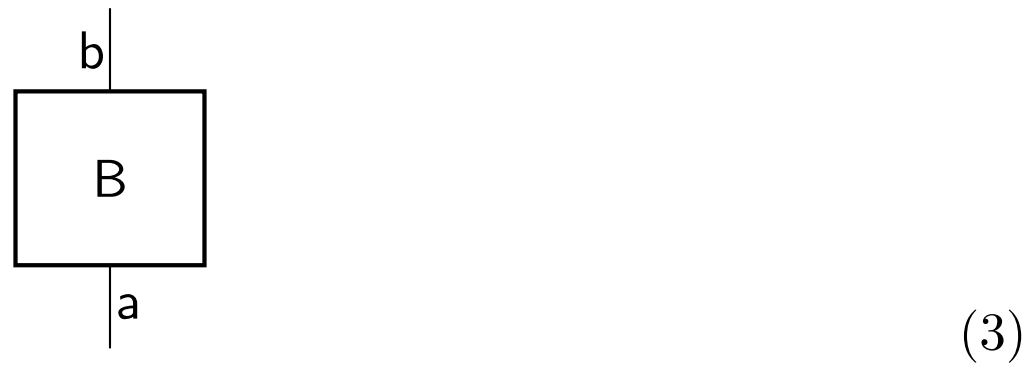

(3)

is drawn using the code

```
\begin{compose}
```

```
\crectangle{A}{2.5}{2.3}{0,0} \csymbol{B}
\thispoint{d}{0,-4.5} \thispoint{u}{0,4.5}
\jointbnoarrow[right]{d}{0}{A}{0} \csymbol{a}
\jointbnoarrow[left]{A}{0}{u}{0} \csymbol{b}
\end{compose}
```

This package can be found at

https://github.com/LucienHardy/compoZition.

along with supplementary material including a set of illustrative examples.

One more comment. Many texts use uppercase letters for the wire labels (which are systems in our application). This convention seems to come from category theory. Penrose [1971], on the other hand, uses lower case letters. The latter makes more sense if you think of these objects as originating from tensors since smaller symbols work better as subscripts and superscripts. We take a Penrosian approach here because we will sometimes use symbolic notation and because the systems are, in some sense, less substantial than the operations which we represent by uppercase letters. This small notational difference does represent shift in interpretation and, one might add, motivates different kinds of symbolic expressions.

## 2.2 Operational Probabilistic Theories

Operationalism is a powerful way to think about physics. The idea is to look at the world from our own point of view. We both act on the world and observe the world. Experiments are like this. We set up an experiment by placing a bunch of apparatuses in a certain arrangement adjusting settings on these apparatuses appropriately (this is us acting on the world). Then we readout results from these apparatuses (this is us observing the world).

A basic concept here is the idea of an *operation* which corresponds to one use of an apparatus. We will consider apparatuses that have apertures (holes) in them through which notional systems may pass. Apertures which systems come out of are *outputs* and apertures which systems go into are called *inputs*. We can represent an operation as follows

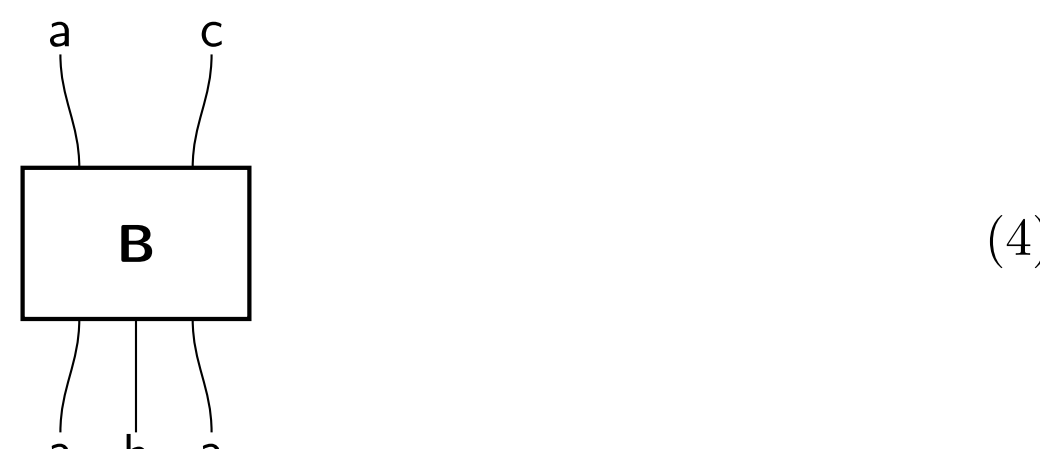

(4)

Inputs are represented by wire attachment points at the bottom of the box and outputs are at the top. The labels on the wires represent the type of system associated with the aperture. Defining an operation involves *proscribing* the how the apparatus is used. We say which apertures are to be used as inputs and for what type of system and, similarly, which are to be used as outputs and for type of system. We are free to fire small rocks into an aperture meant for electrons but we cannot expect the theory we develop to work in such a situation (probably the apparatus would break).

For these introductory remarks on operationalism, we are considering the case where the outputs are after the inputs so we have simple operations. We will discuss complex operations (which we will represent by circles) below. We can align the apertures such that systems can pass between the apparatuses. An example of an experiment, in this way of thinking, might look like this

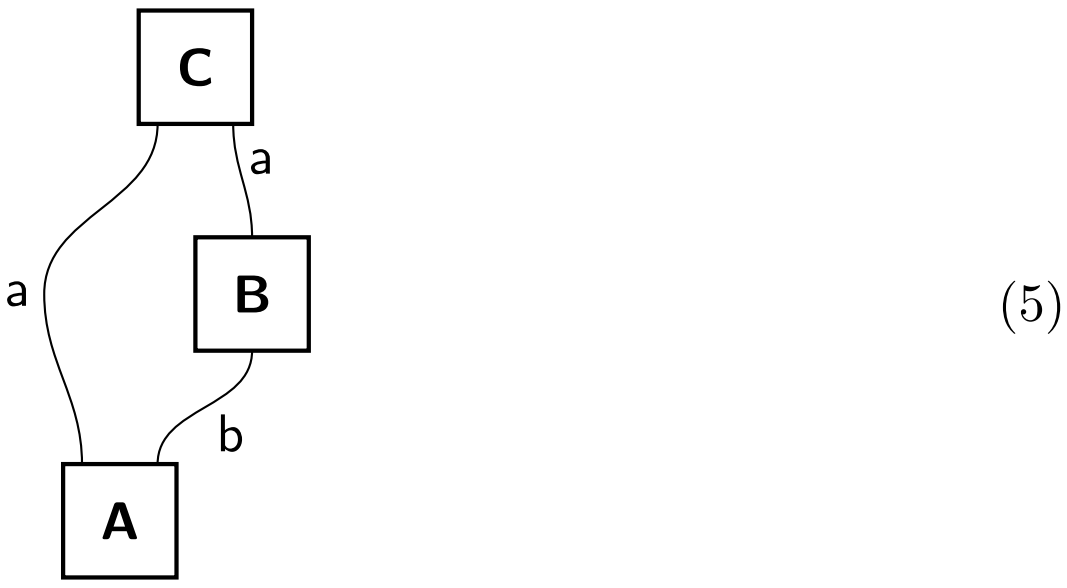

(5)

The wires represent where the apertures are aligned and the label on the wire represents the type of system passing through the aperture. Time goes up. There are no open wires left over so we will call this a *circuit*. The circuit must be a directed acyclic graph (DAG) meaning that there are no closed loops as we trace forward along the wires between the operations.

The operational picture described here is not completely general. For example, it is ill suited to deal with high-school statics problems - such as whether ladder leaning against a wall will remain static or slip. It is, however, well suited to deal with Quantum Theory which is the main driving interest in this book. We can, incidentally, envisage more general operational pictures - see Hardy [2009b].

What we have provided so far is a *descriptive language*. Experiments can be *described* as circuits. To actually make predictions we need to specify the quantity or quantities which we intend to make predictions for. For Quantum Theory this quantity is *probability*. A central (and deep) assumption in this work is that, for a given circuit, the joint probability for the readouts on the operations depends only on the description of the given circuit. Once we have this we assumption we can introduce the $p(\cdot)$ function. This acts on a real value weighed sum of circuits to return the weighted sum of the probabilities for these circuits. We will see that the $p(\cdot)$ function allows us to represent operations as mathematical objects living in a linear space enabling us to do calculations for probabilities for general circuits. We are thereby able to set up *Operational Probabilistic Theories* (which we call OPTs). Within these we can set up Operational Quantum Theory. However, it is interesting that, even before we go to Quantum Theory, we can formulate many ideas and prove many results in the OPT framework. A productive attitude to take is to do as much as possible within the more general OPT framework. Then, if we need to go to the Quantum framework to state an idea or prove a result then this teaches us something about Quantum Theory (this attitude was expounded in Barnum et al. [2006, 2007, 2012] and Barnum and Wilce [2011]). For example, we are able to state our causality assumptions and prove theorems about causality within the OPT framework. However, to prove Stinespring-style dilation theorems, we need to go into the Quantum framework.

Operationalism goes back to Bridgman [1927] who was particularly struck by the need to specify the manner in which spacial-temporal quantities are measured in Special Relativity (see Chang [2009]). What attitude should we adopt to operationalism here? There are, broadly, two approaches we might take. We could think of operationalism as a *fundamental philosophy* whereby we take the very meaning of physical concepts to come from an operational account of how they are measured. Taking this a step further, in a logical positivist direction (see Carnap [1956]), we might even deny that there is any reality beyond this operational description. There is some merit to this way of thinking within Quantum Theory as it offers a way of justifying not going beyond the basic Quantum framework enabling us to be happy with what we have got. Nevertheless, there are many criticisms of this attitude to operationalism (see Chang [2009]). In particular, it has been argued that there are deep philosophical problems with taking this fundamental attitude to operationalism. For example, it has been argued that all our measurements are, ultimately, theory laden (see Hanson [1958]) and so the very quantities we use in an operational description depend on theories we have developed, making them unsuitable for such a purist operational way of thinking. Furthermore, such an attitude is unnecessarily limiting. In physics we do, in fact, often find ourselves going beyond the operational picture we had started with. The attitude towards operationalism I would support is to regard it as an *opportunistic methodology* for the development of new theories. Then operational constructs are more akin to the scaffolding used in erecting a new building. Important for the construction process but, ultimately, removable. This attitude does not have the drawbacks of the fundamental philosophy approach. We can adopt operationalism for opportunistic purposes - because we believe it is a way to make progress in physics at this particular juncture in history. At a different juncture (or even in parallel) we may take an ontological approach whereby we start by positing some underlying picture of reality and build up to predictions from this. The idea of operationalism as a methodology is less limiting than the idea of it as a fundamental philosophy.

The operational approach to Quantum Theory has deep roots going back to Heisenberg's original matrix mechanics paper (see Heisenberg [1968]) conceived in Helgoland in 1925. There he says

> The present paper seeks to establish a basis for theoretical quantum mechanics founded exclusively upon relationships between quantities which in principle are observable.

The modern operational approach uses a mathematical framework which weds this operational way of thinking to probability theory. This is the framework for *Generalized Probabilistic Theories* (GPTs). This terminology was introduced by Barrett [2007]. Barrett based his approach on the framework introduced by Hardy [2001] (which was used to set up an axiomatic derivation of Quantum Theory). This framework is actually a finite dimensional reinvention of a much earlier framework going back to Mackey [1963] with many variants (see Ludwig [1954, 1985 and 1987, 2012], Foulis and Randall [1979], Beltrametti and Bugajski [1997], Gudder et al. [1999], Holevo [1982]). An operator algebraic approach is taken by Alfsen and Shultz [2003]. GPTs continue to be a very active field of research - see Barnum et al. [2014], Janotta and Hinrichsen [2014], Barnum and Wilce [2015], Wilce [2017], Schmid et al. [2021], Plávala [2023], Selby et al. [2023] and recent experimental progress in Mazurek et al. [2021]. A video course can be found by Wilce [2024] on PIRSA. Fritz [2026] writes about GMTs (generalized measurement theories) in Fritz [2026].

An even stronger marriage between probabilities and operationalism is afforded if we take this attitude a step further by taking the pictorial operational description of experiments as a foundation for generalized probabilistic theories (see Hardy [2009b]). This leads to the Operational Probabilistic Theories framework discussed in Sec. 2.1. This program has been implemented by Chiribella, D'Ariano, and Perinotti [2010, 2011] and by Hardy [2013a, 2011b, 2012]. The OPT program is strongly motivated by the pictorial approach (often referred to as process theory) of Coecke and others as discussed in Sec. 2.1. Indeed, the OPT program represents the beginning of a convergence between the operational and process theory communities (the present work takes this convergence even further). One selling point of the OPT/process theory approach is that calculations are represented by pictures that *look just like* the experiment they are a calculation for (this was called the *composition principle* in Hardy [2013b]). This is a deep point and only possible in the first place because both these approaches take the compositional structure of experiments seriously. There is more to it than that, though. The principle is only possible if nature is inclined to go along with it. In this book we will see that the much studied assumption of tomographic locality (see Sec. 9.7) is a sufficient requirement for the composition principle to work in the circuits and networks we study.

## 2.3 Physicality conditions

A central concept in operational theories is what we will call *physicality*. We impose physicality conditions on operations to ensure (i) that the probabilities associated with circuits are between 0 and 1, and (ii) to ensure causality conditions are met. Curiously, the causality conditions automatically impose the subunity property (that probabilities are less than or equal to 1). Thus, physicality conditions consist of (a) positivity conditions and (b) causality conditions.

These conditions appear in textbook versions of standard Operational Quantum Theory as the condition that (a) superoperators are completely positive and (b) superoperators are trace non increasing (see Nielsen and Chuang [2000] for example). Chiribella, D'Ariano, and Perinotti [2010] showed how to relate the trace non-increasing property to a version of causality (that choices in the future should not influence the past).

We will extend and adapt this idea to define causality conditions that apply to the time symmetric perspective. In the time symmetric context the positivity conditions are, essentially, the same as in the time forward case. However, the causality conditions become double causality conditions as we need one for the forward time direction and one for the backward time direction. In the simple case the forward causality condition is essentially the same as that of Chribella et al. and the backward causality condition is the time reverse. In the complex case, the conditions are more complicated as we need both a forward and a backward causality condition for each way of bisecting the associated causal graph into future and past parts. Here we borrow from the work on Quantum Combs by Chiribella et al. [2009a].

In the first part of each book (the operational part) we will prove *composition theorems*. The basic point here is to prove that, when we wire together any number of physical operations, the resulting network is also physical. We will prove these composition theorems for the positivity and causality properties separately since the conditions under which the theorems hold differ slightly. These theorems are more difficult to prove in the case of complex operations where we have to get involved in the intricacies of combining causal diagrams. Correspondingly, the complex case is more interesting.

## 2.4 Simple and complex causal structure

Operations have *simple causal structure* if all the inputs are *before* all the outputs. Circuits then consist of operations wired together in a directed acyclic graph (DAG) meaning that there are no closed loops as we trace forward along the wires between the operations. We represent simple operations by rectangles as we saw in (4).

Operations have *complex causal structure* if some of the outputs on the operation can be regarded as being before some of the inputs. It is easy to build complex operations out of simple operations. For example, the object

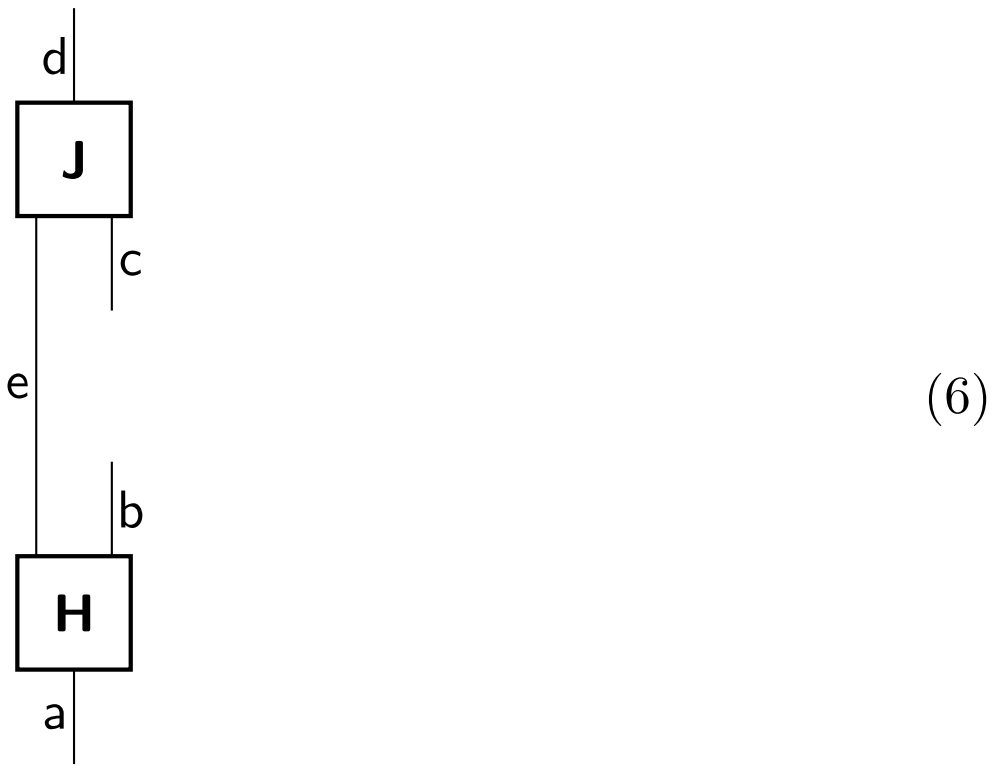

(6)

has the property that output b is before input c. The specification of a complex operation consists of two parts

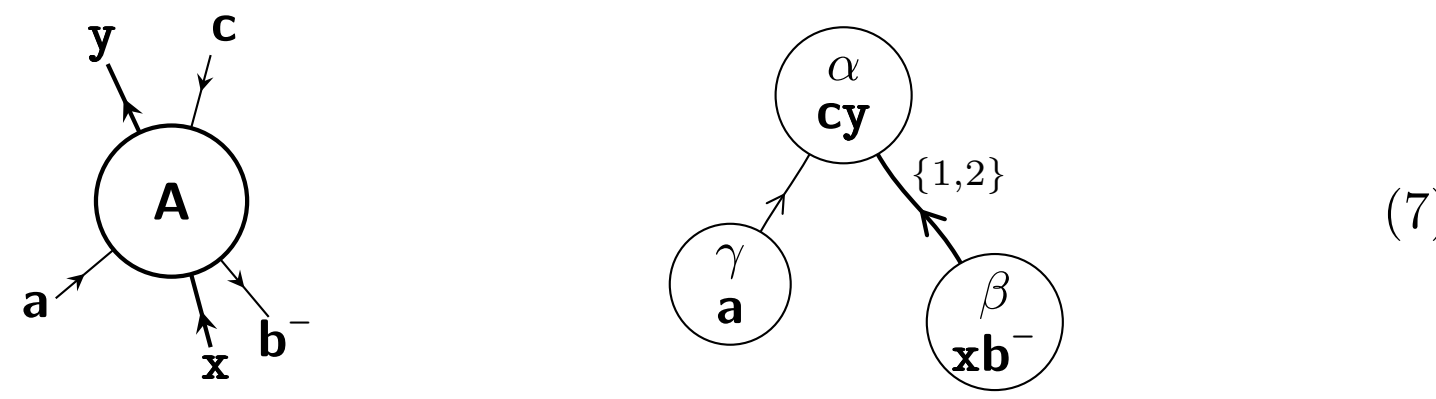

(7)

On the right we have the complex operation "itself" represented by a circle, and on the right we have the *causal diagram* which specifies the causal relationship between the inputs/outputs. Both parts are essential for the full description of a complex operation. We will not assume that all complex operations can be modelled by networks of simple operations (such as that in (6)) neither will be able to prove that this is the case. Thus, complex operations are objects in their own right. We will go into much more detail on how the notation for a complex operation works in Part IV of this book. Though it is worth mentioning in this introduction that a system label such as $\mathbf{a}$ has two parts. We write $\mathbf{a} = \mathbf{a}^+\mathbf{a}^-$ where $\mathbf{a}^+$ represents a system moving in the direction of the arrow on the wire and $\mathbf{a}^-$ represents a system moving against the direction of the arrow (this notation was first introduced in Hardy [2016]). This compact notation is very useful as we will see. It is motivated by Quantum Field Theory where we can have two arbitrary regions of spacetime meeting at a boundary (this is motivated by Oeckl [2003] where he introduced the general boundary formulation - more discussion on this below). Consider the case where, along at least part of this boundary, the normal is spacelike.

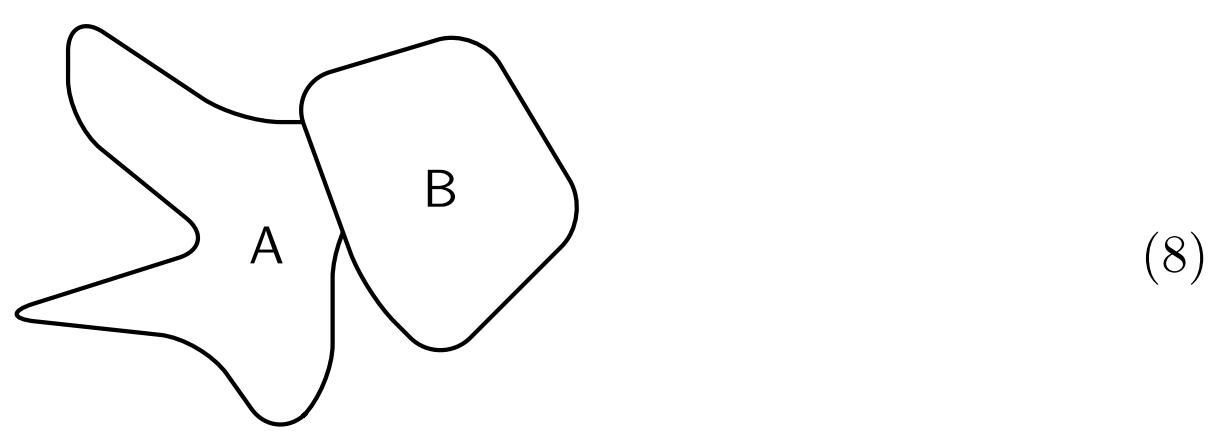

(8)

In this case we can think of quantum systems passing in both directions across this part of the boundary. If we associate operations with each of the regions of spacetime then we have complex operations joined by a system $\mathbf{a}$ with positive and negative parts.

(9)

The arrow is there to establish a convention - systems travel in both directions along the wire labelled by **a**. We can join complex operations to form a circuit

(10)

For the simple case we were able to ensure there are no closed causal loops by imposing that the circuit is a DAG (see (5)). In the complex case the condition that there are no closed causal loops is obtained by joining the causal diagrams associated with the operations the circuit. As we will see, the condition that there are no closed loops is that the causal diagram for a circuit simplifies to the empty set.

We will discuss how to model field theories in the general operational framework in terms of complex operations. In so doing we continue to have finite dimensional state spaces so we should not regard this as a full modelling suitable for Quantum Field Theory which requires infinite dimensional Hilbert spaces. Nevertheless, we can gain insight into spatial-temporal aspects of such theories. In this vein, we will discuss “impossible measurements” due to Sorkin [1993]. We will show that they are ruled out within complex operational theory when we impose physicality conditions (which, as mentioned above, ensure probabilities are between 0 and 1 and causality is satisfied). Interestingly, we can rule out these impossible measurements at the level of the operational theory without having to go to the full Quantum Theory.

The idea of considering operations with complex causal structure has arisen in the literature in multiple guises. As mentioned above, Oeckl [2003] introduced the general boundary formalism where states are associated with the boundary of general regions of space time. Such boundaries can be viewed as having inputs and outputs where some outputs are before some inputs and so we have complex causal structure here. The causaloid framework introduced in Hardy [2005] is, essentially, a GPT framework for situations in which we have indefinite causal structure. It was shown there how to formulate Quantum Theory within the causaloid framework by first specialising to a fixed background causal structure. In that formulation, the objects that arise naturally have complex causal structure. Within the Aharonov school, the idea of *multiple-time measurements* is set up (see Aharonov et al. [2009]). These multiple-time measurements have outputs that are before inputs and so we have complex causal structure. The conceptual seed for this approach is Aharonov, Bergmann, and Lebowitz [1964] on pre-and post-selected ensembles (which we will discuss below in Sec. 2.7). In Gutoski and Watrous [2007] the notion of *quantum strategies* are introduced as part of a general theory of quantum games. A quantum strategy by one player (Alice, say) has inputs and outputs ready to receive and send a sequence of quantum systems to other placers. In this case, some of the outputs are before some of the inputs and so we have complex causal structure. Chiribella, D’Ariano, and Perinotti [2009a] consider similar objects which they call *quan-* *tum combs*. These are part of a broader conceptual development wherein they conceive of these as being basic elements in a broader operational structure of supermaps and, more generally, higher-order maps (see also Chiribella et al. [2008b,a, 2010, 2011], Bisio and Perinotti [2019]). This has become a big field in Quantum Theory. See Jenčová [2024] and Taranto et al. [2025] for recent reviews.

The causaloid framework led to the duotensor and operator tensor approaches (see Hardy [2013a, 2011b, 2012, 2015]) and a formulation of operational General Relativity (see Hardy [2016]). Oeckl developed his ideas further in a series of papers. From our point of view Oeckl [2013] is particularly interesting wherein he develops the *positive formalism* wherein positive operators are associated with the regions of spacetime.

The approach taken in this book is to take the spatial-temporal attitude found in the work by Oeckl and in the causaloid framework as cited above (but where this is spacial-temporal structure is abstracted and associated with causal diagrams), and then leverage the mathematical tools developed by Gutowski and Waltrous concerning recursive normalisation conditions and the conceptual and mathematical tools developed by Chiribella, D'Ariano, and Perinotti that take these recursive normalisation conditions and relate them to causality.

There is a strand of work on indefinite causal order that has grown out of some of the above mentioned work (in particular Chiribella et al. [2009b] and Oreshkov et al. [2012]). This work also fits in naturally with this view of complex causal structure and, in a sense, goes beyond it. In this case there are outputs that are neither before or after some inputs. We will discuss this strand of work later in Sec. 2.11.

The last decade or so has seen the development of many other approaches to causal structure within the quantum foundations community (see Causalworlds [2024] for a recent conference in the CausalWorlds series). There are many strands of work within this community that are rather intertwined and relate to some of the themes in this book.

A significant strand of work on quantum causality has grown out of the program initiated by Leifer and Spekkens [2013], who proposed that quantum theory might be understood as a form of Bayesian inference that is neutral with respect to causal direction. This perspective helped set the stage for later work by Wood and Spekkens [2015], which showed that Bell-inequality-violating correlations cannot be accommodated within classical causal models without a failure of the faithfulness assumption, thereby motivating the development of intrinsically quantum causal models. The program was subsequently developed into an explicit framework for quantum causal models by Allen, Barrett, Horsman, Lee, and Spekkens [2017], who introduced quantum analogues of common causes and causal structure. More recently, Schmid, Selby, and Spekkens [2020] have clarified the operational foundations of this approach, providing a systematic framework for causal modelling in general operational theories and showing how classical and quantum causal models arise as special cases. Complementary progress has also been made on the causal inference side through the development of the inflation technique by Wolfe et al. [2019], which provides powerful constraints for determining whether observed correlations are compatible with a proposed causal structure. Taken together, this line of research seeks to extend the central ideas of causal modelling and causal discovery into the quantum domain while identifying precisely where the classical framework must be generalized.

A second strand of work, pursued by Jonathan Barrett and collaborators in Oxford, investigates how causal structure can be formulated directly in terms of the structure of quantum operations themselves. In this approach, causal relations are not introduced through classical variables or external causal graphs, but instead emerge from the way quantum systems and operations compose. Barrett, Lorenz, and Oreshkov developed a framework of quantum causal models in which the causal relations between laboratories are defined at the level of quantum processes, allowing one to analyse causal structure directly in terms of quantum transformations Barrett et al. [2019]. In subsequent work they extended this framework to allow cyclic quantum causal models, showing how quantum processes generated by unitary dynamics can give rise to causal structures that need not correspond to simple directed acyclic graphs Barrett et al. [2021]. This line of research thus seeks to characterize causal structure intrinsically within quantum theory itself, identifying how the structure of quantum operations constrains the possible patterns of causal influence. In further work Vanrietvelde et al. [2021] propose the framework of routed quantum circuits which generalises standard circuits and allows for an interesting analysis of the quantum switch due to Chiribella et al. [2013]. Subsequent work along these lines appears in Ormrod et al. [2023] and Vanrietvelde et al. [2025].

A third strand, developed by Vilasini, and collaborators, studies the relation between operational or information-theoretic notions of causality and the causal structure of spacetime. A useful starting point is the framework introduced by Vilasini and Colbeck [2022a] for cyclic and fine-tuned causal models compatible with relativistic constraints, which already makes explicit the distinction between signalling constraints and deeper causal relations. This was sharpened in subsequent work showing that the impossibility of superluminal signalling in Minkowski spacetime does not by itself exclude causal loops (see Vilasini and Colbeck [2022b]). Vilasini and Renner [2024a] then developed this further by analysing how cyclic information-theoretic causal structures can or cannot be embedded in acyclic spacetimes , and by proving more general limits on the realization of quantum processes in spacetime (see Vilasini and Renner [2024b]). This research program is especially valuable for clarifying the relation between operational quantum causality and relativistic causal structure.

The approach in this book concerns measurements with a finite set of readouts. A counterpoint to this is the work of Jackson and Caves [2023b,a] who have proposed an instrument-centered formulation of continuous quantum measurement. In their framework the Kraus operators undergo stochastic evolution generated by infinitesimal positive transformations, yielding the usual stochastic master equations for conditioned quantum states as derived dynamics and revealing a geometric structure of measurement processes.

## 2.5 A mostly time symmetric perspective

We will, for the most part, take a time symmetric perspective, though we also discuss the time forward perspective (which, though unstated, is the usual perspective within operational approaches) and the time backward perspective. For simple operational theories we can prove these perspectives are empirically equivalent. It is not clear whether this is true for complex operational theories. Whilst the time forward perspective accords more naturally with our contingent view on the world, it seems that the time symmetric perspective is more fundamental.

An analogy may be useful here. Consider inertial dynamics near the surface of the earth. We can adopt the point of view that a frame of reference fixed at the surface of the earth is inertial and regard gravity as an external force. This is analogous to the time forward frame perspective. It accords with our contingent view on the world. Alternatively, we can adopt the point of view that a freely falling frame of reference is inertial. Then we do not need to regard gravity as an external force. This latter point of view is the more fundamental perspective as it leads to General Relativity. The time symmetric perspective is analogous to the freely falling frame of reference. We could also consider an frame of reference accelerating down at $2g\text{ms}^{-2}$ with respect to the ground. This is analogous to the time backward perspective. In fact, in General Relativity, an even more fundamental point of view is that we should work with general coordinate systems. By analogy, this suggests we should consider general temporal frames. We introduce these in Sec. 14 though we do not develop the ideas further (leaving open a research project to develop this analogy with General Relativity further).

The key difference between the time forward, time symmetric, and time backward perspectives is to do with readouts. A time forward operation (in the simple picture) looks like this

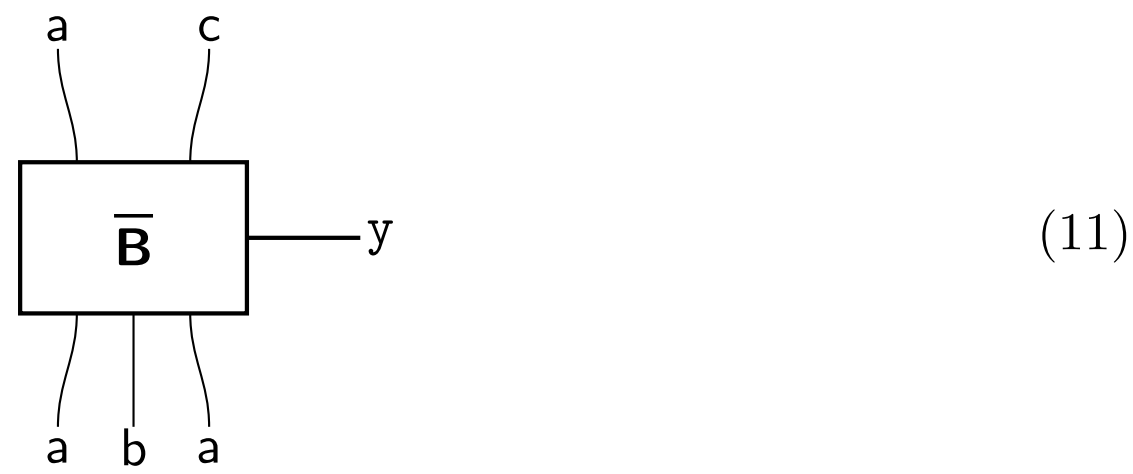

(11)

It has inputs (going in the bottom) and outputs (coming out the top) and it has a classical *outcome* (coming out the right hand side) which we can use to provide a readout. The outcome is available after, but not before the operation happens. This is clearly time asymmetric. A time symmetric operation looks

like this

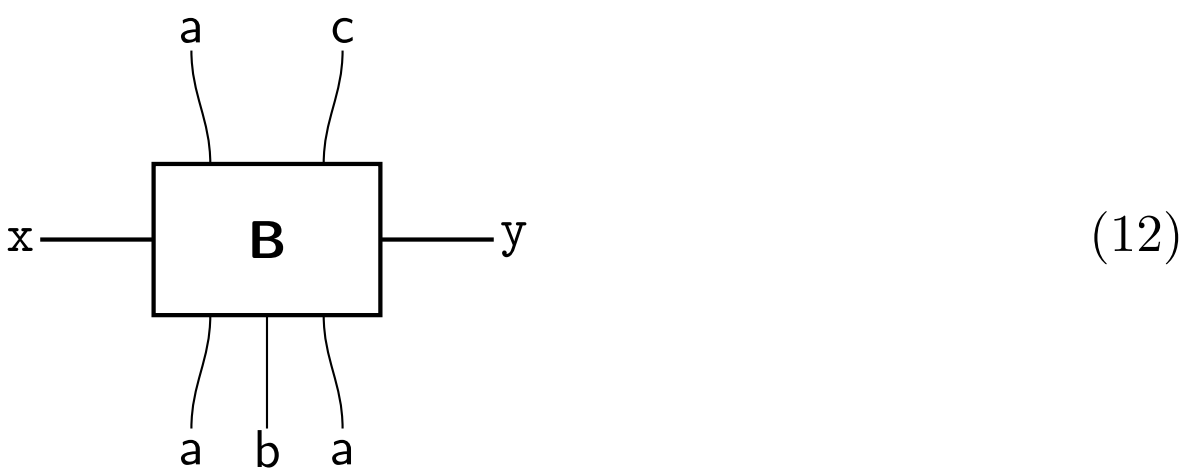

This has both an classical outcome (coming out the right hand side) and a classical *income* (goint in the left hand side) which we can also use to provide a readout. The income represents information available before, but not after the operation. Adding an income is clearly necessary to make the framework time symmetric. However, since the income is only available before while the outcome is only available after, this raises a basic question - can we bring all the readouts (from both incomes and outcomes) together to some localised spatial-temporal region where we can analyse them (and, perhaps, publish the results of this analysis). This is an interesting question and will be discussed in some depth in Sec. 13.7. To complete the story of the key difference between the perspectives we should mention that a time backward operation looks like this

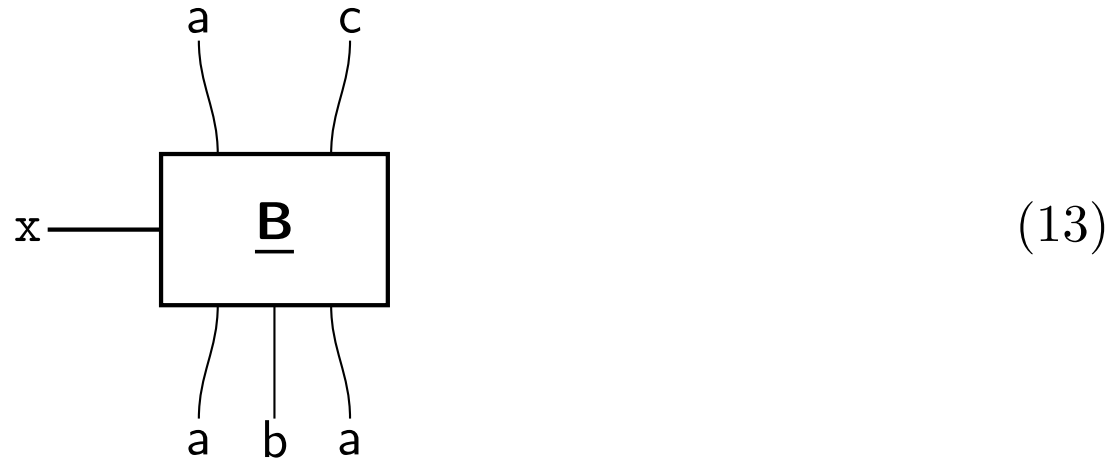

This has only an income but no outcome.

This key difference concerning incomes and outcomes impacts on the kind of causality conditions we impose. In the time forward case we impose a time forward but not a time backward causality condition. In the time backward case we impose a time backward but not a time forward causality condition. And, in the time symmetric case, we impose both time forward and time backward causality conditions. An illustrative consequence of these causality conditions is the shape of the convex spaces associated with states (associated with preparations) and effects (associated with measurement outcomes - these are the natural time reverse of preparations). For the time forward case these spaces look like

this

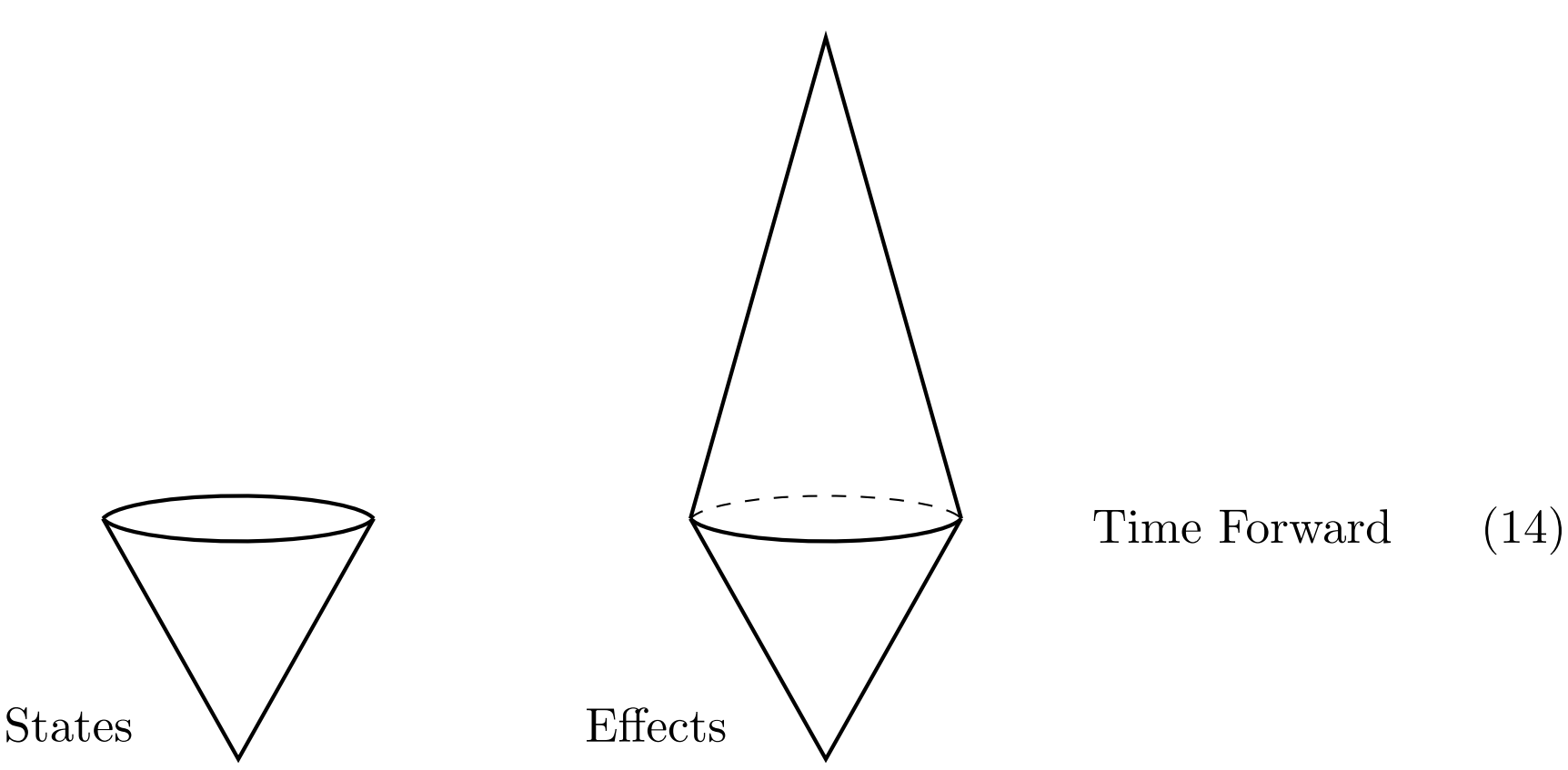

(14)

The main point is that the states live in a cone with a cap representing the normalised states, whilst the effects live in a double cone where the apex is the effect where we ignore the measurement outcome (this is called the *deterministic effect*). The shape of the cross-section shown here as a circle will be different for different system types and different theories. For the time backward case we have

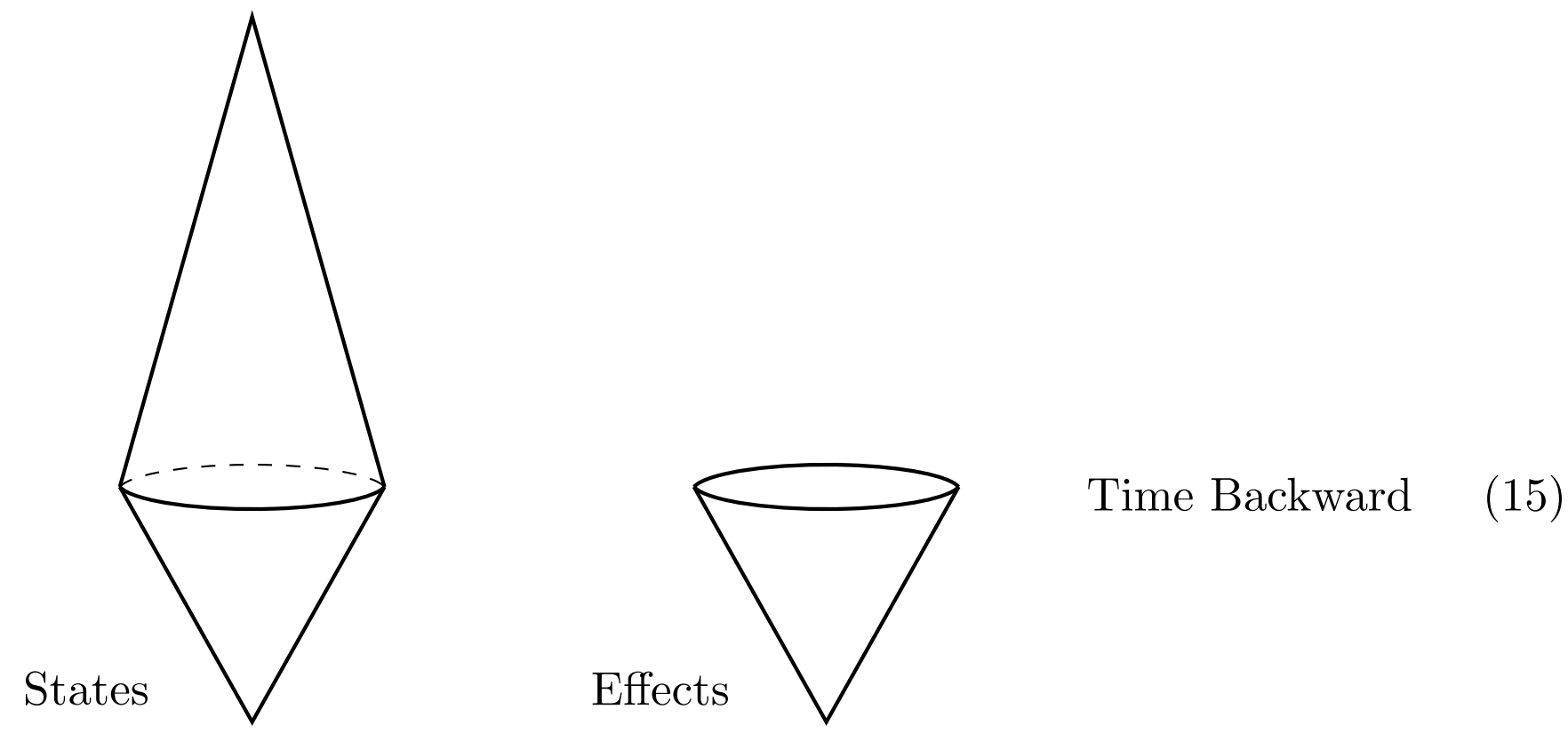

(15)

For the time symmetric case we have

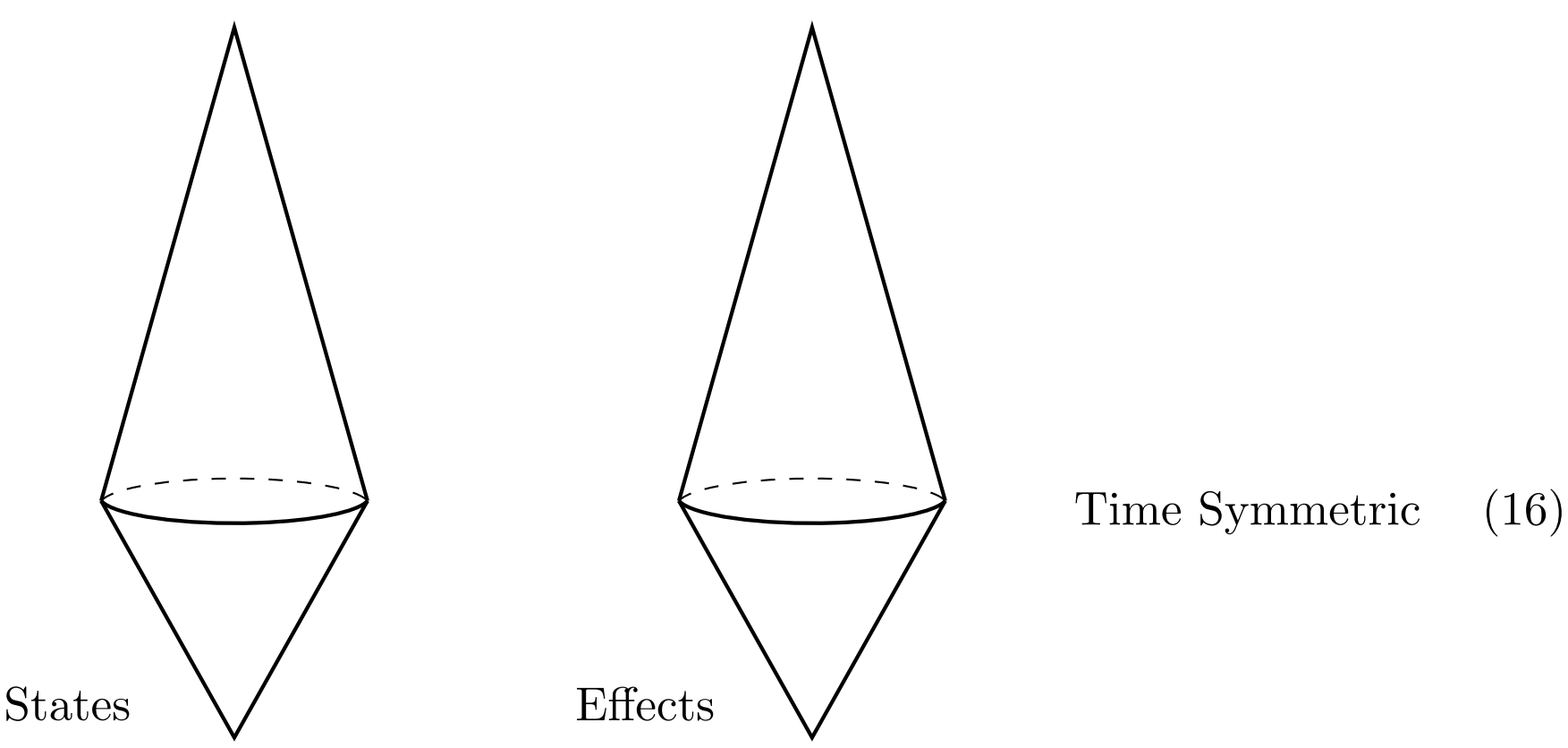

Time Symmetric (16)

Importantly, both the state and effect space have the same shape (as is clearly necessary in a time symmetric theory). This does not mean the time symmetric perspective makes different empirical predictions to the time forward (or time backward) perspectives. It is really about how we package probabilities. In particular, the state space, in the forward case, is spanned by probabilities that are effectively conditioned on incomes whilst, in the time symmetric case, the state space is spanned by joint (not conditional) probabilities. So we repackage the probabilities in different ways as we go between the different perspectives. In Sec. 11.10 we will prove equivalence between the time forward and time symmetric perspectives (and, by inference, the time backward perspective).

We call the systems associated with the inputs and outputs *physical systems*, or just *systems*. We call the systems associated with the incomes and outcomes *pointers* (as this terminology is common in the Quantum Foundations literature for the special systems associated with measurement readouts). We need to introduce a special box associated with taking a readout of a pointer. We call these boxes *readout boxes* and represent them as

$$\texttt{x} \; — \boxed{x} — \; \texttt{x} \qquad (17)$$

in the simple theory and

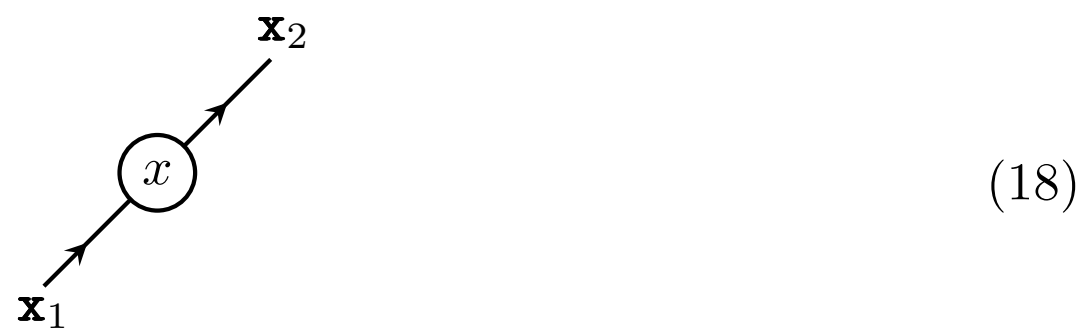

(18)

in the complex theory.

A distinction that is important in this book is the distinction between *deterministic operations* (which we represent by bold uppercase letters $\mathsf{\mathbf{A}}$, $\mathsf{\mathbf{B}}$, ...) and *nondeterministic operations* (which we represent by non-bold symbols $\mathsf{A}$, $\mathsf{B}$, ...). An nondeterministic operation is one where there is one or more implicit

readouts. For example

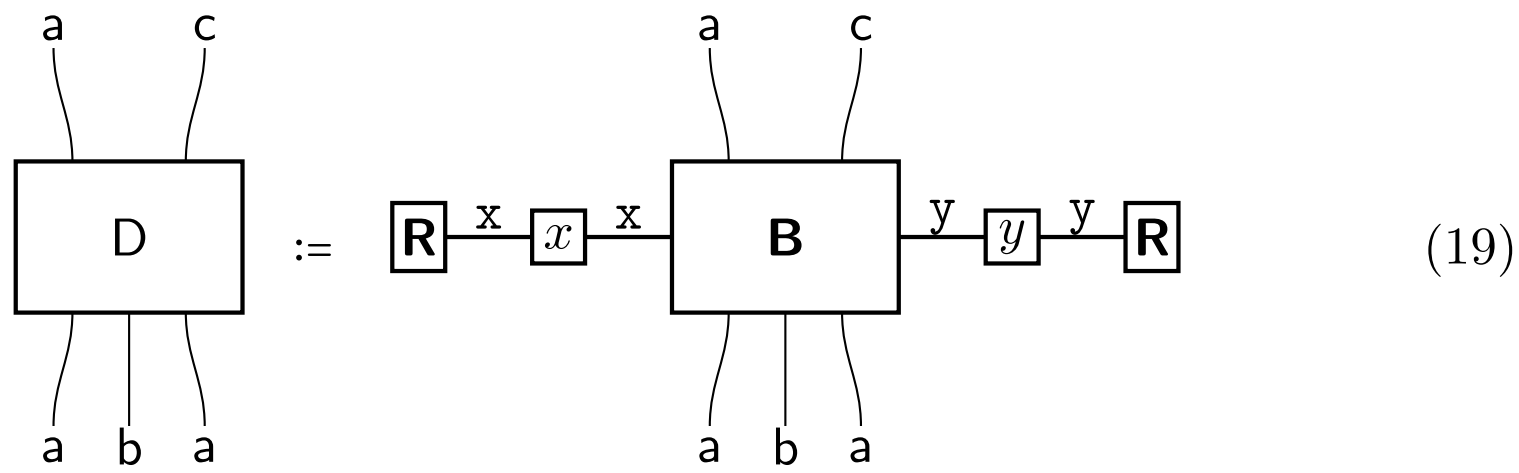

(19)

Here **B** is such that there are no implicit readouts but D clearly does have some.

In the time symmetric theory the following deterministic operations are unique

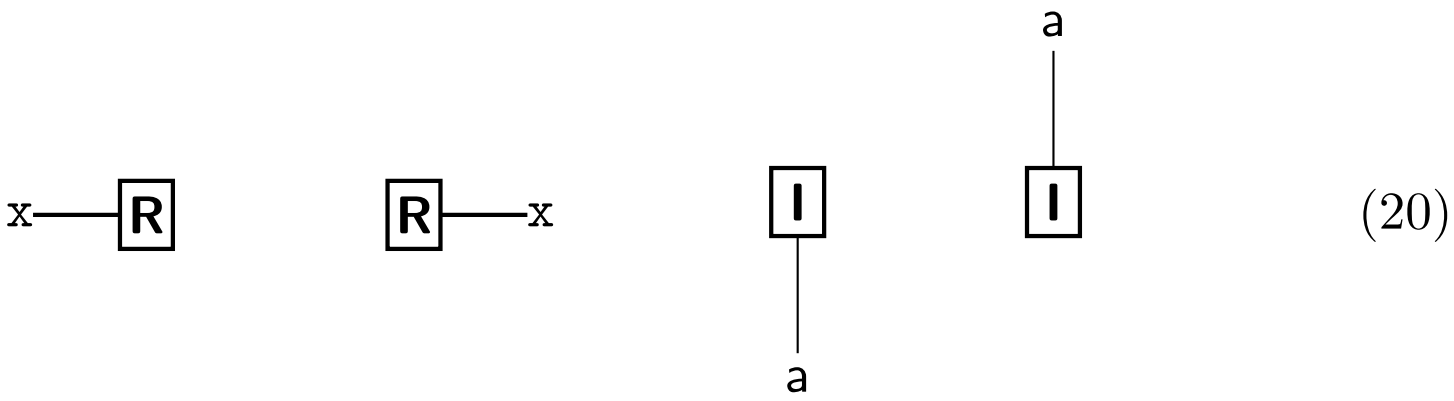

(20)

up to equivalence classes. For systems this is clear since they correspond to the top of the state and effect spaces shown in (16). Something similar must be true for the pointers cases. These objects play an important role in the theory. In some sense, they correspond to ignoring what comes after/before. It is worth commenting that the objects in (20) are often denoted by

x a a x (21)

by other authors.

We will motivate and, for the most part, assume that **R** operations satisfy the *flatness assumption*

$$\text{prob}\left( \boxed{\mathsf{R}} \overset{\mathrm{x}}{—} \boxed{x} \overset{\mathrm{x}}{—} \boxed{\mathsf{R}} \right) = \frac{1}{N_{\mathrm{x}}} \tag{22}$$

where $N_{\mathrm{x}}$ is the number of values $x$ takes.

## 2.6 Double causality conditions

Causality conditions constitute an important part of this work. In the time symmetric case we have *double causality conditions* consisting of a *time forward* *causality condition* and a *time backward causality condition*. In the simple theory these conditions are, respectively,

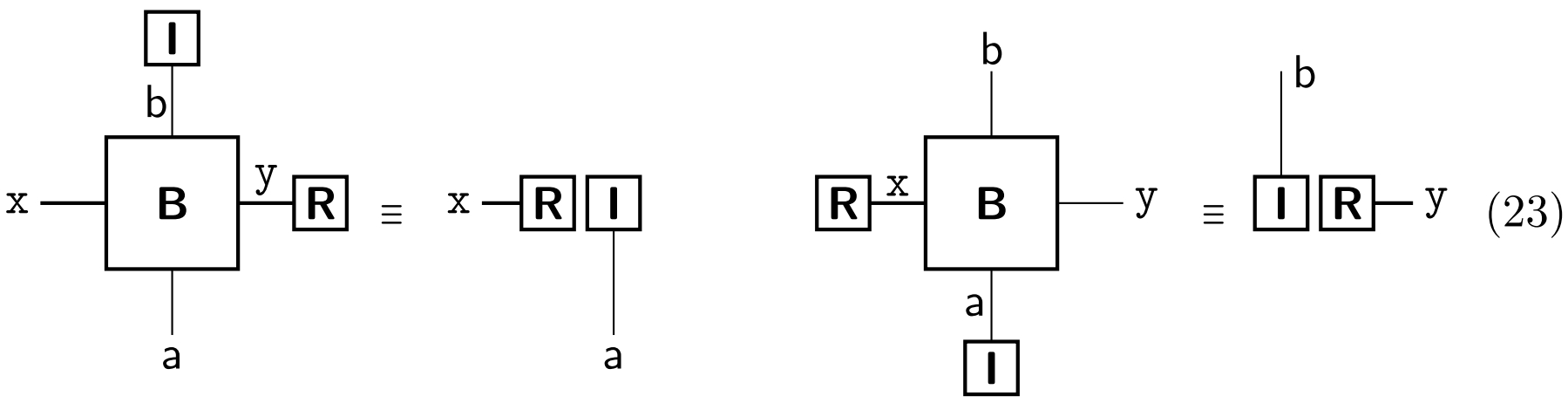

In the complex theory the conditions are that we have

(24)

for every *synchronous partition*, $p$, of the causal diagram (a synchronous partition is the analogue of a spacelike hypersurface for a causal diagram). Clearly the notation in (24) needs to be explained further. This will be done in Part IV. The forward causality condition (in both the simple and the complex case) imposes that choices in the future cannot influence readout probabilities in the past without conditioning on the future. The backward causality condition imposes that choices in the past cannot influence readout probabilities in the future without we conditioning on the past.

In the time forward theory we just have time forward causality conditions and, in the time backward theory we just have time backward causality conditions.

These double causality conditions are, from a conceptual point of view, an application of ideas about causality expressed in Chiribella et al. [2010, 2011] to the time symmetric situation. From a mathematical point of view, these conditions leverage the recursive normalisation conditions discussed by Gutoski and Watrous [2007] and Chiribella et al. [2009a] (particularly in the complex case).

## 2.7 Historical twists and turns in the Quantum time symmetry story

Let us, for a moment, think about the history of time symmetry in Quantum Theory. This is an interesting story with many twists and turns. We will begin with an oversimplified narrative to set up the story. Schoedinger's equation (see Schrödinger [1926]) is clearly time symmetric. However, it was soon realised that wave function projection or collapse is required. This idea is first intimated in the writing of Heisenberg [1925] and formalised in Dirac [1930] and von Neumann [1932] (see the nice discussion by Sudbery [2024]). The idea that the wavefunction evolves, is projected, evolves, is projected, and so on is clearly time asymmetric. Thus it appears that Quantum Theory (at least when we include measurements) is time asymmetric and, indeed, this would be the view of many physicists. However, there are a few more important twists and turns in the story. Watanabe [1955] showed how it is possible to infer probabilities about the past using a "rectodictive wave function" in the same way as we infer probabilities about the future using the usual "predictive wave function". His formalism was not fully time symmetric though because his formula for retrodiction depends on probabilities under the control of the earlier agent (see Di Biagio and Rovelli [2024] for a recent comment on this work). In 1964 Aharanov, Bergman, and Lebowitz (ABL) Aharonov et al. [1964] considered ensembles at some intermediate time which are both preselected and postselected. The resulting equation for calculating probabilities at this intermediate time (the ABL rule) is time symmetric. Interestingly this rule is nonlinear since it has a nontrivial denominator which stems from conditioning on the pre- and post-selection. The approach of Watanabe and ABL pertains to pure states with unitary evolution between measurements. Chiribella, D'Ariano, and Perinotti [2010, 2011] considered the more general theory of mixed states within standard Operational Quantum Theory. The showed that there is a basic causality condition (which, here, I dub the *Pavia causality condition* on account of the name of the town where this work was initiated) that deterministic effects are unique. This condition ensures that we cannot signal backward in time. This shows up in figure on the right of (14) above where we see that there is a unique effect at the top of the double cone. On the other hand, there is no condition ruling out signalling forward in time in standard Operational Quantum Theory (which is from the point of view of time forward perspective). Consequently deterministic states are not unique. This shows up in the figure on the left on (14) where we see there is more than one normalised state (living at the top of the cone). The next twist in the story comes from the work of Di Biagio, Donà, and Rovelli [2020] (which is, in some sense, the spiritual precursor to the approach of the present author). They showed that standard Operational Quantum Theory is set up as a theory that can be used to predict the future based on information about the past (this is the time forward perspective) and, further, there is an empirically equivalent way of setting up the theory as a theory that can be used to predict to past based on information about the future (this is the time backward perspective). In this approach we choose to condition on the past (the time forward perspective) or condition on the future (the time backward perspective). Finally we come to the time symmetric perspective presented in this book (first presented in Hardy [2021], though only for the case of simple causal structure) where we condition on neither past incomes or future outcomes. We define time symmetry to mean that there is a time reverse map on any operator such that the probability of any circuit under this time reverse map is unchanged (to time reverse a circuit we replace each operation by its time reverse which the effect of rotating the circuit through 180°. Thus, for any process, there is a time reversed process with the same probability. The time symmetric perspective is linear because we do not condition on the past incomes or future incomes. In some sense it is the most linear of the frameworks since it is linear independently of any causality conditions. The probability in the time forward perspective is conditioned on the past (incomes) and consequently has a denominator (when written out in terms of joint probabilities). However, this denominator is constant by virtue of the forward causality condition. Likewise, the time backward perspective is linear by virtue of the backward causality condition. We might ask whether this is the end of the story of time symmetry in Quantum Theory. Well, first of all, this story is oversimplified and we will return to this below. But secondly, there are two reasons to think that there will be more chapters even in this version of the story. First, the idea of general temporal frames discussed in Sec. 14 offers a chance to formulate Operational Probabilistic Theories (like Operational Quantum Theory) with respect to a general temporal frame analogous to generally covariant field equations in General Relativity. Second, our experience of the world is clearly not time symmetric. Eggs break in the forward time direction and put themselves back together in the backward time direction. It is possible that the world is, fundamentally, time symmetric, but that we see it as time asymmetric for contingent reasons. This is argued for in Sec. 13.8 and Sec. 13.9. Even if this argument is true (and it may not be), we need to be able to use Quantum Theory in the contingent situation we find ourselves in. One way to do this (following a suggestion by Di Biagio (private communication, 2020)) is to treat preselection as a resource. Tentative steps in this direction are taken in Sec. 13.6.

The story above is very much oversimplified because it is intended to set up the narrative in the present book. Along side the approaches already discussed, other lines of work have sought to restore time symmetry to quantum theory by thinking about retrocausality. An early version of this can be found in the work of de Beauregard [1953], who proposed that the Einstein–Podolsky–Rosen correlations might be understood via influences propagating both forward and backward in time. Cramer [1986] later developed the transactional interpretation, a time-symmetric account in which quantum processes involve a standing-wave "handshake" between retarded and advanced solutions. More recently, Price [1996, 2008, 2012] has argued on general grounds that if the fundamental dynamical laws are time-reversal invariant, then the temporal asymmetry introduced by measurement should not simply be taken as fundamental, and that a realist interpretation may require some form of retrocausality. This line of thought was sharpened by Leifer and Pusey [2017], who showed that, under natural assumptions, a genuinely time-symmetric ontological model of quantum theory cannot reproduce the quantum statistics without retrocausal influences. Additionally, Wharton [2010] and, more recently, Wharton and Argaman [2020] have advocated spacetime or "all-at-once" formulations in which quantum phenomena are understood in terms of global boundary constraints rather than dynamical evolution from past to future, thereby offering locally mediated retrocausal models compatible with Bell's theorem. Although differing in technical framework and emphasis, these approaches share a common motivation: to take seriously the time symmetry of the underlying equations and to question whether the standard measurement postulate, with its built-in temporal asymmetry, is fundamental.

An earlier and conceptually distinct approach to operational time symmetry from that taken in this book was developed by Oreshkov and Cerf [2015, 2016]. In their formulation, one begins from the standard operational framework in which probabilities are expressed conditionally—typically as probabilities of outcomes given preparations—and investigates how a time-reversal transformation can act consistently on such conditional probabilities. To achieve this, they enlarge the notion of operations to include both pre- and post-selection, and derive an operational analogue of Wigner's theorem (see Wigner [1931]), characterizing admissible symmetry transformations in quantum theory. By contrast, in our formulation time symmetry is implemented at a more primitive level by taking joint probabilities as fundamental, so that no temporal conditioning direction is built into the theory from the outset. Forward- or backward-causal descriptions then arise only as constrained perspectives on an underlying time-neutral probabilistic structure. In this sense, while Oreshkov and Cerf show how a fundamentally conditional operational framework can be rendered time symmetric, the approach in this book begins from a time-symmetric probabilistic foundation and derives time forward and time backward theories as special cases. Selby et al. [2022] discuss these two approaches (and introduce a third approach).

A related recent strand of work examines time symmetry through the symmetry structure of quantum evolutions themselves. Chiribella, Aurell, and Życzkowski [2021], extended Wigner's theorem from states to completely positive maps, characterising the symmetry transformations of quantum channels and analysing how time reversal can act at the level of operations. Their results show that symmetries of quantum evolutions decompose into unitary or antiunitary transformations on input and output spaces, while also revealing constraints on implementing time reversal within the standard causal framework. This line of work connects naturally with broader geometric investigations of quantum state spaces, where self-duality and Hilbert–Schmidt structure play a central role in rendering quantum theory unusually symmetric among probabilistic theories.

Chiribella and Liu [2022] look at operations wherein the direction of time is indefinite. This approach is generalised in Apadula et al. [2026] to higher order transformations. These works take a time symmetric approach. The latter work touches on some of the same themes as discussed here concerning complex operations (though, in this book, we do not consider indefinite causal structure).

This work of Chiribella and collaborators connects with an older strand of work – on bistochastic quantum maps (or unital maps) going back to Landau and Streater [1993]. These are completely positive trace preserving maps that also preserve the maximally mixed state. The condition of unitality is strongly connected to the condition of double causality studied here. There are, however, a number of differences in attitude and scope. First, double causality applies to operations within the operational theory (even before we come to Quantum Theory) which have incomes and outcomes. Unital maps are regarded as a restriction on the full set of quantum maps whereas operations satisfying double causality are conceived of within a time symmetric formalism which is shown, in the simple case, to be equivalent to the standard time forward formalism. Further, there are subtleties concerning gauge parameters introduced here and the associated notion of the natural conjuposition group which do not appear in the study of unital maps.

A further recent contribution to the study of time symmetry in quantum theory is due to Parzygnat and collaborators. Parzygnat and Fullwood [2023] develop a framework in which time-reversal symmetry is closely linked to quantum Bayesian updating and "states over time, placing forward and backward inference on a more equal footing. In related work with Buscemi Parzygnat and Buscemi [2023], Parzygnat formulates axioms for retrodiction in categorical terms, identifying structural conditions under which time-reversal symmetry can be achieved once a prior is specified. These works extend the discussion of time symmetry into the domain of probabilistic inference, complementing structural and operational approaches based on symmetry of transformations or joint-probability formulations.

There is a long tradition of relating time asymmetry to the thermodynamic arrow of time. Although this will not be a central theme in this book (though the discussion in Sec. 13.9 touches on this), the topic has generated a vast literature spanning both physics and philosophy. Foundational analyses of the thermodynamic arrow and its relation to Quantum Theory, statistical mechanics and cosmology can be found in Albert [2000], Price [1996], and Winsberg [2012], while broader discussions of time and cosmology appear in Carroll [2010]. The so-called thermal time hypothesis provides another perspective linking time flow with thermodynamics Connes and Rovelli [1994]. Philosophical and foundational investigations of temporal asymmetry and the status of time in physical theory are developed in works such as Horwich [1987] and Maudlin [2007], while broader reflections on the nature and possible fundamentality of time in physics and cosmology can be found in Smolin [2013] and Ismael [2021].

Another area where time symmetry and Quantum Theory meet is in the CPT theorem of Quantum Field Theory. This is a symmetry under charge conjugation, parity transformation, and time reversal. Early results on this go back to Schwinger [1951], Lüders [1954], Pauli [1955], and Bell [1955] (and see Blum and de Velasco [2022] for a review of the history). Importantly, this suggests that when we perform time reversal, we should change properties of the system as well. In this book we will, generally, not do that to keep the discussion simple. However, as discussed at the end of Sec. 6.4, it is easy to expand the definition of time symmetry we work with to allow this.

## 2.8 Operational Quantum Theory

Operations describe objects in the laboratory. By means of the $p(\cdot)$ function mentioned above we are able to put operations into a linear space. We can set up a *correspondence* between operations and *operators*. For the simple case we have

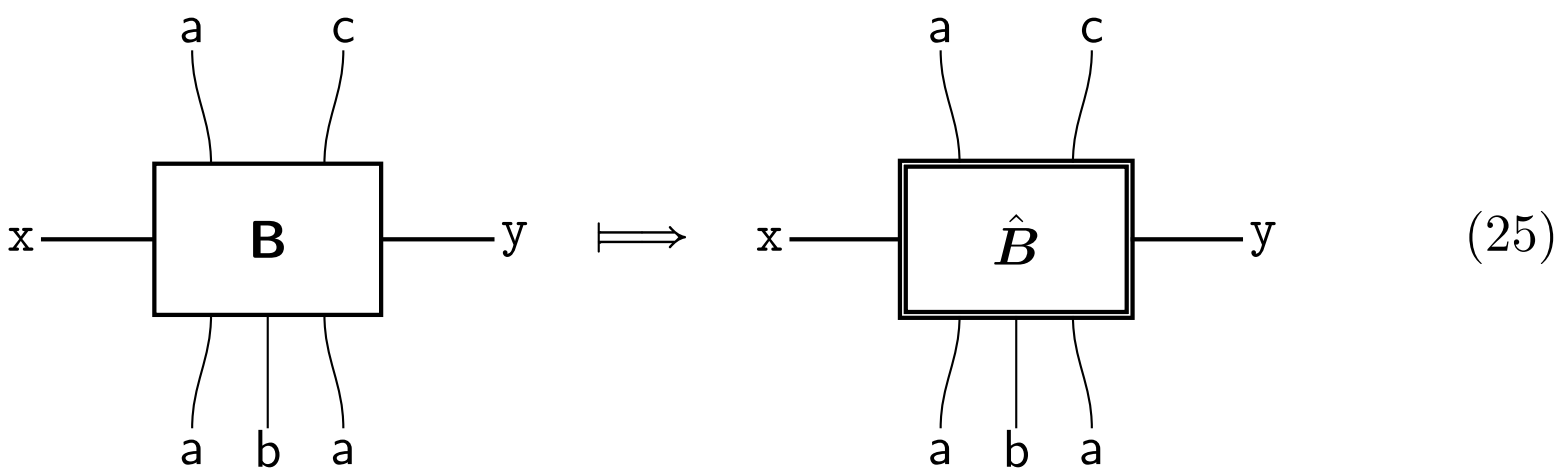

(25)

where $\Longmapsto$ means *corresponds to*. In the case that we think of **B** as labelling an equivalence class of operations we can think of the correspondence as being in either direction. An operator (which I sometimes call an *operator tensor* is a mathematical object living in a space determined by the associated input and output system types (this space derives from underlying Hilbert spaces associated with the systems). Correspondence is aided by the machinery of *fiducial elements*, fiducial matrices, and *duotensors* which are introduced in Sec. 9. The duotensor framework relies on the assumption of tomographic locality (which has many forms discussed in Sec. 9.7).

If we have a circuit comprised of operations then we can calculate the probability associated with that circuit by means of a calculation consisting of operators

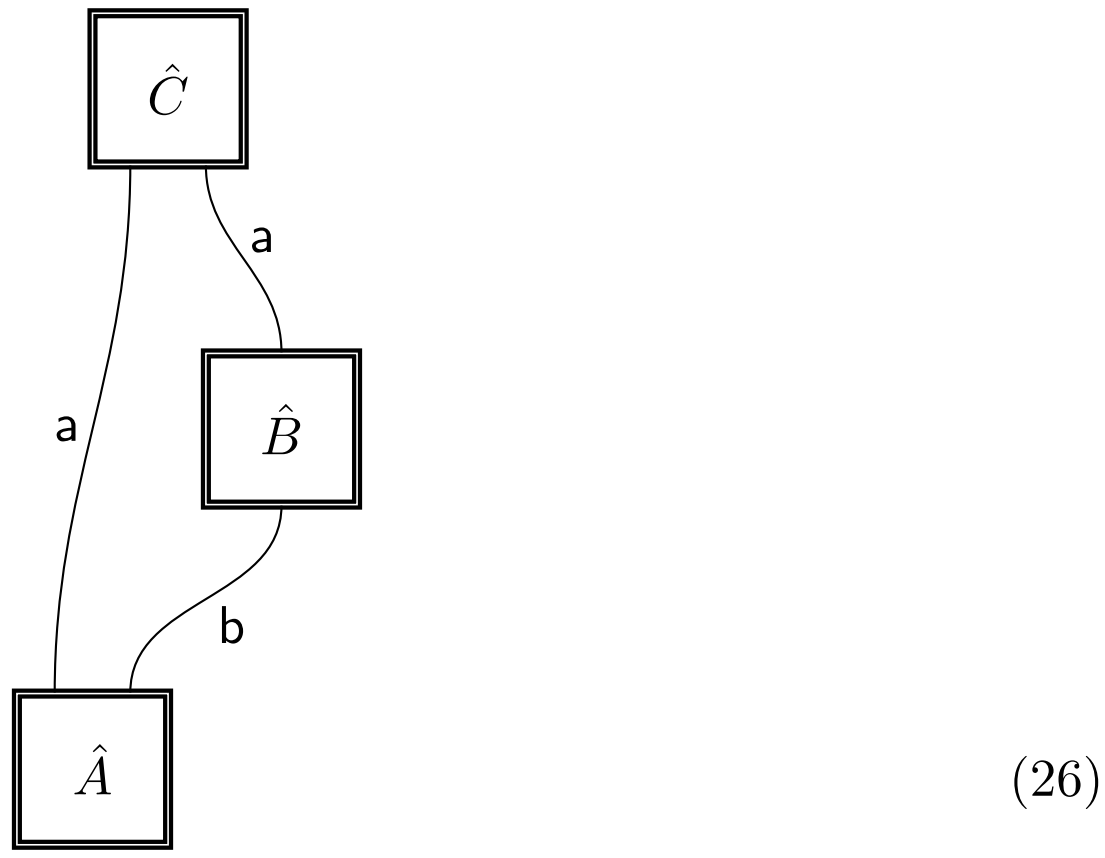

(26)

Here the closed wires indicate taking the partial trace over the associated spaces.

Operations must satisfy physicality conditions inherited from the physicality conditions on operators. These physicality conditions guarantee that circuits have probabilities between 0 and 1 and, further, that appropriate causality conditions are satisfied.

We also have a correspondence between complex operations and complex

operators

$$\mathsf{B} \Longrightarrow \hat{B} \tag{27}$$

The complex operator must satisfy physicality conditions inherited from the physicality conditions on the complex operations and they guarantee that we get probabilities between 0 and 1 and that causality is respected when we do the operator calculation for circuits.

## 2.9 Hilbert objects

Here we provide a few introductory remarks about Hilbert objects. We will illustrate these remarks with simple Hilbert objects (which are represented by rectangles). Similar remarks apply to complex Hilbert objects (which are represented by semi-circles).

Hilbert objects are built out of *left Hilbert objects* and *right Hilbert objects* which belong, respectively, to left Hilbert spaces and right Hilbert spaces. We use the "doubling up" notation

$$\mathsf{a} \; = \; \mathsf{a} \; \mathsf{a} \tag{28}$$

where the small black rectangles - which we call *bobbles* - indicate whether the associated Hilbert space is a left Hilbert space (bobble on left) or a right Hilbert space (bobble on right).

Operators can be built out of objects living in a Hilbert space which we call *Hilbert objects*. For example,

$$\hat{B} \; = \; B \overset{k}{—} B \tag{29}$$

where the line labelled by $k$ represents a sum over $k$. This is the Kraus decomposition where

$$B \in \mathcal{H}_{\mathsf{a}} \otimes \mathcal{H}^{\mathsf{b}} \qquad\qquad B \in \overline{\mathcal{H}}_{\mathsf{a}} \otimes \overline{\mathcal{H}}^{\mathsf{b}} \tag{30}$$

We call the object on the left a *left Hilbert object* and the object on the right a *right Hilbert object*.

Hilbert objects can be represented as a sum over orthonormal bases as follows

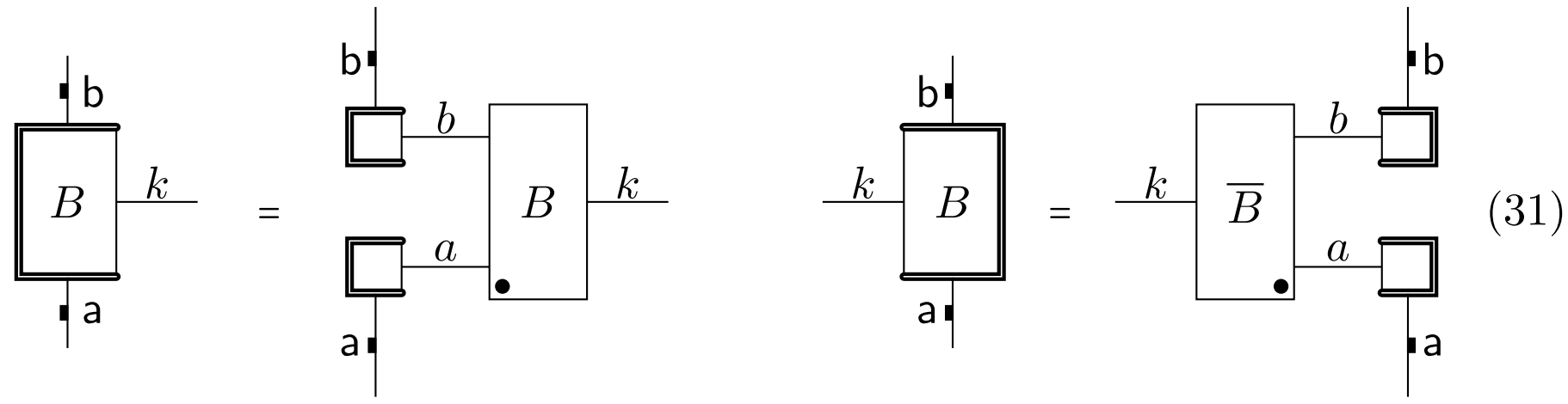

In these expressions the empty boxes are basis elements and we sum over $a$ and $b$. The rectangle $B$ is a hypermatrix (the position of the dot indicates between the two cases indicates taking the *horizontal transpose*. These hypermatrices can be transformed under the *conjuposition group* which involves taking vertical and horizontal transposes and adjoints. Vertical transposes involve flipping the hypermatrix vertically while horizontal transposes involve flipping it horizontally. Adjoints involve, of course, conjugation.

If we expand a Hilbert object using orthonormal bases then apply the conjuposition group to the expansion matrices we induce a group of transformations on the Hilbert object itself. We call this the *normal conjuposition group* (because it is based on orthonormal expansions).

It turns out that, for the time symmetric perspective, there is a more natural orthogonal basis than the orthonormal basis. We call this the orthophysical basis. This basis is normalised in a different and physically more natural way from the orthonormal basis. We can expand Hilbert objects in terms of the orthophysical basis then apply conjupositions to the new expansion hyper-matrices. This gives us the *natural conjuposition group*. The natural vertical adjoint maps an operator to its time reverse.

We will develop *mirror notation*. There are normal and physical mirrors which relate to orthonornal and ortho-physical bases respectively. As an example, we can write

(32)

Mirror notation simplifies expressions and, further, the mirror theorem powers some of the subsequent definitions and proofs.

Isometries, coisometries, and unitaries are familiar in standard treatments of Quantum Theory (coisometries are not so familiar, but we need them because they are the time reverse of isometries). In this work we have normal and natural versions of each of these objects (depending on whether we use orthonormal or ortho-physical bases to define them). Natural unitaries are actually the correct object to use for formulating Quantum Theory from the time symmetric perspective.

Another object crops up in our study of Hilbert objects. This is the *maxometry* which also comes in normal and natural versions. A natural unitary is a special case of a natural maxometry. If we double up a natural unitary we obtain a physical operator (i.e. it satisfies the physicality conditions). However, if we double up a natural maxometry we do not, in general, obtain a physical operator. These maxometric operators cannot be realized in the laboratory.

In standard Operational Quantum Theory the Stinespring dilation theorem is important because it tells us how to model any operation using an ancilla and a unitary thereby telling us how to implement them in the laboratory. It is, however, time asymmetric because the ancilla is prepared in a pure state and then traced over at the end. In the time symmetric case we have a time symmetric dilation theorem. However, as presented, we have to use a natural maxometry rather than a natural unitary. This means that the time symmetric dilation does not provide a prescription for modeling any operation in the laboratory because natural maxometries cannot, in general, be realized in the laboratory. In general the time symmetric dilation for any operator is non-unique. This leaves open the possibility that there always exists a time symmetric dilation where the maxometry is, in fact, unitary. The existence, or not, of such a dilation remains as an open problem.

We can also define complex Hilbert objects. These look like this

$$\underset{\mathsf{a}}{\overset{\mathsf{b}}{B}} \in \mathcal{H}_{\mathsf{a}_1} \otimes \mathcal{H}^{\mathsf{b}_2} \qquad\qquad \underset{\mathsf{a}}{\overset{\mathsf{b}}{B}} \in \overline{\mathcal{H}}_{\mathsf{a}_1} \otimes \overline{\mathcal{H}}^{\mathsf{b}_2} \tag{33}$$

Rather than flipping vertically when we form the conjuposition group we reverse the arrows (which interchanges input and output and so amounts to the same thing as a vertical flip). The black and white arrow heads play a similar role to the left and right bobbles. Complex operations have complex causal structure and this makes the dilation theorems much more interesting. We prove a number of *causal dilation theorems* for complex operators corresponding to physical operations with various classes of causal diagrams. These dilations, like in the simple case, employ natural maxometries and so the remarks above about whether they provide a prescription for implementing them in the laboratory remain open. We do not succeed in proving a general dilation theorem for any causal diagram and this remains as an important open question.

## 2.10 Historical and modern quantum operationalism

Operational quantum theory has a long history emerging from several converging developments in the mid-twentieth century. The density operator formalism, introduced by von Neumann Von Neumann [1927], provided a representation of quantum states as positive trace-one operators, capable of describing both mixtures and subsystems of entangled systems. Stinespring's dilation theorem Stinespring [1955] and Kraus's operator-sum representation Kraus [1971], Kraus et al. [1983] clarified that physically admissible state transformations are completely positive maps, realizable as unitary interactions with an environment. At the same time, the framework of generalized observables (POVMs) was developed, notably by Davies Davies [1976], extending measurement theory beyond projection-valued measures. These elements were unified by Davies and Lewis Davies and Lewis [1970], who introduced quantum instruments—objects that associate each measurement outcome with a completely positive map—thereby combining probability assignment and state update within a single operational structure. This synthesis forms the foundation of modern operational quantum theory and underlies later developments such as quantum combs and higher-order transformations.

Whilst theoretical physicists in many branches of physics stick with an amplitude or Hilbert space point of view, this operational formulation has been very impactful within the Quantum Information community. In this sense, it has become what I will often refer to as the standard textbook presentation. See, for example, the textbooks of Preskill [2024] and Nielsen and Chuang [2000].

The modern pictorial approaches to Quantum Theory, which this book concerns, are built on top of the historical framework described above. This includes the pictorial approach discussed earlier (as expounded in Coecke and Kissinger [2017]) and versions of Quantum Theory formulated in the Operational Probabilistic Theory framework such as in Chiribella et al. [2009a, 2011] and in Hardy [2011b].

The time symmetric version of quantum theory, developed here, is also built on top of this historical framework. It is worth mentioning, however, that it goes beyond this framework in that the physicality conditions become time symmetric and so the spaces of allowed states, effects, and transformations are modified. The clearest signature of this is that the space of allowed states becomes a double cone as described earlier.

Another contemporary branch of work built on top of the historical foundation of Operational Quantum Theory develops quantum theory as a framework for probabilistic inference. A starting point was the proposal by Leifer and Spekkens [2013] to formulate quantum theory as a causally neutral theory of Bayesian inference, emphasizing that the formalism need not presuppose a fixed direction of causal conditioning. Spekkens' epistemic programme reinforced this perspective by treating quantum states as states of knowledge constrained by operational principles (see Spekkens [2007]). Building on these ideas,Schmid et al. [2020] introduced the framework of causal-inferential theories, which systematically distinguishes causal structure from inferential structure within operational models. In this approach, quantum theory is understood as a particular instance of a broader class of operational theories that reconcile causal and inferential relations without reducing one to the other. Although this is a distinct approach from that in the present book, there are some similar themes.

## 2.11 Quantum Gravity as motivation

The primary motivation for formulating operational probabilistic theories in a time symmetric way comes from the problem of Quantum Gravity. The problem of Quantum Gravity is to find a theory that reduces, in appropriate limits, to General Relativity on the one hand, and Quantum Theory (really Quantum Field Theory) on the other hand. Of course, such a theory would have to enjoy some kind of empirical verification to be adopted as the correct theory of Quantum Gravity.

There are two reasons to think that time symmetry may help in solving this problem. The first reason is that General Relativity is a time symmetric theory and so it is likely to help solve the problem of Quantum Gravity if the Quantum Theory we use is also (see discussion on this point by Di Biagio et al. [2020]). The second reason comes from the remarks concerning linearity of the time symmetric formalism discussed in Sec. 2.7. As argued in Hardy [2009a], a theory of Quantum Gravity will likely have indefinite causal structure (because we expect to have something like a superposition of different metric fields which means a superposition of different causal structures). This will force the causality conditions to be modified. Time forward and time backward approaches are linear only by virtue of the causality condition (since it is the causality condition that makes the denominator in the expression for the conditional probability equal to a constant). The time symmetric approach is linear because we work with joint rather than conditional probabilities. It does not need the causality conditions to guarantee this linearity. We are in a much better position to proceed if we have a linear theory, or at least, if we have a domain where the theory is linear independent of any causality assumptions.

The framework developed here has definite causal structure because (1) we have a notion of systems moving between operations in time, and (2) we impose causality conditions. There has been much work developing frameworks with indefinite causal structure and (relatedly) indefinite causal order. In Hardy [2005, 2007, 2009a] the *causaloid formalism* was developed which set up a general probabilistic theory in operational terms for a situation in which there can be indefinite causal structure (see also Markes and Hardy [2009] for a discussion of entropy in the framework, Sakharwade and Hardy [2024] for a diagrammatic formulation of the causaloid framework, and Sakharwade [2021] for general discussion). Chiribella, D'Ariano, Perinotti, and Valiron [2013] developed an operator framework for indefinite causal order and proposed the quantum switch and, independently, Oreshkov, Costa and Bruckner Oreshkov, Costa, and Brukner [2012] developed a similar operator framework and proposed causal order inequalities (which, they proved, are violated by theories having indefinite causal order within their operator framework). There has been a considerable amount of research on theories having indefinite causal structure and order. In Hardy [2009a] it was speculated that such a phenomena might lead to faster than quantum computing. Subsequent work has given concrete substance to this possibility. Chiribella [2012] showed how indefinite causal order could beat Quantum Theory in a discrimination task. Quantum-controlled orderings were shown to yield advantages in oracle and communication-complexity tasks Araújo et al. [2014], Guérin et al. [2016], including enhanced communication through channels that would otherwise be useless Ebler et al. [2018]. More recently, exponential separations in quantum query complexity have been established in models explicitly exploiting indefinite causal order Kristjánsson et al. [2024], and it has been shown that simulating such higher-order processes within standard fixed-order circuit architectures can require substantial computational overhead Bavaresco et al. [2025]. Although these results are typically formulated in oracle or higher-order transformation settings rather than standard decision-problem complexity classes, they provide concrete evidence that relaxing definite causal structure can enlarge the operational power of quantum information processing.

One approach to indefinite causal structure is to try to tame it locally. To see how we might do this note that we can tame non-inertial behaviour locally by going to a freely falling coordinate system. Indeed Einstein's Equivalence Principle says that we can always find a coordinate system such that, in the vicinity of any given point, we have inertial behaviour. The Quantum Equivalence Principle was proposed in Hardy [2020] (see also Hardy [2019] and Hardy [2018]) - that we can always find a quantum coordinate system such that, in the vicinity of any given point, we have definite causal structure. To make this work we need to say what a quantum coordinate system is. The idea of a quantum coordinate system was motivated by the discussion of quantum reference frames in Giacomini et al. [2019] (the idea quantum reference frames go back to Aharonov and Susskind [1967] and Aharonov and Kaufherr [1984]) but it has to be a different idea since quantum reference frames are instantiated by physical systems whilst coordinate systems in General Relativity are not. A quantum coordinate system pertains to a bunch of manifolds which appear in a quantum superposition wherein we put a coordinate system on any one manifold then identify points between these manifolds (which induces coordinate systems on the other manifolds). We can transform these quantum coordinate systems by transforming the coordinate system on one of the manifolds and also by re-identifying points between the manifolds. This allows us to remove indefinite causal structure in the vicinity of a point as required. We can think of quantum coordinate transformations as corresponding, at an abstract level, to quantum diffeomorphisms. Interestingly, exactly this idea of quantum diffeomorphisms was proposed earlier in Anandan [1997, 2002] (though not in the context of eliminating indefinite causal structure locally as discussed). More recent work on the equivalence principle in a quantum setting using quantum reference frame transformations appears in Giacomini and Brukner [2020, 2022]. Recent work on quantum coordinates and quantum diffeomorphisms appear in de la Hamette et al. [2025] and Kabel et al. [2025]. One idea tying this together with the project in this book is that we might be able to extend the quantum equivalence principle by including temporal reference frame transformations such that, in the vicinity of a point, we can transform to a frame of reference such that we have both definite causal structure and time symmetry. Both inertial behaviour and time symmetry afford us a certain amount of linearity in the equations which could be useful in going to a theory of Quantum Gravity.

It can be argued that Quantum Theory is best understood as an operational theory. What about General Relativity? Certainly Einstein was motivated by operational considerations in setting up this theory. In setting up Special Relativity (a necessary precursor to GR) he took a very operational approach to the measurement of space and time intervals. And the Equivalence Principle can be understood as an operational equivalence wherein free fall and floating in space appear the same locally. However, once formulated, General Relativity is a theory of fields living on manifolds and so has a distinct ontological flavour. However, we need to be careful about attributing ontological meaning to the values of these fields at any given spacetime point. In fact, the only "beables" (this is John Bell's term for ontologically real quantities Bell [1987]) are properties that are invariant under general coordinate transformations (or, if we want to be more abstract, under diffeomorphisms). This leaves us with a theory where the operational implications are somewhat obscured. In Hardy [2016] it was shown how to formulate General Relativity as an operational theory with operations connected by wires. The key idea, borrowed and adapted for operational purposes from Westman and Sonego [2009] (see also Westman and Sonego [2008]), is that we can set up a space of scalar fields obtained from the fundamental tensor fields. Then, when we plot a solution to the GR field equations point by point into this space. The resulting surface does not move when we perform a diffeomorphism. This space, then, becomes our operational space and small regions within it can be interpreted as operations. It is possible that this kind of operational approach to General Relativity will be needed when we set up a theory of Quantum Gravity.

Whilst Quantum Gravity provides the main motivation for this excursion into time symmetry, we will not attempt to construct a theory of Quantum Gravity in this book, nor especially discuss this most important of problems in physics. There are many approaches in the literature. The two most developed approaches are string theory (see Polchinski [1998]) and loop quantum gravity (see Smolin [2004] and Rovelli [2007]). The spin foams approach grew out of the loop quantum gravity approach (see Perez [2013]). There are many other approaches including the causal set approach (see Sorkin [2005]) and the causal dynamical triangulations approach (see Loll [2020]).

# 3 The zoo of operational frameworks

## 3.1 18 operational frameworks

We can consider 18 related operational frameworks

$$(\mathrm{TF}, \mathrm{TB}, \mathrm{TS}) \times (\mathrm{S}, \mathrm{C}) \times (\mathrm{OPT}, \mathrm{OCT}, \mathrm{OQT})$$

One example is TSSOQT which is Time Symmetric (causally) Simple Operational Quantum Theory. A particular theory can be written $txo$.

The $o$ indicates whether we are looking at the general case or restricting to the Classical case or to the Quantum case.

$$o = \begin{cases} \text{OPT} & \text{Operational Probabilistic Theory} \\ \text{OCT} & \text{Operational probabilistic Classical Theory} \\ \text{OQT} & \text{Operational Quantum Theory} \end{cases}$$

We will not actually develop Operational probabilistic Classical Theory here (though it is fairly clear how to write down such a theory). We could also consider hybrid classical quantum theories, or alternative probabilistic theories within the operational framework (for example, boxworld in Barrett [2007]).

The $t$ indicates the *temporal frame of reference* where

$$t = \begin{cases} \text{TF} & \text{time forward} \\ \text{TB} & \text{time backward} \\ \text{TS} & \text{time symmetric} \end{cases}$$

are the options. These are the different temporal perspectives we discussed previously.

The $x$ indicates what type of operations we are looking at where

$$x = \begin{cases} \text{S} & \text{causally Simple (or Square)} \\ \text{C} & \text{causally Complex (or Circle)} \end{cases}$$

are the options.

We will mostly be interested in just four of these 18 frameworks, namely

$$\begin{array}{ccc} \text{TSSOPT} & \longrightarrow & \text{TSSOQT} \\ \updownarrow & & \updownarrow \\ \text{TSCOPT} & \longrightarrow & \text{TSCOQT} \end{array} \tag{34}$$

The right pointing arrows indicate correspondence. The up/down pointing arrows can be understood as follows. The down direction indicates that the complex case can be motivated from networks in the simple case. The up direction indicates that we can model the simple case using the complex framework.

We will also look at the TF and TB cases. In principle we expect to be able to convert between TF, TS, and TB using the usual rules concerning conditional and joint probabilities. Then we expect to have

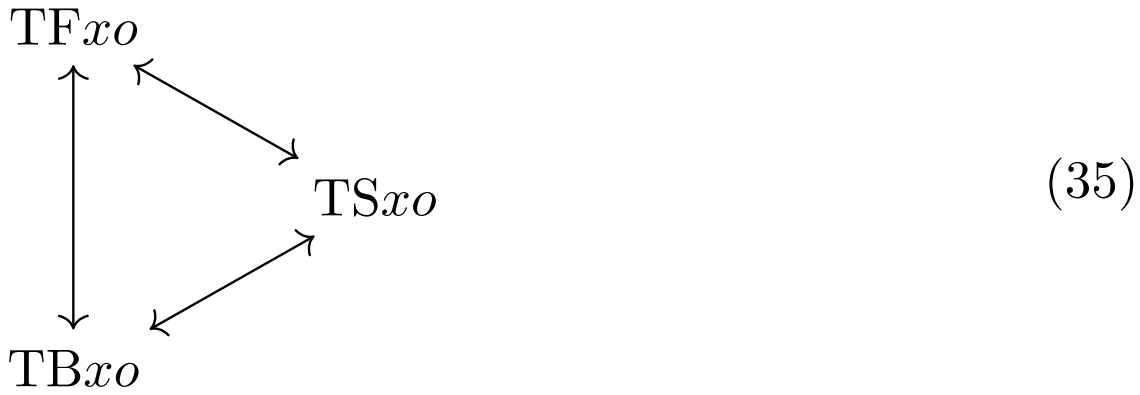

(35)

We prove that this works well in the simple framework but leave it as an open problem for the complex framework.

## 3.2 The structure of operational frameworks

An operational framework, $tx$OPT, can be thought of as having a certain structure.

$tx$OPT =

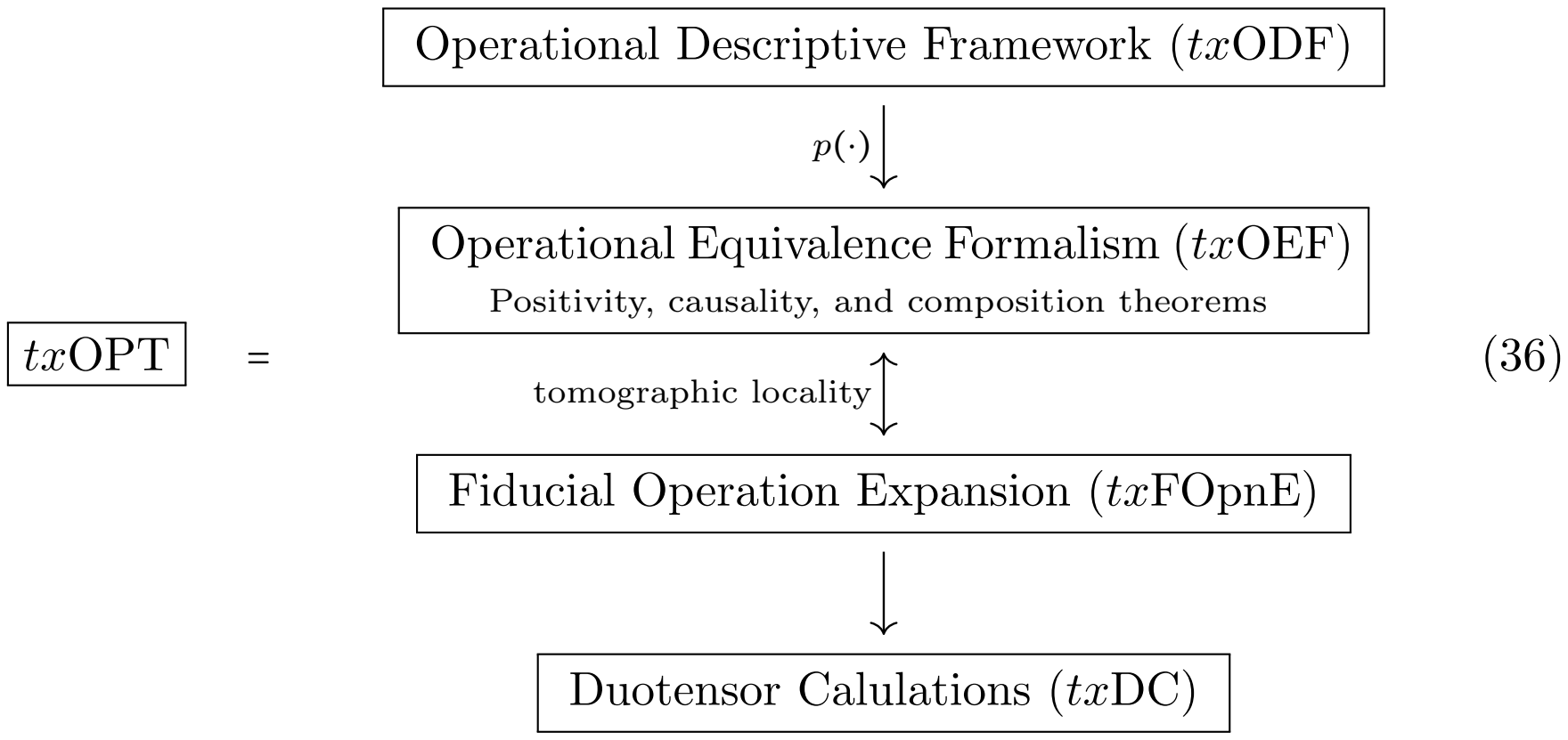

(36)

We can use this structure to go to the classical or the quantum case.

First consider the quantum case. We obtain Operational Quantum Theories, $tx$OQT, by appending structure to $tx$OPT which constrains us to the quantum case. This works as follows.

$tx$OQT =

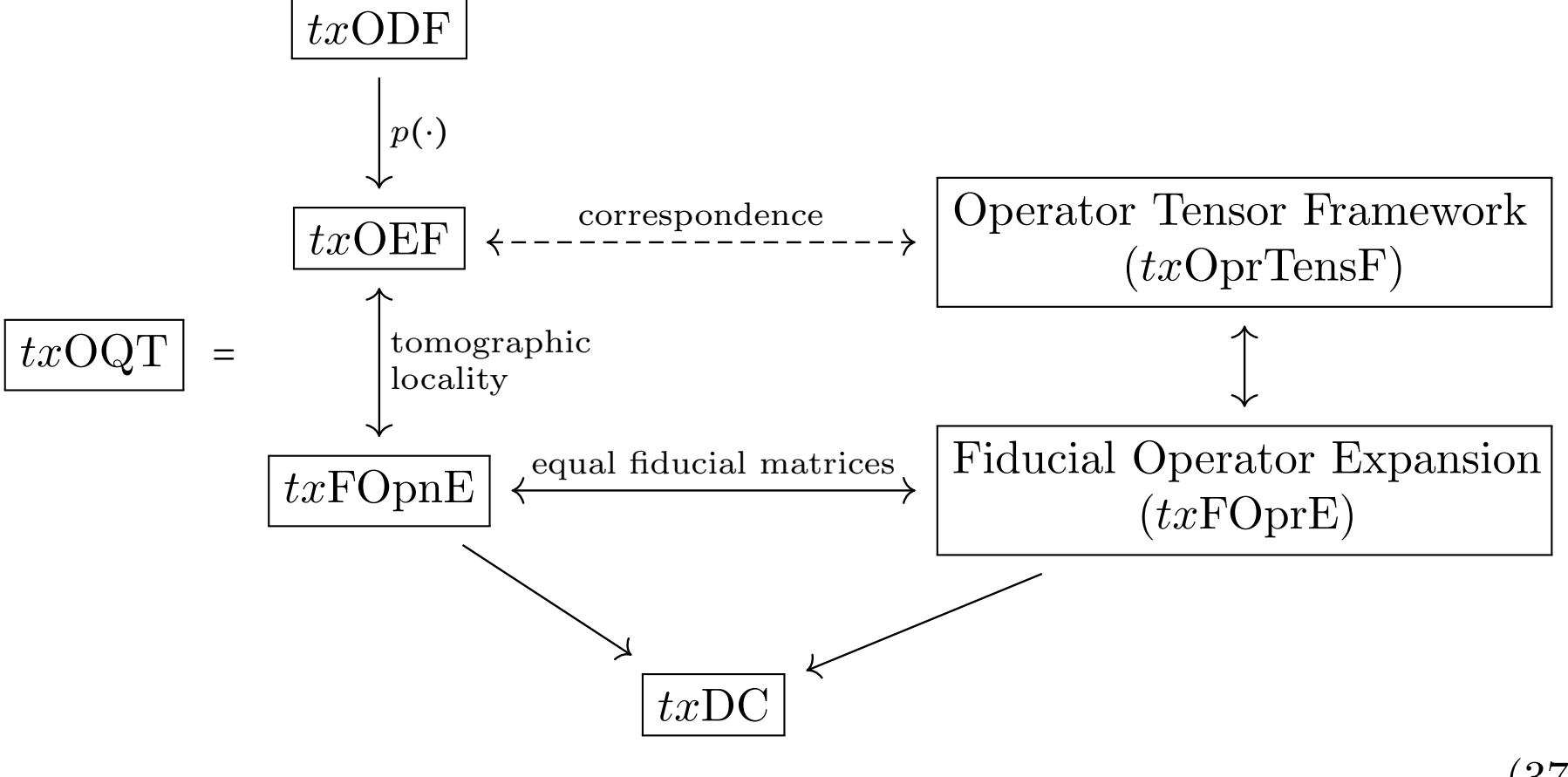

(37)

Compare with (36). The objective is to set up a correspondence from (equivalence classes of) operations to operators. This is shown by the dotted line. This correspondence is achieved by choosing a set of fiducial operations on the one hand and a set of fiducial operators on the other. The fiducial operators are chosen to be positive operators that span the space of operators acting on the relevant Hilbert space. If this Hilbert space is of dimension $N$ then there are $N^2$ such fiducial operators. By combining fiducial elements into circuits we can obtain probabilities that form a matrix - the fiducial matrix. We find a set of fiducial operations in $tx$FOpnE that have the same fiducial matrix as the fiducial operators chosen in $tx$FOprE for each system type (that we can do this is an empirical fact in the Quantum case). Then we form a weighted expansion with these fiducial operators having the same weights (represented by duotensors) as the weights in the fiducial expansion of the corresponding operation. This sets up the correspondence between (equivalence classes of) operations and operators. We can prove, using duotensor calculations, that we get the same probabilities for any circuit if we replace operations by corresponding operators.

We will not dedicate much space to the classical case in this book. It is, however, very straightforward. Similar to the quantum case, we can obtain Classical Probabilistic Theories, $tx$OCT, by appending additional structure that constrains us to the classical case.

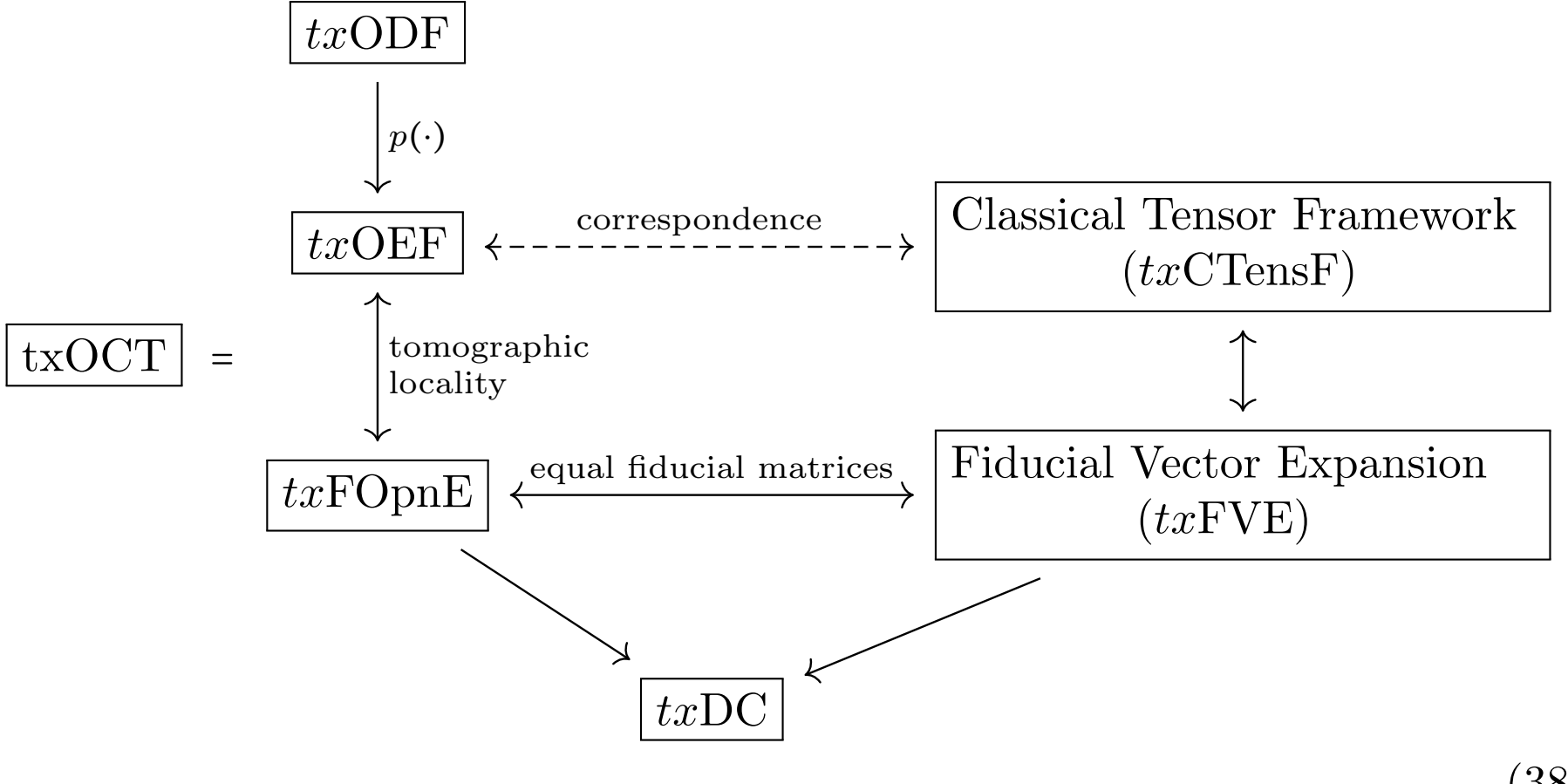

(38)

Compare with (36). The classical tensors are really just tensors. The objective is to set up a correspondence between (equivalence classes of) operations and tensors. This is shown by the dotted line. This correspondence is achieved by choosing a set of fiducial operations on the one hand and a set of fiducial vectors on the other. The fiducial vectors are chosen to be a spanning set of vectors in the corresponding allowed classical probability space. There are $N$ such vectors where $N$ is the number of classical pure states. By combining fiducial elements into circuits we can obtain probabilities that form a matrix - the fiducial matrix. We find a set of fiducial operations in $tx$FOpnE that have the same fiducial matrix as the fiducial vectors chosen in $tx$FVE for each system type. Then we form a weighted expansion with these fiducial operators having the same weights (represented by duotensors) as the corresponding weights in the fiducial expansion of operation. This sets up the correspondence between (equivalence classes of) operations and tensors. We can prove, using a duotensor calculation, that we get the same probabilities for any circuit if we replace operations by corresponding tensors.

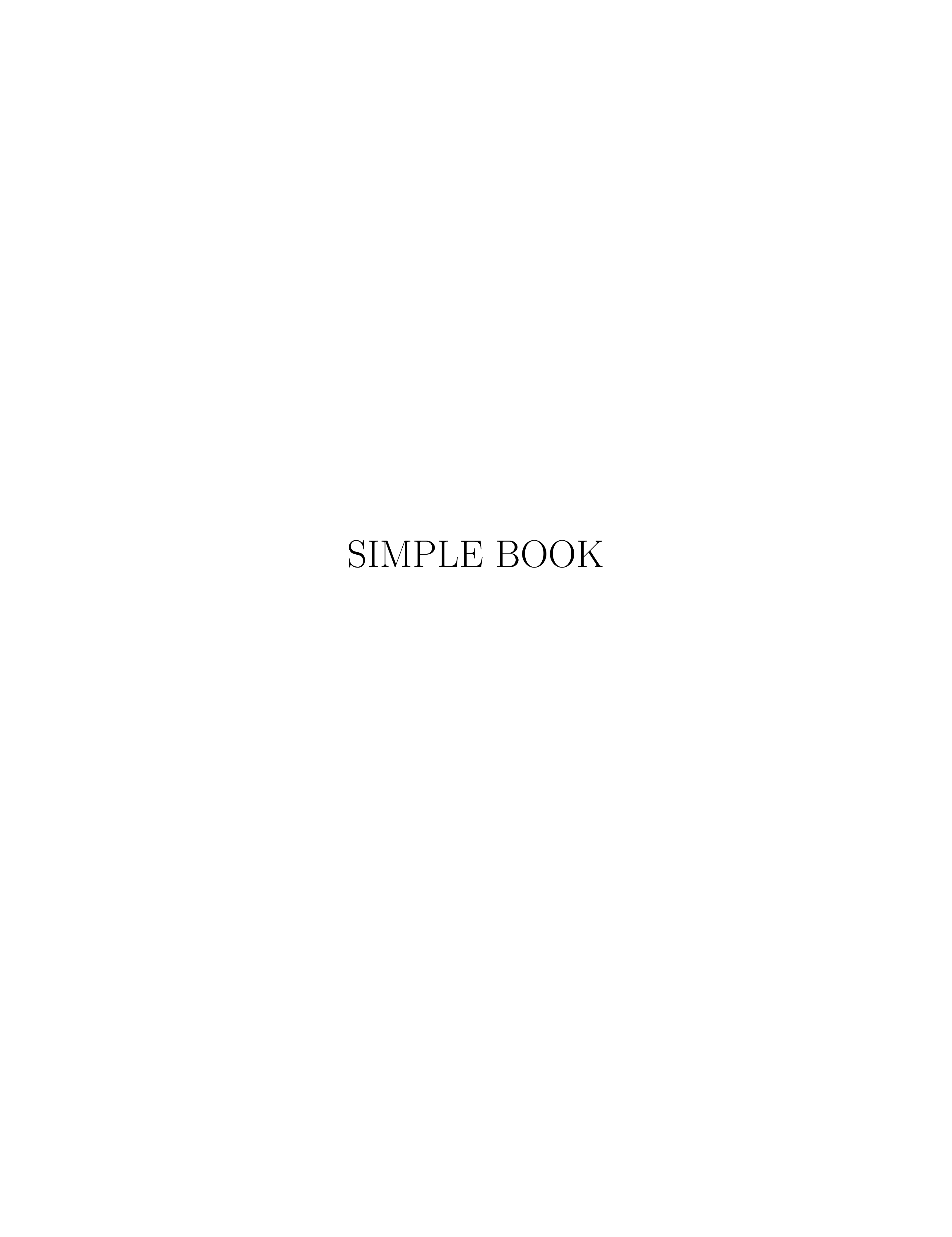

SIMPLE BOOK

# Part I
# Causally Simple Operational Probabilistic Theories

## 4 Introduction to the simple case

In this part we will discuss causally simple operational probabilistic theories (denoted $t$SOPT). This is the general operational theory that can be specialised to Operational Quantum Theory or Classical probabilistic Operational Theory as indicated in the top rung of (34)

$$t\text{SOCT} \longleftarrow t\text{SOPT} \longrightarrow t\text{SOQT} \tag{39}$$

We will mostly focus on the time symmetric temporal perspective ($t$ = TS). We will show in Sec. 11 then show how to convert our results to the time forward case and, in Sec. 12, the time backward case.

We will treat the case of TSSOPT (time symmetric simple operational probability theory) by going through the elements in the flowchart (36). For $tx$ = TSS, the elements of this are

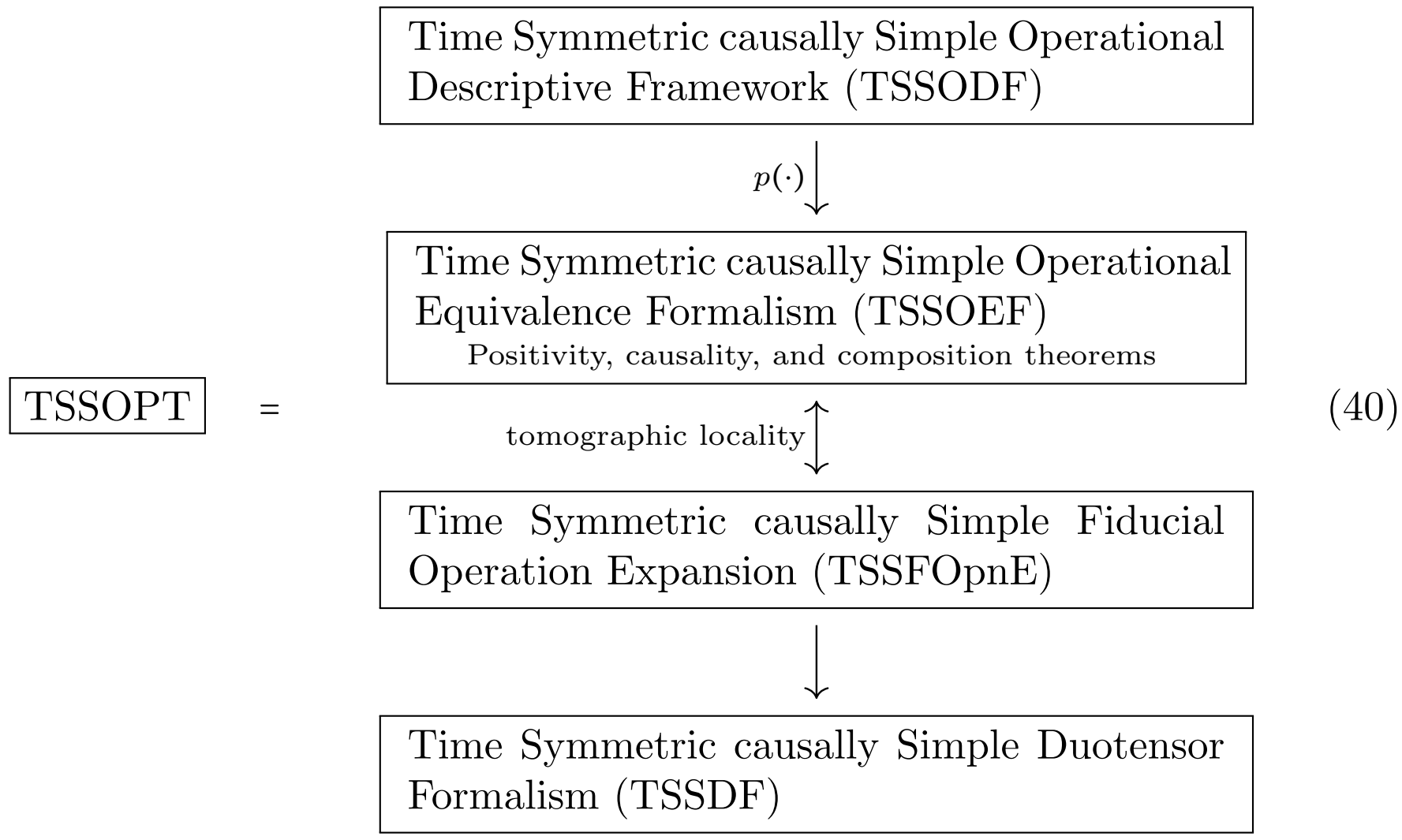

(40)

In what follows we will discuss these elements. The composition theorems (mentioned in the second rung above) concern proving that, when we wire together physical operations the resulting composite is also physical.

# 5 Time Symmetric Causally Simple Operational Descriptive Framework

We begin by outlining a framework for *describing* experiments in operational terms. The descriptive framework does not, in itself, allow us to make predictions. Rather, operational description is a necessary component of an operational theory as it enables us to describe that thing we wish to make predictions about.

There are two basic objects in descriptive framework. These are *deterministic operations* and *readout boxes*. We can wire these together (forming directed acyclic graphs so there are no "closed loops") to form networks and circuits (circuits are networks no open wires). Operations correspond to *one use of an apparatus*. A readout box is introduced where we read off some classical information (say a detector click or a pointer position). Deterministic operations are operations where there are no readout boxes. If we form a circuit comprised only of deterministic operations then there are no readouts and the probability associated with this circuit (as we will see later) is equal to 1.

## 5.1 Deterministic causally simple operations

In the time symmetric causally simple operational descriptive framework we want our operations to be *causally simple*. A causally simple operation can be depicted as follows

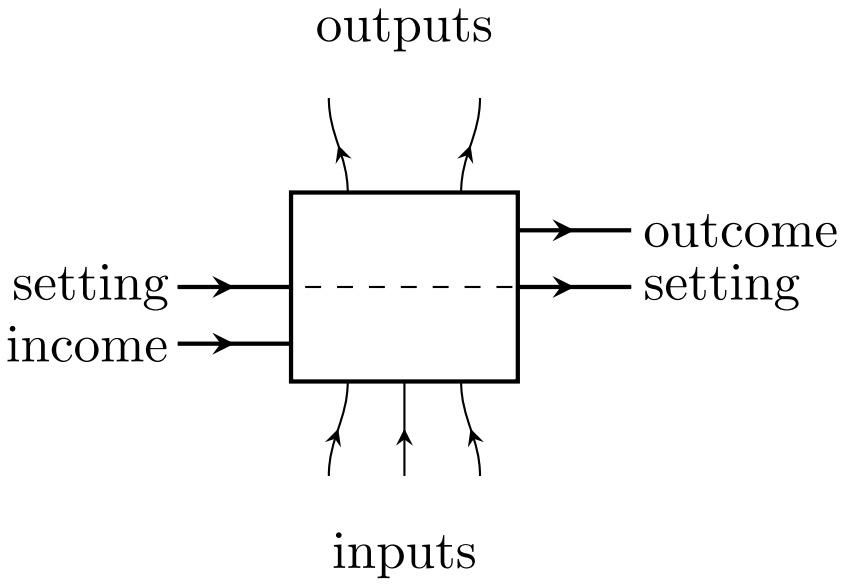

The inputs and outputs are the physical systems the operation acts on. In the quantum case these will be quantum systems. The settings might correspond to knobs on the box which can be put in different positions. Information about the setting is available before and after the apparatus use. The outcomes are classical information that is available after and not before the operation has taken place and may correspond to detector clicks, pointer positions, etc. The incomes are added to make the operation time symmetric Hardy [2021]. Incomes are classical information available before and not after the operation has taken place. An example in quantum theory would be if we send in a quantum system that is constrained to be in one of a give basis set of states. An operation may have multiple inputs, outputs, incomes, and outcomes. In Sec. 13.3 we see that there is no meaningful distinction between the notion of settings and the notion of incomes in time forward theories (where we have preselection). This accounts for why this distinction does not appear in standard textbook treatments of Operational Quantum Theory.

Associated with any wire is a notional system travelling along it. For the sake of having nomenclature, we will refer to those associated with input and output wires as *physical systems* or just as *systems*. And we will refer to those associated with income and outcome wires as *pointer systems* or just as *pointers*.

We call these operations *causally simple* (or just *simple*) because we assume the causal situation is taken to be such that outputs/outcomes on such an operation are all *after* the incomes/inputs. This means that outputs cannot feed into inputs on the same simple operation and, similarly, outcomes cannot feed into incomes on the same simple operation. We represent simple operations by rectangles. Later we will consider the more general case of causally complex operations where the causal structure is potentially more complicated. In this more general case outputs can feed into inputs (and outcomes into incomes) on the same operation. These more general operations will be represented by circles. If it is clear from context that we have a causally simple operation, we will usually drop the "causally simple" as this is a bit cumbersome just referring to it as an "operation".

We will represent (causally simple) deterministic operations as follows

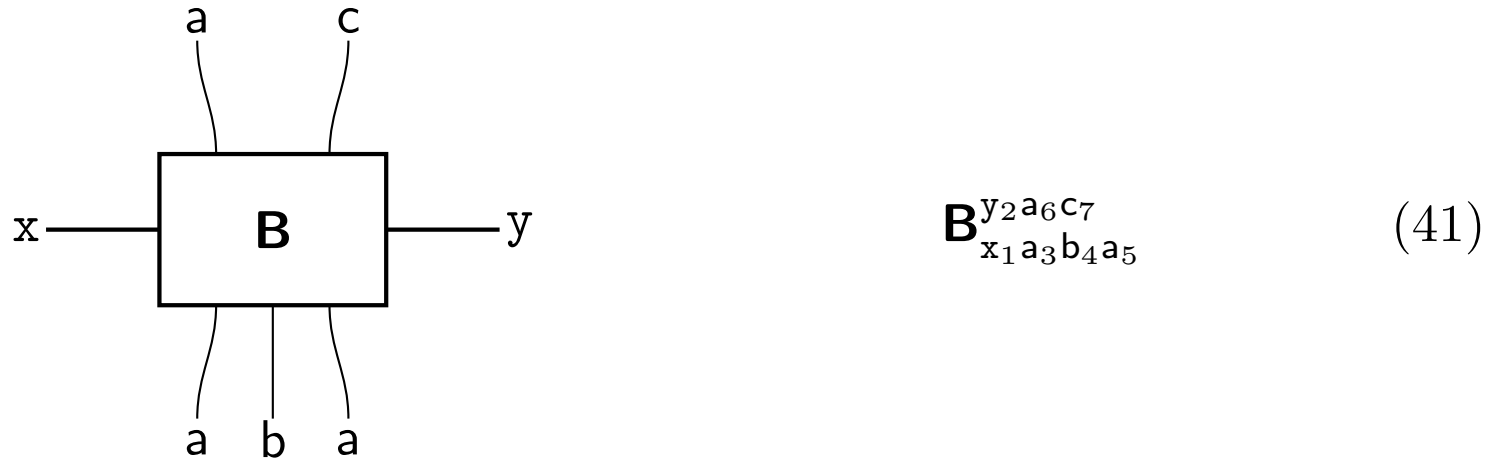

$$\mathbf{B}^{\mathtt{y}_2\mathsf{a}_6\mathsf{c}_7}_{\mathtt{x}_1\mathsf{a}_3\mathsf{b}_4\mathsf{a}_5} \tag{41}$$

This shows an operation $\mathbf{B}$ with diagrammatic representation on the left and symbolic representation on the right. The setting information is no longer shown explicitly but is implicit in the symbol, $\mathbf{B}$ (for a different setting we would use a different symbol, for example $\mathbf{B}'$ or $\mathbf{C}$). Income and outcome types are represented by symbols in LaTeX `\mathtt` font ($\mathtt{x}$ and $\mathtt{y}$ in this instance). An income/outcome of type $\mathtt{x}$ has values denoted by $x = 1, 2, ..$ to $N_{\mathtt{x}}$ (and similarly for $\mathtt{y}$, etc). It is clumsy to refer to "income/outcome types" so we will call these "pointer types" (as, often, in quantum measurement theory we talk about a pointer indicating the readout of a measurement). The input and output types (which we will refer to as "system types") are given by $\mathsf{a}$, $\mathsf{b}$, etc (in LaTeX `\mathsf{a}` font). In the symbolic notation we have integer subscripts on the types to show where the wires (this will be illustrated in (50) below). We will mostly use diagrammatic (pictorial) notation in this book since it is easier to understand.

This operation is deterministic. We use boldface to denote operations that are taken to be *deterministic* - for example $\mathbf{A}$ above is deterministic. A deter-ministic operation can have income and outcome wires but there are no readouts (implicit or otherwise). To make this clear we need to introduce readout boxes.

## 5.2 Readout boxes

If we want to read off the incomes and outcomes we use *readout boxes*

$$\mathtt{x} \,—\, \boxed{x} \,—\, \mathtt{x} \tag{42}$$

The value, $x$, inside the readout box is something that may, or may not happen.

Readout boxes have the following property

$$\mathtt{x} \,—\, \boxed{x} \overset{\mathtt{x}}{—} \boxed{x'} \,—\, \mathtt{x} \qquad \equiv \qquad \begin{cases} \mathtt{x} \,—\, \boxed{x} \,—\, \mathtt{x} & \text{if } \; x = x' \\ \mathtt{x} \,—\, \boxed{0} \,—\, \mathtt{x} & \text{if } \; x \neq x' \end{cases} \tag{43}$$

If $x = x'$ we are reading off the same value twice. If $x \neq x'$ then we get the null box (the box with a "0" in it). This box has the property that if it is added to a circuit then the circuit has zero probability associated with it.

## 5.3 Networks and circuits

We can wire together operations and readout boxes to create *networks*. *Circuits* are networks having no open wires. When we join operations we must (i) match types and (ii) ensure there are no causal loops (wherein we can get back to some given operation by tracing a path through the operations by going from output (or outcome) to input (or income). This means that networks form directed acyclic graphs (DAGs).

Here are three examples of networks

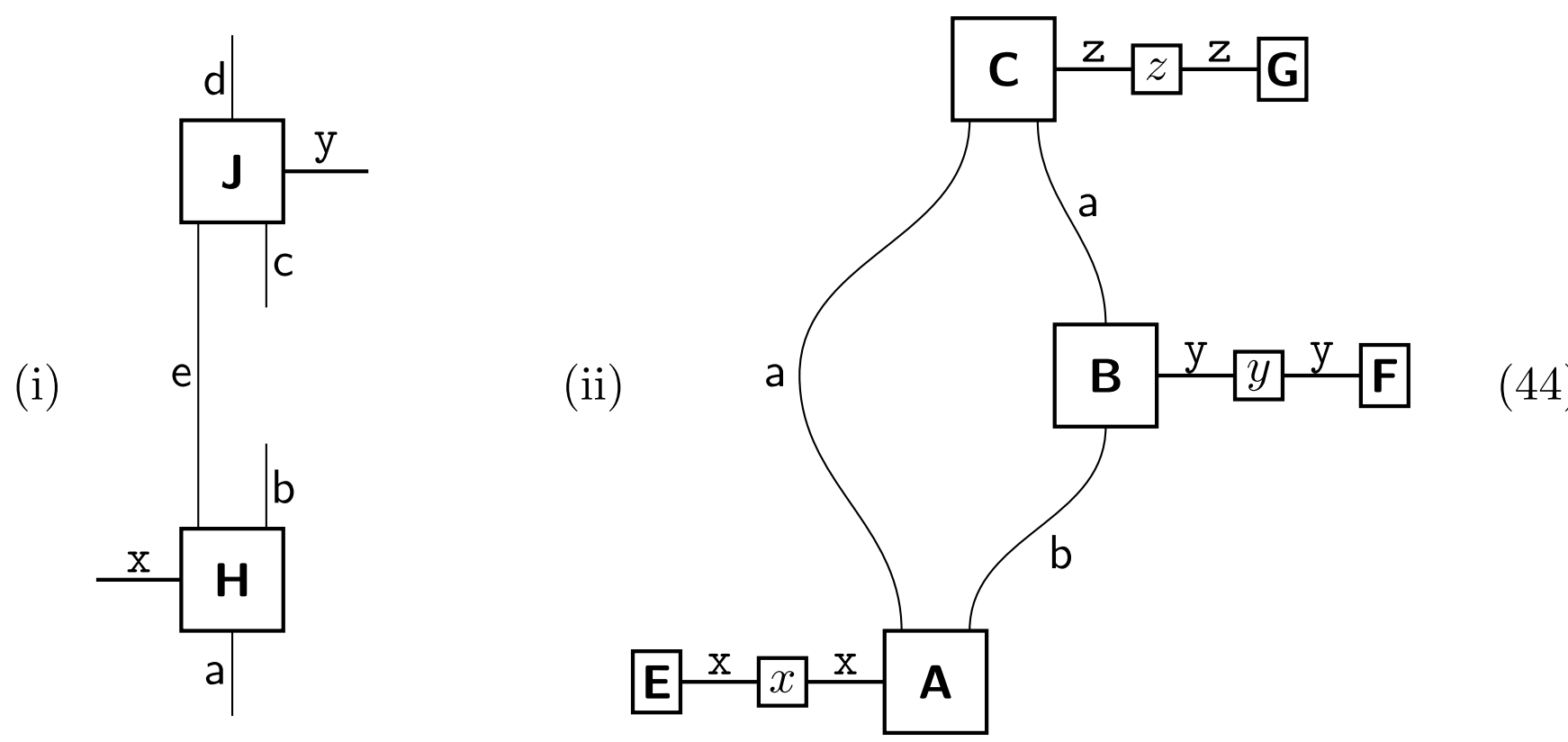

and

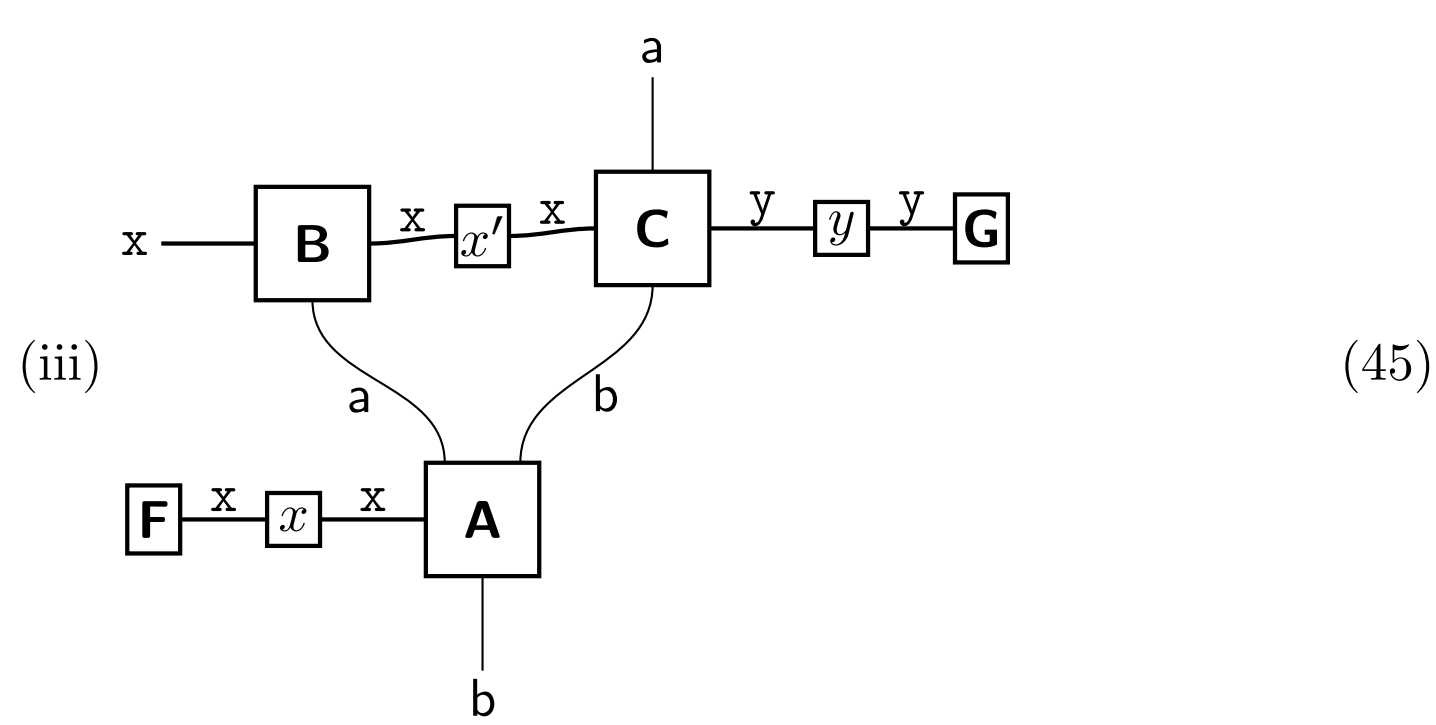

(45)

Example (i) is deterministic since it has no readout boxes. Examples (ii) and (iii) are nondeterministic. Example (ii) is a circuit since it has no open system or pointer wires. Example (iii) has simple causal structure because we can regard all the outputs/outcomes as being after all the inputs/incomes. On the other hand, example (i) has complex causal structure since output b is clearly before input c.

Since nondeterministic networks have some specified readout (for example $xyz$ in (ii)) they may or may not "happen". On the other hand, deterministic networks always happen. This nomenclature was introduced in Chiribella et al. [2009a].

In the case that a network has simple causal structure, we can regard it as a (causally simple) operation as follows

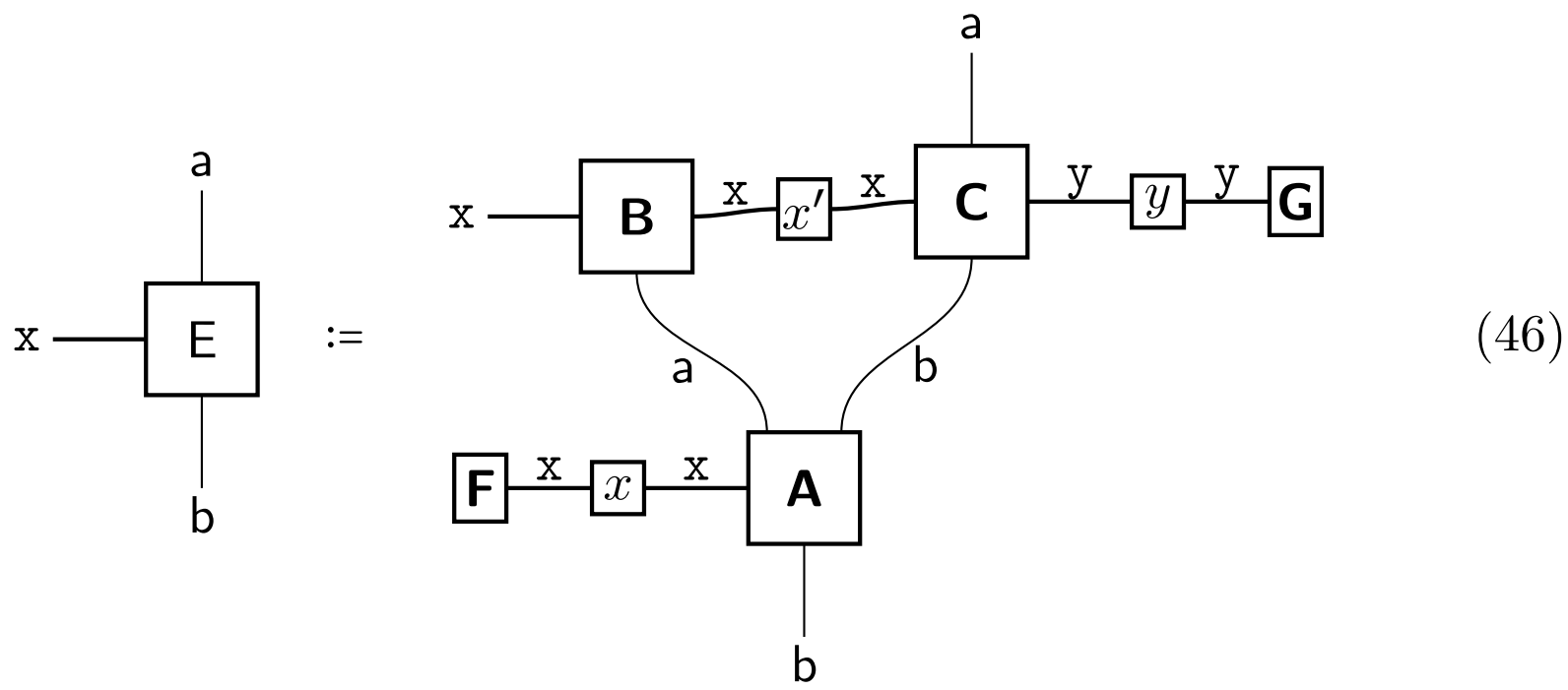

(46)

Since this is an nondeterministic operation we denote it by non-boldface font E. Sometimes we will use the notation, $\mathbf{E}(xx'y)$, for nondeterministic operations where we explicitly include the readouts in parenthesis. Deterministic operation are written in boldface. We will write a general operations which may be deterministic or nondeterministic in non-boldface font.

We cannot write causally complex networks as causally simple operations. In Part IV of this book we will set up the notion of causally complex operations (represented by circles) which can be used to represent causally complex networks.

Circuits (such as example (ii) above) play a special role in the Operational Probabilistic Formalism in that we assume there is a probability for the circuit happening (i.e. the specified readouts being observed). A circuit can be represented by a single box with no open wires where we either choose to make the readouts explicit

$$\boxed{\mathbf{H}(xyz)} \tag{47}$$

or the readouts are implicit

$$\boxed{\mathsf{D}} \tag{48}$$

in the symbol we use to represent the circuit ($\mathsf{D}$ in this case). Symbolically, we can represent a circuit by $\mathbf{H}(xyz)$ or $\mathsf{D}$. A circuit that has no implicit readout boxes (i.e. it is deterministic) would be represented by a boldface letter (e.g. $\mathbf{D}$).

A useful notion is that of a *complete set of operations or networks.* A set of operations (networks) is complete if the readouts for each element are mutually exclusive (only one can happen in any given run) and if all possible readouts are covered by elements of this set. For example, the set

$$\left\{\mathbf{E}(xx'y) : x = 1 \text{ to } N_{\mathsf{x}}, x' = 1 \text{ to } N_{\mathsf{x}}, y = 1 \text{ to } N_{\mathsf{y}}\right\} \tag{49}$$

(where $\mathbf{E}(xx'y)$ is given in (46)) is a complete set.

## 5.4 Composition of deterministic and nondeterministic operations

Determinism is defined in a very primitive fashion - there are no implicit or explicit readouts. This leads us to the following statements.

> **Composition of deterministic and nondeterministic operations.** The following hold true:
>
> 1. If we compose two or more deterministic operations then we obtain a deterministic network (which is a deterministic operation if it has simple causal structure).
> 2. If a network is deterministic then each of its components are deterministic.
> 3. If we compose two or more operations where any one of them nondeterministic then we have a nondeterministic network.
> 4. If a network is nondeterministic then at least one of its components must be nondeterministic.

These statements follow from the fact that the only primitive nondeterministic operation is the readout box (any nondeterministic operation must have a readout box in its composition somewhere).

## 5.5 Labelling of wires

We can think of each wire as having an integer label. We can use this to set up symbolic notation. For example we can write

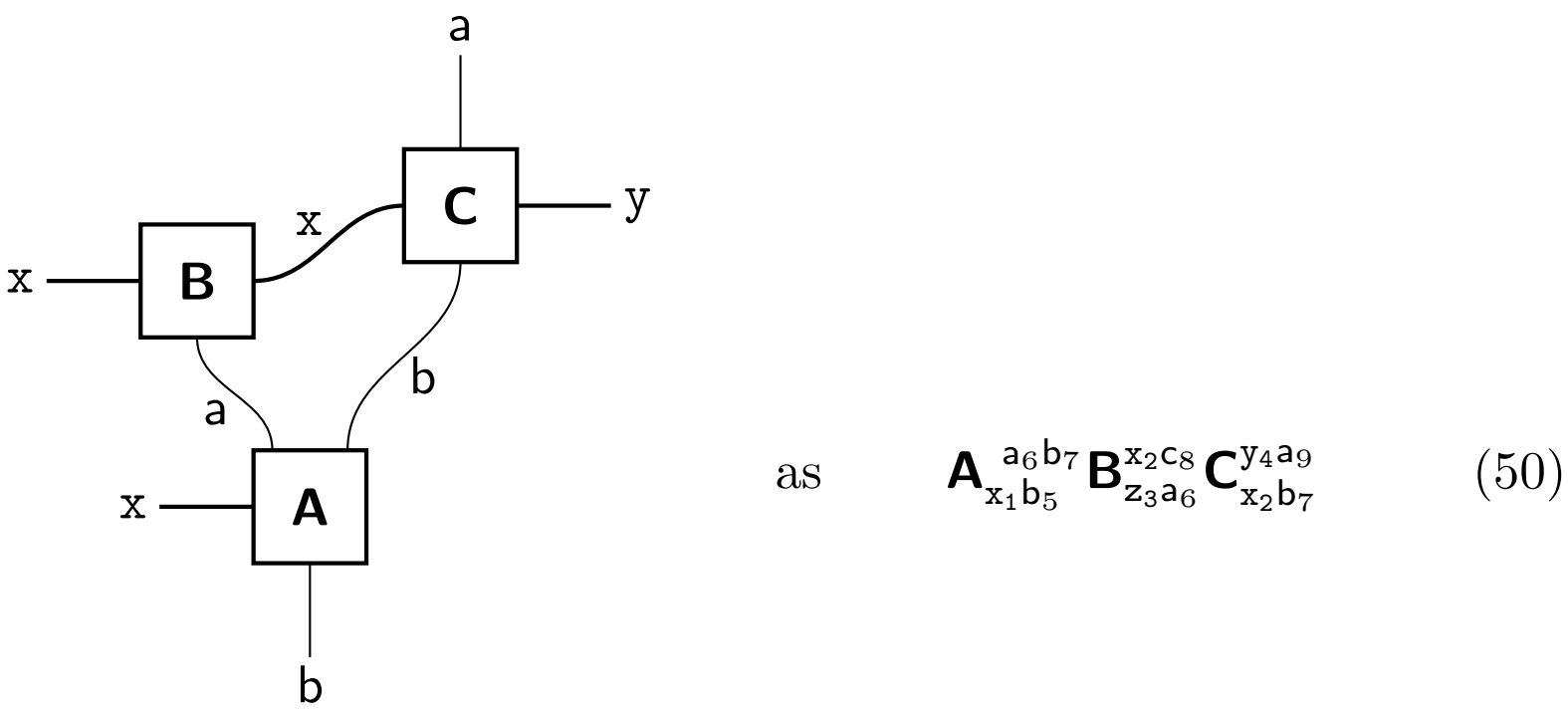

$$\text{as} \qquad \mathbf{A}^{\mathsf{a}_6\mathsf{b}_7}_{\mathsf{x}_1\mathsf{b}_5}\mathbf{B}^{\mathsf{x}_2\mathsf{c}_8}_{\mathsf{z}_3\mathsf{a}_6}\mathbf{C}^{\mathsf{y}_4\mathsf{a}_9}_{\mathsf{x}_2\mathsf{b}_7} \tag{50}$$

Diagrammatic notation is on the left and symbolic notation is on the right. In the symbolic notation on the right the integer shows where the wires go. For example $\mathsf{b}_7$ joins an ouput on $\mathbf{A}$ to an input on $\mathbf{C}$. In symbolic notation we have to explicitly provide this label. Given its immediacy, we will for the most part stick with the diagrammatic notation.

In the diagrammatic notation it is not necessary to display the integer since we can see where the separate wires are. However, there is an issue if we are equating expressions. Consider an equivalence like

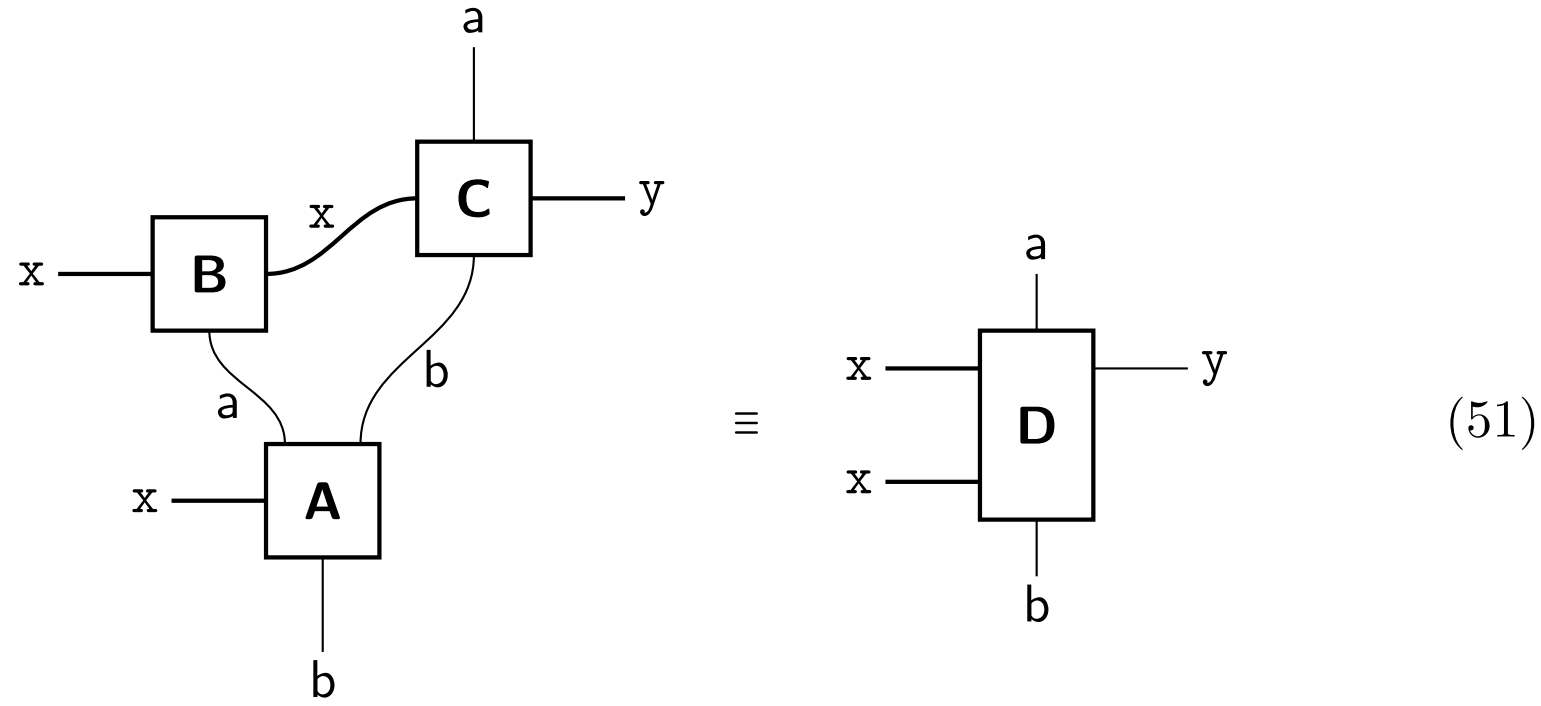

(51)

Equivalence expresses the idea that we can plug the object on one side in the place of the object on the other side in some bigger network and we have equivalent behaviour (for us this will mean equivalent probabilistic behaviour). It is important, then, to know which open wires in the diagram on the left of the equation correspond to which open wires on the right of the equation (so we know how to do this substitution). To identify open wires we impose that the wires enter/leave the diagram in corresponding positions. Indeed, in the above example it is clear which open wire is which as we go across the diagram. In all the examples we will consider it will be clear which are the corresponding open wires is across an equivalence. In the case this correspondence is not clear we can explicitly provide the integer labels.

A few notes are pertinent here. It is not necessary that the closed wires correspond across an equivalence (indeed, there may be a different number of wires and different types for the closed wires across an equivalence). The integer labels, if we do provide them, are just labels and the equivalence has the same meaning if we use different labels (as long as corresponding wires continue to have the same label). In addition to equivalences we will consider equations (using =), definitions (using :=) and various types of inequality. These same comments about identifying corresponding wires pertain in these cases as well.

# 6 Time Symmetric Simple Operational Equivalence Formalism

## 6.1 The circuit probability assumption

Having provided a time symmetric operational descriptive framework, we need a formalism for making predictions. Since we are working within a probabilistic framework we start with the following assumption

> **Circuit probability assumption:** Every circuit (wherein all inputs and outputs and all incomes and outcomes are closed) has a probability associated with it that depends only on the specification of that circuit.

This is taken from Hardy [2021] (a similar assumption was given in earlier works Hardy [2011b]).

Consider a circuit $\mathsf{E}$ obtained by wiring together a bunch of operations including, possibly, readout boxes and conditional boxes. Let $x_{\mathsf{E}}$ represents the incomes and outcomes associated with readout boxes in this circuit. Let the description of this circuit (the wiring and names of the operations) be given by $s_{\mathsf{E}}$. The circuit probability assumption says that

$$\text{prob}(x_{\mathsf{E}}|s_{\mathsf{E}}, \text{extraneous conditions}) = \text{prob}(x_{\mathsf{E}}|s_{\mathsf{E}}) \tag{52}$$

This leads to the important theorem

> **Disjoint circuits factorisation property**. If the circuit probability assumption holds then the probability for a circuit with disjoint parts is the product of the probabilities for each of these disjoint parts.

To prove this consider a circuit $\mathsf{EF}$ comprised of disjoint parts $\mathsf{E}$ and $\mathsf{F}$. The probability for this circuit is

$$\begin{aligned}\text{prob}(\mathsf{EF}) &= \text{prob}(x_{\mathsf{E}}, x_{\mathsf{F}} | s_{\mathsf{E}}, s_{\mathsf{F}}) \\ &= \text{prob}(x_{\mathsf{E}} | x_{\mathsf{F}}, s_{\mathsf{E}}, s_{\mathsf{F}}) \text{prob}(x_{\mathsf{F}} | s_{\mathsf{E}}, s_{\mathsf{F}}) \\ &= \text{prob}(x_{\mathsf{E}} | s_{\mathsf{E}}) \text{prob}(x_{\mathsf{F}} | s_{\mathsf{F}}) \\ &= \text{prob}(\mathsf{E}) \text{prob}(\mathsf{F})\end{aligned}$$

where we have used the circuit probability assumption to remove the extraneous conditions in the second line.

We make the following additional assumptions

**Circuit reality.** The probability of a circuit is a real number.

$$\text{prob}\left( \boxed{\mathsf{B}} \right) \in \mathbb{R} \tag{53}$$

**Circuit positivity.** The probability of a circuit is greater than, or equal to, zero

$$0 \le \text{prob}\left( \boxed{\mathsf{B}} \right) \tag{54}$$

**Circuit subunity.** The probability of a circuit is less than, or equal to, one

$$\text{prob}\left( \boxed{\mathsf{B}} \right) \le 1 \tag{55}$$

**Circuit determinism.** The probability of a deterministic circuit (any circuit comprised only of deterministic operations) is equal to one

$$\text{prob}\left( \boxed{\mathbf{B}} \right) = 1 \tag{56}$$

**Complete circuit sets.** If we have a complete set of mutually exclusive circuits, $\{\mathsf{B}(x) : \forall x\}$, where $x$ denotes the readouts, then the sum of the associated probabilities add to one

$$\sum_x \text{prob}\left( \boxed{\mathbf{B}(x)} \right) = 1 \tag{57}$$

These assumptions are necessary so that probabilities have the usual interpretation.

## 6.2 The $p(\cdot)$ function and equivalence

We will now define the $p(\cdot)$ function which enables us to set up a notion of equivalence between networks. This function is a linear extension of the prob$(\cdot)$ function on circuits. We define the $p(\cdot)$ function as follows

$$p(\alpha\mathsf{A} + \beta\mathsf{B} + \gamma\mathsf{C} + \dots) = \alpha\text{prob}(\mathsf{A}) + \beta\text{prob}(\mathsf{B}) + \gamma\text{prob}(\mathsf{C}) + \dots \tag{58}$$

where $\mathsf{A}$, $\mathsf{B}$, $\mathsf{C}$, ... are circuits (operations wired together with no open wires left over) and $\alpha$, $\beta$, $\gamma$, ... are real numbers (they can be negative). The $p(\cdot)$ function is the linear extension of the prob$(\cdot)$ function to expressions consisting of a linear sum of circuits.

If we have a network, $\mathsf{E}^{\mathtt{y}_3\mathsf{b}_4}_{\mathtt{x}_1\mathsf{a}_2}$, then it can be matched with other networks of the form $\mathsf{H}^{\mathtt{x}_1\mathsf{a}_2}_{\mathtt{y}_3\mathsf{b}_4}$ (so that the types match) to generate a circuit. In order that this is possible it is necessary that $\mathsf{H}$ has a matching causal structure. We call the network $\mathsf{H}$ in this case (where we have matching causal structure and can generate a circuit) a *complement network* to $\mathsf{E}$.

We can consider more general expressions such as

$$\text{exprn} = \alpha + \beta\mathsf{E}^{\mathtt{y}_3\mathsf{b}_4}_{\mathtt{x}_1\mathsf{a}_2} + \gamma\mathsf{F}^{\mathtt{y}_3\mathsf{b}_4}_{\mathtt{x}_1\mathsf{a}_2} + \delta\mathsf{G}^{\mathtt{y}_3\mathsf{b}_4}_{\mathtt{x}_1\mathsf{a}_2} + \dots \tag{59}$$

where $\mathsf{E}$, $\mathsf{F}$, $\mathsf{G}$, ... are networks and we have collected all the incomes, outcomes, inputs, and outputs into $\mathtt{x}$, $\mathtt{y}$, $\mathsf{a}$, and $\mathsf{b}$ respectively. Further, $\alpha$, $\beta$, $\gamma$, and $\delta$ are real numbers (they can be negative). Note in particular that the first term, $\alpha$, is just a number. We demand that each network in the linear expansion associated with a circuit will plug into the same set of *complement networks*. We say that two expressions, $\text{exprn}_1$ and $\text{exprn}_2$, are *equivalent*, i.e.

$$\text{exprn}_1 \equiv \text{exprn}_2 \tag{60}$$

if and only if

$$p(\text{exprn}_1\mathsf{H}) = p(\text{exprn}_2\mathsf{H}) \quad \text{for all complement networks} \quad \mathsf{H} \tag{61}$$

This requires that each network in the expansion of both expressions is of the same causal structure (so they will plug into the same set of complement networks).

The numerical term, $\alpha$, in the expression given in (59) can only appear in equivalences between expressions when each network that does appear in the expression is, in fact, a circuit. In fact, we can prove the important and useful property that

$$\mathsf{A} \equiv \text{prob}(\mathsf{A}) \quad \text{for any circuit} \quad \mathsf{A} \tag{62}$$

This is because, using the disjoint circuit factorisation property in Sec. 6.1,

$$p(\mathsf{AH}) = p(\mathsf{A})p(\mathsf{H}) = p(\text{prob}(\mathsf{A})\mathsf{H}) \qquad \text{for all circuits } \mathsf{H} \tag{63}$$

This example illustrates that equivalence is a broader notion than equality. A circuit can be *equivalent* to a number, under the definition give here, but it is clearly not *equal* to a number since circuits are objects in the laboratory and numbers are objects in a mathematical space.

## 6.3 The $p(\cdot)$ function and inequality

We can also use the $p(\cdot)$ function do obtain a notion of inequality. Consider expressions as defined in Sec. 6.2. We say

$$\text{exprn}_1 \leqq \text{exprn}_2 \tag{64}$$

if and only if

$$p(\text{exprn}_1 \mathsf{H}) \leq p(\text{exprn}_2 \mathsf{H}) \quad \text{for all complement networks } \mathsf{H} \tag{65}$$

We read the symbol $\leqq$ as *less than or equivalent.*

The properties of circuit positivity and circuit subunity in Sec. 6.1 can be written as

$$0 \leqq \boxed{\mathsf{B}} \leqq 1 \tag{66}$$

This makes sense because (by circuit reality) a circuit is equivalent to a real number.

The notions of equivalence and inequality are linked by the following simple theorem.

$$\text{If } \text{exprn}_1 + \text{exprn}_2 \equiv \text{exprn}_3 \text{ and } 0 \leqq \text{exprn}_2 \text{ then } \text{exprn}_1 \leqq \text{exprn}_3 \tag{67}$$

This is easily proved using the definitions of $\equiv$ and $\leqq$.

In Sec. 7.11.2 we will define $\underset{T}{\leqq}$ which means *less than or equivalent with respect to testers*. Testers are a particular subset of complement networks and hence it immediately follows

$$\text{exprn}_1 \leqq \text{exprn}_2 \quad \Rightarrow \quad \text{exprn}_1 \underset{T}{\leqq} \text{exprn}_2 \tag{68}$$

The reverse implication is not generally true (though it will hold for when $\text{exprn}_{1,2}$ are real weighted sums of simple operations). In Quantum Theory we can prove this reverse implication. We do this in the context of complex operational quantum theory (see Sec. 76).

## 6.4 What we mean by time symmetry

We want TSSOPT's to be manifestly time symmetric. Here we define what we mean by such time symmetry. We say that a simple operational probability theory is time symmetric if the following three conditions are met

1. Every allowed operation, $\mathsf{B}$, is associated with a time reversed operation, $\underset{\sim}{\mathsf{B}}$, which is also allowed. The time reversed operation has inputs and outputs are reversed and incomes and outcomes are reversed as illustrated below

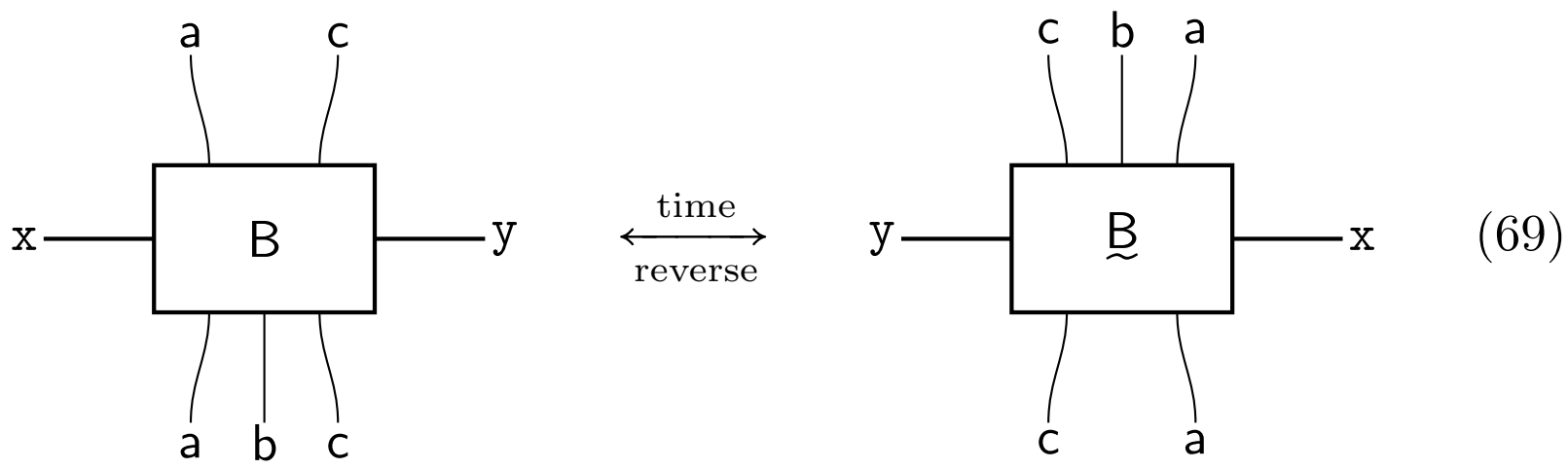

(69)

Note the diagrammatic representation of the time reverse is obtained by rotating by 180°.

2. Readout boxes are equal to their own time reverse.

$$\mathtt{x} \text{---}\boxed{x}\text{---} \mathtt{x} \quad \xleftrightarrow[\text{reverse}]{\text{time}} \quad \mathtt{x} \text{---}\boxed{x}\text{---} \mathtt{x} \tag{70}$$

3. We can obtain the time reverse of any circuit by rotating the whole circuit by 180° and putting tilde's under the symbols. We demand that the probability of the time reversed circuit is equal to the probability of the original circuit. For example

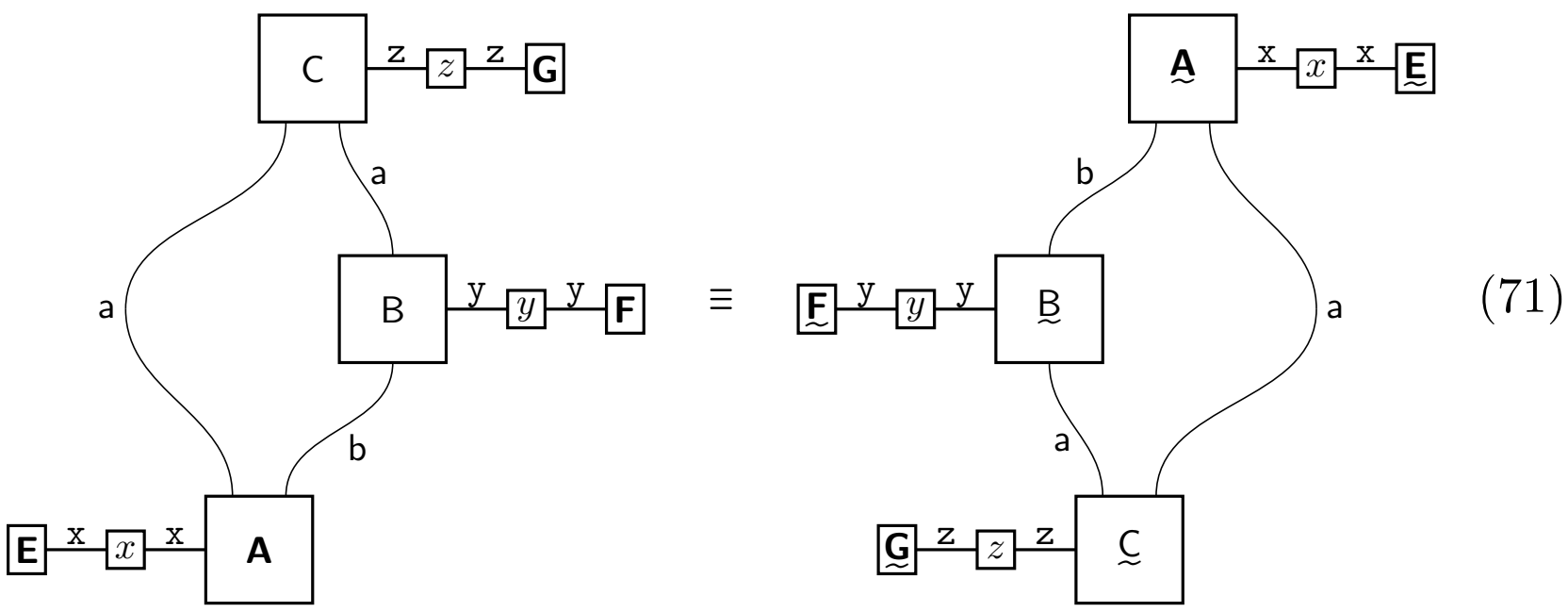

Note that equivalent circuits have the same probability because any circuit is equivalent to its probability (see (62)).

For the first property to be satisfied the constraints on allowed operations must be time symmetric. In Sec. 7 we will discuss these constraints - which we call physicality conditions. They are, indeed, time symmetric.

This is just one way of defining time symmetry. There are generalisations and alternative definitions.

We could allow a generalisation where we also put tildes under the system type symbols (so $\mathsf{a}$ becomes $\underset{\sim}{\mathsf{a}}$) when we form the time reversed operation. Thus, system types would come in pairs, $\mathsf{a}$ and $\underset{\sim}{\mathsf{a}}$. This would provide a framework to investigate Charge Parity Time reversal (CPT) symmetry. (see Schwinger [1951], Lüders [1954], Pauli [1955], and Bell [1955]). The charge and parity would be taken care of by tilde under the type symbols.

There are other definitions of time symmetry. For example, we might define time symmetry to be something that is imposed at the ontological level (see Price [2012] and Leifer and Pusey [2017]).

# 7 Physicality conditions in time symmetric simple operational probabilistic theories

## 7.1 Introduction

We want to be sure that the following conditions are met

**Positivity.** Circuits have probability that is greater than or equal to zero.

**Subunity.** Circuits have probability that is not greater than one.

**Causality.** Appropriate causality constraints are satisfied.

In operational theories these conditions are satisfied by imposing *physicality* conditions on operations consisting of a positivity condition and a causality condition (which, curiously, also imposes subunity). In time symmetric simple operational probabilistic theories we have these two conditions are (a) *tester positivity* which guarantees that probabilities are non-negative, and (b) *double causality* which both imposes certain time symmetric causality conditions and subunity. To express these conditions in full we will need to set up various prerequisites.

In this section we will

1. First we set up the *double causality conditions* for deterministic operations. To do this we will lay out some preliminary causality considerations in Sec. 7.2 and Sec. 7.5. Then, in Sec. 7.6 we will state and prove the double causality theorem which provides causality conditions for deterministic operations.

2. Next we consider positivity. To do this we lay out some preliminaries: we prove the *midcome identity* and the *control wire identity*, and we show how these identities can be used to write a general circuit as a positive weighted sum over *regularised circuits*. Then we state the *tester positivity* condition and prove a composition theorem: if tester positivity is satisfied for each operation in a network then the whole network also satisfies tester positivity. It follows from this theorem that if the network is a circuit, then it has non-negative probability.

3. Finally, we prove a theorem concerning *general double causality conditions* which hold for operations that may be non-deterministic. To prove this theorem we need the notion of tester positivity.

Finally, we provide some conditions for determinism which are useful in later discussions.

## 7.2 Deterministic pointer preparations and results

A *pointer preparation* is an operation having only outcomes.

$$\boxed{\mathsf{E}}\!\!-\!\!-\!\!-\,\mathtt{x} \tag{72}$$

The pointer type, $\mathtt{x}$, can be composite. In the forward in time direction, we can think of this as preparing the pointer type, $\mathtt{x}$, in some probability distribution, $p_{\mathsf{E}}(x)$.

A *pointer result* is an operation having only incomes

$$\mathtt{x}\!-\!\!\boxed{\mathsf{F}} \tag{73}$$

Note that we use the word "result" as the time reverse of "preparation" (just as "effect" is used as the time reverse of "state"). The pointer type, $\mathtt{x}$, can be composite. In the backwards in time direction, this can be thought of as corresponding to some probability distribution, $p_{\mathsf{F}}(x)$, induced by $\mathsf{F}$.

Consider completing a pointer preparation into a circuit by a deterministic result as follows

$$\boxed{\mathsf{E}}\!-\!\!\!-\!\!\boxed{\mathbf{F}} \tag{74}$$

The interpretation of this is that we have some readouts in the past (implicit in $\mathsf{E}$), and then we have no readouts in the future (since $\mathbf{F}$ is deterministic). If, instead, of deterministic result $\mathbf{F}$ we apply deterministic result $\mathbf{F}'$ then we would have

$$\boxed{\mathsf{E}}\!-\!\!\!-\!\!\boxed{\mathbf{F}'} \tag{75}$$

Now, according to an appropriate notion of causality, the choice of $\mathbf{F}$ or $\mathbf{F}'$ in the future should not affect the probability of the readouts on $\mathsf{E}$ in the past. Thus we require

$$\text{prob}\left(\boxed{\mathsf{E}}\!-\!\!\!-\!\!\boxed{\mathbf{F}}\right) = \text{prob}\left(\boxed{\mathsf{E}}\!-\!\!\!-\!\!\boxed{\mathbf{F}'}\right) \tag{76}$$

Further, this should be true for any $\mathsf{E}$. This means

$$\mathtt{x}\!-\!\!\boxed{\mathbf{F}} \;\equiv\; \mathtt{x}\!-\!\!\boxed{\mathbf{F}'} \tag{77}$$

by the definition of equivalence given in Sec. 6.2. This is true for any $\mathbf{F}$ and $\mathbf{F}'$. In other words the deterministic pointer result, for a given pointer type, is unique (up to equivalence). This is a powerful result (which first appears in Chiribella et al. [2010, 2011] for general system types - and which we referred to as the Pavia causality condition in Sec. 2.7). We will reserve the symbol $\mathbf{R}$ for the deterministic results so we write

$$\mathtt{x}\!-\!\!\boxed{\mathbf{R}} \tag{78}$$

for deterministic results (we will also use $\mathbf{R}$ for deterministic preparations - see below). Strictly, deterministic results can be different even though they are equivalent (since equivalence is not the same concept as equality). However, for the most part, we are only interested in the equivalence class. Note that, for simplicity, we will use the symbol $\mathbf{R}$ regardless of the pointer type.

Now we can apply time symmetry. By our definition of time symmetry in Sec. 6.4, we have circuits

$$\boxed{\underset{\sim}{\mathbf{F}}}\!-\!\!\!-\!\!\boxed{\underset{\sim}{\mathsf{E}}} \qquad \text{and} \qquad \boxed{\underset{\sim}{\mathbf{F}'}}\!-\!\!\!-\!\!\boxed{\underset{\sim}{\mathsf{E}}} \tag{79}$$

satisfying

$$\text{prob}\left(\boxed{\underset{\sim}{\mathbf{F}}}\!-\!\!\!-\!\!\boxed{\underset{\sim}{\mathsf{E}}}\right) \equiv \text{prob}\left(\boxed{\underset{\sim}{\mathbf{F}'}}\!-\!\!\!-\!\!\boxed{\underset{\sim}{\mathsf{E}}}\right) \tag{80}$$

This must be true for any pointer result $\underset{\sim}{\mathsf{E}}$ so all deterministic pointer preparations, $\underset{\sim}{\mathsf{F}}$, (for a given pointer type) are equivalent and we can represent them by the same symbol. We will write deterministic preparations as

$$\boxed{\mathsf{R}}\!-\!\!\!-\,\mathtt{x} \tag{81}$$

(strictly we should write $\underset{\sim}{\mathsf{R}}$ but it is always clear whether we have a preparation or result).

Deterministic pointer results and preparations play a big role in this work. We will frequently refer to them as **R** *boxes*.

Note that we have

$$\text{prob}\left(\boxed{\mathsf{R}}\!\overset{\mathtt{x}}{-\!\!\!-\!\!\!-}\!\boxed{\mathsf{R}}\right) = 1 \tag{82}$$

because this is a deterministic circuit.

It follows immediately from the above properties that

$$\boxed{\mathsf{R}}\!-\!\!\!-\!\!\!-\,\mathtt{xy} \quad \equiv \quad \begin{array}{l} \boxed{\mathsf{R}}\!-\!\!\!-\!\!\!-\,\mathtt{y} \\ \boxed{\mathsf{R}}\!-\!\!\!-\!\!\!-\,\mathtt{x} \end{array} \qquad\qquad \mathtt{xy}\,-\!\!\!-\!\!\!-\!\boxed{\mathsf{R}} \quad \equiv \quad \begin{array}{r} \mathtt{y}\,-\!\!\!-\!\!\!-\!\boxed{\mathsf{R}} \\ \mathtt{x}\,-\!\!\!-\!\!\!-\!\boxed{\mathsf{R}} \end{array} \tag{83}$$

To prove this note that we are free to form the network

$$\begin{array}{r} \mathtt{y}\,-\!\!\!-\!\!\!-\!\boxed{\mathsf{R}} \\ \mathtt{x}\,-\!\!\!-\!\!\!-\!\boxed{\mathsf{R}} \end{array} \tag{84}$$

with two deterministic pointer results. Further, this network must be deterministic since each of the component operations are (see remarks in Sec. 5.4). Thus, it is equivalent to any other deterministic result for the composite pointer type `xy`. This proves the right equivalence in (83). The left equivalence follows similarly. This clearly generalises to more than two components for the pointer system.

## 7.3 The flatness assumption

Now introduce the *flatness assumption* which imposes particular properties on these deterministic preparations and results. The flatness assumption is true in Quantum Theory. The flatness assumption is not necessary to set up the formal causality and positivity properties we are seeking in remainder of this section. However, this assumption is important if we are to give sensible interpretation of deterministic pointer preparations and results as we will explain below. Furthermore, the flatness assumption is essential in proving the theorem that time symmetric, time forward, and time backward formulations of Operational Probabilistic Theory are equivalent (see Sec. 11). This theorem is central to the project upon which this book is based.

Now we state the assumption.

**The flatness assumption** states that

$$\text{prob}\left(\boxed{\mathsf{R}}\overset{\mathtt{x}}{\text{——}}\boxed{x}\overset{\mathtt{x}}{\text{——}}\boxed{\mathsf{R}}\right) = \frac{1}{N_{\mathtt{x}}} \tag{85}$$

for all pointer types.

This assumption follows if we assume permutation invariance, i.e. that

$$\text{prob}\left(\boxed{\mathsf{R}}\overset{\mathtt{x}}{\text{——}}\boxed{\pi(x)}\overset{\mathtt{x}}{\text{——}}\boxed{\mathsf{R}}\right) \tag{86}$$

is independent of $\pi$ (where $\pi$ permutes the $x$). To see this we need to use the complete circuit sets property in Sec. 6.1. This permutation invariance property puts each readout on an equal footing. It would be interesting to consider cases where permutation invariance property does not hold. Indeed, were $x$ a continuous rather than discrete quantity (as we assume here) then we would have to work harder to make sense of permutation invariance.

Importantly, the flatness assumption is consistent with the factorisation property in (83). This is clear since, using the flatness assumption, we obtain

$$\begin{aligned}
\frac{1}{N_{\mathtt{xy}}} &= \text{prob}\left(\boxed{\mathsf{R}}\overset{\mathtt{xy}}{\text{——}}\boxed{xy}\overset{\mathtt{xy}}{\text{——}}\boxed{\mathsf{R}}\right) \\
&= \text{prob}\left(\boxed{\mathsf{R}}\overset{\mathtt{x}}{\text{——}}\boxed{x}\overset{\mathtt{x}}{\text{——}}\boxed{\mathsf{R}}\right)\text{prob}\left(\boxed{\mathsf{R}}\overset{\mathtt{y}}{\text{——}}\boxed{y}\overset{\mathtt{y}}{\text{——}}\boxed{\mathsf{R}}\right) \\
&= \frac{1}{N_{\mathtt{x}}N_{\mathtt{y}}}
\end{aligned}$$

using the fact that probabilities factorise over disjoint circuits. Since $N_{\mathtt{xy}} = N_{\mathtt{x}}N_{\mathtt{y}}$, we have consistency. We could explore non-flat choices for $\mathsf{R}$. In so doing, we would have to be sure that factorisation is respected.

Now we turn to the interpretation of the flatness assumption. Here we will provide some intuitive ideas. We will cash out this intuition properly in Sec. 9.9 and Sec. 10.2.1 using duotensors. We can do this by "thinking in a time forward" or "thinking in a time backward fashion". If we think in a time forward fashion, then it is natural to interpret the deterministic result as taking the marginal over the probabilities associated with the different $x$. To represent this intuition we can

$$\text{associate} \qquad \mathtt{x}\text{——}\boxed{\mathsf{R}} \qquad \text{with} \qquad \begin{pmatrix} 1 \\ 1 \\ \vdots \\ 1 \end{pmatrix}^{\mathsf{T}} \qquad \text{"forward thinking"} \tag{87}$$

because the vector on the right simply implements a sum as demanded by the notion of marginalisation (note the $\mathsf{T}$ superscript on the right takes the transpose so we have a row vector which is natural for results as they have a wire coming in from the left). By this intuition, the deterministic result is associated with a flat vector (meaning that all entries are equal). Then, if we think in the time backward fashion, then we expect we can

$$\text{associate} \quad \boxed{\mathsf{R}}\!\!-\!\!\!-\!\!\!-\mathtt{x} \quad \text{with} \quad \begin{pmatrix} 1 \\ 1 \\ \vdots \\ 1 \end{pmatrix} \qquad \text{“backward thinking”} \tag{88}$$

by time symmetry. So the deterministic preparation is also associated with a flat vector. Now, the forward and backward thinking approaches deal with normalisation differently (again, note that this is dealt with properly in Sec. 9.9 and Sec. 10.2.1). However, it is clear at an intuitive level that, if we think forward in time we should normalise the vector associated with the preparation (so that the probabilities sum to 1). Then we

$$\text{associate} \quad \boxed{\mathsf{R}}\!\!-\!\!\!-\!\!\!-\mathtt{x} \quad \text{with} \quad \frac{1}{N_{\mathtt{x}}}\begin{pmatrix} 1 \\ 1 \\ \vdots \\ 1 \end{pmatrix} \qquad \text{“forward thinking”} \tag{89}$$

Now if we “plug” (89) and (87) for the forward way of thinking into the left hand side of (85) and interpret the readout box as projecting onto the given value of $x$ then we obtain the right hand side (i.e. take the scalar product of these vectors after the mentioned projection). The intuition here is that we prepare a flat distribution, project, then take the marginal. On the other hand, if we think backward, then we should normalise the vector associated with the result. Then we

$$\text{associate} \quad \mathtt{x}-\!\!\!-\!\!\!-\!\!\boxed{\mathsf{R}} \quad \text{with} \quad \frac{1}{N_{\mathtt{x}}}\begin{pmatrix} 1 \\ 1 \\ \vdots \\ 1 \end{pmatrix}^{\mathsf{T}} \qquad \text{“backward thinking”} \tag{90}$$

Plugging (90) and (88) (for the backward way of thinking) into the left hand side of (85) gives, again, the right hand side. Now the intuition is the time reverse - we have a flat distribution “coming from the future” and then, in the past, we take the marginal. If we are simply interested in calculating the probability, we do not care whether we employed forward thinking or backward thinking. There is a certain sense in which the normalisation is a global thing which we can choose to allocate to the preparation (in accord with our usual time forward intuition) or the result (in accord with backward thinking). In fact, there is a kind of gauge freedom here. This will crop up when we formulate Quantum Theory (see Sec. 24).

Incidentally, the symbol **R** has been chosen as it is the first letter of the word “Random” since, under the flatness assumption, deterministic preparations and results correspond to the completely random distribution.

## 7.4 Deterministic system preparations and results

A *system preparation* is an operation having only outputs.

$$\begin{array}{c} \mathsf{a} \\ | \\ \boxed{\mathsf{A}} \end{array} \tag{91}$$

The system type, $\mathsf{a}$, can be composite.

A *system result* is an operation having only inputs

$$\begin{array}{c} \boxed{\mathsf{B}} \\ | \\ \mathsf{a} \end{array} \tag{92}$$

The system type, $\mathsf{a}$, can be composite.

We can repeat the analysis given for pointer preparations and results in this case to show that, if we assume causality and time symmetry, there is (up to equivalence) a unique deterministic preparation and a unique deterministic result. Thus, if a future choice of deterministic result does not influence probabilities in the past then we must have

$$\text{prob}\left(\begin{array}{c} \boxed{\mathbf{B}} \\ \mathsf{a}\,| \\ \boxed{\mathsf{A}} \end{array}\right) = \text{prob}\left(\begin{array}{c} \boxed{\mathbf{B}'} \\ \mathsf{a}\,| \\ \boxed{\mathsf{A}} \end{array}\right) \tag{93}$$

for all $\mathsf{A}$ which means that $\mathbf{B}$ and $\mathbf{B}'$ must be equivalent for any such deterministic result. Thus, up to equivalence classes, there is a unique deterministic result, for any given system type, which we write as

$$\begin{array}{c} \boxed{\mathbf{I}} \\ | \\ \mathsf{a} \end{array} \tag{94}$$

By time symmetry this means that we have

$$\text{prob}\left(\begin{array}{c} \boxed{\underset{\sim}{\mathsf{A}}} \\ \mathsf{a}\,| \\ \boxed{\underset{\sim}{\mathbf{B}}} \end{array}\right) = \text{prob}\left(\begin{array}{c} \boxed{\underset{\sim}{\mathsf{A}}} \\ \mathsf{a}\,| \\ \boxed{\underset{\sim}{\mathbf{B}'}} \end{array}\right) \tag{95}$$

for all results, $\underset{\sim}{\mathsf{A}}$. Consequently there is a unique deterministic system preparation, for any given system type, which we will write the unique (up to equivalence class) deterministic preparation as

$$\begin{array}{c} \mathsf{a} \\ | \\ \boxed{\mathsf{I}} \end{array} \tag{96}$$

Strictly, we should write this as $\underset{\sim}{\mathsf{I}}$, but for simplicity we will simply write $\mathsf{I}$ for both deterministic system preparations and results. We use the symbol $\mathsf{I}$ as it corresponds to the first letter of the word "Ignore". We will call the deterministic preparation in (96) an *ignore preparation* and the deterministic result in (94) an *ignore result*. . We will sometimes call these objects $\mathsf{I}$ boxes.

Ignore preparations and results have the important property that

$$\begin{array}{c} \boxed{\mathsf{I}} \\ \mathsf{a}\,| \\ \boxed{\mathsf{I}} \end{array} \equiv 1 \tag{97}$$

since this is a deterministic circuit (recall that circuits are equivalent to their probabilities - see (62)).

We can also prove the factorisation properties

$$\begin{array}{c} \boxed{\mathsf{I}} \\ | \\ \mathsf{ab} \end{array} \equiv \begin{array}{c} \boxed{\mathsf{I}} \\ | \\ \mathsf{a} \end{array} \begin{array}{c} \boxed{\mathsf{I}} \\ | \\ \mathsf{b} \end{array} \qquad\qquad \begin{array}{c} \mathsf{ab} \\ | \\ \boxed{\mathsf{I}} \end{array} \equiv \begin{array}{c} \mathsf{a} \\ | \\ \boxed{\mathsf{I}} \end{array} \begin{array}{c} \mathsf{b} \\ | \\ \boxed{\mathsf{I}} \end{array} \tag{98}$$

This follows immediately from uniqueness of the ignore preparation/result (since the factorised form on the right of these equivalences is one way to implement the corresponding ignore preparation/result).

The uniqueness of the deterministic preparation (up to its equivalence class) is essentially the Pavia causality condition (see Chiribella et al. [2010, 2011]) that we discussed earlier and the argument above is adapted from that work.

## 7.5 Comments on use of causality in a time symmetric setting

In the foregoing analysis we invoked the *basic forward causality principle* that choices in the future should not affect probabilities in the past to prove that we have a unique deterministic result (for both pointers and systems) for a given type. To fully understand how time symmetry is working here we will see that we have to be a bit more careful in our statement of this principle.

Then we used time symmetry to obtain that, similarly, we have a unique deterministic preparation for a given type. In fact, looking at (80) and (95), we can see that we are using the *basic backward causality principle* which is that choices in the past should not affect probabilities in the future. In a time symmetric setting, we are forced to adopt the basic backward causality principle if we have the basic forward causality principle. However, the basic backward causality principle is very counterintuitive (unlike the forward one). We know from our experience that choices in the past *do* affect probabilities in the future.

The key reason for this is that, typically, we are conditioning on the past (but we do not typically condition on the future). There are deep issues here in accounting for why it should be that we prefer to condition on the past. We will discuss this issue in Sec. 13.8. However, it is worth saying a few words at this point to elucidate that the time symmetric framework is consistent with our usual point of view.

Generally, operational theories (such as quantum theory) have been developed in a time forward temporal frame even though, one might argue, the time symmetric frame is more fundamental. We will discuss time forward operational theories in Sec. 11. We will show that it is possible to transform between a time forward, a time symmetric and, indeed, a time backward point of view. Importantly, these different formulations are equivalent.

For the moment, let us explore how we can use conditioning to send information to the future, or to the past. We introduce the symbols

$$\langle x|\!\!-\!\!\mathtt{x} \qquad\qquad \mathtt{x}\!\!-\!\!|x\rangle \tag{99}$$

to indicate preselecting on income $x$ and postselecting on outcome $x$ respectively (we will discuss these symbols more in Sec. 11). These are not allowed preparations and results in the time symmetric theory. However, preselecting on $x$ is an allowed operation in the time forward formulation of operational theories. Similarly, postselecting on $x$ is an allowed operation in the time backward formulation.

Now consider

$$\langle x|\!\!-\!\!\mathtt{x}\!\!-\!\!\left[\, \boxed{x'}\!\!-\!\!\mathtt{x}\!\!-\!\!\boxed{\mathbf{R}} \,\right] \tag{100}$$

We can regard the preselection of $x$ as being in the past whilst the readout box followed by the **R** box (both inside the dashed box) are taken to be in the future. In this scenario, we are able to signal from the past to the future since $x'$ must be equal to $x$. Had we chosen a different $x$ we would see a different readout at $x'$. Better statements for the causality principles above are the following

> **Basic forward causality principle.** We cannot signal from the future to the past unless we condition on outcomes in the future.
>
> **Basic backward causality principle.** We cannot signal from the past to the future unless we condition on incomes in the past.

The "unless" part of the second principle is illustrated in (100). We can consider the time reverse of (100)

$$\left[\, \boxed{\mathbf{R}}\!\!-\!\!\mathtt{x}\!\!-\!\!\boxed{x'} \,\right]\!\!-\!\!\mathtt{x}\!\!-\!\!|x\rangle \tag{101}$$

We can argue that the readout in the $x'$ box depends on the choice of $x$ in the postselect box. This illustrates the "unless" part of the first principle (we can signal backward in time if we preselect on the future).

These principles rule out signalling when we do not condition on later outcomes (for the forward principle) or earlier incomes (for the backward principle). If we do allow conditioning on later outcomes (or earlier incomes) then we can signal to the past (or future) by varying a setting as we will now see. Recall that in Sec. 5.1 we took the setting to be implicit in the symbol, **B**, associated with the operation. Thus, a choice of setting is represented by a choice of **B** versus **B**$'$ say. Such choices are not capable of facilitating signalling without conditioning on an income (for forward in time signalling) or an outcome (for backward in time signalling). To illustrate this, note that we must have

$$\begin{array}{c} \boxed{\mathsf{Y}}\overset{y}{\text{——}} \\ |\,y \\ \boxed{\mathsf{B}} \\ |\,x \\ \boxed{\mathsf{R}}\overset{x}{\text{——}}\boxed{\mathsf{X}} \end{array} \equiv \begin{array}{c} \boxed{\mathsf{Y}}\overset{y}{\text{——}} \\ |\,y \\ \boxed{\mathsf{B}'} \\ |\,x \\ \boxed{\mathsf{R}}\overset{x}{\text{——}}\boxed{\mathsf{X}} \end{array} \equiv \boxed{\mathsf{R}}\overset{y}{\text{——}} \tag{102}$$

because the deterministic pointer preparation is unique. This means that the choice of **B** versus **B**$'$ does not signal to the future when we do not preselect on the past. Similarly, we must have

$$\begin{array}{c} \boxed{\mathsf{Y}}\overset{y}{\text{——}}\boxed{\mathsf{R}} \\ |\,x \\ \boxed{\mathsf{B}} \\ |\,x \\ \overset{x}{\text{——}}\boxed{\mathsf{X}} \end{array} \equiv \begin{array}{c} \boxed{\mathsf{Y}}\overset{y}{\text{——}}\boxed{\mathsf{R}} \\ |\,x \\ \boxed{\mathsf{B}'} \\ |\,x \\ \overset{x}{\text{——}}\boxed{\mathsf{X}} \end{array} \equiv \overset{x}{\text{——}}\boxed{\mathsf{R}} \tag{103}$$

because the deterministic result is unique. Thus we cannot signal to the past if we do not preselect on the future. On the other hand, we can have

$$\begin{array}{c} \boxed{\mathsf{Y}}\overset{y}{\text{——}} \\ |\,y \\ \boxed{\mathsf{B}} \\ |\,x \\ \langle x\rangle\overset{x}{\text{——}}\boxed{\mathsf{X}} \end{array} \not\equiv \begin{array}{c} \boxed{\mathsf{Y}}\overset{y}{\text{——}} \\ |\,y \\ \boxed{\mathsf{B}'} \\ |\,x \\ \langle x\rangle\overset{x}{\text{——}}\boxed{\mathsf{X}} \end{array} \tag{104}$$

(We can have this in the sense that it is not forbidden by our causality principles. Further, Quantum Theory does, indeed, admit such examples.) This means that, if we preselect on a past income, we can signal from the past to the future through our choice of setting **B** or **B′**. Similarly, we can have

(105)

which means that, if we postselect on a future outcome, we can signal from the future to the past through our choice of setting **B** or **B′**.

The foregoing discussion makes it clear that the basic forward and backward causality principles are consistent with our usual intuitions. What we have not done, however, is account for why, as we experience the world, preselection is very natural whilst postselection is not. We will provide a more detailed discussion of preselection and postselection in Sec. 13.

## 7.6 Double causality theorem

Now we have enough preliminaries in place to state and prove the following

**Double causality theorem.** Any deterministic operation

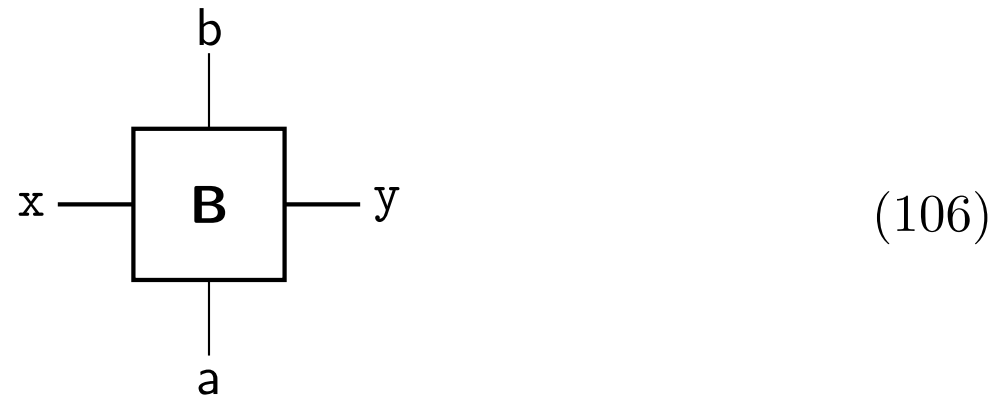

(106)

satisfies

***Forward causality***

(107)

if the basic forward causality principle holds, and

***Backward causality***

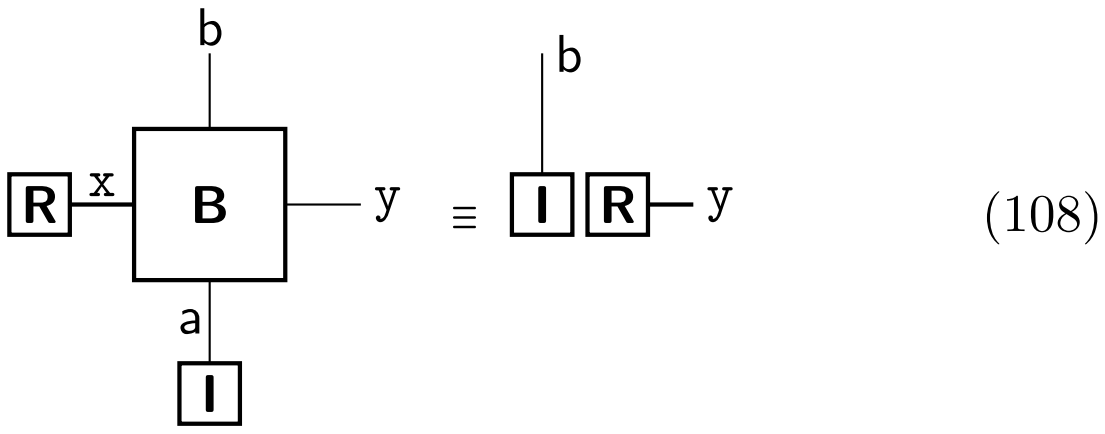

if the basic backward causality principle holds.

We will prove forward causality. The proof for backward causality follows similarly. First consider the circuits

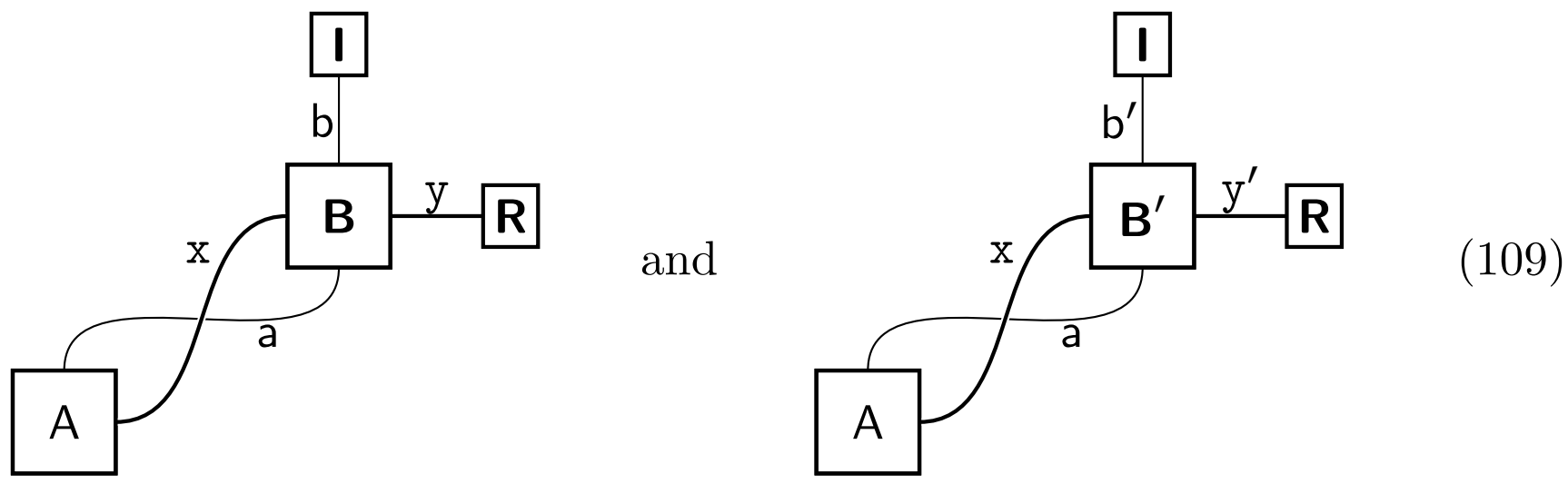

where we regard the operation A as being in the past and the rest of each circuit as being in the future. The the basic forward causality principle tells us that whether we choose **B** or **B′** should not change the probability in the past so we must have

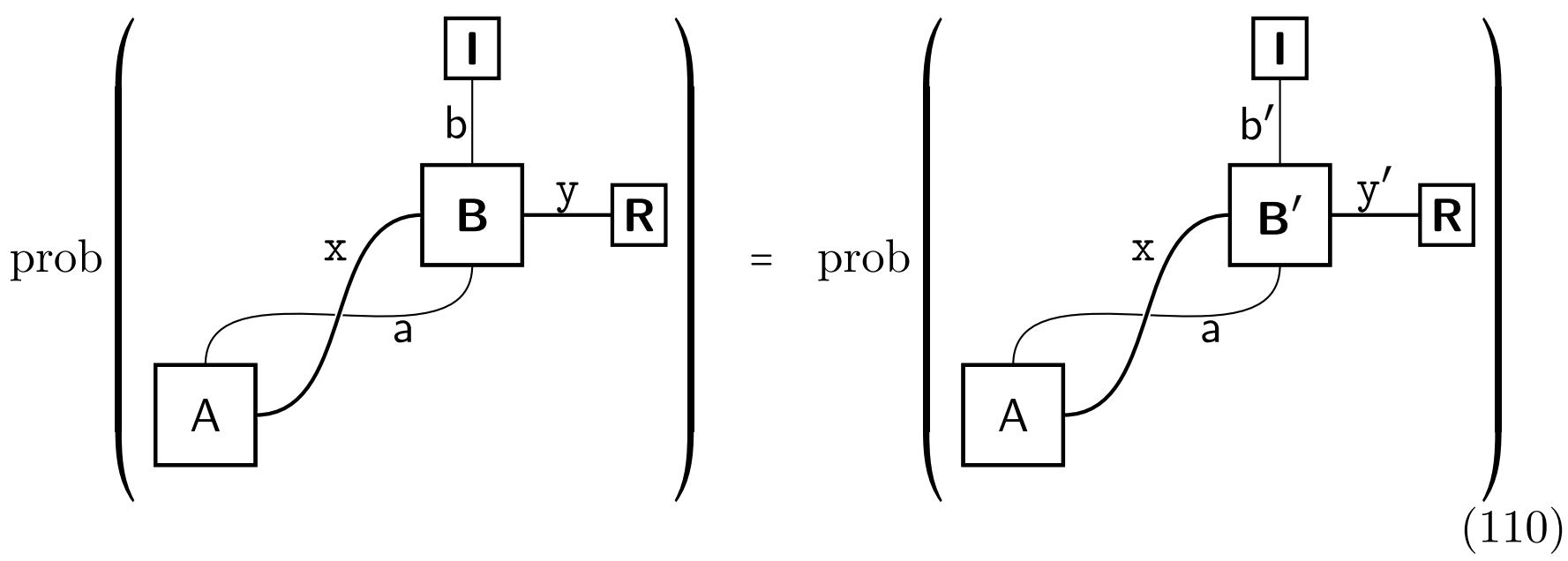

for all A. This means that

I, b, x, B, y, R, a ≡ I, b′, x, B′, y′, R, a (111)

for all $\mathsf{B}$ and $\mathsf{B}'$ (by the definition of equivalence in Sec. 6.2). One particular choice for $\mathsf{B}'$ is where we put $\mathsf{b}' = \mathsf{a}$ and $\mathsf{y}' = \mathtt{x}$ with

x

a

(112)

In this case (111) gives us directly the forward causality condition in (107). The backward causality condition in (108) follows similarly.

Interestingly, many properties follow from double causality. Consider Table 1. It shows many special cases of double causality where we make some of the systems and or pointer types null. Row 1 shows the full double causality condition. The right entry in row 4 shows that a deterministic operation having only outcomes/outcomes can be written in terms of an $\mathsf{R}$ preparation and an $\mathsf{I}$ preparation. The left entry of row 4 shows something similar for a deterministic operation having only incomes/inputs. Row 8 shows that deterministic system results and preparations are unique. Row 9 shows that deterministic pointer results and preparations are unique.

We can also regard the following equation as a special case of double causality

$$\boxed{\mathsf{B}} \equiv \boxed{1} \qquad (113)$$

where the box with a 1 in it simply represents the number 1 (we know this to be true since deterministic circuits have probability equal to 1 and circuits are equivalent to their probabilities according to (62)). This works because the 1 box can be inserted on the right hand side of the forward or backward causality condition. Then, if we treat all the wires as null we get (113).

The forward causality condition is, essentially, equivalent to a condition found in the work of Chiribella et al. [2010, 2011]. In the Quantum setting, this forward causality condition is equivalent to the property that deterministic operations are trace preserving.

## 7.7 Simple double causality composition theorem

As explained in Sec. 5.3, some networks (like network (i) in (44)) have complex causal structure (since outputs/outcomes can be fed into inputs/incomes whilst maintaining a DAG structure) whilst others (like network (iii) in (46)) have simple causal structure. We can, however, restrict ourselves to using any network, even if it has complex causal structure, in *simple form*. This means that when we plug the network into larger networks, we require that outputs/outcomes on the given network are after incomes/inputs. We picture this by imagining that we pull all inputs down, all outcomes up, all incomes to the left, and all

| | | |
|---|---|---|
| 1 | I b x B y R a ≡ x R I a | b R x B y a I ≡ b I R y |
| 2 | x B y R a ≡ x R I a | b R x B y ≡ b I R y |
| 3 | I b x B y R ≡ x R | R x B y a I ≡ R y |
| 4 | x B a ≡ x R I a | b B y ≡ b I R y |
| 5 | I b B y R a ≡ I a | b R x B a I ≡ b I |
| 6 | I b B a ≡ I a | b B a I ≡ b I |
| 7 | x B y R ≡ x R | R x B y ≡ R y |
| 8 | B a ≡ I a | b B ≡ b I |
| 9 | x B ≡ x R | B y ≡ R y |

Table 1: This tabulates many of the special cases of forward causality (left column) and backward causality (right column).

outcomes to the right. Network (i) in simple form looks like this

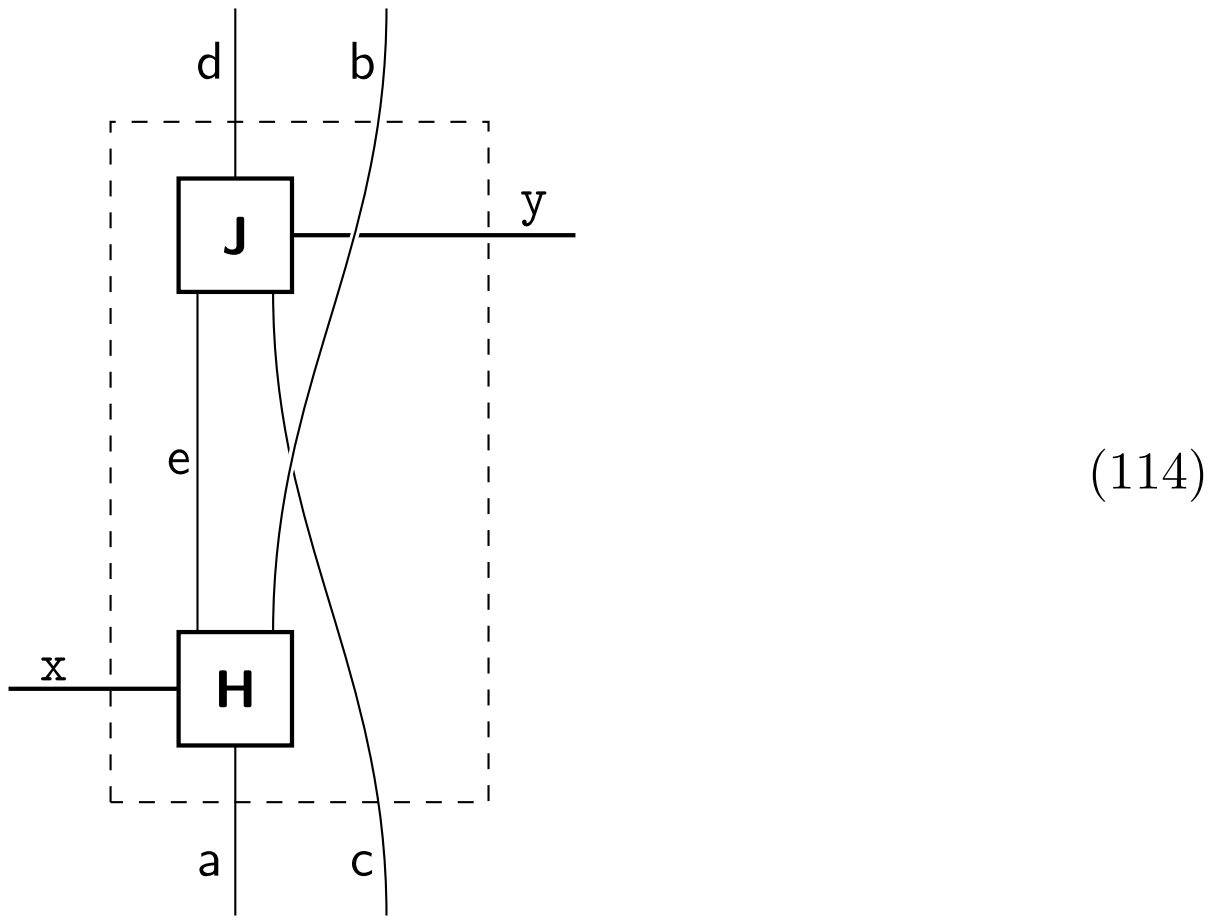

(114)

If we restrict to using this network in this way then the object inside the dashed box can be regarded as an operation.

In this section we will ask whether a network used in simple form satisfies the double causality conditions if the operations that comprise it do. Since we are restricting the network to simple form for this analysis, we call the causality conditions *simple double causality* - this being comprised of *simple forward causality* and *simple backward causality*. If we allow use of the network more generally (not in simple form) then there are more causality conditions. Discussion of the full set of causality conditions for networks will be subsumed under the discussion of complex operations in Part IV later in the book. For now we focus on the simple double causality conditions.

We can now prove the following composition theorem

> **Simple double causality composition theorem** Any deterministic network formed by wiring together two or more deterministic operations will (i) satisfy simple forward causality if each of the component operations satisfy forward causality and (ii) satisfy simple backward causality if each of the component operations satisfy backward causality.

First consider wiring together two deterministic operations, **A** and **B** as follows

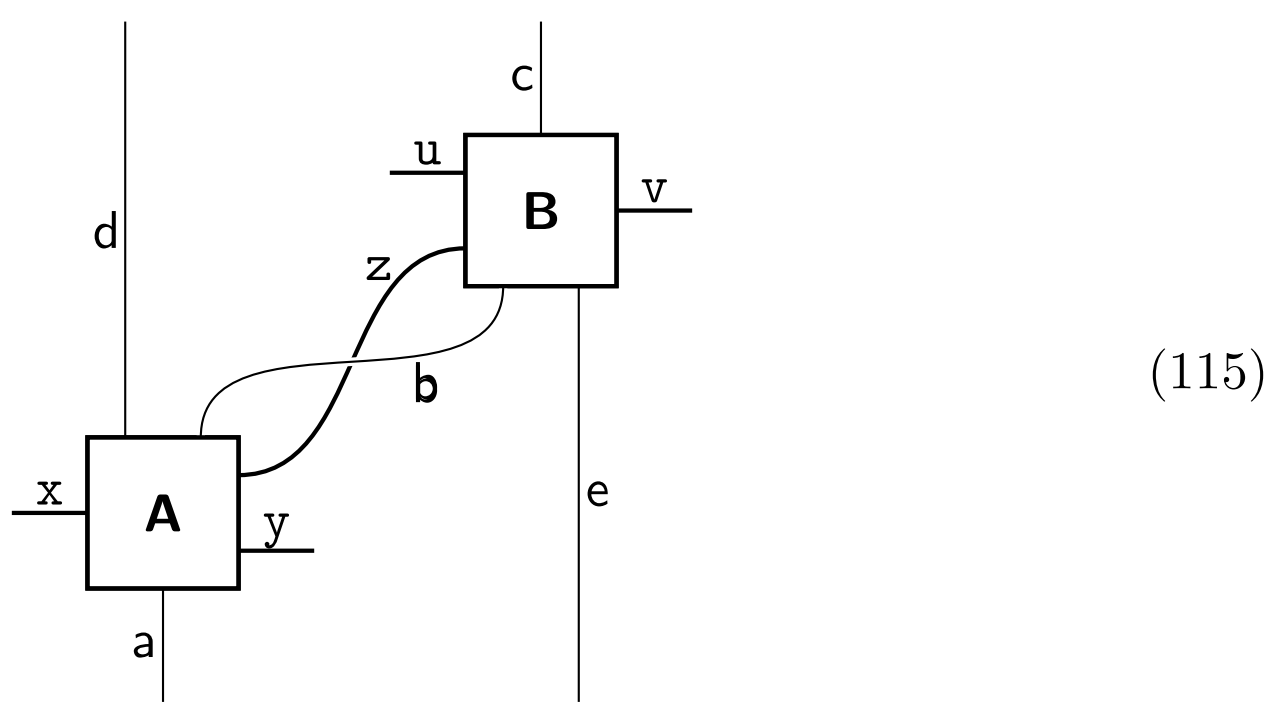

(115)

This is the most general way of wiring two deterministic operations together to form network since either one of the operations must be first (we choose **A**) or they are in parallel. We can delete any of the wires to deal with special cases that arise (note that the "in parallel" case is obtained by deleting wires b and z). It is easy to see that this satisfies simple double causality if **A** and **B** each satisfy double causality. Simple forward causality for this network is the statement that

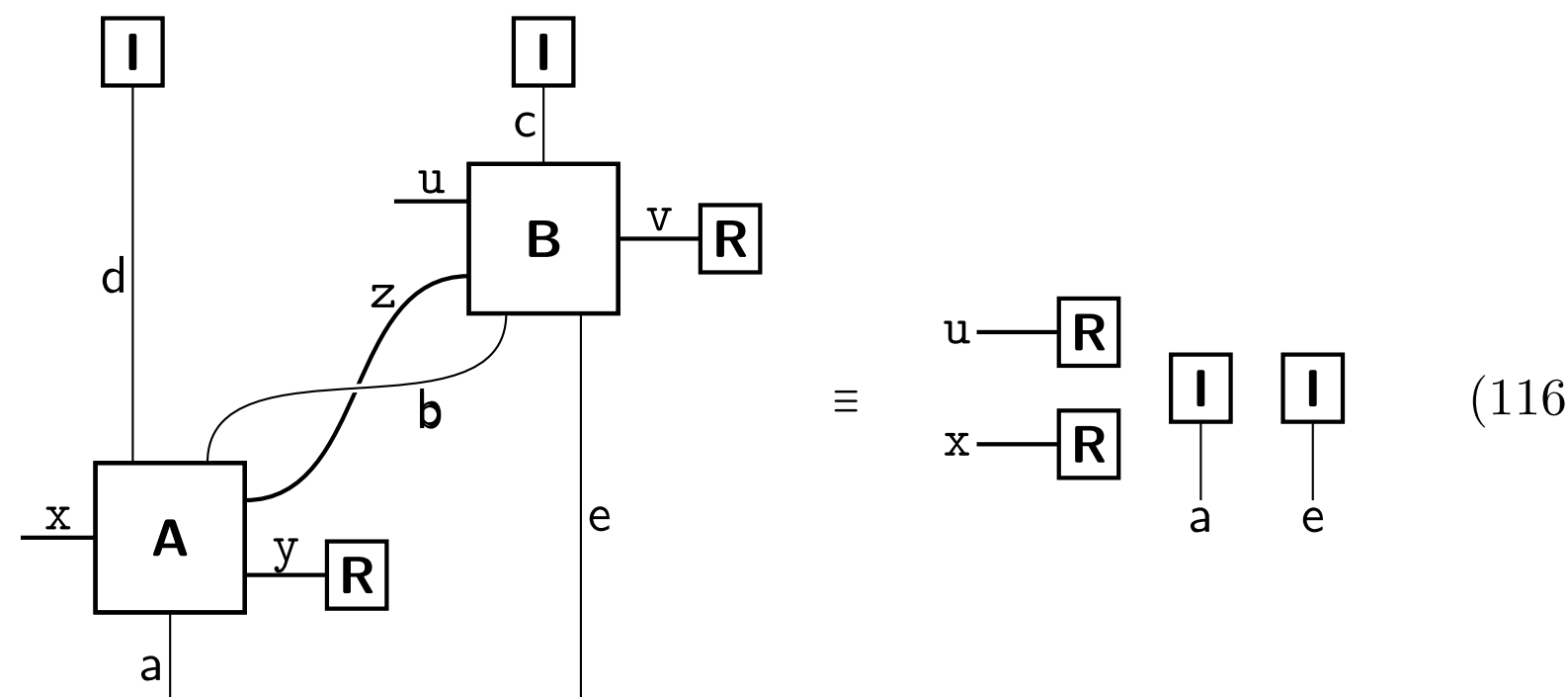

(116)

This follows by applying forward causality first to **B** then to **A** and using the fact that deterministic pointer and system results factorise. This proves (i) for two operations. Now consider wiring together more than two operations. First note that we can impose a causal order on the operations in the network (this is non-unique). Let the $n$-network consist of the first $n$ operations in this causal order. Let there be $N$ operations in the network. We place **I** boxes on all open outputs and **R** boxes on all open outcomes. Now we can apply forward causality to the $N$th operation. This will remove the $N$th operation leaving a $(N-1)$-network whose outputs and outcomes are closed by **I** and **R** boxes respectively. Next we apply forward causality to the $N-1$ operation and so on. In the end, we will be left with only **R** pointer result boxes and **I** system result boxes. This proves (i). Point (ii) follows by following a similar procedure in the reverse time direction.

If this theorem is applied to a deterministic network that is, in fact, a circuit then we simply prove that it has probability equal to 1 (using (113) regarded as a special case of either forward or backward causality).

It is worth noting the following consequence of the above theorem

> **Double causality composition theorem for simple causal structure.** Any deterministic network having simple causal structure that is formed by wiring together two or more deterministic operations will (i) satisfy forward causality if each of the component operations satisfy forward causality and (ii) satisfy backward causality if each of the component operations satisfy backward causality.

This is clear since, if a network has simple causal structure then the simple double causality conditions are just the double causality conditions.

The simple double causality theorem above is rather limited since it only pertains to using networks in simple form. In Part IV, where we introduce complex operations, we will define a double causality conditions that pertain to situations where there is complex causal structure and we will prove that networks satisfy this that double causality is theorem.

## 7.8 Mixed, extremal, homogeneous, heterogeneous, and pure cases

The fact that we have a unique (up to equivalence) deterministic system preparation (result) means that the space of allowed system preparations (results) must have a unique "top" element. The space illustrated in Fig. 1 is a possible candidate for such a space. Indeed, the space of *both* preparations and results for quantum theory in the time symmetric temporal frame has a double cone structure like this. This contrasts with the case of time forward operational probabilistic theories (such as the usual formulation of Quantum Theory) where the space of preparations is a cone with a flat top (representing the deterministic states) whilst the space of results is a double cone like in Fig. 1 whilst, in time backward operational probabilistic theories this is reversed.

When we have a double cone structure like this, the usual taxonomy of pure and mixed do not cover all the interesting cases. We define the following types of system preparation (result).

**Mixed.** A system preparation (result) is mixed if it can be written as being equivalent to a convex sum of distinct preparations (results).

**Extremal.** A system preparation (result) is extremal if it is not mixed.

**Heterogeneous.** A system preparation (result) is heterogeneous if can be written as being equivalent to a sum of distinct non-parallel preparations (results).

**Homogeneous (or pure-parallel).** A Preparation (result) is homogeneous (or pure-parallel) if it is not heterogeneous.

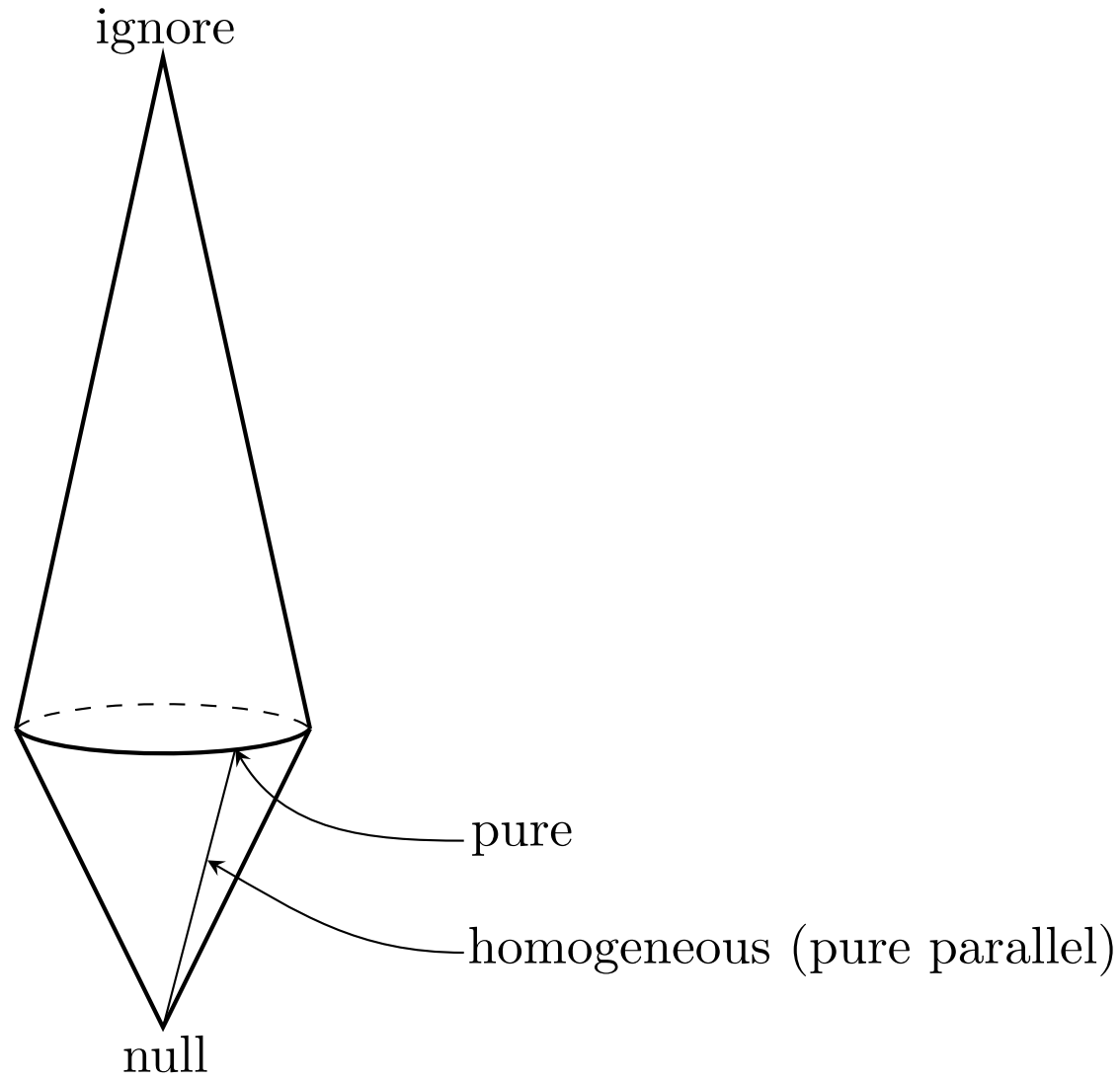

Figure 1: Different types of preparation (or result). The notions of pure and pure parallel (or homogeneous) are illustrated.

**Pure.** A preparation (result) is pure if it is homogeneous and extremal.

Figure 1 illustrates these different types of preparation (result). The space of preparations (results) may look different from this. However, in the time symmetric temporal frame both the set of preparations and the set of results have this double cone appearance. Note that the top (inverted) cone will, in general, be taller than the bottom cone in such spaces. In this figure the pure and homogeneous cases are identified. The extremal cases include the pure, the null, and the "ignore" cases. All other cases are mixed (this includes all cases in the interior, and those on the surface of the double cone except for the pure, null, and ignore). The heterogeneous cases are all the cases which are not homogeneous. In the figure this is all the cases in the interior plus the cases that are on the surface of the top (inverted) cone except for the pure states (this includes the ignore case).

The full space is given by the convex hull of the extremal cases. These are the null, ignore, and pure cases. Thus, if we know the null preparation (result), the ignore preparation (result), and the pure preparations (results) then we can deduce the full space of preparations (results). The null case is just the "zero" preparation (result) - i.e. one that causes any circuit it is in to have probability zero. It is naturally represented by the origin in our space. The real work is in identifying the ignore and pure preparations (results).

We note the following properties which follow from the basic geometry outlined above.

> **Double purity theorem.** The following statements hold. *Forward purity.* Any system preparation is equivalent to a positive weighted sum of pure preparations. *Backward purity.* Any result is equivalent to a positive weighted sum of pure results.

Note that we do not assume these positive weights sum to 1 (and, indeed, this is not true in general). This is clear since any preparation (result) is either (i) heterogeneous in which case it is equivalent to a sum of pure-parallel preparations (results) and therefore equivalent to a positive weighted sum of pure preparations (results) or homogeneous in which case it is proportional to a pure preparation (result).

## 7.9 Pure pointer preparations and results

We take the following to be a defining property of pointer systems

> **Pointer classicality.** The set
>
> $$\left\{ \boxed{\mathsf{R}} \overset{\mathtt{x}}{\text{---}} \boxed{x} \overset{\mathtt{x}}{\text{---}} \;:\text{for all } x \right\} \tag{117}$$
>
> is a complete set of pure preparations. The set
>
> $$\left\{ \overset{\mathtt{x}}{\text{---}} \boxed{x} \overset{\mathtt{x}}{\text{---}} \boxed{\mathsf{R}} \;:\text{for all } x \right\} \tag{118}$$
>
> is a complete set of pure results.

This imposes the necessary classicality property on pointers (since we do take them to be classical - see Sec. 5.1). Incidentally, recall that complete set (as defined in Sec. 5.3) is one whose readouts are mutually exclusive and cover all possibilities).

It follows from the double purity theorem (from Sec. 7.8) and the pointer classicality assumption above that we can write any pointer preparation as

$$\boxed{\mathsf{E}} \text{---}\mathtt{x} \quad \equiv \quad \sum_{x=1}^{N_{\mathtt{x}}} \boxed{\alpha^x} \; \boxed{\mathsf{R}} \overset{\mathtt{x}}{\text{---}} \boxed{x} \overset{\mathtt{x}}{\text{---}} \qquad 0 \leq \alpha_x \leq 1 \quad \forall\; x \tag{119}$$

If $\mathtt{x}$ is a composite system - say $\mathtt{x} = \mathtt{yz}$ - then we have

$$\boxed{\mathsf{E}} \begin{matrix} \overset{\mathtt{y}}{\text{---}} \\ \overset{\mathtt{z}}{\text{---}} \end{matrix} \quad = \quad \sum_{yz} \boxed{\alpha^{yz}} \begin{matrix} \boxed{\mathsf{R}} \overset{\mathtt{y}}{\text{---}} \boxed{y} \overset{\mathtt{y}}{\text{---}} \\ \boxed{\mathsf{R}} \overset{\mathtt{z}}{\text{---}} \boxed{z} \overset{\mathtt{z}}{\text{---}} \end{matrix} \qquad 0 \leq \alpha^{yz} \leq 1 \quad \forall\; yz \tag{120}$$

using the factorisation property in (83). Obviously a similar result holds when we have more than two pointer systems.

It follows from the double purity theorem and the pointer classicality assumption any pointer result can be written as

$$\mathtt{x}\text{---}\boxed{\mathsf{F}} \quad = \quad \sum_{x=1}^{N_{\mathtt{x}}} \overset{\mathtt{x}}{\text{---}} \boxed{x} \overset{\mathtt{x}}{\text{---}} \boxed{\mathsf{R}} \; \boxed{\beta_x} \qquad 0 \leq \beta_x \leq 1 \quad \forall\; x \tag{121}$$

Furthermore, if we have a composite system we obtain

$$\begin{array}{c}\texttt{y}\\ \texttt{z}\end{array}\!\boxed{\mathsf{F}} \;=\; \sum_{yz} \begin{array}{c}\texttt{y}\,\boxed{y}\,\texttt{y}\,\boxed{\mathbf{R}}\\ \texttt{z}\,\boxed{z}\,\texttt{z}\,\boxed{\mathbf{R}}\end{array}\boxed{\beta_{yz}} \qquad 0 \le \beta_{yz} \le 1 \quad \forall\ yz \tag{122}$$

using the factorisation property in (83).

## 7.10 Pointer tomographic process locality

The above results concerning composite systems for preparations and results allow us to prove an important theorem.

> **Pointer tomographic process locality theorem.** It follows from pointer classicality that any operation
>
> $$\begin{array}{c}\texttt{y}\\ \vdots\\ \texttt{x}\end{array}\boxed{\mathsf{M}}\begin{array}{c}\texttt{v}\\ \vdots\\ \texttt{u}\end{array} \tag{123}$$
>
> having only incomes and outcomes is fully characterised the probabilities
>
> $$\text{prob}\left(\begin{array}{c}\boxed{\mathbf{R}}\,\texttt{y}\,\boxed{y}\,\texttt{y}\\ \vdots\\ \boxed{\mathbf{R}}\,\texttt{x}\,\boxed{x}\,\texttt{x}\end{array}\boxed{\mathsf{M}}\begin{array}{c}\texttt{v}\,\boxed{v}\,\texttt{v}\,\boxed{\mathbf{R}}\\ \vdots\\ \texttt{u}\,\boxed{u}\,\texttt{u}\,\boxed{\mathbf{R}}\end{array}\right) \tag{124}$$

This means that two operations, M and M′, are inequivalent only if they yield different joint probabilities as calculated by local process tomography using (124). To prove this we start by noting that the most general circuit containing M can be written in the form

$$\boxed{\mathsf{E}}\begin{array}{c}\texttt{z}\\ \texttt{y}\\ \vdots\\ \texttt{x}\end{array}\boxed{\mathsf{M}}\begin{array}{c}\texttt{z}\\ \texttt{v}\\ \vdots\\ \texttt{u}\end{array}\boxed{\mathsf{F}} \tag{125}$$

Pointer classicality allow us to expand E and F as in (120) and (122) giving

$$\sum_{x\ldots y,u\ldots v,z,z'} \boxed{\alpha^{x\ldots yz}} \begin{array}{c}\boxed{\mathbf{R}}\,\texttt{z}\,\boxed{z}\;\texttt{z}\;\boxed{z'}\,\texttt{z}\,\boxed{\mathbf{R}}\\ \boxed{\mathbf{R}}\,\texttt{y}\,\boxed{y}\,\texttt{y}\;\boxed{\mathsf{M}}\;\texttt{v}\,\boxed{v}\,\texttt{v}\,\boxed{\mathbf{R}}\\ \vdots \qquad\qquad \vdots\\ \boxed{\mathbf{R}}\,\texttt{x}\,\boxed{x}\,\texttt{x}\;\phantom{\boxed{\mathsf{M}}}\;\texttt{u}\,\boxed{u}\,\texttt{u}\,\boxed{\mathbf{R}}\end{array} \boxed{\beta_{u\ldots vz'}} \tag{126}$$

using (43) and (85) for z, we obtain

$$\sum_{x\ldots y,u\ldots v,z} \boxed{\alpha^{x\ldots yz}} \;\boxed{\tfrac{1}{N_{\texttt{z}}}}\; \begin{array}{c}\boxed{\mathbf{R}}\,\texttt{y}\,\boxed{y}\,\texttt{y}\\ \vdots\\ \boxed{\mathbf{R}}\,\texttt{x}\,\boxed{x}\,\texttt{x}\end{array}\boxed{\mathsf{M}}\begin{array}{c}\texttt{v}\,\boxed{v}\,\texttt{v}\,\boxed{\mathbf{R}}\\ \vdots\\ \texttt{u}\,\boxed{u}\,\texttt{u}\,\boxed{\mathbf{R}}\end{array} \boxed{\beta_{u\ldots vz}} \tag{127}$$

If we do the same for $\mathsf{M}'$ then we obtain a different probability iff the local process tomographic probabilities (as in (124)) are different. This proves the theorem.

## 7.11 Midcome and control wire identities

Here we provide *midcome* and *control* wire identities that follow from pointer classicality (the reason for these names is clear from the discussion at the beginning of Sec. 7.11.1).

First, we have the we have the $\omega$-*midcome identity*

$$\overset{\mathtt{x}}{—}\boxed{x}\overset{\mathtt{x}}{—} \quad \equiv \quad \overset{\mathtt{x}}{—}\boxed{x}\overset{\mathtt{x}}{—}\boxed{\mathbf{R}}\boxed{\omega_{\mathtt{x}}(x)}\boxed{\mathbf{R}}\overset{\mathtt{x}}{—}\boxed{x}\overset{\mathtt{x}}{—} \tag{128}$$

where

$$\boxed{\omega_{\mathtt{x}}(x)} \quad := \quad \left( \boxed{\mathbf{R}}\overset{\mathtt{x}}{—}\boxed{x}\overset{\mathtt{x}}{—}\boxed{\mathbf{R}} \right)^{-1} \tag{129}$$

It is easilly verified that this follows from pointer classicality using the pointer tomographic process locality theorem and (43). The $\omega_{\mathtt{x}}(x)$ are determined by the theory.

If we adopt the flatness assumption (85) then we obtain the *midcome identity*

$$\overset{\mathtt{x}}{—}\boxed{x}\overset{\mathtt{x}}{—} \quad \equiv \quad \overset{\mathtt{x}}{—}\boxed{x}\overset{\mathtt{x}}{—}\boxed{\mathbf{R}}\boxed{N_{\mathtt{x}}}\boxed{\mathbf{R}}\overset{\mathtt{x}}{—}\boxed{x}\overset{\mathtt{x}}{—} \tag{130}$$

This identity is true time symmetric operational quantum theory.

Another useful identity that follows from pointer classicality is the *control wire identity*

$$\overset{\mathtt{x}}{———} \quad \equiv \quad \sum_x \overset{\mathtt{x}}{—}\boxed{x}\overset{\mathtt{x}}{—} \tag{131}$$

We can obtain this using the pointer tomographic process locality theorem to check the equivalence of the two sides. The control wire identity expresses a certain classicality since it says that, if we ignore the readout on a pointer wire, we simply have the pointer wire (contrast this with the a wire associated with a quantum system where this would not be true).

### 7.11.1 Regularised circuits

Here we will show that any circuit can be written as being equivalent to a positive weighted sum of what we call *regularised circuits*. This is very useful the discussion of positivity.

First consider

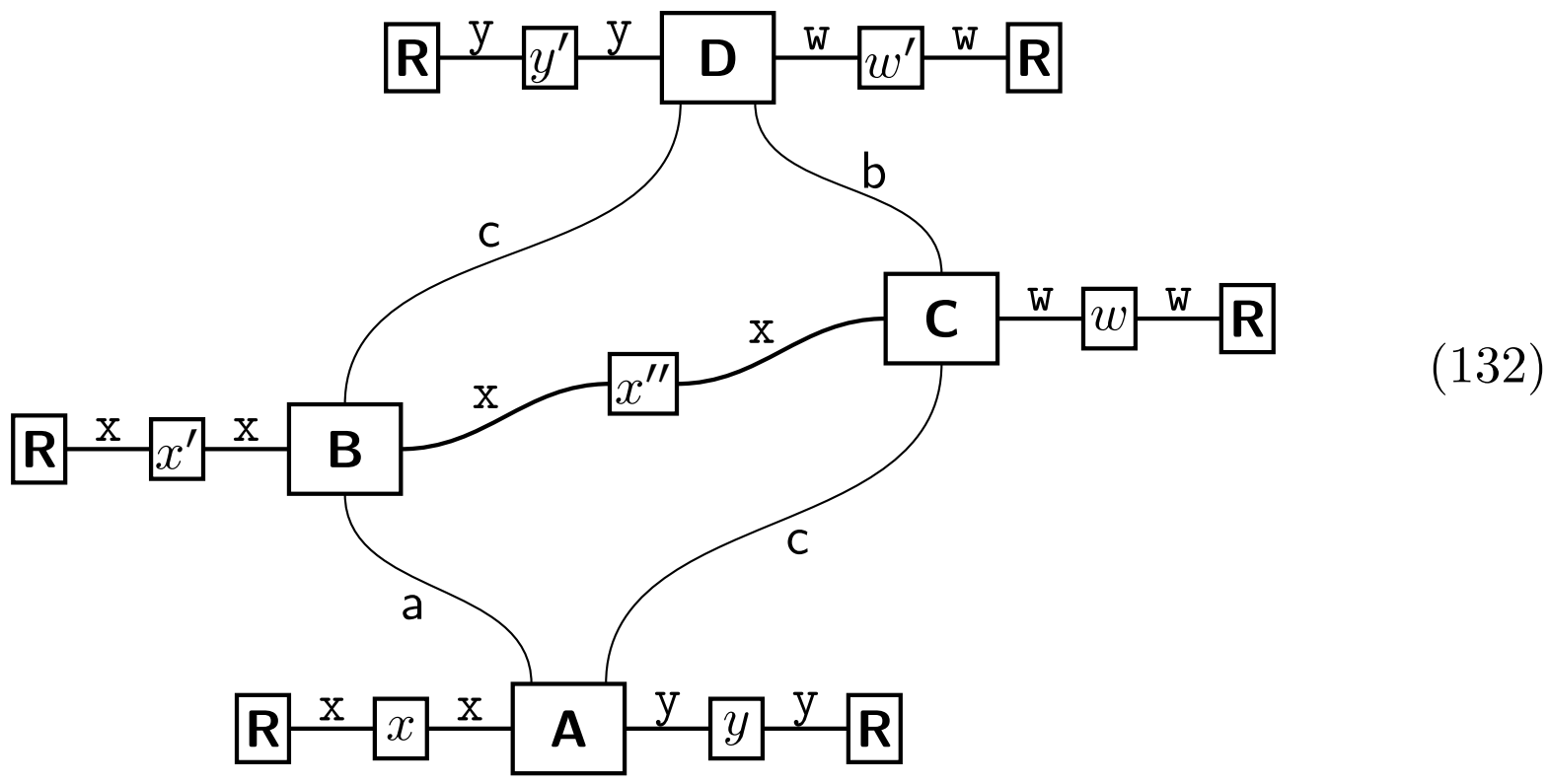

(132)

The $x$ readout box is an income to **A**. The $w$ readout box is an outcome from **C**. The $x''$ readout box is clearly a little different since it does not have an **R** box on either end. We will call this a *midcome* (between **B** and **C**).

We could also have a pointer wire going directly from an outcome on one operation to an income on another operation without a readout box (an example would be if the readout box, $x''$, in (132) were omitted). We will call such a wire a *control wire*.

Consider, a general circuit. We replace all control wires using (131), midcomes wires using the $\omega$-midcome identity (128) (or (130) if we have flatness), and non-deterministic pointer preparations and results using (119) and (121). After all these replacements we will have a positive weighted sum over circuits having no control wires, midcomes, or non-determinstic preparations and results - namely circuits such as

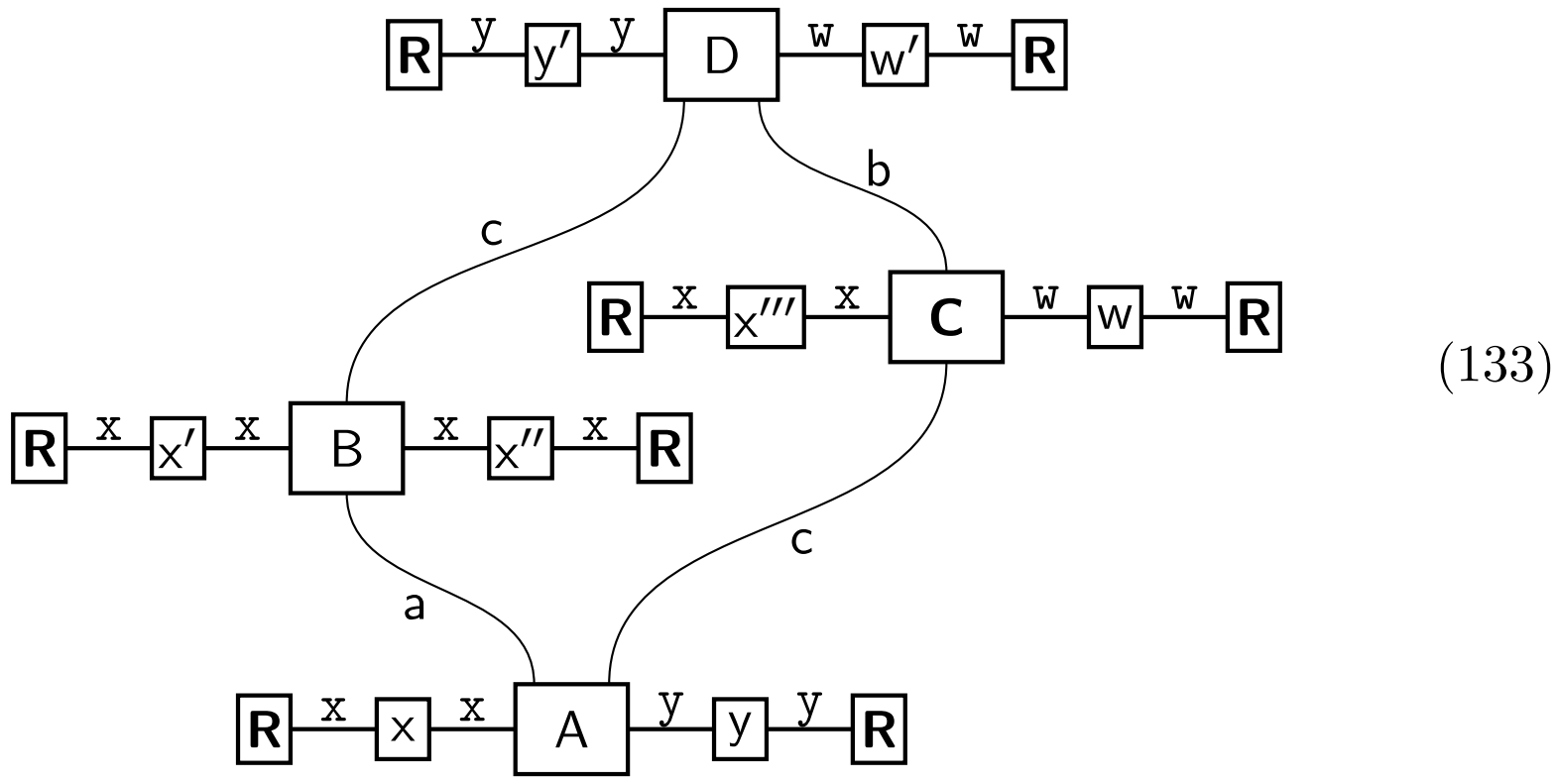

(133)

Let us call such circuits *regularised circuits*. The key result here is that any circuit can be written as a positive weighted sum of regularized circuits.

This is useful when proving positivity (see Sec. 7.12) - if we can prove that regularised circuits have non-negative probability then it follows that general circuits do. It is also useful for considerations to do with circuit reality in the context of Quantum Theory - if regularised circuits have real probabilities then general circuits do (see Sec. 20).

It is useful to define *fully regularised circuits* to be ones in the form

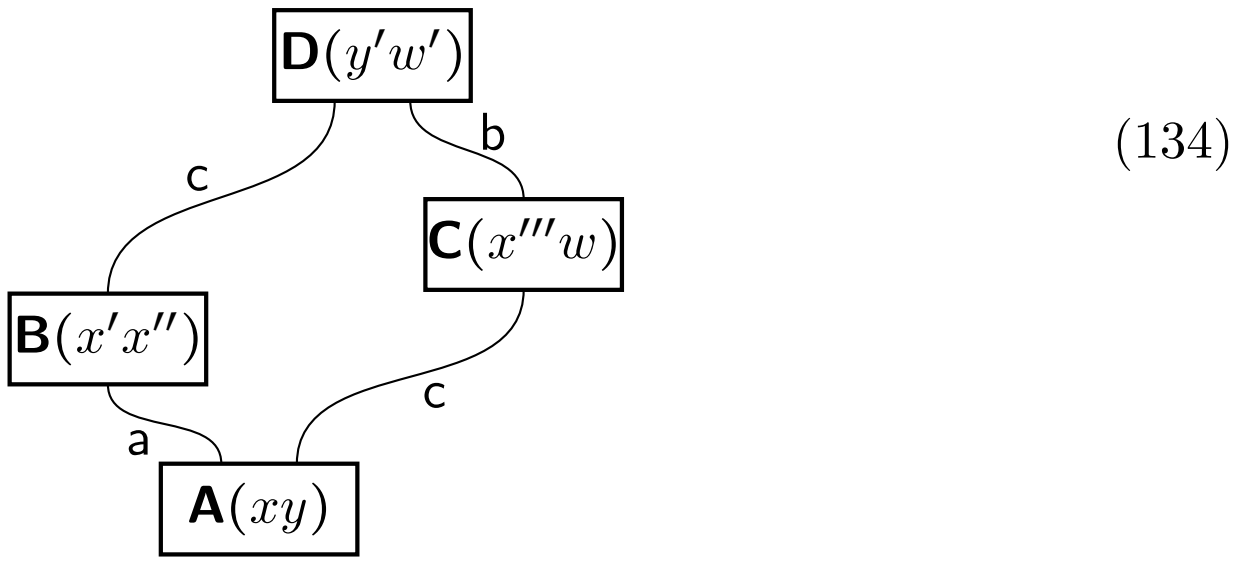

(134)

where we have absorbed all readout and **R** boxes and adopted appropriate notation.

#### 7.11.2 Testers and inequalities

Here we introduce the notion testers. We will illustrate our remarks with the following network

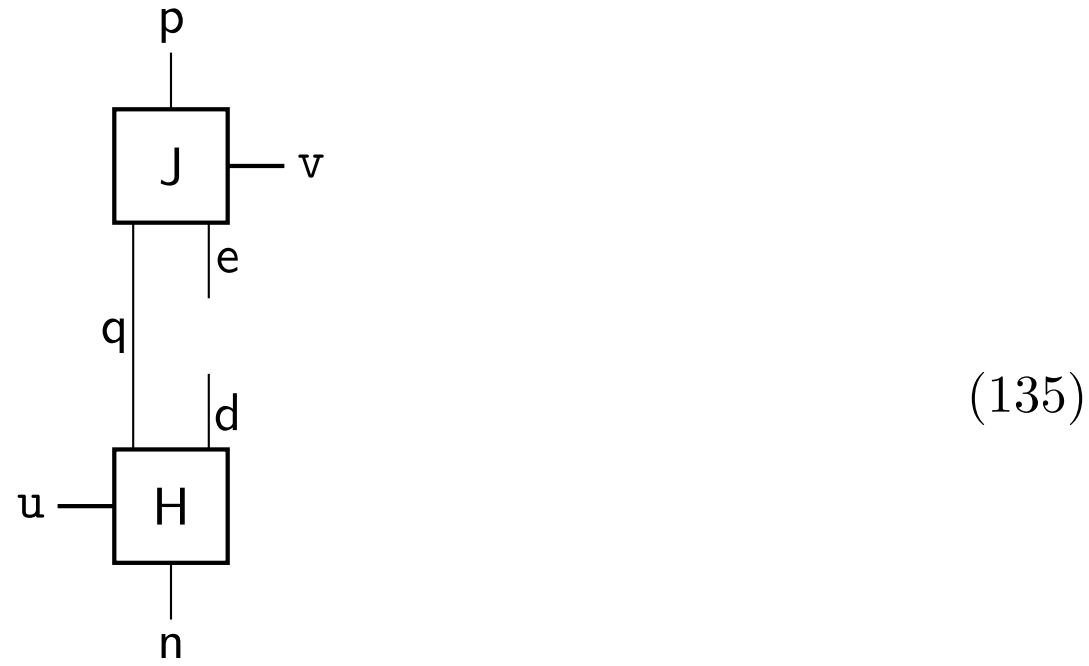

(135)

We can complete any such network into a circuit by means of a complement network (see Sec. 6.2). We will consider a subset of such complement networks, which we call *testers*, having three simplifying properties:

1. We close incomes and outcomes on the network with **R** boxes and readout boxes.

2. We pull all inputs on the network down to meet a single preparation and pull all outputs on the network up to meet a single result.

3. The preparations and result are pure. We also include an ancillary system between the preparation and result (this is h in the examples below).

If we apply a tester to the network in (135) we obtain a circuit of the form

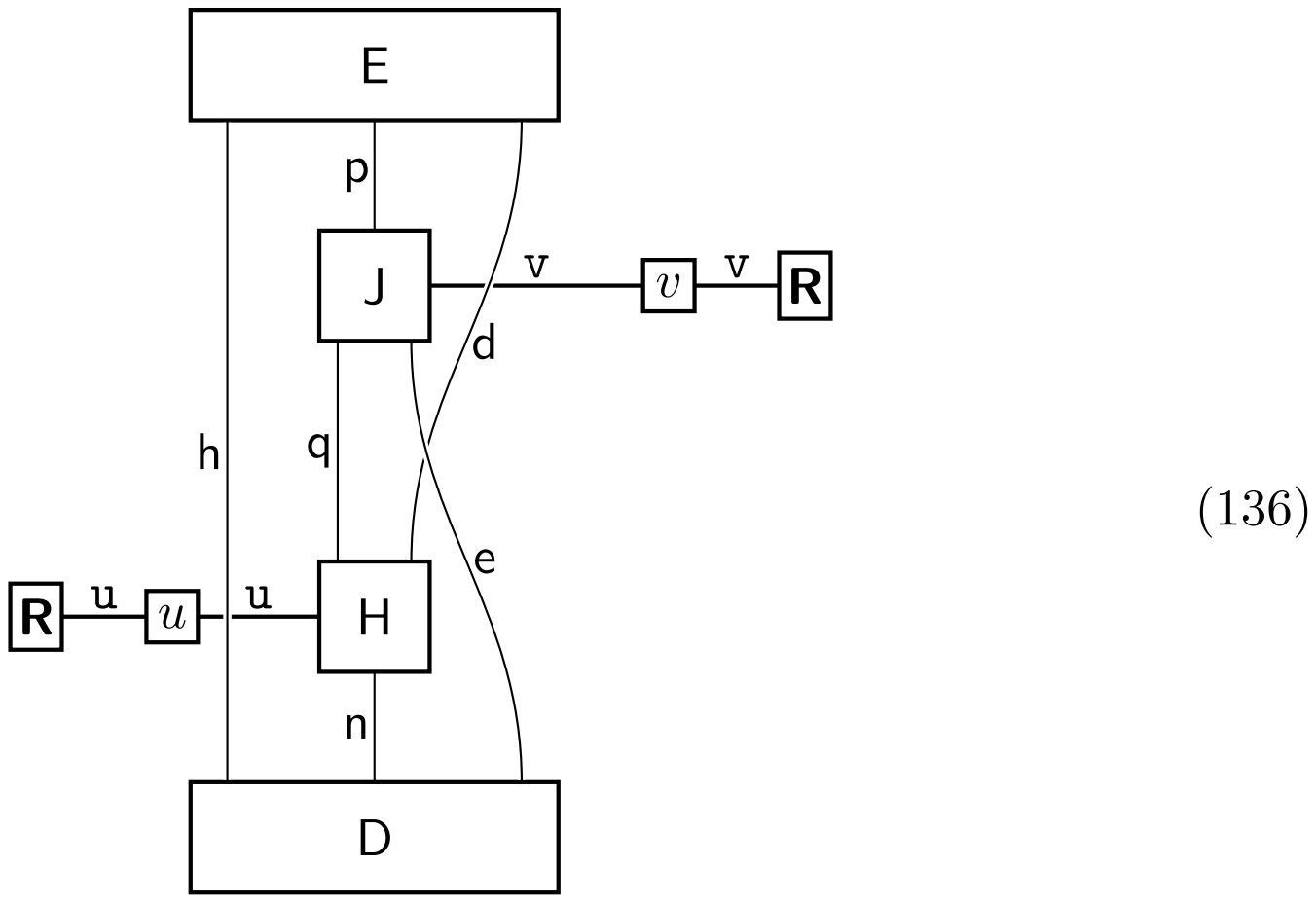

(136)

where D is a pure preparation and E is a pure result. In general, a tester consists of a pure preparation, a pure result, and a bunch of **R** boxes and readout boxes which are used to close incomes and outcomes. Testers can be thought of as that set of complement networks that are possible when we (i) use the given network in simple form (see (114) in Sec. 7.7 and the discussion therein) and (ii) treat it as if it were in a regularised circuit. Point (ii) is pertinent since we are interested in inequalities as will be discussed below.

We say that

$$\text{exprn}_1 \underset{T}{\leqq} \text{exprn}_2 \tag{137}$$

iff

$$p(\text{exprn}_1 T) \leq p(\text{exprn}_2 T) \tag{138}$$

for all testers, $T$, that complete each of the terms in $\text{exprn}_k$ into a circuit. Compare this with the inquality, $\leqq$, defined in Sec. 6.3 where we consider *all* complement networks (not just testers).

The following simple theorem

$$\text{If} \;\; \text{exprn}_1 + \text{exprn}_2 \equiv \text{exprn}_3 \;\; \text{and} \;\; 0 \underset{T}{\leqq} \text{exprn}_2 \;\; \text{then} \;\; \text{exprn}_1 \underset{T}{\leqq} \text{exprn}_3 \tag{139}$$

relates tester inequality and equivalence and follows immediately from the definitions of these notions.

Since testers are a subset of complement networks it follows that

$$\text{exprn}_1 \leqq \text{exprn}_2 \quad \Rightarrow \quad \text{exprn}_1 \underset{T}{\leqq} \text{exprn}_2 \tag{140}$$

(as discussed in Sec. 6.3).

For the special case where these expressions are a real weighted sum of operations (rather than general networks) we can prove the reverse implication, that

$$\text{exprn}_1 \leqq \text{exprn}_2 \quad \Leftarrow \quad \text{exprn}_1 \underset{T}{\leqq} \text{exprn}_2 \tag{141}$$

First we will prove

$$\mathsf{B}^{\mathsf{b}_3\mathsf{y}_4}_{\mathsf{a}_1\mathsf{x}_2}[1] \leqq \mathsf{B}^{\mathsf{b}_3\mathsf{y}_4}_{\mathsf{a}_1\mathsf{x}_2}[2] \quad \Leftarrow \quad \mathsf{B}^{\mathsf{b}_3\mathsf{y}_4}_{\mathsf{a}_1\mathsf{x}_2}[1] \underset{T}{\leqq} \mathsf{B}^{\mathsf{b}_3\mathsf{y}_4}_{\mathsf{a}_1\mathsf{x}_2}[2] \tag{142}$$

where $\mathsf{B}[1]$ and $\mathsf{B}[2]$ are two operations. If we put $\mathsf{B}[l]$ (for $l = 1$ or $l = 2$) into a general circuit there may be midcomes and/or control wires. We can write this circuit as a positive weighted sum of regularised circuits where, importantly, the weights (consisting of various $N_{\mathsf{z}}$ factors coming from the midcome identity (130) as well as the non-negative $\alpha^x$ and $\beta^x$ factors coming from (119) and (121) that are accumulated during the regularisation process described in Sec. (7.11.1)) are the same for the $l = 1$ and $l = 2$ cases. Any regularised circuit containing $\mathsf{B}[l]$ can be put in the form

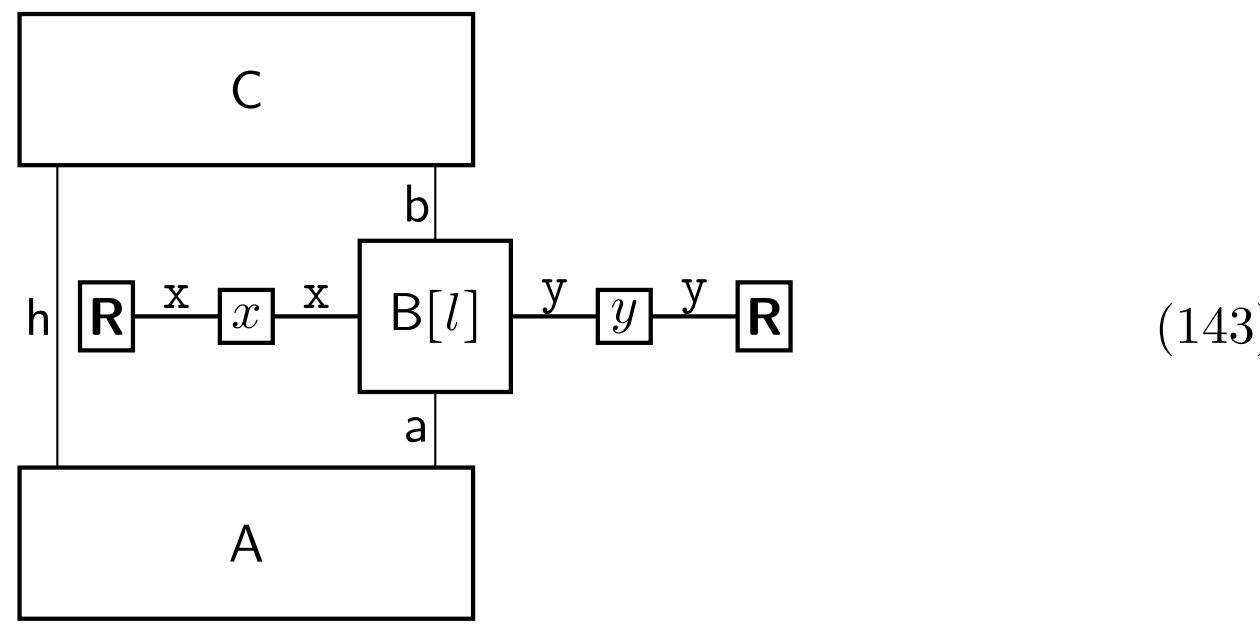

(143)

for appropriately defined $\mathsf{A}$ and $\mathsf{C}$. By the double purity theorem, we can write the preparation $\mathsf{A}$ as a positive weighted sum of pure preparations and $\mathsf{C}$ as a positive weighted sum of pure results. Hence, the circuit in (143) is a positive weighted sum of circuits of the form

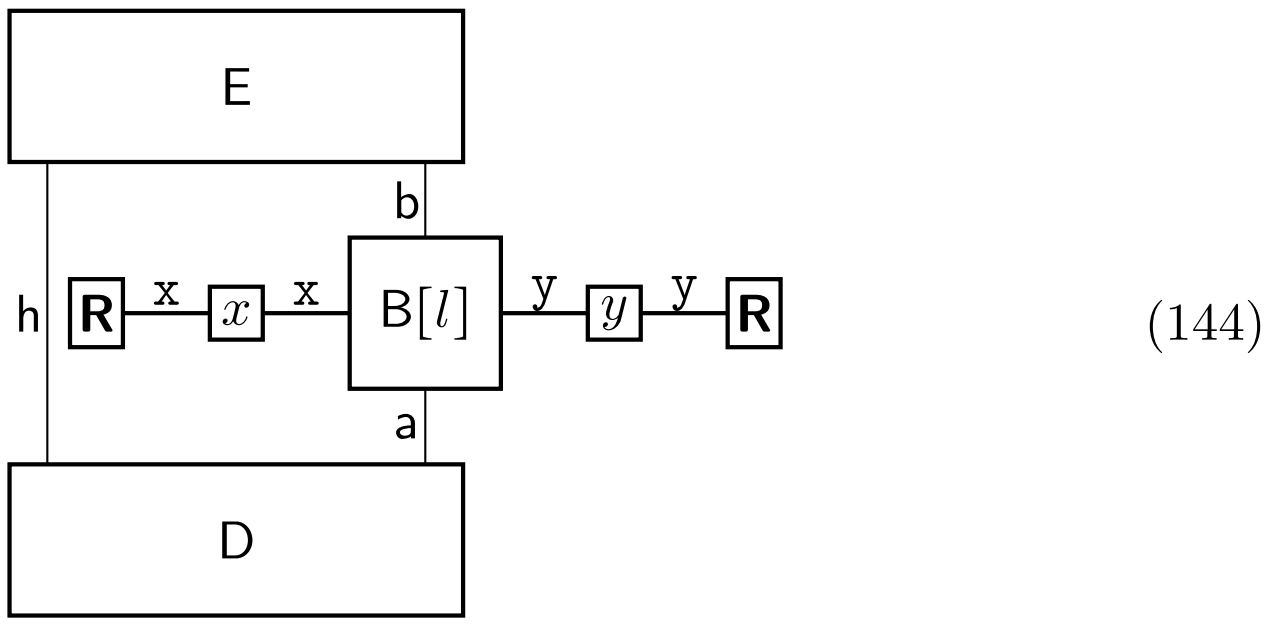

(144)

where $\mathsf{D}$ and $\mathsf{E}$ are pure. Furthermore, the weights in this sum do not depend on $l$. This is a tester applied to $\mathsf{B}[l]$. Since all the weights are positive and independent of $l$, (142) follows. To see, finally, that (141) holds for a weighted sum of operations we can rewrite it in equivalent form as

$$\text{exprn}_1^+ - \text{exprn}_2^- \leqq \text{exprn}_2^+ - \text{exprn}_1^- \quad \Leftarrow \quad \text{exprn}_1^+ - \text{exprn}_2^- \underset{T}{\leqq} \text{exprn}_2^+ - \text{exprn}_1^- \tag{145}$$

where $\text{exprn}_k^\pm$ collects all the terms in $\text{exprn}_k$ with ±ve coefficients. The inequalities (145) each have a positive weighted sum of operations on each side. By linearity of the $p(\cdot)$ function, the truth of (145) follows from (142).

In Sec. 77 we prove that the implication (141) follows in quantum theory even when the expressions are real weighted sums of arbitrary networks (not just simple operations). This proof will be given in the context of complex operational quantum theory.

### 7.11.3 Complete set sums

The control wire identity can be used to set up an equivalence between complete sets of operations/networks (see Sec. 5.3) and the associated deterministic operation/network. Consider the example

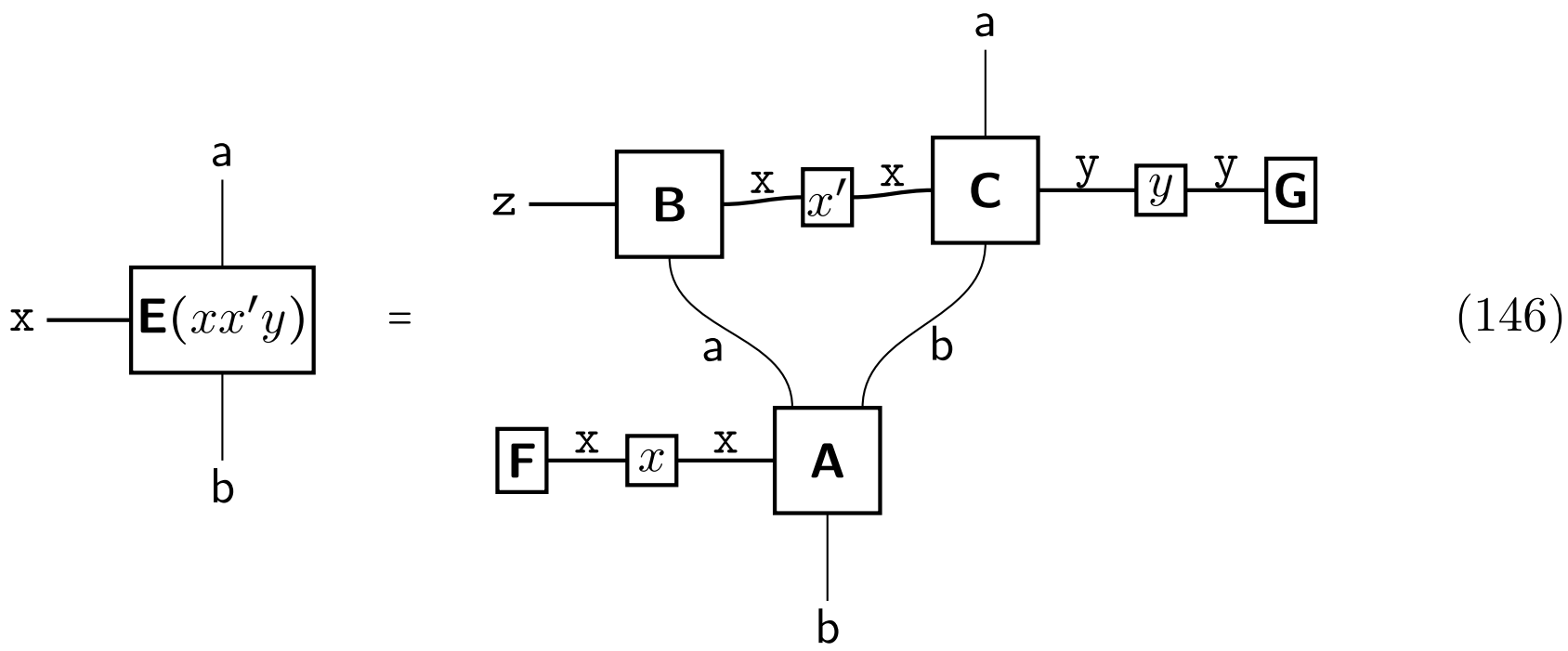

(146)

Then, using the control wire identity, we have

$$\sum_{xx'y} \; {}_{\mathsf{x}}\boxed{\mathbf{E}(xx'y)}^{\mathsf{a}}_{\mathsf{b}} \;\equiv\; {}_{\mathsf{x}}\boxed{\mathbf{E}}^{\mathsf{a}}_{\mathsf{b}} \tag{147}$$

where $\mathbf{E}$ is simply the network on the right hand side of (146) with the readout boxes omitted. In general, if we sum over a complete set of operations/networks we obtain the corresponding deterministic operation/network.

## 7.12 Positivity

In this subsection we will provide a condition on operations such that, if satisfied, then circuits will have a non-negative probability.

### 7.12.1 Tester positivity

Now consider a regularised circuit which contains an operation B (this may be deterministic or nondeterministic). We can put this circuit in the following form

(148)

where we have collected all other operations in the circuit into the preparation A or into the result C. Given the double purity property (see Sec. 7.8), a necessary and sufficient condition for circuit of this form to have positive probability is that

prob ( ) $\geq 0$ (149)

for all systems h, all pure preparations D, and all pure results E. This is clear since, using double purity, all circuits of the form (148) can be written as a positive weighted sum of circuits of the form in (149). We can write this condition as

$0 \underset{T}{\leq}$ (150)

where we are considering testers (see Sec. 7.11.2) of the following type

(151)

for all systems $h$, all pure preparations D and all pure results E. We call the condition in (150) *tester positivity* or $T$-positivity.

We read the symbol $\leqq_T$ as saying *less than or equivalent to, with respect to testers.* Compare this with $\leqq$ discussed in Sec. 6.3. To verify $\leqq$ we must check for all complement networks whereas to verify $\leqq_T$ we only need to check complement networks of the special form in (151). For operations, these amount to the same thing because of the midcome and control identities discussed in Sec. 7.11.

In Part II we will introduce the symbol $\leq_T$ which has a different meaning. It means *less than or equal to, with respect to (operator) testers.*

We have not proven, at this stage, that if all operations in a circuit individually satisfy tester positivity the circuit will have non-negative probability. To prove this we need to think about composing operations and make an assumption about pure preparations and results.

### 7.12.2 Positivity for preparations and results

Before considering composing operations, we will consider applying tester positivity to preparations and results.

For a result the tester positivity condition looks like this

prob 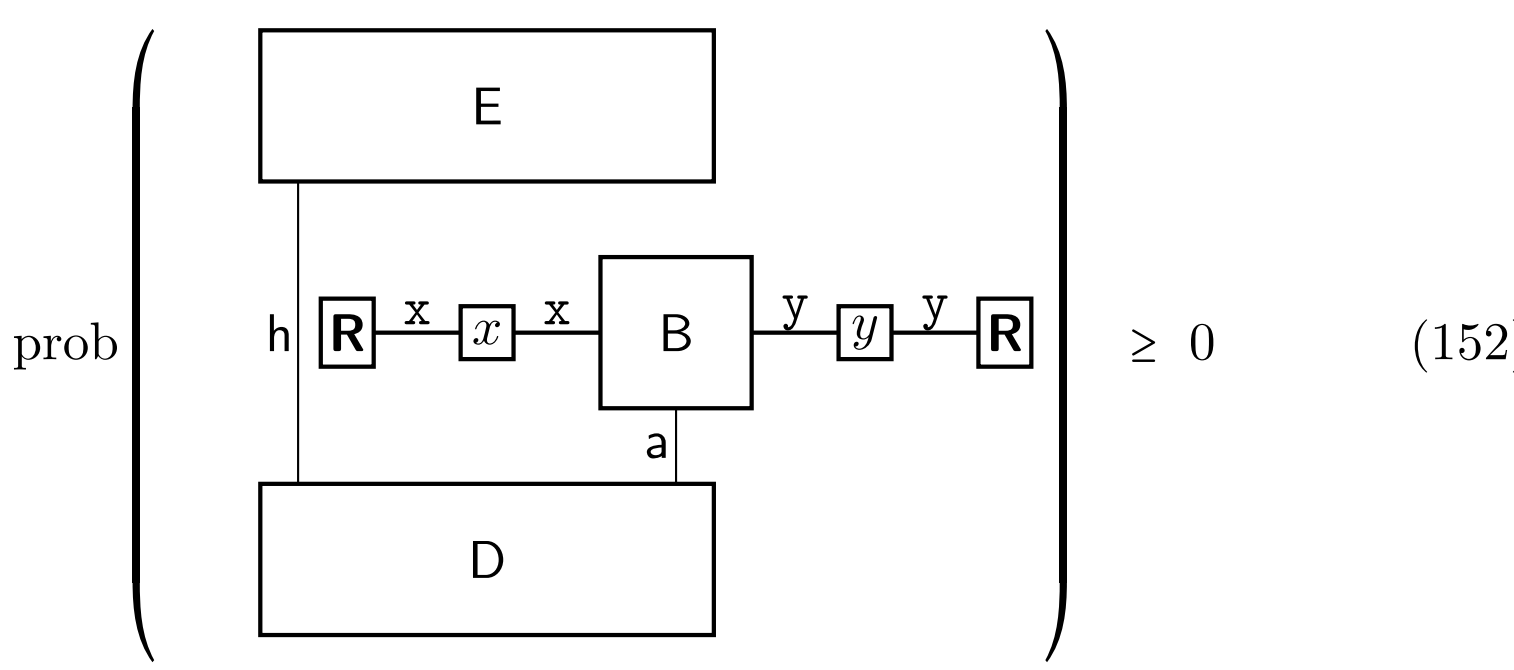

$\geq 0$ (152)

(i.e. b is null) for all pure D and E. The network consisting of D and E comprises a preparation for system a. By forward purity we can expand this as being equivalent to a positive weighted sum of pure preparations and consequently we obtain the condition

$$\text{prob}\left(\begin{array}{c} \boxed{\mathbf{R}} \overset{\mathtt{x}}{\text{—}} \boxed{x} \overset{\mathtt{x}}{\text{—}} \boxed{\mathsf{B}} \overset{\mathtt{y}}{\text{—}} \boxed{y} \overset{\mathtt{y}}{\text{—}} \boxed{\mathbf{R}} \\ \mathsf{a}\,| \\ \boxed{\mathsf{F}} \end{array}\right) \geq 0 \tag{153}$$

for all pure $\mathsf{F}$. This is tester positivity for results.

By analogous reasoning we obtain

$$\text{prob}\left(\begin{array}{c} \boxed{\mathsf{G}} \\ \mathsf{b}\,| \\ \boxed{\mathbf{R}} \overset{\mathtt{x}}{\text{—}} \boxed{x} \overset{\mathtt{x}}{\text{—}} \boxed{\mathsf{B}} \overset{\mathtt{y}}{\text{—}} \boxed{y} \overset{\mathtt{y}}{\text{—}} \boxed{\mathbf{R}} \end{array}\right) \geq 0 \tag{154}$$

for all pure $\mathsf{G}$ as the tester positivity condition for preparations.

In proving the positivity theorems that follow we will assume that all pure preparations and pure results satisfy tester positivity and also that the particular operations in question satisfy tester positivity.

### 7.12.3 Positivity composition theorem

Tester positivity is clearly a necessary condition if we want to have non-negative probabilities. We will show that it is a sufficient condition. To prove this, first we prove a composition theorem.

> **Positivity composition theorem.** Assume all pure preparations and pure results satisfy tester positivity. If we wire together two or more operations to form a network (so there are no closed causal loops) where operation each satisfies the tester positivity condition, then the resulting network also satisfies the tester positivity condition.

Start by considering just two operations, $\mathsf{A}$ and $\mathsf{B}$, which satisfy tester positivity. The condition that network obtained by joining $\mathsf{A}$ and $\mathsf{B}$ satisfies tester positivity

can be written

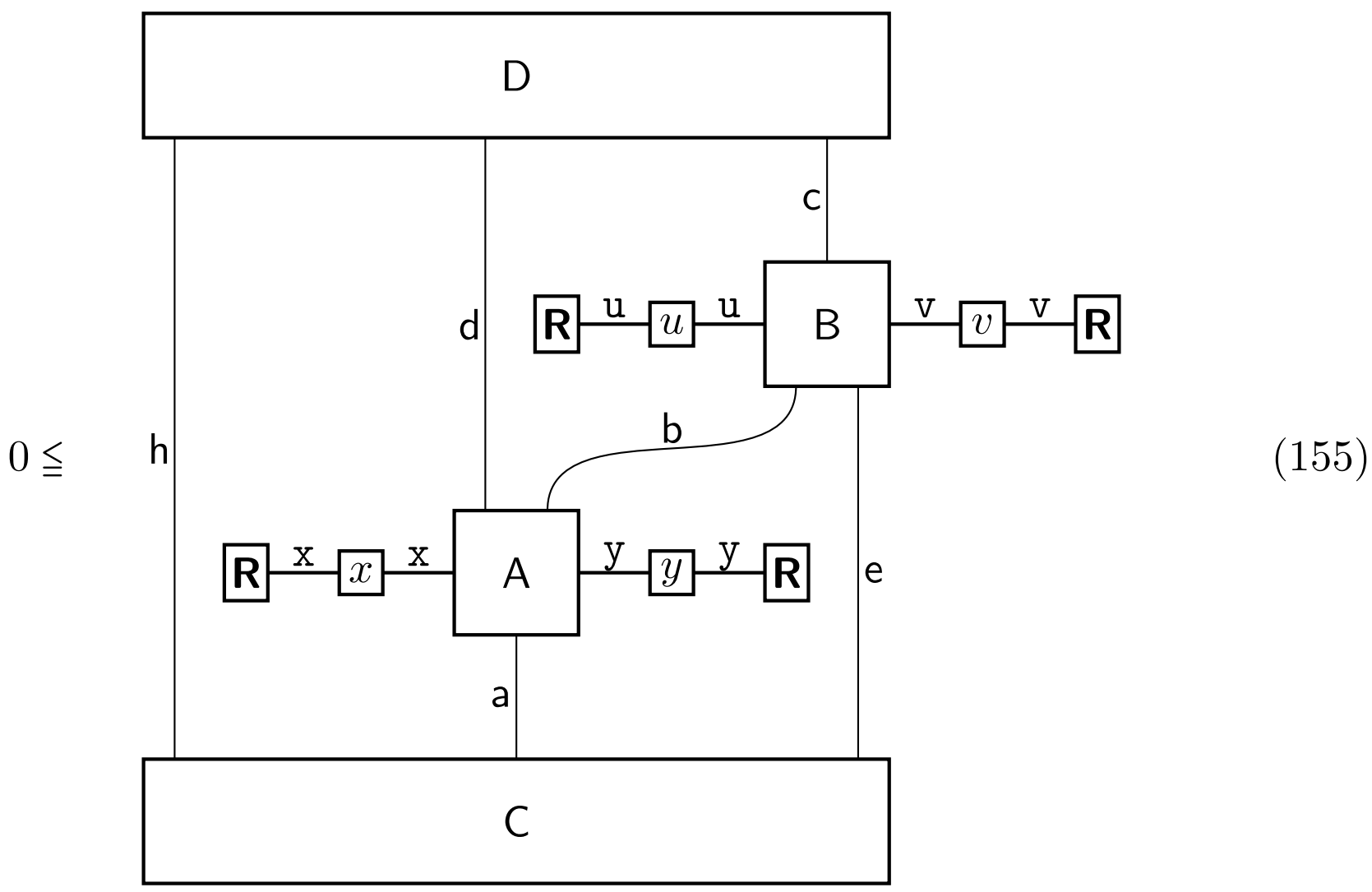

(155)

for all pure C and D. Note this way of joining A and B captures the general situation since one of the two operations must come first (and we show the case where A comes first) or they are in parallel (and this case is captured by omitting the b system). Furthermore, if there are midcomes, control wires, or non-deterministic pointer preparations or results they can be removed by regularising the circuit as described in Sec. 7.11.1 leaving us with this condition for positivity. Consider the system preparation consisting of just C and A

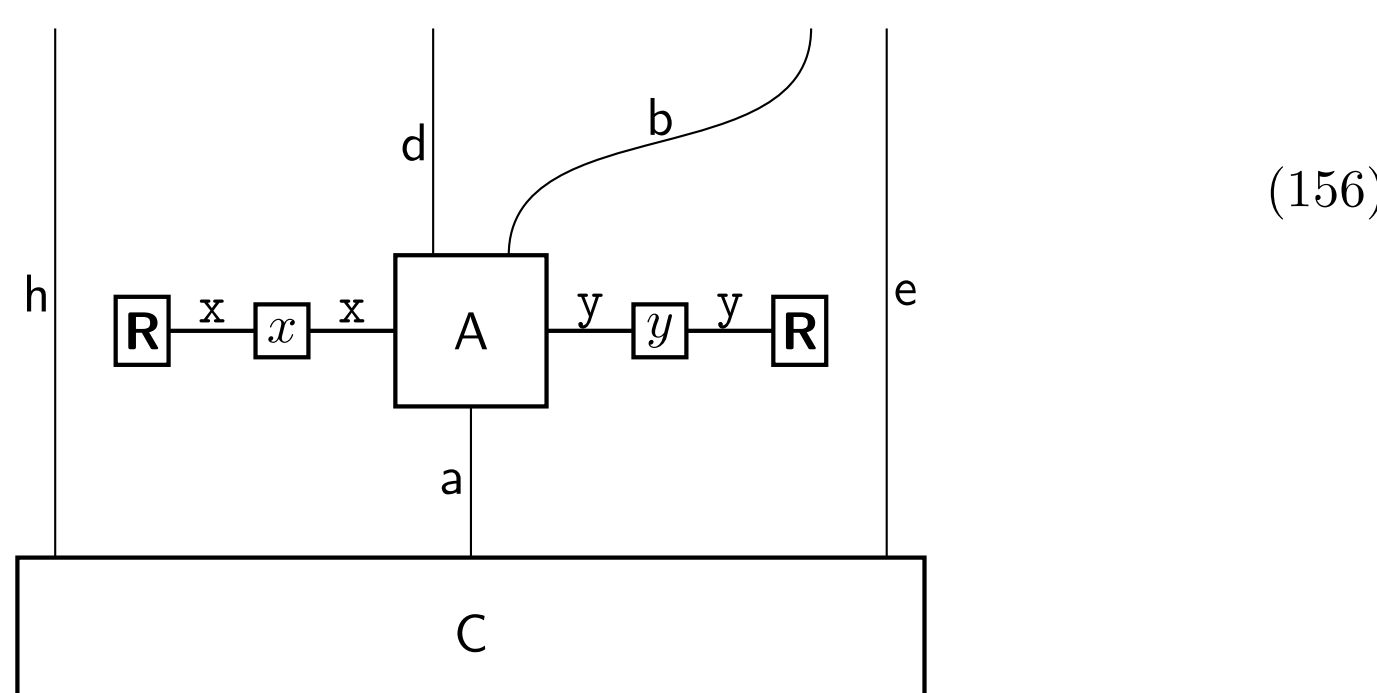

(156)

It is clear that this preparation satisfies tester positivity since this condition is the same as the tester positivity condition for A. Similarly the system result

consisting of B and D

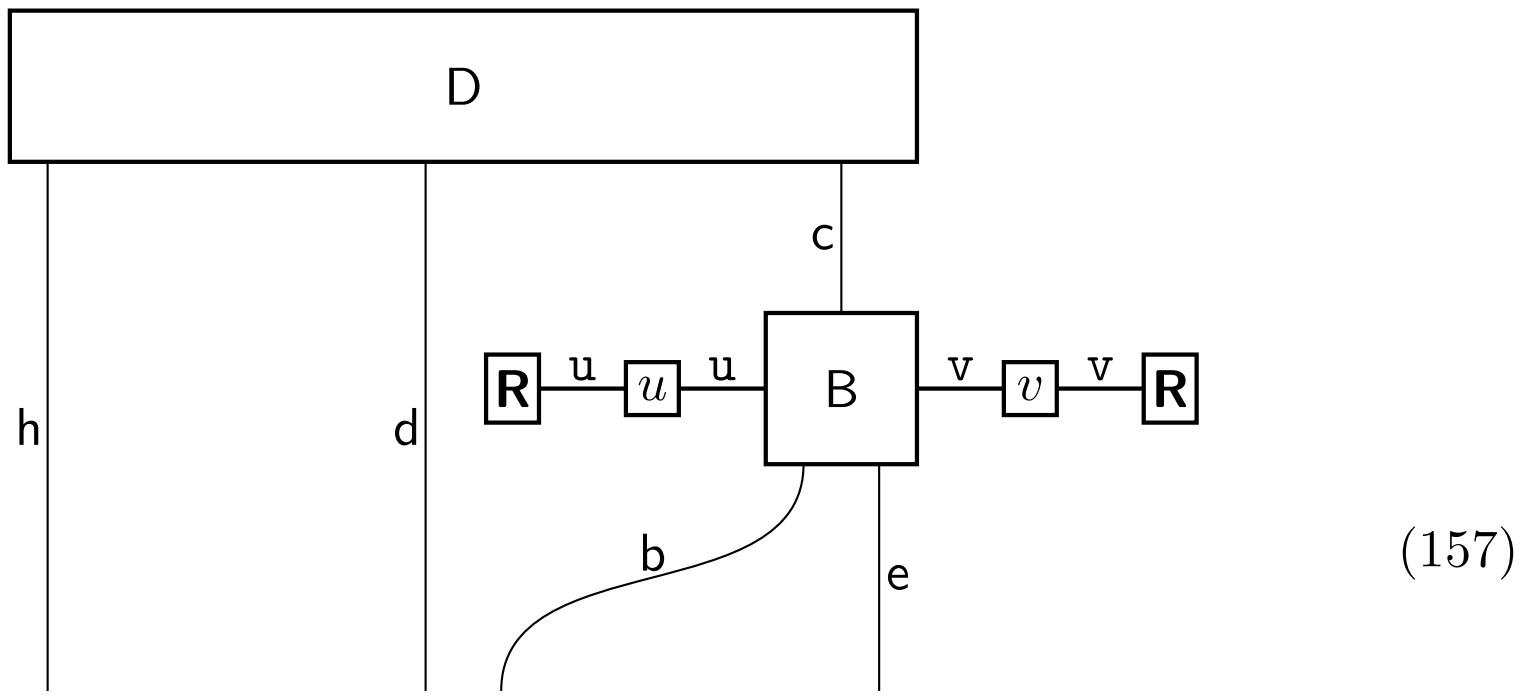

satisfies tester positivity as this condition is the same as the condition that B satisfies tester positivity. By double purity we can write the system preparation in (156) as a positive weighted sum of pure system preparations and we can write the system result in (157) as a positive weighted sum of pure system results. Since, in the theorem statement, we assume that all pure preparations and pure results satisfy tester positivity, this means that the circuit in (155) has a nonnegative probability. Hence

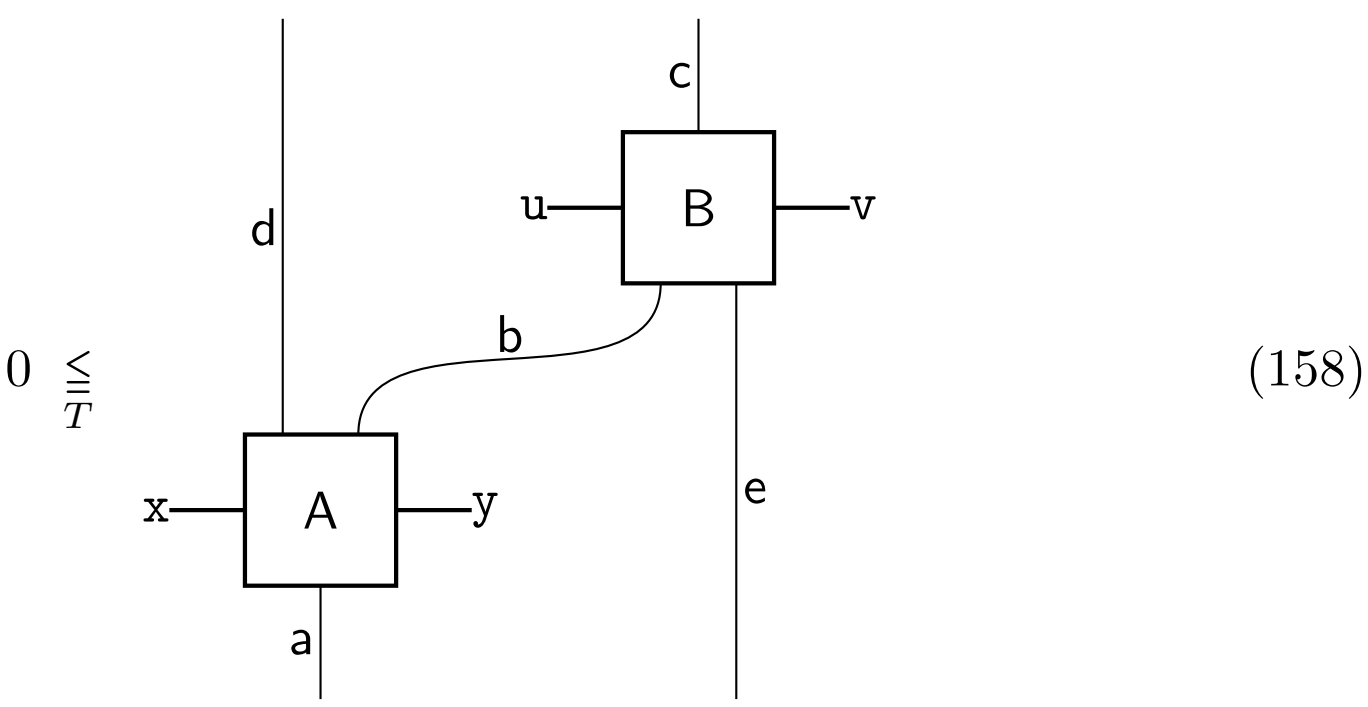

This proves the theorem for two operations. Now consider where we have multiple operations joined together (i.e. constituting a network) where each operation satisfies tester positivity. Since networks have no causal loops (they are directed acyclic graphs) we can put a temporal order on all the operations from first to last (this temporal order is not, in general, unique but any such temporal order will do). Consider the network consisting of the first $n$ operations in the temporal order. Call this the $n$-network. For the purposes of applying tester positivity can regard this network as an operation by pulling all the inputs down and all the outputs up. Now we add the $(n+1)$th operation to the appropriate outputs of the $n$-network (it must add only to outputs because of the way we have imposed a temporal order). If the $n$-network satisfies positivity then the $(n+1)$-network will by the positivity composition theorem. Since the $n = 1$ network is an operation we obtain the theorem by induction.

### 7.12.4 Circuit positivity theorem

Finally, we can prove that circuits have non-negative probabilities.

> **Circuit positivity theorem.** If every operation in a circuit satisfies tester positivity and if all pure preparations and pure results satisfy tester positivity, then the circuit has non-negative probability.

This follows immediately from the positivity composition theorem in Sec. 7.12.3 since a circuit is a special case of a network. To be more explicit, assume the circuit has $N$ operations and using the terminology of the previous theorem. We could decompose it as a $(N-1)$-network which would be a preparation and the final $N$th operation which would be a result. The $(N-1)$-network is positive by the previous theorem. The $N$th operation could be written as a positive weighted sum over pure results (by the double purity property). Thus, the circuit has non-negative probability.

## 7.13 General double causality

Now we provide causality conditions for operations that may not be deterministic. To do this we need to employ the idea of tester positivity.

If we have a complete set of operations

$$\{\mathsf{B}(u) : \forall u\}$$

(meaning the readouts $u$ are mutually exclusive and exhaustive as discussed in Sec. 5.3) then

$$\sum_u \mathsf{B}(u) = \mathsf{B}$$

where $\mathsf{B}$ is deterministic (as discussed in Sec. 5.3). Thus, we have double causality conditions

$$\sum_u \; \text{x} \text{—} \mathsf{B}(u)_{\mathsf{a}}^{\mathsf{b}\,\to\,\mathsf{I}} \text{—y—} \mathsf{R} \;\equiv\; \text{x—}\mathsf{R}\;\; \mathsf{I}\text{—a} \qquad\qquad \sum_u \; \mathsf{R}\text{—x—} \mathsf{B}(u)^{\mathsf{b}}_{\mathsf{a}\,\to\,\mathsf{I}} \text{—y} \;\equiv\; \mathsf{I}\text{—b}\;\; \mathsf{R}\text{—y} \tag{159}$$

using the causality conditions (107, 108) for deterministic operations.

We can also write down double causality conditions for a single nondeterministic operation. A nondeterministic operation, $\mathsf{B}$, is a member of a complete set. Consequently, every non-deterministic operation, $\mathsf{B}$, can be associated with at least one deterministic operation, $\mathsf{B}$ (the sum of the elements of such a complete set as discussed in Sec. 7.11.3). We can define the complement operation, $\mathsf{B} - \mathsf{B}$. Since $\mathsf{B}$ is not necessarily unique, neither is the complement, $\mathsf{B} - \mathsf{B}$. We have

**General double causality theorem.** Consider an operation, B with an associated deterministic operation, **B**, such that (i) **B** satisfies the double causality conditions and (ii) the complement, $\mathsf{B}[c] \equiv \mathbf{B} - \mathsf{B}$, satisfies tester positivity, then it follows that B satisfies the following two conditions

***General forward causality for nondeterministic operations***

(160)

***General backward causality for non-deterministic operations***

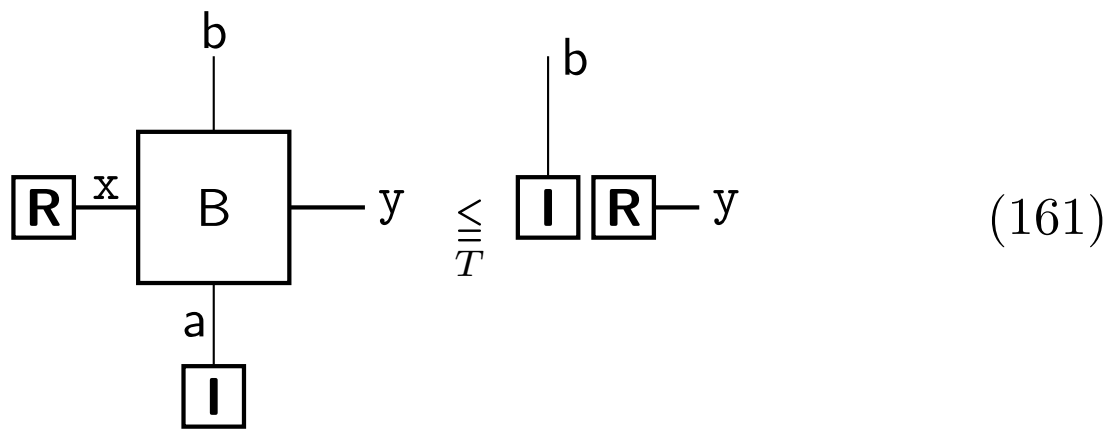

(161)

for any operation, B. Note that if B is deterministic and satisfies the double causality conditions then the inequalities (160) and (161) are saturated and then we have

(162)

(this follows from either of these inequalities when saturated).

The meaning of the inequality, $\leqq_T$, was discussed in Sec. 6.3 (when we close both sides with the same tester template to form a circuit then the corresponding numerical inequality is satisfied for the circuit probabilities, for any such tester). Saturation of this inequality means equivalence ($\equiv$). The proof of this theorem is very simple. Since $\mathsf{B} + \mathsf{B}[c] \equiv \mathbf{B}$, we have

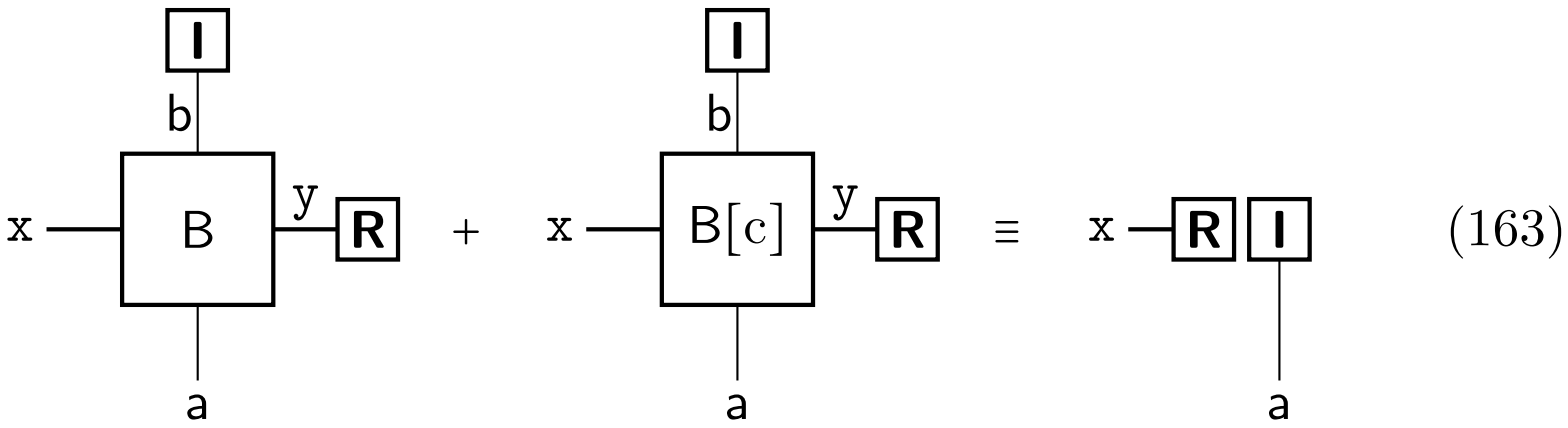

(163)

Since $\mathsf{B}[c]$ satisfies tester positivity it can easily be proven that the second term on the left hand side also does. We can apply the same tester to each term in (163) and the second term is positive. This is true for all testers. Thus we obtain (160). We obtain (161) similarly.

We have provided double causality conditions for deterministic operations. We have also provided general double causality conditions which work for non-deterministic operations. In the latter case the double causality conditions for deterministic operations appears as a special case (wherein the inequalities are saturated). We take the attitude that the deterministic double causality conditions are more fundamental. This is because the basic objects of the theory are deterministic operations and readout boxes. Nondeterministic operations can be built out of deterministic operations and readout boxes. It is worth noting that readout boxes satisfy general double causality (this is very easy to prove from the definition of **R** boxes).

### 7.13.1 Simple general double causality composition theorem

We can redo the simple double causality composition theorem (see Sec. 7.7) but now for operations that may be non-deterministic. We then have the following theorem

> **Simple general double causality composition theorem** Any network formed by wiring together two or more operations will (i) satisfy simple general forward causality if each of the component operations satisfy general forward causality and (ii) satisfy simple general backward causality if each of the component operations satisfy general backward causality. Here forward and backward causality are given by (160) and (161).

The proof of this is similar to the proof in the deterministic case except now we have $\underset{T}{\leqq}$ rather than $\equiv$ between subsequent steps in the proof. First consider composing two operations, $\mathsf{A}$ and $\mathsf{B}$. One must come first (unless they are in parallel). We choose A and then the general form of the composite is

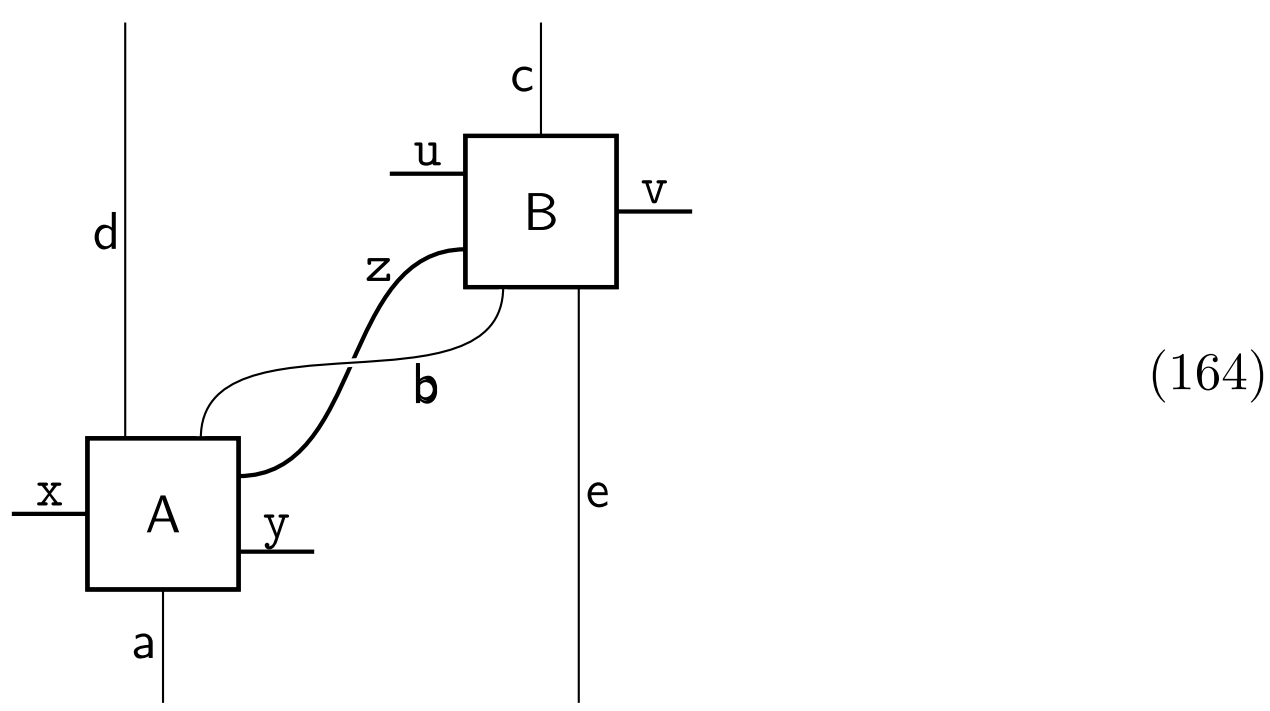

(164)

We can delete wires to deal with special cases (such as then they are in parallel). We can first apply (160) to B and then to A obtaining

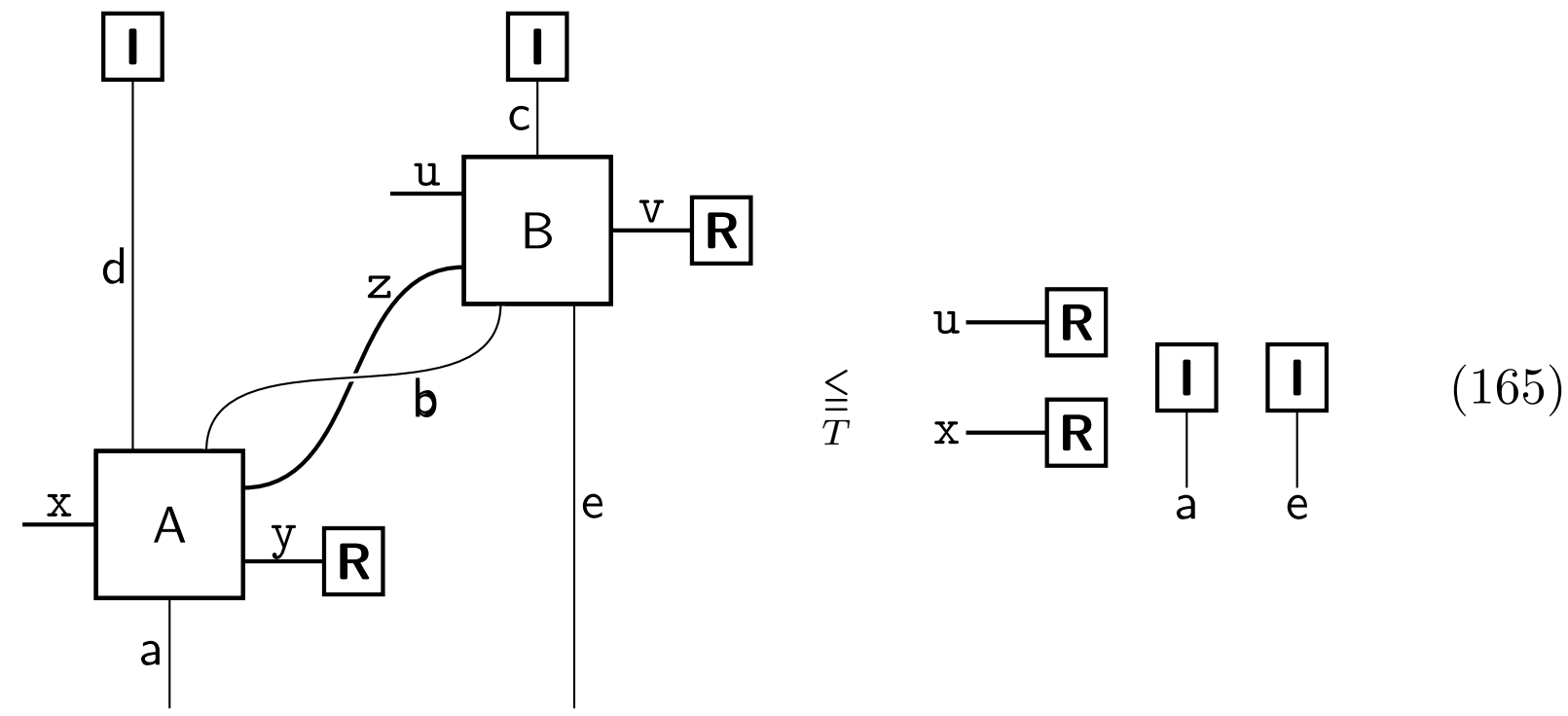

(165)

This proves that the composite object satisfies simple general forward causality. Simple general backward causality can be proved similarly. Now consider more than two operations. We can employ the same steps as in proving the deterministic causality composition theorem in Sec. 49.4.11. We impose a total order and, now, we have an inequality rather than equivalence after each step.

If the network has simple causal structure then the following theorem follows

> **General double causality composition theorem for simple causal structure.** Any network having simple causal structure that is formed by wiring together two or more operations will (i) satisfy general forward causality if each of the component operations satisfy general forward causality and (ii) satisfy general backward causality if each of the component operations satisfy general backward causality. Here general forward and general backward causality are given by (160) and (161).

This is clear since, if a network has simple causal structure then the simple general double causality conditions are just the general double causality conditions.

## 7.14 Subunity of circuit probabilities

The following follows immediately:

> **Circuit subunity theorem.** If every operation in a circuit satisfies general double causality then the circuit has probability less than or equal to one.

To prove this impose a total order on the operations in the circuit. Then we can start by applying general forward causality to last operation. This yields a $\leqq_T$ relationship with a new circuit. Reimpose a total causal order on the remaining operations and apply general forward causality to last operation. We can repeat this until we have only operations with no outputs and outcomes. We can apply general backward causality to these operations. This will yield a circuit composed of only $\mathsf{R}$'s and $\mathsf{I}$'s that is equivalent to one. Thus we prove subunity.

## 7.15 Simple physicality composition theorem

We say an operation is *physical* if it satisfies tester positivity and double causality. We say a network satisfies *simple physicality* if it satisfies tester positivity and simple double causality.

In Sec. 7.12.3 we proved a positivity composition theorem and we proved simple double causality composition theorems (see Sec. 7.13.1 for the deterministic case and Sec. 7.13.1 for the general case).

It is convenient to put all of this together as follows

> **Simple physicality composition theorem.** Under a certain assumption (that pure preparations and pure results satisfy tester positivity), any network built from physical operations will satisfy simple physicality.

By *simple physicality* we mean that tester positivity and simple (general) double causality are satisfied. This theorem subsumes all the composition theorems we have so far stated. However, it is logically course-grained. For example, to prove backwards causality for the network we need only assume backwards causality for the components (we do not to assume forward causality or that pure preparations and results satisfy tester positivity).

We can also state the following for networks having simple causal structure.

> **Physicality composition theorem for simple causal structure.** Under a certain assumption (that pure preparations and pure results satisfy tester positivity), any network having simple causal structure and built from physical operations will satisfy physicality.

This follows because, for a network having simple causal structure, physicality and simple physicality are the same thing.

## 7.16 Conditions for determinism

An operation is deterministic if it is has no readout boxes (implicit or explicit). This is something that can be determined by going into the laboratory and inspecting the piece of apparatus that implements the operation. It would be useful, however, to have a necessary and sufficient condition for determinism that can be written in equation form. It is hard, however, to separate such conditions from causality conditions. To be able to prove a theorem, we will assume that all operations satisfy the (inequality) double causality conditions in (160,161). A necessary condition for an operation to be deterministic is that it passes the determinism test in (168). It follows from this that deterministic operations must, in fact, satisfy the double causality conditions in (107,108) - these are for the deterministic case. Thus, under the assumption that operations satisfy general (i.e. inequality) double causality, a necessary and sufficient condition for determinism is that they satisfy the (deterministic) double causality conditions. We can, however, prove a slightly tighter result than this as we will see in the following theorem.

> **Determinism conditions theorem.** If we assume that all operations satisfy the general double causality conditions in (160, 161) then any one of the following three conditions comprise a necessary and sufficient test for an operation, B, to be deterministic
>
> 1. The forward causality condition
>
> 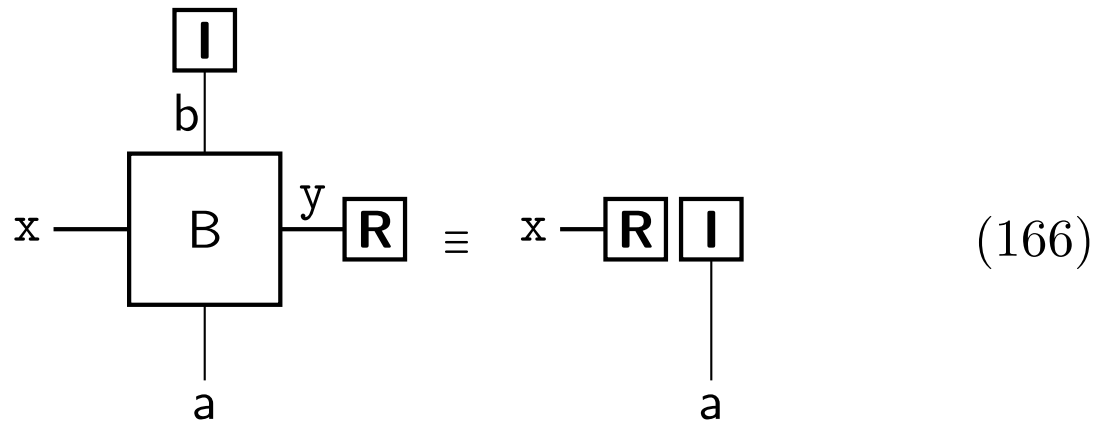
>
>
> (166)
>
> is satisfied.
>
> 2. The backward causality condition
>
> 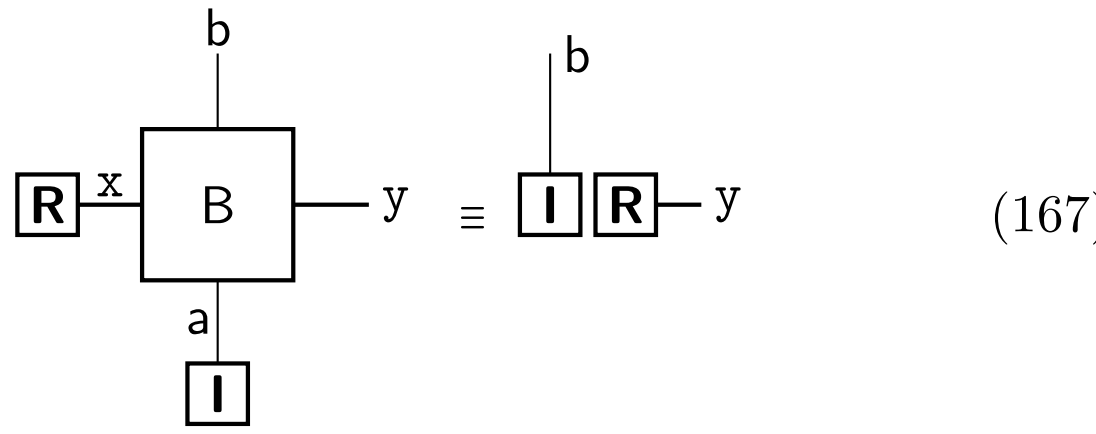
>
>
> (167)
>
> is satisfied.

3. The *determinism test condition*

$$\begin{array}{c} \boxed{\mathbf{I}} \\ | \\ \mathsf{b} \\ \boxed{\mathbf{R}} \overset{\mathsf{x}}{\text{—}} \boxed{\;\mathsf{B}\;} \overset{\mathsf{y}}{\text{—}} \boxed{\mathbf{R}} \\ \mathsf{a} \\ | \\ \boxed{\mathbf{I}} \end{array} \equiv 1 \tag{168}$$

is satisfied.

If an operation, B, is known to be deterministic we would conventionally represent it by bold font, **B**.

Conditions 1. and 2. are necessary since passing the determinism test is a necessary condition for determinism. Condition 3. is necessary since this is the determinism test (elaborating on this - the **R** and **I** elements are deterministic and so, if B is deterministic, this is a deterministic circuit which must have probability equal to 1). Condition 3. is sufficient for the following reasons. Consider embedding B in any circuit comprised of deterministic operations. Each of these deterministic operations must satisfy double causality. If the circuit is not in the form given in (168), then it must be the case that at least one of the other operations can be replaced by **R** and **I** boxes according to either forward or backward causality. We can keep replacing operations until we have a circuit in the form in (168). Thus, under the assumptions in the theorem statement, this condition is sufficient for $B$ to be deterministic. Finally, we see that, applying **R** and **I** boxes to both sides of the forward (or backward) causality condition (so we have circuits on both sides) gives (168). Hence 1. and 2. are also sufficient. Note that these three conditions are not equivalent if we do not impose general double causality, since then we can imagine the existence of operations that satisfy any one but neither of the other conditions.

An operation is deterministic if it has no readout boxes (explicit or implicit). The determinism test above can be regarded as a witness for determinism. It is useful to define the *physical norm*

$$|\mathsf{B}|_{\text{phys-norm}} = \text{prob}\left( \begin{array}{c} \boxed{\mathbf{I}} \\ | \\ \mathsf{b} \\ \boxed{\mathbf{R}} \overset{\mathsf{x}}{\text{—}} \boxed{\;\mathsf{B}\;} \overset{\mathsf{y}}{\text{—}} \boxed{\mathbf{R}} \\ \mathsf{a} \\ | \\ \boxed{\mathbf{I}} \end{array} \right) \tag{169}$$

The physical norm is, of course, equal to 1 for any deterministic operation.

# 8 Double maximality

## 8.1 The double maximality property

Here we define a useful property that a physical theory may have - namely *maximality*. Both classical probability theory and Quantum Theory has this property. Here we define the appropriate concept for the time symmetric frame of reference as follows

> **Double maximality** is the property that, for each pointer type, $\mathtt{x}$, we can associate a system type $\mathsf{x}$ such that there exist deterministic physical operations
>
> (170)
>
> for which we have the equivalence
>
> $\equiv$ (171)
>
> Furthermore, this is maximal in that there does not exist a similar equivalence for the same physical type, $\mathsf{x}$, but for another pointer type, $\mathtt{z}$, having $N_{\mathtt{z}} > N_{\mathtt{x}}$. The right hand side of (171) is just the identity for the type $\mathtt{x}$. We call the operations in (170) *maximal operations*.

When we have double maximality we can associate an integer, $N_{\mathsf{x}}$, with the physical type $\mathsf{x}$ where $N_{\mathsf{x}} := N_{\mathtt{x}}$ (pay attention to the font of the subscripts) where $N_{\mathtt{x}}$ appears in the definition of the maximal operators above.

We call this property "double" maximality because there is the forward property that we can, in some sense, send information from past to future and the backward property that we can send information from future to past. We are not really sending information but establishing correlation in either direction. In the time forward temporal frame we will only be able to send information forward in time (see Sec. 11.11) and in the backward time temporal frame we will only be able to send information backward in time.

Both classical probability theory and Quantum Theory have the property of double maximality. In the case of Quantum Theory the maximal elements are associated with projectors onto an orthonormal basis set.

## 8.2 Physicality of maximal elements

Since maximal elements are physical and deterministic they satisfy positivity and the double causality conditions as we will now discuss.

The positivity conditions are that

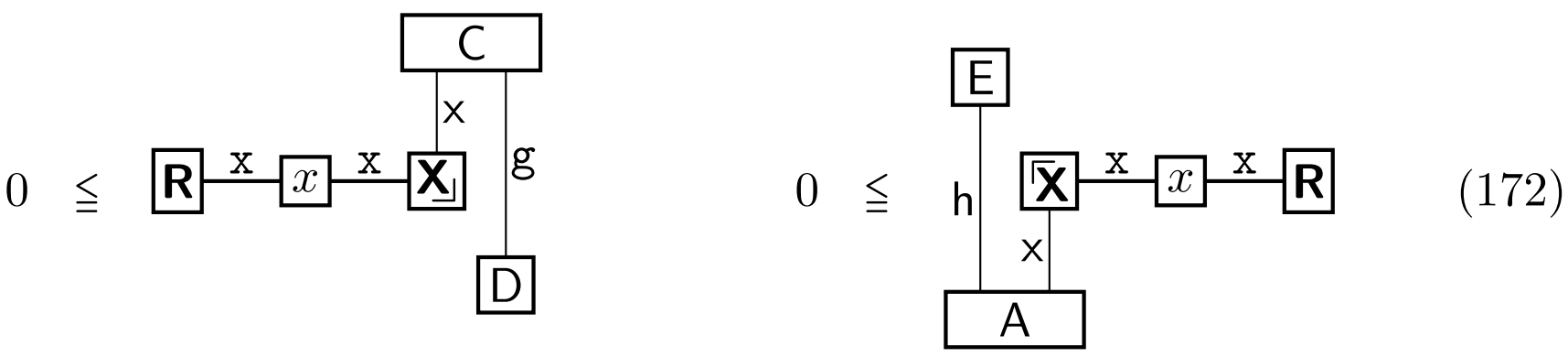

for all pure A, C, D, and E. This condition immediately gives us the following theorem

**Maximal on pure theorem.** We have

$$0 \underset{T}{\leqq} \quad \text{C} \; \text{x} \; \text{g} \; \mathbf{R} \overset{\text{x}}{—} \boxed{x} \overset{\text{x}}{—} \mathbf{X} \qquad\qquad 0 \underset{T}{\leqq} \quad \text{h} \; \bar{\mathbf{X}} \overset{\text{x}}{—} \boxed{x} \overset{\text{x}}{—} \mathbf{R} \; \text{x} \; \text{A} \qquad (173)$$

for all pure A and C where **X** is maximal.

This theorem follows from (172) and the definition of $T$-positivity. We will use this in the proof of the maximal representation theorem below.

The double causality conditions are

$$\mathbf{R} \overset{\text{x}}{—} \mathbf{X} \;(\text{x}) \equiv \mathbf{I} \;(\text{x}) \qquad\qquad \bar{\mathbf{X}} \overset{\text{x}}{—} \mathbf{R} \;(\text{x}) \equiv \mathbf{I} \;(\text{x}) \qquad (174)$$

$$\text{x} — \mathbf{X} \;(\text{x}\; \mathbf{I}) \equiv \text{x} — \mathbf{R} \qquad\qquad \bar{\mathbf{X}} — \text{x} \;(\text{x}\; \mathbf{I}) \equiv \mathbf{R} — \text{x} \qquad (175)$$

## 8.3 Maximal representation theorem

Double maximality allows a useful representation of any operation in terms of maximal operations and an operation having only physical types.

**Maximal representation theorem.** If we have the double maxi-mality property then any operation, $\mathsf{B}$, can be written as

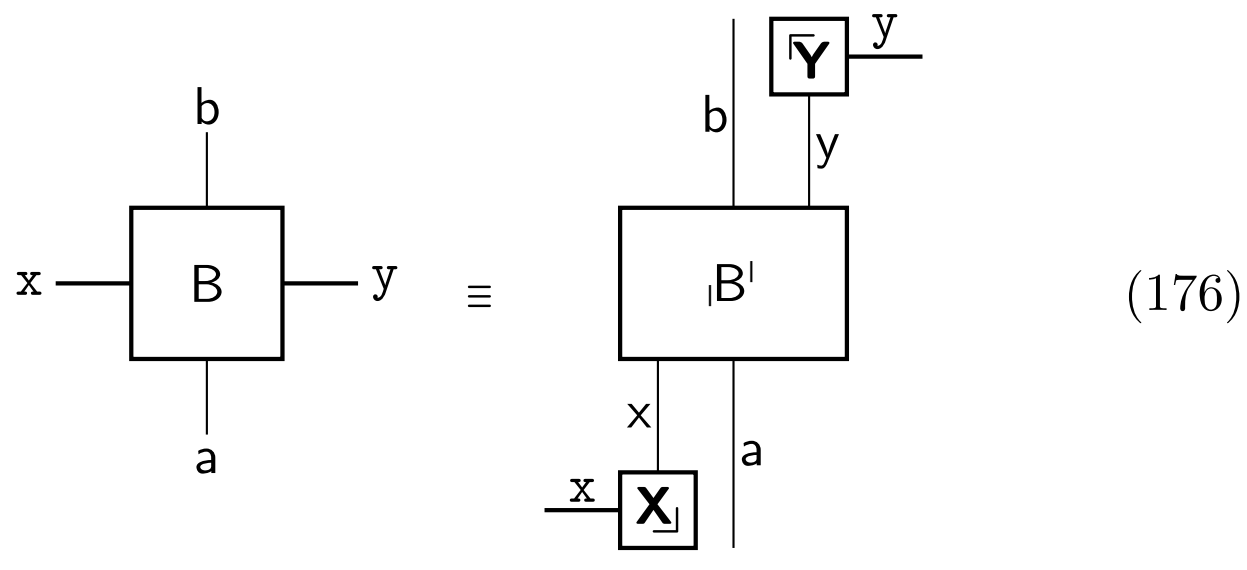

(176)

where

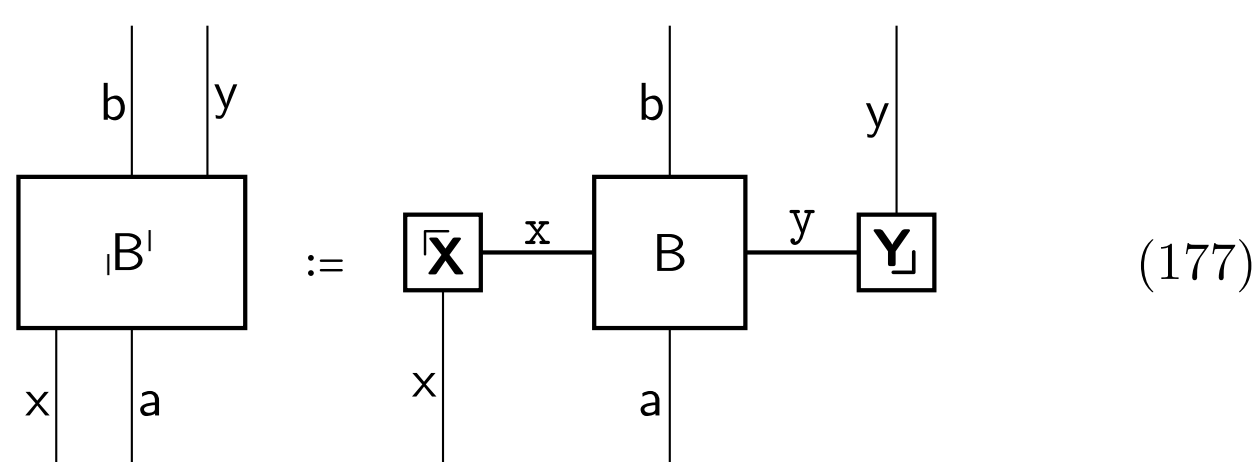

(177)

Furthermore: (i) $_{|}\mathsf{B}^{|}$ satisfies tester positivity if and only if $\mathsf{B}$ satisfies tester positivity, (ii) $_{|}\mathsf{B}^{|}$ satisfies general double causality if and only if $\mathsf{B}$ satisfies general double causality, (iii) $_{|}\mathsf{B}^{|}$ has the same physical norm as $\mathsf{B}$ (see physical norm definition in (169)). As a consequence, $_{|}\mathsf{B}^{|}$ is a physical operator with a given physical norm if and only if $\mathsf{B}$ is physical with the same physical norm.

First note that (176) is easily proved by putting (177) into the right hand side of (176) and applying (171). We can show that $_{|}\mathsf{B}^{|}$ satisfies general double causality (i.e. (160) and (161)) if and only if $\mathsf{B}$ satisfies general double causality using (174, 175). This proves (ii). Now consider (i). Tester positivity of $\mathsf{B}$ is the statement that

$0 \underset{T}{\leq}$

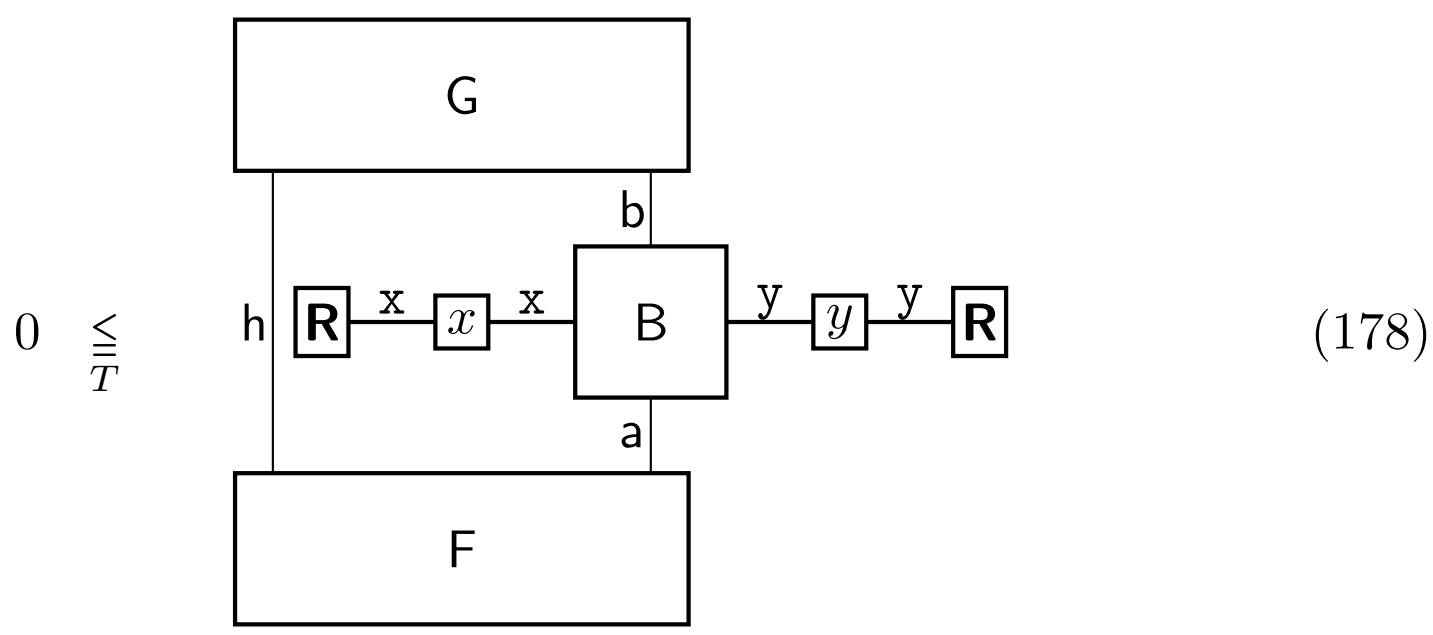

(178)

for all $x$, $y$, $F$, and $G$. Note that, when we stated the tester positivity condition in (150) we were able to restrict to the tester in (151) built from pure preparations for $F$ and pure results for $G$. However, in the proof below, it is convenient to drop this restriction (recall that, by the double purity theorem, all preparations (results) can be written as a positive weighted sum of pure preparations (pure results). One particular choice of $F$ and $G$ gives us the following statement

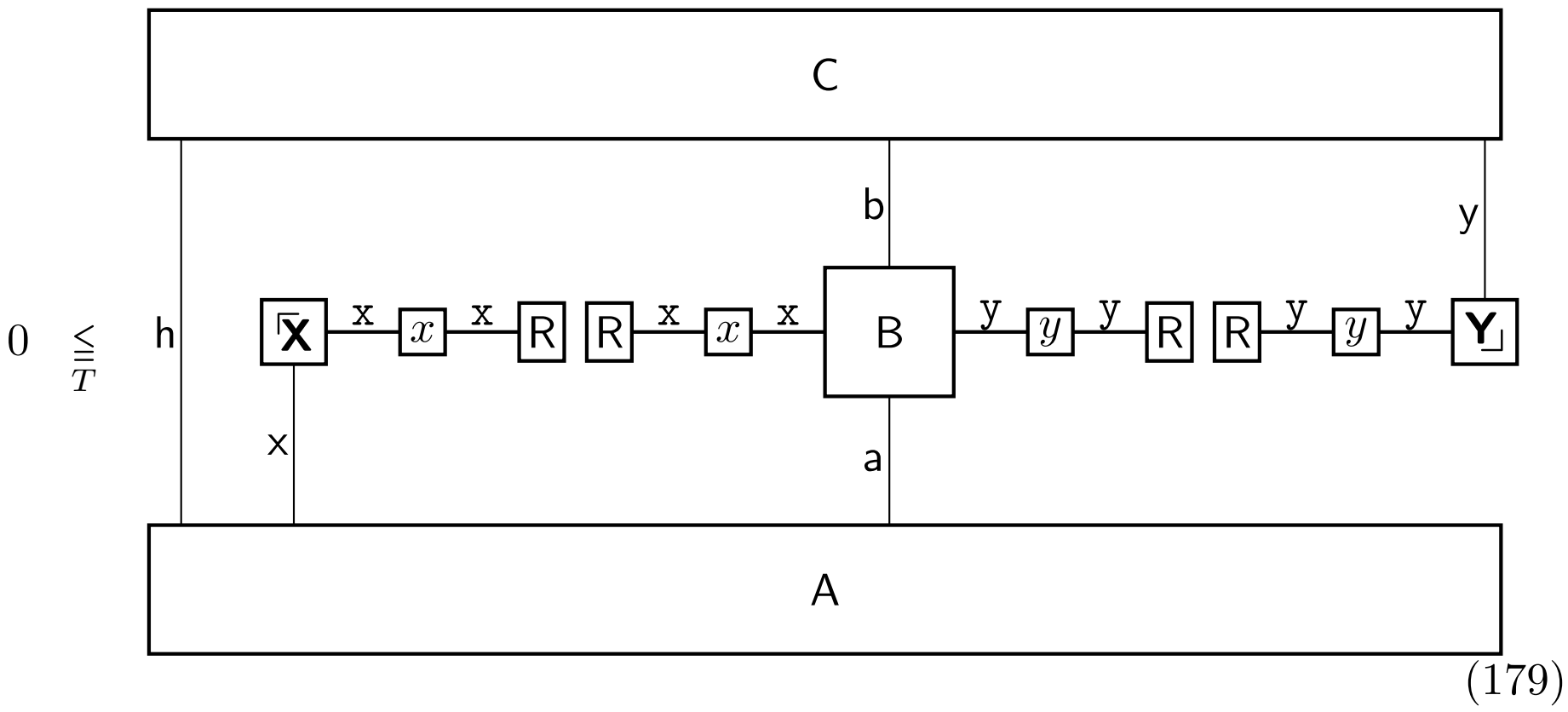

(179)

for all $x$, $y$, $\mathsf{A}$, and $\mathsf{B}$ (we have "pulled" a maximal result, readout box, and $R$ box out of each of $\mathsf{F}$ and $\mathsf{G}$ which we can do since the latter encompass all allowed preparations and results respectively). If we sum this over $x$ and $y$, then, using (130) and (131) we obtain

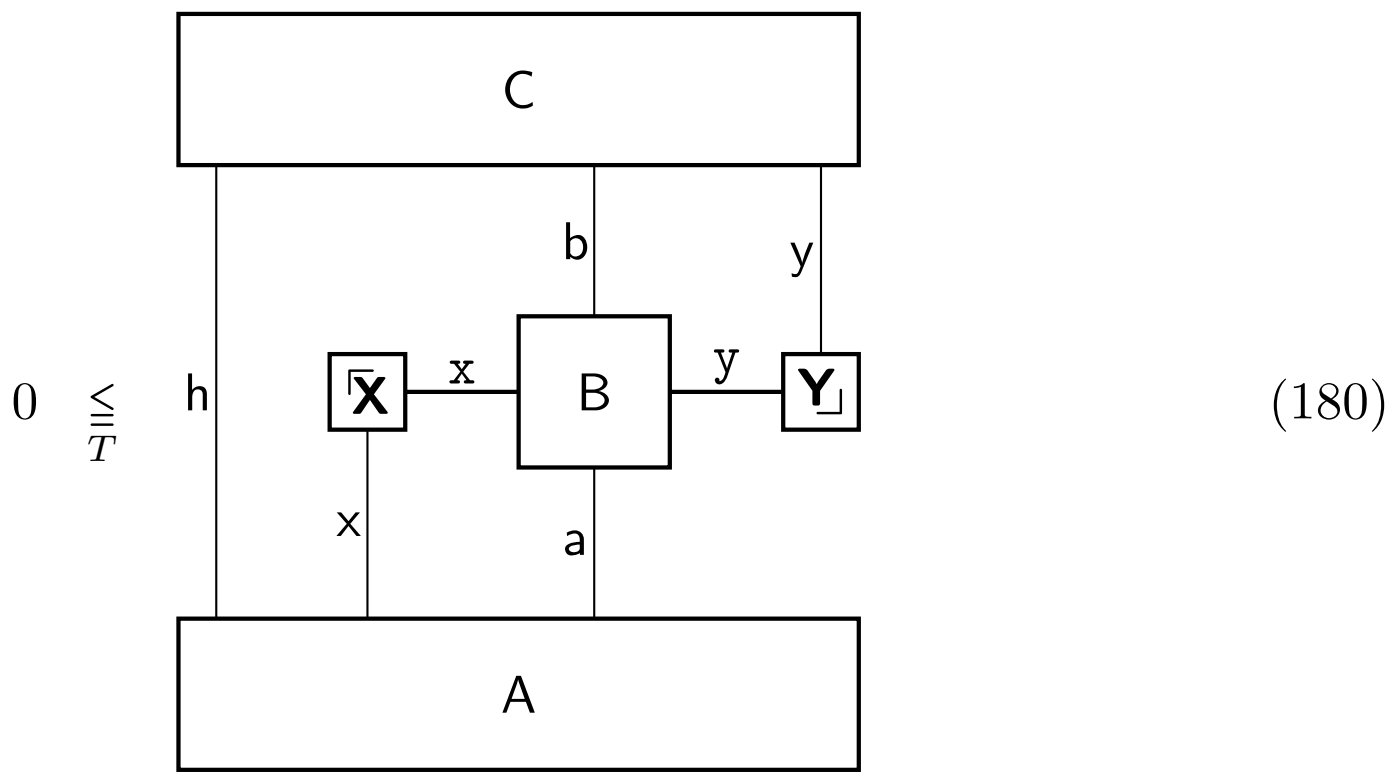

(180)

for all $\mathsf{A}$, and $\mathsf{B}$ which is the condition for $T$-positivity of $\llcorner\mathsf{B}\urcorner$. To prove tester positivity of $\llcorner\mathsf{B}\urcorner$ implies tester positivity of $\mathsf{B}$ we start with (180). By appropriate choice of A and B, this implies for all $x$, $y$, F and G. By use of (171) this gives us (178) for all F and G. Statement (iii) in the theorem concerning the physical norm (defined in (169)) follows using (174, 175). Finally, physicality consists of the conditions that tester positivity and general double causality are satisfied so it follows that $_\llcorner\mathsf{B}^\urcorner$ is physical if and only if B is physical (and, from point (iii), that they have the same physical norm).

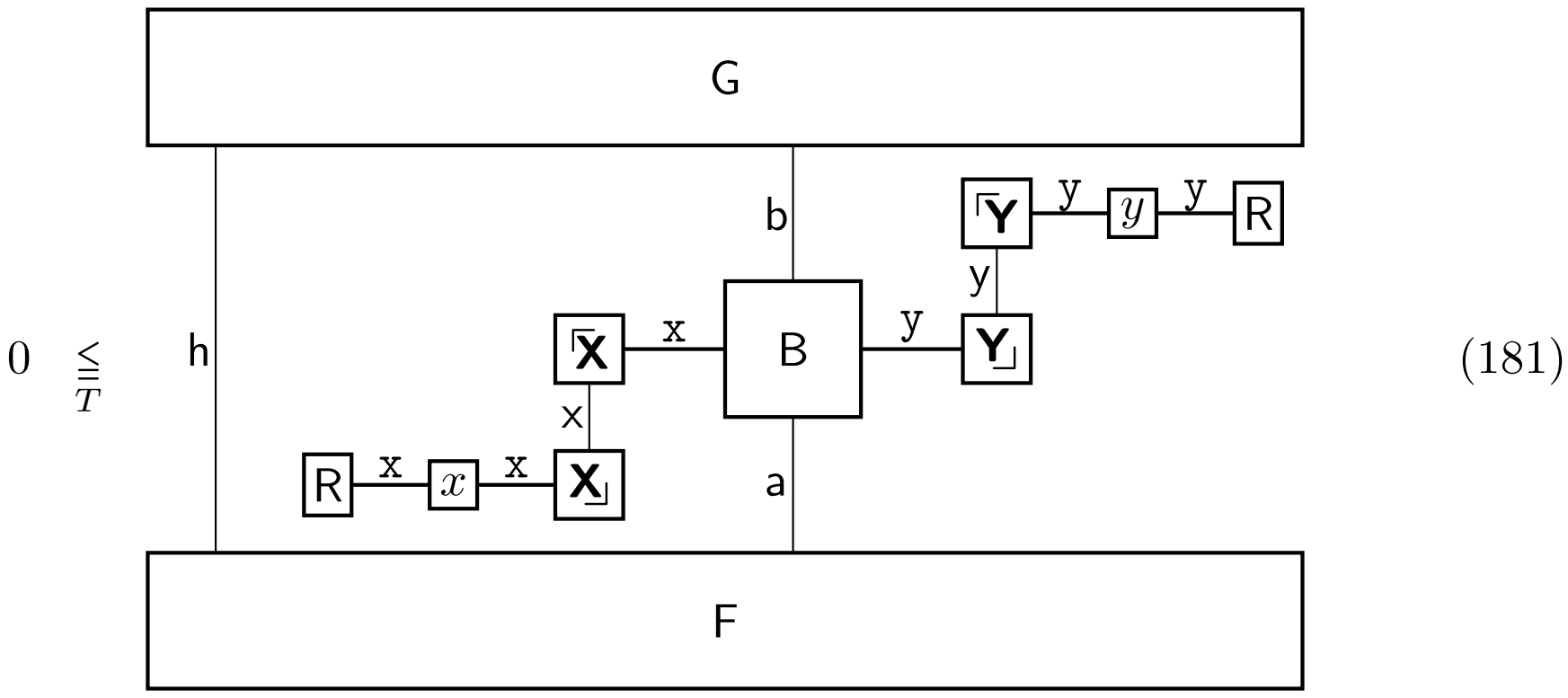

## 8.4 Preparations and results from maximal elements

We can form the following preparations and results from maximal elements

R —x— $x$ —x— X (with output wire x) R (182)

We will call these maximal preparations and results.

We can prove the following theorem

> **Maximal preparations and results theorem.** Any physical preparation or result of the form
>
> 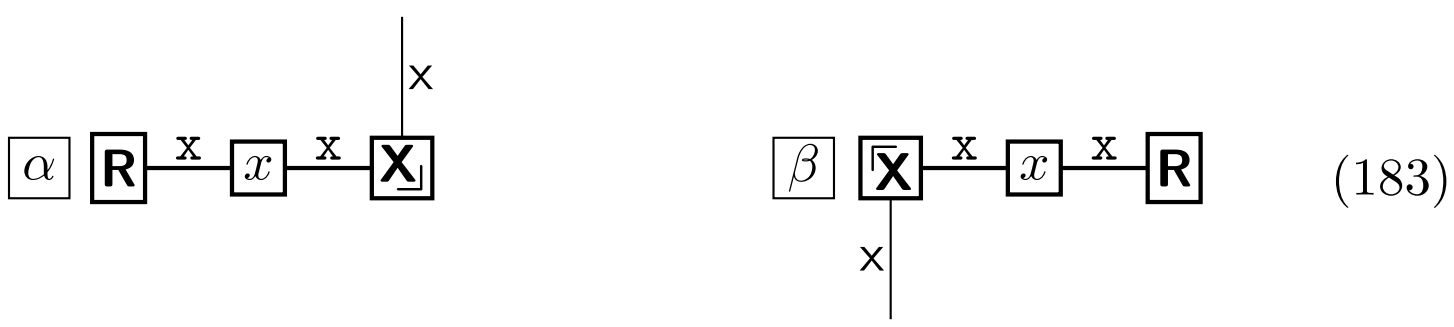
>
>
> (for non-negative $\alpha$ and $\beta$) must have $\alpha, \beta \leq 1$.

In words, there do not exist physical "longer" preparations (results) that are proportional to the maximal preparations (results). We will prove this for the preparation case (the result case follows similarly). Imposing general backward causality (using (161)) we obtain

$$\boxed{\alpha}\ \boxed{\mathbf{R}} \overset{\mathsf{x}}{—} \boxed{x} \overset{\mathsf{x}}{—} \boxed{\mathbf{X}_{\lrcorner}}^{\,\mathsf{x}} \ \underset{T}{\leqq}\ \boxed{\mathbf{I}}^{\,\mathsf{x}} \ \equiv\ \boxed{\mathbf{R}} \overset{\mathsf{x}}{—} \boxed{\mathbf{X}_{\lrcorner}}^{\,\mathsf{x}} \tag{184}$$

where the equality on the right comes from (174). If we apply the maximal result (the right expression in (182)) to both sides of this we obtain

$$\begin{matrix} & & & \boxed{\mathbf{X}} \overset{\mathsf{x}}{—} \boxed{x} \overset{\mathsf{x}}{—} \boxed{\mathbf{R}} \\ & & & \big|\,\mathsf{x} \\ \boxed{\alpha}\ \boxed{\mathbf{R}} \overset{\mathsf{x}}{—} \boxed{x} \overset{\mathsf{x}}{—} & & & \boxed{\mathbf{X}} \end{matrix} \ \leq\ \begin{matrix} & \boxed{\mathbf{X}} \overset{\mathsf{x}}{—} \boxed{x} \overset{\mathsf{x}}{—} \boxed{\mathbf{R}} \\ & \big|\,\mathsf{x} \\ \boxed{\mathbf{R}} \overset{\mathsf{x}}{—} & \boxed{\mathbf{X}} \end{matrix} \tag{185}$$

Using (171) and (43) we see that $\alpha \leq 1$.

In view of the above theorem, it is striking that we have

$$\begin{matrix} & \boxed{\mathbf{X}} \overset{\mathsf{x}}{—} \boxed{x'} \overset{\mathsf{x}}{—} \boxed{\mathbf{R}} \\ & \big|\,\mathsf{x} \\ \boxed{\mathbf{R}} \overset{\mathsf{x}}{—} \boxed{x} \overset{\mathsf{x}}{—} & \boxed{\mathbf{X}} \end{matrix} \ \equiv\ \frac{1}{N_{\mathsf{a}}}\delta_{xx'} \tag{186}$$

In other words, the probability of a maximal preparation followed by the associated maximal result is equal to $\frac{1}{N_{\mathsf{a}}}$. In the time symmetric frame we cannot have longer physical preparations and results that take this probability up to 1.

# 9 Fiducial operation expansion

## 9.1 Introduction

The assumption of *decomposition locality* is that we can expand an arbitrary operation in terms of fiducial elements on the wires. Diagrammatically, this means we can write

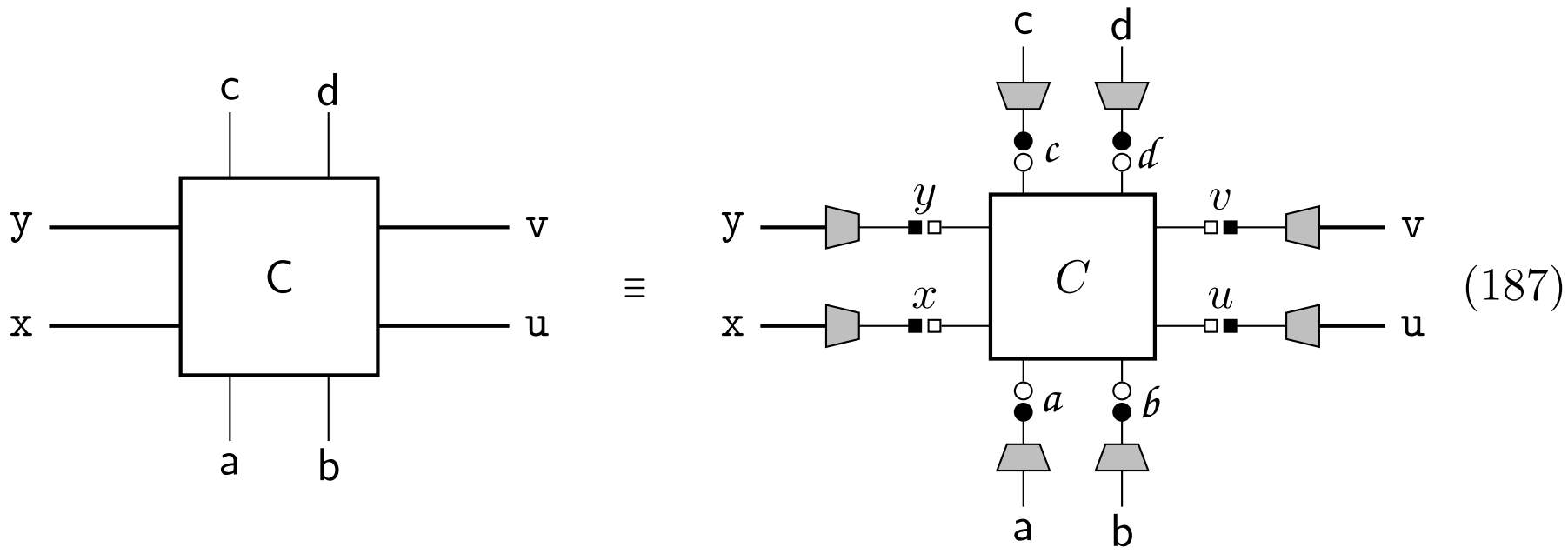

where the matched black and white dot pairs represent a sum over the associated label (e.g. $x$). The little shaded boxes are the fiducial elements. The $C$ box in the middle whose legs have white dots on is an example of a *duotensor*. It is possible to change the colour of any or all of these dots to black. All these elements (fiducials, duotensors, black and white dots) will be explained in this section.

We will see that decomposition locality is equivalent to the property of *process tomographic locality* wherein an operation can be fully characterised (up to its equivalence class) by making local measurements on each of the incomes/outcomes/inputs/outputs. Further, these assumptions are equivalent to what is usually called *tomographic locality* - that it is possible to fully characterise a preparation (up to its equivalence class) by making local measurements on the outcomes and outputs. We will call this assumption *forward tomographic locality*. A similar equivalent assumption is *backward tomographic locality* defined for results.

There is another assumption which we call *wire decomposition locality* which is also equivalent to decomposition locality. This very simple assumption pertains to the decomposition of a wire in terms of fiducials.

We begin by introducing fiducial elements for pointer types and system types. Along the way we will define the *fiducial matrix* and its inverse which can be used to change the colours of dots.

## 9.2 Fiducial pointer elements

We say a set of pointer preparations for a given pointer type is fiducial if (i) an arbitrary pointer preparation for this pointer type can be written as being equivalent a weighted sum of these pointer preparations and (ii) the set is minimal in the sense that there does not exist any other set having fewer elements for which the same is true. We know from (119) that there are $N_{\mathtt{x}}$ elements in a fiducial set (where $N_{\mathtt{x}}$ is the number of values of $x$). Thus we can label fiducial elements by $x$. Diagrammatically we denote the fiducial preparations associated with $\mathtt{x}$ by

$$x \blacksquare\!\!-\!\!-\!\!\lhd\!\!-\!\!-\, \mathtt{x} \tag{188}$$

The wire with the square black dot, labeled by $x$, indicates which element of the fiducial set we have. We can expand an arbitrary pointer preparation as a weighted sum over fiducial elements

$$\boxed{\mathsf{E}}\!\!-\!\!-\!\!-\!\!-\, \mathtt{x} \quad \equiv \quad \boxed{E}\!-\!\overset{x}{\blacksquare\square}\!-\!\!\lhd\!\!-\!\!-\, \mathtt{x} \tag{189}$$

where

$$\boxed{E}\!-\!\!\square x \tag{190}$$

is our first example of a *duotensor*. It provides the weights in the sum over fiducials in (189) and can be associated with the pointer preparation. The adjacent white and black square dot in (189) indicate that we are taking the sum over $x$.

We can similarly introduce fiducial results

$$\mathsf{x} \text{---}\square\text{---}\blacksquare\, x \tag{191}$$

such that an arbitrary pointer result can be expanded as follows

$$\mathsf{x}\text{------}\boxed{\mathsf{F}} \quad\equiv\quad \mathsf{x}\text{---}\square\text{---}\blacksquare\square\text{---}\boxed{F} \tag{192}$$

Here we have a duotensor associated with the result providing the weights in the expansion.

The natural choice for the fiducial pointer elements is

$$\boxed{\mathbf{R}}\overset{\mathsf{x}}{\text{---}}\boxed{x}\overset{\mathsf{x}}{\text{---}} \qquad\qquad \overset{\mathsf{x}}{\text{---}}\boxed{x}\overset{\mathsf{x}}{\text{---}}\boxed{\mathbf{R}} \tag{193}$$

for each given value of $x$. By the pointer classicality property (introduced in Sec. 7.9) these constitute a complete set of pure preparations and results respectively. We know from (119) and (121) that these can be used to expand arbitrary pointer preparations and results.

## 9.3 Fiducial matrices

An important object, associated with any given pointer type, is the *fiducial matrix* defined as

$$\blacksquare\overset{x}{\text{------}}\blacksquare \;=\; \text{prob}\left( x\,\blacksquare\text{---}\square\overset{\mathsf{x}}{\text{---}}\square\text{---}\blacksquare\, x \right) \tag{194}$$

The fiducial matrix is non-singular since we the fiducial elements are chosen to be minimal (see Sec. 9.2). Recall that a circuit is equivalent to its own probability (see (62)). Hence we can write

$$\blacksquare\overset{x}{\text{------}}\blacksquare \;\equiv\; x\,\blacksquare\text{---}\square\overset{\mathsf{x}}{\text{---}}\square\text{---}\blacksquare\, x \tag{195}$$

We denote the inverse of the fiducial matrix by

$$\square\overset{x}{\text{------}}\square \tag{196}$$

This means we have

$$\square\text{------}\square\blacksquare\text{------}\blacksquare \;=\; \square\text{------}\blacksquare \tag{197}$$

and

$$\blacksquare\text{------}\blacksquare\square\text{------}\square \;=\; \blacksquare\text{------}\square \tag{198}$$

where $\square\text{---}\blacksquare$ and $\blacksquare\text{---}\square$ are the identity matrix.

We can use the fiducial matrix or its inverse to change the colour of the dots as follows

$$\text{---}\blacksquare\square\text{---}\square = \text{---}\square \qquad \text{---}\square\blacksquare\text{---}\blacksquare = \text{---}\blacksquare$$

$$\square\text{---}\square\blacksquare\text{---} = \square\text{---} \qquad \blacksquare\text{---}\blacksquare\square\text{---} = \blacksquare\text{---}$$

This is clearly consistent since we can undo the colour change by application of the fiducial or its inverse as appropriate. We have

$$\text{---}\blacksquare\,\square\text{---} \;=\; \text{---}\blacksquare\,\square\text{---}\square\,\blacksquare\text{---}\blacksquare\,\square\text{---} \;=\; \text{---}\square\,\blacksquare\text{---} \tag{199}$$

by inserting the identity from (197). Consequently we can write

$$\text{---}\blacksquare\,\square\text{---} \;=\; \text{---}\square\,\blacksquare\text{---} \;:=\; \text{------} \tag{200}$$

The interpretation here is that summing over a black/white dot pair is the same as summing over a white/black dot pair. Consequently we can denote either simply by a wire as on the right. If we do this then, to evaluate the sum we must insert either a black/white dot pair or a white/black dot pair (we get the same answer either way).

If we choose the fiducial elements to be given by (193) and assume flatness for the $\mathbf{R}$'s (i.e. the property in (85)) then, using (43), we see that the fiducial matrix is proportional to the identity matrix

$$\blacksquare\overset{x}{\text{------}}\blacksquare \;=\; \frac{1}{N_{\mathsf{x}}}\begin{pmatrix} 1 & & & \\ & 1 & & \\ & & \ddots & \\ & & & 1 \end{pmatrix} \tag{201}$$

and so is the inverse of the fiducial matrix

$$\square\overset{x}{\text{------}}\square \;=\; N_{\mathsf{x}}\begin{pmatrix} 1 & & & \\ & 1 & & \\ & & \ddots & \\ & & & 1 \end{pmatrix} \tag{202}$$

So, with these choices for the fiducial element (and assuming flatness) changing a square dot from black to white simply corresponds to multiplication by $N_{\mathsf{x}}$ while changing it from white to black corresponds to multiplication by $1/N_{\mathsf{x}}$.

## 9.4 Fiducial system elements

We can proceed in an exactly analogous manner for physical systems. Here we use round rather than square dots. Fiducial system preparations and fiducial system results are denoted by

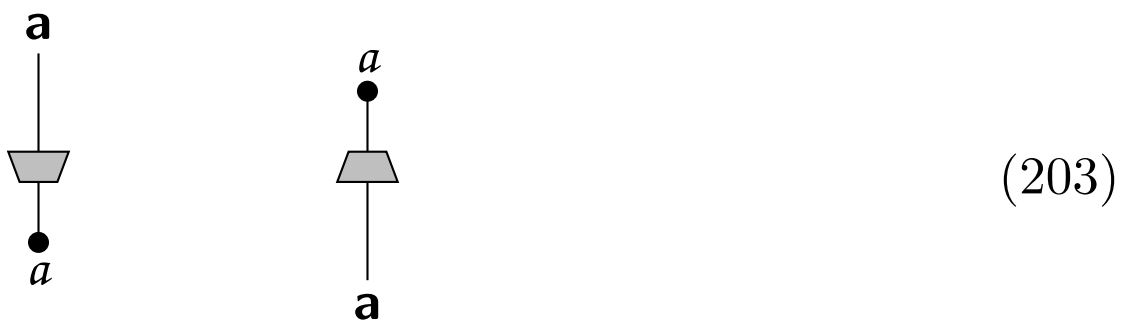

(203)

where $a = 1$ to $K_{\mathsf{a}}$ labels the fiducial elements. The integer, $K_{\mathsf{a}}$ is the number of fiducial elements required for this system type. In Quantum Theory, for example, we have $K_{\mathsf{a}} = N_{\mathsf{a}}^2$ where $N_{\mathsf{a}}$ is the dimension of Hilbert space for this type of system.

We can expand general system preparations and system results as a weighted sum over fiducial elements as follows

$$\text{a} \;\; \text{A} \quad \equiv \quad \text{a} \;\; a \;\; A \qquad\qquad \text{B} \;\; \text{a} \quad \equiv \quad \text{B} \;\; a \;\; \text{a} \tag{204}$$

The weights are given by duotensors as in the pointer case.

## 9.5 Fiducial matrices

We can define a *fiducial matrix* as follows

$$a \quad = \quad \text{prob}\left( a \;\; \text{a} \;\; a \right) \quad \equiv \quad a \;\; \text{a} \;\; a \tag{205}$$

This object was originally introduced in Hardy [2001] to aid the reconstruction of Quantum Theory. It was used in a structurally deeper way in Hardy [2013a] to set up the duotensor formalism. There is was called the *hopping metric*. Fuchs and Schack [2010] use a similar object (though defined using an informationally complete quantum measurements) is used which is later called the *Born matrix* in the QBism literature (see Stacey [2022]). In non-classical theories (such as Quantum Theory) the fiducial matrix cannot be proportional to the identity. Further, the inverse fiducial matrix

$$a \tag{206}$$

can have negative entries.

Clearly we have

$$= \qquad \text{and} \qquad = \tag{207}$$

where the object on the right hand side of each equation is the identity matrix.

We can use the fiducial matrix to change the colour of dots (as in the pointer case). Further, we have

$$= \qquad := \tag{208}$$

for the same reason that we have (200).

## 9.6 Simple circuits

The simplest circuit we can consider is of the form

(209)

We can calculate the probability for this circuit by substituting in (204) to give

(210)

where we shown explicitly four different ways of writing down the scalar expression. Since circuits are equivalent to their probabilities we see the probability is given by the expression on the right which is entirely in terms of duotensors. We see that the calculation for the probability of this simple circuit is given by a mathematical expression that looks the same as the circuit it is a probability for. We will see that is true for general circuits if we adopt the assumption of decomposition locality. This was called *the composition principle* in Hardy [2013b].

## 9.7 Tomographic locality

A well known assumption in the literature on operational probabilistic theories goes by the name *tomographic locality*. This is the assumption that we can fully characterise preparations (up to equivalence classes) by making separate measurements on the outputs and outcomes. Here we will call this *forward tomographic locality* to emphasise that it is time asymmetric. It turns out that this assumption is equivalent (as long as we stick to the case of finite fiducial dimension) to several other assumptions most of which are time symmetric. In particular, forward tomographic locality is equivalent to the assumption of decomposition locality (see (204)) which we encountered in the introduction to this section. In this subsection we will state these equivalent assumptions and prove that they are equivalent.

We will refer colloquially to these equivalent assumptions as "tomographic locality" (though each of these assumptions will have its own name). We will discuss the history and recent work on tomographic locality in Sec. 9.8.

### 9.7.1 Fiducial dimension super-multiplicativity

The fiducial dimension associated with a given system is defined to be the (minimum) number of fiducial preparations (or results) required so that we can write down a general preparation (or result) as in (204). So, for a system of type $\mathsf{a}$ it is equal to $K_{\mathsf{a}}$ and for a pointer of type $\mathtt{x}$ it is equal to $N_{\mathtt{x}}$.

A basic result for fiducial dimensions follows from the factorisation property for disjoint circuits (which is, itself, a consequence of the very basic circuit probability assumption - see Sec. 6.1).

> **Fiducial dimension super-multiplicativity.** The fiducial dimension of a composite system is greater than or equal to the product of the fiducial dimensions of its components.

We will write the fiducial dimension of a composite system $\mathsf{abxy}$ (for example) as $K_{\mathsf{abxy}}$. With this example, the assumption of fiducial dimension super-multiplicativity says that

$$K_{\mathsf{abxy}} \geq K_{\mathsf{a}} K_{\mathsf{b}} K_{\mathtt{x}} K_{\mathtt{y}} \tag{211}$$

For notational homogeneity we will write as $N_{\mathtt{x}} = K_{\mathtt{x}}$ for pointer types (this means that the font really matters as, in general, $K_{\mathtt{x}} \neq K_{\mathsf{x}}$). We will prove the above theorem for the case of a composite system, $\mathsf{ab}$. It is then obvious how to generalise this proof to arbitrary composite systems. Consider a product preparation

(212)

We can complete this into a circuit by a product result

(213)

This circuit factorises into two disjoint circuits and therefore, by the aforementioned factorisation property, so do the probabilities. Hence

(214)

(using the result in (210)). There are $K_{\mathsf{a}} K_{\mathsf{b}}$ linearly independent objects of the form

(215)

Hence we need at least $K_{\mathsf{a}}K_{\mathsf{b}}$ fiducial results to characterise a general preparation since we need this many in the special case of product preparations. It is easy to generalise this proof to arbitrary many systems including pointers. This proves the above theorem.

The assumption that the bound in the above assumption is saturated - so the fiducial dimension of a composite system is greater than or equal to the product of the fiducial dimensions of its components - will be called *fiducial dimension multiplicativity*. This assumption is equivalent to decomposition locality. This will be proven in Sec. 9.7.6.

### 9.7.2 Decomposition locality

Now we have defined fiducial elements we return to the decomposition locality assumption mentioned, by way of motivation, in the introduction.

> **Decomposition locality.** We can expand an arbitrary operation as being equivalent to a weighted sum over the fiducial elements as illustrated in

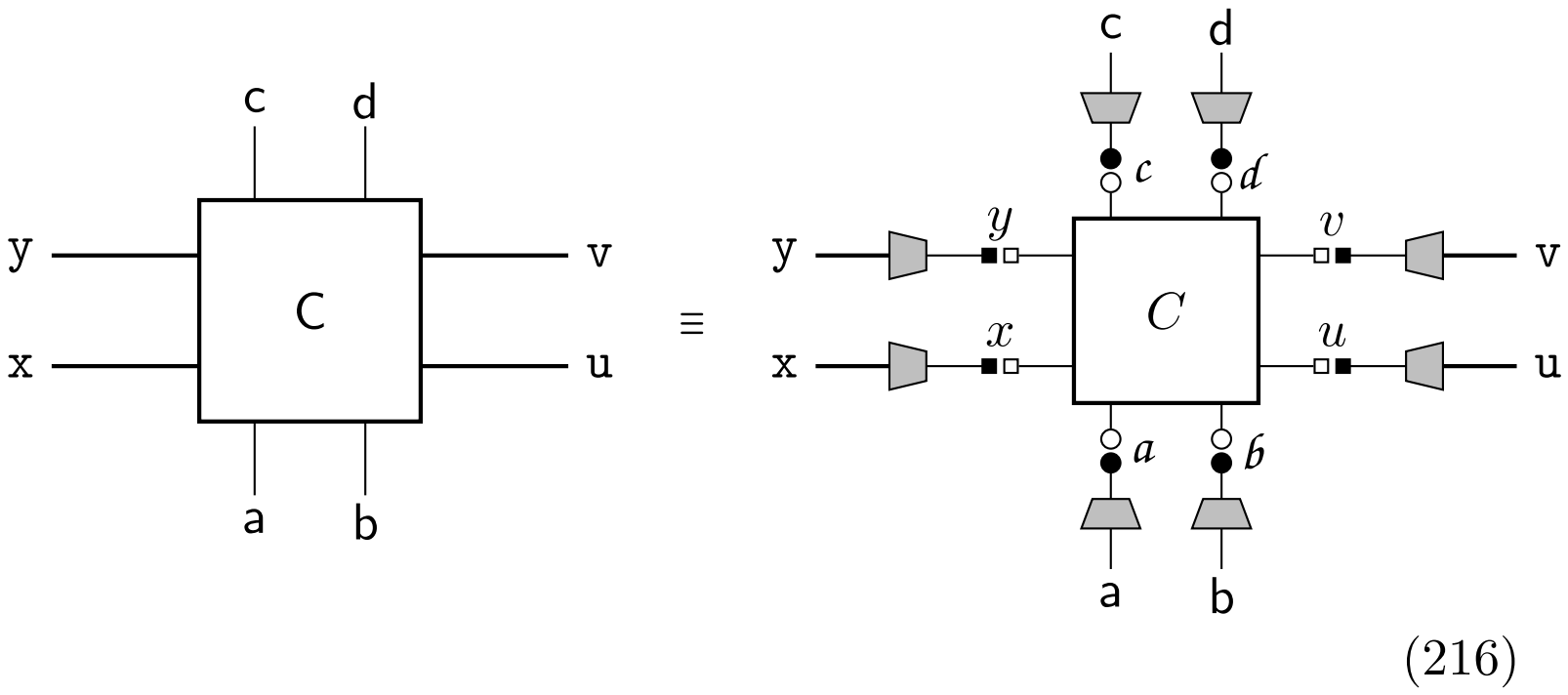

(216)

> Note that the duotensor providing the weights has all white dots.

This assumption was originally introduced in Hardy [2011b], though with only system fiducial elements. There it was called *full decomposability*.

The name duotensor is apt because the duotensor transforms as a tensor when we change fiducial sets and the “duo-” prefix is appropriate because we have a black or white dot on each wire. We can change the colour of any dot using the fiducial matrix or its inverse.

### 9.7.3 Process tomographic locality

We can turn (216) into a circuit by attaching fiducial elements to each input, output, income, and outcome. Using (194) and (205) we obtain

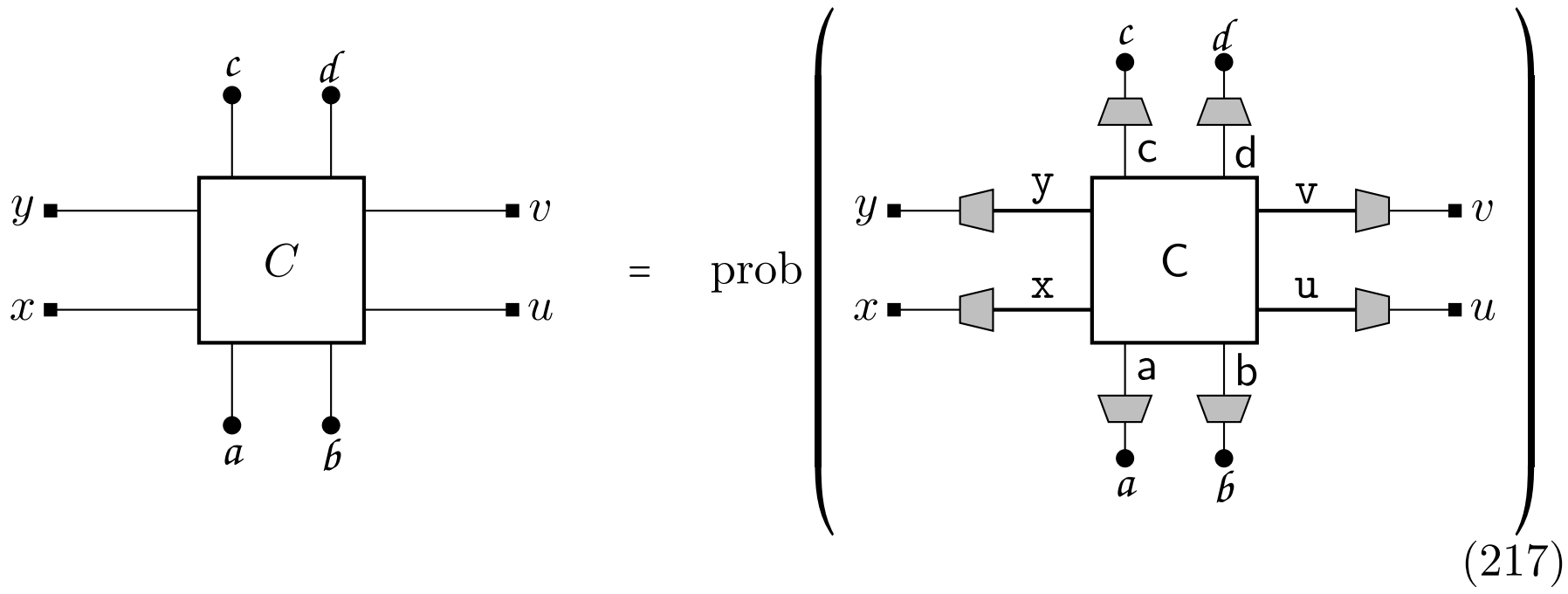

(217)

Thus, we obtain the duotensor with all black dots (therefore a duotensor with all black dots has probabilities as entries). The probability matrix on the right hand side of (217) corresponds to doing local process tomography. Now we can introduce the following assumption

> **Process tomographic locality.** This is the assumption that local process tomography is sufficient to fully characterise operations (up to equivalence classes).

Thus, two operations are equivalent if and only if they have the same (all black dots) duotensor.

It is easy to see that tomographic locality and decomposition locality are equivalent assumptions. This follows because we can change all the dots on the duotensor from black to white. Then the duotensor can be used as expansion coefficients as in (216). To see that decomposition locality and process tomographic locality are equivalent note that, according to the former, two operations are equivalent if and only if they have the same duotensor.

Duotensors with all black dots always correspond to probabilities. Duotensors with all white dots correspond to the weights in a fiducial expansion. We can also have duotensors with mixed black and white dots. If we have a duotensor having only incomes and outcomes and where, further, the incomes are all white (square) dots and the outcomes are all black (square) dots then this corresponds to conditional probabilities of seeing the outcome conditioned on the given income. Once we have inputs and outputs as well then this interpretation breaks down since we can have duotensors with negative entries (certainly this is true in the quantum case).

### 9.7.4 Wire decomposition locality

Surprisingly, there is an assumption equivalent to decomposition locality which only involves wires. In decomposition local theories it turns out we can write

(218)

and, further, this can be used to prove decomposition locality. In fact, we can start by assuming a bit less than this. We state the following assumption

> **Wire decomposition locality.** Wires can be decomposed in terms of fiducial elements as follows
>
> (219)
>
> for some $V$ and $W$ for all x and a.

We will prove this is equivalent to decomposition locality. By thinking of wires as operations

(220)

it is clear that wire decomposition locality follows immediately from decomposition locality (simply because the former is an example of the latter). To prove that decomposition locality follows from wire decomposition locality we need, first, to show that the duotensors $V$ and $W$ are equal to the inverse fiducial matrices (so then (218) follows). To prove this note

(221)

Hence, using (195),

(222)

and so

$$x \,\square\!-\!\boxed{W}\!-\!\square\, x \quad \equiv \quad \square\!\overset{x}{\text{———}}\!\square \tag{223}$$

(where we have applied (197) on the left and (198) on the right). This gives us the simple wire decompositions in (218). Now we have these we can write

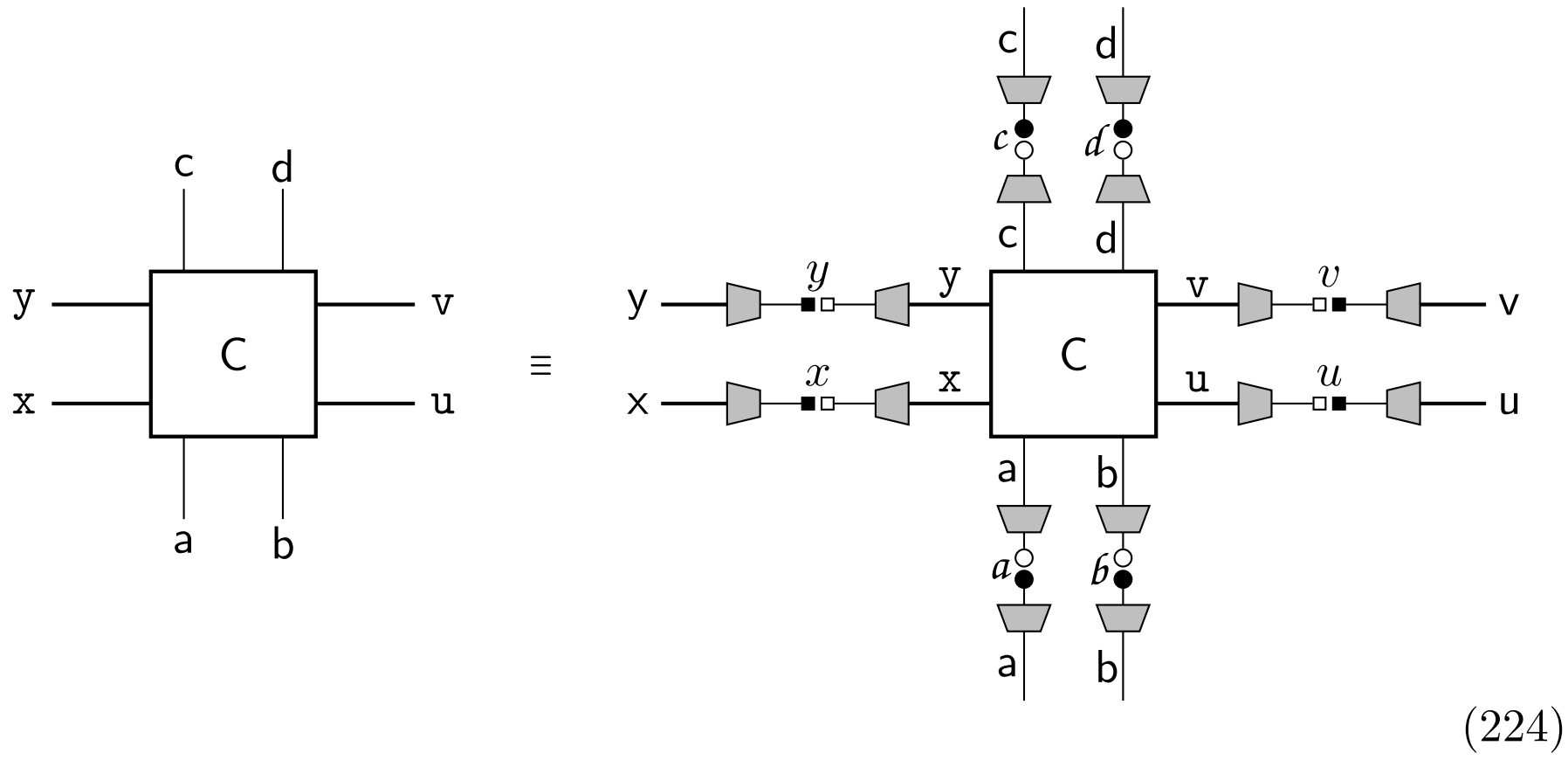

(224)

This is, indeed, the assumption of decomposition locality where the object with white dots in the middle is equivalent to the duotensor with white dots in (216). This completes the proof that wire decomposition locality and decomposition locality are equivalent assumptions.

Wire decomposition locality provides a useful way to obtain a calculation for the probability of a circuit. This will be discussed in Sec. 10.2.

### 9.7.5 Forward and backward tomographic locality

Usually, the assumption of tomographic locality invoked in the literature is actually a time asymmetric assumption. Given the emphasis of this book on time symmetry we will rename this assumption calling it forward tomographic locality.

> **Forward tomographic locality.** This is the assumption that we can fully characterise preparations (up to equivalence classes) by doing local tomography on the outputs and outcomes.

Interestingly this assumption is equivalent to the process tomographic locality assumption even though it only appears to assume "half as much". We will prove this below. Forward tomographic locality tells us that the preparation

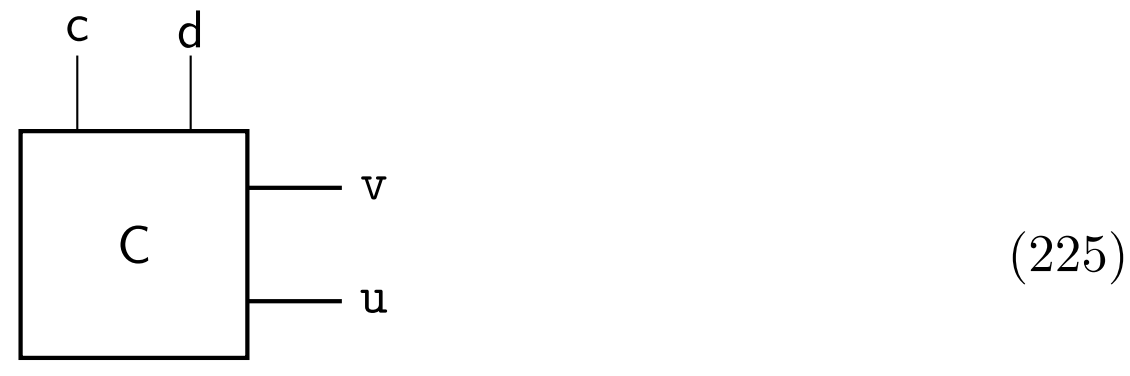

(225)

is fully characterised by the probabilities

$c$ $d$ $C$ $v$ $u$ = prob ( $c$ $d$ c d C v u $v$ $u$ ) (226)

Forward tomographic locality is equivalent to the following

> **Forward decomposition locality.** This is the assumption that we can expand an arbitrary preparation as being equivalent to a weighted sum over the fiducial preparations as illustrated in the example

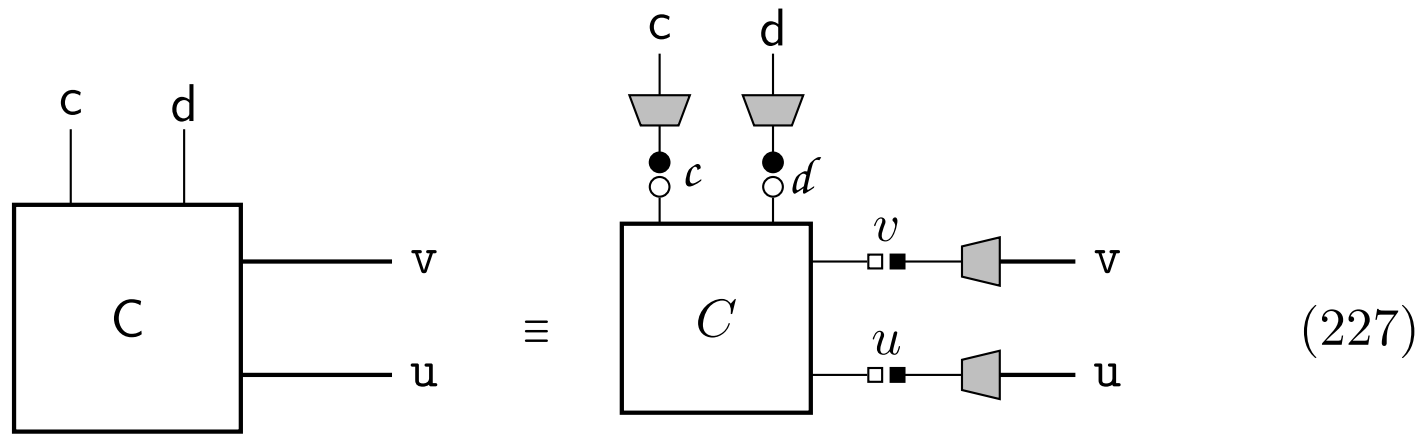

(227)

The above two assumptions are equivalent for the same reason that decomposition locality and process tomographic locality are equivalent - we can change the dots in (226) from black to white (using the inverse fiducial matrix). In this way we obtain the duotensor

$c$ $d$ $C$ $v$ $u$ ≡ $c$ $d$ c d C v u $v$ $u$ (228)

We can use this to provide the expansion coefficients as in the right hand side of (227). Then we see that two operations are equivalent if and only if they have the same duotensor thus proving equivalence of the above two assumptions. If we substitute (228) into (227) we obtain

c d C v u ≡ c d $c$ $d$ c d C v u $v$ $u$ v u (229)

as a useful way of writing down the forward decomposition assumption.

Now we will prove that forward decomposition locality is equivalent to decomposition locality (which amounts to the same thing as proving that forward tomographic locality is equivalent to process tomographic locality). First note that it is immediately clear that decomposition locality implies forward tomographic locality since the former, applied to preparations, yields the latter. Now we need to prove the other direction. To keep the diagrams simple we will only draw system wires (it is easy to include pointer wires too but, since we have adopted the convention that they come out the sides of the boxes, it makes the diagrams messy to include them). We will prove that the forward decomposition locality implies that the operation

(230)

satisfies decomposition locality (we just draw two input and two output wires but the proof obviously generalises to any numbers). According to forward decomposition locality as expressed in (229), we can write

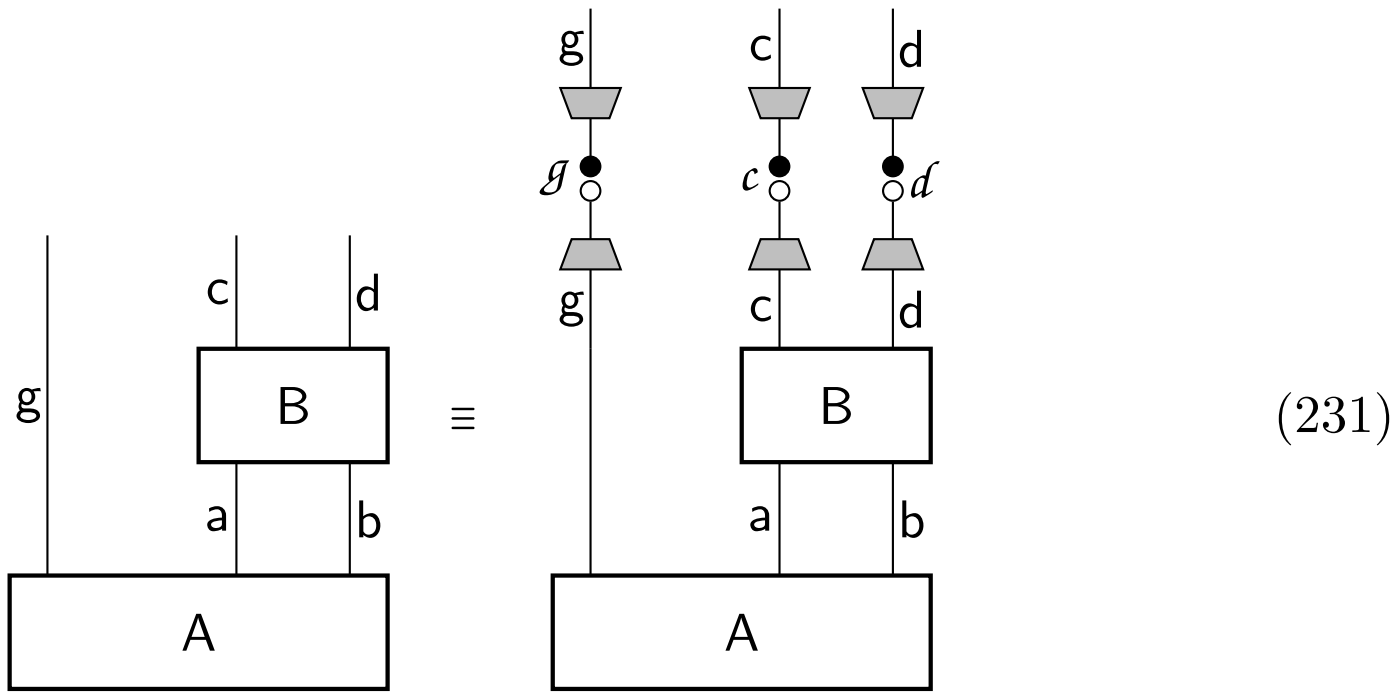

(231)

Using (229) again we can also write

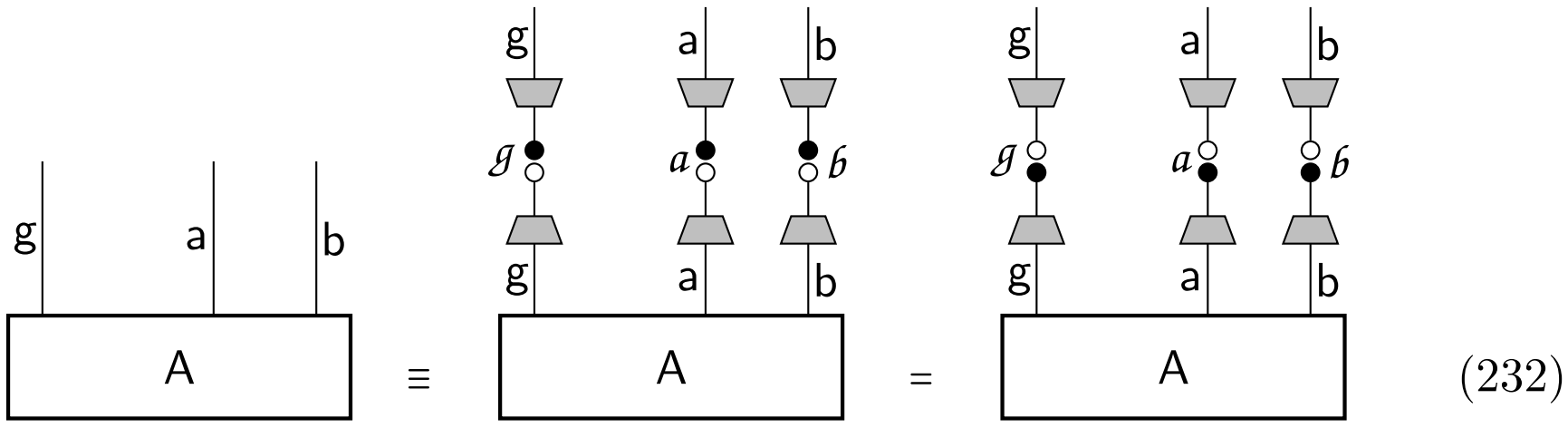

(232)

(where we have swapped the order of the black and white dots in the last line (using (208)). Using these results we obtain

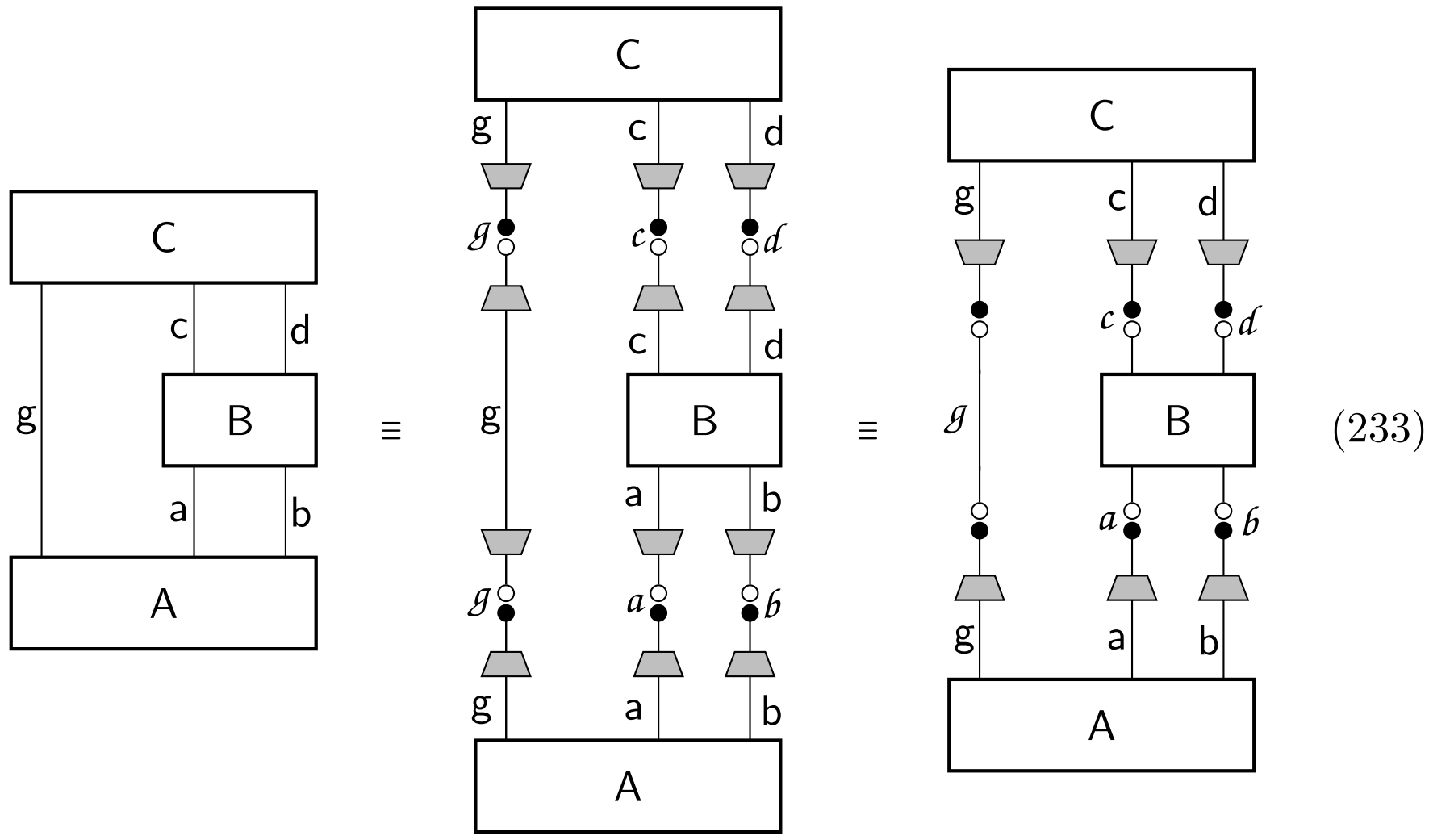

(233)

by substituting (232) into (231), then substituting the resulting expression into the circuit on the left. This is the most general circuit that B can appear in. Hence, we can consistently use

(234)

The equivalence on the right is decomposition locality as given in (230). The equivalence on the left is wire decomposition locality in the form in (218) which can, indeed, be regarded as an application of decomposition locality. This proves that decomposition locality follows from forward decomposition locality. It is curious that we can derive decomposition locality and wire decomposition locality as a "package deal".

We can also state backward versions for these assumptions.

> **Backward tomographic locality.** This is the assumption that we can fully characterise results (up to equivalence classes) by doing local tomography on the inputs and incomes.

which is equivalent to

**Backward decomposition locality.** This is the assumption that we can expand an arbitrary result as being equivalent to a weighted sum over the fiducial results as illustrated in the example

$$\text{(235)}$$

These assumptions can, clearly, be shown to be equivalent to process tomographic locality by the same techniques as we just employed with the forward cases.

### 9.7.6 Fiducial dimension multiplicativity

Recall, from Sec. 9.4, that $K_{\mathsf{a}}$ is the number of fiducial elements associated with a system of type $\mathsf{a}$ (and therefore $K_{\mathsf{a}}$ is the dimension of the space of preparations or of results). We can make the following assumption

**Fiducial dimension multiplicativity** is the property that

$$K_{\mathsf{ab}} = K_{\mathsf{a}} K_{\mathsf{b}} \tag{236}$$

It can be shown that this assumption is equivalent to the other assumptions. To prove this first we note that

$$K_{\mathsf{ab}} \geq K_{\mathsf{a}} K_{\mathsf{b}} \tag{237}$$

as was shown in Sec. 9.7.1. We can prove that this inequality is saturated using forward tomographic locality assumption. We want to prove that we only need $K_{\mathsf{a}} K_{\mathsf{b}}$ product results to characterise general preparations. To prove this we will assume the contrary and prove a contradiction. Consider doing tomography to characterise a preparation $\mathsf{A}$ and assume that we need at least one more fiducial product result, $E_{\mathsf{a}} F_{\mathsf{b}}$ (in addition to the $K_{\mathsf{a}} K_{\mathsf{b}}$ fiducial results we already have). Then we have

$$\text{(238)}$$

where we make the first step by doing tomography on the preparation comprised of operations $\mathsf{A}$ and $\mathsf{F}$, and we make the second step by doing tomography on the system comprised of $\mathsf{A}$ and the fiducial result acting on $\mathsf{a}$. Thus, we see that the additional product result, $\mathsf{EF}$ is fully characterised by a sum over the existing $K_{\mathsf{a}}K_{\mathsf{b}}$ product fiducial results and so is not necessary. This completes the proof. We can extend this proof multiplicative property to composite systems consisting of any (finite) number of both system and pointer types. For example we have

$$K_{\mathsf{abxy}} = K_{\mathsf{a}}K_{\mathsf{b}}K_{\mathsf{x}}K_{\mathsf{y}} \tag{239}$$

This can be proven using the above techniques. Recall that, the fiducial dimension for pointer systems is $K_{\mathsf{x}} = N_{\mathsf{x}}$. Consequently fiducial dimension multiplicativity gives

$$N_{\mathsf{xy}} = N_{\mathsf{x}}N_{\mathsf{y}} \tag{240}$$

when applied to a composite pointer system.

### 9.7.7 Operation locality

An interesting consequence of decomposition locality is that implies that we can associate a separate mathematical object (a duotensor in our case) with each operation. Let us illustrate this with an example. Consider an operation $\mathsf{A}$ which can be regarded as being comprised of an operation $\mathsf{B}$ and an operation $\mathsf{C}$. Then, by decomposition locality, we have

(241)

This property, which we call *operation locality* (see Hardy [2011a]), allows us to separately model each operation by a duotensor. Decomposition locality implies this very natural property holds. The converse does not appear to be true. Thus, we can imagine that there are theories that satisfy operation locality but not decomposition locality.

## 9.8 History and modern work on tomographic locality

The property now called tomographic locality has an interesting prehistory. It was implicit in the standard formulation of quantum mechanics from its early development. In particular, the tensor-product structure introduced in the complex Hilbert space formulation of quantum theory already implies that the parameters specifying a composite state are determined by correlations between subsystem measurements Jordan et al. [1934]. Operational approaches to quantum theory developed later also emphasized describing states through probabilities of measurement outcomes (the tomographic point of view). Notable examples include the operational frameworks of Ludwig Ludwig [1985 and 1987], the measurement-theoretic formalism of Davies and Lewis Davies and Lewis [1970], and later work by Wootters [1986] and by Fivel [1994] where hints of the tomographic locality principle are found. Investigations of alternative formulations of quantum mechanics further highlighted the importance of the composite system structure: Stueckelberg showed that quantum theory can be formulated in real Hilbert space with an additional operator playing the role of the imaginary unit Stueckelberg [1960], while work on Jordan-algebraic formulations revealed that constraints on tensor products play a key role in singling out complex quantum theory Hanche-Olsen [1983]. Operational considerations about determining quantum states from measurement statistics were also explored by Wootters, who analyzed how many probabilities are required to specify a quantum state and emphasized the accessibility of global states through local measurement correlations Wootters [1990]. In this paper Wootters describes the version of tomographic locality which below fiducial dimension multiplicity (although he subtracts 1 for normalisation which slightly complicates the formulae). The principle itself, in various guises, was used explicitly as an axiom in many of the reconstruction of Quantum Theory of the present century (see Hardy [2001], Dakic and Brukner [2009], Masanes and Müller [2011], Chiribella et al. [2011], and many more) and is discussed by Barrett [2007] in his article on General Probability Theories.

Recent work has explored both the role of tomographic locality and the consequences of relaxing it. Within operational frameworks for generalized probabilistic theories, Hardy [2011a] showed that locally tomographic theories admit a particularly simple circuit representation and also described how more general theories beyond tomographic locality can be accommodated by enlarging the mathematical structure (for example through tensor sums). Hardy and Wootters [2012] investigated theories satisfying a weaker compositional property, introducing the notion of *bilocal tomography*, with real-vector-space quantum theory providing a canonical example. A variety of mechanisms leading to violations of tomographic locality have since been studied, including superselection-restricted theories such as fermionic quantum theory by D'Ariano et al. [2014], symmetry-restricted constructions by Centeno et al. [2025], and generalized probabilistic theories with modified composition rules allowing even locally classical systems to exhibit entanglement and other quantum-like phenomena by D'Ariano et al. [2020, 2021]. Related work has also explored operational theories with minimal dynamical structures and altered composition rules (see Rolino et al. [2025]). More recently, Baldijão et al. [2026] give a fascinating analysis of entanglement in tomographically nonlocal theories within the General Probability Theories framework deriving the tensor sum between local and nonlocal parts and showing that the failure of tomographic locality gives rise to distinct forms of entanglement and clarifying their operational consequences. Together these investigations illustrate that violations of tomographic locality arise naturally in a broad class of operational theories and provide useful foils for understanding the structural role of this principle in quantum theory.

## 9.9 Duotensors for special cases

There are a few special cases for which we want to write down the associated duotensor. We will use the natural choice of pointer fiducial elements in (193).

Given this natural choice of pointer fiducials (and assuming flatness for the **R**'s as in (85)) we immediately have

$$\boxed{\boldsymbol{R}}\!-\!\!-\!\blacksquare\, x \;=\; \frac{1}{N_{\mathtt{x}}}\begin{pmatrix}1\\1\\\vdots\\1\end{pmatrix} \qquad\qquad x\,\blacksquare\!-\!\!-\!\boxed{\boldsymbol{R}} \;=\; \frac{1}{N_{\mathtt{x}}}\begin{pmatrix}1\\1\\\vdots\\1\end{pmatrix}^{\top} \tag{242}$$

since the probability for each $x$ is $1/N_{\mathtt{x}}$. Note the $\top$ superscript on the right takes the transpose so we have a row vector (which is natural when the wire is on the left). Applying the inverse of the fiducial matrix we obtain

$$\boxed{\boldsymbol{R}}\!-\!\!-\!\square\, x \;=\; \begin{pmatrix}1\\1\\\vdots\\1\end{pmatrix} \qquad\qquad x\,\square\!-\!\!-\!\boxed{\boldsymbol{R}} \;=\; \begin{pmatrix}1\\1\\\vdots\\1\end{pmatrix}^{\top} \tag{243}$$

This is what we expect when we are ignoring the income/outcome (i.e. when we marginalise).

The above duotensors connect with the vectors introduced in Sec. 7.3 to represented forward and backward thinking. We will discuss them below in Sec. 10.2.1.

A readout box can be expanded in terms of fiducials as follows

$$\mathtt{x}\;\text{---}\!\rhd\!\text{---}\overset{x}{\blacksquare\square}\text{---}\boxed{x}\text{---}\overset{x}{\square\blacksquare}\text{---}\!\lhd\!\text{---}\;\mathtt{x} \tag{244}$$

Readout boxes satisfy (43). From this it is easy to show that

$$x\,\square\!-\!\boxed{x}\!-\!\blacksquare\, x \;=\; \begin{pmatrix}0 & & & & \\ & 0 & & & \\ & & \ddots & & \\ & & & 1 & \\ & & & & \ddots\end{pmatrix} \tag{245}$$

where the 1 is in the $x$ position along the diagonal. Note that the $x$'s on the wires indicate the variable type to be summed over (when this duotensor is matched with other duotensors) whilst the $x$ inside the box indicates a particular value of this variable. Here we have given the duotensor when we have a white dot on the left and a black dot on the right as this has the natural interpretation of inputting the pointer, looking to see if the pointer has the particular value, $x$ (in which case the matrix returns a 1 for the probability), then outputting the system. We can obtain the duotensor with two white dots by applying the inverse fiducial matrix - this introduces a factor $N_{\mathtt{x}}$.

# 10 Time symmetric duotensor calculations

We can use duotensors to calculate the probability associated with a circuit. We can do this either (i) by replacing all the operations in the circuit with their local decompositions or (ii) by replacing all the wires in the circuit with their wire decompositions. These two techniques lead to the same calculation.

## 10.1 Using decomposition locality

Lets consider (i) first. Consider, for example, the circuit

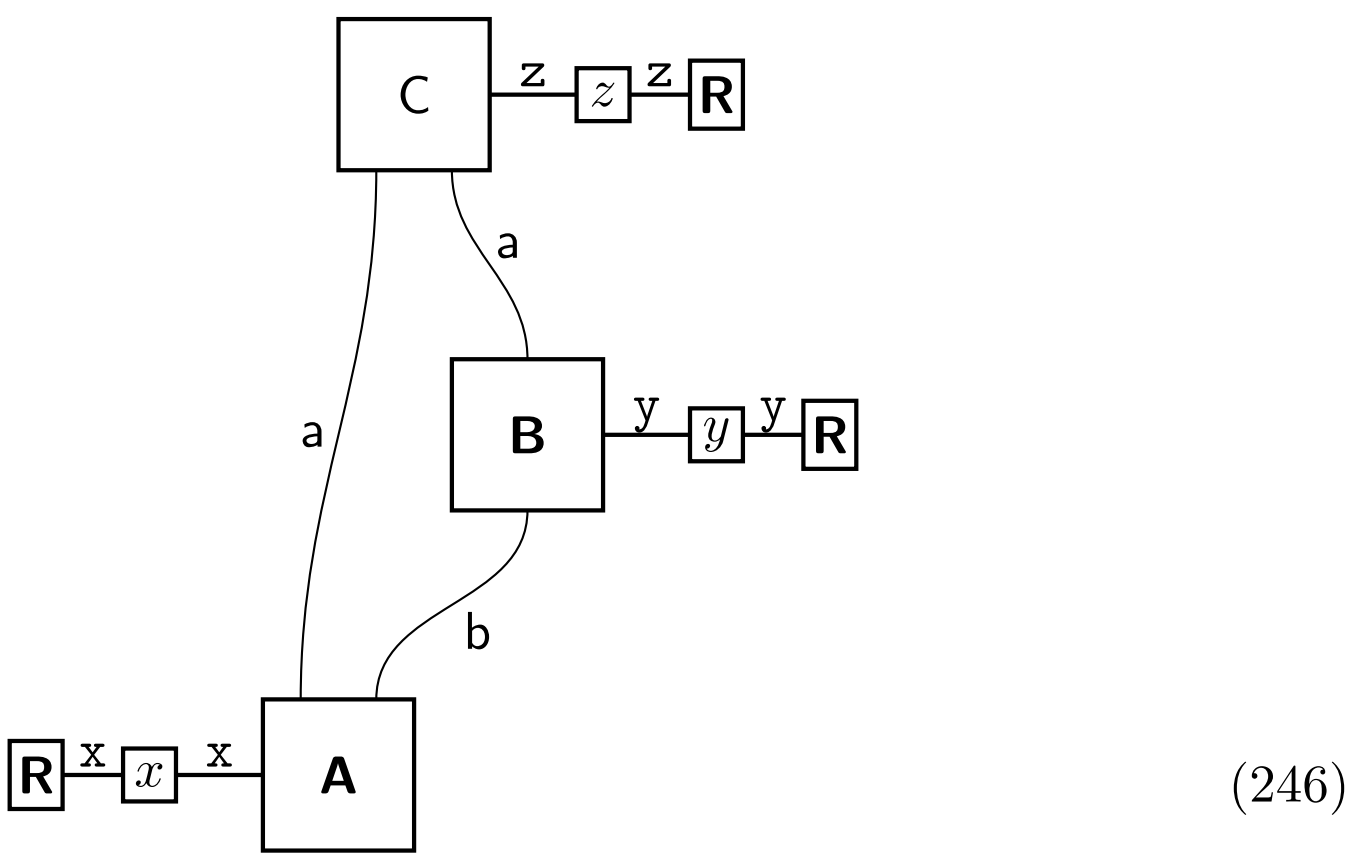

(246)

If we replace each operation by its local decomposition we obtain that the equivalent expression

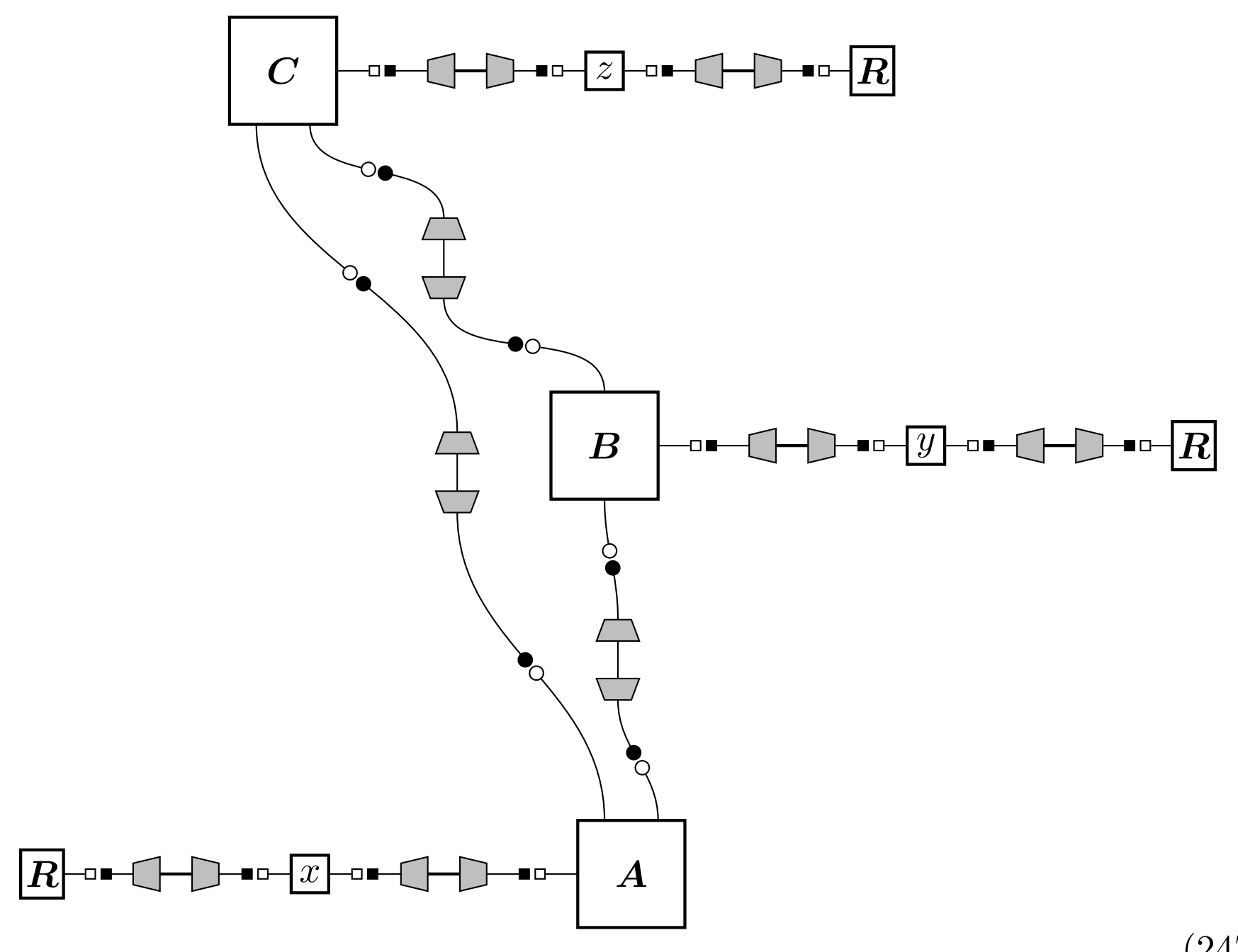

(247)

Note that the bold font, $\boldsymbol{A}, \boldsymbol{B}, \boldsymbol{C}, \boldsymbol{R}$ for the duotensors indicates that these duotensors are associated with deterministic operations. Replacing the fiducial circuits with the equivalent fiducial matrices we obtain the equivalent diagram

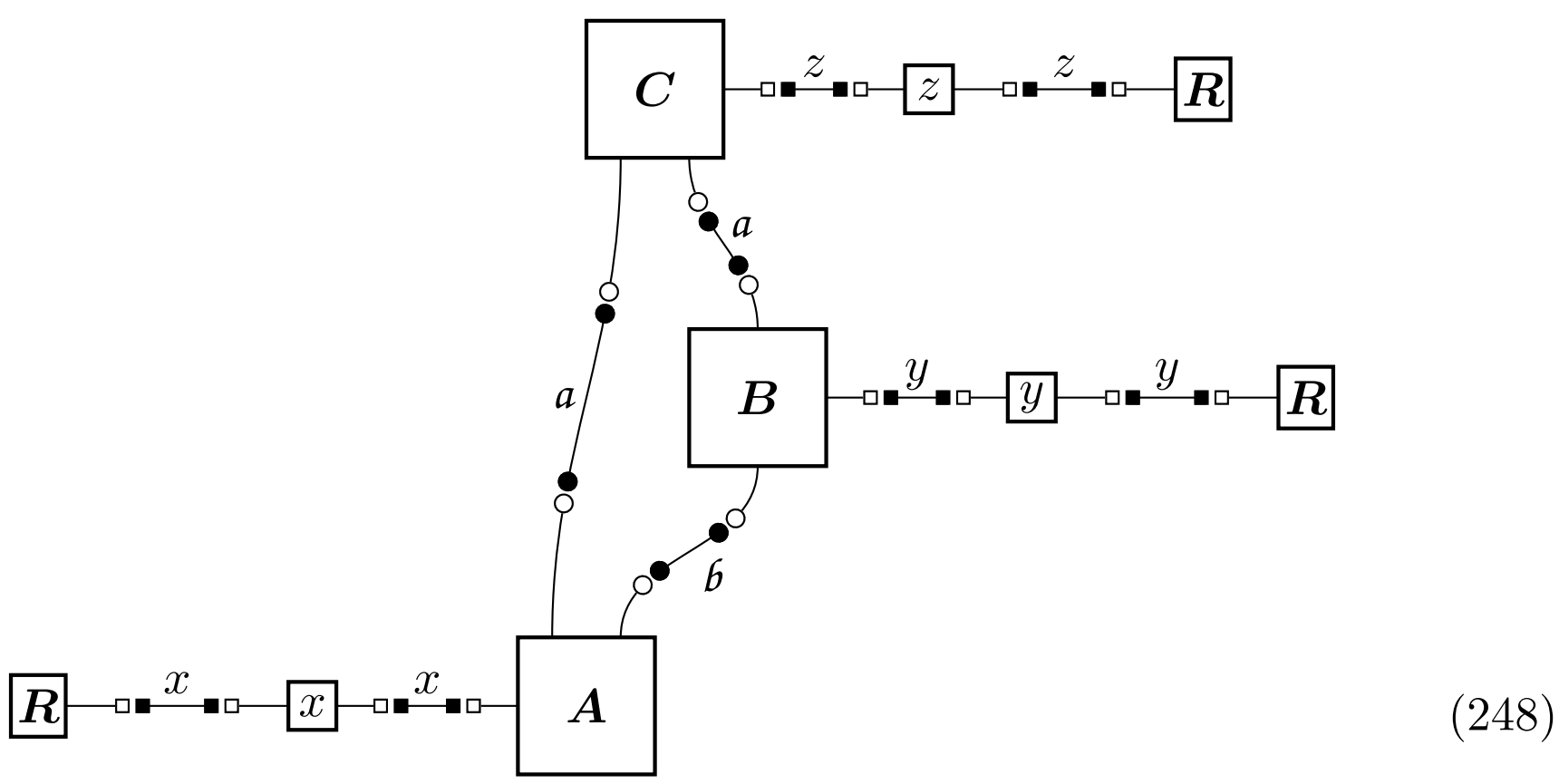

(248)

Finally, we can replace black/white dot pairs with a wire. So we obtain

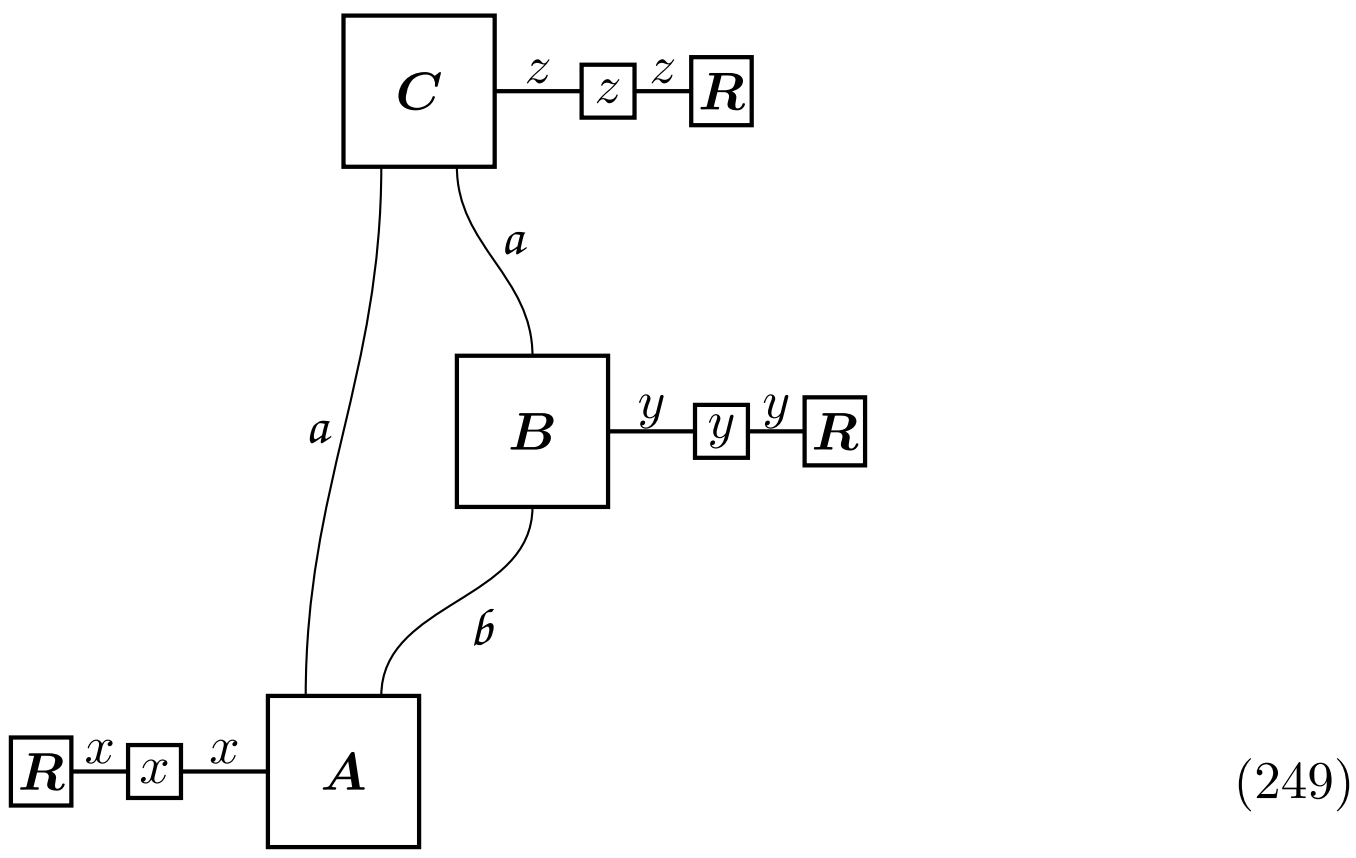

(249)

Do do this calculation we need to reinsert black and white dots so we know which form for each duotensor we are using.

The key take home point is that the calculation for a circuit looks the same as the circuit itself. This clearly works for any circuit.

## 10.2 Using wire decomposition

Now consider, instead, replacing the wires in (246) by their decompositions. Then we obtain

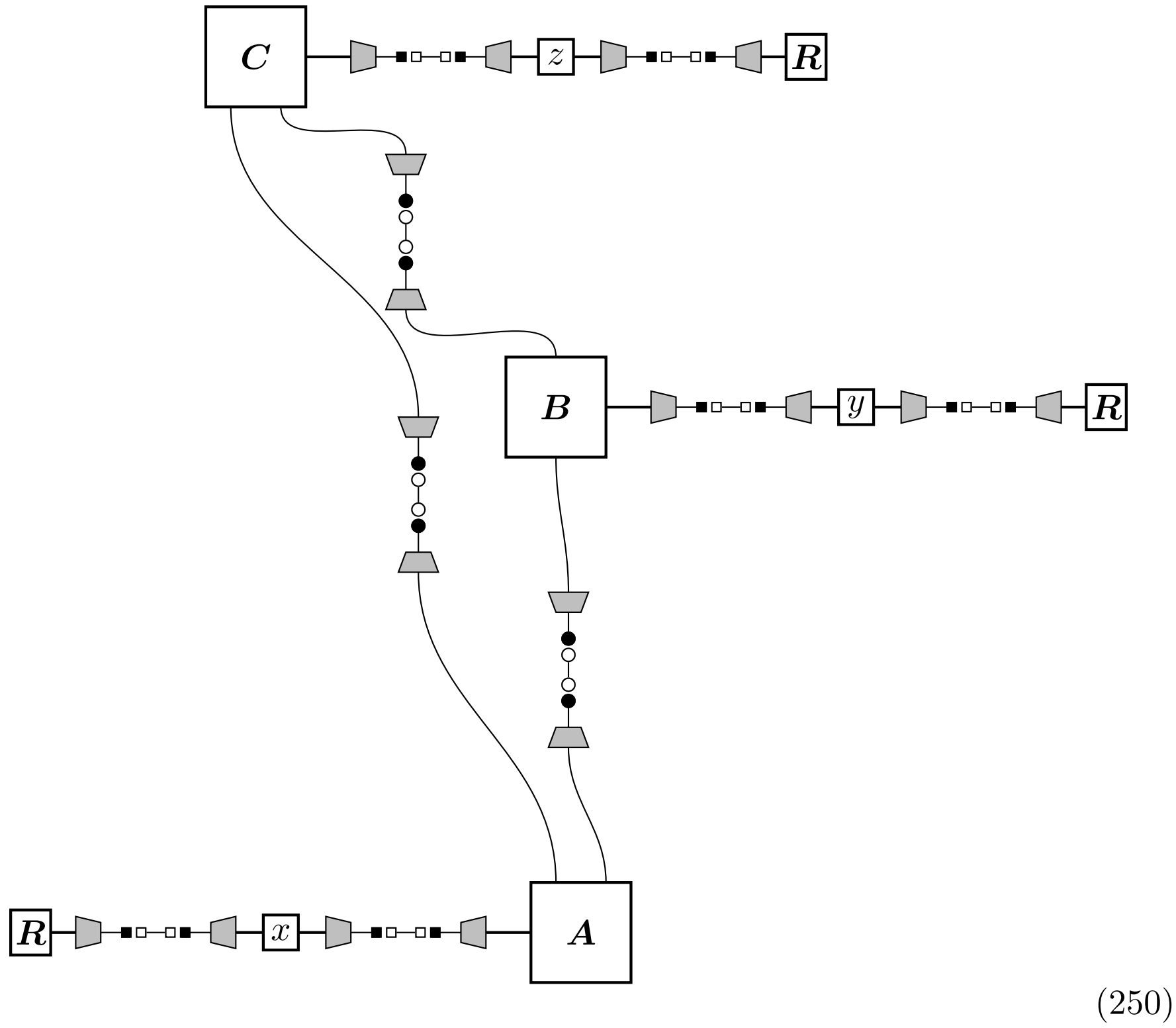

(250)

where we have inserted black and white dots to bring out the natural structure. This is equivalent to the calculation

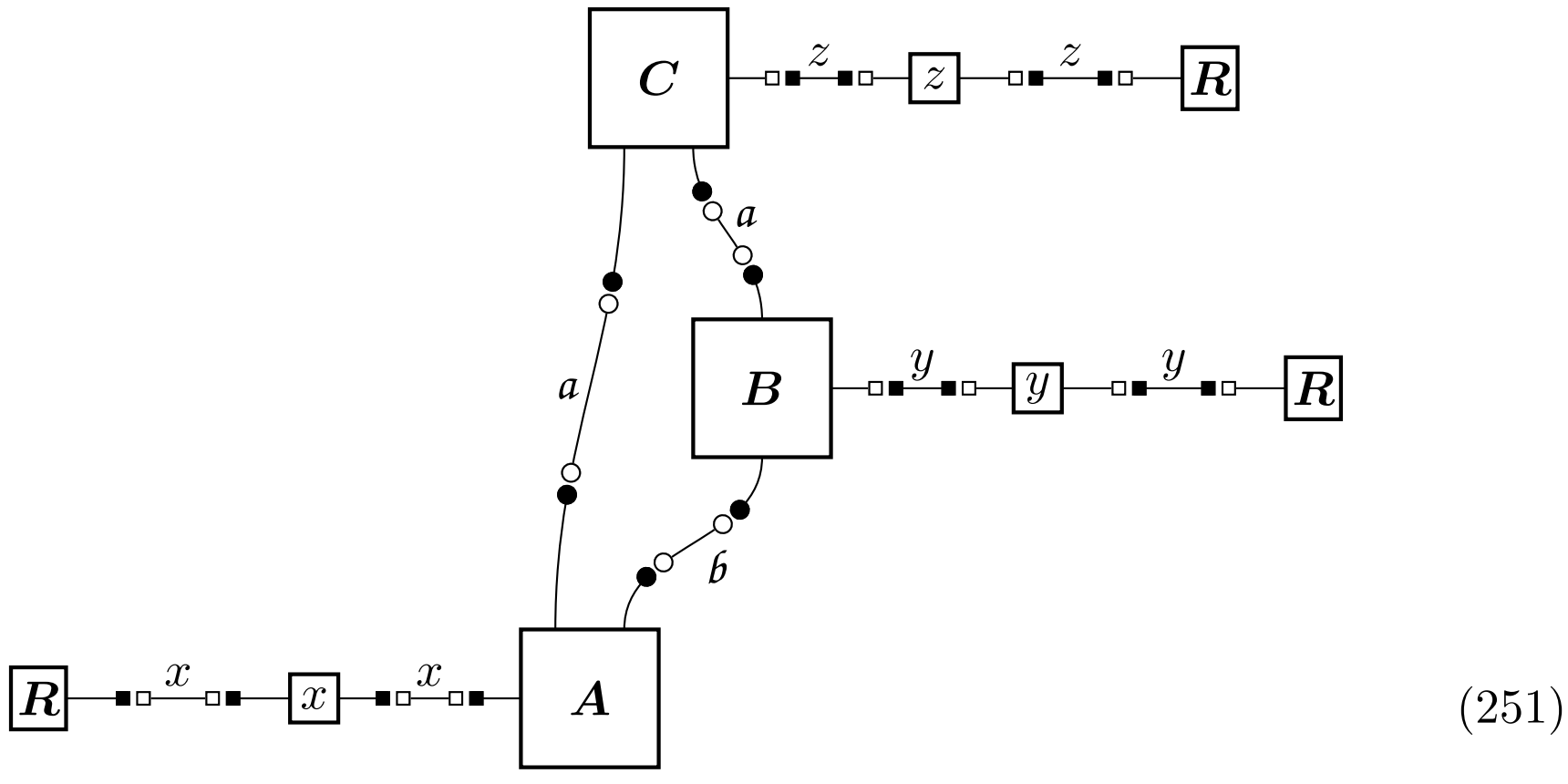

(251)

which can be written as in (249) as before. This shows how wire decomposition is a simple and powerful way to convert a circuit into an equivalent duotensor calculation.

### 10.2.1 Forward and backward thinking

It is worth reconnecting here with the intuitive arguments in Sec. 7.3 for the flatness assumption involving "forward thinking" and "backward thinking".

We can calculate

$$\text{prob}\left( \boxed{\mathsf{R}} \overset{\mathtt{x}}{\text{——}} \boxed{x} \overset{\mathtt{x}}{\text{——}} \boxed{\mathsf{R}} \right) = \boxed{R}\text{—}\overset{x}{\blacksquare\square}\text{—}\boxed{x}\text{—}\overset{x}{\blacksquare\square}\text{—}\boxed{R} = \frac{1}{N_{\mathtt{x}}} \tag{252}$$

This corresponds to forward thinking as discussed in Sec. 7.3 since we send in a normalised distribution with the preparation and then marginalise with the result. Alternatively, we can calculate

$$\text{prob}\left( \boxed{\mathsf{R}} \overset{\mathtt{x}}{\text{——}} \boxed{x} \overset{\mathtt{x}}{\text{——}} \boxed{\mathsf{R}} \right) = \boxed{R}\text{—}\overset{x}{\square\blacksquare}\text{—}\boxed{x}\text{—}\overset{x}{\square\blacksquare}\text{—}\boxed{R} = \frac{1}{N_{\mathtt{x}}} \tag{253}$$

This corresponds to backward thinking. Generally, if we do a duotensor calculation with all the square black dots to the left and the round black dots to the bottom in matched black/white dot pairs then the calculation fits naturally with a forward thinking approach because such a calculation can be thought of as evolving a (possibly subnormalised) probability distribution forward in time. If the black dots are to the right/top then we have a backward thinking approach which can be thought of as evolving a probability distribution backward in time. Importantly, we get the same answer either way. Indeed, we can also do the calculation where some pairs have the black dot on the left (or bottom) and some pairs have the black dot on the right (or top). We still get the same answer.

There is a difference between the forward thinking and backward thinking approaches described above (which provide different ways to calculate the joint probability over incomes and outcomes) and the time forward and time backward perspectives (which provide ways to calculate the conditional probability where we condition on incomes or outcomes respectively). We will now discuss these perspectives.

# 11 Time forward Simple Operational Probabilistic Theories

## 11.1 Introduction

The standard textbook way of thinking about operational probabilistic theories (such as Quantum Theory) is in the time forward temporal frame. In this book we are, for the most part, sticking with the time symmetric perspective. It is important to appreciate that we can transform between the time symmetric, time forward, and time backward perspectives.

If we assume flatness for deterministic pointer preparations and results (this is the property in (85)) then we can prove that the forward, backward, and time symmetric formulations are equivalent. Equivalence of time forward and time symmetric perspectives is proven in Sec. 11.10. Equivalence of time backward and time symmetric perspectives can be proven similarly.

These different perspectives refer to the same physics but packaged in different ways. This “repackaging” affects the physicality constraints on operations and, thereby, the shape of the associated spaces. For example, in the time symmetric formalism, both preparations/results live in a double cone shaped space (as illustrated in Fig. 1) wherein there is a unique ignore preparation/result. In the time forward formalism, results still live in a double cone shaped space but preparations live in a single cone shaped space capped off by a plane of normalised states. In this single cone shaped space, pure preparations are normalised. And, in the time backward formalism, preparations live in a double cone shaped space while results live in a single cone shaped space. In this section we will show how to transform between the time symmetric and time forward simple operational probability theories and show that they are equivalent.

## 11.2 Operations in the time forward frame

The time forward temporal frame of reference is the standard formulation of Quantum Theory found in textbooks and used in most papers on Quantum Theory. An operation has outcomes but no incomes

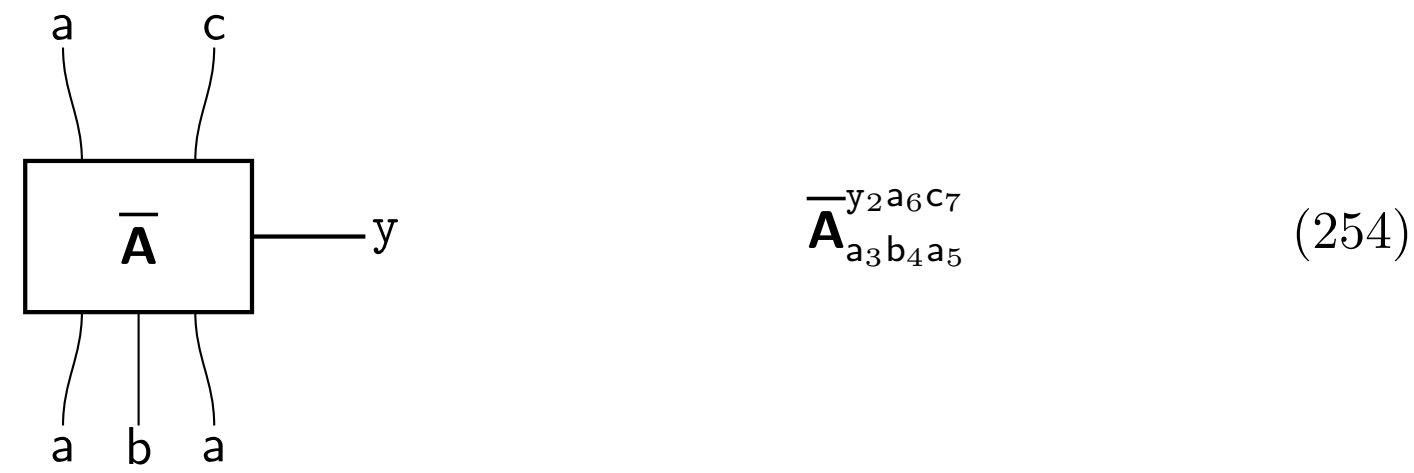

$$\bar{\mathsf{A}}^{y_2 a_6 c_7}_{a_3 b_4 a_5} \tag{254}$$

(Compare with (41).) The bar over the letter $\mathsf{A}$ indicates that we are in the time forward setting. Starting from this we could set up all the machinery indicated in flowchart (36) from scratch for the time forward simple case along the same lines as we did for the time symmetric simple case. In so doing, we would be able to avail ourselves of the all the forward properties. However, we would have to think afresh about using the backward properties since most of these are inapplicable. Instead we will show how to model time forward operations in terms of time symmetric ones. We will also see how to model time symmetric operations in terms of time forward ones so we can prove that the time forward and time symmetric formalisms are equivalent.

## 11.3 Preselection and outcome flag boxes

First we will define preselection and flag boxes. A preselection box is represented as

$$\langle\boxed{x}\!\text{—}\,\mathtt{x} \tag{255}$$

It indicates we preselect on the value $x$ for this income. This means

$$\text{prob}\left(\langle\boxed{x}\overset{\mathtt{x}}{\text{—}}\boxed{\mathtt{x}}\overset{\mathtt{x}}{\text{—}}\boxed{\mathbf{R}}\right) \;=\; 1 \tag{256}$$

It is convenient to also define a outcome "flag" box,

$$\overset{\mathtt{x}}{\text{—}}\boxed{x}\rangle \;:=\; \overset{\mathtt{x}}{\text{—}}\boxed{x}\overset{\mathtt{x}}{\text{—}}\boxed{\mathbf{R}} \tag{257}$$

This is where we read off the outcome then "dispose" of the pointer system. so, clearly,

$$\text{prob}\left(\langle\boxed{\mathtt{x}}\overset{\mathtt{x}}{\text{—}}\boxed{\mathtt{x}'}\rangle\right) \;=\; \delta_{xx'} \tag{258}$$

The reason for choosing the flag shape is that it fits neatly over the preselection shape (and this helps us display the important *midcome identity* in Sec. 11.6). The name is further appropriate because of the common English expression of "flag goes up" meaning *to signal attention* which is appropriate for an outcome.

The preselect box is not allowed in the time symmetric theory since it is deterministic

$$\text{prob}\left(\langle\boxed{x}\overset{\mathtt{x}}{\text{—}}\boxed{\mathbf{R}}\right) \;=\; 1 \tag{259}$$

(this follows by summing (258) over $x'$ and using the control wire identity (131) for $x'$) and yet it violates backward causality for deterministic operations since

$$\langle\boxed{x}\overset{\mathtt{x}}{\text{—}} \;\not\equiv\; \boxed{\mathbf{R}}\overset{\mathtt{x}}{\text{—}} \tag{260}$$

(look at right column of row 9 of Table 1). In fact, since we have a different preselection box for each $x$, these boxes violate the primitive property in the time symmetric formulation that there is a unique deterministic pointer preparation as discussed in Sec. 7.2.

The outcome flag box is allowed in the time symmetric theory since it is built from allowed objects. We should also note that it is not deterministic since

$$\text{prob}\left(\boxed{\mathbf{R}}\overset{\mathtt{x}}{\text{—}}\boxed{x}\rangle\right) \;=\; \frac{1}{N_{\mathtt{x}}} \tag{261}$$

follows from the definition of the **R** boxes in (85) (contrast this with (259) for the preselection box).

By summing (257) over $x$ and using the control wire identity (131) we obtain

$$\sum_{x}^{N_{\mathtt{x}}} \overset{\mathtt{x}}{\text{—}}\boxed{x}\rangle \;\equiv\; \overset{\mathtt{x}}{\text{—}}\boxed{\mathbf{R}} \tag{262}$$

We will use this in Sec. 11.10.

The main application of the preselection box and outcome flag box for us is to allow us to convert between time forward operations and time symmetric operations as we will see in Sec. 11.4 and Sec. 11.7.

## 11.4 Modelling TF with TS

We can think of an operation in the time forward case as arising from conditioning on incomes as they arise on operations in the time symmetric case. Then we can model an operation $\overline{\mathbf{B}}$ in the time forward theory in terms of an operation, $\mathsf{B}$ from the time symmetric theory by conditioning on the incomes of the latter as follows

$$\overset{b}{\underset{a}{\boxed{\overline{\mathsf{B}}}}}\!-\!\mathrm{y} \quad \Leftarrow\!\bar{\square}\!=\!\square \quad \langle x \,\overset{\mathrm{x}}{-}\, \overset{b}{\underset{a}{\boxed{\mathsf{C}}}}\!-\!\mathrm{y} \tag{263}$$

We use the $\Leftarrow\!\bar{\square}\!=\!\square$ symbol to indicate when the time forward operation on the left is modelled by the time symmetric operation on the right. The $\overline{\mathsf{B}}$ is given in bold as it represents what we will call a *forward deterministic* operation. This is where we condition on incomes and have no implicit readout boxes for outcomes.

To readout the outcomes we can use the outcome flag box or, equivalently (see (257)), a readout box followed by an $\mathsf{R}$ box

$$\overset{b}{\underset{a}{\boxed{\overline{\mathsf{B}}}}}\,\overset{\mathrm{y}}{-}\, y\rangle \quad \Leftarrow\!\bar{\square}\!=\!\square \quad \langle x \,\overset{\mathrm{x}}{-}\, \overset{b}{\underset{a}{\boxed{\mathsf{C}}}}\,\overset{\mathrm{y}}{-}\,\boxed{y}\,\overset{\mathrm{y}}{-}\,\boxed{\mathsf{R}} \tag{264}$$

We can take the outcome, $y$, to be implicit. Then we would have

$$\overset{b}{\underset{a}{\boxed{\overline{\mathsf{D}}}}} \quad \Leftarrow\!\bar{\square}\!=\!\square \quad \langle x \,\overset{\mathrm{x}}{-}\, \overset{b}{\underset{a}{\boxed{\mathsf{C}}}}\,\overset{\mathrm{y}}{-}\,\boxed{y}\,\overset{\mathrm{y}}{-}\,\boxed{\mathsf{R}} \tag{265}$$

where the outcome, $y$ is implicit in the $\overline{\mathsf{D}}$ symbol (we now use a non-bold font for $\overline{\mathsf{D}}$).

An important special case are operations in the TF frame for which, when modelled in the TS setting, there is no preselection on incomes. We denote such cases by an asterisk subscript. We have

$$\overset{b}{\underset{a}{\boxed{\overline{\mathsf{B}}_*}}}\!-\!\mathrm{y} \quad \Leftarrow\!\bar{\square}\!=\!\square \quad \overset{b}{\underset{a}{\boxed{\mathsf{C}}}}\!-\!\mathrm{y} \tag{266}$$

Thus, in such cases, $\overline{\mathsf{B}}_* = \mathsf{C}$ and we say there is *no implicit preselection.* This is a distinction that can only be appreciated by thinking about how time forward operations are modelled in the time symmetric formalism - it is not a notion which is, in itself, native to the time forward formalism.

## 11.5 Circuits in the time forward case

A circuit in the time forward case looks like this

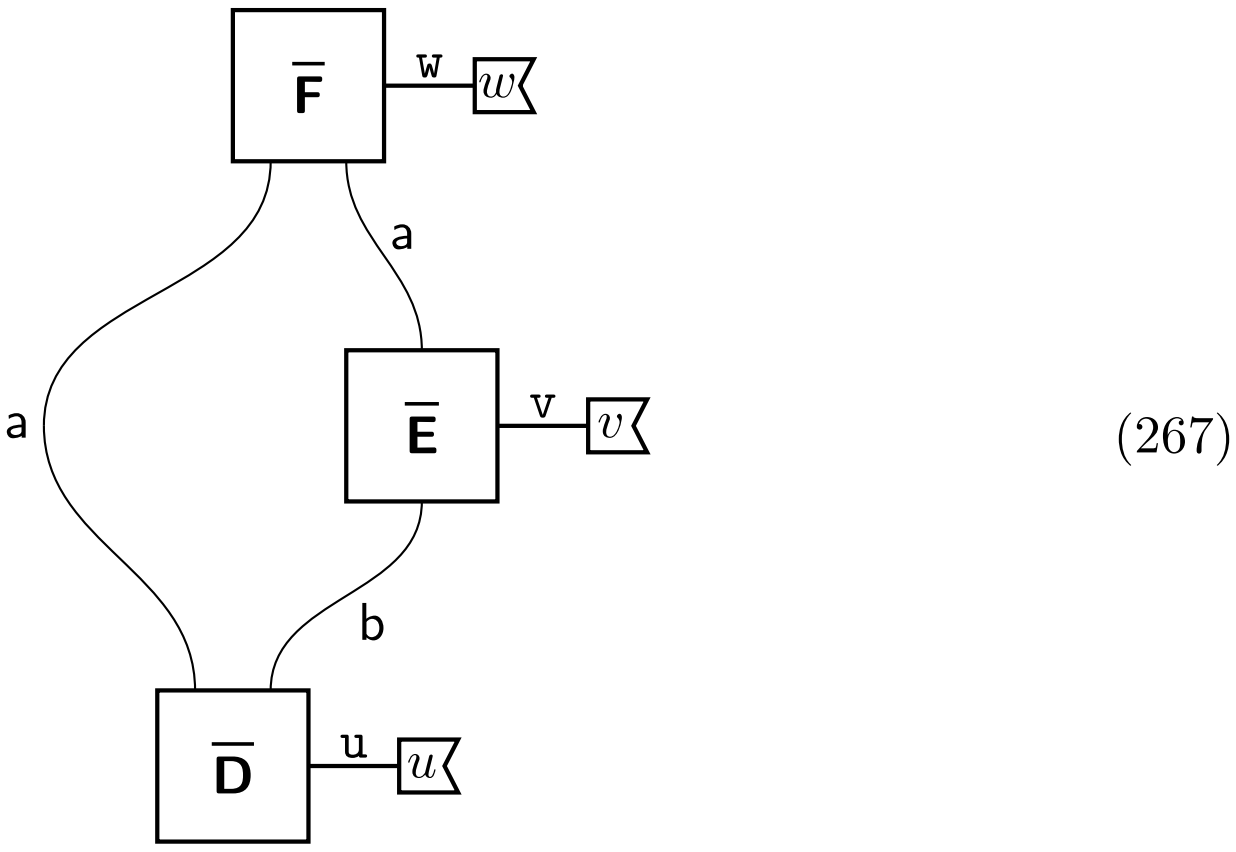

$$(267)$$

By regarding each operation as resulting from preselecting on incomes, we can write this circuit as follows

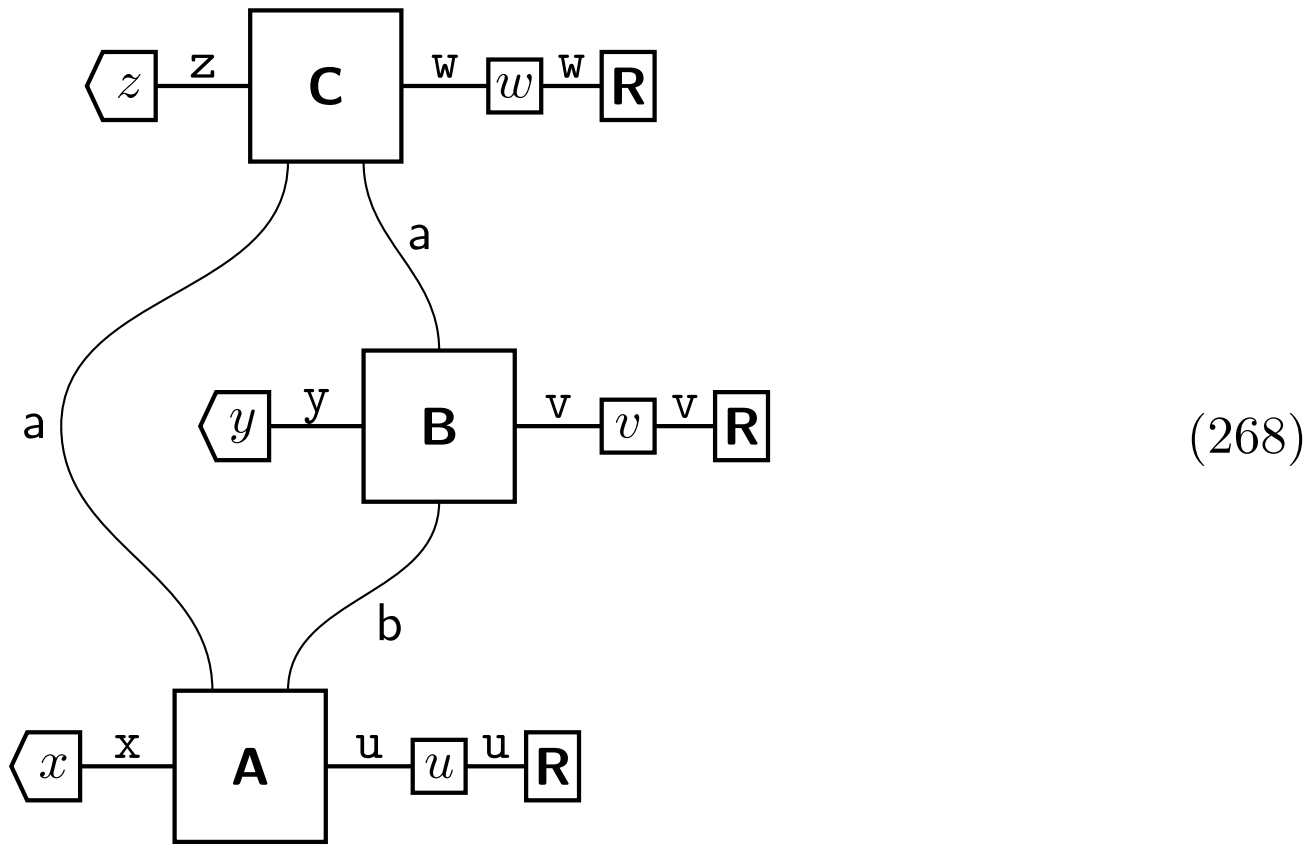

$$(268)$$

where we are now using time symmetric modeling for the operations. Since we preselect on all incomes there is no possibility of midcomes or control wires in the sense of Sec. 7.11. However, the same effect could be achieved by other means in a rich enough theory (for example, if we had a hybrid classical quantum theory then control effects could be achieved through the exchange of classical systems between the operations).

For the circuit in (268), we want to calculate the conditional probability $\text{prob}(uvw|xyz)$. This is given by

$$\text{prob}(u, v, w|x, y, z) = \frac{\text{prob}(x, y, z, u, v, w)}{\text{prob}(x, y, z)} \tag{269}$$

which we can write as

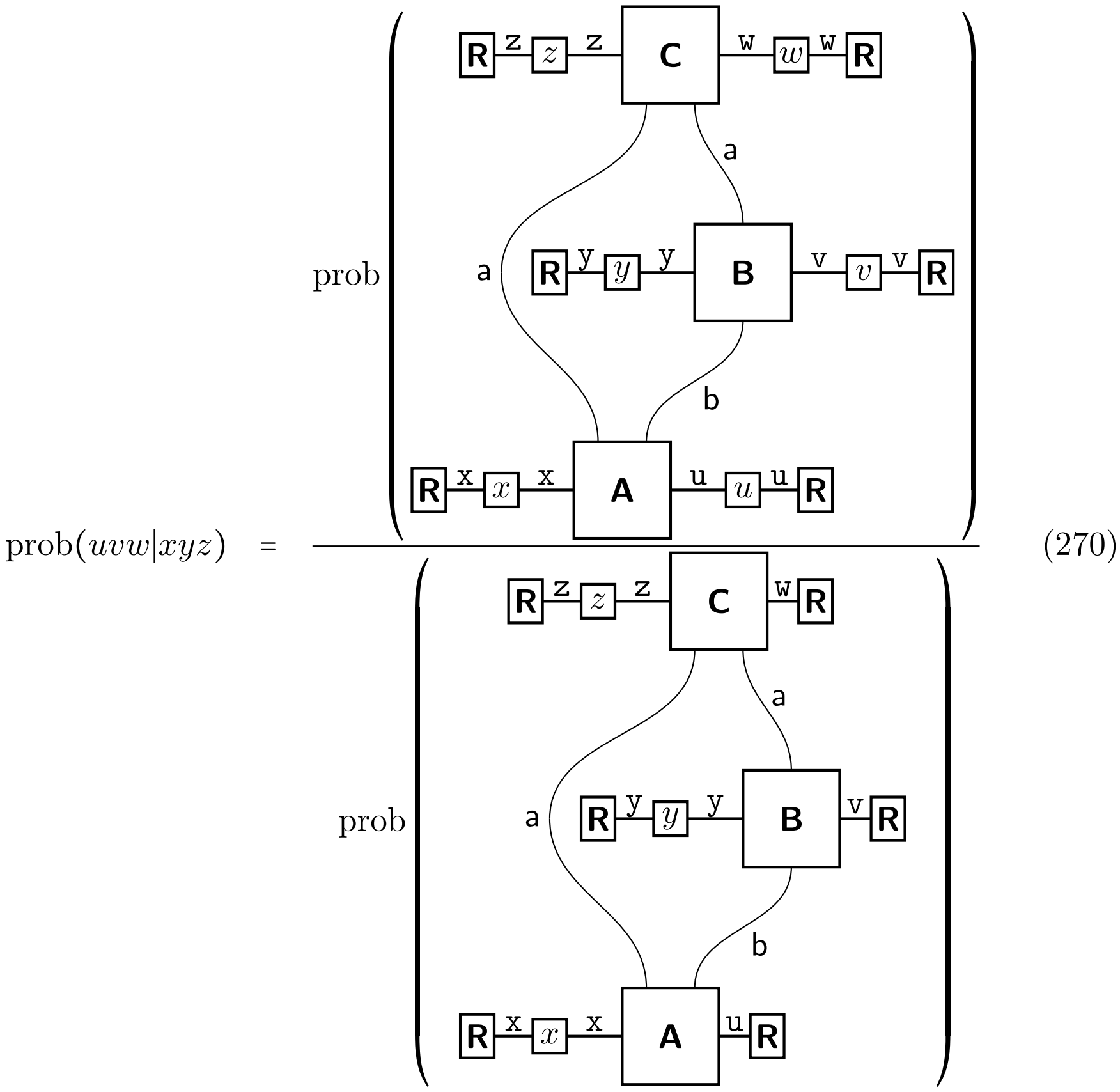

(270)

Applying forward causality to the circuit in the denominator, starting at **C**,

then progressing through **B**, then **A**, it is easy to see that

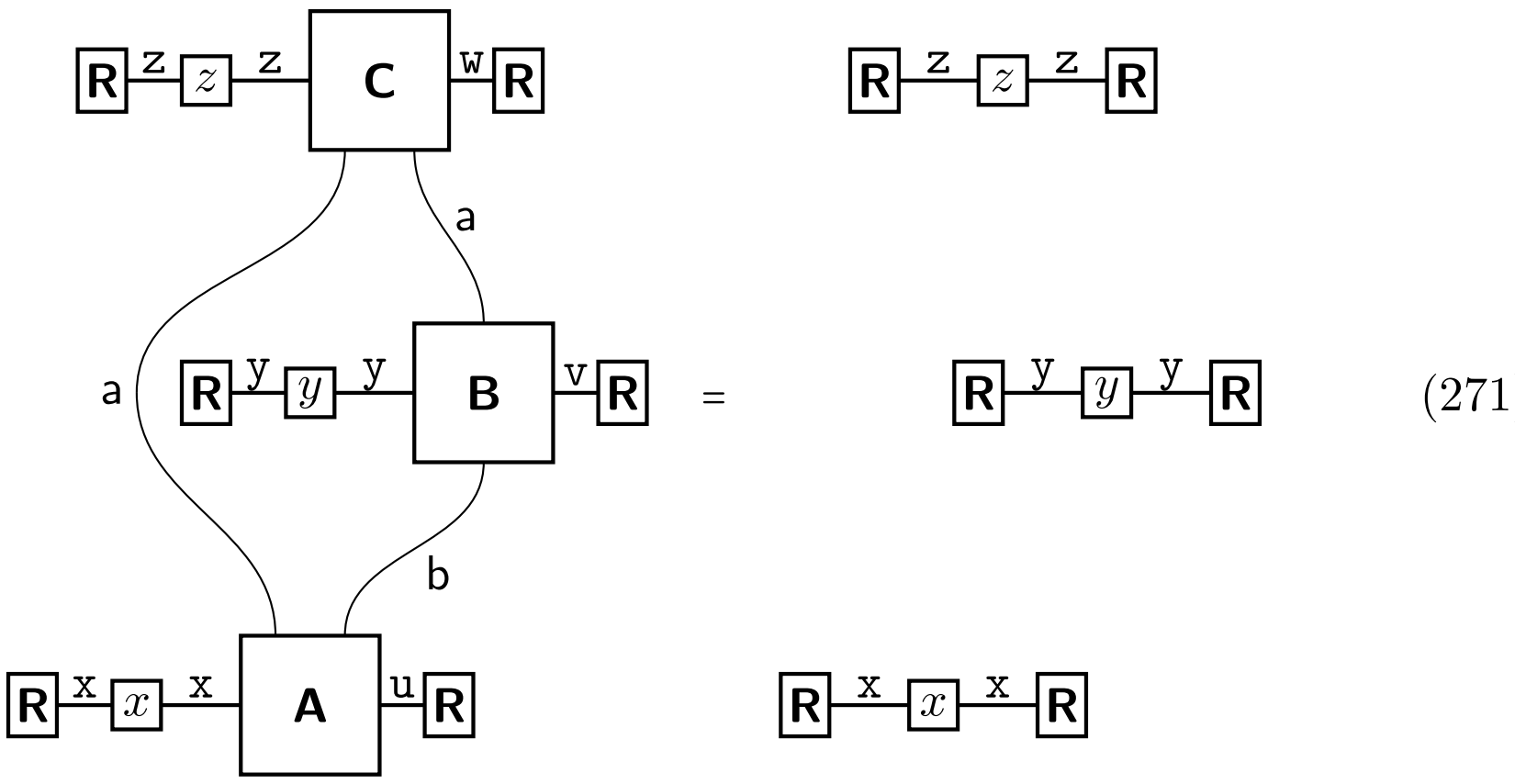

This clearly works for any circuit of this form (where we have **R** boxes after the open outcomes). Now, if we assume flatness (85) then this is equal to $\frac{1}{N_{\mathtt{x}}N_{\mathtt{y}}N_{\mathtt{z}}}$. We can write this as

$$\text{prob}(x, y, z) = \frac{1}{N_{\mathtt{x}}N_{\mathtt{y}}N_{\mathtt{z}}} \tag{272}$$

This gives rise to a normalisation factor of $N_{\mathtt{x}}N_{\mathtt{y}}N_{\mathtt{z}}$ when substituted into (270). We can associate each factor in this product with the associated preselection element. Thus we can use the *replacement rule*

$$\langle x \,\overset{\mathtt{x}}{—} \quad \equiv \quad \boxed{N_{\mathtt{x}}}\ \boxed{\mathsf{R}}\overset{\mathtt{x}}{—}\boxed{x}\overset{\mathtt{x}}{—} \tag{273}$$

The probability $\text{prob}(uvw|xyz)$ is given by using the substitution in (273) in (268) and is equal to

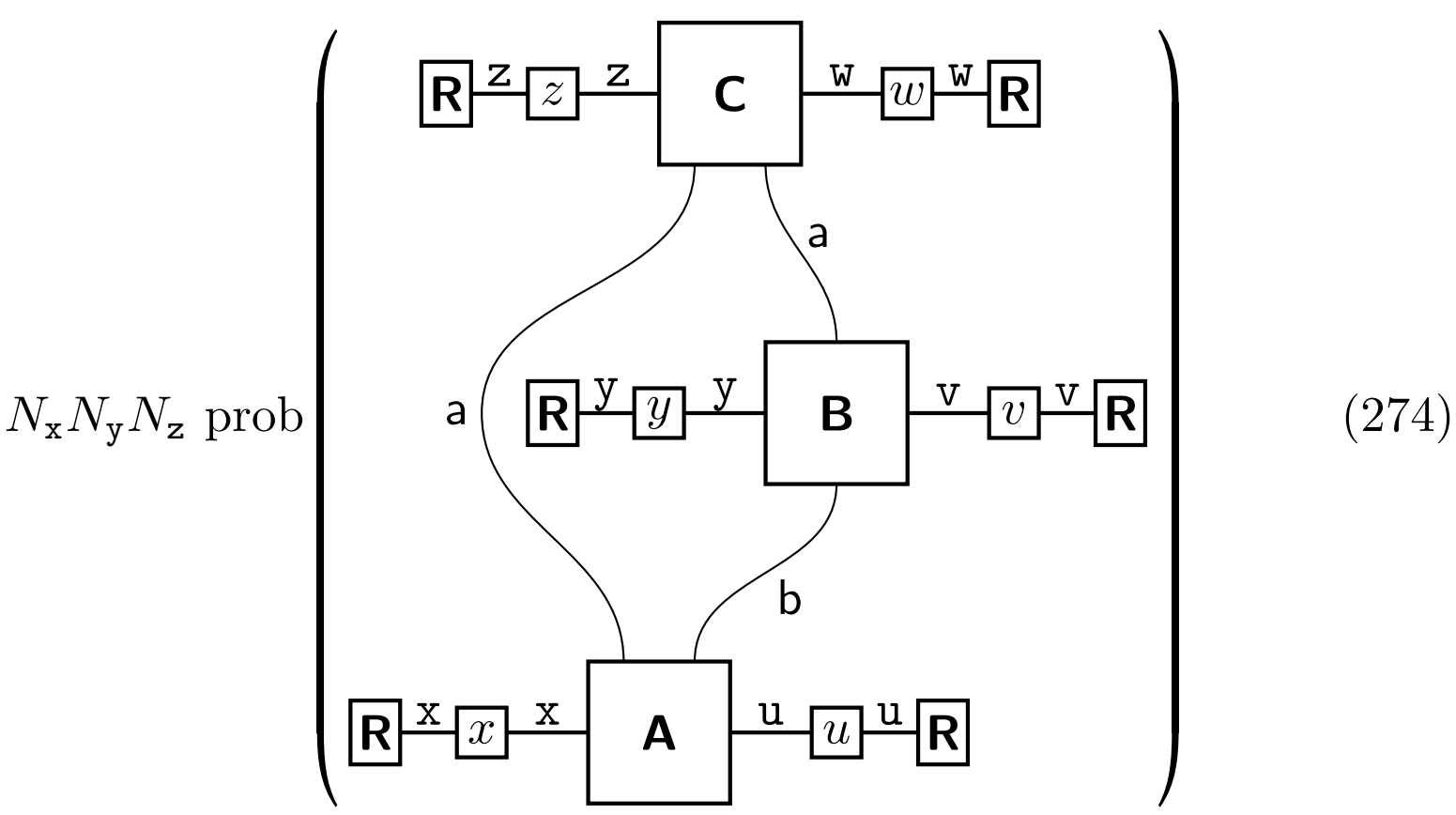

This establishes that there is a simple connection between the time forward and time symmetric cases.

There is an important subtlety here which we express as a warning:

**The replacement rule warning**: The replacement rule in (273) is only proven to work when used in a circuit where all incomes are preselected and, importantly, no outcomes are postselected.

If we were to postselect on some or all outcomes while still preselecting on incomes then this replacement rule would not work (because the application of forward causality above would be blocked and denominator would not be constant). One way forward is to regard the preselection box as an allowed operation in the time forward theory then to regard the replacement rule as a way of modelling this in the time symmetric theory

$$\langle x\!\mid\overset{\mathtt{x}}{\rule{2em}{0.4pt}} \qquad \Leftarrow\!\bar{\square}\!\!=\!\!\square \quad \boxed{N_{\mathtt{x}}}\ \boxed{\mathsf{R}}\overset{\mathtt{x}}{\rule{1em}{0.4pt}}\boxed{x}\overset{\mathtt{x}}{\rule{1em}{0.4pt}} \tag{275}$$

When we model a TF situation we heed the above warning so this will work. This way of modelling preselection has a natural interpretation if we think in a think in a time forward fashion (see discussion in Sec. 7.3 and Sec. 10.2.1). The **R** box followed by the $x$ box can be thought of as preparing an $x$ with probability $\frac{1}{N_{\mathtt{x}}}$. Then the $N_{\mathtt{x}}$ factor in (275) renormalises this probability to 1.

It is also worth making the related comment that the factor, $N_{\mathtt{x}}N_{\mathtt{y}}N_{\mathtt{z}}$, is a global factor arising from a probability calculation for a full circuit (the one in the denominator in (270)). By using the replacement rule in (273) we are, in a sense, deciding to attribute the $N_{\mathtt{x}}$ factor to the $\mathtt{x}$ income, the $N_{\mathtt{y}}$ factor to the $\mathtt{y}$ income, and the $N_{\mathtt{z}}$ factor to the $\mathtt{z}$ income.

This connection between the time symmetric and time forward pictures can be understood as the application of the standard rule for calculating conditional probabilities from joint probabilities

$$\text{prob}(u,v,w|x,y,z) = \frac{\text{prob}(x,y,z,u,v,w)}{\text{prob}(x,y,z)} \tag{276}$$

where this exercise is greatly simplified as the denominator takes the form in (272). We can also use this procedure to go from the time forward frame to the time symmetric frame.

## 11.6 The forward seatbelt identity

Recall the midcome identity

$$\overset{\mathtt{x}}{\rule{2em}{0.4pt}}\boxed{x}\overset{\mathtt{x}}{\rule{2em}{0.4pt}} \quad \equiv \quad \overset{\mathtt{x}}{\rule{1em}{0.4pt}}\boxed{x}\overset{\mathtt{x}}{\rule{1em}{0.4pt}}\boxed{\mathsf{R}}\ \boxed{N_{\mathtt{x}}}\ \boxed{\mathsf{R}}\overset{\mathtt{x}}{\rule{1em}{0.4pt}}\boxed{x}\overset{\mathtt{x}}{\rule{1em}{0.4pt}} \tag{130 revisited}$$

from Sec. 7.11 (this follows if we assume flatness). Using (257) and the replacement rule, (273), we obtain

$$\overset{\mathtt{x}}{\rule{1.5em}{0.4pt}}\boxed{x\rangle}\boxed{\langle x}\overset{\mathtt{x}}{\rule{1.5em}{0.4pt}} \quad \equiv \quad \overset{\mathtt{x}}{\rule{2em}{0.4pt}}\boxed{x}\overset{\mathtt{x}}{\rule{2em}{0.4pt}} \tag{277}$$

We call this the *forward seatbelt identity* because of the resemblance to clicking the two parts of a seatbelt together). It has the following interpretation: reading from left to right, reading out the outcome $x$ then preselecting the outcome $x$ is the same as simply reading out $x$ as the pointer system passes by. We adopted the flag shape to make this equation look natural. It is easy to obtain a backward seatbelt identity. Since the replacement rule comes with a warning, that same warning applies to using the seatbelt identity. We will use it to prove equivalence between TS and TF approaches. In TF approaches we always heed the replacement rule warning. Since we use the midcome identity to obtain the forward seatbelt identity, we are also assuming flatness.

Using the control wire identity (131) we obtain

$$\sum_{x=1}^{N_{\mathsf{x}}} \; \overset{\mathsf{x}}{—}\boxed{x}\!\langle\!\langle\boxed{x}\overset{\mathsf{x}}{—} \quad \equiv \quad \overset{\mathsf{x}}{———} \tag{278}$$

by summing (277) over $x$. This equation is useful when modeling time symmetric operations with time forward operations.

## 11.7 Modelling TS with TF

We wish to prove that time symmetric and time forward simple operational probabilistic theories are equivalent. Thus we need a way to go from time forward to time symmetric operations. The key to doing this is to note that, when we modelled $\overline{\mathsf{B}}$ using $\mathsf{C}$ as in (263) we are actually picking out just one income on $\mathsf{C}$. We could model other time forward operations by preselecting other outcomes. Thus, associated with any time symmetric operation $\mathsf{C}$ there are $N_{\mathsf{x}}$ time forward operations which we will write

$$\left[\overline{\mathsf{C}}(x)\right]_{\mathsf{a}}^{\mathsf{b}}\!—\mathsf{y} \quad \Leftarrow\!\square\!\!-\!\!\bar{\square} \quad \langle\boxed{x}\overset{\mathsf{x}}{—}\left[\mathsf{C}\right]_{\mathsf{a}}^{\mathsf{b}}\!—\mathsf{y} \tag{279}$$

where $\overline{\mathsf{B}}$ is associated with just one (say for $x=1$) so $\overline{\mathsf{B}} = \overline{\mathsf{C}}(1)$. Hence, to model a time symmetric operation, $\mathsf{C}$, we need a set of time TF operations. We do this modeling as folows

$$\mathsf{x}—\left[\mathsf{C}\right]_{\mathsf{a}}^{\mathsf{b}}\!—\mathsf{y} \quad \Leftarrow\!\square\!\!-\!\!\bar{\square} \quad \sum_{x} \overset{\mathsf{x}}{—}\boxed{x}\!\langle \left[\overline{\mathsf{C}}(x)\right]_{\mathsf{a}}^{\mathsf{b}}\!—\mathsf{y} \tag{280}$$

where the $\Leftarrow\!\square\!\!-\!\!\bar{\square}$ symbol stands for when a TS operation expression is modelled by a set of TF operation. If we substitute (279) for $\overline{\mathsf{C}}(x)$ into the right hand side of (280) and use the forward seatbelt identity summed over $x$ (as in (278)) then we verify it returns $\mathsf{C}$ as required (we are implicitly using flatness since the forward seatbelt identity is based on the midcome identity which needs flatness).

## 11.8 Positivity in the time forward frame

If we were to start afresh in setting up the time forward formalism then we would need to define a $p(\cdot)$ function on time forward circuits, $\overline{\mathsf{E}}$ (these being circuits in which we preselect on all the incomes). For the sake of having notation, we could call this the $\overline{p}(\cdot)$ function. For circuits having preselected incomes $\mathtt{xyz}$ like that discussed in discussed in Sec. 11.5 we would have $\overline{p}(\cdot) = N_{\mathtt{x}} N_{\mathtt{y}} N_{\mathtt{z}} p(\cdot)$. We get the normalisaton factor $N_{\mathtt{x}} N_{\mathtt{y}} N_{\mathtt{z}}$. In general, we would taking a weighted sum over circuits each of which may have different preselected incomes. In this case the normalisation factor would be different. However, these normalisation factors can be absorbed into the weights in the weighted sum. Consequently we retain the linearity structure. This means that pure preparations/results in the time symmetric theory map to pure preparations/results in the time forward theory. Hence the tester positivity condition for operations in the time forward frame is

$$0 \underset{\overline{T}}{\leq} \overline{\mathsf{B}}_{\mathsf{a}}^{\mathsf{b}\,\mathsf{y}} \tag{281}$$

where the $\overline{T}$ stands for "tester" indicating that the given operation is positive with respect to any tester of the following type

b
h
y
$y$
y
**R**
a
$\overline{\mathsf{E}}$
$\overline{\mathsf{D}}$

(282)

for all pure preparations $\overline{\mathsf{D}}$ and pure results $\overline{\mathsf{E}}$. The positivity composition theorem, network positivity theorem, and positivity theorem all go through for the time forward frame. This can simply be derived from the time symmetric cases by multiplying by the normalisation factors.

## 11.9 Causality in the time forward frame

The double causality condition on a deterministic operation, **C**, in the time symmetric frame is

(283)

In the time forward frame we must condition on all incomes. For the backwards causality condition (on the right) we cannot condition on the income on **C** because it is already closed. Hence, this condition is not generally applicable in the time forward frame.

The exception to this being when **C** has no income. For such a case, $\mathbf{C} = \overline{\mathbf{B}}_*$ (using the asterisk notation introduced in Sec. 11.4). We have

If no implicit preselection (284)

This special case tells us that when there is no preselection implicit in $\overline{\mathbf{B}}$ then, on sending in the ignore preparation, the outcome is uncorrelated with the output. Since (as noted at the end of Sec. 11.4) the notion of implicit preselection is not native to the time forward formalism, the time backward causality condition in (284) also is not native to the time forward formalism.

The time forward causality condition in (283) (on the left) is generally applicable in the time forward frame. We will see how to write this down for time forward operations. First we preselect on the incomes to give

(285)

and so we obtain the following causality condition for a deterministic operation, $\overline{\mathbf{B}}$, in the time forward case

(286)

In the deterministic case we have the forward causality condition (286) and the backward causality condition (284) which only applies for the special case where $\overline{\mathbf{B}}$ has no implicit preselection. Together these constitute the $t$ =TF causality conditions.

The general causality conditions for an operation, $\overline{\mathsf{B}}$, which may be non-deterministic, are

$$\mathbf{I} \overset{\mathsf{b}}{\text{—}} \underset{\mathsf{a}}{\overline{\mathsf{B}}} \overset{\mathsf{y}}{\text{—}} \mathbf{R} \ \underset{\overline{T}}{\leq} \ \underset{\mathsf{a}}{\mathbf{I}} \tag{287}$$

and, only in the special case where there is no implicit preselection in $\overline{B}$,

$$\text{If no implicit preselection} \qquad \overset{\mathsf{b}}{\underset{\mathsf{a}}{\overline{\mathsf{B}}_*}} \text{—}\, \mathbf{I} \ \overset{\mathsf{y}}{\text{—}} \ \underset{\overline{T}}{\leq} \ \overset{\mathsf{b}}{\mathsf{I}}\ \mathsf{R} \overset{\mathsf{y}}{\text{—}} \tag{288}$$

where these inequalities are saturated for the deterministic case. This inequality is obtained using the tester positivity condition along the same lines as in Sec. 7.13

We can obtain a causality composition theorem and a network causality theorem by considering the forward cases of the time symmetric versions of those theorems.

In the time forward case the constraints arising from causality are different and so the space of allowed operations is different. Most notably, pure system preparations are normalised in the time forward case because we divide by the normalisation factor $1/N_{\mathtt{x}}$ when we condition. They are not normalised in the time symmetric case.

## 11.10 Equivalence of time forward and time symmetric approaches

In the time forward formalism, a physical operation must satisfy $\overline{T}$ positivity (discussed in Sec. 11.8) and general forward causality (as discussed in Sec. 11.9). In the time symmetric formalism, a physical operation satisfies $T$ positivity (see Sec. 7.12 and general double causality (as discussed in Sec. 11.9). We can now prove a theorem that is central to this book because it proves that the time symmetric approach is equivalent to the usual time forward approach. A quantum version of this theorem appears in Mrini and Hardy [2024].

**TS and TF equivalence theorem:** This theorem assumes flatness and has two parts.

**Modeling TS by TF.** Any physical operation in the time symmetric theory can be modelled by a set of physical operations in the time forward theory.

**Modeling TF by TS.** Any physical operation in the time forward theory can be modelled by a physical operation in the time symmetric theory.

Note that we have used the forward seatbelt identity in setting up the elements of this proof in the forgoing chapters which means we need flatness. We also use flatness below to obtain (290). It is sufficient to consider deterministic operations since nondeterministic operations can be modelled on deterministic ones by making some readout boxes implicit. First we show that any physical TS deterministic operation can be modelled by a set of physical TF deterministic operations. We have already put in place the machinery necessary to prove this works. If we start with a TS deterministic operation, $\mathbf{C}$, we can generate a set of deterministic TF operations, $\overline{\mathbf{C}}(x)$ using (279). If $\mathbf{C}$ is physical (in the TS sense) then it follows from the considerations in Sec. 54.3 and Sec. 11.9 that each of $\overline{\mathbf{C}}(x)$ is physical (in the TF sense). Finally, we can model $\mathbf{C}$ in terms of $\overline{\mathbf{C}}(x)$ by the expression in (280). This proves we can model any physical operation in the time symmetric theory by a set of physical operations in the TF theory. Now we will prove that any physical deterministic operation, $\overline{\mathbf{B}}$, in the TF theory can be modelled by a physical deterministic operation, $\mathbf{C}$, in the TS theory. We start by putting

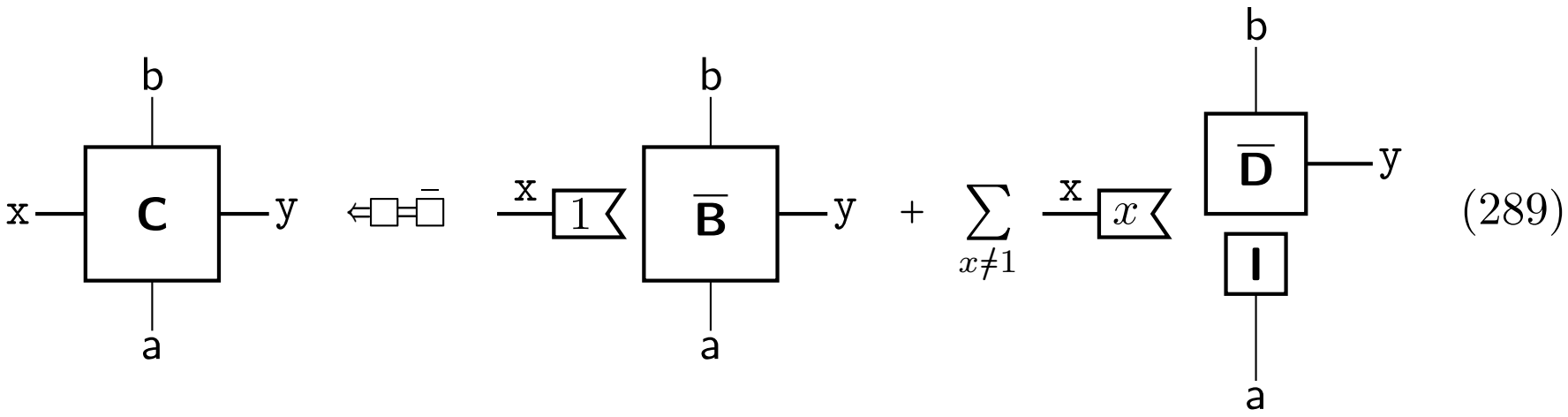

(289)

Note that we are choosing $\overline{\mathbf{C}}(1) = \overline{\mathbf{B}}$ and then we are choosing the remaining $\overline{\mathbf{C}}(x)$ (for $x \neq 1$) to all be equal. It is clear that we can obtain $\overline{\mathbf{B}}$ from $\mathbf{C}$ by preselecting on $x = 1$. The question that remains is whether we can choose a physical $\overline{\mathbf{D}}$ (in the TF sense) such that $\mathbf{C}$ is physical (in the TS sense). We will now prove that this is the case. If we apply backward causality to this we obtain

(290)

(using flatness). Motivated by this we define $\overline{\mathsf{D}}$ as follows

$$\overline{\mathsf{D}}^{\,\mathsf{b}}_{\;\mathsf{y}} \;:=\; \boxed{\frac{N_{\mathsf{x}}}{N_{\mathsf{x}}-1}}\; \mathsf{I}^{\mathsf{b}}\; \mathsf{R}_{\mathsf{y}} \;-\; \boxed{\frac{1}{N_{\mathsf{x}}-1}}\; \overline{\mathsf{B}}^{\,\mathsf{b}}_{\;\mathsf{y}} \circ_{\mathsf{a}} \mathsf{I} \tag{291}$$

(we obtain this if we replace the $\Leftarrow\!\square\!\Rightarrow\!\bar{\square}$ in (290) with a := and rearrange terms). We need to be sure that $\overline{\mathsf{D}}$ is physical. First we see that $\overline{T}$-positivity is guaranteed if we choose $N_{\mathsf{x}}$ to be big enough. We will show in Sec. 11.12 below that, in spectral theories (of which Quantum Theory is an example), $N_{\mathsf{x}} = N_{\mathsf{y}}N_{\mathsf{b}}$ is big enough. Second, if we apply forward causality to $\overline{\mathsf{D}}$, as defined above, we obtain

$$\mathsf{I} \circ_{\mathsf{b}} \overline{\mathsf{D}} \circ_{\mathsf{y}} \mathsf{R} \;\equiv\; 1 \tag{292}$$

using the fact that $\overline{\mathsf{B}}$ satisfies forward causality. This is, in fact, the forward causality condition for a deterministic operation having no input or income (such as $\overline{\mathsf{D}}$). Hence $\overline{\mathsf{D}}$ is a physical operation in the time forward theory. Further, $\overline{T}$-positivity of $\overline{\mathsf{B}}$ follows from physicality for $\overline{\mathsf{B}}$. Then we see that $T$-positivity of $\mathsf{C}$ defined in (289) follows from $\overline{T}$ positivity of $\overline{\mathsf{B}}$ and $\overline{\mathsf{D}}$. Furthermore, if we plug the definition of $\overline{\mathsf{D}}$ into (289) and apply backward causality we see that both sides agree proving that $\mathsf{C}$ satisfies backward causality (this is because of the way $\overline{\mathsf{D}}$ was defined). It remains to prove that $\mathsf{C}$ satisfies forward causality. This is straight forward. If we apply forward causality to the right hand side of (289) we obtain

$$\overset{\mathsf{y}}{\longrightarrow}\boxed{1}\!\!\langle\; \mathsf{I}_{\mathsf{b}} \;+\; \sum_{x\neq 1} \overset{\mathsf{y}}{\longrightarrow}\boxed{x}\!\!\langle\; \mathsf{I}_{\mathsf{b}} \;=\; \overset{\mathsf{y}}{\longrightarrow}\mathsf{R}\;\; \mathsf{I}_{\mathsf{b}} \tag{293}$$

where we have used the fact that $\overline{\mathsf{B}}$ satisfies forward causality (since it is physical) and (292). We use (262) to show that the expression on the left of (293) is equal to the expression on the right hand side. Consequently $\mathsf{C}$ satisfies forward causality. Since $\mathsf{C}$ is $T$ positive, satisfies backward and forward causality, it is physical. This proves any deterministic physical operation in the time forward theory can be modelled by a deterministic physical operation in the time symmetric theory. The case of nondeterministic operations can be modelled by deterministic operations by making some outcomes implicit and so the theorem is proven in general.

The proof that physical operations in TS can be modelled by physical operations in TF is relatively easy. The reason for this is that we start with a physical operation in TS which is, therefore, constrained to satisfy both forward causality and backward causality (as well as $T$-positivity). We use this to obtain a set of operations in TF which inherit the property of forward causality (as well as $T$-positivity) and so are already physical in the TF sense. We then use these operations to model the TS operation we started with which is, therefore, automatically physical (in the TS sense). This proof is easy because physical operations in TS are, on the face of it, more constrained. The proof that physical operations in TF can be modelled by physical operations in TS is harder since we need to construct an TS operation out of operations which we can prove to be physical in TF, then prove that the TS operation so constructed is physical (in the TS sense). In the case of complex operations, the easy proof (that physical TS complex operations can be modelled by physical TF complex operations) goes through in the same way as in the simple case. However, the hard proof (that physical TF complex operations can be modelled by physical TS complex operations) does not appear to work (see Sec. 53.6).

## 11.11 Forward maximality

In the time symmetric temporal frame we can introduce the assumption of double maximality. In the time forward temporal frame we have to condition on incomes so we have, instead, the property

> *Forward maximality* is the property that, for each pointer type, $\mathtt{x}$, we can associate a physical type $\mathsf{x}$ such that there exist operations
>
> $$\tag{294}$$
>
> for which we have the equivalence
>
> $$\equiv \tag{295}$$
>
> Furthermore, this is maximal in that there does not exist a similar equivalence for the same physical type, $\mathsf{x}$, but for another pointer type, $\mathtt{z}$, having $N_{\mathtt{z}} > N_{\mathtt{x}}$.

We call the object on the left in (294) a *maximal preparation*, and we call the object on the right a *maximal measurement*. The full set of maximal preparations are perfectly distinguishable by means of the corresponding maximal measurement. Interestingly, we can use the circuit in (295) to send information forward in time but there is no corresponding circuit (in the time forward frame) that sends information backwards in time.

## 11.12 Spectrality

Now we will introduce a notion of spectrality and define spectral theories (of which operational classical probability theories and operational quantum the-ory) are examples. We will define these notions with respect to the time forward temporal frame though they clearly extend to the time backward and time symmetric frames. We define spectrality for preparations. A similar definition can be given for results. Our primary reason for introducing spectrality in this book is to prove that we can choose $N_{\mathsf{x}} = N_{\mathsf{y}} N_{\mathsf{b}}$ in the proof in Sec. 11.10 above (such that $\overline{\mathbf{D}}$ defined in (291) satisfies $\overline{T}$-positivity). However, spectrality is actually of more general interest and can play a role in reconstructions of Quantum Theory. This idea is ultimately rooted in the spectral decomposition of density operators in quantum mechanics and in earlier mathematical work on ordered vector spaces and Jordan algebras describing quantum state spaces Jordan et al. [1934], Alfsen and Shultz [2003]. As an explicit assumption in operational reconstructions it is mentioned (though not used) in Hardy [2004] in a discussion of generalized probabilistic theories. It is explicitly used in some reconstructions of quantum theory – for example Dakic and Brukner [2009], Wilce [2011], and Barnum et al. [2014] (see also Barnum and Wilce [2014]). More recently spectrality has been studied systematically in the GPT framework, for example by Chiribella and Scandolo [2015] who gave operational axioms guaranteeing diagonalization of states and explored its consequences for entropy and thermodynamics beyond quantum theory.

First we define the following property.

**Spectrality property:** A preparation, $\overline{\mathbf{A}}$ is said to be *spectral* when we can write

$$\overline{\mathbf{A}}^{\mathsf{c}} \;\equiv\; \overline{\mathbf{E}} \overset{\mathsf{c}}{—} \mathbf{Z}_{\lrcorner}^{\mathsf{c}} \;\equiv\; \sum_{c=1}^{N_{\mathsf{c}}} \overline{\mathbf{E}} \overset{\mathsf{c}}{—} c\langle\; \rangle c \overset{\mathsf{c}}{—} \mathbf{Z}_{\lrcorner}^{\mathsf{c}} \tag{296}$$

for some maximal $\mathbf{Z}_{\lrcorner}$, where we have

$$0 \;\leqq\; \overline{\mathbf{E}} \overset{\mathsf{c}}{—} c\langle \qquad \text{for all } c \tag{297}$$

We call (297) the spectral positivity property.

In words, this is saying that the preparation can be written as being equivalent to a positive weighted sum over maximal elements. Note that the second equivalence in (296) follows from (278).

We have the following simple theorem for spectral operators

**Spectral $\overline{T}$-positivity theorem:** Spectral preparations satisfy $\overline{T}$-positivity.

This follows immediately from the definition of $\overline{T}$-positivity in Sec. 11.8.

**Spectral subunity theorem:** The property

$$\overline{\mathbf{E}} \overset{\mathsf{c}}{—} c\langle \;\leqq\; 1 \tag{298}$$

is required for any spectral operator, $\overline{\mathbf{A}}$ expanded as in (296), that satisfies forward causality.

Spectral subunity is also easy to prove. Forward causality for a deterministic preparation, $\overline{\mathsf{A}}$, is the property

$$\begin{array}{c}\boxed{\mathsf{I}}\\ |\,\mathsf{c}\\ \boxed{\overline{\mathsf{A}}}\end{array} \equiv 1 \tag{299}$$

We have

$$\begin{array}{c}\boxed{\ulcorner\mathsf{Z}}\overset{\mathsf{c}}{\text{——}}\boxed{c'}\\ |\,\mathsf{c}\\ \boxed{c}\overset{\mathsf{c}}{\text{——}}\boxed{\mathsf{Z}\lrcorner}\end{array} \equiv \delta_{cc'} \tag{300}$$

using (295) and (258). It follows from this and (299)

$$\sum_c \boxed{\overline{\mathsf{E}}}\overset{\mathsf{c}}{\text{——}}\boxed{c} \equiv 1 \tag{301}$$

Hence, given spectral positivity (297), we have *spectral subunity*

$$\boxed{\overline{\mathsf{E}}}\overset{\mathsf{c}}{\text{——}}\boxed{c} \leqq 1 \tag{302}$$

as required.

We define spectral theories as follows

> **Spectral theories:** A time forward operational probabilistic theory is a spectral theory if every system preparation has the spectral property.

It is well known that quantum theory has the spectral property since any density operator can be diagonalised in some basis. Operational classical probability theories also have the spectrality property.

Now we prove that the preparation, $\overline{\mathsf{D}}$, defined in (291) and used in the proof of the important theorem in Sec. 11.10, can be made to satisfy $\overline{T}$-positivity if $N_{\mathsf{x}} = N_{\mathsf{by}}$. Given the definition of $\overline{\mathsf{D}}$ in (291) we have

$$\boxed{\overline{\mathsf{D}}}^{\,\mathsf{b}}\overset{\mathsf{y}}{\text{——}}\boxed{\mathsf{Z}\lrcorner}^{\,\mathsf{y}} \equiv \boxed{\frac{N_{\mathsf{x}}}{N_{\mathsf{x}}-1}}\ \boxed{\mathsf{I}}^{\,\mathsf{b}}\ \boxed{\mathsf{I}}^{\,\mathsf{y}} - \boxed{\frac{1}{N_{\mathsf{x}}-1}}\ \begin{array}{c}\boxed{\overline{\mathsf{B}}}^{\,\mathsf{b}}\overset{\mathsf{y}}{\text{——}}\boxed{\mathsf{Z}\lrcorner}^{\,\mathsf{y}}\\ |\,\mathsf{a}\\ \boxed{\mathsf{I}}\end{array} \tag{303}$$

where we have used (174) in the first term after the equivalence. This is equal to

$$\boxed{\overline{\mathsf{D}}}^{\,\mathsf{b}}\overset{\mathsf{y}}{\text{——}}\boxed{\mathsf{Z}\lrcorner}^{\,\mathsf{y}} \equiv \boxed{\frac{N_{\mathsf{x}}}{N_{\mathsf{x}}-1}}\ \boxed{\mathsf{R}}\overset{\mathsf{by}}{\text{——}}\boxed{\mathsf{Z}\lrcorner}^{\,\mathsf{by}} - \boxed{\frac{1}{N_{\mathsf{x}}-1}}\ \boxed{\overline{\mathsf{E}}}\overset{\mathsf{by}}{\text{——}}\boxed{\mathsf{Z}\lrcorner}^{\,\mathsf{by}} \tag{304}$$

where we have used (98) and (174) on *by* in the first term after the equivalence and spectrality on the second term. It follows that expression on the left of (304) is spectral with respect to $\mathsf{Z}_\lrcorner$ if we have spectral positivity

$$0 \;\leqq\; \boxed{\frac{N_{\tt x}}{N_{\tt x}-1}}\; \boxed{\mathsf{R}}\overset{\tt by}{\rule[0.5ex]{1.5em}{0.4pt}}\boxed{by}\!\langle \;-\; \boxed{\frac{1}{N_{\tt x}-1}}\; \boxed{\overline{\mathsf{E}}}\overset{\tt by}{\rule[0.5ex]{1.5em}{0.4pt}}\boxed{by}\!\langle \tag{305}$$

We have

$$\boxed{\overline{\mathsf{E}}}\overset{\tt by}{\rule[0.5ex]{1.5em}{0.4pt}}\boxed{by}\!\langle \;\leqq\; 1 \tag{306}$$

by spectral subunity. It follows that the expression on the left in (304) is guaranteed to be spectral with respect to $\mathsf{Z}_\lrcorner$ if $N_{\tt x} = N_{\tt by}$ (where we use (261) on ${\tt by}$). It then follows that this expression satisfies $\overline{T}$-positivity (using the spectral $\overline{T}$-positivity theorem above). It is then a simple matter to prove that $\overline{\mathsf{D}}$ satisfies $\overline{T}$-positivity. We do this by noting that we can recover $\overline{\mathsf{D}}$ by hitting the expression on the left of (304) with $^\ulcorner\mathsf{Z}$ and using double maximality. Then we use the fact that

$$0 \underset{\overline{T}}{\leqq} \boxed{\overline{\mathsf{F}}}\ \text{(outputs b, y)} \quad\Rightarrow\quad 0 \underset{\overline{T}}{\leqq} \boxed{\overline{\mathsf{F}}}\ \text{(output b; output y into } \boxed{^\ulcorner\mathsf{Z}} \text{ with output y)} \tag{307}$$

for any preparation, $\overline{\mathsf{F}}$. This follows simply from the definition of $\overline{T}$-positivity in Sec. 11.8 (as the inequality on the right is, effectively, more constrained).

One final comment on $\overline{\mathsf{D}}$. We showed in Sec. 11.10 that $\overline{\mathsf{D}}$ can be constructed such that it satisfies positivity and forward causality. It is possible a particular theory has other constraining principles on operations and we do not know that $\overline{\mathsf{D}}$ satisfies them as well. If we assume we are working in spectral theories then we can do better. It is reasonable to assume that we can realise any spectral preparation (that is any positive weighted mixture of some set of maximal preparations where the sum of the weights does not exceed unity). We were able to show that the object on the left in (304) is spectral (if $N_{\tt x} = N_{\tt by}$) and so it can be prepared (the existence of $\overline{\mathsf{D}}$ then follows from the above arguments).

## 11.13 Operation expansion and duotensors

Operator fiducial expansions and duotensor calculations are set up in the same way as the time symmetric case. The difference is that the fiducial elements must correspond to time forward operations (so incomes are conditioned on). The natural choice for the fiducial pointer elements is

$$\rangle\!\boxed{x}\overset{\tt x}{\rule[0.5ex]{1.5em}{0.4pt}} \qquad\qquad \overset{\tt x}{\rule[0.5ex]{1.5em}{0.4pt}}\boxed{x}\overset{\tt x}{\rule[0.5ex]{1.5em}{0.4pt}}\boxed{\mathsf{R}} \tag{308}$$

Compare this with (193) for the time symmetric case. If we make this choice for the pointer fiducial elements then the pointer fiducial matrix is simply the identity (rather than having a $1/N_{\tt x}$ factor as in the time symmetric case. Then the inverse pointer fiducial matrix is also the identity. System fiducial preparations will be preselected in the time forward case and so correspondingly, the system fiducial matrix will have different normalisation from the time symmetric case.

Duotensor calculations work the same way as in the time symmetric case.

# 12 Time backward simple operational probabilistic theories

Time backward Simple Operational Theory is just the time reverse of the time forward case. Thus, operations have no outcomes and can be obtained by postselecting on outcomes from the time symmetric case

(309)

where we have *postselected* on $y$ on the right hand side. Note, we denote backward deterministic operations in the time backward case by **B̲**.

The tester positivity condition is

(310)

with respect to testers of the form

(311)

for all pure D̲ and E̲.

The causality conditions on deterministic operations are

b b
R x B ≡ I (312)
a
I

and

I
b
If no implicit postselection x B* ≡ x R I (313)
a a

(where $\underline{\mathsf{B}}^*$ denotes that there is no implicit postselection). For operations that may be non-deterministic, the general causality conditions are

b b
R x B $\underset{T}{\leqq}$ I (314)
a
I

and

I
b
If no implicit postselection x B* $\underset{T}{\leqq}$ x R I (315)
a a

Saturation of these inequalities gives the deterministic case.

In the time backward frame we have backwards maximality - this being the time reverse of forward maximality (discussed for the time forward frame).

The fiducial expansion of operations is the time reverse of the time forward case. The natural choice of pointer fiducial elements is

R x $x$ x        x $x$ (316)

The pointer fiducial matrix is equal to the identity again.

# 13 Classical copying, preselection, and coalescence

Typically we imagine that all the data collected from an experiment can be brought one place so it can be analysed (and, perhaps, a summary published). In our present context, this means that we need to be able to collect the readouts in one place. To do this, we need to be explicit about the wiring that would be used to actually bring all this information to one place. We immediately hit a problem in the time symmetric theory. Such theories do not allow copying of classical information either forward in time, or backward in time. Copying forward in time becomes possible if we have preselection boxes (which are allowed in the time forward theory). Copying backward in time becomes possible if we have postselection boxes (which are allowed in the time backward theory).

Imagine we want to copy the classical information on a pointer wire

x (317)

forward in time. To do this we would have to insert a box into the wire as follows

x forward copy x x (318)

What does it mean for this box to copy in the forward time direction? A natural definition would be that it has the following property

$x$ x forward copy x x $\equiv$ $x$ x $x$ x (319)

for all $x$. If we have such a box we can concatenate it to produce multiple copies.

We can define backward copying boxes by flipping the above diagrams horizontally.

## 13.1 No classical copying in time symmetric theory

Here we will see that, assuming flatness, there is no forward or backward copying of pointer information (which is classical) in the time symmetric theory. Here we prove the impossibility of a copying machine that works for all $x$. Later we will prove an even stronger theorem (though based on more assumptions) that it is not possible even to copy a given value of $x$ (so there is no preselection amplification).

> **No pointer copy operation in time symmetric theory.** Consider time symmetric operational probabilistic theories with flatness
>
> **No forward-copying.** There is no deterministic operation, $\mathsf{A}$, satisfying
>
> $x$ x $\mathsf{A}$ x x $\equiv$ $x$ x $x$ x (320)
>
> for all $x$ that also satisfies backward causality.

**No backward-copying.** There is no deterministic operation, **B**, satisfying

$$\equiv \qquad (321)$$

for all $x$ that also satisfies forward causality.

This means that there are no deterministic physical operations capable of copying pointer information forward or backward in time.

It follows from (320) that

$$\equiv \qquad (322)$$

We will prove the no forward copying case (the no backward copying case follows similarly). Consider the left hand side of (322). This is a circuit with no postselection. Hence we are heeding the preselection warning (see Sec. 11.5). Since we also have flatness, we can employ the forward seatbelt identity (see Sec. 11.6). We can append an **R** box followed by a flag and sum over $x$ as follows

$$\sum_{x=1}^{N_x} \quad \equiv \quad \equiv \qquad (323)$$

where the second expression follows from the summed seatbelt identity (278) and the third expression follows using backward causality. The expression on the right is equivalent to $\frac{1}{N_x^2}$ by flatness. However, if we append an **R** box followed by a flag and sum over $x$ for the expression on the right hand side of (322) we obtain

$$\sum_{x=1}^{N_x} \qquad (324)$$

Using (258), the seatbelt identity, (43), and flatness we see that this is equivalent to $\frac{1}{N_x}\delta_{x'x''}$ which is different from the expression we got from the left hand side. This proves the theorem.

## 13.2 Classical copying with preselection and with postselection

Now we will see that, if we actually have preselection boxes, then we can copy classical information forward in time. Similarly, if we have postselection boxes, we can copy classical information backward in time.

**Pointer copying with preselection and postselection.** This theorem consists of two parts.

**Time forward-copying possible with preselection.** There exists a deterministic operation, **C**, satisfying the physicality conditions such that

(325)

for all $x$.

**Time backward-copying possible.** There exists a deterministic operation, **E**, satisfying the physicality conditions such that

(326)

for all $x$.

Thus preselected physical deterministic operations can forward-copy but not backward-copy.

The proof of the first point is by example. Assume that **C** in (325) implements a permutation such that $1x \to xx$ (it is always possible to find a permutation that does this and, further, permutations satisfy the conditions for determinism and physicality). Then we have the forward copying property in (325). Note that we can model this permutation operation using a duotensor with black square dots on the left and white square dots on the right. Then the duotensor has 0's and 1's as entries as appropriate to implement the permutation matrix.

We might wonder whether postselection can be used to enable forward copying, or preselection used to enable backward copying (so, in either case, the selection is "opposed" to the direction of copying). The following theorem states that this is not the case.

**Pointer copying with opposed pre/post-selection.** Consider time symmetric theories with flatness.

**Time forward-copying with postselection not possible.** If we assume flatness, then there is no deterministic operation, **F**, satisfying

(327)

for all $x$ that also satisfies backward causality.

**Time backward-copying with preselection not possible.** If we assume flatness, then there is no deterministic operation, **D**, satisfying

(328)

for all $x$ that also satisfies forward causality.

Thus postselected physical deterministic operations can backward-copy but not forward-copy.

The special case of this theorem when $N_{\mathsf{y}} = 1$ returns the no pointer copying theorem we obtained in Sec. 13.1. We will prove that postselection cannot be used to enable forward copying (the proof of the backward case is similar). First, consider

(329)

For postselection to enable forward copying as in (327) we require that $x = x' = x''$ if $y = 1$. Now add an **R** box followed by a $x$ flag box, and sum over $x$, as follows

(330)

To obtain the expression on the right we have used the forward seatbelt identity (which is allowed since we have assumed flatness and do not have postselection) and then we used forward causality. Now, following on from our previous remarks, if (327) is to implement forward copying then, after we sum over $x$, we still require that $x' = x''$ if $y = 1$. We can see, however, that this is not possible the right hand side of (330) shows that $y$, $x'$, and $x''$ are uncorrelated. This proves the theorem.

## 13.3 On incomes and settings

Recall the following figure

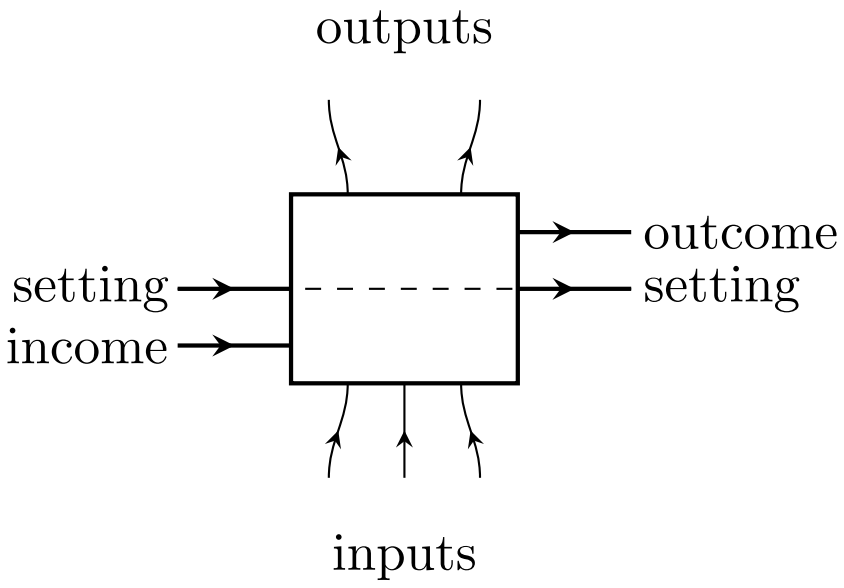

(from Sec. 5.1). On the left we have both incomes and settings. What is the difference between them? The key difference is that the setting is available both before and afterwards (for example, the position of a knob can be read off both before and after the operation has happened). The income, on the other hand, is only available before. However, if we have preselection, then we can copy the income so that this copy becomes available afterwards. Thus, in theories with preselection, there is no meaningful distinction between the notions of incomes and settings. This explains why we do not have this distinction in standard time forward operational Quantum Theory. It is, of course, similarly true that there is no meaningful distinction between outcomes and settings in theories having postselection.

## 13.4 No preselection or postselection amplification

If we have a preselected pointer preparation can we amplify this so we have two preselected pointer preparations? Here we prove that we cannot, and similarly that we cannot amplify postselected preparations either.

> **No preselection or postselection amplification in time symmetric theory.** Consider time symmetric operational probabilistic theories with flatness
>
> **No forward amplification.** There is no deterministic operation, **A**, satisfying
>
> $x$ x — **A** — x, x $\equiv$ $x$ x, $x$ x (331)
>
> for some given $x$ that also satisfies backward causality and tester positivity.

**No backward amplification.** There is no deterministic operation, **B**, satisfying

$$\text{(332)}$$

for some given $x$ that also satisfies forward causality and tester positivity.

This means that there are no deterministic physical operations capable of amplifying either forward or backward in time.

This theorem statement is similar to the no copying theorem statement in Sec. 13.1. The differences are that here we only demand the transformation works for some given $x$ (this is sufficient to amplify a preselection/postselection), and we need to use tester positivity to prove this result (unlike the no copying result). We provide a proof by contradiction of this theorem for the preselection case with $x = 1$. So, preselection amplification would be a process satisfying

$$\text{(333)}$$

This implies

$$\sum_{x=1}^{N_{\mathtt{x}}} \qquad \equiv \qquad + \quad \alpha \qquad \text{(334)}$$

where

$$\alpha \equiv \sum_{x=2}^{N_{\mathtt{x}}} \qquad \text{(335)}$$

Now, by tester positivity, $\alpha$ must be non-negative. Further, we can use (323) and (261) to simplify the left hand side of (334) and (258) and (261) to simplify the first term on the right hand side. Putting all this together, we obtain

$$\frac{1}{N_{\mathtt{x}}^2} \geq \frac{1}{N_{\mathtt{x}}} \tag{336}$$

But this is not possible for any non-trivial case (where $N_{\mathtt{x}} > 1$). This proves, by contradiction, that we cannot amplify preselection as in (333).

## 13.5 On formal and strong preselection

The time symmetric formulation of operational theories is formally equivalent to the time forward formulation (as was shown in Sec. 11.10). In particular, we can calculate the conditional probabilities of time forward operational theories from joint probabilities calculated within the time symmetric theory. An example was given in Sec. 11.5 where we calculate the pertinent conditional probability as follows

$$\text{prob}(u, v, w|x, y, z) = \frac{\text{prob}(u, v, w, x, y, z)}{\text{prob}(x, y, z)} \tag{337}$$

The joint probabilities in the numerator and denominator on the right can both be calculated from the time symmetric formulation.

Let us think more carefully about the notion of preselection. Imagine we want to preselect on a particular income so we have a preparation like

$$\langle x \overset{\mathsf{x}}{\longrightarrow} \boxed{\mathbf{X}}^{\;\mathsf{x}} \tag{338}$$

What does it mean to preselect $x$ here? At a very basic level we are always free to talk about the conditional probability, $\text{prob}(-|x, -)$, of something conditioned on a particular value of $x$. As just discussed, we can always use the joint probabilities calculated in the time symmetric formalism to calculate such conditional probabilities. Let us call this *formal preselection*.

However, the notion of preselection is more meaningful if

- we can *purposely* select a particular income, $x$, and
- we are able to *locally compare* preselected income values with readouts of the experiment in some small region of spacetime.

Let us call this *strong preselection* to contrast it with the weaker notion of formal preselection. The reason for the second point is that we want to be able to bring all the data we have collected together so we can analyse it. The localised region of spacetime will, one imagines, be *after* the regions of spacetime where we perform the experiment.

It is worth commenting that the notion of strong preselection, as defined above, invokes notions that are prevalent in physics but which we do not normally discuss in a head-on fashion. First, the idea that we "purposely select" something raises issues to do with free will that are beyond the scope of this book. However, it is worth noting that, if we have a source of pointer preparations which are preselected in some given state (say "1"), then we can use permutation operations (which are free operations) to transform this given pointer preselection to any other value. So let $\mathbf{A}$ be such a permutation operation (that maps $1 \to x$). Then

$$\langle 1 \overset{\mathsf{x}}{\longrightarrow} \boxed{\mathbf{A}} \overset{\mathsf{x}}{\longrightarrow} \; = \; \langle x \overset{\mathsf{x}}{\longrightarrow} \tag{339}$$

Thus, a notion of free will is built into the choice of the operation, **A**. We can think of the choice of operation as being determined by the setting. Curiously, the setting is a time symmetric notion (it represents classical information available before and after the operation). Second, strong preselection requires us to think about making all classical information (the readouts) available at a certain place in spacetime (so it can be analysed). In real experiments this is often done by electrical wiring taking the information to some box where it can be correlated and processed.

We can also define a notion of *strong postselection* wherein

- we can purposely select a particular outcome, $y$, and
- we are able to compare postselected outcomes with readouts of the experiment in some localised region of spacetime.

This is the time reverse of strong preselection. The localised region of spacetime here will, one imagines, be *before* the regions of spacetime where we perform the experiment.

## 13.6 Strong preselection or postselection as a resource

Resource theory has been developed both at an abstract level (see Coecke et al. [2016], Fritz [2017], and Gonda and Spekkens [2023]), and applied to many particular cases Chitambar and Gour [2019]. Di Biagio (private communication, 2020) suggested that we could have a resource theory in which preselection is treated as a resource. The key idea in resource theories is to have operations that are free and operations that are regarded as resources (so not free). There are, then, some consistency requirements so that we cannot generate, or amplify resources using the free operations.

In a preselection resource theory we have

**Free operations.** Deterministic physical operations and readout boxes from the time symmetric theory for all a, b, y, z, and x, and for any $x$. Since nondeterministic operations can be built from deterministic ones and readout boxes, this includes all physical operations in the time symmetric theory.

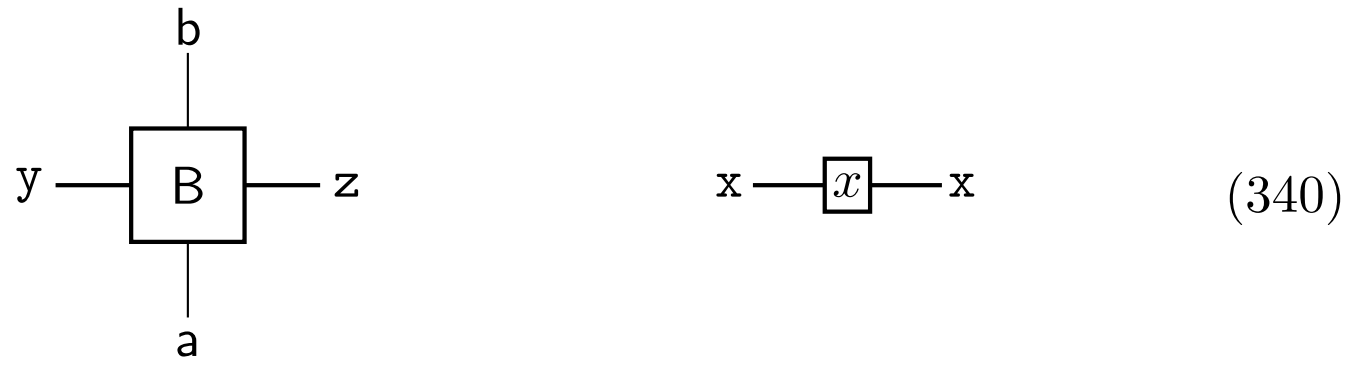

**Resources.** Preselection boxes such as

$$x \;\text{—}\; \mathtt{x} \qquad (341)$$

for any system type, x, and for any $x$. We will interpret these as providing strong preselection as defined in Sec. 13.5.

We have already proven that we cannot amplify preselection using deterministic time symmetric physical operations so we have the kind of consistency we require for a resource theory. Also note that we can use pointer preselection boxes to prepare quantum systems in normalised states.

We can, similarly, define a postselection resource theory in which postselection boxes become the resource instead of preselection boxes.

The resource theory approach is useful since we can use it to ask about conversion between different resources and develop resource measures. We will not develop the resource theory of preselection or postselection beyond the above comments in this book.

## 13.7 Encapsulating a time symmetric circuit with external preselection

A principle contention of this book is that the time symmetric point of view is fundamental. This is driven, in particular, by the ambition of finding a theory of Quantum Gravity. A theory of Quantum Gravity must reduce to General Relativity and Quantum Theory in appropriate limits. Therefore, an appropriate methodology is to examine conceptual differences between the two theories. General Relativity is time symmetric while Quantum Theory (in its usual operational formulation) appears to be time asymmetric. We have seen that we can, in fact, repackage such time asymmetric operational formulations into a time symmetric formulism (and, as we will see, Quantum Theory fits into this new time symmetric formulism). Thus, this can be taken as a step towards a theory of Quantum Gravity. However, we do have the problem that we cannot copy the classical information collected at the readouts for later analysis without strong preselection (or postselection).

It is, however, possible to encapsulate a time symmetric circuit (inside a dashed box) whilst passively copying the readout information using the resource of some external preselection boxes (outside the dashed box) so that this readout information can all be taken to some given location in the future for analysis. This is best illustrated by example. Thus, consider the circuit

(342)

We can calculate $\text{prob}(x, y)$ using the machinery of the time symmetric formal-ism. We can encapsulate this as follows

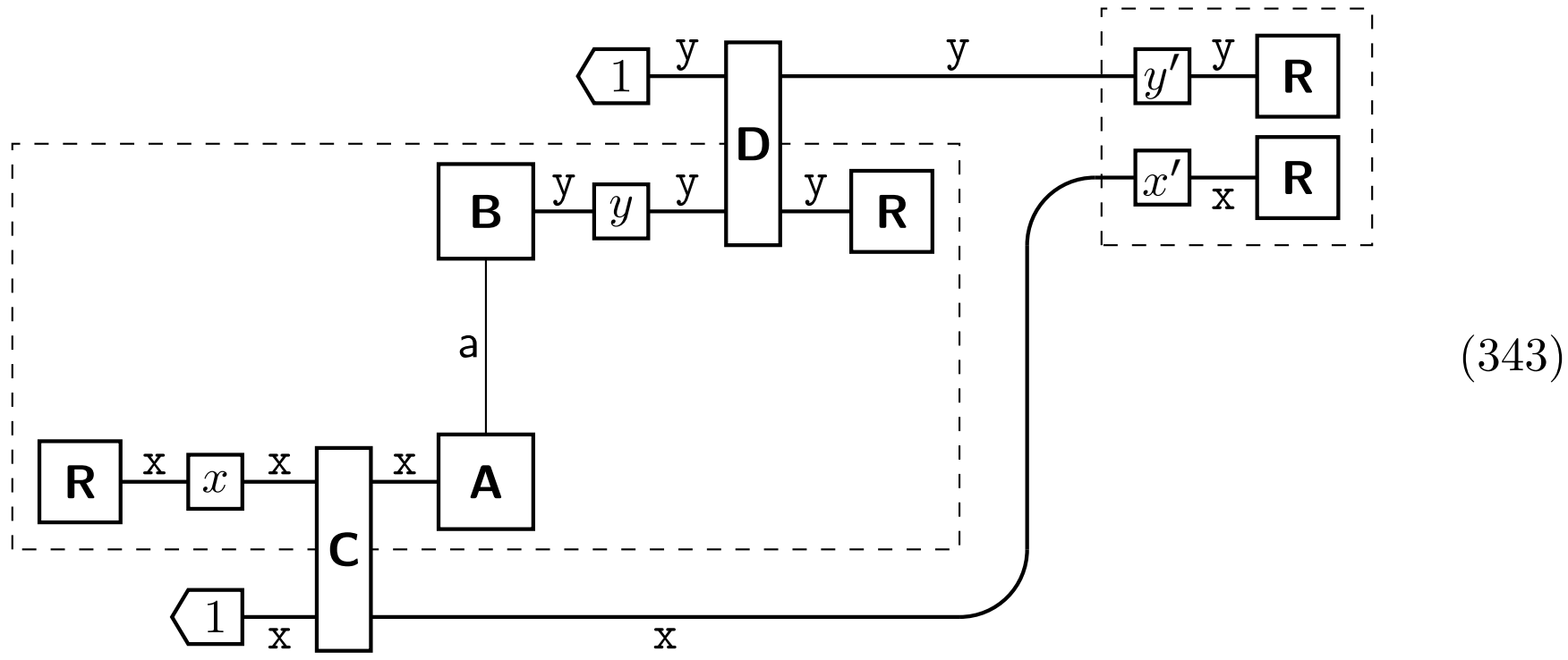

(343)

Here **C** and **D** are permutation transformations chosen to allow forward-copying of x and y respectively with the shown preselections (as discussed in Sec. 13.2). Thus, the readout $x$ is copied to the readout $x'$ (so these are equal). Similarly, the readout $y$ is copied to the readout $y'$. The forward-copying process does not disturb the pointer being copied. Thus we can passively monitor the time symmetric circuit in (342) by adding the extra structure in (343). The probability $\text{prob}_{\text{encap}}(x', y')$ read-off at the end is the same as $\text{prob}(x, y)$ for $x' = x$ and $y' = y$. The larger dashed box shows the original circuit "encapsulated". The operations inside the box are time symmetric (and further, the permutation operations **C** and **D** are passive). Outside this encapsulated circuit, we allow preselection. The smaller dashed box shows that all the information has been brought to a small spacetime region. We could, in this smaller dashed box, have some processing of the data by means of more operations which might involve the use of further preselection.

This encapsulation means that we can perfectly well view experiments as being time symmetric and still collect data if we have a time asymmetric passive monitoring process using a strong preselection resource. However, if we want to maintain that the world is actually time symmetric at some deep fundamental level, then we need to account for the existence of strong preselection. This is the issue we will now turn our attention to.

## 13.8 Why can we preselect?

There is, we might say, a "wind" of strong preselection blowing from the past to the future that allows the forward copying of information and provides us with memories of the past, but not of the future. In our present context, it is this wind that provides the time asymmetry in the world we live in. Where does this wind come from?

Certainly, to backtrack a little, it is *our experience* of the world that we have strong preselection in the sense described in Sec. 13.5, but that we do not have strong postselection. Why is this? This is a deep problem and it is not the intent in this book to give a definitive account of where the appearance of strong preselection comes from. However, in this subsection, we will discuss some possible ways we might go about accounting for the existence of strong preselection. In the next subsection we will flesh out one possible approach to doing this (involving what we will call *coalescence*) in more detail.

The time symmetric formalism sits very unnaturally with the notion of strong preselection since the only deterministic pointer preparations are $\mathsf{R}$ which correspond to flat distributions over $x$ (suggesting we are not able to purposely select a particular income) and, further, we cannot copy incomes (so we cannot compare them with later outcomes).

The time forward operational formalism (which is, essentially, the standard approach to Quantum Theory in textbooks) does fit naturally with strong preselection since we can have pure deterministic pointer preparations (which suggests we can purposely select a particular income, $x$) and, further, we can copy readouts so they can be transported to later times (so we are able to compare $x$ with later outcomes). However, given the formal equivalence with the time symmetric formalism (proved in Sec. 11.10), this natural fit does not actually prove we have strong preselection rather than just formal preselection.

If we are to account for this strong preselection we will have to think more carefully. Here we will raise a few possibilities that offer some hope of providing an understanding the origin of strong preselection from within the time symmetric point of view.

**Time asymmetric physics somewhere.** It is possible that the time symmetric physicality conditions breakdown somewhere. One possibility is that they breakdown in places where we have strong Quantum Gravity effects (such as in the early universe). In such a case, the early universe would provide us with ample strong preselection which we can regard as a resource.

**We are already highly preselected.** By the time we come to perform an experiment, it has to be the case that we have experimentalists and apparatuses. We are already necessarily conditioning on this so that we can even begin to talk about our experiment.

**Special role for mind.** A radical possibility is that conscious mind can play a special role in observing readouts. For example, it is possible that it could allow awareness of some readouts $x$ and $y$ even though they are at different times - conscious mind might allow a (perhaps only slightly) nonlocal in time appreciation of outcomes (this very much fits in with the work by the philosopher Zhou [2023a,b] on the specious present). A more prosaic approach would be to model mental processes to see how they fit into the operational model in a more detailed way (this idea has been pursued in Milburn and Shrapnel [2018] and Evans et al. [2021]). The fact that we are a contingent consequence of a long chain of events might imply that we, ourselves, are the necessary source of the strong preselection resource required to forward-copy.

It would be ironic if, as per the first point, we need Quantum Gravity to account for the existence of strong preselection since the motivation for the present work is to formulate Quantum Theory in a time symmetric way so it is more like General Relativity with the idea that this will aid the discovery of a theory of Quantum Gravity.

It is possible that we need to appeal to some theory of mind here. However, we need to be careful not to fall into the trap of trying to account for one mystery with another mystery. In the end, science requires clear explanations. There are other ways, besides those mentioned above, to bring mind into the discussion. One more way to bring mind into the picture is through the interpretation of probability. According to one point of view (strongly advocated for by Fuchs [2010, 2017]) probabilities are beliefs and therefore mental in origin. In such a case, we are free to believe in preselected pointer states. It is not clear whether this point of view is necessarily time asymmetric. An point of view that is explicitly time asymmetric at the ontological level is that of Dowker [2022a,b] where spacetime "atoms" are born in the forward time direction, these birth events being the neural correlates of conscious experience. There is also, of course, a line of thinking that goes back to Wigner [1961] speculating that collapse of the wavefunction is caused by conscious mind. This point of view has been developed in recent decades by Hameroff and Penrose [2014]. It is also an explicitly time asymmetric point of view at the ontological level. In this book we are interested in time symmetry at the operational level. An interesting question is how the question of time symmetry/asymmetry at the operational and ontological levels interface with one another.

## 13.9 Coalescence - transgressing the boundaries

Here we will pursue one approach to accounting for the existence of strong preselection within a broader time symmetric point of view. We do this by suggesting a mechanism to account for the fact that we are already highly preselected by virtue of the fact that apparatuses and experimentalists are a precondition for doing operational physics (this is the second point in Sec. 13.8 above). Where do such apparatuses and experimentalists come from in the first place? The operational framework we have set up treats systems and operations on a different footing. The systems could be the fundamental particles of physics. A highly composite system made out of very many such particles might *coalesce* into a sufficiently macroscopic physical object that we might want to call an operation. Let us represent this *schematically* by

$$\mathbf{I}^{\mathsf{c}\mathsf{d}\mathsf{e}\cdots\mathsf{f}} \;\rightsquigarrow\; p\; \mathbf{B}_{\mathsf{a}\mathsf{x}}^{\mathsf{b}\mathsf{y}} \;+\; (1-p)\; \mathbf{J}^{\mathsf{c}\mathsf{d}\mathsf{e}\cdots\mathsf{f}} \tag{344}$$

We can read this schematic process as follows. We start with a maximally mixed preparation, represented by $\mathbf{I}^{\mathsf{cde}\ldots\mathsf{f}}$ on the left. Then we can imagine a process whereby a piece of apparatus (associated with the operation $\mathbf{B}$) might form from the systems $\mathsf{c}$, $\mathsf{d}$, ..., $\mathsf{f}$. This will only happen when these systems have the right correlations - hence the process is probabilistic. We imagine that $p$ is very small. Since $p$ is small, $\mathbf{J}$ is almost the same as $\mathbf{I}$. Such a process transgresses the boundaries between system and operation and so would take us outside the operational framework we have set up (Sokal [1996]). It is clear the schematic process illustrated in is outside the operational framework because the types are not matched between the different terms (we have used the symbol $\rightsquigarrow$ to indicate this). Consequently it is possible that neither $\mathbf{B}$ nor $\mathbf{J}$ is constrained by the time symmetric physicality conditions. Consequently, this coalescence process is not described by the time symmetric operational probabilistic theory we have set up. It is, therefore, possible that preselection operations will coalesce out of this maximally mixed "primeval soup". This, then, provides a tentative origin for time symmetry violating operations. To strengthen this claim we would need to provide a theoretical description of how coalescence could actually produce preselection operations in the broader context of a time symmetric theory. This does not seem unlikely since the process we have described is probabilistic and works in a forward time direction. There is a danger here however. We do not want it to be the case that, in subverting the time symmetric operational theory, we allow formation of any kind operations kind of process since this might lead to an uncontrolled violation of the basic cherished principles we have used to set up the whole framework in the first place.

The schematic process in (344) describes the probabilistic formation of a single operation. To do operational physics this is not enough. We need coalescence of a whole world of apparatuses, and experimentalists (along with laboratories and, perhaps, a few theorists too).

The idea that strong preselection comes about through coalescence in the broader context of a time symmetric theory is a radical idea. In such a view, we can maintain that the initial state of the universe is maximally mixed (in accord with flatness) and so has maximal entropy (as opposed to the usual point of view that the universe started in a very low entropy state). However, we would, by virtue of the fact that coalescence of apparatuses and ourselves is necessary for us to begin to do physics experiments, only be observing a world conditioned on this very coalescence. If the laws of physics are time symmetric, then this coalescence could happen forward in time or backward in time (in which case it would feel like it was forward in time) or both if we have a many worlds type ontology. We can imagine a branching structure where coalescences forward in time give rise to a forward in time branching structure whilst coalescences backward in time give rise to a backward in time branching structure (so that the overall picture remains time symmetric). In a full theory of Quantum Gravity spatiotemporal properties (such as the metric and the size of the universe) would have to be part of the state (they could not be treated as a fixed background). Understanding how this might fit in with a truly time symmetric picture represents an interesting challenge. For example, one supposes that the universe would have to be both expanding and contracting "at the same time".

We have described coalescence above as a process - something that happens over time eventually leading to experimentalists and apparatuses. Certainly it is reasonable to expect that this will happen. However, if the initial state really is maximally mixed, then some components of this initial mixture will already contain the necessary correlations between the fundamental particles for fully formed experimentalists and apparatuses. So, there is already (with small probability) a world of classical objects sufficient to do operational physics in the initial state. We might, by the way, question the very notion of an "initial state". We can imagine an infinite past where coalescing branches feel like they have a finite history from the inside.

There is, in this way of thinking, a radical many worlds type of interpretation in which the universe is always in a maximally state (whether regarded in the forward or backward time direction) rather than a pure state as in the standard many worlds interpretation (see Everett [1957]). This would be an "everything everywhere all at once" interpretation Kwan and Scheinert [2022]. When viewed from within, an operational world may have coalesced within some branches allowing us to account for appearances. It is not entirely clear such a point of view is coherent. Tools of the sort developed by Saunders [1998], Deutsch [1999], and Wallace [2003, 2012] might be useful in attempting to making sense of this more radical many worlds style interpretation. see Saunders et al. [2010] for a discussion of the many worlds interpretation by many authors (including some critical voices). For those of us not entirely comfortable with the many worlds ontology, one might hope for a "one world" ontology consistent with this time symmetric coalescence picture. That remains a challenge for future work.

There is a point to be discussed concerning the applicability of operational thinking. On the one hand, Quantum Theory is well understood in operational settings where it can be used to calculate probabilities. On the other hand, cosmological considerations drive us to think outside the operational framework (for example, Quantum Cosmologists often speak about the wavefunction of the universe where there are no external apparatuses or experimentalists Hartle and Hawking [1983]). If we stick within a fully operational point of view, then it is not clear how we can address cosmological issues. One way to come at this is to think about the notion of "now". In an operational setting we can foliate circuits into synchronous partitions (this is a set of wires for which no wire in the set can be arrived at by tracing forward through the circuit from any other wire in the set). Synchronous partitions are basically the space-like hypersurfaces of circuits. We can associate a notion of "now" with some such synchronous partition. This notion, however, is highly dependent on having operations (i.e. experimental apparatuses). It is easy to think about apparatuses in a terrestrial laboratory but harder to think about this in, say, the early universe. What, then, does it mean to speak of the notion of "now" for the universe as a whole when we do not have the operational backdrop of instruments defining foliations as describe above? What is required here is an ontology which can inform how operational notions are grounded in a bigger picture of reality. The notion of a consistent "now" across all branches may not survive in this deeper ontology. Indeed, this would seem to follow from the idea of quantum coordinates in Hardy [2020] where a general quantum coordinate transformation (or a quantum diffeomorphism) can reassign the notion of "now". A lesson from history is that operational thinking can help us on route to making progress in our understanding of the underlying ontology, but we might have to go beyond these operational notions at some point. Operationalism, from this point of view, might be regarded as a kind of conceptual scaffolding - useful while an ontology is under construction but ultimately to be removed. Though, of course, even after the scaffolding has been removed, such an ontological theory still has to make contact with the operational world so we can use it to make predictions we can test in real experiments.

# 14 General temporal frames

## 14.1 Introduction

We have considered just three temporal frames: time symmetric, time forward, and time backward. This discrete situation contrasts sharply with the situation in General Relativity where, in a very different context, general coordinate systems provide for a continuum of frames. We will see that we can, in fact, define a continuum of temporal frames in operational theories. Such frames do, in general, lead to a nonlinear expression for the probability because the denominator in the expression for the relevant conditional probability does not simplify to a constant. This is unlike the case of time symmetric, time forward, and time backward temporal frames where we have linearity (in the latter two cases this is because the double causality conditions ensure that this denominator does simplify to a constant).

We will return to the comparison with general coordinates from General Relativity in Sec. 14.6.

## 14.2 General case

The temporal frame we have is determined by what we attach to the incomes and outcomes. In general we can consider a network with open pointer wires such as

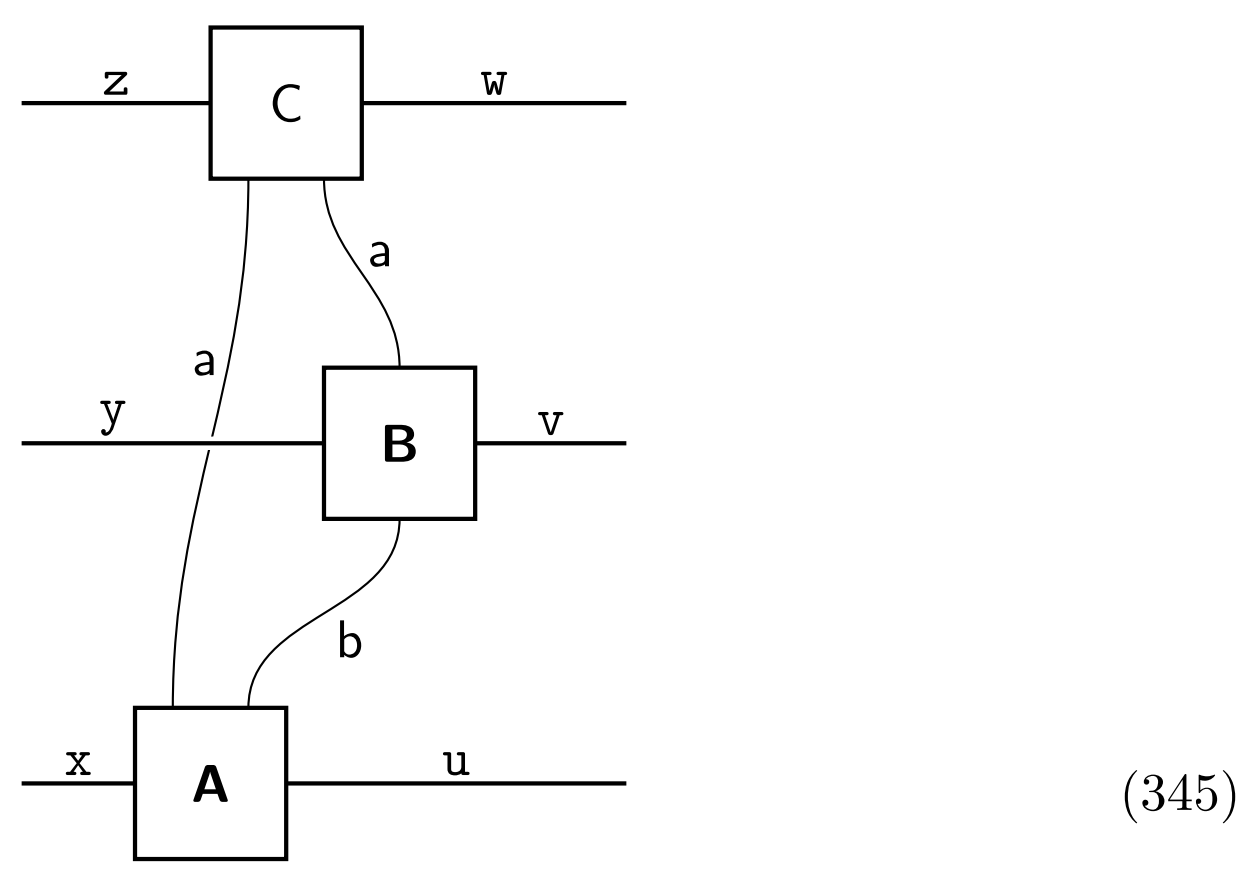

(345)

(where all the system wires are closed). We can *view* this network by appending a network (call this the *reference network*) to the open pointer wires so we have a circuit. We allow preselection and postselection boxes in the reference fragment. We can think of any such view as constituting a general temporal reference frame. For example we could close (345) into a circuit as follows.

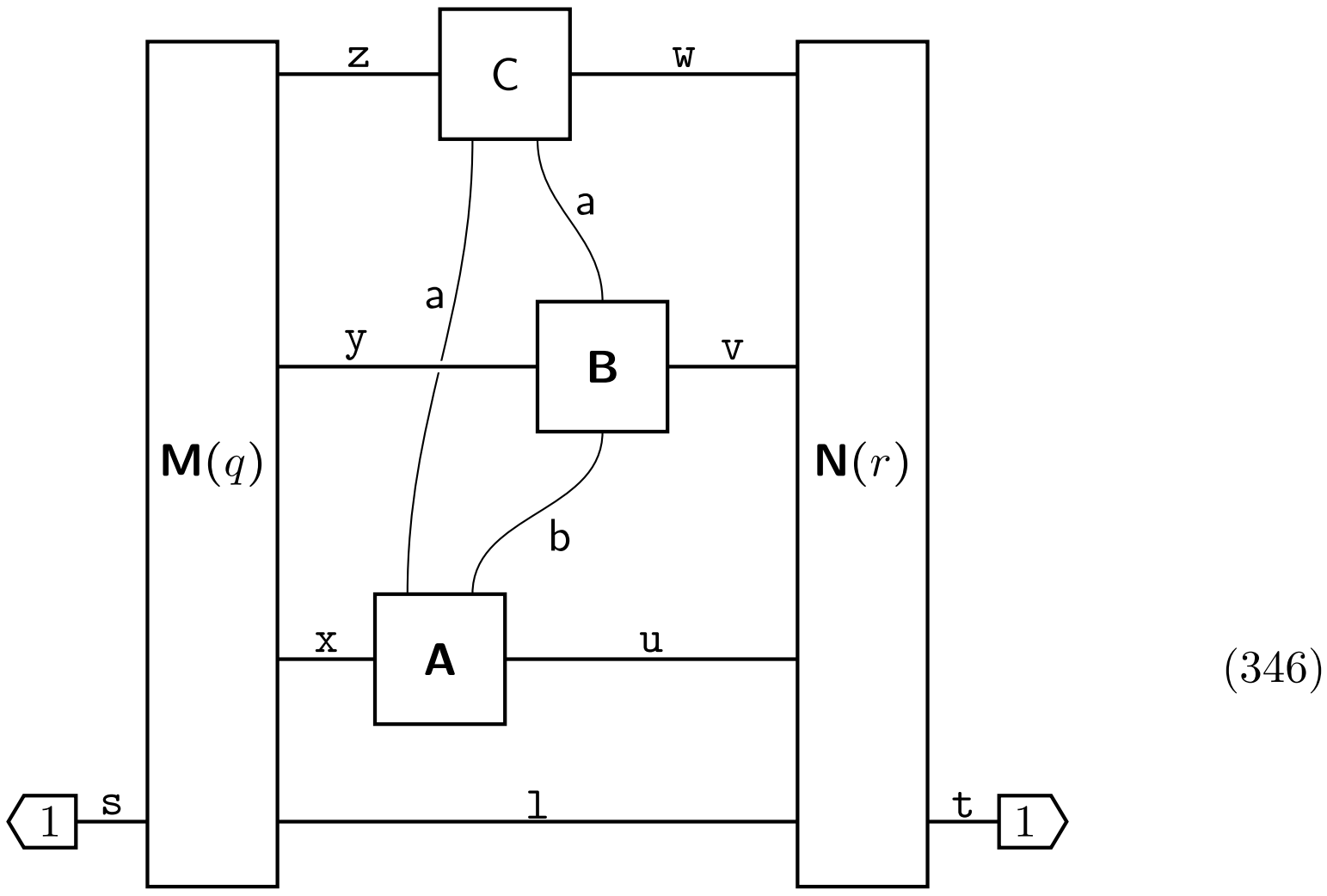

(346)

(Note this is not the most general way to view this circuit because, for example, the outcome wire from **A** could be fed into **B**).

## 14.3 Locally implementable case

There exist temporal frames that can be implemented locally by attaching boxes separately to each income and outcome. We will call these *locally implemented temporal frames*.

We can locally close these open incomes and outcomes in various ways. The most general bit of apparatus we can place on an income or outcome are

1 s **Q**$(q)$ x t 1      x 1 m **S**$(r)$ n 1      (347)

respectively. A locally implemented general temporal frame to view the frag-

ment in (345) looks like

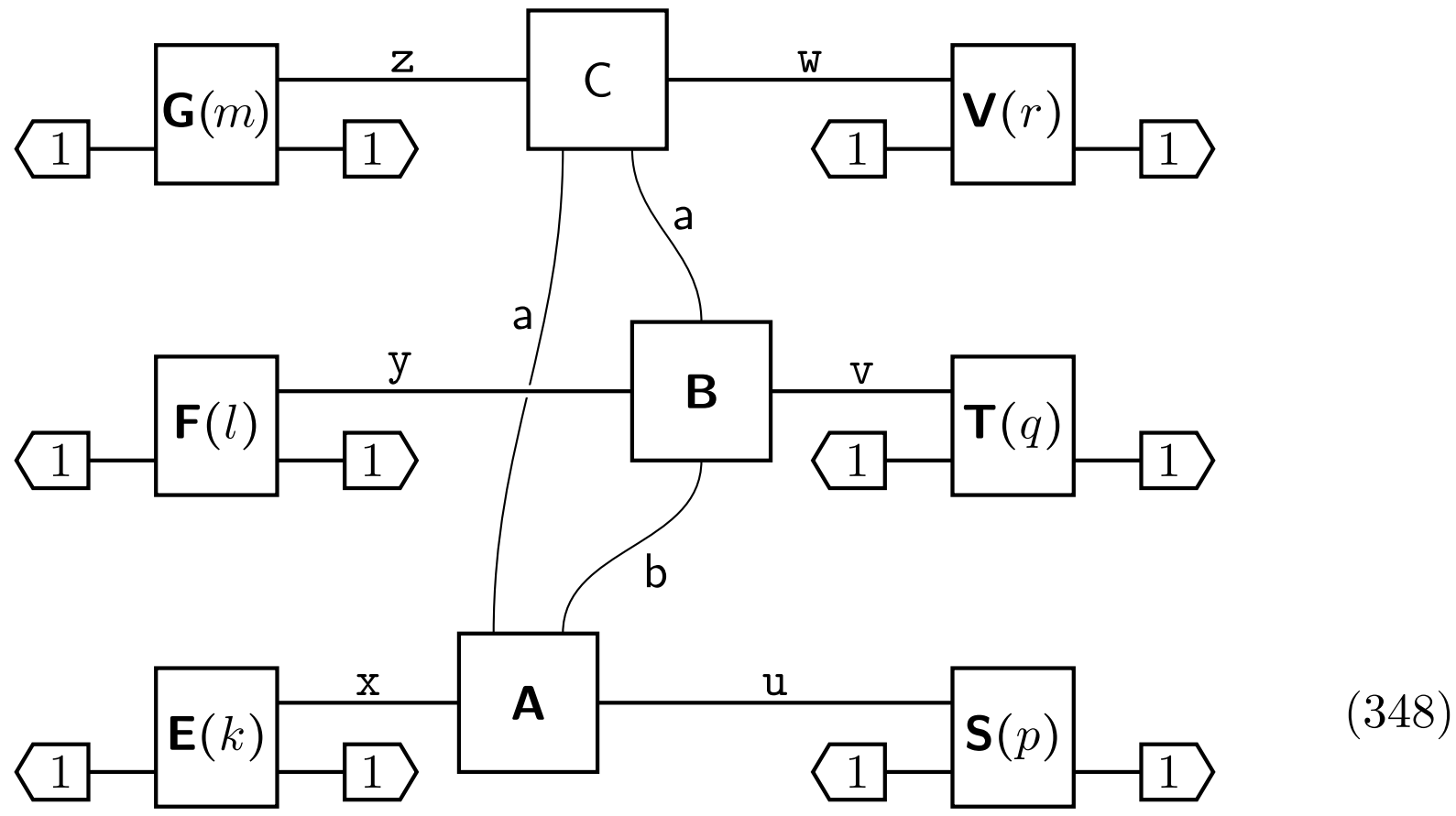

$$(348)$$

(we have omitted some wire labels). The probability associated with this circuit is given by

$$\text{prob}(k, m, l, p, q, r | 1, 1, \ldots, 1) = \frac{\text{prob}(k, m, l, p, q, r, 1, 1, \ldots, 1)}{\text{prob}(1, 1, \ldots, 1)} \tag{349}$$

where the 1's refer to the pre- and post-selection boxes. The denominator is given by summing over $k, m, l, p, q, r$, and is equal to the probability for

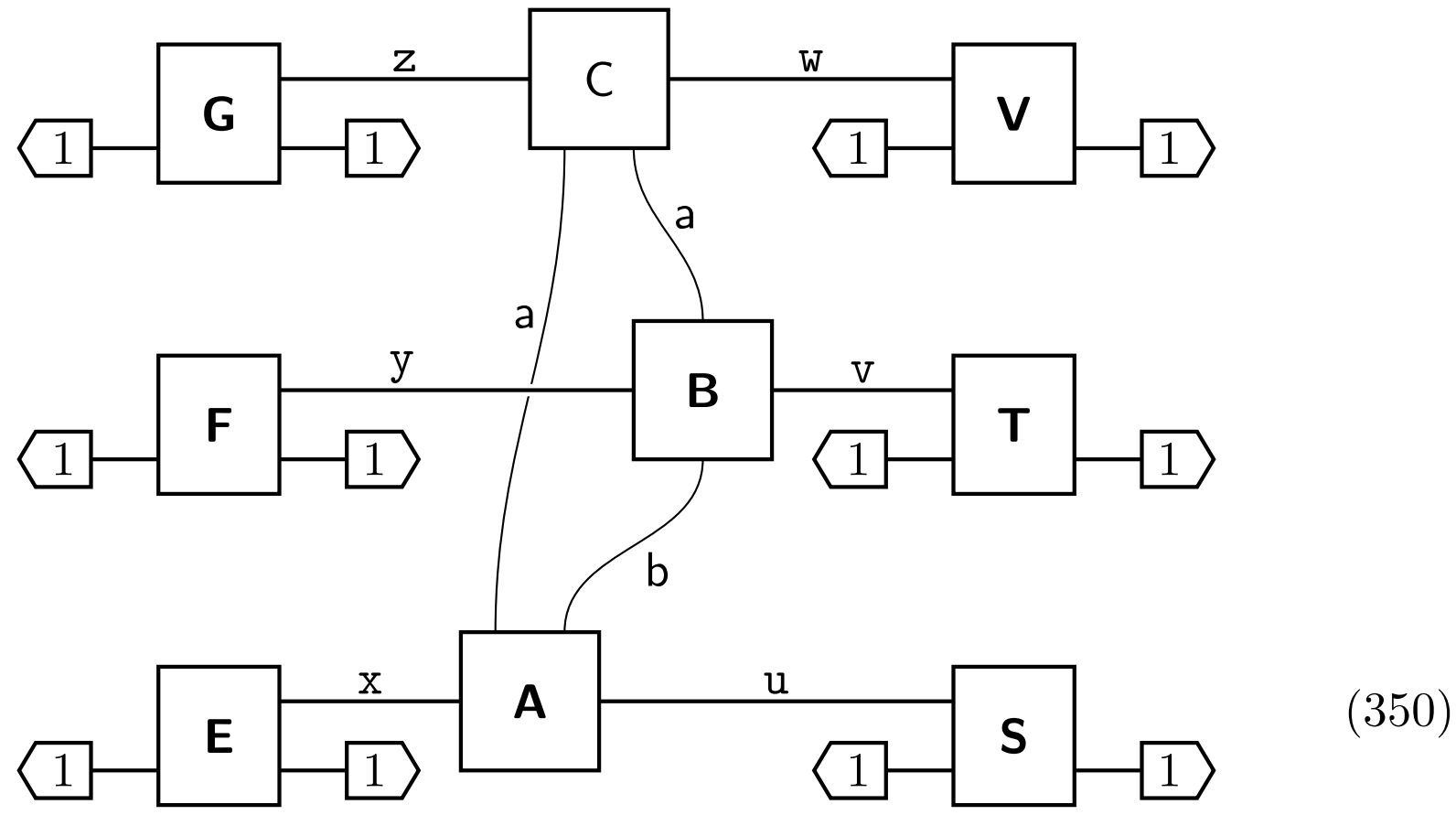

$$(350)$$

This denominator does not, in general, simplify. Compare with the probability shown in (270) for the forward frame. In that case the denominator does simplify using forward causality as shown in (271). It is clear that the time symmetric, time forward, and time backward temporal frames are special cases of this general temporal frame. Since $\mathsf{S}(u)$ and $\mathsf{T}v$ are general operations they are specified by real numbers and consequently there are a continuum of such frames.

The time symmetric, time forward, and time backward temporal frames use the same type of apparatus to close each of the incomes and the same type of apparatus to close each of the outcomes. In this sense they are global (by analogy with a global inertial reference frame).

## 14.4 The ABL rule

A special case of interest of a non-global frame is illustrated in the example

$$
\begin{array}{ccccccccc}
 & & & & \mathsf{A} & \overset{\mathrm{s}}{—} & \boxed{1}\rangle & & \\
 & & & & \mathrm{b}\,| & & & & \\
\boxed{\mathsf{R}} & \overset{\mathrm{x}}{—} & \boxed{x} & \overset{\mathrm{x}}{—} & \mathsf{B} & \overset{\mathrm{y}}{—} & \boxed{y} & \overset{\mathrm{y}}{—} & \boxed{\mathsf{R}} \\
 & & & & \mathrm{a}\,| & & & & \\
 & & \langle\boxed{1} & \overset{\mathrm{r}}{—} & \mathsf{L} & & & &
\end{array}
\tag{351}
$$

Here we preselect to form the preparation, postselect to form the result, and have **R** and readout boxes on the middle operation. The probability $p(x, y)$ is given by

$$
\mathrm{prob}(x, y|1, 1) = \frac{\mathrm{prob}\left(\begin{array}{ccccccccc}
 & & & & \mathsf{A} & \overset{\mathrm{s}}{—} & \boxed{1} & \overset{\mathrm{s}}{—} & \boxed{\mathsf{R}} \\
 & & & & \mathrm{b}\,| & & & & \\
\boxed{\mathsf{R}} & \overset{\mathrm{x}}{—} & \boxed{x} & \overset{\mathrm{x}}{—} & \mathsf{B} & \overset{\mathrm{y}}{—} & \boxed{y} & \overset{\mathrm{y}}{—} & \boxed{\mathsf{R}} \\
 & & & & \mathrm{a}\,| & & & & \\
\boxed{\mathsf{R}} & \overset{\mathrm{r}}{—} & \boxed{1} & \overset{\mathrm{r}}{—} & \mathsf{L} & & & &
\end{array}\right)}{\mathrm{prob}\left(\begin{array}{ccccccccc}
 & & & & \mathsf{A} & \overset{\mathrm{s}}{—} & \boxed{1} & \overset{\mathrm{s}}{—} & \boxed{\mathsf{R}} \\
 & & & & \mathrm{b}\,| & & & & \\
 & & & \boxed{\mathsf{R}}\overset{\mathrm{x}}{—} & \mathsf{B} & \overset{\mathrm{y}}{—}\boxed{\mathsf{R}} & & & \\
 & & & & \mathrm{a}\,| & & & & \\
\boxed{\mathsf{R}} & \overset{\mathrm{r}}{—} & \boxed{1} & \overset{\mathrm{r}}{—} & \mathsf{L} & & & &
\end{array}\right)}
\tag{352}
$$

The denominator does not simplify under double causality. This is the situation considered by Aharonov, Bergmann, and Lebowitz (ABL) Aharonov et al. [1964] except (i) we have added the income, $x$ on the middle box to make it more time symmetric, and (ii) the calculation here works for general states (not just pure states).

## 14.5 General system temporal frames

It is worth mentioning a way to develop this point of view further (though this is a little outside the main thrust of our discussion). The network we started with in (345) has only pointer wires open. We could consider how to view a fragment with both pointer and system wires open. This leads to a notion of general system temporal frames. For example, we could view the network

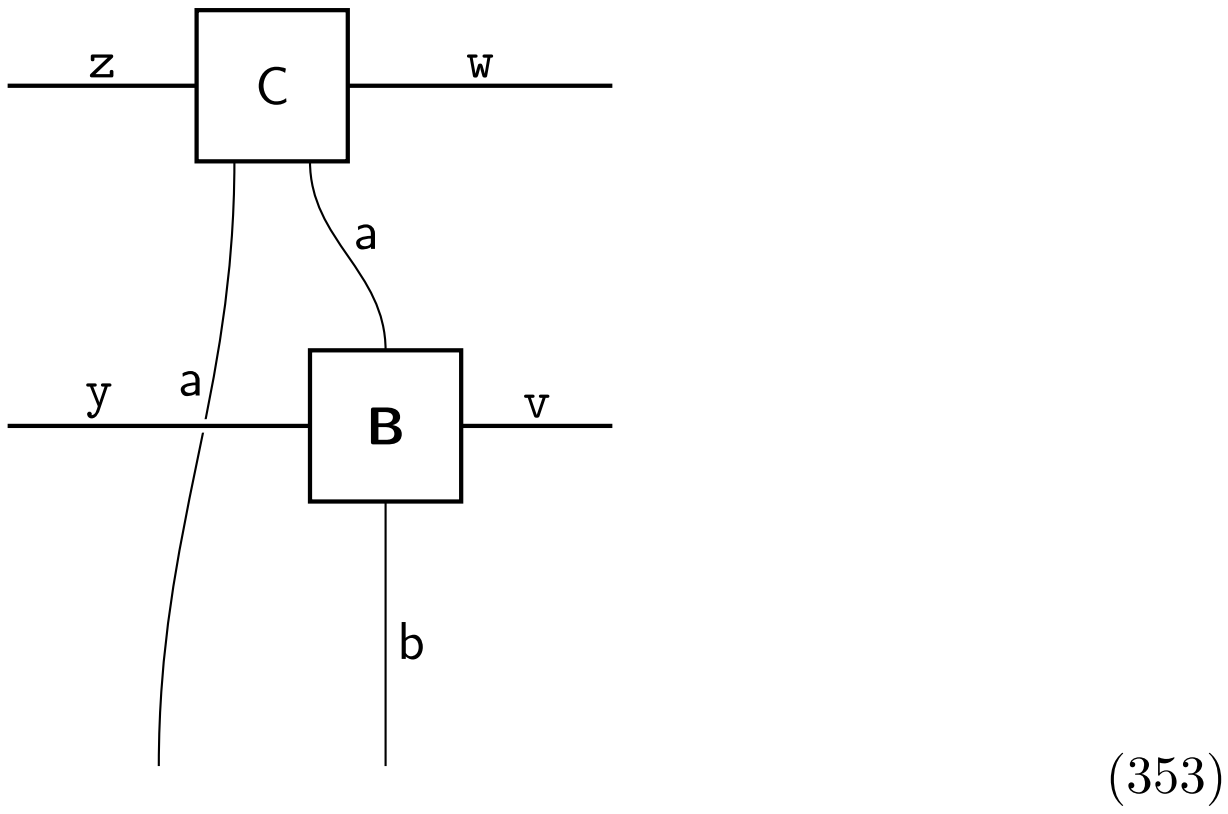

(353)

as follows

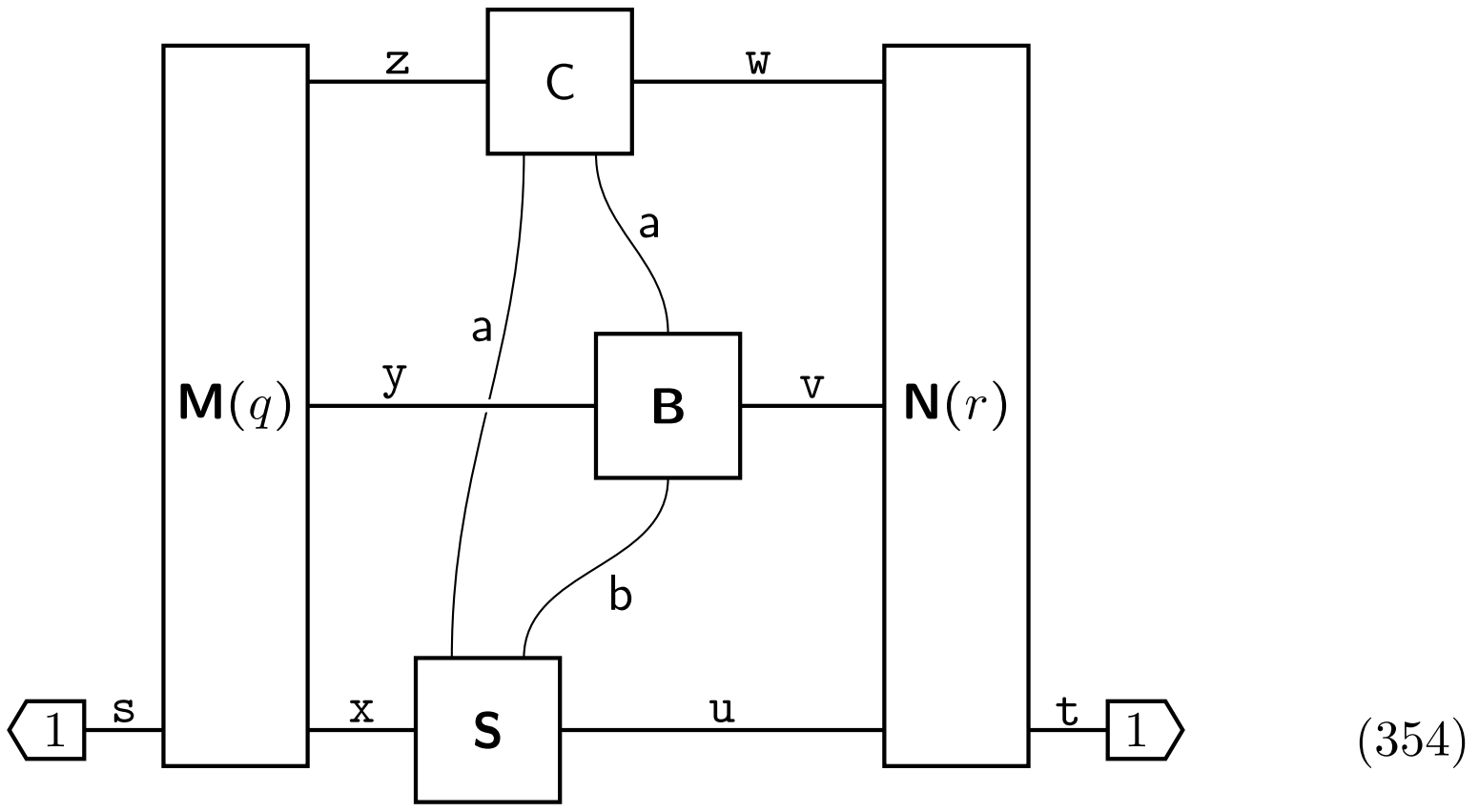

(354)

In the case that the systems are quantum systems we have a notion of *general quantum temporal frames* (since the perspective is, in part, controlled by quantum systems). This notion is not obviously related to Quantum Reference Frames as introduced by Aharonov and Susskind Aharonov and Susskind [1967] and Aharonov and Kaufherr Aharonov and Kaufherr [1984].

## 14.6 Should we put all temporal frames on the same footing?

In General Relativity there is a dual attitude. On the one hand we put all coordinate systems on an equal footing through the principle of general covariance. This plays an important role in determining the form of the equations. On the other hand, local inertia frames play a special role. This special role is witnessed in the principle of minimal coupling whereby we replace partial derivatives in Special Relativistic field equations by covariant derivatives as a means of guaranteeing that there is a reference frame in which inertial physics continues to hold locally and, also, enabling us to obtain general relativistic field equations from special relativistic ones. It is also witnessed in the vielbein formalism (preferred by some) where the basic object (*vielbeins* usually denoted by $e^a_\mu$) is the transformation between inertial coordinates to general coordinates at the given point.

We might similarly argue that, on the one hand we should formulate operational theories such as Quantum Theory so that all temporal frames are on the same footing, and on the other hand treat the time symmetric case as special. It is reasonable to argue that the time symmetric frame is the special one because there is no denominator. Thus, we always have linearity. As it happens, the time forward and time backward temporal frames also turn out to be linear, but only because the forward and backward causality conditions respectively make it so. If we generalise to a new theory where these conditions do not hold then we will no longer have linearity.

One might argue that the time forward temporal frame (that appears in textbooks) is the fundamental one because it relates most closely to how we experience the world. The counter argument to this is that such experience is contingent on less fundamental features of the world. In particular it relies on having an abundant resource of preselection which may be a consequence of the rather contingent phenomenon of coalescence. A similar situation pertains in General Relativity. We might argue that a reference frame stationed at the ground is the fundamental one since this relates most closely to how we experience the world. However, this is based on rather contingent aspects of our situation in the world - that we are standing on a big mass that formed (one might even say coalesced) at some earlier point in history).

Here we have presented the idea that we use up a preselection resource to view the world in the time forward temporal frame. An interesting question, clearly outside the scope of this book, would be to ask what the resource is in General Relativity we use when we want to view the world in a non-inertial reference frame.

# Part II

# Causally Simple Operational Quantum Theory

## 15 Introduction

In the previous part of this paper we set up the theory of causally simple operational probabilistic theories ($t$SOPT). Now we leverage this structure to set up causally simple operational quantum theory ($t$SOQT). The structure of $t$SOQT is obtained by appending to the structure of $t$SOPT in the following way

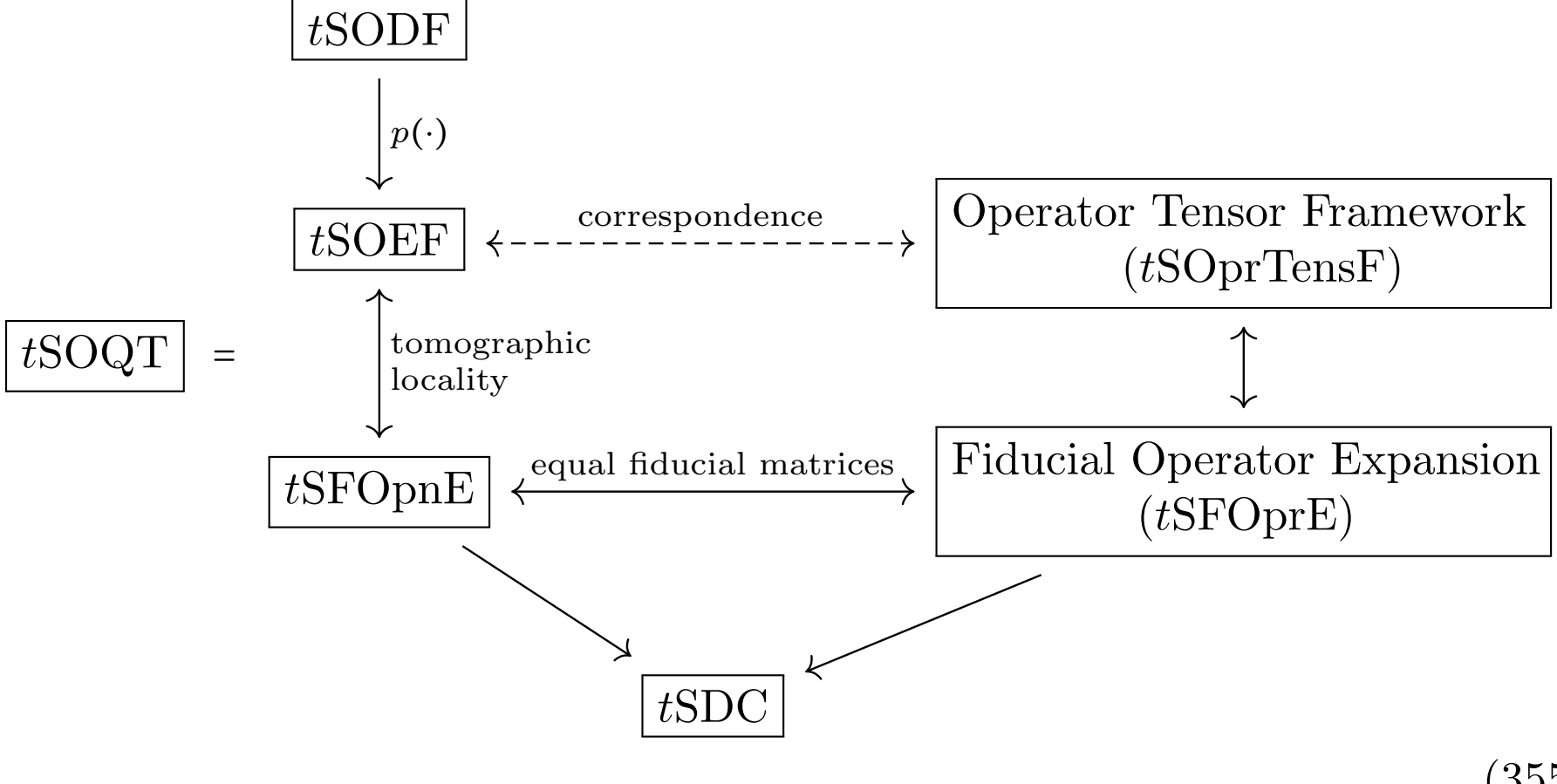

(355)

This is the flowchart in Sec. 3 with $x$ =S (as we are treating operations with simple causal structure). The lowest entry on the flowchart, $t$SDC, standing for $t$-simple duotensor calculations) helps tie the quantum structure on the right with the operational probabilistic structure on the left. We will mostly take a time symmetric perspective here with $t$ =TS as we did in the previous part of this book.

Central to this approach are *operator tensors* (which we will often just call *operators* though they do have a tensorial-style structure which plays an important role). Here the operators have simple causal structure so we will call them *simple operators* if we need to disambiguate from those having complex causal structure (which will be introduced in Part V). We will set up the operator tensor framework wherein operators tensors play an analogous role to operations. Then we will show how operator tensors can be expanded in terms of fiducial elements analogously to how this can be done with operations. Then we see that, when the fiducial matrices coming from the OPT side of the flowchart (on the left) are equal to the fiducial matrices coming from the OQT side, we are able to establish a correspondence between operations and operator tensors such that we can calculate probabilities for circuits by means of an operator tensor calculation.

The physicality constraints (positivity and causality) on operations get translated into constraints on operator tensors. We will develop the necessary mathematical prerequisites to be able to formulate these positivity and causality constraints on operator tensors.

# 16 Simple Operator tensors

## 16.1 Purity correspondence assumption

In Sec. 19 we will set up a notion of correspondence between operations and operators. It is useful to start with a notion of correspondence just for pure preparations and pure results (purity was defined in Sec. 7.8). We will say that the operator corresponding to a preparation is a *state* and the operator corresponding to a result is an *effect*.

The idea of purity plays an important role in setting up $t$-SOPT. In particular, pure preparations and results are used to state the property of tester positivity. Thus, we need a statement about what operators these correspond to. We make the following assumption.

**Purity correspondence assumption.** We have the following

*Pure preparations.* Pure system preparations, $\mathsf{A}^{\mathsf{a}_1}$, correspond to rank one operators, $\hat{A}^{\mathsf{a}_1} = |A\rangle^{\mathsf{a}_1}\langle A|$ (where $|A\rangle^{\mathsf{a}_1}$ is not necessarily normalised to 1). Furthermore, all appropriately normalised rank one operators, $\hat{A}^{\mathsf{a}_1} = |A\rangle^{\mathsf{a}_1}\langle A|$, have a pure preparation, $\mathsf{A}^{\mathsf{a}_1}$, that they correspond to.

*Pure results.* Similarly, pure system results, $\mathsf{C}_{\mathsf{a}_1}$, correspond to rank one operators, $\hat{C}_{\mathsf{a}_1} = |C\rangle_{\mathsf{a}_1}\langle C|$ (where, again, $|C\rangle_{\mathsf{a}_1}$ is not necessarily normalised to 1). Furthermore, all appropriately normalised rank one operators, $\hat{C}_{\mathsf{a}_1} = |C\rangle_{\mathsf{a}_1}\langle C|$, have a pure result, $\mathsf{C}_{\mathsf{a}_1}$, that they correspond to.

We will say that we have a pure state when this corresponds to a pure preparation. Likewise, we have a pure effect when this corresponds to a pure result.

Here we used the bra and ket notation since it will be familiar to most readers. It is, however, antiquated from the point of view in this book. The best notation for these objects is diagrammatic and will be introduced in Part III. Even if we stick with symbolic notation there are reasons not to use bra and ket notation explained in Sec. 28.2

We will see in Sec. 25 that, in the time symmetric frame, the normalisation of these pure states and effects is such that the maximum probability given by a pure preparation followed by a pure result is, in fact, equal to $\frac{1}{N_{\mathsf{a}}}$ (and not equal to 1 as we have in the standard time forward frame where we condition on incomes). This follows from the maximal preparations and results theorem in Sec. 8.4.

Note that, since homogeneous preparations and results are proportional to pure preparations and results, we can also assert that these also correspond to rank one projectors.

It follows from the purity correspondence assumption and the double purity theorem (in Sec. 7.8) that any state (effect) can be written as a positive weighted sum of rank one operators. Thus it follows that any preparation followed by any result yields, under correspondence to the corresponding state and effect, a real non-negative number as is required for this number to represent a probability. This fact provides strong motivation for the purity correspondence assumption.

## 16.2 States and effects

According to the purity correspondence assumption, the state associated with a pure preparation is written $|A\rangle^{\mathsf{a}_1}\langle A|$. We can write this as $|A\rangle^{\mathsf{a}_1}\otimes{}^{\mathsf{a}_1}\langle A| \in \mathcal{H}^{\mathsf{a}_1}\otimes\overline{\mathcal{H}}^{\mathsf{a}_1}$ where

$$|A\rangle^{\mathsf{a}_1} \in \mathcal{H}^{\mathsf{a}_1} \qquad\qquad {}^{\mathsf{a}_1}\langle A| \in \overline{\mathcal{H}}^{\mathsf{a}_1} \tag{356}$$

(Note that, in the time symmetric frame, $|A\rangle^{\mathsf{a}_1}$ will not be normalised.) In general, a state is mixed and we can write it as

$$\hat{A}^{\mathsf{a}_1} \in \mathcal{H}^{\mathsf{a}_1} \otimes \overline{\mathcal{H}}^{\mathsf{a}_1} \tag{357}$$

We write $\mathcal{L}^{\mathsf{a}_1} := \mathcal{H}^{\mathsf{a}_1} \otimes \overline{\mathcal{H}}^{\mathsf{a}_1}$. Now, a mixed state is a positive weighted sum of pure states and is, consequently, Hermitian. Thus, we actually have

$$\hat{A}^{\mathsf{a}_1} \in \mathcal{V}^{\mathsf{a}_1} \subset \mathcal{L}^{\mathsf{a}_1} \tag{358}$$

where $\mathcal{V}^{\mathsf{a}_1}$ is the space of Hermitian operators in $\mathcal{L}^{\mathsf{a}_1}$. We write

$$\boxed{\mathsf{A}}^{\,\mathsf{a}} \quad \Longmapsto \quad \boxed{\boxed{\hat{A}}}^{\,\mathsf{a}} \;\in\; \mathcal{V}^{\mathsf{a}_1} \tag{359}$$

for the diagrammatic representation of states corresponding to preparations. The symbol, $\Longmapsto$, means *corresponds to*.

Similarly, according to the purity correspondence assumption, associated with a pure result is an effect written $|B\rangle_{\mathsf{a}_1}\langle B|$. We can write this as ${}_{\mathsf{a}_1}\langle B|\otimes|B\rangle_{\mathsf{a}_1} \in \mathcal{H}_{\mathsf{a}_1}\otimes\overline{\mathcal{H}}_{\mathsf{a}_1}$ (we have reversed the order of the ket and bra to be consistent with the discussion in Sec. 28 later) where

$${}_{\mathsf{a}_1}\langle B| \in \mathcal{H}_{\mathsf{a}_1} \qquad\qquad |B\rangle_{\mathsf{a}_1} \in \overline{\mathcal{H}}_{\mathsf{a}_1} \tag{360}$$

In general, an effect is mixed and we can write it as

$$\hat{B}_{\mathsf{a}_1} \in \mathcal{H}_{\mathsf{a}_1} \otimes \overline{\mathcal{H}}_{\mathsf{a}_1} \tag{361}$$

We write $\mathcal{L}_{\mathsf{a}_1} := \mathcal{H}_{\mathsf{a}_1} \otimes \overline{\mathcal{H}}_{\mathsf{a}_1}$. Now, actually quantum effects are represented by Hermitian operators. Thus, we actually have

$$\hat{B}_{\mathsf{a}_1} \in \mathcal{V}_{\mathsf{a}_1} \subset \mathcal{L}_{\mathsf{a}_1} \tag{362}$$

where $\mathcal{V}_{\mathsf{a}_1}$ is the set of Hermitian operators in $\mathcal{L}_{\mathsf{a}_1}$. Diagrammatically, we write

$$\boxed{\mathsf{B}}_{\,\mathsf{a}} \quad \Longrightarrow \quad \boxed{\hat{B}}_{\,\mathsf{a}} \in \mathcal{V}_{\mathsf{a}_1} \tag{363}$$

for the diagrammatic representation of effects corresponding to results.

## 16.3 Operator tensors

The basic object in the simple operator tensor framework is the simple operator tensor. This can be represented diagrammatically and symbolically by

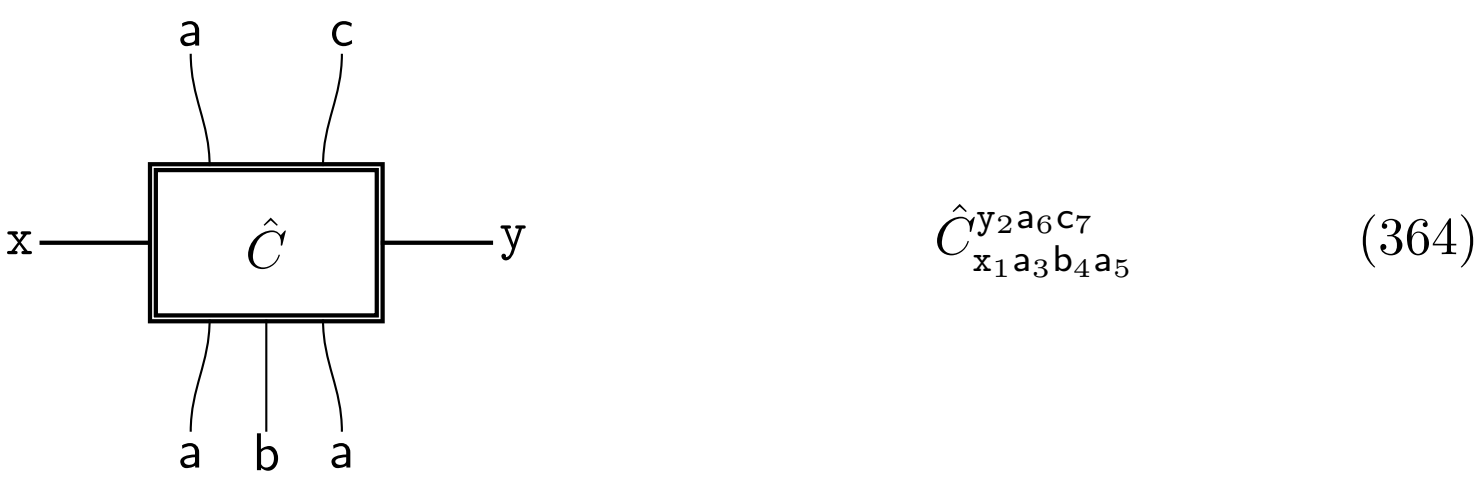

$$\hat{C}^{\mathsf{y}_2\mathsf{a}_6\mathsf{c}_7}_{\mathsf{x}_1\mathsf{a}_3\mathsf{b}_4\mathsf{a}_5} \tag{364}$$

The double border in the diagrammatic notation tells us that we have a operator tensor rather than an operation. For the deterministic case we will use bold $\hat{\boldsymbol{C}}$ as we did with operations. A simple operator tensor is an operator with some tensorial structure added. The example above is an element of the space

$$\mathcal{P}_{\mathsf{x}_1} \otimes \mathcal{L}_{\mathsf{a}_3} \otimes \mathcal{L}_{\mathsf{b}_4} \otimes \mathcal{L}_{\mathsf{a}_5} \otimes \mathcal{P}^{\mathsf{y}_2} \otimes \mathcal{L}^{\mathsf{a}_6} \otimes \mathcal{L}^{\mathsf{a}_6} \otimes \mathcal{L}^{\mathsf{c}_7} \tag{365}$$

The spaces $\mathcal{L}$ are as defined above. The income space, $\mathcal{P}_{\mathsf{x}_1}$ is a real vector space of dimension $N_{\mathsf{x}}$ and the outcome space, $\mathcal{P}^{\mathsf{y}_2}$, is a real vector space of dimension $N_{\mathsf{y}}$.

The subscripts and superscripts of a simple operator tensor tell us what space that operator tensor belongs to. In the case that there are no inputs or outputs (no system wires) then we will omit the hat over the $C$ symbol.

For brevity, rather than using the lengthy nomenclature "simple operator tensors", we will usually use "operator tensors" or just "operators" (except when we need a longer name for disambiguation purposes).

We will see in Sec. 20 that tester positivity implies that these operator tensors are actually Hermitian so the above example will actually be an element of

$$\mathcal{P}_{\mathsf{x}_1} \otimes \mathcal{V}_{\mathsf{a}_3} \otimes \mathcal{V}_{\mathsf{b}_4} \otimes \mathcal{V}_{\mathsf{a}_5} \otimes \mathcal{P}^{\mathsf{y}_2} \otimes \mathcal{V}^{\mathsf{a}_6} \otimes \mathcal{V}^{\mathsf{a}_6} \otimes \mathcal{V}^{\mathsf{c}_7} \tag{366}$$

The spaces $\mathcal{V}$ are as defined above.

# 17 Networks and circuits

We can wire together operators to create operator networks. For example

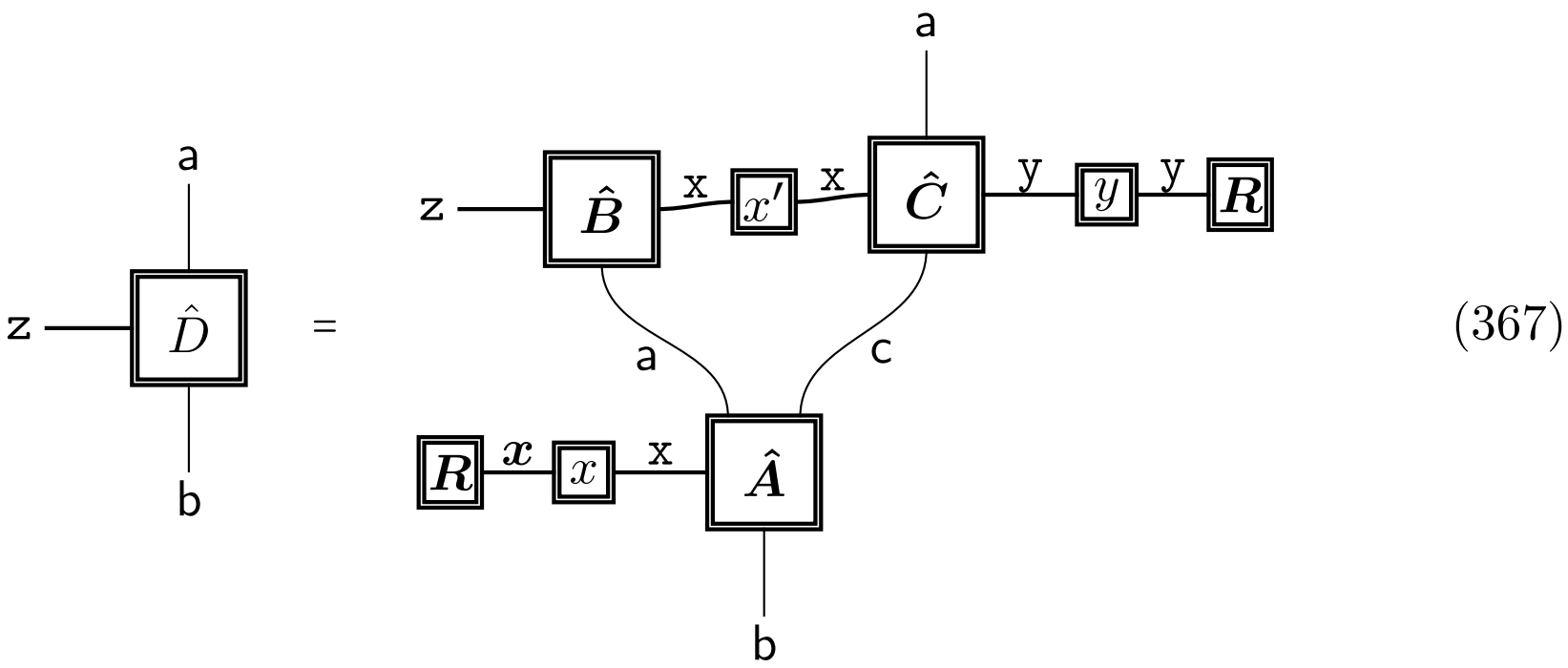

$$(367)$$

The expression on the right is a calculation. To understand how these calculations work it is best to start with simple examples.

Consider the operator network

$$\hat{A}^{\mathsf{a}_1}\hat{B}_{\mathsf{a}_1} \tag{368}$$

(diagrammatic notation on the left, symbolic notation on the right). The repeated index, $\mathsf{a}_1$, indicates taking the trace of $\hat{A}\hat{B}$ i.e.

$$\hat{A}^{\mathsf{a}_1}\hat{B}_{\mathsf{a}_1} = \text{trace}(\hat{A}\hat{B}) \tag{369}$$

Clearly we also have

$$\hat{A}^{\mathsf{a}_1}\hat{B}_{\mathsf{a}_1}\hat{C}_{\mathsf{b}_2} = \text{trace}(\hat{A}\hat{B})\hat{C}_{\mathsf{b}_2} \tag{370}$$

Hence, by linearity, an expression like

$$\hat{A}^{\mathsf{a}_1}\hat{D}_{\mathsf{a}_1\mathsf{b}_2} \tag{371}$$

indicates taking the partial trace with respect to the space associated with $\mathsf{a}_1$. This will become more explicit when we expand operators in terms of fiducial operators.

We can tell a similar story for pointer systems. The object

$$E^{\mathsf{x}_1}F_{\mathsf{x}_1} \tag{372}$$

means we take the dot product between the vectors $E$ and $F$. We can extend this linearly to examples like

$$E^{\mathsf{x}_1}G_{\mathsf{x}_1\mathsf{y}_2} \tag{373}$$

where now we take the "partial dot product" with respect to the space associated with $\mathtt{x}_1$. This is just the usual rule for abstract tensors. We can make this explicit when we expand in terms of fiducial vectors.

Linearity allows us to consider cases having both system and pointer indices such as that in (367).

If we have a circuit (wherein no wires are left open) such as

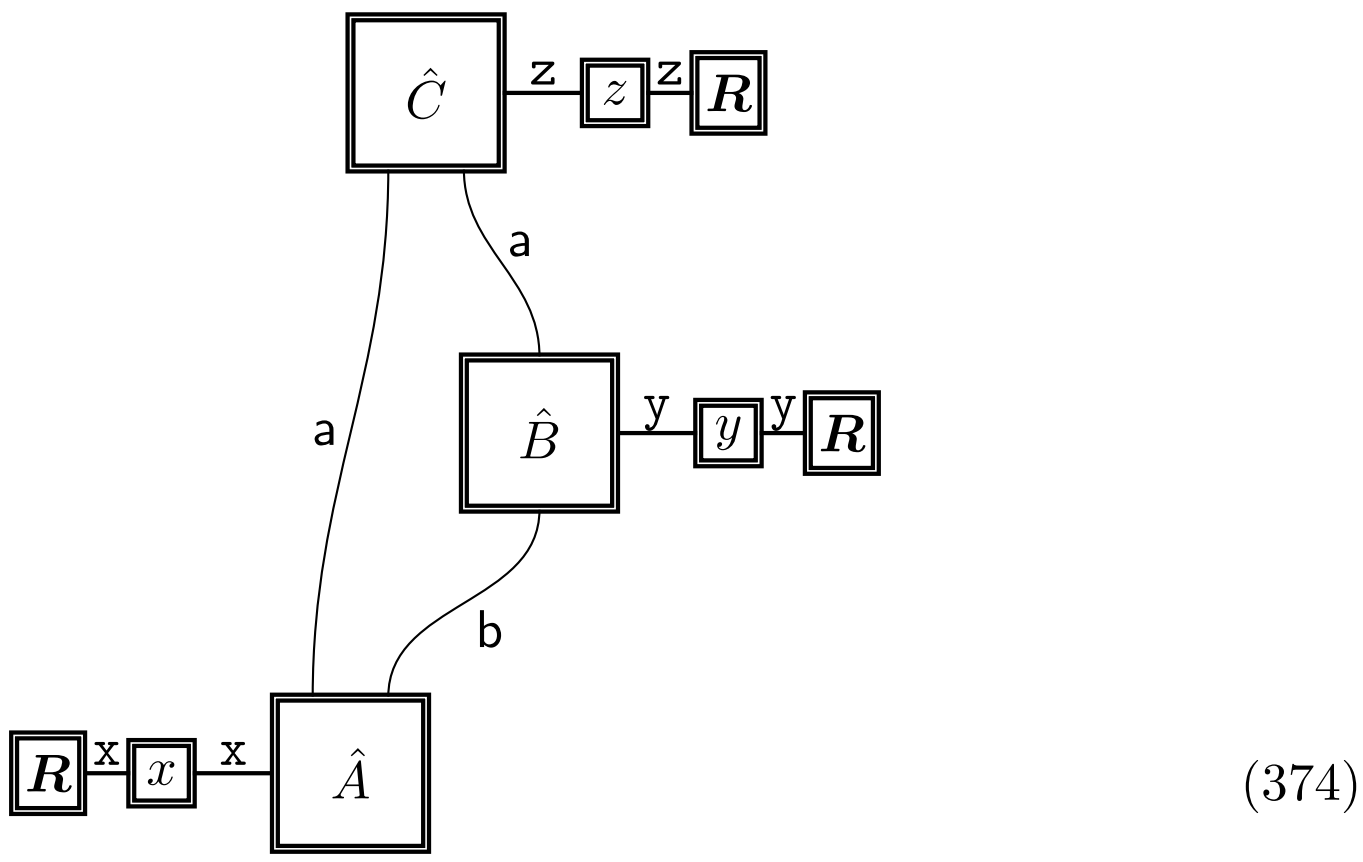

(374)

then this will equate to a number. Further, since we restrict to Hermitian spaces, $\mathcal{V}$, this number will be real. In fact, we want this number to be equal to the probability for this circuit and so we will have to impose further constraints (analogous to those in the case of operations) to get a real number between 0 and 1.

In the case of operations, we have equivalences between expressions (for example, a circuit comprised of operations is equivalent to its probability). In the case of operators we have *equality* (for example, an operator circuit is equal to its probability).

# 18 Special operator tensors

We have the following special operators. First

$$\mathtt{x} \text{—} \boxed{x} \text{—} \mathtt{x} \qquad (375)$$

is an operator corresponding to a readout box.

Second, we have operations

$$\boxed{\boldsymbol{R}} \text{—} \mathtt{x} \qquad\qquad \mathtt{x} \text{—} \boxed{\boldsymbol{R}} \qquad (376)$$

corresponding to a deterministic pointer preparation and a deterministic pointer result.

Third, we have maximal operators

$$\text{x} \!-\!\! \boxed{\hat{X}}^{\,\text{x}} \in \mathcal{P}_{\text{x}_1} \otimes \mathcal{V}^{\text{x}_2} \qquad\qquad \boxed{\hat{X}}\!-\!\text{x}_{\,\text{x}} \in \mathcal{P}^{\text{x}_1} \otimes \mathcal{V}_{\text{x}_2} \tag{377}$$

We have indicated the spaces to which these maximal elements belong. Consider the object on the left. We can think of this as an object whereby a classical income is correlated with a quantum output. The classical incomes correspond to a set of orthogonal elements in the vector space $\mathcal{P}^{\text{x}_1}$. The quantum outputs correspond to a set of projectors in $\mathcal{V}_{\text{x}_2}$ onto a basis of the corresponding Hilbert space.

And, fourth, we have

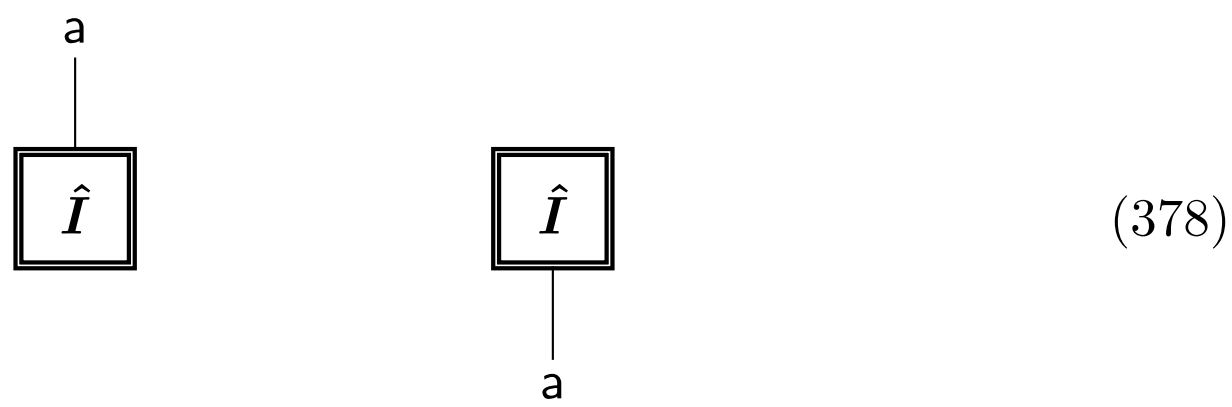

(378)

corresponding to ignore preparations and ignore results.

We will return to these special operators later in Sec. 24.

# 19 Correspondence from operations to operators

We wish to set up a correspondence between operations and operators.

> **Correspondence.** We say that we have established a correspondence from operations to operators if, under this correspondence, any circuit comprised of operations has probability equal to the evaluation of the corresponding circuit comprised of corresponding operators

If we have established correspondence then we have, for example,

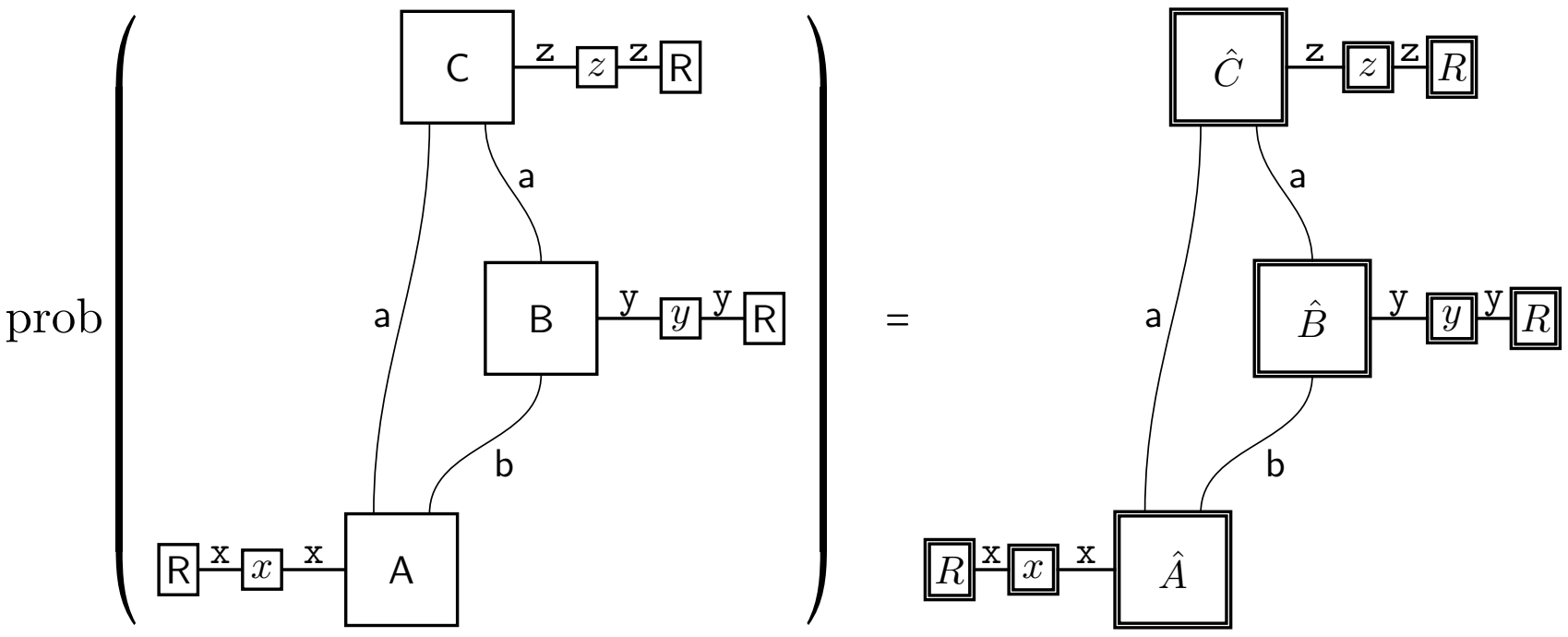

where we use the convention that corresponding objects are represented by the same letter (so that $\mathsf{A}$ corresponds to $\hat{A}$, $\mathsf{B}$ corresponds to $\hat{B}$, etc.).

The map from operations to corresponding operators can be many-to-one and we denote it by $\Longmapsto$. There is, however, a one-to-one map between equivalence classes of operations and corresponding operators.

Correspondence from operations to operators induces a correspondence from networks comprised of operations to networks comprised of operators.

In Sec. 23.1 we will prove a correspondence theorem which establishes the form of this correspondence using fiducial elements. In Sec. 23.3 we will prove the correspondence rule - that equivalence in the operation theory goes over to equality in the operator theory (with something similar for inequalities). Thus, we can take relationships over from the operation theory into the operator theory. The correspondence theorem and the correspondence rule, as we will prove them, rely on the assumption of tomographic locality.

We have represented operations that we assert to be both deterministic and physical by bold font (i.e. $\mathbf{B}$). We will represent operators that, we assert, correspond to deterministic and physical operations by bold also (i.e. $\hat{\boldsymbol{B}}$). We will omit the "hat" For operators having only incomes and outcomes (i.e. no system input or output) - for example we write $\boldsymbol{R}$ rather than $\hat{\boldsymbol{R}}$.

# 20 Circuit reality and Hermiticity

We will see that the combination of circuit reality (see Sec. 6.1) and the purity correspondence assumption (see Sec. 16.1) imply that operator tensors are Hermitian. We will prove this by decomposing operators into Hermitian and anti-Hermitian parts. This approach is fairly intuitive though does require some more technical proofs (which we put off to later when we have developed necessary tools).

There is a second approach to proving that operator tensors are Hermitian using duotensor machinery. This is explained in Sec. 23.2.

For convenience, we will write

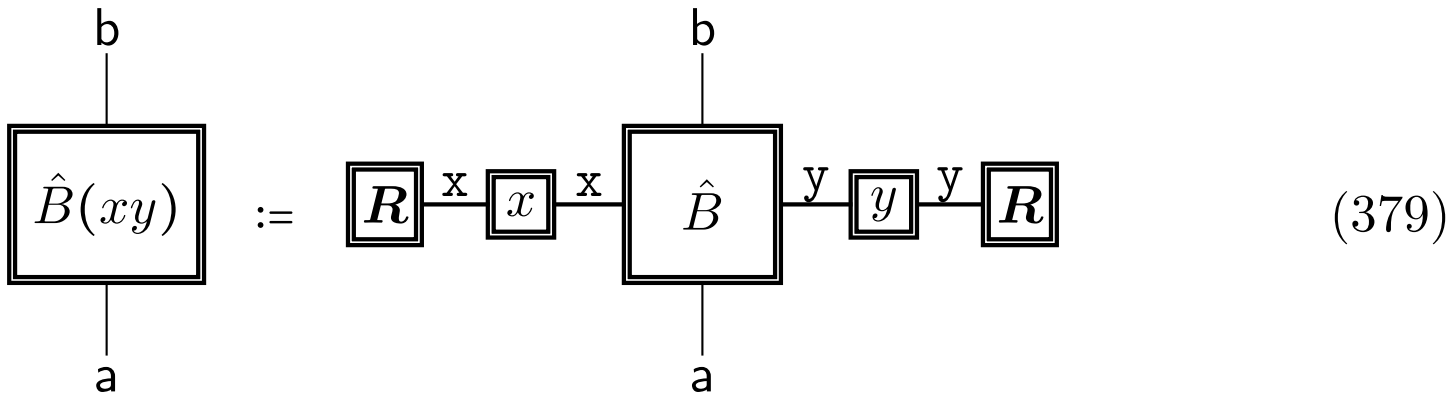

(379)

Any circuit can be written in the form (148). Under correspondence we require

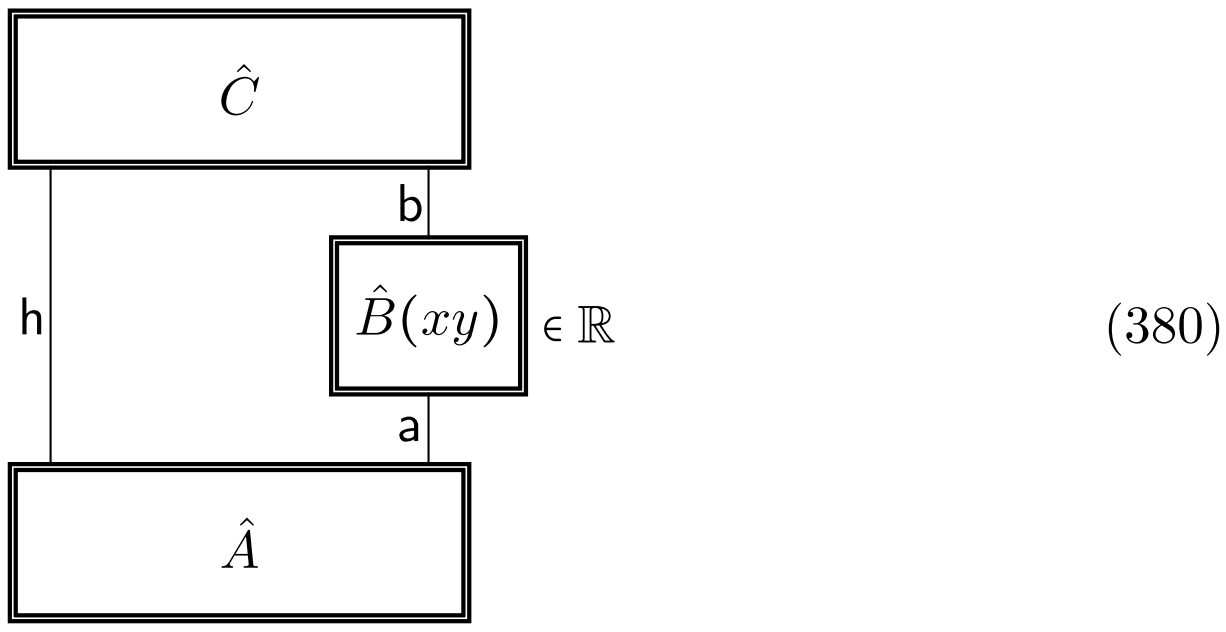

$$\in \mathbb{R} \tag{380}$$

by circuit reality (the probability for a circuit must be real as discussed in Sec. 6.1).

Any operator $\hat{B}$ can be written as

$$\hat{B} = \hat{B}_H + \hat{B}_A \quad \text{where} \quad \begin{matrix} \hat{B}_H = \frac{1}{2}(\hat{B} + \hat{B}^\dagger) & \text{is Hermitian} \\ \hat{B}_A = \frac{1}{2}(\hat{B} - \hat{B}^\dagger) & \text{is anti-Hermitian} \end{matrix} \tag{381}$$

This means we must have

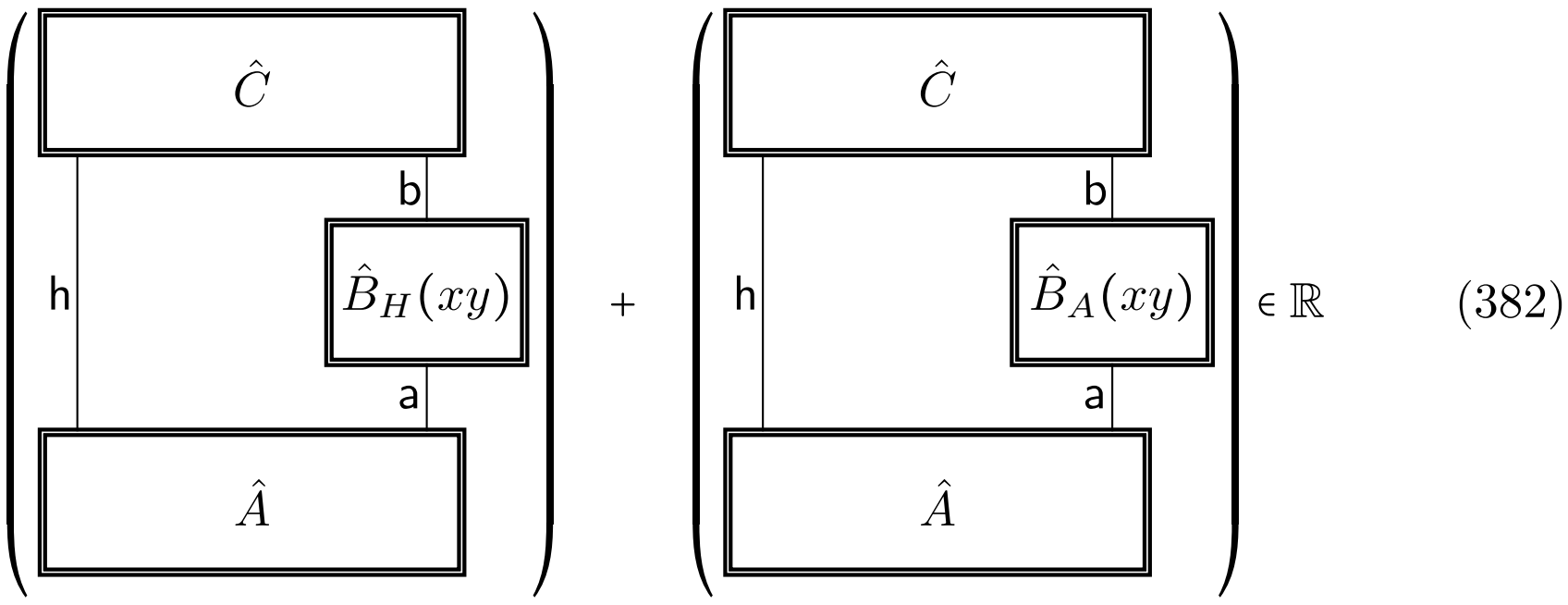

$$\in \mathbb{R} \tag{382}$$

We show in Sec. 42.1 that operator circuits comprised of only Hermitian operators must be real. This means that the term on the left must be real. In Sec. 42.3 we show that operator circuits in the form of the term on the right (with $\hat{B}_A(xy)$) must be pure imaginary for some choices of rank one $\hat{A}$ and $\hat{C}$ unless $\hat{B}_A(xy) = 0$ (the proof of this uses the fact that the eigenvalues of anti-Hermitian matrices must be pure imaginary or zero). The purity correspondence assumption (see Sec. 16.1) requires that, up to normalisation, every rank one state, $\hat{A}$ and rank one effect, $\hat{C}$, have corresponding preparations and results. Consequently circuit reality and the purity correspondence assumption imply that $\hat{B}(xy)$ is Hermitian. Recall the definition of $\hat{B}(xy)$ in (379). For each value of $xy$ we get a Hermitian matrix. It follows that

$$\underset{\mathsf{a}}{\overset{\mathsf{b}}{\mathsf{x}\,\boxed{\hat{B}}\,\mathsf{y}}} \quad \in \quad \mathcal{P}_{\mathsf{x}} \otimes \mathcal{P}^{\mathsf{y}} \otimes \mathcal{V}_{\mathsf{a}} \otimes \mathcal{V}^{\mathsf{b}} \tag{383}$$

# 21 Fiducial Operator Expansion

In Sec. 9 we saw how to expand operations in terms of fiducial operations. This was made possible by assuming decomposition locality (or, equivalently, tomographic locality). In this Section, we will see that we can expand (Hermitian) operator tensors in terms of fiducial operators in an analogous manner. This is made possible by the fact that operator tensors in quantum theory are Hermitian.

## 21.1 Fiducial operators

We introduce fiducial sets of operators for system outputs and inputs

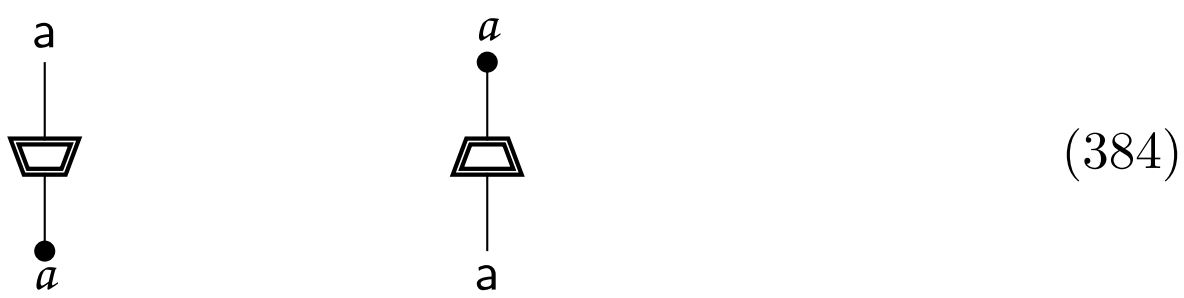

(384)

On the left we have fiducial operator preparations. We can choose any set of Hermitean operators that span the space, $\mathcal{V}^{\mathsf{a}_1}$. Since the space is Hermitian, we have $a = 1$ to $N_{\mathsf{a}}^2$. On the right we can choose any set of fiducial operator results which span the space $\mathcal{V}_{\mathsf{a}_1}$ and, of course, $a = 1$ to $N_{\mathsf{a}}^2$.

We also introduce fiducial sets of vectors for pointer preparations and results

$$x \blacksquare\!\!-\!\!\rhd\!\!-\ \mathsf{x} \qquad\qquad \mathsf{x}\ -\!\!\lhd\!\!-\!\!\blacksquare\, x \tag{385}$$

For the fiducial pointer preparations (on the left) we can choose any set of vectors that span the vector space $\mathcal{P}^{\mathsf{x}_1}$. We have $x = 1$ to $N_{\mathsf{x}}$. Similarly, the fiducial pointer results (on the right) can be chosen to be any set of vectors that span the vector space $\mathcal{P}^{\mathsf{x}_1}$.

## 21.2 Fiducial matrices

We can use these fiducial elements to define fiducial matrices

$$(386)$$

The inverses of the fiducial matrices are

$$(387)$$

Reasoning in the same way as in Sec. 9.3, we have

$$(388)$$

and, reasoning in the same way as in Sec. 9.5, we have

$$(389)$$

Consequently we can insert pairs of black and white dots in wires (round for system wires and square for pointer wires) as we like.

## 21.3 Same fiducial matrices as for operations

It is important for what follows that we can make choices for fiducial elements such that we get the same fiducial matrices for operations as for operators.

The natural choice for the fiducial pointer preparations and results is that they are orthogonal but not normalised so that

$$= \frac{1}{N_{\mathtt{x}}} \begin{pmatrix} 1 & & & \\ & 1 & & \\ & & \ddots & \\ & & & 1 \end{pmatrix} \qquad (390)$$

This is equal to the fiducial matrix obtained with the natural choice of fiducial operations (193) as discussed earlier in Sec. 9.3. Thus, for pointers, we can get the same fiducial matrices in the operation and operator cases. The inverse of the fiducial matrix in (390) is

$$= N_{\mathtt{x}} \begin{pmatrix} 1 & & & \\ & 1 & & \\ & & \ddots & \\ & & & 1 \end{pmatrix} \qquad (391)$$

(we saw this inverse before in (202)).

What about the system case? If it is true that nature is modelled by Quantum Theory then it has to be the case that we can find fiducial sets in the operation case that have the same fiducial matrix as some fiducial sets in the operator case. A specific example for a spanning set of $N_{\mathsf{a}}^2$ operators which could serve as fiducial elements are the following

$$|a\rangle_{\mathsf{a}}\langle a| \qquad |aa'\rangle_{\mathsf{a}}\langle aa'| \qquad |iaa'\rangle_{\mathsf{a}}\langle iaa'| \tag{392}$$

for $a = 1$ to $N_{\mathsf{a}}$ and $a > a'$ where

$$|aa'\rangle_{\mathsf{a}} = \frac{1}{\sqrt{2}}(|a\rangle_{\mathsf{a}} + |a'\rangle_{\mathsf{a}}) \qquad |iaa'\rangle_{\mathsf{a}} = \frac{1}{\sqrt{2}}(|a\rangle_{\mathsf{a}} + i|a'\rangle_{\mathsf{a}}) \tag{393}$$

We can choose these as both fiducial preparation and result operators. Recall that pure states and effects are not normalised for the time symmetric theory. Consequently, the basis vectors being used here are orthogonal but not normalised to 1. If the world is modelled by Quantum Theory then there will exist fiducial preparation and result operations that give rise to the same fiducial matrix as we obtain from this spanning set of operators (if they are suitably normalised).

## 21.4 Fiducial operator expansion

We can expand a Hermitian operator using fiducial elements in the obvious way. For example,

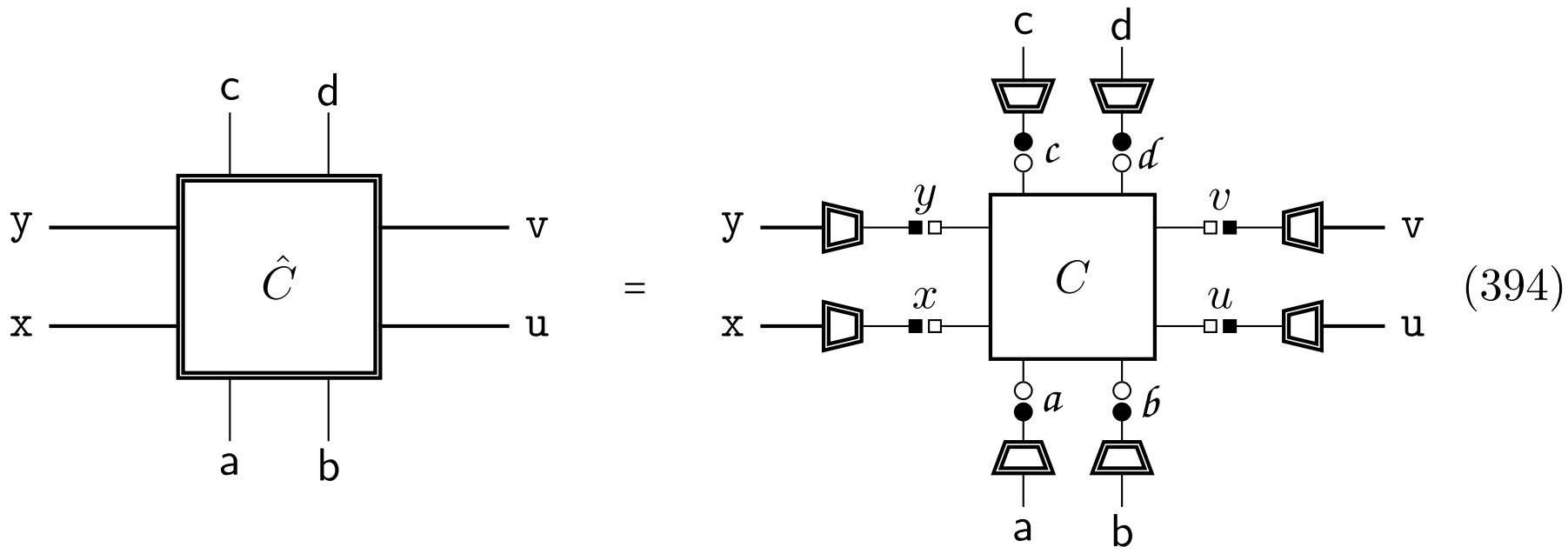

(394)

Compare this with (216). In particular, note that we have equality rather than equivalence. The duotensor with white dots provides the expansion coefficient. Note that the entries in this duotensor are all real. This is possible because we are expanding a Hermitian operator in terms of Hermitian fiducials (Hermitian operators form a real vector space). This provides the route to establish *correspondence* between operations and operators as we will discuss later. This works for Hermitian operators because as the system fiducials are Hermitian and the space for the operator tensor, $\hat{C}$, is

$$\mathcal{P}_{\mathsf{x}} \otimes \mathcal{P}_{\mathsf{y}} \otimes \mathcal{P}^{\mathsf{u}} \otimes \mathcal{P}_{\mathsf{v}} \otimes \mathcal{V}_{\mathsf{a}} \otimes \mathcal{V}_{\mathsf{b}} \otimes \mathcal{V}^{\mathsf{c}} \otimes \mathcal{V}^{\mathsf{d}} \tag{395}$$

so the space is a tensor product of the fiducial spaces. This tensor product structure corresponds to tomographic locality in the case of operations (see discussion in Sec. 9.7.2). An interesting counter situation is where we start with Hermitian matrices then restrict the entries to be real. In this case we do not have this tensor product structure (see Hardy and Wootters [2012]).

### 21.5 Wire expansions

We can expand system and pointer wires in terms of fiducials as follows

$$\mathsf{x} \quad = \quad \mathsf{x} \; x \; \mathsf{x} \qquad (396)$$

$$\mathsf{x} \quad = \quad \mathsf{x} \; x \; \mathsf{x} \qquad (397)$$

The proof of this is analogous to the proof in Sec. 9.7.4 for wires in the case of operations.

## 22 Duotensor calculations

We can evaluate operator networks and circuits using fiducial expansions or using the wire expansion. This follows analogous steps to the case of networks and circuits built out of operations (though now we have equality rather than equivalence between expressions). For example, we can evaluate the operator circuit

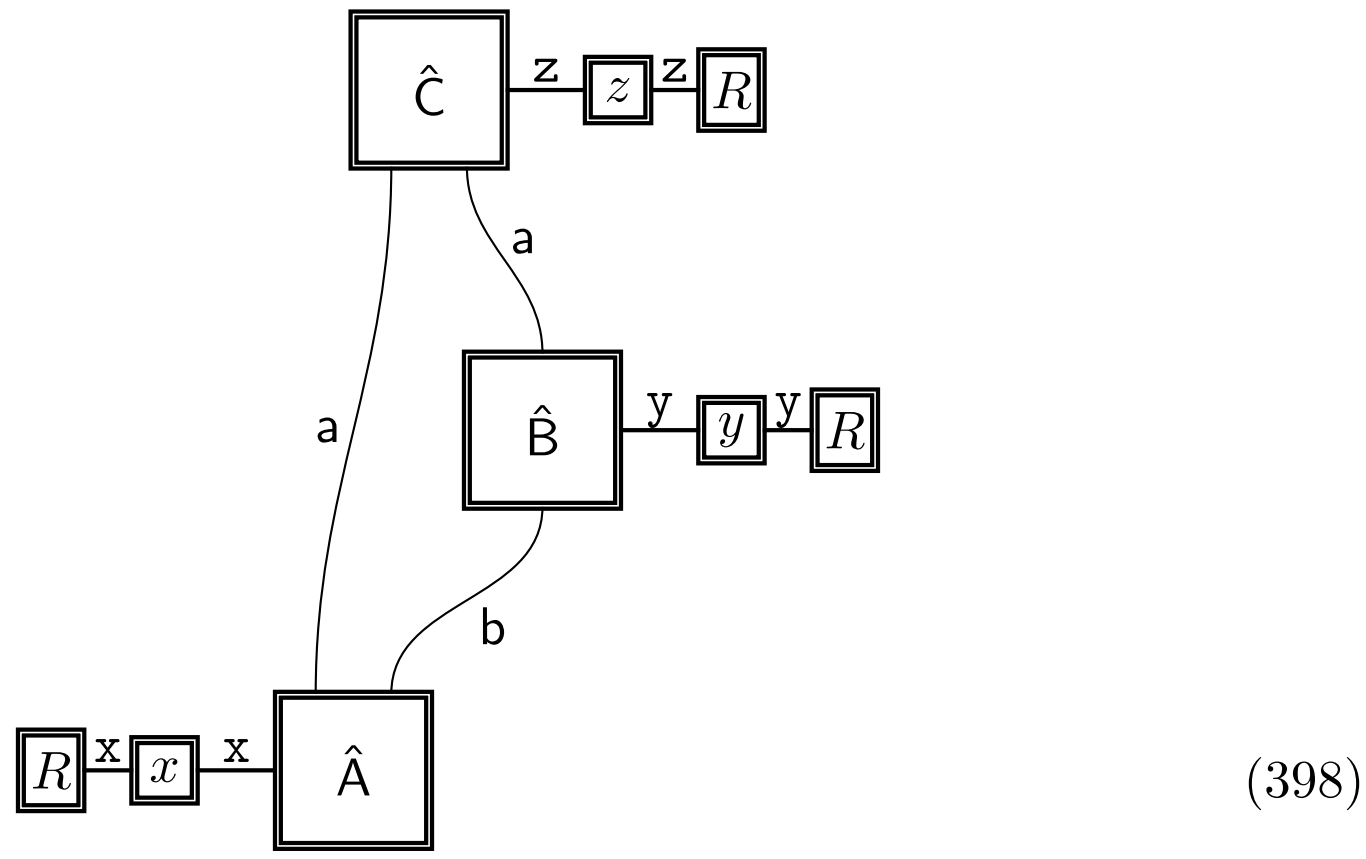

(398)

We can evaluate this either by expanding the operators in terms of fiducials or by using fiducial expansions of the wires. Consider the former approach.

Replacing operators with fiducial expansions we get

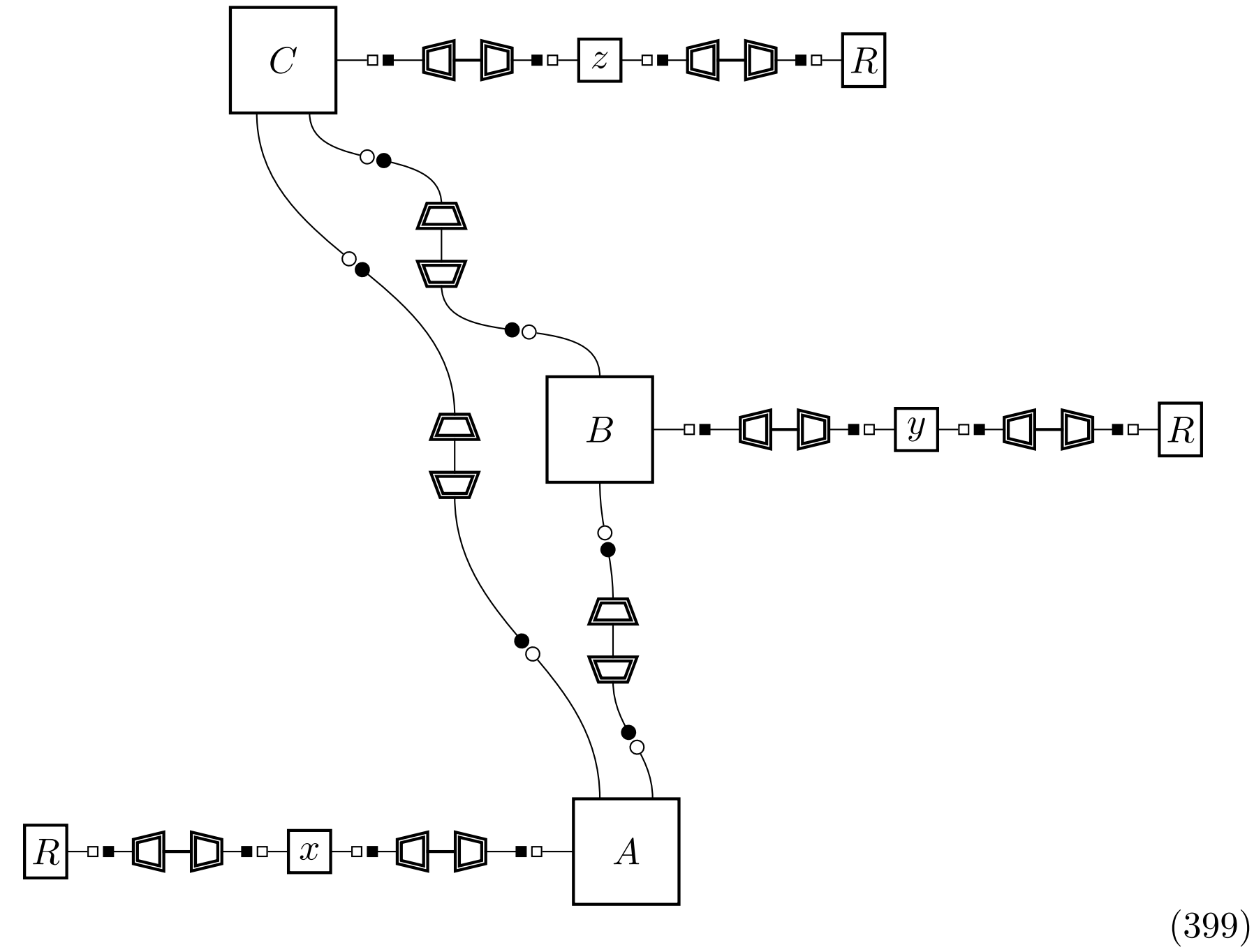

(399)

Recognising the fiducial matrices in this expression gives

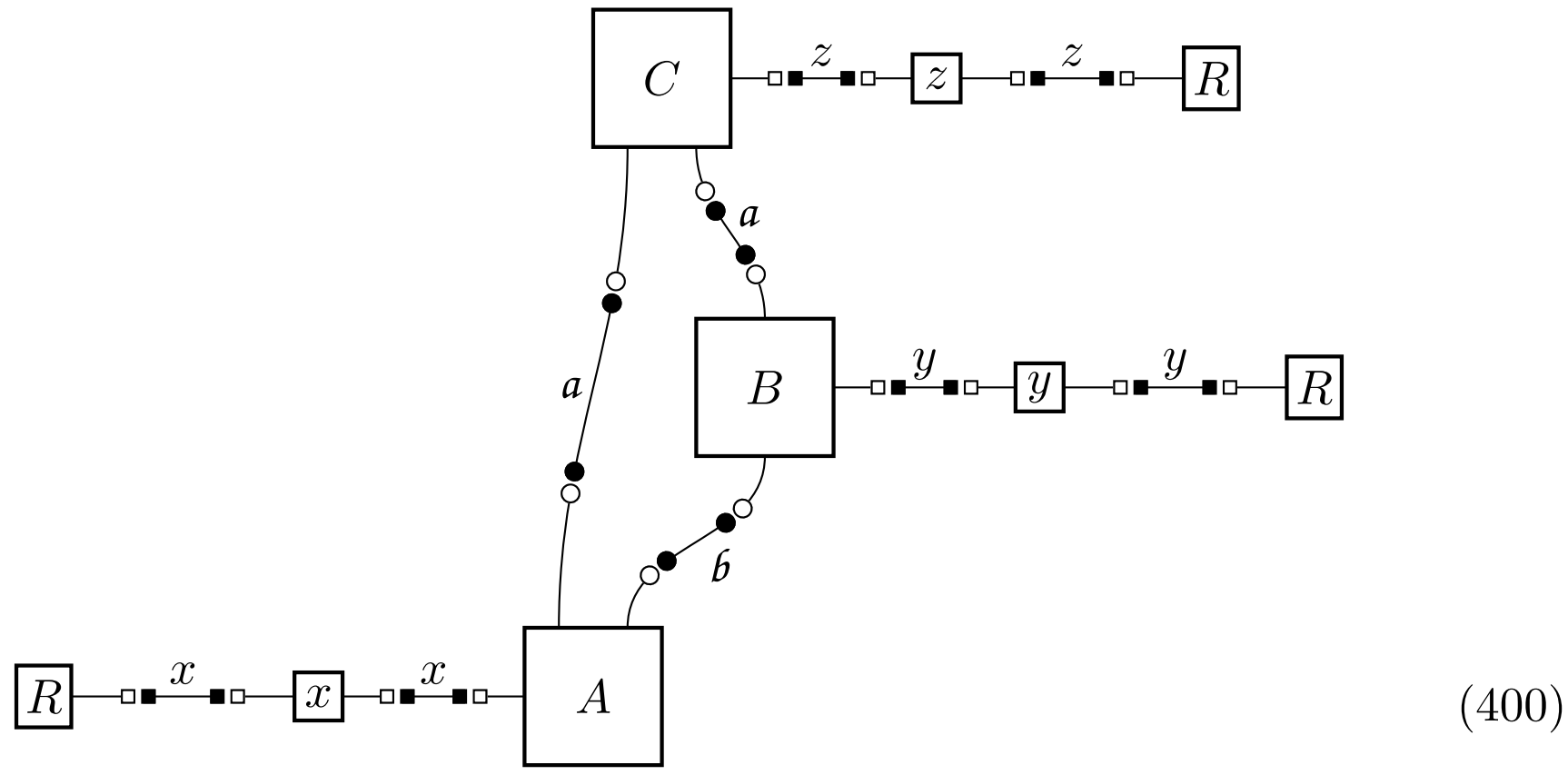

(400)

Finally we can delete black white pairs of dots (as we can reinsert them) so we obtain

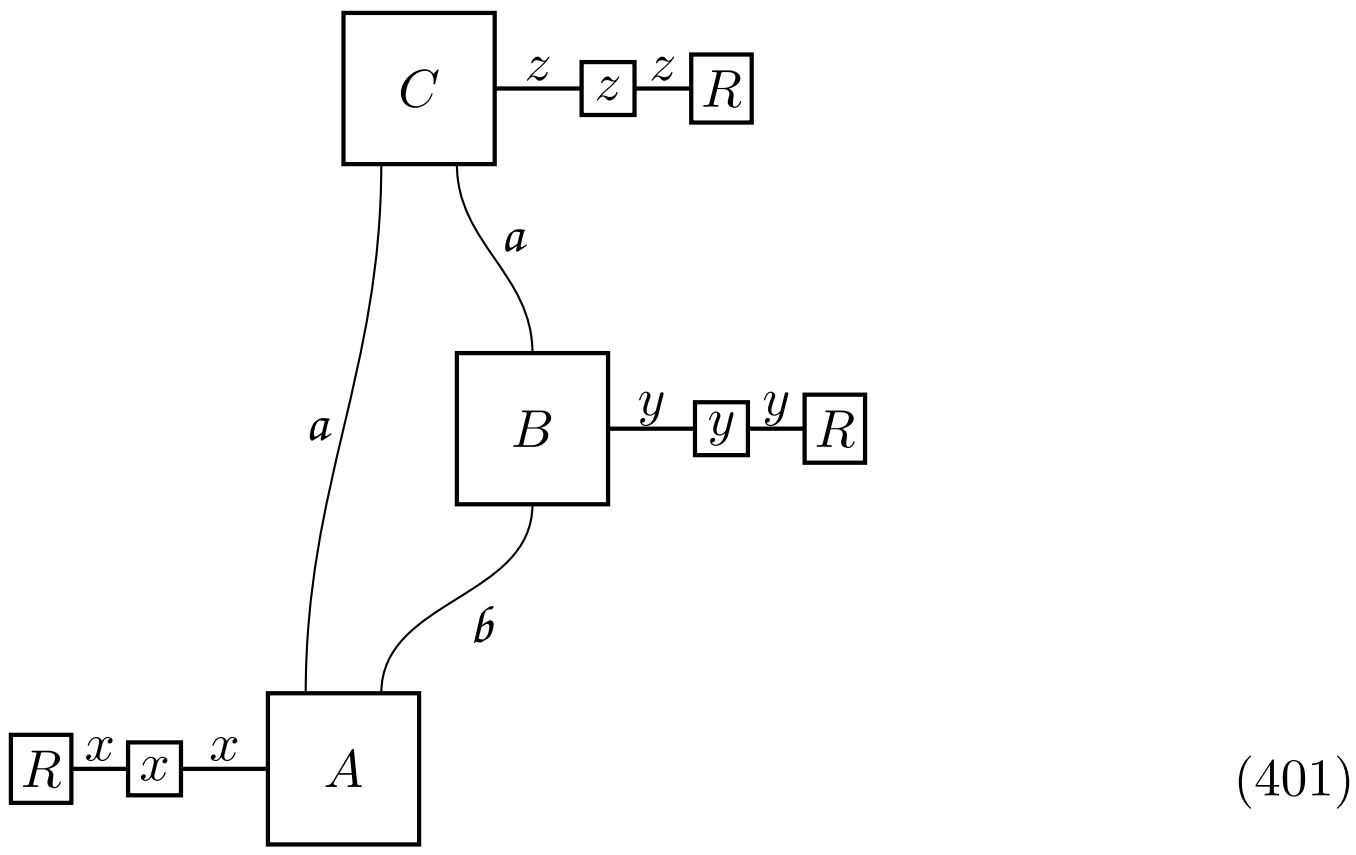

(401)

Importantly, if the fiducial matrices are the same as in the operation case, then we obtain the same answer as in the operation case.

# 23 Correspondence

## 23.1 Correspondence theorem

In Sec. 19 we provided a notion of correspondence between operations and operators (such that under this correspondence, any circuit comprised of operations has a probability equal to the evaluation of the corresponding circuit comprised of corresponding operators).

Given what we have established, it is clear that such a correspondence is given as follows

> **Correspondence theorem.** If the fiducial matrices coming from the fiducial operations are the same as the fiducial matrices coming from fiducial operators, i.e.
>
> $x$ ▪—◧ x ◨—▪ $x$ ≡ $x$ ▪—◧ x ◨—▪ $x$ := ▪—$x$—▪ (402)
>
> and
>
> $\mathcal{X}$ ● x ● $\mathcal{X}$ ≡ $\mathcal{X}$ ● x ● $\mathcal{X}$ := $\mathcal{X}$ ●—● (403)

then the operation

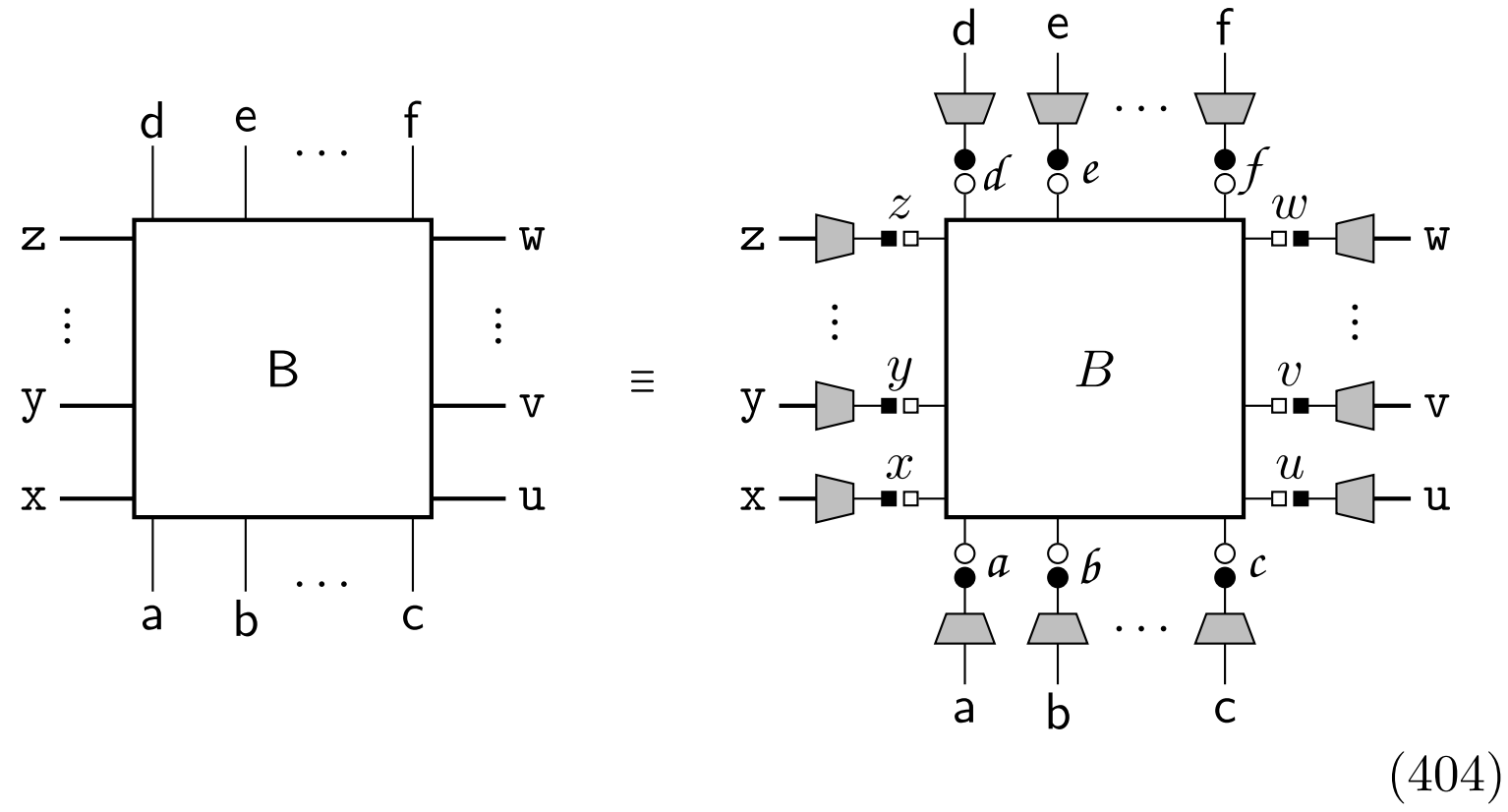

(404)

corresponds to the operator

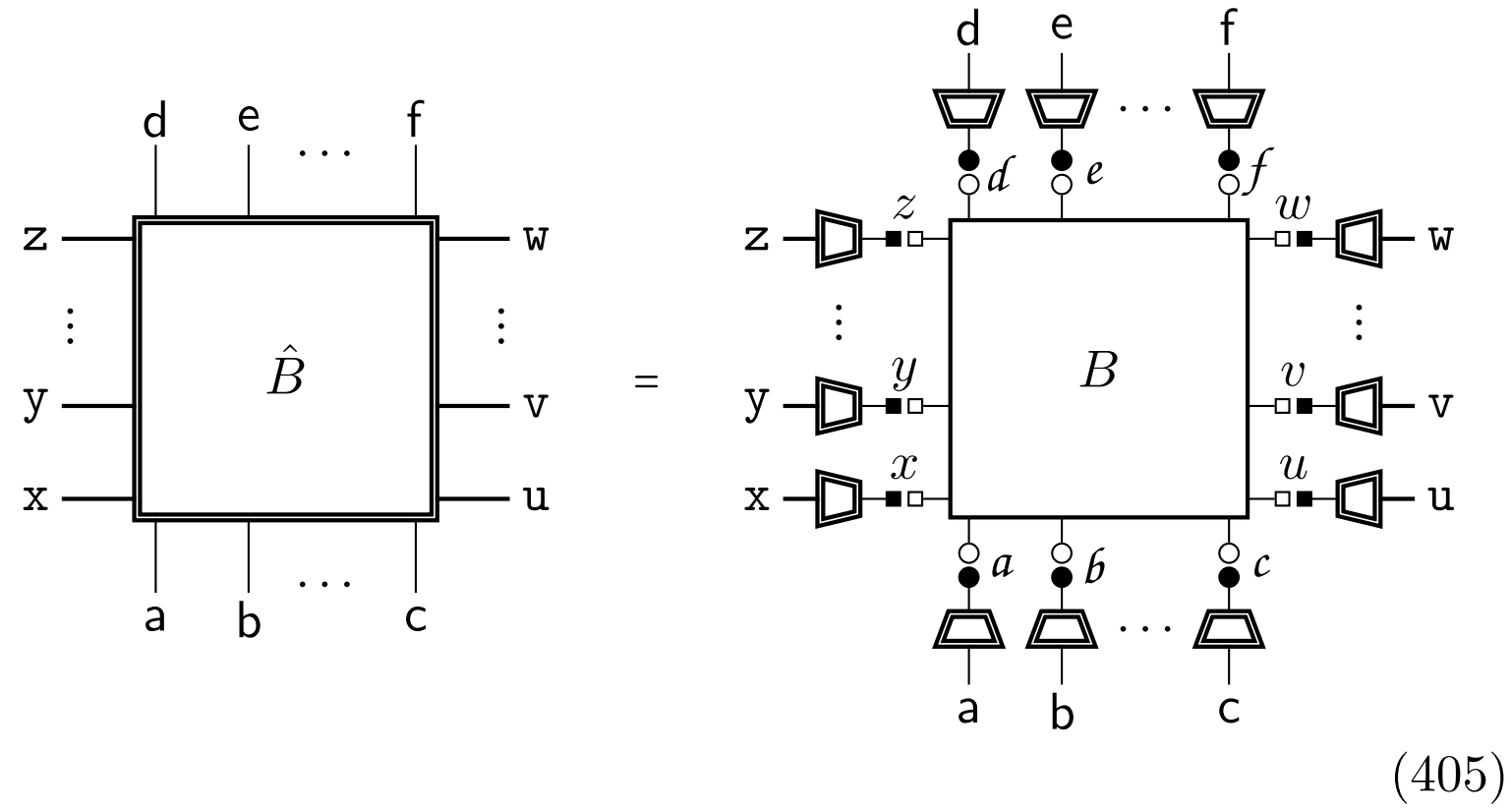

(405)

where, importantly, we have the same duotensor in these two expansions.

The proof of this theorem is straightforward. Start with any operation circuit (such as the example in (246)). Replace each operation with the fiducial expansion,as in (404) (this yields (247) in our example). This is equivelent to a duotensor calculation (this would be (248) in our example). Since the fiducial matrices are equal this is equal to expression with operator fiducials in place of the operation fiducials (as in (399)). Then finally, we can recognise that we have operator expansions (as in (405)) and replace these with equal operators. This returns the same circuit but now with operators rarther than operations (i.e. that in (374)). Since this is equivalent to the original circuit they must have equal probabilities.

Note that we established in Sec. 21.3 that we can have the same fiducial matrices in the operation and operator case (as long as the world is actually described by Quantum Theory). This correspondence is indicated by a dotted line on the flowchart (355). The proof of the correspondence theorem using duotensors as just outlined is evident on this flowchart. The correspondence map is between equivalence classes of operations and operators since operations having the same duotensor expansion belong to the same equivalence class (they give the same probabilities in any circuit).

## 23.2 Circuit reality and correspondence

We proved that operator tensors should be Hermitian in Sec. 20 using circuit reality and the purity correspondence assumption by considering Hermitian and anti-Hermitian operators. We can also prove Hermiticity using duotensors and the fact that fiducial operators are Hermitian.

First, circuit reality implies that the fiducial matrices in (194) and (205) consist of real number entries (and therefore their inverses also consist of real number entries). Furthermore, circuit reality implies that the duotensor in (217) with all black dots consists of all real entries. If we change the colour of some or all of these dots to white dots (using the corresponding inverse fiducial matrix) then we still have all real entries. Consequently, duotensors have all real entries.

Second, as argued in Sec. 16.1, it follows from the purity correspondence principle, that preparations and results correspond to Hermitian operator tensors. Hence, fiducial operators must be Hermitian.

Thus, under correspondence, the operator tensor associated with a general operation, $\hat{B}$, is given in (405). Given the above facts, this is a sum over real weighted tensor products of Hermitian operators, and is therefore Hermitian.

## 23.3 Correspondence rule

In Sec. 6.2 we introduced expressions like

$$\text{exprn} = \alpha + \beta \mathsf{E}_{\mathtt{x}_1 \mathsf{a}_2}^{\mathtt{y}_3 \mathsf{b}_4} + \gamma \mathsf{F}_{\mathtt{x}_1 \mathsf{a}_2}^{\mathtt{y}_3 \mathsf{b}_4} + \delta \mathsf{G}_{\mathtt{x}_1 \mathsf{a}_2}^{\mathtt{y}_3 \mathsf{b}_4} + \dots \tag{406}$$

(where $\alpha$, $\beta$, ... are real) corresponding to real weighted sums of networks. We call such expressions *operation expressions*. We can take any such expression and convert it into the corresponding sum of operators. In the above example we would have

$$\widehat{exprn} = \alpha + \beta \hat{E}_{\mathtt{x}_1 \mathsf{a}_2}^{\mathtt{y}_3 \mathsf{b}_4} + \gamma \hat{F}_{\mathtt{x}_1 \mathsf{a}_2}^{\mathtt{y}_3 \mathsf{b}_4} + \delta \hat{G}_{\mathtt{x}_1 \mathsf{a}_2}^{\mathtt{y}_3 \mathsf{b}_4} + \dots \tag{407}$$

We call such expressions *operator expressions*. Then we say

$$\text{exprn} \Longmapsto \widehat{exprn} \tag{408}$$

(recall that $\Longmapsto$ means *corresponds to*).

We can have equivalence and inequality relationships between such expressions. Equivalence is denoted by $\equiv$ (see Sec. 6.2). We have defined inequality with respect to general complement networks, denoted by $\leqq$ in see Sec. 6.3. We also defined inequality with respect to testers. This is denoted by $\leqq_T$ (see Sec. 7.11.2).

It is worth recalling (as discussed in Sec. 6.2) that, whenever we have an equivalence or inequality relationship between expressions, all terms in the expressions must have the same types so that they can be closed into circuits by complement networks. This implies that any numerical term such as $\alpha$ in (406) can only be non-zero when the other terms correspond to circuits.

It is sufficient for our purposes to consider only the tester inequality, $\leqq_T$. We can prove the following

> **Correspondence rule.** Equivalences and inequalities, which are respectively of the form
>
> $$\text{exprn}_1 \equiv \text{exprn}_2 \qquad\qquad \text{exprn}_3 \underset{T}{\leqq} \text{exprn}_4 \tag{409}$$
>
> between operation expressions go over to equations and inequalities of the form
>
> $$\widehat{exprn}_1 = \widehat{exprn}_2 \qquad\qquad \widehat{exprn}_3 \underset{T}{\leq} \widehat{exprn}_4 \tag{410}$$
>
> between operator expressions under the correspondence
>
> $$\text{exprn}_i \Longmapsto \widehat{exprn}_i$$
>
> for $i = 1, 2, 3, 4$.

The inequality relationship, $\underset{T}{\leq}$, between operator expressions is with respect to operator testers as we will see. To prove the correspondence rule, first consider two equivalent operation expressions. We have

$$\text{exprn}_1 \equiv \text{exprn}_2 \quad\Longleftrightarrow\quad p(\text{exprn}_1\mathsf{H}) = p(\text{exprn}_2\mathsf{H}) \quad \forall \text{ complement networks } \mathsf{H} \tag{411}$$

Under correspondence we obtain

$$p(\widehat{exprn}_1\hat{H}) = p(\widehat{exprn}_2\hat{H}) \quad \forall \text{ complement operator networks } \hat{H} \tag{412}$$

since circuits take the same probability under correspondence. Recall that $p(x) = x$ for any real number, $x$. Consequently we simply have

$$\widehat{exprn}_1\mathsf{H} = \widehat{exprn}_2\mathsf{H} \tag{413}$$

Since this must be true for all $\hat{H}$ we obtain

$$\widehat{exprn}_1 = \widehat{exprn}_2 \tag{414}$$

as required. To prove the correspondence rule for inequalities we follow the same path. By definition we have

$$\text{exprn}_1 \underset{T}{\leqq} \text{exprn}_2 \quad\Longleftrightarrow\quad p(\text{exprn}_1\mathsf{T}) \leq p(\text{exprn}_2\mathsf{T}) \quad \forall \text{ testers } \mathsf{T} \tag{415}$$

Under correspondence we obtain

$$p(\overparen{exprn}_1 \hat{T}) \leq p(\overparen{exprn}_2 \hat{T}) \quad \forall \text{ testers } \hat{T} \tag{416}$$

where $\hat{T}$ is an *operator tester*. We write this as

$$\overparen{exprn}_1 \underset{T}{\leq} \overparen{exprn}_2 \tag{417}$$

(i.e. the definition of $\underset{T}{\leq}$ in (417) is given in (416)).

The step from (413) to (414) for equality invokes the fact that complement operator networks, $\hat{H}$, form an overcomplete basis. In fact, it follows from local process tomography that we can form a complete basis just using complement networks consisting of fiducial elements (this is what drives the duotensor approach and the correspondence theorem). Thus, it is actually only necessary to check that (413) holds for a small subset of all possible complement operator networks. By contrast, the step from (416) to (417) for inequality is by definition. To be sure that (417) holds we need to check (416) holds for *every* operator tester $\hat{T}$ (a complete tomographic subset will not do).

We need to give a notational word of caution. We could say that We could also consider a correspondence whereby

$$\text{exprn}_1 \leqq \text{exprn}_2 \tag{418}$$

corresponds to

$$\overparen{exprn}_1 \leq \overparen{exprn}_2 \qquad \text{(we will not use this notation)} \tag{419}$$

where the latter would be defined with respect to all complement networks (not just testers). This would be fine except for the fact that the notation, $\leq$, between operators already has a well established and different meaning in the mathematical literature. Namely, $\hat{A} \leq \hat{B}$ is usually taken to mean that the eigenvalues of $\hat{B} - \hat{A}$ are all non-negative (or, equivalently, that $\langle\psi|\hat{A}|\psi\rangle \leq \langle\psi|\hat{B}|\psi\rangle$ for all $|\psi\rangle$). We will stick with the standard meaning of $\leq$ and so we will not write (419). In Sec. 43 we will see that we can use this standard inequality relationship if we first take the "input twist" (see (935)). The twist operation will be defined later.

The correspondence rule is important since it means we can take any equivalence or inequality in the operation theory and find a corresponding equality or inequality in the operator theory. For example, the double causality conditions for deterministic operations

(420)

immediately give us double causality conditions for deterministic operators

$$ \tag{421} $$

by the correspondence rule. Further, the condition

$$ \equiv \ 1 \tag{422} $$

(see Sec. 7.16) which tests to see if an operation, $\mathsf{B}$, is deterministic becomes

$$ \equiv \ 1 \tag{423} $$

to test if the operator, $\hat{B}$, is deterministic. In fact, by correspondence with the conditions in Sec. 7.16, we also know that an operator is deterministic if it satisfies either of the conditions in (421).

# 24 Special operator tensors and normalisation gauge

We discussed special operator tensors in Sec. 18. They are summarised in Table 2. Here we will show how to represent them using correspondence. We will see that there is a gauge freedom due to the fact that probability is a property of a circuit as a whole and not the component parts. This leads to the introduction of *normalisation gauge parameters* $\alpha_{\mathsf{a}}$ and $\alpha^{\mathsf{a}}$ associated with physical systems, and $\beta_{\mathsf{x}}$ and $\beta^{\mathsf{x}}$ associated with pointer systems.

The natural choice of fiducial pointer preparation in the operational theory is given in (193). This is that, for each $x$, the fiducial preparation is equal to an **R** box followed by readout box with that value of $x$. The corresponding operator, for the given value of $x$, must be a vector in $\mathcal{P}^{\mathtt{x}_1}$. Thus, we can write

$$x\blacksquare\!-\!\!\sqsupset\!\!-\mathtt{x} \quad \xrightarrow[x]{\text{given}} \quad \beta^{\mathtt{x}}\begin{pmatrix} 0 \\ 0 \\ \vdots \\ 1 \\ \vdots \\ 0 \end{pmatrix} \tag{424}$$

where the 1 is in the $x$ position (we will discuss $\beta^{\mathtt{x}}$ below). Since the fiducial element has a wire on both sides, it should be represented as a matrix which we obtain by placing $N_{\mathtt{x}}$ column vectors next to each other. Hence we obtain

$$x\blacksquare\!-\!\!\sqsupset\!\!-\mathtt{x} \;=\; \beta^{\mathtt{x}}\begin{pmatrix} 1 & & \\ & 1 & \\ & & \ddots \end{pmatrix} \qquad \mathtt{x}-\!\!\sqsubset\!\!-\!\blacksquare x \;=\; \beta_{\mathtt{x}}\begin{pmatrix} 1 & & \\ & 1 & \\ & & \ddots \end{pmatrix} \tag{425}$$

|  |  |  |
|---|---|---|
| $\boxed{\boldsymbol{R}}$ — $\mathtt{x}$ and $\mathtt{x}$ — $\boxed{\boldsymbol{R}}$ | $\boldsymbol{R}^{\mathtt{x}_1}$ and $\boldsymbol{R}_{\mathtt{x}_1}$ | Deterministic preparation and result operators |
| $\mathtt{x}$ — $\boxed{x}$ — $\mathtt{x}$ | $O^{\mathtt{x}_2}_{\mathtt{x}_1}[x]$ | Readout box operator |
| $\mathtt{x}$ — $\boxed{\hat{X}}$ — $\mathtt{x}$ and $\boxed{\hat{X}}$ — $\mathtt{x}$, $\mathtt{x}$ | $\hat{\boldsymbol{X}}^{\mathtt{x}_2}_{\mathtt{x}_1}$ and $\hat{\boldsymbol{X}}^{\mathtt{x}_1}_{\mathtt{x}_2}$ | Maximal output and maximal input operators |
| $\mathtt{x}$, $\boxed{\hat{I}}$ and $\boxed{\hat{I}}$, $\mathtt{x}$ | $\hat{\boldsymbol{I}}^{\mathsf{a}_1}$ and $\hat{\boldsymbol{I}}_{\mathsf{a}_1}$ | Ignore output and input operators |

Table 2: Operators that play a special role. We provide diagrammatic and symbolic notation for these operators.

The fiducial result operation on the right is obtained similarly (though we should think of it as placing $N_{\mathtt{x}}$ row vectors to form a matrix). We have the normalisation condition

$$\beta^{\mathtt{x}}\beta_{\mathtt{x}} = \frac{1}{N_{\mathtt{x}}} \tag{426}$$

With this normalisation condition, we get the correct fiducial matrix (390). The $\beta^{\mathtt{x}}$ and $\beta_{\mathtt{x}}$ are normalisation gauge parameters. They can take any real values for each pointer system type, $\mathtt{x}$ consistent with the normalisation condition (426). The key thing about the two matrices in (425) is that they are invertible since we want the fiducial preparation operators corresponding to the columns (associated with each $x$ to be linearly independent (and similarly for the rows of for the fiducial results). They are proportional to the identity in (425) because we have made the natural choice of fiducials in (193).

We can change the colour of the square dot in (425) from black to white using the inverse of the fiducial matrix (202). This gives

$$x\,\square\!-\!\!\triangleleft\!\!-\,\mathtt{x} \;=\; \beta^{\mathtt{x}} N_{\mathtt{x}} \begin{pmatrix} 1 & & \\ & 1 & \\ & & \ddots \end{pmatrix} \qquad \mathtt{x}\,-\!\!\triangleright\!\!-\!\square\,x \;=\; \beta_{\mathtt{x}} N_{\mathtt{x}} \begin{pmatrix} 1 & & \\ & 1 & \\ & & \ddots \end{pmatrix} \tag{427}$$

The $N_{\mathtt{x}}$ factor comes from the inverse of the fiducial matrix.

We can write

$$\boxed{\boldsymbol{R}}\!-\!\mathtt{x} \;=\; \boxed{\boldsymbol{R}}\!-\!\overset{x}{\square\blacksquare}\!-\!\triangleleft\!-\!\mathtt{x} \tag{428}$$

Using (425) and (243) we obtain the equation on the left below

$$\boxed{\boldsymbol{R}}\!-\!\mathtt{x} \;=\; \beta^{\mathtt{x}} \begin{pmatrix} 1 \\ 1 \\ \vdots \\ 1 \end{pmatrix} \qquad \text{and} \qquad \mathtt{x}\!-\!\boxed{\boldsymbol{R}} \;=\; \beta_{\mathtt{x}} \begin{pmatrix} 1 \\ 1 \\ \vdots \\ 1 \end{pmatrix} \tag{429}$$

The equation on the right is obtained similarly. Note we have used correspondence because we take the duotensor (243) used in the expansion of an operation and use it to construct the operator. Note that the normalisation gauge parameters, $\beta^{\mathtt{x}}$ and $\beta_{\mathtt{x}}$, enter into how we represent these pointer preparations and results. Given the normalisation condition (426) we see that

$$\boxed{\boldsymbol{R}}\overset{\mathtt{x}}{-\!\!-}\boxed{\boldsymbol{R}} \;=\; 1 \tag{430}$$

as required (by correspondence with the right equation in (85)).

The readout box can be expanded as

$$\mathtt{x}\!-\!\boxed{x}\!-\!\mathtt{x} \;=\; \mathtt{x}\!-\!\triangleright\!-\!\overset{x}{\blacksquare\square}\!-\!\boxed{x}\!-\!\overset{x}{\square\blacksquare}\!-\!\triangleleft\!-\!\mathtt{x} \;=\; \begin{pmatrix} 0 & & & & \\ & 0 & & & \\ & & \ddots & & \\ & & & 1 & \\ & & & & \ddots \end{pmatrix} \tag{431}$$

where the 1 is in position $x$ along the diagonal. We have used (245) and (425) in the last step. We have used (426) to remove the $\beta^{\mathtt{x}}\beta_{\mathtt{x}}$ factor.

We require maximal operators (see Table 2) have the double maximality property

$$\begin{matrix} & \boxed{\hat{\boldsymbol{X}}}\!-\!\mathtt{x} \\ & \mid \mathtt{X} \\ \mathtt{x}\!-\! & \boxed{\hat{\boldsymbol{X}}} \end{matrix} \;=\; \overset{\mathtt{x}}{-\!\!-\!\!-\!\!-} \tag{432}$$

by correspondence with (171). If we sandwich the expressions on each side of this equation between a fiducial pointer preparation and a fiducial pointer result we obtain

$$\text{x} \qquad = \qquad \overset{x}{\blacksquare\!\!-\!\!-\!\!\blacksquare} \tag{433}$$

We associate maximal operators with projectors, $|x\rangle_{\mathtt{x}_1}\langle x|$ on to a basis set, $|x\rangle_{\mathtt{x}_1}$. Thus, for each value of $x$ we have

$$\xrightarrow[x]{\text{given}} \alpha^{\mathtt{x}}|x\rangle^{\mathtt{x}_1}\langle x| \qquad\qquad \xrightarrow[x]{\text{given}} \alpha_{\mathtt{x}}|x\rangle_{\mathtt{x}_1}\langle x| \tag{434}$$

where we have the normalisation condition

$$\alpha^{\mathtt{x}}\alpha_{\mathtt{x}} = \frac{1}{N_{\mathtt{x}}} \tag{435}$$

This normalisation condition ensures that we get back the fiducial matrix (390) when we plug (434) into (433). The normalisation gauge parameters $\alpha^{\mathtt{x}}$ and $\alpha_{\mathtt{x}}$ can take any real values consistent with (435). We can expand the maximal elements in terms of the fiducials as follows

$$= \qquad\qquad\qquad = \tag{436}$$

Here we are summing over $x$ where the black and white dot meet. In Sec. 18 we commented that maximal elements have a correlated pointer and system. We can see this explicitly in (436). Since the duotensor notation may be less familiar it is worth representing the object in more standard notation as

$$= \sum_{x=1}^{N_{\mathtt{x}}} \beta_{\mathtt{x}}\alpha^{\mathtt{x}}N_{\mathtt{x}}\ \mathbf{n}_{\mathtt{x}_1} \otimes |x\rangle^{\mathtt{x}_2}\langle x| \tag{437}$$

and

$$= \sum_{x=1}^{N_{\mathtt{x}}} \beta^{\mathtt{x}}\alpha_{\mathtt{x}}N_{\mathtt{x}}\ \mathbf{n}^{\mathtt{x}_1} \otimes |x\rangle_{\mathtt{x}_2}\langle x| \tag{438}$$

where the $\mathbf{n}_{\mathtt{x}_1}[x]$ and $\mathbf{n}^{\mathtt{x}_1}$ are orthonormal vectors in $\mathcal{P}_{\mathtt{x}_1}$ and $\mathcal{P}^{\mathtt{x}_1}$ respectively. The notation on the right hand sides of (437, 438) is quite foreign to the approach in this book but it helps in seeing how maximal elements are correlations between a pointer and system. The constants $\beta_{\mathtt{x}}\alpha^{\mathsf{x}}N_{\mathtt{x}}$ and $\beta^{\mathtt{x}}\alpha_{\mathsf{x}}N_{\mathtt{x}}$ come from (427) and (434).

The ignore operators (see Table 2) can be written in terms of the maximal operators

$$\boxed{\hat{I}}^{\mathsf{x}} \;\equiv\; \boxed{R}\overset{x}{\text{-}\square\blacksquare\text{-}}\boxed{\hat{X}}^{\mathsf{x}} \qquad\qquad \boxed{\hat{I}}_{\mathsf{x}} \;=\; \boxed{\hat{X}}_{\mathsf{x}}\overset{x}{\text{-}\blacksquare\square\text{-}}\boxed{R} \tag{439}$$

Using (243) and correspondence along with (434), we obtain

$$\boxed{\hat{I}}^{\mathsf{x}} = \alpha^{\mathsf{x}}\hat{\mathbb{1}}^{\mathsf{x}_1} \qquad\qquad \boxed{\hat{I}}_{\mathsf{x}} = \alpha_{\mathsf{x}}\hat{\mathbb{1}}_{\mathsf{x}_1} \tag{440}$$

where

$$\hat{\mathbb{1}}^{\mathsf{x}_1} = \sum_{x=1}^{N_{\mathtt{x}}} |x\rangle^{\mathsf{x}_1}\langle x| \qquad\qquad \hat{\mathbb{1}}_{\mathsf{x}_1} = \sum_{x=1}^{N_{\mathtt{x}}} |x\rangle_{\mathsf{x}_1}\langle x| \tag{441}$$

are the identity operators in $\mathcal{V}^{\mathsf{x}_1}$ and $\mathcal{V}_{\mathsf{x}_1}$ respectively.

The normalisation condition (435) now guarantees that

$$\boxed{\hat{I}}\;\big|^{\mathsf{x}}\;\boxed{\hat{I}} \;=\; 1 \tag{442}$$

which we require to be true by correspondence with (97).

These normalisation gauge parameters, $\alpha^{\mathsf{x}}$, $\alpha_{\mathsf{x}}$, $\beta^{\mathtt{x}}$, and $\beta_{\mathtt{x}}$ are absent in the usual treatments of Quantum Theory. If we set $\alpha_{\mathsf{x}} = 1$ (so $\alpha^{\mathsf{x}} = \frac{1}{N_{\mathtt{x}}}$ in (440)) then we obtain the usual expressions used in such standard treatments of Quantum Theory wherein ignore preparation corresponds to the maximally mixed preparation, $\mathbb{1}^{\mathsf{x}_1}/N_{\mathtt{x}}$, while and the ignore result corresponds to the trace operation, $\mathbb{1}_{\mathsf{x}_1}$. In such standard treatments of Quantum Theory it looks like we have a time asymmetry even at the very basic level of how we represent ignore preparations and results. It is instructive here to see that introducing the normalisation gauge parameters allows us to think of these two objects as being on the same footing.

## 25 Maximal states and maximal effects are pure

In Sec. 8.4 we defined maximal preparations and results in equation (182). The corresponding definitions are

$$\boxed{\boldsymbol{R}} \overset{\mathsf{x}}{—} \boxed{x} \overset{\mathsf{x}}{—} \boxed{\hat{\boldsymbol{X}}\lrcorner}^{\,\mathsf{x}} \qquad\qquad {}_{\mathsf{x}}\boxed{\ulcorner\hat{\boldsymbol{X}}} \overset{\mathsf{x}}{—} \boxed{x} \overset{\mathsf{x}}{—} \boxed{\boldsymbol{R}} \tag{443}$$

Since we have defined maximal elements to correspond to a projection onto a basis set (in Sec. 24) it follows that maximal states and effects are proportional to pure states and effects (since the latter are represented by rank one operators as discussed in Sec. 16.1). In fact, we can see that maximal states (effects) are equal to pure states (effects). This follows immediately under correspondence from the maximal preparations and results theorem in Sec. 8.4. Further, all pure states (effects) are equal to maximal states (effects). This follows since both are represented by rank one operators and every rank one operator of appropriate length belongs to at least one set of maximal states (results). We state these simple facts as a theorem

> **Purity and maximality theorem.** Maximal states (effects) are pure states (effects) and pure states (effects) are maximal states (effects). Pure states are represented by rank one projectors, $\hat{A}^{\mathsf{a}_1} = |A\rangle^{\mathsf{a}_1}\langle A|$. Pure effects are represented by rank one projectors, $\hat{C}_{\mathsf{a}_1} = |C\rangle^{\mathsf{a}_1}\langle C|$. Pure states and effects satisfy the normalisation
>
> $$\begin{array}{c}\boxed{\hat{\boldsymbol{I}}}\\ |\,\mathsf{a}\\ \boxed{\hat{A}}\end{array} = \frac{1}{N_{\mathsf{a}}} \qquad\qquad \begin{array}{c}\boxed{\hat{C}}\\ |\,\mathsf{a}\\ \boxed{\hat{\boldsymbol{I}}}\end{array} = \frac{1}{N_{\mathsf{a}}} \tag{444}$$
>
> This means that pure states are necessarily nondeterministic.

The normalisation in (444) follows since we have established that pure states are in the form (443) above. We can append an ignore operator directly to these equations then use the operator equations obtained under correspondence with (175) to simplify.

In Sec. 7.8 we discussed the double cone space of states (and effects). Now we can see why we need this double cone structure. Since the pure states, for example, cannot be deterministic they must be closer to the null state.

## 26 Physicality conditions

We say an operation is physical if (i) it satisfies tester positivity and (ii) it satisfies $t$-causality (so double causality $t$ =TS, forward causality for $t = TF$, and backward causality for $t$ =TB). Under correspondence, this maps onto a physicality condition on operators. An operator is physical if (i) it satisfies $T$-positivity and (ii) it satisfies $t$-causality. Let us examine these physicality conditions on operators in the $t$ =TS case.

$T$-positivity is the condition that

b

$$0 \underset{T}{\leqq} \quad \mathsf{x} - \boxed{\hat{B}} - \mathsf{y} \tag{445}$$

a

where $\leqq_T$ is with respect to any tester of the form

$$\hat{E}$$

b

$$\mathsf{h} \quad \boxed{\boldsymbol{R}} \overset{\mathsf{x}}{-} \boxed{x} \overset{\mathsf{x}}{-} \qquad\qquad \overset{\mathsf{y}}{-} \boxed{y} \overset{\mathsf{y}}{-} \boxed{\boldsymbol{R}} \tag{446}$$

a

$$\hat{D}$$

where $\hat{D}$ and $\hat{E}$ are rank one projectors. This follows under correspondence from (150) and the purity correspondence assumption (in Sec. 16.1). Note that, since we are testing for positivity, it is sufficient to consider only testers wherein $\hat{D}$ and $\hat{E}$ are normalised rank one projectors. If we are in the $t$ =TF case we omit the income $\boldsymbol{R}$ and $x$ boxes on the left. If we are in the $t$ =TB case we omit the outcome $x$ and $\boldsymbol{R}$ boxes on the right. There are two additional mathematically equivalent ways of stating this $T$-positivity condition on operators which we will discuss in detail in Sec. 43. One says that $T$-positivity of $\hat{B}$ implies that it is a positive operator (in the usual sense) under the *input twist* (which will be defined). The other says that $T$-positive operators can be written in the *twofold* form

b b b

$$\boxed{\hat{B}} \; = \; \boxed{B} \overset{l}{-\!-\!-} \boxed{B} \tag{447}$$

a a a

where this new diagrammatic notation will be explained.

Operators that correspond to deterministic operations are called *deterministic operators*. For the $t$ =TS case we have the double causality conditions

$$\text{(diagram)} \quad = \quad \text{(diagram)} \qquad\qquad \text{(diagram)} \quad = \quad \text{(diagram)} \tag{448}$$

on any deterministic operator, $\hat{\boldsymbol{B}}$ (and, indeed, any operator satisfying these conditions is necessarily deterministic). This follows under correspondence from (107 108). In the case we have a nondeterministic operator, we have the general double causality conditions

$$\text{(diagram)} \quad \underset{T}{\leq} \quad \text{(diagram)} \qquad\qquad \text{(diagram)} \quad \underset{T}{\leq} \quad \text{(diagram)} \tag{449}$$

where the inequalities are saturated in the deterministic case. This follows under correspondence from (160 161).

In the $t$ =TF case we have the general forward causality condition

$$\text{(diagram)} \quad \underset{T}{\leq} \quad \text{(diagram)} \tag{450}$$

The inequality is saturated in the deterministic case. This follows under correspondence from (287).

In the $t$ =TB case we have the general backward causality condition

$$\text{(diagram)} \quad \underset{T}{\leq} \quad \text{(diagram)} \tag{451}$$

where the inequality is saturated in the deterministic case. This inequality follows from (314) using correspondence.

## 27 Axioms for Simple Operational Quantum Theory

We can now state axioms for simple operational quantum theory.

**Axioms for Simple Operational Quantum Theory.**

0 All realisable circuits are directed acyclic graphs (DAG's)

1 All realisable operations are physical.

2 Every realisable operation has a physical operator corresponding to it.

3 Every physical operator has a realisable operation corresponding to it.

Here we use the word "realisable" to indicate circuits and operations that can be realised in the real world (the laboratory). Axiom 0 prohibits circuits that would have closed time like loops. Axiom 1 guarantees that probabilities for circuits are between 0 and 1 and that $t$-causality is satisfied. Quantum Theory concerns operators associated with Hilbert spaces. Axiom 2 enforces this. Axiom 2 also enforces that realisable operations are physical (by correspondence). It is logically possible that some physical operators have no corresponding realisable operation. Axiom 3 rules this out. Physicality is understood with respect to the temporal frame of reference we are working in.

- *If we are in $t$=TS frame* then, for operations, the physicality conditions are that $T$-positivity (150), general forward cauality (160), and general backward causality (161) are satisfied. For operators we use the corresponding conditions.

- *If we are in the $t$ =TF frame* then, for operations, the physicality conditions are that $T$-positivity (281) and the general forward causality in (286, 287) are satisfied For operators we use the corresponding conditions.

- *If we are in the $t$ =TB frame* then, for operations, the physicality conditions are that $T$-positivity (310) and general backward causality in (312, 314) are satisfied We use the corresponding conditions for operators.

The corresponding conditions for operators, in each case, is obtained in the obvious way as, for example, discussed in Sec. 26.

If, as in standard treatments, we work in the TF temporal frame, then these axioms are equivalent to the assertion that operations are associated with trace non-increasing completely positive maps. The complete positivity part is equivalent to tester positivity. The trace non-increasing part is equivalent to the forward causality condition (287). The other temporal frames are equivalent.

# Part III
# Simple Hilbert Objects

## 28 Hilbert Objects

In order to further study $T$-positivity in Quantum Theory and in order to set up a dilation theorem (along the lines of the famous Stinespring dilation theorem in Quantum Theory) we need to see how operators can be related to objects that live in Hilbert space. In text book presentations an operator can be written

$$\hat{A} = \sum_l |\varphi_l\rangle\langle\varphi_l| \tag{452}$$

We will provide a diagrammatic representation. This pictorial way of thinking about Quantum Theory has a long history which we discussed in Sec. 2.1. The particular tricks of representing what we will call Hilbert objects diagrammatically, representing them in Hilbert squares and then doubling up to get operators can be found in the work of Abramsky and Coecke [2004] and Selinger [2007]. The book by Coecke and Kissinger [2017] provides a good overview and more detailed description of the history. The diagrammatic notation used here is a little different to that used in these references for two reasons. First, we will root the diagrammatic notation within a study of the *conjuposition group* of hypermatrices (which we introduce in Sec. 29.1). Second, in order to make contact with the double border boxes used to represent operators, we will write

b b b

$\hat{A}$ = $A$ $l$ $A$ (453)

for an operator of the form in (452) when it corresponds to a preparation (we can think of "pushing" the two boxes on the right hand side together to form a box like that on the left). More generally, we will write

b b b

$\hat{B}$ = $B$ $k$ $B$ (454)

a a a

for a completely positive map expressed in terms of Kraus operators, $B_k$

$$\sum_k B_k \ \cdot \ B_k^\dagger \tag{455}$$

(which acts on a density matrix, $\hat{\rho}$ as $\sum_k B_k \hat{\rho} B_k^\dagger$).

The essential element in this description is what we will call a *Hilbert object*. In the two examples above we have (on the right of the equation) a left Hilbert object and a right Hilbert object. The left Hilbert object in (454) is the "left half" of an operator tensor and the right Hilbert object is the "right half" and so they are suggestively represented by box with a double border all around except on one side so it looks like they are meant to be joined together to form an operator tensor. In this section we will look at Hilbert objects in some detail. In subsequent sections we will explore their properties.

In this Part we introduce the mathematical machinery of Hilbert objects for the purpose of formulating Quantum Theory. The Hilbert objects introduced here are for the purpose of formulating causally simple operational quantum theory. They are represented by rectangles. We might call them *simple Hilbert objects* though mostly we will just call them *Hilbert objects* for brevity. Later we will introduce a theory of Hilbert objects for causally complex quantum theory. These might be called *complex Hilbert objects*. They are represented by semicircles.

## 28.1 Doubling up notation

We use the following doubling up notation for wires

$$\mathsf{a}\Big| \quad = \quad \mathsf{a}\,\|\,\mathsf{a} \tag{456}$$

We see an example of this in (454). We call the small black rectangles on the wires "bobbles".

## 28.2 Basic Hilbert objects

There are four basic Hilbert space objects we need. These correspond to bras and kets and to inputs and outputs as follows

$$\begin{array}{ll} {}_{\mathsf{a}_1}\langle B| \in \mathcal{H}_{\mathsf{a}_1} \qquad & |B\rangle_{\mathsf{a}_1} \in \overline{\mathcal{H}}^{\mathsf{a}_1} \\ \\ |A\rangle^{\mathsf{a}_1} \in \mathcal{H}^{\mathsf{a}_1} \qquad & {}^{\mathsf{a}_1}\langle A| \in \overline{\mathcal{H}}^{\mathsf{a}_1} \end{array} \tag{457}$$

The bra/ket notation is, from our point of view, rather antiquated. It is better to use diagrammatic notation for the same objects as follows

$$\begin{array}{ll} \boxed{B}\,\mathsf{a} \ \in \mathcal{H}_{\mathsf{a}_1} \qquad & \boxed{B}\,\mathsf{a} \ \in \overline{\mathcal{H}}_{\mathsf{a}_1} \\ \\ \mathsf{a}\,\boxed{A} \ \in \mathcal{H}^{\mathsf{a}_1} \qquad & \mathsf{a}\,\boxed{A} \ \in \overline{\mathcal{H}}^{\mathsf{a}_1} \end{array} \tag{458}$$

As we will see, this notation goes naturally with the double border box notation we have adopted for operator tensors. We will, below, consider left and right Hilbert objects which have both inputs and outputs.

We will say that the objects on the left of (457, 458) are *left Hilbert objects* belonging to a *left Hilbert space* and objects on the right are *right Hilbert objects* belonging to a *right Hilbert space*. Left and right Hilbert objects using the same letter (e.g. $A$ above) are *horizontal adjoints* of one another. This is just as, in bra-ket notation $|A\rangle^{\mathsf{a}_1}$ and ${}^{\mathsf{a}_1}\langle A|$ are adjoints. Here we add the word "horizontal" because we flip horizontally in the diagrammatic notation. We will see in Sec. 32 that the horizontal adjoint here is an example of what we will call the normal conjuposition group. Its full name will be the *normal horizontal adjoint*. We will also define the *normal vertical adjoint* (which involves flipping vertically) along with other transformations on Hilbert objects.

As we noted above, the bra/ket notation is rather antiquated. In particular, a left Hilbert object is a bra if it has an input and a ket if it has an output. We will frequently want to notate objects (like Kraus operators) that live in the tensor product of left input and left output Hilbert objects (these are general left Hilbert objects). Objects in this tensor product space cannot be notated by either a bra or a ket and so we have, in some sense, a notational failure. For the most part we will employ the diagrammatic notation in (458). However, it is also useful to have *new symbolic notation* that is better tuned to the framework being presented here. To this end, we suggest

$$
\begin{array}{ll}
{}_{\mathsf{a}_1}[\![B| \in \mathcal{H}_{\mathsf{a}_1} & \qquad |B]\!]_{\mathsf{a}_1} \in \overline{\mathcal{H}}^{\mathsf{a}_1} \\[2ex]
{}_{\mathsf{a}_1}[\![A| \in \mathcal{H}^{\mathsf{a}_1} & \qquad |A]\!]_{\mathsf{a}_1} \in \overline{\mathcal{H}}^{\mathsf{a}_1}
\end{array}
\tag{459}
$$

This new symbolic notation better models the diagrammatic notation. In spite of the antiquated nature of the bra/ket notation, we will sometimes use it below rather than the new symbolic notation because of its familiarity and consequent usefulness from a pedagogical point of view.

## 28.3 Scalar products

Inputs feed into outputs and bras and kets naturally combine to give us the scalar products

$$
{}_{\mathsf{a}_1}\langle B|A\rangle^{\mathsf{a}_1} \qquad\qquad {}^{\mathsf{a}_1}\langle A|B\rangle_{\mathsf{a}_1}
\tag{460}
$$

Represented diagrammatically, these scalar products are

$$
\begin{array}{ccc}
\boxed{B} & \qquad & \boxed{B} \\
|\,\mathsf{a} & & |\,\mathsf{a} \\
\boxed{A} & & \boxed{A}
\end{array}
\tag{461}
$$

The scalar product on the left is the complex conjugate of the scalar product on the right. In terms of the new symbolic notation the scalar product is notated

$$ {}^{\mathsf{a}_1}[\![A|\,{}_{\mathsf{a}_1}[\![B| \qquad\qquad |A]\!]^{\mathsf{a}_1}|B]\!]_{\mathsf{a}_1} \tag{462}$$

Note that, in the new symbolic notation, we do not form a scalar product by a $\langle$bra$|$ket$\rangle$ structure. Rather, the superscript $\mathsf{a}_1$ is matched by a subscript $\mathsf{a}_1$.

We will write

$$|A|^2 := \begin{array}{c}\boxed{A}\\ \vert\ \mathsf{a}\\ \boxed{A}\end{array} \tag{463}$$

where we say $|A|$ is the *norm* of $A$.

## 28.4 Sums of Hilbert objects

We can take sums of Hilbert objects. For example, in standard (antiquated) symbolic notation we might have

$$\begin{array}{ccc} {}_{\mathsf{a}_1}\langle B| = \sum_k \alpha_k \, {}_{\mathsf{a}_1}\langle D[k]| & \Leftrightarrow & |B\rangle_{\mathsf{a}_1} = \sum_k \bar{\alpha}_k |D[k]\rangle_{\mathsf{a}_1} \\ |A\rangle^{\mathsf{a}_1} = \sum_l \beta_l |E[l]\rangle^{\mathsf{a}_1} & \Leftrightarrow & {}^{\mathsf{a}_1}\langle A| = \sum_l \bar{\beta}_l \, {}^{\mathsf{a}_1}\langle E[l]| \end{array} \tag{464}$$

Note that on the right we have taken the complex conjugate of the coefficients as well as flipping kets to bras. We can write the same four equations diagrammatically

$$\begin{array}{ccccc} \begin{array}{c}\boxed{B}\\ \vert\ \mathsf{a}\end{array} = \begin{array}{c}\boxed{D}\overset{k}{\text{—}}\boxed{\alpha}\\ \vert\ \mathsf{a}\end{array} & \Leftrightarrow & \begin{array}{c}\boxed{B}\\ \vert\ \mathsf{a}\end{array} = \begin{array}{c}\boxed{\bar{\alpha}}\overset{k}{\text{—}}\boxed{D}\\ \vert\ \mathsf{a}\end{array} \\ \begin{array}{c}\vert\ \mathsf{a}\\ \boxed{A}\end{array} = \begin{array}{c}\vert\ \mathsf{a}\\ \boxed{E}\overset{l}{\text{—}}\boxed{\beta}\end{array} & \Leftrightarrow & \begin{array}{c}\vert\ \mathsf{a}\\ \boxed{A}\end{array} = \begin{array}{c}\vert\ \mathsf{a}\\ \boxed{\bar{\beta}}\overset{l}{\text{—}}\boxed{E}\end{array} \end{array} \tag{465}$$

When we flip the diagram to obtain right Hilbert objects from left Hilbert objects, two things are happening. First, we are taking the complex conjugate of the coefficients (denoted by the bar over $\overline{\alpha}$ and $\overline{\beta}$). Secondly, we are taking what we will call the "horizontal transpose" since we are flipping the boxes with the $\alpha$ and $\beta$'s (so now the wire is on the otherside). Generally, in matrix manipulations, conjugate plus transpose equals adjoint. Thus we will call this the *horizontal adjoint*. This is discussed in more detail in Sec. 29.1 where we will also see that there is a vertical transpose and a corresponding vertical adjoint. In anticipation of later terminology it is worth mentioning that the transformations in (465) can be implemented by expanding the Hilbert objects with respect to a basis. The normal way of doing this is to use an orthonormal basis. In this case we will refer to the *normal horizontal adjoint* of the Hilbert object, the *normal vertical adjoint* of the Hilbert object, and so on. However, there is a more natural way of implementing these maps (for the time symmetric case) wherein the basis elements are not normalised to 1 but, instead, normalised in a more physically useful way. When we use this "ortho-physical" basis we will refer to the *natural horizontal adjoint*, the *natural vertical adjoint*, and so on. It so happens that the normal horizontal adjoint is actually the same as the natural horizontal adjoint and so we can simply refer to this as the horizontal adjoint. The same is not true for the vertical adjoint or, indeed, the horizontal transpose.

In the new symbolic notation we notate these as follows

$$
\begin{aligned}
{}_{\mathsf{a}_1}[\![B| = \alpha_k \;\; {}_{\mathsf{a}_1}[\![D|^k \qquad &\Leftrightarrow \qquad |B]\!]_{\mathsf{a}_1} = \bar{\alpha}^k \;\; {}_k|D]\!]_{\mathsf{a}_1} \\
{}_{\mathsf{a}_1}[\![A| = \beta_l \;\; {}_{\mathsf{a}_1}[\![E|^l \qquad &\Leftrightarrow \qquad |A]\!]^{\mathsf{a}_1} = \bar{\beta}^l \;\; {}_l|E[l]]\!]^{\mathsf{a}_1}
\end{aligned}
\tag{466}
$$

where summation over the labels $k$ and $l$ is implicit. Further, $\bar{\alpha}^k$ is the complex conjugate of $\alpha_k$ for each given value of $k$.

We can build up left or right Hilbert objects that belong to a tensor product space. For example,

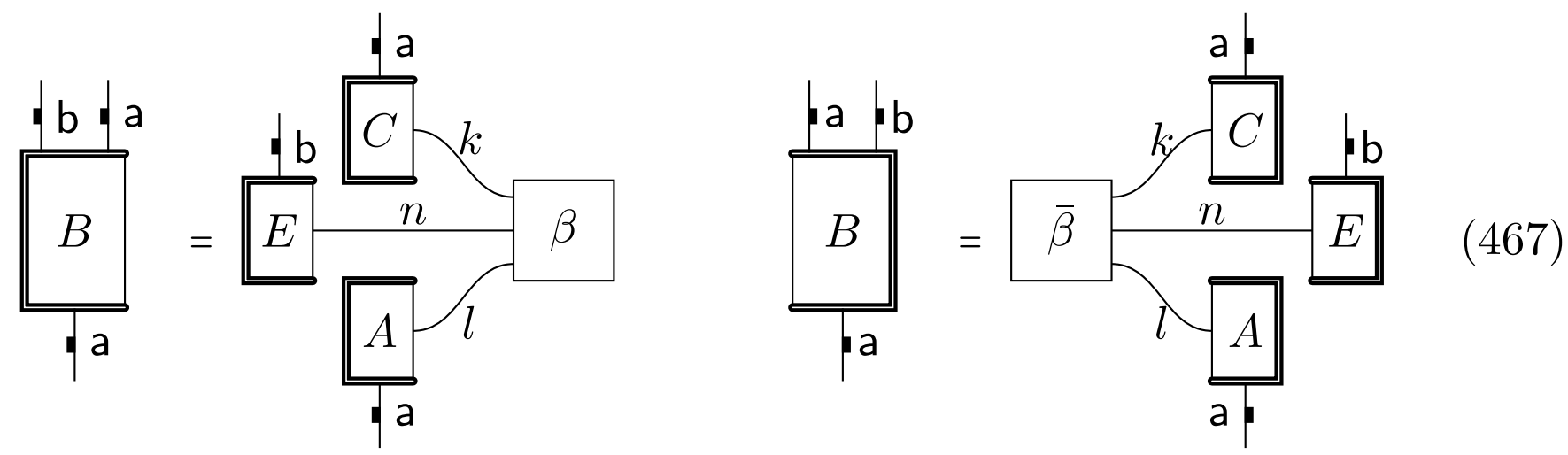

(467)

The object on the left is a left Hilbert object since it belongs to a belongs to a left Hilbert space, namely $\mathcal{H}_{\mathsf{a}_1} \otimes \mathcal{H}^{\mathsf{a}_2} \otimes \mathcal{H}^{\mathsf{b}_3}$. The object on the right is a right Hilbert object as it belongs to a right Hilbert space, namely $\overline{\mathcal{H}}_{\mathsf{a}_1} \otimes \overline{\mathcal{H}}^{\mathsf{a}_2} \otimes \overline{\mathcal{H}}^{\mathsf{b}_3}$. The objects in (467) are normal horizontal adjoints of one another.

## 28.5 Left and right Hilbert networks

We can form *left Hilbert networks* by wiring together left Hilbert objects and *right Hilbert networks* by wiring together right Hilbert objects. For example

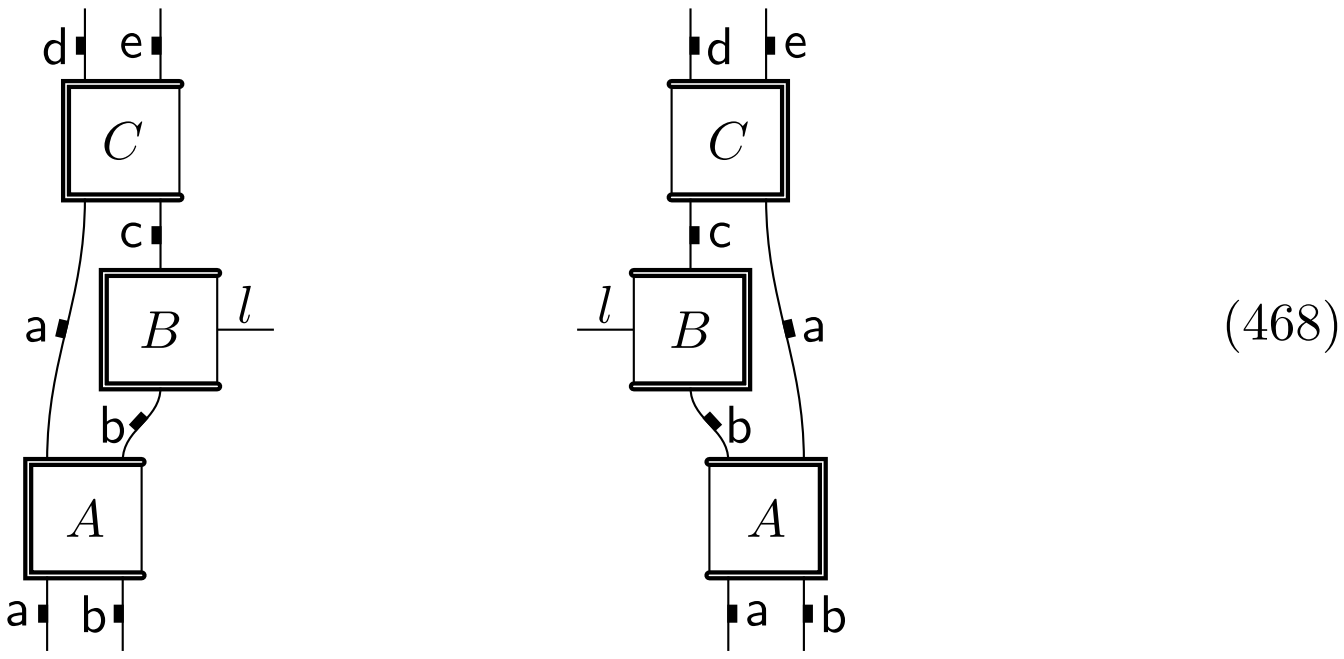

(468)

A left (or right) Hilbert network is, itself, a left (or right) Hilbert object since it belongs to the left (or right) Hilbert space associated with the open system wires left over.

If there are no system wires left open then we will say we have a *left or right Hilbert circuit*. For example

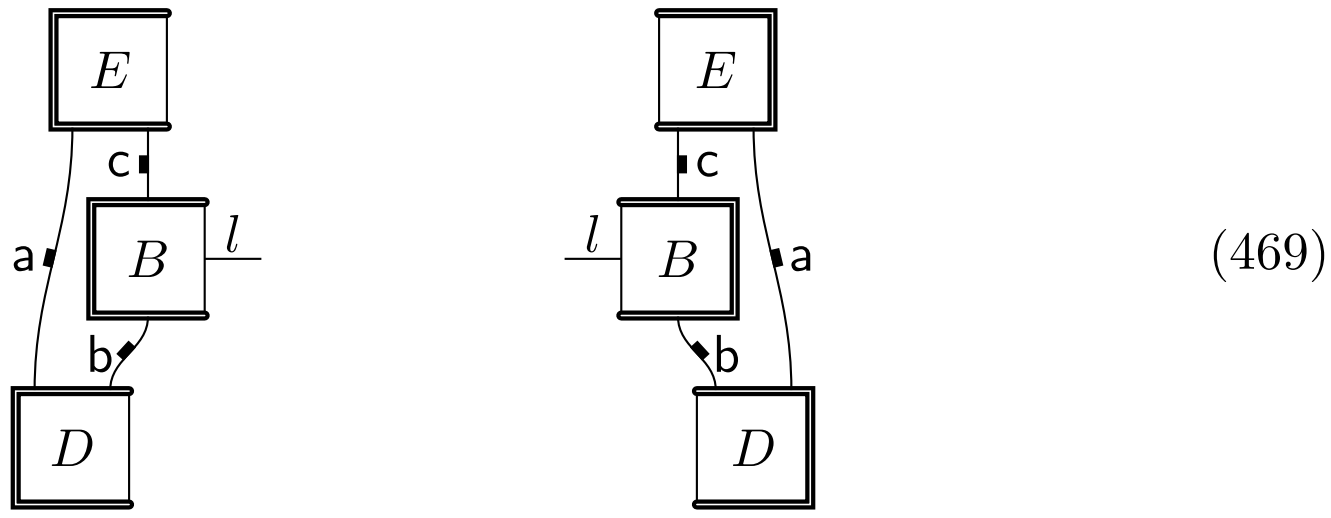

(469)

provides a left Hilbert circuit (on the left) and a right Hilbert circuit (on the right). These Hilbert circuits have no open system wires (though they can have open label wires like the $l$ wire in these examples). A Hilbert circuit is a list of complex numbers (labeled by $l$ in the above examples) and when we flip a Hilbert circuit, we get the complex conjugates of these numbers. We can think of this as the horizontal adjoint since the wire labeled by $l$ ends up pointing in the other direction.

## 28.6 Left and right composite Hilbert objects

We can construct left Hilbert objects in an even more general way by allowing multiple left Hilbert objects and hypermatrices. For example

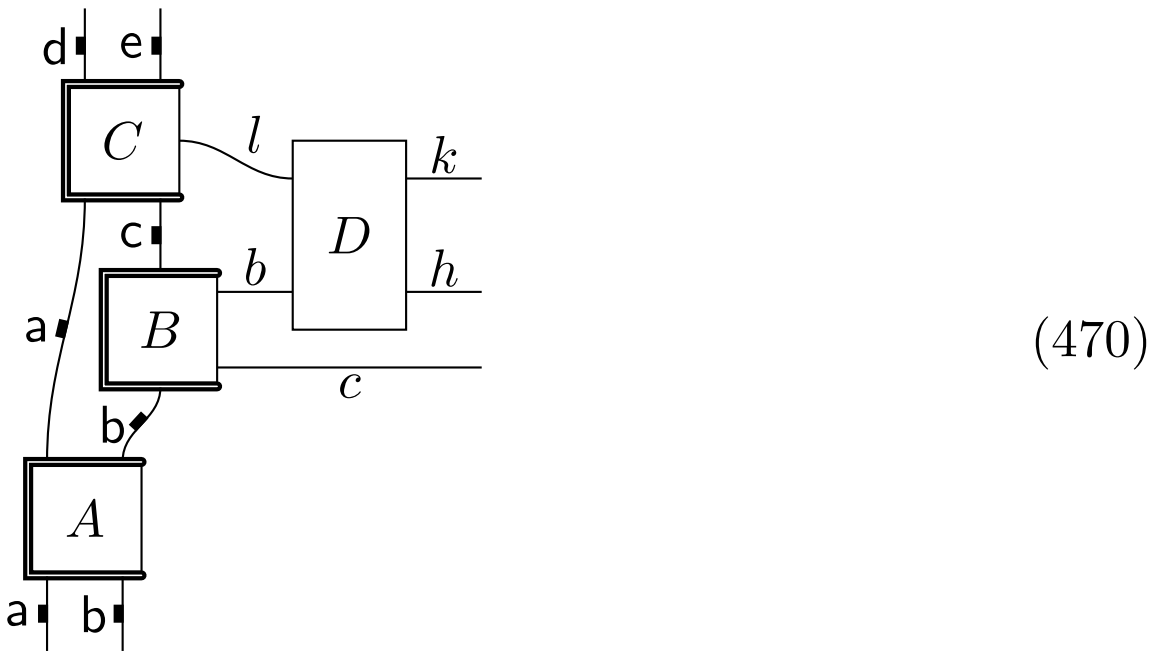

(470)

Right composite Hilbert objects can be formed similarly.

## 28.7 General Hilbert objects

We can form general Hilbert objects such as

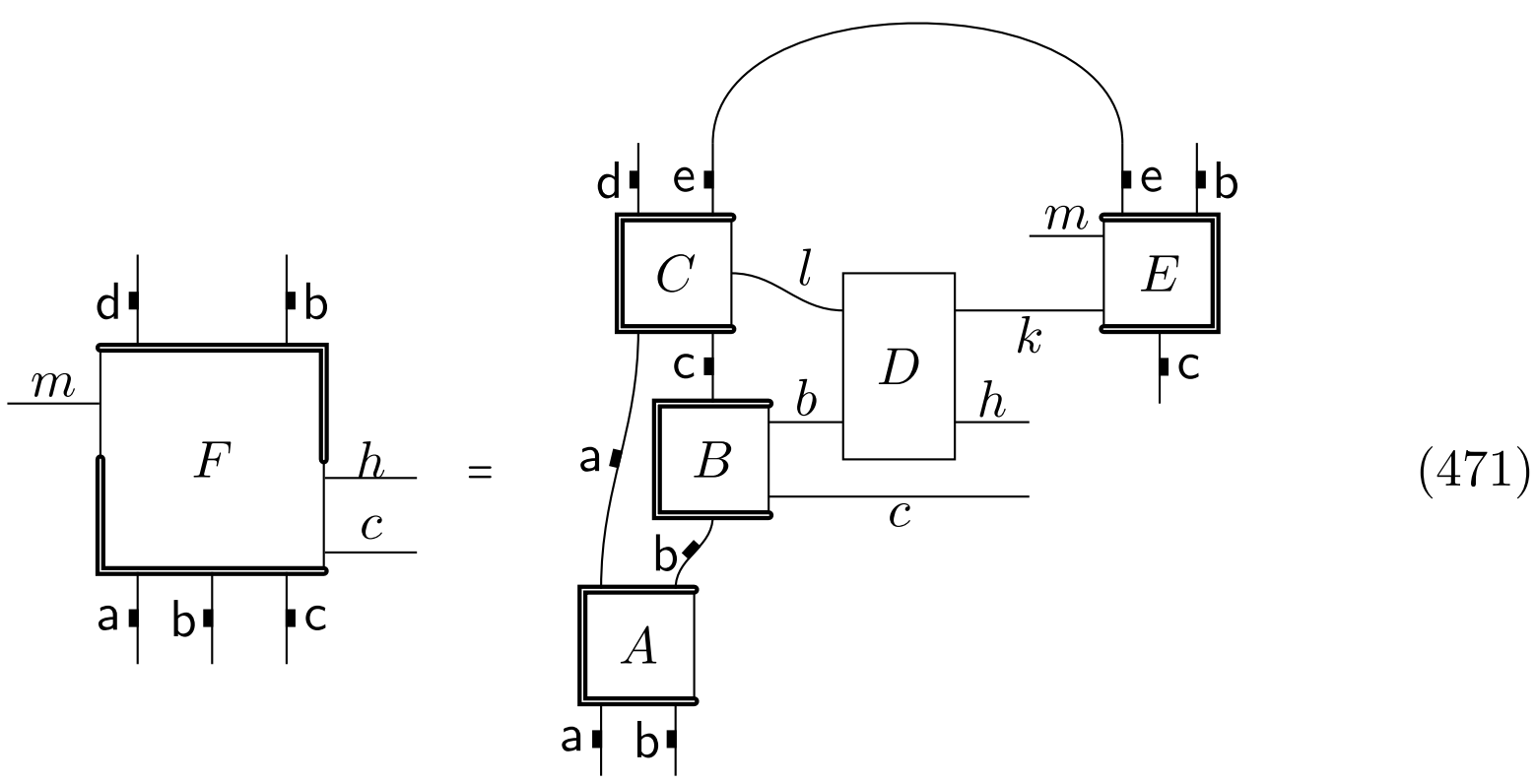

(471)

General Hilbert objects may have both left and right Hilbert objects in its composition. Left Hilbert objects are a special case of general Hilbert objects (as are right Hilbert objects). If a general Hilbert object has both left and right wires then we will call it a *hybrid Hilbert object*. We are especially interested in operator tensors which, as we will see in Sec. 31, are a special case of such hybrid Hilbert objects (where left and right wires are all matched up in pairs).

Hybrid Hilbert objects crop up when we connect left and right Hilbert objects using the vertical twist operations which will defined in (513). In particular, see the example of using a twist mirror in (see (751)). In general, it useful to be able to break diagrams up in any manner we wish. General fragments of a diagram will be general Hilbert objects (and may be hybrid).

# 29 Hypermatrices

Hypermatrices are higher dimensional matrices. We have seen some examples already. For example, in (467) we had the boxes

$$k \quad n \quad l \quad \beta \qquad\qquad \overline{\beta} \quad k \quad n \quad l \tag{472}$$

which appeared as expansion coefficients for a left and right Hilbert object respectively. These are hypermatrices. More generally, a hypermatrix may have wires on both sides of the box such as

$$b \quad d \quad M \quad a \quad c \tag{473}$$

It has horizontal structure (as there can be wires on the left and the right sides) and vertical structure (since there can be multiple wires on each side). We will be especially interested in when these appear as expansions with respect to bases (such as in (530) below). In such a case we can transform the basis and, thereby, transform the hypermatrix. Then hypermatrices become tensors. However, this tensorial property is not important for defining transpose, conjugate, and adjoints which is what we will turn to next.

## 29.1 Conjupositions of hypermatrices

We will be interested in various transformations on hypermatrices that involve moving the elements around and also, maybe, taking the conjugate of elements. We will call these *conjupositions* (the word "conjuposition" is a portmanteau of the words "conjugation" and "transposition"). We will see that there are eight conjuposition transformations of interest and that they form a group.

In the example in (472) above, in going from the hypermatrix on the left to that on the right, we have flipped the box horizontally (we will call this the horizontal transpose below) and we have taken the complex conjugate. This is clearly a kind of adjoint - we will call it the horizontal adjoint. There are several other types of transformations on hypermatrices which we will discuss below.

First we will consider transposition. Consider the following hypermatrix

$$b \quad d \quad M \quad a \quad c \tag{474}$$

We have placed a small circular dot to help us track the different objects that result under transposition as will be clear below.

We can define the following three transposes of this hypermatrix.

d M b c a := d M b c a — horizontal transpose $H$ (475)

b M d a c := a M c b d — vertical transpose $V$ (476)

c M a d b := c M a d b — transpose $T$ (477)

We also add the identity transformation (denoted by $I$) which leaves $M$ unchanged. It is useful to include the $I$ since then $\{I, H, V, T\}$ is a group of transformations with identity element $I$. Each transformation is its own inverse since $HH = VV = TT = I$. Clearly have $T = HV = VH$ - the transpose corresponds to taking the vertical and horizontal transpose in either order. In fact we also have $H = TV = VT$ and $V = TH = HT$. These are the properties of the Klein-four group, $K_4$. This is to be expected since $K_4$ is the symmetry group of a (non-square) rectangle). Indeed if we "gently pull on" the wires for the expressions on the right in above definitions we can see that the rectangle will flip over and agree with the rectangle on the left. Including the original $M$ we have four hypermatrices generated by taking transposes. These transposes also work when there are more than two wires on each side of the box. The vertical transpose simply inverts the order of the wires.

We can also take the complex conjugate of all the components in the hypermatrix, $M$. We can denote the new expansion box by $\overline{M}$. This doubles the number of objects we can obtain to eight.

$$\begin{array}{ccc} b & & d \\ & M & \\ a & & c \end{array} \qquad \begin{array}{ccc} b & & d \\ & \overline{M} & \\ a & & c \end{array} \tag{478}$$

$$\begin{array}{ccc} d & & b \\ & M & \\ c & & a \end{array} \qquad \begin{array}{ccc} d & & b \\ & \overline{M} & \\ c & & a \end{array} \tag{479}$$

$$\begin{array}{ccc} b & & d \\ & M & \\ a & & c \end{array} \qquad \begin{array}{ccc} b & & d \\ & \overline{M} & \\ a & & c \end{array} \tag{480}$$

$$\begin{array}{ccc} c & & a \\ & M & \\ d & & b \end{array} \qquad \begin{array}{ccc} c & & a \\ & \overline{M} & \\ d & & b \end{array} \tag{481}$$

The conjuposition, $\overline{I}$, is the conjugation transformation. It transforms between the left and right hand hypermatrices in each row above. We can denote performing conjugation and transposition by putting a bar over the corresponding transposition symbol. Thus, we have the full set of eight transformations

$$\mathcal{C} = \{I, \overline{I}, H, \overline{H}, V, \overline{V}, T, \overline{T}\} \tag{482}$$

We call these the *conjuposition transformations*. We have $\overline{H} = \overline{I}H = H\overline{I}$, $\overline{V} = \overline{I}V = V\overline{I}$, and $\overline{T} = \overline{I}T = T\overline{I}$. We give each of the conjupositions a name as shown in Table 3. These transformations form a group, the *conjuposition group* (which we will denote by $\mathcal{C}$). This group is acting "behind the scenes" to generate the Hilbert cube which we will discuss in Sec. 32.8. This group has various subgroups some of which (again, behind the scenes) play an important role. For example, the subset,

$$\mathcal{S}_{BI} = \{I, \overline{H}, \overline{V}, T\} \tag{483}$$

forms a subgroup (the Klein four group) and is behind the Hilbert square discussed in Sec. 32.5. This subgroup turns out to be basis independent and is particularly important for us. Another subgroup is

$$\mathcal{S}_{\text{side}} = \{I, \overline{I}, \overline{V}, V\} \tag{484}$$

This subgroup is behind the side Hilbert squares discussed in Sec. 32.9 and plays an important role in the book of Coecke and Kissinger [2017] - as will be discussed in Sec. 32.9. One other subgroup is of particular interest. This is

$$\mathcal{S}_{\text{nonconjugating}} = \{I, H, V, T\} \tag{485}$$

This subgroup does not use conjugation. There are many other subgroups - including two element subgroups of the form $\{I, e\}$ where $e$ is any element (since $ee = I$ for $\mathcal{C}$).

| $I$ | identity | $\overline{I}$ | conjugate |
|---|---|---|---|
| $H$ | horizontal transpose | $\overline{H}$ | horizontal adjoint |
| $V$ | vertical transpose | $\overline{V}$ | vertical adjoint |
| $T$ | transpose | $\overline{T}$ | adjoint |

Table 3: The conjuposition transformations on a hypermatrix.

## 29.2 Conjupositions of hypermatrix networks

We can wire together a bunch of hypermatrices forming a hypermatrix network such as

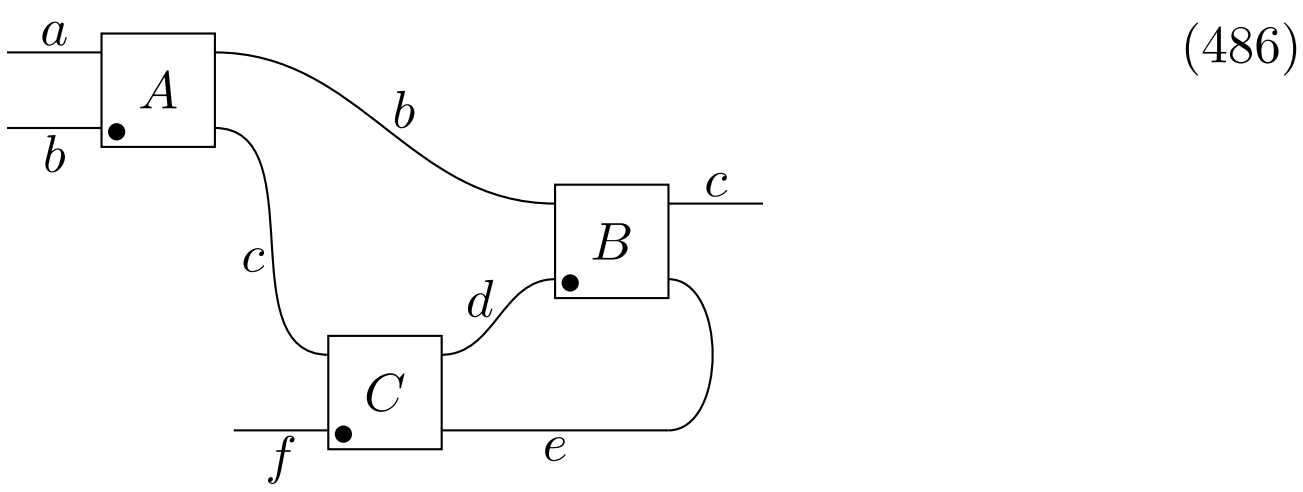

(486)

We can consider applying conjuposition transformations on such hypermatrix networks as a whole. Interestingly, we can implement such conjupositions by applying them separately to each hypermatrix in the network and wiring those matrices together appropriately. In our graphical notation, this simply amounts to flipping the whole diagram horizontally or vertically (or both for the transpose case) and applying complex conjugates if appropriate. For example, the transpose, $T$, of the hypermatrix network in (486) is

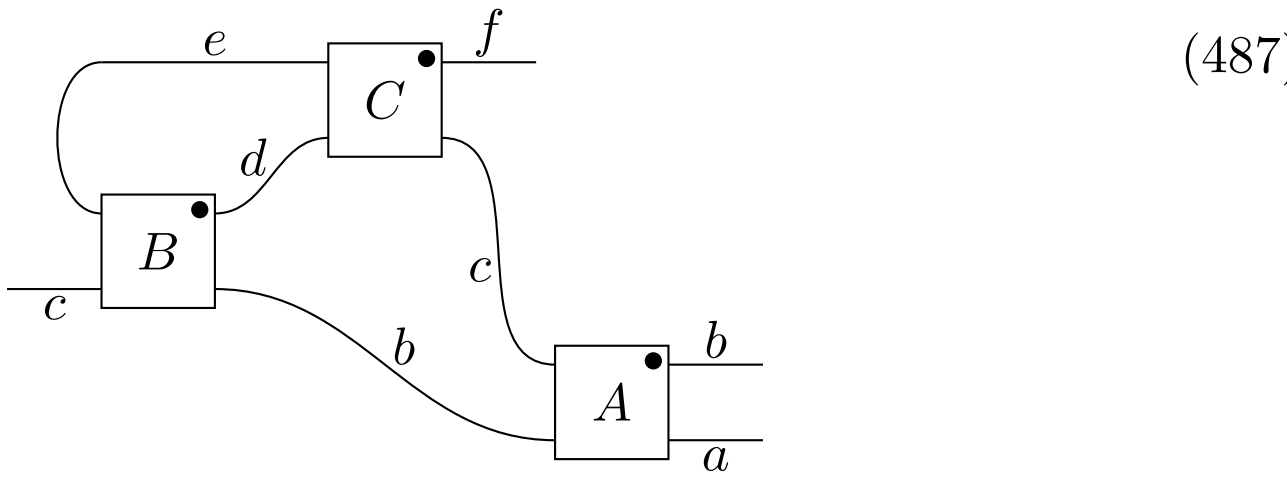

(487)

This diagram is obtained by flipping the network in (486) horizontally then vertically. The horizontal adjoint, $\overline{H}$, of the expression in (486) is given by flipping hozitontally and conjugating

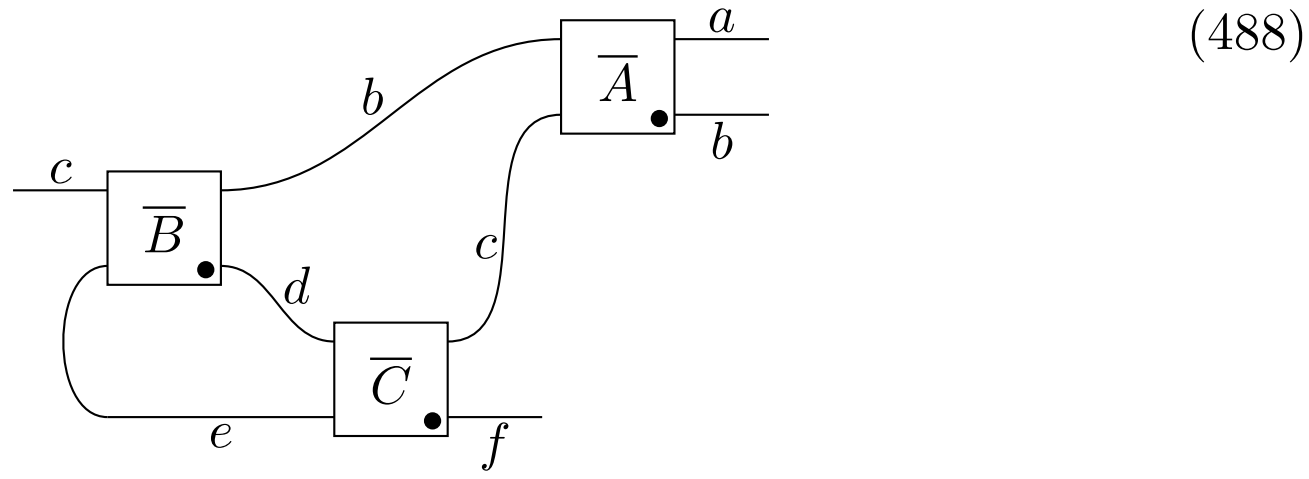

(488)

The proof that this technique works for general hypermatrix networks and any element of the conjuposition group is straightforward and outlined below. First we will consider some simple examples.

For the horizontal transpose we have

a C b D c H = c D b C a (489)

while, for the horizontal adjoint, we have

a C b D c $\overline{H}$ = c $\overline{D}$ b $\overline{C}$ a (490)

The proofs for these two cases are similar. Let us prove the horizontal transpose property first. Using (475) the horizontal transpose of $CD$ in (489) is equal to

c C b D a = c D b C a (491)

We see that the objects inside the dotted boxes are horizontal transposes and so we get (489). Equation (490) is proven in a similar fashion where we take the complex conjugate of the elements and use the fact that

$$\left(\sum_i a_i b_i\right)^* = \sum_i a_i^* b_i^* \tag{492}$$

Such sums appear in matrix multiplication. If we take the complex conjugate of a product of matrices we get the same thing as if we take the product of the complex conjugate of those matrices (wherein each element is replaced by its complex conjugate).

The two simple proofs above employ two tricks. First, we can rearrange the diagram as in going from the left hand side of (491) to the right hand side to see that the horizontal transpose is applied to each element. This trick also works for the vertical transpose. Second, we can apply complex conjugation afterwards using the property in (492). Using these tricks it is easy to show that any hypermatix network can be transformed under any element of the conjuposition group by flipping and conjugating it appropriately. If the network contains only two hypermatrices then we prove this using a technique similar to that in the simple example above. If the network contains more than two hypermatrices then we can employ a third trick. First, we group all but one of the hypermatrices together and regard them as one hypermatrix and then apply the above technique. Then we iterate by pulling one more hypermatrix out and so on.

## 29.3 Padding matrices

Here we introduce some useful notation. We define a *padding matrix* as follows

$$\overset{k}{\text{---}}\!\diamond\!\overset{m}{\text{---}} \;=\; \begin{cases} 1 & \text{if } \; k = m \\ 0 & \quad \text{else} \end{cases} \tag{493}$$

where $k = 1$ to $K$ and $m = 1$ to $M$ with $K \leq M$. We will call this the *padding matrix*. For the case where $K = 3$ and $M = 5$ the padding matrix looks lke this

$$\overset{k}{\text{---}}\!\diamond\!\overset{m}{\text{---}} \;=\; \begin{pmatrix} 1 & 0 & 0 & 0 & 0 \\ 0 & 1 & 0 & 0 & 0 \\ 0 & 0 & 1 & 0 & 0 \end{pmatrix} \tag{494}$$

The padding matrix is useful if we want "pad" a labeled set of objects with some zero elements. For example

$$\underset{a}{\overset{b}{\boxed{B}}}\overset{k}{\text{---}}\!\diamond\!\overset{m}{\text{---}} \tag{495}$$

Now, for $m = k$, this just returns the original set, and for $m > k$ this returns zero elements.

Sometimes it is useful to use the $f$-padding matrix. This is defined as follows

$$\overset{k}{\text{---}}\!\diamond_{f}\!\overset{m}{\text{---}} \;=\; \begin{cases} 1 & \text{if } \; m = f(k) \\ 0 & \quad \text{else} \end{cases} \tag{496}$$

where $f$ is a one-to-one (i.e. injective) function. The matrix associated with this has a single 1 in each row but not necessarily each column (and 0's otherwise). Clearly the padding matrix in (493) is a special case. We can think of a $f$-padding matrix as an appropriate permutation matrix (which simply permutes) followed by a padding matrix.

# 30 Orthonormal bases and associated objects

A basic tool for analysing Hilbert spaces is an orthonormal basis set. Once we have introduced these we can use them to define further objects. In particular, we have identity decompositions, yanking equations, and twist operations. Orthonormal bases can be used to expand Hilbert objects where the expansion weights are provided by a hypermatrix. We can then perform transformations on these objects by both transforming the bases and, using a conjuposition, transforming the expansion hypermatrix. This leads to the *normal conjuposition group*. Further important concepts that come out of orthonormal bases are the notion of a normal isometry, a normal coisometry, a normal unitary (these are standard notions from the literature where the word "normal" is omitted) and a normal maxometry (this is a new notion that will play an important role).

We will see that orthonormal bases do not reflect the underlying symmetry of time symmetric Quantum Theory. Instead we should use what we will call *ortho-physical* bases (and a complement we call *ortho-deterministic bases* that are derived from ignore operators). These bases are more natural for the underlying physics and, correspondingly, we can define the natural conjuposition group, natural unitaries, and so on.

Although orthonormal bases are not natural (in the above sense) for the underlying physics, they are mathematically simple and play an important role in setting up the necessary natural concepts. Hence we will start with them.

## 30.1 Orthonormal bases

We can choose some orthonormal basis set for each state space in (457) as follows

$$\begin{array}{ll} \left\{ {}_{\mathsf{a}_1}\langle a| : a = 1 \text{ to } N_{\mathsf{a}} \right\} & \left\{ |a\rangle_{\mathsf{a}_1} : a = 1 \text{ to } N_{\mathsf{a}} \right\} \\ \left\{ |a\rangle^{\mathsf{a}_1} : a = 1 \text{ to } N_{\mathsf{a}} \right\} & \left\{ {}^{\mathsf{a}_1}\langle a| : a = 1 \text{ to } N_{\mathsf{a}} \right\} \end{array} \tag{497}$$

These can be represented diagrammatically

$$(498)$$

where $a = 1$ to $N_a$. The orthogonality conditions

$${}^{\mathsf{a}_1}\langle a|a'\rangle_{\mathsf{a}_1} = \delta_{aa'} \qquad {}_{\mathsf{a}_1}\langle a|a'\rangle^{\mathsf{a}_1} = \delta_{aa'} \qquad \text{for all} \quad a, a' \tag{499}$$

can be written

$$(500)$$

For the most part it will be possible to use just one orthonormal basis set (for each system type) which denote with the white-filled empty boxes as above. We can introduce further basis sets using shading (following Coecke and Kissinger). In our notation this is as follows

(501)

where $U$ is a unitary matrix - i.e. it satisfies

(502)

(503)

It is clear that the shaded basis will also satisfy the orthogonality relations

(504)

We can prove this using (500) and the unitarity properties

(505)

which follow from (503) using the definition of the horizontal transpose in (475). Interestingly, we can invert the expressions in (501) expressing the white bases in terms of unitaries acting on the shaded ones. Then, we can prove these white bases are orthonormal if the shaded ones are using the other unitary equation (502).

## 30.2 Identity decompositions

There are ten ways to join the basis elements in (30.1) together at their label wires. Four of these ten ways correspond to decompositions of the identity. These are basis independent and discussed in this section. The remaining six ways correspond to transpose operations. These are not basis independent and are discussed in Sec. 30.4.

The first two decompositions of the identity are

$$(506)$$

These come from the decompositions of the identity $\sum_a |a\rangle_{\mathsf{a}_1}\ {}^{\mathsf{a}_2}\langle a|$ and $\sum_a |a\rangle^{\mathsf{a}_1}\ {}_{\mathsf{a}_2}\langle a|$ respectively. The proof that this acts as the identity can be seen by how it acts. To prove the equation on the left of (506) consider adding a $|A\rangle^{\mathsf{a}_1}$ box to the bottom

$$(507)$$

(where we have used the orthonormal relations (498)). Thus, this acts as the identity on $|A\rangle^{\mathsf{a}_1}$. We could, instead, add a box to the top with a similar result. One consequence of this argument is that (506) is independent of the choice of orthonormal basis.

The other two decompositions of the identity are

$$(508)$$

(from $\hat{\mathbb{1}}_{\mathsf{a}_1} = \sum_a |a\rangle_{\mathsf{a}_1}\langle a|$) and

$$(509)$$

(from $\hat{\mathbb{1}}^{\mathsf{a}_1} = \sum_a |a\rangle^{\mathsf{a}_1}\langle a|$). These provide us with a *cap* (right expression in (508)) and *cup* (right expression in (509)). These are very useful. It is important that these caps and cups are independent of the choice of basis. To see this note that

$$(510)$$

follows from (501) and the unitarity relationship (502) (and its conjugate).

## 30.3 Yanking equations

We can use the identity equations in Sec. 30.2 and the orthogonality conditions in (499) to prove "yanking" equations

(511)

These equation are useful in what follows. The proof of the yanking equation on the left is as follows

(512)

(we have taken some liberties with the positions of the incoming and outgoing wires to keep this proof compact). The first step uses the cap and cup identity decompositions (508, 509), the second step uses the orthogonality conditions (500), the third step is a graphical manipulation (we can move boxes around as long as we maintain their orientation and keep wires attached to the same places), and the final step uses the decomposition of the identity on the left of (506). The other yanking equation can be obtained similarly.

The mathematical structure behind cups, caps, and yanking equations goes back to Kelly and Laplaza [1980]. The diagrammatic representation is introduced in Joyal and Street [1991]. Cups and caps are, though, already implicit in Penrose [1971]. They were introduced into the study of Quantum Theory within the process theoretic approach of Coecke and collaborators in the original paper by Abramsky and Coecke [2004]. The vivid and apt terminology "yanking" equations became standard later is found in the book of Coecke and Kissinger [2017]. The standard approach in this work is to take cups to correspond to maximally entangled states (rather than the identity as we do) and to take caps to correspond to maximally entangled effects (rather than the identity). Then the yanking equation represents quantum teleportation. Note that Coecke and Kissinger [2017] discuss (see pages 149-150 therein) in formal language, yanking equations that are, essentially, in the form of (511) above. Also, note that we will discuss yanking equations that correspond to teleportation in the sense of Coecke et al. (see the comment below (519) in Sec. 30.4).

## 30.4 Twist objects

In Sec. 30.2 we defined four identity relations by joining together basis objects at their label wires. Further, these are basis independent. We can define a further six objects - the twist objects - so called because they "twist" the side of the bobble (the small black rectangle indicating whether we have a left or right wire) as the bobble "slides" through. These twist objects are not basis independent, however. We will denote the twist objects by a $w$ inside a circle (since *twist* has a "$w$" in it and we want to reserve $T$ for "transpose").

The six twist objects are as follows. The *vertical twist objects* are

(513)

The cup twist objects are

(514)

The cap twist objects are

(515)

We see, using (501) that the twist objects are basis dependent (since the unitaries in (501) do not cancel using the unitary relations in (502) and (503)). We can convert between any pair of these twist objects by appropriate application of the caps in (508) and (509). For example

(516)

Futhermore, if two $w$-circles meet they "annihilate". For example we have

(517)

This also happens more generally when we have twists on cups and/or caps. For example

(518)

(we can think of the twists as having "opposite sense" so they cancel). Properties of these types, as illustrated by the examples in (516), (517), (518), are easily derived by inserting the definitions of these objects in terms of bases. One interesting example of this $w$-annihilation property is

$$= \quad \tag{519}$$

where we have, further, used the yanking equation (511) to simplify the right hand side after the $w$-annihilation. This is, in fact, the yanking equation used in the book of Coecke and Kissinger [2017] – involving maximally entangled states and effects – which they relate to quantum teleportation (Bennett et al. [1993]).

Applying an appropriate twist object flips a left Hilbert object into a right Hilbert object and vice versa. For example,

$$= \quad = \quad = \quad \tag{520}$$

In the second expression we have expanded the left Hilbert object in terms of a basis and used (513). The dotted box in the third expression indicates the object that is defined as $A^H$ in the fourth expression. The $H$ stands for horizontal transpose (discussed in Sec. 29.1).

## 30.5 Ignore operators in terms of orthonormal bases

We saw in Sec. 24 that ignore operations, denoted with $\hat{I}$'s, are proportional to the identity operations, denoted with $\mathbb{1}$ (see (440)).

$$\hat{I} = \alpha^{\mathsf{a}}\hat{\mathbb{1}}^{\mathsf{a}_1} \qquad \hat{I} = \alpha_{\mathsf{a}}\hat{\mathbb{1}}_{\mathsf{a}_1} \tag{521}$$

The normalisation gauge parameters, $\alpha_{\mathsf{a}}$ and $\alpha^{\mathsf{a}}$ are chosen such that $\alpha_{\mathsf{a}}\alpha^{\mathsf{a}} = \frac{1}{N_{\mathsf{a}}}$ (see (435) in Sec. 24) which guarantees

$$\equiv \quad 1 \tag{522}$$

This is required by correspondence with (97).

Using (440) with (509, 508)

$$(523)$$

and

$$(524)$$

where the normalisation condition is

$$|\gamma^{\mathsf{a}}|^2 |\gamma_{\mathsf{a}}|^2 = \frac{1}{N_{\mathsf{a}}} \tag{525}$$

This guarantees that a deterministic circuit built from a deterministic preparation and a deterministic result has probability equal to 1. These $\gamma_{\mathsf{a}}$'s are related to the $\alpha_{\mathsf{a}}$'s by

$$\alpha^{\mathsf{a}} = |\gamma^{\mathsf{a}}|^2 \qquad \text{and} \qquad \alpha_{\mathsf{a}} = |\gamma_{\mathsf{a}}|^2 \tag{526}$$

The normalisation condition (435) on the $\alpha$'s then agrees with (525). In Sec. 33 we will have reason to impose the further constraint that $\gamma^{\mathsf{a}}\gamma_{\mathsf{a}} \in \mathbb{R}$.

## 30.6 Basis expansion of left and right Hilbert objects

We can expand any left Hilbert object or right Hilbert object in terms of a basis. This provides a way to define the horizontal adjoint. Consider

$$(527)$$

We say that these objects are horizontal adjoints of one another. We will see, in Sec. 32.2, that this definition is independent of the choice of orthonormal basis (in fact, it is also independent of whether we use an orthonormal basis or an unormalised orthogonal basis - we will see this in Sec. 33.3).

It is useful to adopt the notation $\ulcorner B$ and $B\urcorner$ for the expansion matrices,

instead of $B$ and $\overline{B}$, so that (527) can be written as

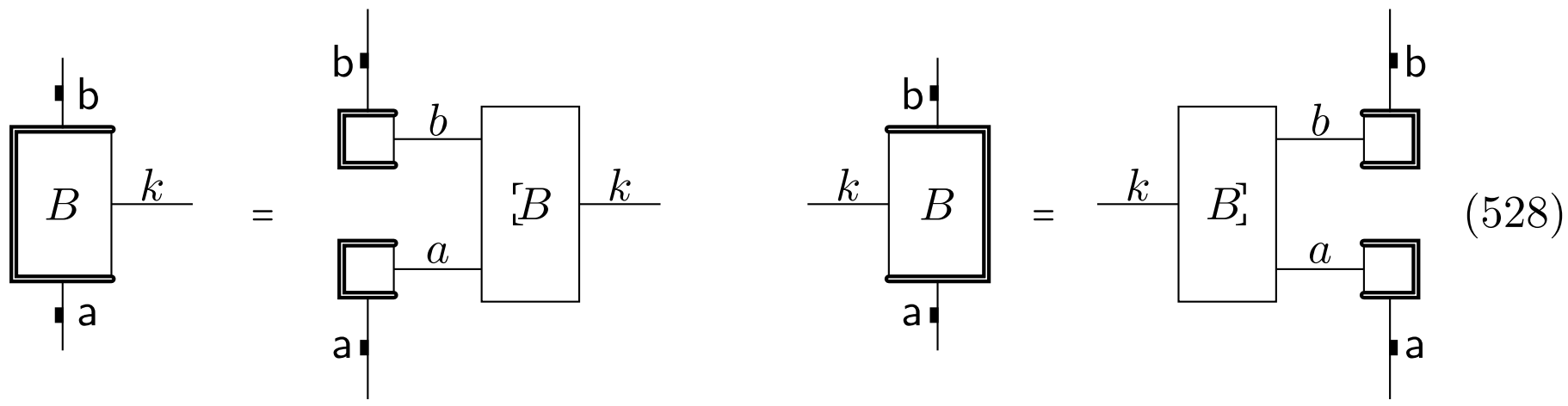

This notation better reflects the horizontal symmetry.

# 31 Operator tensors

## 31.1 Expansion in terms of left and right Hilbert objects

We have already introduced operator tensors in Sec. 16 introducing them as elements of a space determined by the wires coming out (as illustrated in the example in (364) which is an element of the space given in (365). Here we will discuss operator tensors having only inputs and outputs (no incomes or outcomes). An example of an operator tensor is

(529)

Here the $\alpha$ box represents a matrix with entries labeled by $ll'$ (recall that, in the diagrammatic notation, we label each leg by $l$ since we can see they are in different places from the diagram). We can always expand an operator tensor in terms of the orthonormal basis we have chosen as follows

(530)

We use the notation $\ulcorner C \urcorner$ for the expansion hypermatrix of an operator tensor in terms of orthonormal bases. This will help with disambiguation later when we want to expand left and right Hilbert objects using $\ulcorner C$ and $C \urcorner$ respectively.

In Sec. 5.5 we discussed how we need to identify open wires across equivalences, equations, and inequalities. We can think of each wire having an integer label where wires that correspond across an equation have the same integer label. However, rather than explicitly displaying the integer label explicitly we use the position at which the wire enters/leaves the diagram to tell us which wires correspond. Now we have introduced an extra complication as wires can be doubled up. Look, for example, at (530) above. If the $\mathsf{b}$ wire on the left side of the equation has integer label, 3, then we can label the two $\mathsf{b}$ wires on the right side of the equation as 3-left and 3-right. Note further, that on the left of the equation the output wires are in the order $\mathsf{ba}$ reading from left to right. We adopt the convention of maintaining this same order for the left Hilbert object wires but reversing this order for the right Hilbert object wires when we double up. This is, indeed, what we have done in (530) above. Given this convention, it is not actually necessary to explicitly label the wires to be able to identify corresponding wires.

## 31.2 Normal operator tensors

Consider an operator tensor $\hat{B}$ with basis expansion

(531)

In the usual terminology, we say a matrix, $M$, is *normal* iff $MM^\dagger = M^\dagger M$ (where the $\dagger$ denotes adjoint). Here we have hypermatrices (see Sec. 29.1) and so we need to use the right notion of adjoint. We will say the operator tensor, $\hat{B}$, is normal iff $\ulcorner B \lrcorner \ulcorner B \lrcorner^{\overline{H}} = \ulcorner B \lrcorner^{\overline{H}} \ulcorner B \lrcorner$ where $\overline{H}$ represents the horizontal adjoint as discussed in Sec. 29.1. Hermitian and anti-Hermitian matrices are special cases of normal matrices of particular interest to us (they will be discussed below). A standard result in the theory of matrices is that a matrix is unitarily diagonalisable if and only if it is normal (see Horn and Johnson [2013]). This means it can be written

(532)

The diamond shaped box, $\lambda$, represents a diagonal matix with eigenvalues, $\lambda_l$. These eigenvalues can, in general, be complex for a normal matrix. Now, since we are using the horizontal adjoint, we should call the unitarity property horizontal also. The $\underleftarrow{B}$ box is a horizontal unitary matrix and the $\underrightarrow{B}$ box is its horizontal adjoint (see Sec. 29.1). Explicitly, we have

$$\underleftarrow{B}^{l}_{ab} = (\underrightarrow{B}^{ab}_{l})^* \tag{533}$$

Since $\underleftarrow{B}$ is a square matrix, $l$ runs from 1 to $N_{\mathsf{a}}N_{\mathsf{b}}$. The horizontal unitarity property of $\underleftarrow{B}$ can be expressed as follows

(534)

We can write

(535)

by substituting (532) into (530) where

(536)

We will call the object on the left a *left eigenvector* and the object on the right a *right eigenvector*. The horizontal unitarity conditions in (534) are equivalent to the following properties on these left and right eigenvectors.

(537)

and

(538)

These horizontal unitarity conditions have to do with diagonalisation of a normal matrix (which is defined with respect to the horizontal adjoint) as discussed above. They are different from the unitary conditions on $U$ to be discussed in Sec. 37 (which have to do with the vertical adjoint). Condition (538) can be interpreted as the condition that these eigenvectors are orthonormal.

We will have cause to consider left Hilbert circuits containing the left eigenvector, $\underleftarrow{B}$. In particular, we can show that there exists a choice for $D$ and $E[l]$ such that

(539)

where the box with an $l$ in it on the right indicates selecting on this particular value of $l$ and $\nu$ is a constant (which we include so $D$ and $E$ can correspond to physical operators). This is true if we put

(540)

and

(541)

where $\alpha$ and $\beta$ are constants such that $\alpha\beta = \nu$ (where $\nu$ is the constant in (539)). Substituting (540) and (541) into the left hand side of (539), using one of the horizontal unitarity properties for $\underleftarrow{B}$ in (534), and using the basis orthogonality relations (500) gives the right hand side of (539). It is worth noting that $E[l]$, as given in (540), can be defined in terms of the right eigenvector, $\underrightarrow{B}$, as follows

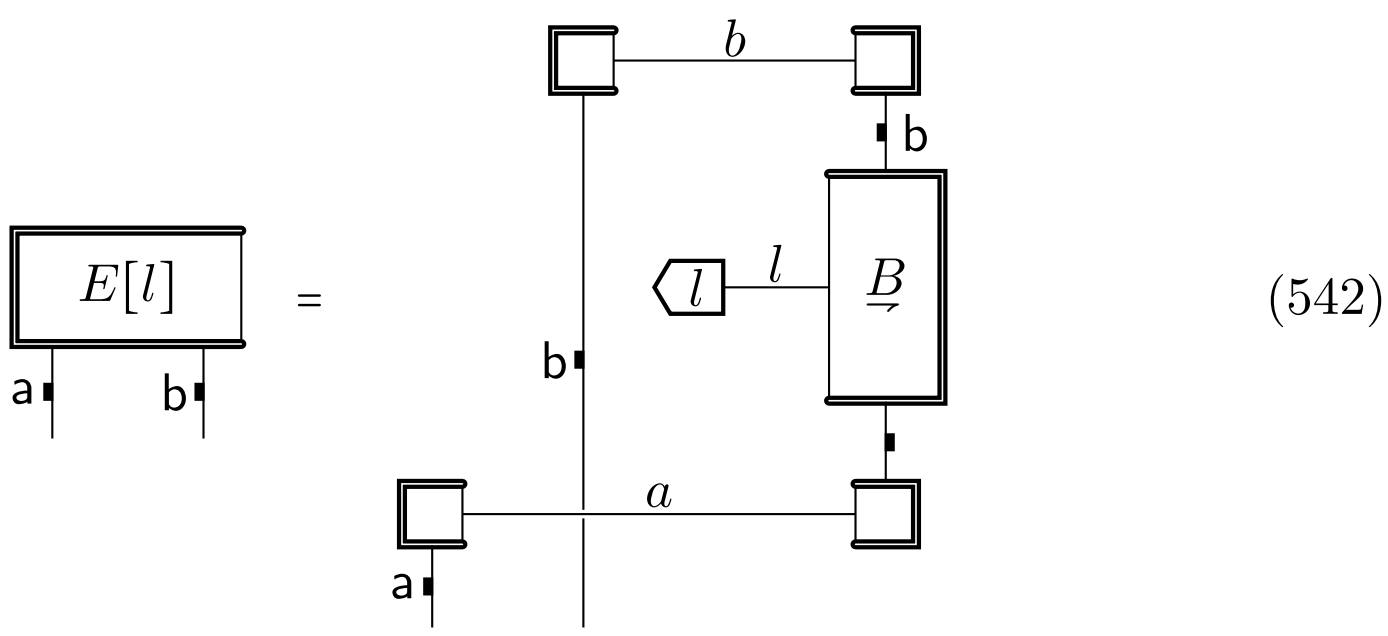

(542)

If we do this then the result in (539) follows using (538).

## 31.3 Hermitian operator tensors

Of particular interest are *Hermitian operator tensors*. These have the property that the orthonormal basis expansion matrix, $[B]$, is Hermitian - i.e. it satisfies $[B] = [B]^{\overline{H}}$ (where $\overline{H}$ indicates the horizontal adjoint as discussed in Sec. 29.1). It follows from this defining property that (i) Hermitian matrices, $[B]$, are normal and (ii) that $[B]$ has real eigenvalues. Hence, we can write

$$\hat{B}\,{}^{\mathsf{b}}_{\mathsf{a}} = \underleftarrow{B}\,{}^{\mathsf{b}}_{\mathsf{a}} \overset{l}{-} \lambda \overset{l}{-} \underrightarrow{B}\,{}^{\mathsf{b}}_{\mathsf{a}} \tag{543}$$

where, now, the eigenvalues $\lambda_l$ are real.

## 31.4 Anti-Hermitian operators

As we noted in Sec. 20, any operator can be written as the sum of a Hermitian and an anti-Hermitian operator.

An anti-Hermitian operator tensor is one for which the orthonormal bais expansion matrix, $B$, is anti-Hermtian. Anti-Hermitian matrices (also called "skew" matrices) are matrices having the property that $B = -B^{\dagger}$ (where $\dagger$ is the horizontal adjoint). It follows immediately from this defining property that (i) the matrix, $B$, is normal, and (ii) the eigenvalues of $B$ are pure imaginary or zero. Consequently, we can put anti-Hermitian operator tensors in the form

$$\hat{B}\,{}^{\mathsf{b}}_{\mathsf{a}} = \underleftarrow{B}\,{}^{\mathsf{b}}_{\mathsf{a}} \overset{l}{-} i\mu \overset{l}{-} \underrightarrow{B}\,{}^{\mathsf{b}}_{\mathsf{a}} \tag{544}$$

where the $\mu_l$ are real.

The operator tensors in Quantum Theory are Hermitian. However, we need to consider anti-Hermitian operator tensors to prove that circuit reality implies Hermiticity (see Sec. 20 and Sec. 42.3).

## 31.5 Twofold operators

Twofold operators are ones which can be written as

(545)

where

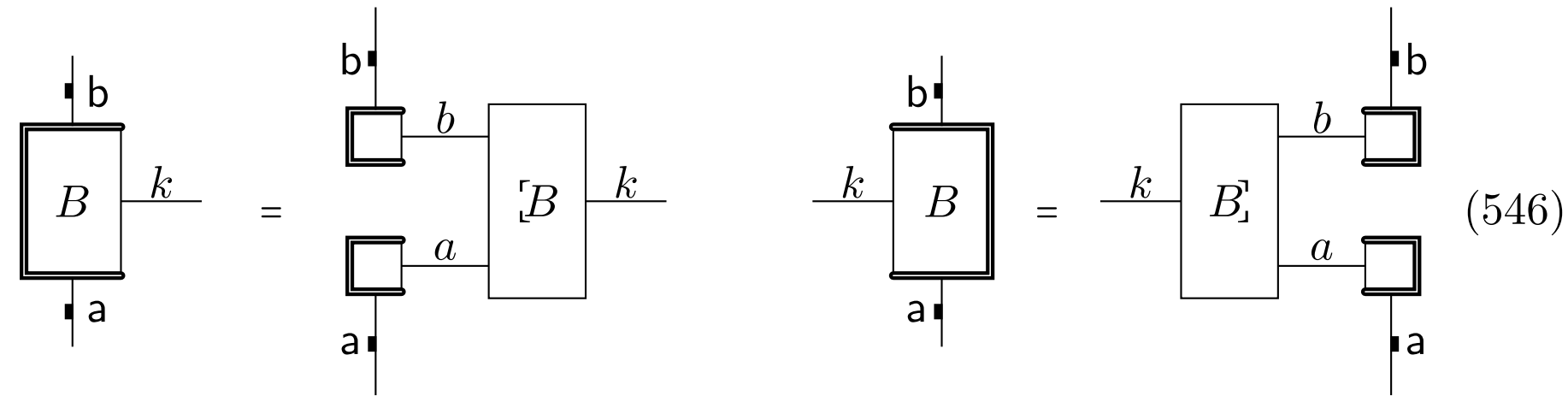

(546)

such that the matrices ${}^{\ulcorner}_{\llcorner}B$ and $B^{\urcorner}_{\lrcorner}$ are horizontal adjoints. If we substitute (546) into (545) we see that the expansion matrix for $\hat{B}$ is

(547)

It is easy to see that ${}^{\ulcorner}_{\llcorner}B^{\urcorner}_{\lrcorner}$, when given in this way, is Hermitian with nonnegative eigenvalues. First note that if we apply (490) to the right hand side of (547) get back the same object. Consequently, ${}^{\ulcorner}_{\llcorner}B^{\urcorner}_{\lrcorner} = \overline{{}^{\ulcorner}_{\llcorner}B^{\urcorner}_{\lrcorner}}^{H}$ so ${}^{\ulcorner}_{\llcorner}B^{\urcorner}_{\lrcorner}$ is Hermitian. Further, note that

$\geq 0$ for all $V$ (548)

Since $B^{\urcorner}_{\lrcorner}$ is the left adjoint of ${}^{\ulcorner}_{\llcorner}B$, the left hand side is a sum of the form $\sum_l A_l \overline{A}_l$ and hence is nonnegative. Thus, ${}^{\ulcorner}_{\llcorner}B^{\urcorner}_{\lrcorner}$ has positive eigenvalues.

For any given operator, $\hat{B}$, the twofold form is not unique. There are two useful theorems concerning this which we will now discuss.

First we have the following theorem.

**Unitary freedom in twofold representation.** We have

(549)

where $k = 1$ to $K$ and $m = 1$ to $M$, and we consider the case where $K \leq M$, if and only if

(550)

where $U$ is a horizontal unitary.

Note we have used the "cut diamond" padding matrix notation (introduced in (493) whose elements are equal to 1 for $k = m$ and 0 otherwise. This theorem is equivalent to the standard theorem in Quantum Theory - that we can find a unitary transformation between sets of Kraus operators representing the same completely positive map (after padding the set of smaller rank with zeros). See, for example, Nielsen and Chuang Nielsen and Chuang [2000] (Theorem 8.2 therein). We will leave the proof of this theorem to Sec. 68.7 where we consider operators with complex causal structure since the notation we adopt there makes this proof more straightforward. In Sec. 68.7 we also prove two preliminary theorems and a corollary concerning isometric and unitary maps between a general decomposition and one built out of orthogonal Hilbert objects. These are used to prove the above theorem. Those theorems are also applicable here.

Before stating the second theorem we give the following definition

**The rank of twofold operator.** The rank of a twofold operator tensor

(551)

is the minimum size of the set $\{l\}$ over all such twofold decompositions. We will write this as $\text{rank}(\hat{B})$.

Note that this definition only works for operators with no income or outcome wires (to the sides). We will extend the definition of rank for the case where there are incomes and outcomes by using the maximal representation in Sec. 44.2.

We can state the following theorem

**Minmax theorem for twofold representation.** The rank of a

twofold operator

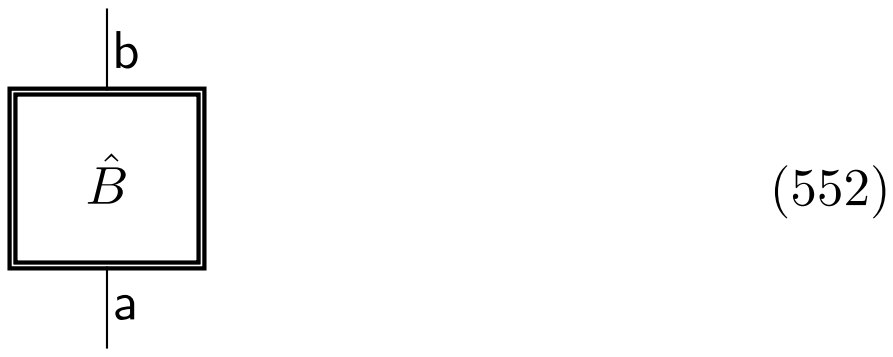

(552)

is equal to the number of nonzero eigenvalues of the expansion matrix, $\ulcorner B \lrcorner$, and satisfies $\text{rank}(\hat{B}) \leq N_{\mathsf{a}}N_{\mathsf{b}}$. Further, there exist twofold operators which necessarily saturate this upper bound.

We can write $\hat{B}$ in the form in (551). Let us assume that $l = 1$ to $L$. By definition, $\text{rank}(\hat{B}) \leq L$. Each term in this sum is a rank one operator (and therefore Hermitian). The $L$ terms taken together span a subspace of Hermitian operators. Let this subspace have dimention $K$. We can equally span this subspace by $K$ Hermitian operators built from a set of orthonormal Hilbert objects labeled by $k = 1$ to $K$. These can be regarded as the subset of the eigenvectors introduced in the discussion of normal operator tensors in Sec. 31.2. If we weight each of these eigenvectors by $\sqrt{\lambda_k}$ (these eigenvalues are nonnegative in the twofold positive case) then we obtain back $\hat{B}$. Clearly, in such a construction, we always have $K \leq L$ and, therefore the rank is equal to the number of nonzero eigenvalues. To prove $\text{rank}(\hat{B}) \leq N_{\mathsf{a}}N_{\mathsf{b}}$ we note that $\hat{B}$ is Hermitian with nonnegative eigenvalues. Since Hermitian operators are normal, we can write them in the form in (543). The $\underleftarrow{B}$ and $\underrightarrow{B}$ matrices are unitary and therefore square. Hence, as noted in Sec. 31.2, $l$ runs from 1 to $N_{\mathsf{a}}N_{\mathsf{b}}$ and so $L = N_{\mathsf{a}}N_{\mathsf{b}}$. Since we have $\lambda_l \geq 0$ we can absorb these eigenvalues (as $\sqrt{\lambda_l}$'s) into the hypermatrices $\underleftarrow{B}$ and $\underrightarrow{B}$ whilst maintaining the property that the latter are horizontal adjoints and so we are in the form in (551) with $L = N_{\mathsf{a}}N_{\mathsf{b}}$. Clearly, some of the eigenvalues could be zero and so we have $\text{rank}(\hat{B}) \leq N_{\mathsf{a}}N_{\mathsf{b}}$. On the other hand, if all the eigenvalues are nonzero then we cannot have $\text{rank}(\hat{B}) < N_{\mathsf{a}}N_{\mathsf{b}}$ and so we necessarily saturate the upper bound. An example of the latter is $\hat{B}_{\mathsf{a}}^{\mathsf{b}} = \hat{\boldsymbol{I}}_{\mathsf{a}}\hat{\boldsymbol{I}}^{\mathsf{b}}$).

It is worth noting that, although we can always write the operator above in twofold form with $L \leq N_{\mathsf{a}}N_{\mathsf{b}}$, we can also write down decompositions where $L$ is greater than $N_{\mathsf{a}}N_{\mathsf{b}}$. We can do this by padding the sets of left and right Hilbert objects used in the decomposition with zeros, or we can simply have sets of left and right Hilbert objects that are linearly dependent.

Since the rank is less than or equal to $N_{\mathsf{a}}N_{\mathsf{b}}$ we can always write

$$\hat{B}_{\mathsf{a}}^{\mathsf{b}} = B_{\mathsf{a}}^{\mathsf{b}\,ab} \; B_{ab\,\mathsf{a}}^{\mathsf{b}} \qquad (553)$$

where we have replaced the $l$ label wire by $a$ and $b$ label wires. Then we have

$$(554)$$

These objects are *horizontal adjoints* of one another. This form is quite natural and we will use it to illustrate taking what we will call *normal conjupositions* of left and right objects in what follows.

We will say that operator tensors that can be written in twofold form (as in (551, 553)) satisfy *twofold positivity*. We will see in Sec. 43 that this is equivalent to the notion of tester positivity discussed in Sec. 26 when there are no incomes or outcomes (the case where there are incomes and outcomes is easily dealt with there). This notion of twofold positivity is essentially the same as the notion of ⊗-positivity in Coecke and Kissinger [2017]. This concept was introduced, in a diagrammatic setting, by Selinger Selinger [2007] though is really just the representation of Kraus [1971]. A notational difference is that here we use the horizontal label wires (labeled $a$ and $b$) to indicate the sum rather than a cup or cap. We can introduce cups and/or caps using the equations

$$(555)$$

(with similar equations for $b$). These equations are obtained through the definitions in (508), (509), and the orthonormality equations (500).

## 31.6 Homogeneous operator tensors

We call any operator tensor that can be written in the form

$$(556)$$

*homogeneous*. Homogeneous operators are necessarily twofold operators. Twofold operators that cannot be written in this form are called *heterogeneous* since then they must be written as a sum of homogeneous operator tensors that are not proportional to one another (as in (551) where $l$ runs over at least two values). Clearly homogeneous operators are have rank equal to 1 whereas heterogeneous operators have rank greater than 1.

The terminology *homogeneous* was introduced in Sec. 7.8 for preparations and results in contradistinction to heterogeneous preparations and results. A heterogeneous preparation (or result) is one that can be written as the sum of distinct non-parallel preparations (results). A homogeneous preparation (or result) is one that is not heterogeneous. In the case of operator tensors corresponding to preparations and results homogeneous operators are, indeed, in the form in (556) (this is clear from the discussion in Sec. 16.1).

Maximal states and maximal effects are clearly homogeneous because they are pure (as proven in Sec. 25). Unitary operator tensors to be discussed in Sec. 37.3 are also homogeneous.

# 32 Normal conjupositions of Hilbert objects

## 32.1 Eight normal conjupositions

Here we describe the *normal conjuposition group* which acts on Hilbert objects. This is obtained by expanding the Hilbert object using orthonormal bases and using the conjuposition group on the expansion hypermatrix. We call this the "normal" conjuposition group because we use orthonormal bases. Though the word "normal" also has the connotation of indicating that this is the usual way things are done - this is true in that usual notions of adjoint and transpose, as used in standard Quantum Theory, are to be found amongst the normal conjuposition group. We will see later that the normal conjuposition group is not quite the right symmetry group for time symmetric Quantum Theory. That honour belongs to the *natural conjuposition group* (introduced in Sec. 33 which is based on orthogonal (but not normal) bases chosen to reflect the underlying physics.

It is convenient to illustrate our remarks using the expansion introduced in (554) (as these have $ab$ label wires whose behaviour under conjupositions is interesting). Since there are eight conjupositions acting on the expansion hypermatrices, we obtain eight Hilbert objects (four left and four right) starting with any given left or right Hilbert object. Written down explicitly, these are

(557)

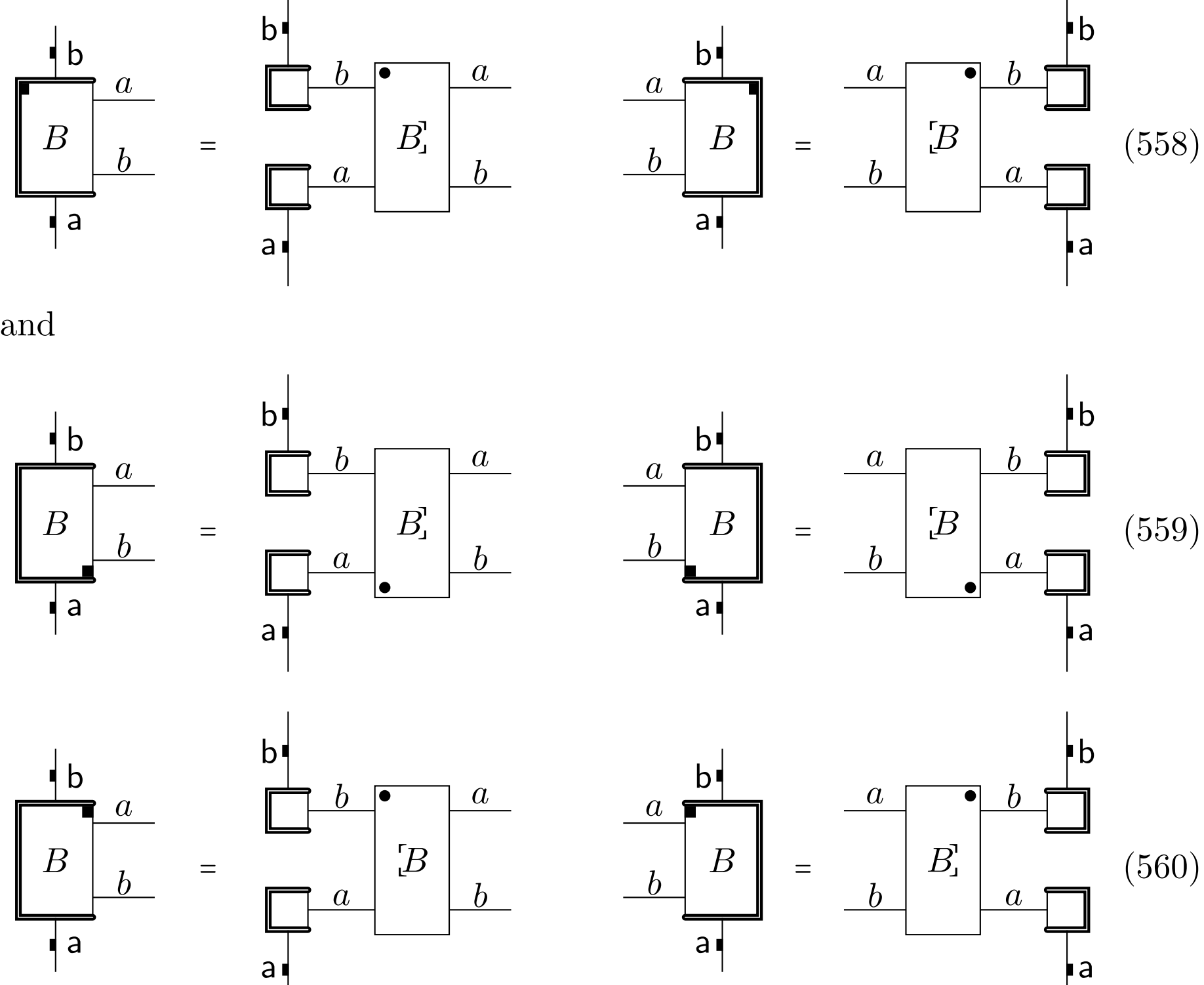

We have included small black squares on the Hilbert objects. These will help us track normal conjupositions. Two hypermatrices with ${}^{\ulcorner}_{\llcorner}B$ and $B^{\urcorner}_{\lrcorner}$ and having the small circle dot in the same position are conjugates. Thus ${}^{\ulcorner}_{\llcorner}B$ and $B^{\urcorner}_{\lrcorner}$ play the same role as $B$ and $\overline{B}$ in the notation of Sec. 29.2. In going from the left Hilbert object in (557) to the left Hilbert object in (558) we are taking the vertical adjoint of the expansion matrix. Thus we call this transformation between Hilbert objects the *normal vertical adjoint*. In going from the left Hilbert object in (557) to the right Hilbert object (557) we are taking the horizontal adjoint of the expansion matrix and so we call this transformation between Hilbert objects the *normal horizontal adjoint*. It turns out that the normal horizontal adjoint is equal to horizontal adjoint taken with respect to orthogonal bases that are not normal and so, in this particular case, we can call this the *horizontal adjoint*. A similar statement is not true for the other normal conjupositions.

We will denote the normal conjupositions by the same symbols as we used for the conjuposition group (i.e. $\{I, \overline{I}, H, \overline{H}, V, \overline{V}, T, \overline{T}\}$).

We will begin by considering each normal conjuposition in the subgroup, $\mathcal{S}_{BI} = \{I, \overline{H}, \overline{V}, T\}$. The effect of each of these conjuposition operations turns out to be basis independent (as we will prove).

## 32.2 The normal horizontal adjoint of Hilbert objects

Consider again the left and right Hilbert objects in (557).

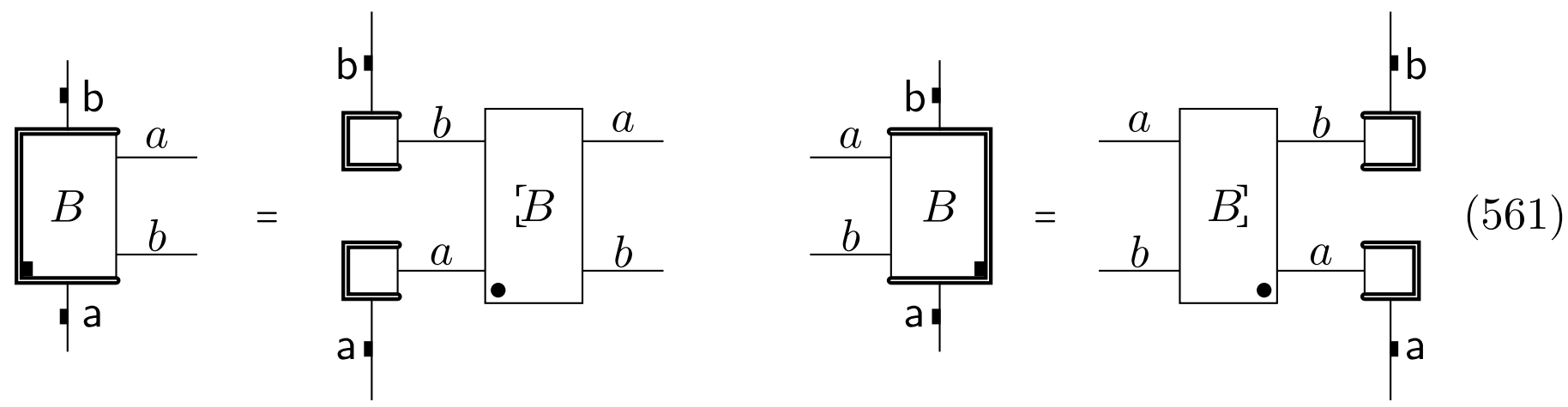

As pointed out above, these are normal horizontal adjoints. The notation is such that, if we flip the diagram over horizontally, we take the normal horizontal adjoint. We will see below that flipping it vertically takes the normal vertical adjoint.

Now we make an important observation. The horizontal adjoint is basis independent. To see this note that, using the method in Sec. 30.1, we can transform our expansion of a left object to a new basis by applying unitaries, $U$ and $V$, as follows

(562)

Using (490) we see that the horizontal adjoint of

(563)

is

(564)

Thus, the normal horizontal adjoint of the left Hilbert object in (562) can be obtained by taking the horizontal adjoint of the expansion matrix in the shaded

basis and gives

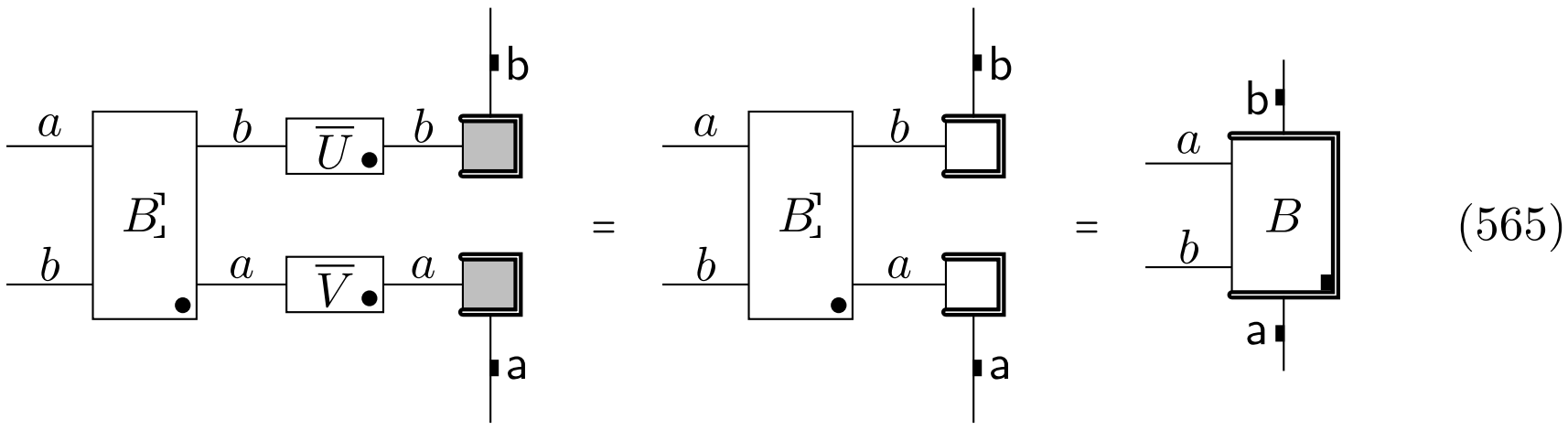

In the second expression we have undone the effect of the unitaries to give back the original basis. Thus, we see that taking the normal horizontal adjoint of a left Hilbert object is independent of the basis we use. The same is true, of course, for taking the normal horizontal adjoint of a right Hilbert object. Note that when we take the horizontal adjoint $U$ and $V$ are transformed to $U^{\overline{H}}$ and $V^{\overline{H}}$ respectively. This is essential for the above proof to work since this allows us to undo the effect of the unitaries and get back the original (unshaded) bases (see (501) where we saw that, if left bases transform under $U$, then right bases transform under $U^{\overline{H}}$).

It is worth noting here that the normal horizontal transpose is not basis independent. We could try to follow through the same reasoning as above but with the horizontal transpose (rather than horizontal adjoint) to attempt to prove that this is also basis independent. This proof is blocked because we get $U^H$ and $V^H$ after the horizontal transpose but we need $U^{\overline{H}}$ and $V^{\overline{H}}$ to transform back to the unshaded basis element.

## 32.3 The normal transpose of Hilbert objects

The normal transpose can be defined in terms of conjuposing expansion hypermatrices. Alternatively, we can define it using caps and cups. We will show these two definitions are equivalent.

First consider the transpose where we suppress the $ab$ labels on the side. The normal transpose of a left Hilbert object returns a right Hilbert object and can be written as follows

(566)

The usefulness of the small square dots is apparent here. With this square dot notation, we can "derive the equation above by "sliding" the box along the wire then using the yanking equation (the right equation in (511)) to straighten out the wire. This *sliding manoeuvre* is discussed in Coecke and Kissinger [2017].

We can, of course, take the normal transpose of a right Hilbert object in the same way. For example,

(567)

The sliding manoeuvre works in this case as well.

If we include the *ab* label wires then the definition of the normal transpose of a left object is as follows:

(568)

We can see that the sliding interpretation still works. As we slide the left object over, the *ab* wires become untwisted. Note that, in this work, there is no mathematical meaning to whether a wire passes over or under another wire. Hence, we can allow wires to "pass through" one another as they untwist.

Next we want to prove equivalence of this definition of the normal transpose to the definition in terms of expansion matrices. First we note the normal transpose of the elements in a basis set are as follows

(569)

(570)

These are obtained using (508), (509) and the orthonormality relations (500). Now we can apply the definition of the normal transpose to the expanded form

giving

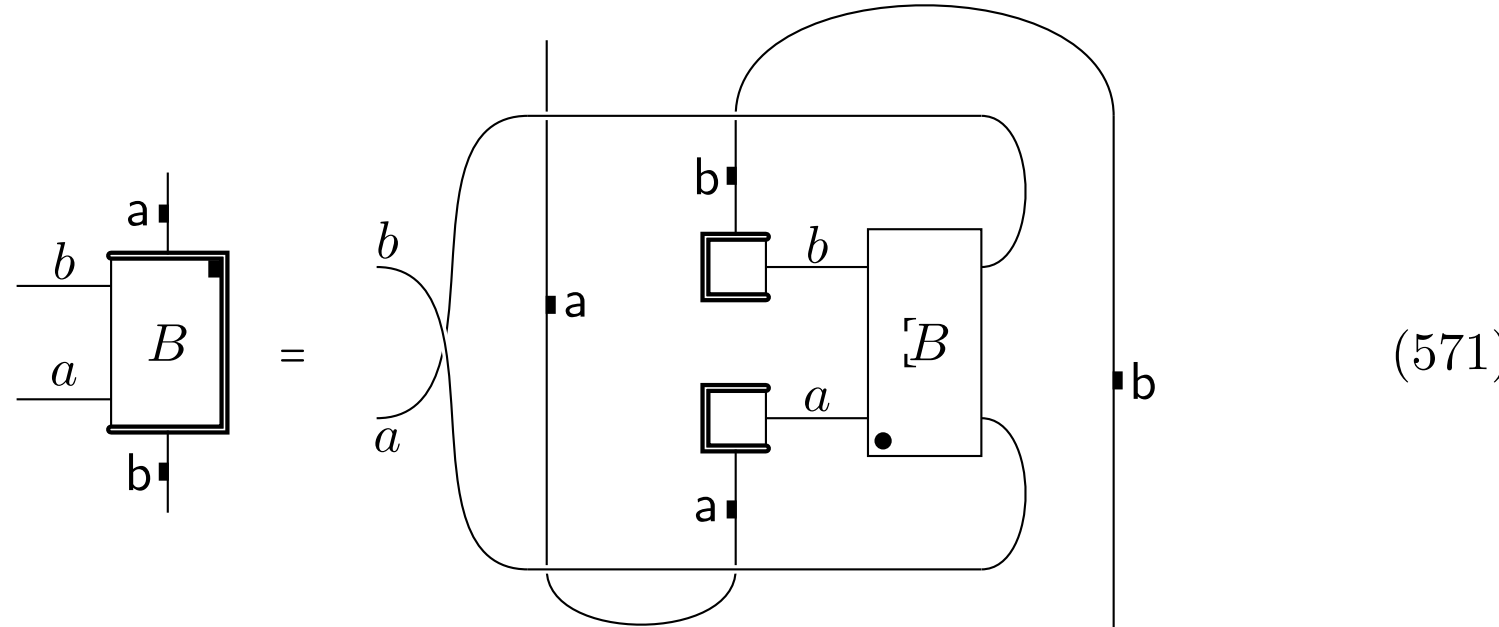

(571)

Using (569) and (570) and the definition of the transpose of a hypermatrix in (477), we obtain

(572)

Note the position of the circular dot in (572) compared with (571). This shows that the normal transpose of a left object returns a right object whose expansion matrix is the transpose of the expansion matrix of the left object we started with. Note, in particular, that we have ${}^{\ulcorner}_{\llcorner}B$ both before (in (571)) and after (in (572)) indicating that we are not taking the conjugate of the elements.

We can, similarly, take the normal transpose of a right object obtaining a left object with transposed expansion matrix. The normal transpose of the right object in (561) is

(573)

Note that we have $B^{\urcorner}_{\lrcorner}$ both before and after the transformation as expected (but that the circular dot has moved).

It is important to note that the normal transpose of a left (or right) Hilbert object is basis independent (as long as we are transforming between orthonormal bases). This is clear from the definition in (566) (or (568)) since the cup and cap objects used (originally defined in (509) and (508)) are basis independent as noted in Sec. 30.2.

It is worth mentioning a simple derivation of the sliding manoeuvre. By putting a cap on (566) (the definition of the transpose) then applying the yanking equation (511) gives the left equation below

(574)

This left equation can be viewed as sliding the $B$ left object over so it becomes a right object. The equation on the right is obtained by applying a cup to the definition of $\mathsf{C}^T$ and using the yanking equation. These two equations capture the basic sliding manoeuvre.

## 32.4 The normal vertical adjoint of Hilbert objects

For hypermatrices we can obtain the vertical adjoint, $\overline{V}$, by performing $T\overline{H}$ (see Sec. 29.1). Thus we want this to be true for Hilbert objects as well. We have already seen how to do the normal horizontal adjoint (see Sec. 32.2) and normal transpose (see Sec. 32.3) of Hilbert objects so we can obtain the normal vertical adjoint. This goes as follows

normal horizontal adjoint → normal transpose → (575)

Thus, we see that, diagrammatically, the normal vertical adjoint is obtained by vertically flipping the diagram. Comparing (572) with the left object in (561) we see that, when we take the normal vertical adjoint of a left object, we take the vertical adjoint of the expansion matrix.

The normal vertical adjoint is basis independent. This follows from the fact that we can obtain the normal vertical adjoint by performing the normal horizontal adjoint then the normal transpose, each of which are basis independent.

## 32.5 The Hilbert Square

The normal horizontal adjoint is obtained by flipping horizontally in this diagrammatic representation. Rotating through 180° takes the normal transpose. Flipping vertically takes the normal vertical adjoint. This can be illustrated as

follows

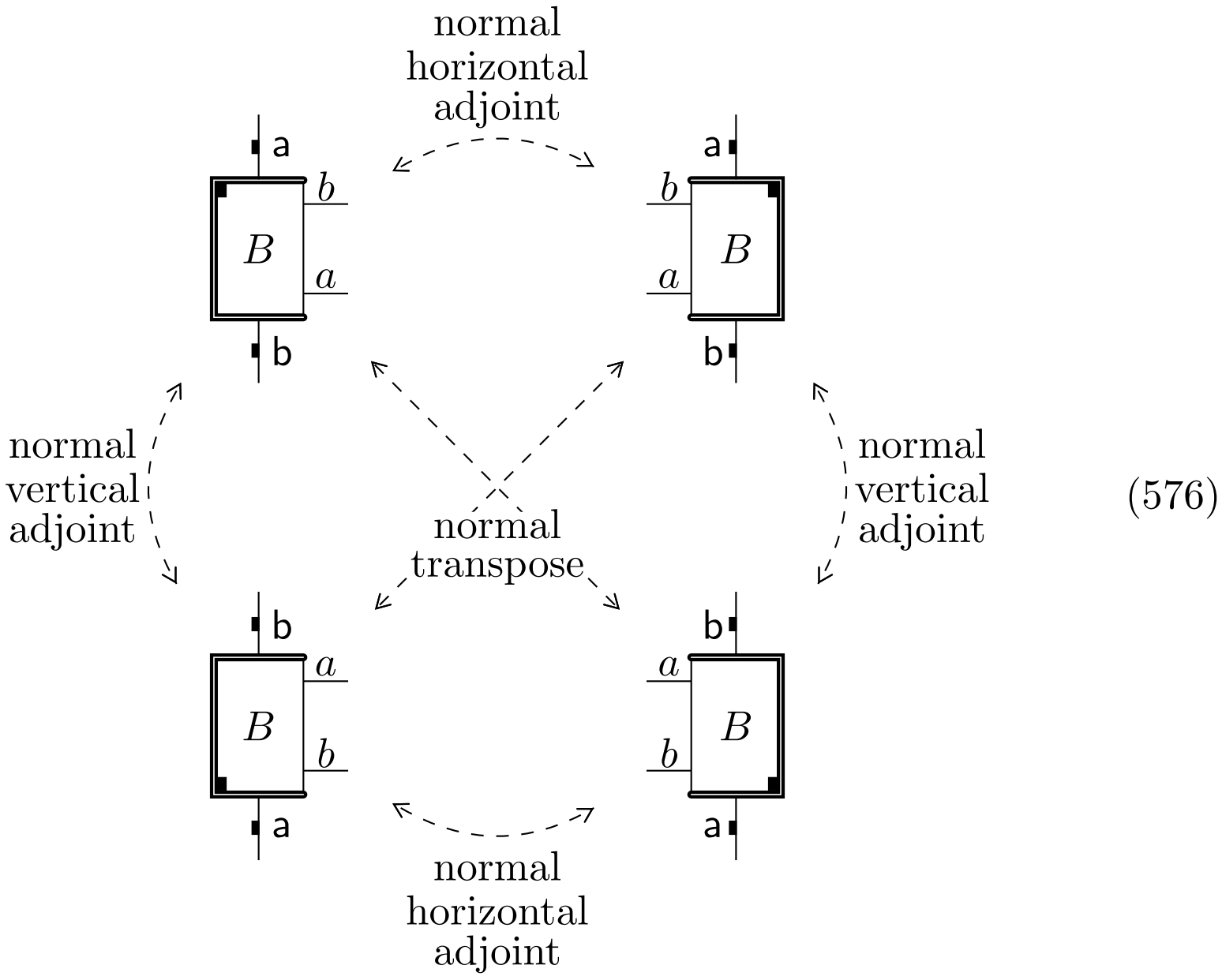

(576)

We call this *the Hilbert square*. The transformations between these objects is basis independent. The underlying group of transformations is $\mathcal{S}_{BI} = \{I, \overline{H}, \overline{V}, T\}$ which is a subgroup of the normal conjuposition group.

For completeness, here are the basis expansions of the four Hilbert objects in the Hilbert square.

(577)

(578)

We can verify that the expansion matrices are related by the appropriate elements of $\mathcal{S}_{BI} = \{I, \overline{H}, \overline{V}, \overline{T}\}$.

The Hilbert square (576) above is motivated by a similar looking diagram in the book of Coecke and Kissinger [2017] – page 110 therein. However, the square they consider is not basis independent. Rather, it is generated by the subgroup $\mathcal{S}_{\text{side}} = \{I, \overline{I}, V, \overline{V}\}$. We will see that the square considered by Coecke and Kissinger appears on the left (and right) side of the Hilbert cube in Sec. 32.8. We will discuss in more detail the connection between the approach discussed by Coecke and Kissinger and the approach in this book in Sec. 32.9 so far as these squares are concerned.

## 32.6 The normal horizontal and vertical transposes of Hilbert objects

In Sec. 30.4 we defined the twist objects using orthonormal bases. We can use twist objectss to transform a left Hilbert object into a right Hilbert object as follows

(579)

This equation can serve as the definition of the normal horizontal transpose (as we will see). We indicate the new object by the placement of the small black square dot (note this position was not used in the Hilbert square (576)). Since the twist objectss are basis dependent, this transformation is basis dependent. We can see, by expanding out in a basis, that this transformation is the normal horizontal transpose. The expression on the right in (579) is

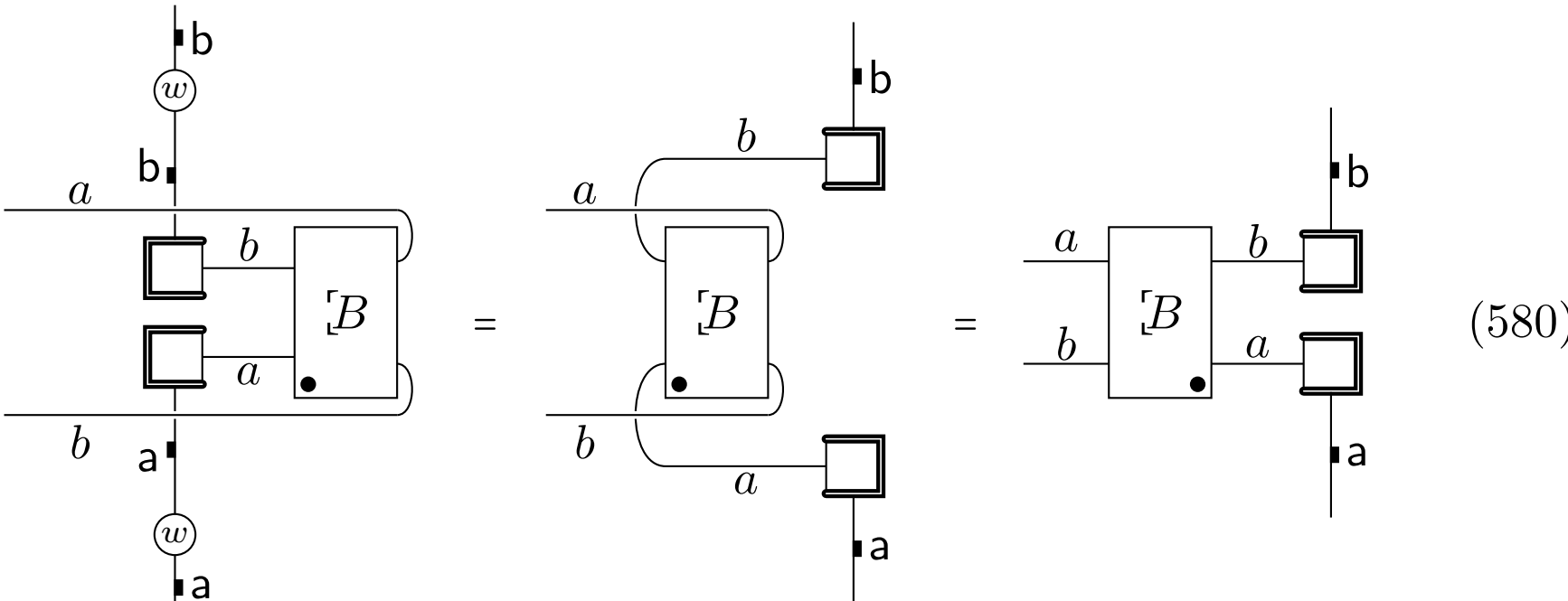

where we have used (513) in the first step and (475) in the second step. This is, indeed, the normal horizontal transpose, $H$.

We can use the cup and cap twist objects to implement another transformation as follows

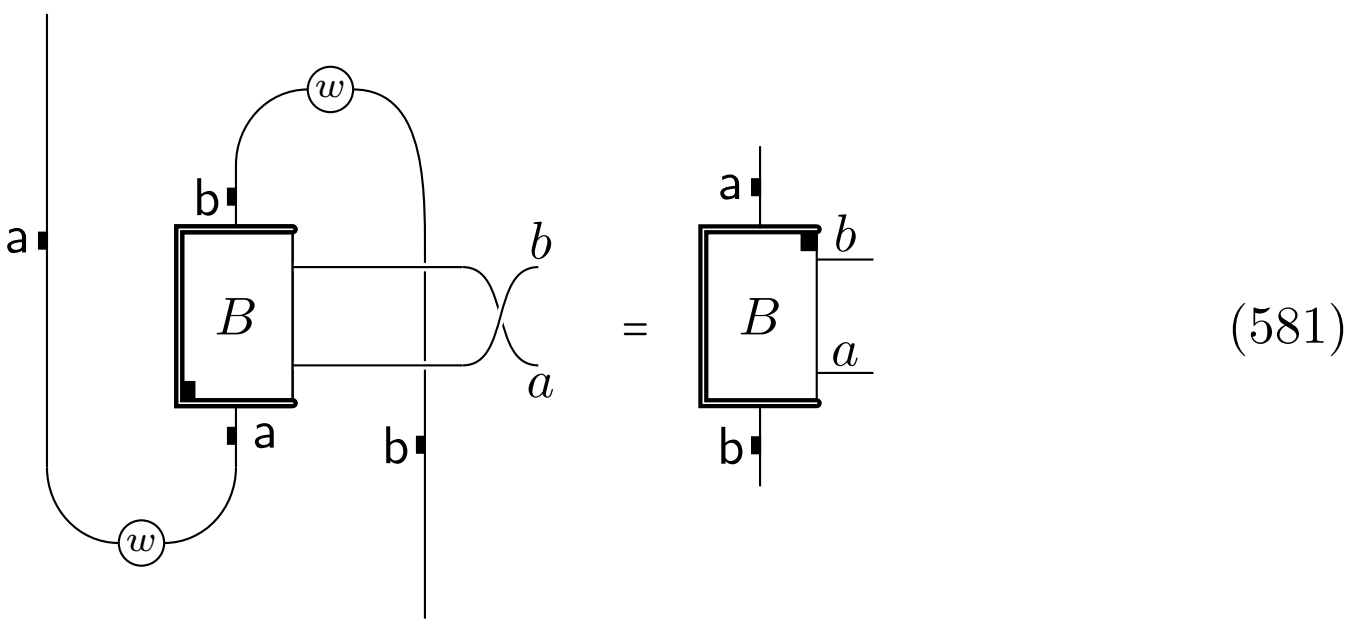

(581)

It can easily be verified, by expanding using bases, that this is the normal vertical transpose.

The normal horizontal and vertical transposes are nonconjugating. They generate the subgroup

$$\mathcal{S}_{\text{nonconjugating}} = \{I, H, V, T\} \tag{582}$$

This group is important because each of these transformations can be implemented on a left or right Hilbert object by using caps, cups, and twists (they can also be implemented on general Hilbert objects). If we have a composite Hilbert object then we can we can use these caps/cups/twists to "get at" just one of the components and implement a nonconjugating conjuposition on that component alone. This allows us to transform between equivalent equations. This is discussed in Sec. 32.13.

## 32.7 Shadow Hilbert squares

If we choose some bases (for a and b) then we can define the twist object and use it to perform the normal horizontal transpose on each element in a Hilbert square. This will give us four more Hilbert objects which form their own square:

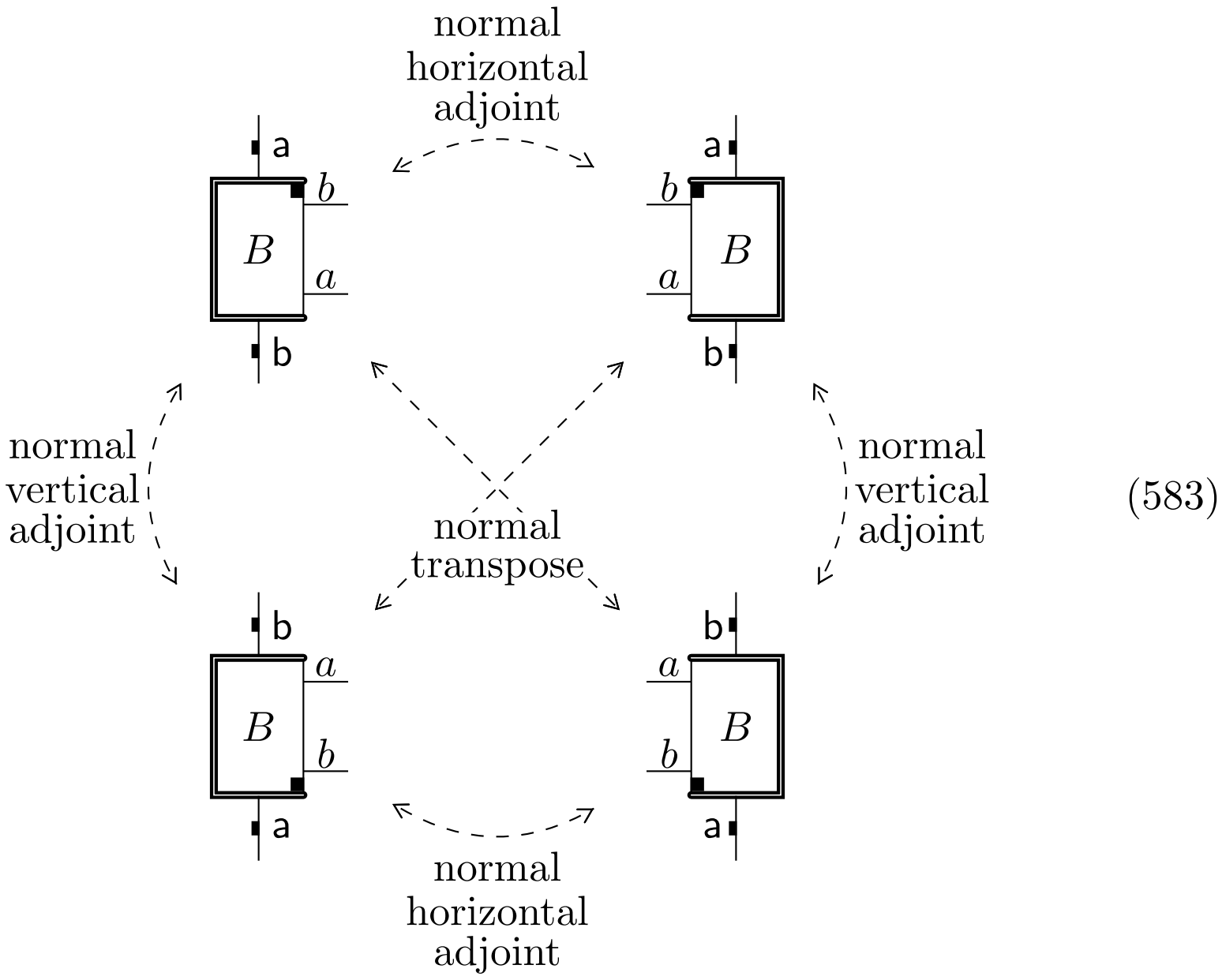

We will say this is the *shadow* of the original Hilbert square - we call it the *shadow Hilbert square* with respect to the bases choice we have made. The shadow Hilbert square is, in fact, a Hilbert square in its own right. If we choose a different basis then we will get a different shadow.

The back face of the Hilbert cube to be discussed in Sec. 32.8 is chosen to be the shadow (with respect to some bases choice) of the front face (which is our starting Hilbert square). For completeness we display here the basis expansions of the four elements of the shadow Hilbert cube.

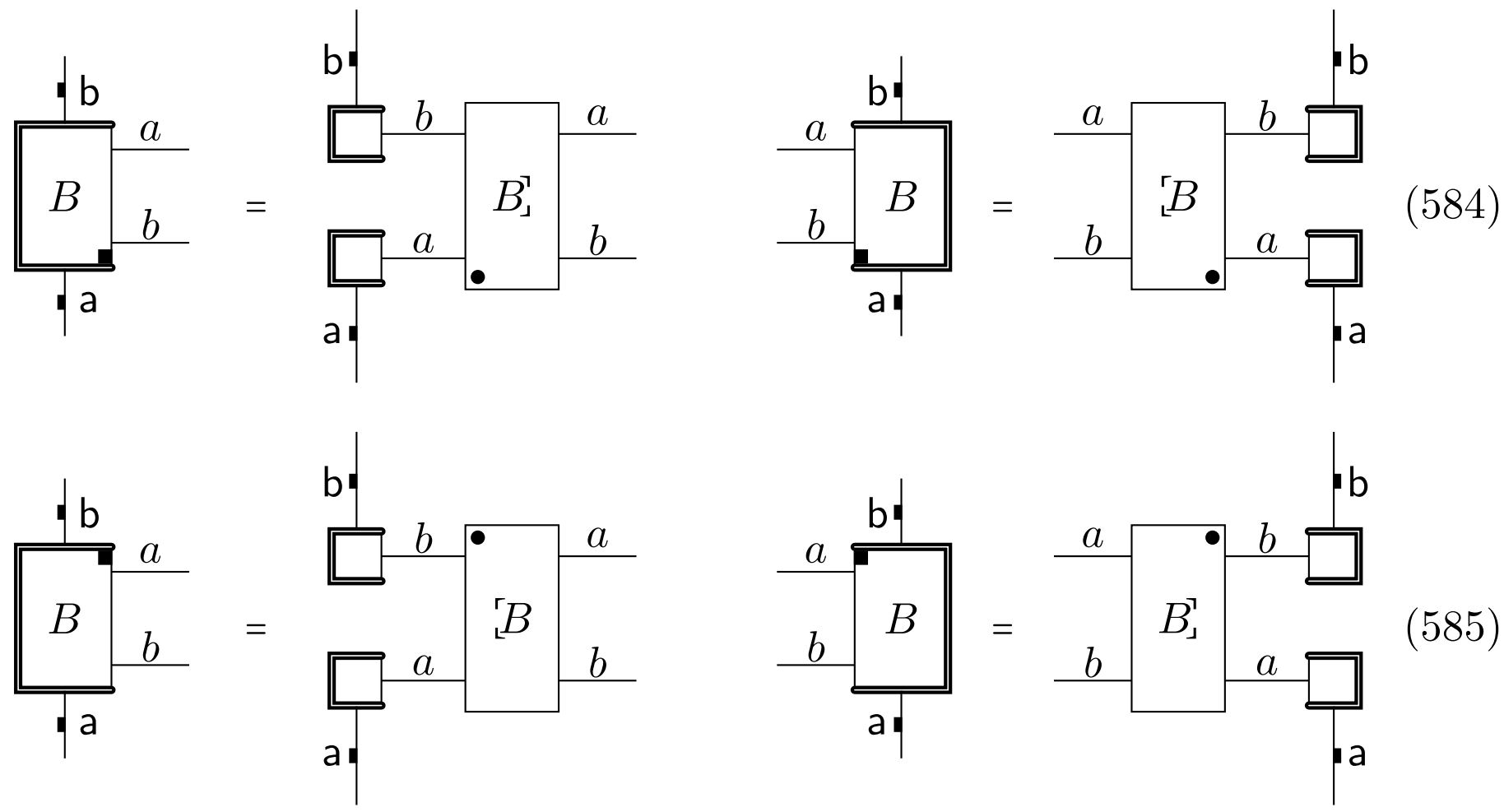

Observe how the $H$ transformation in (579) takes us from the left object in (577) to the right object in (584). Note that all eight expansion hypermatrices obtained under the conjuposition group $\mathcal{C} = \{I, \overline{I}, H, \overline{H}, V, \overline{V}, T, \overline{T}\}$ appear in (577, 578) and (584, 585). To see this compare the expansion hypermatrices in these eight equations above with the expansion matrices in (478-481) bearing in mind that the conjugate represented by $M$ or $\overline{M}$ there is represented by ${}^{\ulcorner}_{\llcorner}B$ or $B^{\urcorner}_{\lrcorner}$ here.

## 32.8 The Hilbert cube

If we start with a Hilbert square (as in (576)) and choose a particular orthonormal basis then we can generate the shadow Hilbert square (as in (583)). The resulting objects (whose expansions appear in (577, 578, 584, 585) can be arranged to form a cube as follows

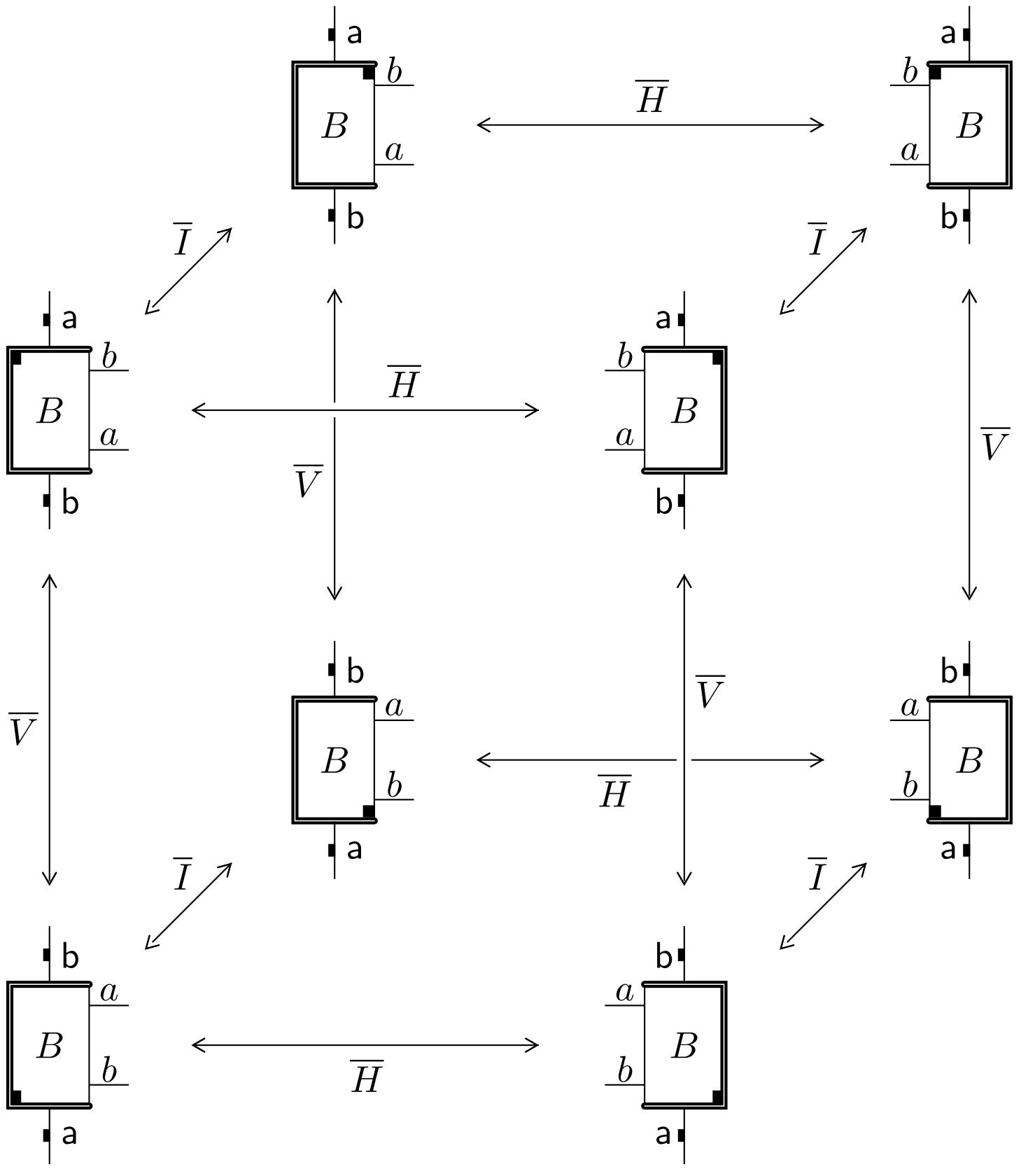

(586)

We have not shown the transformations associated with the diagonals to prevent overcrowding in the diagram. There are two diagonals in each of the six faces plus four interior diagonals. Here are some comments.

1. The cube is generated by the operations in the normal conjuposition group

$$\mathcal{C} = \{I, \overline{I}, H, \overline{H}, V, \overline{V}, T, \overline{T}\}$$

   Every edge and every diagonal (including the interior diagonals) is labeled with one of these group elements.

2. We have labelled explicitly the operations associated with the edges. This is enough to deduce all the other operations (associated with the various diagonals) since the edges provide a path between any pair of Hilbert objects.

3. The operations associated with the diagonals of the top and bottom faces are the horizontal transpose, $H$, as was shown above (see (579) and (580)). This allows us to deduce that the operations associated with the edges that connect the front and back squares is the conjugate, $\overline{I}$, as labeled (since $\overline{H}H = \overline{I}$). This is a basis dependent transformation.

4. We can think of the $\overline{H}$, $\overline{V}$, and $\overline{I}$ operations as inducing movement along the $x$, $y$, and $y$ axes respectively where $x$ is horizontal, $y$ is vertical, and $z$ is the depth axis. We could have arranged the Hilbert objects differently to have $H$, $V$, and $\overline{I}$, respectively along these axes (which would make it easier to read off the transformations associated with the diagonals). However, by associating the basis independent operations, $\overline{H}$ and $\overline{V}$, with the $x$ and $y$ axes, we are able to have the Hilbert square and its shadow at the front and back, which is more useful for us (since the elements of these two squares are related by basis independent transformations).

5. The diagonals on the front and back faces are the transpose $T$ (as we have already seen since the Hilbert square and the shadow Hilbert square are of the form shown in (576)).

6. The diagonals on the side faces are the normal vertical transpose, $V$. This operation can be effected by applying a twist cup and twist cap to the input and output wires of the given left or right Hilbert object as illustrated in (581).

7. The diagonals passing through the interior of the cube (between opposite corners) are the normal adjoint, $\overline{T}$. This is a basis dependent transformation.

8. The front face of the cube is the original Hilbert square (see (576)). The back face is the shadow Hilbert square (see (583)). These are associated with the basis independent normal conjupositions $\mathcal{S}_{BI} = \{I, \overline{H}, \overline{V}, T\}$. There are many more squares appearing in the Hilbert cube. However, only the Hilbert square and shadow Hilbert square are associated with basis independent transformations.

9. We will call the faces on the side the *side Hilbert squares*. On the left side is the *left Hilbert square* and on the right side is the *right Hilbert square*. The operations associated with the side Hilbert squares are $\mathcal{S}_{\text{side}} = \{I, \overline{I}, V, \overline{V}\}$. We will discuss the side Hilbert squares in more detail below (Sec. 32.9).

10. Another square of interest is the one constituted by the front left two objects and the back right two objects a *twist Hilbert square* since it is associated with the group $\mathcal{S}_{\text{nonconjugating}} = \{I, H, V, T\}$ of operations that can be implemented with caps, cups, and twists (see the discussion at the end of Sec. 32.6. There is another twist square - corresponding to the front right and back left elements. The twist squares do not have conjugation and hence they are not sufficient to formulate Quantum Theory (since we need to take the product of amplitudes and their conjugates to get probabilities). However, the nonconjugating group is of interest because it can implemented on one component of a composite Hilbert object allowing us to find equivalent equations (see Sec. 32.13).

11. We can also consider a square constituted of the top front and the bottom back objects. We will call this a *slanted Hilbert square*. Such a square is associated with the group $\{I, \overline{H}, V, \overline{T}\}$. There is another slanted Hilbert square associated with the front bottom and back top elements. Since these squares do not have the normal vertical adjoint operation they are of limited use in Quantum Theory (the normal vertical adjoint is used in the study of unitary operations - see Sec. 37.

In Appendix A we discuss a possible generalisation of the Hilbert cube - the Hilbert hypercube.

## 32.9 The side Hilbert squares

The Hilbert square was motivated by a similar looking square (they call it a "quartet") in the book of Coecke and Kissinger [2017] – see page 110 therein. However, as we will see, their square is actually a side Hilbert square (though, of course, they do not call it that).

The side Hilbert squares are interesting because, whilst they require basis dependent transformations, they enable us to do normal conjugation in a way that is sufficient to formulate Quantum Theory. The left Hilbert square looks

like this

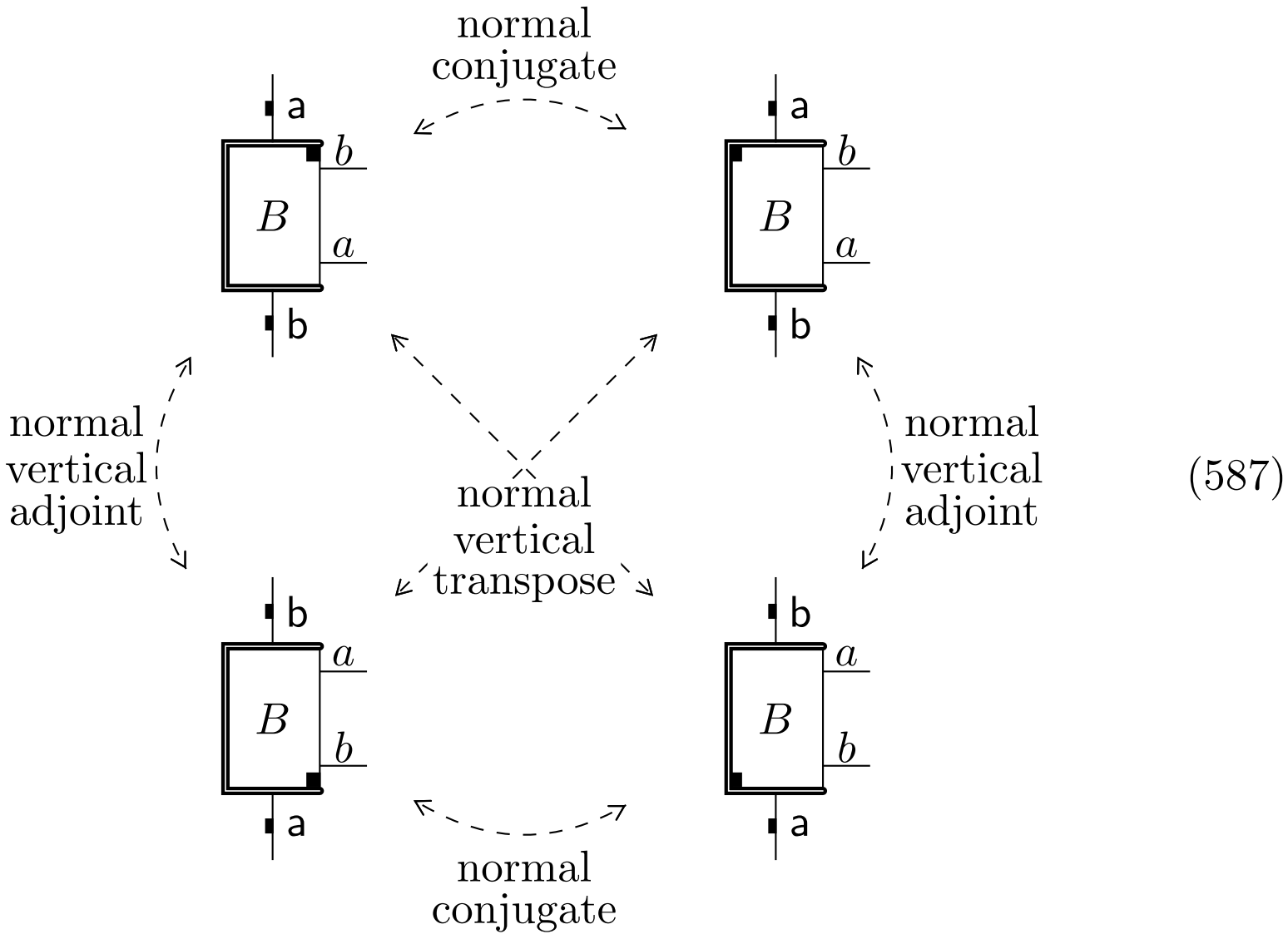

(587)

The diagonals are associated with the normal vertical transpose, $V$, as we discussed in Sec. 32.8. The normal vertical transpose can be implemented using twist cups and caps as shown in (581).

This seem like a good point to reflect on some of the differences between the present approach and that of Coecke and Kissinger [2017] in their book which nicely summarises the main approach taken in the process theory community. First note that they start at with more abstract processes which are not, necessarily, instantiated by what we have called Hilbert objects here. However, these abstract processes can be instantiated by Hilbert objects and so we can make a comparison.

Coecke and Kissinger do not have the horizontal label wires (labeled here by $a$, $b$, $\ldots$). Rather, they put summation symbols, $\sum_{ab}$, in their diagrams. Consequently they are not motivated to define horizontal and vertical transposes or horizontal and vertical adjoints as we have done here. We can, however, easily translate their terminology into the present terminology. Their conjugate is the same as the normal conjugate, $\overline{I}$, here. Their "transpose" is the same as the normal vertical transpose, $\overline{V}$, here. Their "adjoint" is the same as the normal vertical adjoint, $\overline{V}$, here.

The aforementioned figure on page 110 of Coecke and Kissinger [2017] illustrates a "quartet" of processes that are related by conjugate, "transpose", and "adjoint" transformations. These are the same as the transformations in the group $\mathcal{S}_{\text{side}} = \{I, \overline{I}, V, \overline{V}\}$ (where we now include $I$ for completeness). Hence, as commented above, their quartet is effectively one of the side Hilbert squares (it does not matter which, though the left one is more natural since it associates kets with preparations - see (457)). Their cup and cap are the basis dependent twist cup and twist cap in (514) and (515). In the advanced material in Chapter 4.6 (pg 145) of Coecke and Kissinger do, in fact, discuss basis independent cups and caps equivalent to (509) and (508) used here and indicate reasons for their preference for the basis dependent cups and caps.

How it is possible to have formulations of the same theory (Quantum Theory) based on seemingly different objects - the Hilbert square in this book and the left Hilbert square in Coecke and Kissenger's book? To understand this note that, in the formulation of Quantum Theory in this book we calculate probabilities by evaluating circuits built out of operator tensors. These operator tensors must be in twofold form (so that the probabilities calculated are nonnegative - see Sec. 43). Consider an operator tensor that is in twofold form

(588)

This is built out of left and right Hilbert objects taken from the Hilbert square. If we act on the right Hilbert spaces with twist objects then we obtain

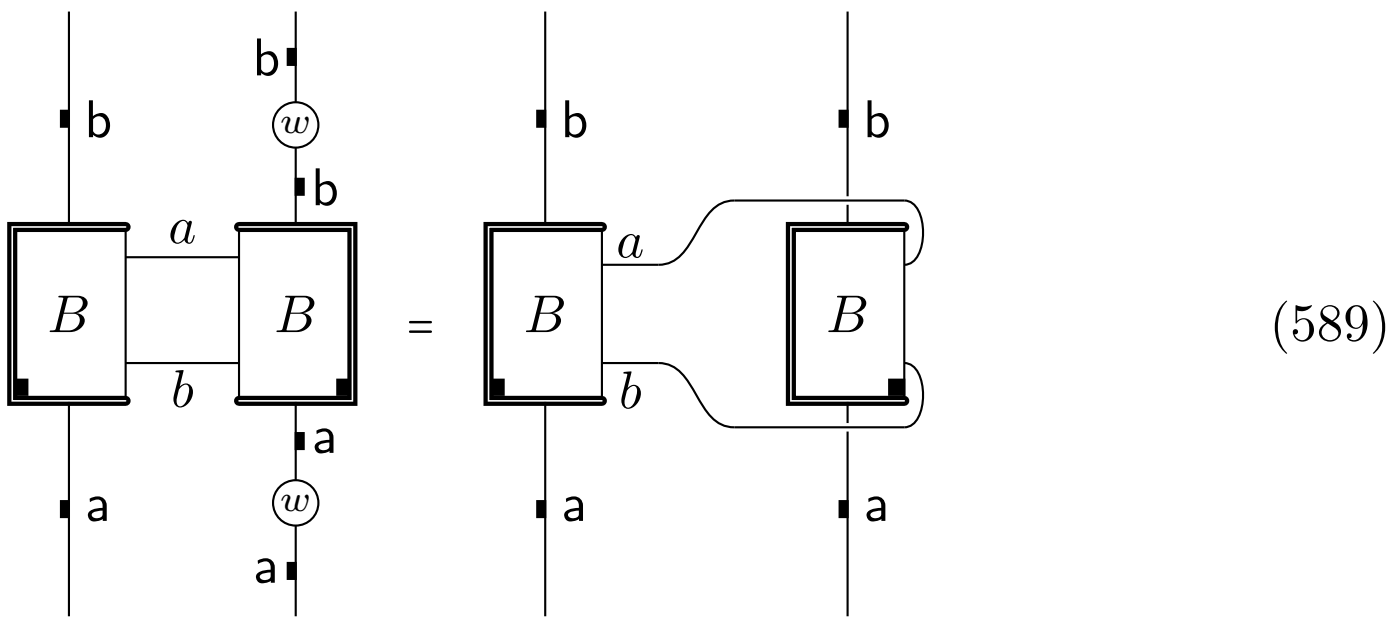

(589)

This object is built out of left Hilbert objects taken from the left Hilbert square. Let us call it a *left operator tensor*. If we build a circuit out of these left operator tensors then the twists cancel (because of the $w$-annihilation property as discussed in Sec. 30.4) and so we will get the same answer as if we had built the circuit out of operator tensors. This means we can use either operator tensors or left operator tensors in formulating Quantum Theory. Coecke and Kissinger are, effectively, using left Hilbert operators. Of course, it is worth commenting that the diagrammatic notation in this book has been deliberately chosen to work with the basis independent, operator tensor, Hilbert square approach rather than the basis dependent, left operator tensor, left Hilbert square approach. It need not be this way and, indeed, the diagrammatic notation of Coecke and Kissinger works well with their approach.

## 32.10 Normal conjupositions of basis elements

We can consider taking normal conjupositions the basis elements themselves. If we want to do this then it is useful to include square dots as follows

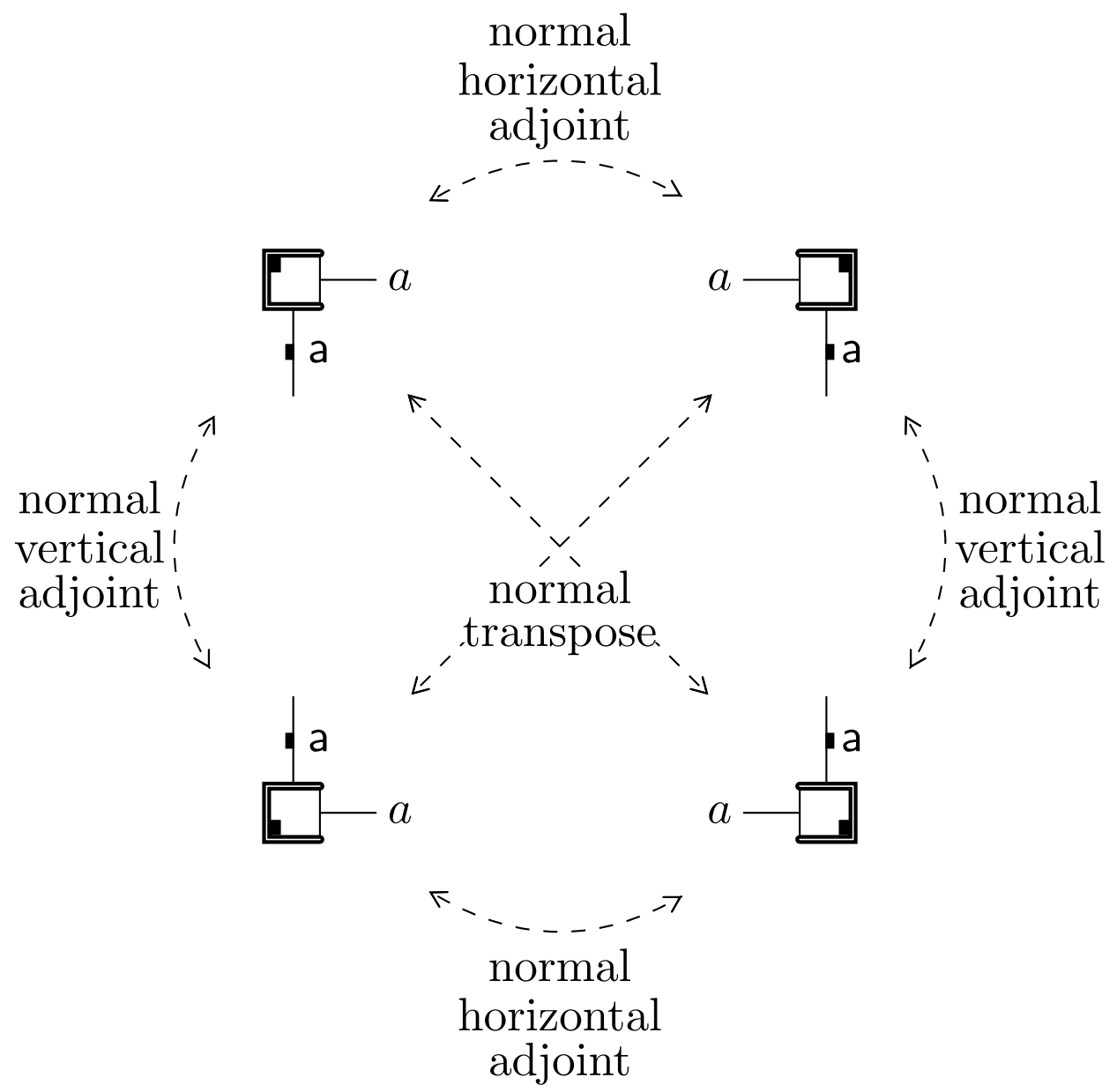

(590)

where these are taken to be the same objects as those appearing in (498). The normal horizontal transpose, $H$, of these objects can be obtained by applying twists (as shown in (579)). Consider, for example

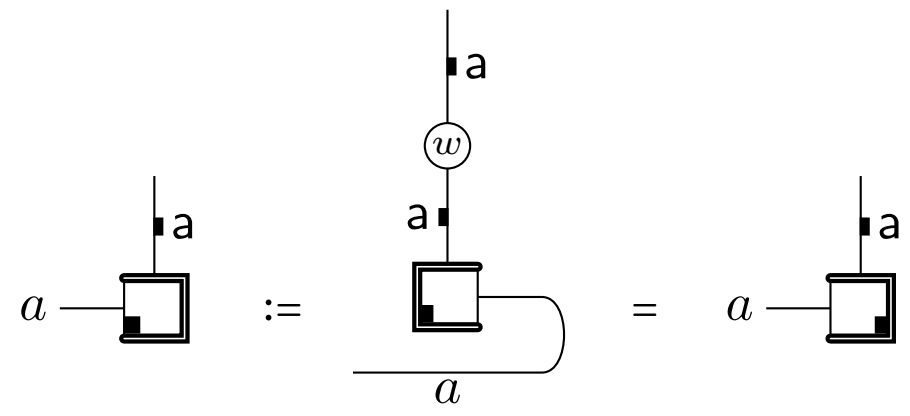

(591)

where we have used the definition of the twist object in (513) to obtain the equality. Thus, applying $H$ to this basis element is the same as applying $\overline{H}$. Similar remarks apply to taking $H$ or $\overline{H}$ for the other basis elements shown in (590). This is not surprising as the twist transformation uses the very basis we are transforming (we can think of there being no off-diagonal elements so it does not make any difference whether we apply $H$ or $\overline{H}$). This means that the shadow to the Hilbert square in (590) has the same elements in it. In turn, this means that basis elements are left unchanged under the action of $\overline{I}$ (as illustrated in (591)).

We leave it as an exercise to show that if we start with a Hilbert square built out of elements of a different basis (such as shown in (501)) than the one used to define the twist object then there are off-diagonal elements and so we do not have this degeneracy with respect to $H$ and $\overline{H}$.

## 32.11 Normal Conjupositions of left and right composite Hilbert objects

It is often useful to apply an element of the normal conjuposition group to a left or right composite Hilbert object. Such transformations are straightforward because of the way hypermatrix networks behave under the conjuposition as discussed in Sec. 29.2. For example, it is easy to show that

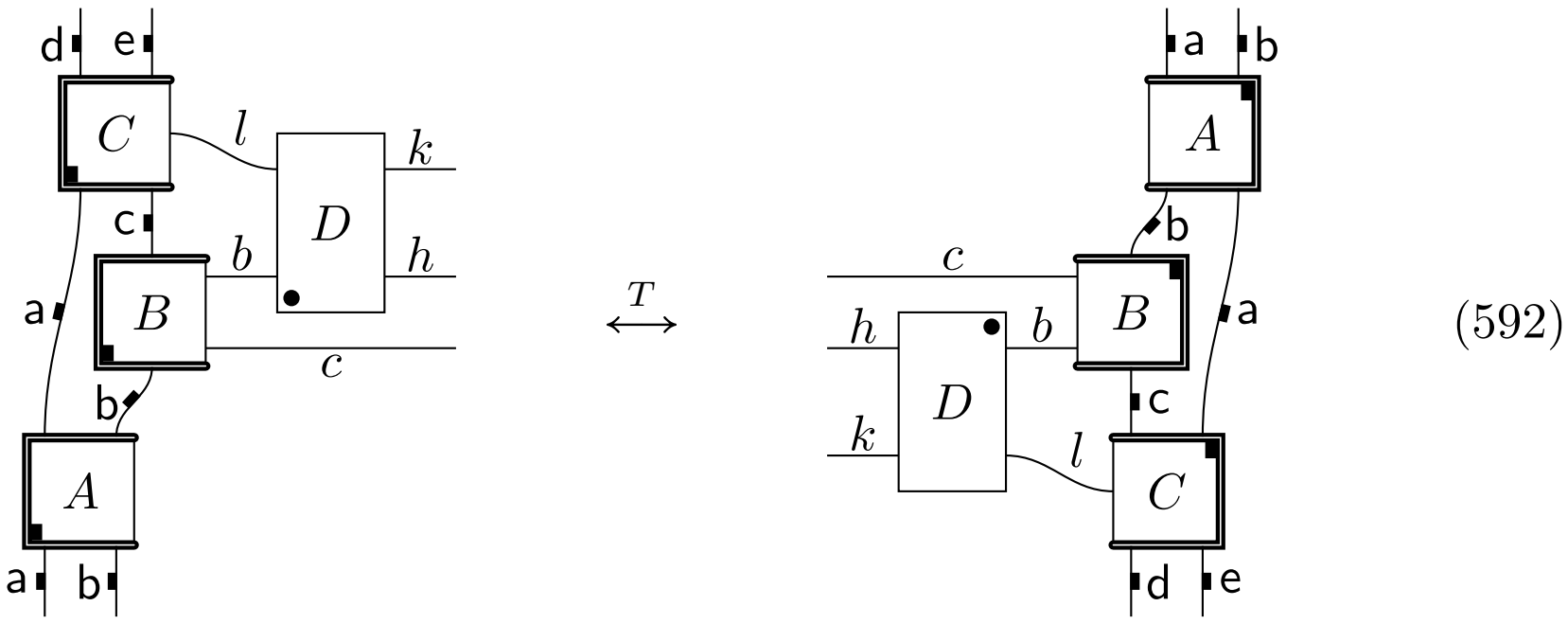

(592)

This is what we would expect for the $T$ operation - each object making up the left Hilbert network is transformed under $T$ resulting in the right Hilbert network. In general, the any conjuposition of such a composite object preserves the compositional form whilst transforming the object as in this example.

It is instructive to show how to use the results of Sec. 29.2 (showing how hypermatrix networks transform under the conjuposition group) on a simple example. We have

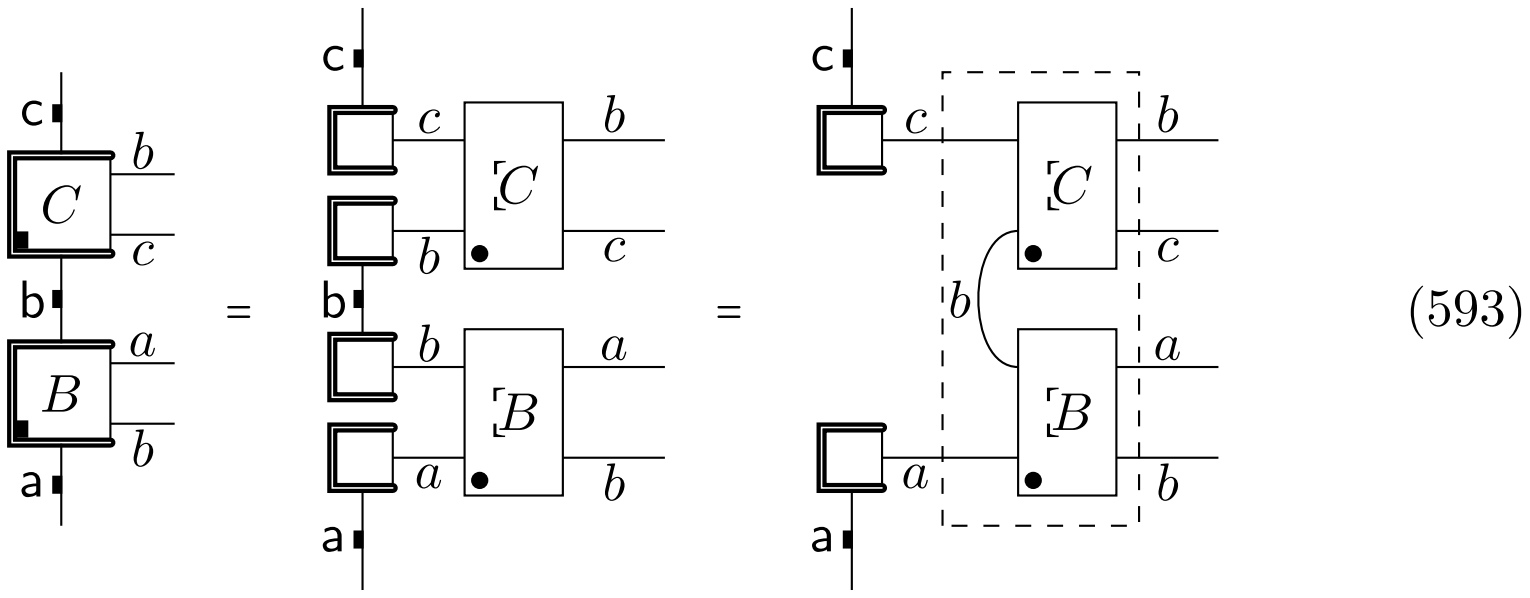

(593)

The hypermatrix inside the dashed box is the expansion hypermatrix for this left Hilbert network. We can, then, conjupose this left composite Hilbert object by performing the corresponding conjuposition transformation on this expansion hypermatrix and then appending the appropriate bases elements as illustrated by the examples in Sec. 32.5 and Sec. 32.7. The conjuposition of the hypermatrix inside the dashed box is obtained by flipping/rotating and conjugating this object as described in Sec. 29.2. One comment worth making here is that this works because the scalar product of the orthonormal basis elements is taken to be real (see (500)). Were this scalar product complex then the hypermatix inside the dashed box would not conjupose appropriately.

The eight elements of the conjuposition group acting on this composite Hilbert object give a Hilbert cube as follows

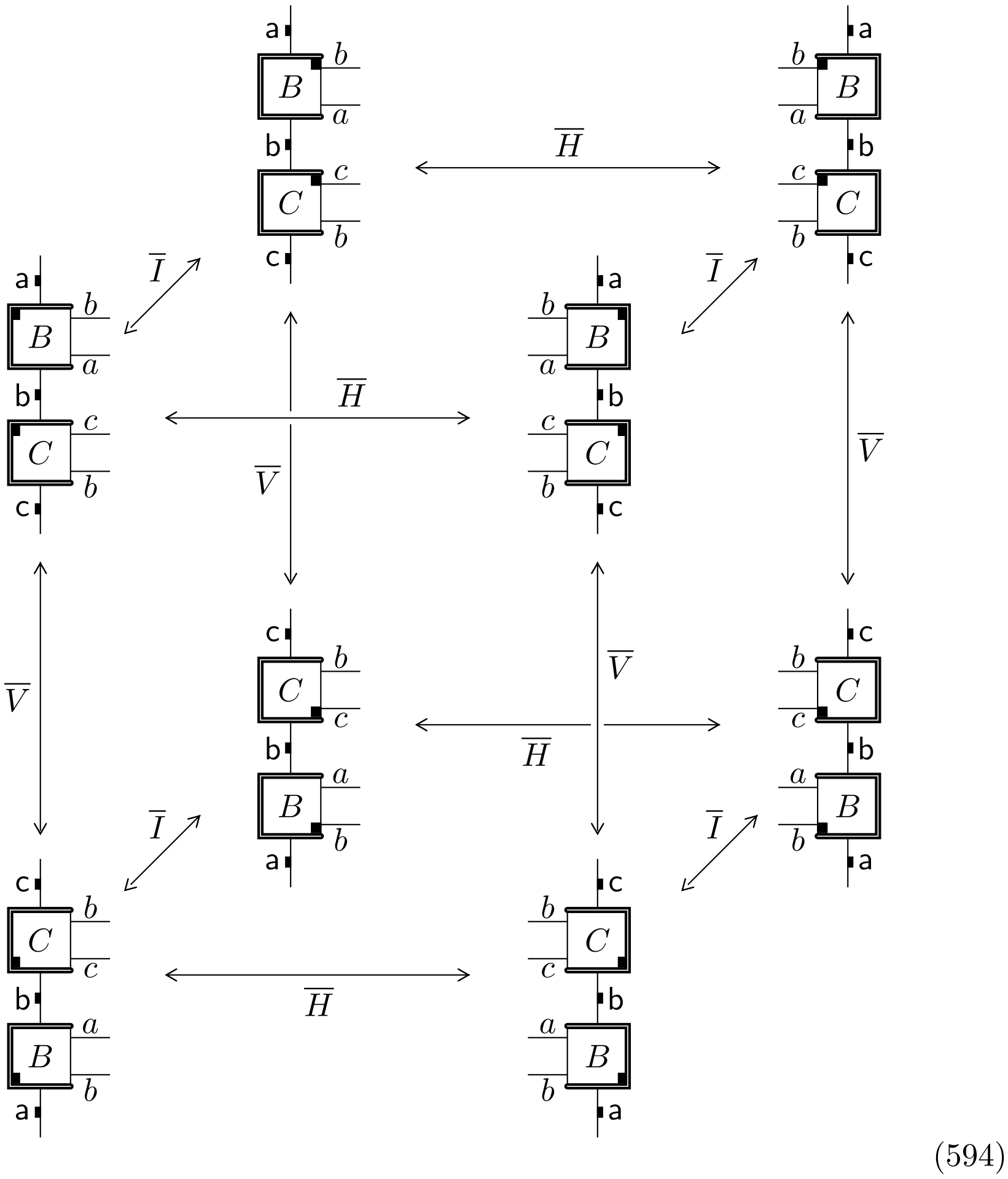

(594)

It is clear here that Hilbert networks transform in the way we would expect.

An important application of the results above is that, given any equation (or inequality) between composite Hilbert objects, we can conjupose it (using any element of the conjuposition group) to generate an equivalent equation (or inequality). For example,

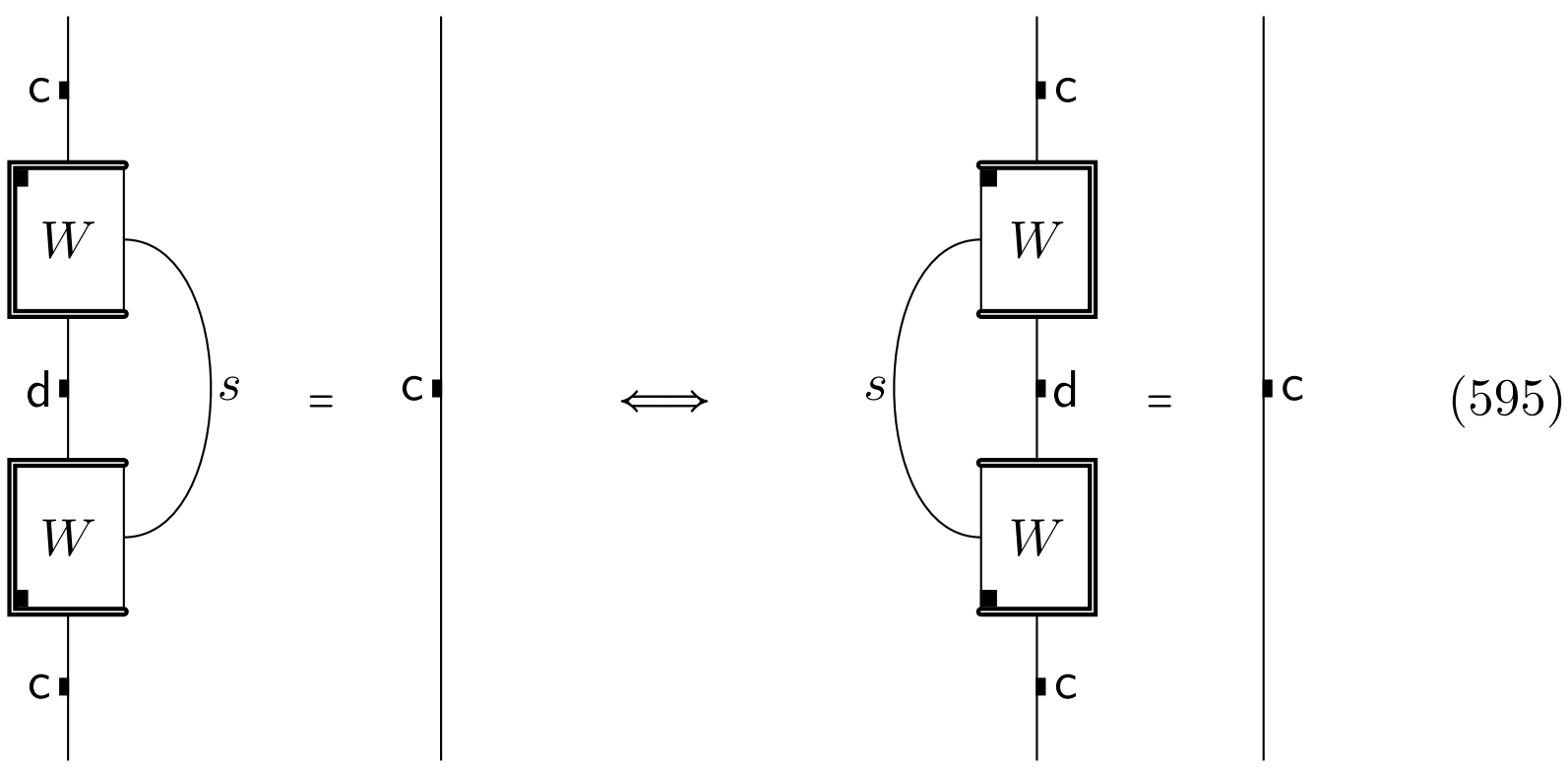

$$(595)$$

(note that this equation appears in Appendix B as (1899)). We obtain this equivalence by applying the $H$ operation. We will discuss how to form equivalent equations in greater generality in Sec. 32.13.

## 32.12 Normal conjuposition of general Hilbert objects

We can form general Hilbert objects such as

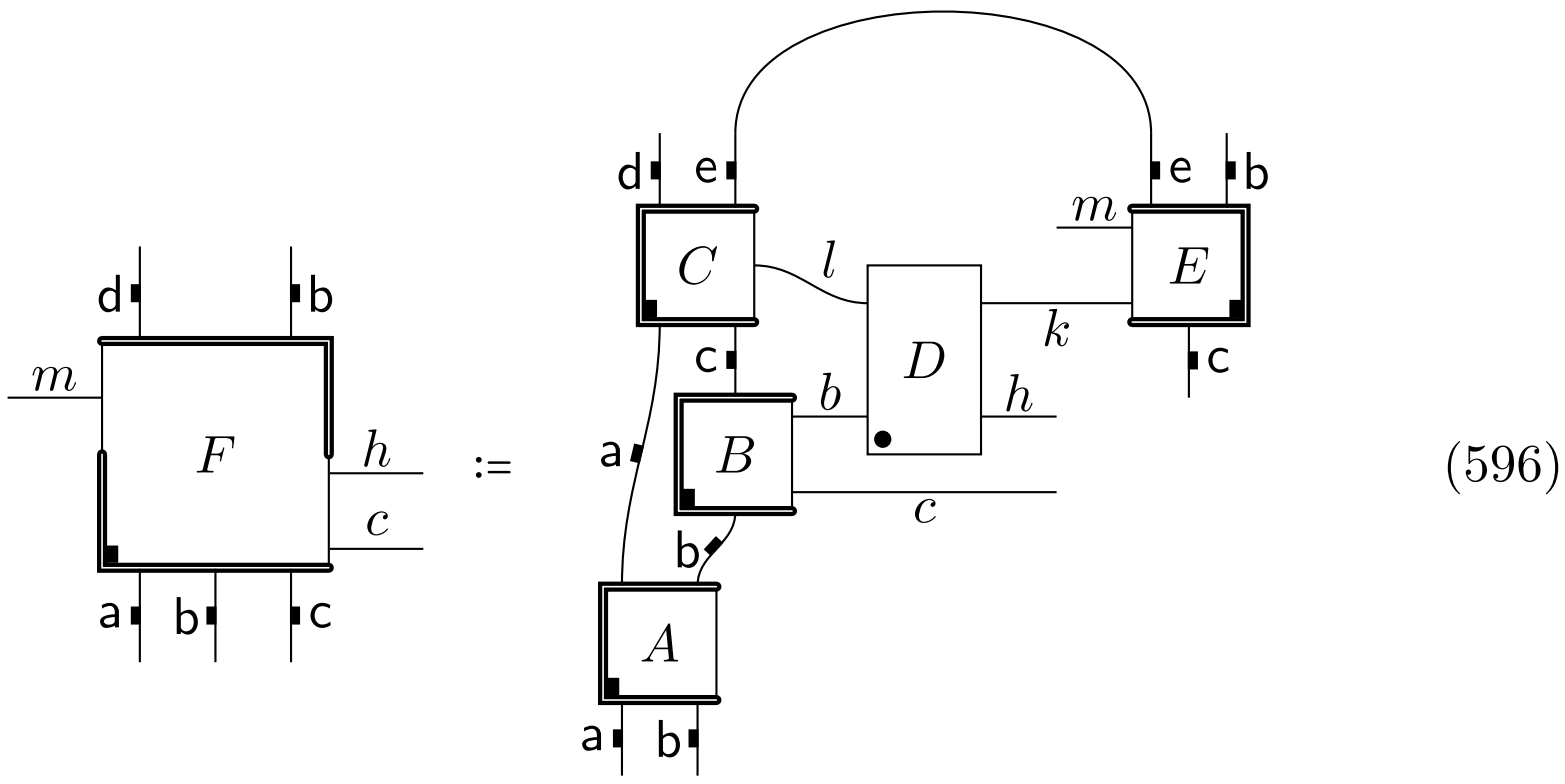

$$(596)$$

This example is hybrid since it has both left and right elements (see Sec. 28.7).

We can take normal conjupositions of general Hilbert objects in the obvious way. For example, if we apply $T$ to (596) we obtain

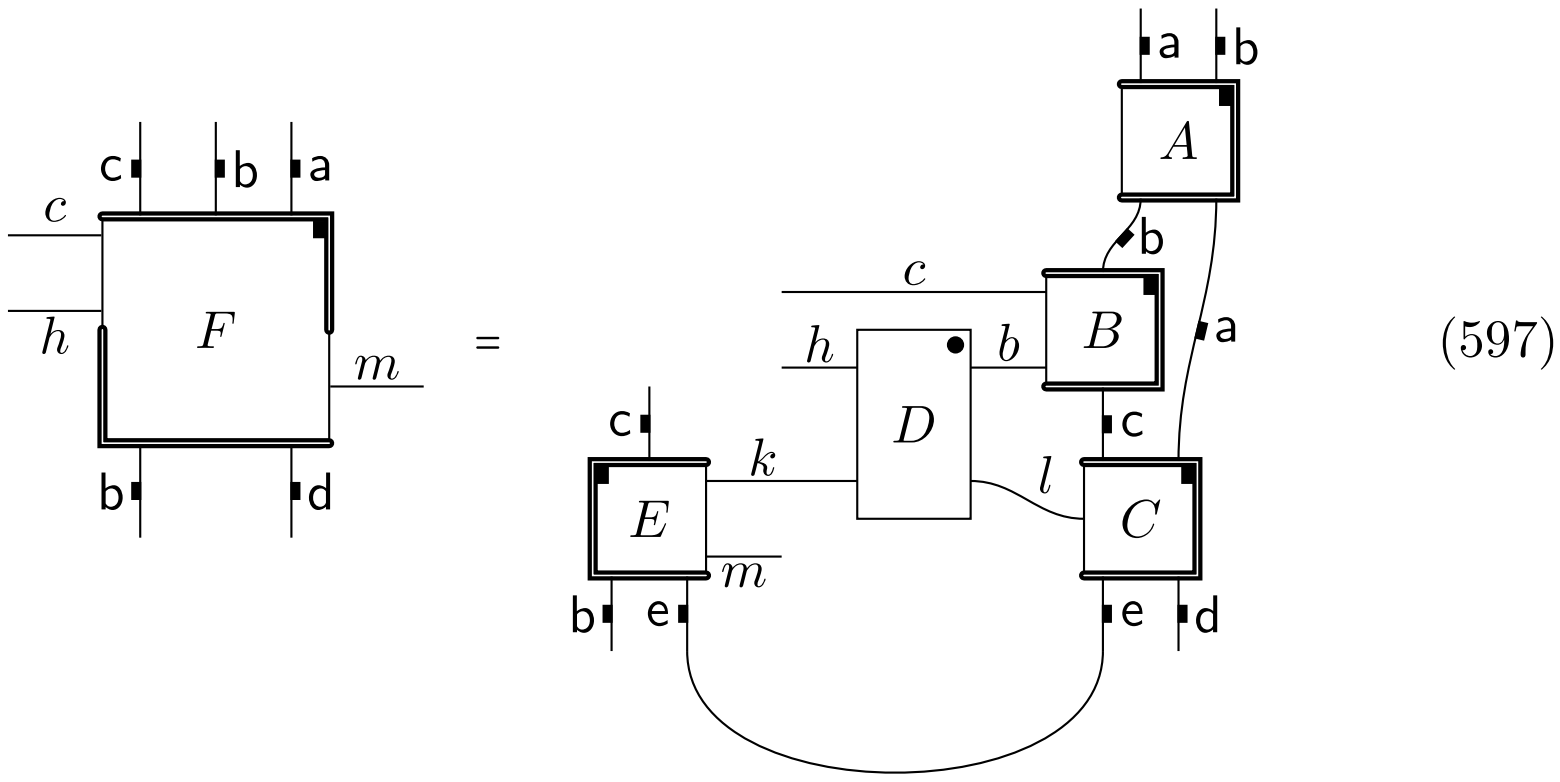

(597)

This can be proven by expanding out the component Hilbert objects with respect to bases and then using results from Sec. 29.2 to conjupose the resulting hypermatrix network.

We can apply any of the eight normal conjugations to a general Hilbert object. The resulting object is obtained by flipping horizontally, vertically, or rotating by 180° and, furthermore, moving the square dot as appropriate. Applying the normal conjupositions to some general Hilbert object results in a

Hilbert cube of such objects

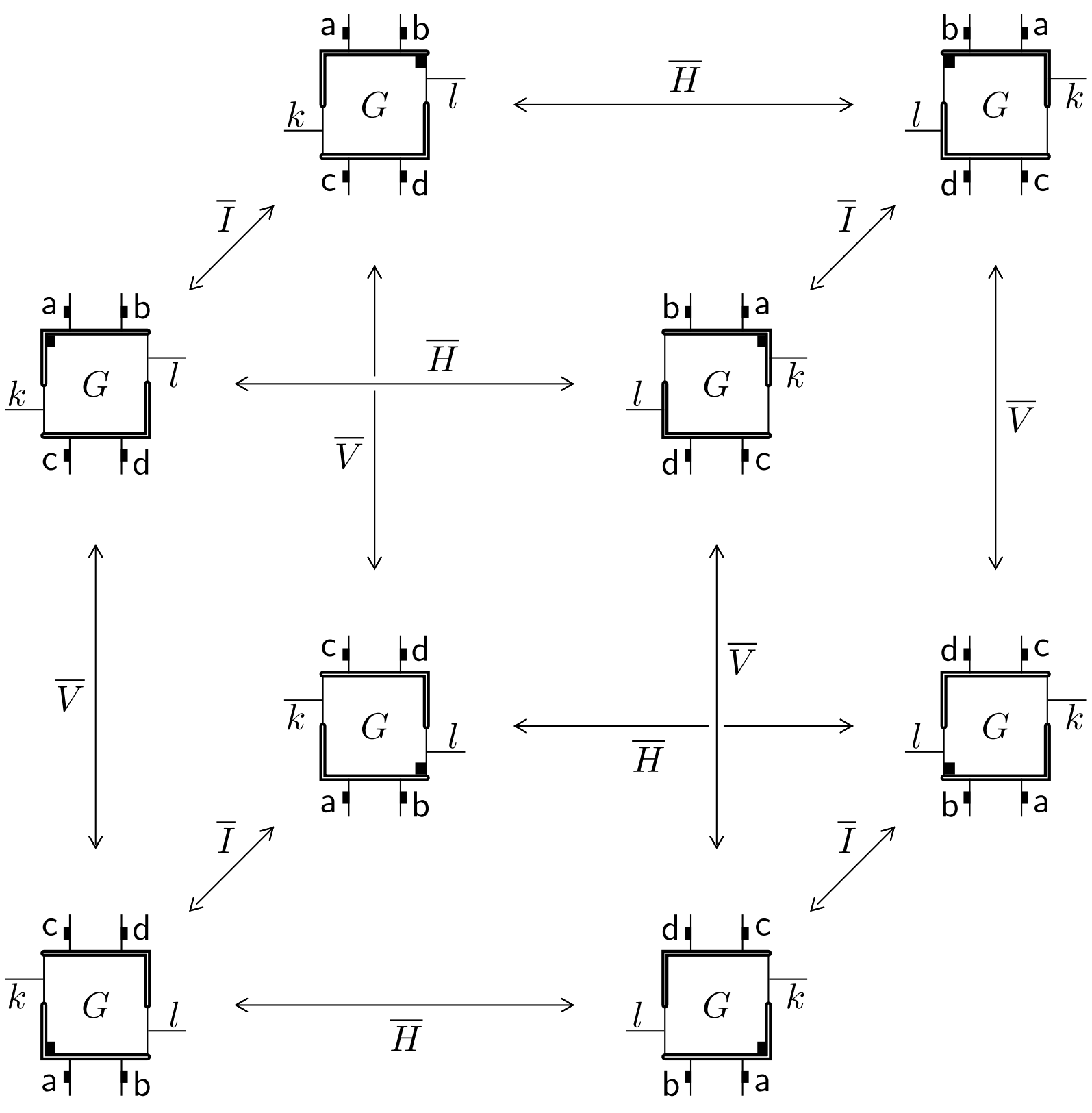

(598)

We will mostly work in terms of left Hilbert objects, right Hilbert objects, or operator tensors. All of which are special cases of general Hilbert objects. The special notation for general Hilbert objects is, however, useful when we want to prove general theorems as we will see next.

## 32.13 Equivalent equations

If we have an equation (or inequality) relating general Hilbert objects then we can taken the normal conjuposition of each side of the equation (or inequality) with any given element of the normal conjuposition group resulting in an equivalent equation (or inequality). In fact, we can go further and apply normal conjupositions in the subgroup, $\mathcal{S}_{\text{nonconjugating}}$, to only part of the expression appearing on each side of an equation (or inequality). We will just discuss the case of equations - the case for inequalities follows similarly.

Consider equations of the general form

(599)

Such equations can be converted to equivalent equations by applying the same normal conjuposition on each side. For example we obtain

(600)

by applying the normal adjoint, $\overline{T}$, to both sides. Thus every equation belongs to a set of eight equivalent equations (though, note, if we have certain symmetries then some of these equations will be the same - an pertinent example is when our general Hilbert objects are actually Hermitian operator tensors).

Next consider an equation where the expression on each side is composite such that the open wires on each component on the left match with a component on the right. An example is the equation

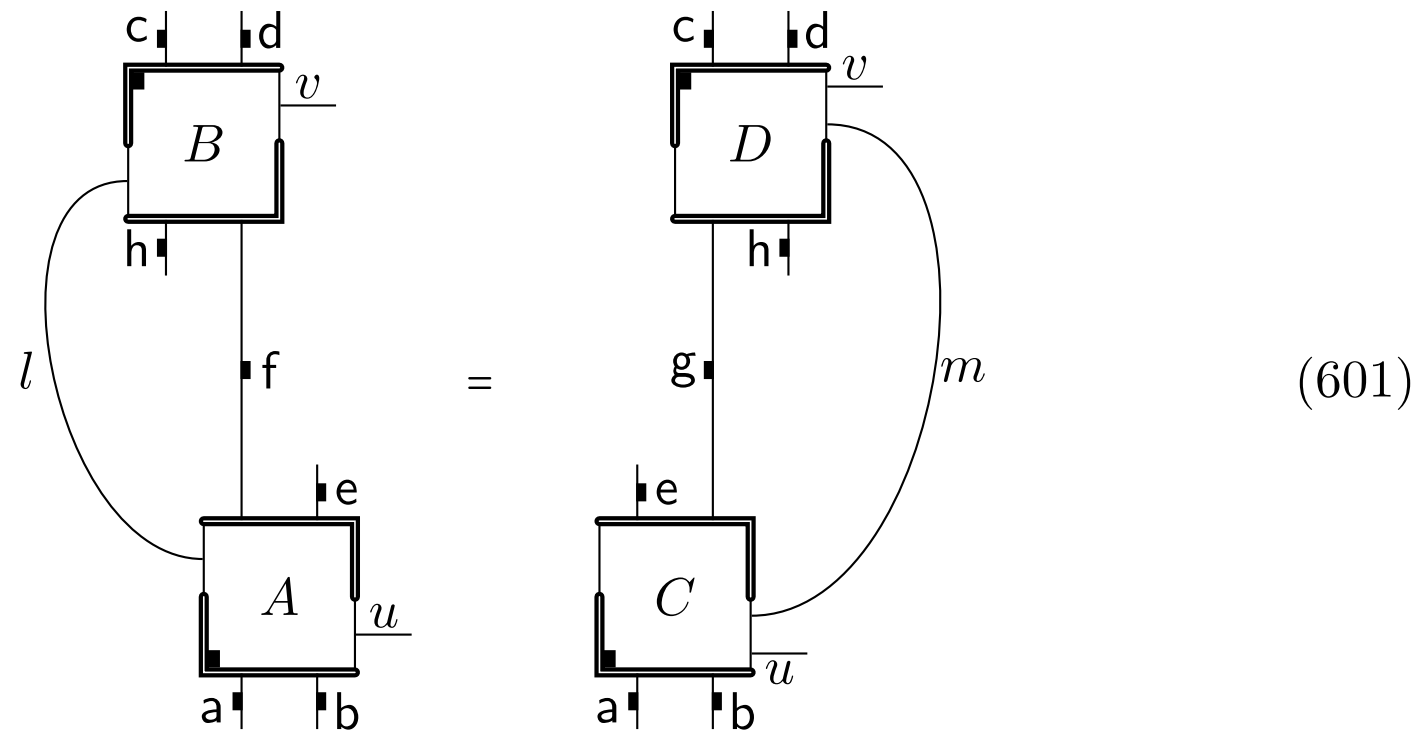

(601)

Here we see that the $A$ and $C$ components have the same open wires on them, as do the $B$ and $D$ components. The wires joining the components can be different from one side of the equation to the other (as in this example). The next move is to insert decompositions of the identity on each of the system wires connecting the components. In our example this works as follows

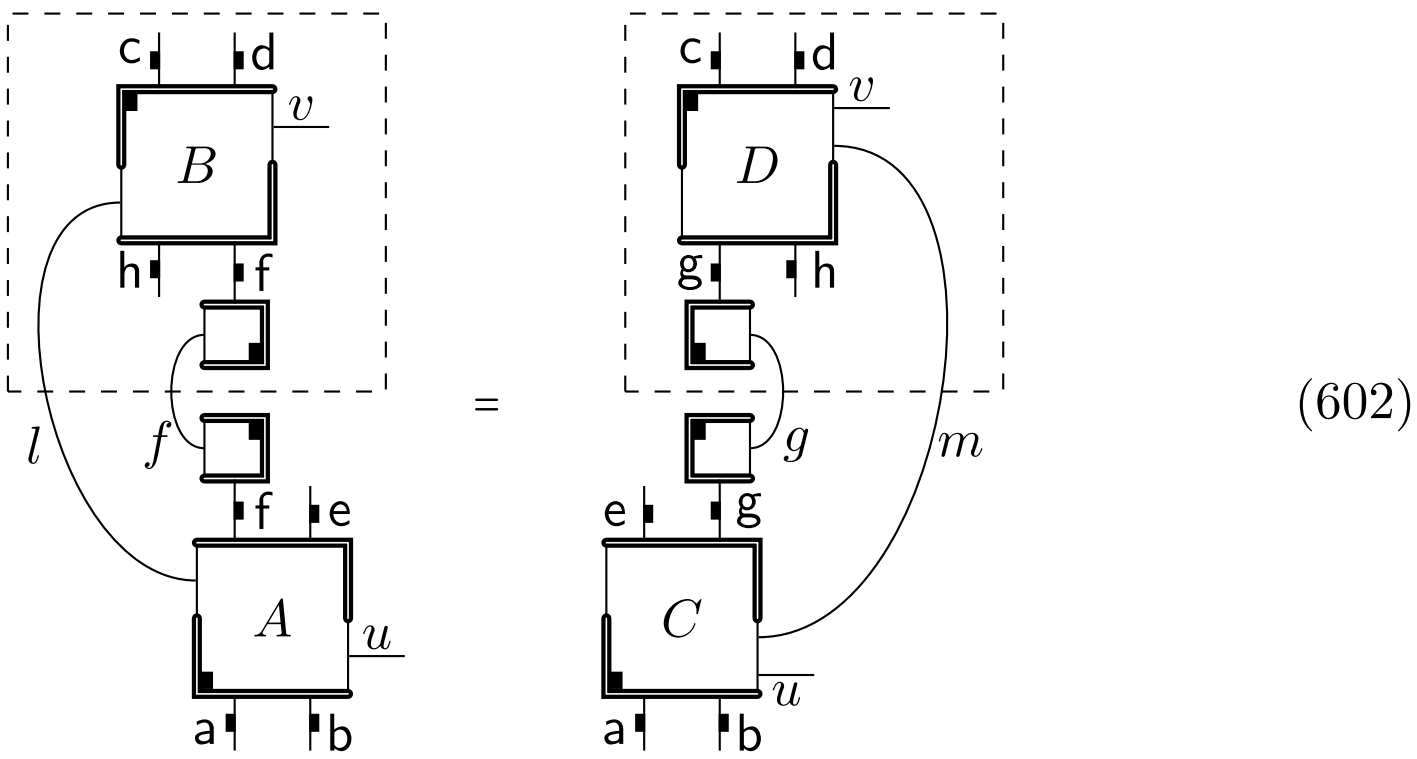

(602)

Now each side of the equation has two components (where we count the basis elements in with the general Hilbert object it is attached to by a system wire) that are connected only by label wires. We can apply caps, cups, and twists to each of the open wires on either (or both) of the components and, thereby, implement any element of the subgroup, $\mathcal{S}_{\text{nonconjugating}}$ on the component. Consider the normal vertical transpose, $V$. This is implemented by applying twist caps and cups (see example in (581)). Thus, if we apply twist caps and cups to the open wires on the component inside the dashed box on each side of the equation then we will effect the normal vertical transpose, $V$, on this object on each side of the equation resulting in the equation

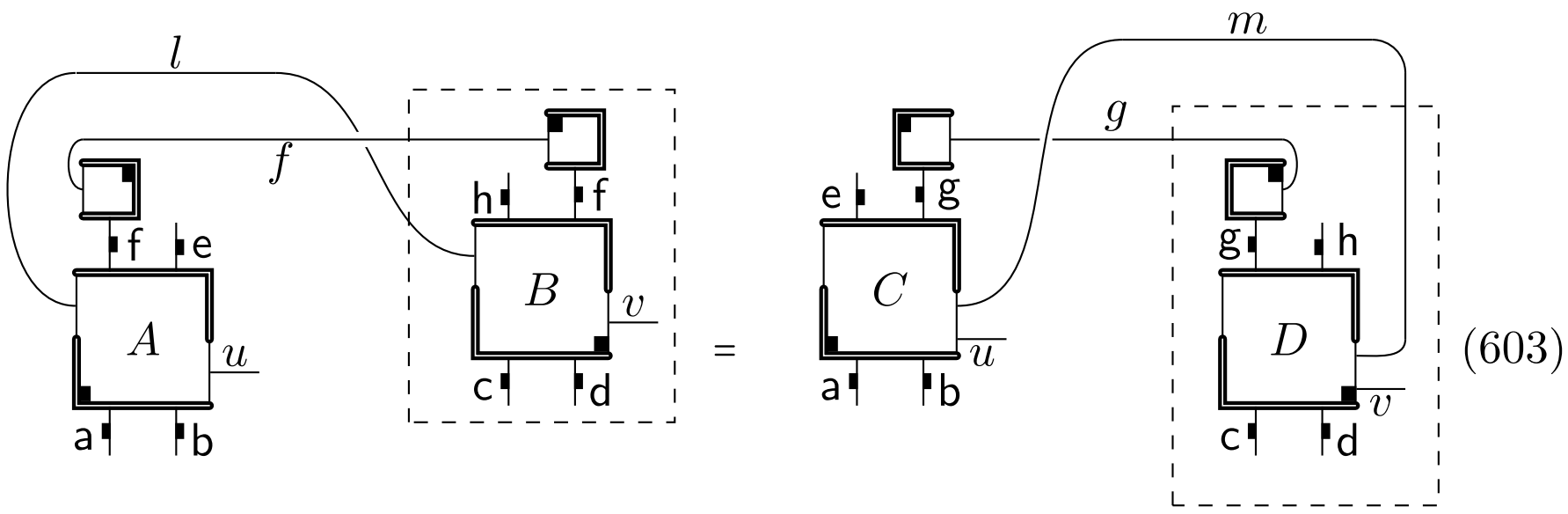

(603)

To complete the manipulation, we can use the twist cap notation (see (515)) giving

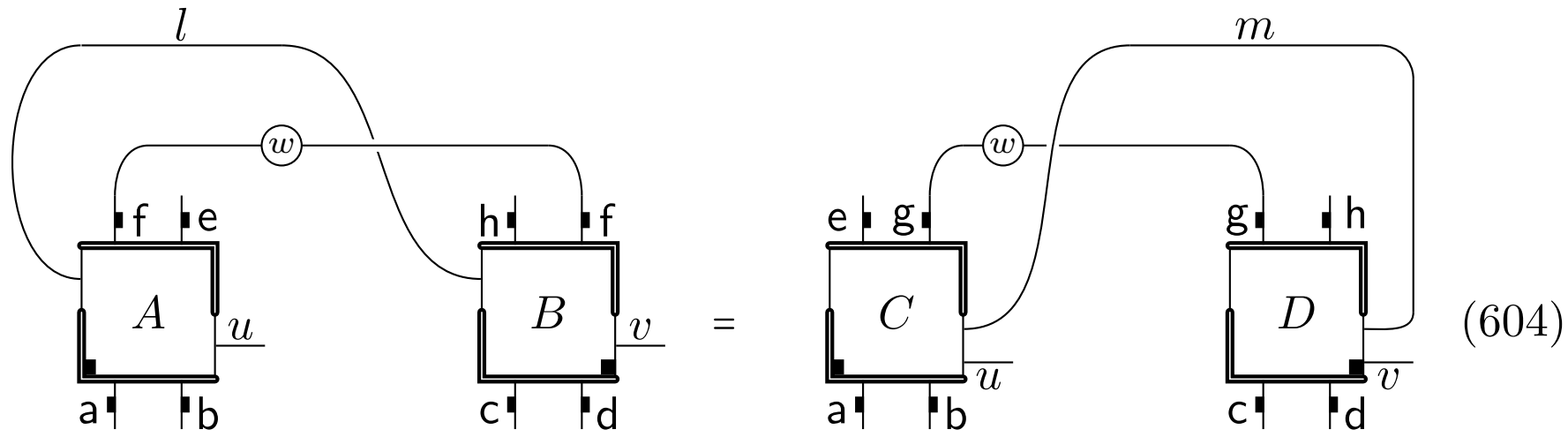

(604)

(note we are using the result in Sec. 32.10 that the basis elements are unchanged under the action of $\overline{I}$ so the position of the black dots on the basis elements does not matter). The twist, $w$ will appear on connecting wires when we implement $V$ (as in the above example) or $H$, but not when we implement $T$ (since the transpose can be effected with just cups and caps as in (568)).

To summarize, we can apply any element of the conjuposition group to the whole expression on both sides and get a new equation. Further, if we have an equation with the same composite form on both sides then we can apply any nonconjugating normal conjuposition (by applying caps/cups/twists as appropriate) to any of the components as detailed above. Finally, it is worth noting that we could apply partial normal nonconjugating transformations by only applying caps/cups/twists to some of the wires in the given component.

## 32.14 Normal conjuposition of operator tensors

We can perform normal conjupositions on operator tensors. Normal conjupositions in the subgroup $\{I, \overline{H}, \overline{V}, T\}$ acting on operator tensors are basis independent and, consequently, are of most interest to us. We will provide explicit expressions for these.

We can perform a normal conjuposition by expanding in an orthonormal basis (for each system type) as follows

(605)

(this equation appeared as (531) though now we have introduced a black square and circle so we can follow the transformations). The $I$ transformation leaves this unchanged of course.

The $\overline{H}$ transformation flips (605) over horizontally and takes the conjugate of the entries of hypermatrix $\ulcorner B \urcorner$ as follows

(606)

The black square has been moved horizontally to indicate the horizontal adjoint being performed. If the operator tensor is Hermitian then it is unchanged after the $\overline{H}$ operation (see Sec. 31.3).

The $\overline{V}$ transformation flips (605) vertically and takes the conjugate as follows:

(607)

The black square has been moved vertically to indicate the normal vertical adjoint being performed.

The normal transpose, $T$, is given by rotating (605) through 180° (or we can think of this as flipping horizontally and vertically) to give

(608)

The small black square has been moved to the opposite corner (consistent with rotating through 180°.

It is interesting to note that if we have a Hermitian operator tensor then its normal vertical adjoint, ($\overline{V}$, yields the same object as its normal transpose, $T$. To see this compare (607) and (608).

Any operator tensor (Hermitian or not) can be written in the form

(609)

(the left and right Hilbert objects are not, in general, uniquely determined by $\hat{B}$). It is instructive to see how to implement the normal conjupositions by acting on this form as well. Here we will simply show how the normal conjupositions work for the cases of $\overline{H}$, $\overline{V}$, and $T$ We leave it as an exercise to prove that these yield the same result as above when we act on the form expanded in terms of bases.

First, the normal horizontal adjoint is given by

(610)

We see that we simply flip the diagram horizontally.

The normal vertical adjoint is given by

(611)

We simply flip the diagram vertically.

The normal transpose is given by

(612)

Here we simply rotate the diagram through 180°.

The normal transpose of an operator tensor can be defined using cups and caps. First define

(613)

Then we can write

(614)

for transpose of an operator tensor. To see that this works consider applying the cap and cup defined in (613) to $\hat{B}$ as defined in (609). This gives

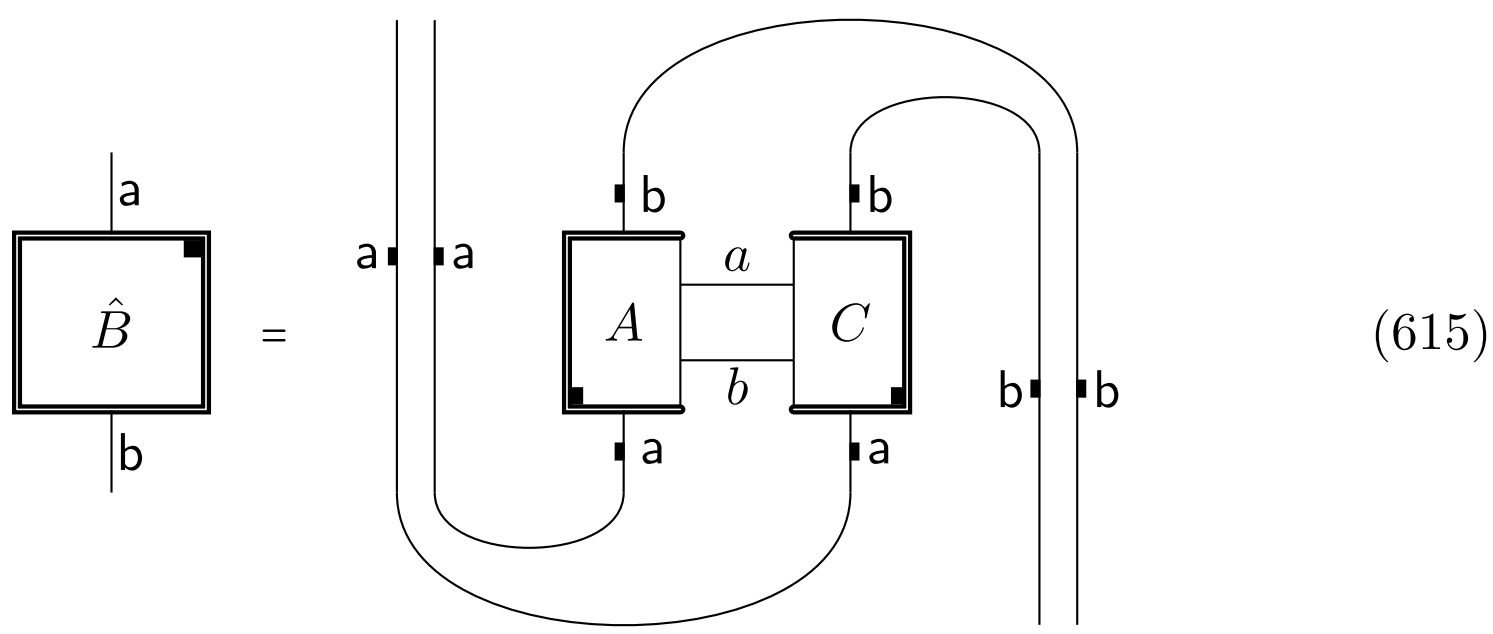

(615)

Here we are taking the transpose of both $A$ and $C$. If we slide $A$ and $C$ over the cap (see the discussion on "sliding" in Sec. 32.3), then straighten out wires appropriately, we get (612) as required. We can apply the cap and cup in (613) directly to the expansion of $\hat{B}$ in (605). If we do this, it is easy to verify that we get expression in (608).

## 32.15 Normal conjupositions of operator tensor networks

We can join operator tensors together to form operator tensor networks. For example

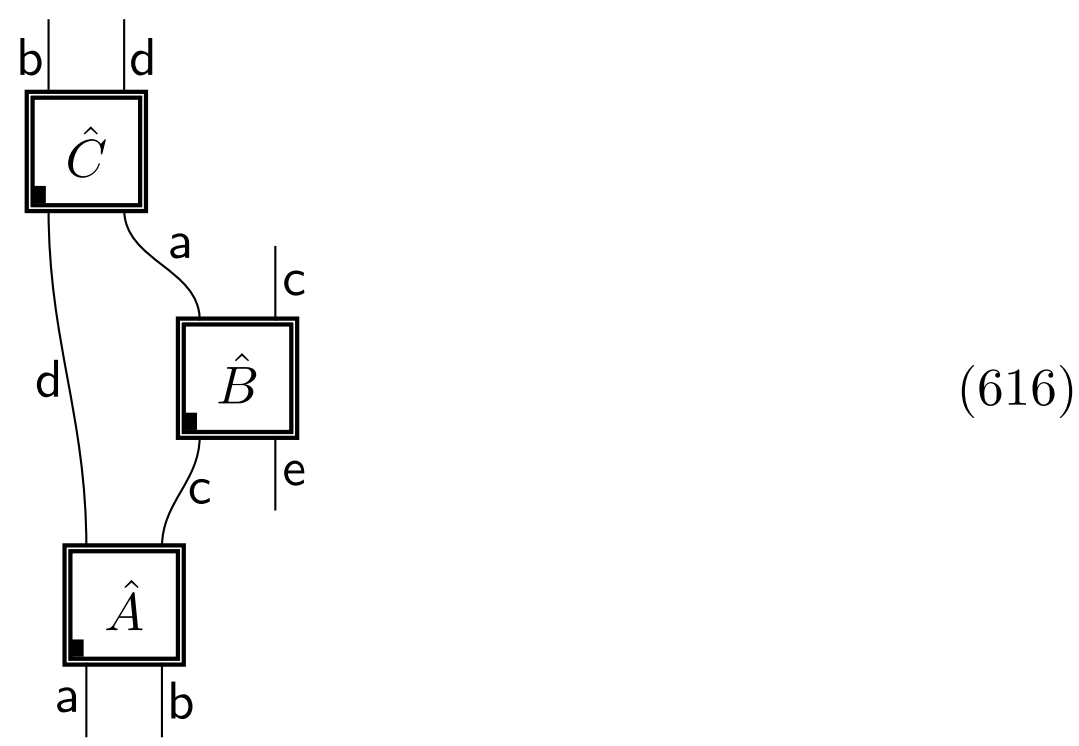

(616)

We can conjupose this operator tensor network under any element of the normal conjuposition group. For example, under a transpose, $T$, this becomes

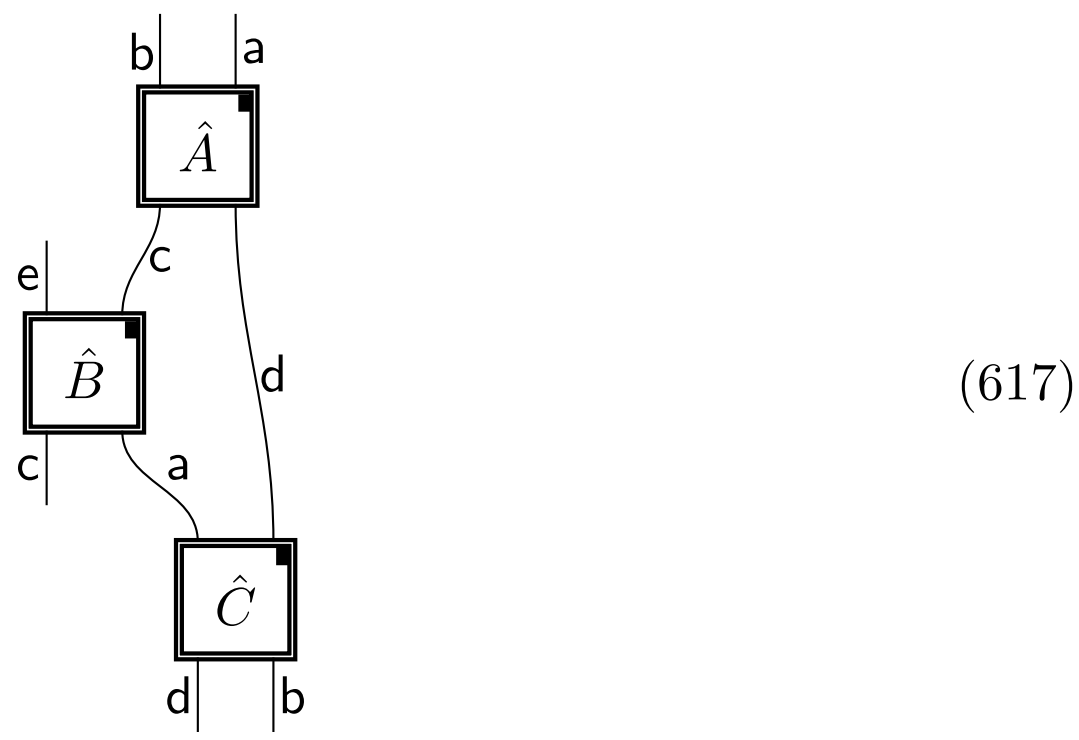

(617)

We can prove this by expanding each operation out in terms of bases and proceeding as before. Let us illustrate this with a simple example. We have

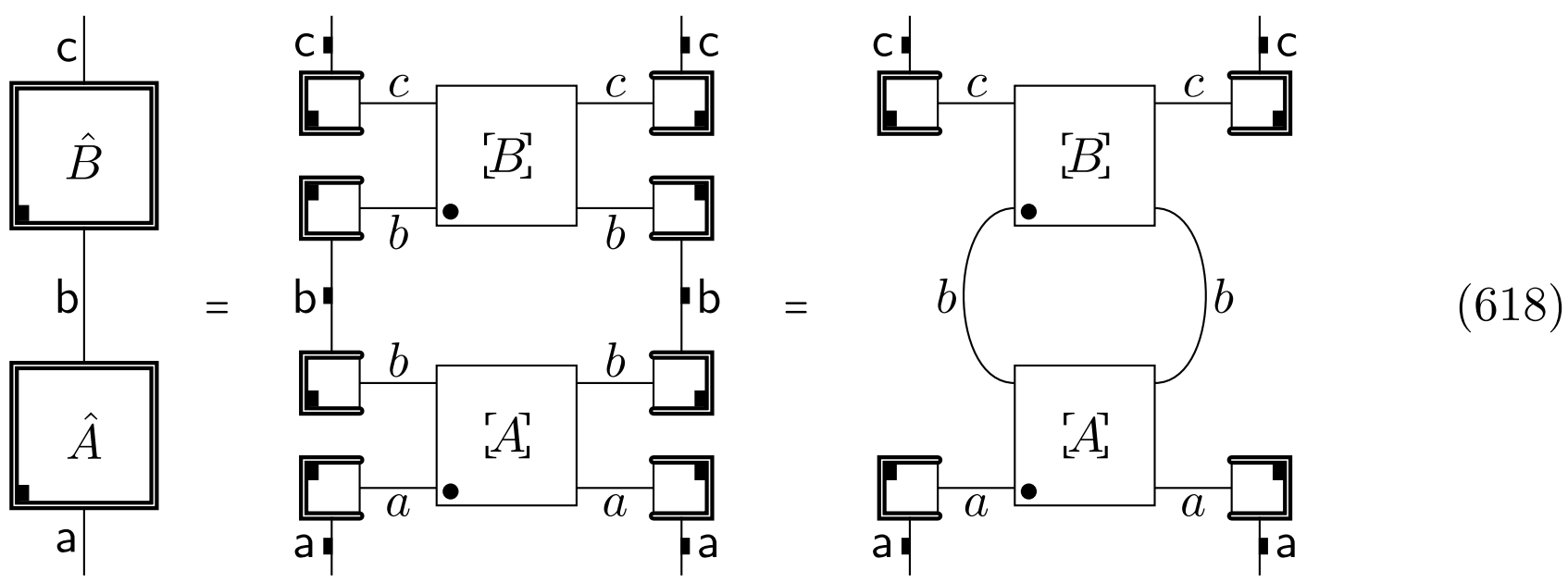

(618)

We can take the normal conjuposition of the object on the right by conjuposing the composite hypermatrix then appropriately reattaching the basis elements. The conjuposed composite hypermatrix will have the same form (see Sec. 29.2). For example, consider the normal transpose of the above example is given as follows

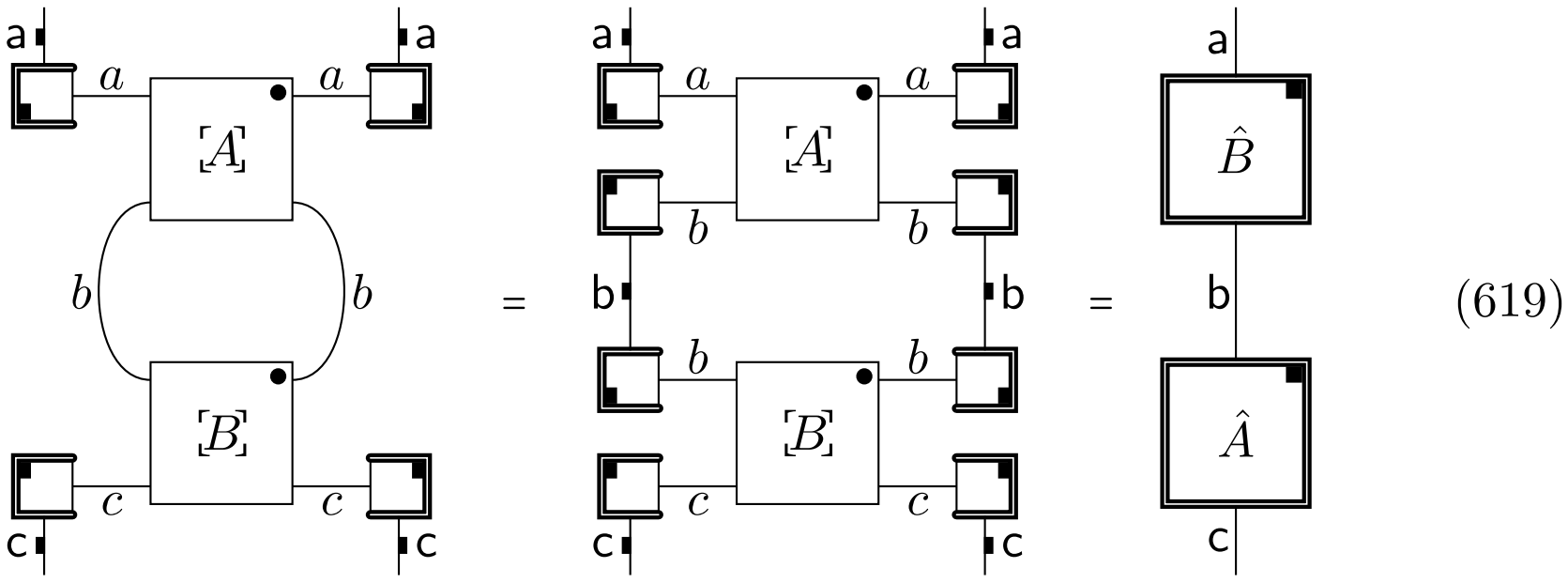

(619)

This expansion technique works for any element of the conjuposition group and for any network of operator tensors. An alternative way to prove this for the special case of the transpose, $T$, is to use (614).

If we have an equation relating operator tensor networks then we can perform some given conjuposition of this equation to obtain an equivalent equation.

## 32.16 On the physical meaning of normal conjupositions

In Quantum Theory operations correspond to operator tensors (as discussed in Sec. 19). This affords us the opportunity to think about normal conjupositions in physical terms. We will see that something is not quite right which will lead us to the idea of *natural conjupositions* in Sec. 33.

We want any given operator circuit to equate to a probability $p$. For example

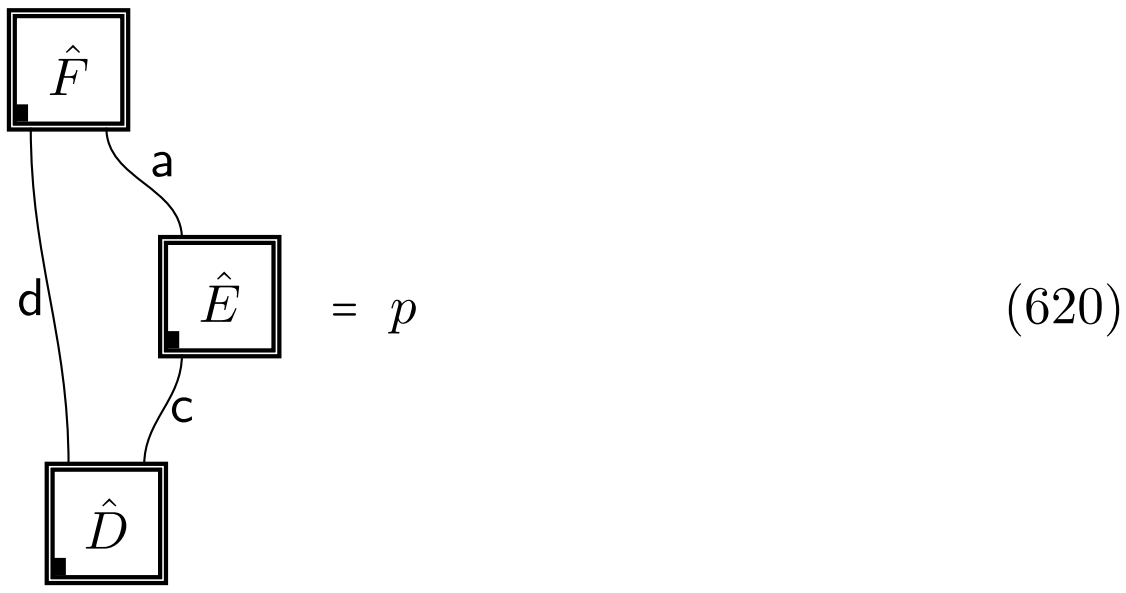

We can take the normal horizontal adjoint of any such equation. Then we obtain

$$\text{[circuit with } \hat{F}, \hat{E}, \hat{D} \text{ connected by wires a, c, d]} = \overline{p} \qquad (621)$$

By circuit reality (see Sec. 6.1) we want $p = \overline{p}$. This is guaranteed if the operator tensors are all equal to their horizontal adjoint - i.e. they are Hermitian as discussed in Sec. 31.3. To prove that such circuits equate to a real number only if the operator tensors are all Hermitian requires some assumptions and this is discussed elsewhere in this book (see Sec. 20, Sec. 42.1, and Sec. 23.2).

The basis independent normal conjupositions are $\{I, \overline{H}, \overline{V}, T\}$. Since they are basis independent, we should ask what they mean physically. Given that, in Quantum Theory, we impose Hermiticity on the operator tensors, $\overline{H}$ and $I$ have the same effect and $\overline{V}$ and $T$ have the same effect. Hence, we effectively have only one nontrivial basis independent conjuposition (let us take it to be the transpose $T$). If we have an equation involving operator tensors then we can obtain an equivalent equation by applying $T$. In particular, if we have a circuit equal to a probability, $p$, then the transposed circuit has the same probability,

$p$. For example, under $T$ the equation (620) becomes

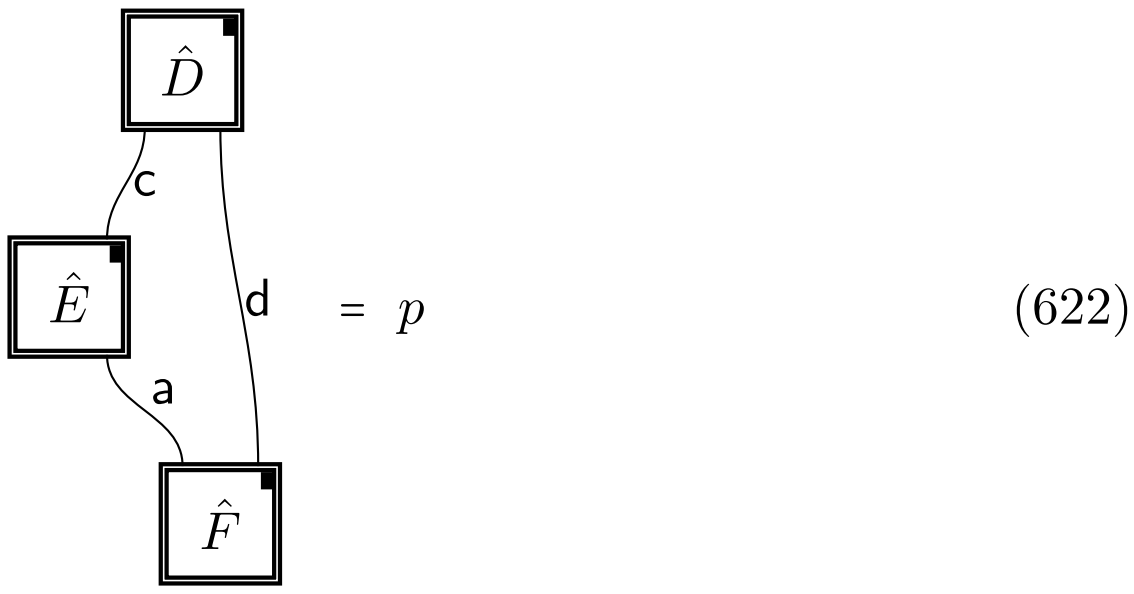

(622)

Here we have taken the transpose of each operator tensor and thereby inverted the circuit. The clear temptation is to interpret this as the time reverse and say that the normal transpose, $T$, acting on an operation gives us the time reverse of that operation. However, this is not quite right as we will now see.

If the normal transpose is to represent the time reverse then it should send the ignore preparation operator to the ignore result operator. However we actually have

(623)

The transpose does not get the gauge factors right. Now, this can easily be fixed by an appropriately rescaled transpose. In Sec. 33 we show how to motivate such a rescaling by working with an orthogonal basis whose elements are not normalised.

# 33 Natural conjupositions

We just saw (in Sec. 32.16) that the normal transpose does not transform between the ignore preparation operator and ignore result operator (see (623)) as we would require were the normal transpose a symmetry of the physics - there is a scaling issue. The same is true of the normal vertical adjoint for the same reason. This means that the normal conjupositions do not quite not represent the physical symmetry group we are seeking. Here we will set up the *natural conjuposition group* (that acts on Hilbert objects) which is the correct symmetry group for the time symmetric version of Quantum Theory (TSSOQT) we are studying here. This group does transform ignore operators appropriately. It also transforms the physicality conditions appropriately (both the $T$-positivity and the (general) double causality conditions) and so it is the appropriate group to implement the time reverse discussed in Sec. 6.4. We will discus physicality conditions and implementing the time reverse later in Sec. 33.16.

Natural conjupositions are enacted by expanding Hilbert objects and operator tensors in terms of what we will call *ortho-physical bases* and then applying the corresponding conjuposition (in the sense of 29.2) to the resulting expansion hypermatrix, then reattaching the appropriate ortho-physical bases. This will be illustrated in (636, 637) and (643, 585) below.

## 33.1 Ortho-physical bases

First we define the following scaled basis objects

$$ a,\ \mathsf{a} \quad := \quad \boxed{\gamma_{\mathsf{a}}}\ a,\ \mathsf{a} \qquad\qquad a,\ \mathsf{a} \quad := \quad a,\ \mathsf{a}\ \boxed{\overline{\gamma}_{\mathsf{a}}} $$

$$ \mathsf{a},\ a \quad := \quad \boxed{\gamma^{\mathsf{a}}}\ \mathsf{a},\ a \qquad\qquad \mathsf{a},\ a \quad := \quad \mathsf{a},\ a\ \boxed{\overline{\gamma}^{\mathsf{a}}} \tag{624} $$

We will call these an *ortho-physical basis set*. They are denoted by a small triangle marker (rather than the small optional square marker (see (590)) for the orthonormal basis set).

The gauge normalisation condition is

$$ |\gamma^{\mathsf{a}}\gamma_{\mathsf{a}}|^2 = \frac{1}{N_{\mathsf{a}}} \tag{625} $$

However, for physical conjuposition to behave appropriately on composite objects (see Sec. 33.14) we need to go further and impose that $\gamma^{\mathsf{a}}\gamma_{\mathsf{a}}$ is real. Thus, we have the condition

$$ \gamma^{\mathsf{a}}\gamma_{\mathsf{a}} = \frac{1}{N_{\mathsf{a}}^{\frac{1}{2}}} = \overline{\gamma}^{\mathsf{a}}\overline{\gamma}_{\mathsf{a}} \tag{626} $$

We call this the *ortho-physical gauge normalisation condition*. It implies the gauge normalisation condition (625). Further, it gives the ortho-physical conditions

$$ a,\ \mathsf{a},\ a \quad = \quad \boxed{\frac{1}{\sqrt{N_{\mathsf{a}}}}}\Big(a \qquad\qquad a,\ \mathsf{a},\ a \quad = \quad a\Big)\boxed{\frac{1}{\sqrt{N_{\mathsf{a}}}}} \tag{627} $$

(these are analogous to the orthonormality conditions in (500)). The condition that $\gamma^{\mathsf{a}}\gamma_{\mathsf{a}}$ is real imposes that the constants in (627) are real.

We can transform the basis of ortho-physical bases in the same way as we did for orthonormal bases

$$(628)$$

(compare this with (501)) where $U$ is a unitary matrix - i.e. it satisfies

$$(629)$$

$$(630)$$

It is easy to show that the new (shaded) bases satisfy the same ortho-physicality conditions as the original ortho-physical bases do (see (627)).

## 33.2 Objects constructed from ortho-physical basis

In Sec. 30.2 and Sec. 30.4 we defined multiple objects using orthonormal bases. These were various decompositions of the identity including the cap and cup, and various twist objects. We can define similar objects using ortho-physical basis elements.

We can use the ortho-physical basis to decompose the identity as follows

$$(631)$$

Again, we have used the ortho-physical gauge normalisation condition in (626). This equation should be compared with (506).

Comparing with (523) we see that the ignore operators are expressed in a straightforward way in terms of these

$$(632)$$

and similarly (comparing with (524)) ignore preparations are expressed in a straightforward way using the ortho-physical basis

$$(633)$$

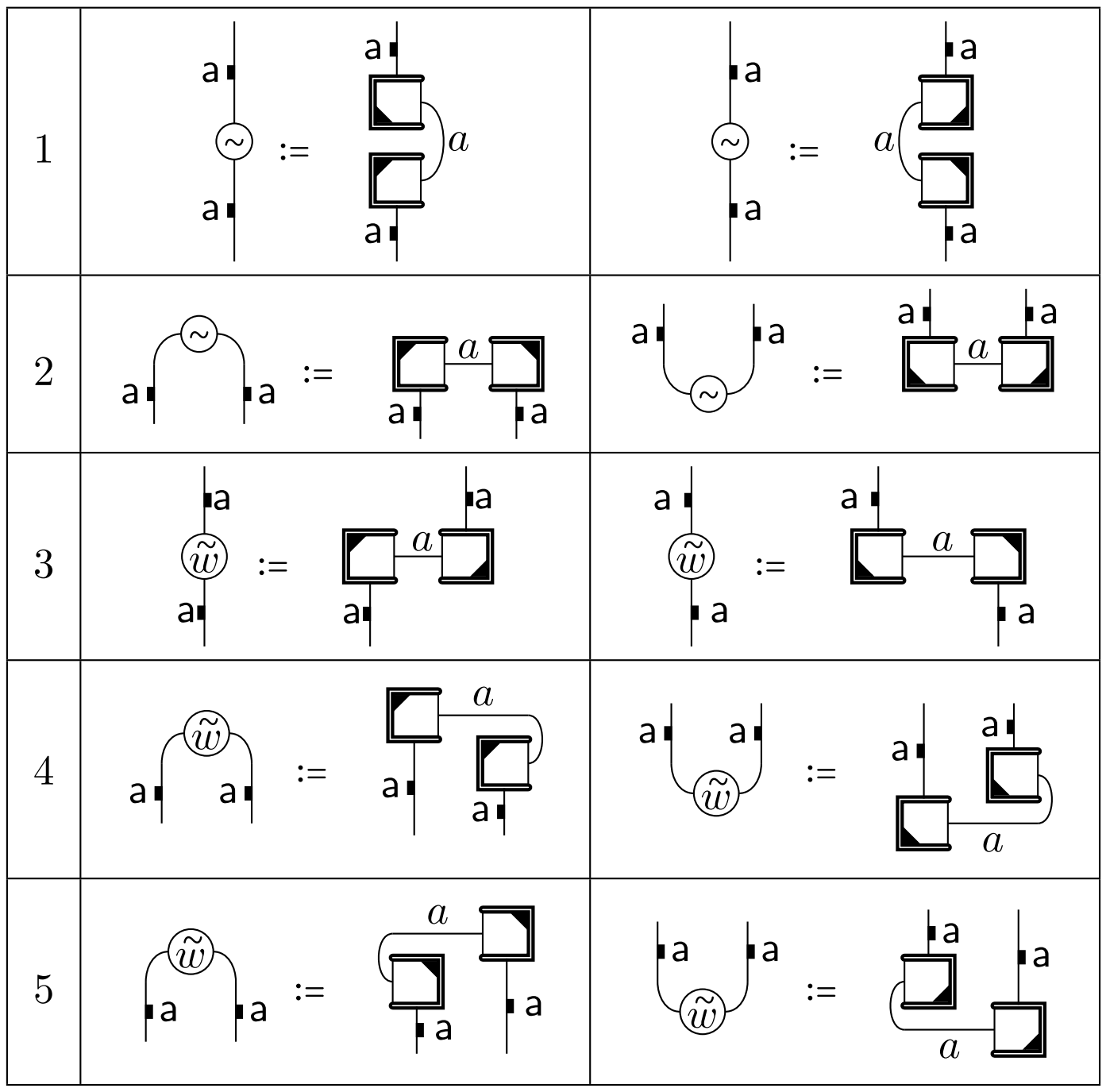

Table 4: This tabulates all the objects that can be built by joining ortho-physical basis elements at their label wires.

Reading off these equations we see that expansion matrix of the ignore operator in the ortho-physical basis is simply the identity. We define

(634)

We call these *physical caps* and *physical cups* respectively. We have

(635)

so physical caps and physical cups are just the ignore operators in a different guise.

In Table 4 we define all objects that can be formed by joining two basis elements at their label wires. Each of these objects is proportional to the corresponding object built out of orthonormal basis elements. We have already discussed the objects in rows 1 and 2 of the table in Sec. 33.1 (where these constants of proportionality were given). The objects in rows 3, 4, and 5 are *physical twist* objects (here *twist* is indicated by the use of the $w$ as "w" appears in the word "twist").

For normal conjupositions we saw that identity decompositions including caps and cups, and also twists play two roles. First they allowed us to perform the nonconjugating normal conjupositions in $\mathcal{S}_{\text{nonconjugating}} = \{I, H, V, T\}$ (as discussed in Sec. 32.6). Second, they will allow us to implement the theory of mirrors (see Sec. 34. We will see that the objects in Table 4 do allow us to set up the theory of physical mirrors. However, they are not normalised in the right way for implementing non-conjugating physical conjupositions. For that purpose we need to define special caps, cups, and twists defined in Sec. 33.10. Further, we will need special caps and cups to prove a *physical mirror theorem* (see Sec. 35.3).

## 33.3 Natural conjupositions of Hilbert objects

We define natural conjupositions by conjuposing the expansion hypermatrix we obtain using ortho-physical bases to expand the given Hilbert object. To see how this works consider

(636)

(637)

(compare with (577, 578)). The above four objects form the *natural Hilbert square*. As we are expanding in the ortho-physical basis, we use the notation ⦅$B$ instead of ⌈$B$ and $B$⦆ instead of $B$⌉. Going between ⦅$B$ and $B$⦆ involves conjugation (as with ⌈$B$ and $B$⌉).

It is interesting to compare the effect of applying a natural conjuposition with the effect of applying the corresponding normal conjuposition. To make this comparison we can substitute in the expressions in (624) expressing (636) and (637) in terms of $\gamma$ factors and orthonormal bases.

Going from the left object to the right object in (636) is a *natural horizontal adjoint* transformation which we denote by $\underset{\sim}{\overline{H}}$. On the left we have a $\gamma_{\mathsf{a}}\gamma^{\mathsf{b}}$ coefficient (when we substitute in (624)). On the right we have a $\overline{\gamma}_{\mathsf{a}}\overline{\gamma}^{\mathsf{b}}$ coefficient. If we had, instead, taken the normal horizontal adjoint, $\overline{H}$, of the left object then the $\gamma_{\mathsf{a}}\gamma^{\mathsf{b}}$ coefficient would be conjugated becoming $\overline{\gamma}_{\mathsf{a}}\overline{\gamma}^{\mathsf{b}}$. Thus we get the same right object. Thus the natural horizontal adjoint is the same as the normal horizontal adjoint. That is

$$\underset{\sim}{\overline{H}} = \overline{H} \tag{638}$$

(indeed, note that the above reasoning holds for any orthogonal basis normalised or not). We will see that a similar result does not hold for the other conjupositions.

Going from the left object in (636) (which has coefficient $\gamma_{\mathsf{a}}\gamma^{\mathsf{b}}$) to the left object in (637) (which has coefficient $\gamma^{\mathsf{a}}\gamma_{\mathsf{b}}$) is a *natural vertical adjoint* transformation which we denote by $\underset{\sim}{\overline{V}}$. This is not the same as the $\overline{V}$ transformation. If we perform $\overline{V}$ on the left object in (636) then the coefficient is conjugated and becomes $\overline{\gamma}_{\mathsf{a}}\overline{\gamma}^{\mathsf{b}}$. Thus, if we first perform $\overline{V}$ then apply the conversion factor

$$\kappa_{\mathsf{a}}^{\mathsf{b}}\left[\overline{V} \to \underset{\sim}{\overline{V}}\right] = \frac{\gamma_{\mathsf{b}}\gamma^{\mathsf{a}}}{\overline{\gamma}^{\mathsf{b}}\overline{\gamma}_{\mathsf{a}}} \tag{639}$$

then we obtain the natural vertical adjoint, $\underset{\sim}{\overline{V}}$. Note that we have indicated by subscripts and superscripts on the $\kappa$ symbol that the input and output of the original object are $\mathsf{a}$ and $\mathsf{b}$ respectively. We should also indicate that this is applied to a left (rather than right) Hilbert object. However, interestingly, if we apply the ortho-physical gauge normalisation condition (626) we can show

$$\kappa_{\mathsf{a}}^{\mathsf{b}}\left[\overline{V} \to \underset{\sim}{\overline{V}}\right] = \left|\frac{\gamma^{\mathsf{a}}}{\gamma^{\mathsf{b}}}\right|\sqrt{\frac{N_{\mathsf{a}}}{N_{\mathsf{b}}}} \in \mathbb{R} \tag{640}$$

Since this conversion factor is real we obtain the same conversion factor if we compare $\overline{V}$ and $\underset{\sim}{\overline{V}}$ acting on a right Hilbert object.

Going from the left object in (636) to the right object in (637) is the natural transpose which we denote by $\underset{\sim}{T}$. It is easy to show that

$$\kappa_{\mathsf{a}}^{\mathsf{b}}\left[T \to \underset{\sim}{T}\right] = \frac{\overline{\gamma}_{\mathsf{b}}\overline{\gamma}^{\mathsf{a}}}{\gamma^{\mathsf{b}}\gamma_{\mathsf{a}}} \tag{641}$$

If we impose the ortho-physical gauge normalisation condition (626) then we obtain

$$\kappa_{\mathsf{a}}^{\mathsf{b}}\left[T \to \underset{\sim}{T}\right] = \left|\frac{\gamma^{\mathsf{a}}}{\gamma^{\mathsf{b}}}\right|\sqrt{\frac{N_{\mathsf{a}}}{N_{\mathsf{b}}}} \in \mathbb{R} \tag{642}$$

(which is the same as the conversion factor for the natural vertical adjoint). Since this is real, we get the same conversion factor if we start with a right Hilbert object.

We can also write

(643)

(644)

for the shadow Hilbert square objects.

We can transform between the eight objects above by conjuposing the expansion matrix. This gives the full natural conjuposition group. We will denote elements of the natural conjuposition group as follows

$$\mathcal{S}_{\rm phys} = \{I, \overline{\underset{\sim}{I}}, \underset{\sim}{H}, \overline{\underset{\sim}{H}}, \underset{\sim}{V}, \overline{\underset{\sim}{V}}, \underset{\sim}{T}, \overline{\underset{\sim}{T}}\} \tag{645}$$

As noted above, $\overline{\underset{\sim}{H}} = \overline{H}$. Also, clearly the identity transformation, $I$, is the same in the two cases.

We might wonder whether $\underset{\sim}{H} = H$. This is not generally the case unless $\gamma^{\mathsf{b}}\gamma_{\mathsf{a}}$ is real or imaginary. Indeed, examining (636, 637, 643, 644) we see that the conversion factor is

$$\kappa^{\mathsf{b}}_{\mathsf{a}}\left[\text{L to R}, H \to \underset{\sim}{H}\right] = \frac{\overline{\gamma}^{\mathsf{b}}\overline{\gamma}_{\mathsf{a}}}{\gamma^{\mathsf{b}}\gamma_{\mathsf{a}}} \tag{646}$$

This conversion factor is complex (unless $\gamma^{\mathsf{b}}\gamma_{\mathsf{a}}$ is real or imaginary) and of magnitude 1. We have indicated that this is a left (L) to right (R) transformation. If consider starting with the right Hilbert object in (584) then we obtain, instead,

$$\kappa^{\mathsf{b}}_{\mathsf{a}}\left[\text{R to L}, H \to \underset{\sim}{H}\right] = \frac{\gamma^{\mathsf{b}}\gamma_{\mathsf{a}}}{\overline{\gamma}^{\mathsf{b}}\overline{\gamma}_{\mathsf{a}}} \tag{647}$$

which is the complex conjugate of the expression in (646).

We can obtain the $\overline{\underset{\sim}{I}}$ transformation on the left object in (584) by performing the $\overline{I}$ transformation then multiplying by the conversion factor

$$\kappa^{\mathsf{b}}_{\mathsf{a}}\left[\text{L to L}, \overline{I} \to \overline{\underset{\sim}{I}}\right] = \frac{\gamma^{\mathsf{b}}\gamma_{\mathsf{a}}}{\overline{\gamma}^{\mathsf{b}}\overline{\gamma}_{\mathsf{a}}} \tag{648}$$

while, if we do this on the right object in (584), the conversion factor is the complex conjugate

$$\kappa^{\mathsf{b}}_{\mathsf{a}}\left[\text{R to R}, \overline{I} \to \overline{\underset{\sim}{I}}\right] = \frac{\overline{\gamma}^{\mathsf{b}}\overline{\gamma}_{\mathsf{a}}}{\gamma^{\mathsf{b}}\gamma_{\mathsf{a}}} \tag{649}$$

We have that $\overline{\underset{\sim}{I}} = \overline{I}$ if $\gamma^{\mathsf{b}}\gamma_{\mathsf{a}}$ is real.

One special case of particular interest is when the output system is of the same type as the input system for our Hilbert object (so $\mathsf{b} = \mathsf{a}$). In this case we can see, by inspecting the above equations that, if the ortho-physical gauge condition (626) holds, then

$$\kappa^{\mathsf{a}}_{\mathsf{a}}\left[\text{all cases}\right] = 1 \tag{650}$$

This means that if we have the same output as input system types, a standard conjupositions has the same effect as the corresponding physical conjuposition (if the ortho-physical gauge condition holds).

## 33.4 Gauge fixing of $\gamma$'s

There is a natural temptation to fix the the $\gamma$'s. We could adopt the convention that

$$\gamma^{\mathsf{a}} = 1 \tag{651}$$

for all $\mathsf{a}$. Then we have $\gamma_{\mathsf{a}} = \frac{1}{\sqrt{N_{\mathsf{a}}}}$. As discussed at the end of Sec. 24, this is the convention in standard Quantum Theory (whereby application of the ignore operator corresponds to the mathematical trace operation). However, this convention is time asymmetric and obscures the underlying time-symmetry in the time symmetric formulism.

We could, instead, choose a time symmetric gauge fixing condition such as

$$\gamma^{\mathsf{a}} = \gamma_{\mathsf{a}} = \frac{1}{N_{\mathsf{a}}^{\frac{1}{4}}} \tag{652}$$

This convention is explicitly time-symmetric. However, it stronger than we need for our purposes and may obscure interesting physics. Further, we do not need to adopt this convention for our formalism to be time symmetric.

It is more interesting to consider *partial* gauge fixing conditions. In fact, we have already introduced the ortho-physical gauge normalisation condition in (626), that $\gamma^{\mathsf{a}}\gamma_{\mathsf{a}} = \frac{1}{\sqrt{N_{\mathsf{a}}}}$. This is useful since it allows for physical conjuposition to preserve the compositional form of composite objects (as will be discussed in Sec. 33.14). Further, it makes the conversion factor in (639) and (641) real and independent of whether we start with a left or right Hilbert object. For these reasons we will mostly use the ortho-physical gauge normalisation condition (though it is often interesting to look at the functional forms without using this condition).

We can go a bit further than the ortho-physical gauge normalisation condition. In particular, consider the following

> **The $\gamma(N)$ gauge**. We demand that, for any system type $\mathsf{a}$, we have
>
> $$\gamma^{\mathsf{a}} = \gamma(N_{\mathsf{a}}) \quad \text{and} \quad \gamma_{\mathsf{a}} = \frac{1}{\sqrt{N_{\mathsf{a}}}}\gamma^{-1}(N_{\mathsf{a}}) \tag{653}$$
>
> where $\gamma(N)$ is any (non-zero) complex-valued function of the integer, $N$, satisfying the completely multiplicative property
>
> $$\gamma(MN) = \gamma(M)\gamma(N) \tag{654}$$
>
> for all $M$ and $N$.

The completely multiplicative property is necessary so that composite systems behave appropriately. If we have a composite system $\mathsf{cd}$ then we can regard it as a single system with $\gamma^{\mathsf{cd}} = \gamma(N_{\mathsf{cd}})$ or we can regard it as a composite system whereby the $\gamma$ factors multiply giving $\gamma^{\mathsf{c}}\gamma^{\mathsf{d}} = \gamma(N_{\mathsf{c}})\gamma(N_{\mathsf{d}})$. Equating these two and using the fact that $N_{\mathsf{cd}} = N_{\mathsf{c}}N_{\mathsf{d}}$ gives the completely multiplicative property.

The $\gamma(N)$ gauge satisfies the ortho-physical gauge normalisation condition (626). Further, it guarantees the following very useful property

$$\kappa_{\mathsf{a}}^{\mathsf{b}}\left[\overline{V} \to \underset{\sim}{\overline{V}}\right] = \kappa_{\mathsf{a}}^{\mathsf{b}}\left[T \to \underset{\sim}{T}\right] = 1 \qquad \text{if} \quad N_{\mathsf{a}} = N_{\mathsf{b}} \tag{655}$$

as is clear from (640, 642). This property turns out to be interesting when we look at unitaries. In Sec. 38 we will see that, under the $\gamma(N)$ gauge, natural unitaries (defined with respect to natural conjupositions) are also normal unitaries (defined with respect to normal conjuposions). This offers a path to formulate time symmetric Quantum Theory using normal conjupositions. However, the natural conjuposition group is much more natural and so it is better, in the end, not to employ this partial gauge fixing.

Before we finish this topic, it is worth noting we can go a bit further and require that the function, $\gamma(N)$, above is real valued. We will call this the *real* $\gamma(N)$ *gauge*. If we do this then

$$\kappa_{\mathsf{a}}^{\mathsf{b}}\left[\text{L to R}, H \to \underset{\sim}{H}\right] = \kappa_{\mathsf{a}}^{\mathsf{b}}\left[\text{R to L}, H \to \underset{\sim}{H}\right] = 1 \tag{656}$$

and

$$\kappa_{\mathsf{a}}^{\mathsf{b}}\left[\text{L to L}, \overline{I} \to \underset{\sim}{\overline{I}}\right] = \kappa_{\mathsf{a}}^{\mathsf{b}}\left[\text{R to R}, \overline{I} \to \underset{\sim}{\overline{I}}\right] = 1 \tag{657}$$

This is clear from (646, 647, 648, 649). The real $\gamma(N)$ gauge guarantees that, when $N_{\mathsf{a}} = N_{\mathsf{b}}$, natural conjupositions have the same effect as normal conjupositions.

It is worth noting that both the time forward gauge fixing convention in (651) and the time symmetric gauge fixing convention in (652) are consistent with the real $\gamma(N)$ gauge.

## 33.5 The natural Hilbert cube

We can arrange the Hilbert objects obtained by applying the natural conjuposition group to some given object as a *natural Hilbert cube*. If we start with a left (or right) Hilbert object, we obtain

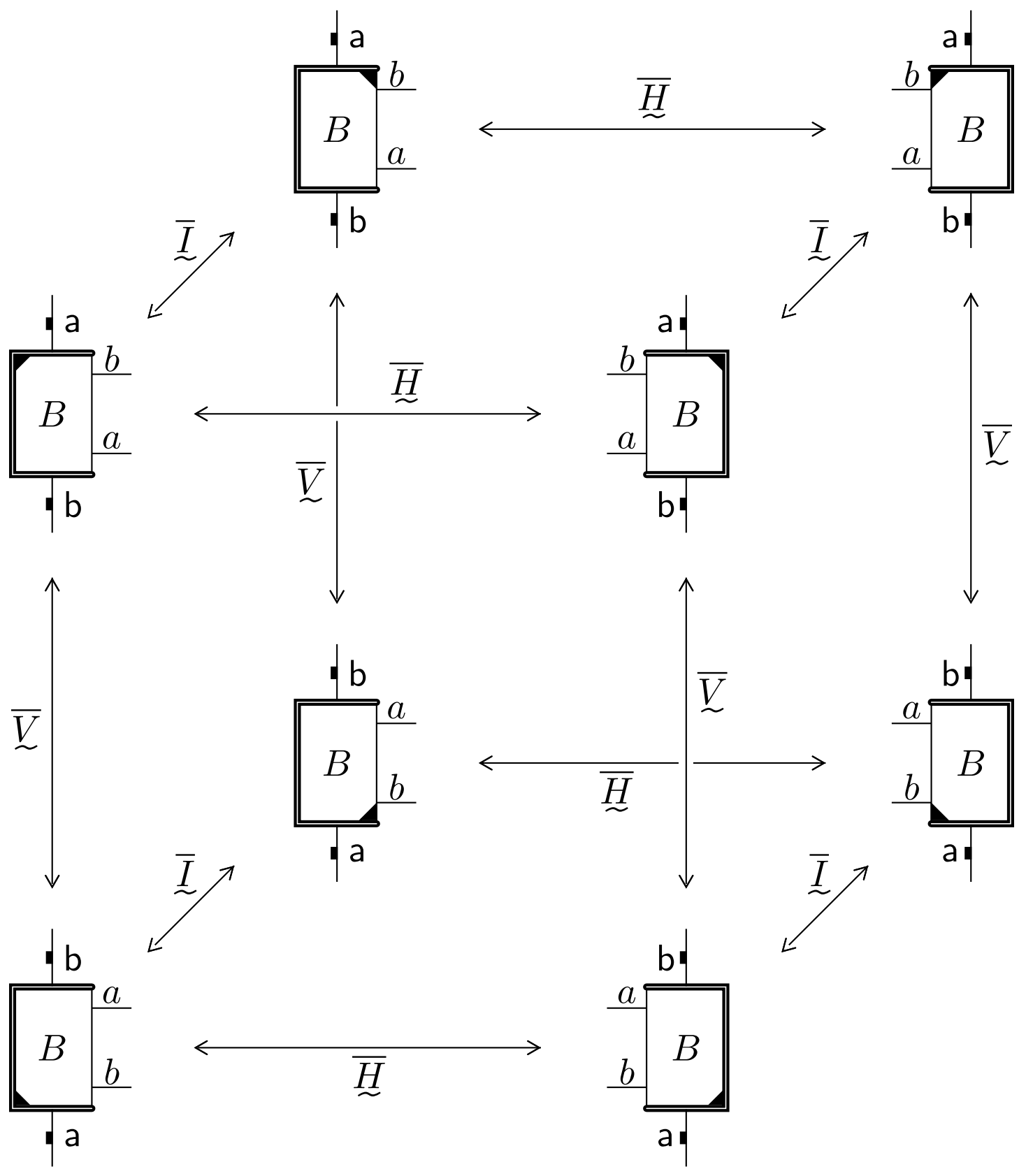

(658)

(compare with (586)). The natural conjupositions transform between the elements of this Hilbert cube. In Sec. 32.8 we made various comments on the normal Hilbert cube, in particular about the many squares to be found in that cube. We can make similar comments about the natural Hilbert cube. The counterparts to the squares considered there are associated with corresponding subgroups of the natural conjuposition group with similar properties.

## 33.6 Natural conjupositions of operator tensors

The natural conjupositions of operator tensors works in the same way as normal conjuposition of operator tensors (as discussed in Sec. 32.14) except, now, we expand with respect to ortho-physical bases. Consider

(659)

The natural vertical adjoint of this is given by

(660)

The other natural conjupositions are defined similarly.

We can also implement natural conjupositions directly if the starting operator tensor is written in the form

(661)

then we can apply some natural conjuposition to each of the left and right objects composing this operator tensor to effect the given natural conjuposition on the operator tensor. This gives the same result as expanding in terms of a ortho-physical basis and conjuposing the expansion matrix (we leave it as an exercise to the reader to prove this). For example

(662)

The other natural conjupositions can be obtained similarly.

## 33.7 Natural conjupositions of ignore operators

We are now in a position to see that the ignore operators transform appropriately under natural conjupositions. Indeed, it is clear from the above remarks that

$$\text{(663)}$$

(where we have now included the triangle markings on the $\hat{\boldsymbol{I}}$'s for clarity). We can see this either by considering the equation on the left as providing the expansion of the ignore operation result in the ortho-physical basis. Then the expansion matrix is simply the identity and so the natural transpose is given by the equation on the right. Or we can apply (662).

This proves that natural conjupositions fix the problem we identified with normal conjupositions in Sec. 32.16.

## 33.8 Ortho-deterministic bases

Here we introduce the *ortho-deterministic basis* which is, in some sense, the complement of the ortho-physical basis. This new basis is defined as follows

$$\text{(664)}$$

where the $\eta$'s are chosen such that

$$\text{(665)}$$

This means

$$\gamma^{\mathsf{a}}\eta_{\mathsf{a}} = 1 \qquad\qquad \eta^{\mathsf{a}}\gamma_{\mathsf{a}} = 1 \tag{666}$$

If we use the ortho-physical gauge normalisation condition in (626) then we obtain

$$\eta_{\mathsf{a}} = \sqrt{N_{\mathsf{a}}}\gamma_{\mathsf{a}} \qquad\qquad \eta^{\mathsf{a}} = \sqrt{N_{\mathsf{a}}}\gamma^{\mathsf{a}} = 1 \tag{667}$$

Thus, the ortho-deterministic bases are given by multiplying the ortho-physical bases by $\sqrt{N_{\mathsf{a}}}$.

It is worth noting that combining ortho-deterministic and ortho-physical bases gives us useful identities

$$\text{a} \quad a \quad = \quad a \quad = \quad \text{a} \qquad\qquad a \quad = \quad a \quad = \quad \text{a} \tag{668}$$

These follow from (666).

It is natural to ask why we used the ortho-physical rather than the ortho-deterministic basis to define the natural conjuposition group. Interestingly, with one small assumption, it does not actually matter which we use. We will prove the following

> **Natural conjuposition group equivalence.** If we assume the ortho-physical gauge normalisation condition (626), then the action of conjuposition group on Hilbert objects defined with respect to the ortho-deterministic basis is the same as the of the conjuposition group on Hilbert objects when defined with respect to the ortho-physical basis - i.e. the natural conjuposition group.

To see this recall (from Sec. 33.3) that the conversion factor for $V$ to $\utilde{V}$ is given by

$$\kappa^{\mathsf{b}}_{\mathsf{a}}\left[\overline{V} \to \overline{\utilde{V}}\right] = \frac{\gamma_{\mathsf{b}}\gamma^{\mathsf{a}}}{\overline{\gamma}^{\mathsf{b}}\overline{\gamma}_{\mathsf{a}}} \tag{669}$$

If we follow similar logic to that in Sec. 33.3 for the ortho-deterministic bases then we obtain the conversion factor

$$\kappa^{\mathsf{b}}_{\mathsf{a}}\left[\overline{V} \to \overline{V}_{\text{det}}\right] = \frac{\eta_{\mathsf{b}}\eta^{\mathsf{a}}}{\overline{\eta}^{\mathsf{b}}\overline{\eta}_{\mathsf{a}}} \tag{670}$$

Now, since the $\eta$'s are related to the $\gamma$'s by (667), this is actually the same conversion factor as in (669) (recall that we used the ortho-physical gauge normalisation condition to obtain (667)). Similar remarks pertain to the other conversion factors. Thus, the conjuposition group obtained from the ortho-deterministic bases is the same as that obtained from the ortho-physical basis (under the ortho-physical gauge normalisation condition). We will always assume the ortho-physical gauge normalisation condition when necessary so we only have one new conjuposition group associated with the ortho-physical and the ortho-deterministic bases.

## 33.9 Interpretation of bases

We can provide an interpretation of ortho-physical bases and ortho-deterministic bases by using these elements to build twofold operators. Thus, for the ortho-physical basis we can build preparation and result operators

$$\hat{X}_{\text{phys}}(a) \;:=\; \cdots \qquad\qquad \hat{Y}_{\text{phys}}(a) \;:=\; \cdots \tag{671}$$

These operators are not deterministic since

$$\frac{\hat{\boldsymbol{I}}}{\hat{X}_{\text{phys}}(a)} = \frac{1}{N_{\mathsf{a}}} \qquad\qquad \frac{\hat{Y}_{\text{phys}}(a)}{\hat{\boldsymbol{I}}} = \frac{1}{N_{\mathsf{a}}} \tag{672}$$

However, they are physical since they are in twofold form (so satisfy $T$-positivity) and we can see that they satisfy the double causality conditions (160,161) under correspondence. The latter gives

$$\hat{X}_{\text{phys}}(a) \underset{T}{\leq} \hat{\boldsymbol{I}} \qquad\qquad \hat{Y}_{\text{phys}}(a) \underset{T}{\leq} \hat{\boldsymbol{I}} \tag{673}$$

It is easy to see that these conditions are satisfied. Thus, the name orthophysical basis is warranted. The state and effect in (671) are clearly homogeneous. In fact they are, further, pure (see Fig. 1 for an illustration of these concepts). This follows from the theorem in Sec. 25. In particular, compare (672) above with (444).

Similarly, using the deterministic basis we can build preparation and result operators

$$\hat{X}_{\text{det}}(a) \;:=\; \cdots \qquad\qquad \hat{Y}_{\text{det}}(a) \;:=\; \cdots \tag{674}$$

These operators are deterministic since

$$\frac{\hat{\boldsymbol{I}}}{\hat{X}_{\text{det}}(a)} = 1 \qquad\qquad \frac{\hat{Y}_{\text{det}}(a)}{\hat{\boldsymbol{I}}} = 1 \tag{675}$$

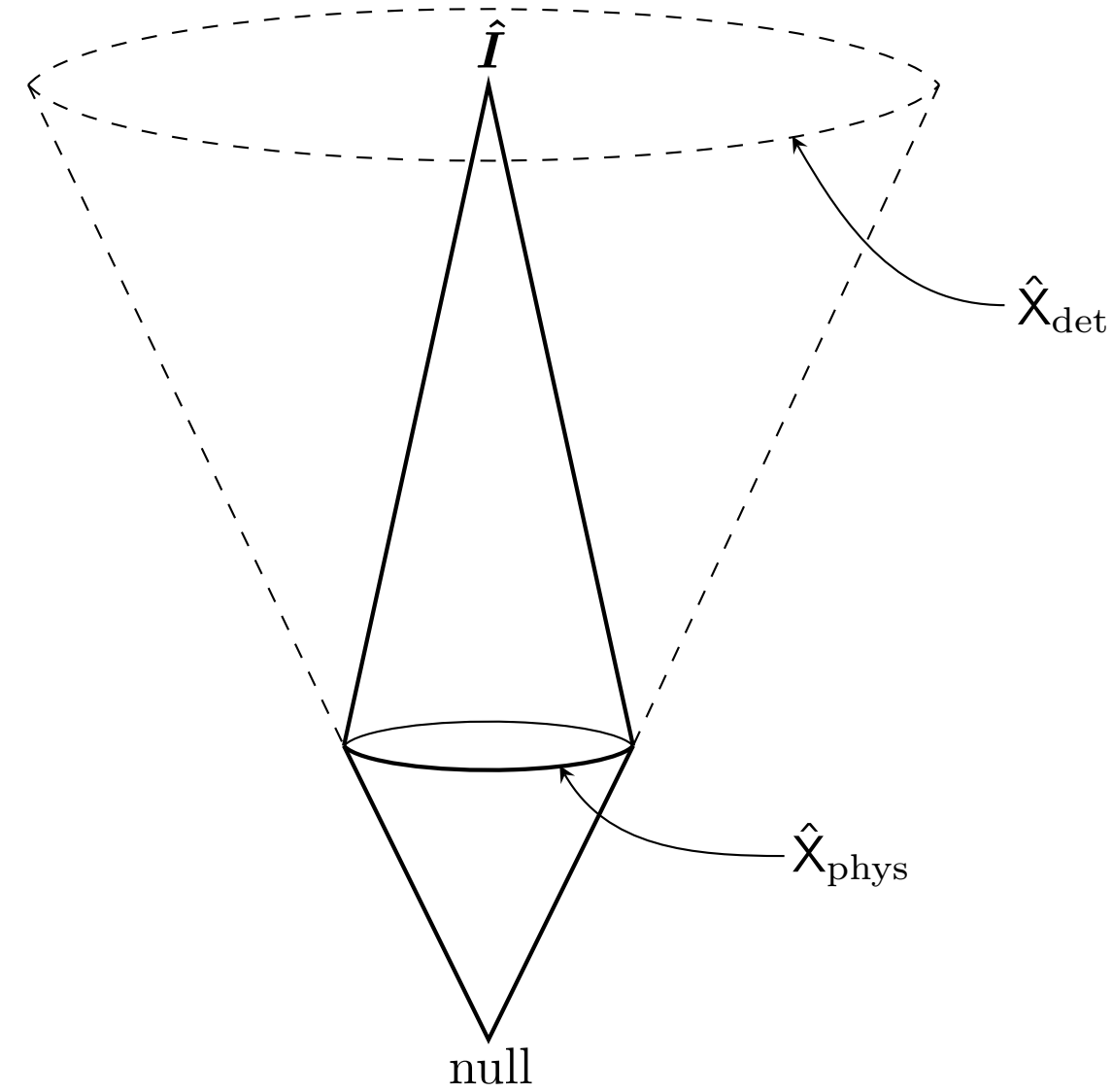

Figure 2: This diagram illustrates the convex space for preparation operators. The dashed fragment is not part of the allowed physical space. The preparation $\hat{X}_{\mathrm{phys}}$ is physical but not deterministic. The preparation $\hat{X}_{\mathrm{det}}$ is deterministic but not physical. A similar diagram pertains for the results $\hat{Y}_{\mathrm{phys}}$ and $\hat{Y}_{\mathrm{det}}$.

However, they are not physical since

$$\overset{\mathsf{a}}{\boxed{\hat{X}_{\mathrm{det}}(a)}} \neq \overset{\mathsf{a}}{\boxed{\hat{\boldsymbol{I}}}} \qquad\qquad \underset{\mathsf{a}}{\boxed{\hat{Y}_{\mathrm{det}}(a)}} \neq \underset{\mathsf{a}}{\boxed{\hat{\boldsymbol{I}}}} \tag{676}$$

in violation of the backward and forward causality conditions respectively (which have the consequence that a physical deterministic preparation/result is necessarily equal to the ignore preparation/result). A geometric interpretation is provided in Fig. 2.

## 33.10 Special caps, cups, and twists

We can obtain useful objects when we join an ortho-deterministic basis element to an ortho-physical element at the label wire. We will call these objects *the specials*. There are twenty ways of doing this. However, it turns out it does not matter which element is which (since the ortho-deterministic element simply contributes an overall factor of $\sqrt{N_{\mathsf{a}}}$. Removing this degeneracy, we obtain ten distinct specials. These ten specials are shown in Table 5. The decompositions of the identity in row 1 were already discussed. We will call the remaining objects *special caps*, *special cups*, and *special twists*. The objects in rows 4 and 5 are *twisted special caps and cups* and are, in fact, what would normally be called maximally entangled states and effects (with a particular normalisation).

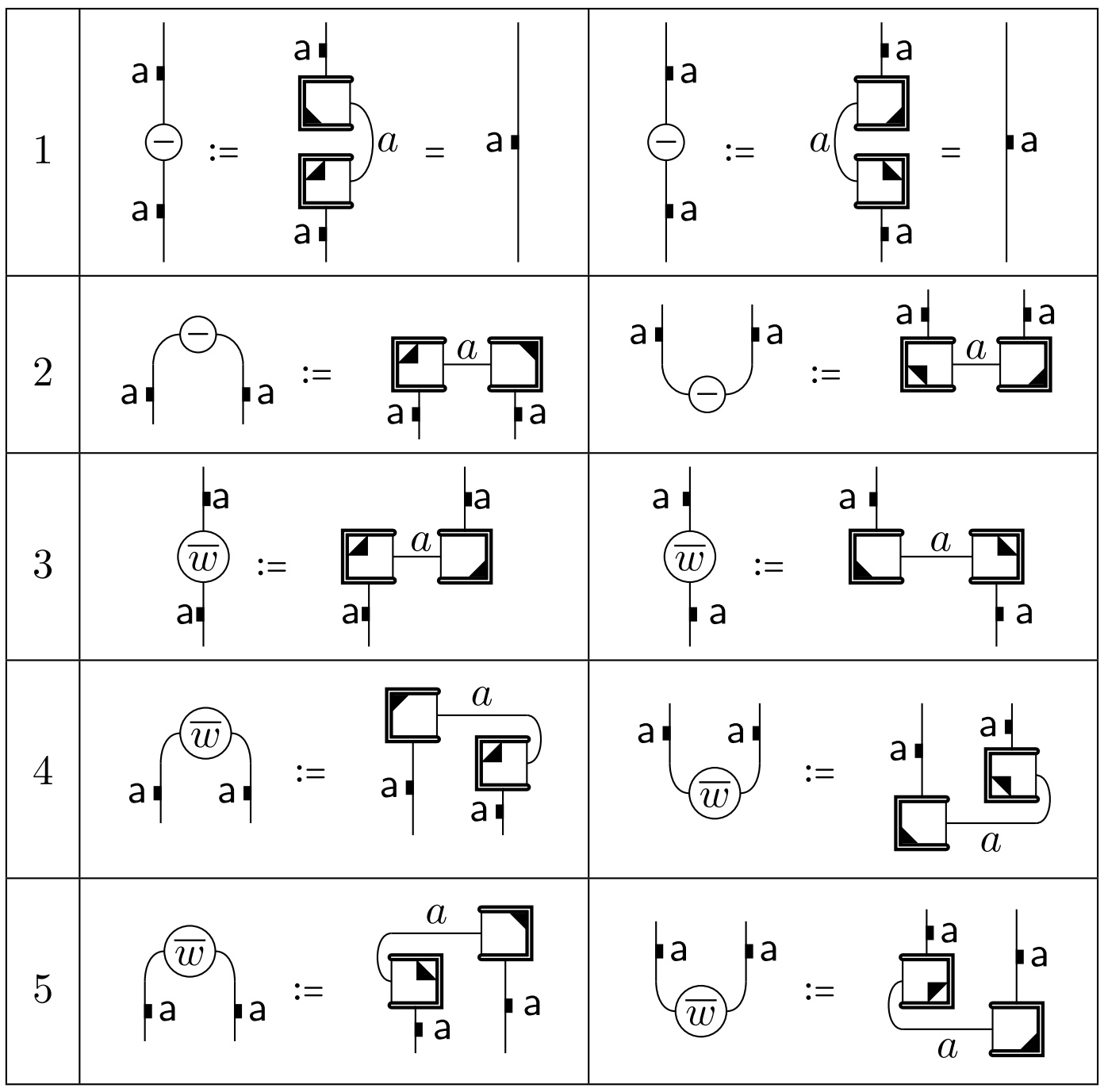

Table 5: This tabulates *the specials* - all the objects that can be built by joining an ortho-physical basis element to an ortho-deterministic basis element at the label wire. An ortho-deterministic element is equal to $\sqrt{N_{\mathsf{a}}}$ times the corresponding ortho-physical element. Consequently, we get the same result for the objects in this table if we interchange the position of the ortho-physical and the ortho-deterministic element in each case. In row 1 we have used (666) to simplify.

The special caps, cups, and twists can be used to implement the nonconjugating natural conjupositions (just as (standard) caps, cups, and twists were used to implement the (standard) nonconjugating conjupositions as discussed in Sec. 32.3) and Sec. 32.6).

The special cap is

(677)

The special cup is

(678)

We are using the ortho-physical gauge normalisation condition in (626) to obtain the equalities above. Note that, if we do not use this condition, then these special caps and cups have a "sense" (either clockwise or anticlockwise).

The special caps and cups have the following properties

(679)

and

(680)

It is convenient to define

(681)

since sometimes our cups and caps may bend so that the bobbles are on the inside.

The special cups and caps satisfy yanking equations as follows

(682)

These are easily proven using the definitions above.

The special twist objects are

(683)

and

(684)

They have the following properties

(685)

and

(686)

These properties are useful below.

It is easy to see that we have the special twist annihilation properties

(687)

using the defining equations for these special twists above. We can also form yanking equations using the special twists

(688)

Here we are using the twisted special cap and cup in row 4 of Table 5. We have a similar yanking equation using the objects in row 5 of this table.

## 33.11 Using special caps, cups, and twists for nonconjugating transformations

We saw in Sec. 32.6, we can implement nonconjugating normal conjupositions using caps, cups, and twists. Here we will see that we can implement the nonconjugating natural conjupositions ($\mathcal{S}_{\text{nonconjugating}} = \{I, \underset{\sim}{H}, \underset{\sim}{V}, \underset{\sim}{T}\}$) using special caps, special cups, and special twists. In fact, these special caps, cups, and twists are guaranteed to do the job since they convert between ortho-physical basis elements appropriately as we saw in Sec. 33.10. It is, however, instructive to see how they work. Since $\underset{\sim}{V} = \underset{\sim}{T}\underset{\sim}{H}$ it is enough to show how just $\underset{\sim}{T}$ and $\underset{\sim}{H}$ work.

First consider the natural transpose. By analogy with (568) (for the standard case) we have

(689)

Comparing this with (568) it is easy to show that the natural transpose, $\underset{\sim}{T}$ is given by multiplying the normal transpose, $T$, by

$$\kappa_{T\leftarrow\underset{\sim}{T}} = \eta_{\mathsf{b}}\overline{\gamma}_{\mathsf{b}}\eta^{\mathsf{a}}\overline{\gamma}^{\mathsf{a}} = \frac{\overline{\gamma}_{\mathsf{b}}\overline{\gamma}^{\mathsf{a}}}{\gamma^{\mathsf{b}}\gamma_{\mathsf{a}}} \tag{690}$$

using the definitions of the ortho-physical and ortho-deterministic bases and using (666) to prove the second equality. We obtained this result previously (in Sec. 33.3) by inspecting the hypermatrix expanded forms of these Hilbert objects.

Expanding out the left Hilbert object being naturally transposed gives

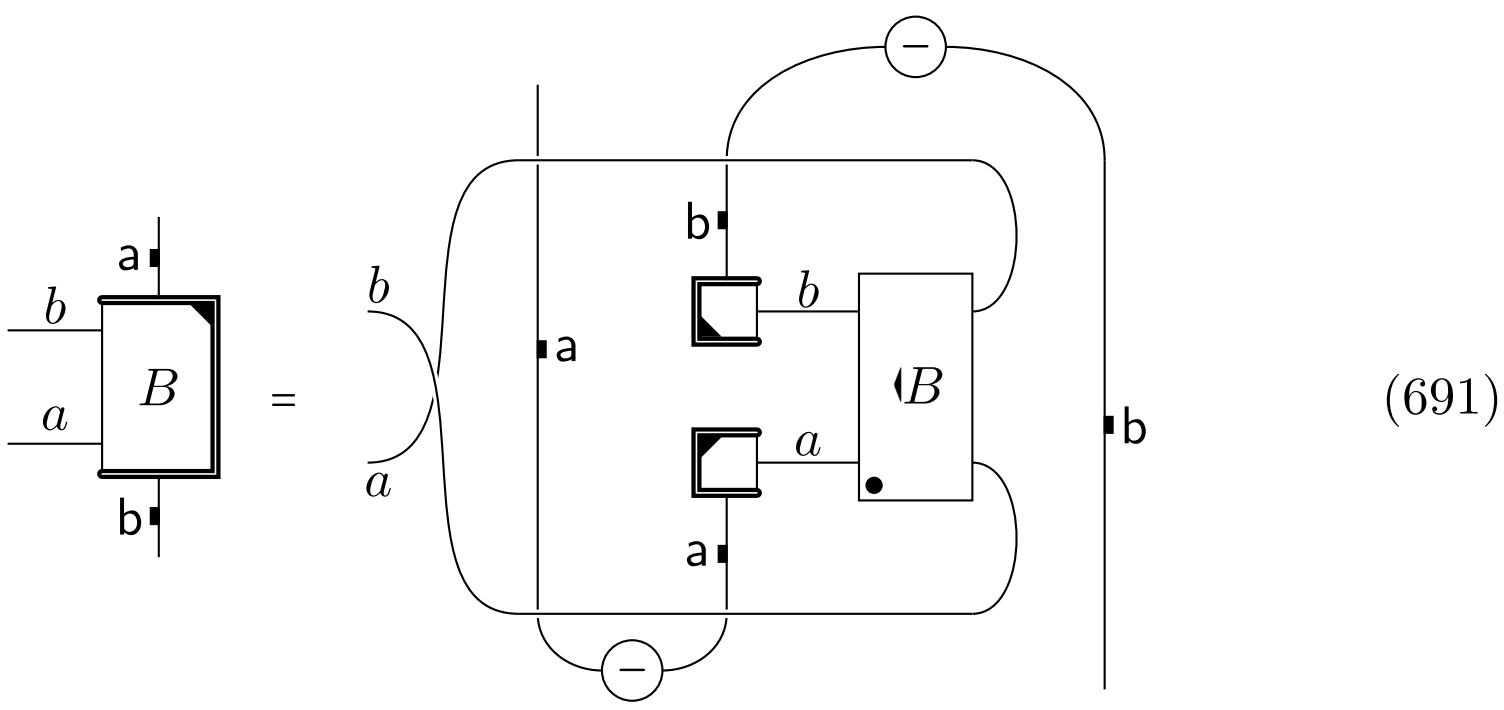

(691)

Using the definitions of special caps and cups in (677, 678) and the definition of the transpose of a hypermatrix in (477) we obtain

(692)

which is the correct expansion for the naturally transposed object.

The natural horizontal transpose is given by

(693)

This is analogous to (579) and can be regarded as the definition of the natural horizontal transpose. It is easy to verify this agrees with the definition of the natural horizontal transpose in terms of hypermatrix expansions by following analogous steps to those in (580) (where orthonormal bases is replaced ortho-physical bases and twists are replaced by physical twists). Also note that, if $\mathsf{b} = \mathsf{a}$, then the coefficients coming from the special twists cancel and (693) agrees with (579) (so $\underset{\sim}{H}$ agrees with $H$ in this case).

## 33.12 Basis independence in the natural conjuposition group

We saw how to perform basis transformation for the ortho-physical bases in (628). We can prove that the transformations in $\{I, \underset{\sim}{\overline{H}}, \underset{\sim}{\overline{V}}, \underset{\sim}{T}\}$ are basis independent. This is obvious for $I$.

Basis independence of $\underset{\sim}{\overline{H}}$ follows using exactly analogous reasoning as was used to prove basis independence for $\overline{H}$ in Sec. 32.2.

Basis independence of the natural transpose follows from the fact that we can implement it using special cups and caps (as discussed in Sec. 33.11). Further, it is easy to show that these special cups and caps are basis independent.

The natural vertical adjoint is obtained by $\underset{\sim}{T}\underset{\sim}{\overline{H}}$ and is, consequently also basis independent.

## 33.13 Natural conjupositions of ortho-physical basis elements

For orthonormal bases we saw that normal conjugation, $\overline{I}$, leaves basis elements unchanged (see (591) and the discussion below). Similarly, natural conjugation, $\overline{\underset{\sim}{I}}$, leaves ortho-physical basis elements unchanged. To see this note that

(694)

where the first step is an application of using the special twist to take the physical horizontal transpose (as in (693)) and the second step is from the way the special twist is defined (see (683)). Similar remarks follow for the other ortho-physical basis elements.

One application of this observation is that we can transform between a physical cap and a physical cup by any of the natural conjupositions having a vertical component:

$$\xleftrightarrow{\underset{\sim}{V},\overline{\underset{\sim}{V}},\underset{\sim}{T},\overline{\underset{\sim}{T}}} \tag{695}$$

(this is clear looking at the definition of physical caps and cups in (634)). Further, physical caps are invariant under $\underset{\sim}{H}, \overline{\underset{\sim}{H}}, \overline{\underset{\sim}{I}}$, as are physical cups.

Given that the special caps and cups are proportional to physical caps and cups (677, 678)) with constant of proportionality equal to $N_{\mathsf{a}}^{\frac{1}{1}}$, we also have

$$\xleftrightarrow{\underset{\sim}{V},\overline{\underset{\sim}{V}},\underset{\sim}{T},\overline{\underset{\sim}{T}}} \tag{696}$$

(meaning that any element of the set on one side can be transformed by any of the transformations above the arrow to any element of the set on the other side). This is useful in the theory of physical mirrors (see Sec. 35).

## 33.14 Natural conjuposition preserves compositional structure

In Sec. 32.11 we saw how normal conjuposition of composite objects preserves their compositional form. It is highly desirable that natural conjupositions do the same. Consider the simple example

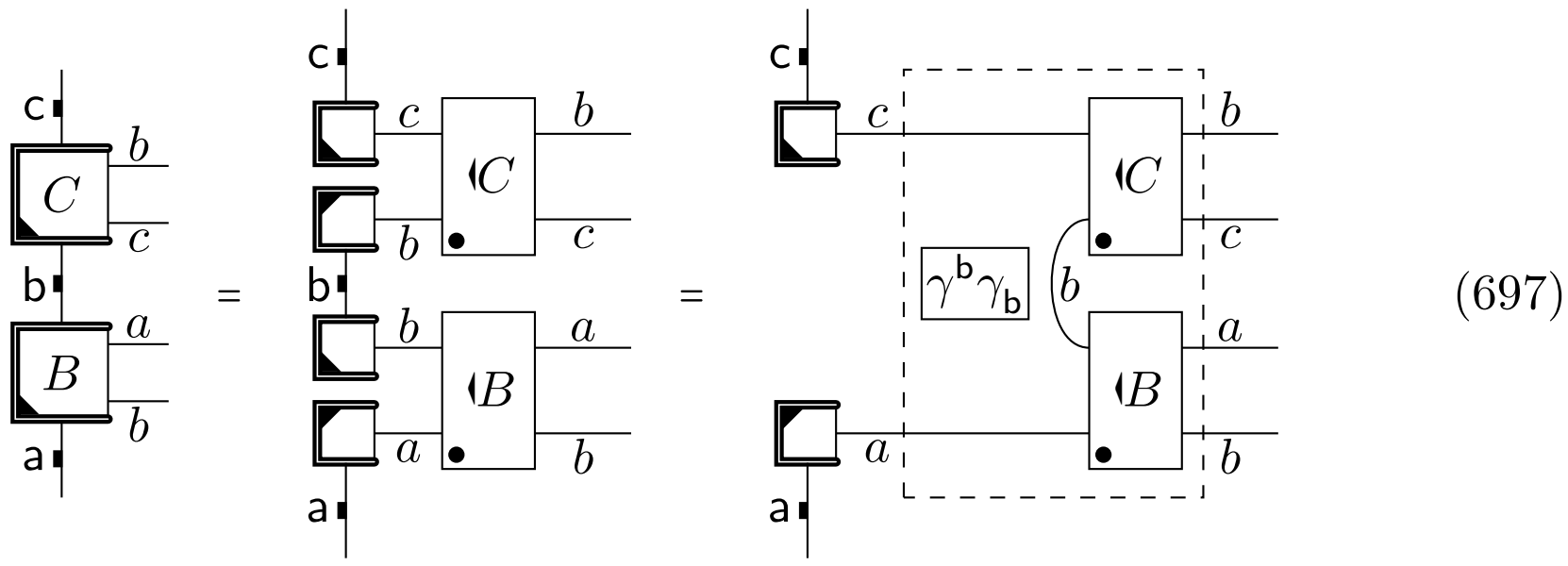

(compare with (593)). If take any natural conjuposition of this then we simply take the corresponding conjuposition of the hypermatrix in the dashed box and reexpress in the appropriate ortho-physical bases. However, this only works if $\gamma^{\mathsf{b}}\gamma_{\mathsf{b}}$ is real. This provides the motivation for the ortho-physical gauge normalisation condition in (626). With this condition, it is easy to see that natural conjuposition preserves the form of composite objects in general (by arguing along the same lines as in Sec. 32.11). In fact, it is clear this works for general composite Hilbert objects.

The form of operator networks is also preserved under natural conjupositions. It is instructive to see how this works with a simple operator network. We have

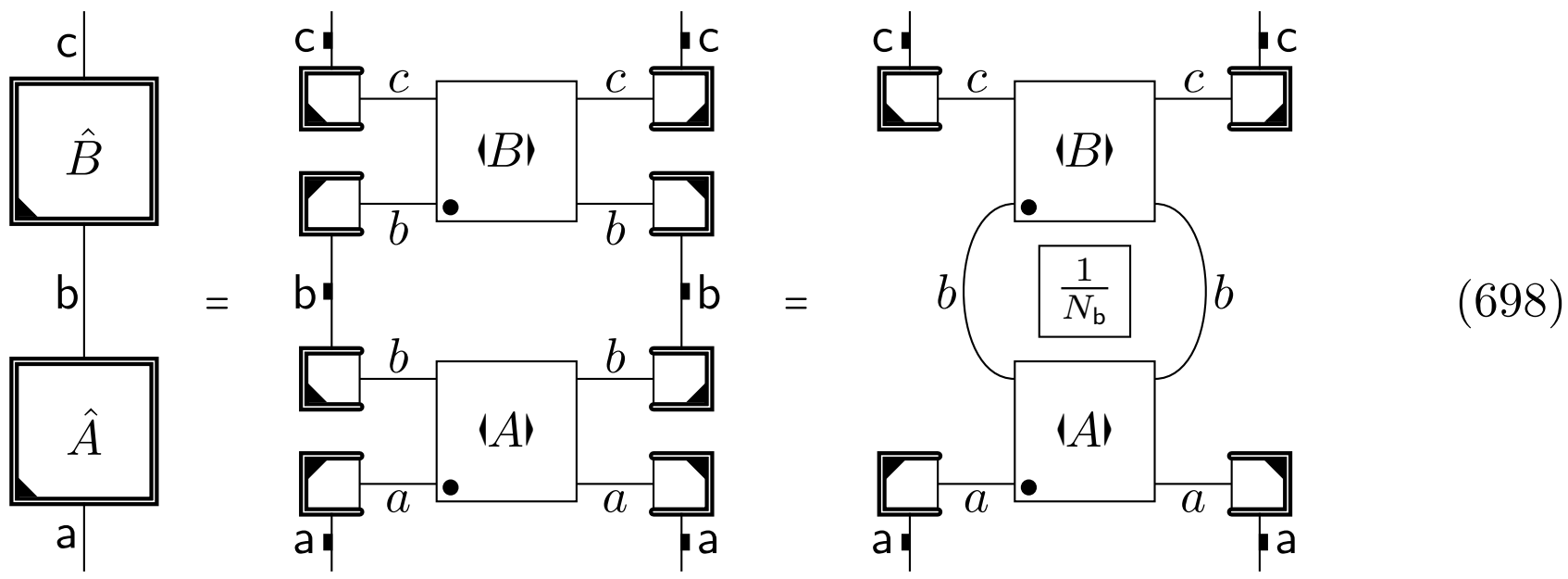

where the $\frac{1}{N_{\mathsf{b}}}$ constant comes from (631). We can implement natural conjupositions by applying the corresponding conjuposition to the expansion matrix then reexpressing in terms of the ortho-physical bases. For example, the natural transpose is given by taking the transpose of the expansion hypermatrix (using the results of Sec. 29.2 this has the same compositional form) and then reexpressing in terms of the ortho-physical bases as follows

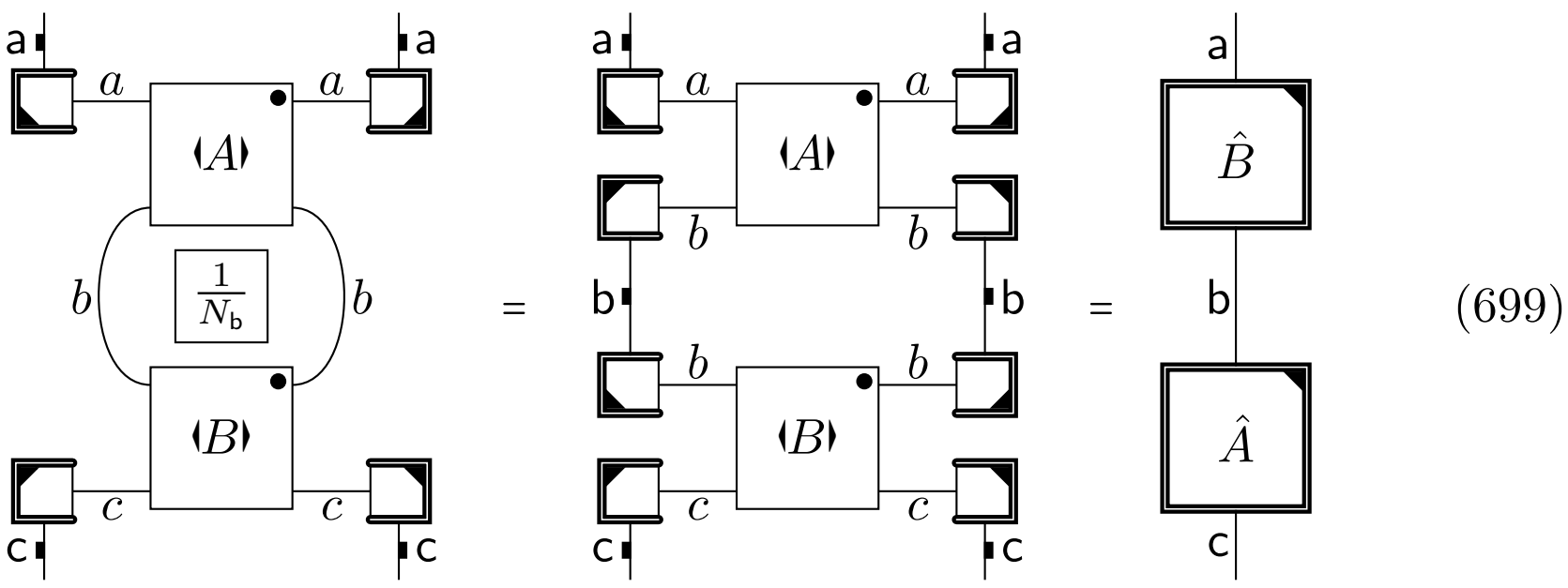

This works for any natural conjuposition and for any operator network.

The above examples are special cases of natural conjupositions of general Hilbert objects such as

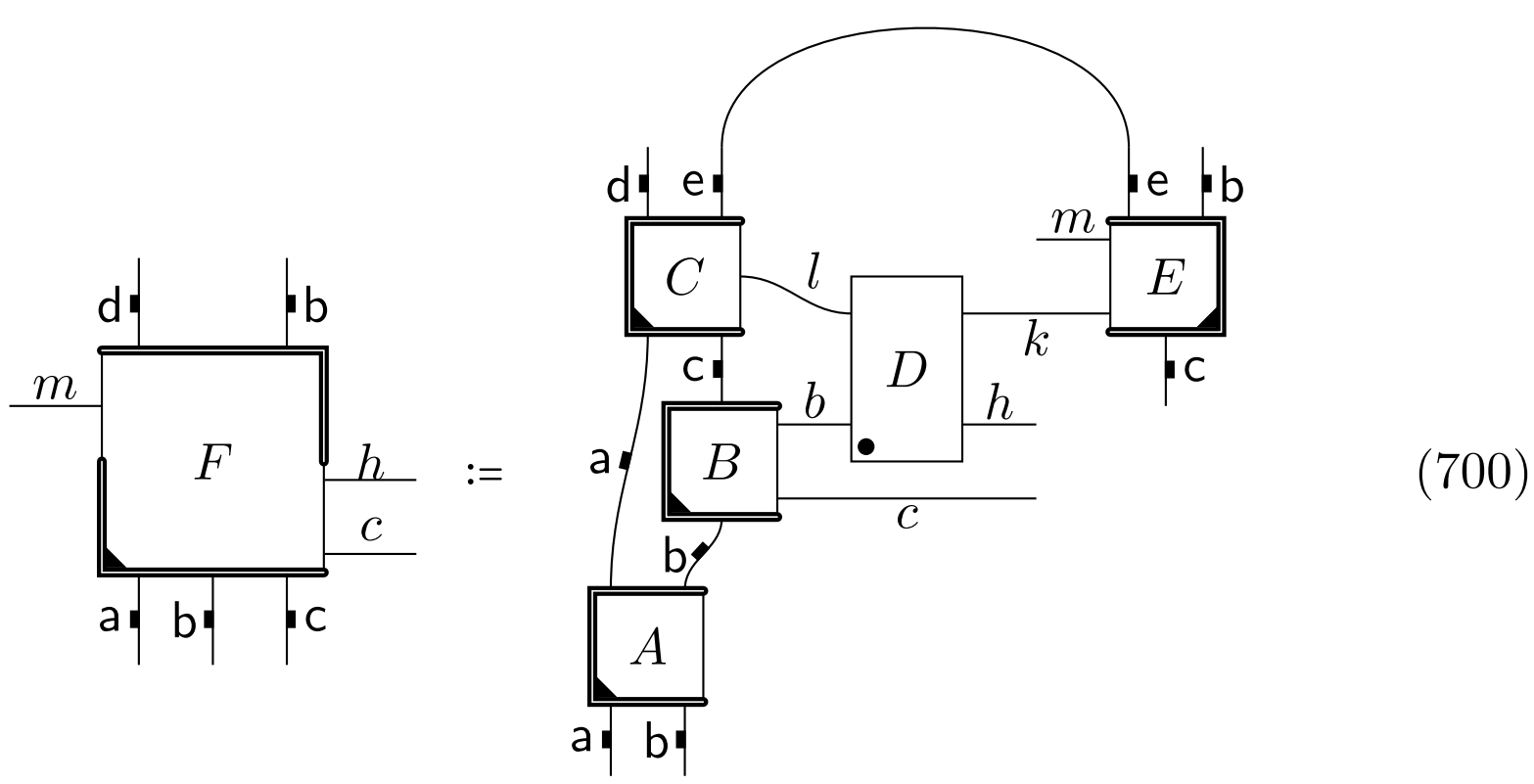

As long as we assume the ortho-physical gauge (626), natural conjuposition preserves the general form. For example, the natural transpose of (700) is

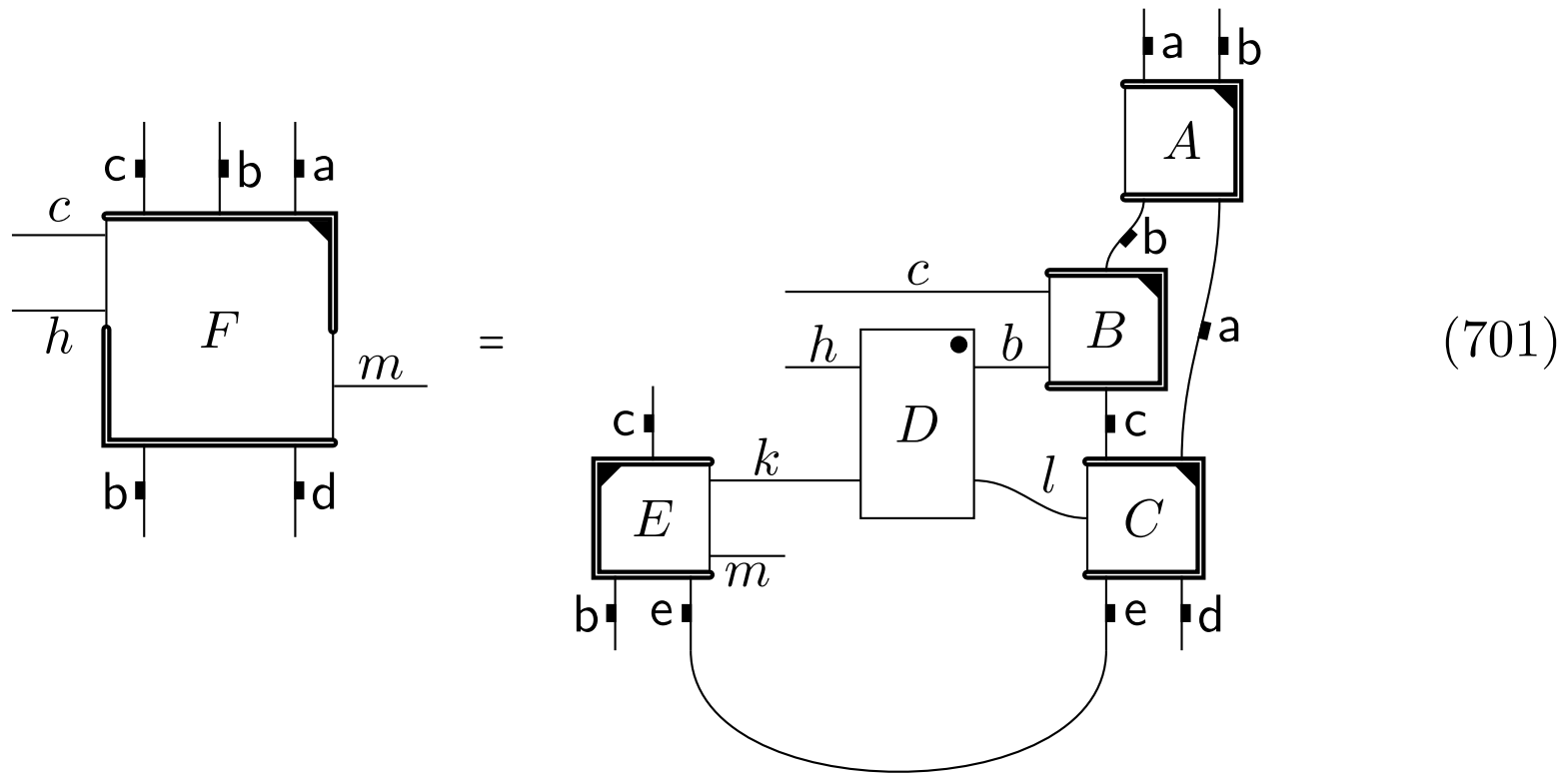

General Hilbert objects form a natural Hilbert cube under natural conjupositions

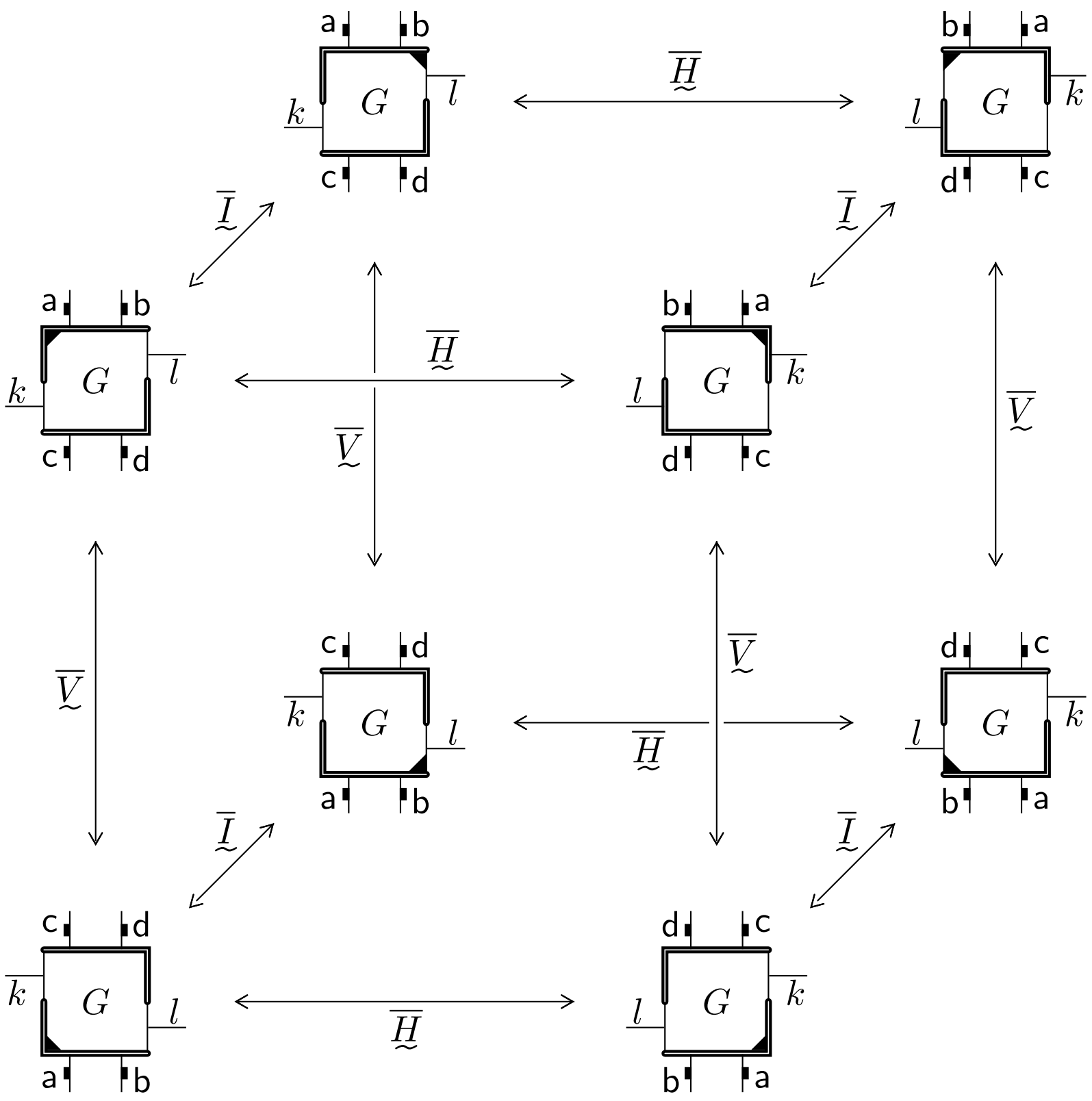

(702)

To ensure the compositional form is preserved here we must impose the orthophysical gauge when we take natural conjupositions.

## 33.15 The natural transpose of an operator tensor

In Sec. 32.14 we saw we could take the normal transpose of an operator tensor by attaching doubled up caps and cups (see (614)). Here we will see that we can take the natural transpose of an operator by attaching doubled up special caps and cups.

Let us first define these *special system caps and cups* obtained by doubling up special caps and cups with bobbles (defined in (677, 678, (681)))

(703)

The special caps and cups with bobbles are defined in (677, 678, (681)). It is easy to show that

(704)

using (526, 525, 666). It is clear from (440) that these convert between ignore preparation and result operators.

Note that the constants in (704) are reciprocals. Hence we obtain the following

(705)

(compare with (705)).

Now we can take the natural horizontal transpose as follows

(706)

This works because the special system cap and cup attach separately to the left and right Hilbert object that comprise the operator tensor (this was illustrated for the normal case in (615).

What happens if we have incomes and outcomes? For example

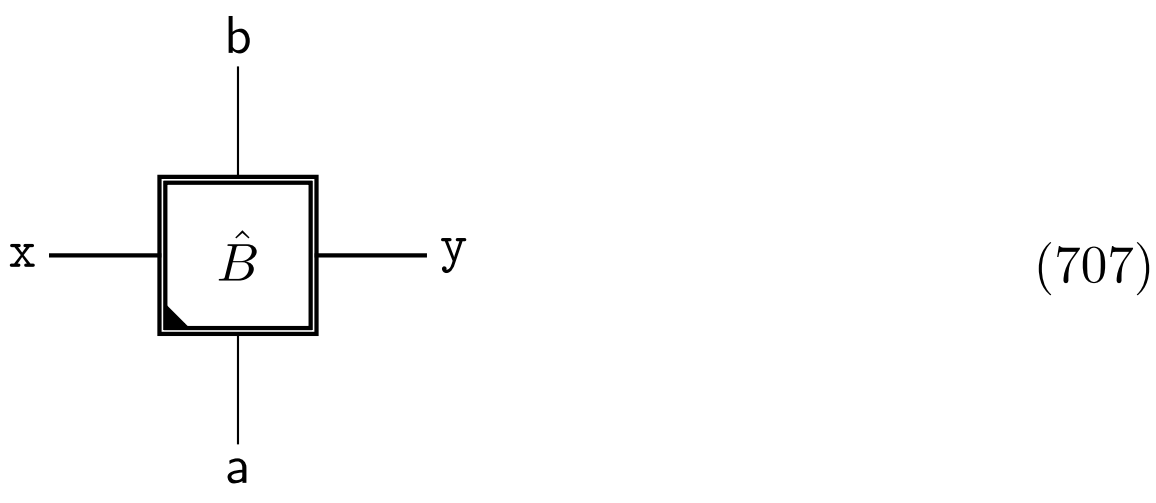

(707)

We can take the natural adjoint of this by attaching appropriate special caps and cups to the income and outcome wires as well as we will now see. First note that the operator B lives in the tensor product space

$$\mathcal{P}_{\mathtt{x}_1} \otimes \mathcal{P}^{\mathtt{y}_2} \otimes \mathcal{V}_{\mathsf{a}_3} \otimes \mathcal{V}^{\mathsf{b}_4} \tag{708}$$

We can do the natural transpose on the $\mathcal{V}_{\mathsf{a}_3} \otimes \mathcal{V}^{\mathsf{b}_4}$ part as before by attaching the special caps and cups. To time reverse the $\mathcal{P}_{\mathtt{x}_1} \otimes \mathcal{P}^{\mathtt{y}_2}$ we need appropriate transformations on incomes and outcomes. To get the normalisation right we can impose that this transformation converts between the preparation and result $\boldsymbol{R}$ boxes. We can do this using the *special pointer cap and cup*

$$\mathtt{x}\,\supset_{\ominus}\,\mathtt{x} \;=\; \boxed{\frac{\beta_{\mathtt{x}}}{\beta^{\mathtt{x}}}}\;\mathtt{x}\,\supset\,\mathtt{x} \qquad\qquad \mathtt{x}\,{}_{\ominus}\!\subset\,\mathtt{x} \;=\; \boxed{\frac{\beta^{\mathtt{x}}}{\beta_{\mathtt{x}}}}\;\mathtt{x}\,\subset\,\mathtt{x} \tag{709}$$

where

$$\mathtt{x}\,\supset\,\mathtt{x} \;:=\; \begin{pmatrix}1 & 0 & \cdots & 0\\ 0 & 1 & \cdots & 0\\ \vdots & \vdots & \ddots & \vdots\\ 0 & 0 & \cdots & 1\end{pmatrix} \qquad \in \mathcal{P}_{\mathtt{x}_1} \otimes \mathcal{P}_{\mathtt{x}_2} \tag{710}$$

and

$$\mathtt{x}\,\subset\,\mathtt{x} \;:=\; \begin{pmatrix}1 & 0 & \cdots & 0\\ 0 & 1 & \cdots & 0\\ \vdots & \vdots & \ddots & \vdots\\ 0 & 0 & \cdots & 1\end{pmatrix} \qquad \in \mathcal{P}^{\mathtt{x}_1} \otimes \mathcal{P}^{\mathtt{x}_2} \tag{711}$$

Inspecting (429), it is easy to verify that the special pointer cap and cup convert between preparation and result **R** operator boxes. Thus, we have the correct coefficients in (709). We have the special pointer yanking equation

$$\mathtt{x}\,\supset_{\ominus}\,\mathtt{x}\,{}_{\ominus}\!\subset\,\mathtt{x} \;=\; \overline{\quad\mathtt{x}\quad} \tag{712}$$

since the constants in (709) are reciprocals.

We can now define the natural transpose for an operator with incomes and

outcomes as follows

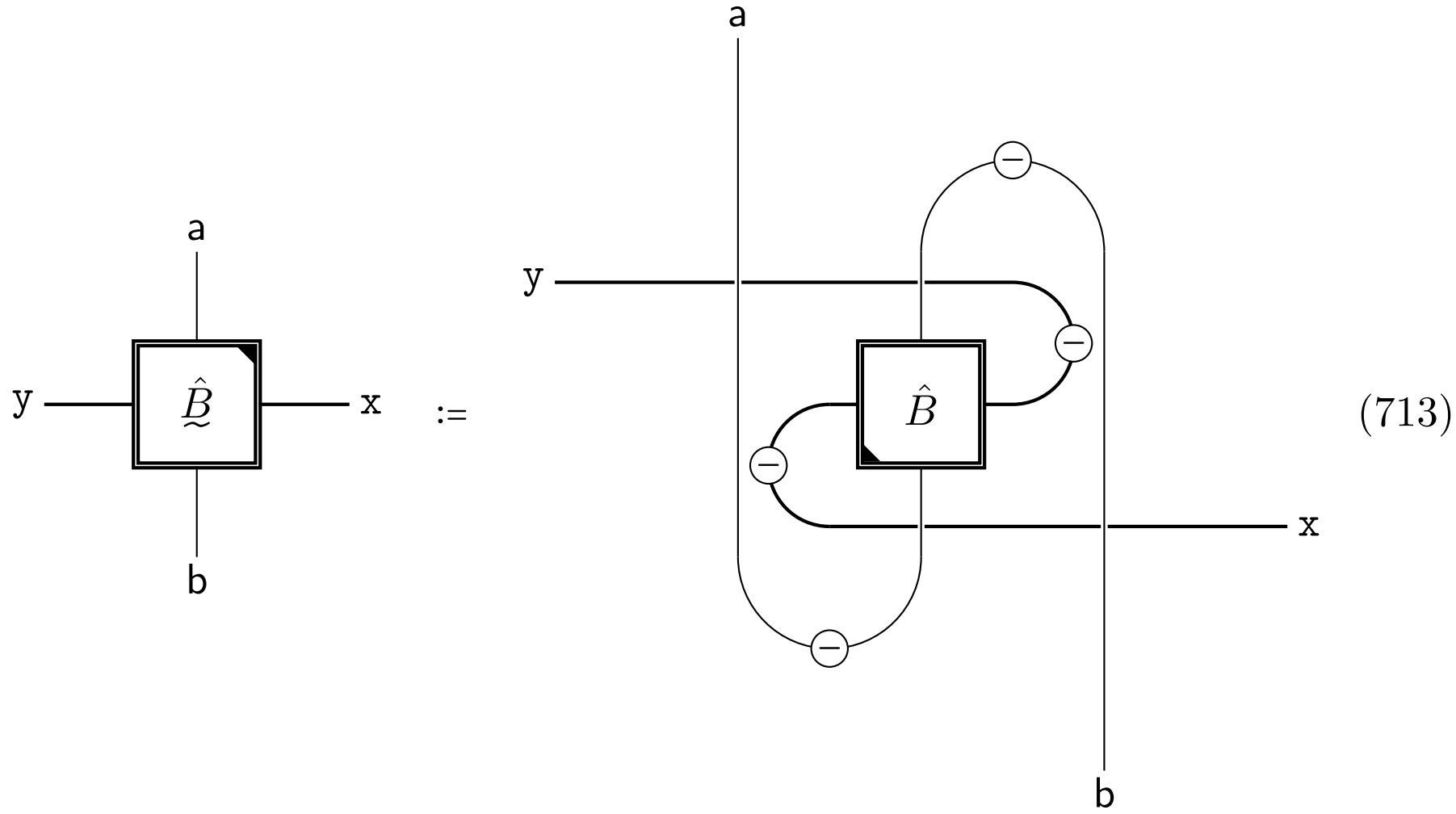

Note that we have extended the notation with the triangular marks to this case where we have incomes and outcomes. Also note that, if there is more than one wire coming out of any side, then we place the special caps or cups so that the order is reversed (consistent with rotation through 180°). This definition is such that, if we take the natural transpose of a network of operator tensors the compositional structure is preserved - the new network is obtained simply by rotating the original through 180°. This is easily proven by (i) putting special pointer and system caps and cups on each open wire and (ii) by replacing closed wires with the cap-cup wires given by the special yanking equations (705) and (712). Then we can identify the natural transpose of each of the original operator tensors.

## 33.16 Physical interpretation of natural conjupositions

In Sec. 32.16 we saw that the basis independent part of the normal conjuposition group does not quite represent the right symmetry group for time symmetric simple operational quantum theory (TSSOQT) because the normal transpose (and the normal vertical adjoint) fail to transform ignore preparation operators to ignore result operators (see (623)). We saw in Sec. 33.7 that the natural transpose *does* transform between ignore preparation and result operators. This is clearly a necessary condition for the natural transpose to be a useful symmetry. We will start by restricting ourselves to operator tensors having only inputs and outputs. We will see that the physicality conditions are invariant under the natural conjuposition group. We will first restrict ourselves to the basis independent subgroup

$$\{I, \overline{\underset{\sim}{H}}, \overline{\underset{\sim}{V}}, \underset{\sim}{T}\} \tag{714}$$

as this is the most interesting, and then we will comment on the other natural conjupositions. Since the physicality conditions express the constraints on allowed operators, this means that the natural conjupositions constitute a symmetry group of the physics here (TSSOQT). We will see that we can identify the natural transpose, $\underset{\sim}{T}$, as enacting the time reverse. This interpretation goes over to the case when we allow the operator tensor to have incomes and outcomes.

Consider an arbitrary operator tensor. If it has incomes and outcomes we know how to take take $I$ and $\underset{\sim}{T}$ but not $\overline{\underset{\sim}{H}}$ and $\overline{\underset{\sim}{V}}$. For the moment we will proceed with an operator without incomes and outcomes. In fact we can justify this restriction either by using the maximal representation of an operator shown later in (952) or, in a suitably regularised circuit, by absorb the $\boldsymbol{R}$ boxes and readout boxes into the operator as shown in (379). Thus, we will consider an operator tensor

(715)

without incomes or outcomes. We will return to the case where there are incomes and outcomes below. We have included the small triangle marking so we can track physical conjupositions. Our convention for placement of these markings is such that operator in (715) maps to

(716)

under the action of $I$, $\overline{\underset{\sim}{H}}$, $\overline{\underset{\sim}{V}}$, and $\underset{\sim}{T}$ respectively (these are the elements of the group in (714)).

The physicality conditions on such an operator are as follows. First, that it can be written in twofold positive form as follows

(717)

Second, that the general double causality conditions

(718)

hold (we get these by correspondence with (160, 161) ignoring incomes and outcomes). The inequality is saturated for the deterministic case. The twofold positivity condition (717) is proven in Sec. 43 to be equivalent to the tester positivity condition in Sec. 26.

First, let us consider the action of the group elements in (714) on the twofold condition. We obtain the four conditions

$$\text{(719)}$$

and

$$\text{(720)}$$

These tell us that (i) each of the operators in (716) satisfies twofold positivity (if any given one does) and (ii) that

$$\text{(721)}$$

Since $\underset{\sim}{\overline{H}} = \overline{H}$, this second point tells us that these operators are Hermitian (i.e. invariant under the horizontal adjoint).

Next we will consider the general double causality conditions in (718). First note that we have

$$\text{(722)}$$

Note in particular that the group in (714) maps between these elements - see discussion in Sec. 33.7. We can apply each of the elements of the group in (714) to the double physicality conditions. This is easy to do since the compositional structure is preserved under natural conjupositions (as discussed in Sec. 33.14). We obtain a total of four sets of general double causality conditions. However, given the equalities in (721), we effectively only get one additional set of general double causality conditions, namely

$$\begin{array}{c} \mathsf{a} \\ \hat{B} \\ \mathsf{b} \\ \hat{I} \end{array} \underset{T}{\leq} \begin{array}{c} \mathsf{a} \\ \hat{I} \end{array} \qquad\qquad \begin{array}{c} \hat{I} \\ \mathsf{a} \\ \hat{B} \\ \mathsf{b} \end{array} \underset{T}{\leq} \begin{array}{c} \hat{I} \\ \mathsf{b} \end{array} \tag{723}$$

This condition applies to the natural transpose of the $\hat{B}$ operator tensor we started with.

The crucial insight from the above results is that

$$\begin{array}{c} \mathsf{b} \\ \hat{B} \\ \mathsf{a} \end{array} \tag{724}$$

is physical if and only if

$$\begin{array}{c} \mathsf{a} \\ \hat{B} \\ \mathsf{b} \end{array} \tag{725}$$

is physical. This makes $\underset{\sim}{T}$ the perfect candidate for the time reversal map (see the discussion of time reversal symmetry in Sec. 6.4) since it is guaranteed to map between physical operators. In particular we write

$$\begin{array}{c} \mathsf{b} \\ \hat{B} \\ \mathsf{a} \end{array} \underset{\text{reverse}}{\overset{\text{time}}{\longleftrightarrow}} \begin{array}{c} \mathsf{a} \\ \hat{B} \\ \mathsf{b} \end{array} \tag{726}$$

The requirements for the theory to be time symmetric (as discussed in Sec. 6.4) were that such a map exists and that, if we use it to form the time reverse of any circuit, we get the same probability. To verify the second point, consider

the circuit

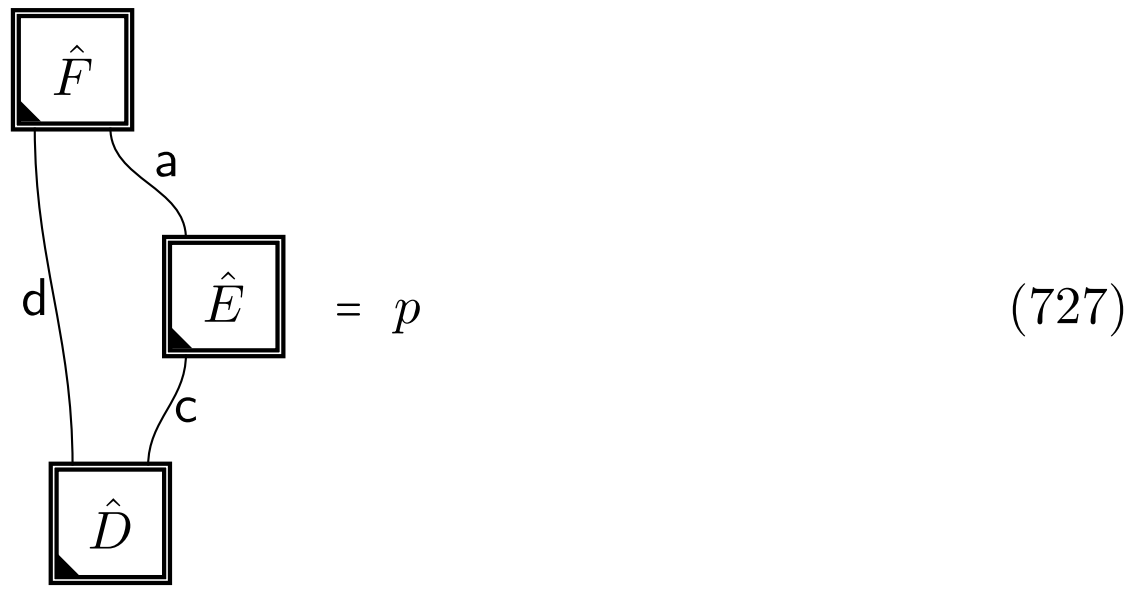

(where $p$ is the associated probability for this circuit). Thus, under $\underset{\sim}{T}$ we obtain

$$\hat{D} \quad \text{c} \quad \hat{E} \quad \text{d} \quad \text{a} \quad \hat{F} \quad = p \tag{728}$$

(this follows from the reasoning illustrated in going from (698) to (699)). In general, we rotate a circuit through 180° when we implement the physical transpose. Note also that $p$, on the right hand side is a scalar and so is unchanged under $\underset{\sim}{T}$ (we can think of it as a box with no wires). Thus, we satisfy the second condition for a theory to be time symmetric.

The group in (714) has four elements. If we restrict to physical operators then $\overline{\underset{\sim}{H}}$ has the same action as $I$, and $\overline{\underset{\sim}{V}}$ has the same action as $\underset{\sim}{T}$. Thus $\overline{\underset{\sim}{H}}$ leaves the physics unchanged. This symmetry (which is the same as $\overline{H}$) is required so that probabilities are real (see discussion in Sec. 32.16). The natural vertical adjoint, $\overline{\underset{\sim}{V}}$, or equivalently the natural transpose, $\underset{\sim}{T}$, can be regarded as implementing the time reverse map. It does change the physical situation in that it returns the time reversed situation. We might argue that, in a time symmetric theory, there is no way to distinguish between the time forward direction and the time backward direction and, therefore, that the physical transpose also leaves the physical situation unchanged. This is a difficult debate to settle. However, it is worth noting that if we have two disjoint circuits and we apply either $\overline{\underset{\sim}{H}}$ or $\underset{\sim}{T}$ to only one of the two circuits then the resulting diagrams look different in the two cases.

Now we will consider the case where there are incomes and outcomes. We defined the natural transpose for this case in (713). We will show here that (i) if such an operator tensor is physical then its natural transpose is also physical and (ii) the probability for an operator circuit on which we apply the natural transpose (by rotating through 180°) is unchanged. Hence we can consistently

put

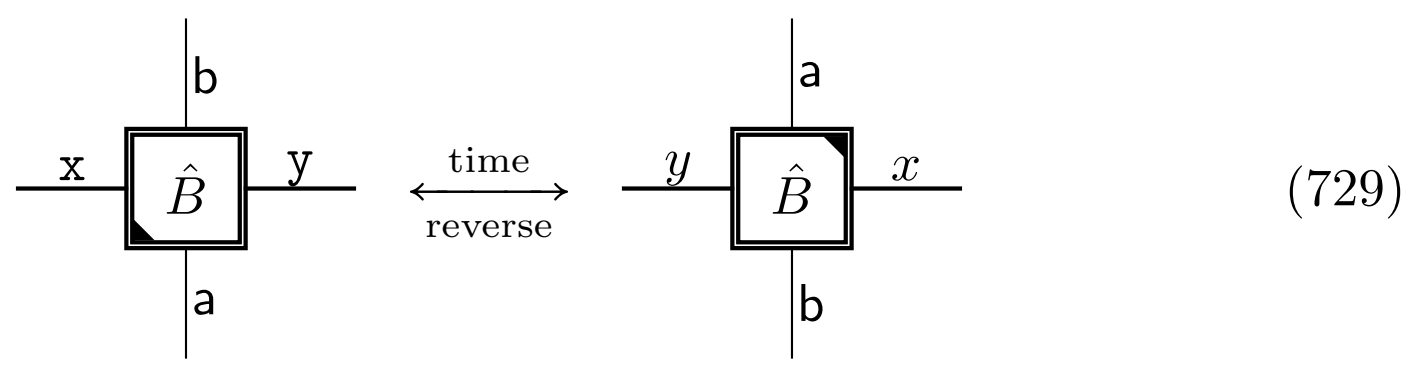

$$(729)$$

To prove (i) we need to consider positivity and double causality. The condition for $\hat{B}$ to be positive is that it satisfies the tester positivity condition

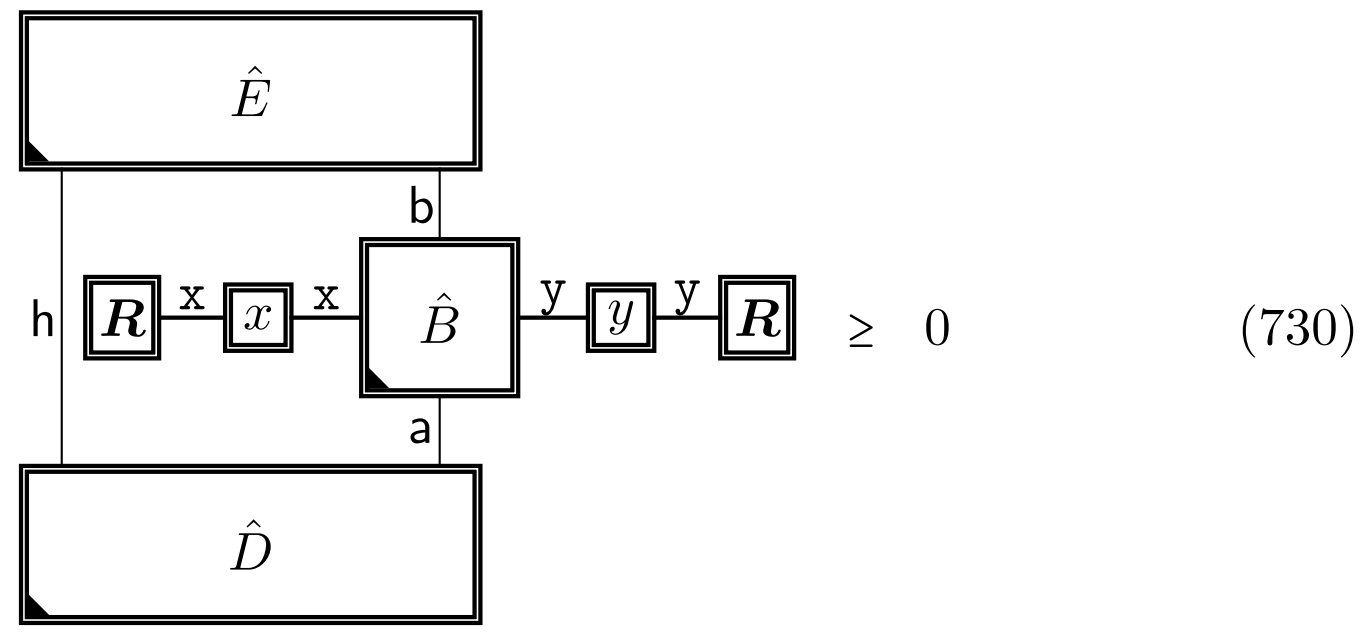

$$(730)$$

for all pure $\hat{D}$ and $\hat{E}$ (this is equivalent to the twofold positivity condition as shown in Sec. 43.2). If we apply the natural transpose to this condition then we simply rotate the circuit through 180° (since the compositional structure is unchanged - see comments at the end of Sec. 33.15). The resulting inequality is simply the tester positivity condition for $\hat{B}^{\underset{\sim}{H}}$ (note that pure preparation operators map to pure result operators (and vice versa) under the natural transpose because they are homogeneous - i.e. represented by rank one projectors). Next consider the double causality condition. If we rotate the double causality conditions for $\hat{B}$ through 180° then we simply get the double causality conditions for $\hat{B}^{\underset{\sim}{H}}$. Thus $\hat{B}^{\underset{\sim}{H}}$ is physical if $\hat{B}$ is. Now consider (ii). This follows immediately from the fact that the natural transpose is implemented by rotating a circuit through 180° (leaving the compositional form unchanged) and that the natural transpose leaves scalars (such as the probability $p$) unchanged.

Our remarks above were restricted to the basis independent elements of the natural conjuposition group. What about the other elements? If we apply $\underset{\sim}{\overline{I}}$ then we are taken to the shadow square (the back face of the Hilbert cube) which are then related by the basis independent subgroup. We will denote the elements of the shadow square (obtained by applying $\underset{\sim}{\overline{I}}$ to the corresponding elements in (716)) as follows

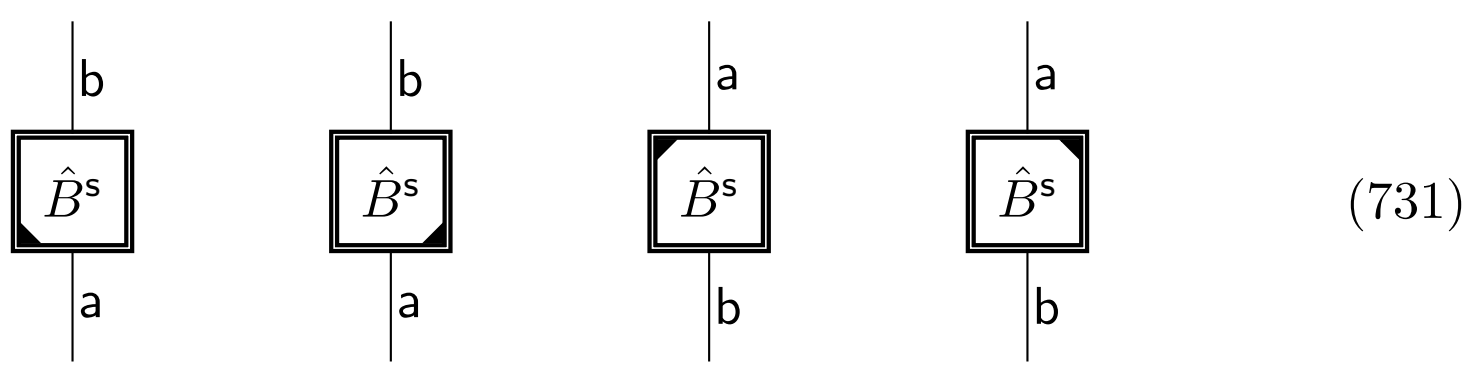

$$(731)$$

(where the $\mathsf{s}$ superscript stands for "shadow"). The effect of $\overline{\underset{\sim}{I}}$ is to take the complex conjugate of the expansion matrix (when we expand in terms of ortho-physical basis elements (here we are using the fact that $\overline{\underset{\sim}{I}}$ leaves basis elements unchanged - see (694)). If we have a circuit built out of Hermitian operators, and we apply $\overline{\underset{\sim}{I}}$ to each operator, then the circuit will evaluate to the same (real) number. It is also easy to see that the positivity condition and causality conditions continue to be satisfied. This does suggest that an alternative time reverse transformation is $\overline{\underset{\sim}{T}}$. It is, however, basis dependent, and so we prefer $\underset{\sim}{T}$. While we are here, it is worth noting that we could also apply partial natural conjupositions. For example, we could apply the natural conjuposition only to the right Hilbert object on an operator written in twofold form (this is the natural analogue of the example given in (589)). We could also formulate Quantum Theory in terms of those objects.

# 34 Normal mirrors

Many of the diagrammatic equations we are interested in are symmetric either about a vertical axis or about a horizontal axis (and a few are symmetric about both). We can provide notation using mirrors to remove the corresponding diagrammatic redundancy. We will have vertical mirrors and horizontal mirrors. The definition of how a mirror "reflects" requires us to adopt a basis. There are normal mirrors (using a orthonormal basis), physical mirrors (using an ortho-physical basis, and even deterministic mirrors (using an ortho-deterministic basis).

In this section we will set up the theory of *normal mirrors*. We will have vertical and horizontal normal mirrors. Each normal mirror has an element of the normal conjuposition group associated with it. In the first place we will consider *regular normal mirrors* associated with $\overline{H}$ and $\overline{V}$. Later we will consider *twist normal mirrors* associated with $\overline{I}$, and $\overline{T}$. Normal mirror notation enables us to formulate a *normal mirror theorem* which can be used to convert between equivalent equations.

In Sec. 35 we will consider physical mirrors which are associated with the natural conjuposition group. Physical mirrors are more useful for the actual physics we want to describe. We will mention deterministic mirrors in Sec. 35.7 though we do not develop the theory of these mirrors since they do not have a clear application to Quantum Theory.

These mirrors do not represent actual mirrors (which are, themselves, no strangers to Quantum Theory - particularly in Quantum Optics). In case any such ambiguity should arise, we could call the mirrors here *Hilbert mirrors* since their job is to perform certain mathematical "reflections" on Hilbert objects.

## 34.1 Regular normal mirror notation

A regular normal mirror associated with $\overline{H}$ is defined such that

(732)

for any left Hilbert object. In words, the mirror produces an image under $\overline{H}$ and any label wires are attached to this image appropriately. Two applications of this are

(733)

and

(734)

where we have, in the rightmost expression in each case, extended our notation to allow system wires to connect to normal mirrors having the given meaning. Note we have used (508).

Using this extended notation we can, for example, write

(735)

The expression on the left of this equation is symmetric about a vertical axis. The expression on the right of this equation has a "mirror" which is "reflective" on its left side. This mirror produces a "reflection" (the normal horizontal adjoint, $\overline{H}$) which is a right Hilbert object. We interpret both the left object and its reflection as being part of the mathematical expression. Thus the two sides of the equation are the same. We can, instead, have a mirror which is reflective on its right side:

(736)

Now we have a right object which reflects to produce a left object.

It is not necessary to have wires that attach to the mirror. For example, for a homogeneous operator we can write

(737)

In the first expression we have the notation given in (556) for a homogenous operator.

If we have a mirror that is reflective on the left side then we assume that everything on the left side is reflected even if we do not extend the mirror up and down to the full extent of the diagram. For example,

(738)

Although it is fine to extend the mirror, it is a little more aesthetically pleasing to keep the mirror short. In such a case, if we "get close" to the mirror we can, of course, see the full "reflection" and so this is in keeping with the reflection analogy. We use this example to illustrate how a constant (inside the thin bordered box) becomes the complex conjugate when reflected. When we go from an expression written in mirror notation (like the left hand side of (738)) to one where the mirror has been removed and the reflection has been drawn in explicitly we will we are “reflecting out” the expression (this nomenclature is reminiscent of “multipling out” brackets).

We will associate a horizontal regular normal mirror with $\overline{V}$. The action of this mirror is defined through the following property

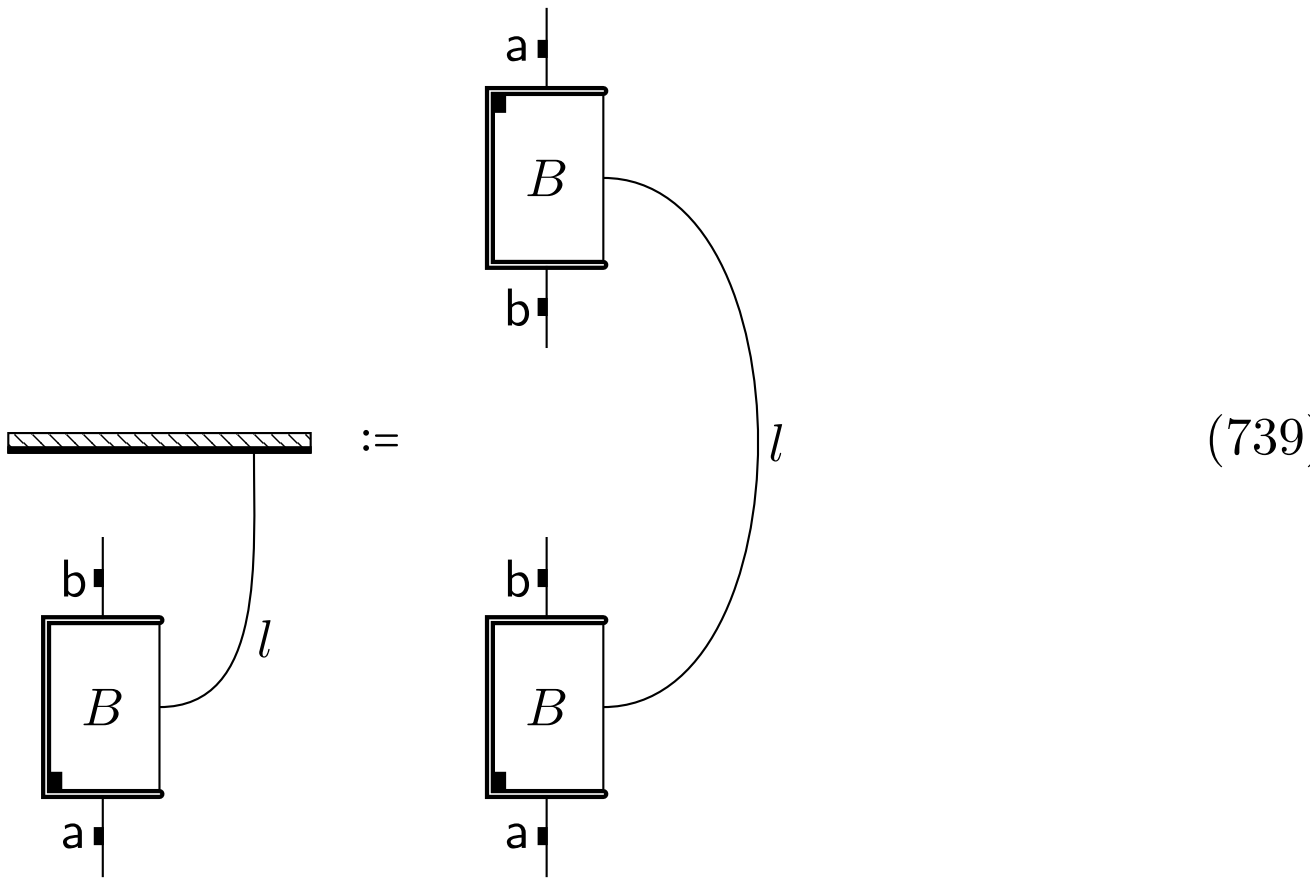

(739)

The mirror forms an image under $\overline{V}$ and any label wires joining the mirror are attached to this image as appropriate.

Special applications of this is the following

(740)

and

(741)

To make the second step of (740) we have used (506). To make the second step of (741) we have used (506. 500, 508, 509). In the last step in each of (740) and (741) we have extended the notation allowing a system wire to connect with the mirror having the given meaning.

This extended notation allows us to use mirror notation in expressions that are symmetric about a horizontal axis. For example

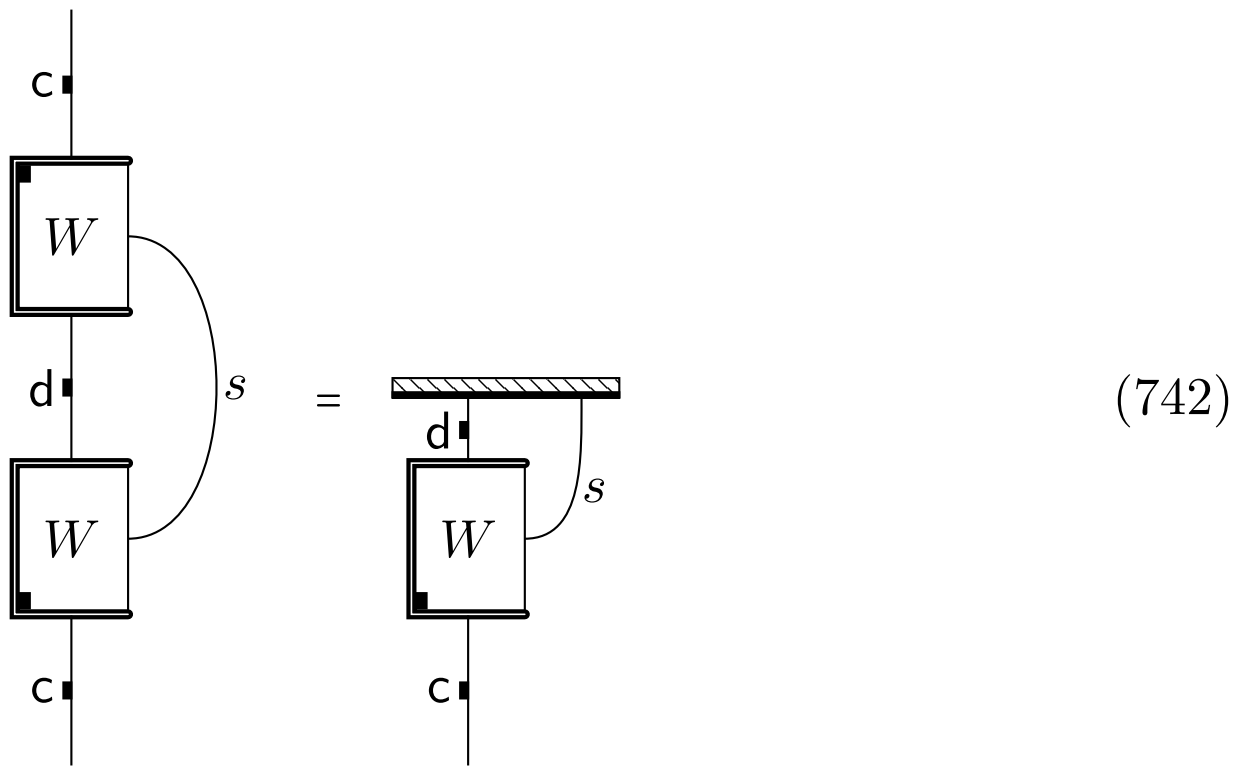

(742)

The mirror here produces an image that is the normal vertical adjoint ($\overline{V}$). If there where constant terms then their reflection would be the complex conjugate (this is true for horizontal mirrors as well).

We can write down expressions that have both a vertical and a horizontal regular normal mirror in them. Here is an example.

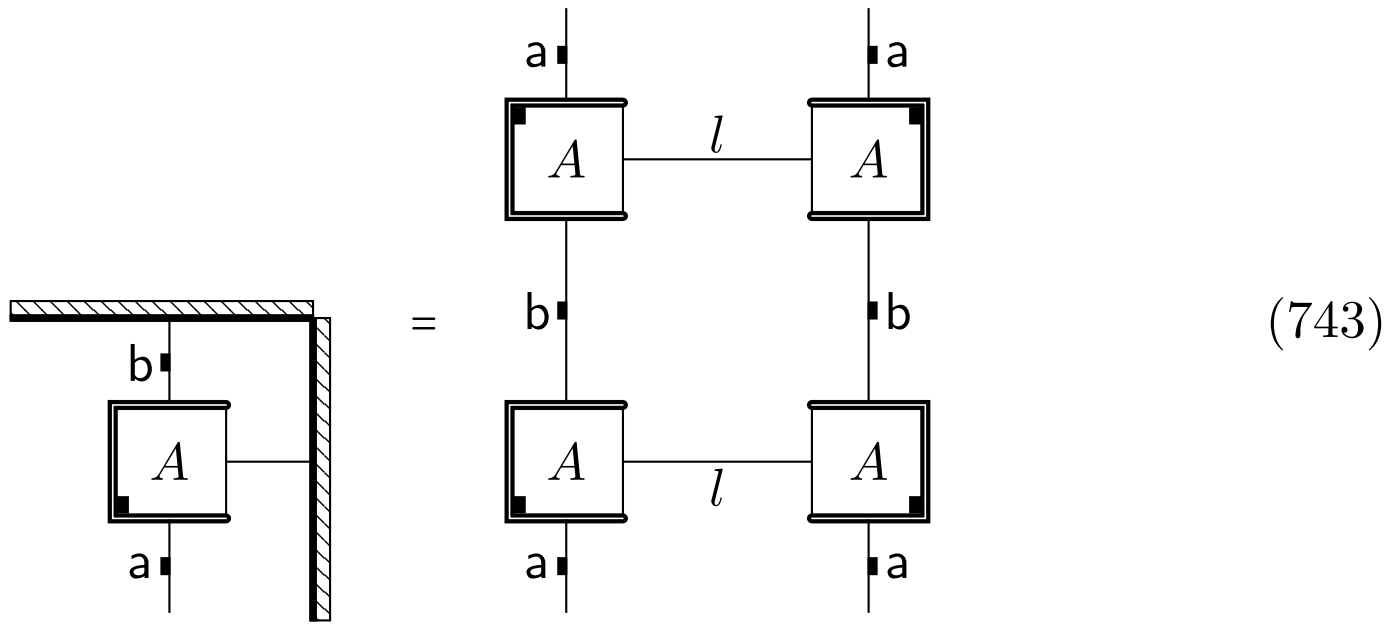

(743)

Indeed, two mirrors placed like this will produce two reflections (one to the right, one above) and also a third image which is the original one rotated. This third image can be thought of as acting with $\overline{VH}$ (where we reflect horizontally then reflect this reflection vertically) or with $\overline{HV}$ (where we reflect vertically then reflect this reflection horizontally). Since $\overline{VH} = \overline{HV}$ we get the same image either way (equal to the transpose, $T$, of the original).

## 34.2 Smoke and mirrors

The normal mirrors discussed above (in Sec. 34.1) create images under $\overline{H}$ and $\overline{V}$. We can attempt to define more normal mirrors that work under other elements of the normal conjuposition group. With this in mind, consider Table 6 where we have partitioned the elements of this group into those that are conjugating (i.e. have a bar over the symbol) or not, and those that are basis independent or basis dependent. We call mirrors that are both conjugating and basis independent

|  | basis independent | basis dependent |
|---|---|---|
| conjugating | $\overline{H}$, $\overline{V}$ | $\overline{I}$, $\overline{T}$ |
| non-conjugating | $I$, $T$ | $H$, $V$ |

Table 6: Partitioning the normal conjuposition group with mirrors in mind.

*regular*. These are normal mirrors that act under $\overline{H}$ and $\overline{V}$ which we already considered in Sec. 34.1. We can also define normal mirrors that are conjugating but basis dependent. These are mirrors that act under $\overline{I}$ and $\overline{T}$. We will call these *normal twist mirrors* as they employ twist objects. Since they are conjugating they offer an alternative way to formulate the equations of Quantum Theory. We will discuss them below. It is possible to define a third pair of mirrors - those associated with normal horizontal and vertical transposition ($H$ and $V$). Since these mirrors do not involve conjugation they do not appear to be useful for formulating Quantum Theory (where probabilities are formed by the sums over products of the form $c\overline{c}$) and so we will not discuss in any detail. Finally, there are the transformations $I$ and $T$. These transformations are basis independent - however it does not appear to be possible to define mirrors associated with these transformations without using a basis making such mirrors rather odd objects. We leave it as an exercise for the reader to draw examples for the non-conjugating mirrors.

Let us set up normal twist mirrors (associated with $\overline{I}$ and $\overline{T}$). We can imagine these mirrors have a chamber filled with "smoke" allowing these unusual reflections. The vertical normal twist mirror creates an $\overline{I}$ reflection and can be defined through the equation

(744)

Note that the label wire attaches appropriately to the image.

Particular applications of this definition are the following

(745)

and

(746)

We are using (514) and (515) to get to the third expression in each case. We introduce the extended notation defined in the final expression allowing a system wire to attach to a twist mirror with the meaning that, when reflected out, we insert a twist (the circle with $w$). See (515) for definition of this circle notation. Importantly, note we have used the fact (noted in Sec. 32.10) that the $\overline{I}$ acting on a basis element leaves that basis element unchanged (see (591) for an example).

Using this extended notation we can define objects such as

(747)

The "reflection" in this mirror is reminiscent of René Magritte's painting, *La reproduction interdite* depicting a man looking in the mirror where the "reflected" image is his own back. A caveat is that the reflection in (747) does show the small black square reflected as one would expect in a mirror. For this mirror we must insert a twist object for every system wire that passes through the mirror (as depicted in the above example).

A horizontal normal twist mirror creates an image under $\overline{T}$ while respecting the connections of label wires as follows

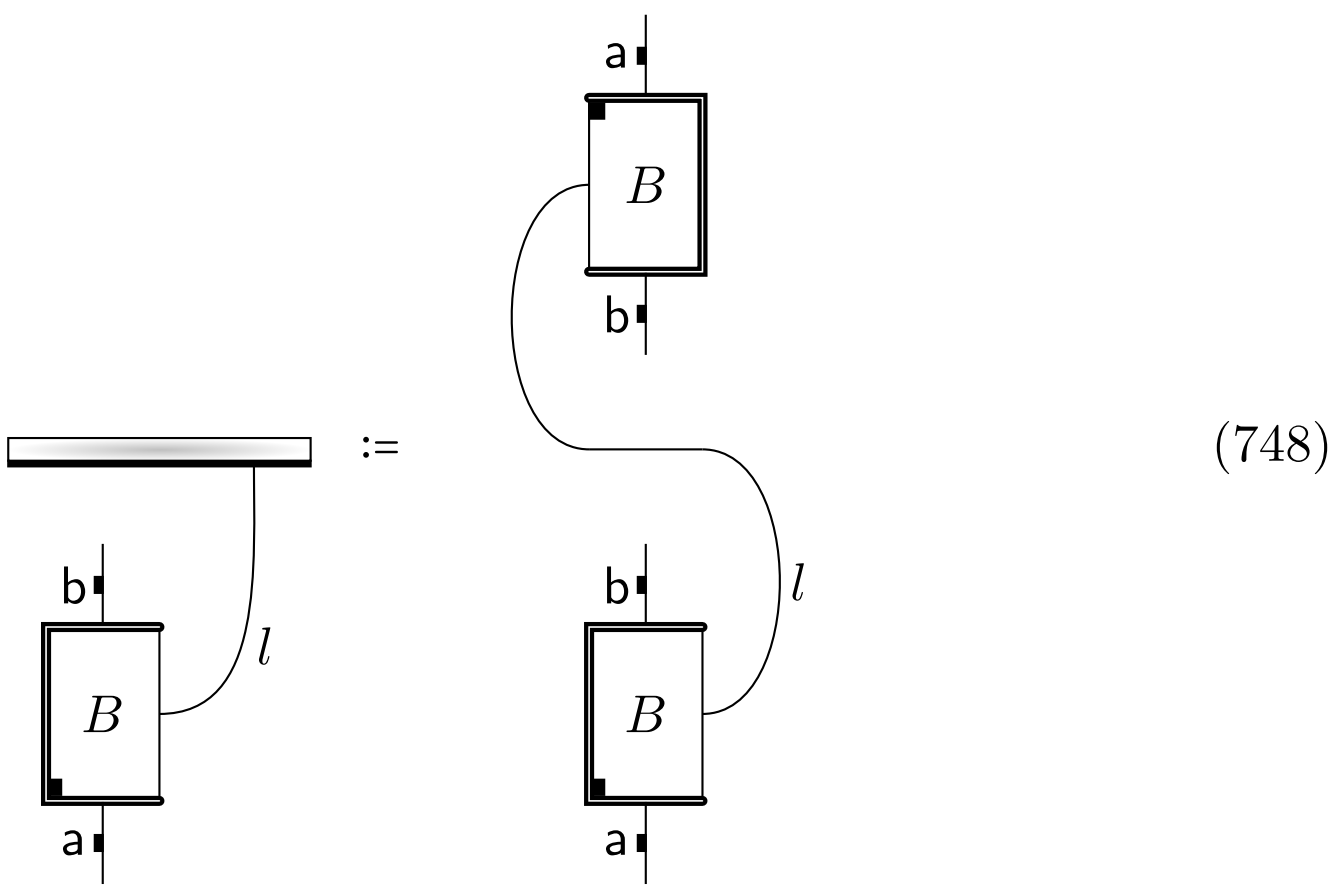

(748)

Particular applications of this to basis elements are as follows

(749)

and

(750)

We have used (513, 591) to obtain (749). To obtain (750) we use (500, 513, 591). We have defined extended notation in the last term allowing a system wire to be attached to a mirror.

An example for this horizontal normal twist mirror is

(751)

The "reflection" is the original object acted up on by $\overline{T}$. Every system wire passing through the mirror acquires a twist object as depicted.

It should be noted that twist mirrors, like natural mirrors conjugate any complex numbers upon reflection.

Since the "reflective coating" can be on either side we have a total of eight conjugating mirror types (four regular and four twist).

## 34.3 Normal mirrors for general Hilbert objects

Till now we have considered reflecting either left or right Hilbert objects. We can, however, easily extend this notation to reflect general Hilbert objects. For example, it is natural to write

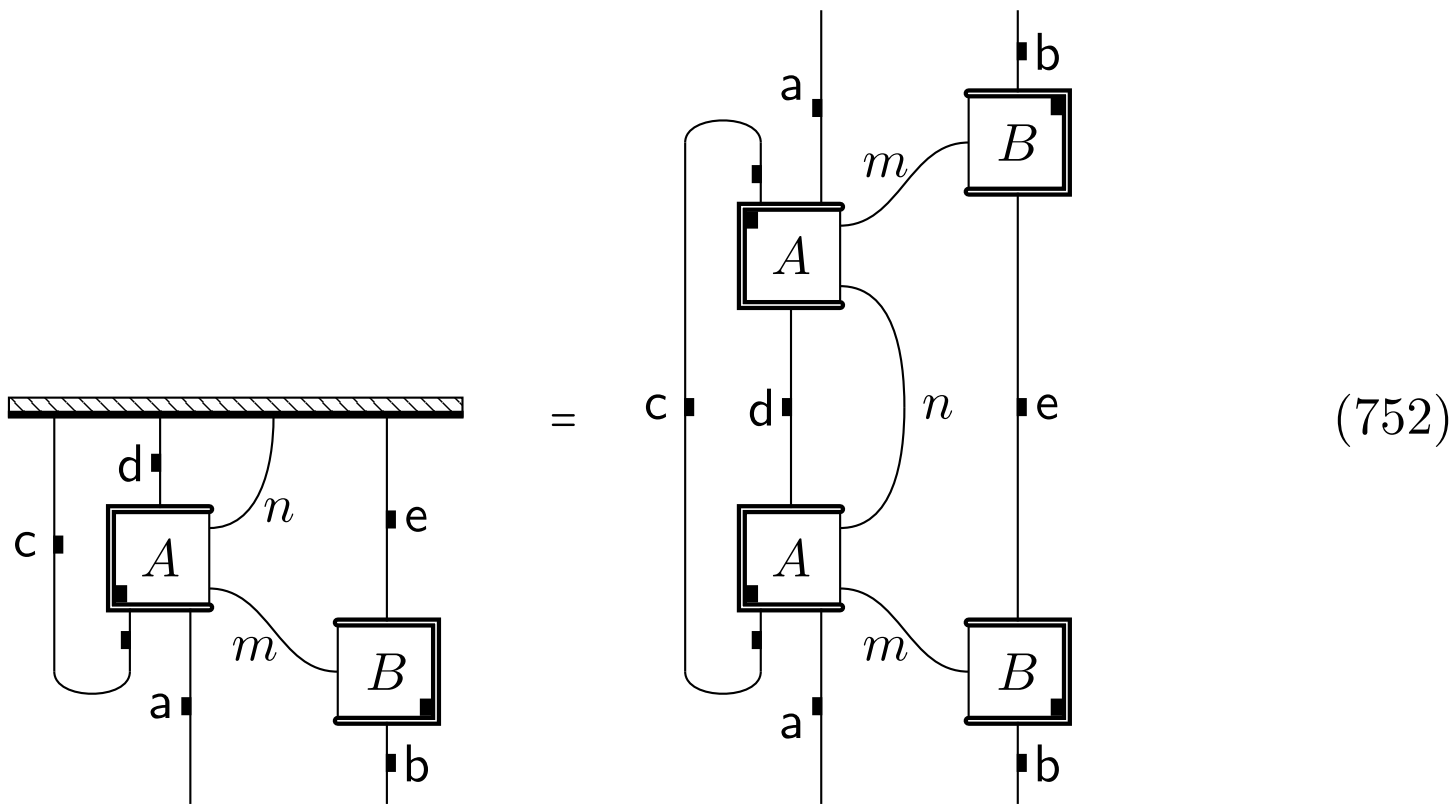

(752)

Thus, there is no need to have only left (or only right) objects being reflected. For the most part, since we are interested in using mirror notation to exploit left/right or up/down symmetry, we will not be interested in extending the mirror notation in this way. However, we do wish to prove a *normal mirror theorem* and, for this we do need to consider reflecting general Hilbert objects (composed of both left and right Hilbert objects).

We will provide defining equations for our mirrors that work for general objects.

In place of the defining equation (732) we have

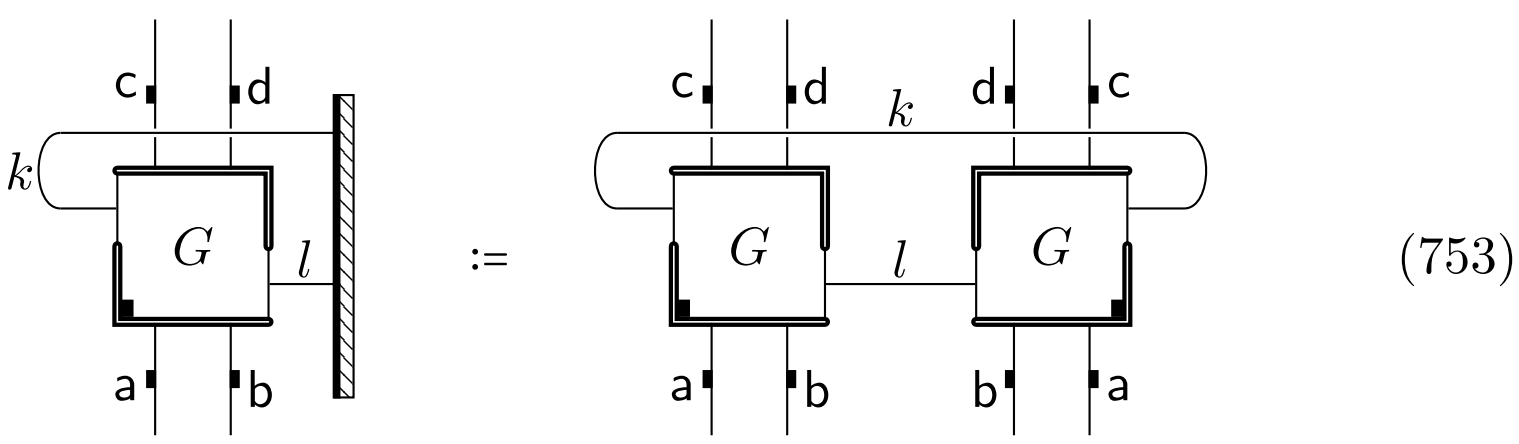

(753)

From this defining equation we can obtain (733) and (734) as well as

(754)

and

(755)

so we can extend our notation to allow left and right system wires to attach to such mirrors.

In place of (739) we have

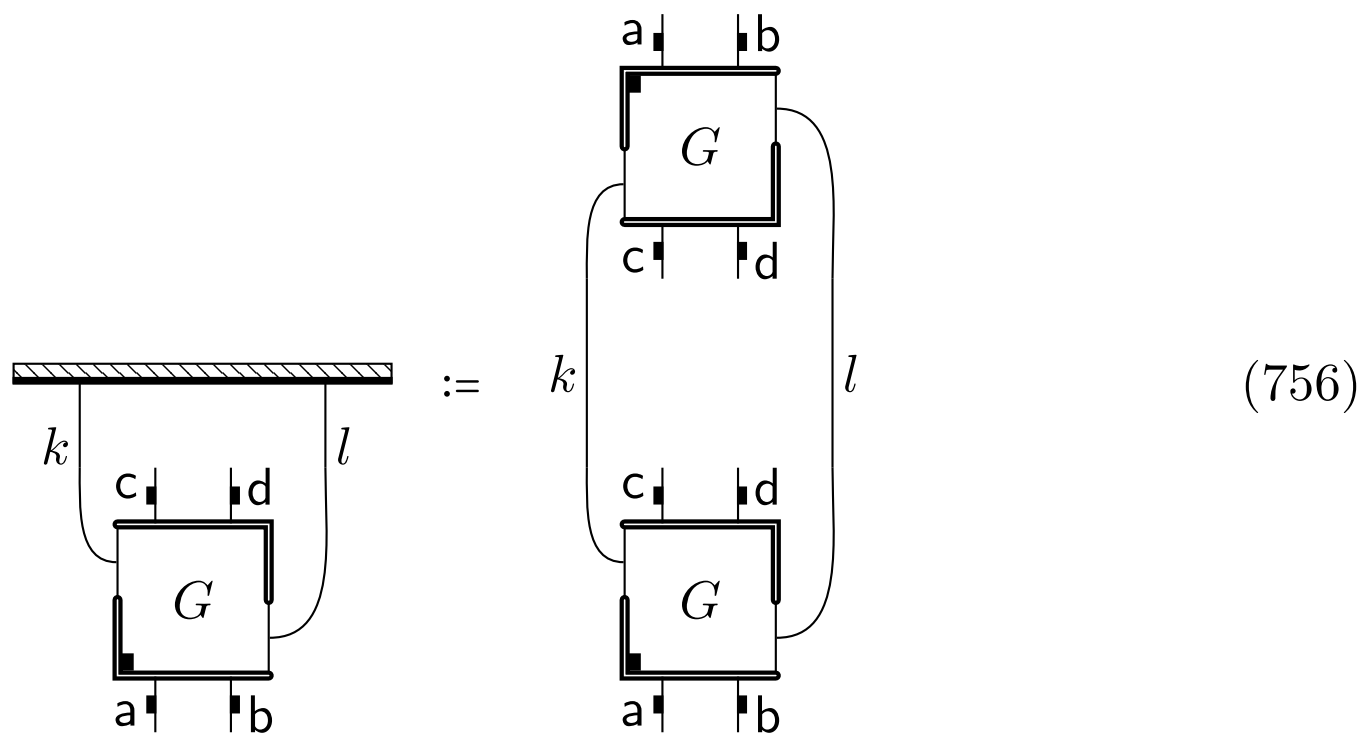

(756)

From this defining equation we can still obtain (740) and (741) as special cases as well as two more equations formed by flipping these equations horizontally.

We can also define the action of twist mirrors on general Hilbert objects. In place of the definition (744) we have

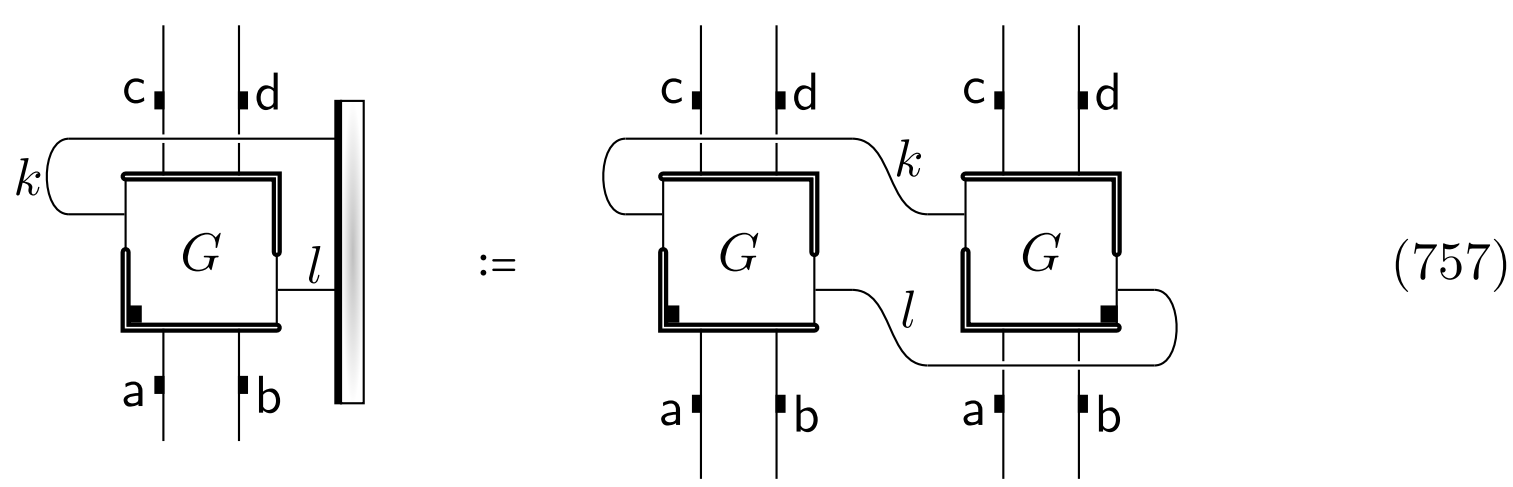

(757)

Particular applications of this definition are (745, 746) as well as

(758)

and

(759)

These allow us to extend the notation and attach system wires to the mirrors.

The definition in (748) is replaced by

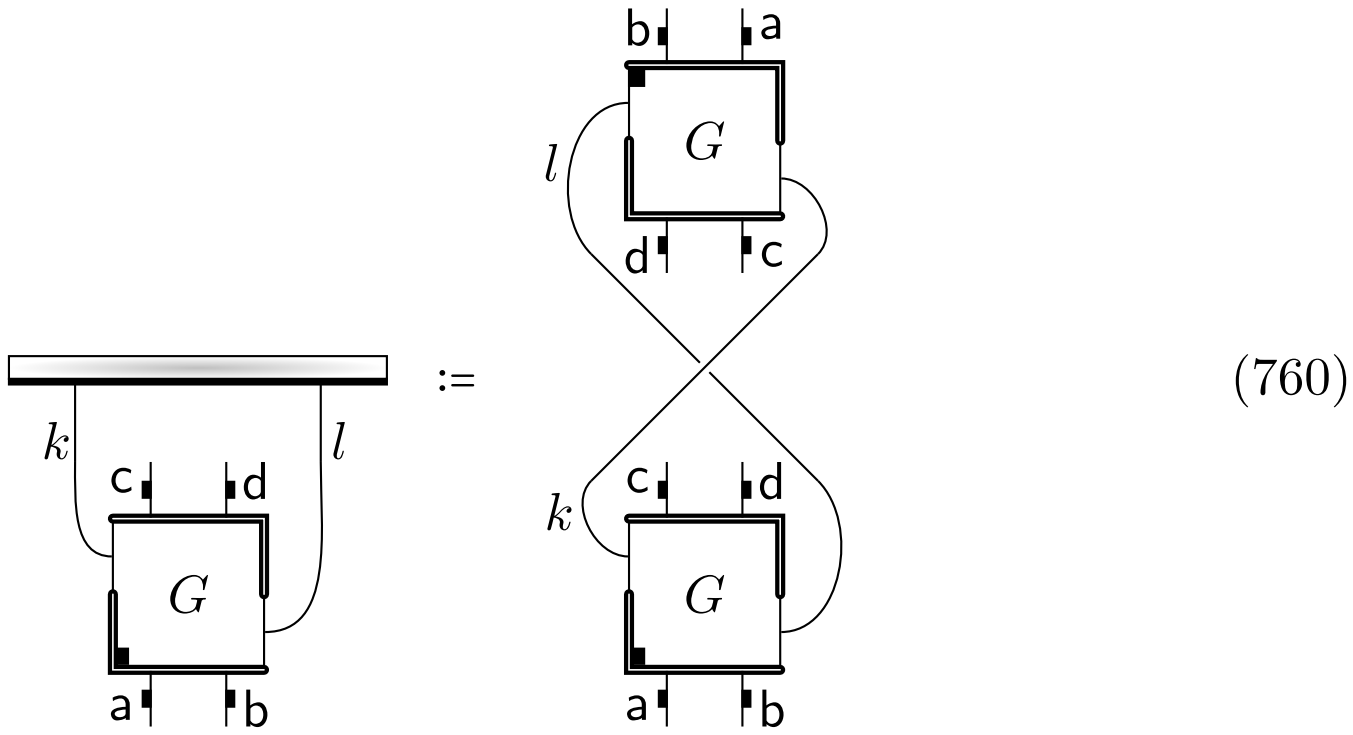

(760)

Particular applications of this definition are (749, 750) and horizontal reflections of these equations. These allow us to extend the notation so we can attach system wires to the mirror.

There is an interesting property of the four above mirror expressions in (753, 756, 757, 760). Namely that we can transform between these four expressions by performing a nonconjugating transformation *only on the image part* of the expression as follows

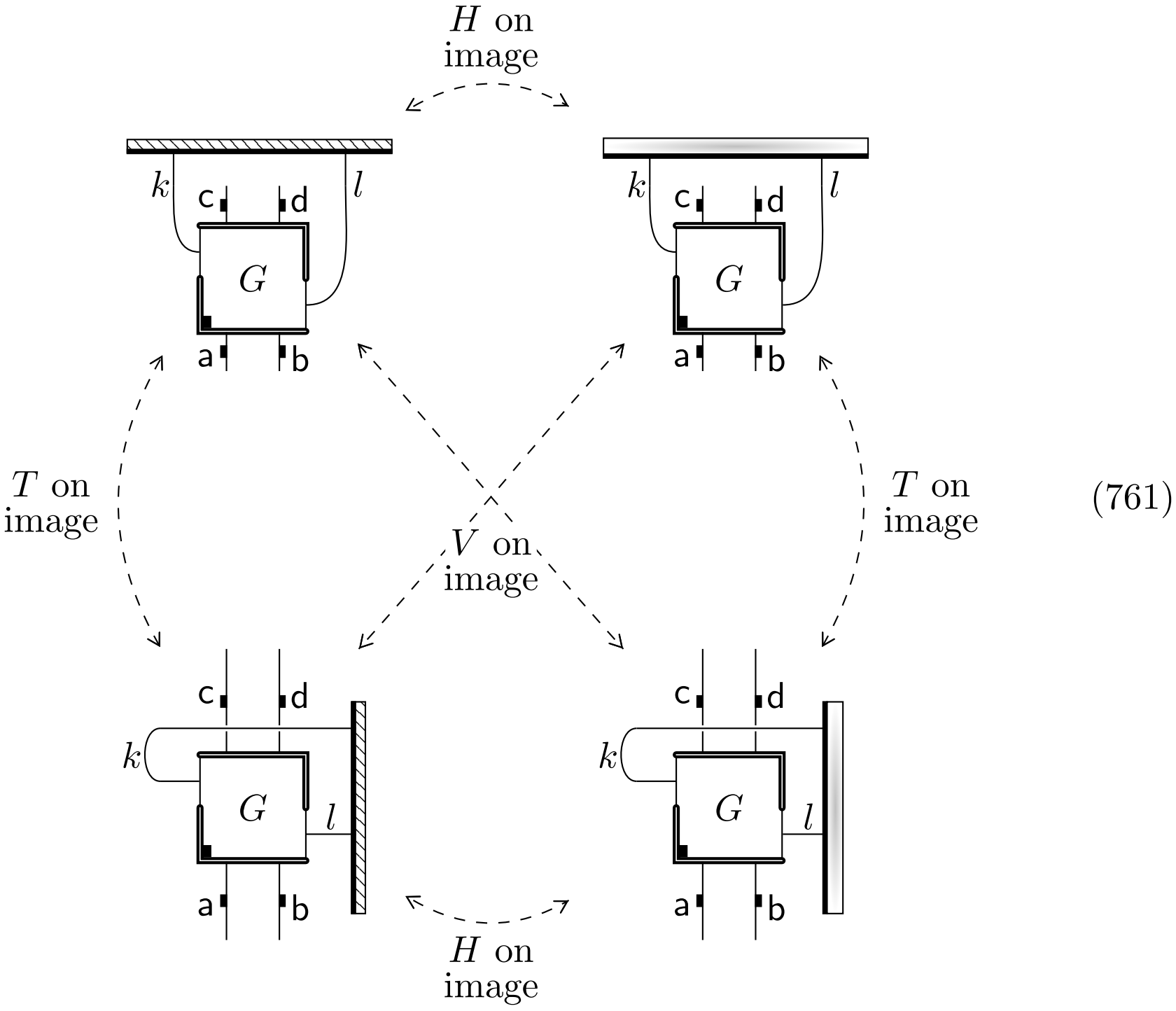

(761)

This is easily verified by reflecting out the expressions and applying the given transformations to just the image. These nonconjugating transformations are the members of the group $\mathcal{S}_{\text{nonconjugating}} = \{I, H, V, T\}$ discussed in Sec. 32.6. Recall that these are the transformations that can be applied to only part of an expression by means of caps/cups/twists. A similar statement to that depicted in (761) is true if we apply the nonconjugating transformations to just the object (i.e. not the image) part of the reflected out expressions.

## 34.4 Mirror equation equivalences

In Sec. 32.13 we showed how to form equivalent equations either by applying the same conjuposition to both sides of the equation or, if both of the equation have the same compositional form, by applying a nonconjugating transformation to corresponding components on both sides of the equation. Equations expressed in mirror form provide a particularly natural way to implement such transformations which we will now explore. In Sec. 34.5 we will prove a curious theorem - the mirror theorem - which takes advantage of this. First we will look at some examples of mirror equations and how they are equivalent to other mirror equations.

We will write all mirror equations in *balanced mirror equation form* by which we mean we have the same kind of mirror on each side.

First consider the equation

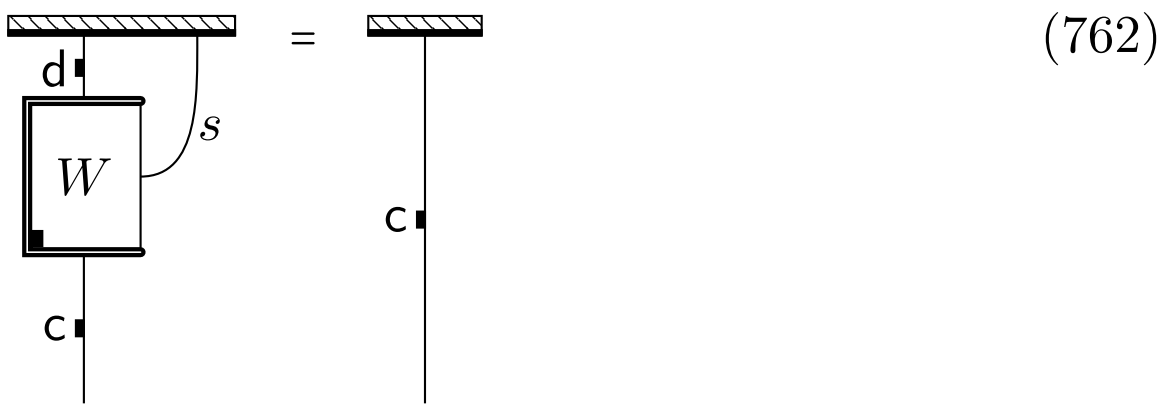

(762)

This equation is equivalent to

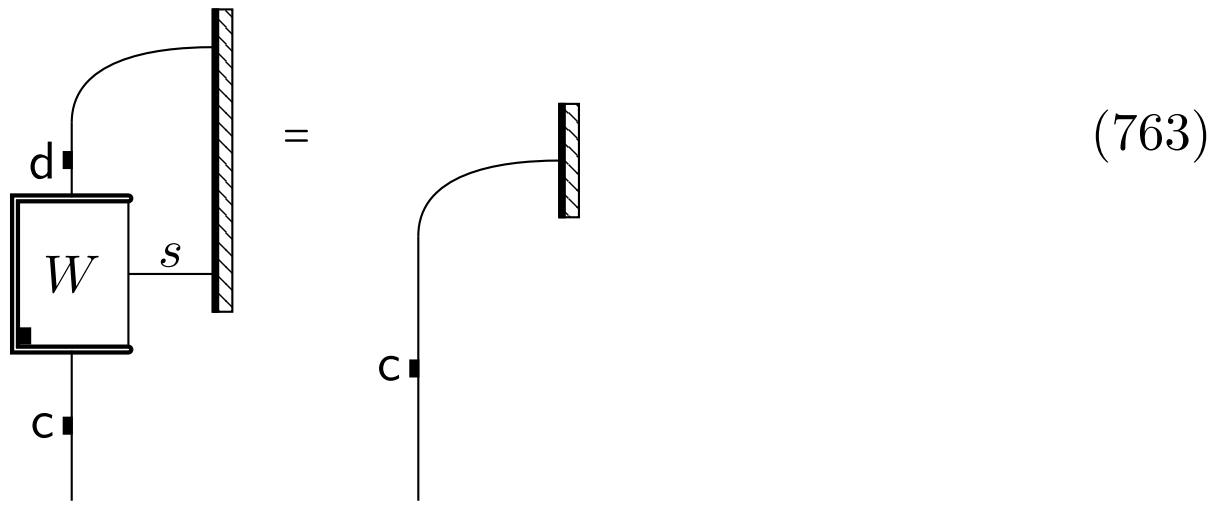

(763)

where the mirror has been rotated by 90°. To prove this, first complete the reflections in (762) taking us to the form given in (1899). Then we add a cap at the top (to the c system) on each side of the equation. Then we use the sliding manoeuvre (see Sec. 32.3). The resulting equation is symmetric about a vertical axis and can be written in the form shown in (763). We can run this proof backwards so (762) and (763) are equivalent. It is clear that, in general, if we have an equation with left (or right) objects reflected in a horizontal mirror (as in (762)) on each side, we can form an equivalent equation with vertical mirrors on each side (as in (763)). The proof of equivalence simply works by adding caps (or cups) as necessary and sliding.

This also works if we consider twist mirrors. Thus, the equation

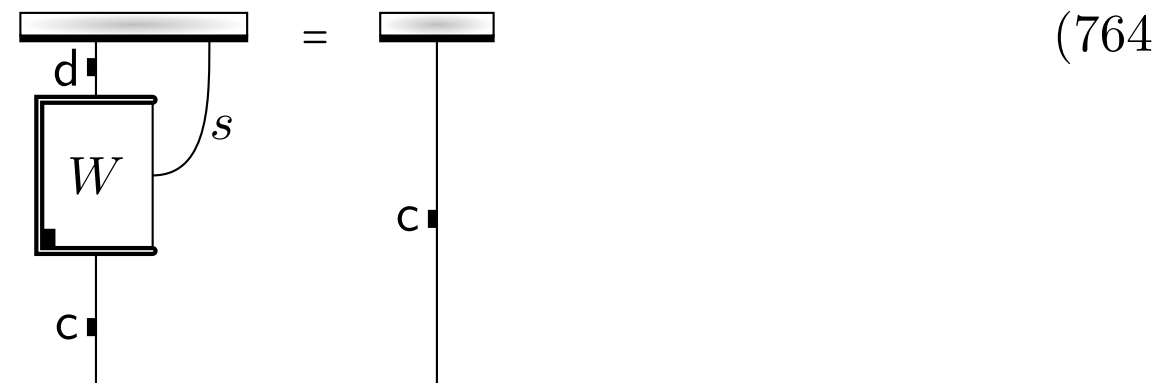

(764)

is equivalent to

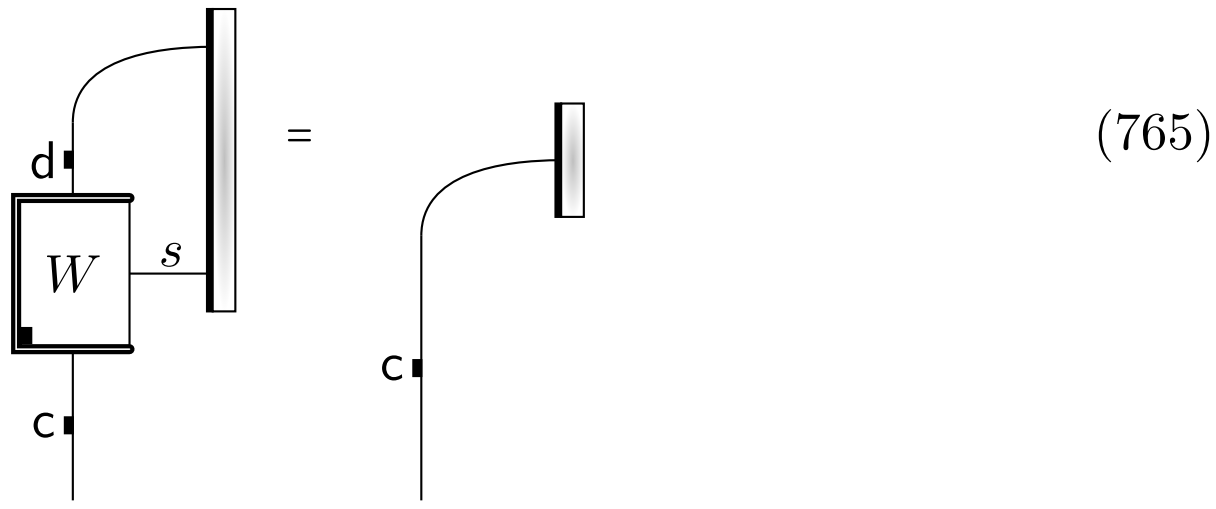

(765)

where the twist mirror has been rotated by 90°. The proof of this follows by similar techniques as for the natural mirrors - we add a cap and employ the sliding manoeuvre. This also clearly works in general for such examples.

In fact, it can be shown all the equations (762), (763), (764), and (765) are equivalent. This is because if we replace natural mirrors by twist mirrors, we also get equivalent equations. This is easily proven for the above example but we will consider a slightly more complicated example to illustrate some key points.

We can prove that the equation

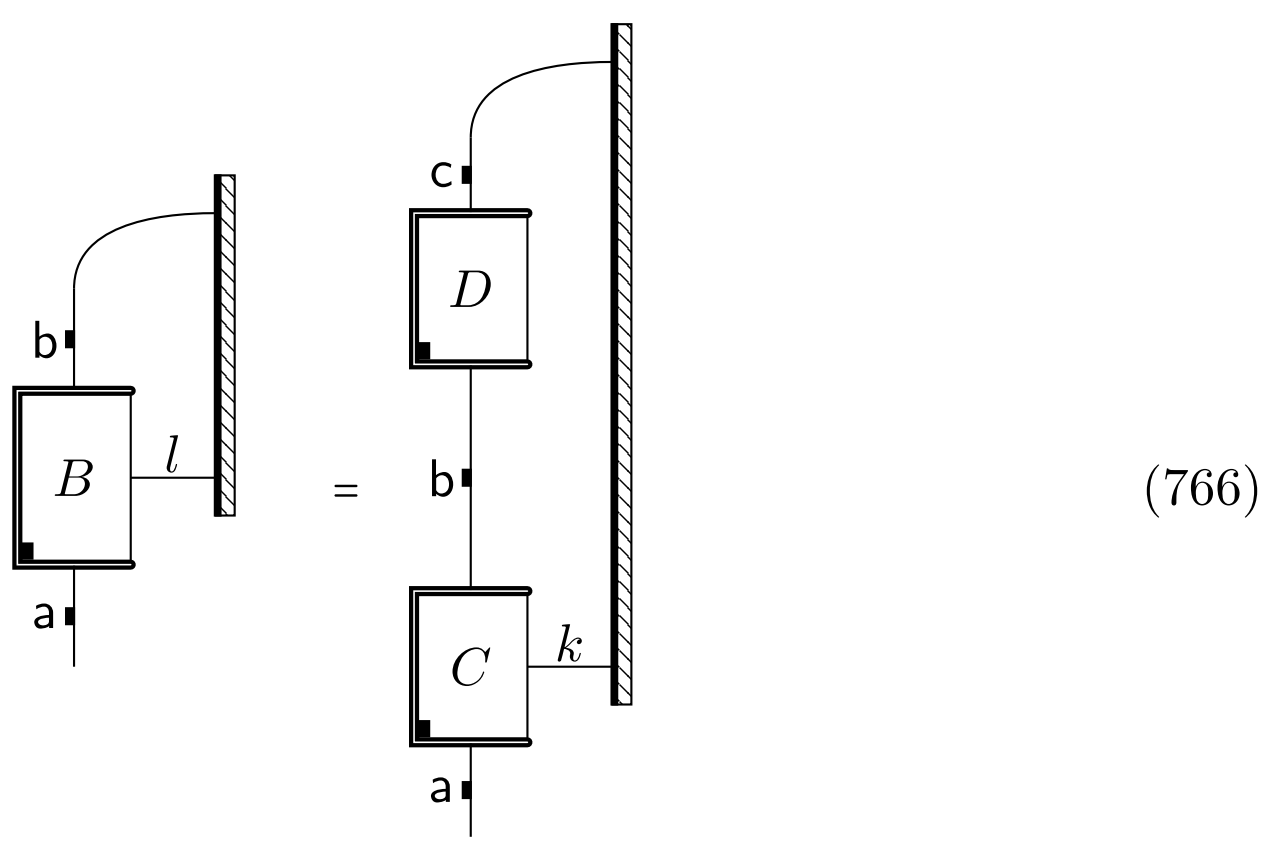

(766)

is equivalent to the equation

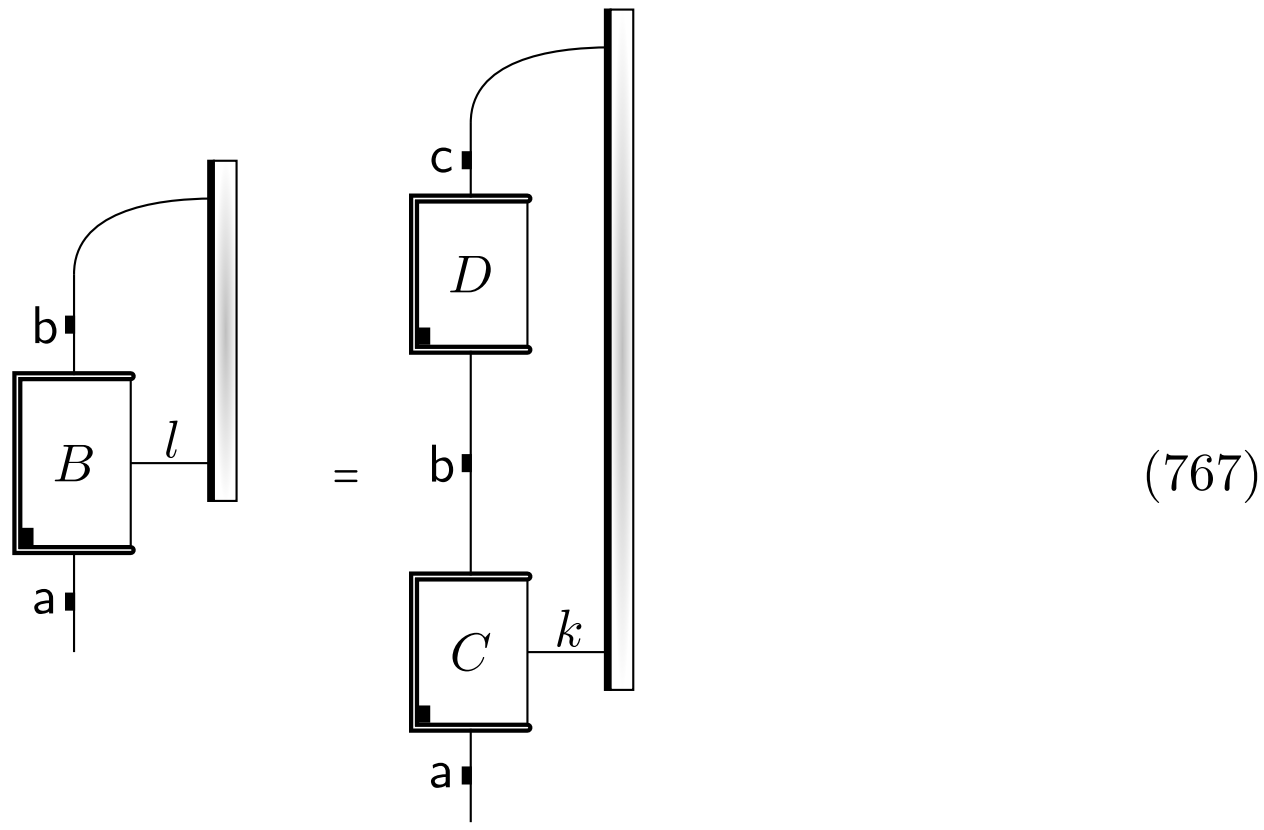

(767)

To prove this first expand out (766). We obtain

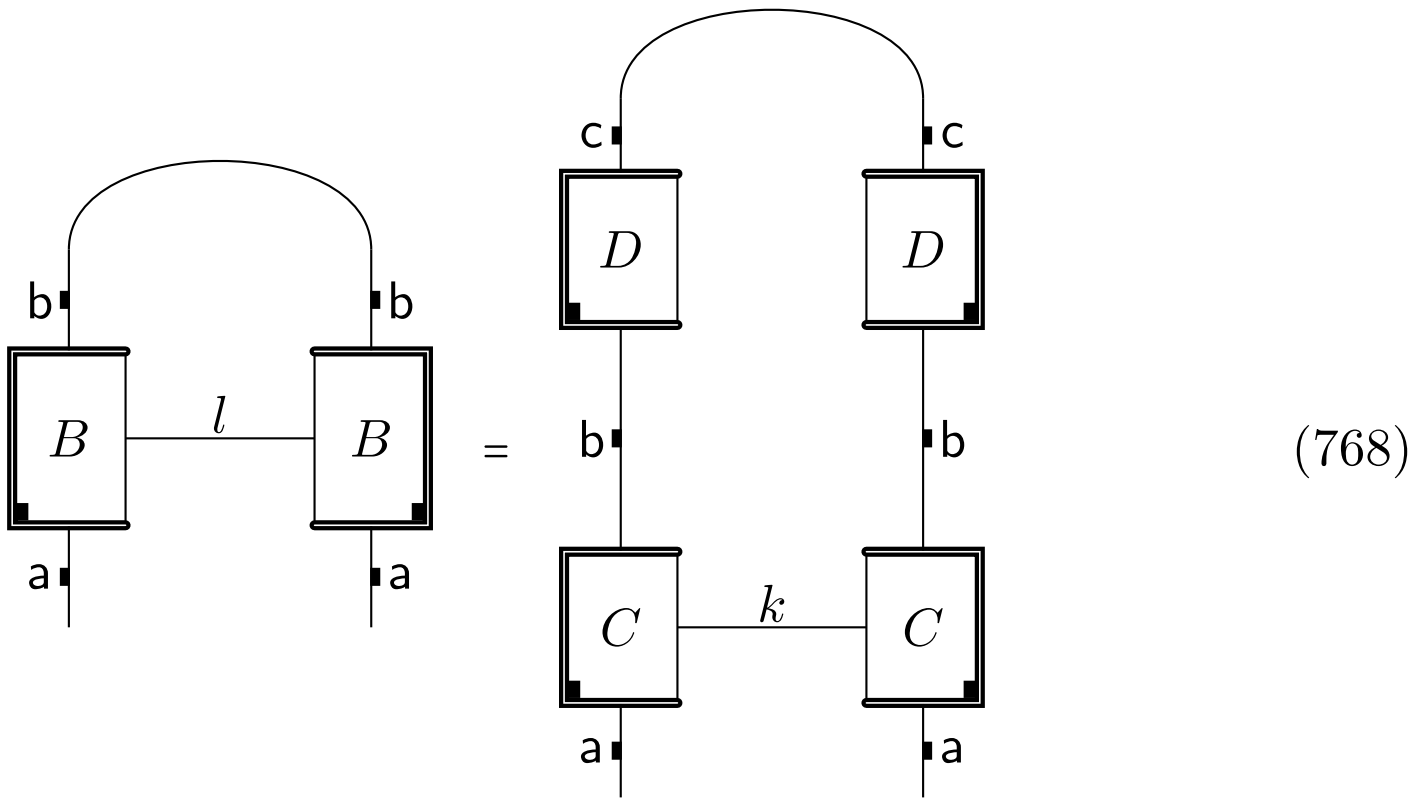

(768)

Now we insert pairs of twists in each closed right Hilbert space system wire (we can do this since they annihilate by (517)) and we apply a twist to each open wire on the right to give

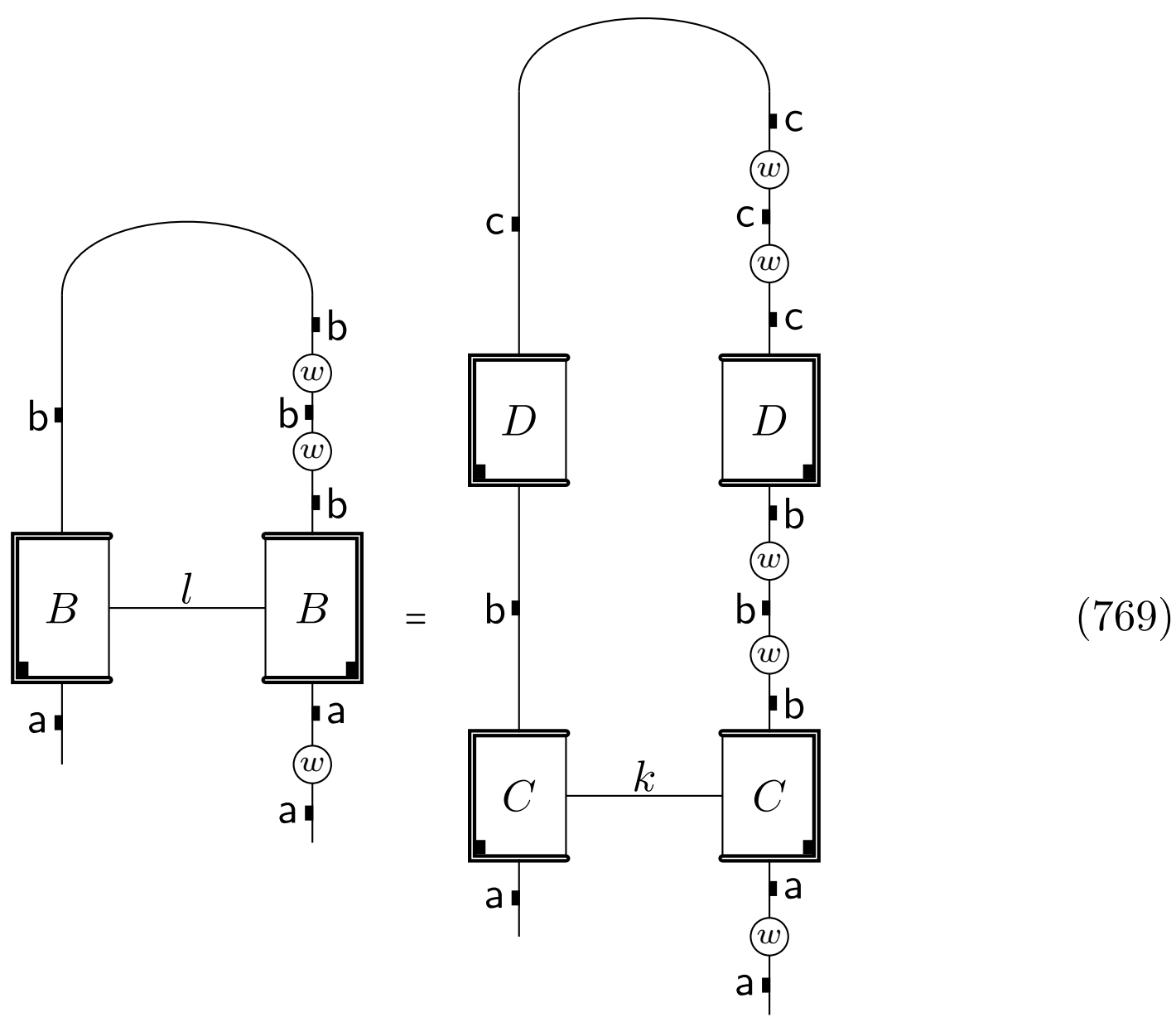

(769)

Using (579) this gives (767) as required. These steps are reversible so (766) and (767) are equivalent equations. Note that the expression on (for example) the left hand sides of these equations are not equal but can be obtained from one another by applying a twist object on the open wires.

## 34.5 Normal mirror theorem

We will now prove a mirror theorem which says we can always rotate and replace normal mirrors in balanced equations to obtain equivalent equations. Any balanced mirror equation having vertical regular normal mirrors on both sides can be put in the form

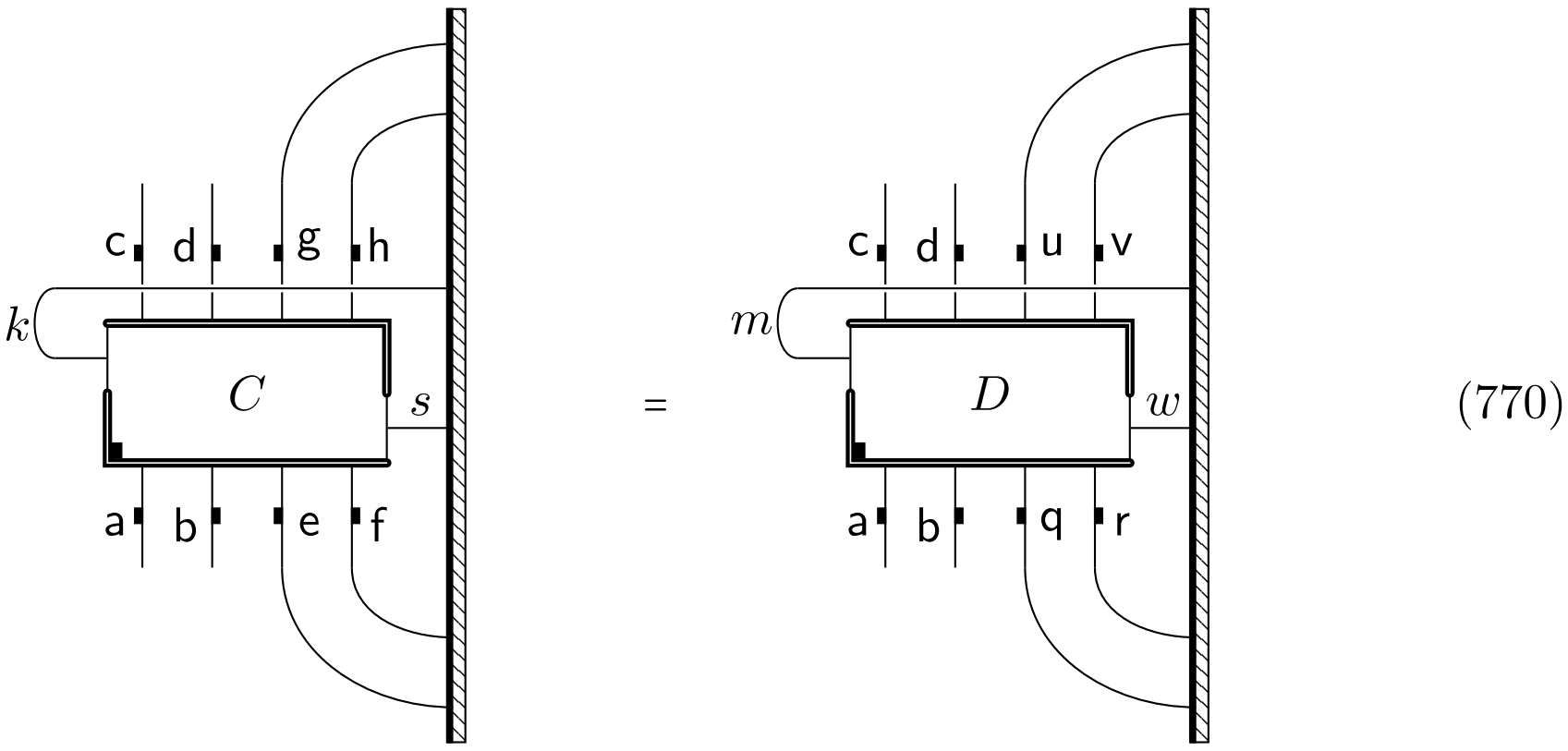

(770)

The two kinds of manipulation discussed in Sec. 32.13 can be applied to such equations. First, we can apply any element of the normal conjuposition group to each side. For example, if we apply $V$ we end up with the new equation

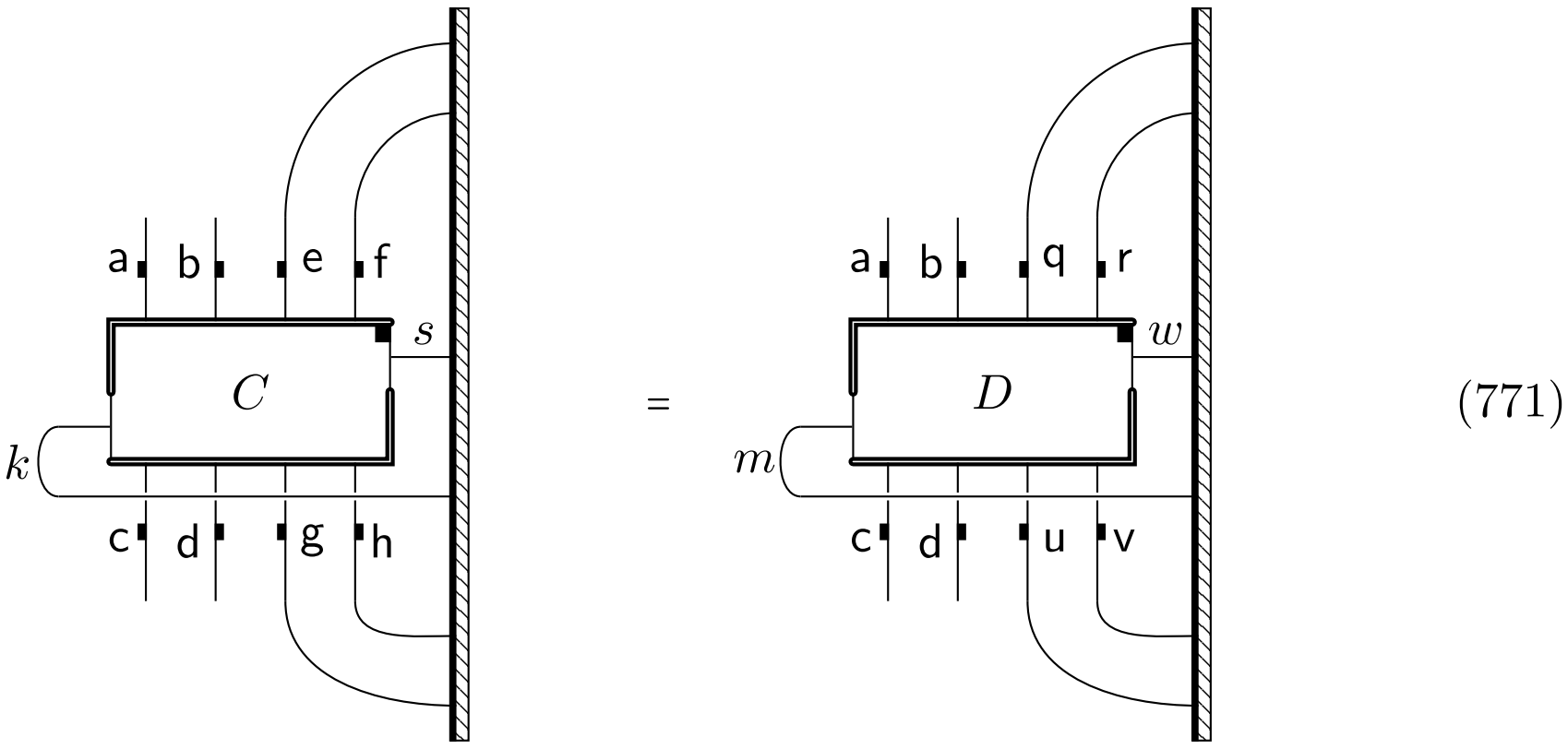

Second, we can apply a nonconjugating transformation to just the image. To see how to do this it is useful to first put (770) in a form where only label wires attach to the mirror. We do this by inserting basis elements on each system wire that attaches to the mirrors (using (740,741)). This gives

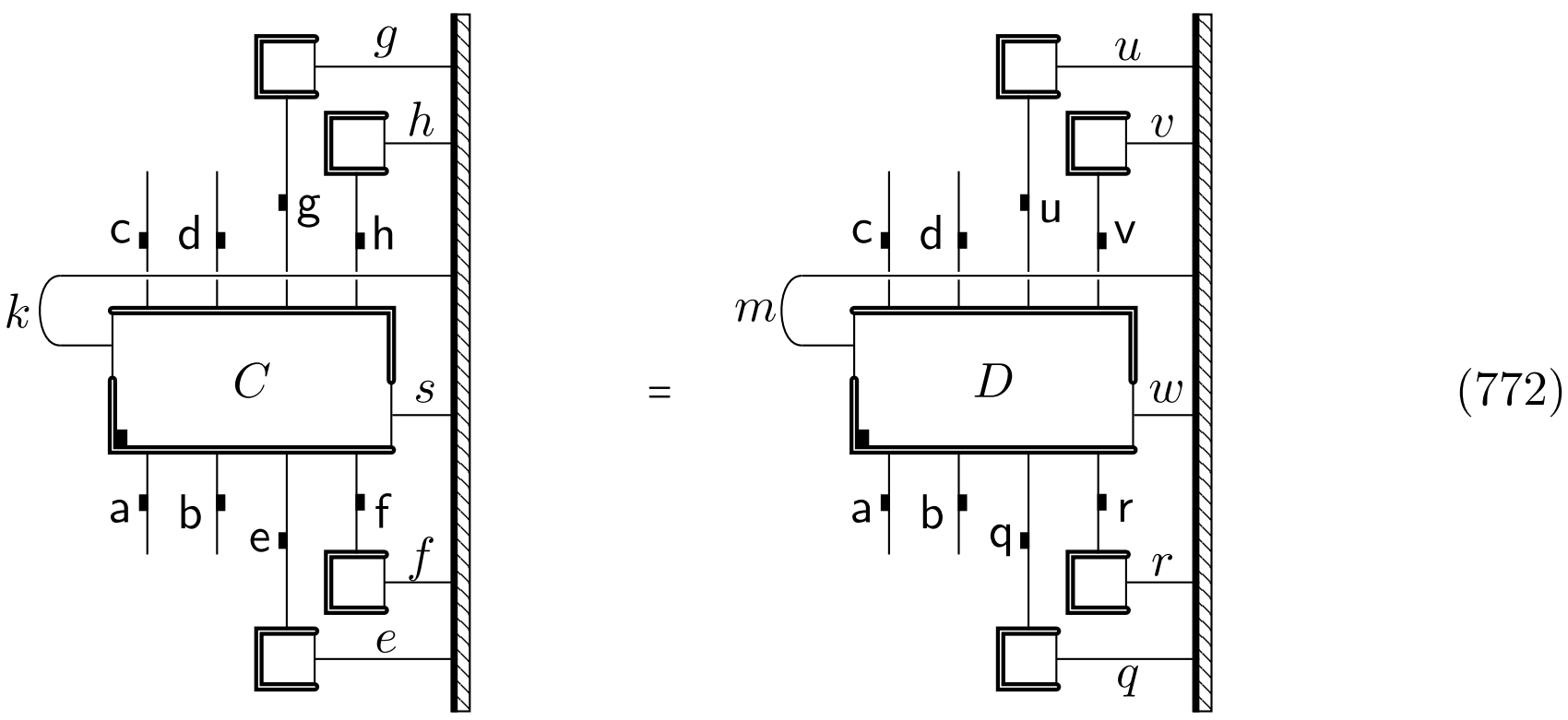

These equations are of the form

(773)

where $l = ghs$ and $n = uvw$, $A$ contains $C$ and the basis boxes, and $B$ contains $D$ and the basis boxes.

Now, as stated above, we can perform a nonconjugating transformation to just the image part of the expression on each side of the equation thereby obtaining an equivalent equation. We saw in (761) that any nonconjugating transformations acting only on the image will convert the expression into a similar expression with a different mirror. This gives the following three additional equations.

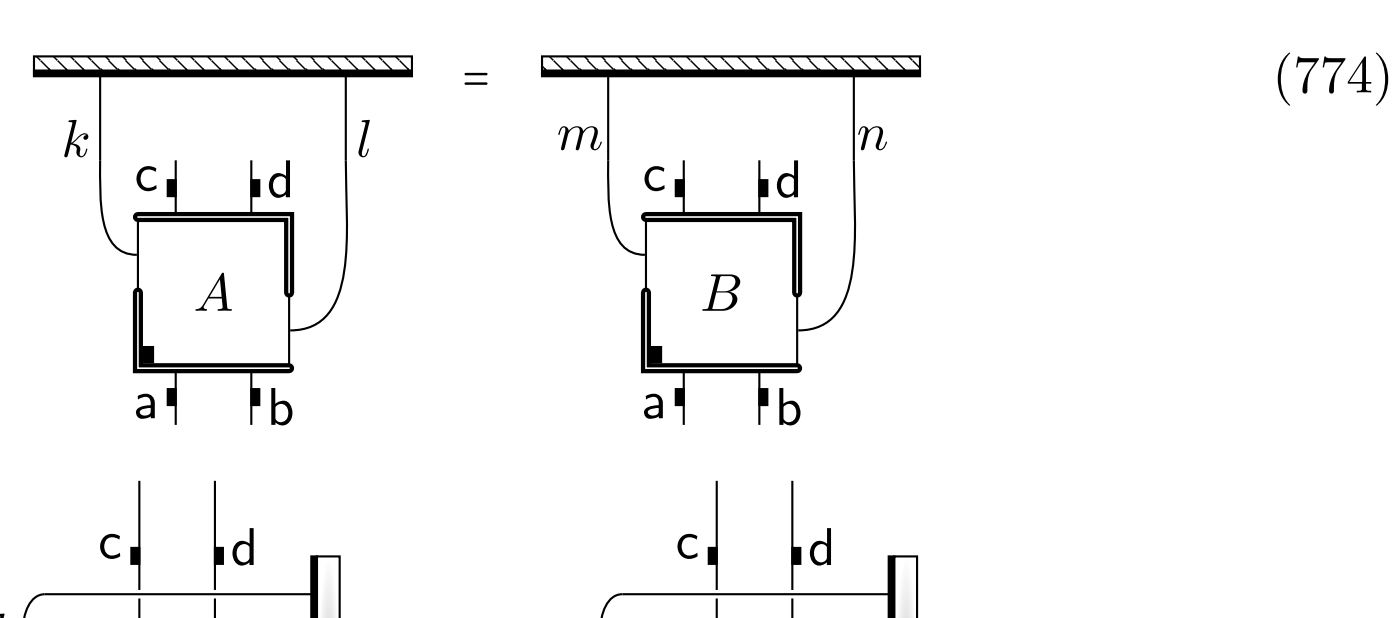

(774)

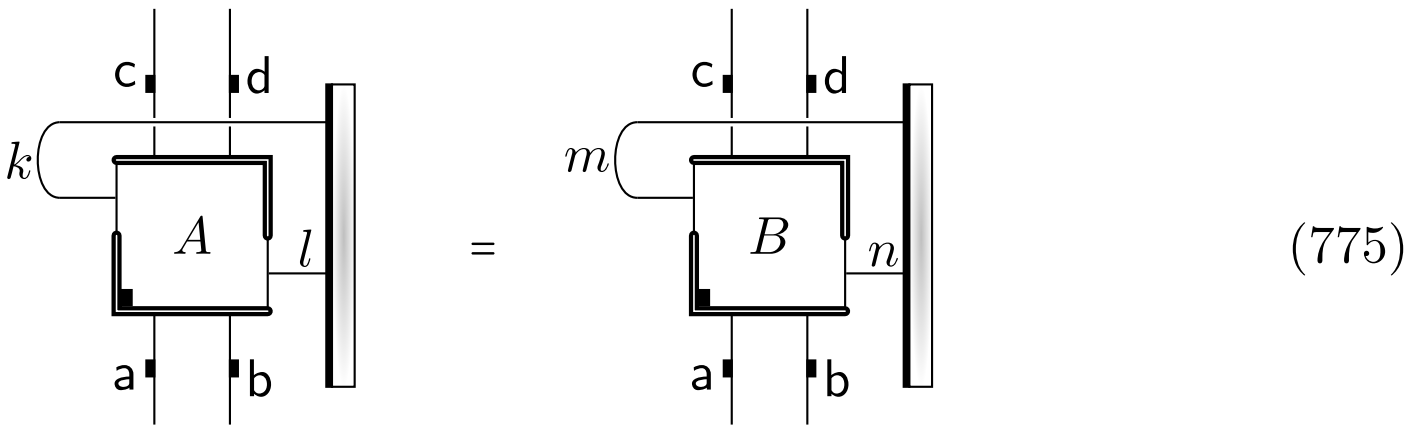

(775)

and

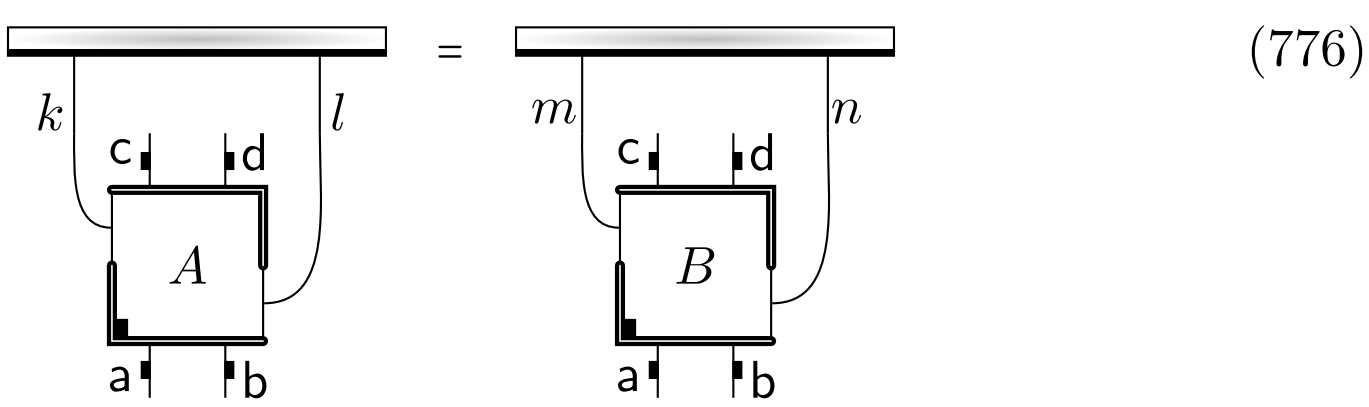

(776)

Finally we can, if we wish, reattach the system wires to the mirrors. If we do this with (775) then we get an equation like (770) but where the natural mirrors have been replaced by smoke mirrors. If we do this with (774) then we obtain

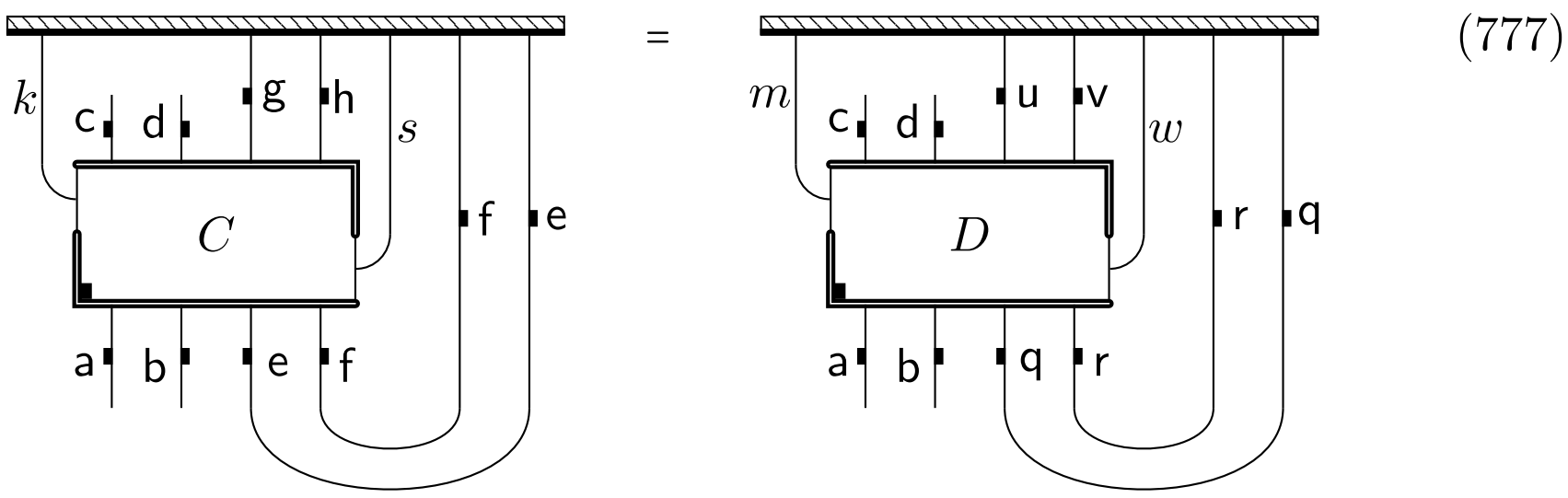

(777)

If we do this with (776) we obtain an equation like (770) but where the natural mirrors have been replaced by smoke mirrors.

It is worth noting that the wires that are attached to the mirror are "dragged" with the mirror when we rotate it. In the above examples we considered rotating the mirrors in (770) through 90° anticlockwise. We could also rotate the mirrors clockwise through 90° (this is clear as the results in (761) extend to that case). Then we have mirrors whose reflective surfaces are on top. We could also rotate so that we have mirrors whose reflective surfaces are on the right. With these further comments in mind we can finally state the normal mirror theorem.

> **Normal mirror theorem.** Consider an equation, expressed such that the left hand side has a general Hilbert object, $C$, reflected in a conjugating normal mirror and the right hand side has another general Hilbert object, $D$, reflected in the same kind of mirror. This equation is equivalent to any equation formed such that the left hand side is expressed in terms of any normal conjuposition acting on $C$ reflected in any conjugating normal mirror and the right hand side has the same normal conjuposition acting on $D$ reflected in the same kind of normal mirror. Further, we require that the attached wires in the original equation are "dragged" so they attach to the mirror on each side of the equation in the corresponding way.

This theorem clearly follows from the foregoing discussion.

A normal mirror can be horizontal or vertical, have or not have smoke, and have the reflective coating on either side. Hence we have eight possible types of mirror. We have eight normal conjupositions. Thus, it would appear that we have 64 equivalent equations. However, there is clearly double counting since any equation formed by reflecting to the right can clearly be formed by reflecting to the left instead (and similarly for reflecting up or down). Hence, we actually only have 32 equations in any equivalence set formed using the mirror theorem. In any application we will probably only be interested in a small number of these.

Let us illustrate the mirror theorem with an example.

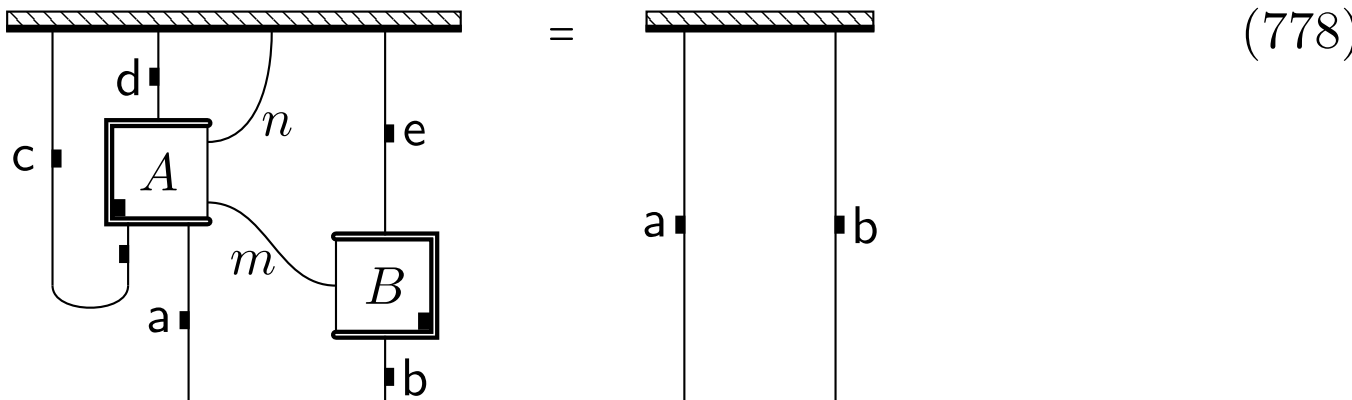

(778)

is equivalent to

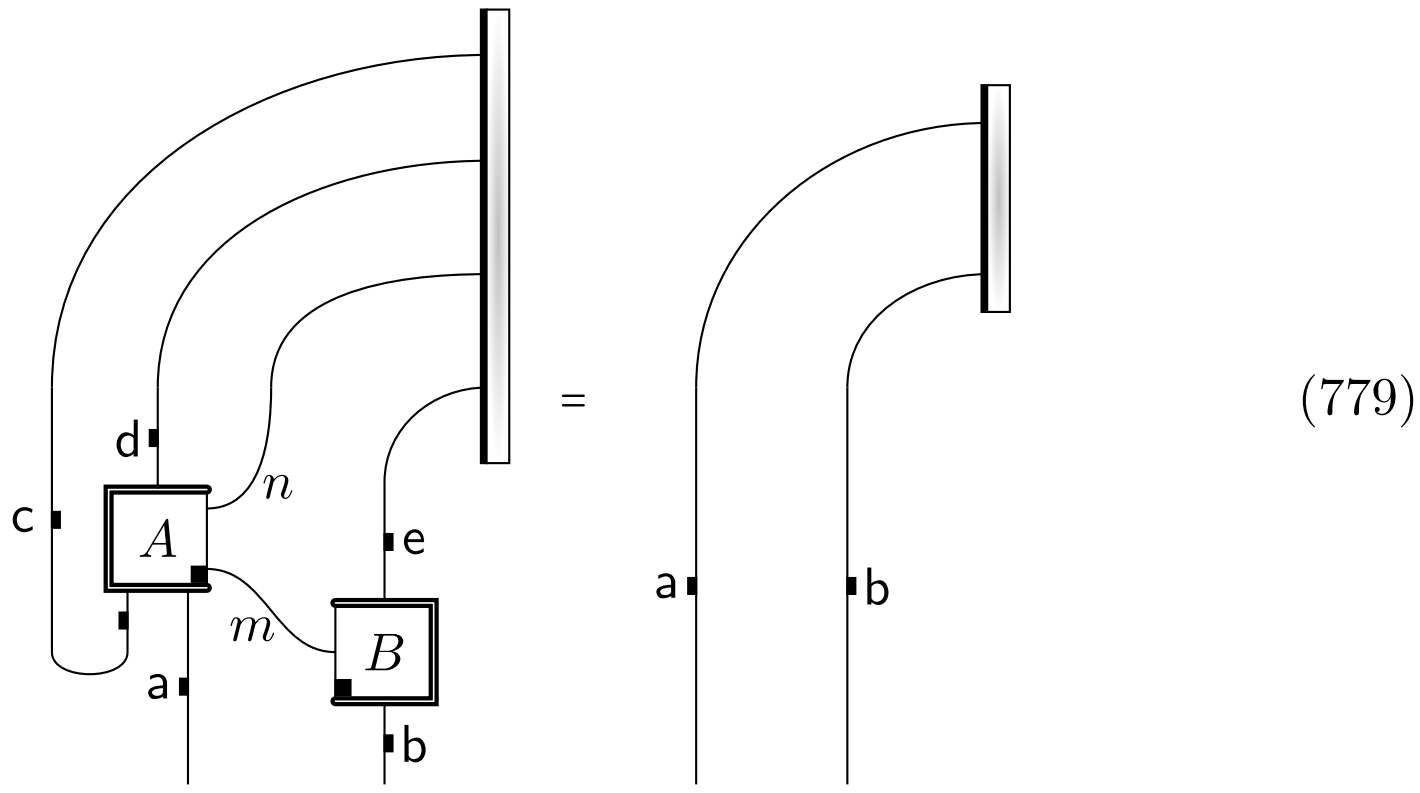

(779)

where, first of all, we have applied $\overline{I}$ to the whole expression on each side, and subsequently we have applied $V$ to the image (through the process outlined above). This last move has the effect of replacing the regular mirror in (778) with a twist mirror rotated through 90°.

This mirror theorem applies to equalities, not inequalities. We could try to set up a mirror theorem for inequalities. The problem is that we require a notion of positivity to define what we mean by inequalities. However, the notion of positivity we currently have (twofold positivity or one of the equivalent notions) is not invariant under the image transformations we consider in the mirror theorem. If we could find a more general notion of positivity that is invariant under these image transformations then we would be able to set up a mirror theorem with inequalities. A general notion of positivity of this sort would be interesting in its own right.

## 34.6 Scalar products and norms

Consider an expression like

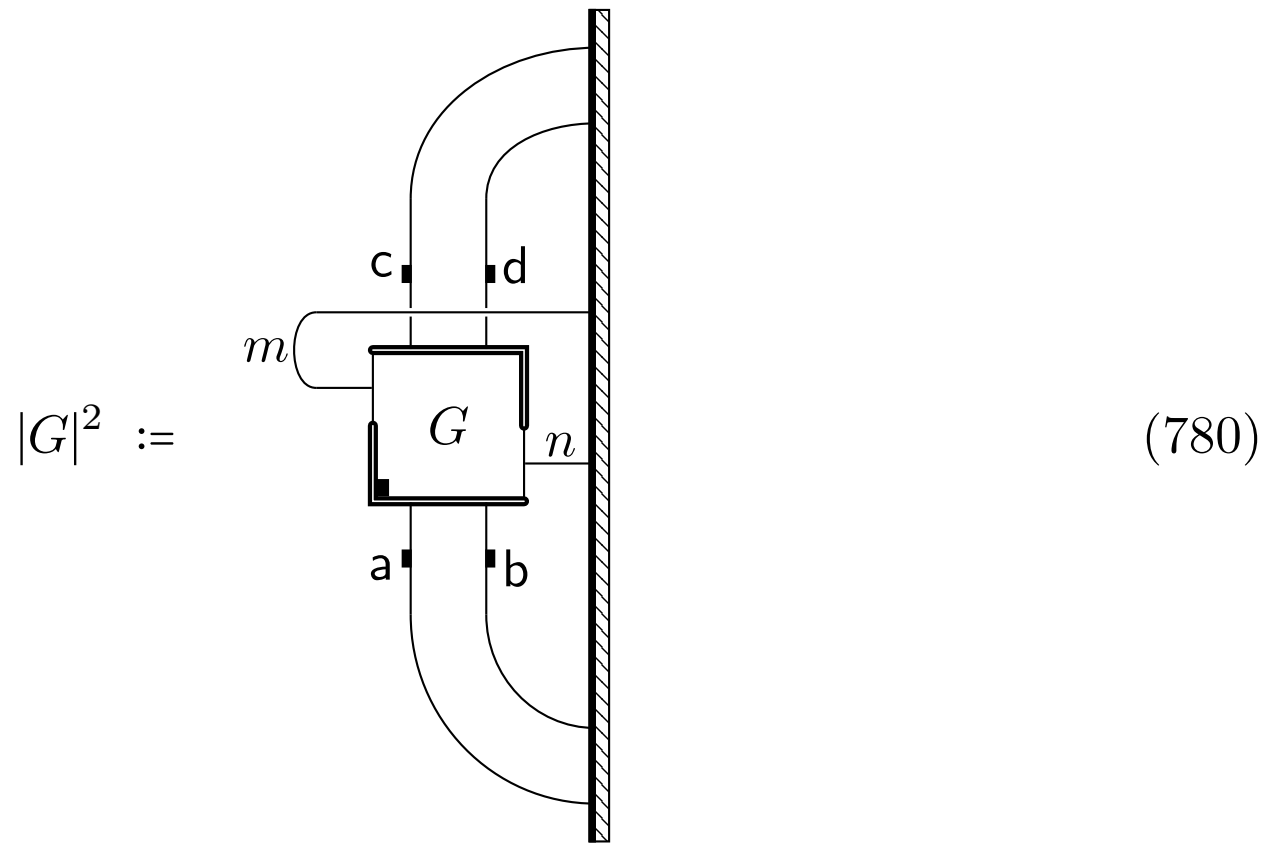

(780)

having no open wires. We can think of this as the scalar product (of $G$ with its horizontal adjoint). We call $|G|$ the *norm* of $G$. This is a generalisation of the norm introduced in (463). By expanding out the general Hilbert object in terms of a basis we can show that this expression is necessarily real and non-negative (because the mirror is conjugating). This justifies writing $|G|^2$. We can write

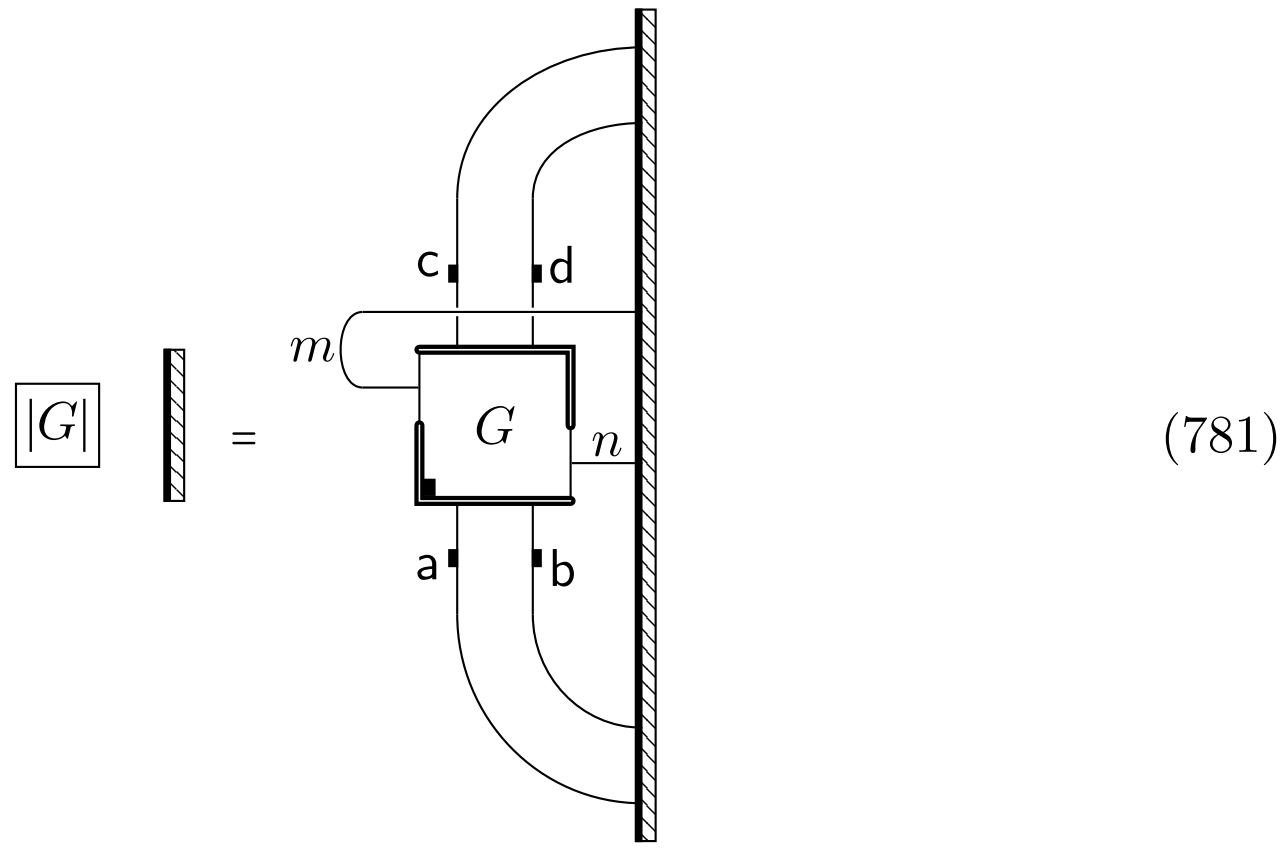

(781)

It immediately follows from the mirror theorem that any mirror transformation of this equation has the same constant on the left hand side. For example

$$= |G|^2 = \qquad (782)$$

$|G|$ $k$ c d $l$ a b $G$ a b

Thus, when we have scalar products like (780) all its normal mirror transformations are equal. Consequently $G$ has the same norm no matter what mirror we use to calculate it.

A simple application of the above ideas is the following theorem

> **Self scalar products non-negative.** The following properties for scalar products hold

$$0 \leq \quad = \quad = \quad = \quad = |B|^2 \qquad (783)$$

$B$ a $B$ a a $B$ $B$ $B$ a $B$ $B$ $B$ a a $B$

The inequality follows as the norm is nonnegative. The subsequent three equalities can be obtained using the mirror theorem applied to

$$|B| \quad = \quad \text{a} \; B \qquad (784)$$

where the left hand side holds as this scalar product is real. We can obtain further equalities by further application of the normal mirror theorem using twist mirrors. We say that $B$ is *normalised* if the scalar product in (783) is equal to 1.

There is a sense in which a given object is, up to normalisation and overall phase, the unique object pointing in a given direction in the associated direction as made precise in the following theorem.

**Scalar product theorem.** If

$$B \;\text{a}\; B \quad = \quad C \;\text{a}\; C \qquad (785)$$

then

$$0 \quad \leq \quad C\,C \;\; \text{a} \;\; \text{a} \;\; B\,B \quad \leq \quad B\,B \;\; \text{a} \;\; \text{a} \;\; B\,B \qquad (786)$$

and

$$C\,C \;\; \text{a} \;\; \text{a} \;\; B\,B \quad = \quad B\,B \;\; \text{a} \;\; \text{a} \;\; B\,B \quad \Leftrightarrow \quad C \quad = \quad e^{i\theta} \; B \qquad (787)$$

for some $\theta$.

This theorem simply restates well known properties of Hilbert vectors in the diagrammatic notation used in this book. We leave its proof as an exercise - though some parts of the proof can be obtained easily using the mirror theorem.

## 34.7 Pulling out a wire

We note that

$$\text{a} \quad = \quad \boxed{N_{\mathsf{a}}^{\frac{1}{2}}} \quad = \quad \boxed{N_{\mathsf{a}}} \tag{788}$$

This is easily obtained using (733) and (734). Thus, if we "pull out a wire" from a mirror, we pick up a factor of $N_{\mathsf{a}}^{\frac{1}{2}}$.

We can apply the mirror theorem to this equation leading to various other equations. For example, it follows from (788) and the mirror theorem that

$$\text{a} \quad \text{a} \quad = \quad \boxed{N_{\mathsf{a}}^{\frac{1}{2}}} \tag{789}$$

Here we have rotated by 90° and replaced the natural mirror with a standard mirror.

In the next section we will introduce *physical mirrors*. These are simpler in that, when we pull out a wire, we pick up a factor of 1 (see Sec. 35.5).

# 35 Physical mirrors

We have seen how to set up a theory of mirrors based on the normal conjuposition group. We can do the same for the natural conjuposition group. There is an important point to note. We use particular properties of the orthogonal basis when we extend our notation to allow system wires to attach to mirrors. For example, in the case of normal mirrors we use the properties of orthonormal bases in (733, 734). We have two orthogonal bases that are associated with the natural conjuposition group - the ortho-physical basis (see Sec. 33.1) and the complementary ortho-deterministic bases (see Sec. 33.8). Thus, we can set two natural theories of mirrors - for *physical mirrors* and for *deterministic mirrors*. The theory of physical mirrors turns out to be the most useful (because, as we will see, when a system wire attaches to a vertical mirror, it gives rise to an ignore operator - see (791, 792). Thus we will study physical mirrors in detail. We briefly touch on the theory of deterministic mirrors in Sec. 35.7.

We define physical mirrors based on the conjugating elements of the natural conjuposition group and prove a physical mirror theorem. We do this in an analogous way to the case of normal mirrors. We will provide the definitions of these mirrors and the mirror theorem here, all expressed in terms of general Hilbert objects. We will see that the expressions in Table 4 play a role in defining what happens when a system wire connects to a physical mirror. We will note an interesting difference between physical and normal mirrors in Sec. 35.5 and comment on the physical significance of physical mirrors in Sec. 35.6.

## 35.1 Regular physical mirrors

The regular physical mirrors are associated with the basis independent transformations $\overline{\underset{\sim}{H}}$ and $\overline{\underset{\sim}{V}}$.

The mirror associated with $\overline{\underset{\sim}{H}}$ is defined through

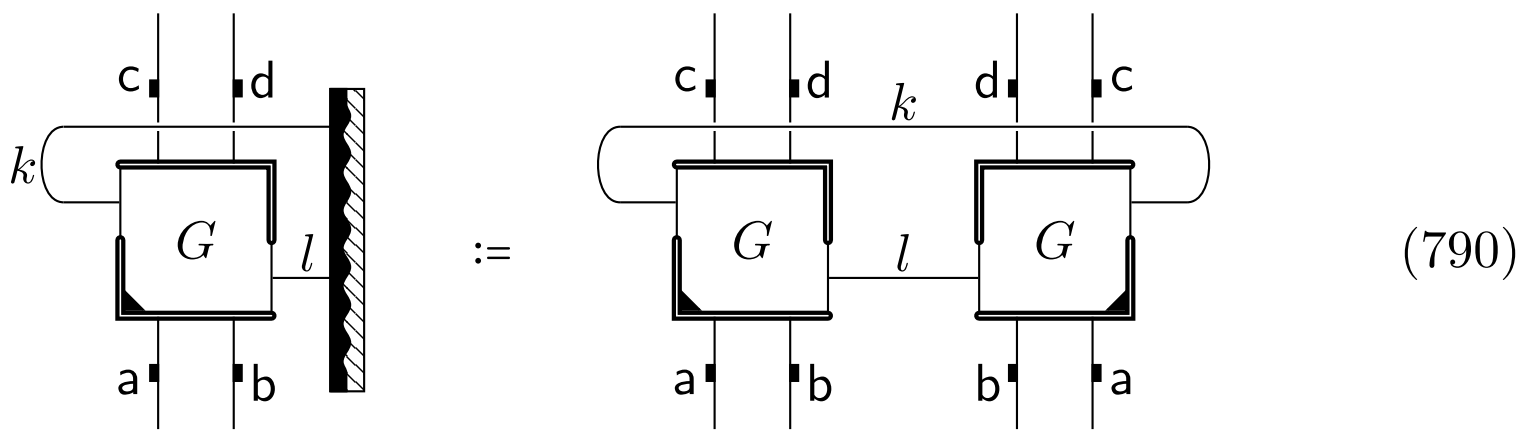

(compare with (753)). The image is the physical horizontal adjoint. If we apply this definition to the ortho-physical basis elements themselves we obtain

(791)

(792)

(793)

and

(794)

This means we can extend our notation (as shown in the final expressions in the above equations) to allow system wires to connect to this mirror (it is at this point that we enter the territory of physical mirrors rather than deterministic mirrors). This employs the objects in row 2 of Table 4.

The mirror associated with $\overline{\underset{\sim}{V}}$ is defined through

(795)

(compare with (756)). The image is the physical vertical adjoint. If we apply this definition to the ortho-physical basis elements themselves we obtain

(796)

This employs the objects in row 1 of Table 4. We have indicated, in the final expression, an extension of the notation allowing system wires to connect to the mirror. We can, similarly obtain the horizontal flip of this

(797)

where, again, we have indicated in the last step an extension of the notation.

There are two further ways of reflecting basis elements in a mirror. First consider

(798)

The second step in (798) uses (677, 678). The third step uses the fact that the special cup transforms into a special cap under the physical vertical adjoint (see (696)) and the horizontal flip of (796) above. We can obtain the horizontally flipped version of this similarly

(799)

The need for special cups in (798) and (799) is curious - compare with the (741) which is the corresponding equation for normal conjupositions. The special cup in (798) slightly complicates the physical mirror theorem (which will be given in Sec. 35.3). Interestingly, when we come to the complex case we will not need special cups because we will have arrows on the wires (see discussion at the end of Sec. 72.1) and then the mirror theorem is correspondingly simpler.

## 35.2 Twist physical mirrors

The twist physical mirrors are associated with the basis dependent transformations, $\underset{\sim}{\overline{I}}$ and $\underset{\sim}{\overline{T}}$.

The mirror associated with $\underset{\sim}{\overline{I}}$ is defined as follows

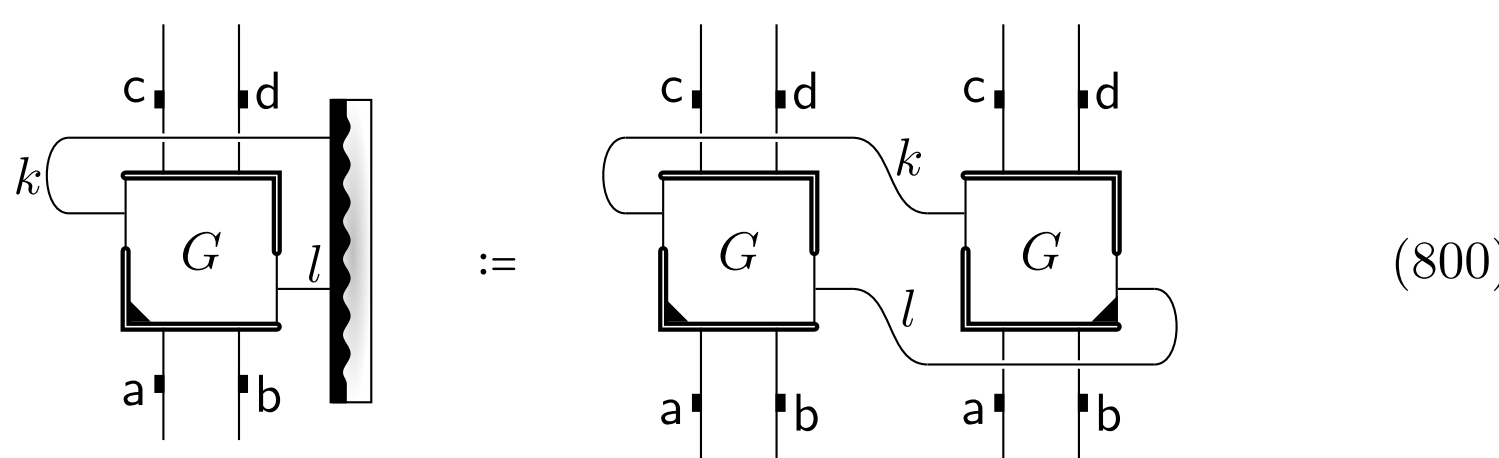

(800)

(compare with (757)). where we obtain

(801)

(802)

(803)

and

(804)

by applying this definition to the ortho-physical basis elements. This employs the objects in rows 4 and 5 in Table 4. This allows us to extend the notation so the system wires can connect to this mirror.

The mirror associated with $\overline{\underset{\sim}{T}}$ is defined through

(805)

(compare with (760)). where we have

(806)

as an application of this definition to ortho-physical bases allowing the indicated extension of the notation so system wires can connect to this mirror. We can also obtain the horizontal flip of this by starting with a right ortho-physical effect. We can obtain the following and

(807)

Here the second step uses the definition of special caps and cups in (677, 678) and the third step uses (796) and the fact that the image of the indicated special cup in the twist mirror is the special cap (since the image is the physical adjoint, $\overline{\underset{\sim}{T}}$, we can use (696)).

## 35.3 Physical mirror theorem

We can prove a physical mirror theorem for physical mirrors (the physical mirrors we have considered are all conjugating ones). This theorem is slightly complicated by the appearance of the special cups in (798) and (807).

This theorem states that

> **Physical mirror theorem.** Consider an equation, expressed such that the left hand side has a general Hilbert object, $C$, reflected in a conjugating physical mirror and the right hand side has another general Hilbert object, $D$, reflected in the same kind of mirror. This equation is equivalent to any equation formed such that the left hand side is expressed in terms of any physical conjuposition acting on $C$ reflected in any conjugating physical mirror and the right hand side has the same physical conjuposition acting on $D$ reflected in the same kind of mirror. Further, we require that the attached wires in the original equation are "dragged" so they attach to the mirror on each side of the equation in the corresponding way and that any cups (or caps) resulting from this dragging are required to be special cups (or caps).

Compare with the (standard) mirror theorem in Sec. 34.5. The proof is along the same lines as the standard mirror theorem (except for the appearance of special cups and caps). The key to the proof is the property

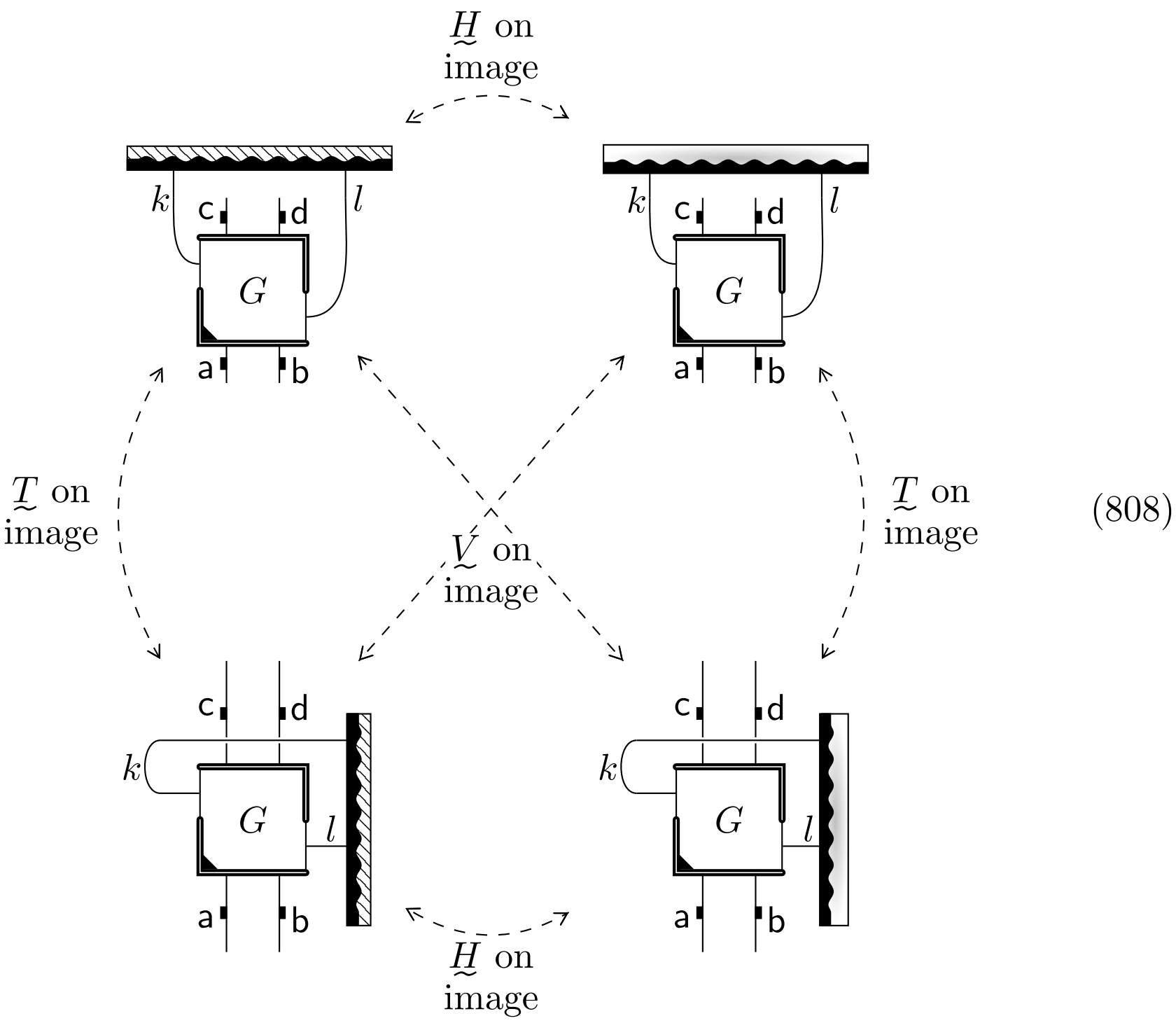

(808)

(compare with (761)). This is easy to prove by reflecting out these expressions and applying the given transformation to just the image. To prove the theorem start by considering a general equation

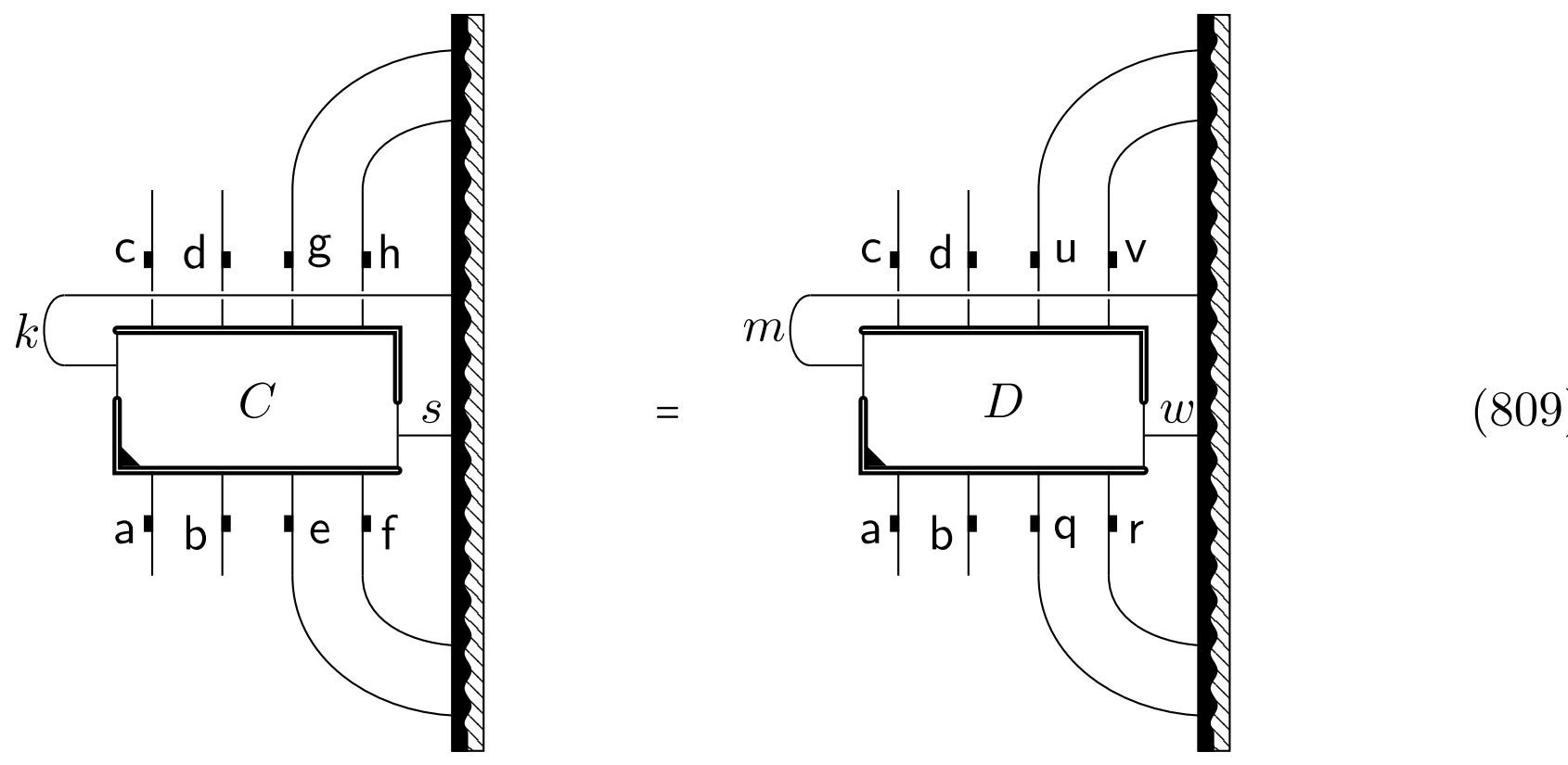

(809)

We can write this as

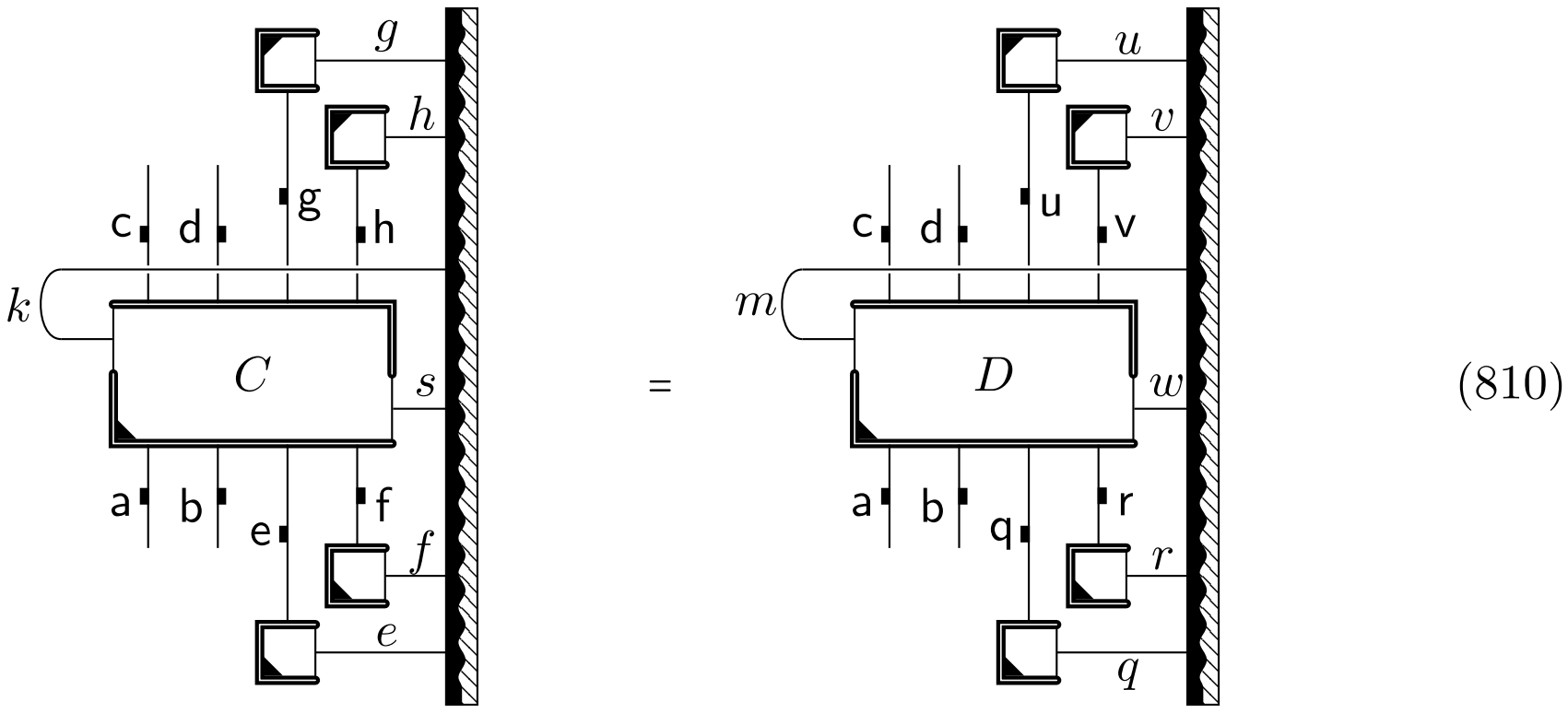

(810)

This is in the form

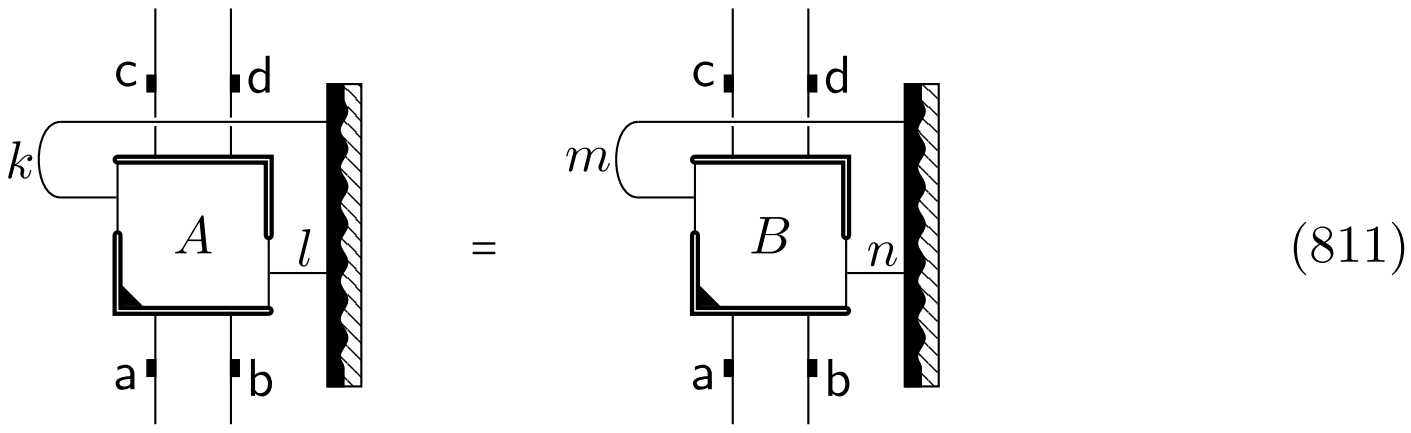

(811)

Using the result (808) above, we can apply a nonconjugating physical conjuposition to the image on both sides of the equation giving three new equivalent equations. We will just consider the equation obtained by applying the physical transpose, $\underset{\sim}{T}$, to the image as the other cases are similar. Then we obtain

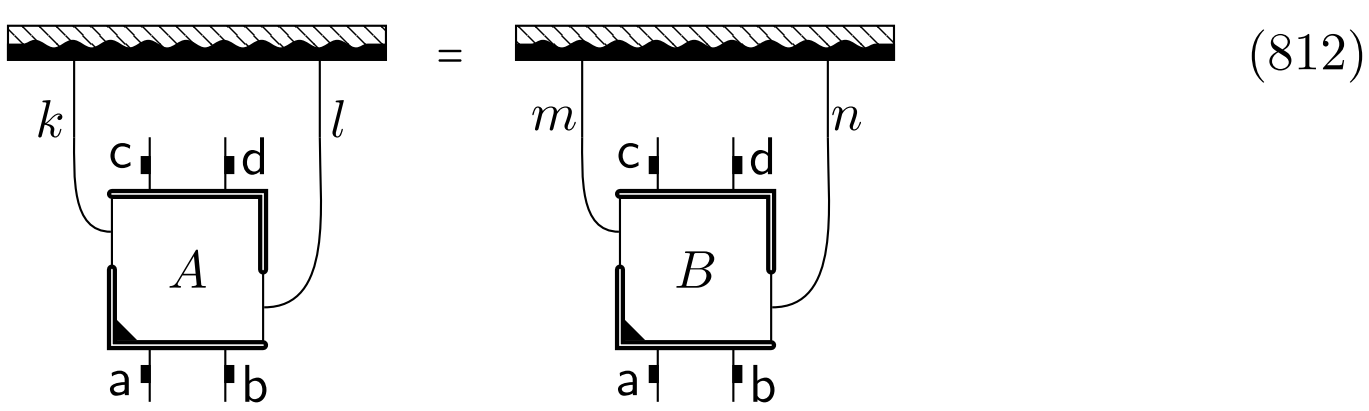

(812)

Referring back to (810) and using (796) and (798) we obtain

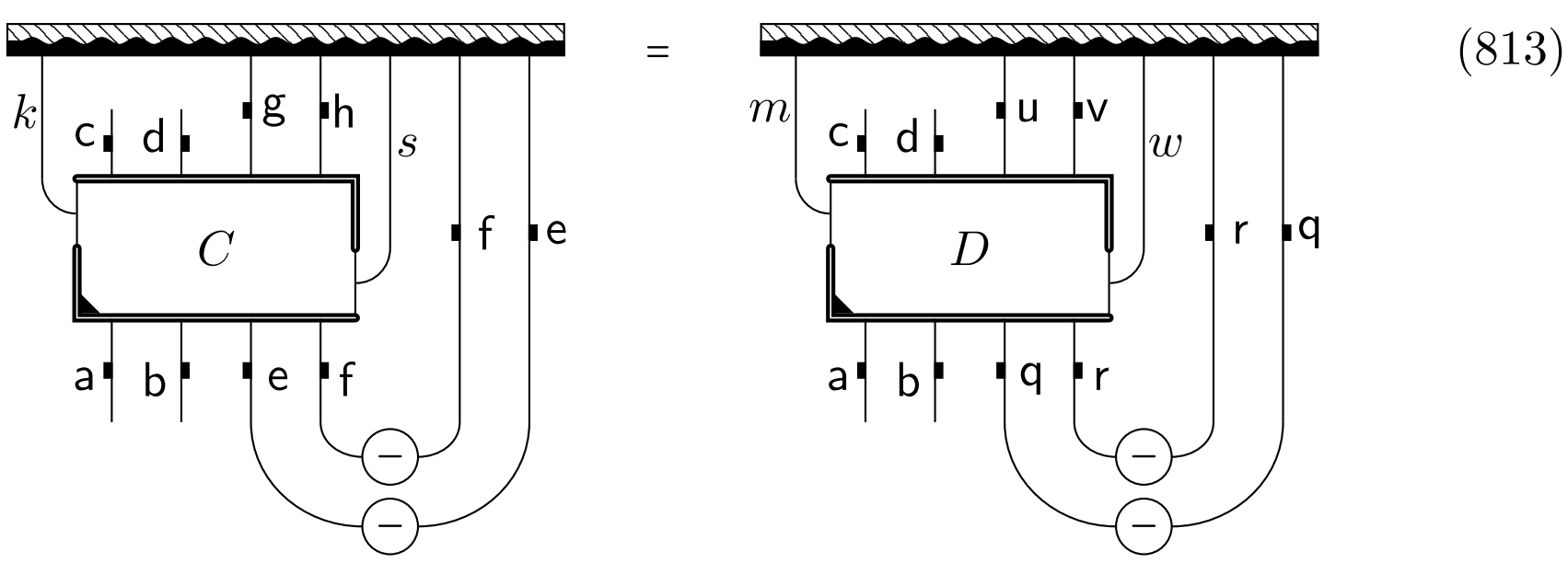

(813)

Note, in particular, the appearance of the special cups coming from (798). If we compare with (809) we can see that these have appeared when we drag the wires and create cups. These special cups (and special caps) can be thought of as necessary because standard caps and cups do not transform into one another under appropriate physical conjupositions while special caps and cups do (see (696). In a certain sense, use of these special caps and cups allows us to write equations in "covariant form".

We can use the mirror theorem to obtain other equations. For example, we can replace the natural physical mirrors in (813) with physical twist mirrors (having smoke) giving

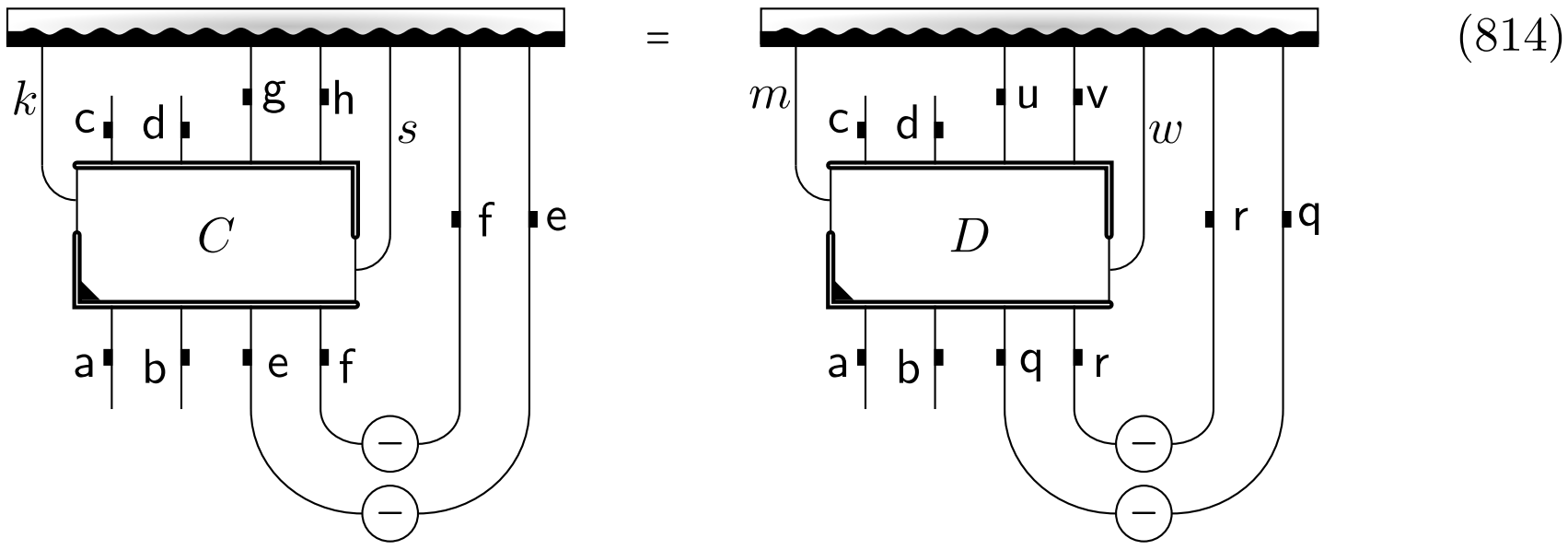

(814)

This equation could have been obtained directly from (809) by a procedure similar to that described above (whereby we arrived at (813) from (809)) where, now, the special cups come from using equation (807).

## 35.4 Physical scalar products and physical norms

In Sec. 34.6 we discussed scalar products and scalar norms with respect to standard conjupositions using standard mirrors. We can do something similar for physical mirrors. We have to be a little more careful because of the appearance of special cups and caps in the physical mirror theorem. We define the *physical norm*, $|\underset{\sim}{G}|$, of a general Hilbert object through the equation

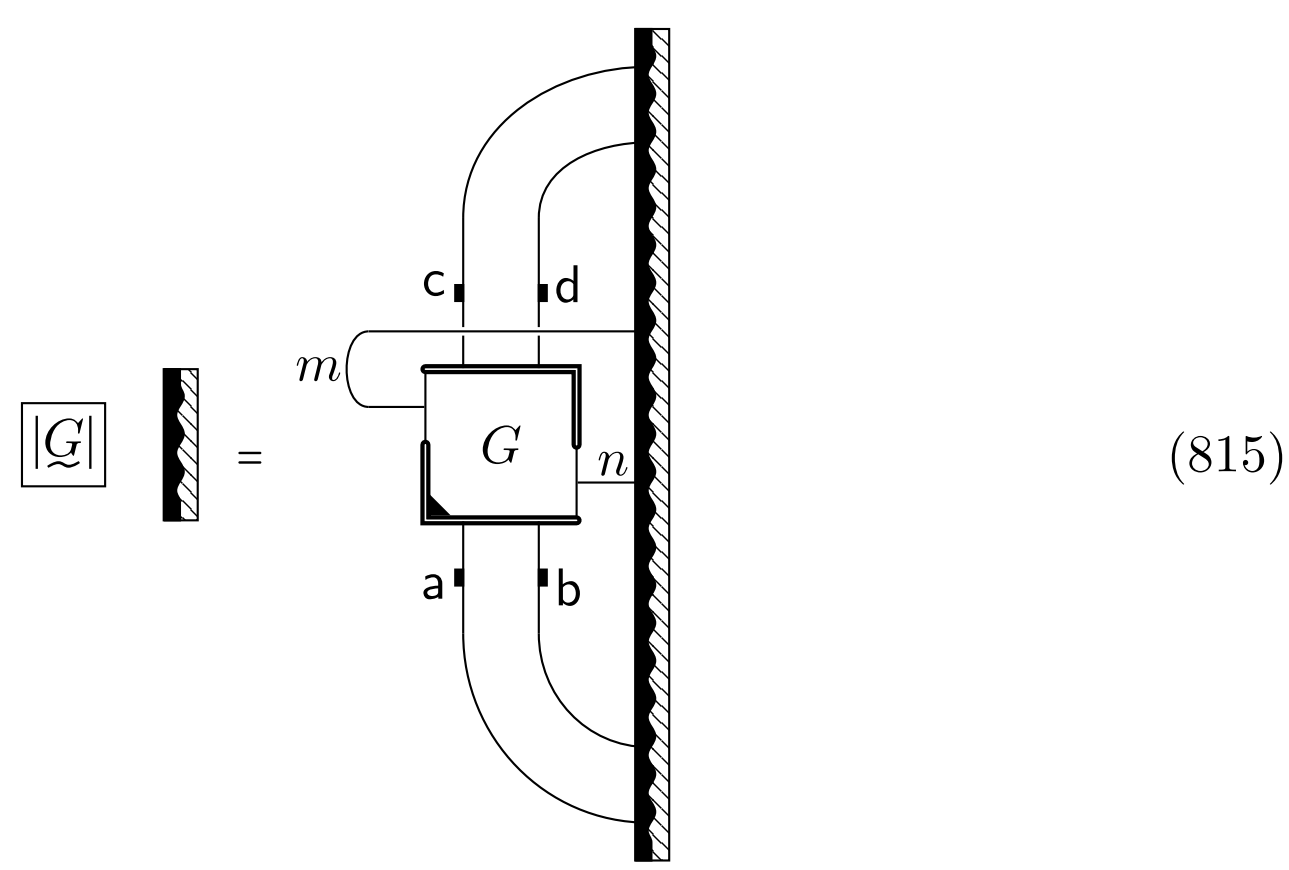

(815)

(the left hand side reflects out to $|\underset{\sim}{G}|^2$ of course). By the physical mirror theorem we can transform this equation to provide other ways of calculating the physical norm. We have to insert special cups and/or caps as necessary.

If $|\underset{\sim}{G}| = 1$ then we say that $G$ is *physically normalised*.

The case where the physical norm of a left (or right) Hilbert object is equal to 1 is of particular interest. Consider

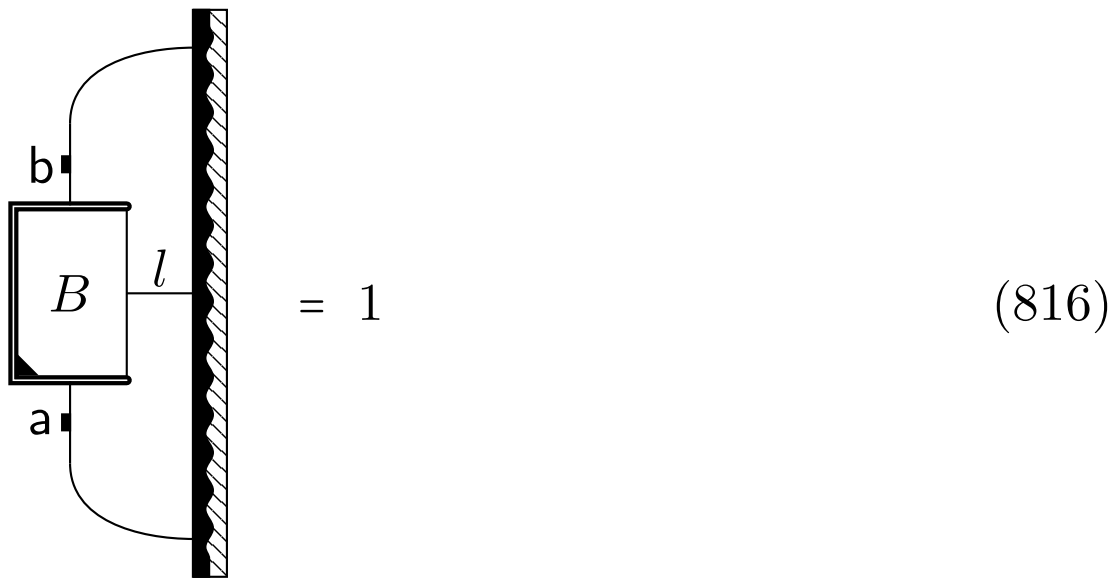

(816)

This is the same as the equation

$\hat{I}$ b $\hat{B}$ a $\hat{I}$ = 1 (817)

(where $\hat{B}$ is the operator tensor formed from $B$). This is a necessary condition for $\hat{B}$ to be deterministic (and, given certain assumptions, it is also a sufficient condition - see Sec. 7.16). Thus, physical normalisation is a natural idea from the point of view of the physics.

## 35.5 Pulling out a wire

We have a physically interesting equation for physical mirrors that does not hold for standard mirrors. This is

a = 1 (818)

We have inserted a 1 on the right hand side to make the meaning clear (we could have omitted the 1 - we take a mirror with no objects reflected to equate to 1). Compare with the case of standard mirrors as discussed in Sec. 34.7 where a factor of $N_{\mathsf{a}}^{\frac{1}{2}}$ is introduced.

If we reflect out the above equation (using (791 and 792) we obtain

(819)

Using (632) and (633) this is just the deterministic circuit

(820)

formed by having an ignore result act on an ignore preparation. So pulling out a wire from a vertical natural physical mirror gives us the basic deterministic circuit.

We can rotate the "pulling out a wire" equation in (818) anticlockwise through 90°. According to the physical mirror theorem (from Sec. 35.3), when we do this we introduce a special cup as follows.

(821)

It is easy to verify this equation is correct from basic definitions.

We can also use physical twist mirrors. By the physical mirror theorem we obtain

(822)

and

(823)

from (818). The right hand sides of these equations equate, of course, to 1.

The message is that we can pull out a wire to the side of any vertical physical mirror without changing the equation (since this just introduces an overall constant factor of 1. We can also pull out a wire to the top or bottom of any horizontal physical mirror but we have to insert a special cap or cup as appropriate. Then we also get an overall constant factor of 1.

## 35.6 Physical significance of physical mirrors

Physical mirrors are based on the natural conjuposition group which is, in fact, more natural for the time symmetric formulation of Quantum Theory. We saw that the time reverse transformation is implemented by the natural transpose, $\utilde{T}$. Further, physical mirrors have the property that a wire joining the mirror gives rise to the ignore operation

a = $\hat{I}$ a    a = $\hat{I}$ a    (824)

(see (791, 792) and ((632) and(633)). We can, therefore, express the condition that an operator, $\hat{B}$, is physical very simply in terms of physical mirrors as follows.

> **Physicality conditions (using physical mirrors).** The operator tensor, $\hat{\boldsymbol{B}}$, is deterministic and physical if we can write
>
> b $\hat{\boldsymbol{B}}$ a = b $\boldsymbol{B}$ $l$ a    (825)
>
> where
>
> 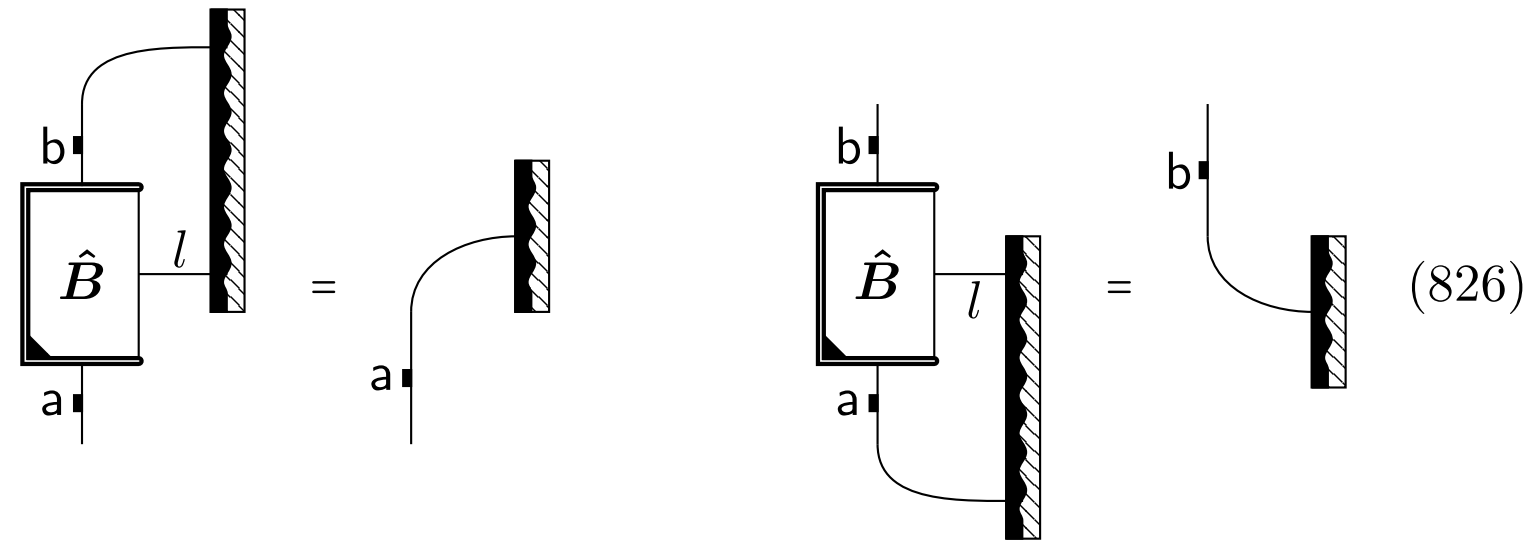
>
>
> The condition in (825) is the twofold positivity condition and the conditions in (826) are the double causality conditions.

Note that the condition for positivity in (825) could equally be written with a normal mirror (since $\overline{H} = \overline{\utilde{H}}$). However, the double causality conditions in (826) could not be written so simply with normal mirrors.

Another respect in which physical mirrors are more natural is that we can pull out a wire from a vertical mirror without changing an expression (as per (818)) because the associated coefficient is 1. For a normal mirror this would incur a coefficient of $N_{\mathsf{a}}$. This difference suggests that natural conjupositions and physical mirrors may help with the limit as $N_{\mathsf{a}} \to \infty$. Indeed, using physical mirrors and natural conjupositions solves the problem discussed by Coecke and Kissinger [2017] – page 239 therein – that a cup closed by a cap evaluates to $\infty$.

The above remarks pertained to physical vertical mirrors. Horizontal mirrors induce transformations that flip the inputs and outputs vertically (either $\overline{\underset{\sim}{V}}$ or $\underset{\sim}{T}$). Do horizontal mirrors also have physical significance. We will make three comments here. First, the image in a horizontal mirror is the time reverse (see Sec. 33.16). Second, these mirrors will turn out to be very useful when we come to define *natural unitaries*. Third, the physical mirror theorem tells us that any equations (or inequalities) expressed with physical vertical mirrors can also be expressed with physical horizontal mirrors and so these different kinds of mirrors are deeply connected.

Ortho-physical bases, natural conjupositions, and physical mirrors are closer to the physics. However, orthonormal bases, normal conjupositions and normal mirrors make for simpler calculations and, further, are useful for defining the natural and physical cases. We will keep both sets of machinery in our repertoire.

## 35.7 Deterministic mirrors

Given any notion of bases, we can define a corresponding conjuposition group and, thereby, define the action of mirrors accordingly. We saw in Sec. 33.8 that ortho-deterministic bases give rise to the same conjuposition group as ortho-physical bases (as long as we assume the ortho-physical gauge normalisation condition (626)). Hence, the defining equations for mirrors in (790, 795, 757, 760) would be the same. However, the notation introduced in (791-794, 796-798, 801-804, 806-798). allowing system wires to attach to mirrors would not be the same. Thus, the theory of deterministic mirrors looks a little different from that of physical mirrors.

Now, interestingly, if we connect an ortho-deterministic basis element to a physical mirror we obtain

a $\qquad a \qquad = \qquad a$ (827)

(to see this use (796) and (665)). Thus, ortho-deterministic bases "play" very nicely with physical mirrors. Also, it is worth noting that a similar "complementary" fact would hold for an ortho-physical basis reflected in a deterministic mirror.

In the case of physical mirrors, when we attach a wire to a vertical mirror, we obtain the corresponding ignore operator which is physically meaningful. The same is not true for deterministic mirrors. In the absences of a clear physical application for deterministic mirrors we omit a full study of them here.

# 36 Normal isometries and coisometries

Consider a Hilbert Square of objects

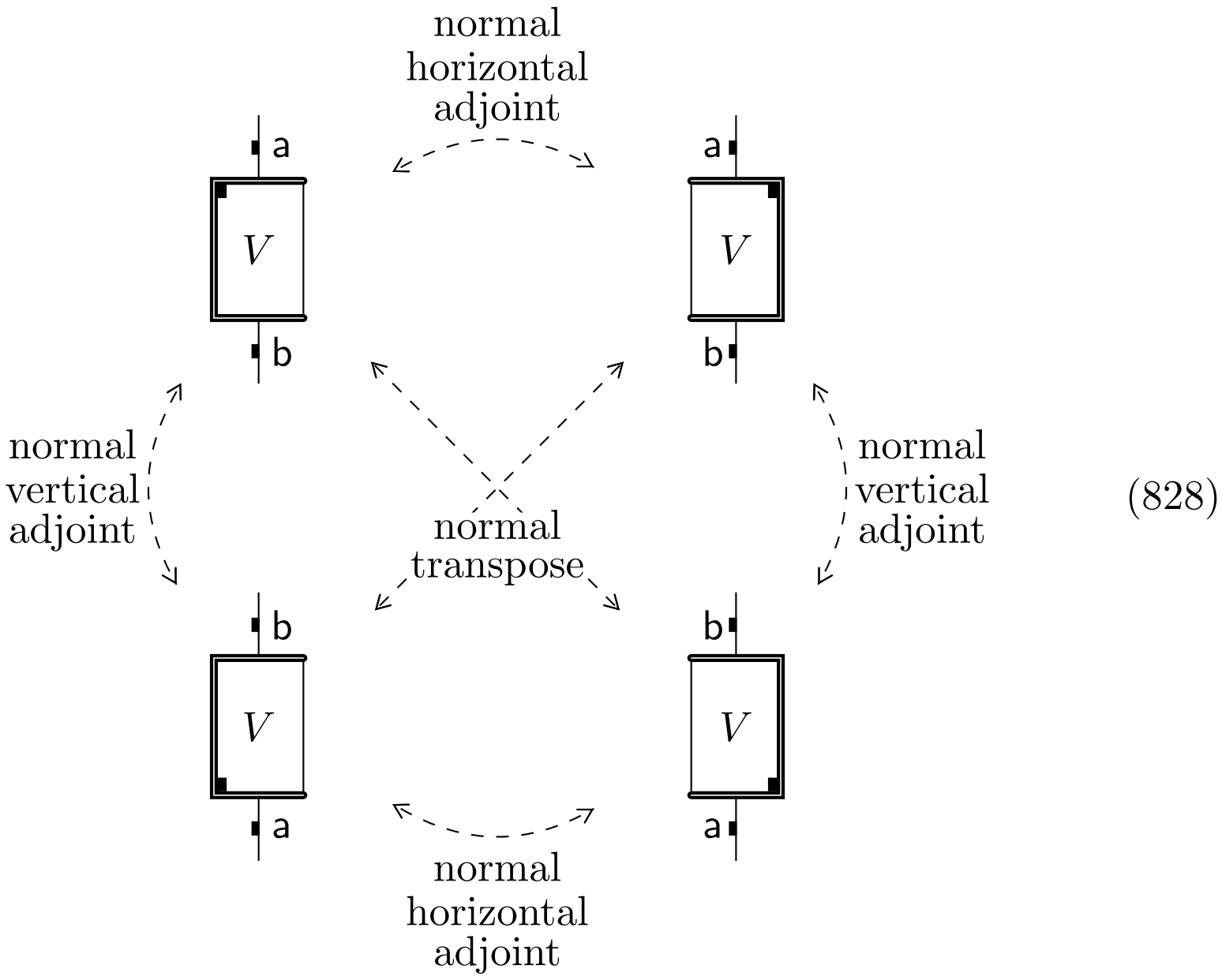

(828)

These elements are related by the subgroup $\{I, \overline{H}, \overline{V}, T\}$ of the normal conjupositions. We will provide definitions for these elements to be what we will call *normal isometries* and *normal coisometries*. These are essentially what are normally called just “isometries” and “coisometries” in standard textbooks. Later we will define natural isometries and natural coisometries based on the corresponding subgroup of the natural conjupositions.

## 36.1 Normal isometries

We define

**Normal isometry.** The left Hilbert object

b

$V$

a

(829)

is a normal isometry iff

(830)

By flipping the above diagram horizontally we obtain the definition for a normal isometry for the corresponding right Hilbert object.

We can write down the normal isometry condition in normal mirror notation as follows

(831)

Normal isometries have the property that they can be slid upwards along the wire and "absorbed" a normal mirror above. By the same token, reading this equation backwards, if we have a system wire joining a normal mirror above we can "pull" any isometry out of the mirror. From this we can write down 32 equivalent equations expressing the normal isometry condition using the mirror theorem (see Sec. 34.5).

For a normal isometry we must have $N_{\mathsf{b}} \geq N_{\mathsf{a}}$. To see this consider what happens if $N_{\mathsf{b}} < N_{\mathsf{a}}$. Then we clearly have the following channel capacity problem. Consider sending in a basis set of states. At the intermediate stage these would be projected down to a set of states spanning a $N_{\mathsf{b}} < N_{\mathsf{a}}$ dimensional Hilbert space and no linear transformation can then recover the full set of $N_{\mathsf{a}}$ basis states. Hence we cannot have $N_{\mathsf{b}} < N_{\mathsf{a}}$.

It is easy to see that the normal isometry condition in (830) is equivalent to the following condition

(832)

This follows from the normal mirror theorem. We can also prove (832) by adding a cap to (830) and then sliding the top object over the cap. We will sometimes refer to this normal isometry condition as the *forward normal unitarity* condition (it is completely equivalent to the normal isometry condition).

## 36.2 Normal coisometries

Consider again the Hilbert square in (828) for $V$'s. We can instead, define coisometries.

> **Normal coisometry.** The left Hilbert object
>
> (833)
>
> is a normal coisometry iff
>
> (834)
>
> By flipping the above diagram horizontally we obtain the definition of normal coisometry for the corresponding right Hilbert object (that obtained by horizontally flipping (833)).

For normal coisometries we have $N_\mathsf{b} \leq N_\mathsf{a}$. This follows by similar channel capacity reasoning to the normal isometric case above. The normal coisometry condition in (834) is equivalent to the following condition

(835)

We will sometimes refer to this as the *backwards normal unitarity* condition (it is completely equivalent to the normal coisometry condition). To prove this condition we add a cup to (834) and slide the lower object through the cup (we can also use the normal mirror theorem to prove this). We can write the normal coisometry condition using a normal mirror as follows

(836)

Normal coisometries have the property that the Hilbert object can be slid down a wire and "absorbed" by a normal mirror below

## 36.3 Normal Isometries, coisometries, and the Hilbert cube

We can prove various results concerning normal isometries and coisometries.

1. If a given Hilbert object is a normal isometry, its normal horizontal adjoint is also a normal isometry. Similarly, if a given Hilbert object is a normal coisometry, its normal horizontal adjoint is also a normal coisometry.

2. If a left object is a normal isometry then its normal vertical adjoint is a normal coisometry (and vice versa).

3. If a left object is a normal (co)isometry then its normal conjugate ($\overline{I}$) is a normal (co)isometry.

Point 1 follows from the fact that the defining conditions for (co)isometries can be flipped horizontally (by applying a $\overline{H}$ transformation). Point 2 follows from inspecting the defining conditions for normal (co)isometries. To prove point 3 we insert twist objects in the appropriate (co)isometry relationship and use the $w$-annihilation property (517). For example, let us begin by assuming

b
$V$
a

is isometric. (837)

From this the condition in (830) follows. Inserting twist objects into (830), we obtain

a
$w$
a
$V$
b
$w$
b
$w$
b
$V$
a
$w$
a

= a (838)

Using (579) this gives

$$\text{(diagram)} = \text{(diagram)} \tag{839}$$

This equation can also be obtained from the normal mirror theorem. It follows from (839) that

is isometric (840)

By point 1 above, it follows that

is isometric (841)

is a normal isometry. These inferences work in the other direction (we have if and only if) so (841) implies (837). Putting together these facts, we see that if any bottom element (the ones with the black square in the lower position) of

the normal Hilbert cube

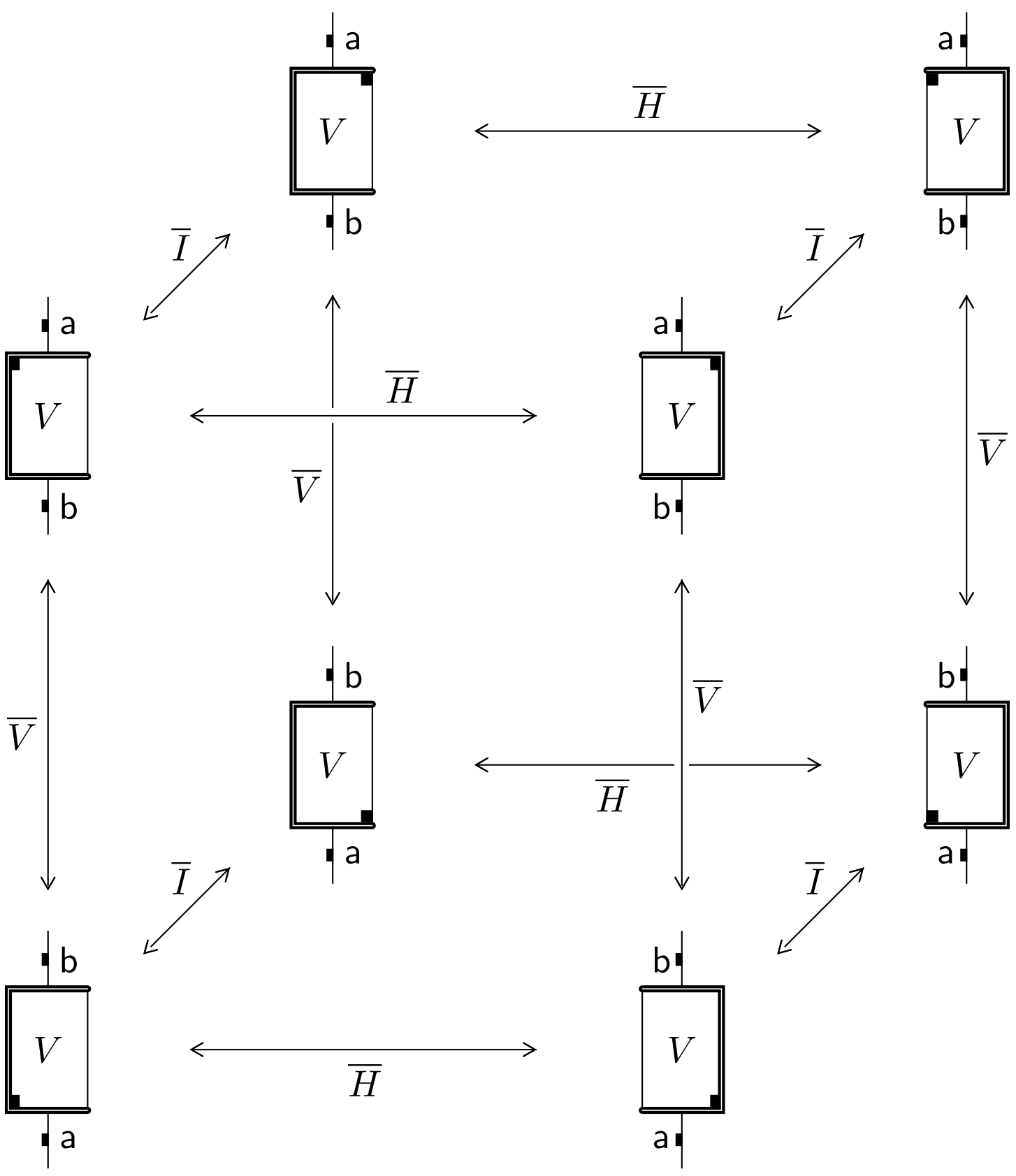

(842)

is a normal isometry then all the bottom elements of the Hilbert cube of $V$'s are normal isometries and all the top elements are normal coisometries. Similarly, if any top element is a normal coisometry then all top elements are normal coisometries and all bottom elements are normal isometries. Similar comments clearly hold for a Hilbert hypercube (see Appendix A). In general, if any element of a Hilbert (hyper)cube is a normal isometry (or coisometry) then all other elements at the same level (top or bottom) are also normal isometries (or coisometries) while all elements at the other level (bottom or top) are normal coisometries (or isometries).

There are 32 equations relating the elements of the Hilbert cube of $V$'s when any given element is an isometry or coisometry which can be derived using the normal mirror theorem (see Sec. 34.5). Looking at the normal Hilbert cube of $V$'s in (842) where the bottom elements are normal isometries and the top elements are normal coisometries, we have

(843)

from which the equivalent relationships

(844)

follow by the normal mirror theorem (we can also derive them directly by adding a cap and a cup (respectively) and sliding as discussed in Sec. 36). Equations (843) and (844) involve only the front elements of the Hilbert cube. We also have the shadow versions of these relationships where the black dot is moved horizontally (for example (839)) for the back of the Hilbert cube. Furthermore, we can have relationships that mix front and back (shadow) elements. For example

(845)

and

(846)

Equations (845) and (846) can be derived by the normal mirror theorem. We can also obtain them directly reasoning as follows. We can obtain (845) by twisting the top object in (830) but leaving the bottom object as is and taking care to balance the $w$-circles on each side using the $w$-annihilation property (see (838) for comparison). We obtain (846) by twisting the right object in the corresponding equation in (844) while leaving the left object untwisted. We can obtain the remaining equations by flipping these equations vertically, horizontally, or by applying $\overline{I}$ (which moves the black square to the other horizontal position).

## 36.4 Normal isometries, coisometries, and bases

Note the following property of normal isometries.

> **Normal isometries and orthonormal bases.** If $V$ is a normal isometry, so that
>
> (847)
>
> then the preparations
>
> (848)
>
> are orthonormal.

The proof of this is simple. Using the normal isometry property we have

(849)

We have used the normal isometry property and orthonomality of the basis elements. This proves orthonomality of the $X$'s. By application of this normal isometry, we produce $N_{\mathsf{a}}$ orthonormal $X$'s. We have $N_{\mathsf{b}} \geq N_{\mathsf{a}}$ since $V$ is a normal isometry. If $N_{\mathsf{b}} > N_{\mathsf{a}}$ then do not obtain a complete orthonormal set for $\mathsf{b}$ by acting on $\mathsf{a}$ with an normal isometry.

We can write the preparation in (848) as follows

(850)

The $f$-padding matrix (represented by the cut diamond shape) was defined in (496) such that, if $a$ labels rows and $b$ labels columns, then there is a single 1 in each row and 0's elsewhere. Appending a basis element to the open $a$ label gives

(851)

Using (506) gives

(852)

This provides a rather pictorial way of understanding how normal isometries work. It is clear from this form that if we send an standard (i.e. unshaded) orthonormal basis element in the bottom, we obtain the shaded basis element as output. Of course, it is also true that, even if we send a general basis element in the bottom, we obtain an orthonormal basis element as output.

There are, of course, similar results concerning normal coisometries and basis sets obtained by "flipping" the normal isometry case upside down.

# 37 Normal unitaries

Now we will consider normal unitary left and right Hilbert objects. We will usually denote unitaries by $U$. Consider a Hilbert square of $U$s as follows.

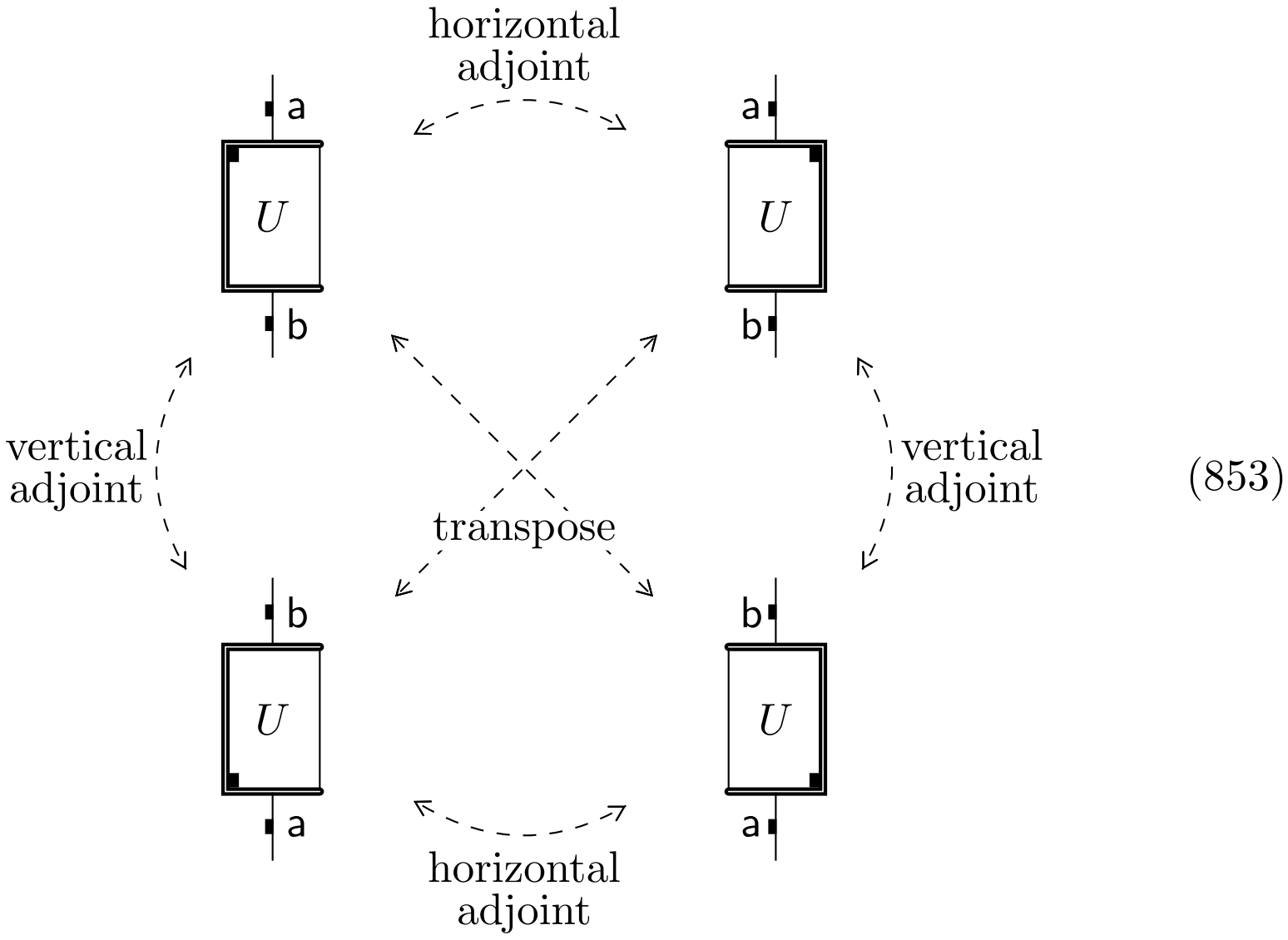

(853)

We will see that if any one of these four Hilbert object is a normal unitary the others are too.

Normal unitaries are defined with respect to the normal conjuposition group. Later we will define natural unitaries with respect to the natural conjuposition group. Natural unitaries turn out to be more natural from a physics point of view.

## 37.1 Equivalent definitions of normal unitaries

The usual textbook definition of unitarity (for matrices) is that a matrix $U$ is unitary iff it is isometric ($UU^\dagger = \mathbb{1}$) and coisometric ($U^\dagger U = \mathbb{1}$). This forces $U$ to be square. Consider left Hilbert space object

b
U
a

(854)

We will provide a number of equivalent definitions for normal unitarity.

The definition closest to the standard textbook definition given above is

**Normal unitarity definition 1.** The left object,

(855)

is a normal unitary if and only if it has the two properties

(856)

The property on the left is **normal isometry** and the property on the right is **normal coisometry**.

We can, similarly, define the right Hilbert object obtained by flipping (854) horizontally to be a normal unitary iff the conditions obtained by flipping (866) horizontally are satisfied.

These defining properties of a normal unitary can be put in an equivalent form using the sliding manoeuvre leading to the next definition.

**Normal unitarity definition 2.** $U$ in (854) is a normal unitary if and only if it has the two properties

(857)

We will call the condition on the left **forward normal unitarity** (it is equivalent to normal isometry) and the condition on the right **backward normal unitarity** (it is equivalent to normal coisometry).

These provide a useful definition of normal unitarity for us as they are more in keeping with the time symmetric attitude taken here (and strongly related to forward and backward causality as we will see). The equations in (857) follow by the normal mirror theorem from (856).

We noted in Sec. 36 that $N_{\mathsf{b}} \geq N_{\mathsf{a}}$ for normal isometries and $N_{\mathsf{b}} \leq N_{\mathsf{a}}$ for coisometries. Since normal unitaries are both normally isometric and normally coisometric, they must satisfy $N_{\mathsf{b}} = N_{\mathsf{a}}$. This simple fact guides us to the next definition for normal unitarity

> **Normal unitarity definition 3 and 4.** $U$ in (854) is a normal unitary if and only if it has the properties that $N_{\mathsf{a}} = N_{\mathsf{b}}$ and it satisfies either forward normal unitarity or backward normal unitarity (or equivalently, normal isometry or normal coisometry).

This definition is equivalent to definition 2 because, if $N_{\mathsf{a}} = N_{\mathsf{b}}$ and normal isometry is satisfied then it follows that normal coisometry is also satisfied. This is equivalent to the usual textbook statement that a square matrix satisfying $UU^\dagger = \mathbb{1}$ is unitary. Let us simply provide the usual textbook proof employing matrix multiplication. If the isometry property $UU^\dagger = 1$ holds and $U$ is square then $U$ and $U^\dagger$ are necessarily invertible. It follows that $U^\dagger U = U^\dagger U(U^\dagger (U^\dagger)^{-1}) = U^\dagger (UU^\dagger)(U^\dagger))^{-1} = U^\dagger (U^\dagger)^{-1} = \mathbb{1}$ (using isometry in the penultimate step). This proves the coisometry property. This proof can straightforwardly be set to pictures and applied to normal unitaries. Also, clearly, we could start with coisometry and prove isometry for square matrices.

These considerations motivate another definition.

> **Normal unitarity definition 5.** $U$ in (854) is a normal unitary if and only if it takes a complete orthonormal basis set for $\mathsf{a}$ to a complete orthonormal basis set for $\mathsf{b}$. In diagrams we require that
>
> (858)
>
> represent a complete orthonormal basis set for $\mathsf{b}$. Here $\pi$ is a permutation matrix.

The time reverse of this condition works as well. We can prove that definition 5 follows from definition 3 above. It follows from the normal isometry of normal unitaries that acting on a basis set with a normal unitary produces an orthonormal set (see (848)). It follows from the condition that $N_{\mathsf{a}} = N_{\mathsf{b}}$ that this is a complete orthonormal basis set. We prove that definition 4 follows from definition 5. Completeness of the basis implies $N_{\mathsf{a}} = N_{\mathsf{b}}$. Further, from (858), we obtain

(859)

(using the fact that the permutation matrix, $\pi$, times its horizontal transpose gives the identity). This is the backward normal unitarity condition.

Finally, we can use the above definition to obtain our last definition

> **Normal unitarity definition 6.** $U$ in (854) is a normal unitary if and only if it can be written in the form
>
> (860)
>
> where the unshaded basis is any given basis and the shaded basis is then fixed by the normal unitary.

We prove this by showing it is equivalent to definition 5 since, by appending (unshaded) basis element to the open label wire of (858) we obtain

(861)

Using (506) gives (860). Further, if send in (unshaded) basis elements into the bottom of (860), we obtain (858). This proves definition 6 is equivalent to definition 5.

## 37.2 Normal unitaries and the normal Hilbert cube

Consider a normal Hilbert cube of $U$s as follows.

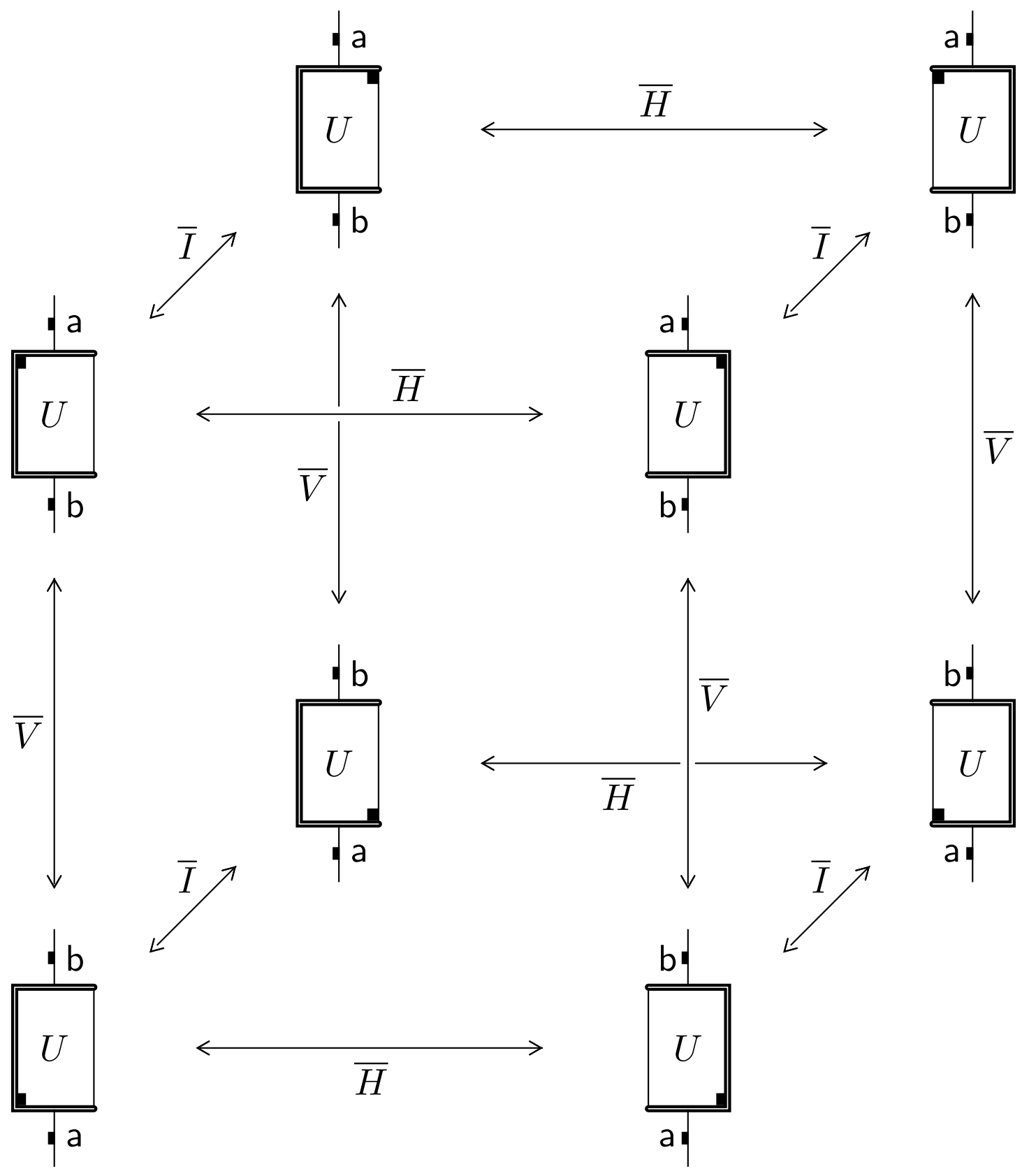

(862)

If any one of the eight objects in this Hilbert cube is a normal unitary it follows that all the other objects are normal unitaries as well. This follows from the results of Sec. 36.3. To see this assume some given element of the above Hilbert cube is a normal unitary. This means it is a normal isometry and consequently all elements at the same level (top or bottom) are also normal isometries and all elements at the other level are normal coisometries. Normal unitarity of this given element also means that it is a normal coisometry. This means that all elements at this given level are also normal coisometries and all elements at the other level are normal isometries. Hence all elements are both normal isometries and normal coisometries - so they are normal unitares.

Normal isometry of any given element of the Hilbert cube provides 32 equations relating elements of the Hilbert cube obtained from the normal mirror theorem. Normal coisometry of that same element provides a further 32 equations. Thus, we have 64 equations deriving from normal unitarity. This includes defining equations, such as those in (856), the equivalent forward and backward unitarity conditions in (857), and various equivalent versions involving the twist operator such as

(863)

which comes from reflecting out an expression with a twist mirror. This particular equation expresses the well known property that a maximally entangled state (represented here by the twist cup, $w$) is invariant when we act on one side with a (normal) unitary and the other side with (what we call here) the normal conjugate transformation, $\overline{I}$, of this normal unitary.

## 37.3 Normal unitary operator tensors

We can define a normal unitary operator tensor by doubling up

(864)

Note that normal unitary operator tensors are homogeneous (see Sec. 31.6). This normal unitary operator tensor has normal vertical adjoint

(865)

(since $\hat{U}$ is Hermitian, this is also equal to its normal transpose). It follows from (856) that normal unitary operator tensors have the properties

(866)

# 38 Natural isometries, coisometries, and unitaries

## 38.1 Need for natural unitaries

In Sec. 36 we discussed normal isometries and coisometries and in Sec. 37 we discussed normal unitaries. Normal unitaries are just the usual unitaries which are central to textbook treatments of (time forward) Quantum Theory. Normal unitarity is defined with respect to normal conjupositions. However, natural conjupositions are actually the correct symmetry group for the time symmetric version of Quantum Theory as we saw in Sec. 33.16. By adapting the mirror definitions of normal isometries, coisometries, and unitaries we will see that we can define *natural isometries*, *natural coisometries*, and *natural unitarities* with respect to natural conjupositions (rather than normal conjupositions).

Natural unitaries are more fundamental than normal unitaries from our point of view (because they are based on the correct physical symmetry). Does this mean we should abandon the standard unitaries? Fortunately, it turns out that if we use the $\gamma(N)$ gauge (see Sec. 33.4) then natural unitaries satisfy the conditions for normal unitaries and vice versa. Since textbooks implicitly assume a gauge fixing that satisfies the $\gamma(N)$ gauge, we can consistently regard natural unitaries as being more fundamental whilst not having to abandon normal unitaries.

## 38.2 Definitions

For the case of normal conjupositions we have

(867)

where the left equation is the normal isometry condition, the right equation is the normal coisometry condition, and both equations taken to gether constitute the normal unitarity conditions. If we replace the normal mirrors in these definitions by physical mirrors as follows

(868)

then we have definitions for *natural isometries* (left equation), *natural coisometries* (right equation), and *natural unitaries* (both equations).

Reflecting out the condition in (868), we obtain

$$\text{(diagram)} \tag{869}$$

respectively. We can use (631) to remove the small circles giving

$$\text{(diagram)} = \sqrt{\frac{N_\mathsf{b}}{N_\mathsf{a}}}\ \mathsf{a} \qquad\qquad \text{(diagram)} = \sqrt{\frac{N_\mathsf{a}}{N_\mathsf{b}}}\ \mathsf{b} \tag{870}$$

Comparing these with (856), we see that the natural isometry (left) and natural coisometry (right) conditions involve a rescaling with respect to the normal isometry and normal coisometry conditions. We can use the same channel capacity reasoning as used in Sec. 36, to argue that for natural isometries (the left equation) we must have $N_\mathsf{b} \geq N_\mathsf{a}$. Similarly, for natural coisometries (the right equation) we can prove $N_\mathsf{a} \geq N_\mathsf{b}$. Hence, the coefficients in (870) are greater than or equal to one.

Now let us focus on natural unitarity. In this case we impose both natural isometry and natural coisometry. Then we must have $N_\mathsf{a} = N_\mathsf{b}$. Hence the conditions for natural unitarity are

$$\text{(diagram)} \tag{871}$$

These conditions are almost the same as the normal unitarity defining conditions in (856). However, rather than taking the normal vertical adjoint (as indicated by the small squares in (856)) we are taking the natural vertical adjoint (as indicated by the small triangles). Thus, natural unitarity is different to normal unitarity.

## 38.3 Gauge fixing

Although natural unitarity is different from normal unitarity, there are two interesting ways in which these two types of unitarity can coincide.

**Same input and output types.** Recall from the discussion at the end of Sec. 33.3 that, if the input and output are of the same type then the normal and natural conjupositions have the same effect. Hence, in this special case the normal vertical adjoint is the same as the natural vertical adjoint and hence the normal unitary is the same as the normal unitary.

**Impose $\gamma(N)$ gauge.** If we impose the $\gamma(N)$ gauge then, as discussed in Sec. 33.4, the natural vertical adjoint, $\overline{\underset{\sim}{V}}$ becomes equivalent to the normal vertical adjoint, $\overline{V}$ in the special case where $N_{\mathsf{a}} = N_{\mathsf{b}}$ (which we have here for natural unitarity). In this case normal and natural unitaries are the same. We have, incidently, invoked the ortho-physical gauge condition (626) when used (631) to remove the small circles in (869). Recall that the ortho-physical gauge condition does, in fact, follow from the $\gamma(N)$ gauge (see Sec. 33.4).

## 38.4 Natural unitary operators

A natural unitary operator tensor is written as

b, $\hat{U}$, a = b, $\boldsymbol{U}$, a; b, $\boldsymbol{U}$, a (872)

(compare with (864)) where the conditions for natural unitarity in (868) are satisfied. Note that we are now using bold font for $\boldsymbol{U}$. This is because, as we will see, natural unitary operators are deterministic. Using the physical mirror theorem (Sec. 35.3) we can convert the natural unitarity conditions in (868) into the equivalent form

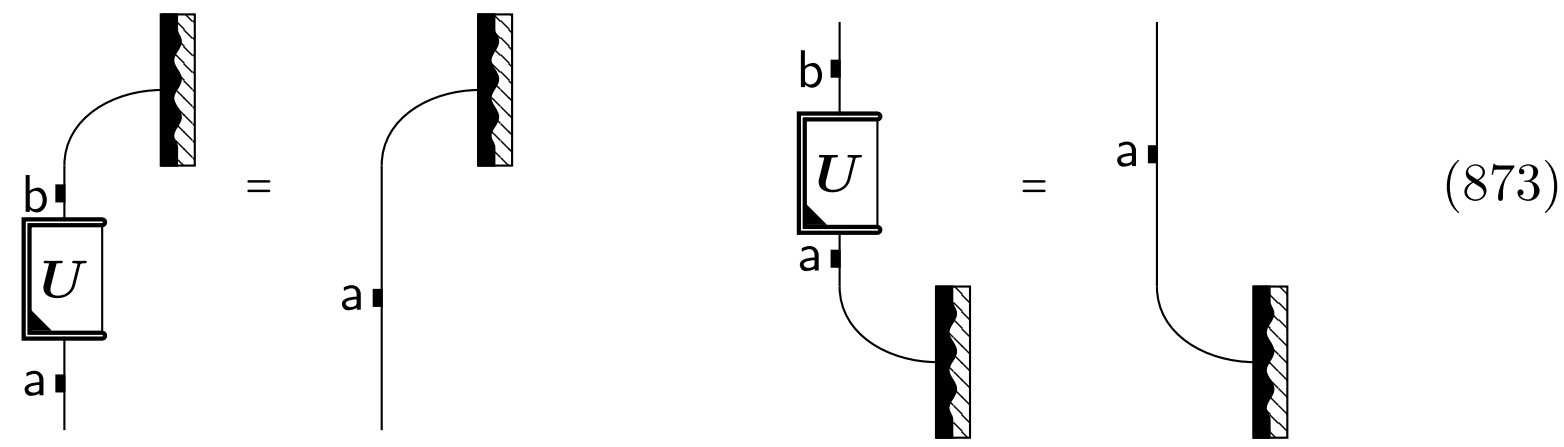

(873)

If we reflect these out and use (634) in conjunction with (632, 633) then we get

$$\hat{\boldsymbol{I}}\ \overset{\mathsf{b}}{|}\ \hat{\boldsymbol{U}}\ \overset{\mathsf{a}}{|} \;=\; \hat{\boldsymbol{I}}\ \overset{\mathsf{a}}{|} \qquad \overset{\mathsf{a}}{|}\ \hat{\boldsymbol{U}}\ \overset{\mathsf{a}}{|}\ \hat{\boldsymbol{I}} \;=\; \overset{\mathsf{a}}{|}\ \hat{\boldsymbol{I}} \qquad \text{where} \qquad \overset{\mathsf{b}}{|}\ \hat{\boldsymbol{U}}\ \overset{\mathsf{a}}{|} \;=\; \overset{\mathsf{b}}{|}\ \boldsymbol{U}\ \overset{\mathsf{a}}{|}\ \ \overset{\mathsf{b}}{|}\ \boldsymbol{U}\ \overset{\mathsf{a}}{|} \qquad (874)$$

These are simply the forward and backward causality conditions for a deterministic homogeneous operator tensor, $\hat{\boldsymbol{U}}$ (recall that homogeneity is the condition that there is no label wire connecting left and right Hilbert objects - see Sec. 31.6). Thus, we have the following important observation

> **Deterministic homogenous physical operators** are necessarily natural unitary operators. Also, natural unitary operator tensors are deterministic physical operators.

The second statement, that *natural* unitary operators are physical and deterministic follows because they satisfy the double causality conditions and $T$-positivity (the latter follows since the natural unitary operator tensor is written in twofold form - see longer discussion in Sec. 43). For this reason, natural rather than normal unitaries are the more fundamental object.

If we assume the $\gamma(N)$ gauge then normal unitaries are also natural unitaries (and hence physical and deterministic). This role as deterministic homogeneous operators accounts for the importance of unitarities in Quantum Theory. It is worth noting, however, that we need to adopt the $\gamma(N)$ gauge if we are to interpret normal unitaries in this way. As mentioned above, *natural unitaries* are the more fundamental objects as they have this interpretation whether or not we adopt the $\gamma(N)$ gauge. In standard treatments of Quantum Theory a gauge fixing is chosen (that $\gamma_{\mathsf{a}} = 1$ and $\gamma^{\mathsf{a}} = \frac{1}{\sqrt{N_{\mathsf{a}}}}$ for all $\mathsf{a}$) that satisfies the $\gamma(N)$ gauge and so the distinction between normal unitaries and natural unitaries is not evident.

## 38.5 Relationship with bases

In (848) we saw that a normal isometry acting on an orthonormal basis set returns an (possibly incomplete) orthonormal set (see also (850)). Further, in (858) we saw that a normal unitary acting on an orthonormal basis returns a (complete) orthonormal basis set. We have similar results for natural isometries and natural unitaries acting on ortho-deterministic basis sets (introduced in Sec. 33.8).

First, for any natural isometry $V$ we have

(875)

(compare with (850)) where $f$ is a $f$-padding matrix (as defined in (496)). We can prove this by verifying the following equality

(876)

using (796) (for a shaded basis) and (665) along with the natural isometry condition (left equation in (868)). By appending an ortho-physical basis element to the open $a$ label wire, we can show that any natural isometry can be written as

(877)

(compare with (852)).

Now consider natural unitaries. We can prove that a (complete) ortho-deterministic basis set is transformed to a complete ortho-deterministic basis set by a natural unitary as follows

(878)

where $\pi$ is a permutation matrix. This follows from (875) for the special case where $N_{\mathsf{b}} = N_{\mathsf{a}}$ (which forces the natural isometry to become a natural unitary). We can also prove that any natural unitary can be written in terms of ortho-physical and ortho-deterministic bases as follows

(879)

We can prove this by attaching a ortho-physical basis to the open label wire $a$ in (878) as follows

(880)

Then, using (668) we obtain (879).

We can use a natural unitary to transform the ortho-physical bases as follows

(881)

It can easily be checked that this holds because $N_{\mathsf{a}} = N_{\mathsf{b}}$ for a natural unitary (so a similar result does not hold, in general, for natural isometries).

# 39 Normal isometries and coisometries from normal unitaries

Here we will show that normal isometries and normal coisometries can be written as projected normal unitaries. We will also see how to write normal isometric and coisometric operators in terms of a normal unitary operator that is both projected and acted upon by $\mathbb{1}$.

Later we will see that similar statements are true for the natural case.

## 39.1 A little normal theorem

Here we prove a useful theorem connecting any normal isometry or coisometry to a normal unitary which we will use to prove subsequent theorems.

**Little normal theorem.** If $V$ is an normal isometry then we can always find a normal unitary, $U$, such that

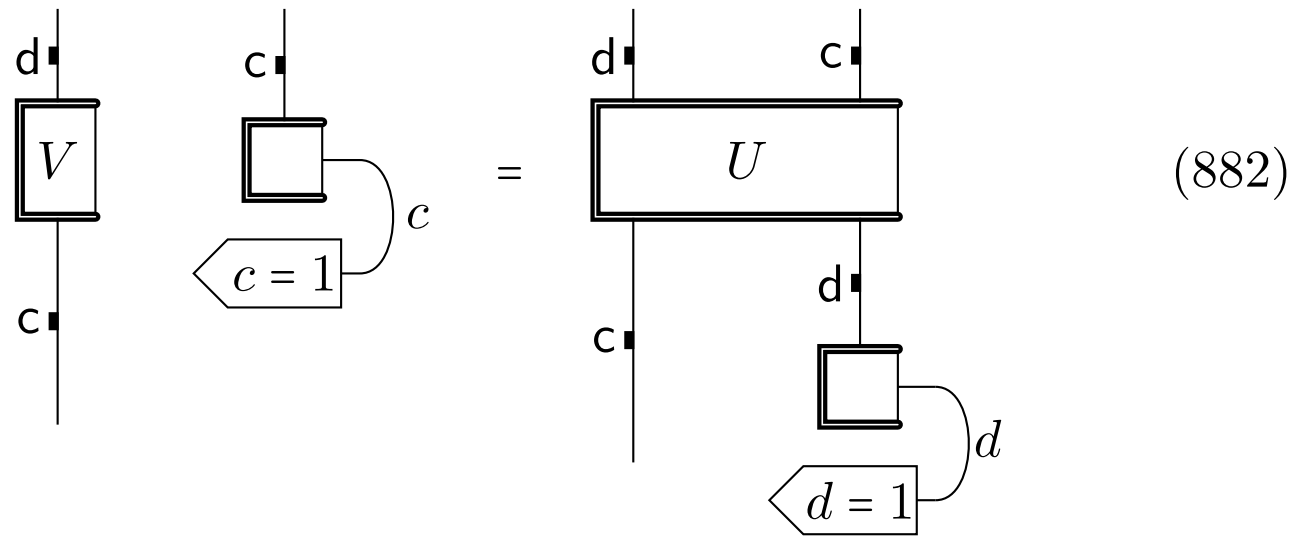

(882)

In general, $U$ is not unique. If $V$ is a normal coisometry then we can find a normal unitary, $U$, such that

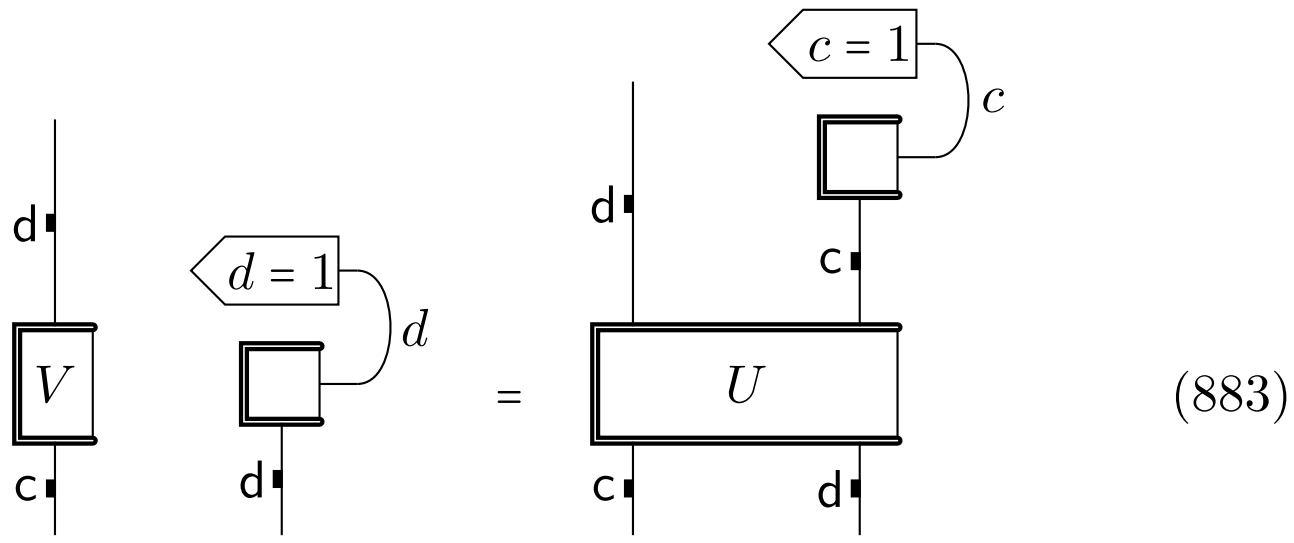

(883)

In general, $U$, is not unique.

The first part of theorem is central to textbook proofs of the Stinespring dilation (the second part is the time reverse). Here we have adapted the diagrammatic proof here from the book of Coecke and Kissinger [2017]. We will prove the normal isometry case. The normal coisometry case has a similar proof. If $V$ is a normal isometry then we can write

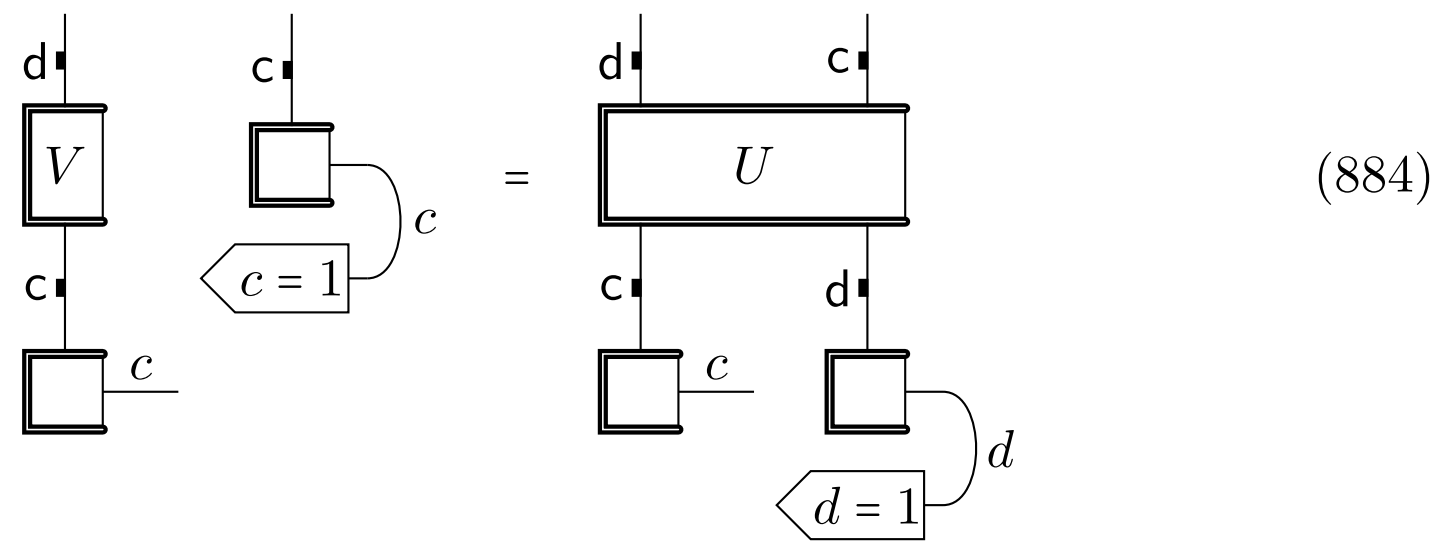

(884)

This is clear since normal unitaries and normal isometries preserve orthonormality (see Sec. 37). The orthonormality preserving property of the normal isometry $V$ guarantees that the left hand side objects are orthonormal for different $c$. So we can choose $U$ acting on those basis elements inputted into $U$ on the right such that we get the basis elements on the left. Then we can complete the normal unitary such that the remaining orthonormal basis elements (for values of $d \neq 1$) that can be inputted into $U$ are mapped to some set of remaining orthonormal basis elements orthogonal to the ones already given by the expression on the left. Next, we can append basis elements for system $\mathsf{c}$ to the $c$ label wire as follows

(885)

Finally, we can use the decomposition of the identity wire in (506) to obtain (884).

## 39.2 Normal (co)isometries as projected unitaries

Since we have adopted a time symmetric perspective, it is interesting to see how normal isometries and coisometries can be regarded as corresponding to projecting on both an input and an output of a normal unitary (we call such objects "projected unitaries").

> **Normal isometries and coisometries as projected normal unitaries.** If $V$ is a normal isometry or a normal coisometry, we can write
>
> (886)
>
> where $U$ is a normal unitary, and $A$ and $C$ are normalised. For a normal isometry, $N_{\mathsf{d}} \geq N_{\mathsf{c}}$, and for a normal coisometry, $N_{\mathsf{d}} \leq N_{\mathsf{c}}$. Not all projected normal unitaries are either normal isometries or normal coisometries.

The inequality relationships between $N_{\mathsf{c}}$ and $N_{\mathsf{d}}$ are simply properties of normal isometries and normal coisometries (discussed in Sec. 37). The above theorem follows almost immediately from the "little normal theorem" in the previous subsection. We will prove this for the case of normal isometries using the first part of the little normal theorem. The case of normal coisometries follows similarly from the second part. Hitting both sides of (882) (from the little normal theorem) with the $c = 1$ basis element on the c output we obtain

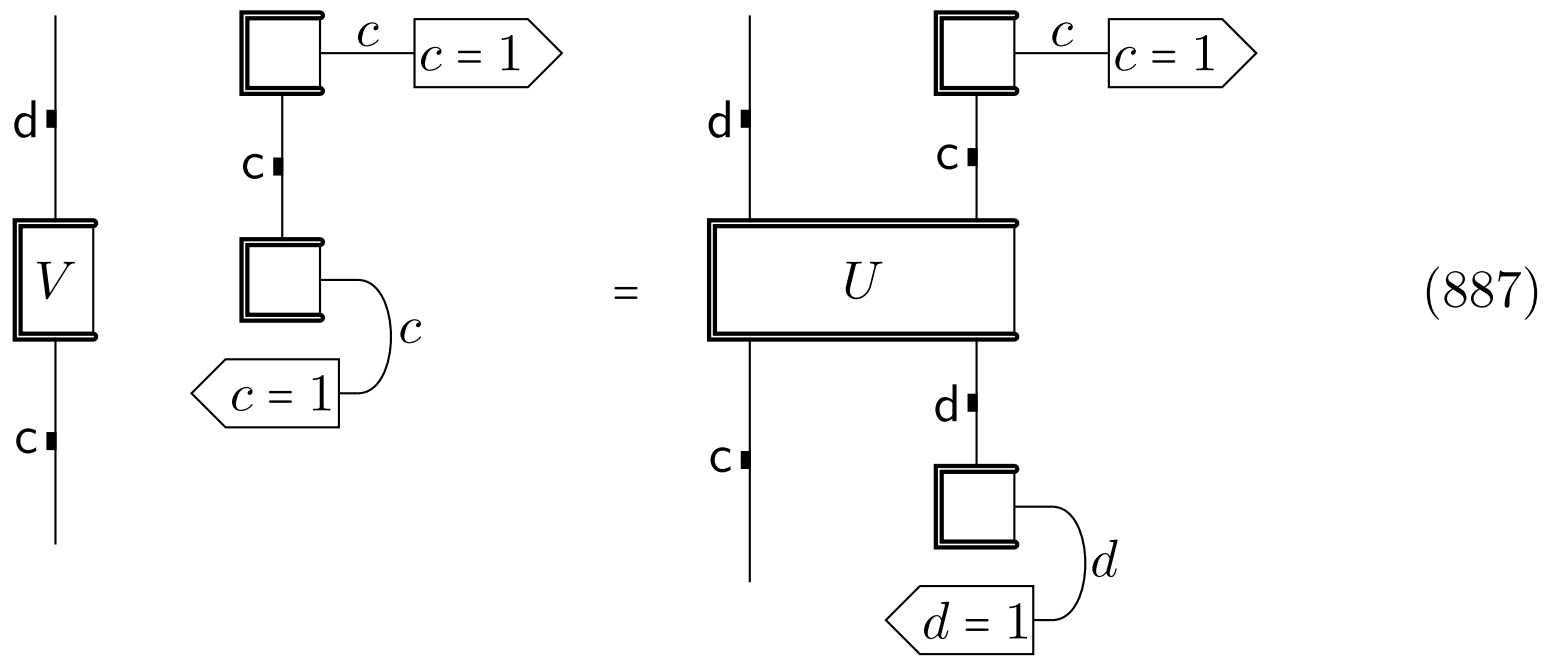

(887)

The factor involving “preselecting” and “postselecting” in the $c = 1$ basis element is equal to 1. This proves (886) as required. To see that not all projected normal unitaries are normal isometries or normal coisometries it is sufficient to provide an example. Consider

(888)

The object inside the dashed box is clearly a normal unitary so this is a projected normal unitary. It is easy to verify that the left object in (888) is neither an normal isometry (it does not satisfy (830) or a normal coisometry (it does not satisfy (834)). In Appendix B we prove necessary and sufficient conditions for a projected normal unitary to be a normal isometry or a normal coisometry.

## 39.3 Normal (co)isometric operators from normal unitary operators

Consider an the normal (co)isometric operator

(889)

where $V$ is normal (co)isometric. We can obtain this operator from a normal unitary in two different ways.

**Normal (co)isometric operator from normal unitary operator.** Any normal isometric operator can be written in the alternative forms

(890)

where $U$ is a normal unitary and $\hat{A}$ and $\hat{B}$ are homogeneous and normalised. Any normal coisometric operator can be written in the alternative forms

(891)

where $U$ is a normal unitary and $\hat{A}$ and $\hat{B}$ are homogeneous and normalised.

Note, $\mathbb{1}$ represents the identity operator defined in (441). The first alternative in each case follows immediately from (886). The second alternative in each case follows straightforwardly from the little normal theorem in Sec. 39.1. To see this consider the case where $V$ as a normal isometry.

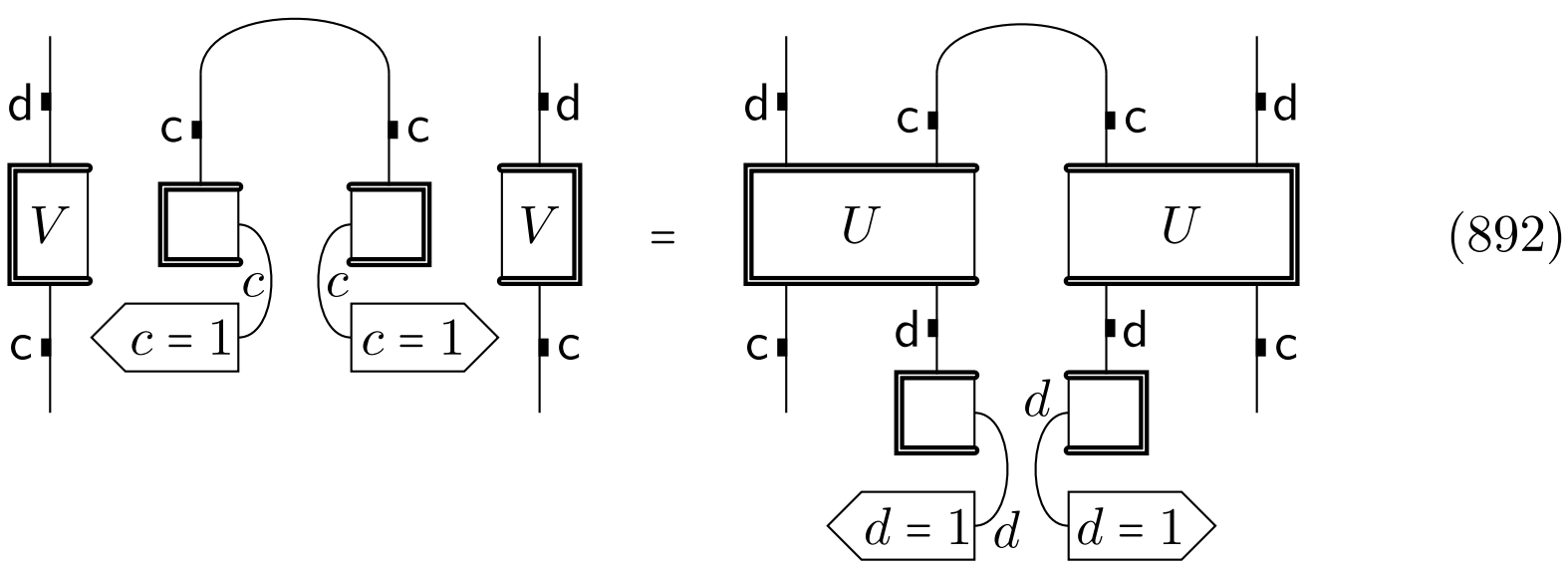

where we have doubled up (882) from the little normal theorem and placed a cap on the output c wires. The left expression is equal to the left expression in (890) and the right expression is the doubled up form of the right expression in (890). This proves the second alternative for the normal isometric case. The normal coisometric case is proved similarly.

# 40 Natural (co)isometries from natural unitaries

In this section we will redo the theorems of the previous section for natural (co)isometries and natural unitaries. We will get more natural objects appearing. For example, rather than having identity operators, $\hat{\mathbb{1}}$, as in (890,891), we will have ignore operators, $\hat{\boldsymbol{I}}$. The theorems here make use of both the ortho-physical basis (introduced in Sec. 33.1) and the ortho-deterministic basis (introduced in Sec. 33.8).

## 40.1 A little natural theorem

First we set up a little theorem (like that in Sec. 39.1).

> **A little natural theorem:** If $V$ is a natural isometry then we can always find a natural unitary, $\boldsymbol{U}$, such that
>
> 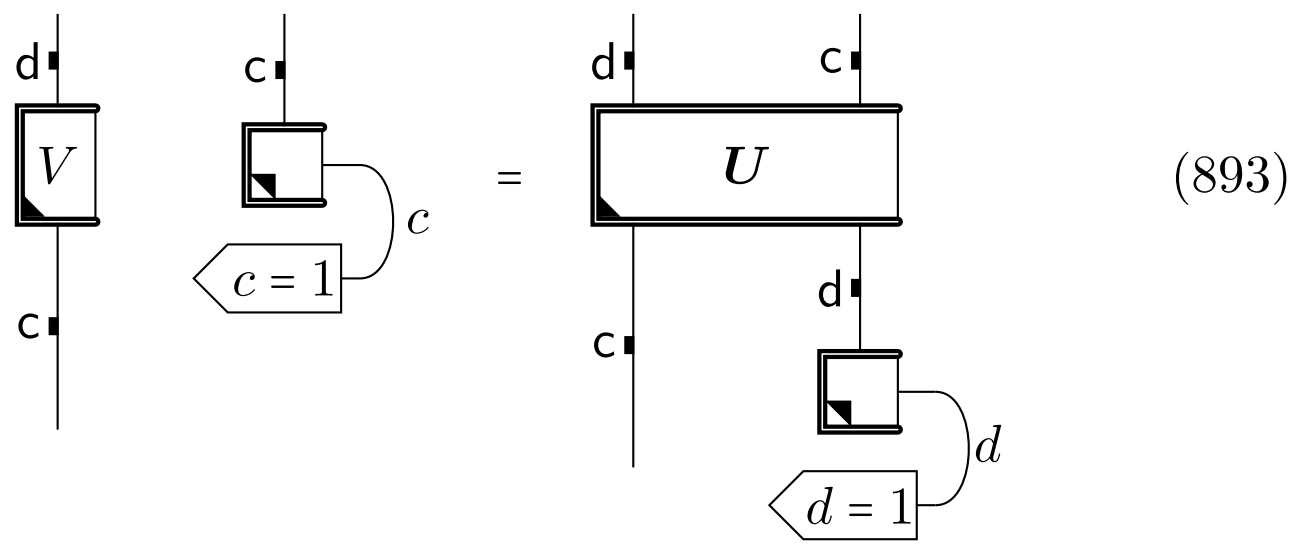
>
> (893)
>
> In general, $\boldsymbol{U}$ is not unique. If $V$ is a natural coisometry then we can always find a natural unitary, $\boldsymbol{U}$, such that
>
> 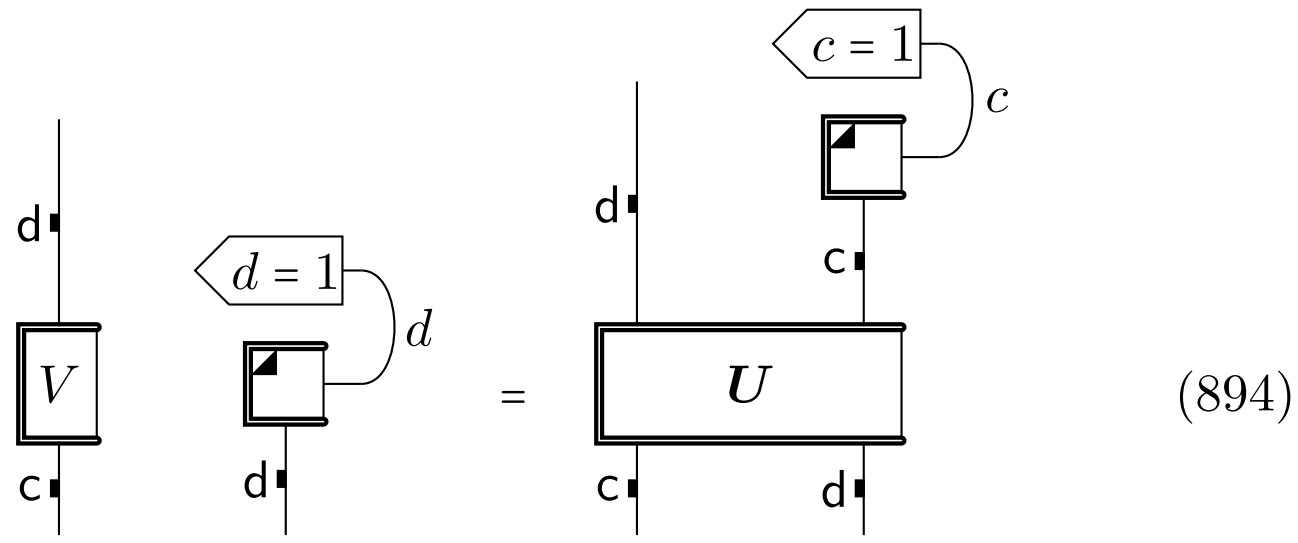
>
> (894)
>
> In general, $\boldsymbol{U}$ is not unique.

The proof of this theorem is along exactly the same lines as the proof of the little normal theorem in Sec. 39.1. Consider the case where $V$ is a natural isometry.

We can write

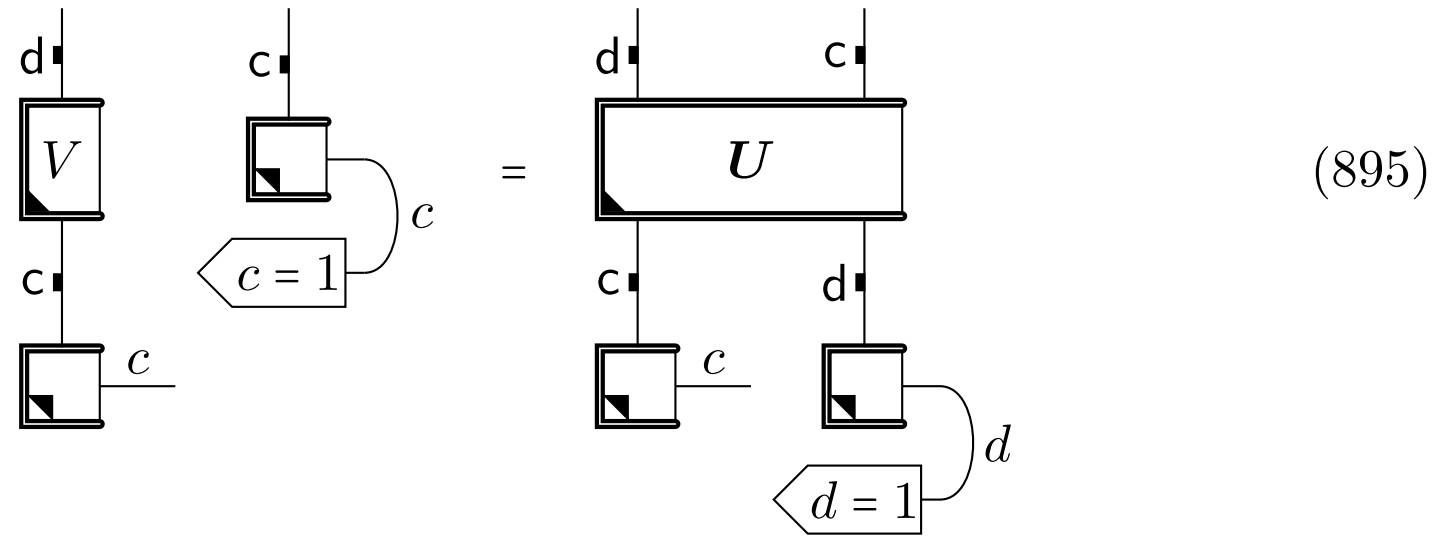

(895)

To prove this recall that natural isometries preserve ortho-deterministic bases (see (875)). Thus, on the left hand side we have $N_{\mathsf{c}}$ ortho-deterministic basis elements. Natural unitaries also preserve ortho-determinstic bases (see (878)). Thus, we can choose $\boldsymbol{U}$ such that it maps its input to the same $N_{\mathsf{c}}$ ortho-deterministic basis elements. This partially determines $\boldsymbol{U}$ - we can simply fill out the remaining $N_{\mathsf{cd}} - N_{\mathsf{c}}$ output set of $\boldsymbol{U}$ with any ortho-deterministic basis elements orthogonal to the above mentioned set of $N_{\mathsf{c}}$ basis elements (thus, $\boldsymbol{U}$ is not unique in general). We can attach an ortho-physical basis to the open $c$ label wire giving

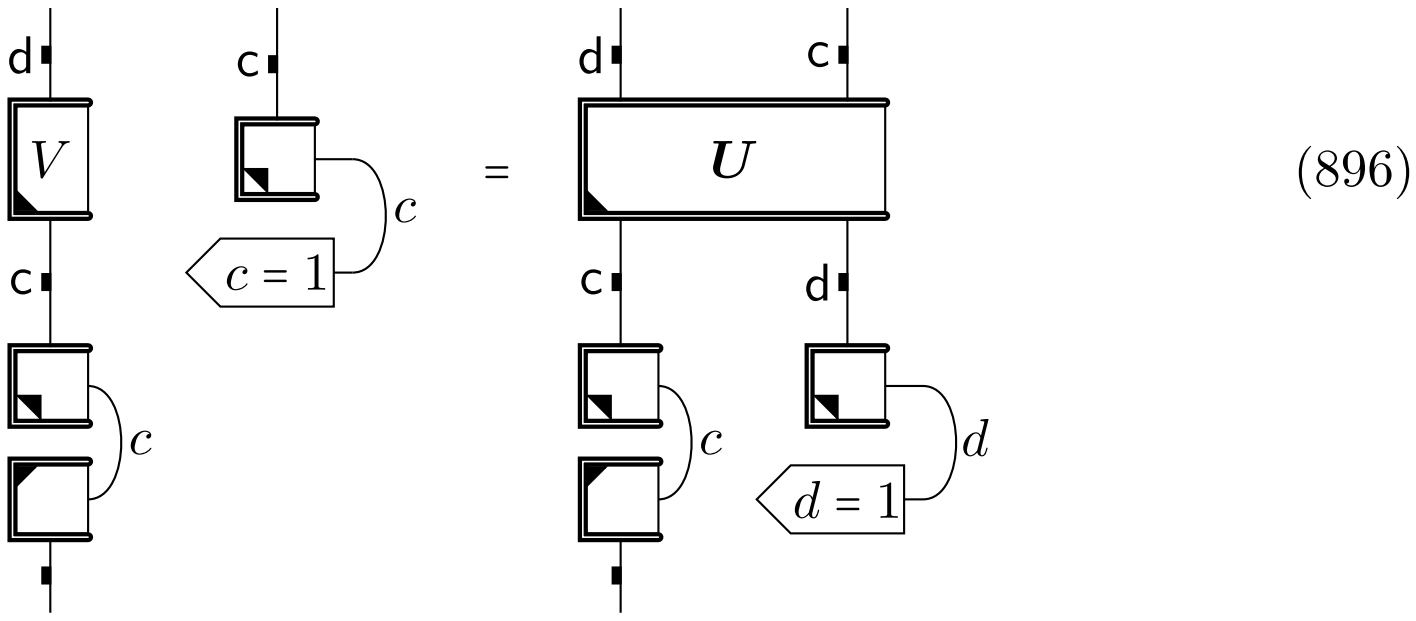

(896)

Then we can use the decomposition of the identity in (668) to obtain (893). The proof of (894) (when $V$ is a physical-mirror coisometry) follows by similar reasoning).

## 40.2 Natural (co)isometries as projected physical unitaries

We can use the little natural theorem to prove a theorem that natural isometries and natural coisometries can be written as projected natural unitaries. This is similar to the result in Sec. 39, though note the projectors are not normalised in the standard way.

**Natural (co)isometries as projected natural unitaries.** If $V$

is a natural isometry then we can write it as

(897)

where $N_{\mathsf{d}} \geq N_{\mathsf{c}}$. If $V$ is a natural coisometry then we can write it as

(898)

where $N_{\mathsf{c}} \geq N_{\mathsf{d}}$. There exist projected natural unitaries that are not proportional to either a natural isometry or a natural coisometry.

This follows immediately from (893) using (665). The example given in (888) suffices to prove the statement in the final sentence of the above theorem.

## 40.3 Natural (co)isometric operators from natural unitary operators

We can use the above results to write natural (co)isometric operators in terms of natural unitary operators in two alternative ways.

**Natural (co)isometric operators from natural unitary operators.** Any natural isometric operator, $\hat{V}$, can be written in the two alternative forms

(899)

where $\hat{\boldsymbol{U}}$ is a natural unitary, and $\hat{A}_{\rm det}$ and $\hat{C}_{\rm phys}$ are homogeneous and satisfy

$$\begin{array}{c}\boxed{\hat{\boldsymbol{I}}}\\ \mathsf{d}\\ \boxed{\hat{A}_{\rm det}}\end{array} = 1 \qquad\qquad \begin{array}{c}\boxed{\hat{C}_{\rm phys}}\\ \mathsf{c}\\ \boxed{\hat{\boldsymbol{I}}}\end{array} = \frac{1}{N_{\mathsf{d}}} \tag{900}$$

Any natural coisometric operator, $\hat{V}$, can be written in the two alternative forms

$$\begin{array}{c}\mathsf{d}\\ \boxed{\hat{V}}\\ \mathsf{c}\end{array} = \begin{array}{cc} \mathsf{d} & \boxed{\hat{C}_{\rm det}}\\ & \mathsf{c}\\ \multicolumn{2}{c}{\boxed{\hat{\boldsymbol{U}}}}\\ \mathsf{c} & \mathsf{d}\\ & \boxed{\hat{A}_{\rm phys}}\end{array} = \begin{array}{cc} \mathsf{d} & \boxed{\hat{C}_{\rm det}}\\ & \mathsf{c}\\ \multicolumn{2}{c}{\boxed{\hat{\boldsymbol{U}}}}\\ \mathsf{c} & \mathsf{d}\\ & \boxed{\hat{\boldsymbol{I}}}\end{array} \tag{901}$$

where $\hat{\boldsymbol{U}}$ is a natural unitary, and $\hat{A}_{\rm phys}$ and $\hat{C}_{\rm det}$ are homogeneous and satisfy

$$\begin{array}{c}\boxed{\hat{\boldsymbol{I}}}\\ \mathsf{d}\\ \boxed{\hat{A}_{\rm phys}}\end{array} = \frac{1}{N_{\mathsf{d}}} \qquad\qquad \begin{array}{c}\boxed{\hat{C}_{\rm det}}\\ \mathsf{c}\\ \boxed{\hat{\boldsymbol{I}}}\end{array} = 1 \tag{902}$$

Note that the operator tensors labeled with the "det" subscripts are not physical and can be built from an element of an ortho-deterministic basis set (see the discussion in Sec. 33.9). The operator tensors labeled with "phys" are physical and can be built from an element of an ortho-physical basis set (this is also discussed in Sec. 33.9).

# 41 Pseudo-physicality and maxometry

Unitaries are central to textbook Quantum Theory. However, in the time symmetric point of view, a more general object - a maxometry - can take the role of unitaries. This is particularly true in the dilation theorem which will be discussed in Sec. 44. In this section we will motivate maxometries through consideration of the notion of pseudo-physicality. Then we will define *natural maxometries* using physical mirrors. These are the main objects we are interested in. Natural unitaries are a special case of natural maxometries. We will also define *normal maxometries*. Normal unitaries are a special case of normal maxometries.

The word "maxometry" is a portmanteau of "maximally" (from maximally entangled) and "(co)isometry" (we will see below why this is appropriate).

## 41.1 Pseudo-physical operators

Consider an operator

d h
$\hat{N}$
c g
(903)

We will say that this is a *pseudo-physical operator* with respect to ancillae g and h operator if

$\hat{I}$
d h
$\hat{N}$
c g
$\hat{I}$
(904)

is physical. In other words, a pseudo-physical operator has the property that, when we ignore its ancillae, it becomes physical. It is possible that a pseudo-physical operator itself is not physical.

We will, further, say $N$ is a *deterministic pseudo-physical operator* if it is both pseudo-physical and deterministic. This is true if and only if it satisfies $T$-positivity and

$\hat{I}$ $\hat{I}$
d h
$\hat{N}$ = $\hat{I}$
c g c
$\hat{I}$

d $\hat{I}$ c
h
$\hat{N}$ = $\hat{I}$
c g
$\hat{I}$ $\hat{I}$
(905)

These are the double causality conditions (421) for the operator in (904).

## 41.2 Natural maxometries

A *Natural maxometric operator* is a homogeneous deterministic pseudo-physical operator. Spelling this out we have the following definition.

**Natural maxometric operators.** We say that

d h d h h d
$\hat{M}$ = $M$ $M$
c g c g g c
(906)

is a *natural maxometric operator* with respect to ancillae g and h if

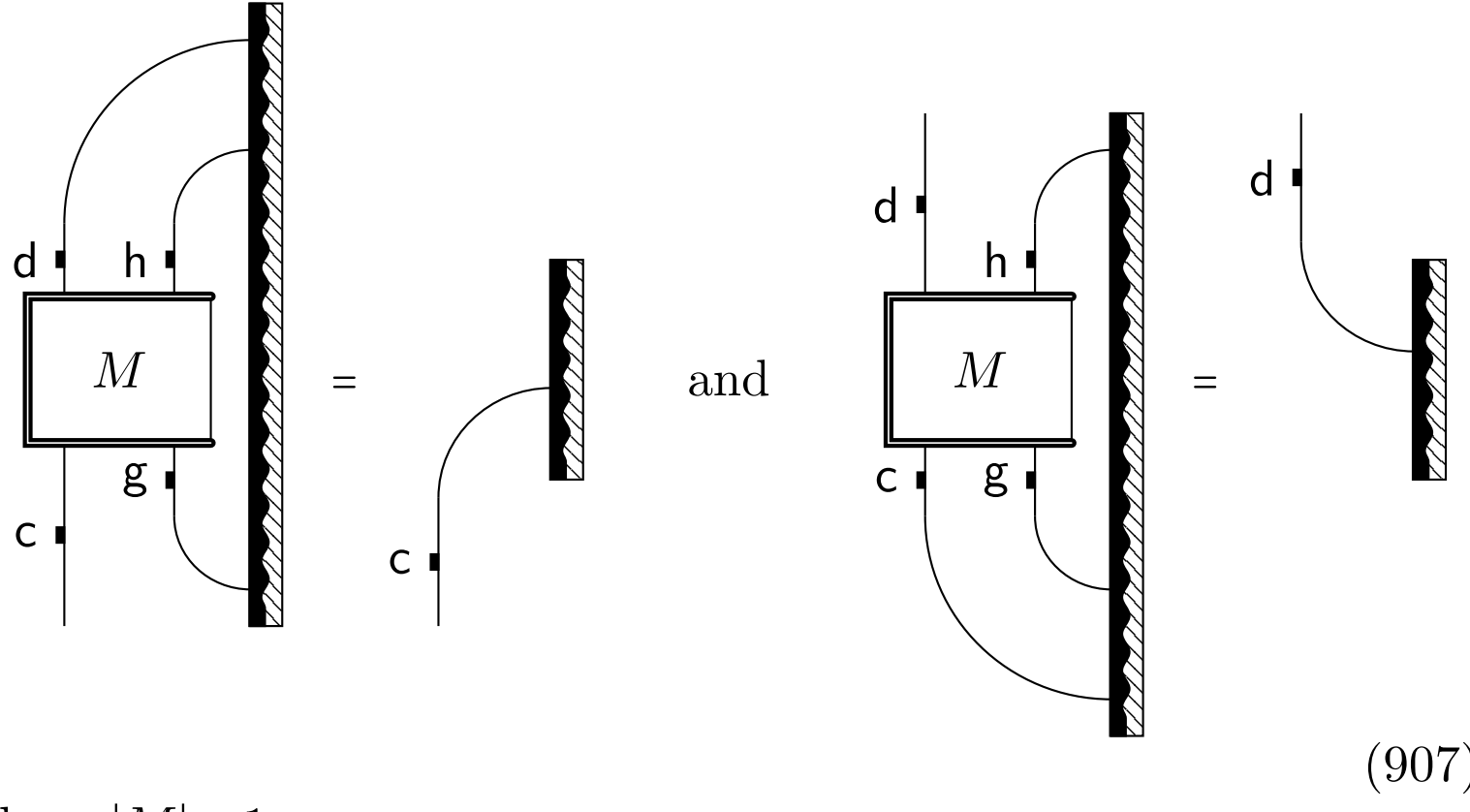

(907)

where $|\underset{\sim}{M}| = 1$.

We obtain these conditions by substituting (906) into (905) (where $\hat{M}$ replaces $\hat{N}$), using (635) and the definition of a physical mirror (see Sec. 35).

Furthermore, we define

**Natural maxometries**. If the conditions in (907) are satisfied then we say that the left object

(908)

is a *natural maxometry* with respect to ancillae g and h.

We have a similar definition for a right object to be a natural maxometry.

Recall from Sec. 38.4 that a deterministic homogeneous physical operator is necessarily a natural unitary operator. This means that, if a natural maxometry, $\hat{M}$, is actually physical (not just pseudo-physical) then it is necessarily a natural unitary operator. Furthermore, we can show that all natural unitaries are natural maxometries. Let

(909)

be a natural unitary. To see that it is a natural maxometry with respect to ancillae g and h note that

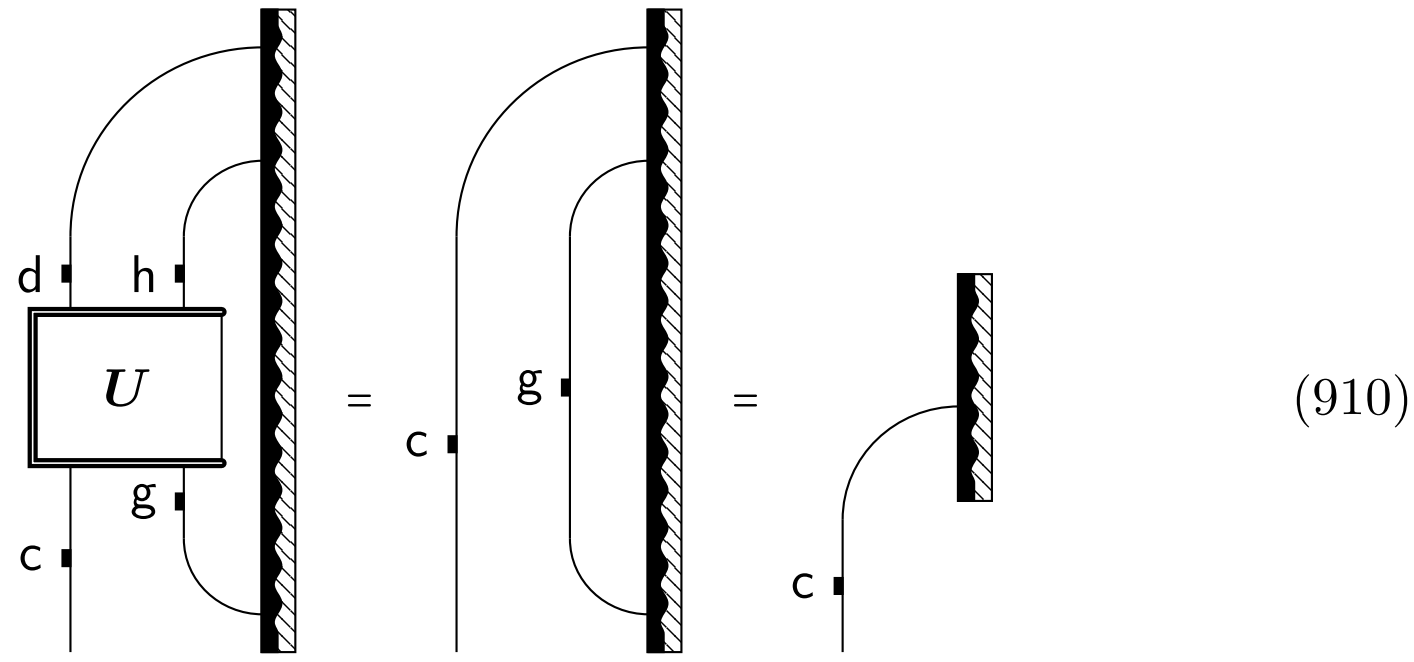

(910)

We have used natural unitarity in the first step (see (873)) and, in the second step we are using the fact that the g wire equates to 1 when reflected in this physical mirror (here we are using the property that we can pull out a wire as illustrated in (818)). Equality of the first and third expressions prove that the first property required for a natural maxometry in (907) is satisfied. We can prove the second property required for natural maxometry in a similar fashion.

However, not all natural maxometries are natural unitaries. Here is a counterexample.

(911)

where

(912)

where we have used the "cut diamond" notation for $f$-padding matrices (see Sec. 29.3). It is easy to verify that the conditions for natural maxometry in (907) are satisfied. It is also clear that the example in (911) does not satisfy the conditions for natural unitarity (look at (873) for example).

We call these objects "maxometries" because the defining properties in (907)

can be reexpressed as the requirement that

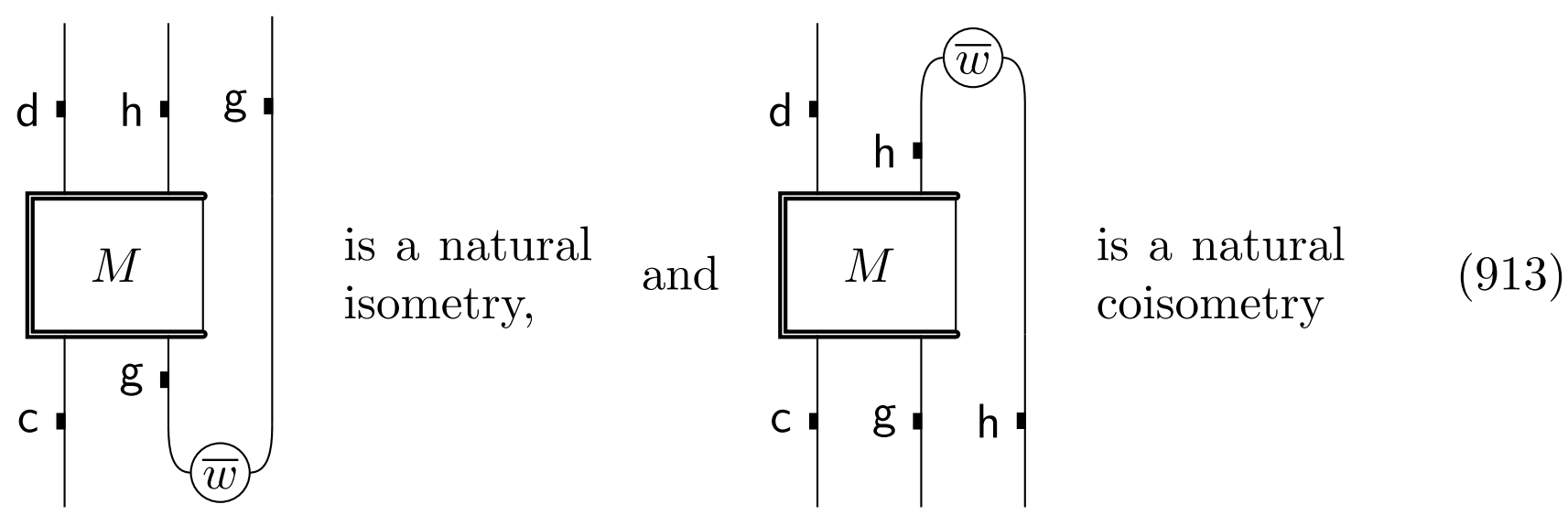

(913)

where

(914)

This is easy to prove. Consider the left condition in (913). We can obtain this by using the physical mirror theorem to rotate the left condition in (907). This inserts a special cup (as per the physical mirror theorem described in Sec. 35.3). We can then insert a special twist in the g wire to obtain the required result. The right condition in (913) is obtained similarly. These objects in (914) are, in fact, the twisted special cup and cap defined in row 4 of Table (5). They are maximally entangled. Thus, if we attach an appropriately normalised maximally entangled object to the input (or output) ancilla of a natural maxometry, we obtain a natural isometry (or a physical-mirror coisometry). This is the motivation for using the word "maxometric".

In fact, the object on the left (913) is, by itself, a natural maxometry. This is because, in addition to being a natural isometry, it also satisfies

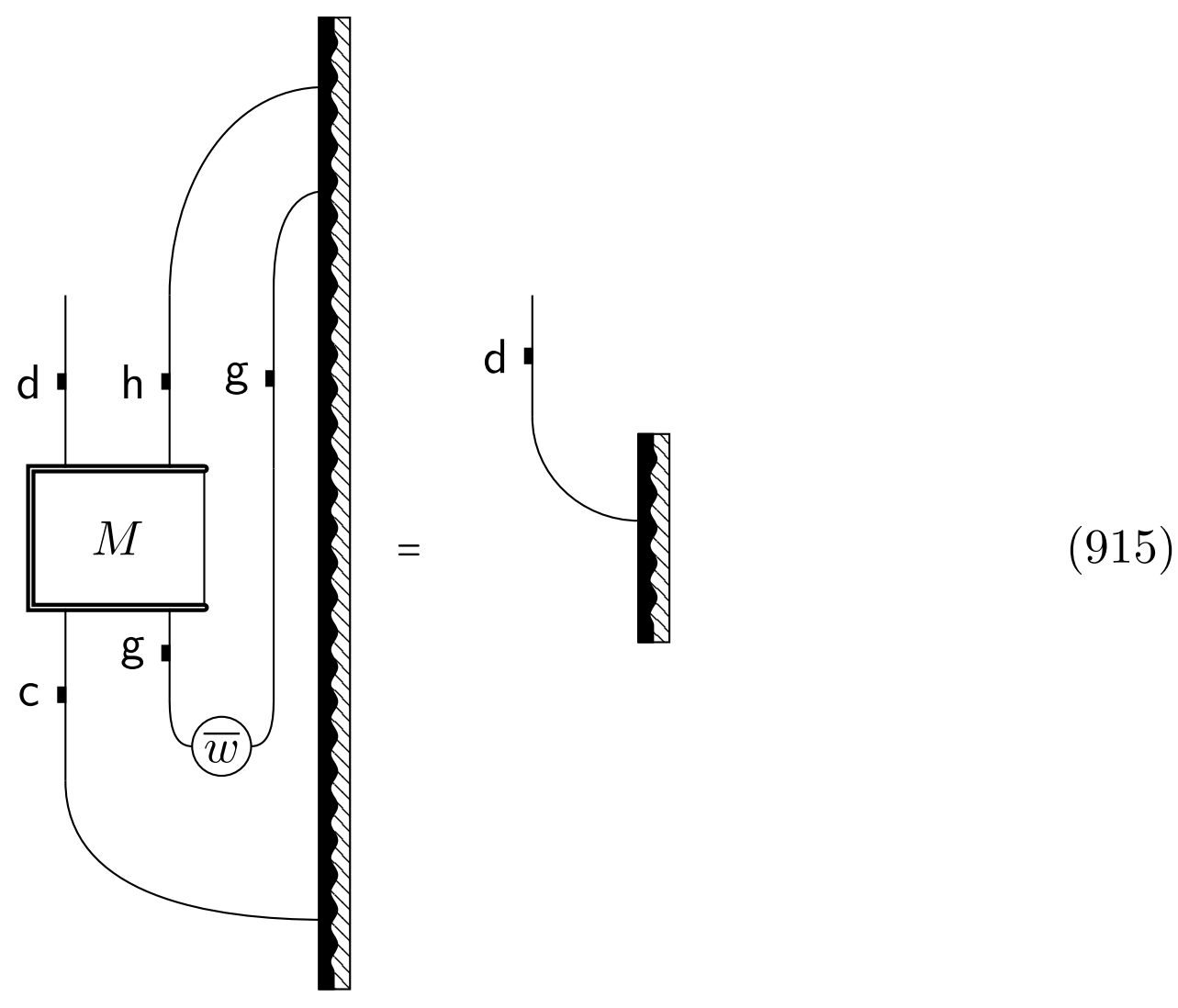

(915)

A similar statement is true about the object on the right of (913). This turns out to be an important point. If we attach a twisted special cap to the input (or output) of a natural maxometry then we obtain a new natural maxometry which is a natural isometry (coisometry).

## 41.3 Normal maxometries

We saw that natural unitaries are special cases of natural maxometries. It is interesting to define a notion of normal maxometries for which normal unitaries are a special case. Thus we have

> **Normal maxometries.** If the left Hilbert object
>
> 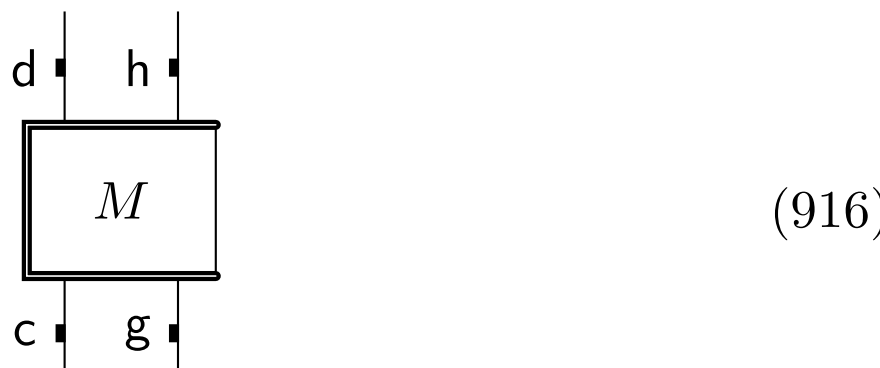
>
>
> (916)
>
> has the properties
>
> 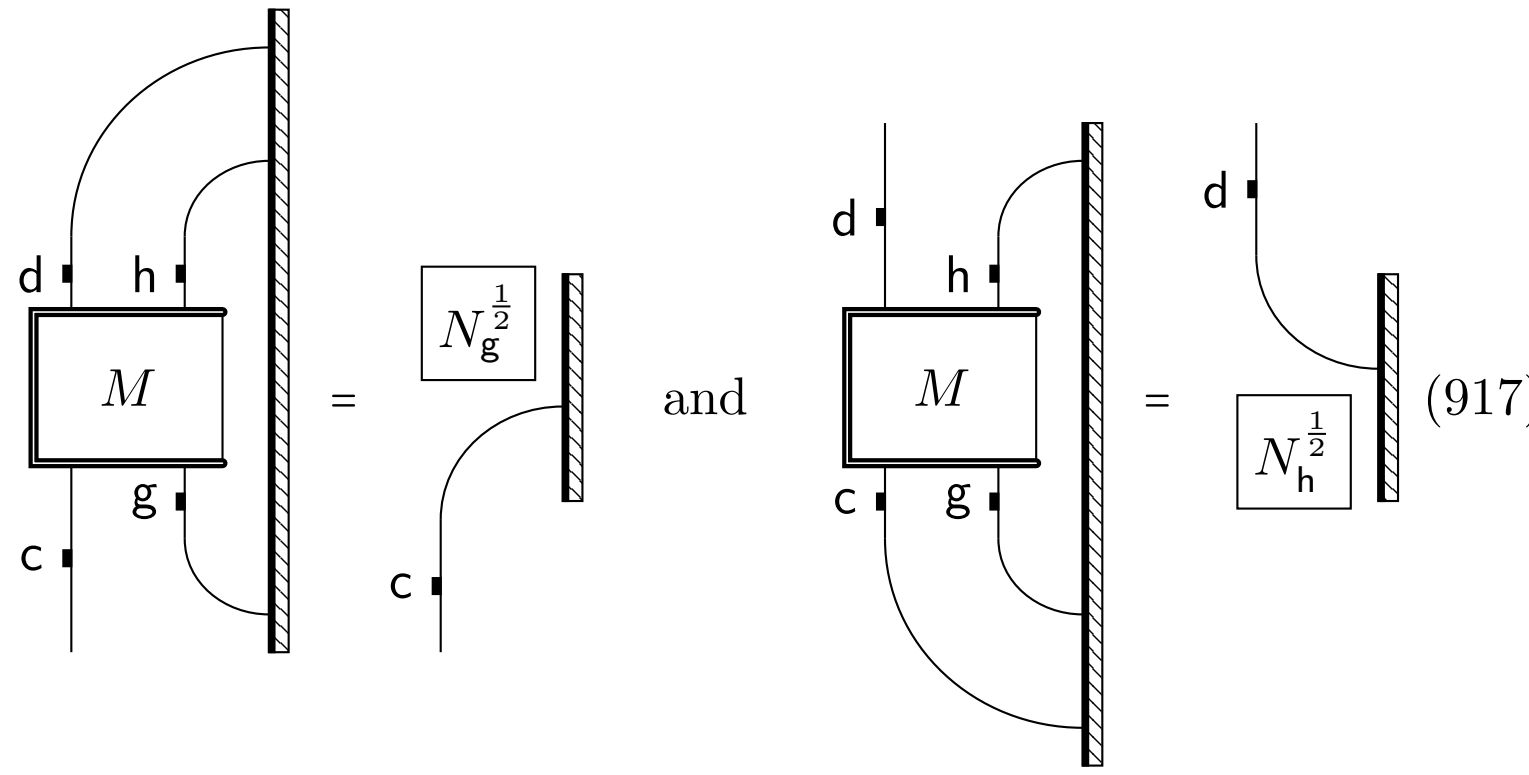
>
>
> (917)
>
> then we say it is a *normal maxometry* with respect to the ancillae g and h. Furthermore, we must have
>
> $$N_{\mathsf{g}} N_{\mathsf{c}} = N_{\mathsf{h}} N_{\mathsf{d}} = |M|^2 \tag{918}$$
>
> where $|M|$ is the norm of the Hilbert object, $M$, as defined in Sec. 34.6.

Note that (918) is easily obtained from (917) by attaching the open wires to the mirrors and using (788).

We can now prove, as mentioned above, that normal unitaries are normal maxometries. Consider a normal unitary

(919)

This is a normal maxometry with respect to ancillae g and h. The condition in the left in (917) is satisfied because

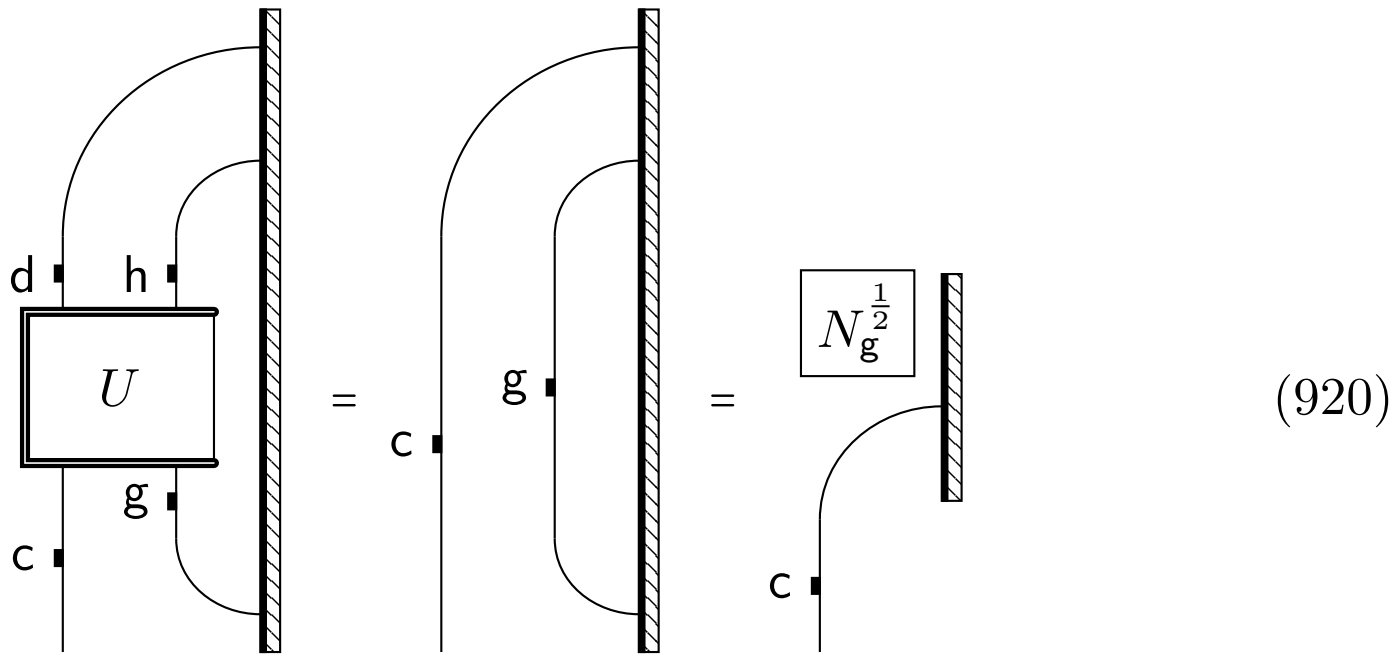

We use normal unitarity in the first step (see (857)) and in the second step we use (788). The condition on the right in (917) is proved similarly.

We note, further, that a normal maxometry is equivalently defined by the properties

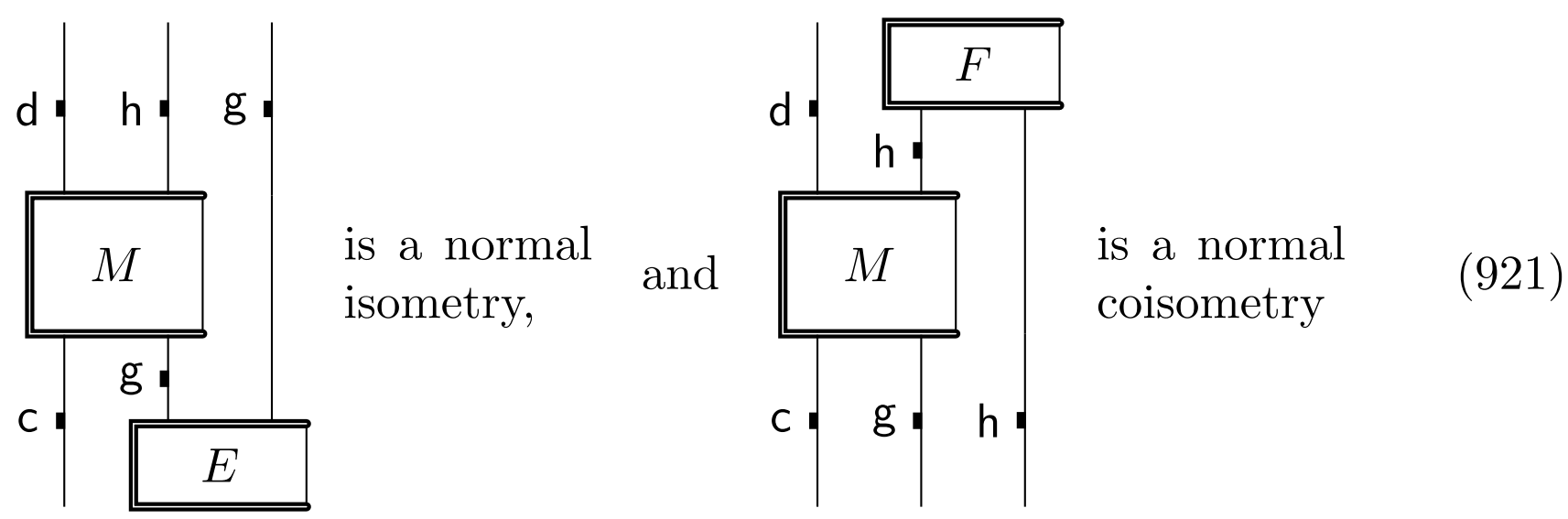

where $E$ and $F$ are, respectively, the maximally entangled state and effect

(922)

To see this note that we can use the normal mirror theorem to rotate the mirrors

in (917) to obtain

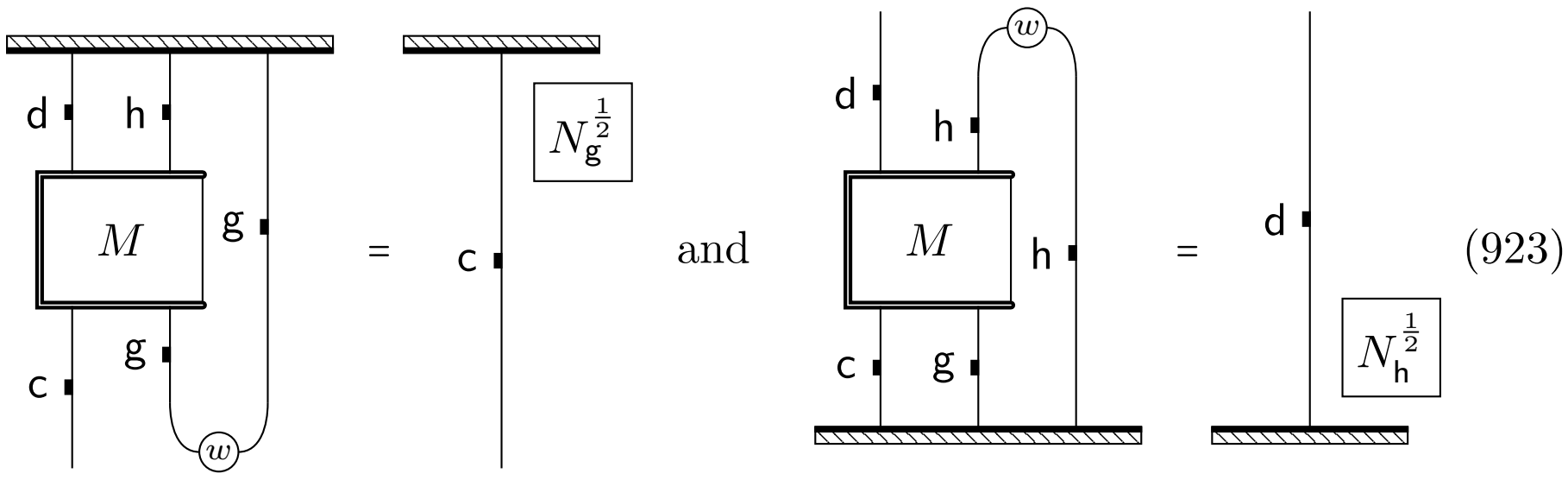

(923)

where we have also inserted twist objects (which we can do as the twist object is annihilated by its reflection in the mirror (see Sec. 30.4)). Since twist objects are maximally entangled clearly (923) expresses the requirement in (921). Here we are using the mirror definitions for a normal isometry (in (831)) and normal coisometry (in (836)). Note that the maximally entangled states in (922) can be expressed with respect to other bases since the twist-annihilation property works regardless of the basis.

Interestingly, if $M$ takes the "square" form

(924)

(so the input ancilla is the same type as the output signal and the output ancilla is the same type as the input signal) then the conditions in (907) for this to be a natural maxometry are the same as the conditions (917) for this to be a normal maxometry.

# 42 Evaluating operator circuits using Hilbert objects

Consider a general circuit. We can write it as a positive weighted sum of fully regularised circuits (as discussed in Sec. 7.11.1). Consider such a fully regularised circuit. We can also consider the corresponding operator circuit. If we can evaluate such fully regularised operator circuits then we can evaluate a general operator circuit. An example of a fully regularised circuit and of the corresponding operator circuit are

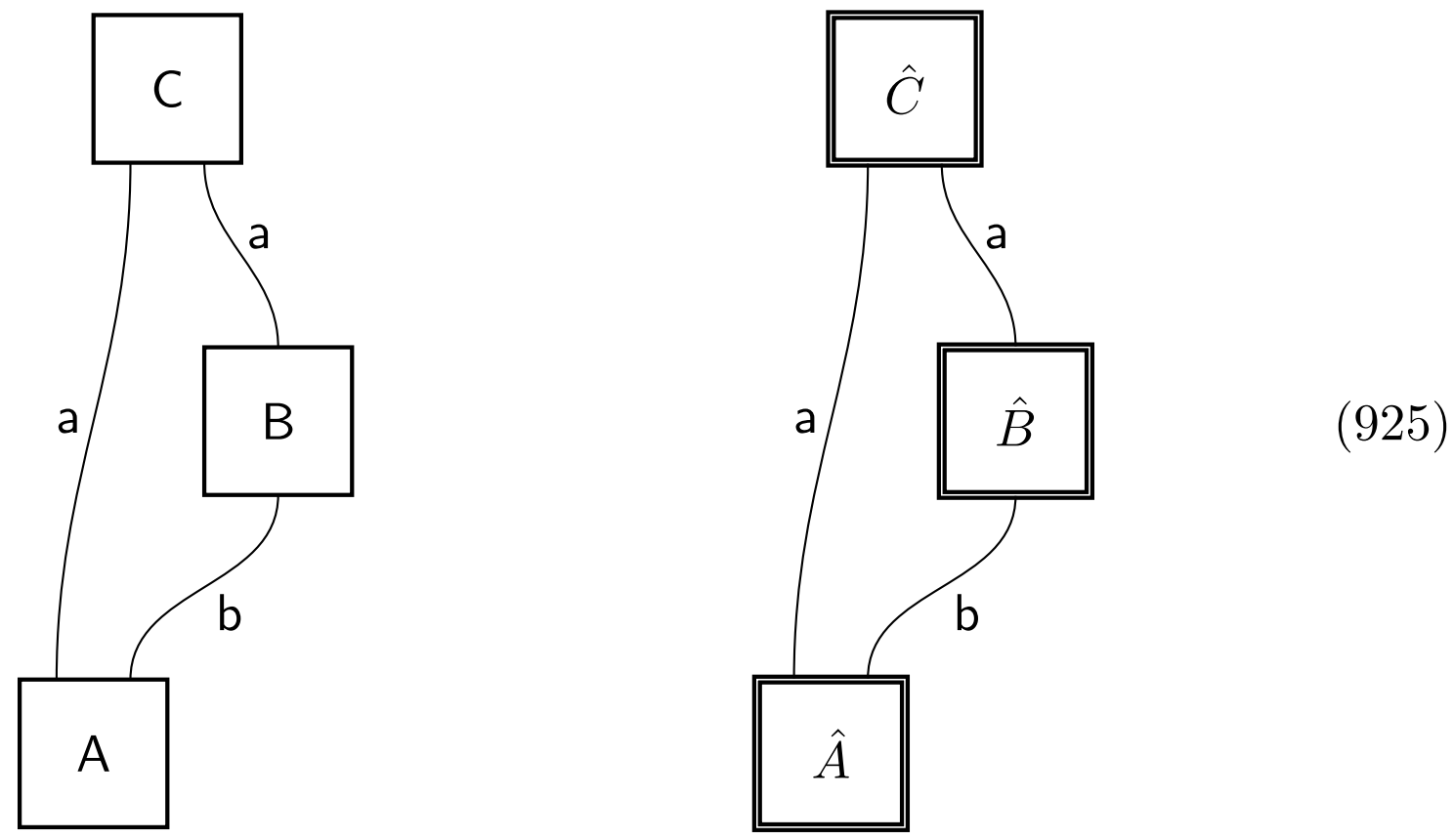

(925)

The probability of the circuit on the left should be equal to the operator circuit on the right. As discussed in Sec. 6.1, these probabilities should satisfy the circuit reality property (the probability is a real number) and the circuit positivity assumption (the probability is non-negative). In this section show (1) that operator circuits with Hermitian operators are real (2) that operator circuits with whose operators are all twofold positive are non-negative, and (3) that if we admit anti-Hermitian operators then some operator circuits will evaluate to a pure imaginary number. The first and third of these results complete the proof in Sec. 20 that circuit reality implies Hermiticity. To prove these results it is sufficient to consider fully regularised circuits (since general circuits correspond to a positive (and therefore, real) weighted sum of regularised circuits).

## 42.1 Operator circuits with Hermitian operators are real

Assume that all operators in a circuit are Hermitian. In this case we can substitute each operator by the form in (543) where the Hermitian nature of the operator is explicit. For example

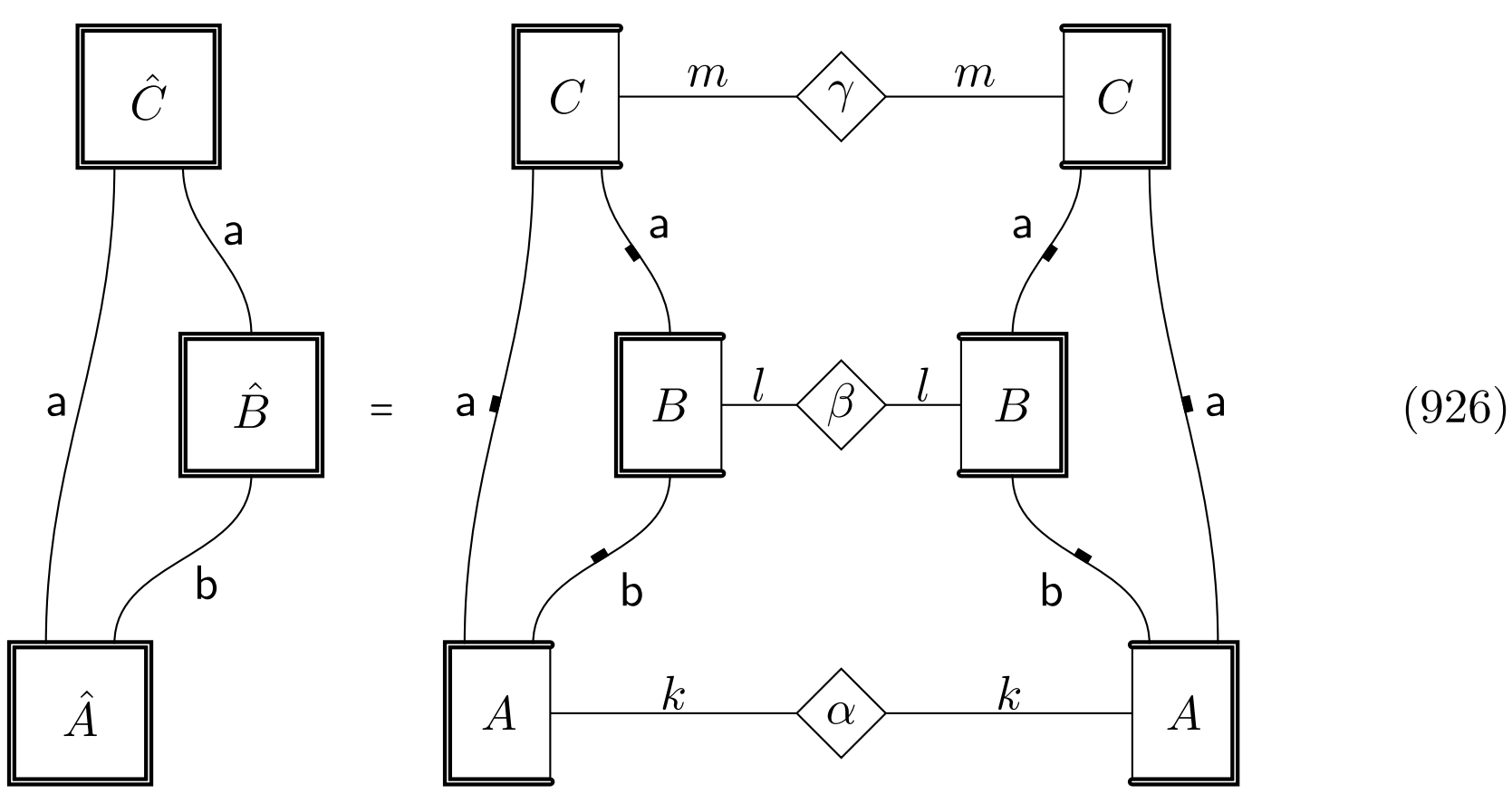

(926)

where the eigenvalues $\alpha_k$, $\beta_l$, and $\gamma_m$ are all real. It is clear that the expression on the left of (926) is real since it is of the form $\sum_{klm} D_{klm}\alpha_k\beta_l\gamma_m\overline{D}_{klm}$. Here $D_{klm}$ is the value of the left Hilbert circuit in (926) while $\overline{D}_{klm}$ is the value of the right Hilbert circuit.

## 42.2 Operator circuits with twofold positive operators are non-negative

If all the operators in an operator circuit are twofold positive (i.e. in the form in (447)) then the operator circuit clearly evaluates to a non-negative real number. For example,

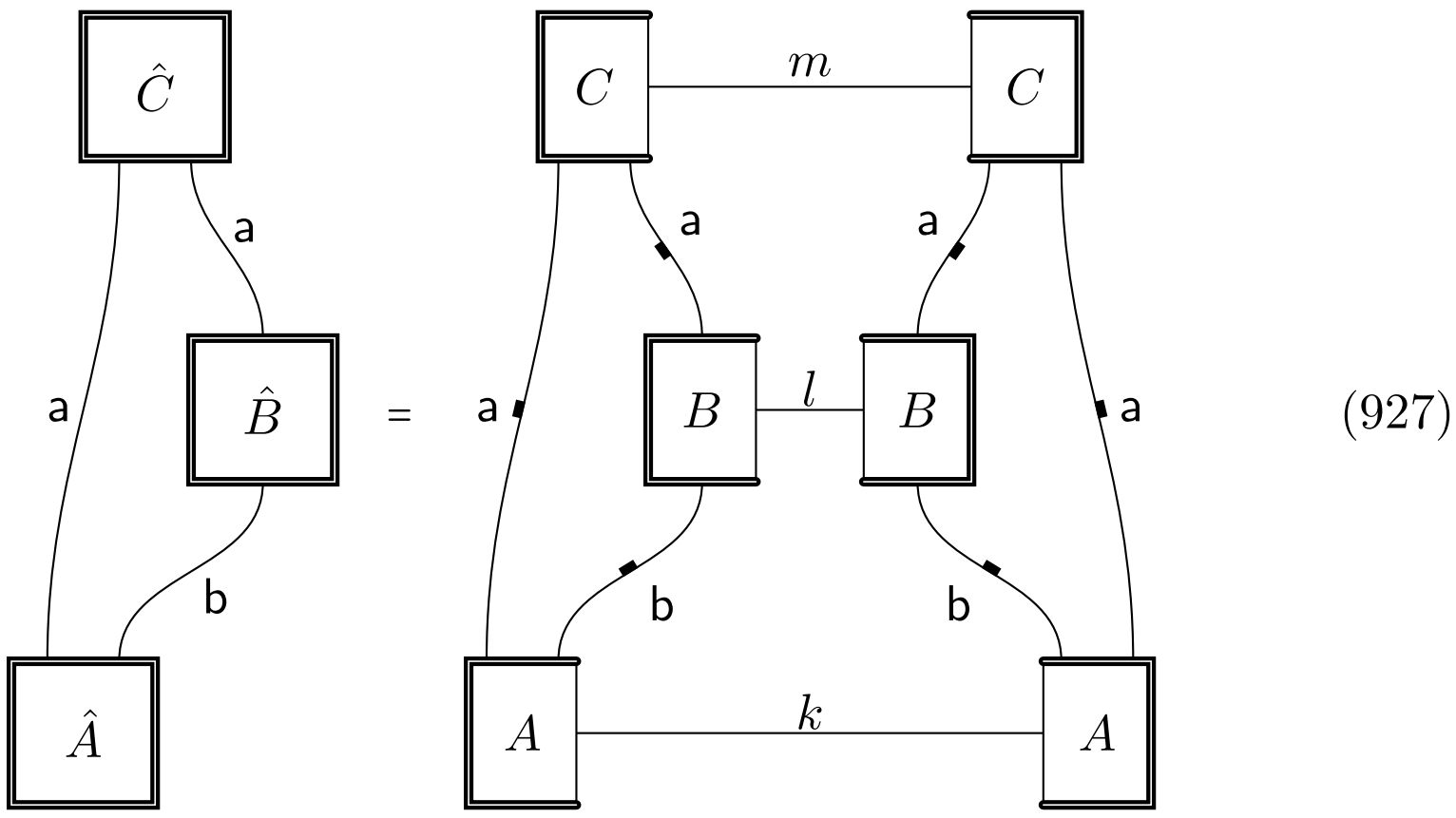

(927)

This is clearly positive since it is of the form $\sum_{klm} D_{klm}\overline{D}_{klm}$. It is worth noting that this can be written in mirror form (using a normal regular vertical mirror or a natural regular vertical mirror).

In the special case where all the operators are homogeneous we lose the sums over $k$, $l$, and $m$ and the operator circuit has the form

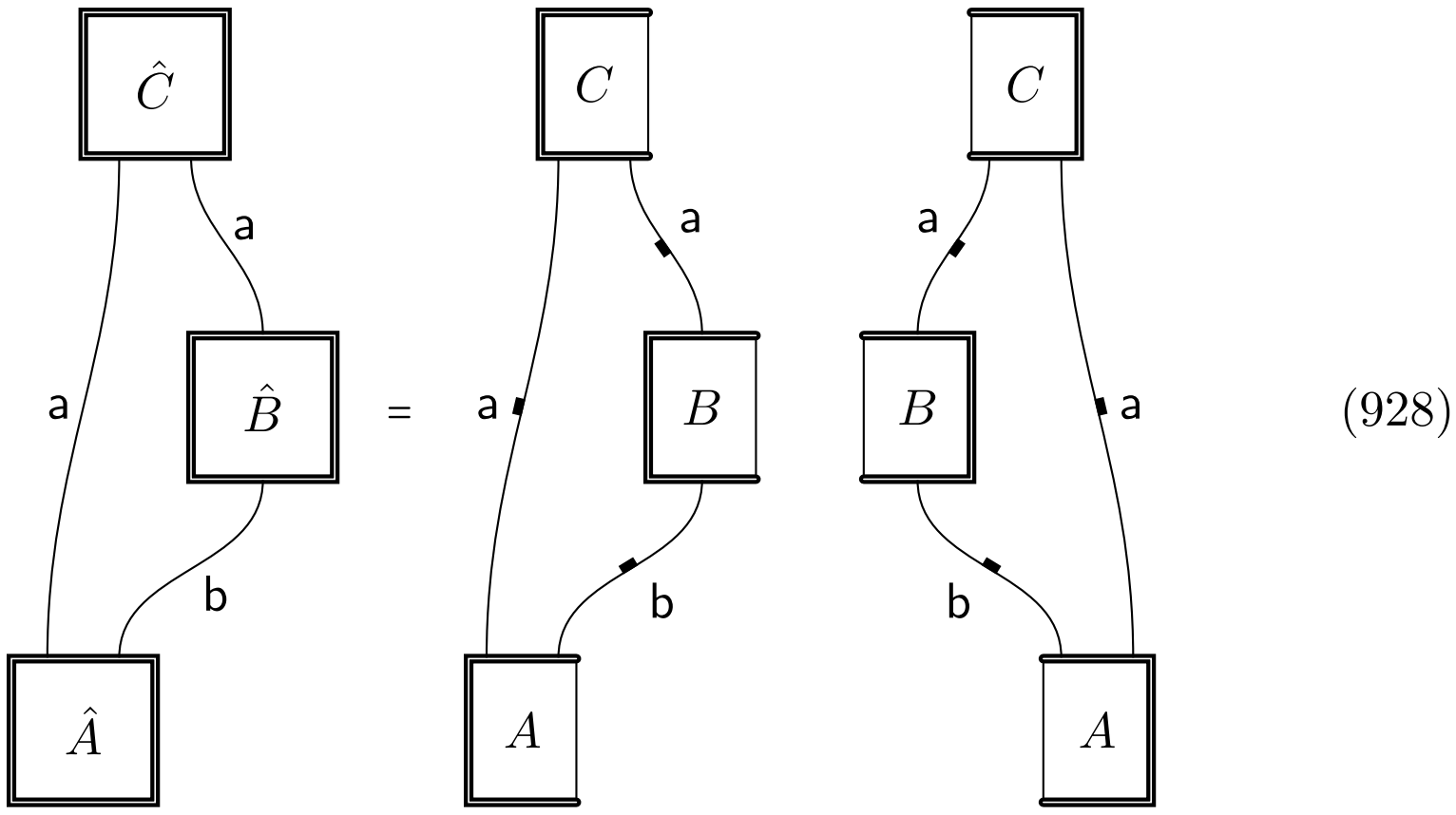

(928)

which is just the product, $D\overline{D}$, of a left and right Hilbert circuit (with no sum).

## 42.3 Operator circuits with an anti-Hermitian operator

Now we are in a position to complete the argument in Sec. 20 that circuit reality implies Hermiticity. First, consider an operator circuit having an anti-Hermitian operator, $B$, in it

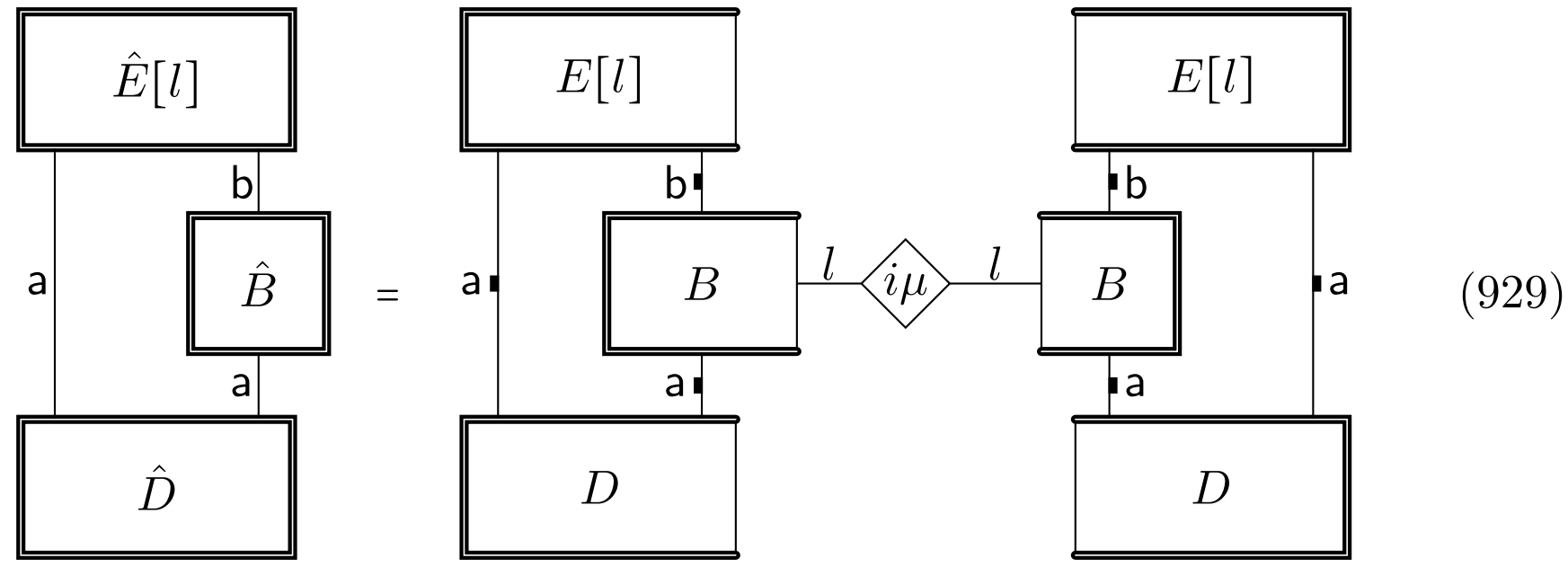

(929)

as discussed in Sec. 20. Since $\hat{B}$ is anti-Hermitian the eigenvalues $i\mu_l$ are pure imaginary, or zero. We we chose $\hat{D}$ and $\hat{E}[l]$ to be homogeneous (they could be pure). On the right we show the form of such an operator circuit when expanded in terms of left and right Hilbert objects. Anti-Hermitian operator tensors are normal. Now, we saw in Sec. 31.2 that when $\hat{B}$ is a normal operator tensor we can find $D$ and $E[l]$ such that

(930)

Thus, the expression in (929) becomes equal to

(931)

For any given $l$, this is equal to $i\nu\overline{\nu}\mu_l$. This can only be real for all $\hat{E}[l]$ if $\mu_l = 0$ for all $l$. Hence, the only way to be sure that the operator circuit in (929) is always real is for the eigenvalues of $\hat{B}_A(xy)$ to all be zero. Referring back to the argument in Sec. 20, this completes the proof that circuit reality implies that the anti-Hermitian part of any operator tensor must be zero and hence operator tensors are necessarily Hermitian.

# 43 Positivity conditions

## 43.1 Three positivity conditions

For convenience, we will write

$$\hat{B}(xy)\ {}^{\mathsf{b}}_{\mathsf{a}} \;:=\; \boldsymbol{R} \overset{\mathsf{x}}{—} \boxed{x} \overset{\mathsf{x}}{—} \hat{B}\ {}^{\mathsf{b}}_{\mathsf{a}} \overset{\mathsf{y}}{—} \boxed{y} \overset{\mathsf{y}}{—} \boldsymbol{R} \tag{932}$$

to simplify the following discussion.

There are three useful notions of positivity we can impose on operators. These are

**Tester positivity** comes from correspondence with the operational theory (as discussed in Sec. 26) is the condition

$$0 \leq \hat{E} \,\Big|\, \mathsf{h} \quad \overset{\mathsf{b}}{\underset{\mathsf{a}}{\hat{B}(xy)}} \,\Big|\, \hat{D} \tag{933}$$

for all rank one $\hat{D}$ and $\hat{E}$. Recall that the purity correspondence assumption (see Sec. 16.1) gives this mathematical condition physical significance.

**Twofold positivity** is the condition

$$\hat{B}(xy)\ {}^{\mathsf{b}}_{\mathsf{a}} \;=\; B(xy)\ {}^{\mathsf{b}}_{\mathsf{a}} \overset{l}{—} B(xy)\ {}^{\mathsf{b}}_{\mathsf{a}} \tag{934}$$

naturally associated with operators when we build them out of Hilbert space elements (as discussed in Sec. 31.5).

**Input Twist positivity** is the condition that, when we apply a twist to input space (but not the output space) we have a positive operator

$$0 \;\leq\; \hat{B}_w(xy) \quad \text{(input } \mathsf{a} \text{, output } \mathsf{b}\text{)} \tag{935}$$

where

$$\hat{B}_w(xy)\,[\mathsf{a}\to\mathsf{b}] \;=\; \hat{B}_w(xy)\,[\mathsf{a}\,\mathsf{a}\to\mathsf{b}\,\mathsf{b}] \;:=\; \hat{B}_w(xy)\,[\mathsf{a}\,\mathsf{a}\to\mathsf{b}\,\mathsf{b}] \text{ with } w \text{ applied to each input } \mathsf{a} \tag{936}$$

If we denote, in symbolic notation, the input twisted operator by $\hat{B}^{\mathsf{b}_2}_{\mathsf{a}^{w}_{1}}(xy)$ then the positivity condition in 935 can be written as $\langle\psi|\hat{B}^{\mathsf{b}_2}_{\mathsf{a}^{w}_{1}}(xy)|\psi\rangle \geq 0$ for all $|\psi\rangle$.

Note that the last condition, input twist positivity, is equivalent to output twist positivity where we apply twists to just the outcomes (see remarks at the end of Sec. 43.2) and so this condition is, like the others, time symmetric. The input twist object could also be called the input normal horizontal transpose operation since the twist implements $H$.

The twist object is based on orthonormal bases. We can, instead, use the physical twist objects (denoted by $\tilde{w}$ and defined in the third row of Table 4) to define a *input physical twist*. It is easy to adapt the proof to be given below to this case. Positivity with respect to the input physical twist is equivalent to that with respect to the input twist.

## 43.2 Operator positivity theorem

We will now prove the following theorem

> **Operator positivity theorem.** For simple operator tensors, the tester positivity, twofold positivity, and input twist positivity conditions are equivalent.

Given this theorem, we can simply refer to $T$-positivity of operators as we have been doing. Before proving the theorem we note that each of these different ways of expressing positivity has its own advantage. Tester positivity emphasises the operational meaning of positivity. Twofold positivity connects in a beautiful way with the theory of Hilbert space and input twist positivity allows us to check positivity by inspecting eigenvalues (since it relates to the standard textbook notion of positivity of operators).

First we will prove that tester positivity and twofold positivity are equivalent (i.e. (933) ⇔ (934)). It is easy to prove that (934) ⇒ (933). If we insert the form on the right hand side of (934) into the expression on the right hand side of (933) we obtain

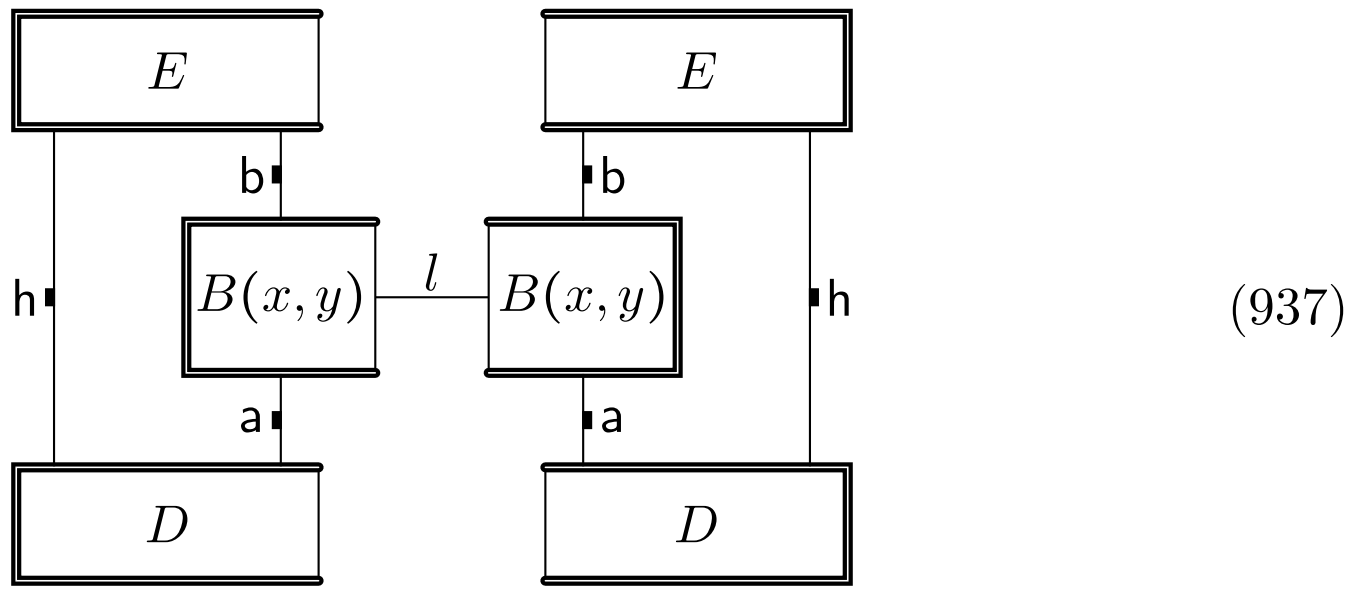

(937)

This is clearly positive since it is the sum (over $l$) over products of a number with its complex conjugate (see Sec. 28.5). Now we want to prove (933) ⇒ (934). First note that tester positivity implies circuit reality (since non-negative numbers are necessarily real). We know from the discussion in Sec. 20 and Sec. 42 that circuit reality implies Hermiticity. Hence, $\hat{B}(xy)$ is Hermitian. Using the form in (543) for a Hermitian operator, inequality (933) implies

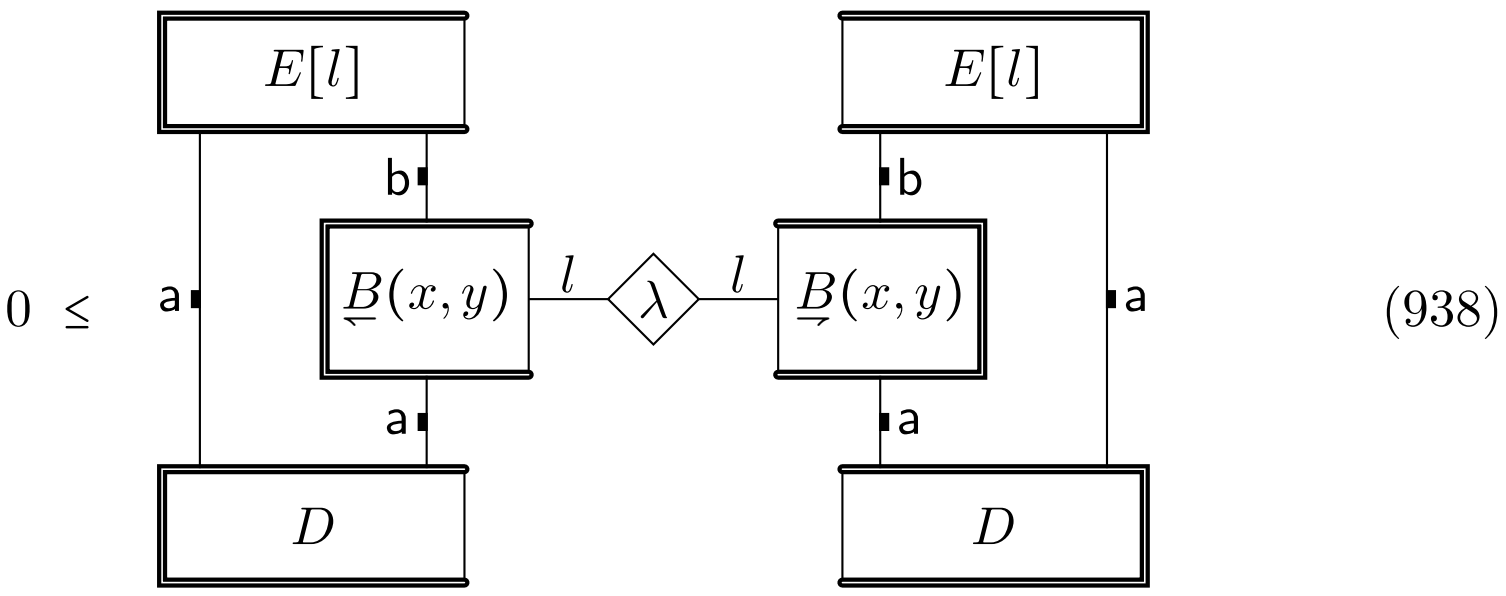

(938)

where the eigenvalues, $\lambda_l$, are real. We have put h = a and we have introduced the $l$ in $E[l]$ so we can use a result from Sec. 31.2. Using (539) and its conjugate (938) becomes

$$0 \leq \; \langle l \rangle \overset{\boxed{\nu}}{\underset{}{—l—}} \langle \lambda \rangle \overset{\boxed{\bar{\nu}}}{\underset{}{—l—}} \langle l \rangle \qquad \forall l \tag{939}$$

which implies that $\lambda_l \geq 0$ for all $l$. Hence (934) follows (since the $\lambda_l$ can be absorbed when positive as $\sqrt{\lambda_l}$ to the left and to the right as discussed in Sec. 31.5).

Now we will prove tester positivity and input twist positivity are equivalent.

Tester positivity can be written as

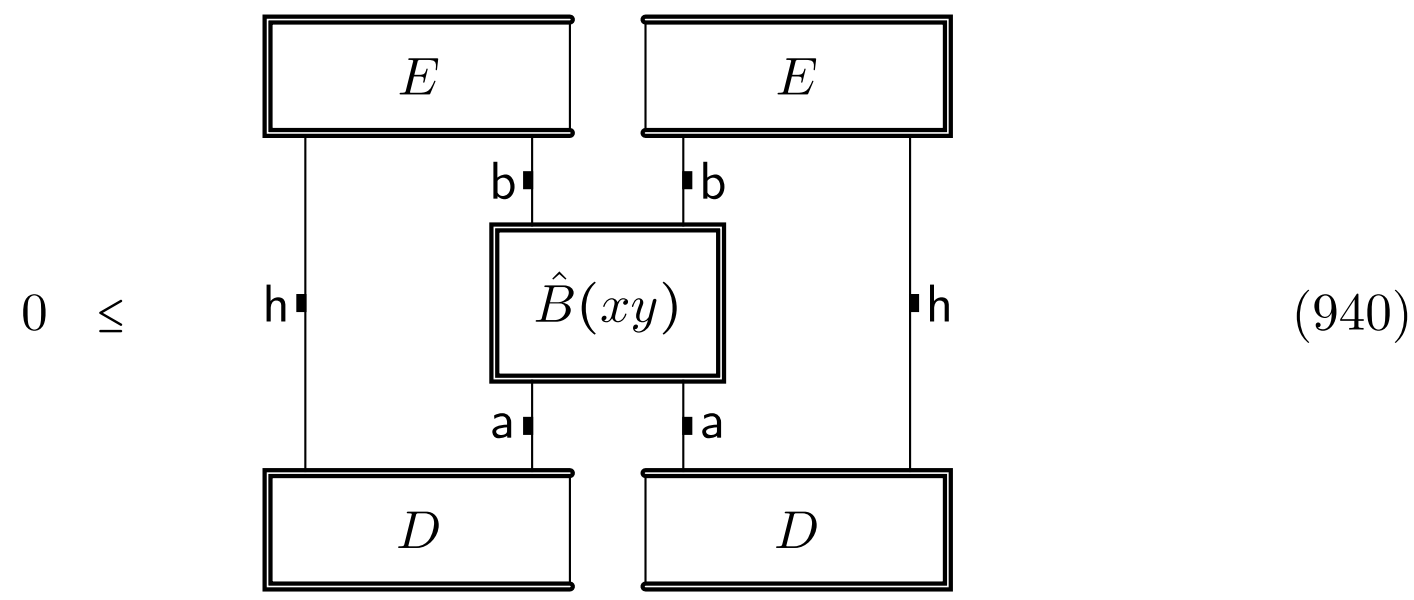

We can expand $D$ and $E$ with respect to a basis for h giving

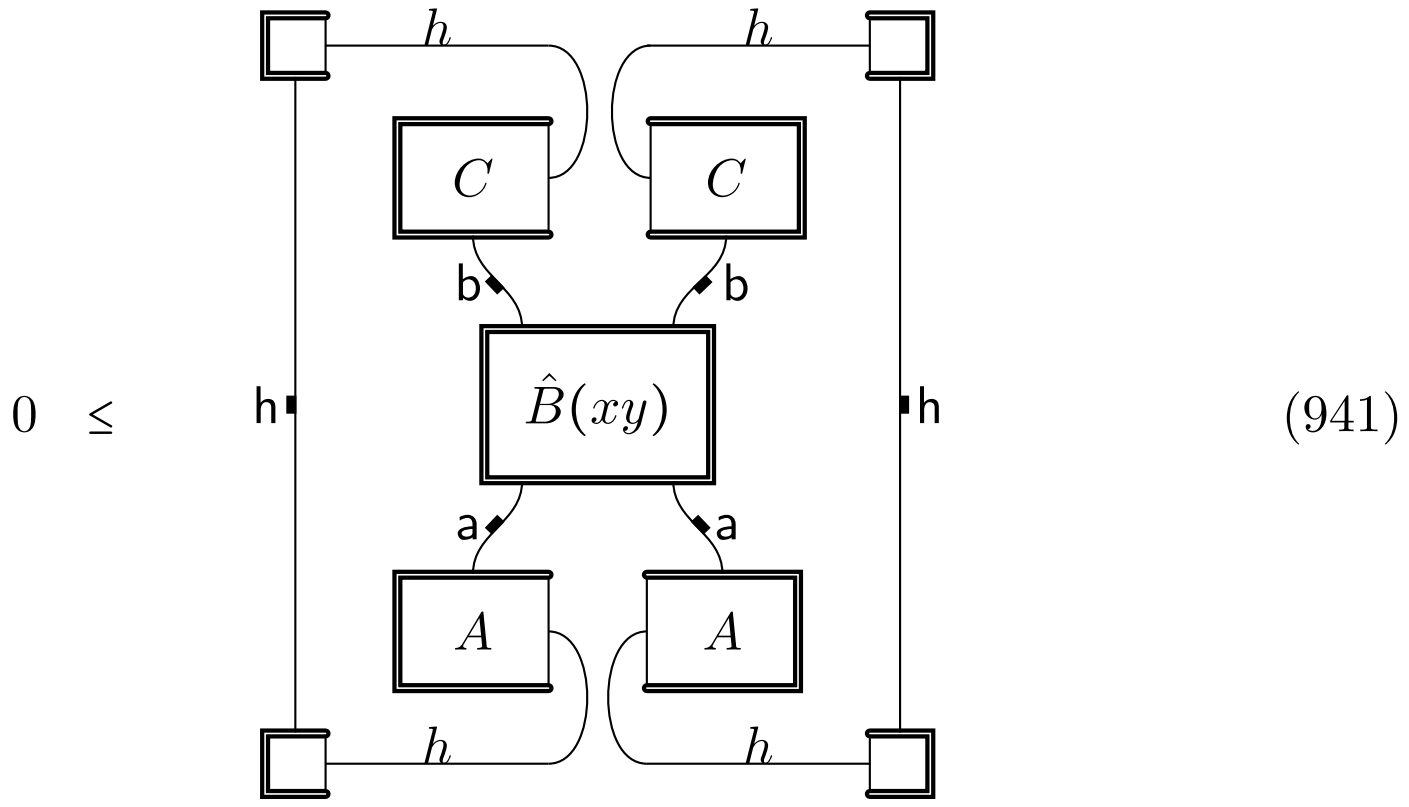

using the orthogonality conditions for the h basis elements we obtain

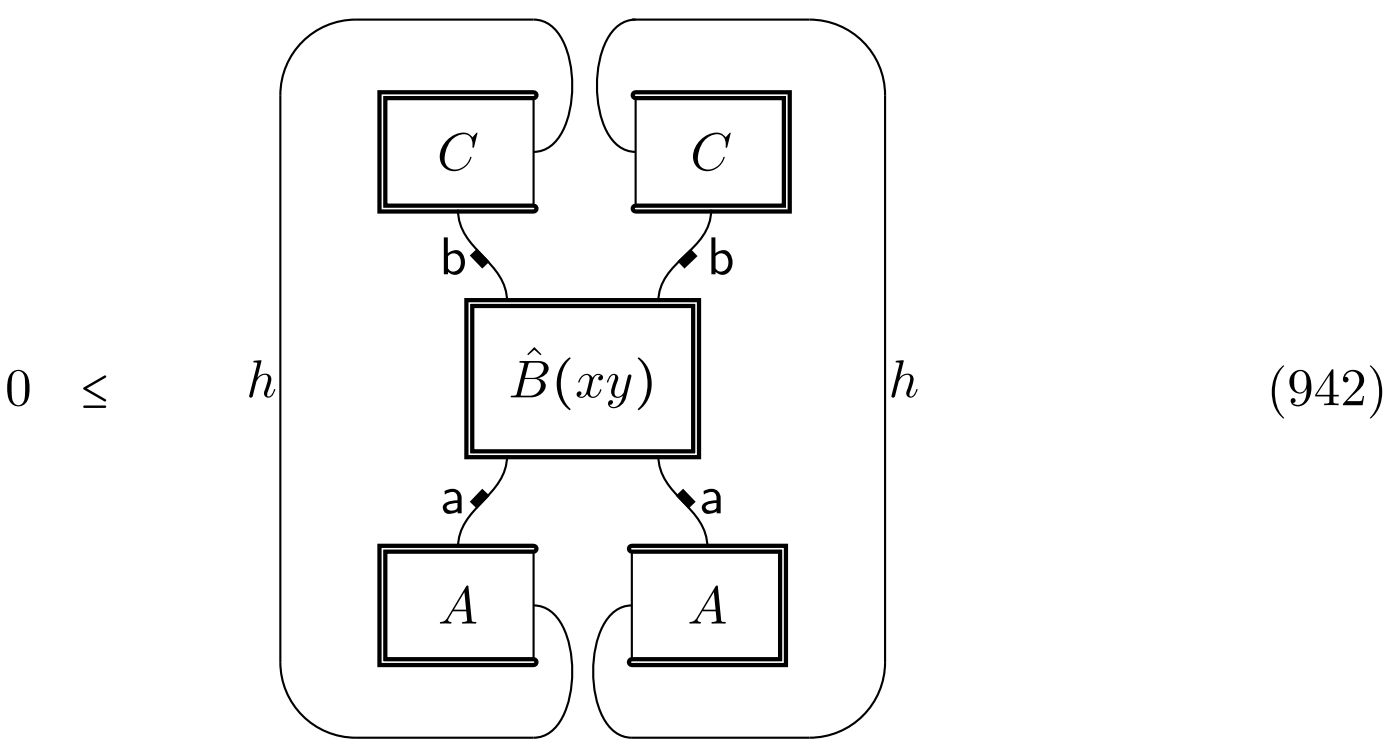

We can write this as

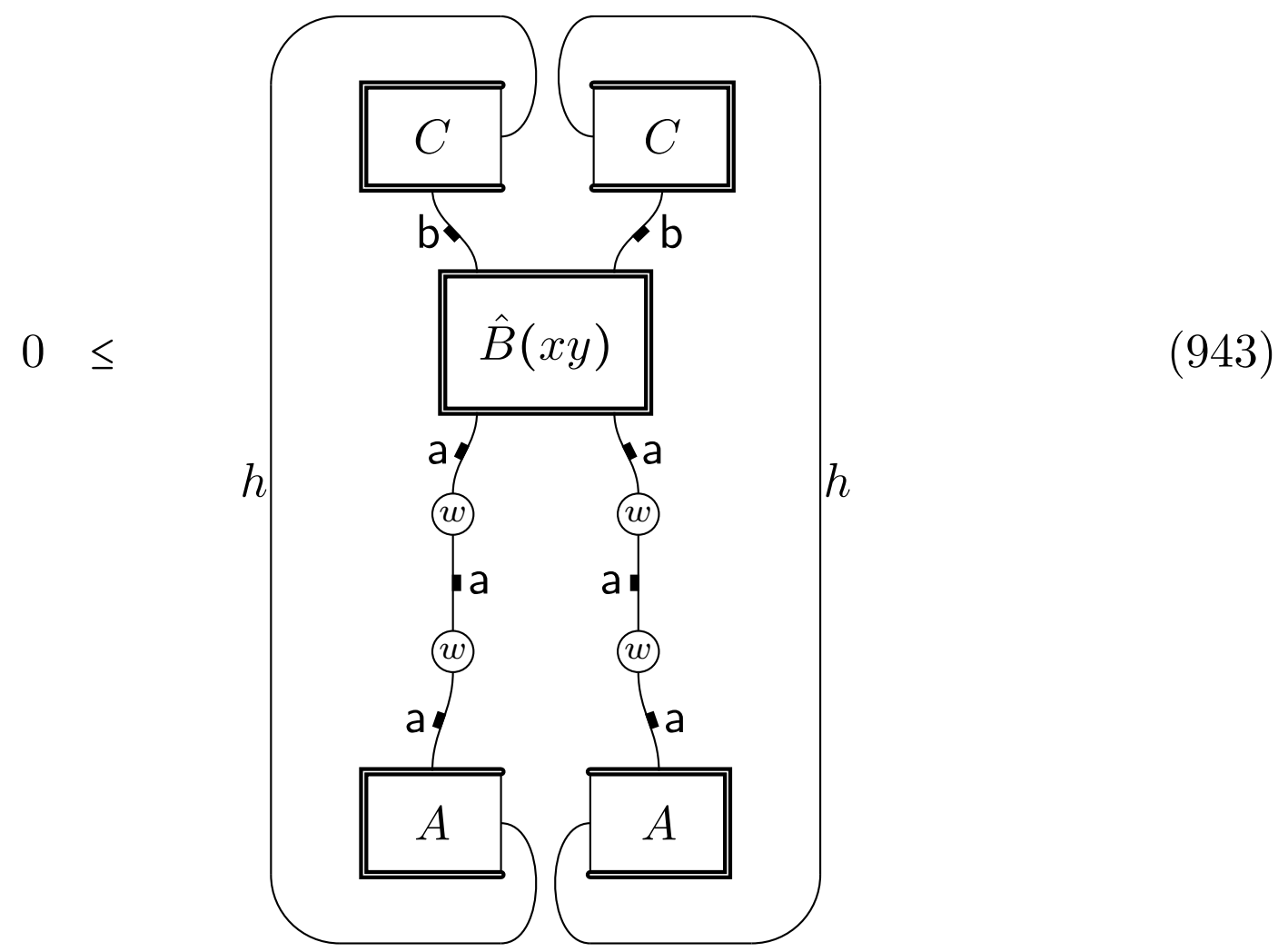

Absorbing the upper twist objects into $B$ and the lower ones into the $A$ boxes we obtain.

$$0 \leq \quad C \quad C \quad \hat{B}_{H_{\text{in}}}(xy) \quad A_H \quad A_H \tag{944}$$

Recall from (579) that applying a twist to every wire implements the normal horizontal transpose. If we just apply it to the input wires then we have the *input twist* which can also be called the *input normal horizontal transpose* (hence the subscript $H_{\text{in}}$ above). On the $A$ boxes we apply the output twist which, since these boxes have no input wires, implements the horizontal transpose (which we denote by $A^H$). The $A^H$ and $C$ boxes on the right are both kets, i.e. $|A_h^H\rangle^{\mathsf{a}_1}$ and $|C_h\rangle_{\mathsf{b}_2}$ (see (457) and (458)). Their combined state can be written $|\psi\rangle_{\mathsf{a}_1}^{\mathsf{b}_2}$ where

$$|\psi\rangle_{\mathsf{a}_1}^{\mathsf{b}_2} = \sum_h |A_h^H\rangle^{\mathsf{a}_1} |C_h\rangle_{\mathsf{b}_2} \tag{945}$$

It is possible choose $|A_h^H\rangle^{\mathsf{a}_1}$ and $|C_h\rangle_{\mathsf{b}_2}$ such that $|\psi\rangle_{\mathsf{a}_1}^{\mathsf{b}_2}$ is proportional to any element of the corresponding Hilbert space. Likewise, we have a bra, ${}_{\mathsf{a}_1}^{\mathsf{b}_2}\langle\psi|$ on the left. Hence (944) is the input twist positivity condition (of the form ${}_{\mathsf{a}_1}^{\mathsf{b}_2}\langle\psi|B_T|\psi\rangle_{\mathsf{a}_1}^{\mathsf{b}_2} \geq 0$). These steps are all reversible and so see that input twist

positivity is equivalent to tester positivity. This completes the proof of the operator positivity theorem.

It is worth commenting that we can use the same proof technique to prove that output normal transpose positivity is equivalent to tester positivity. This also tells us that input tester positivity and output tester positivity are equivalent.

Furthermore, we can consider the case of using physical twists instead. If we replace $w$ by $\tilde{w}$ (so we are implementing the input physical twist) and use the fact that physical twists cancel (except that an overall positive factor, $1/N_{\mathsf{a}}$, gets introduced for each pair of cancelling $\tilde{w}$'s) then we can go through similar steps and prove positivity. Thus, input physical twist positivity is the same condition as input twist positivity.

## 43.3 The Choi-Jamiołkowski operator

Many treatments of Quantum Theory use what is usually called the Choi-Jamiołkowski operator. We will not make direct use of this object in this book. However, it is useful to discuss it to make contact with these other treatments.

The Choi-Jamiołkowski operator is defined as follows

b
$\hat{B}_{CJ}(xy)$
a
:=
a b b a
$\hat{B}(xy)$
a a
$w$ $w$
(946)

We can invert this operation by appending twist caps and applying the yanking identities like that in (519). Thus, we have an isomorphism between operators having inputs and outputs and operators having only outputs (states). We could, instead, have appended twist caps in the first place. This would give an isomorphism between general operators and operators having only inputs (effects). Note that operator in (946) was actually defined by Choi Choi [1975]. Jamiołkowski Jamiołkowski [1972] defined a basis independent operator (that given in (946) with the twists removed).

The Choi-Jamiołkowski operator in (946) can be obtained by putting cups on the input twisted operator (shown in (936)). It is easy to prove from this that input twist positivity is equivalent to positivity of the Choi-Jamiołkowski operator (and so positivity of (946) is an alternative and equivalent way of expressing $T$-positivity.

We could extend the Hilbert cube by appending all the objects obtained by applying non-conjugating operations to either the inputs or outputs. This would provide an even bigger zoo of objects. For the purposes of the present work this does not help us.

# 44 Dilation

## 44.1 Introduction

An important theorem in Quantum Theory is the Stinespring dilation theorem. This first appeared as a mathematical result in Stinespring [1955]. The dilation, as applied to Quantum Theory, first appeared in Hellwig and Kraus [1970] (see also Hellwig and Kraus [1969] for historical context). Interestingly, Hellwig and Kraus did not cite Stinespring nor prove that any completely positive map has such a dilation. This connection appears to have been first made explicitly in Davies [1976]. Modern treatments can be found in many textbooks like Preskill [2024].

In its standard presentation the Stinespring dilation pertains to the time forward frame of reference (to TFSOQT). If we translate it into our diagrammatic language, it says that any deterministic physical operator without outcomes in this time forward theory (one satisfying forward causality and $T$-positivity) can be written as

(947)

where $\hat{K}$ is a normalised pure (homogeneous) state, $\hat{I}$ is the ignore effect, and $\hat{\boldsymbol{U}}$ is a unitary operator (we would want this to be natural in our treatment though this idea does not exist in standard textbook treatments). It is easy to include outcomes in the dilation. Then we have

(948)

We have included a maximal result, $\ulcorner\hat{\boldsymbol{Y}}$, which provides the outcome (a Stinespring dilation with outcomes was first provided in Ozawa [1984]). We have also written the pure state, $\hat{K}$, in terms of preselecting a maximal state, $\hat{\boldsymbol{X}}\lrcorner$ (where the preselection box is defined by correspondence with (273) and so this pure state preparation is by correspondence with the left expression in (294)). This dilation is clearly not time symmetric.

What, then, is the correct dilation for a physical operator in the time symmetric theory (so it satisfies forward and backward causality as well as $T$-positivity)? The temptation is to simply add an ignore state, $\hat{\boldsymbol{I}}$ going into the natural unitary and remove the preselection box. This would give

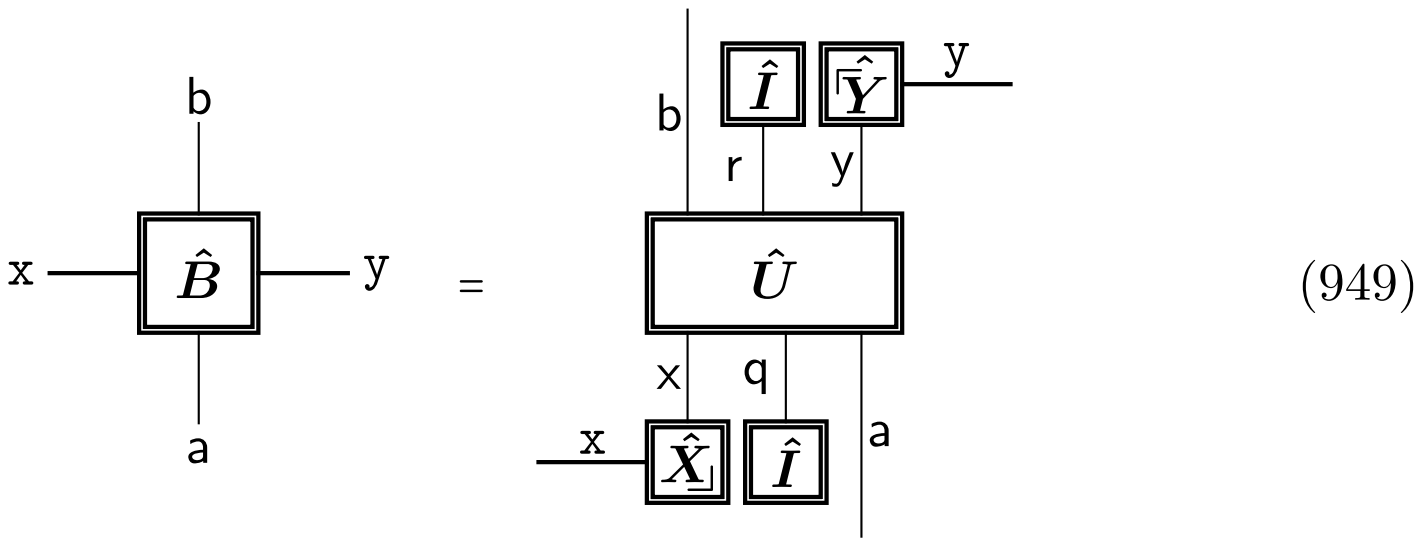

(949)

However, this is not the form we obtain in the dilation theorem to be proved below. Rather, we obtain the form

(950)

Here $\hat{M}$ is a natural maxometric operator with respect to ancilla **q** and **r** (as discussed in Sec. 41.2). This form for the dilation in is clearly time symmetric (in the sense that the time reverse of this is of the same form).

If $\hat{M}$ is actually physical then it is necessarily a physical unitary (since any homogeneous deterministic physical operator is a physical unitary - see Sec. 38). However, it might not be physical. This raises an important question which is *unresolved* in this work. Can any dilation in the form (950) (with a natural maxometric operator) be written in the form (949) (with a physical unitary)? This issue is important because we cannot build $\hat{M}$ in the laboratory if it is non-physical. In the usual time forward picture the Stinespring dilation offers a way to build any physical operator in terms of more basic operators. Even more important than this, it allows for a *peaceful coexistence* between the church of the larger Hilbert space and the church of the smaller Hilbert space. We will explore these issues in Sec. 45.1 through three examples and offer some conjectures. We will discuss peaceful coexistence and churches in Sec. 45.3.

## 44.2 Maximal representation of operators

Before we state and prove the dilation theorem, recall the maximal representation theorem from Sec. 8.3. We have the corresponding theorem for operators.

**Maximal representation theorem (for operators).** Any oper-ator, $\hat{B}$, can be written as

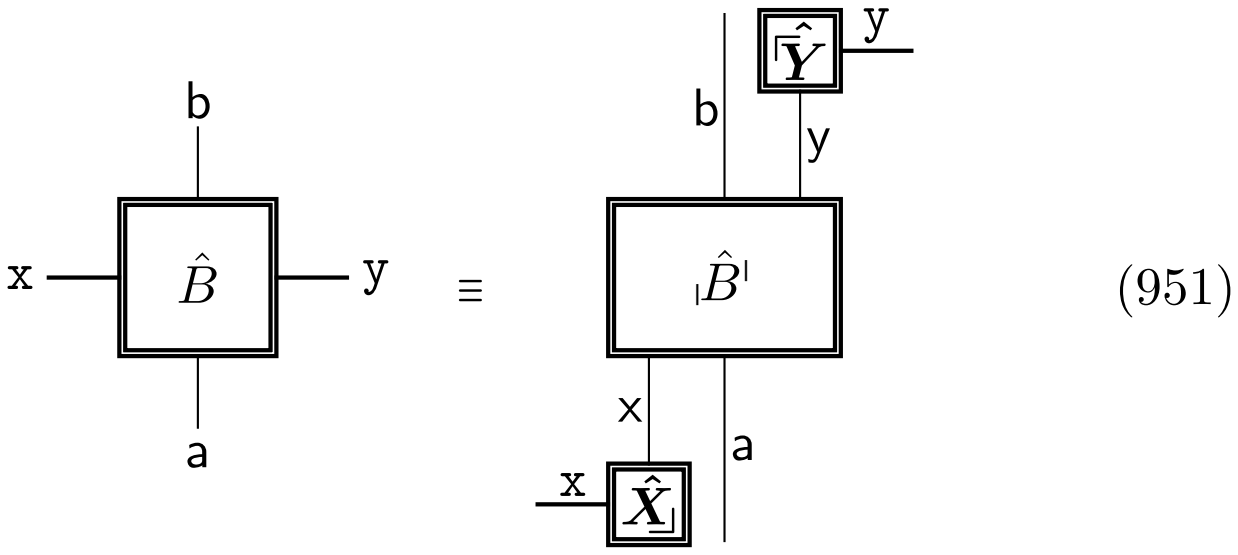

(951)

where

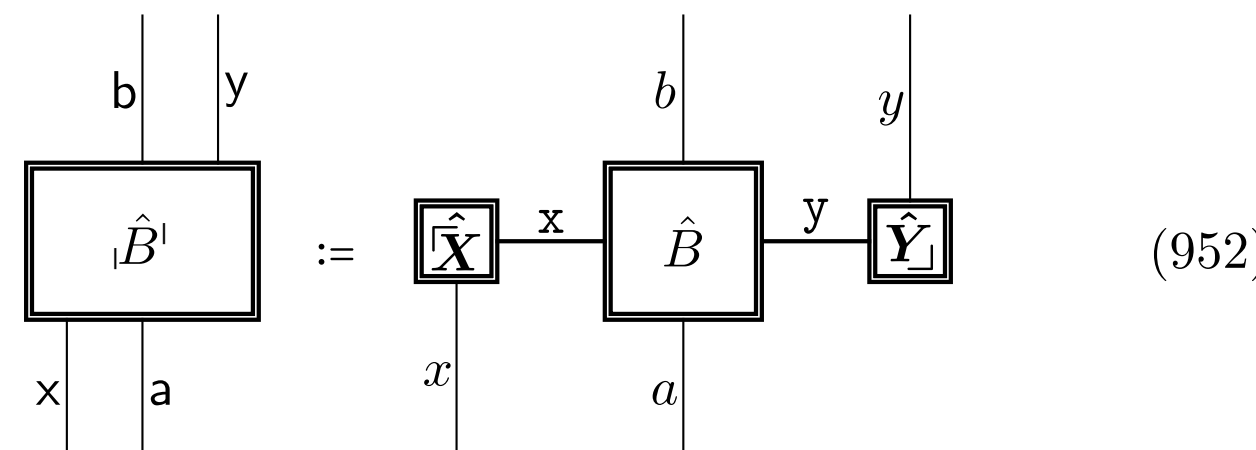

(952)

> Furthermore: (i) $\llcorner\hat{B}\urcorner$ satisfies tester positivity if and only if $\hat{B}$ satisfies tester positivity, (ii) $\llcorner\hat{B}\urcorner$ satisfies general double causality if and only if $\hat{B}$ satisfies general double causality, (iii) $\llcorner\hat{B}\urcorner$ is deterministic if and only $\hat{B}$ is deterministic. As a consequence, $\llcorner\hat{B}\urcorner$ is a physical (and deterministic) if and only if $\hat{B}$ is physical (and deterministic).

This theorem follows by correspondence with the operation case discussed in Sec. 8.3.

Note that the maximal definition allows us to extend the definition of *rank* introduced in Sec. 31.5 to the case where there are incomes and outcomes. The rank of a twofold positive operator, $\hat{B}$, is defined to be equal to the rank of $\llcorner\hat{B}\urcorner$.

## 44.3 Dilation theorem

We are now in a position to state a dilation theorem for the time symmetric case.

> **Dilation theorem.** An operator, $\hat{\boldsymbol{B}}$, is deterministic and physical if and only if can we write it in the form

(953)

where $\hat{M}$ is a natural maxometric operator with respect to ancillae $\mathsf{q}$ and $\mathsf{r}$. These ancillae can be any systems satisfying $N_{\mathsf{q}}N_{\mathsf{r}} \geq \text{rank}(\hat{\boldsymbol{B}})$.

In some expressions below, we will group systems together (e.g. $\mathsf{xa}$) for convenience. First we will provide the "only if" part of the theorem (this is more difficult than the "if" part). We note that it follows from the maximal representation theorem in Sec. 44.2 that $\hat{\boldsymbol{B}}$ is deterministic and physical if and only if ${}_{\llcorner}\hat{\boldsymbol{B}}^{\urcorner}$ is deterministic and physical. So we can turn our attention to ${}_{\llcorner}\hat{\boldsymbol{B}}^{\urcorner}$. If this is physical then it must satisfy $T$-positivity which means we can write it in twofold form (see Sec. 43) as follows

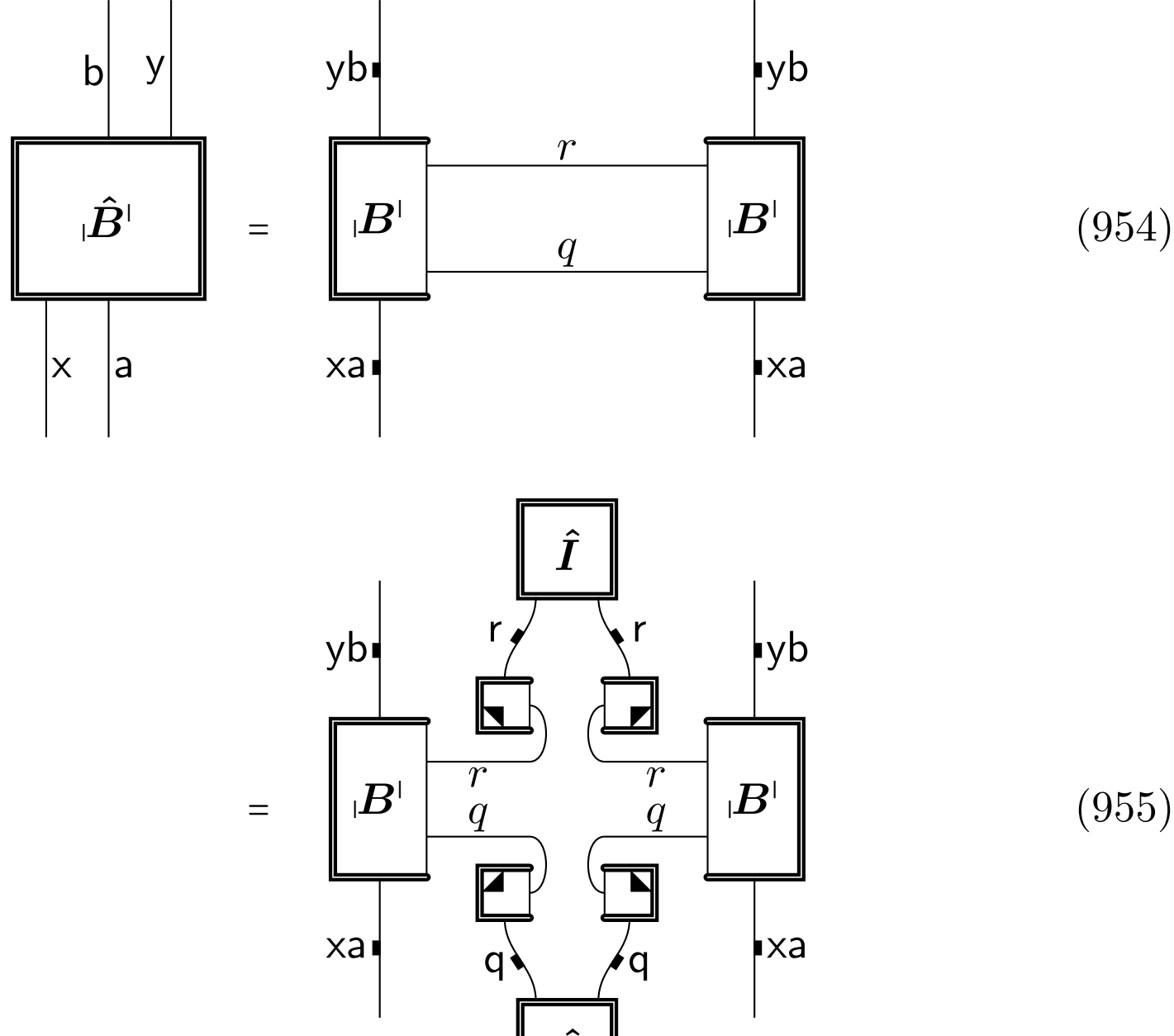

where $N_{\mathsf{q}}N_{\mathsf{r}} \geq \text{rank}({}_{\llcorner}\hat{\boldsymbol{B}}^{\urcorner})$ by the minmax theorem in Sec. 31.5. Note one possible choice is $\mathbf{q} = \mathbf{by}$ and $\mathbf{r} = \mathbf{ax}$ (by the minmax theorem). The final expression follows from the definitions of $\hat{\boldsymbol{I}}$ (see (634, 635)) and the defining property

(665)) of the ortho-deterministic basis. We have

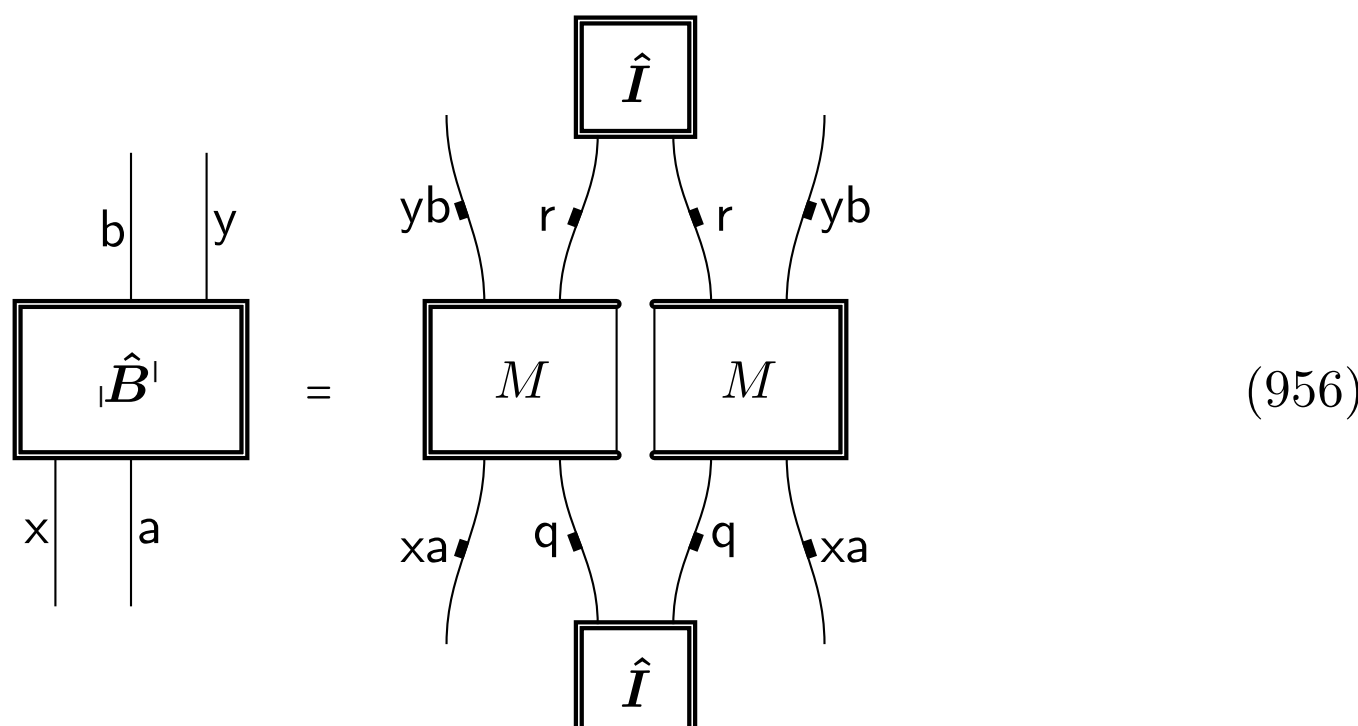

(956)

where

(957)

Since ${}_{\lfloor}\hat{\boldsymbol{B}}^{\rfloor}$ is deterministic and physical we require it satisfies the double causality conditions

(958)

If we insert (956) into these equations we get the conditions for $M$ to be a natural maxometry. This proves that, if $\hat{\boldsymbol{B}}$ is deterministic and physical, it can be written in the dilated form in the theorem. Now we need to provide the "if" part of the proof. That is we need to prove if an operator, $\hat{\boldsymbol{B}}$, that can be written as in (953), where $\hat{M}$ is a natural maxometric operator, then $\hat{\boldsymbol{B}}$ is both deterministic and physical. In fact, this follows immediately by reversing the steps. First, we see that $\hat{\boldsymbol{B}}$, written in the form in (953), necessarily satisfies twofold positivity. Then we see that if $\hat{M}$ is a natural maxometric operator then the double causality conditions for ${}_{\lfloor}\hat{\boldsymbol{B}}^{\rfloor}$ are satisfied. From this the double causality conditions $\hat{\boldsymbol{B}}$ readily follow. This proves the dilation theorem (the requirement that $N_{\mathsf{q}}N_{\mathsf{r}} \geq \text{rank}({}_\lfloor\hat{\boldsymbol{B}}^\rceil)$ follows from the comment below (955)).

Since the natural maxometric operator, $\hat{M}$, is central to the above result it is worth providing the following corollary concerning its relationship with $\hat{\boldsymbol{B}}$.

> **Dilation corollary 1.** The natural maxometric operator $\hat{M}$ used in the dilation of $\hat{\boldsymbol{B}}$ is of the form
>
> (959)
>
> where
>
> (960)
>
> Further, any $M$ of this form where ${}_\lfloor\hat{\boldsymbol{B}}^\rceil$ is physical, is necessarily a natural maxometry.

This is easily proved from (957) that appears in the proof of the dilation theorem above. It is interesting to note that it is consistent with the dilation theorem to choose q = yb and r = xa in which case $M$ takes the "square" form in (924). In this case the condition for $M$ to be a natural maxometry is the same as the condition for it to be a normal maxometry.

It is also worth stating a second corollary concerning the special case where $\hat{\boldsymbol{B}}$ is homogeneous.

> **Dilation corollary 2.** If $\hat{\boldsymbol{B}}$ is rank one (so ${}_\lfloor\hat{\boldsymbol{B}}^\rceil$ is homogeneous) then a necessary and sufficient condition for it to be physical and deterministic is

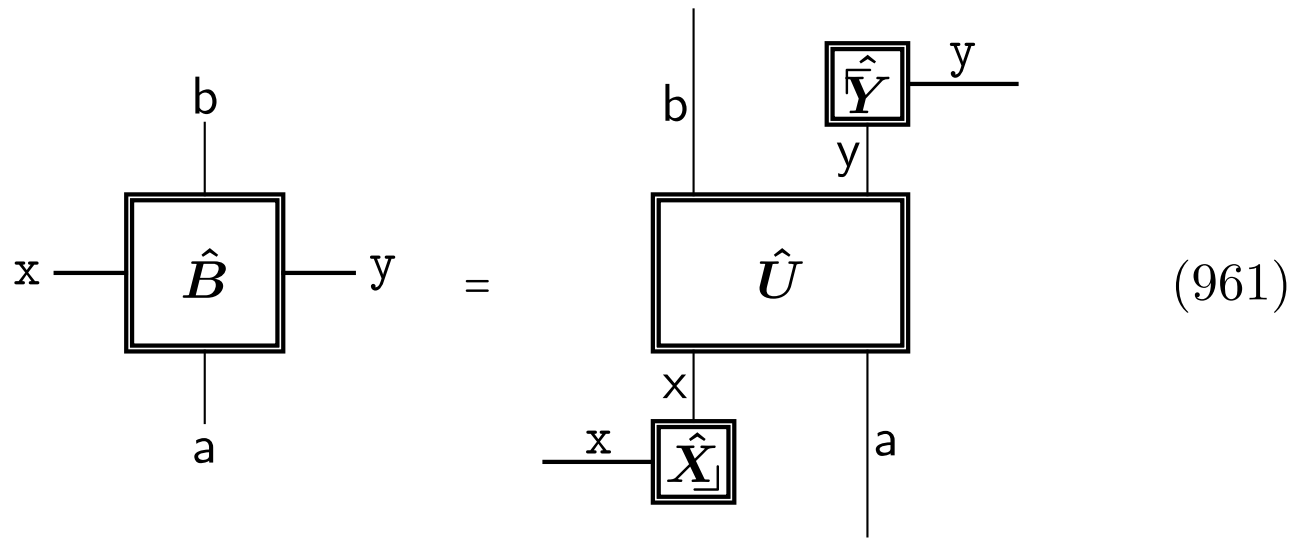

(961)

where $\hat{\boldsymbol{U}}$ is a natural unitary.

This is an immediate consequence of the dilation theorem. If $\hat{\boldsymbol{B}}$ has rank one we can choose $N_{\mathsf{q}} = N_{\mathsf{r}} = 1$ according to the dilation theorem. This means that systems q and r are null and so can be omitted. In this case we do not have any $\hat{\boldsymbol{I}}$'s in the dilation. Further, it means that the natural maxometry, $\hat{M}$, has no ancillae. A maxometry with no ancilla is, in fact, a natural unitary (compare the defining properties of a natural maxometry in (907) with the defining properties of a natural unitary in (873)). This proves we must have (961).

## 44.4 Dilation theorem with sufficient pairs

In the above dilation theorem we gave a single dilation which, by itself, is both necessary and sufficient for an operator to be deterministic and physical. In the complex book, when we come to consider what will be called "causal dilation theorems" (in Sec. 78), we will not be able to provide a single dilation that is both necessary and sufficient by itself. Instead, we will provide a pair of dilations each of which is necessary. Taken together the pair of dilations will be seen to be sufficient - hence we will call these *sufficient pairs*.

Interestingly, we can also formulate the dilation theorem in Sec. 44.3 in terms of a sufficient pair by employing some theorems about natural maxometries we proved earlier. We can do this in two ways - in terms of an isometry and a coisometry, or in terms of unitaries.

First consider the case with (co)isometric operators.

**Sufficient pair dilation theorem -(co)isometric version.** An operator

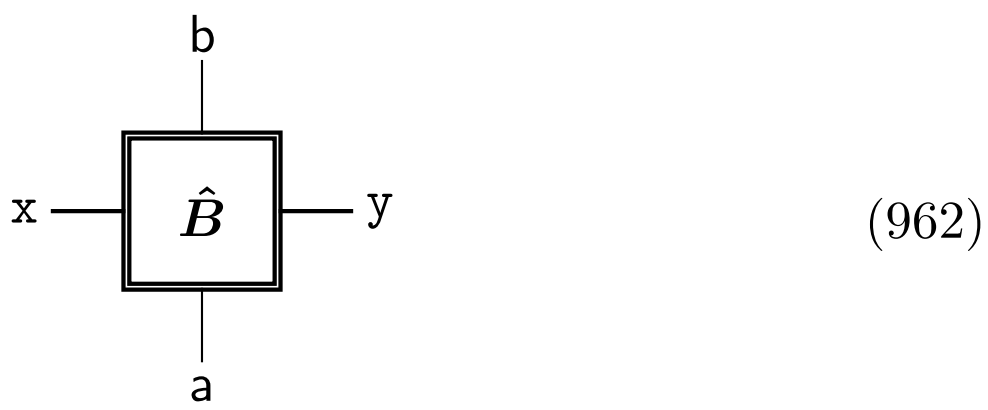

(962)

is deterministic and physical if and only if it can be written being equal to *both*

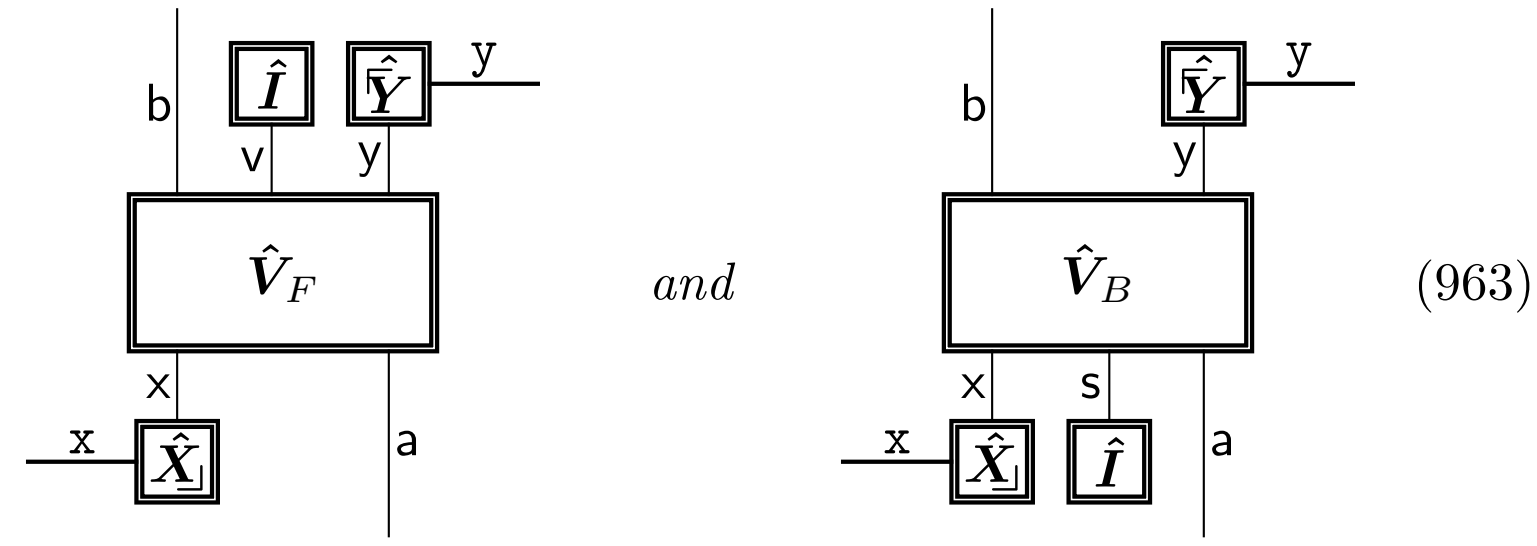

(963)

where $\hat{\boldsymbol{V}}_F$ is a natural isometric operator and $\hat{\boldsymbol{V}}_B$ is a natural coisometric operator.

The proof of this is straight forward. The left expression in (963) follows from the dilation theorem since we can choose $\mathsf{q}$ in (953) to be null (as long as $N_{\mathsf{r}} \geq \text{rank}(\hat{\boldsymbol{B}})$). Then the natural maxometric operator has no input ancilla. Such a maxometric operator is necessarily a natural isometric operator. The right expression in (963) can be proven similarly. To see that these two conditions, taken together, are sufficient for $\hat{\boldsymbol{B}}$ to be deterministic and physical note first that both expressions are clearly $T$-positive since the components are. Then note that the expression on the left satisfies forward causality because of the properties of the isometric operator and, similarly, the expression on the right satisfies backward causality because of the properties of the coisometric operator.

Now consider the theorem with natural unitary operators.

**Sufficient pair dilation theorem.** An operator

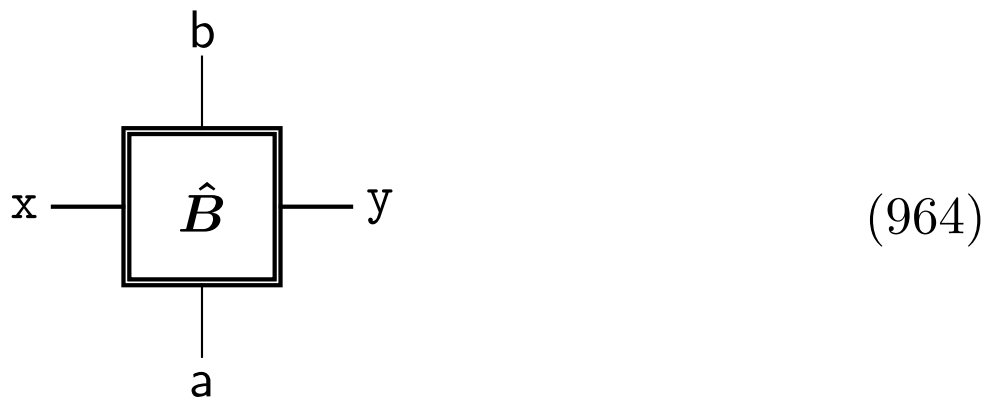

(964)

is deterministic and physical if and only if it can be written being equal to *both*

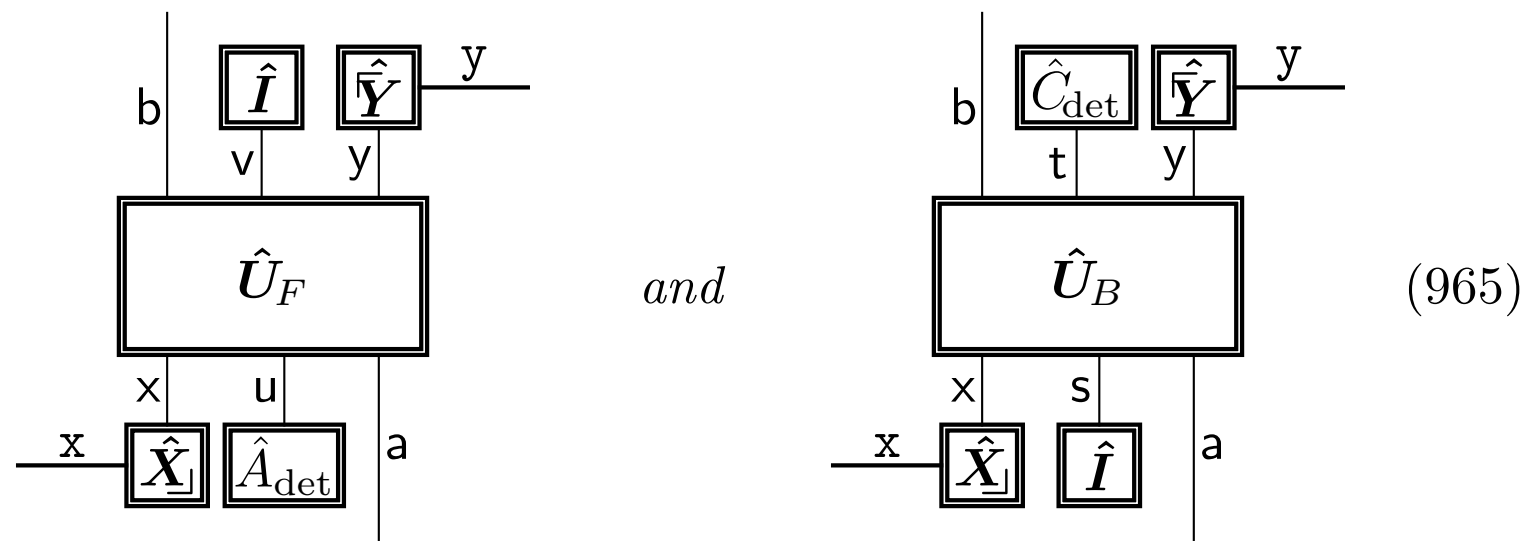

(965)

where $\hat{A}_{\text{det}}$ and $\hat{C}_{\text{det}}$ are homogeneous and deterministic, and where $\hat{\boldsymbol{U}}_F$ and $\hat{\boldsymbol{U}}_B$ are natural unitary operators.

Note that $\hat{A}_{\text{det}}$ and $\hat{C}_{\text{det}}$ are, necessarily, not physical. Consequently, it is not possible to build $\hat{\boldsymbol{B}}$ by means of either composition suggested in (965). In this sense, neither of the dilations in the sufficient pair support the cause of peaceful coexistence discussed in Sec. 44.1. This theorem follows immediately from the previous one by using (899) on the left and (901) on the right. It is clear that both expressions are $T$-positive. Further, the left expression satisfies forward causality and the right expression satisfies backward causality. Hence sufficiency follows. This proves the theorem.

Dilation in terms of sufficient pairs are a useful device. However, we may want to ask "what is the actual dilation?" So long as we are in the time symmetric frame of reference, all the dilations we have considered have elements that are unphysical in the general case. On the other hand, in the time forward theory $\hat{A}_{\rm det}$ is a physically allowed preparation. Thus, if we believe that the time forward frame is, in some deeper ontological sense, the correct perspective, we might claim that the left dilation in (965) is "the correct one". A similar comment pertains for the backward time frame and the right dilation in (965). Ultimately, we would like ontological statements to be independent of the frame we are in and so this issue remains unresolved here. The next section takes up (but does not resolve) this issue.

# 45 Constructability?

## 45.1 Constructible and unconstructible operators

The dilation in (953) uses a natural maxometric operator, $\hat{M}$. In the case that $\hat{M}$ is actually physical then (as shown in Sec. 38) it is actually a natural unitary and we will write it as $\hat{\boldsymbol{U}}$. Thus, there exist a class of deterministic physical operators having the dilation

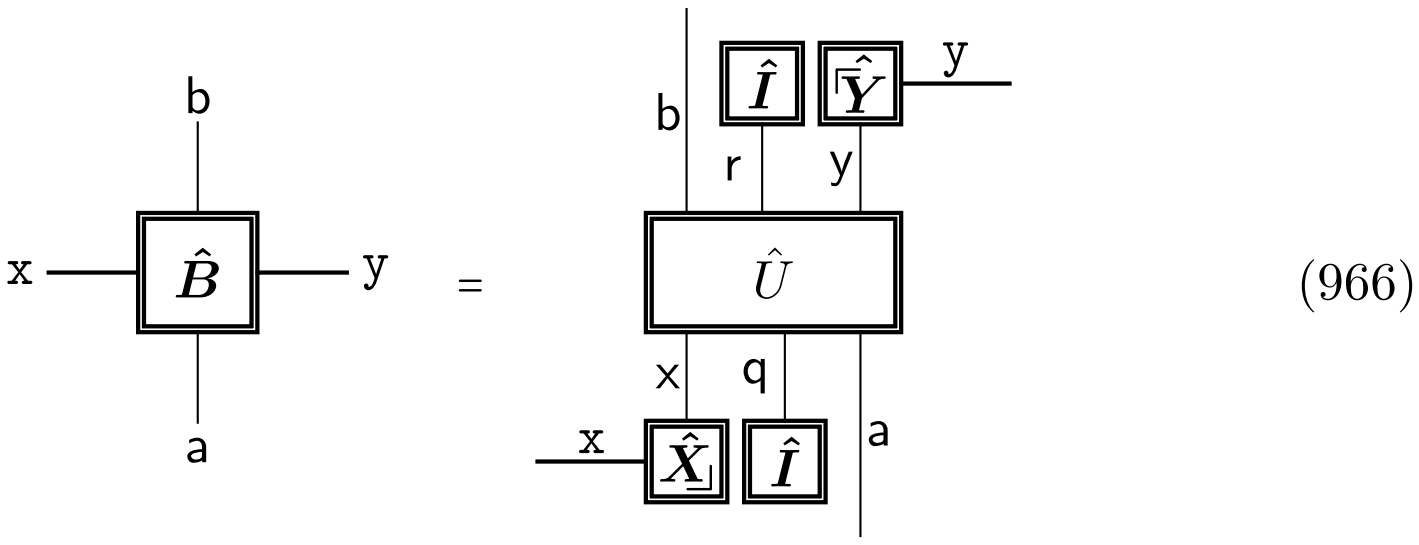

(966)

where $\hat{U}$ is a natural unitary (we already discussed dilations like this in Sec. 44). We call operators having the form in (966) *constructible operators*. Such operators are necessarily deterministic and physical (by the dilation theorem). We will call any deterministic physical operator that cannot be expanded in this way *unconstructible*. For such cases, $\hat{M}$, is unphysical.

Here we will explore whether there exist genuinely unconstructible operators by studying three examples. For simplicity consider deterministic physical operator, $\hat{\boldsymbol{B}}$ without incomes or outcomes. According to the dilation theorem, we can write such a deterministic physical operator in dilated form as

(967)

where we have written the deterministic maxometric operator, $\hat{M}$, in twofold form. We have chosen $\mathsf{r} = \mathsf{a}$ and $\mathsf{q} = \mathsf{b}$ in the dilation (this is consistent with the dilation theorem as stated above). We require $M$ to be a natural maxometry and we are seeking examples where it is not a natural unitary. We note, incidentally, that $M$ is a special case of the example discussed in (924) and, therefore, imposing the conditions that it is a natural maxometry are equivalent to imposing those such that it is a normal maxometry.

We will consider setting $M$, in turn, to each of the following

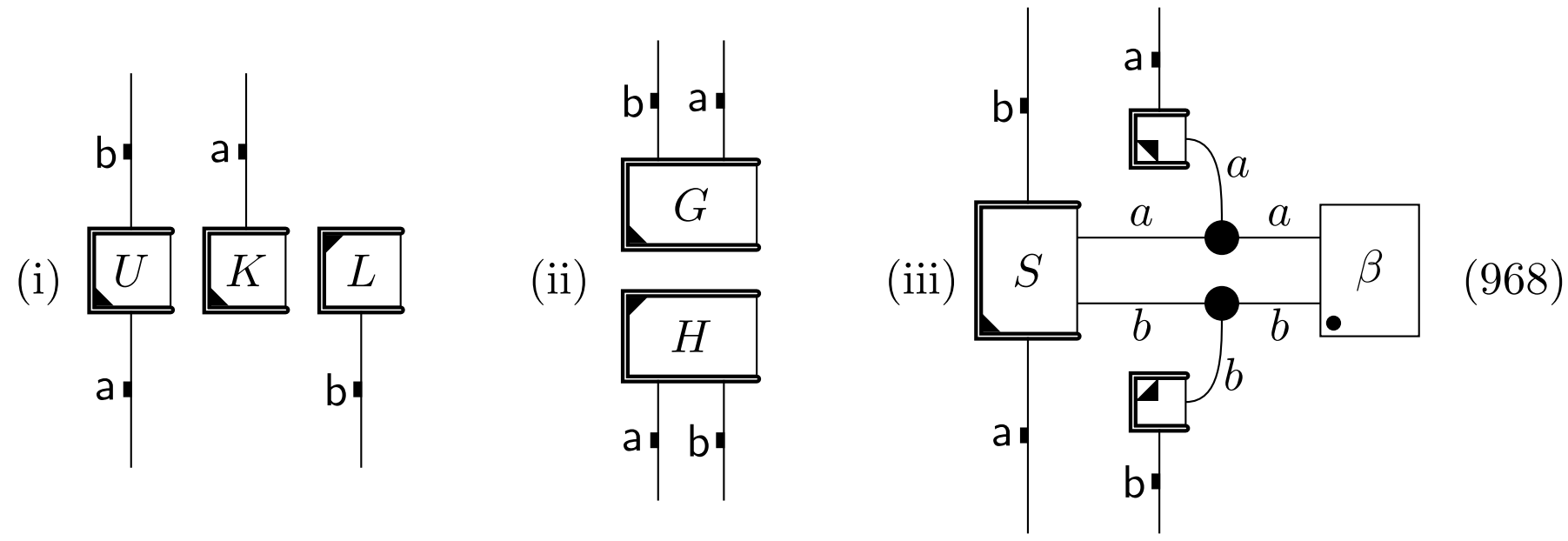

Note that the second example is the same as that considered earlier in (911). We wish to impose that each example is a deterministic maxometry.

**Example (i).** For this example we require that $U$ is a natural unitary (note this means that $N_{\mathsf{a}} = N_{\mathsf{b}}$). This clearly satisfies the conditions for a natural maxometry if we impose $|\underset{\sim}{K}| = |\underset{\sim}{L}| = 1$.

**Example (ii).** This example is a natural maxometry if

(969)

and

$$(970)$$

where we have used (907). Either of these conditions imply that $|\underset{\sim}{G}||\underset{\sim}{H}| = 1$ (this is evident if we close both sides of either (969) or (970) with a ~ circle).

**Example (iii).** For this example we impose that

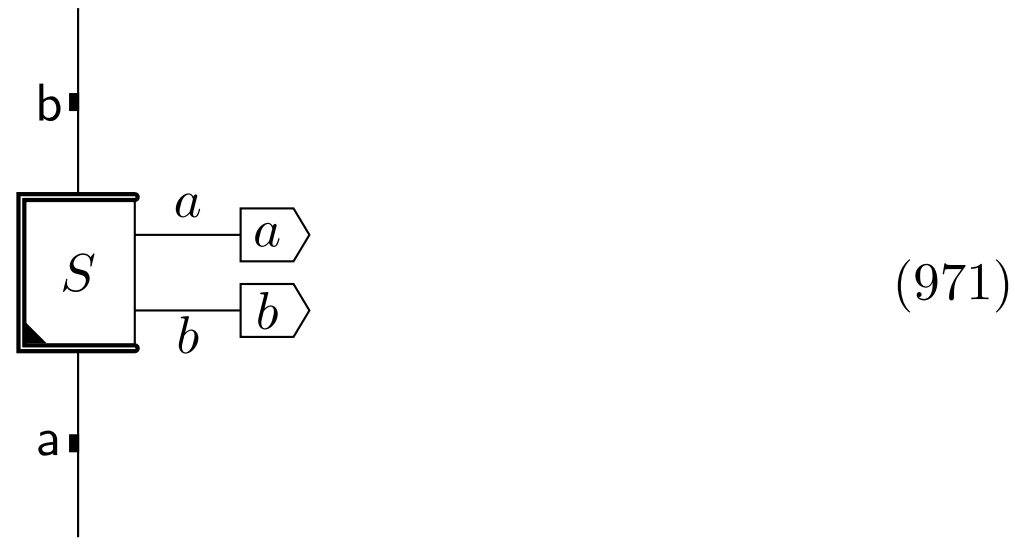

$$(971)$$

is a natural unitary for each choice of $b$ and $a$ (this means that $N_{\mathsf{a}} = N_{\mathsf{b}}$). Then we can verify that the conditions for example (iii) to be a natural maxometry is

$$= 1 \qquad (972)$$

This can be obtained from (907).

With these impositions each of the three examples is a natural maxometry. It is also the case that none of these examples are natural unitaries (except in trivial cases).

Now consider inserting each of the three examples in (968) into (967) in turn. If we insert example (i) in we simply obtain

$$(973)$$

(where we have used the deterministic normalisation conditions $|\tilde{K}| = |\tilde{L}| = 1$ mentioned above). The ancillary terms do nothing and so we are just left the natural unitary. So, in fact, we do have a constructible operator. Thus, not every unphysical natural maxometry leads to an unconstructible operator when inserted into the dilation.

Now consider example(ii). If we insert this into (967) and use (969, 970) we obtain

(974)

(in the last step we have used the fact physical caps and cups provide ignore operators - see (635)). This can be written as

(975)

The object inside the dashed box is a natural unitary operator so this is also in constructible form. We still do not have an example of an unconstructible operator. It is interesting to note that the ignore operators in (975) are not the same ignore boxes from our initial dilation (in (967)). This example shows that a given $\hat{\boldsymbol{B}}$ can have quite different dilations.

Now consider example (iii) in (968). If we insert this example into (967) we obtain

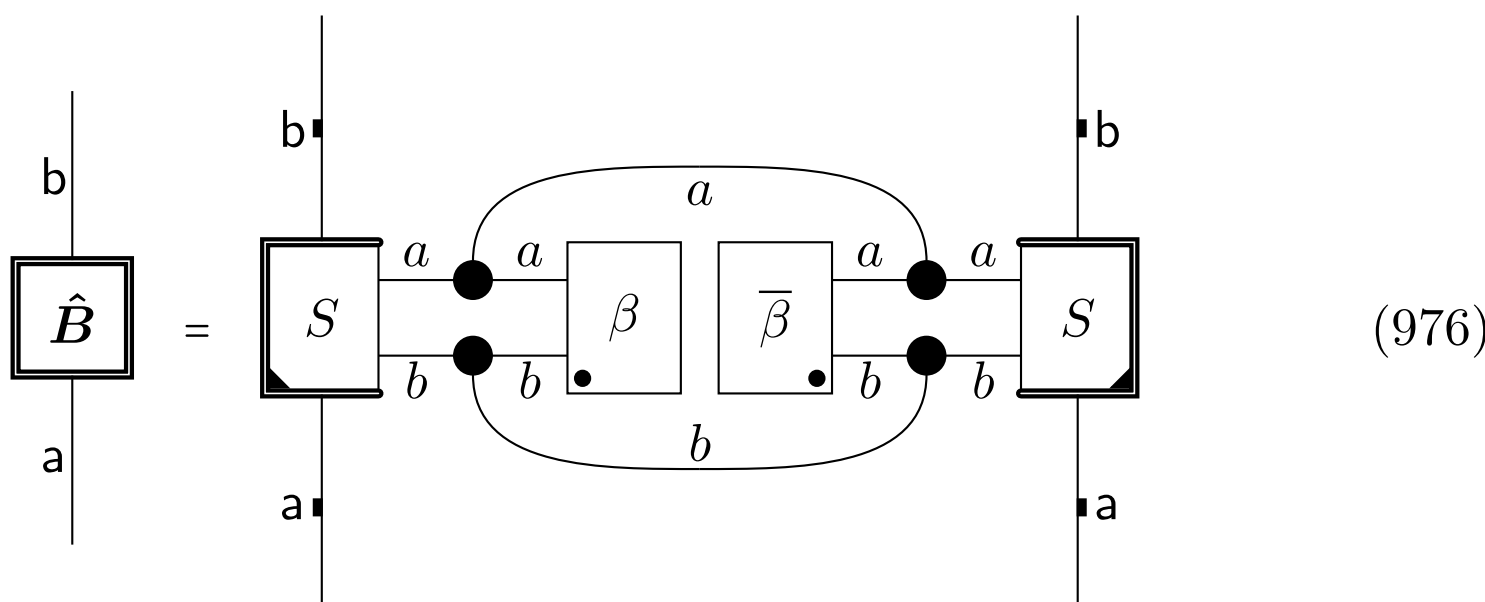

(976)

The $b$ and $a$ label wires joining the two halves ensure that we have the same natural unitary on both sides. Thus, this expression is simply a weighted sum of natural unitaries. Note (1) that, for each $ba$, these weights are positive (equal to $\beta_{ab}\overline{\beta}_{ab}$), and (2) that the sum (over $ba$) of the weights is equal to 1 by virtue of condition in (972) for this to be a deterministic maxometry. This means we can interpret $\hat{B}$ in (976) as convex mixture of natural unitary operations as follows

(977)

where the diamond box is a diagonal matrix and with entries, $\lambda_{ba}$, on the diagonal satisfying $\sum_{ba} \lambda_{ba} = 1$ and $0 \leq \lambda_{ba}$. Is this an example of an unconstructible operator? Clearly, for the special case where one of the diagonal entries of $\lambda$ is equal to 1 (and the remaining entries are equal to 0) is constructible since this is just a physical unitary. Another special case turns out to be constructible. This is when we choose all the diagonal entries to be equal as we will see. Consider using

(978)

(where $S$ is a natural unitary for each $h$). In fact this maxometry is a natural unitary (as can be verified by inserting it into (868)). Inserting (978) into a suitably adapted version of (967) (where the ancillae that we act on with ignore operators are now all systems of type h) and simplifying gives

(979)

This is exactly the diagonal form with equal weights we were seeking (where the weights are equal to $N_{\mathsf{h}}^{-1}$). Furthermore, since the expression in (978) is a natural unitary, this is a constructible operator. Furthermore, the left Hilbert object in (978) is a special case of example (iii) in (968) (if we choose $\beta$ appropriately).

Hence we have proven that two special cases of (977) are constructible operators even though we started with a non-physical natural maxometry (example (iii) in (968)). What about the general case? In fact, as we will show, it is easy to see that if all the diagonal entries of $\lambda$ are rational then the operator is constructible. Furthermore, even when there are irrational entries, we can approximate the operator to arbitrary accuracy by a constructible operator if we allow $N_{\mathsf{h}}$ to be arbitrarily big. It remains unproven whether or not we can exactly model such operators in a construcible way with finite $N_{\mathsf{h}}$. First consider the rational case. In this case we can write all the diagonal entries of $\lambda$ with as fractions with a given denominator (the lowest common denominator) which we set to be equal to $N_{\mathsf{h}}$. We have $h = 1$ to $N_{\mathsf{h}}$. We partition these $N_{\mathsf{h}}$ values of $h$ into sets having ranks equal to the numerators of the of the diagonal entries of $\lambda$. For each $h$ in one of these sets we set $S$ equal the natural unitary associated with that diagonal entry of $\lambda$. In this way we exactly obtain $\hat{\boldsymbol{B}}$. This proves the rational case. For the irrational case (which we will refer to as "example (iii)-irrational" below) we can approximate the diagonal entries with some rational numbers with a given denominator. We can clearly increase this denominator to improve the approximation.

## 45.2 Constructibility conjectures

It remains an open problem to resolve whether unconstrucitble operators actually exist. For the purposes of the foregoing discussion it is useful to state a few conjectures that cover pretty much all the logically possible resolutions.

> **Unconstructibility conjecture.** Unconstructible operators exist.

This conjecture could be proved by exhibiting an example that cannot be put in constructible form. The opposite conjecture is

> **Constructibility conjecture.** Unconstructible operators do not exist.

We see some evidence for this above in that it was shown that operators that looked, on the face of it, to be unconstructible turned out to be constructible. Even for example (iii)-irrational (at end of Sec. 45.1 above) we were able to approximate the operator by a constructible operator arbitrarily well by having ancillae of arbitrarily large dimension. The above conjectures are a little ambiguous - do we allow arbitrarily large ancillae or regard them as unphysical?

We can state conjectures that take into account the dimensions of ancillae. Thus we offer the following conjectures.

> **Finite constructibility conjecture.** All deterministic physical operators, whose systems are of finite dimension, can be exactly modelled as constructible operators with finite ancillae.

Example (iii)-irrational stands as evidence against this. However, it has not been proven that there is no construcible dilation with finite ancillae for this example. Alternatively, we might make the following conjecture.

> **Finite unconstructibility conjecture.** If we allow only finite dimensional ancilla, then there exist deterministic physical operators whose systems are of finite dimension that cannot be exactly modelled as constructible operators.

Example (iii)-irrational is evidence for this conjecture (but not proof of this conjecture). We can also state the following

> **Approximate constructibility conjecture.** Every deterministic physical operator can be modelled as a constructible operator to arbitrary accuracy provided we allow ancillae of arbitrarily large dimension.

Example (iii)-irrational is evidence for (though, of course, not proof of) this conjecture. Contrary to this is the conjecture

> **Strong unconstructibility conjecture.** There exist deterministic physical operators, whose systems are of finite dimension, which cannot be modelled by a constructible operator to arbitrary accuracy even with arbitrarily large dimension ancillae.

Proof of this conjecture would be the strongest statement against constructibility. We do not have a good example to provide evidence for this conjecture.

We might want to prove theorems where the dimensions of the ancillae are constrained to be some function (maybe polynomial) of the dimensions of the systems of the deterministic physical operator under consideration. However, until we have some tenable proof techniques, there is no good reason to conjecture any particular functional dependency.

## 45.3 Peaceful coexistence or not?

According to the "Church of the larger Hilbert space" the whole universe is actually in some pure state and evolves unitarily according to the Schroedinger equation using the fundamental Hamiltonian for the universe (the expression is attributed to John Smolin - see Anonymous [2015]). This is the point of view of a large segment of the theoretical physics community who regard mixed states and mixing evolution as non-fundamental. There is a smaller segment of the theoretical physics community (often more routed in the Quantum Information perspective) who regard mixed states and mixing evolution as the more basic objects (with pure states and pure evolution being special cases). These are the adherents of the "Church of the smaller Hilbert space" (this counter expression is due to Matthew Leifer - see his blog post at Leifer [2006]).

Within time forward operational Quantum Theory, the Stinespring dilation (see (947, 948)) shows how any operation can be associated with an operator that is built out of pure elements (normalised pure states and a unitary operation). For this reason, it offers a route to the peaceful coexistence between the two extreme points of view since it allows us to model any mixing operator that may appear in the smaller church according to the point of view of the larger church. At the very least, it offers a way for small church adherents to reassure large church adherents that there is no "funny business" going on in operational quantum theory.

However, we might argue that the time symmetric point of view is more fundamental than the time forward. Or we might argue that any resolution of this conflict between churches should work equally well for any temporal frame of reference. Then we would want to consider the time symmetric dilation in (953). In this case peaceful coexistence between the churches depends on the whether or not the conjectures hold in Sec. 45.2. In the case that there does not always exist a unitary dilation for physical operations, we have to think hard about these churches and which offers the better path forward.

If unconstructibility holds as a fact, then the dilation theorem in Sec. 44.3 breaks the peaceful coexistence. Indeed, one might say it undermines the church of the larger Hilbert space since a general operation cannot be thought of as being composed of more basic pure elements. This is not surprising from a church of the smaller Hilbert space point of view. Indeed, this point of view is routed in deeper insights that go beyond Hilbert space into the more basic structure of operational probabilistic theories as discussed in Part I of this book. In the operational probabilistic point of view a state, for example, is a list of probabilities and pure states are merely special cases in the convex space.

We should mention the measurement problem which haunts both these churches. In the larger church we have to explain how we see definite outcomes when the wavefunction simply evolves unitarily. In the smaller church we have to account for both the existence of apparatuses and, ultimately, say what the underlying reality is. The discussion on coalescence in Sec. 13.9 touches on some of these issues (though not to the extent that it offers a resolution of the measurement problem). A full discussion of the measurement problem is worthy of a book in its own right so we will not comment further. See Tomaz et al. [2025] for a recent review of the measurement problem.

## 45.4 Other forms for the dilation

In Appendix C it is shown that any natural maxometry can be modelled by a projected natural unitary while the converse is not true. This means we can write the dilated form in (953) as written as

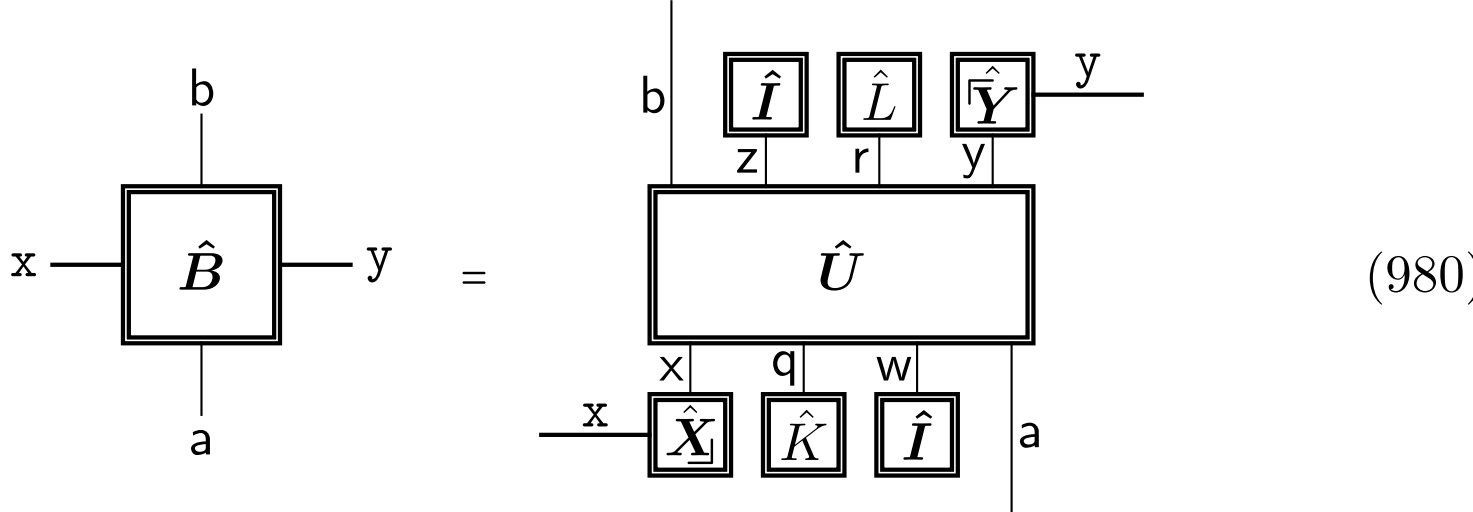

(980)

for appropriate homogeneous $\hat{K}$ and $\hat{L}$. In general, $\hat{K}$ and $\hat{L}$ will be nonphysical. Two comments are important here. First, not all dilations of the form in (980) need correspond deterministic physical operators (even when appropriately normalised). This is because not all projected unitaries are proportional to maxometries. Second, in the maxometric dilation in (953) the natural maxometric operator may be non-physical whilst, in the dilation in (980) the operators $\hat{K}$ and $\hat{L}$ may be non-physical (so we have not got rid of non-physicality in the dilation - we have rather put it somewhere else).

We can write the dilation (980) as

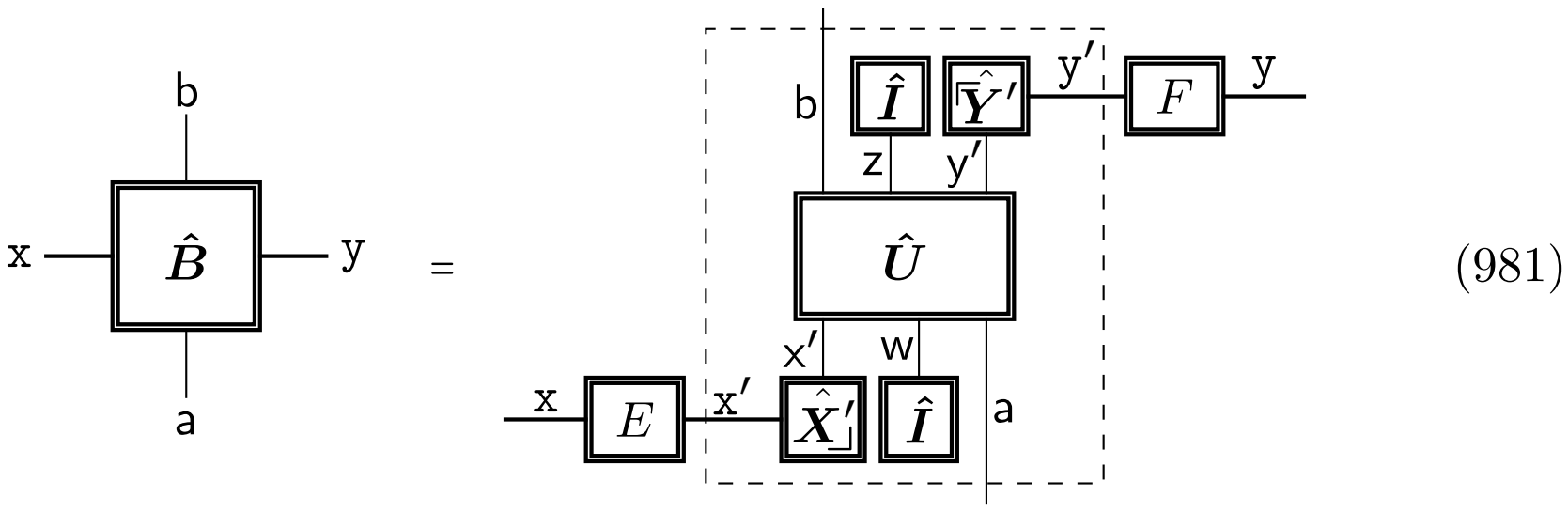

(981)

for appropriate definitions of $E$ and $F$. Now it is the $E$ and $F$ that are non-physical. Once again, note that not all dilations in this form need correspond to deterministic physical operators. This form is interesting because the operator inside the dashed box is a constructible operator. Thus, we can regard a general operator as a kind of classical processing of a constructible one (though the classical processing by $E$ and $F$ is nonphysical). Possible choices for $E$ and $F$ which give back (980) are as follows. $E$ is given by the expression inside the dotted box of

(982)

and $F$ is given by the expression inside the dotted box of

(983)

If $\alpha$ and $\beta$ are greater than 1 then $E$ and $F$ are nonphysical.

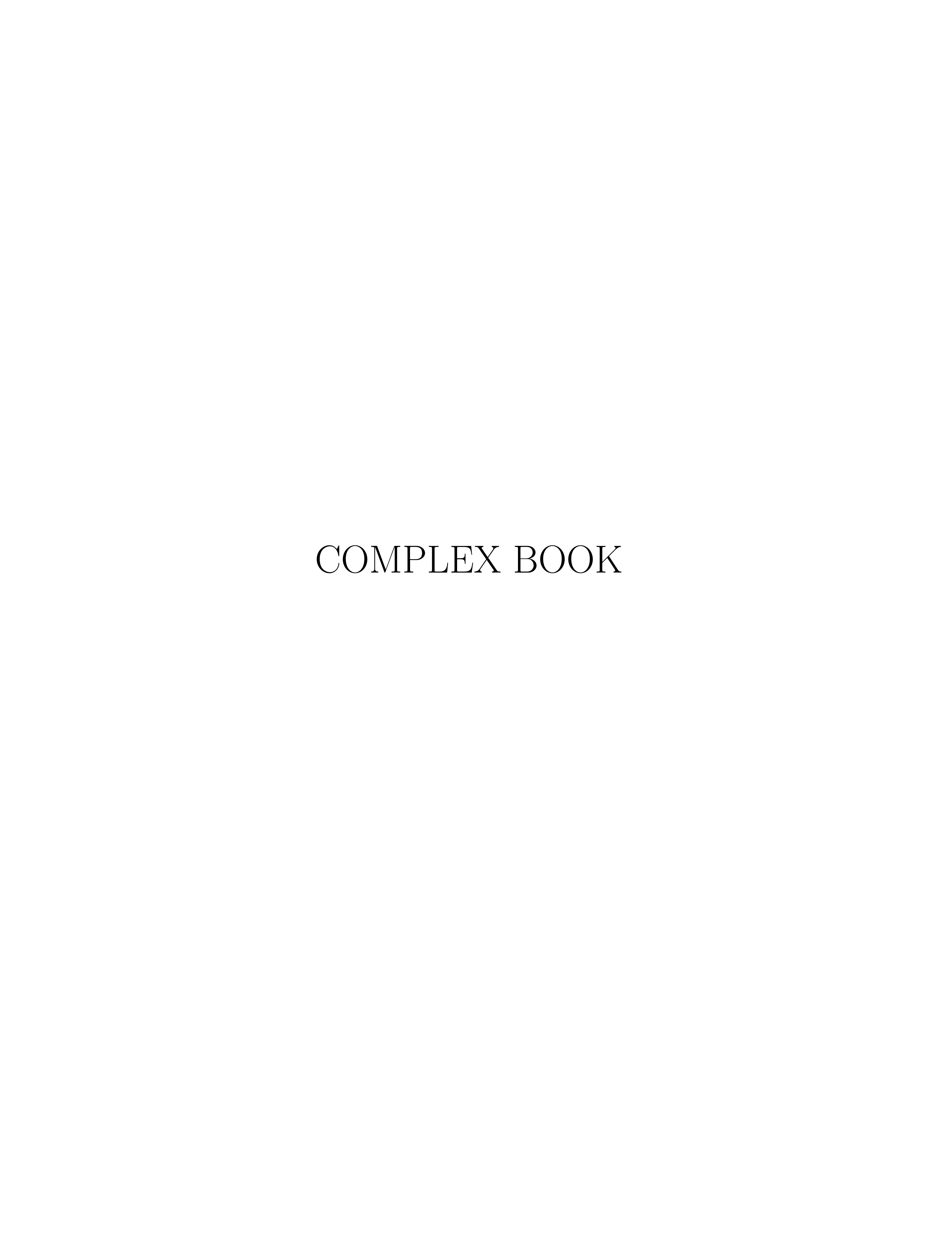

# Part IV
# Causally Complex Operational Probabilistic Theories

## 46 Introduction to the causally complex case

### 46.1 Flowcharts

In this part we will discuss causally complex operational probabilistic theories (denoted $t$COPT). Then in Part V we will discuss causally complex operational quantum theory. We can also consider causally complex operational probabilistic classical theories. Together these three cases constitute the middle rung of (34)

$$t\text{COCT} \longleftarrow t\text{COPT} \longrightarrow t\text{COQT} \tag{984}$$

The word "complex" does not refer to complex numbers but to the complicated causal structure we study in this case. We will mostly focus on the time symmetric temporal perspective ($t$ = TS). We discuss the time forward case in Sec. 53 and the time backward case in Sec. 54. We find that it is possible to model the TS case using the TF theory or the TB theory. However, it remains as an open problem as to whether the TS theory can model the TF or the TB theory in the complex case. The time symmetric theory may be more constrained. This contrasts with the simple case where we were able to prove equivalence.

We will treat case of TSCOPT (time symmetric complex operational probability theory) by going through the elements in the flowchart (36). For $tx$ = TSC,

the elements of this are

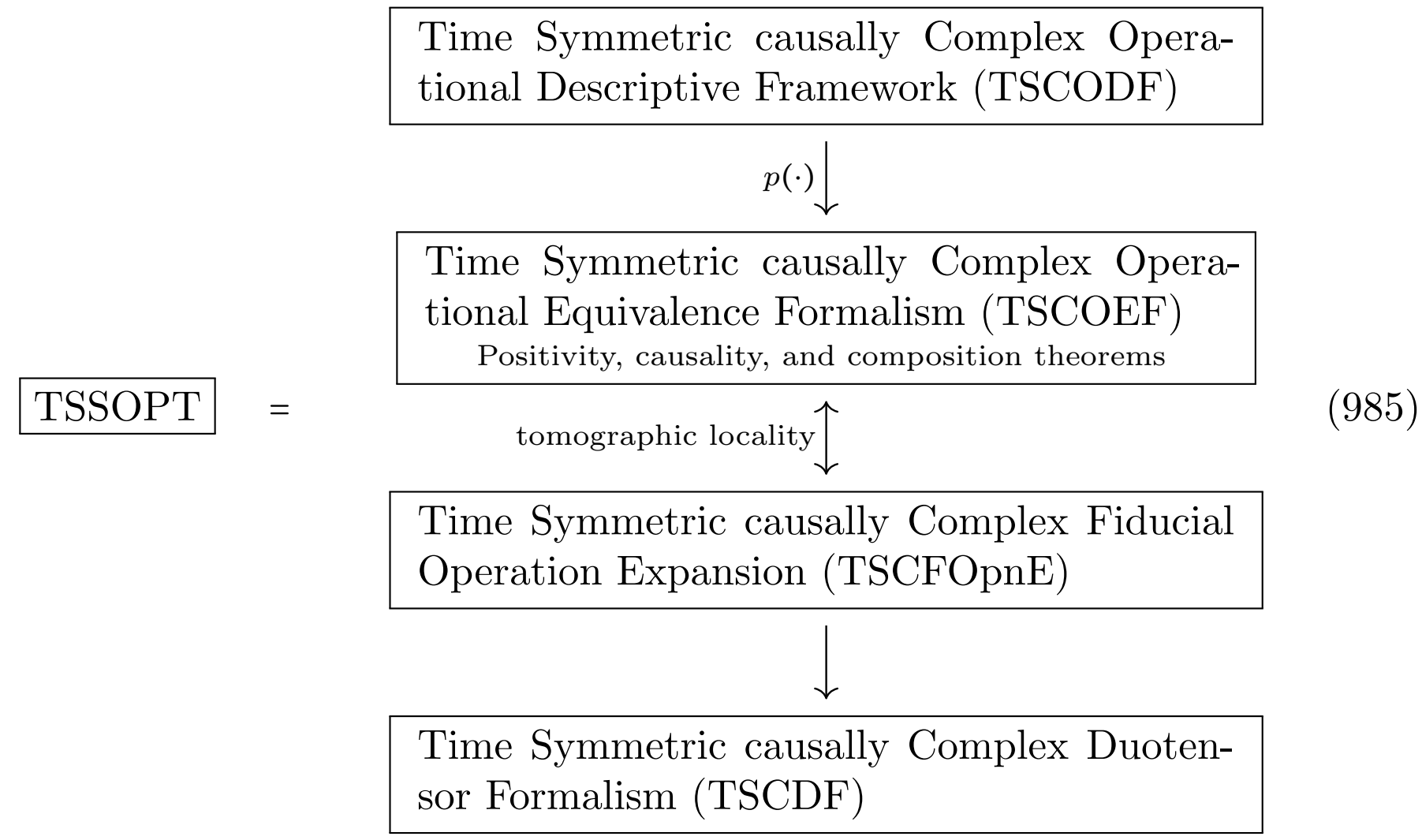

(985)

In what follows we will discuss these elements. Much of the structure builds on, and is analogous to, the simple theory. However, there are some significant differences. The causality principles are more interesting. The composition theorems are more intricate to prove. There are a bigger variety of dilation theorems which we call causal dilation theorems since they pertain to various causal scenarios.

## 46.2 Basic idea

Simple operations have the property that all the inputs/incomes are regarded as being to the past of all the outputs/outcomes. Complex operations do not necessarily have this property. The key idea is that we take simple networks (i.e. those comprised of simple operations) and use them to motivate causally complex operations. A simple network has a certain causal structure. We abstract away this causal structure obtaining a *causal diagram*. The causal diagram forms part of the specification of the complex operation. Further, causal diagrams allow us to write down physicality conditions for complex operations. A simple network (comprised of simple operations that are physical) is an example of a physical complex operation. However, we will not be able to prove that all physical complex operations can be written as physical simple networks. Furthermore, there is evidence when we come to Time Symmetric Complex Operational Quantum Theory that there are physical complex operations that cannot be written as physical simple networks (in particular, see the section on causal dilation in Sec. 78).

A complex operation is specified as follows

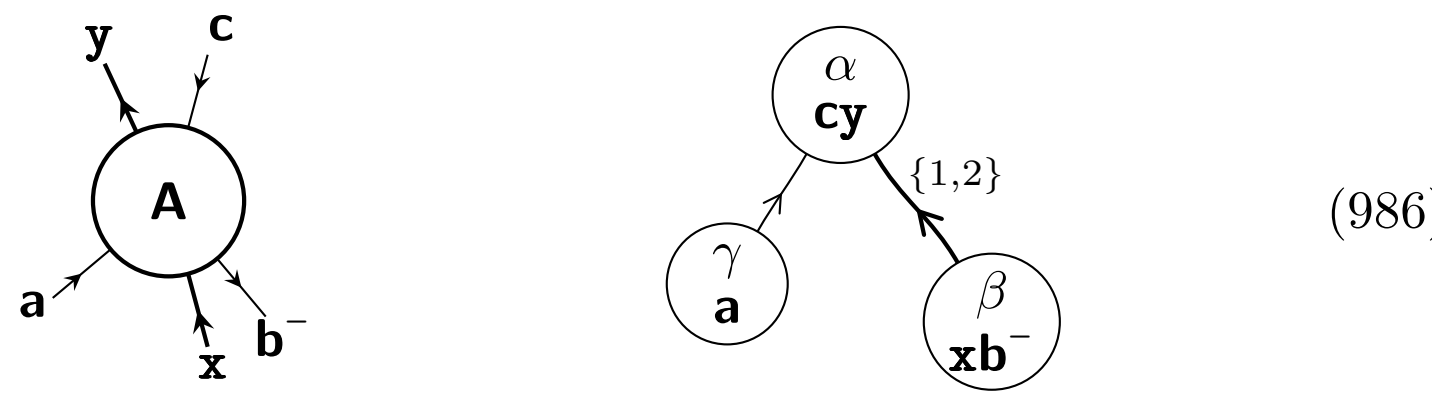

(986)

where the diagram on the left is the *causally complex operation itself* and the diagram on the right is the *causal diagram* whose meaning (which is probably a little obscure to the reader at this point) will be explained in due course. We will illustrate how the above example can be obtained from a simple network in Sec. 47.1. The causal diagram is a essential component of the full description of a causally complex operation.

We use a compact notation wherein $\mathbf{a} = \mathbf{a}^+\mathbf{a}^-$ where $\mathbf{a}^+$ is travelling in the direction of the arrow on the wire (in the complex operation) and $\mathbf{a}^-$ is travelling against the direction of the arrow. Hence,

> The arrows on the causally complex operation itself (on the left of (986)) establish a convention and do not, in themselves, indicate that a system is travelling forward in time along the direction of the arrow.

Motivation for using this compact notation comes from modelling spacetime in the context of field theories as discussed in Sec. 55 where we associate operations with regions of space time (this was also discussed in Sec. 2.4). Then we see how, at a part of the boundary whose normal is spacelike, systems can be travelling in both directions across this boundary which motivates this compact notation. This is illustrated in the example in (1320) for the simple network in (1319).

The arrows in the causal diagram on the right of (986) are *causal links* and they *do* indicate the forward time direction. Open wires in the operation appear as nodes in the causal diagram. As represented here this causal diagram has *explicit causal structure* given by the arrows and *implicit causal structure* represented by $\alpha$, $\beta$, and $\gamma$ (the idea of implicit causal structure will be explained in Sec. 47.12. When the implicit causal structure is taken into account, the causal diagram must be a directed acyclic graph (DAG) so there are no closed timelike loops. The key idea is that the (complex operation, causal diagram) pair represents the open wires and the causal links between them without representing the internal structure that is evident a simple network diagram. Complex operations are causally more complicated than simple operations. A detailed discussion of approaches in the literature that consider complex causal structure was provided in Sec. 2.4. The approach taken in this book is, particularly, born from the approaches of Oeckl [2003], Hardy [2005], Gutoski and Watrous [2007], Chiribella et al. [2009a], and Hardy [2016].

We will say that causally complex operators are *physical* if they satisfy two properties

**Tester Positivity.** This condition (provided in Sec. 49.2.2) will guarantee that circuits have non-negative probabilities.

**$t$-causality.** For $t$ =TS we have (general) double causality conditions (see Sec. 49.4.5). For $t$ =TF we have forward causality conditions plus a backward causality condition that applies only in special cases (see Sec. 53.4). For $t$ =TB we have backward causality conditions plus a forward causality condition that applies only in special cases (see Sec. 54.4). These conditions guarantee (in the appropriate $t$COPT) that we have causality and that probabilities for circuits are not greater than unity.

Only physical operators are allowed in this framework. The positivity requirement is clearly essential. The causality condition could be relaxed in principle - for example if one was interested in indefinite causal structure (see Hardy [2005], Chiribella et al. [2009b], Oreshkov et al. [2012]). However, this condition also guarantees that probabilities are less than or equal to 1. Any modification of causality condition would need to continue to respect this subunity property.

We use simple networks to motivate complex operations. Once we motivate the idea of a complex operation, we can impose physicality conditions (that guarantee positivity and causality as just discussed). The question of whether it is possible to represent any such physical complex operation by a simple network remains unsettled though the evidence points in the direction of this not being possible. Thus, a complex operation is possibly a more general idea than that of a simple network. This leads to the question of do we either (i) allow any physical operation even if it cannot be modelled by a simple network or (ii) impose extra constraints so that complex operations must be modellable by simple networks. It is more natural to take the former attitude and this opens up the exciting possibility that there is new physics in the complex case going beyond what we can implement by simple networks.

The question of whether we can model any physical complex operation by a simple network remains an open question even within complex Quantum Theory. This leads us to the interesting topic of causal dilation theorems for the quantum case as discussed in Sec. 78.

We can join multiple causally complex operations together to form a complex network. For example

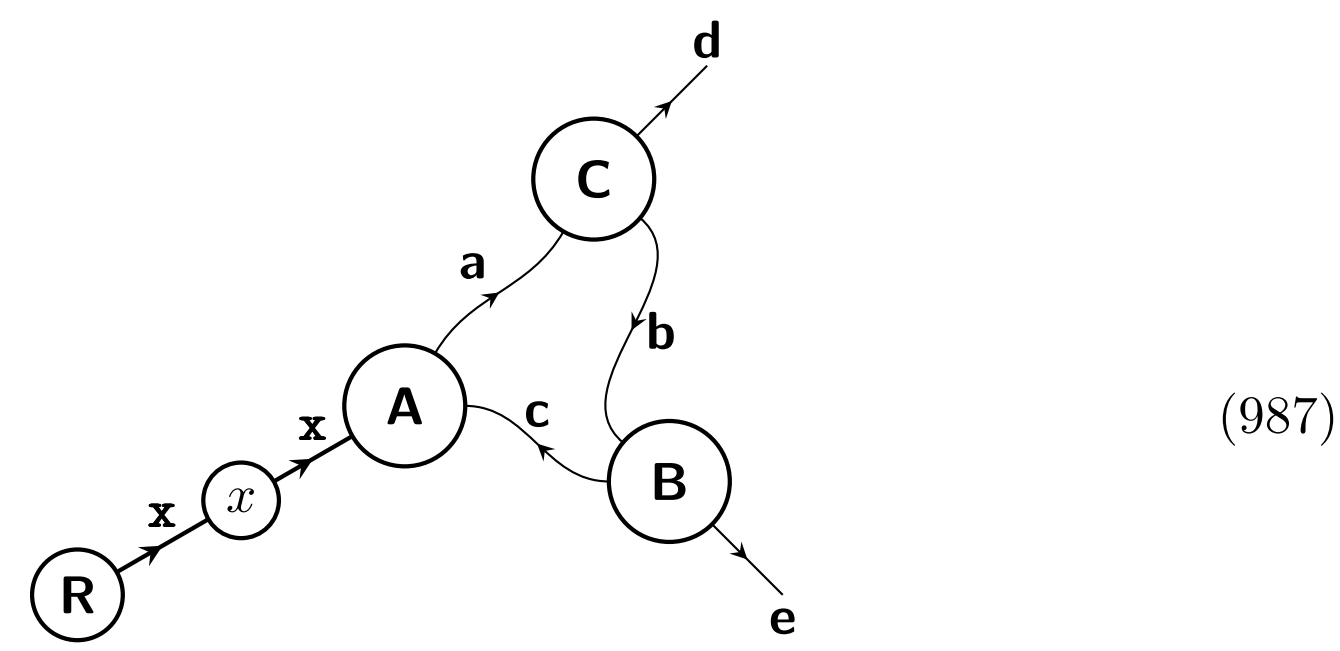

(987)

This diagram does not have to be a DAG with respect to the arrows since, as mentioned above, these arrows are there to establish a convention (they do not, in themselves, correspond to the forward in time direction). This complex network will, itself, have a causal diagram which can be calculated from the causal diagrams of the complex operations it is composed from. The causal diagram associated with a complex network does have to be a DAG (when all the implicit causal structure is taken into account). In the example in (987) the circle with an $x$ in it is the appropriate generalisation of readout boxes to the causally complex case. These will be discussed in Sec. 47.18. The circle with an **R** in it is the appropriate generalisation of the **R** boxes to the causally complex case. These will be discussed in Sec. 48.5.

Any complex network can, itself, be regarded as a complex operation. For example, the complex network in (987) can be represented as the following operation

$$\text{D} \quad \mathbf{d},\ \mathbf{e} \tag{988}$$

since it this has the same open wires. Note that when we join simple operations we do not necessarily end up with a network that can be regarded as simple operation because the resulting causal structure may be complex. However, this is not an issue in the complex case since complex operations are, of course, allowed to have complex causal structure. As in the causally simple case, we use boldface to represent deterministic operations (where there are no implicit readout boxes) and unbolded font to represent nondeterministic operations (as in the example (988) which has an implicit readout box shown explicitly in (987)).

# 47 Descriptive framework

## 47.1 Causally complex operations

We motivate the idea of causally complex operations by looking at simple networks. A causally simple operation has inputs/incomes that are before the outputs/outcomes. A simple network (a network constructed out of simple operations) may, however, have some outputs/outcomes that are *before* some inputs/incomes. For example

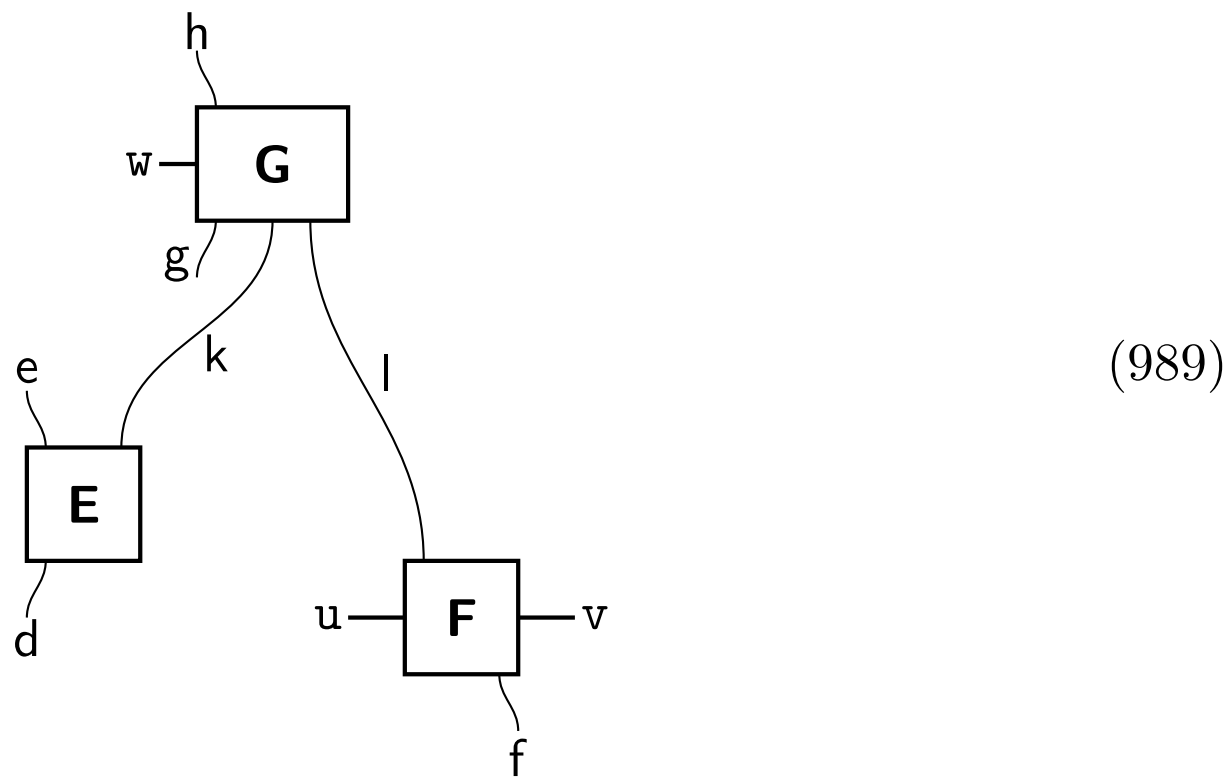

(989)

Here we see that output e is before input g so this is not a simple causal structure. We may want to treat a simple network as a single object in itself. In this case, as we will explain, we can represent the above simple network by a *causally complex operation* as follows

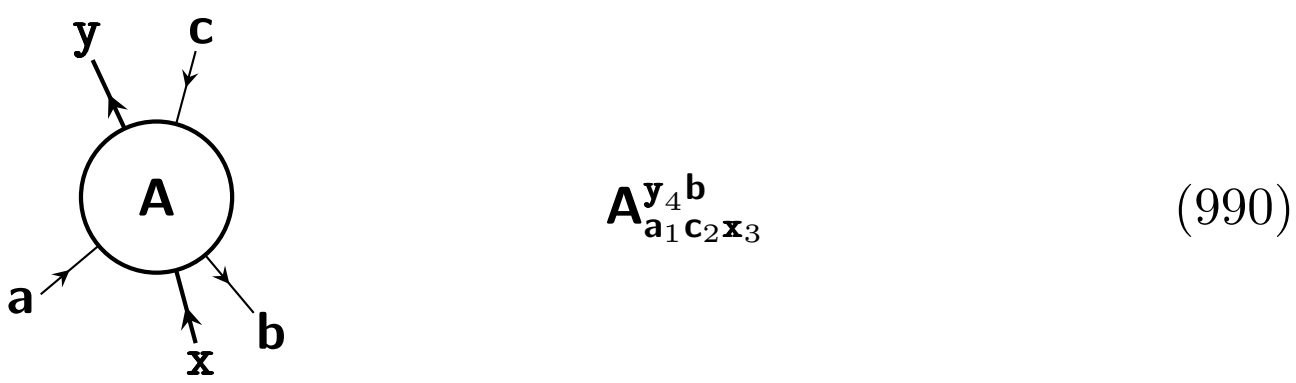

$$\mathbf{A}_{\mathbf{a}_1 \mathbf{c}_2 \mathbf{x}_3}^{\mathbf{y}_4 \mathbf{b}} \tag{990}$$

The diagrammatic notation is on the left and the symbolic notation is on the right. We represent complex operations by circles rather than using rectangles as we do in the causally simple case. Further, we use bold for the types ($\mathbf{a}$, $\mathbf{x}$, ...). We put $\mathbf{a} = \mathbf{a}^+\mathbf{a}^-$ where $\mathbf{a}^+$ indicates a system moving in the direction of the arrow on the $\mathbf{a}$ wire, and $\mathbf{a}^-$ indicates a system moving against the direction of this arrow. To connect with the simple network (989), we will put $\mathbf{a}^+ = \mathsf{d}$ and $\mathbf{a}^- = \mathsf{e}$. Thus, $\mathbf{a}$ is a composite wire consisting of a system travelling in the direction of the wire and a system travelling against the direction of the wire. The arrow on the $\mathbf{a}$ wire establishes a convention with respect to which $\mathbf{a}^\pm$ derive their meaning. A full set of identifications that enable us to represent the simple network in (989) by the causally complex operation (990) is

$$\begin{array}{llll} \mathbf{c}^+ = \mathsf{g} & \mathbf{c}^- = \mathsf{h} & \mathbf{y}^+ = 0 & \mathbf{y}^- = \mathsf{w} \\ \mathbf{a}^+ = \mathsf{d} & \mathbf{a}^- = \mathsf{e} & & \\ \mathbf{b}^+ = 0 & \mathbf{b}^- = \mathsf{f} & \mathbf{x}^+ = \mathsf{u} & \mathbf{x}^- = \mathsf{v} \end{array} \tag{991}$$

Note that $\mathbf{b}^+ = 0$ because there are no systems going in the direction of the arrow on the $\mathbf{b}$ wire. In such a case we can simply label the wire with $\mathbf{b}^-$ as

follows

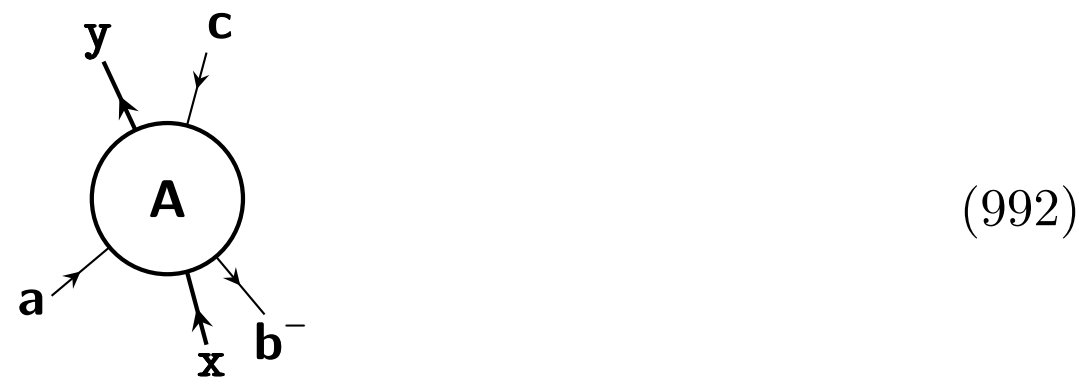

(992)

The minus superscript indicates this system is going against the arrow direction.

Associated with every causally complex operation is a causal diagram. In the case that we started with a simple network we can obtain the causal diagram by looking at this simple network. In the example above we have the following *causal diagram*

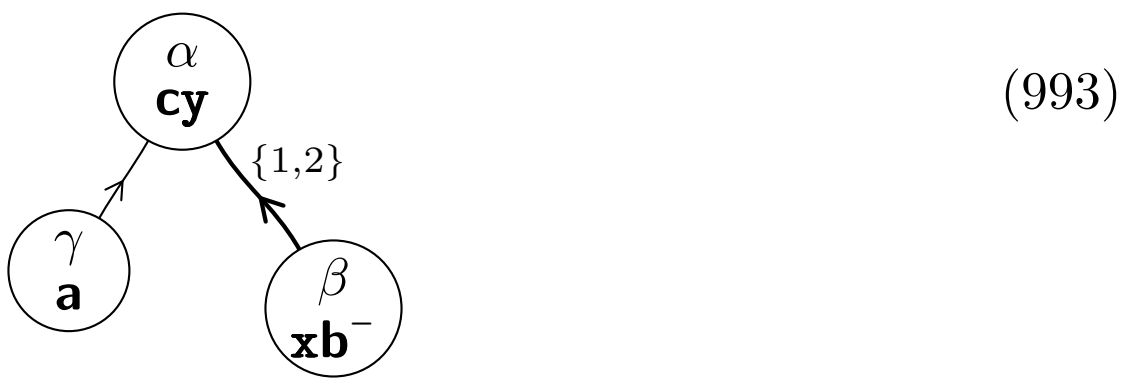

(993)

(We will show, in detail, where this causal diagram comes from in the foregoing discussion - in particular, see Sec. 47.3, Sec. 47.9, and Sec. 47.12.) The nodes (circles) contain two kinds of information. First they contain $\alpha$, $\beta$, and $\gamma$ which denotes the "implicit" causal structure in this node. We will elaborate on this in Sec. 47.12. Second, they contain the labels of the open wires from the simple network (e.g. **cy**) associated with the operation in (992). The arrows indicate causal links. The *explicit* causal structure in this causal diagram is represented by the arrows between nodes. We can obtain the causal diagram from the simple network by drawing an arrow from one node to another whenever there is a forward route through the network from any wire associated with the first node to any wire associated with the second node. For example, there is a route through the network from u (which is associated with node **xb**$^-$) to h (which is associated with node **yc**). These arrows carry a set of labels (a subset of the integers as in $\{1, 2\}$ for example). The rank of this set is the number of causal connections between the two nodes required to connect the implicit causal information between the nodes (this will become clear in Sec. 47.12). If only one causal connection is required then this set can be omitted and we will represent the causal link by a thin line.

The full description of a causally complex operation consists of the complex operation itself and the causal diagram. The above discussion shows how to motivate causally complex operations by looking at simple networks and obtain the causal diagram from the network. However, we do not assume that all causally complex operations are obtained from simple networks. In general, we simply provide the causal diagram part of the specification of the causally complex operation.

We will refer to causally complex operations as *complex operations* or simply as *operations* when it is clear from context that we are talking the causally com-plex case. We regard these operations along with an associated causal diagram as the basic objects in $t$COPT.

## 47.2 Interconvertible forms for complex operations

It is convenient to employ a notational convention whereby we can convert between different ways of representing a complex operation. For example

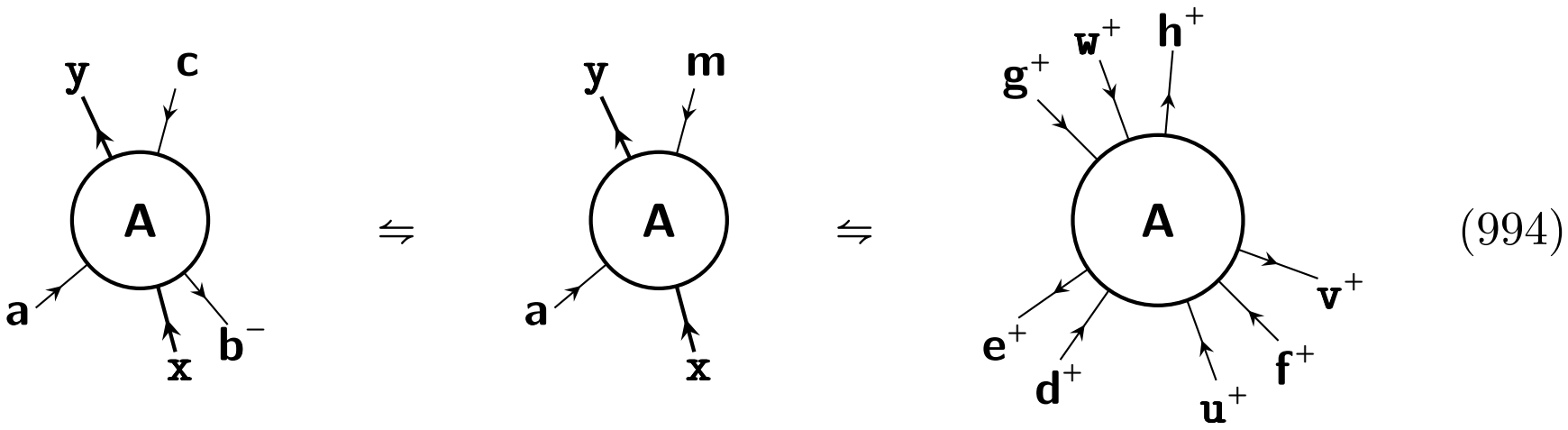

(994)

Here ⇋ means "interconvertible with" wherein we have the same composite system in and out overall. For the above examples this works if

$$\text{out} \qquad \mathbf{a}^-\mathbf{x}^-\mathbf{c}^-\mathbf{y}^+ = \mathbf{a}^-\mathbf{x}^-\mathbf{m}^-\mathbf{y}^+ = \mathbf{e}^+\mathbf{v}^+\mathbf{h}^+ \tag{995}$$

$$\text{in} \quad \mathbf{a}^+\mathbf{x}^+\mathbf{b}^-\mathbf{c}^+\mathbf{y}^- = \mathbf{a}^+\mathbf{x}^+\mathbf{m}^+\mathbf{y}^- = \mathbf{d}^+\mathbf{u}^+\mathbf{f}^+\mathbf{w}^+\mathbf{g}^+ \tag{996}$$

There are some issues with adopting this convention for the following reason. The point on the circle at which a wire is attached is significant. This is particularly the case if we have two or more systems of the same type since it is these points of attachment that label the systems. If we go from the first instance of **A** in (994) to the second and back again then we need to keep a record of where **c** attaches somewhere in the background. This small notational difficulty is a small price to pay for the convenience of being able to freely convert between different forms as in (994).

The relationship ⇋ is not quite equality since we cannot simply swap operations that are interconvertible as the wiring will not match. However, we can assert equality for complex networks like

g → B —b, a, c→ C → w  =  g → B —d, e→ C → w   (997)

(where the B's are interconvertible as are the C's) since the converted wiring is closed within the complex network.

It is useful to have notation for act of converting between different ways of decomposing systems. We will write

a, x, b | c, d, y   (998)

We call the thin rectangle a *converter*. At a converter we demand the total system type going in is the same as that going out. In the above example this condition is

$$\mathbf{axb}^R = \mathbf{cdy}^R \tag{999}$$

where the superscript $R$ indicates swapping the + and − components so $(\mathbf{b}^R)^\pm = \mathbf{b}^\mp$. We can write (999) out more explicitly as

$$\mathbf{a}^\pm \mathbf{x}^\pm \mathbf{b}^\mp = \mathbf{c}^\pm \mathbf{d}^\pm \mathbf{y}^\mp \tag{1000}$$

## 47.3 Causal diagram for simple network

Consider, again, the simple network in (989)

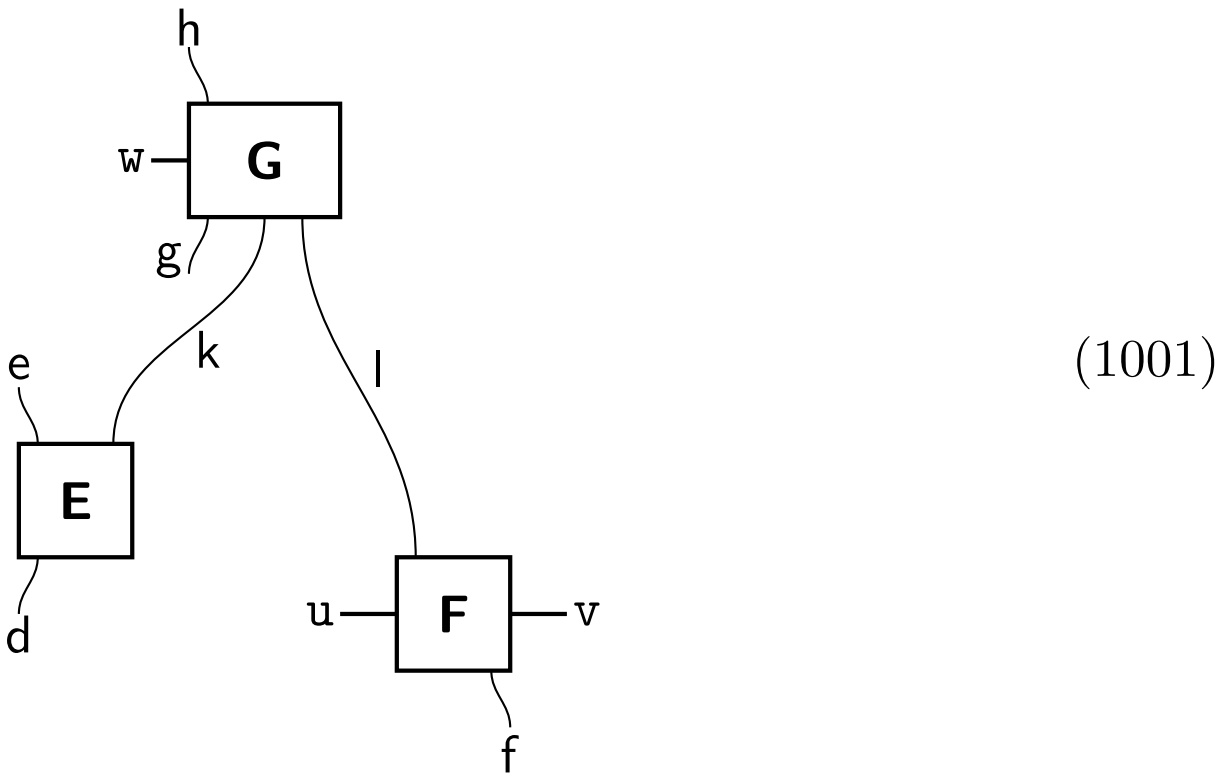

(1001)

We will see in Sec. 47.9 how to deduce the causal diagram for an arbitrary simple network by fusing the causal diagrams for the simple operations that it is comprised of. It is, however, clear by examining the above simple network that its causal diagram is

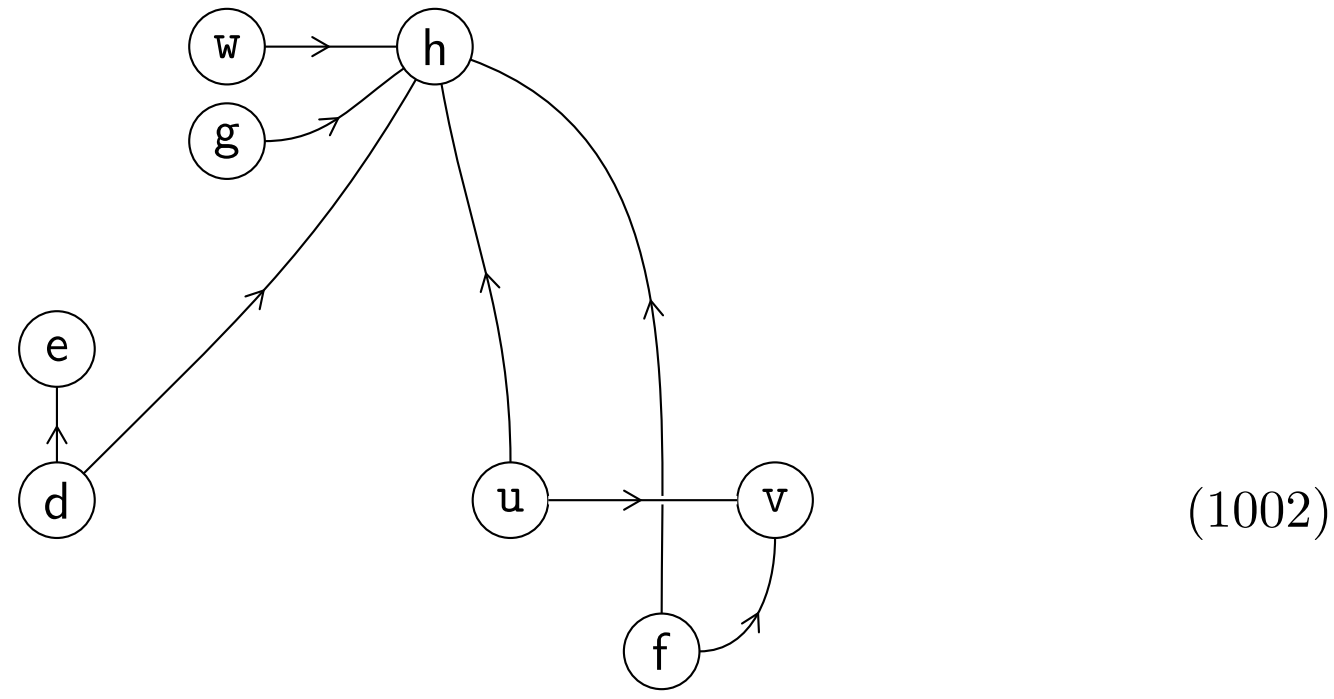

(1002)

We obtain this by tracing forward between the open wires and placing a forward pointing arrow in the causal diagram where we find a connection. We obtain the same diagram by tracing backwards and placing a backward pointing wire where we find a connection (this is an important observation in the context of a time symmetric theory). We will call the nodes here *base nodes* since they have individual wires associated with them rather than composite wires. This means that these wires correspond either to an input/income or to an output/outcome.

One important feature to note is that the base nodes fall into two types which we will label ±. We denote base nodes associated with an output/outcome wire in the simple network with a +. These nodes have causal arrows pointing into them in the causal diagram. We denote nodes associated with an input/income wire in the simple network with a −. These base nodes have causal arrows pointing out of them in the causal diagram. In (1002) h, e, and v are + base nodes and the other nodes are − base nodes. This particular causal diagram has arrows into or out of every base node and consequently we do not need to explicitly include the ±'s to say what type of node we have. It is possible, however, to have a base node with no causal arrows attached. For example,

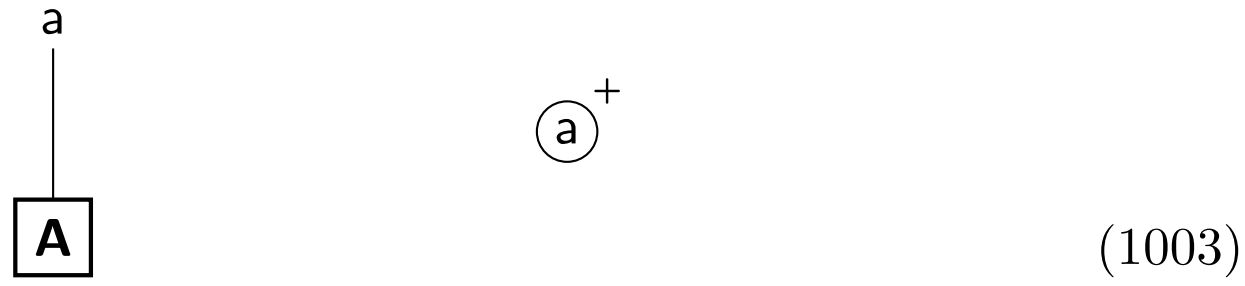

(1003)

The preparation on the right corresponds to a + base node in the causal diagram on the left. We indicate this by a + sign next to the base node.

## 47.4 On the meaning of causal diagrams

We can deduce the causal diagram of a simple network by inspection. However, it is possible some of the boxes in the network have properties that actually disallow certain causal influences were we to investigate further. We could attempt to deduce these properties by embedding the network inside various bigger networks to deduce the causal structure by means of some kind of causal tomography. These considerations lead to the following two possible ways to understand causal diagrams, which we call *hard* and *soft* by virtue of a (obvious but imperfect) analogy with hardware and software. We take the hard approach in this paper. There is, however, much interesting work where the (more nuanced) soft approach is taken.

**Hard approach.** In this approach, we can think of the causal structure as something that is simply given, or something that is deduced from operational considerations corresponding to how a causally complex operation is implemented physically in terms of how its components are wired together or in terms of constraints imposed by some background spatial-temporal causal structure (coming, for example, from a spacetime metric). In the hard approach we think of the causal structure as something that is imposed from without.

**Soft approach.** The causal structure is deduced from operational considerations that probe the inner structure of the complex operation by means of some kind of causal tomography (this would involve completing the complex operation into various circuits to probe the causal influence between input/income wires and output/outcome wires). In the soft approach we think of the causal structure as something that is determined from within.

As long as the underlying physics is local with respect to the hard causal structure then we expect all the causal arrows that appear in the "soft causal diagram" to appear also in the "hard causal diagram". The analogy with hardware and software is that, in the hardware we see how the boxes (e.g. integrated circuits) are wired together while, in the software, we can program these boxes to do different things.

The hard approach is more suited to our current goals since we will be interested in imposing physicality constraints that depend on the given causal diagram and then looking for the full set of complex operations consistent with that causal diagram. Thus, in the hard approach we can think of the causal diagram as something that is imposed rather than something that is determined.

The soft approach, on the other hand, opens us up to more subtle types of causal structure that might arise in a tomographic scenario. To illustrate this, consider the property of *causal atomicity* (see Barrett et al. [2019], Ormrod et al. [2023], van der Lugt and Lorenz [2025], and also Wilson and Vanrietvelde [2021]). This is the property that a system a influences a composite system bc precisely when it influences at least one of b or c. Interestingly, there are actually examples that violate causal atomicity in Quantum Theory (see Appendix A of van der Lugt and Lorenz [2025]). However, causal diagrams, as we have set them up, necessarily satisfy causal atomicity meaning that there are subtleties that causal diagrams cannot accommodate. The causaloid framework (see Hardy [2005]) is an extreme example of this where every set of nodes has a matrix associated with it that conveys the counterpart of causal structure - this goes beyond the idea of causal structure being something that is indicated by the presence or absence of an arrow between a pair of nodes.

The soft approach allows us to discover more subtle types of causal structure such as a violation of causal atomicity. We expect a theory of Quantum Gravity to have indefinite causal structure, a quantitative description of which is dynamically determined by some Quantum analogue of the energy momentum tensor. Thus, we will not have a fixed background causal structure and so the hard approach, in its present form, will not work.

Related distinctions to those above appear in several strands of recent work on causal structure in quantum theory. In the programme of causal identification, one distinguishes between the abstract causal structure of a process and its concrete causal structure (see Friend and Kissinger [2024], Barrett et al. [2025]). The abstract causal structure corresponds to the comb shape or compositional skeleton of the process, specifying the pattern of slots and wires and the possible signalling relations between them, while the concrete causal structure corresponds to the actual causal influences instantiated by a particular physical process and which may be identified through operational probing or tomography. In later work on compositional causal modelling this distinction is reformulated in terms of syntactic and semantic causal structure (see Friend et al. [2025]). Related ideas also arise in work on inferring causal relations in quantum systems from intervention and signalling relations (see Ried et al. [2015]).

## 47.5 Fusing causal diagrams

The ±'s labeling the two types of base node are important since, as we will now see, we can only join a − node to a + node since input/income wires join to output/outcome wires (we cannot join two − base nodes, or two + base nodes). We say the nodes are *matched* if this is true and the wire types (e.g. `e`, `u`, ...) inside are the same. For example, we can join a $\textcircled{\texttt{d}}^{+}$ node and a $\textcircled{\texttt{d}}^{-}$ node.

Consider joining (fusing) the causal diagram in (1002) to the causal diagram on the right below

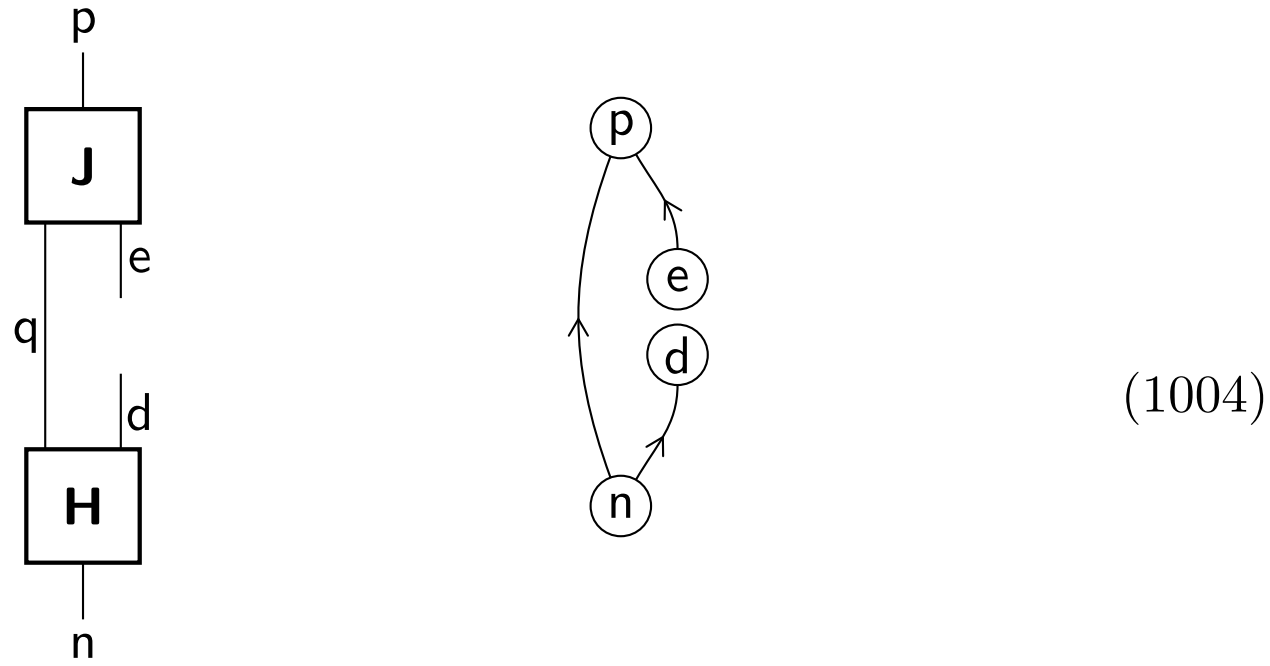

(1004)

at the nodes `d` and `e`. We can do this since these nodes are matched. When we do this we get the following

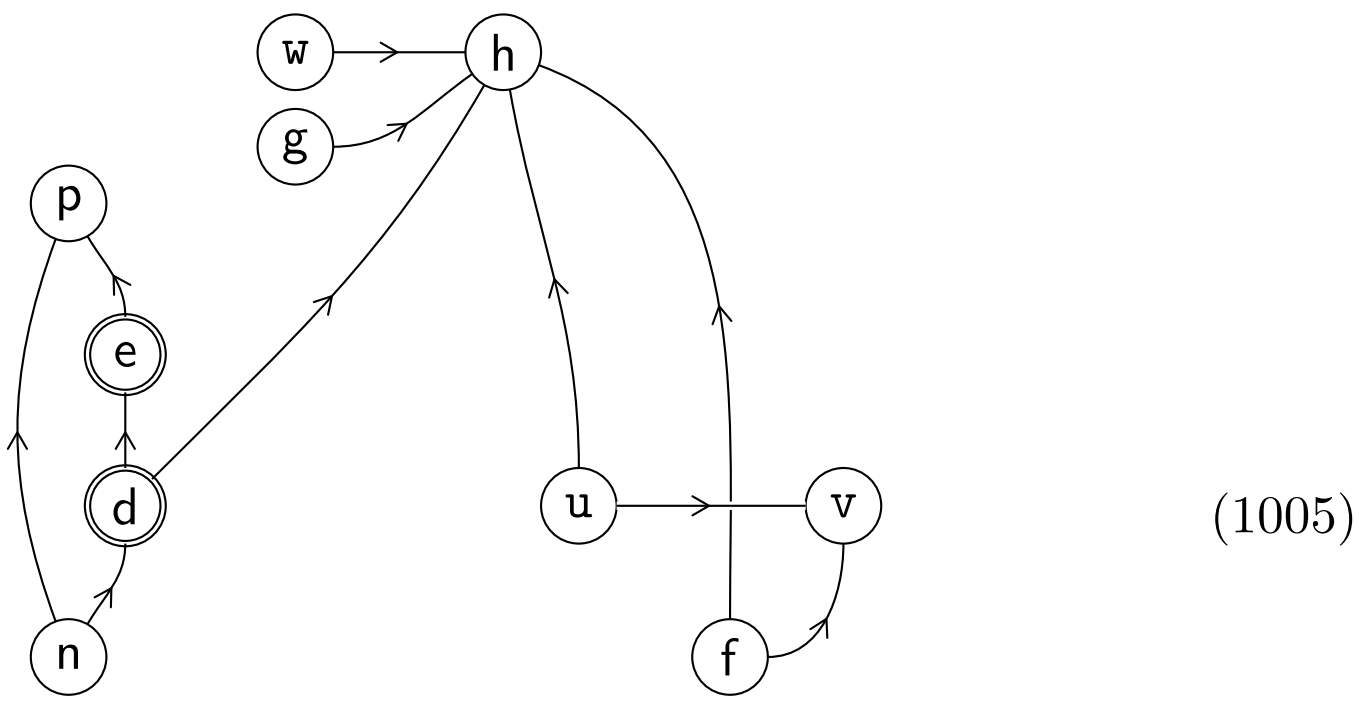

(1005)

where we have double border circles to indicate where the fusion has happened. These overlapping nodes act as a conduit for causal structure. We no longer care what the system type inside is. Thus we can write

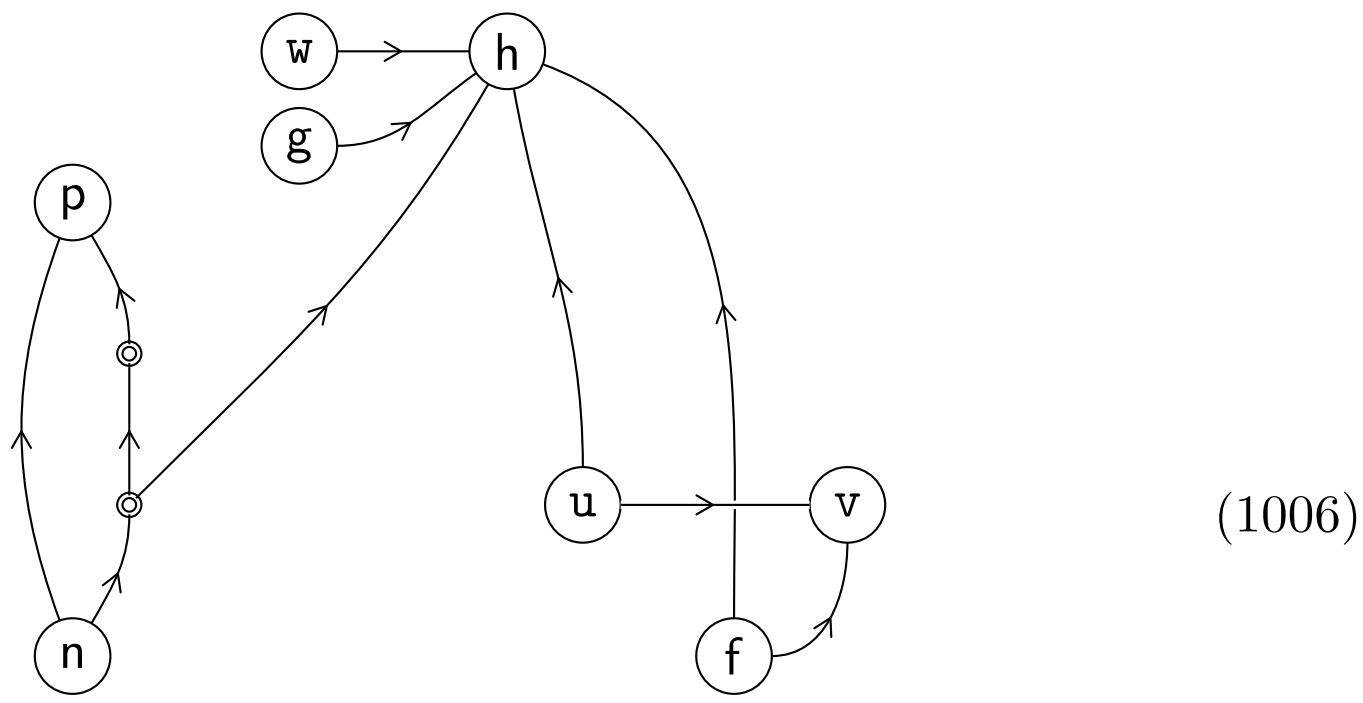

(1006)

where we have introduced the symbol ⊚ which has causal arrows attached. This symbol represents the body of a *causal spider* while the attached arrows are the legs of the spider. It indicates that we have a causal link from every incoming link to every outgoing link. We will discuss causal spiders in Sec. 47.6. It is clear by inspection that this causal diagram simplifies to

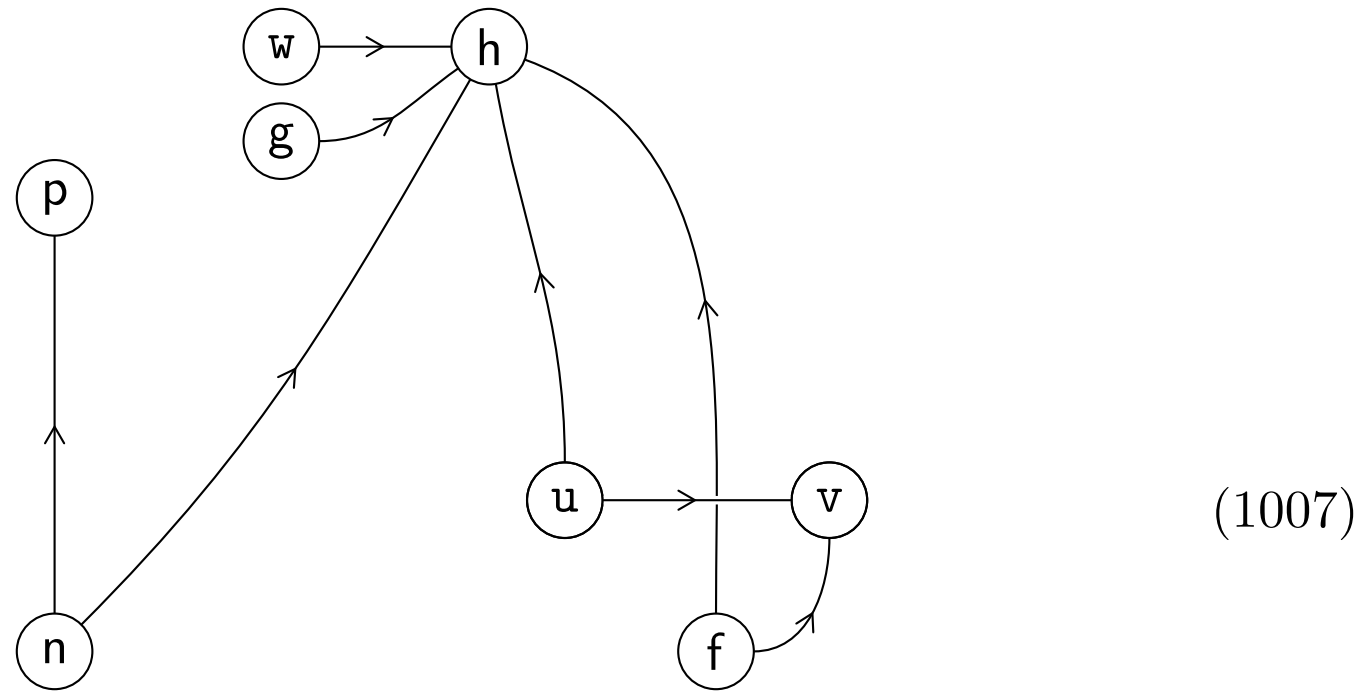

(1007)

because we do not need to duplicate the causal link between n and p so we have deleted one of the two links while keeping the causal link between n and h. We can perform these manipulations in a more local manner by sliding spiders along wires using identities between causal spiders which we will now discuss.

## 47.6 Causal spiders

We will call the small double border circles *causal spiders* (the term "spider" is used in Coecke and Kissinger [2017] for objects which are graphically similar but have a different physical meaning). In general, the $m \to n$ causal spider

$$m \left\{ \;\vdots\; \circledcirc \;\vdots\; \right\} n \qquad (1008)$$

means we have a causal link connecting *every* incoming wire to *every* outgoing wire.

There are a few special cases that are worth noting. First, for the $1 \to 1$ causal spider we have

(1009)

since we just have one causal connection going through the spider. Second, note that if we join a $-$ and a $+$ base node that have no causal arrows (such as when we wire a preparation to a result) then we get a spider body with no legs

(1010)

These carry no causal information and can be deleted (though see the comments concerning the causal diagram for a circuit in Sec. 47.7 below). Third, we can consider one-legged causal spiders

(1011)

These indicate that a particular causal route is closed to the future (left expression) or from the past (right expression). One-legged causal spiders are sometimes useful (e.g. in stating the residua causal diagrams theorem in Sec. 49.4.7).

A $m \to n$ general causal spider can be written in terms of $1 \to n$ and $m \to 1$ causal spiders. For example, a $2 \to 3$ causal spider can be written

(1012)

We will call this the $2 \to 3$ *causal spider dance identity*. This clearly generalises to an arbitrary $m \to n$ causal spider.

Another useful equation is the following

(1013)

We call this the *causal spider tango identity*.

We can, in certain circumstances, slide causal spiders until they hit nodes which absorb them. For example we have

(1014)

since clearly the right side has the same causal meaning. This equation also holds if the arrows are reversed.

To illustrate the use of these equations consider again

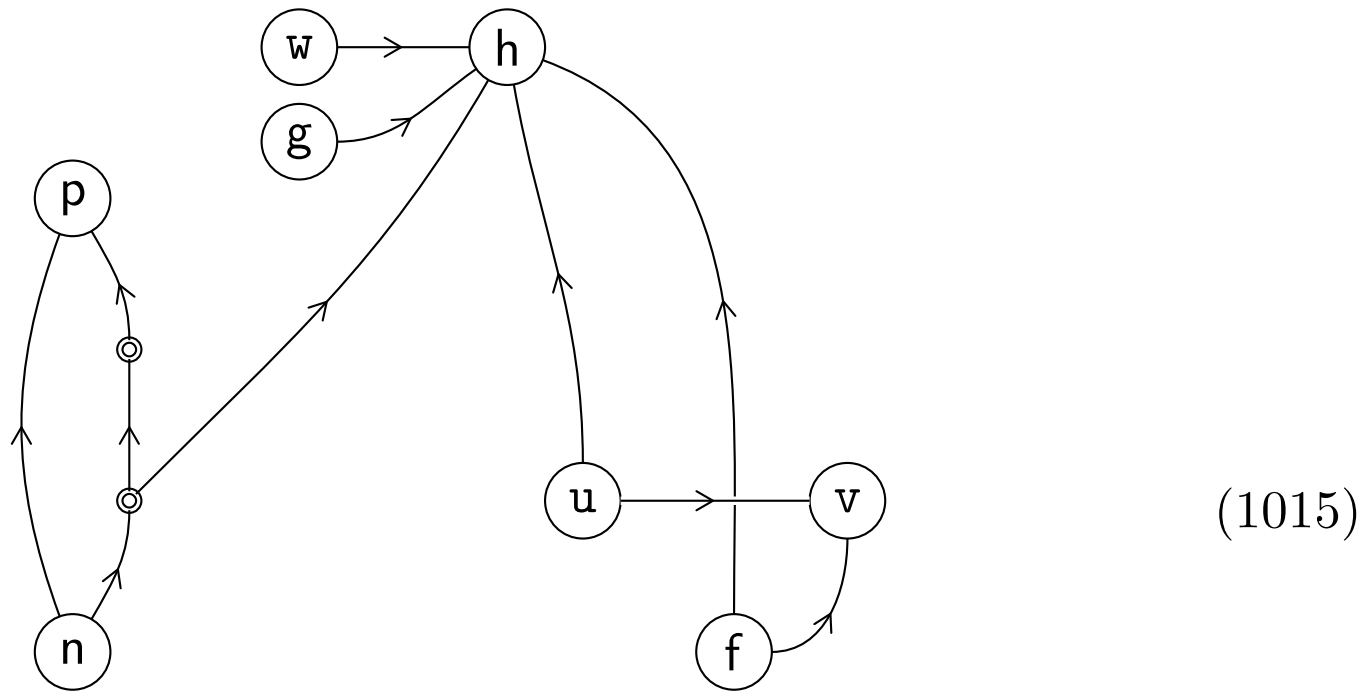

(1015)

This was (1006) above. We can either delete the upper ◎ using (1009) or slide it upwards until it is absorbed by node p using (1014). Then we can slide the lower spider down until it is absorbed by node n using (1014). This gives

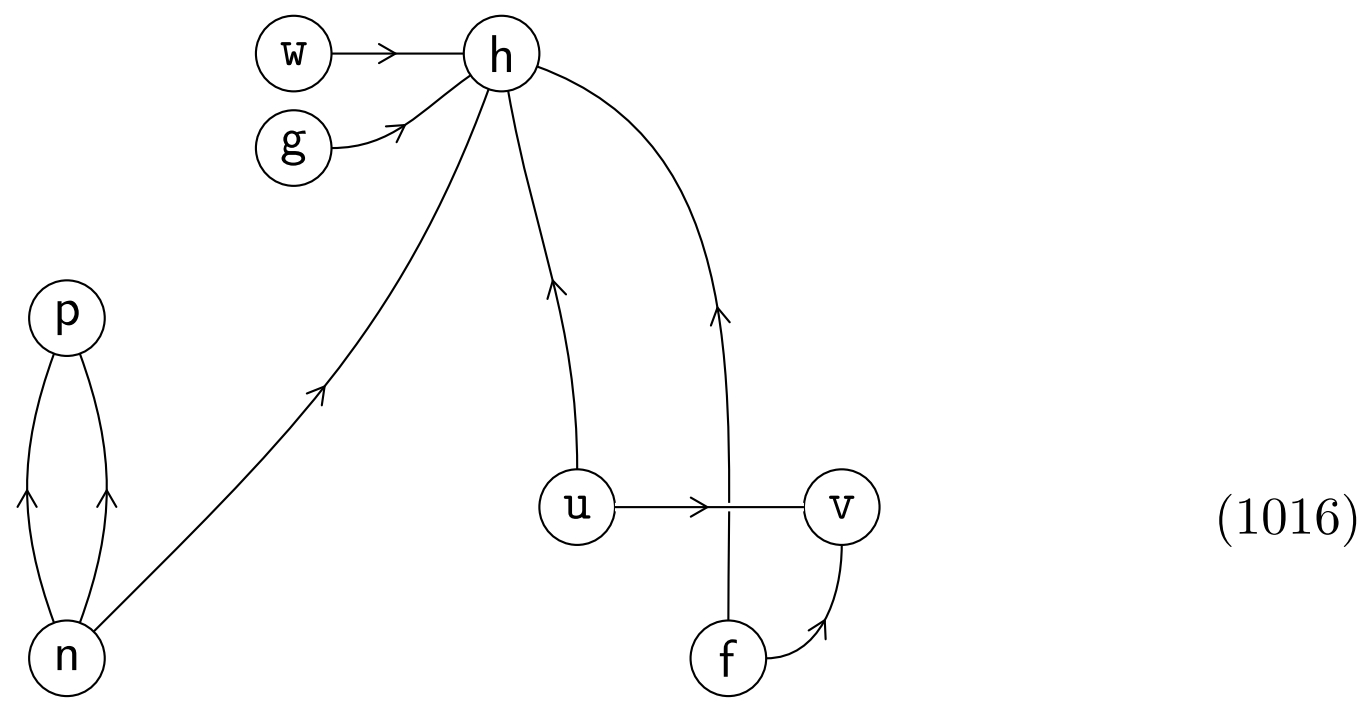

(1016)

Now we can use (1014) in reverse to pull a spider out of node p

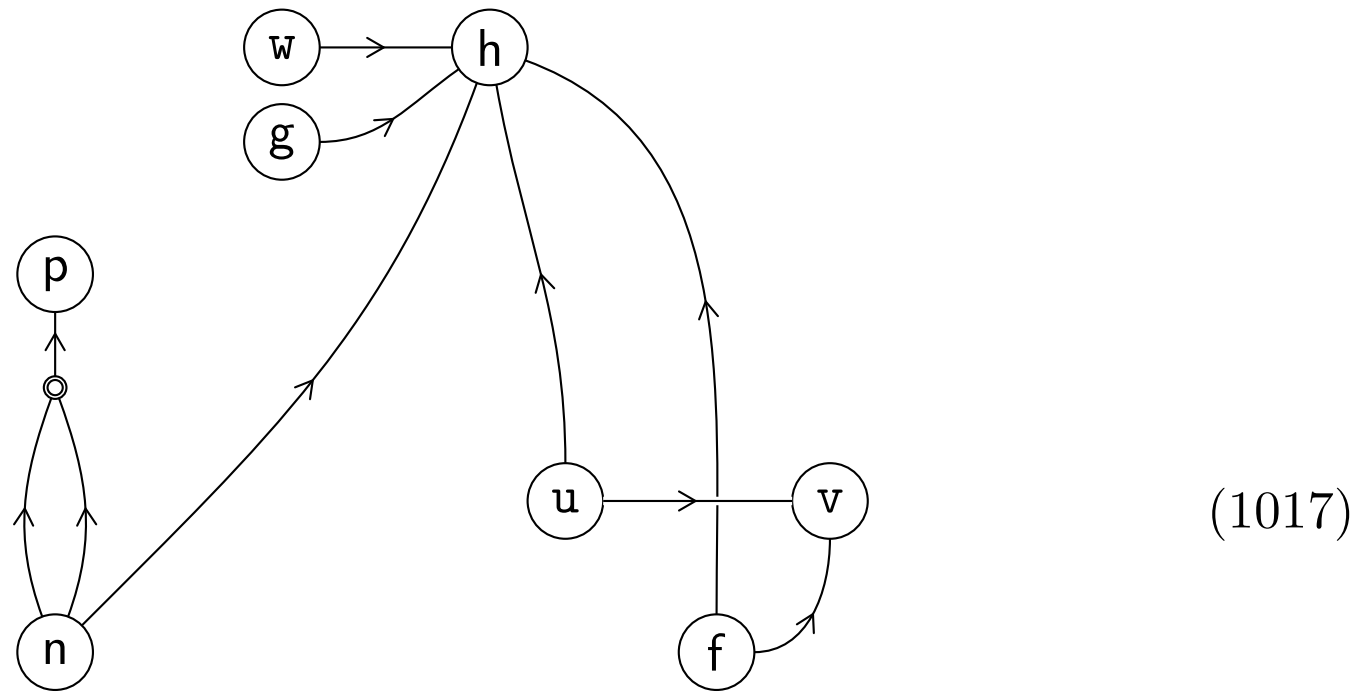

(1017)

Finally, we can slide this causal spider down until it is absorbed into node n so now we only have one causal link between nodes n and p. This finally gives us (1007).

If we have a causal spider with legs that go to multiple nodes, we can first use the causal spider dance identity so that we have only $1 \to n$ and $m \to 1$ spiders. We can then slide these so that they are absorbed by nodes. If these spiders collide we can use the tango identity to remove this obstruction. In this way we can remove the spiders as long as we have a directed acyclic graph (DAG). An example where we do not have a DAG is considered in Sec. 47.10.

## 47.7 Causal diagram for a circuit

A circuit has no open wires. Consequently, its causal diagram will have no nodes and therefore no causal links - it is represented by the "empty" causal diagram. For the sake of having a symbol to represent this case we will represent the causal diagram of a circuit by a spider body

$$\circledcirc \tag{1018}$$

since this is what we obtain when we join a + base node to a − base node. This simplifies to the empty causal diagram.

## 47.8 Causal diagrams for simple operations

Simple operations are used to build simple networks which motivate complex operations. The simple operation

(1019)

has a causal diagram that can we written in various equivalent ways

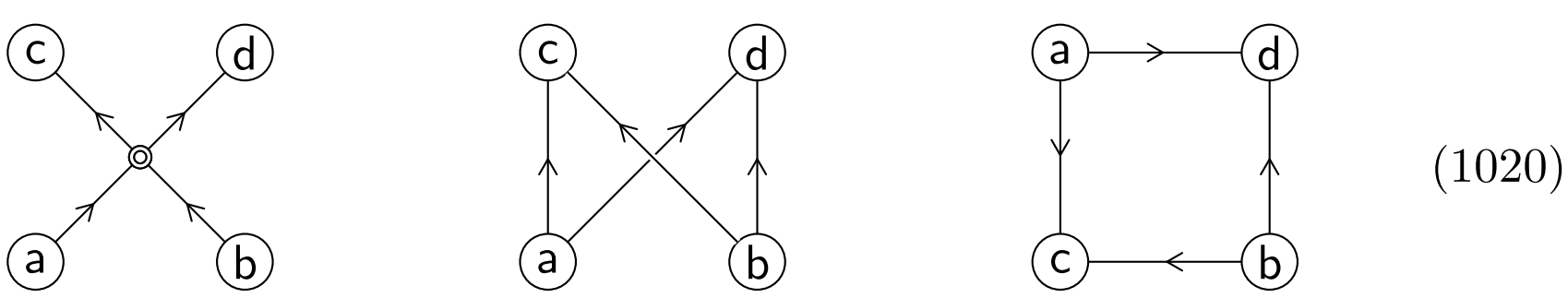

(1020)

Similarly, the causal diagram for the simple operation

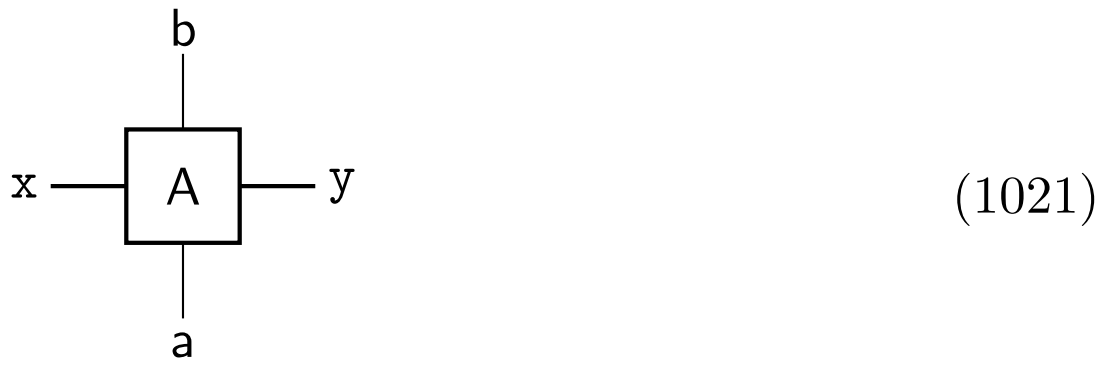

(1021)

can be written in the following equivalent ways

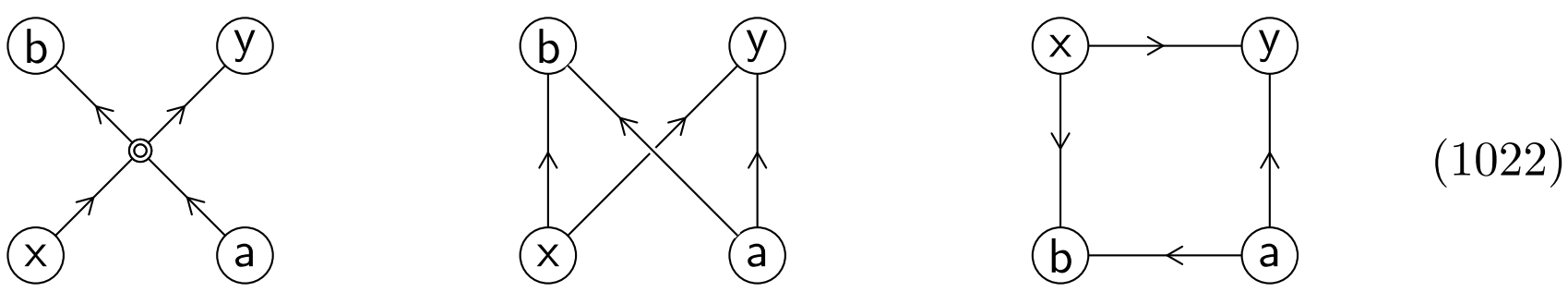

(1022)

In general, a simple operation with $m$ inputs/incomes and $n$ outputs/outcomes can be written by placing the corresponding nodes on the legs of a $m \to n$ causal spider.

The case of simple preparations and simple results is interesting. For a simple preparation of systems a and b we have

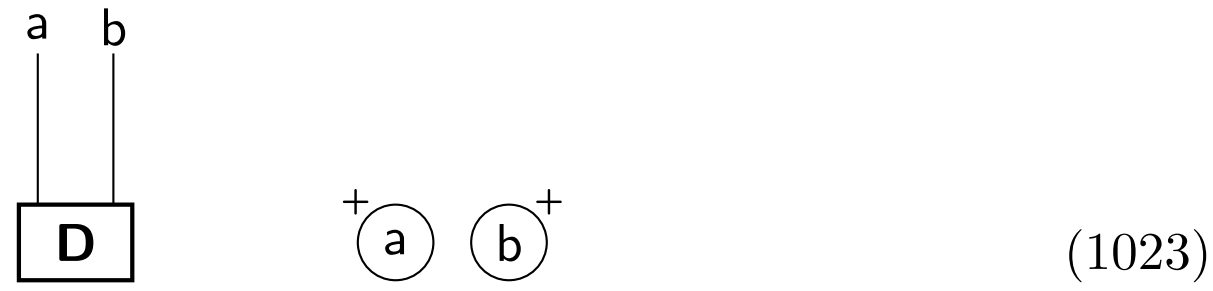

(1023)

Similarly, for a simple result, we have

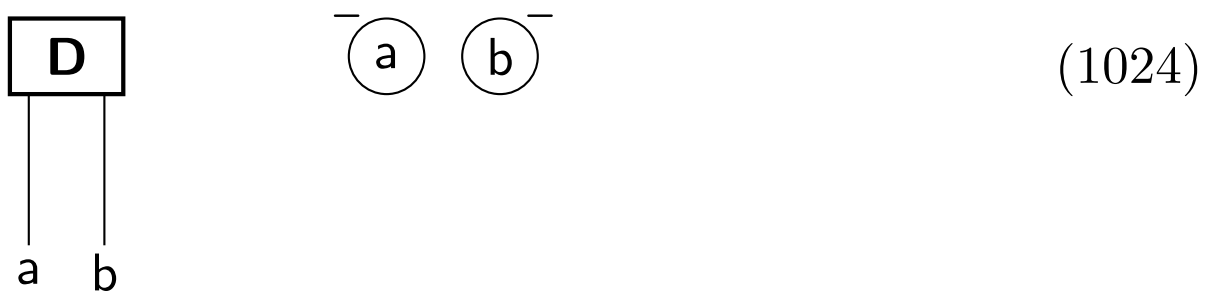

(1024)

As commented earlier, if we combine a preparation and result (like these) to form a circuit then the causal diagram is the empty set.

## 47.9 Deducing the causal diagram of simple network

In Sec. 47.3 we wrote down the causal diagram of the simple network

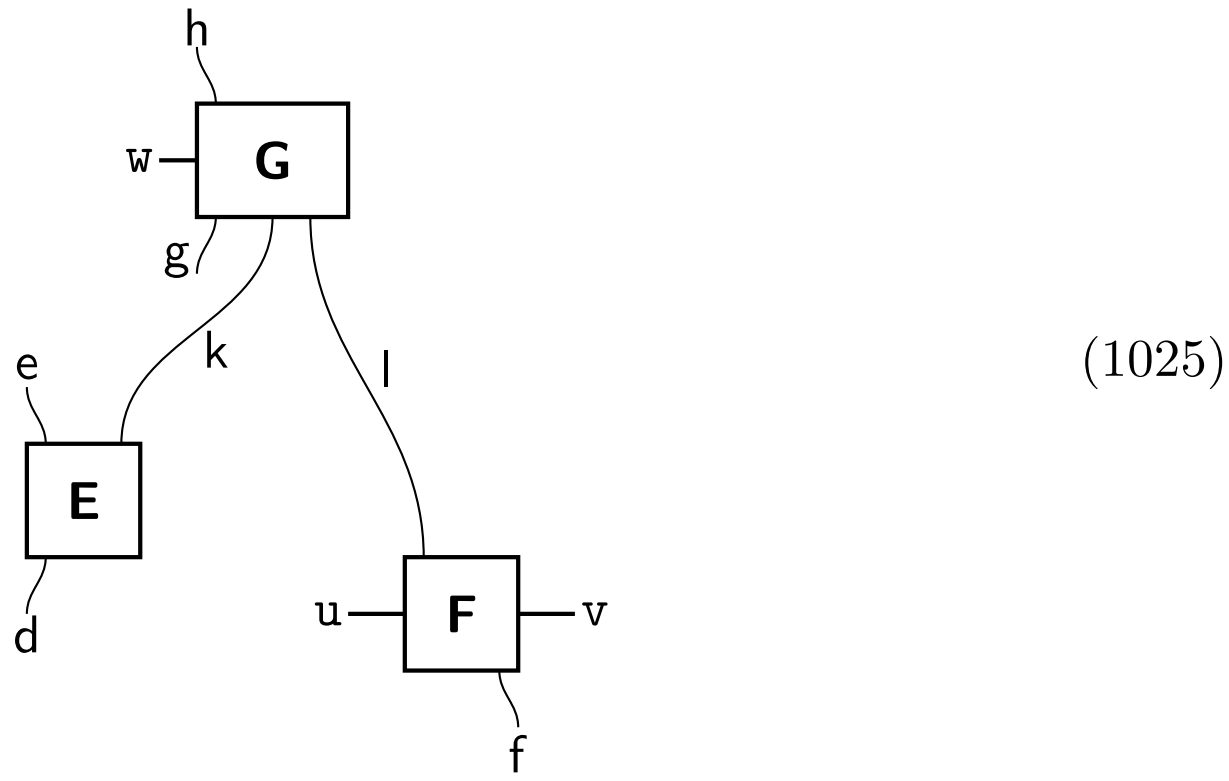

(1025)

by examining the causal structure implicit in the wiring. For this simple example this is fairly obvious. However, for a more complicated simple network it is good to have a systematic way to deduce the causal diagram. We can do this by fusing the causal diagrams for the simple operations that comprise the given simple network. If we fuse together the causal diagrams for **E**, **F**, and **G** then we obtain

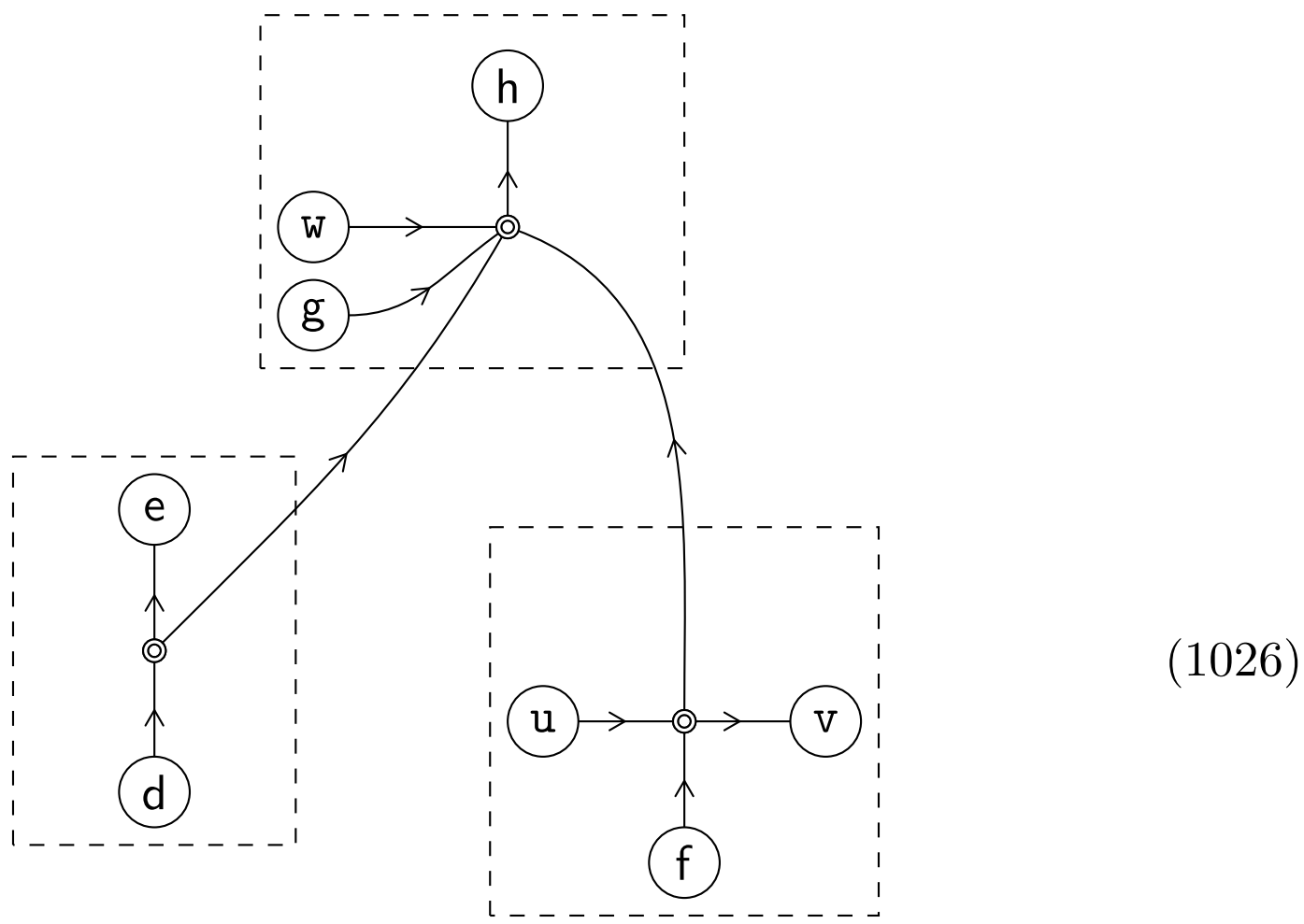

(1026)

where the dashed boxes indicate the causal diagram for each of the simple operations. Now, using causal spider manipulations, we can simplify this to the causal diagram given previously in (1002). This technique clearly works for any simple network.

It is worth noting that, after the first step, the causal diagram looks like the simple network it is associated with (it has the same connectivity between the components coming from the different simple operations). This is clear comparing (1026) with (1025). This is to be expected since causal influence is associated with the system and pointer wires. However, after simplification (using spider manipulations) this immediate resemblance may be obscured. The simple network has, in a certain sense, more information in it because the wires are labeled by their types (d, f, . . . ) and so we have a quantitative sense of how much "information" can flow along this wire. In the causal diagram, however, we draw a wire if there is a causal connection without any such quantitative measure. It would be interesting to investigate causal diagrams wherein the wires were labeled by some such quantitative measures but we will not do that in this book. Since we regard causal diagrams as constraints (this is the "hard" point of view outlined in Sec. 47.4) we allow any amount of "information" to flow along any given wire. This allows us to impose causality constraints effectively since "information" cannot flow where there is no causal arrow.

## 47.10 Disallowed wirings

We do not allow simple networks to have closed timelike loops so the causal diagram must be a DAG. It is interesting to see how we might get non-DAG causal diagrams if we join together operations in the "wrong" way. Consider joining two simple operations **A** and **B** as follows

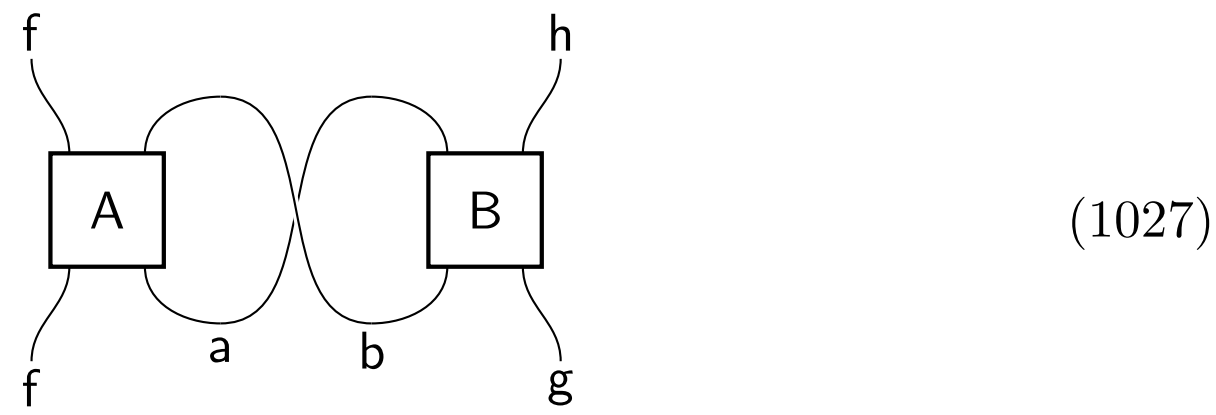

(1027)

This is not an allowed network as it is non-DAG. We can, nevertheless, look at the causal diagram. This is given by joining

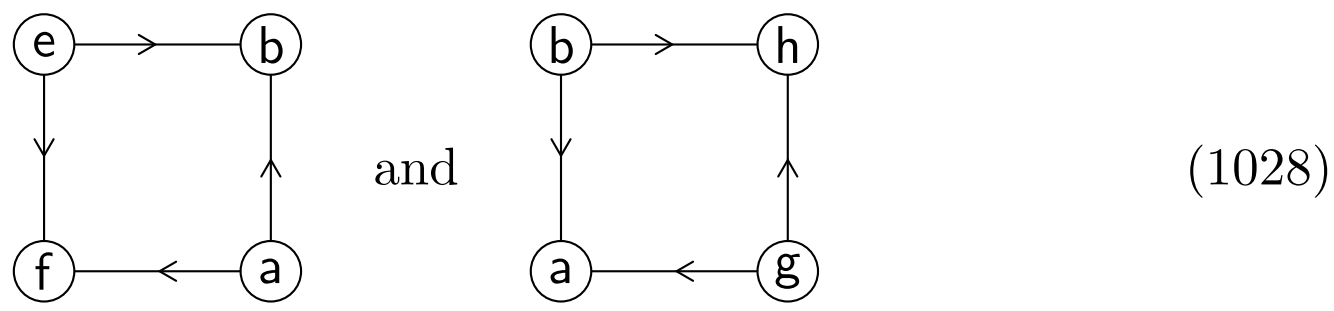

(1028)

at a and b. Then we obtain

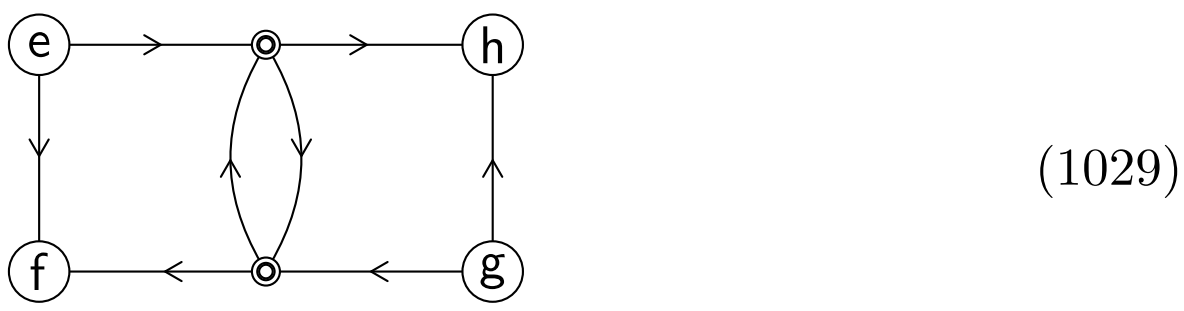

(1029)

This is non-DAG because of the loop in the middle (which comes from the loop a → b → a in the non-DAG simple network above (if we were able to delete either of these two arrows in the network then the causal diagram would become a DAG). We can reduce the valency of the causal spiders using the 2 → 2 causal dance identity and absorbing some spiders into nodes giving

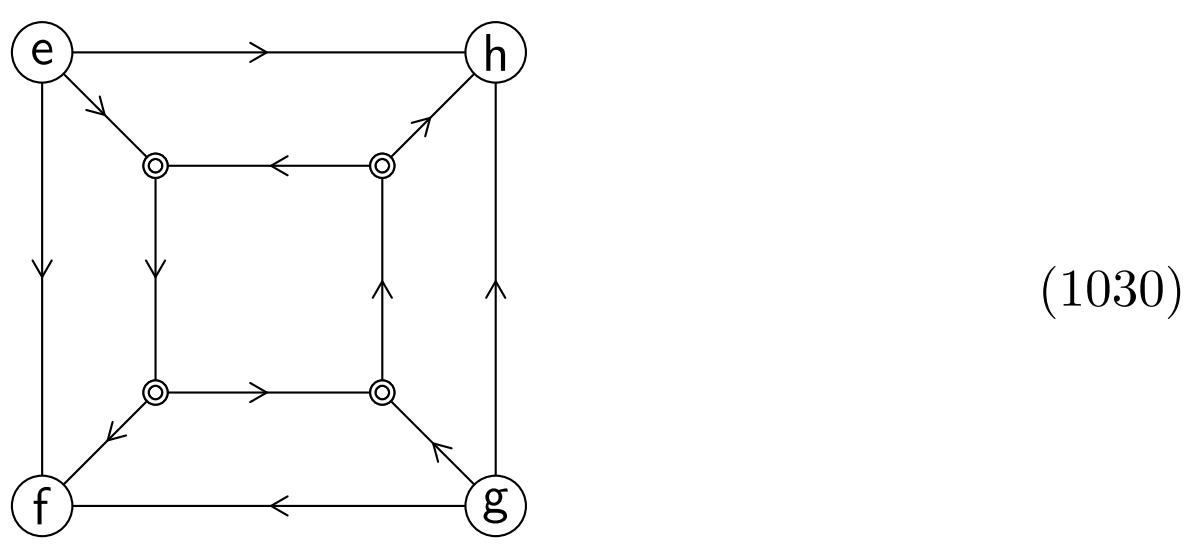

(1030)

However, we are blocked from absorbing the remaining 1 → 2 and 2 → 1 causal spiders and so cannot simplify further. The resulting diagram is remains non-DAG (as we would expect). In this framework, we regard operations with non-DAG causal diagrams as non-physical and disallow them (though, of course, it would be interesting to study this further to see if such causal loops could be consistently included).

## 47.11 Fission on causal diagrams

We may wish to regard a causal diagram as consisting of different parts. We use the notation

(1031)

The notation on the right visually suggests fusion/fission of nodes. We will call them *fusion nodes*. The lower node can be regarded as a + type (as it has a causal arrow pointing into it) and the upper node can be regarded as a − type. Consider an example. We introduce a partition line as indicated by the dashed line on the left then we have

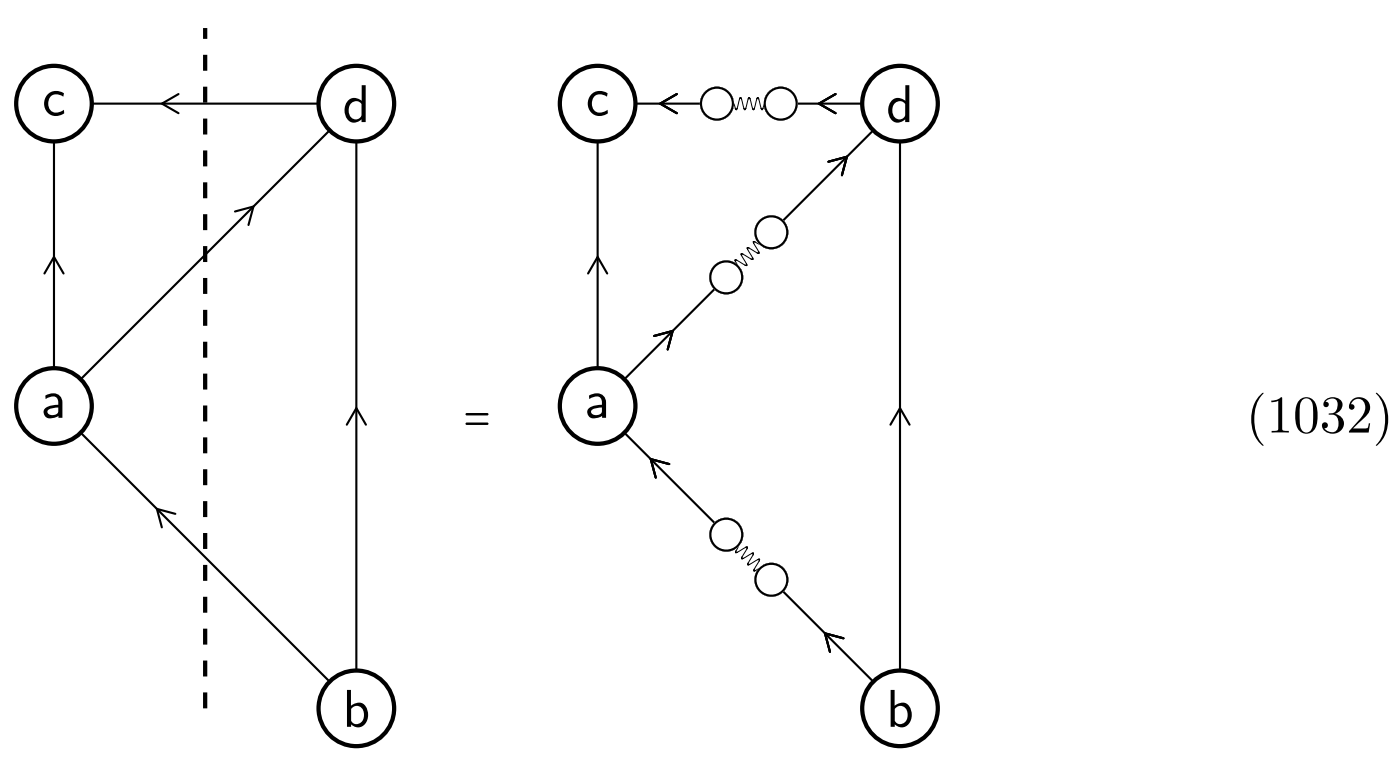

(1032)

In this example, the nodes at the fusion sites belonging to the left hand part of the causal graph are of mixed type (two −'s and one +). Similarly the right hand part also has mixed type nodes at the fusion sites. We will say this is an *asynchronous partition*. Here is an example of a *synchronous partition*.

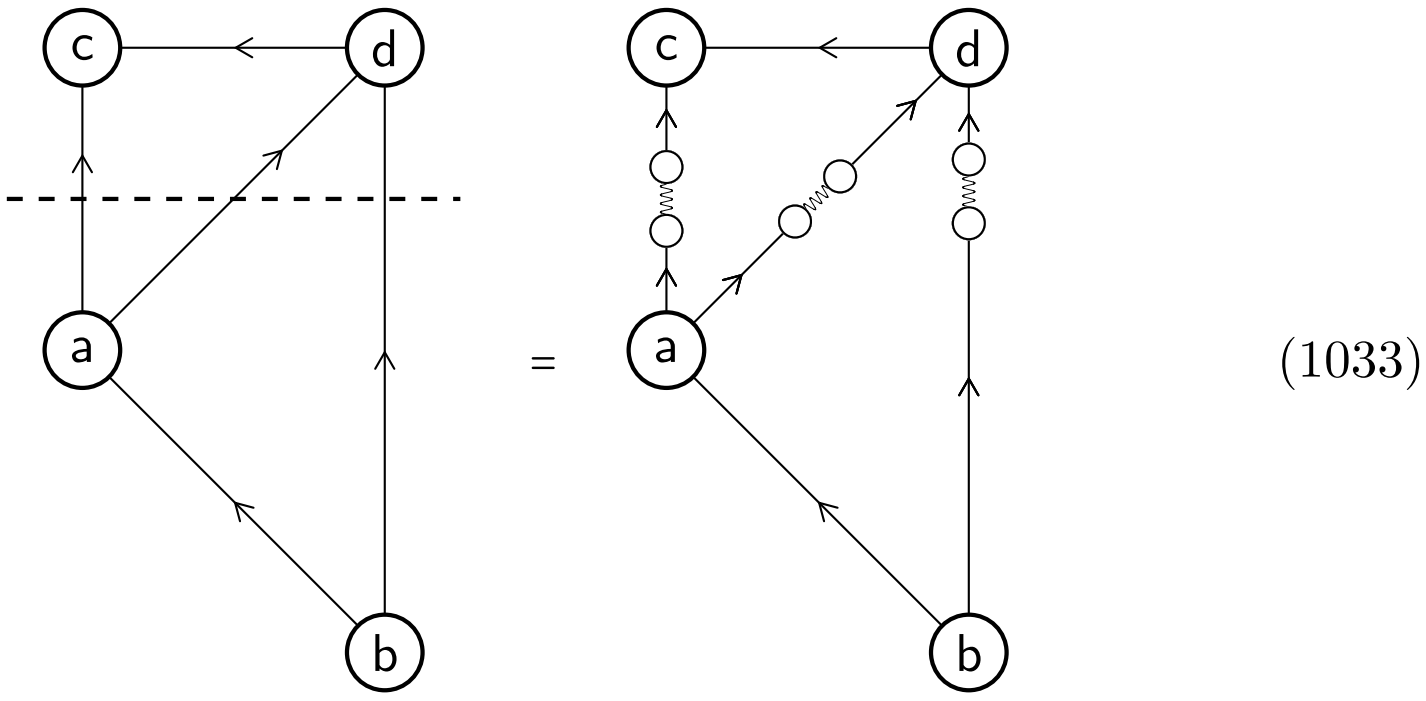

(1033)

The lower part has all + type nodes at the fusion sites. The upper part has all − type nodes at the fusion sites. We can regard the upper part of the causal diagram as being to the future of the lower part. Synchronous partitions are a discrete graphical analogue of space-like hypersurfaces. We will be interested in synchronous partitions when we come to consider positivity and causality for complex operations.

As the final step of fission, we can imagine breaking a causal diagram at the fission sites so we get two separate causal diagrams. Then each diagram will contain some nodes that do not have wire types (a, x, ...) in them. It is useful to allow this in some of the proofs below.

## 47.12 Implicit causal structure in nodes

In the causal diagram in (1002) all the causal structure is explicit. We will define a procedure of course-graining on causal diagrams. This is where we group together the nodes to create new nodes. First we identify the nodes to be course-grained. We can course-grain into disjoint sets of nodes in anyway we want. For example,

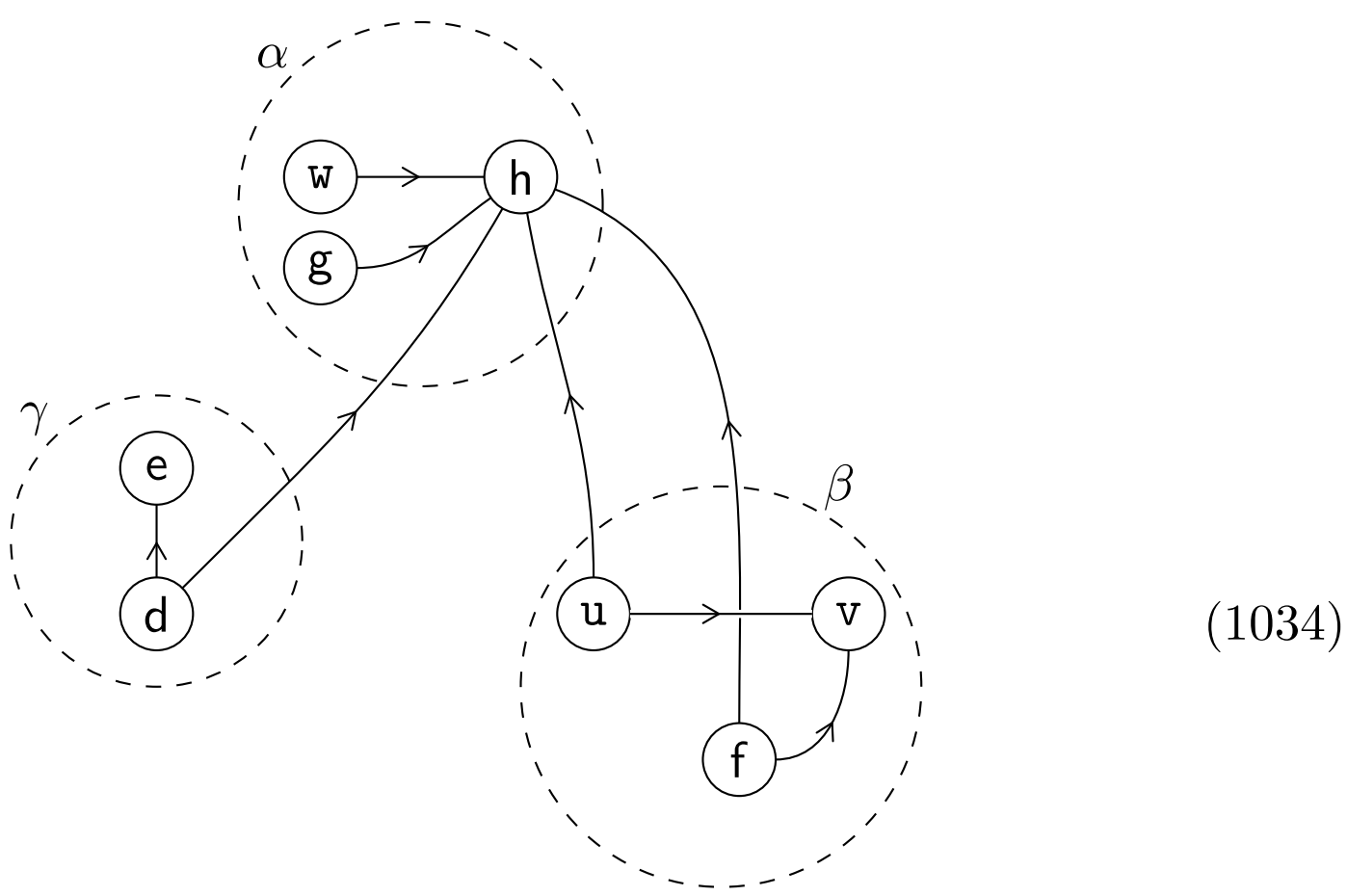

(1034)

Now we can represent what happens inside each of the dashed circles by a single node as follows

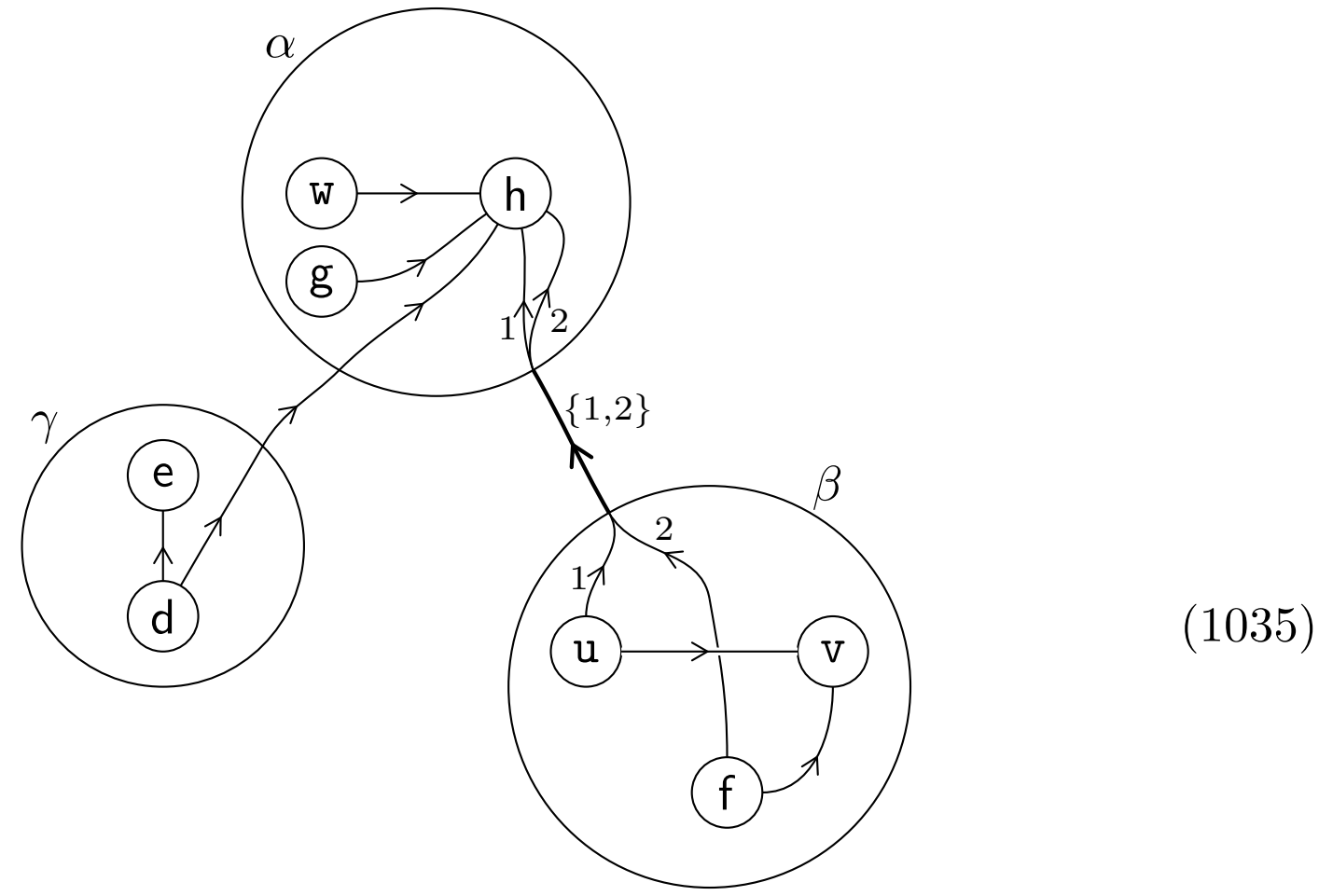

(1035)

Thus, the operation **A** discussed in Sec. 47.1 is represented by

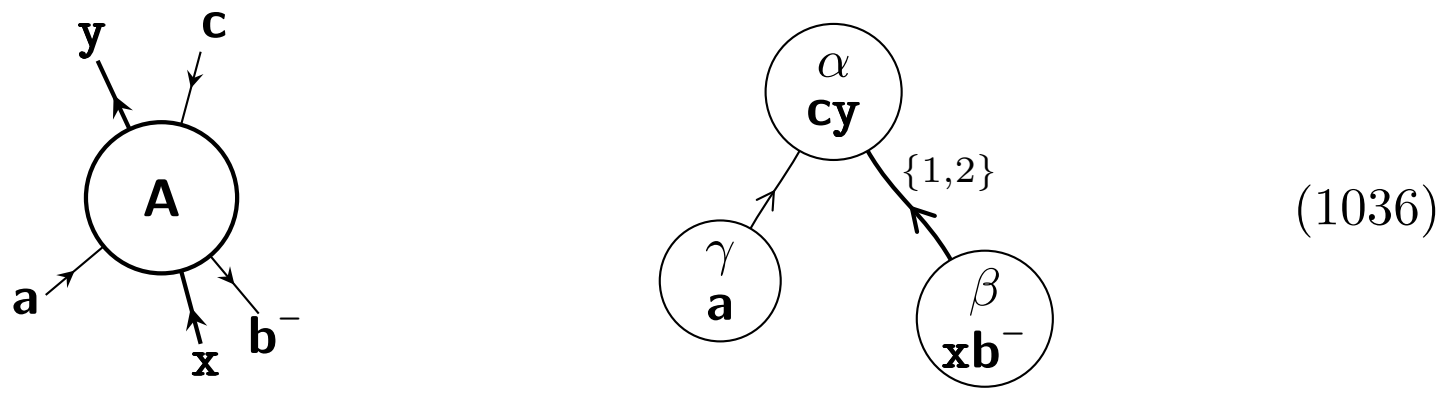

(1036)

where, on the left, we have the operation and on the right we have the causal diagram. The implicit causal structure in each node is designated by $\alpha$, $\beta$, and $\gamma$. The identifications of the types **a**, **b**, **c**, **x**, and **y** in operation **A** are given in (991). The causal link with a $\{1, 2\}$ on it consists of two labeled wires which are needed because the causal structure $\beta$ has two wires coming out of it in the direction of h.

## 47.13 Causal diagrams that appear to be non-DAG

The course-grained causal diagram in (1036) is, itself, a directed acyclic graph (DAG). However, this need not be the case. Consider, instead, course-graining the base causal diagram (1002) as follows

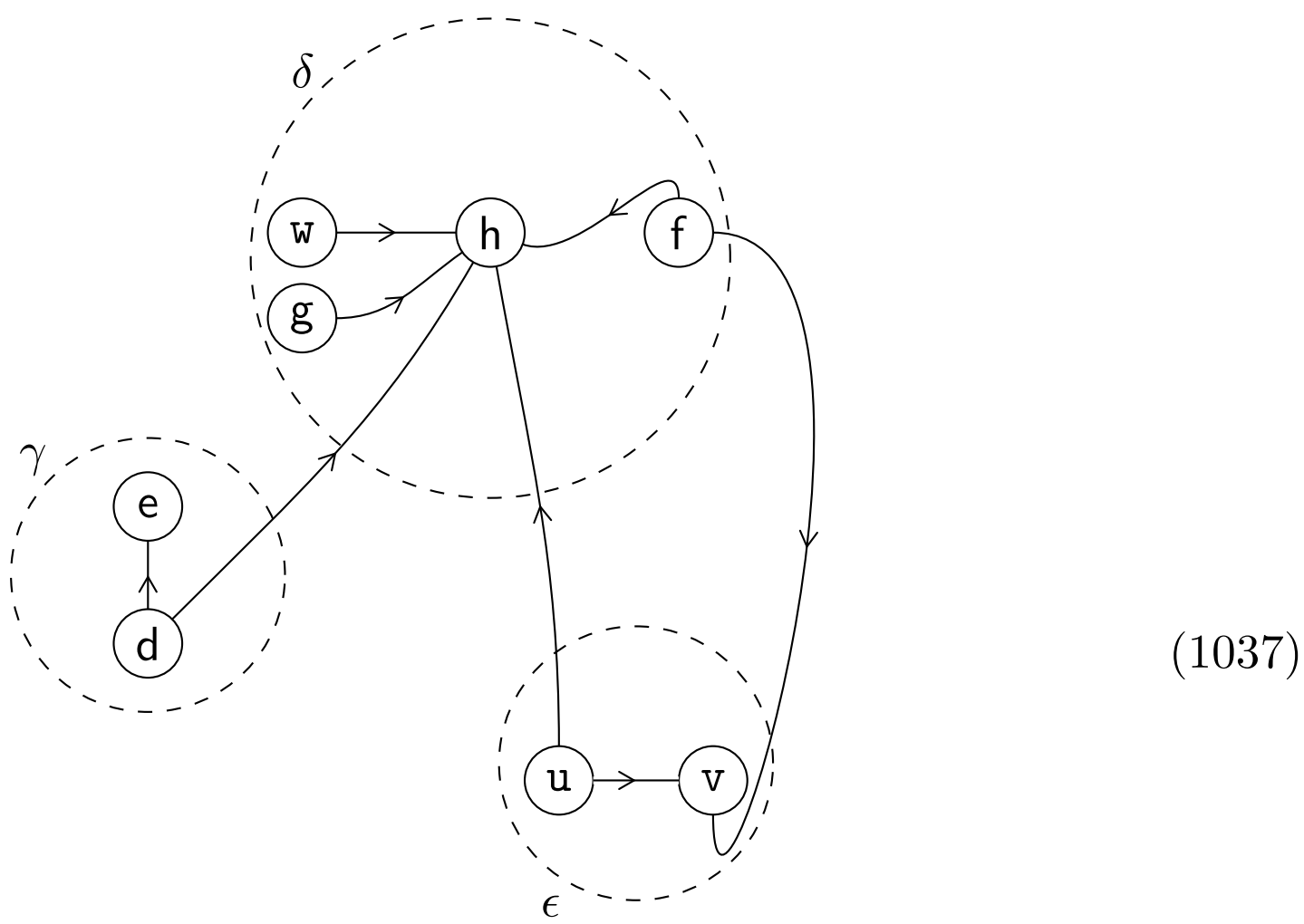

Here we have "moved" f. If we do this then the operation is given by

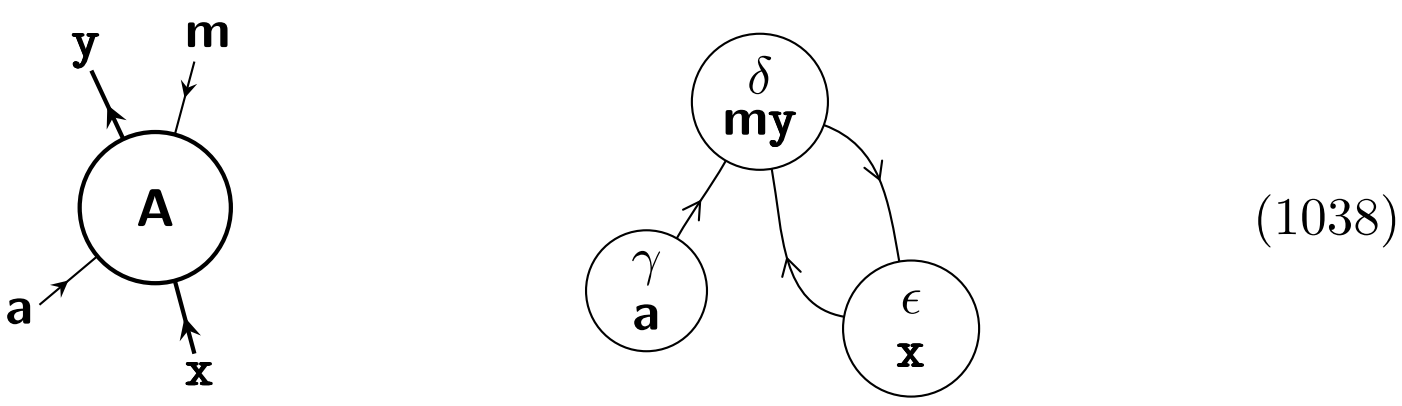

On the right is the course-grained causal diagram. Note that this is not a DAG at this level of course-graining. On the left is the operation where

$$\begin{array}{cccc} \mathbf{m}^+ = \mathsf{gf} & \mathbf{m}^- = \mathsf{h} & \mathbf{y}^+ = 0 & \mathbf{y}^- = \mathtt{w} \\ \mathbf{a}^+ = \mathsf{d} & \mathbf{a}^- = \mathsf{e} & & \\ & & \mathbf{x}^+ = \mathtt{u} & \mathbf{x}^- = \mathtt{v} \end{array} \tag{1039}$$

Although the causal diagram in (1038) is not a DAG at the level of the explicit causal structure, the implicit causal structure in the nodes means that this is actually a DAG at the fine-grained level. This has to be the case since the base causal diagram is a DAG.

## 47.14 Joining causally complex operations

We can join causally complex operations to form causally complex networks. In so doing we need to join both the operation and the causal diagrams.

To start consider the simple network

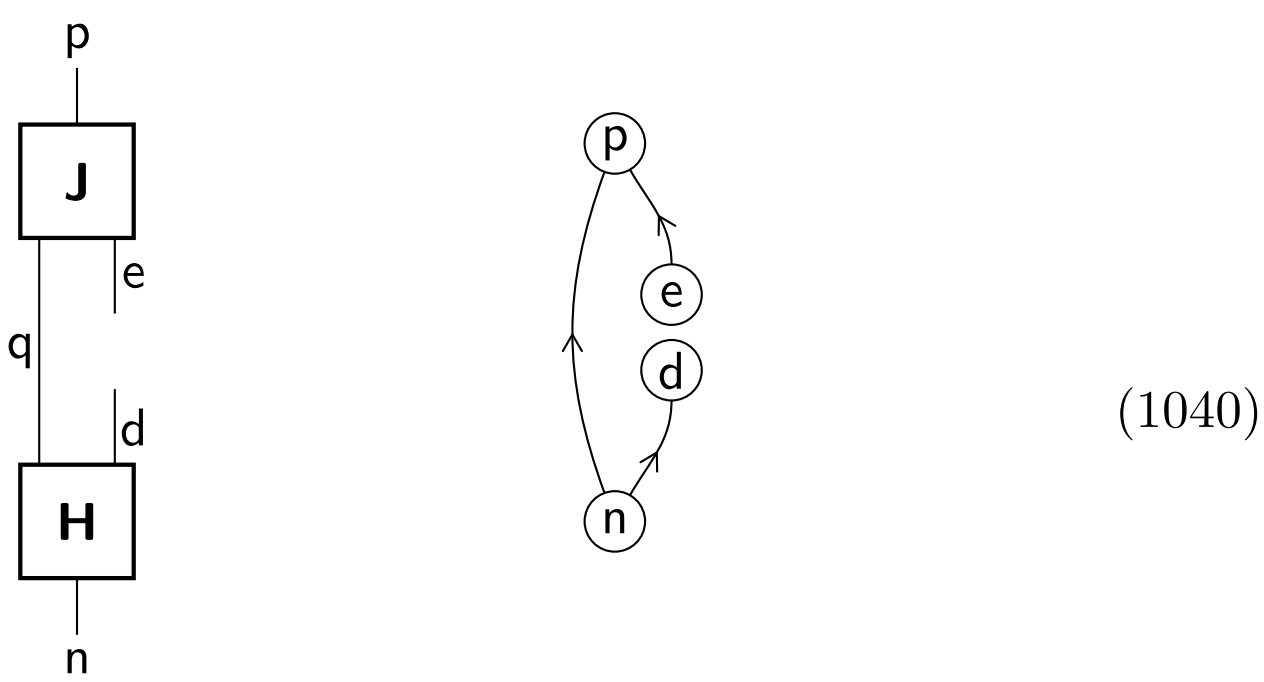

(1040)

where we have shown the causal diagram on the right. This simple network can be joined to the simple network in (1001) at d and e. Thus, the corresponding complex operations also can be joined. We can represent (1040) as a complex operation as follows

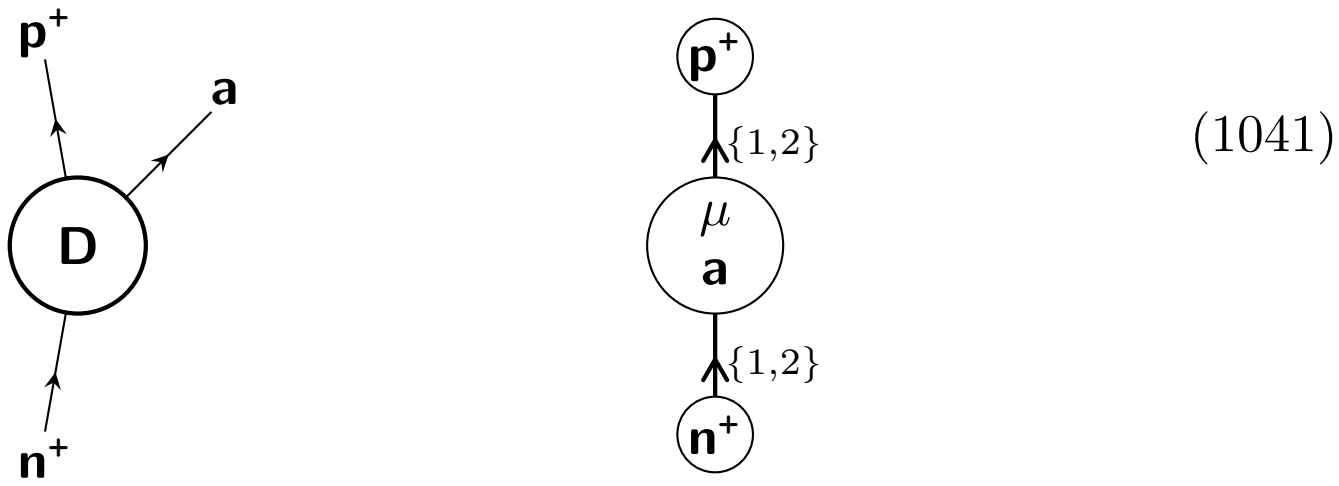

(1041)

where the causal diagram on the left is

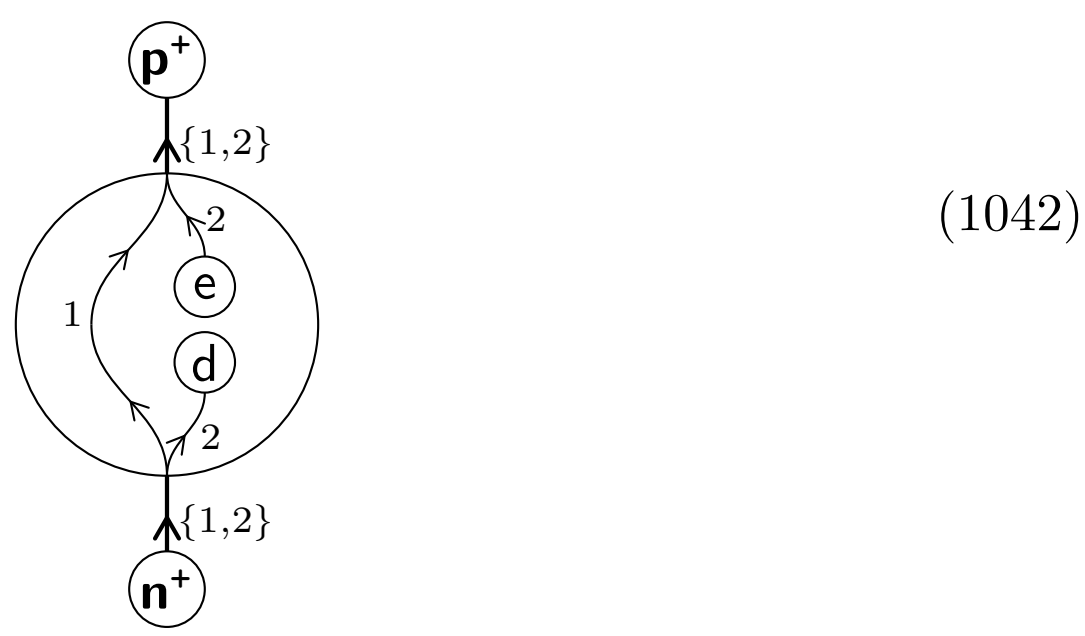

(1042)

showing the implicit causal structure.

Now consider joining **D** in (1041) to **A** in (1036) at **a**. Then we have

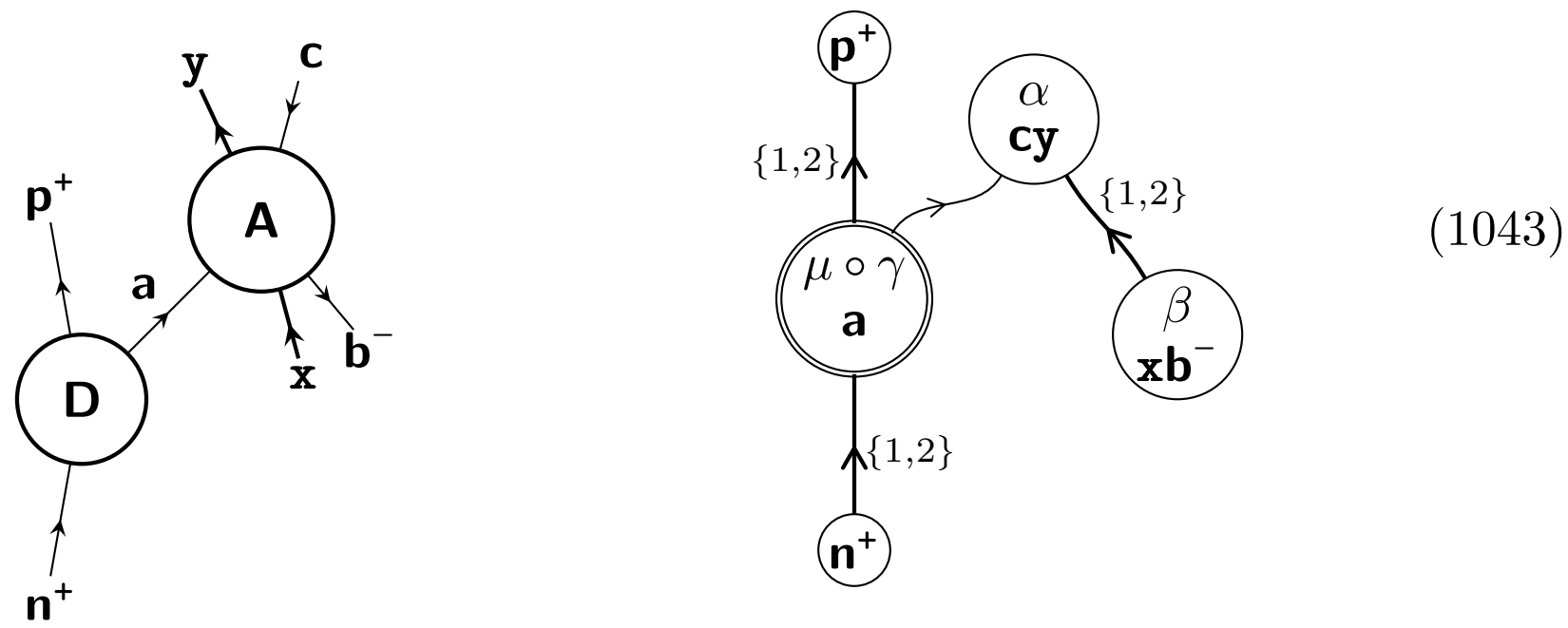

(1043)

The way we join causal diagrams needs some explanation. Since we join the outgoing **a** wire from **D** to the ingoing **a** wire into **A** we need to overlap the nodes in the causal diagram having an **a**. We indicate that these nodes are overlapping by the double border. Since these overlapping nodes have some implicit causal structure we denote the new implicit causal structure in this node by $\mu \circ \gamma$. We will explain in detail below how to calculate the new causal structure in such situations. Ultimately, we want a causal diagram whose nodes correspond only to open wires. Since the **a** wire is now closed, we want to eliminate the double border node.

At the overlapped node in (1043) we are overlapping the following two nodes

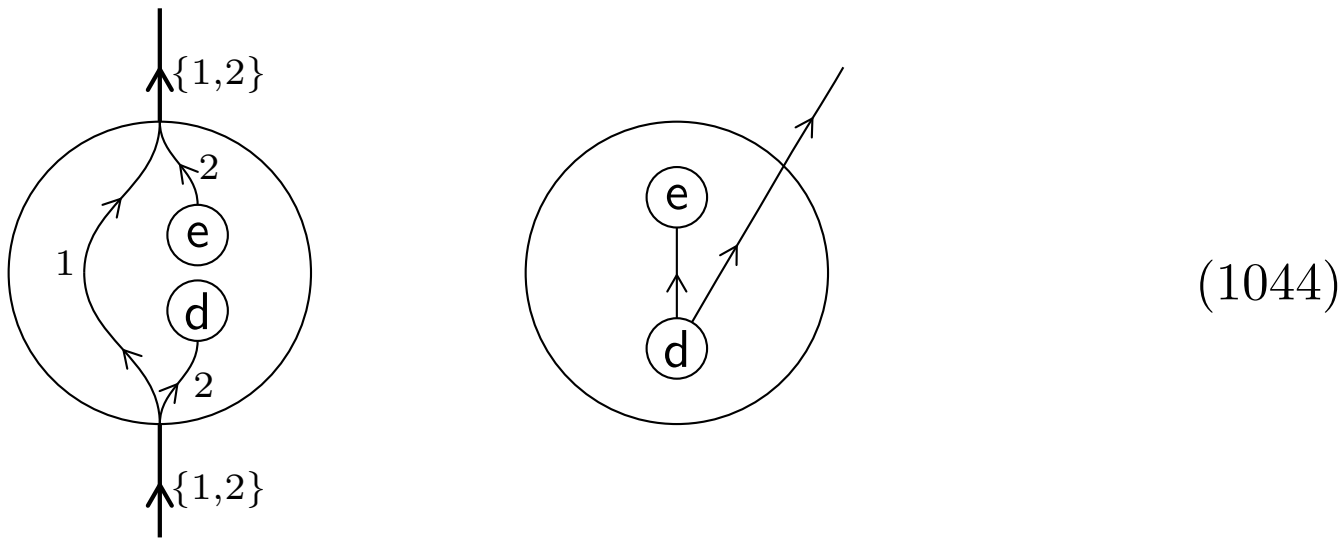

(1044)

where we now show the implicit causal structure. To do this, we literally overlap the nodes on the interior and keep all the causal links giving

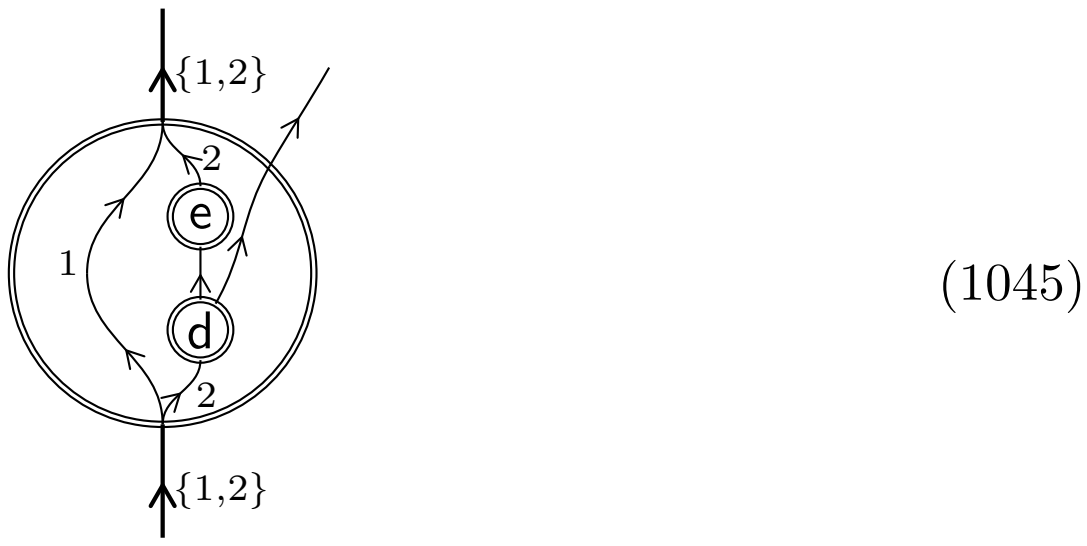

(1045)

Importantly, note that we match − and + type base nodes. We can now simplify this. First, we can drop the large double border circle making the implicit causal structure explicit.

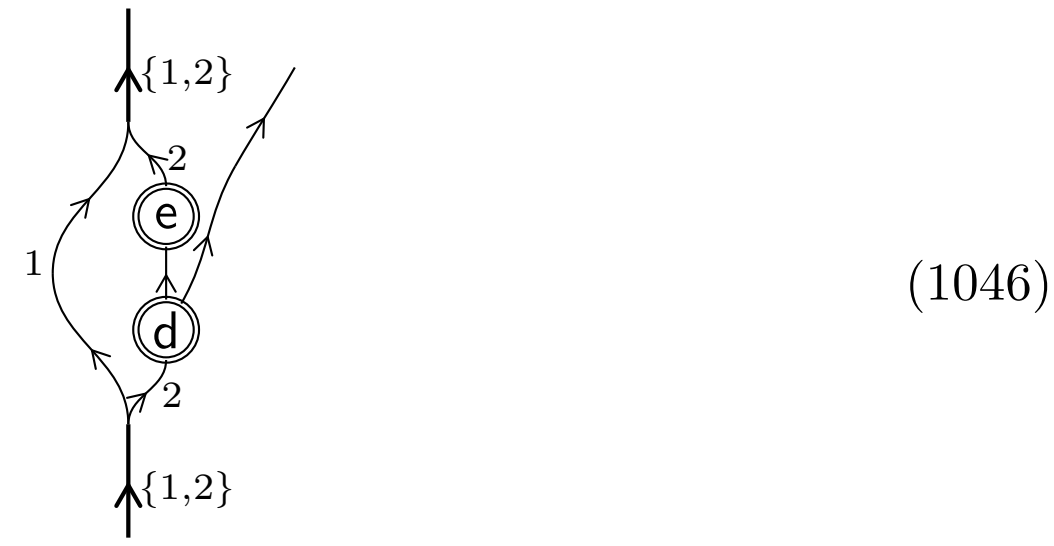

(1046)

Now we have overlapping nodes d and e (corresponding to joining the **a** wire). These overlapping nodes act as a conduit for causal structure and we do not, any longer, care about the system type (d or e) inside. Thus, we write

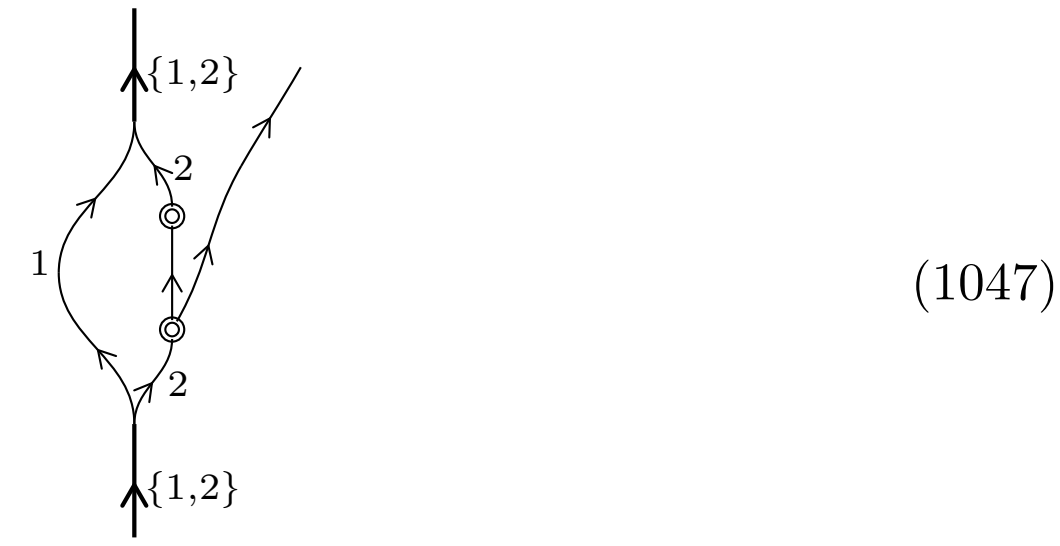

(1047)

To simplify further, we reinsert this into the causal diagram in (1043) giving

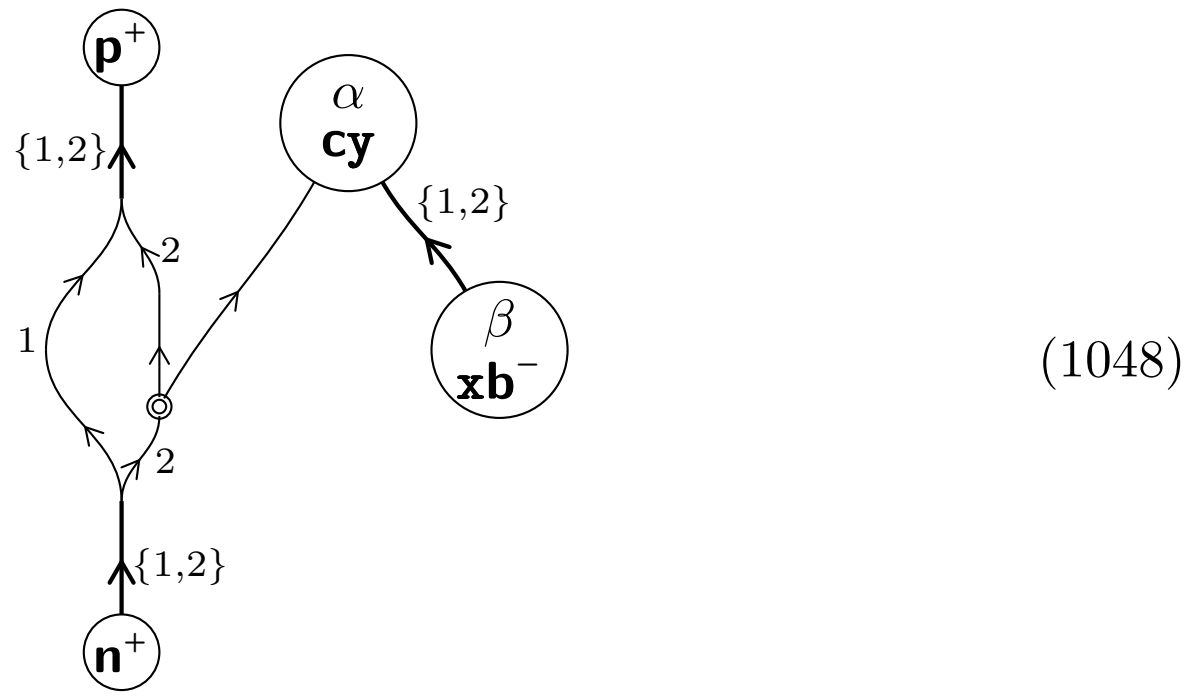

(1048)

We can simplify this further in the same way we simplified (1006) using the causal spider identities as shown in Sec. 47.6 giving

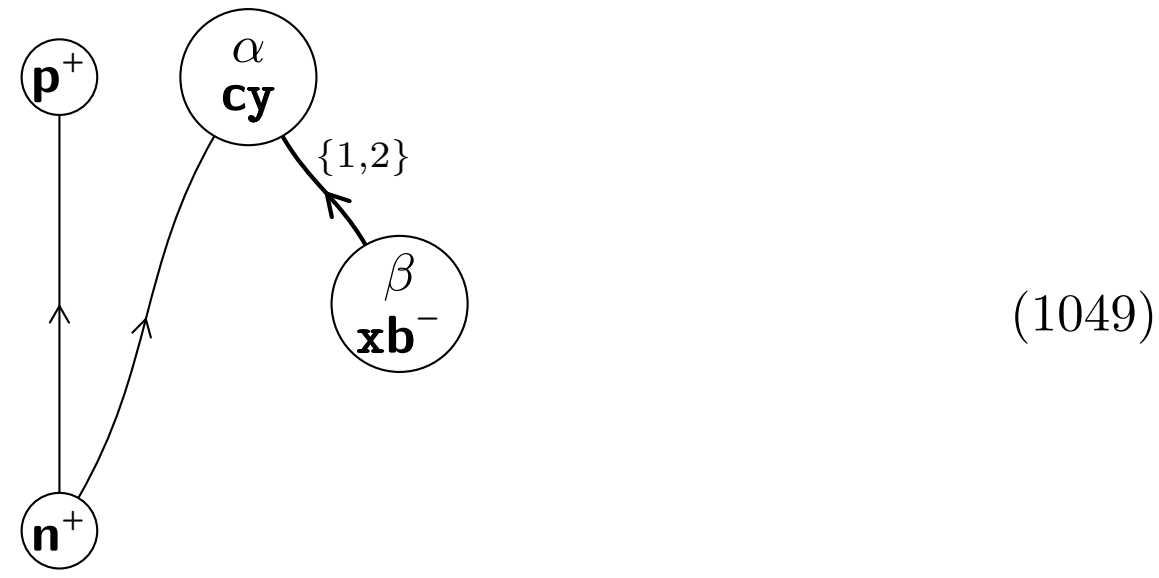

(1049)

If we make the implicit causal structure $\alpha$ and $\beta$ explicit then we get the causal structure in (1007) that we obtain when we join together the simple networks (1001) and (1040) corresponding to complex operations **A** and **D**.

## 47.15 Base representation of operations

We can always give a base representation of a causally complex operation. A base representation is where systems are not composite (and, therefore, are broken up into + and − parts). The resulting *base causal diagram base causal diagram* gives the full causal picture without any implicit causal structure.

For our example of an operation deriving from the simple network in (1001) the base representation is given by

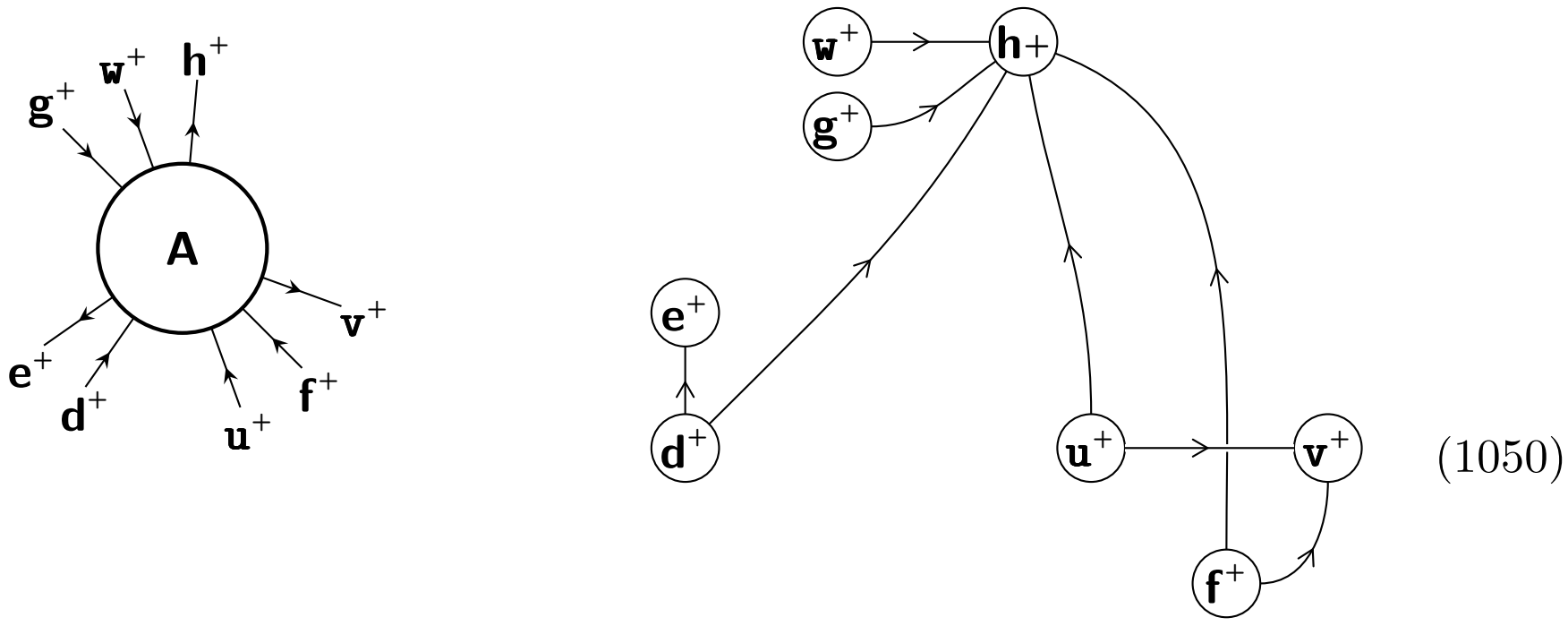

(1050)

where we have set $\mathbf{d}^+ = \mathsf{d}$, $\mathbf{u}^+ = \mathsf{u}$, . . . for the purpose of representing this base operation and its causal diagram. The + superscripts to indicate that these systems move in the direction of the arrow on the wire in the base operation. This is different from the ± introduced in Sec. 47.3 to denote the two types of base node. For example, the $\mathbf{d}^+$ node in the above causal diagram corresponds to an input and so is a − type node.

## 47.16 Simple causal structure, simple form, and synchronous partitions

One type of causal structure that plays an important role in our considerations is *simple causal structure.*

> **Simple causal structure.** We say we have simple causal structure if the base causal diagram has the property that every output is to the future of every input (in the sense that there is an arrow from every input to every output).

As we saw in Sec. 47.15 above, base representation is where our systems are not composite and are broken up into + and − parts (so every node is a base node).

For example, if the base causal diagram is

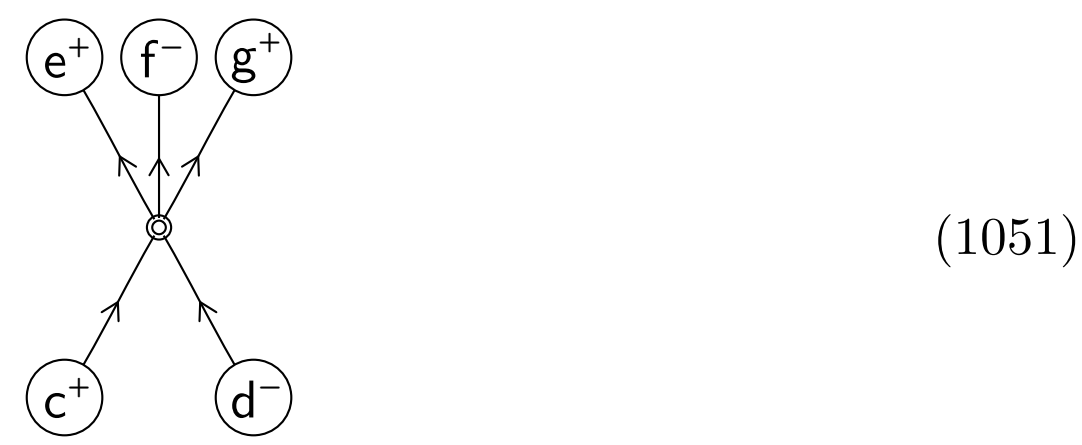

(1051)

then we have simple causal structure. The base nodes with arrows pointing out of them are input base nodes base nodes with arrows pointing into them are output base nodes. Note that, as required for simple causal structure, we have a causal arrow from every input base node to every output base node.

It is useful to be able to notate simple causal structure when the causal diagram is not in base form. If a node is not a base node then, in general, it has some implicit causal structure which we designate by $\alpha$, $\beta$, ... We adopt notation whereby we omit this designation the following simple cases

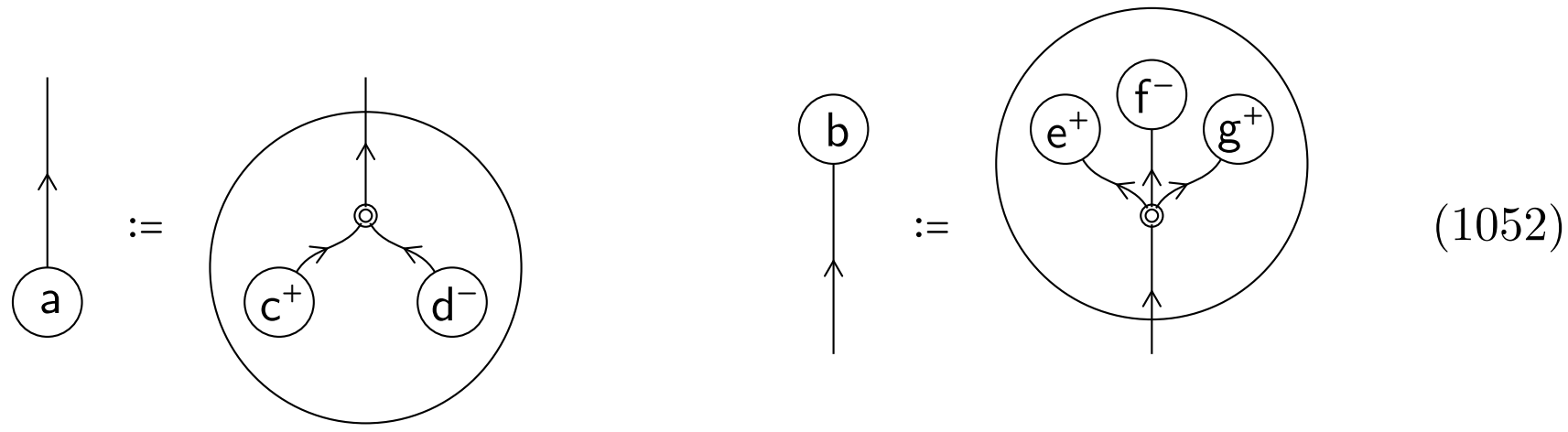

(1052)

On the left all base nodes in the big circle are inputs and on the right they are outputs. The notation on the left generalises in the obvious way for arbitrary composite systems made out of input base nodes (such that base node has a causal arrow pointing to the ⊚ node). This generalisation works when we have both physical and pointer systems (e.g. **ax**). Similarly, the notation on the right generalises in the obvious way for composite systems made out of output base nodes. Using this notation we can write the simple causal structure in (1051) as

(1053)

Note that, using the spider tango identity (1013), this gives (1051). In general, we may have pointer systems also. In this case we can write any simple causal structure as

(1054)

With the above mentioned generalisations of the notation in (1052), we can write any simple causal diagram in this form.

Most causal diagrams will not have simple causal structure. However, any causal diagram can be put into what we will call *simple form*.

> **Simple form of causal diagram.** The simple form of a causal diagram is
>
> 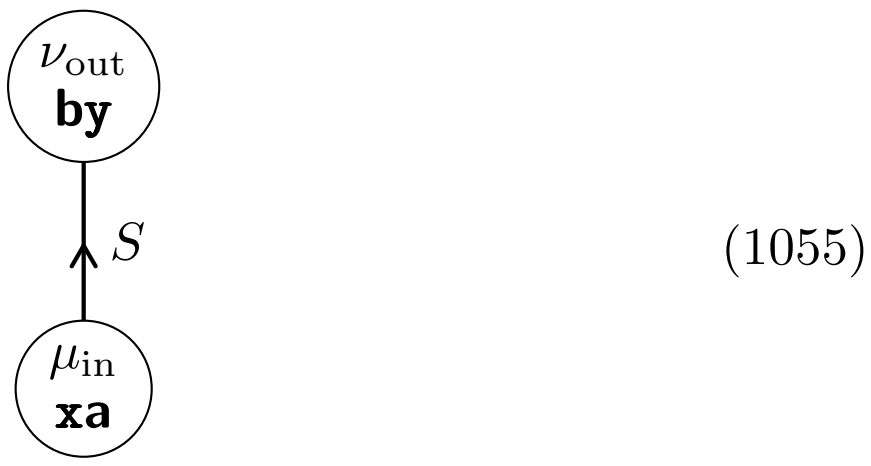
>
>
> (1055)
>
> where $\mu_{\text{in}}$ indicates that the implicit causal structure for **ax** in the lower circle consists of only inputs, and $\nu_{\text{out}}$ indicates that the implicit causal structure for **by** consists of only outputs.

As an example, consider the causal diagram

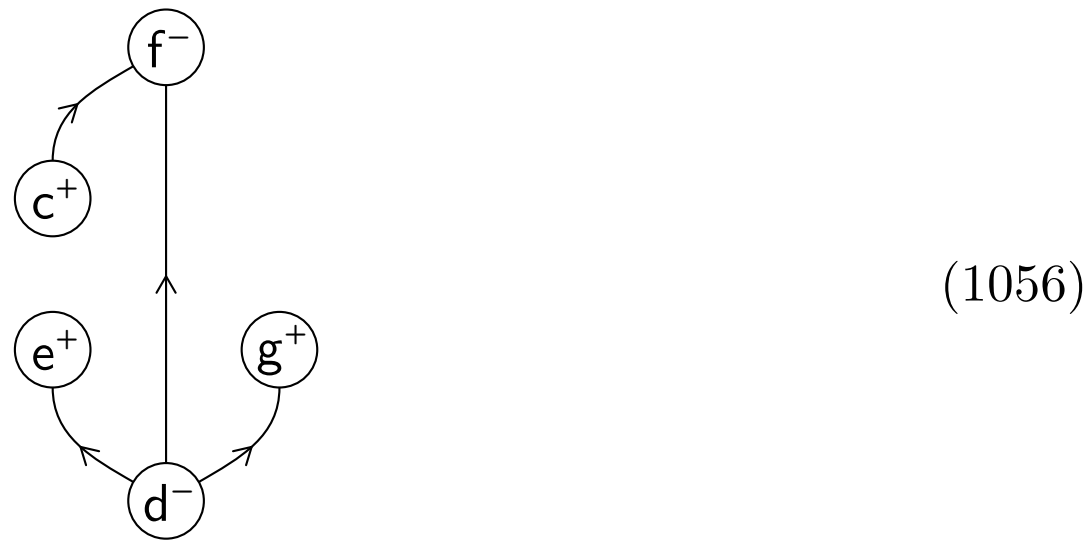

(1056)

We can put this in simple form as follows

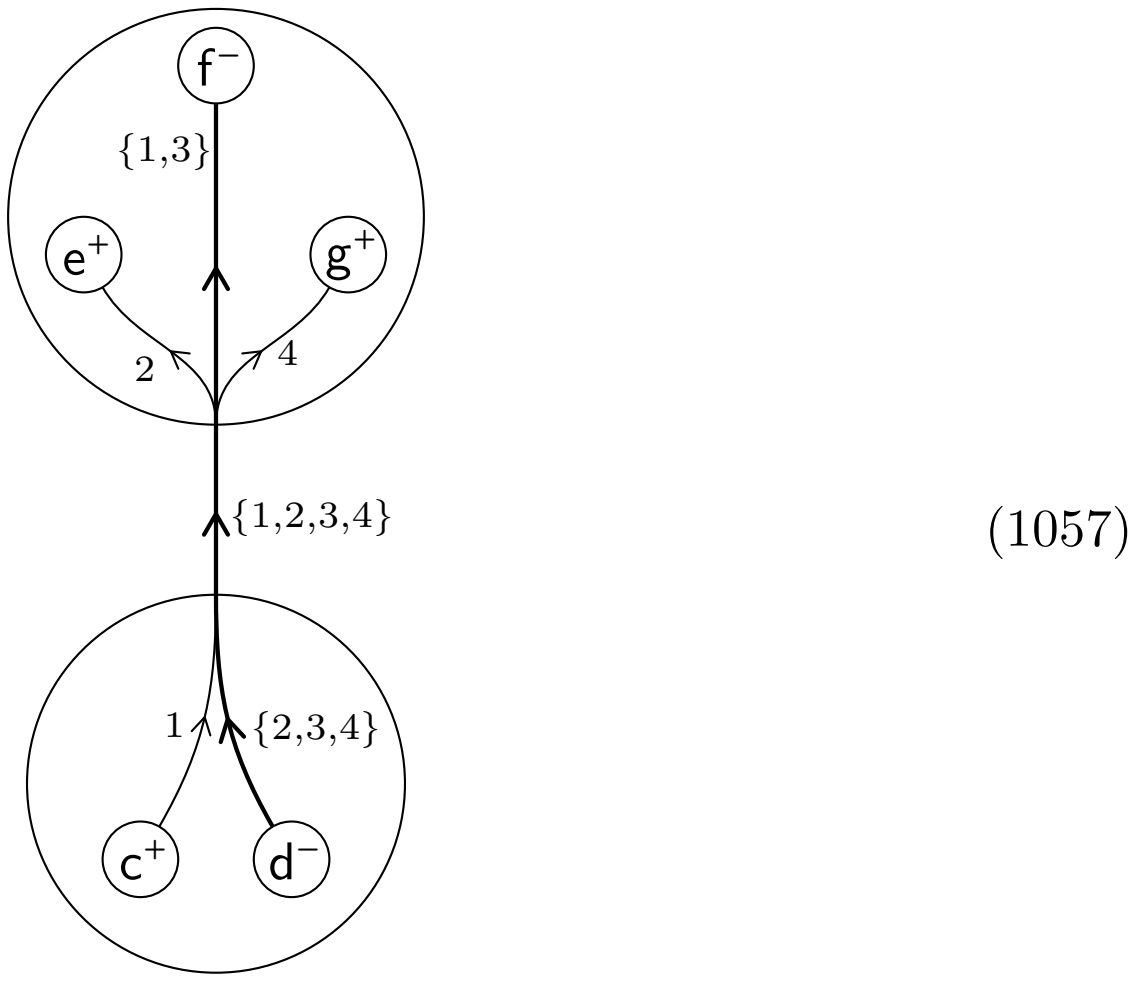

(1057)

We can think of simple form as being where we pull all the inputs down and all the outputs up. The fact that we can always put a causal diagram in simple form relates to the fact that any causal diagram can be obtained from a simple causal diagram by deleting arrows (it can be verified that (1057) is obtained from (1051) by deleting the arrow from $\mathbf{c}^+$ to $\mathbf{e}^+$ and the arrow from $\mathbf{c}^+$ to $\mathbf{g}^+$). If we have an operation of the form

(1058)

then the simple form of the causal diagram is

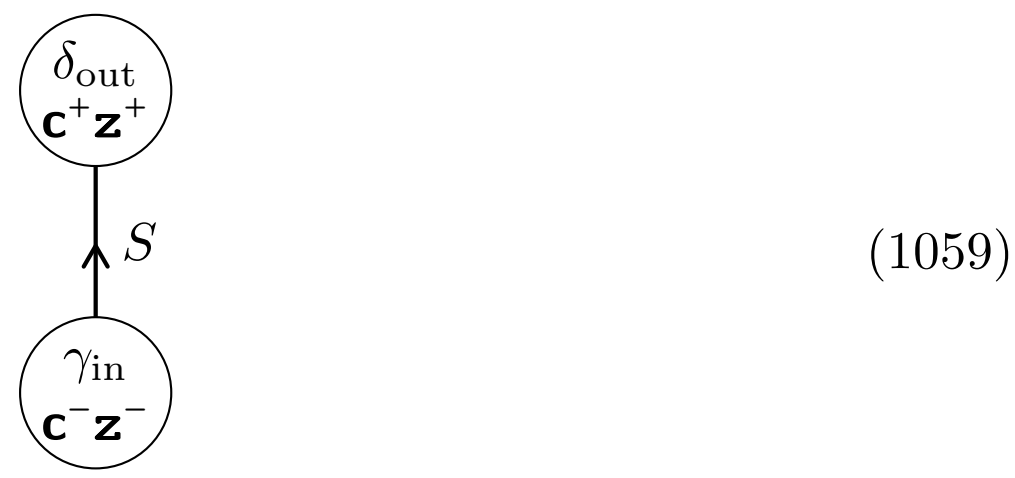

(1059)

All inputs are in the earlier node and all outputs are in the later node.

Finally, we introduce the idea of a *synchronous partition*.

> **Synchronous partition.** A synchronous partition is a partition of the causal diagram into two parts, of the form
>
> 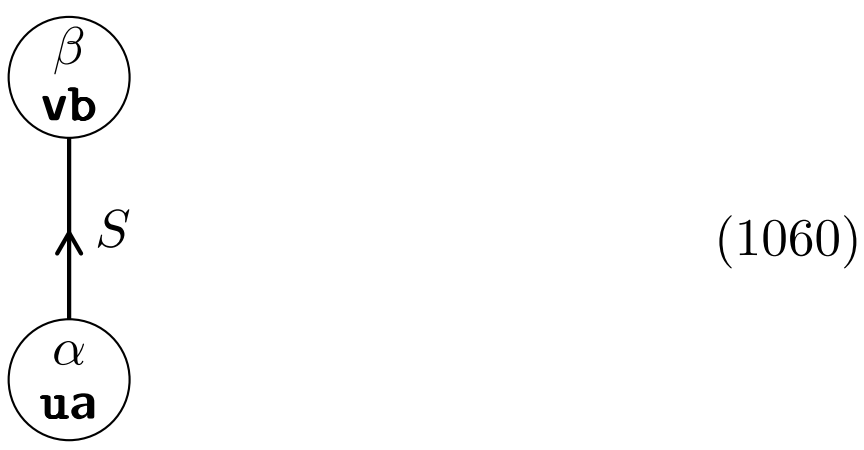
>
>
> (1060)
>
> such that one part can be regarded as being to the future of the other (so any arrows connecting the two parts go from the past to the future part).

Here is an example

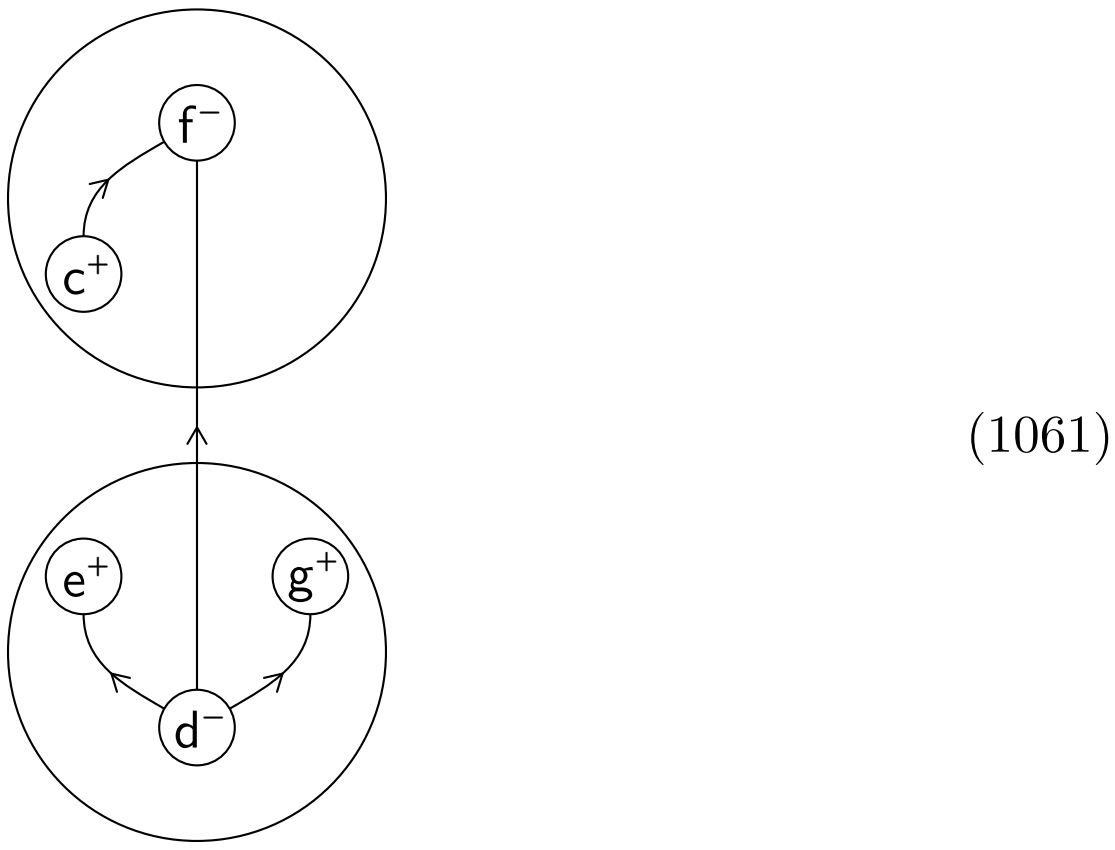

Note that, for a general synchronous partition, both the earlier and later parts can have both input nodes and output nodes in them (as in the above example). In general, a causal diagram will have multiple synchronous partitions. Different synchronous partitions can be thought of as the analogue of spacelike hypersurfaces since they partition the causal diagram into past and future parts. We will often denote a synchronous partition as follows

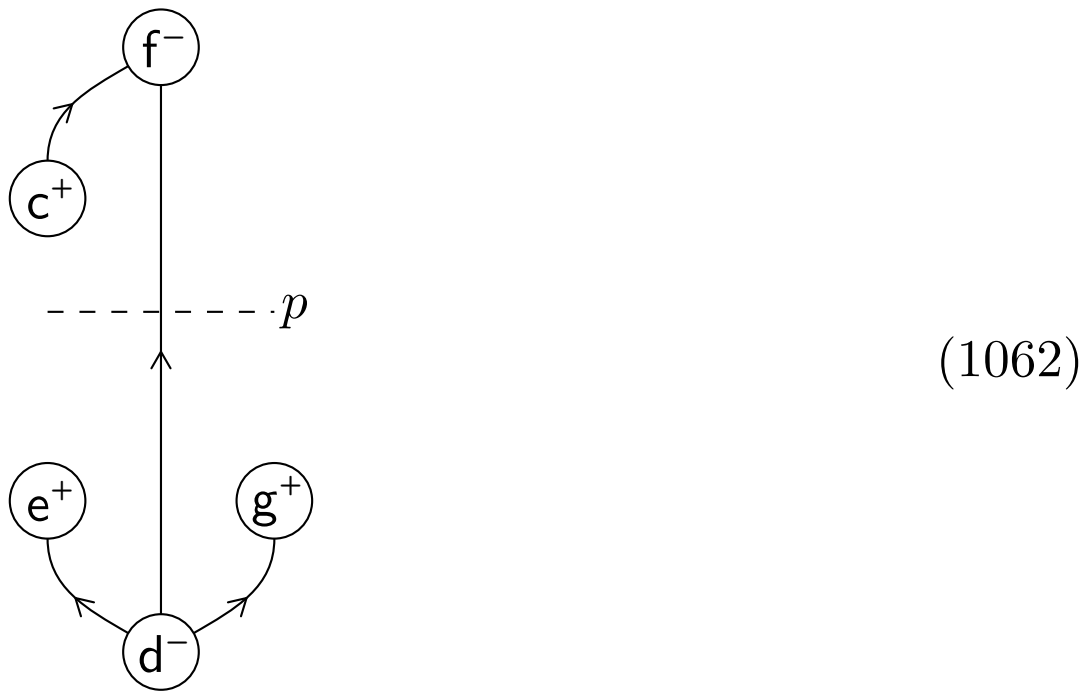

where the line, $p$, cut through the wires associated with the partition.

## 47.17 Preparations and results

A preparation is an operation with only outputs and outcomes. For example

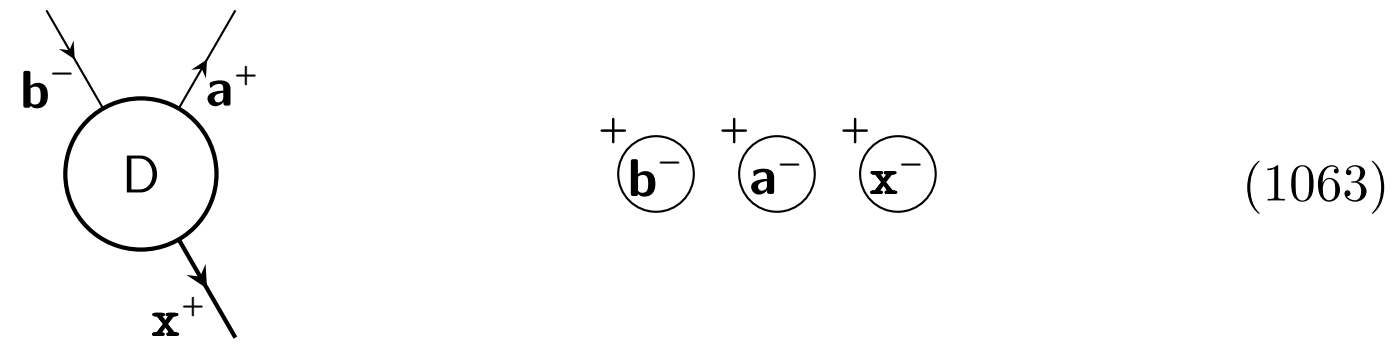

The causal diagram is shown on the right. If it has only system outputs then it is called a system preparation. If it has only pointer outcomes then it is called a pointer preparation.

A result is an operation with only inputs and incomes. For example

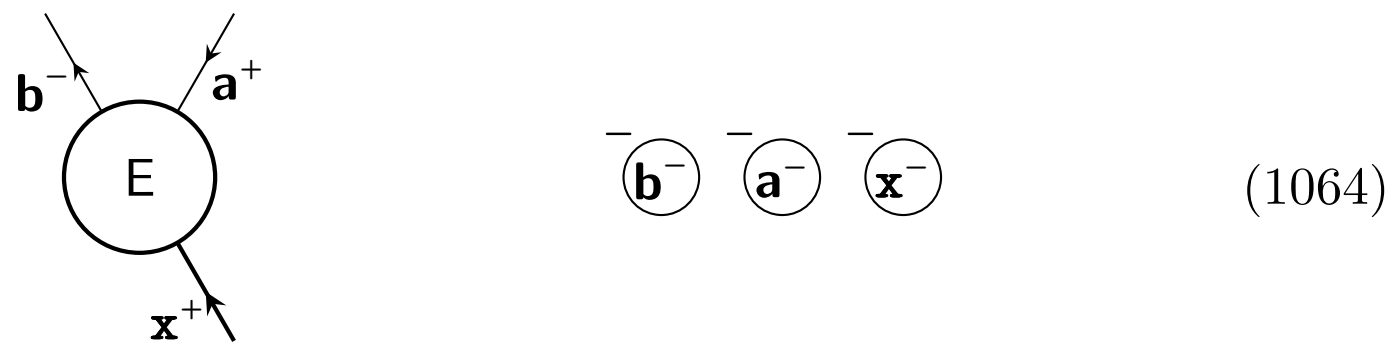

(1064)

The causal diagram is shown on the right. If it has only system inputs then it is called a system result. If it has only pointer incomes then it is called a pointer preparation.

## 47.18 Readout boxes

We represent a readout box as follows

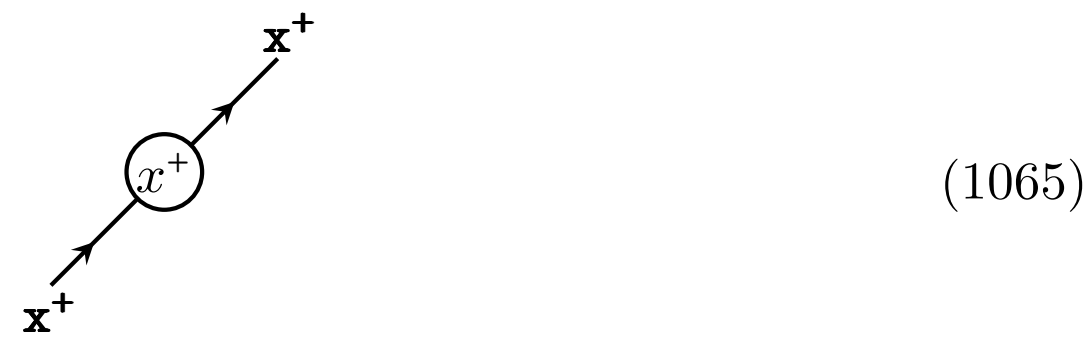

(1065)

This is where we read $x^+$ off the $\mathbf{x}^+$ wire. We can also have

(1066)

This is where we read $x^-$ off a $\mathbf{x}^-$ wire. This pointer system is moving in the opposite direction to the arrow. We can combine these two to have

(1067)

where $x = (x^+, x^-)$ (we can write this as $x = x^+x^-$). In this readout box, we read off both $x^+$ and $x^-$ from the $\mathbf{x}$ wire.

Readout boxes have the property

$$\mathbf{x} \rightarrow (x) \xrightarrow{\mathbf{x}} (x') \rightarrow \mathbf{x} \quad \equiv \quad \begin{cases} \mathbf{x} \rightarrow (x) \rightarrow \mathbf{x} & \text{if } x = x' \\ \mathbf{x} \rightarrow (0) \rightarrow \mathbf{x} & \text{if } x \neq x' \end{cases} \tag{1068}$$

(compare with (43)). The 0 box has the property that, if added to any circuit, then that circuit has probability zero. We have similar properties when we replace $x$ by $x^{\pm}$ and $\mathbf{x}$ by $\mathbf{x}^{\pm}$.

We provide the causal diagram for readout boxes. We have

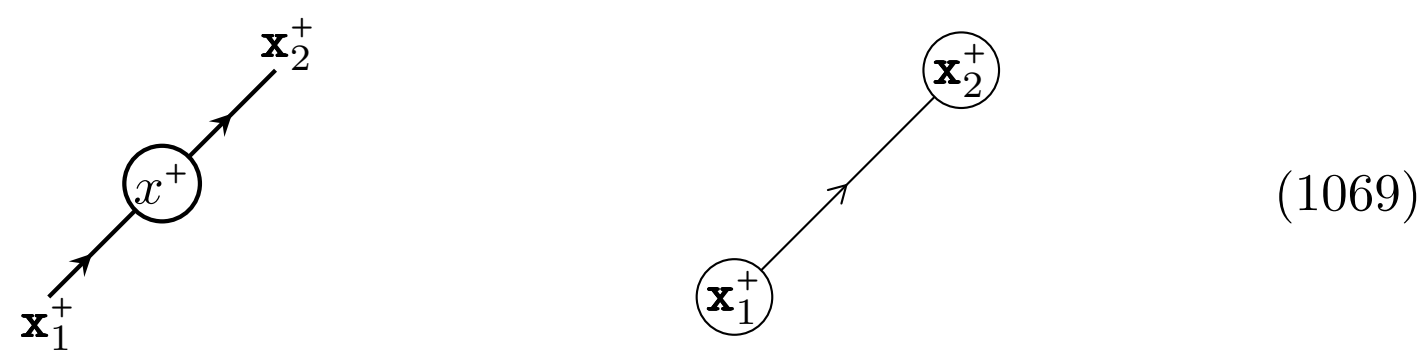

$$(1069)$$

where the readout box on the left has the causal diagram on the right. Note that we have used integer subscripts since we have repeated types and we need to know how the operation on the left relates to the causal diagram on the right. Further, we have

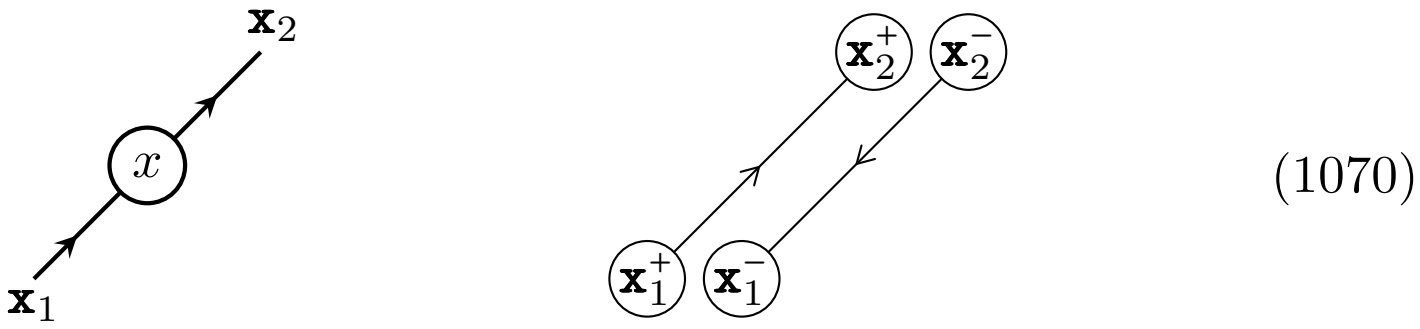

$$(1070)$$

for the more compact notation introduced in (1067).

# 48 The Operational Equivalence Formalism

## 48.1 Probabilities for circuits

As in the simple case, in the complex case we have

> **Circuit probability assumption.** Every circuit (wherein there are no open wires) has a probability associated with it that depends only on the specification of that circuit.

By similar reasoning to that in Sec. 6.1 we can prove that probabilities for disjoint circuits factorise. Further, we assume probabilities are real, non-negative, not greater than one, and that deterministic circuits have probability equal to one.

## 48.2 The $p(\cdot)$ function and equivalence

As in Sec. 6.2 we define a $p(\cdot)$ function as the linear extension of the prob$(\cdot)$ function such that

$$p(\alpha\mathsf{A} + \beta\mathsf{B} + \gamma\mathsf{C} + \dots) = \alpha\text{prob}(\mathsf{A}) + \beta\text{prob}(\mathsf{B}) + \gamma\text{prob}(\mathsf{C}) + \dots \tag{1071}$$

for circuits, $\mathsf{A}$, $\mathsf{B}$, $\mathsf{C}$, dots. Here $\alpha$, $\beta$, $\gamma$, ... are real numbers.

An operation complements another operation or network if the two can be joined together to create a circuit. This requires that their causal diagrams complement each other appropriately.

As in Sec. 6.2, we can use the $p(\cdot)$ function to define equivalence between expressions of the form

$$\text{exprn} = \alpha + \beta \mathsf{E}_{\mathbf{x}_1}^{\mathbf{a}_2} + \gamma \mathsf{F}_{\mathbf{x}_1}^{\mathbf{a}_2} + \delta \mathsf{G}_{\mathbf{x}_1}^{\mathbf{a}_2} + \dots \tag{1072}$$

Two such expressions, $\text{exprn}_1$ and $\text{exprn}_2$, are equivalent when

$$p(\text{exprn}_1 \mathsf{H}) = p(\text{exprn}_2 \mathsf{H}) \quad \text{for all complement operations} \quad \mathsf{H} \tag{1073}$$

We write this equivalence as

$$\text{exprn}_1 \equiv \text{exprn}_2 \tag{1074}$$

As in Sec. 6.2, a circuit is equivalent to its own probability.

## 48.3 The $p(\cdot)$ function and inequality

As in Sec. 6.3, can use the $p(\cdot)$ function to define a notion of inequality as follows. We say

$$\text{exprn}_1 \leqq \text{exprn}_2 \tag{1075}$$

iff

$$p(\text{exprn}_1 \mathsf{H}) \leq p(\text{exprn}_2 \mathsf{H}) \quad \text{for all complement operations} \quad \mathsf{H} \tag{1076}$$

This is a very strong form of inequality since it requires we look at all complement operations.

In Sec. 49.2.3 (see also Sec. 49.2.2) we will define a weaker notion of inequality (denoted $\leqq_T$) with respect to testers which is sufficient for our purposes.

## 48.4 What we mean by time symmetry

In Sec. 6.4 we discussed how to implement time symmetry in the case of causally simple operational probabilistic theories. Here we will discuss how to do this in the causally complex case. Since we have set up our notation in a different way (with arrows and system types having a +ve and a −ve part) we implement time symmetry in a different way as follows

1. Every allowed operation, $\mathsf{B}$, is associated with a time reversed operation, $\utilde{\mathsf{B}}$, which is also allowed. The time reverse operation has all attached arrows reversed and the associated causal diagram has the arrows reversed.

For example

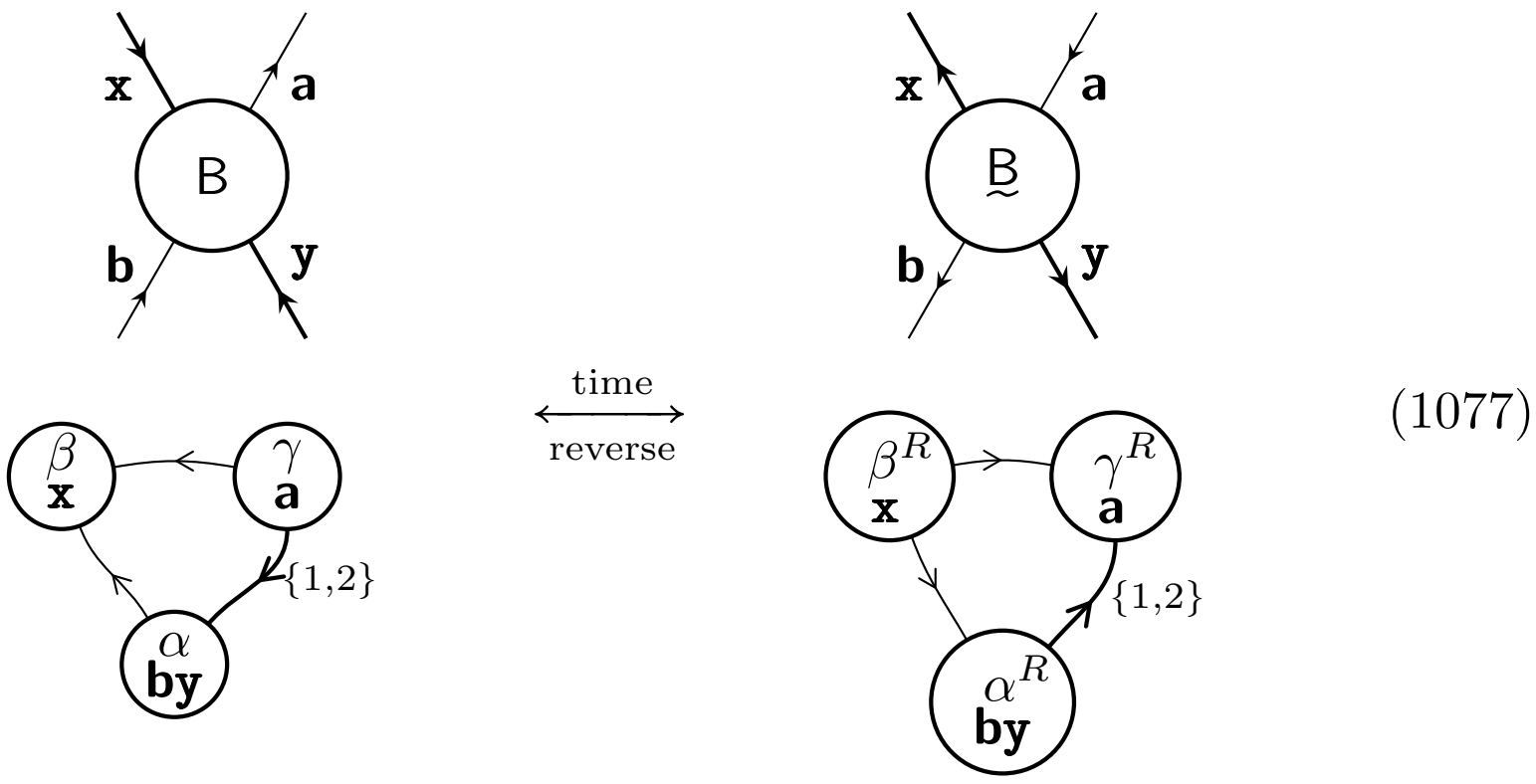

(1077)

Note how all the arrows are reversed. The notation $\alpha^R$ indicates we have reversed the arrows in $\alpha$.

2. The time reverse of a readout box is given by reversing the arrows

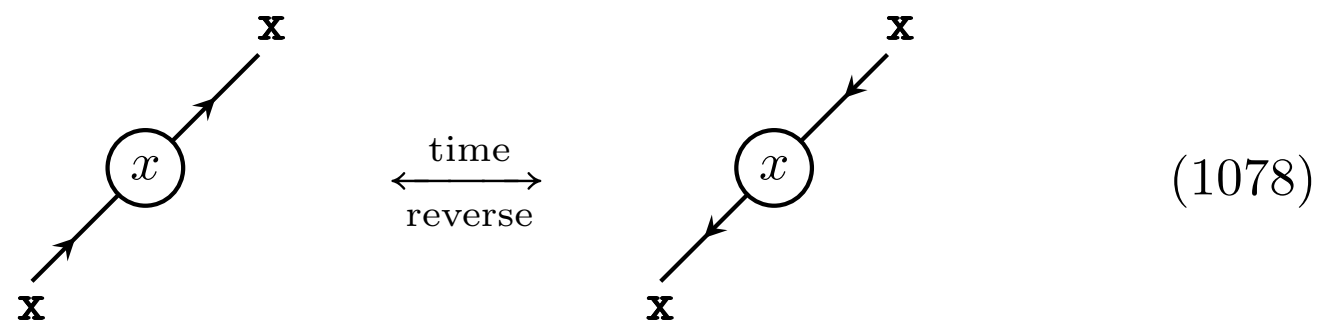

(1078)

3. We can obtain the time reverse of any circuit by reversing the direction of all the arrows and placing a tilde under operation symbol. We demand that the probability of the time reversed circuit is equal to the probability of the original circuit. For example

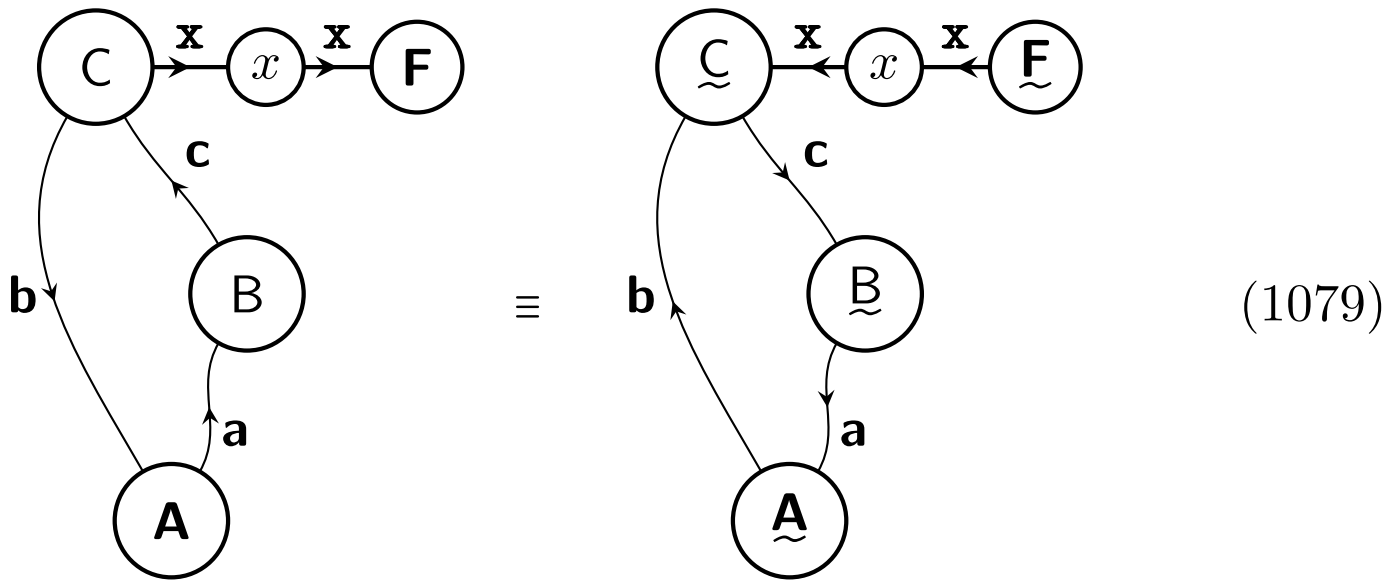

(1079)

(recall that equivalence for circuits implies equal probabilities).

Unlike in the causally simple case, we do not have to rotate the circuit through 180° to implement the time reversal. The absolute time direction of any given +ve or −ve part of any open wire can be read off the causal diagram (though we would need to know something of the detail of the implicit causal structure, $\alpha$, $\beta$, or $\gamma$ in the example above).

## 48.5 Deterministic pointer boxes

As in the simple case, we reserve the symbol **R** for deterministic pointer boxes. We write $\underset{\sim}{\mathbf{R}}$ (the time reverse) as **R** - this will not lead to any ambiguity since we can look at the wires attached to see the role of the given box.

We represent a deterministic pointer preparation box by

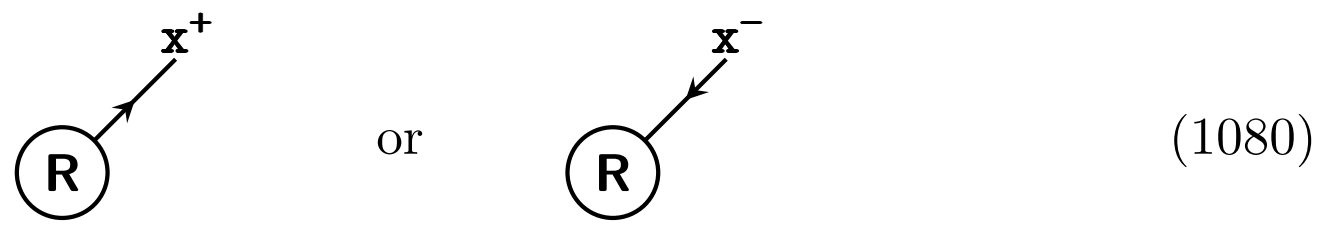

(1080)

and a deterministic pointer result box by

$$\text{R} \; \mathbf{x}^+ \qquad \text{or} \qquad \text{R} \; \mathbf{x}^- \tag{1081}$$

Preparations and results have simple causal structure and, hence, these are the same objects we considered in the simple case in Sec. 7.2 (albeit represented now by a circle rather than a rectangle). It follows from the basic causality principles employed there that these **R** boxes are unique (up to equivalence classes).

We can combine these using the notation

$$\text{R} \; \mathbf{x} \quad \leftrightharpoons \quad \text{R} \; \mathbf{x}^+ \; \text{R} \; \mathbf{x}^- \tag{1082}$$

and

$$\text{R} \; \mathbf{x} \quad \leftrightharpoons \quad \text{R} \; \mathbf{x}^+ \; \text{R} \; \mathbf{x}^- \tag{1083}$$

The objects in (1082, 1083) are deterministic but they are not preparations or results. There are other deterministic operations having a pointer system **x** like this. However, as we will discuss below, if we restrict to the operations having the same causal diagram, then it is reasonable to assume these objects are unique.

This more compact notation is useful. For example, the network from the simple case

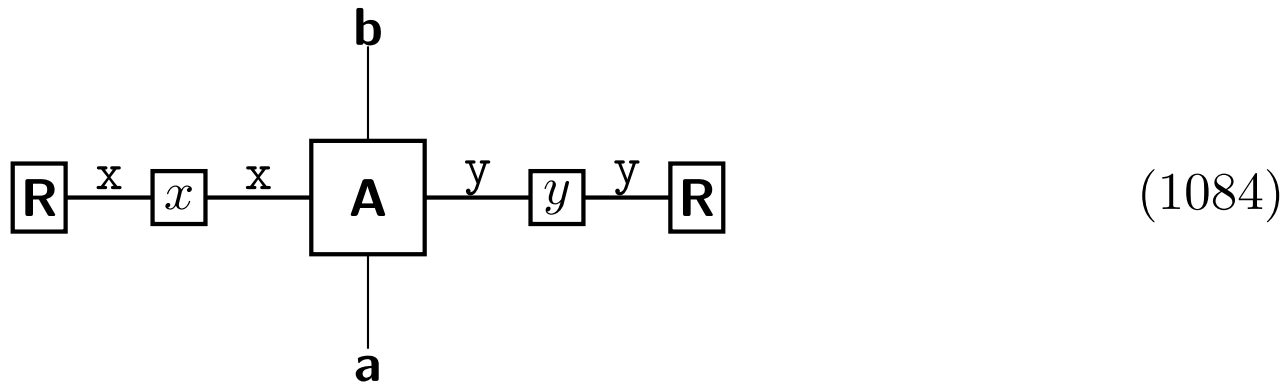

(1084)

can be written as

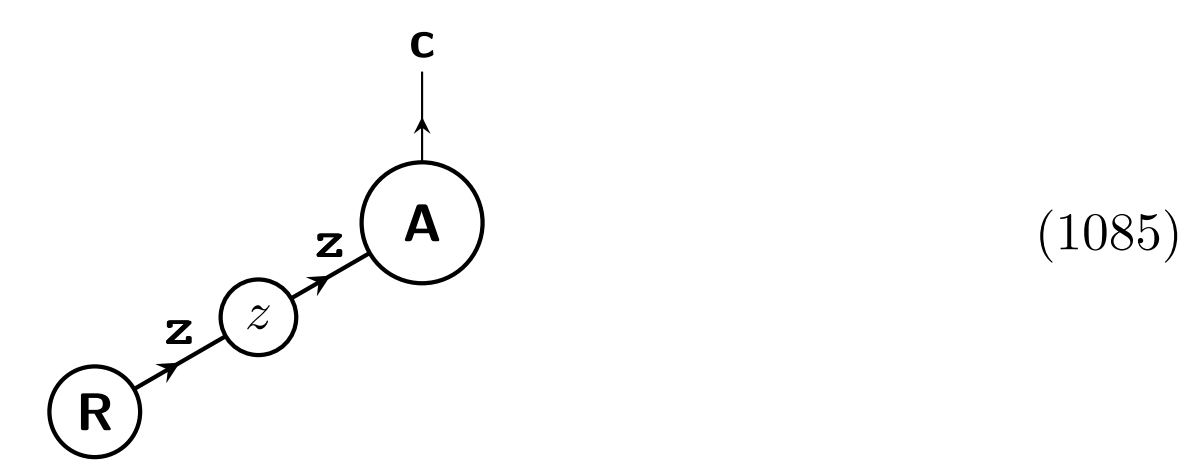

(1085)

where $\mathbf{z}^+ = \mathsf{x}$, $\mathbf{z}^- = \mathsf{y}$, $\mathbf{c}^+ = \mathsf{b}$, and $\mathbf{c}^- = \mathsf{a}$.

We must have

$$\mathbf{R} \xrightarrow{\mathbf{x}^\pm} \mathbf{R} \equiv 1 \qquad\qquad \mathbf{R} \xrightarrow{\mathbf{x}} \mathbf{R} \equiv 1 \qquad (1086)$$

since these **R** boxes are deterministic.

It follows from uniqueness (following the reasoning in Sec. 7.2) that **R** preparations and results factorize for composite systems. In our current notation this property can be written as

$$\mathbf{R} \xrightarrow{\mathbf{x}^\pm\mathbf{y}^\pm} \quad \leftrightharpoons \quad \mathbf{R} \xrightarrow{\mathbf{x}^\pm} \; \mathbf{R} \xrightarrow{\mathbf{y}^\pm} \qquad (1087)$$

Hence we have

$$\mathbf{R} \xrightarrow{\mathbf{xy}} \quad \leftrightharpoons \quad \mathbf{R} \xrightarrow{\mathbf{x}} \; \mathbf{R} \xrightarrow{\mathbf{y}} \qquad (1088)$$

for the compact notation.

Flat pointer boxes have very simple causal diagrams. For example we have

$$\mathbf{R} \xrightarrow{\mathbf{x}^+} \qquad\qquad (\mathbf{x}^+)^+ \qquad (1089)$$

since there is just a single wire. Further, we have

$$\mathbf{R} \xrightarrow{\mathbf{x}} \qquad\qquad {}^-(\mathbf{x}^-) \quad (\mathbf{x}^+)^+ \qquad (1090)$$

for the compact notation introduced in (1082). We earlier asked whether **R** boxes in compact notation are unique. The most natural interpretation of this causal diagram is that we have a pointer preparation and a separate pointer result. Therefore, if an operation having this causal diagram is deterministic, then it is reasonable to assume it is unique. We can actually prove this is true if we make the *no correlation without causation assumption* and the *weak tomographic locality assumption* which are introduced later in Sec. 48.9 (see the *operations factorise when their causal diagrams do theorem* in that section).

We also have cause to define these $\mathbf{R}_+$ and $\mathbf{R}_-$ boxes

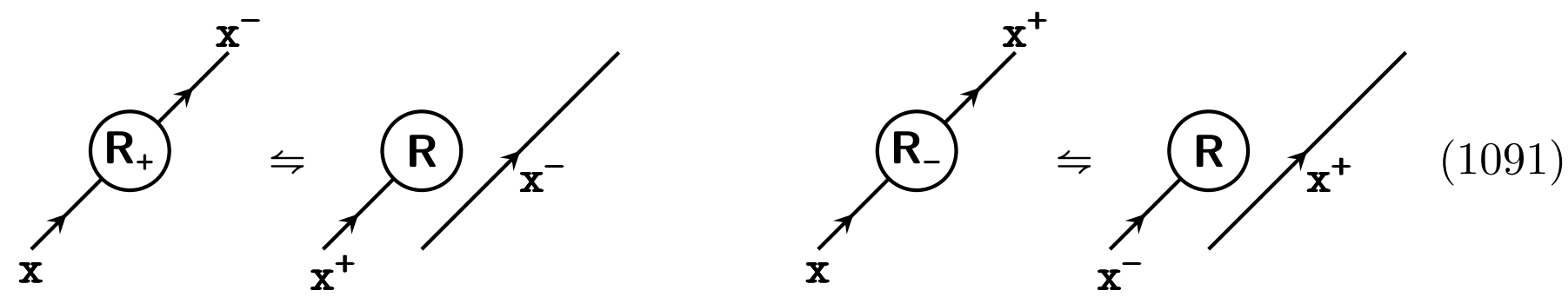

The $\mathbf{R}_+$ operation absorbs $\mathbf{x}^+$ from the in pointing arrow in an $\mathbf{R}$ box while leaving $\mathbf{x}^-$ to pass unhindered. We will use these boxes in stating the causality property. We also define $\mathbf{R}^\pm$ operations as follows

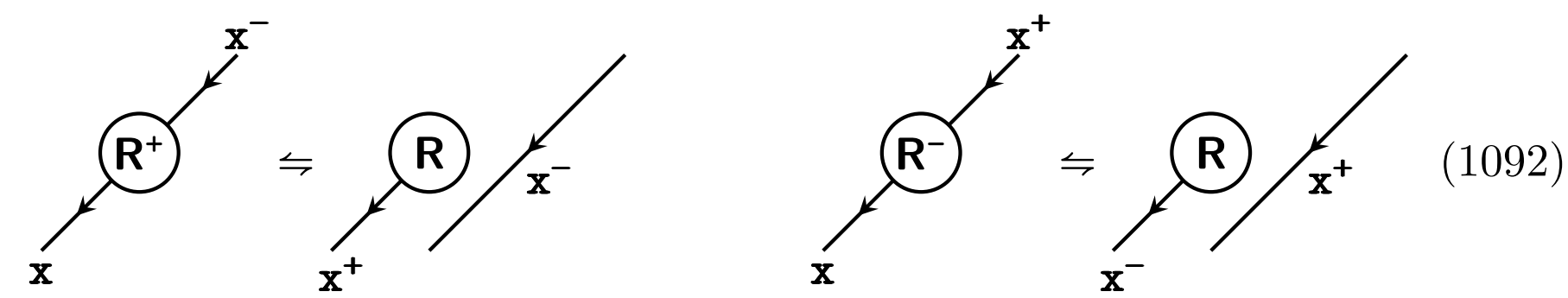

These absorb $\mathbf{a}^\pm$ from the out pointing arrow.

## 48.6 Flatness

In Sec. 7.3 we motivated the flatness assumption by arguing that the time reverse of marginalisation gives a flat distribution over the pointer variable. In our current notation the *flatness assumption* is

$$\equiv \frac{1}{N_{\mathbf{x}^\pm}} \tag{1093}$$

Compare with (85) for the simple case. This assumption becomes

$$\equiv \frac{1}{N_{\mathbf{x}}} \tag{1094}$$

for the compact notation.

As in the simple case, the flatness assumption is necessary to prove some of the theorems that follow.

## 48.7 Ignore boxes

Now we introduce ignore boxes which follow a similar pattern to **R** boxes. We reserve the symbol **I** for ignore boxes. We write $\underset{\sim}{\mathbf{I}}$ (the time reverse) as **I** - this will not lead to any ambiguity since we can look at the wires attached to see the role of the given box.

We define an ignore preparation as

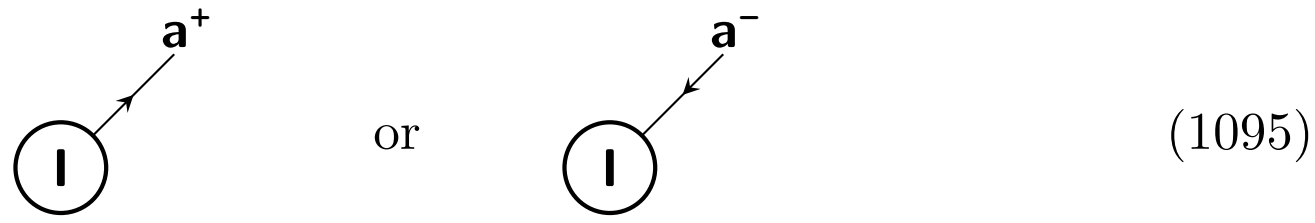

(1095)

and an ignore result as

$$\textsf{I} \nwarrow \mathsf{a}^+ \qquad \text{or} \qquad \textsf{I} \nearrow \mathsf{a}^- \tag{1096}$$

We can combine these as follows

$$\textsf{I} \nearrow \mathsf{a} \qquad \text{or} \qquad \textsf{I} \swarrow \mathsf{a} \tag{1097}$$

where the example on the left is interconvertible with

$$\textsf{I} \nearrow \mathsf{a}^+ \quad \textsf{I} \nearrow \mathsf{a}^- \tag{1098}$$

Ignore boxes have the property

$$\textsf{I} \xrightarrow{\mathsf{a}^\pm} \textsf{I} \ \equiv\ 1 \qquad\qquad \textsf{I} \xrightarrow{\mathsf{a}} \textsf{I} \ \equiv\ 1 \tag{1099}$$

since they are deterministic. Ignore boxes on composite systems factorise as follows

$$\textsf{I} \nearrow \mathsf{ab} \quad \leftrightharpoons \quad \textsf{I} \nearrow \mathsf{a} \quad \textsf{I} \nearrow \mathsf{b} \tag{1100}$$

This follows from the factorisation of ignore boxes in $t$SOPT.

Ignore boxes have very simple causal diagrams. For example we have

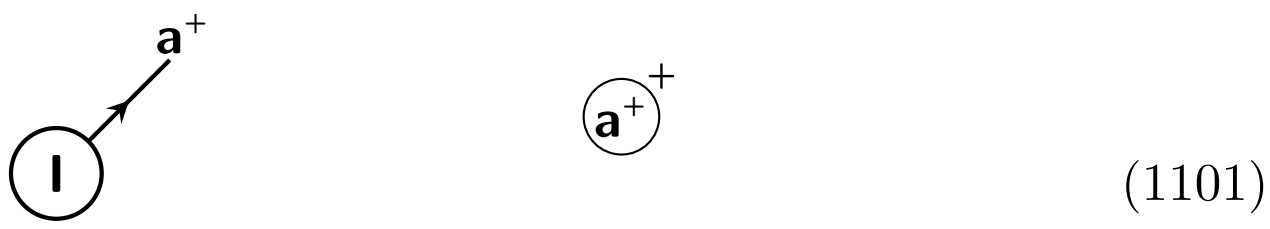

(1101)

since there is just a single wire. Also, we have

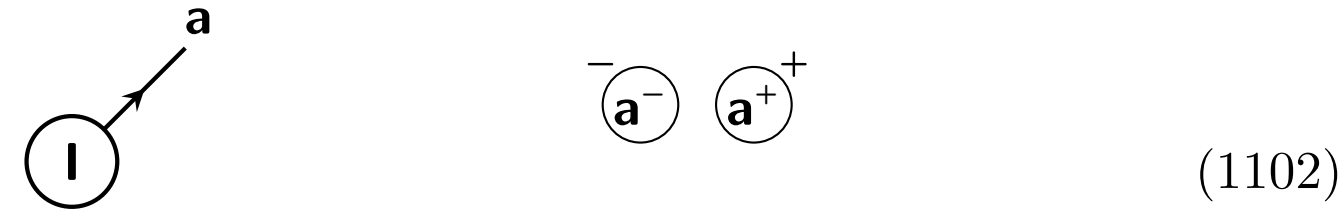

(1102)

for the compact notation introduced in (1097).

We have cause to define $\mathsf{I}_+$ and $\mathsf{I}_-$ boxes as follows

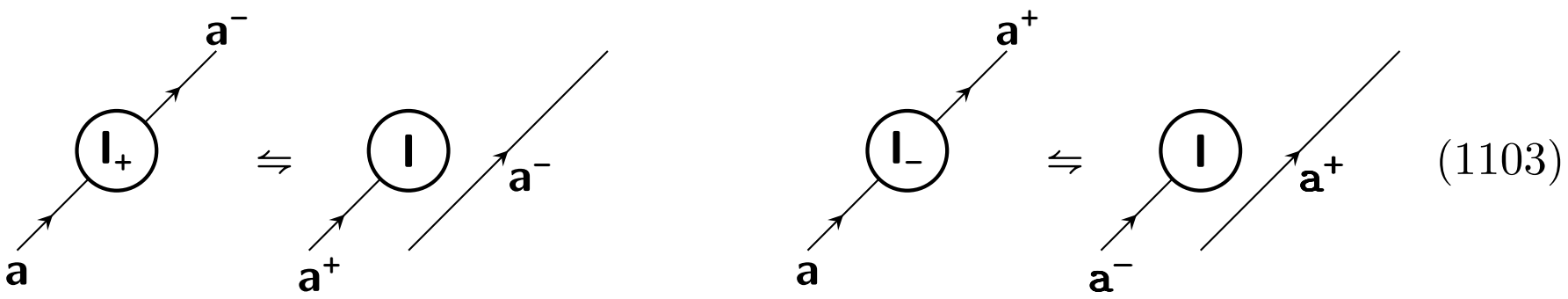

(1103)

The $\mathsf{I}_+$ operation absorbs $\mathbf{a}^+$ from the in pointing arrow. We will use these boxes in stating the causality property. We also define $\mathsf{I}^\pm$ boxes.

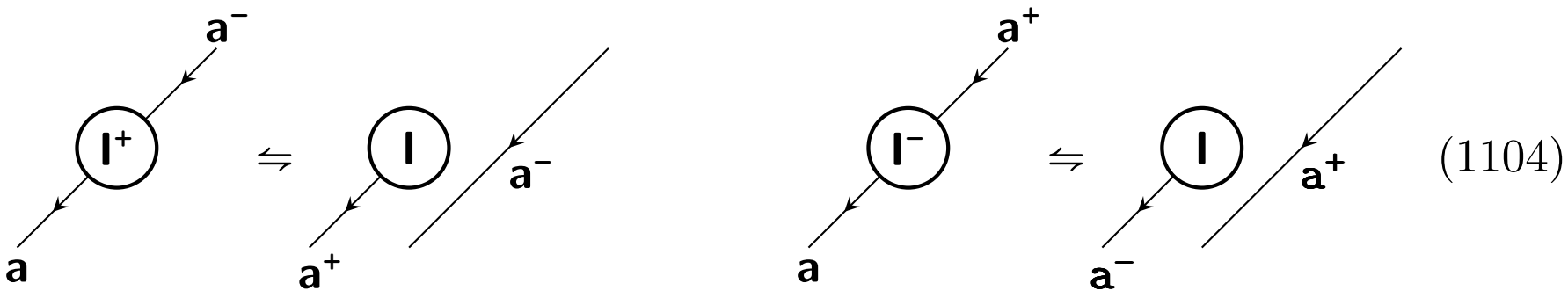

(1104)

The $\mathsf{I}^+$ operation absorbs $\mathbf{a}^+$ from the out pointing arrow.

## 48.8 The determinism test and the physical norm

As in the simple case, a necessary condition for an operation, B, to be deterministic is that it satisfies the determinism test

$$\text{[R} \xleftarrow{\mathbf{x}} \text{B} \xrightarrow{\mathbf{a}} \text{I]} \equiv 1 \qquad (1105)$$

Further, it is useful to define the *physical norm* as

$$|B|_{\text{phys norm}} = \text{prob}\left( \text{R} \xleftarrow{\mathbf{x}} \text{B} \xrightarrow{\mathbf{a}} \text{I} \right) \qquad (1106)$$

Clearly, a necessary condition for B to be deterministic is that it has physical norm equal to 1 so passes the determinism test. We will see in Sec. 49.4.13 that passing the determinism test is also a sufficient condition for an operation to be deterministic if we assume operations satisfy what we will call general double causality.

## 48.9 Disjoint causal diagrams

A special case of interest is when the causal diagram of an operation consists of two (or more) disjoint parts. In such a case it is reasonable to assume that, when we form a circuit by separately addressing the systems associated with these disjoint parts, the associated probabilities factorise. We can make this assumption precise as follows

> **No correlation without causation assumption.** If a deterministic operation has a causal diagram consisting of two disjoint parts, i.e.
>
> 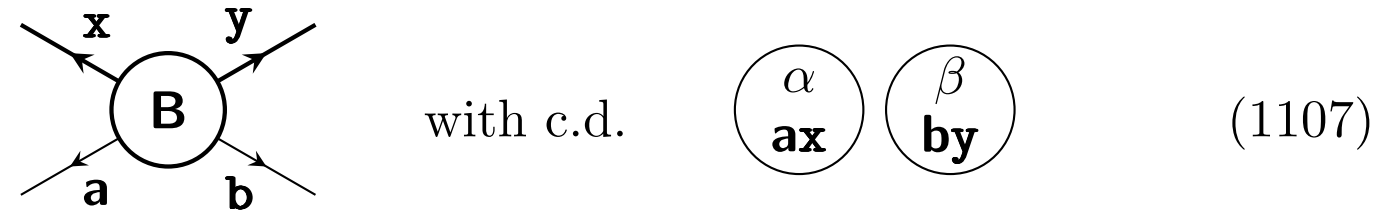
>
>
> then
>
> (1108)
>
> for all A and C.

Note that the equivalence in (1108) implies that the associated probability factorises (since circuits are equivalent to their probabilities). We are, effectively, using the **R** and **I** boxes to provide marginals. Typically, in the case that we have no correlation, the joint probability would be the product of the marginals. Also, note that, A attaches to **xa** which are associated with one disjoint part of the causal diagram of **B** and C attaches to the other disjoint part of **B**'s causal diagram. Thus, there is no possibility of direct cause, common cause, or even postselection between these two parts. In such a circumstance, it is reasonable to expect no correlation This is a reasonable implementation of the sometimes maligned slogan "no correlation without causation". This principle goes back to Reichenbach [1956]. See Beebee et al. [2009] for modern discussion of it. It has been evoked in modern discussions of quantum theory – see Wood and Spekkens [2015]. The above assumption does not preclude inducing a correlation by introducing an operation whose causal diagram connects the two disjoint parts.

For example, it does not impose that the probability for

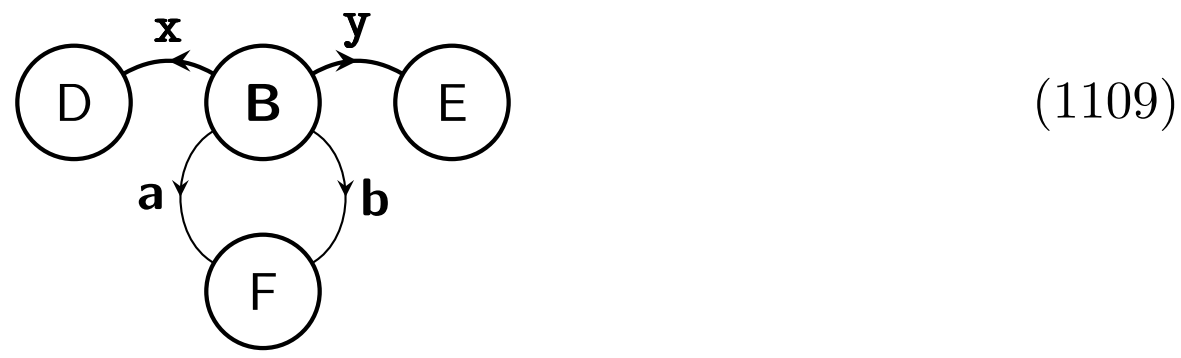

(1109)

factorises. In general, the causal diagram associated with F can connect the disjoint parts of the causal diagram associated with **B** allowing correlations.

For the most part, we will not use the above assumption. However, we will need it when we prove the *many ancestors and many descendants theorems* in Sec. 78.21 for the quantum case.

It is reasonable to ask for a stronger property, namely that the operation itself factorises when its causal diagram factorises. We will prove a theorem showing that this stronger property follows from the above assumption if we also assume tomographic locality. Tomographic locality is the assumption that we can fully characterise an operation by joint probabilities obtained when we attach separate operations on each of the open wires to form a circuit (this assumption was discussed in Sec. 9.7 for the simple case and will be discussed in Sec. 51.3 for the complex case). In fact, for the proof below it is sufficient for the proof below to attach separate operations on a bipartition of the open wires (in particular, we will use the bipartition $(\mathbf{ax}, \mathbf{by})$. For the sake of having a name, we will call this the *weak tomographic locality principle*.

> **The weak tomographic locality principle.** For any bipartition of the open wires (say $(\mathbf{ax}, \mathbf{by})$), there exist sets $\mathsf{A}[k]$ (for $k = 1$ to $K$) and $\mathsf{C}[l]$ (for $l = 1$ to $L$) such that
>
> $$\mathsf{A}[k] \overset{\mathbf{x}}{\underset{\mathbf{a}}{\rightleftarrows}} \mathbf{B}' \overset{\mathbf{y}}{\underset{\mathbf{b}}{\rightleftarrows}} \mathsf{C}[l] \;\equiv\; \mathsf{A}[k] \overset{\mathbf{x}}{\underset{\mathbf{a}}{\rightleftarrows}} \mathbf{B}'' \overset{\mathbf{y}}{\underset{\mathbf{b}}{\rightleftarrows}} \mathsf{C}[l] \qquad \forall k, l \quad (1110)$$
>
> only if $\mathbf{B}' \equiv \mathbf{B}''$.

Equivalence of the expressions in (1110) means that these circuits have the same probabilities. The weak tomographic locality principle is implied by the usual tomographic principle.

With these preliminaries in place we can now state and prove the following theorem.

> **Operations factorise when their causal diagrams do.** If the *no correlation without causation assumption* and the *weak tomographic locality principle* both hold, then a deterministic operation whose causal diagram factorises, i.e.

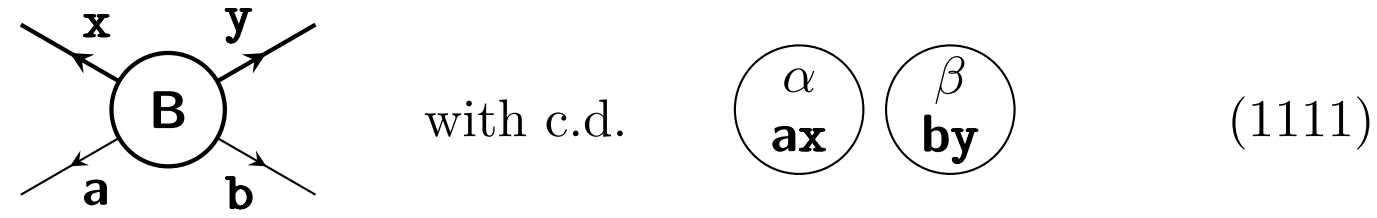

(1111)

can we written in factorised form as

(1112)

where

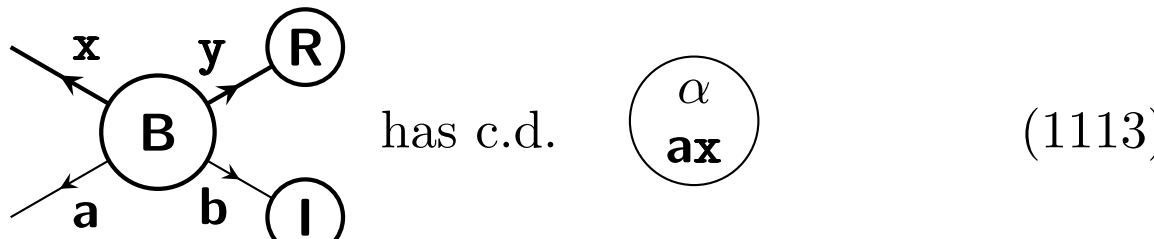

(1113)

and

has c.d. $\beta$ **by** (1114)

Further, if the causal diagram associated with an operator has more than two disjoint parts then the operation factorises into the same number of factors having causal diagrams equal to these disjoint parts.

The proof of this is simple. First note that it follows from the no correlation without causation property that

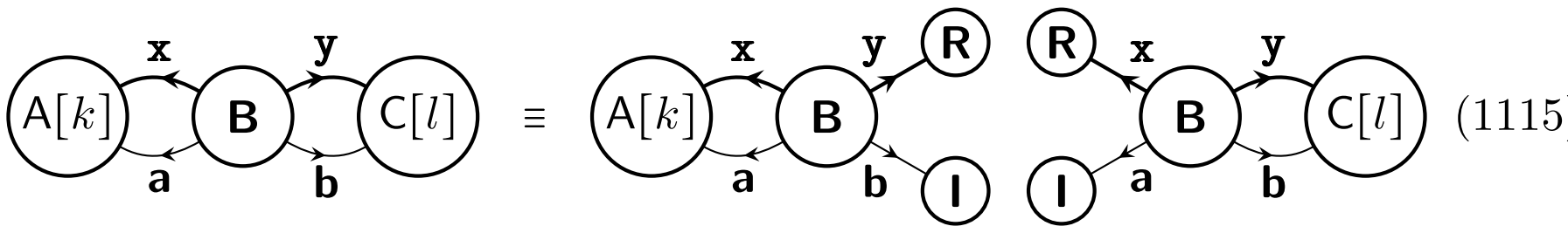

(1115)

Thus, given the weak tomographic locality principle, the factorised operation on the right hand of (1112) is equivalent to **B**. To prove that each of the factors has the corresponding causal diagrams (as in (1113, 1114)) we simply fuse the causal diagrams of the appended **I** and **R** operations using the techniques in Sec. 47.5. To prove that the theorem generalises to the case where the causal diagram has more than two disjoint parts we can group all but one of the causal diagrams and apply the theorem across this grouped/ungrouped partition. This will separate out one factor (that associated with the ungrouped causal diagram). Then we can repeat this process until we have only one of the disjoint causal diagrams on the "grouped" side. This proves the theorem.

To prove the many ancestors and many descendants theorems we use the above factorisation property. However, since quantum theory already satisfies tomographic locality, we will only need the no correlation without causation assumption.

# 49 Physicality conditions

## 49.1 Prerequisites

### 49.1.1 Control and midcome identities

In Sec. 7.11 we stated the control wire identity (131) and the midcome identity (130) in the language of the simple theory. We can also state these in the language of the complex theory. The control wire identity is

$$\xrightarrow{\;\mathbf{x}\;} \quad \equiv \quad \sum_x \xrightarrow{\;\mathbf{x}\;} \textcircled{$x$} \xrightarrow{\;\mathbf{x}\;} \tag{1116}$$

This is easily obtained from (131) by starting with separate $\mathbf{x}^+$ and $\mathbf{x}^-$ control identities and combining them. The midcome identity is

$$\xrightarrow{\;\mathbf{x}\;} \textcircled{$x$} \xrightarrow{\;\mathbf{x}\;} \quad \equiv \quad \xrightarrow{\;\mathbf{x}\;} \textcircled{$x$} \xrightarrow{\;\mathbf{x}\;} \textcircled{$\mathbf{R}$} \overset{\textcircled{$N_{\mathbf{x}}$}}{} \textcircled{$\mathbf{R}$} \xrightarrow{\;\mathbf{x}\;} \textcircled{$x$} \xrightarrow{\;\mathbf{x}\;} \tag{1117}$$

which is obtained in a similar way. The $N_{\mathbf{x}}$ inside the circle is an overall factor. Note that $N_{\mathbf{x}} = N_{\mathbf{x}^+} N_{\mathbf{x}^-}$. To derive the midcome identity we need to assume flatness.

### 49.1.2 Regularised circuits

A generalised circuit may have control wires and midcome wires. We can use the control and midcome identities to write a general circuit as a positive weighted sum of *regularized circuits*. In a regularised circuit all pointer wires go to, or come from, a readout box then an $\boldsymbol{R}$ box such as in the example

$$\text{[circuit: } \mathsf{C} \xleftarrow{\mathbf{z}} \textcircled{$z$} \xleftarrow{\mathbf{z}} \mathbf{R};\quad \mathsf{C} \xrightarrow{\mathbf{a}} \mathsf{B},\ \mathsf{B} \xrightarrow{\mathbf{c}} \mathsf{A},\ \mathsf{A} \xrightarrow{\mathbf{b}} \mathsf{C};\quad \mathbf{R} \xleftarrow{\mathbf{x}} \textcircled{$x$} \xleftarrow{\mathbf{x}} \mathsf{A} \text{]} \tag{1118}$$

(compare with Sec. 7.11.1 for the simple case).

We can go one step further and absorb the readout boxes and $\mathbf{R}$ boxes giving

$$\text{[circuit: } \mathsf{C}(z) \xrightarrow{\mathbf{a}} \mathsf{B},\ \mathsf{B} \xrightarrow{\mathbf{c}} \mathsf{A}(x),\ \mathsf{A}(x) \xrightarrow{\mathbf{b}} \mathsf{C}(z) \text{]} \tag{1119}$$

where $\mathsf{A}(x)$ and $\mathsf{C}(z)$ are defined appropriately. We call this *fully regularised form*.

The regularised form will prove useful when stating the strong tester positivity (see Sec. 49.2.1) and the fully regularised form is useful when evaluating corresponding operator circuits (see Sec. 75) in the Quantum Theory.

### 49.1.3 Complete sets of complex operations

We can build complete sets of complex operations from deterministic operations and readout boxes. For example

(1120)

This is a complete set since the $\mathsf{B}(stz)$ are mutually exclusive and one of them must happen (see Sec. 5.3 for discussion of the simple case). We can use the control identity (1116) to write

(1121)

The network on the right is deterministic. We can write

(1122)

In general, complete sets of operations sum to a deterministic operation (see Sec. 7.11.3 for discussion of the simple case).

### 49.1.4 Purity for complex operations

In Sec. 7.8 we defined mixed, extremal, heterogeneous, homogeneous, and pure cases for system preparations and results in the simple case. We can define system preparations and results here as objects of the form

(1123)

and all the definitions in Sec. 7.8 go through in the same way.

The natural object in the complex case is

$$\text{C} \xrightarrow{\mathbf{a}} \tag{1124}$$

which, in the simple case, could be an operation with an input and an output, or more generally, a network having complicated causal structure. In fact, since our notion of equivalence applies equally to these objects, we can also define mixed, heterogeneous, homogeneous, and pure cases for operations like that in (1124) in the same way as we did in Sec. 7.8. A pure operation is one that is homogeneous and extremal. It is extremal if it cannot be written as being equivalent to a sum of distinct operations. It is homogeneous if it cannot be written as being equal to a sum of distinct non-parallel operations. See Fig. 1 for an illustration of these ideas.

When we proved the positivity composition theorem for simple operations in Sec. 7.12.3 we assumed that pure preparations and pure results are positive. For the positivity composition theorem to be given below for causally complex operations we have a choice. Either (i) we assume that preparations and results are positive (as we did for simple operations) or (ii) we make the stronger assumption that pure operations of the general sort given in (1124) are positive. If we make assumption (ii) then we immediately prove that all operations are positive since any operation can be written as a positive weighted sum of pure ones. It is more interesting to make the weaker assumption (i) since then there is something to prove and, furthermore, we take advantage of the temporal structure in an interesting way. Thus, in proving the positivity composition theorem for complex operations we will continue to assume only that pure preparations and results are positive.

## 49.2 Positivity

### 49.2.1 Strong positivity condition

Consider an arbitrary circuit involving any operation, B. This circuit can be written as

$$\text{B} \overset{\mathbf{x}}{\underset{\mathbf{a}}{\rightrightarrows}} \text{C} \tag{1125}$$

where **x** and **a** may be composite. Therefore the condition on operation B that it gives nonnegative probabilities for any circuit is

$$\text{prob}\left( \text{B} \overset{\mathbf{x}}{\underset{\mathbf{a}}{\rightrightarrows}} \text{C} \right) \geq 0 \qquad \forall \quad \text{C} \tag{1126}$$

Using the control identity (1116) and midcome identities (1117) we can write this circuit as

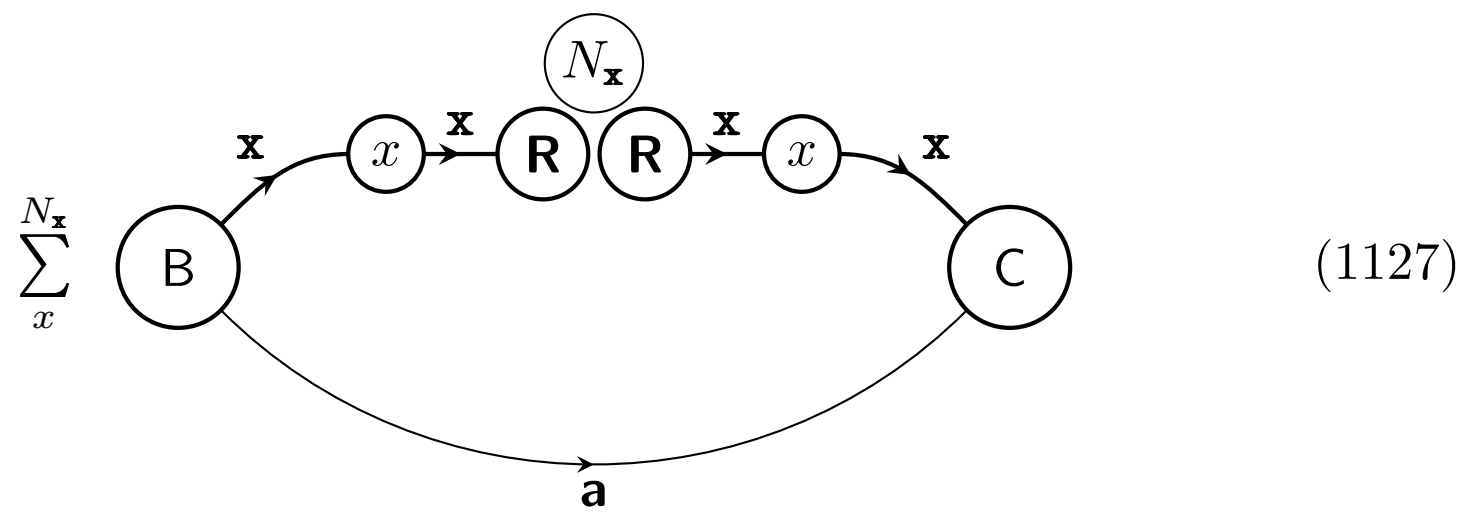

(1127)

(Since we are using the midcome identity, we are assuming flatness.) We can write this as

$$\sum_x^{N_{\mathbf{x}}} \; N_{\mathbf{x}} \; \mathsf{B} \;\; \mathbf{x}: x \to \mathbf{R}, \;\; \mathbf{a}: \mathsf{D}[x] \tag{1128}$$

by absorbing an $\mathbf{R}$ box and an $x$ readout box into $\mathsf{D}[x]$. We have put the circuit in regularized form. Further, we can write $\mathsf{D}[x]$ as being equal to a positive weighted sum of pure operations, $\mathsf{E}$. Hence, the condition becomes

$$\text{prob}\left( \mathsf{B} \;\; \mathbf{x}: x \to \mathbf{R}, \;\; \mathbf{a}: \mathsf{E} \right) \geq 0 \tag{1129}$$

for all pure $\mathsf{E}$. We can write this condition as

$$0 \underset{sT}{\leq} \mathsf{B} \;\; \mathbf{x}, \; \mathbf{a} \tag{1130}$$

where this means that the probability is non-negative with respect to any tester of the form

$$\mathbf{x}: x \to \mathbf{R}, \;\; \mathbf{a}: \mathsf{E} \tag{1131}$$

for pure $\mathsf{E}$. We call testers of this form *strong testers*. We call this *strong tester positivity*.

### 49.2.2 Tester positivity for complex operations

The positivity condition above is very strong since it applies to all pure operations E. In accord with the remarks in Sec. 49.1.4 we will consider a weaker positivity condition that only uses pure preparations and results. This is in keeping with the positivity condition used in the simple case. It turns out this weaker condition is sufficient to prove the positivity theorems we want to prove. Our weaker positivity condition is

$$\text{prob}\left( \mathsf{G} \quad \mathbf{a}^+ \quad \mathbf{g}^+ \quad \mathsf{B} \xrightarrow{\mathbf{x}} x \xrightarrow{\mathbf{x}} \mathbf{R} \quad \mathbf{a}^- \quad \mathsf{F} \right) \geq 0 \qquad (1132)$$

for all pure preparations F and pure results G, and for all systems $g^+$. (Note we have changed the angle at which **x** attaches to B for ease of graphical representation here). Using the thin rectangle converter notation (as defined in (998)) this is the same as

$$\text{prob}\left( \mathsf{G} \quad \mathbf{a}^+ \quad \mathbf{g}^+ \quad \mathbf{a} \quad \mathsf{B} \xrightarrow{\mathbf{x}} x \xrightarrow{\mathbf{x}} \mathbf{R} \quad \mathbf{a}^- \quad \mathsf{F} \right) \geq 0 \qquad (1133)$$

We can write this condition as

$$0 \underset{T}{\leq} \xleftarrow{\mathbf{a}} \mathsf{B} \xrightarrow{\mathbf{x}} \qquad (1134)$$

where $\underset{T}{\leq}$ means we have positivity with respect to any tester of the form

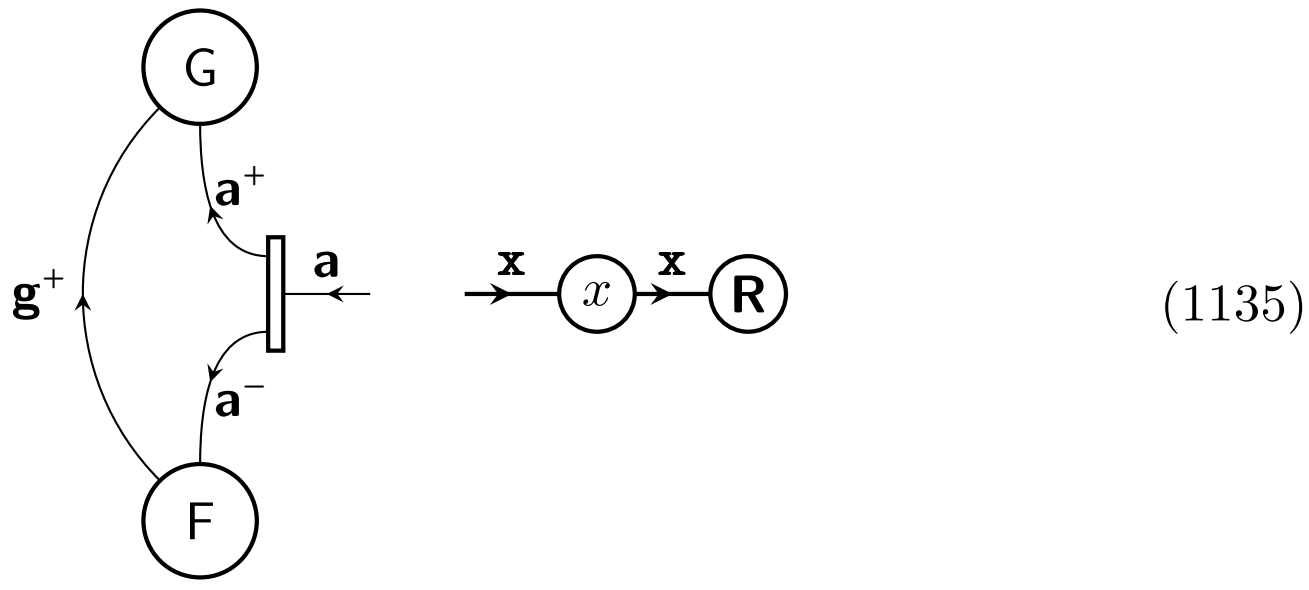

(1135)

where F and G are pure. If a complex operation satisfies (1134) we say it satisfies *tester positivity*. We see, by comparing (1134) with (1130) that tester positivity is a weaker condition than strong tester positivity.

Note that not all circuits involving the complex operation B can be written in the form in (1132). We have simply pulled up all outputs and pulled down all inputs whereas a more general circuit might take advantage of the causally complex structure. This turns out not to matter for the theorems we want to prove (and mirrors the fact that we pulled up all outputs and down all inputs when proving the positivity composition theorem in Sec. 7.12.3).

### 49.2.3 Inequalities

Recall that, from Sec. 48.3, we have by definition

$$\text{exprn}_1 \leqq \text{exprn}_2 \quad \Leftrightarrow \quad p(\text{exprn}_1 \mathsf{H}) \leqq p(\text{exprn}_2 \mathsf{H}) \quad \begin{array}{l}\forall \text{ complement} \\ \text{networks } \mathsf{H}\end{array} \tag{1136}$$

Following the steps in Sec. 49.2.1 we will see that (assuming flatness) it is sufficient to check this for complement networks of the form

(1137)

for pure E. These are *strong testers* (defined previously in Sec. 49.2.1. Thus, we provide the definition

$$\text{exprn}_1 \underset{sT}{\leqq} \text{exprn}_2 \quad \Leftrightarrow \quad p(\text{exprn}_1 \mathsf{S}) \leqq p(\text{exprn}_2 \mathsf{S}) \quad \text{for all strong testers } \mathsf{S} \tag{1138}$$

This is a little simpler than (1136) since we have to check fewer complement networks. To see that (1138) is sufficient note (i) we can rearrange the inequality $\text{exprn}_1 \leqq \text{exprn}_2$ as

$$\text{exprn}_1^+ - \text{exprn}_2^- \leqq \text{exprn}_2^+ - \text{exprn}_1^- \tag{1139}$$

where $\text{exprn}_k^{\pm}$ collects all the terms in $\text{exprn}_k$ with ±ve coefficients. An arbitrary complement network can be written as a positive weighted sum of strong testers. Hence

$$\text{exprn}_1^+ - \text{exprn}_2^- \leqq \text{exprn}_2^+ - \text{exprn}_1^- \quad \Leftarrow \quad \text{exprn}_1^+ - \text{exprn}_2^- \underset{sT}{\leqq} \text{exprn}_2^+ - \text{exprn}_1^- \tag{1140}$$

follows by the linearity of the $p(\cdot)$ function, it is sufficient to check only the strong testers to check the $\leqq$ inequality.

We can restrict the set of complement networks we consider further so that they are of the form

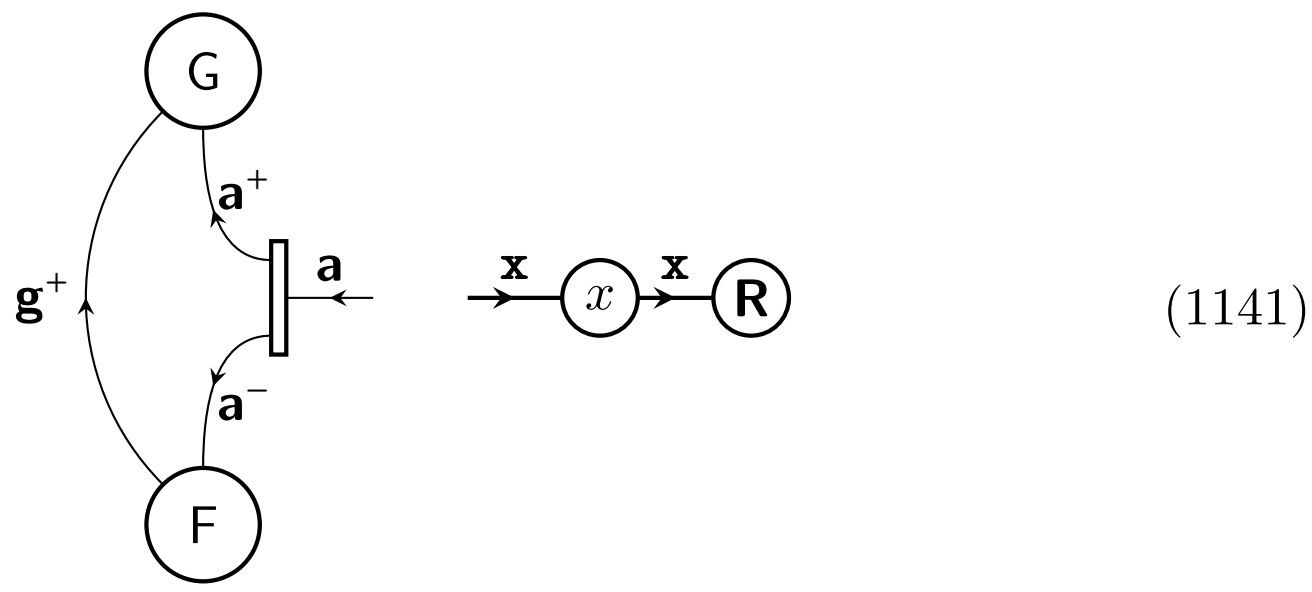

(1141)

where F is a pure preparation and G is a pure result. These are *testers*. They involve splitting **a** into positive and negative parts. This is the same as the testers considered in the simple case (see Sec. 7.11.2) but now written in the complex notation. Now we can define

$$\text{exprn}_1 \underset{T}{\leqq} \text{exprn}_2 \quad \Leftrightarrow \quad p(\text{exprn}_1 \mathsf{T}) \leqq p(\text{exprn}_2 \mathsf{T}) \;\; \text{for all testers } \mathsf{T} \qquad (1142)$$

Here we follow the definition of tester inequality given in the simple case (in Sec. 7.12.1).

Clearly

$$\text{exprn}_1 \leqq \text{exprn}_2 \quad \Rightarrow \quad \text{exprn}_1 \underset{T}{\leqq} \text{exprn}_2 \qquad (1143)$$

It follows by the same reasoning given in Sec. 7.12.1 that

$$\text{exprn}_1 \leqq \text{exprn}_2 \quad \Leftarrow \quad \text{exprn}_1 \underset{T}{\leqq} \text{exprn}_2 \qquad (1144)$$

if all the terms in $\text{exprn}_{1,2}$ consist of operations having simple causal diagrams. In the case of Quantum Theory, we can prove that (1144) holds for arbitrary expressions. We do this in Sec. 77 leveraging the Hilbert space structure and the definition of twofold positivity.

### 49.2.4 RI partial contraction positivity theorem

We will sometimes have cause to partially contract an operation, B, with a **R** and **I**. We prove the following theorem

> **RI partial contraction positivity theorem.** If B satisfies tester positivity then
>
> $$\underset{T}{\geqq} 0 \qquad (1145)$$
>
> for any such contraction with **R** and **I**.

To prove this note that (1145) is true if

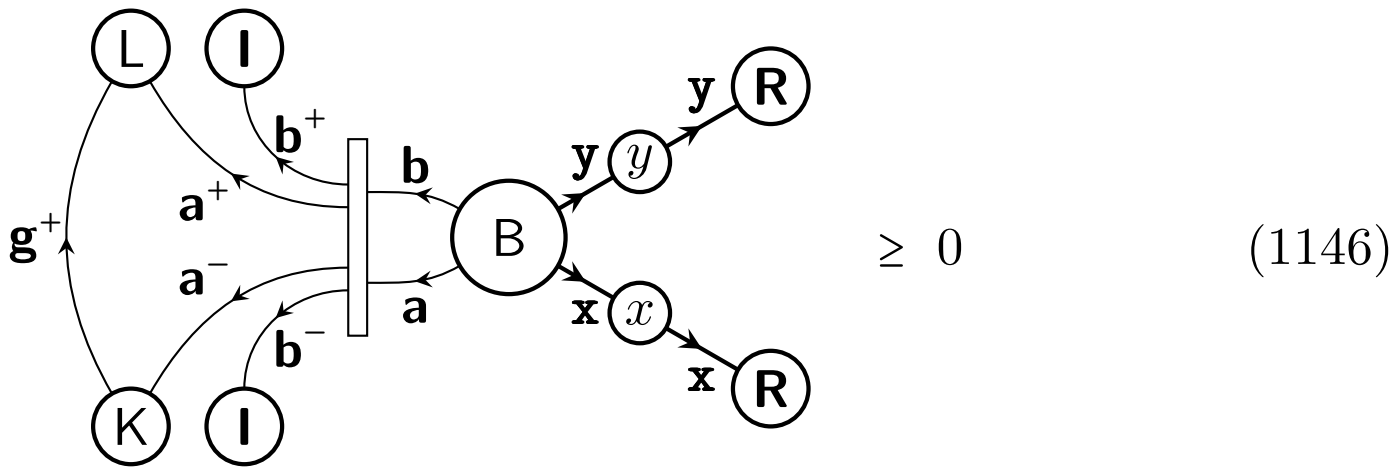

(since, if we sum over $y$, we get (1145)). And, in turn, this is true if for all pure $\mathsf{F}$ and $\mathsf{G}$ since the preparation $\mathsf{K}^{\mathbf{g}^+}\mathbf{I}_{\mathbf{b}^-}$ can be written as positive weighted sum of pure preparations and the result $\mathsf{L}_{\mathbf{g}^+}\mathbf{I}_{\mathbf{b}^+}$ can be written as a positive weighted sum of pure results. The condition in (1147) is the condition for the positivity of $\mathsf{B}$ and so the theorem follows.

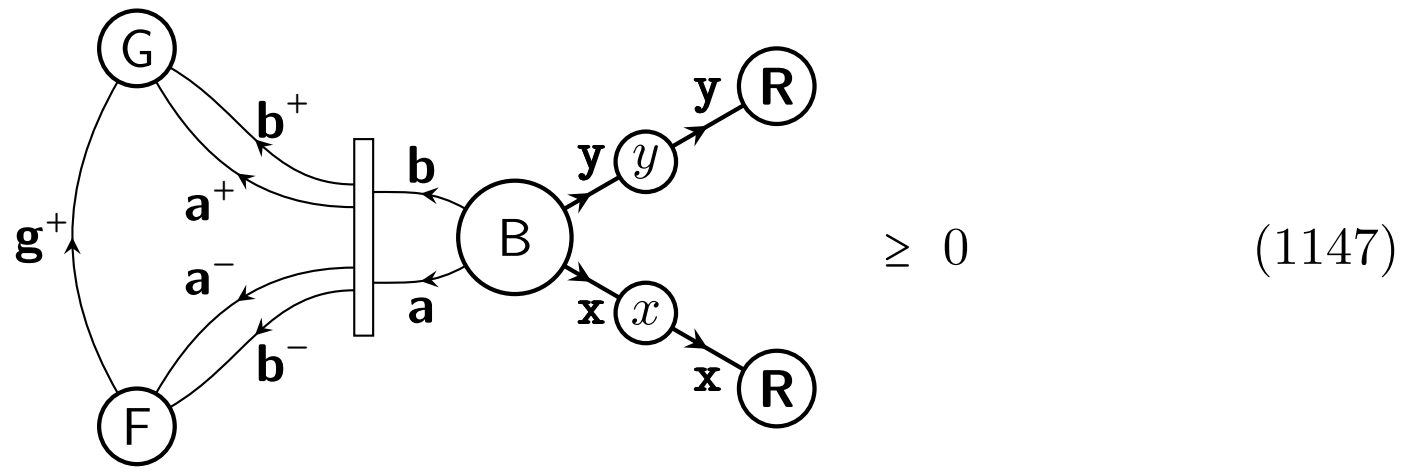

### 49.2.5 Synchronous partition for positivity assumption

Here we will introduce an assumption which we will use in Sec. 49.2.6 to prove an important theorem - namely that we have appropriate positivity under composition of operations.

Consider a bundle of causal links represented by a thick wire with a set, $S$, of wire labels. We can write this as

$$S \quad = \quad \cdots \quad := \qquad (1148)$$

In the second expression we have separated out the individual causal arrows and inserted a fusion nodes as defined in (1031). We have omitted the individual labels on these wires (drawn from $S$). In the final expression we introduce notation for this situation.

We will need the following assumption.

**Synchronous partition for positivity assumption.** For any causally complex operation satisfying tester positivity

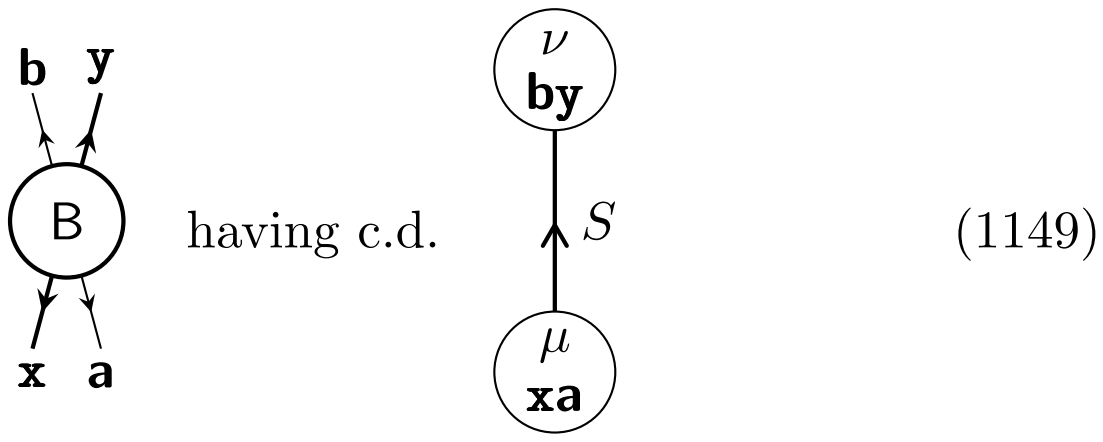

(1149)

(where $S$ is a set of causal link labels) we can write

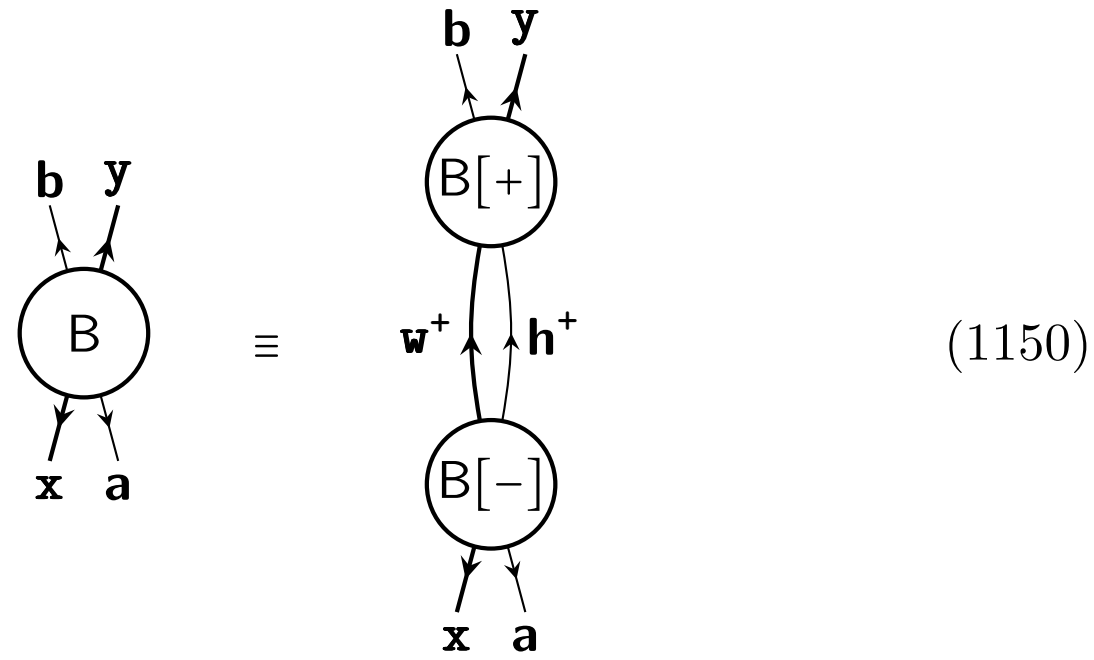

(1150)

for some $\mathbf{h}^+$ and $\mathtt{w}^+$, where

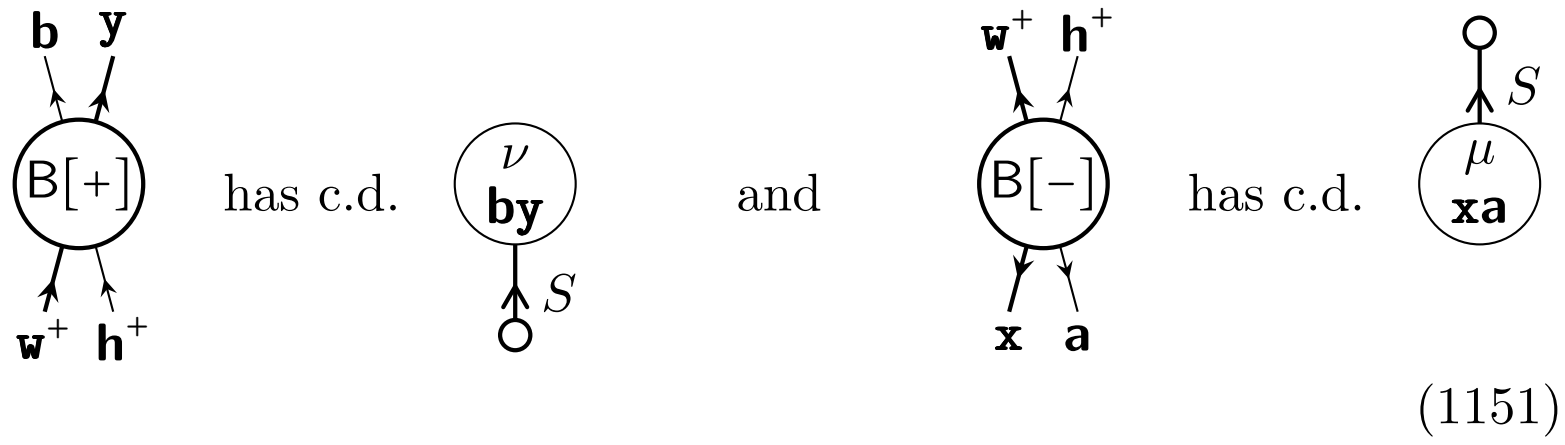

(1151)

where the causally complex operations $\mathsf{B}[+]$ and $\mathsf{B}[-]$ satisfy tester positivity.

Note that $\mathsf{B}[+]$ and $\mathsf{B}[-]$ are not necessarily unique and either $\mathbf{h}^+$ and/or $\mathtt{w}^+$ may be null. Further, note that the nodes in the causal diagrams in (1151) are to be interpreted as the nodes in (1148) (that is they can be written in fanned out into $|S|$ nodes as in the middle expression in (1151)). We have not put any wire type inside these nodes. We know that, overall, these nodes are associated with the wire type $\mathbf{h}^+\mathtt{w}^+$. However, the assumption does not specify how to regard $\mathbf{h}^+\mathtt{w}^+$ as a composite system and divide these components between these $|S|$ nodes in the causal diagram. This does not matter for what follows.

If we assume double maximality (see Sec. 8) then any pointer type wire can be replaced by a system type wire. In this case, we can replace $\mathtt{w}^+$ by $\mathbf{w}^+$. Then we can simply absorb this into the definition of $\mathbf{h}^+$. So, in this case, we only need $\mathbf{h}^+$ between $\mathsf{B}[-]$ and $\mathsf{B}[+]$. Instead, we could remove the $\mathtt{w}^+$ wire using the control and midcome identities (however, this introduces a sum over $w$).

We will see that time symmetric complex operational quantum theory (TSCOQT) satisfies the above synchronous partition for positivity assumption (this is discussed in Sec. 78.7).

### 49.2.6 Positivity under composition

In Sec. 7.12.3 we proved that, if all pure preparations and pure results satisfy tester positivity then any network built out of simple operations, each satisfying tester positivity, will itself be tester positive. Now we will prove a similar theorem for complex operations. This turns out to be a more involved proof than in the simple case.

> **Positivity composition theorem.** Assume all pure preparations and pure results satisfy tester positivity and that the synchronous partition for positivity assumption holds. If we wire together two or more causally complex operations, each satisfying tester positivity, then the resulting network will satisfy tester positivity.

First, consider wiring together two tester positive causally complex operations, $\mathsf{A}$ and $\mathsf{B}$, which have no pointer types. The case where there are pointer types is then easy to deal with and will be dealt with at the end of this theorem proof. The case where there are more than two operations also follows simply. In general, any such situation can be represented by

$$\mathbf{a} \leftarrow \mathsf{A} \xrightarrow{\mathbf{c}} \mathsf{B} \rightarrow \mathbf{b} \tag{1152}$$

To prove that this satisfies tester positivity we need to prove that

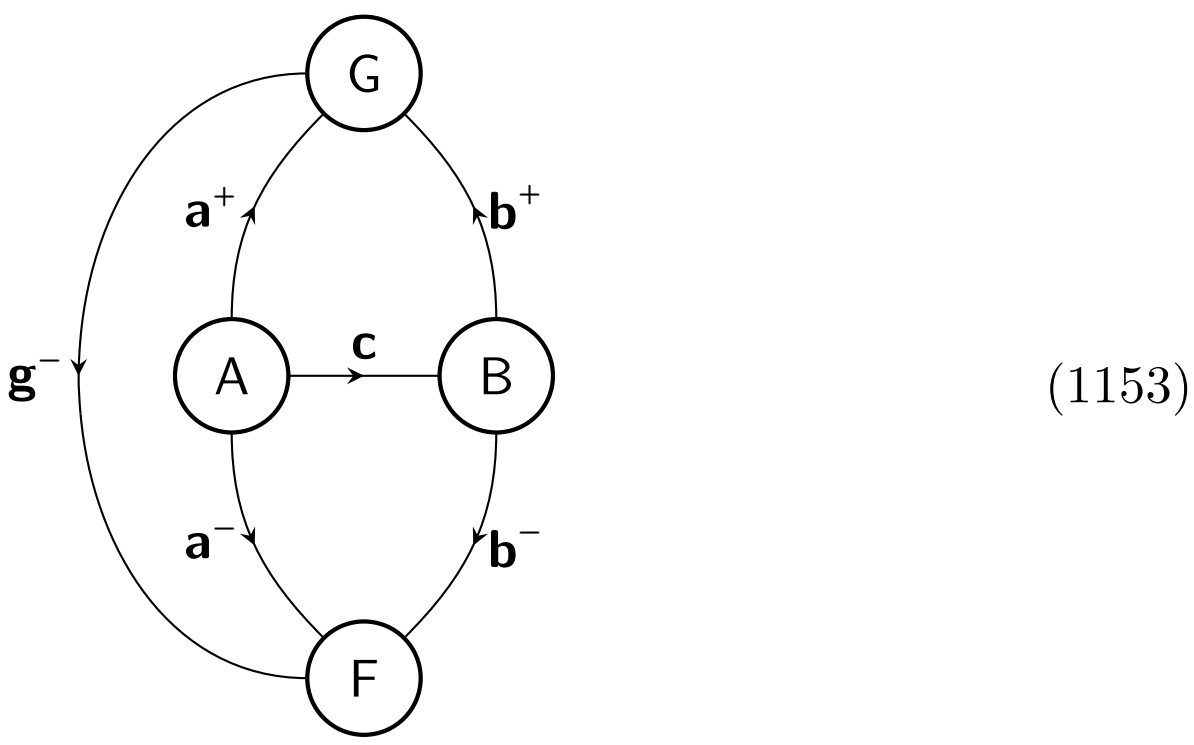

(1153)

is positive where $\mathsf{F}$ and $\mathsf{G}$ are pure. For later convenience, we have the arrow on the $\mathbf{g}^-$ pointing backwards - this system is travelling from $\mathsf{F}$ to $\mathsf{G}$ as indicated by the $-$ sign on the $\mathbf{g}^-$. When we join $\mathsf{A}$ and $\mathsf{B}$ we also join their causal diagrams.

For example, we may have

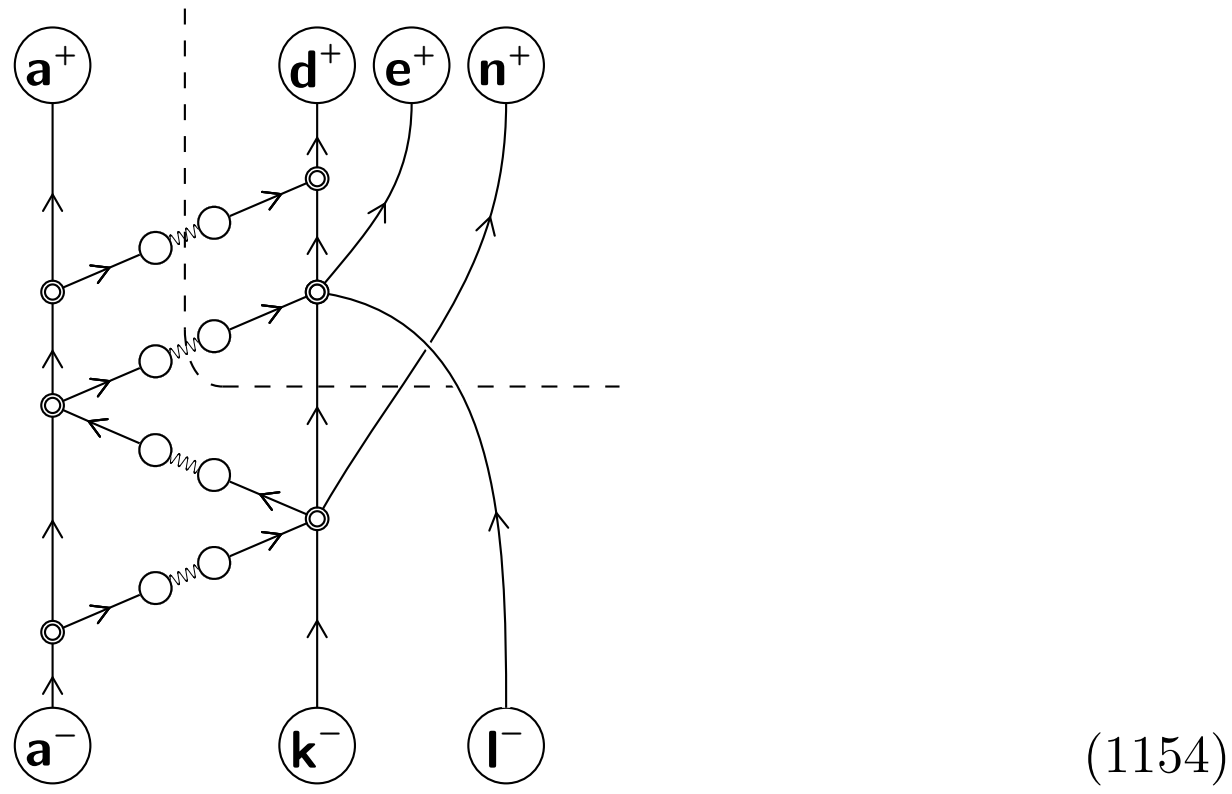

(1154)

(ignore, for the moment, the dashed line). Here the causal diagram on the left for A is fused to the causal diagram on the right for B (at the fusion nodes). To be consistent with (1152) we require $\mathbf{d}^+\mathbf{e}^+\mathbf{n}^+ = \mathbf{b}^+$ and $\mathbf{k}^-\mathbf{l}^- = \mathbf{b}^-$. The central fusion nodes are associated with $\mathbf{c}$ where the two operations are joined together. The particular causal structure in the example could be achieved if $\mathbf{c} = \mathbf{p}^+\mathbf{q}^-\mathbf{r}^+\mathbf{s}^+$ where $\mathbf{p}^+$ is associated with the bottom fusion site, $\mathbf{q}^-$ is associated with the next one up, and so on. These fusion nodes are nodes in a DAG and consequently we can always impose some total order on them (this total order may not be unique). Now consider the dashed line. We can partition the causal diagram along this line. This is a synchronous partition both for the whole causal diagram and also for that part of the causal diagram associated with B. We can always find a synchronous partition line through some number of fusion nodes like this partitioning off part of B or A's causal diagram for any such joining of causal diagrams. Let us assume that we are partitioning of part of B's causal diagram as in this example (the same argument would go through were it A's causal diagram). Since this is a synchronous partition for B's causal diagram, we can apply the synchronous partition for positivity assumption to B. Then (1153) becomes

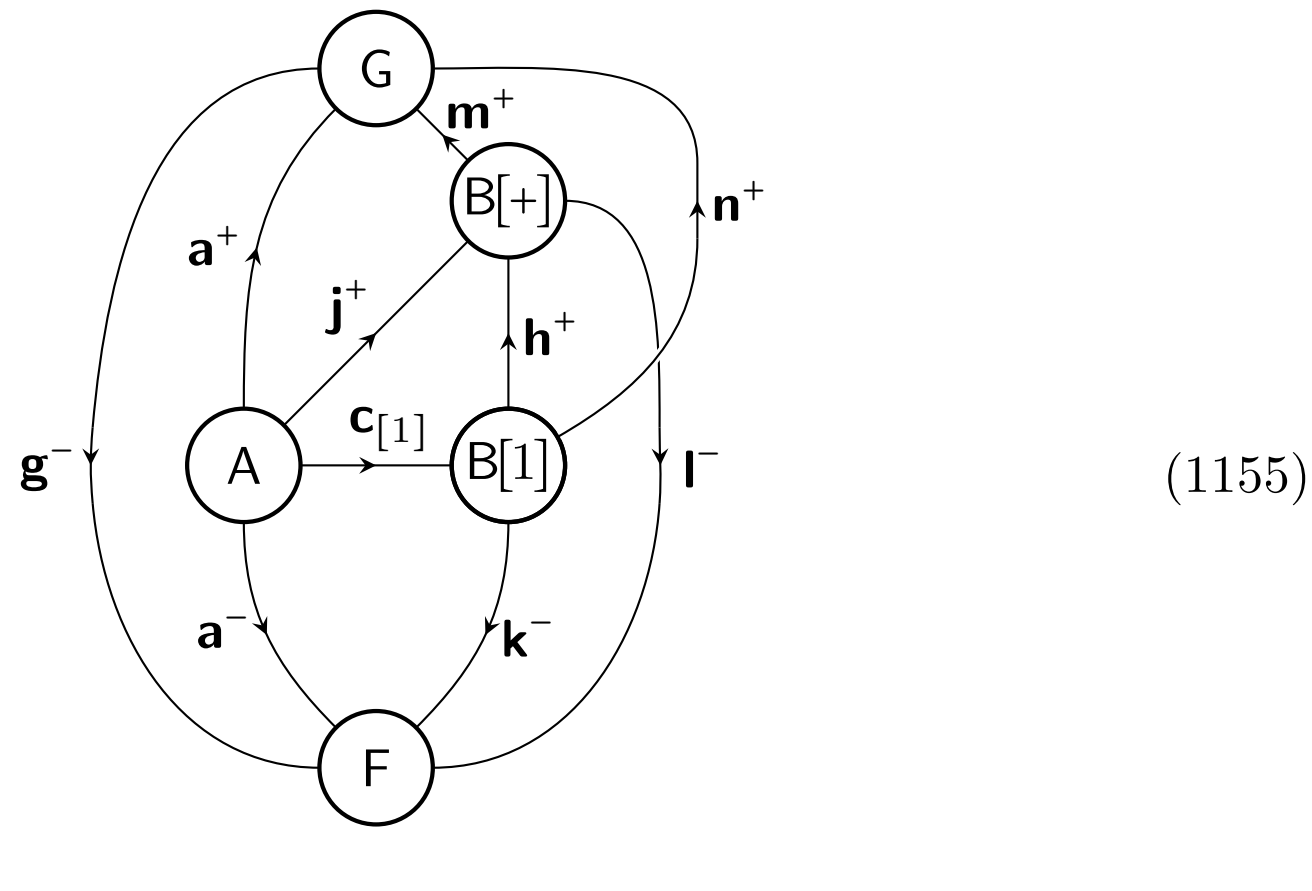

(1155)

Here $\mathsf{k}^-\mathsf{l}^- = \mathsf{b}^-$, $\mathsf{m}^+\mathsf{n}^+ = \mathsf{b}^+$, and $\mathsf{j}^+\mathsf{c}_{[1]} = \mathsf{c}$. The key point to note here is that all the arrows connected to $\mathsf{B}[+]$ have a definite direction (they have a + or a − system on them). This is made possible because of the way we partition the causal diagram in (1154) which guarantees that $\mathsf{j}^+$ has a + on it. We illustrated this with respect to the causal diagram example in (1154). However, it is clear that, for any example, we can find such a synchronous partition that either cuts away part of the causal diagram associated with $\mathsf{A}$ or (as in this case) $\mathsf{B}$. By the synchronous partition for positivity assumption, $\mathsf{B}[+]$ satisfies tester positivity. This means we can absorb $\mathsf{B}[+]$ into $\mathsf{G}$ obtaining a result satisfying tester positivity that is not necessarily pure (but is equal to a positive weighted sum over pure results). Similar reasoning is given in the case of (157) for the simple operator case. Therefore, the expression in (1155) is non-negative if

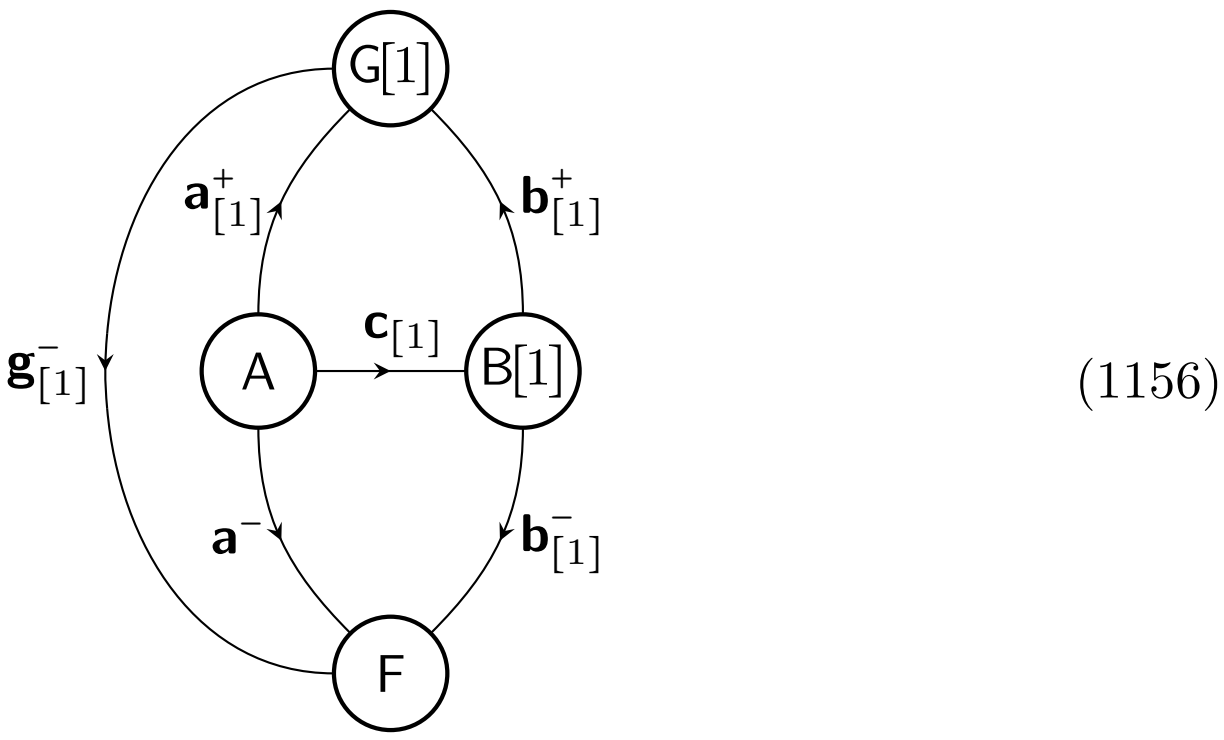

(1156)

is non-negative for all pure $\mathsf{F}$ and $\mathsf{G}[1]$. Here $\mathsf{a}^+_{[1]}\mathsf{j}^+ = \mathsf{a}^+$, $\mathsf{b}^+_{[1]} = \mathsf{h}^+\mathsf{n}^+$, $\mathsf{b}^-_{[1]} = \mathsf{k}$, and $\mathsf{g}^-_{[1]} = \mathsf{g}^-\mathsf{l}^-$. By the synchronous partition for positivity assumption, $\mathsf{B}[1]$ satisfies tester positivity and hence we are back where we started (compare with (1153)) but where the causal diagram is smaller. In our example, this smaller causal diagram is the following

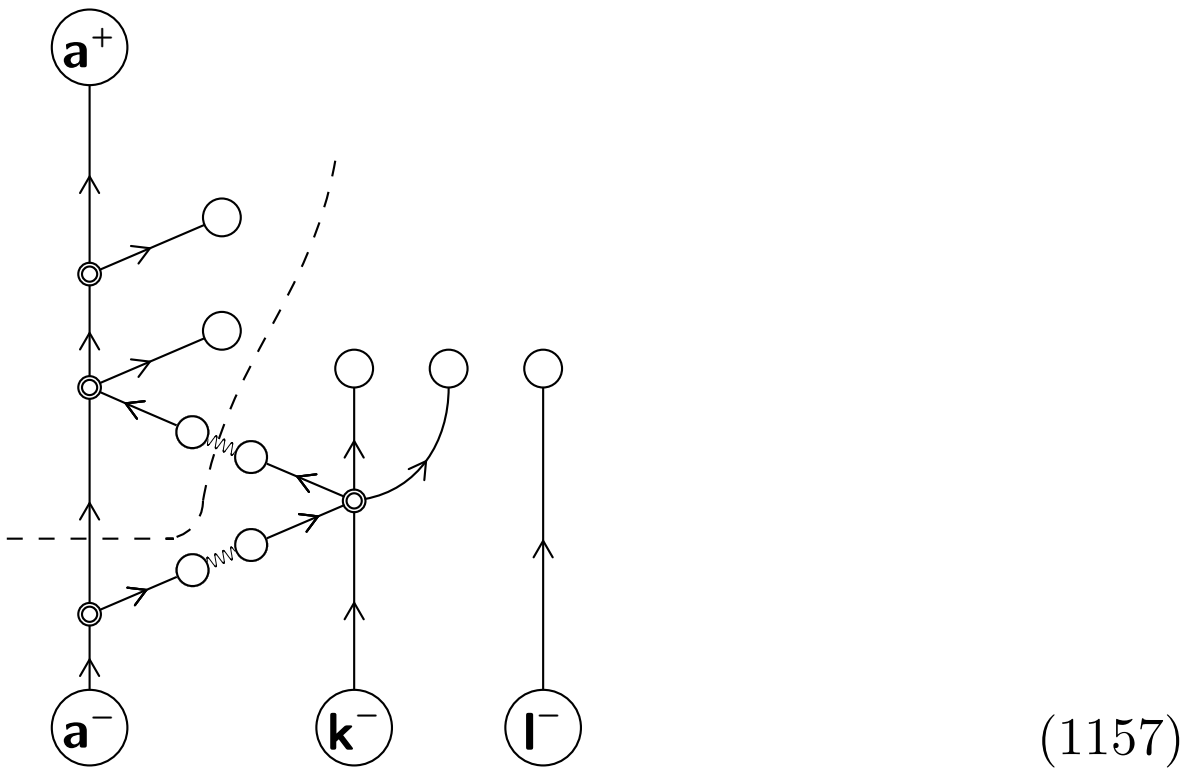

(1157)

We can iterate this process taking a "bite" out of $\mathsf{A}$ (in the example, this bite is associated with the synchronous partition shown by the dashed line in (1157) above), yielding $\mathsf{A}[2]$, then take a bite out of $\mathsf{B}[1]$ yielding $\mathsf{B}[3]$ and so on until all of $\mathsf{A}$ and $\mathsf{B}$ have been absorbed into $\mathsf{G}$ so the process terminates. When it does terminate, we will have a circuit consisting of pure result, $\mathsf{G}[N]$, acting on a pure preparation, $\mathsf{F}$. Since, by assumption, pure preparations and results are tester positive, this circuit must have non-negative probability. This proves that (1153) has non-negative probability which proves the theorem in the case where there are no pointer types. In fact, we can terminate the process when $\mathbf{c}[n]$ is null in the + direction or in the − direction since it is then clear the resulting expression has non-negative probability (so, in our example, we can see in (1157) that we are already there after two bites). To deal with the case where there are pointer types we can pursue one of two alternative strategies. The simplest strategy is simply to redraw the above diagrams where each physical system wire is accompanied by a pointer wire. This will not change the graphical nature of the proof. A more involved (and more revealing) proof is to use the control and midcome identities to replace pointer wires by expressions involving a sum over circuits involving pointer boxes and $\mathbf{R}$ boxes. We need to prove that each term in this sum over circuits is positive so we can focus on one such term. Some of these result boxes and $\mathbf{R}$ boxes can be absorbed into $\mathsf{F}$ and $\mathsf{G}$. The ones attached to $\mathsf{A}$ can be absorbed to give $\mathsf{A}'$. The pointer boxes attached to $\mathsf{B}$ can be absorbed to give $\mathsf{B}'$. Thus, to prove the composition theorem for two operations we have the same problem but with $\mathsf{A}'$ and $\mathsf{B}'$. To prove the composition theorem for more than two operations that are wired together, we simply order these operations (in any way) then start by joining together the first two to form a new operation that satisfies tester positivity, then we join the third operation, and so on.

## 49.3 Positivity of circuits

Circuits have probabilities and we require that these probabilities are non-negative. This property follows from the positivity composition theorem.

> **Circuit positivity theorem.** If every operation in a complex circuit satisfies tester positivity, if the synchronous partition for positivity assumption holds, and if all pure preparations and pure results are $T$-positive, then the circuit has non-negative probability.

Consider a circuit made out of $N$ complex operations satisfying tester positivity. We can join any $N-1$ of them forming a complex operation $\mathsf{A}$. Call the $N$th operation $\mathsf{B}$. Now consider joining these two operations. The situation is that shown in (1152) in the Sec. 49.2.6 but where, now, $\mathbf{a}^{\pm}$ and $\mathbf{b}^{\pm}$ are null. We can go through the steps of the proof of positivity composition theorem (taking "bites" out of the causal diagram) where we now treat $\mathbf{a}^{\pm}$ and $\mathbf{b}^{\pm}$ as null. In the final step we prove that the circuit has non-negative probability.

## 49.4 Causality

In this subsection we set up the double causality conditions for causally complex deterministic operations. We will state these conditions in Sec. 49.4.2. We offer two alternative ways to obtain these causality conditions from more basic assumptions. The first alternative is provided in Sec. 49.4.5. There we show the double causality conditions they can be obtained from the *weak synchronous partition assumption for causality* to be introduced in Sec. 49.4.4. The second alternative is provided in Sec. 49.4.6. There we show how the double causality conditions can be obtained from the *no correlations with out causation* assumption from Sec. 48.9 along with the assumption of tomographic locality (actually, we only need the weak tomographic locality assumption introduced in the same section).

The double causality condition involves an object we call the *residuum*. We will prove some useful theorems concerning these residua.

We will prove a general double causality conditions for complex operations that may not be deterministic. In addition we prove a composition theorem showing that networks built from operations satisfying (general) double causality to, themselves, satisfy (general) double causality. This theorem is essential if the theory is to be well behaved under composition.

### 49.4.1 Simple double causality conditions

In Sec. 7.7 we applied the forward and backward causality conditions to simple deterministic networks. Since simple networks act as models for causally complex operations, we can apply these same causality conditions here adapted to our circle rather than square notation. We will refer to them as *simple double causality conditions* here.

> **Simple double causality condition for deterministic complex operations.** Consider a deterministic causally complex operation
>
> $$\xleftarrow[\mathsf{z}]{} (\mathsf{B}) \xrightarrow[\mathsf{c}]{} \qquad (1158)$$
>
> The simple forward causality condition is
>
> $$\xleftarrow[\mathsf{z}^-]{} (\mathsf{R}_+) \xleftarrow[\mathsf{z}]{} (\mathsf{B}) \xrightarrow[\mathsf{c}]{} (\mathsf{I}_+) \xrightarrow[\mathsf{c}^-]{} \quad \equiv \quad \xleftarrow[\mathsf{z}^-]{} (\mathsf{R}) \quad (\mathsf{I}) \xrightarrow[\mathsf{c}^-]{} \qquad (1159)$$
>
> and the simple backward causality condition is
>
> $$\xleftarrow[\mathsf{z}^+]{} (\mathsf{R}_-) \xleftarrow[\mathsf{z}]{} (\mathsf{B}) \xrightarrow[\mathsf{c}]{} (\mathsf{I}_-) \xrightarrow[\mathsf{c}^+]{} \quad \equiv \quad \xleftarrow[\mathsf{z}^+]{} (\mathsf{R}) \quad (\mathsf{I}) \xrightarrow[\mathsf{c}^+]{} \qquad (1160)$$
>
> where $\mathsf{R}_\pm$ and $\mathsf{I}_\pm$ are defined in (1091) and (1103) respectively.

Note that the causal diagram for $\mathsf{B}$ above can be put in simple form as shown in (1059). These simple causality conditions are not the full set of causality conditions we need for deterministic operations. We will provide those in Sec. 49.4.5.

### 49.4.2 The double causality conditions

For the causally complex operational probabilistic theories we obtain a bigger set of causality conditions - we have one time forward and one time backward causality condition pertaining to each synchronous partition of the operation. This is necessary since causally complex operations are, in general, causally more complicated than the simple case. These causality conditions are strongly related to (and, in essence, this is the same idea as) the recursive causality conditions given by Chiribella et al. [2009a] for quantum combs which first appeared as recursive normalisation conditions for quantum strategies in the work of Gutoski and Watrous [2007] (the causality interpretation being due to CDP). Our discussion here is more general than quantum combs/strageties because (i) we consider general operations (we will discuss the specific case of Quantum Theory later), and (ii) causally complex operations are causally more general objects than quantum combs (wherein the teeth are causally sequential). Furthermore, we are working in the time symmetric temporal frame whereas earlier work is in the time forward frame (we show how to get the forward frame from the time symmetric one in Sec. 53.4). Additionally, the diagrammatic notation is different. A preliminary version of the causality conditions presented here were first given in Hardy [2018] (see also Hardy [2023]).

For the moment we will simply state the double causality conditions. Later we will provide two different ways of obtaining them from more basic assumptions (in Sec. 49.4.5 and Sec. 49.4.6). In Sec. 49.4.8 we will show that the simple double causality above are special cases of the full set of double causality conditions (corresponding to when the synchronous partition used is fully before or fully after the causal diagram).

Now we will state the double causality conditions.

**Double causality conditions.** The double causality conditions apply to synchronous partitions of any deterministic complex operation, $\mathsf{B}$. For any synchronous partition, $p$, we can write

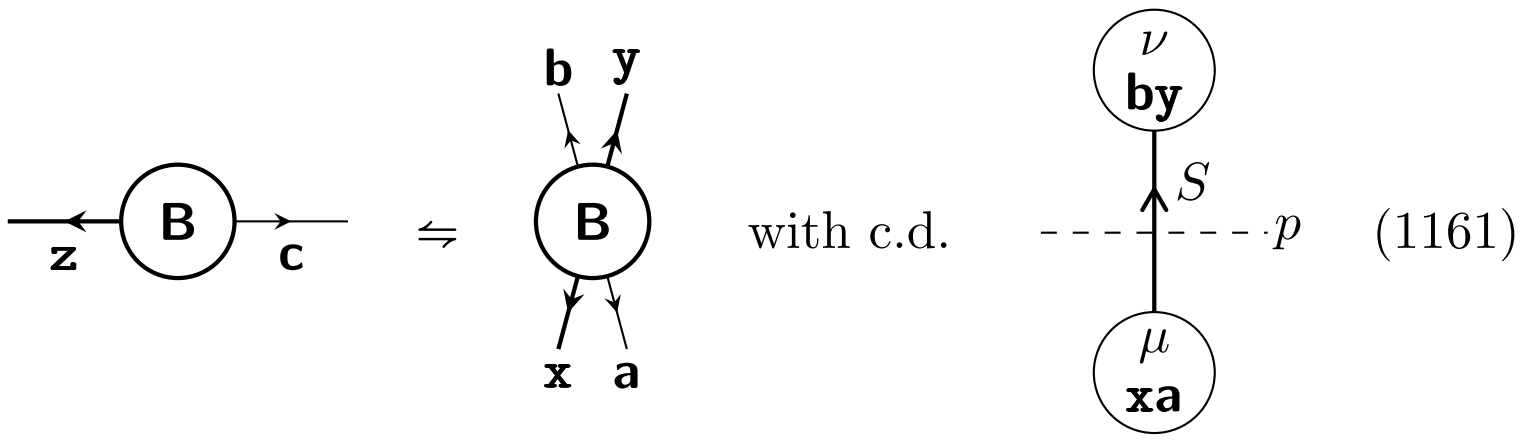

The double causality conditions are satisfied when the *forward causal-*

*ity condition*

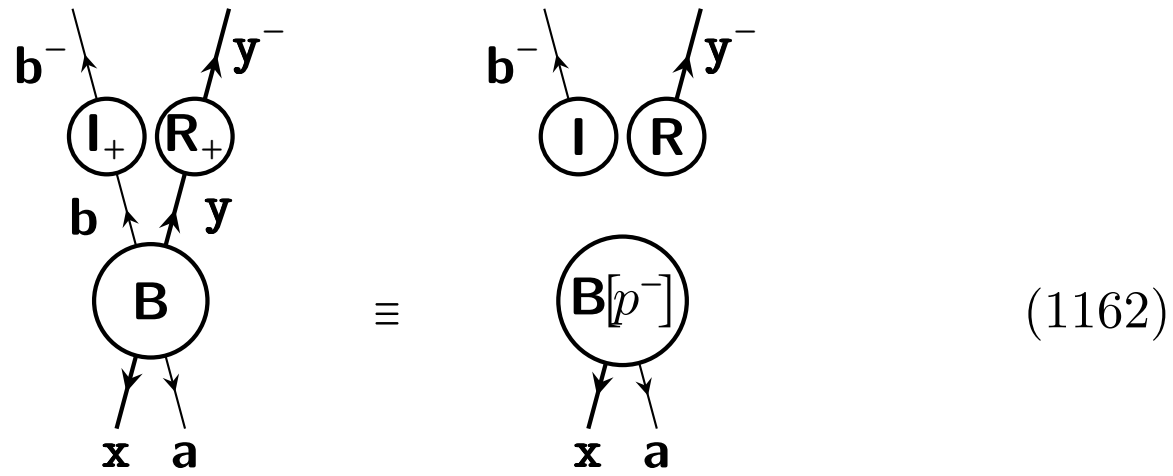

(1162)

and the *backward causality condition*

(1163)

are satisfied for every synchronous partition, $p$. It is useful to have terms for the "anatomy" of the right hand sides of these equivalences. We call $\mathbf{B}[p^-]$ (1162) the *past residuum* with respect to $p$, and we call $\mathbf{B}[p^+]$ in (1163) the *future residuum* with respect to $p$. Further, we call the $\mathbf{R}$'s and $\mathbf{I}$'s that appear on the right hand sides $\mathbf{R}$-*crumbs* and $\mathbf{I}$-crumbs (or just *crumbs*).

Recall that $\mathbf{R}_\pm$ was defined in (1091) and $\mathbf{I}_\pm$ was defined in (1103). The reason for the terminology "crumbs" is that hitting the operation to the future/past of some $p$ with $\mathbf{R}_\pm$ and $\mathbf{I}_\pm$ (as on the left hand sides of these equivalences) is a bit like taking a "bite" out of it (the crumbs being left over).

In Sec. 49.4.7 we will prove some useful theorems concerning the residua, $\mathbf{B}[p^\pm]$.

### 49.4.3 Synchronous partition for causality assumption

In Sec. 49.2.5 we introduced the synchronous partition for positivity assumption which we used to prove positivity under composition in Sec. 49.2.6. That particular assumption can be proven to hold in Quantum Theory (as we will discuss in Sec. 78.7). Here we will introduce another synchronous partition assumption - this time for causality. We use this to find double causality conditions for the causally complex case in Sec. 49.4.5. We are not able to prove that this assumption holds in Quantum Theory (TSCOQT) - though we can conjecture that it does. However, a weaker form of the assumption does hold in Quantum Theory as we will prove in Sec. 78.7. Furthermore, this weaker form is sufficient to obtain the same double causality conditions. We introduce the weaker form of the assumption in Sec. 49.4.4.

The synchronous partition assumption with simple causality is the following.

**Synchronous partition for causality assumption.** For any deterministic causally complex operation

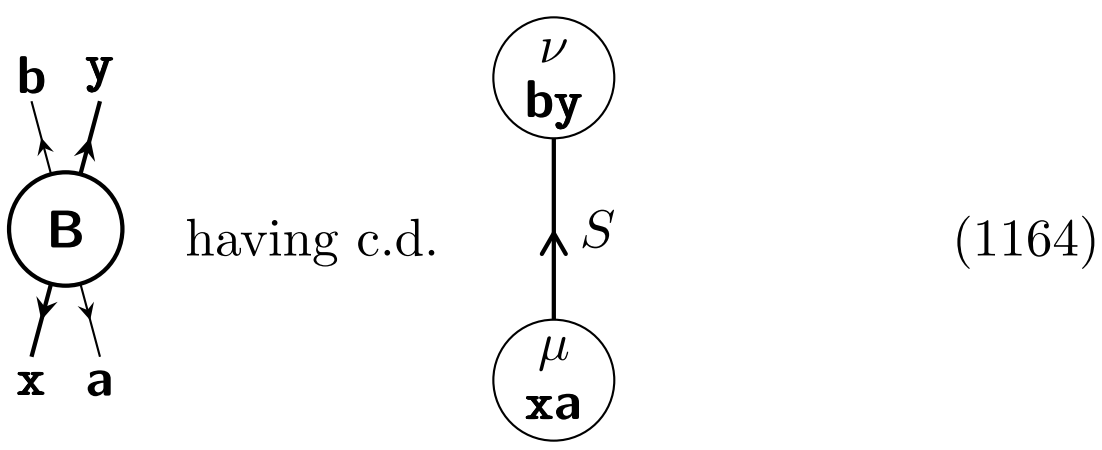

(1164)

satisfying simple double causality, we can write

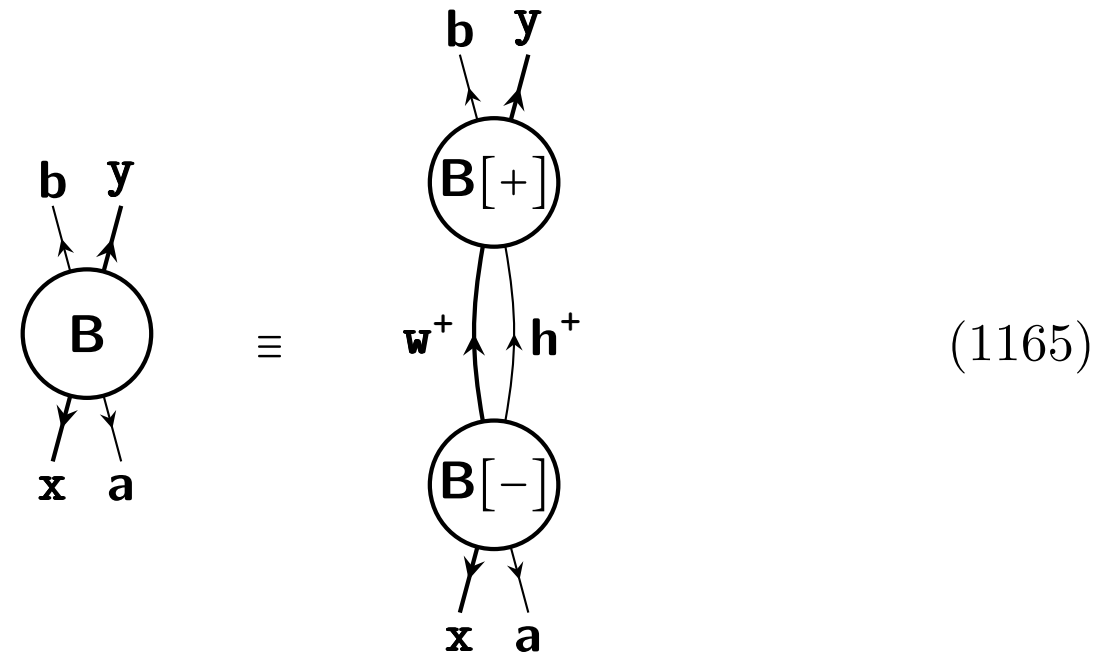

(1165)

for some $\mathbf{h}^+$ and $\mathtt{w}^+$, where

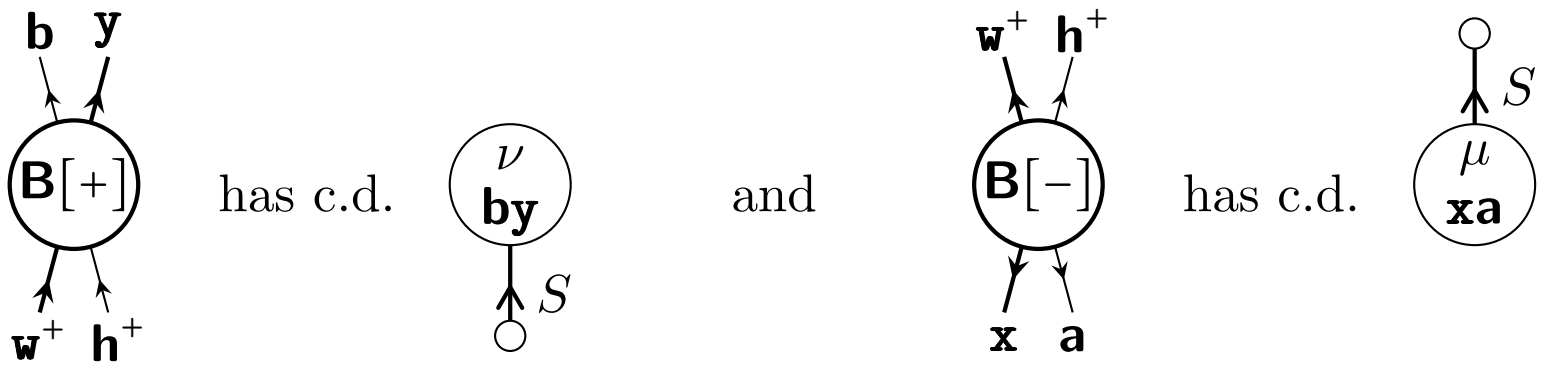

(1166)

and where the causally complex operations $\mathbf{B}[+]$ and $\mathbf{B}[-]$ satisfy the simple double causality condition for deterministic operators.

Clearly this assumption has the same form as the synchronous partition assumption with tester positivity stated in Sec. 49.2.5 where tester positivity is replaced by satisfying the simple double causality condition and where we restrict to deterministic operations.

### 49.4.4 Weak synchronous partition assumption for causality

Now we state the aforementioned weaker assumption.

**Weak synchronous partition assumption with simple double**

**causality.** For any deterministic causally complex operation

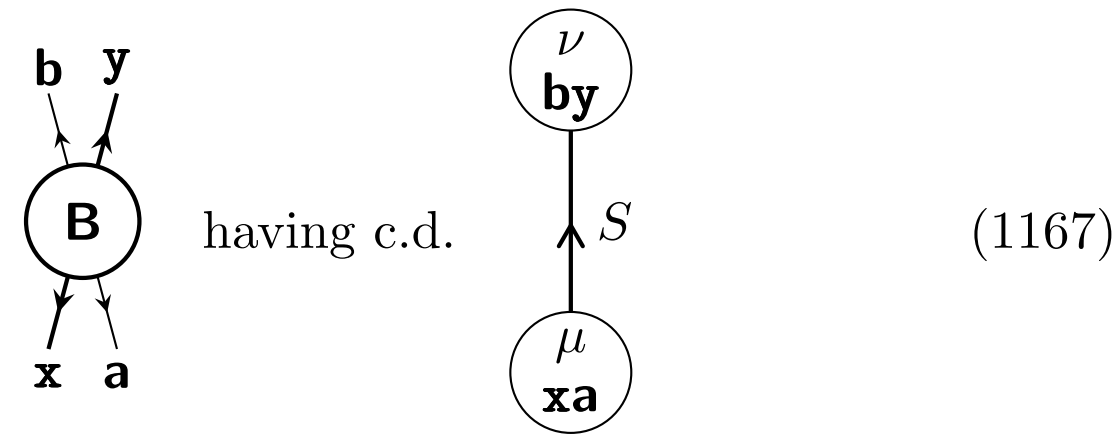

satisfying simple double causality, we can write

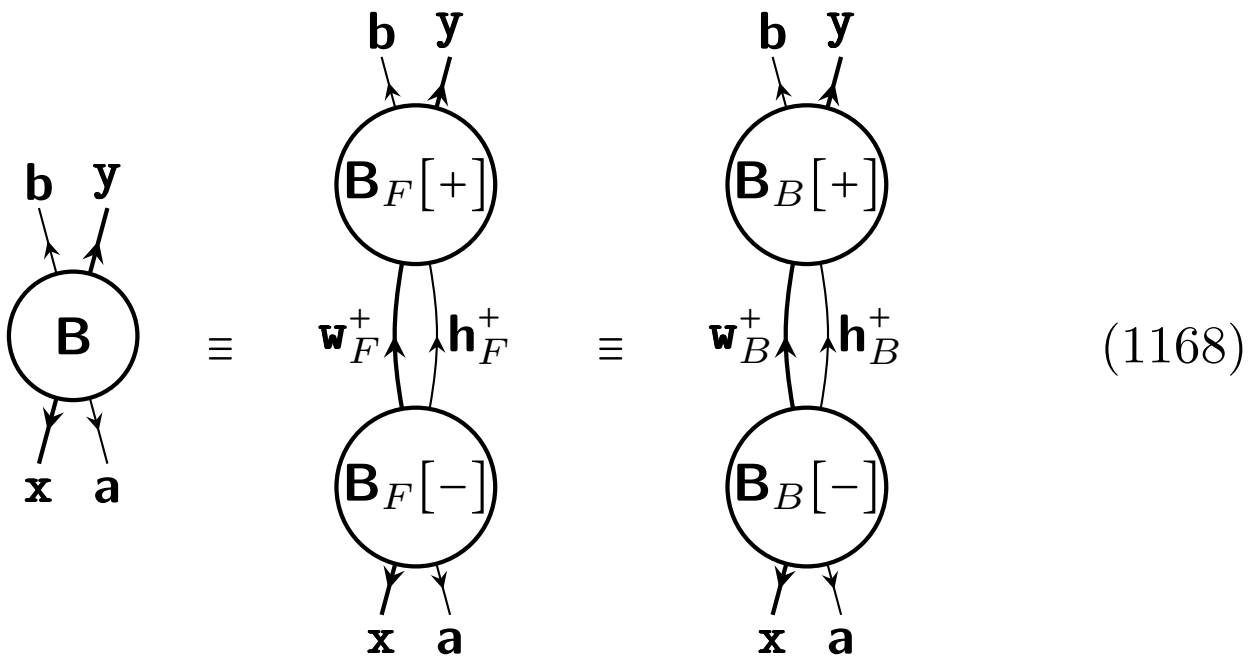

for some $\mathbf{h}^+_{F,B}$ and $\mathbf{w}^+_{F,B}$, where

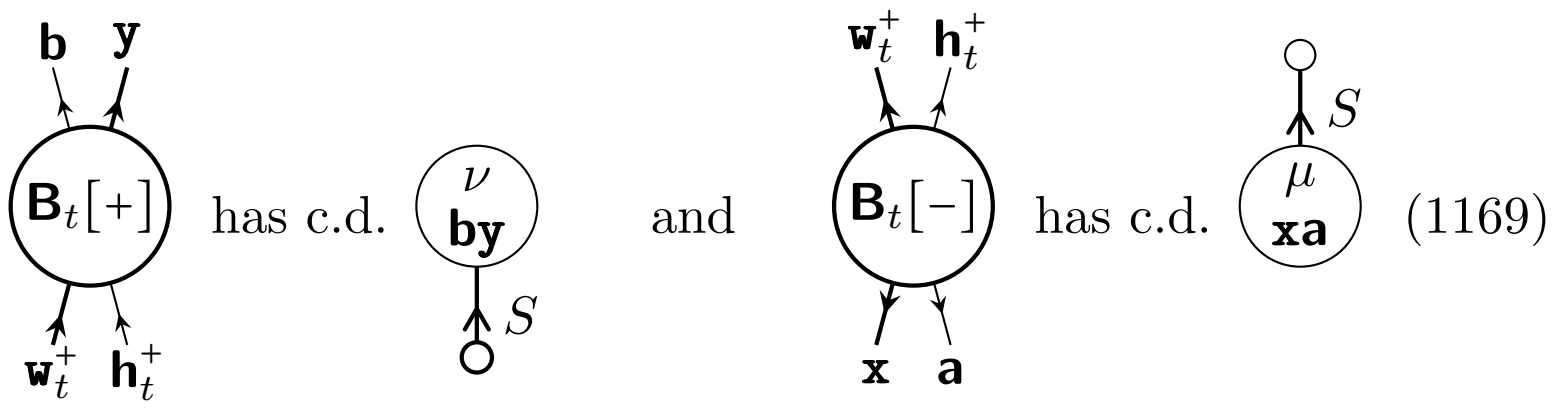

(for $t = F, B$) and where $\mathbf{B}_F[+]$ satisfies simple forward causality (for a deterministic operator) and $\mathbf{B}_B[-]$ satisfies simple backward causality (for a deterministic operator).

This assumption is clearly weaker than the synchronous partition with simple double causality assumption.

### 49.4.5 Double causality theorem I

We will prove the following theorem.

**Double causality theorem I**. Assume the weak synchronous partition assumption for causality holds. Consider a deterministic causally complex operation **B** satisfying simple double causality such that

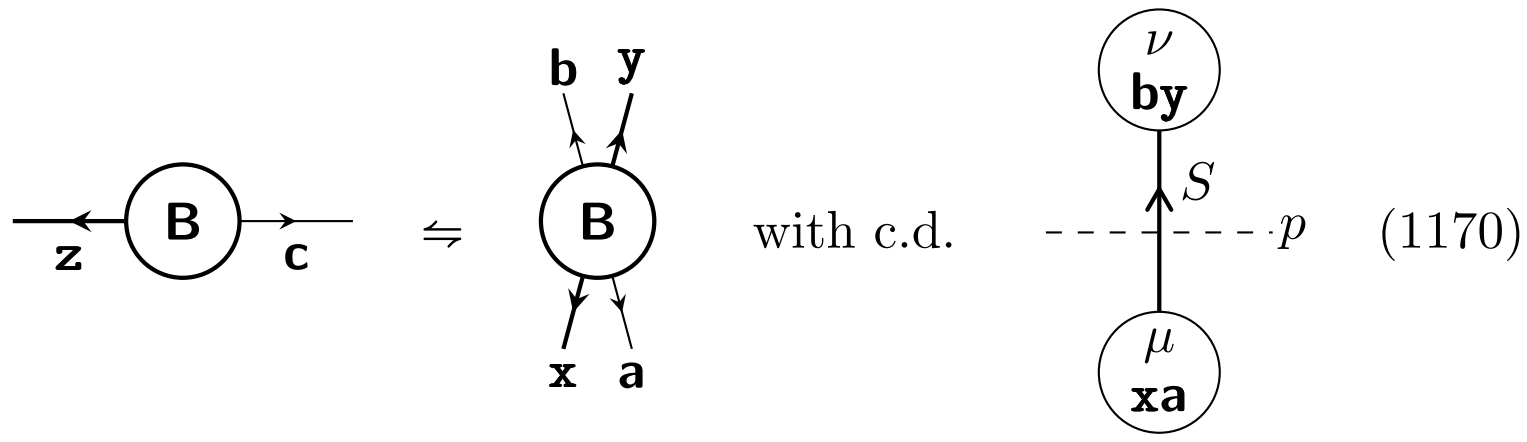

(1170)

Then the *forward causality* condition

(1171)

follows. The *backward causality* condition

(1172)

also follows.

We will prove the forward causality condition. The backward causality condition follows similarly. First we note that it follows from the weak synchronous partition for causality assumption that $\mathbf{B}_F[+]$ satisfies simple forward causality. Hence

(1173)

(To see this note that all systems on open wires are moving towards $\mathbf{B}_F[+]$ so we have a result.) Since **B** satisfies simple double causality we can use the synchronous partition assumption (with simple double causality) to obtain the first step below.

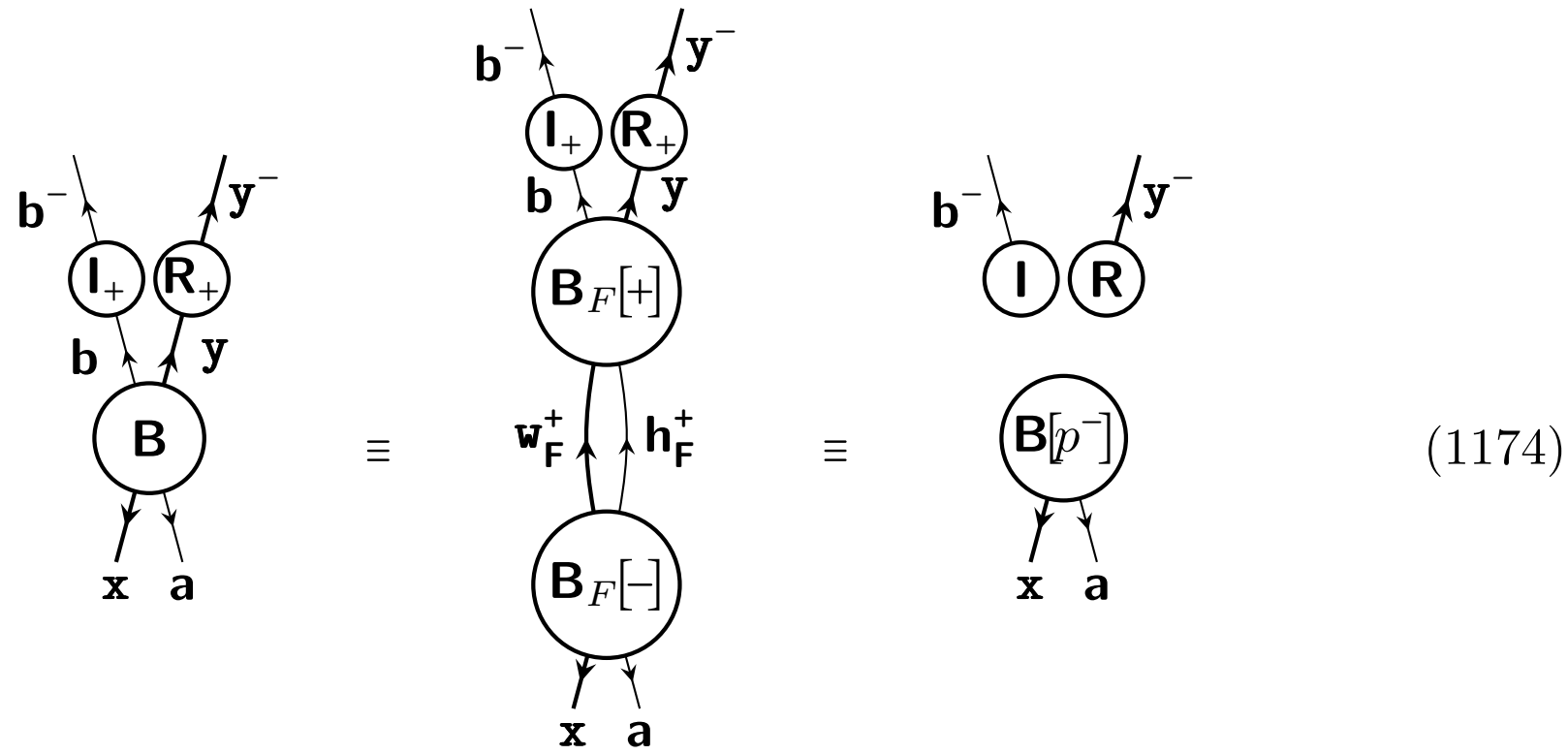

(1174)

By the synchronous partition assumption (with simple double causality) $\mathbf{B}_F[+]$ satisfies simple forward causality and hence we can use (1173) obtaining the second step above where

$$\mathbf{B}[p^-] \equiv \mathbf{B}_F[-] \text{ with } \mathbf{R} \text{ on } \mathbf{w}_F^+ \text{ and } \mathbf{I} \text{ on } \mathbf{h}_F^+ \quad (1175)$$

This proves forward causality condition. The backward causality condition is proved similarly.

### 49.4.6 Double causality theorem II

Here we offer an alternative proof of double causality that is, in some ways, more compelling since it uses the very basic assumption of no correlations without causation from Sec. 48.9 instead of the weak synchronous partition assumption. However, it only applies to theories that satisfy weak tomographic locality (as defined in Sec. 48.9). Since Quantum Theory and Classical Probability Theory are two such theories this is still of interest.

We state the theorem as follows

> **Double causality theorem II**. Assume no correlation without causation and weak tomographic locality. Consider a deterministic causally complex operation $\mathbf{B}$ satisfying simple double causality such

that

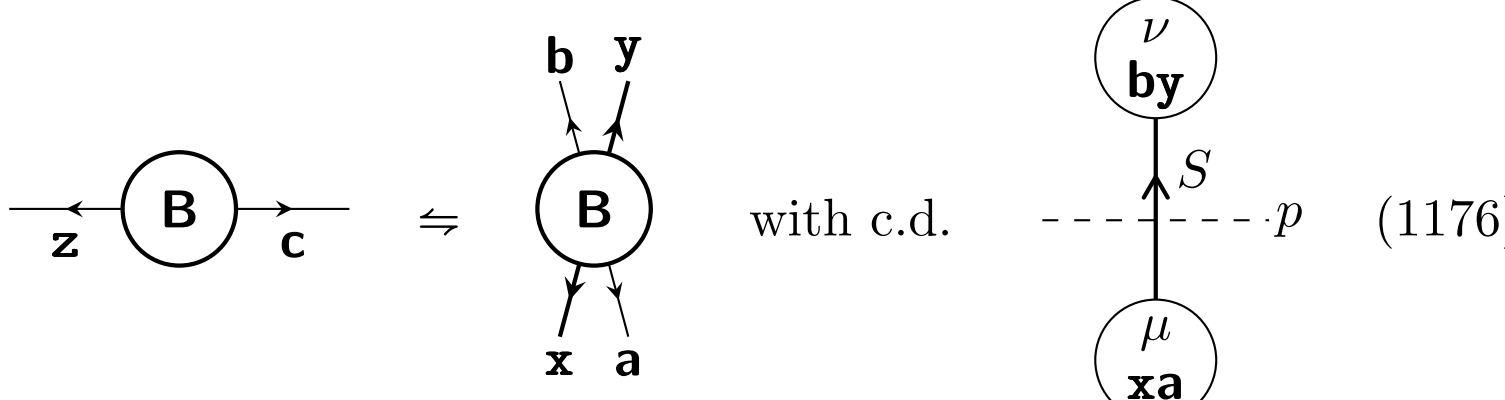

(1176)

Then the *forward causality* condition

$$\equiv \qquad (1177)$$

follows. The *backward causality* condition

$$\equiv \qquad (1178)$$

also follows.

Consider the forward case. From the fact that **B** satisfies simple forward causality we can obtain

$$\equiv \qquad (1179)$$

by first applying simple double causality to **B** then closing the open $\mathbf{a}^-$ and $\mathbf{x}^-$

wires with **I** and **R** respectively. Next we note that

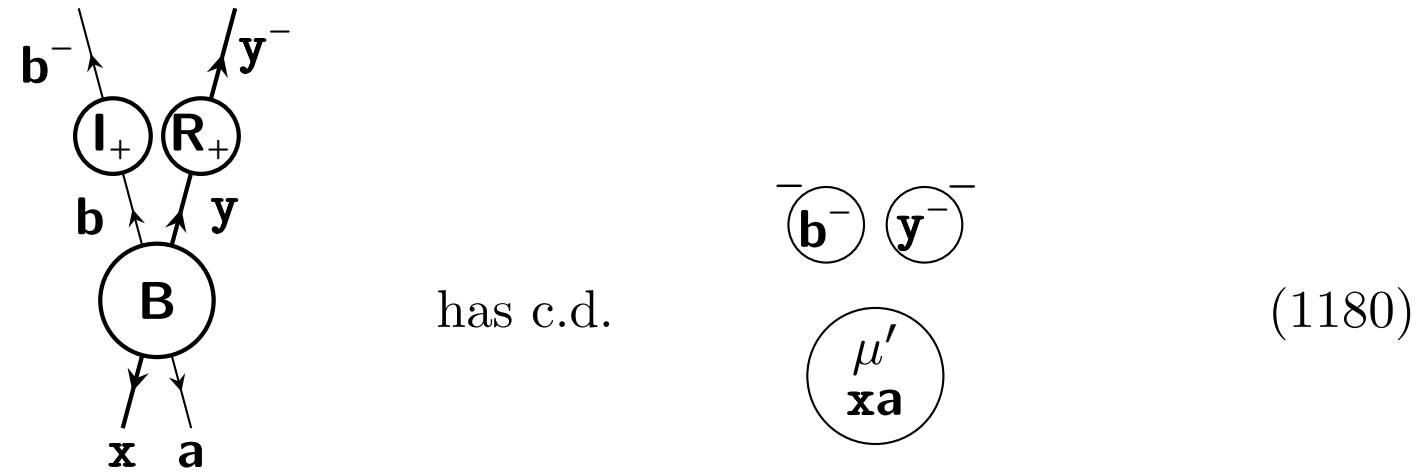

(1180)

This is true because $\mathbf{b}^-$ and $\mathbf{y}^-$ are, respectively, an open input and an open income which means they can only have causal arrows coming out of them. However, given the form of the causal diagram in (1176) and the fact that we have closed $\mathbf{b}^+$ and $\mathbf{y}^+$, there is no where for any such causal arrows to go. Thus, the causal diagram must take this form. Since the causal diagram has disjoint parts we can apply the *operations factorise when their causal diagrams do* theorem from Sec. 48.9 (where **b** becomes $\mathbf{b}^-$ and **y** becomes $\mathbf{y}^-$). Using (1179) this immediately gives (1177) where

(1181)

The backward case follows similarly. This proves the theorem.

### 49.4.7 Five residua theorems

The double causality conditions have the past residuum, $\mathbf{B}[p^-]$, and future residuum, $\mathbf{B}[p^+]$. We will prove five theorems concerning these residua. The first four theorems which enable us to treat them as deterministic physical operations with a known causal diagram. The fifth theorem is a simple observation concerning residua.

First we prove that residua can be calculated from the original operation

**Residua form theorem.** The past and future residua with respect to $p$ can be calculated from **B** as follows

(1182)

Furthermore, the residua are deterministic.

To prove the form for **B**$[p^-]$ note

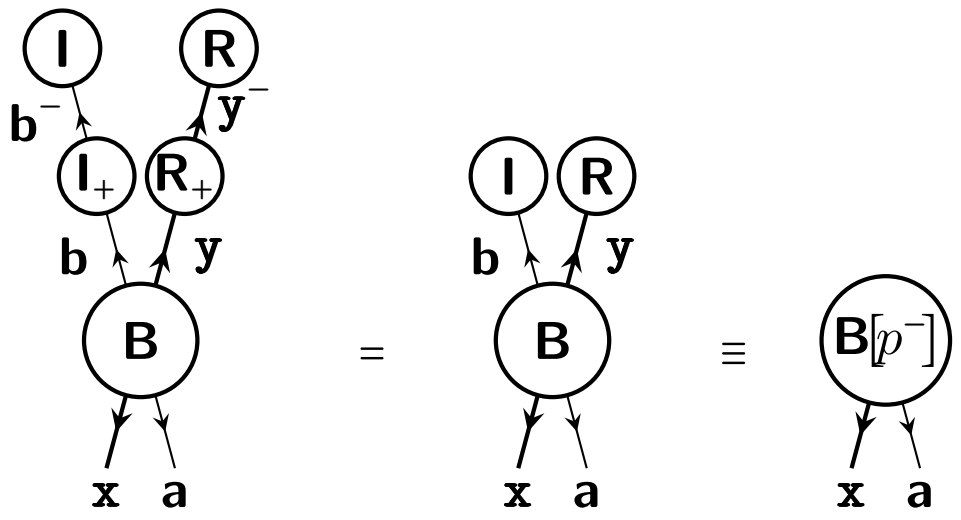

(1183)

where the middle expression follows from the left expression by the definition of $\mathbf{R}_+$ and $\mathbf{I}_+$. We have used (1091) and (1099) (in addition to (1162)) to go from the expression on the left to that on the right. This gives the form for the past residuum. The form for the future residuum is obtained similarly. That the residua are deterministic follows immediately from (1182) and the fact that **B**, **I**, and **R** are deterministic.

Next we prove

> **Residua tester positivity theorem.** If **B** satisfies tester positivity and the double causality conditions ((1162) and (1163)) then the residua, **B**$[p^\pm]$, also satisfy tester positivity.

The proof of this is straightforward. Using (1182) above, the theorem follows immediately from the **R I** partial contraction positivity theorem in Sec. 49.2.4.

Now we prove that the residua have the expected causal diagrams.

> **Residua causal diagrams theorem.** Given an operation

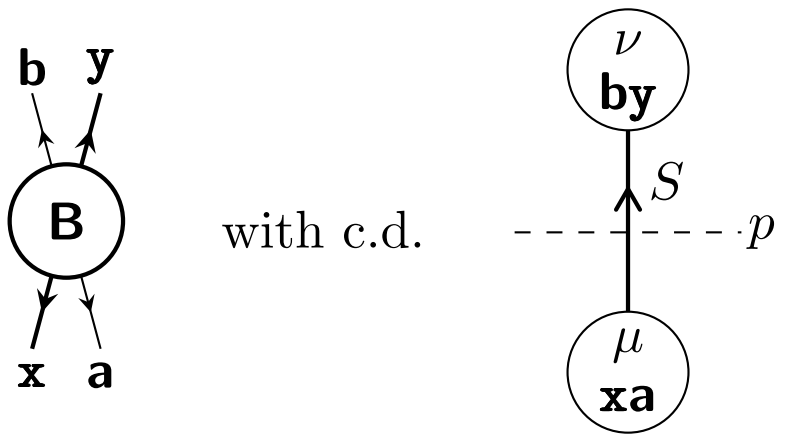

(1184)

> we have

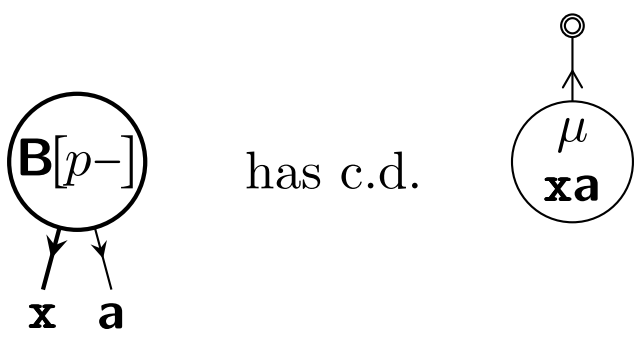

(1185)

for the past residuum (with respect to $p$), and for the future residuum (with respect to $p$).

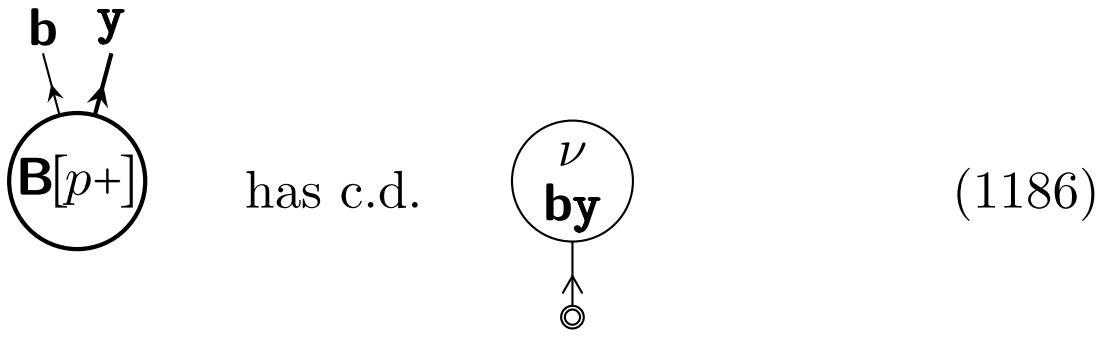

Note that we have used one-legged causal spider notation introduced in (1011) to close causal links. The proof of this is straightforward. Given the form for the past residuum in (1182), we can obtain the causal diagram for $\mathbf{B}[p^-]$ by contracting the causal diagram in (1184) with the causal diagrams for $\mathbf{I}$ and $\mathbf{R}$ (see Sec. 67.6 and Sec. 48.7). This gives the causal diagram in (1185). By similar reasoning we obtain (1186) for the future residuum.

Next we prove a theorem concerning the double causality of the residua.

**Residua causality theorem.** This theorem has three parts

**Forward.** If $\mathbf{B}$ satisfies forward causality then the past residua, $\mathbf{B}[p^-]$ also satisfy forward causality.

**Backward.** If $\mathbf{B}$ satisfies backward causality then the future residua, $\mathbf{B}[p^+]$ also satisfies backward causality.

**Double.** If $\mathbf{B}$ satisfies double causality then the residua, $\mathbf{B}[p^\pm]$ also satisfy double causality.

The three parts of this theorem are logically distinct but, clearly, related. To see this note that, in what follows, to prove that $\mathbf{B}[p^-]$ satisfies backward causality we need to consider both forward causality (to get hold of $\mathbf{B}[p^-]$) and backward causality. First we prove that $\mathbf{B}[p^-]$ satisfies forward causality if $\mathbf{B}$ does (this proves the "Forward" part of the theorem). Assume that we have two synchronous partitions, $p$ and $q$ where $q < p$ (i.e. $q$ does not cross $p$ and is earlier than $p$). Put

b y B x a ⇋ b y e B v u d (1187)

where $p$ lies between nodes **by** and **ev** while $q$ lies between **ev** and **ud**. Applying forward causality on synchronous partition $p$ gives

$$\equiv \tag{1188}$$

Applying the forward causality on synchronous partition $q$ instead gives

$$\equiv \tag{1189}$$

We can equate the right hand sides of (1188) and (1189). Next, we hit both sides of this with $\mathbf{I}_{\mathbf{b}^-}\mathbf{R}_{\mathbf{y}^-}$ yielding

$$\equiv \tag{1190}$$

This shows that $\mathbf{B}[p^-]$ satisfies forward causality with respect to $q$ completing the proof of the "Forward" part of the theorem. The "Backward" part of the theorem follows similarly. To prove the "Double" part of the theorem we need to prove that $\mathbf{B}[p^-]$ satisfies backward causality with respect to $q$ assuming that $\mathbf{B}$ satisfies double causality. First, applying forward causality on $p$, we obtain

$$\equiv \tag{1191}$$

Now, instead, applying backward causality on $q$, we obtain

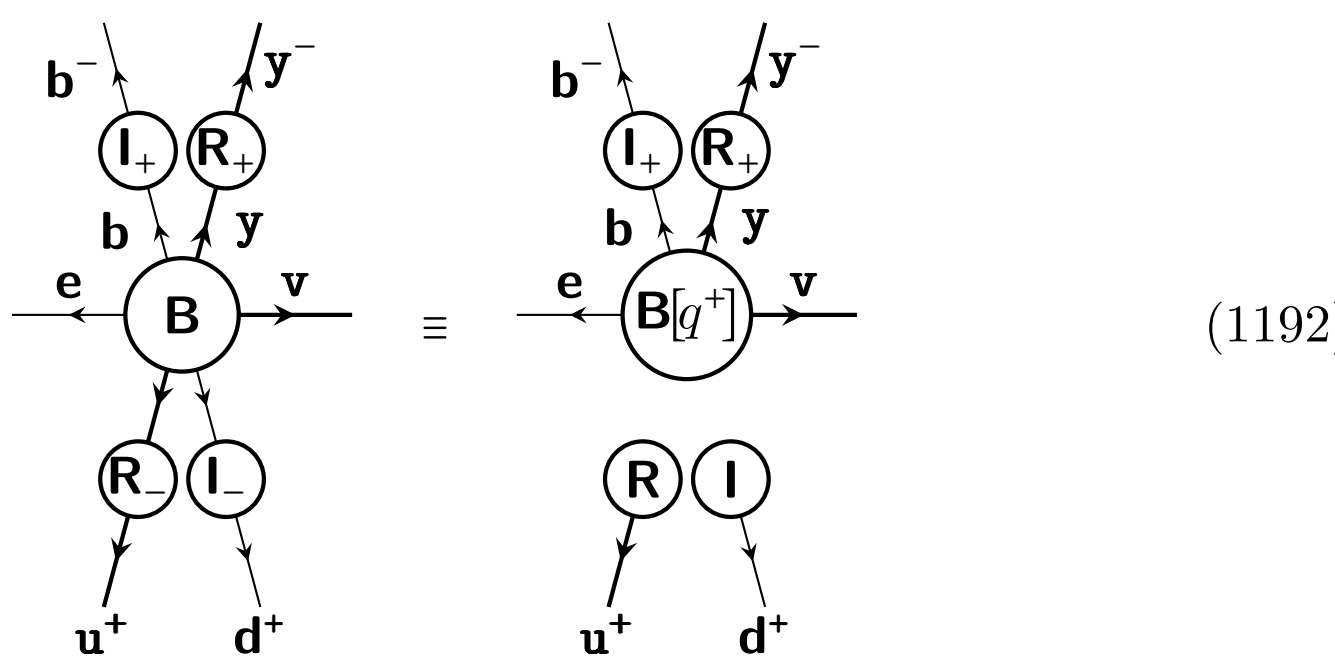

(1192)

Equating the right hand sides of (1191) and (1192) and hitting with **R**'s and **I**'s as appropriate, we obtain

(1193)

This proves that $B[p^-]$ satisfies backward causality (where we can read off the resulting future residuum from the right hand side). We can prove, similarly, that $\mathbf{B}[p^+]$ satisfies forward causality if $\mathbf{B}$ satisfies double causality. This, together with our earlier proofs of the "Forward" and "Backward" parts of the theorem proves the "Double" part of the theorem.

The above theorems allow us to think of the causality conditions as iterative. We apply them to $\mathbf{B}$, then obtain a residuum, $\mathbf{B}[p^\pm]$, which must also satisfy these conditions (with a known causal diagram). So we can apply the causality conditions on the residua yielding new residua and so on. This yields a set of conditions with respect to some foliation into ordered synchronous partitions). Indeed, this iterative attitude is taken in Gutoski and Watrous [2007] for quantum strategies and Chiribella et al. [2009a] for quantum combs. However, we can take a different attitude where we think of the double causality conditions (1162, 1163) as imposing a forward and backward condition for every synchronous partition, $p$ (even though these may not belong to an ordered foliation).

Finally, we prove a theorem which allows us to deduce some properties of an operation from its residuum.

**Forward and backward causality of an operation from its**

**residua.** This theorem consists of a forward case and a backward case.

**Forward case.** If an operation, **B**, satisfies deterministic forward causality at $p$ and, further, the resulting past residuum, $\mathbf{B}[p^-]$, satisfies deterministic forward causality for all synchronous partitions, then **B** satisfies deterministic forward causality for all $q \leq p$.

**Backward case.** If an operation, **B**, satisfies backward causality at $p$ and, further, the resulting future residuum, $\mathbf{B}[p^+]$, satisfies deterministic backward causality for all synchronous partitions, then **B** satisfies deterministic backward causality for all $q \geq p$.

Here the notation $q \leq p$ means that the synchronous partition, $q$, does not intersect any wires of the causal diagram that are to the future of the wires intersected by $p$. This theorem is essentially obvious. We will prove the forward case. The past case follows by similar reasoning. The condition for forward causality of **B** at $p$ being satisfied is given in (1737) above. The condition that $B[p^-]$ satisfies forward causality at some synchronous partition, $q$, is given in (1190) above. If we substitute the latter into the former then we obtain (1189) which is the condition for **B** satisfying forward causality at $q$ (which is earlier than $p$). This holds for all $q \leq p$ and so we prove the theorem.

### 49.4.8 Simple double causality as special case

The full set of double causality conditions are generated by applying (1162) and (1163) to every synchronous partition of the causal diagram. We will see that the simple double causality condition in (1159) and (1160) are special cases.

First consider

$$\overset{\mathbf{z}}{\longleftarrow} \textcircled{\mathbf{B}} \overset{\mathbf{c}}{\longrightarrow} \tag{1194}$$

We can apply forward causality to a synchronous partition, $f$, in which the nodes **c** and **z** are to the future. We obtain

$$\overset{\mathbf{z}^-}{\longleftarrow} \textcircled{\mathbf{R}_+} \overset{\mathbf{z}}{\longleftarrow} \textcircled{\mathbf{B}} \overset{\mathbf{c}}{\longrightarrow} \textcircled{\mathbf{I}_+} \overset{\mathbf{c}^-}{\longrightarrow} \quad \equiv \quad \overset{\mathbf{z}^-}{\longleftarrow} \textcircled{\mathbf{R}} \quad \textcircled{\mathbf{I}} \overset{\mathbf{c}^-}{\longrightarrow} \tag{1195}$$

We obtain this by applying (1162) with **by** set to **cz** and with **a** and **x** each set to the null system. Note that, when **a** and **x** are null, $\mathbf{B}[f-]$ becomes a circuit (i.e. it has no open wires). Since **B** is deterministic, $\mathbf{B}[f-]$ is deterministic. A deterministic circuit is equivalent to 1 and so $\mathbf{B}[f-]$ drops out. The condition in (1195) is the simple forward causality condition (1159).

We can obtain the simple backward causality in (1160) in a similar fashion but now applying the backward causality condition (1163) to the synchronous partition, $b$, with the nodes **c** and **z** are to the past. Then we apply (1163) with **ax** set to **cz** and where **b** and **y** each set to the null system.

### 49.4.9 Double causality conditions for simple causal structure

Here we consider what happens when we apply the double causality conditions to the case where we have simple causal structure (see Sec. 47.16). For example,

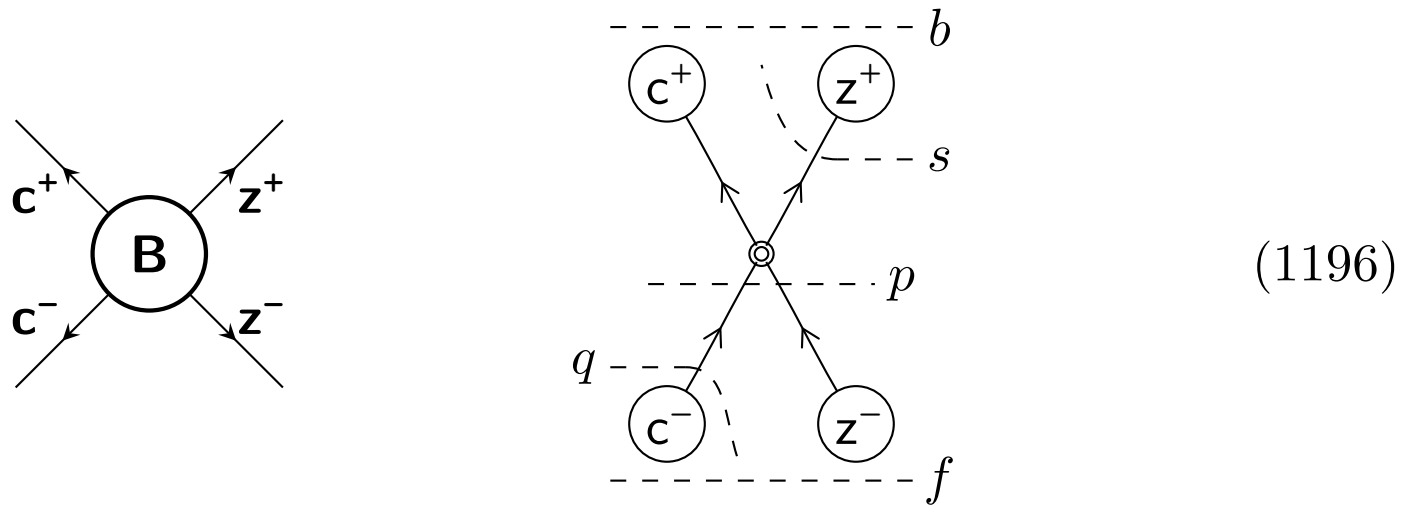

(1196)

There exist multiple synchronous partitions of the causal diagram. We will consider the examples $f$, $b$, $p$, $q$, and $s$ shown above. We already know from Sec. 49.4.8 that, if we apply forward causality to $f$, we obtain the simple forward causality condition. In our present notation this is

(1197)

If we apply backward causality to $b$, we obtain the simple backward causality condition.

(1198)

Here we will see that, when we apply the double causality conditions to other synchronous partitions, we do not obtain any new conditions when the causal structure is simple.

If we apply the forward causality condition with respect to the partition that has incomes in the past and outcomes in the future (this is $p$) we obtain

(1199)

If we use the fact that $B[p^-]$ is deterministic and has only incomes then this condition reduces to the simple forward causality condition in (1197).

Now consider applying the forward causality with respect to partition $q$. This gives

(1200)

Now, $\mathsf{B}[q-]$ is a deterministic result (it has only an income). Consequently it is the ignore result and so we get back the simple forward causality condition.

If we apply forward causality with respect to $s$ then we obtain

(1201)

which can be regarded as a mere definition (rather than an actual condition).

It is easy to see that, regardless of which synchronous partition we use, the forward causality conditions give us either simple forward causality or a mere definition, and the backward causality conditions give us either simple backward causality or a mere definition. Thus, in general, when we have simple causal structure, the double causality conditions reduce to the simple double causality conditions.

### 49.4.10 General double causality

In Sec. 49.1.3 we saw that a complete set of operations sum to a deterministic operation. Thus, if we have a complete set of operations, $\mathsf{B}(u)$, then we can write down the double causality conditions as follows. Forward causality is

(1202)

and backward causality is

(1203)

for any complete set of operations $\mathsf{B}(u)$ and any synchronous partition, $p$.

A nondeterministic operation, $\mathsf{B}$, is a member of a complete set of operations. We can show, under certain positivity assumptions, that it satisfies inequality double causality conditions (this proof proceeds along the same lines as for the simple case in Sec. 7.13). Since $\mathsf{B}$ is a member of (at least one) complete set of operations we can associate (at least one) deterministic operation, **B**, with it (the sum of the elements of such a complete set). Then we can define B$[c]$ = **B** − B as the complement of B. If we assume that B$[c]$ satisfies tester positivity and **B** satisfies the double causality conditions (1162, 1163) then we can prove the following *general double causality conditions*. The *general forward causality condition* is

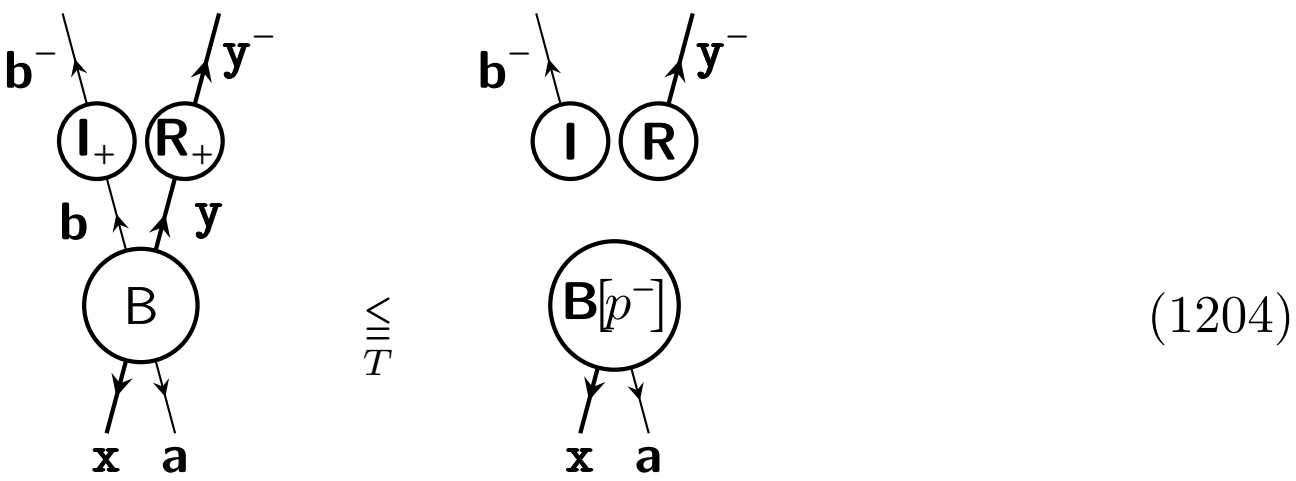

(1204)

and the *general backward causality condition* is

$$\textbf{b}\ \textbf{y} \quad \mathsf{B} \quad \textbf{x}\ \textbf{a} \quad \textbf{R}_-\ \textbf{I}_- \quad \textbf{x}^+\ \textbf{a}^+ \quad \underset{T}{\leq} \quad \textbf{b}\ \textbf{y} \quad \textbf{B}[p^+] \quad \textbf{R}\ \textbf{I} \quad \textbf{x}^+\ \textbf{a}^+ \tag{1205}$$

for suitable operations $\textbf{B}[p^\pm]$. To prove (1204) note that

$$\textbf{b}^-\ \textbf{y}^- \ \ \textbf{I}_+\ \textbf{R}_+ \ \ \textbf{b}\ \textbf{y} \ \ \mathsf{B} \ \ \textbf{x}\ \textbf{a} \quad + \quad \textbf{b}^-\ \textbf{y}^- \ \ \textbf{I}_+\ \textbf{R}_+ \ \ \textbf{b}\ \textbf{y} \ \ \mathsf{B}[c] \ \ \textbf{x}\ \textbf{a} \quad \equiv \quad \textbf{b}^-\ \textbf{y}^- \ \ \textbf{I}\ \textbf{R} \ \ \textbf{B}[p^-] \ \ \textbf{x}\ \textbf{a} \tag{1206}$$

since **B** satisfies (1162). We are assuming that B$[c]$ satisfies tester positivity. It follows from the **R I** partial contraction theorem in Sec. 49.2.4 that the second term in (1206) satisfies tester positivity. From this (1204) immediately follows. The backwards condition in (1205) follows similarly.

### 49.4.11 Causality composition theorem

Now we are in a position to prove the following.

> **Causality composition theorem (deterministic case).** Consider joining two or more deterministic causally complex operations. Then the following statements are true:

**Forward.** If each of these operations satisfies forward causality (1162) then the resulting network will also satisfy forward causality.

**Backward.** If each of these operations satisfies backward causality (1163) then the resulting network will also satisfy backward causality.

First we consider joining together just two operations having no pointer types. The cases where we have pointer types and where we have more than two operations are easy to deal with and will be discussed at the end of this proof. We will prove the forward case. The backward case follows similarly. Consider joining two deterministic operations, without pointer types, each of which satisfies forward causality. Any such situation can be represented as

a — A —h— B — b (1207)

We want to show that this satisfies forward causality with respect to any synchronous partition, $p$. In particular, let us consider the case where

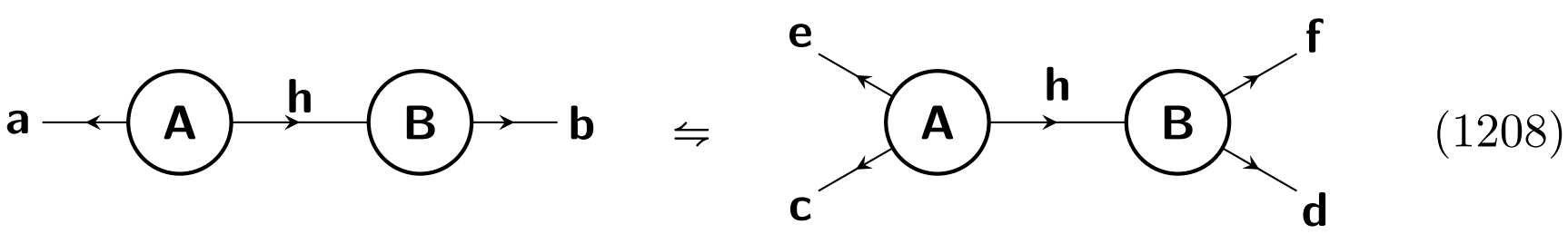

(1208)

with

$\nu$ **ef** — $S$ — $p$ — $\mu$ **cd** (1209)

To illustrate our remarks, consider the following example of a causal diagram

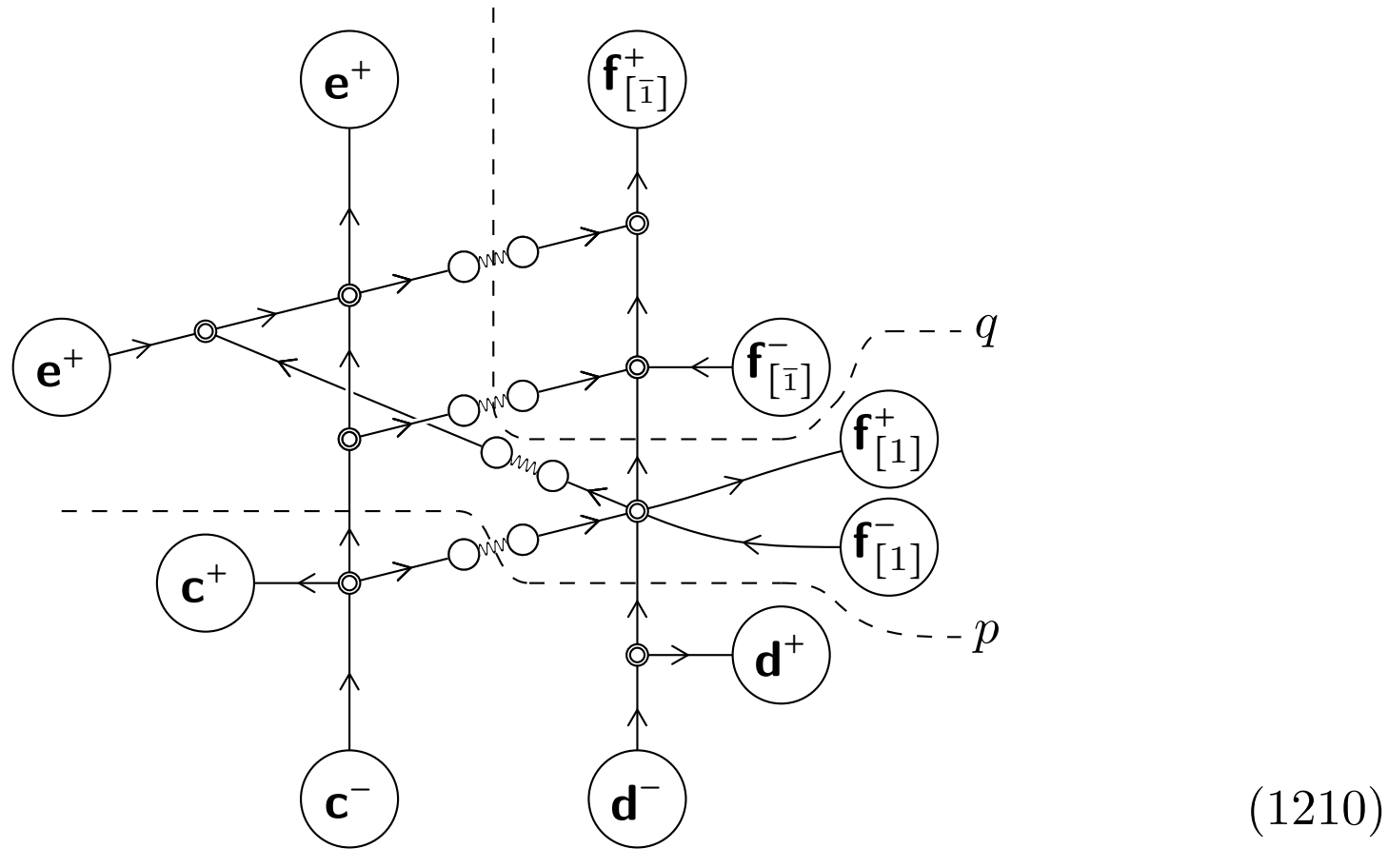

(1210)

where $\mathbf{f}^{\pm}_{[1]}\mathbf{f}^{\pm}_{[\bar{1}]} = \mathbf{f}^{\pm}$. In this example, a synchronous partition $p$, consistent with (1209) is shown. Further, in this example, the causal diagram to the left of the fusion nodes is associated with $\mathbf{A}$ and the causal diagram to the right of the fusion nodes is associated with $\mathbf{B}$. The statement that forward causality is satisfied by the network formed from $\mathbf{A}$ and $\mathbf{B}$ is the following

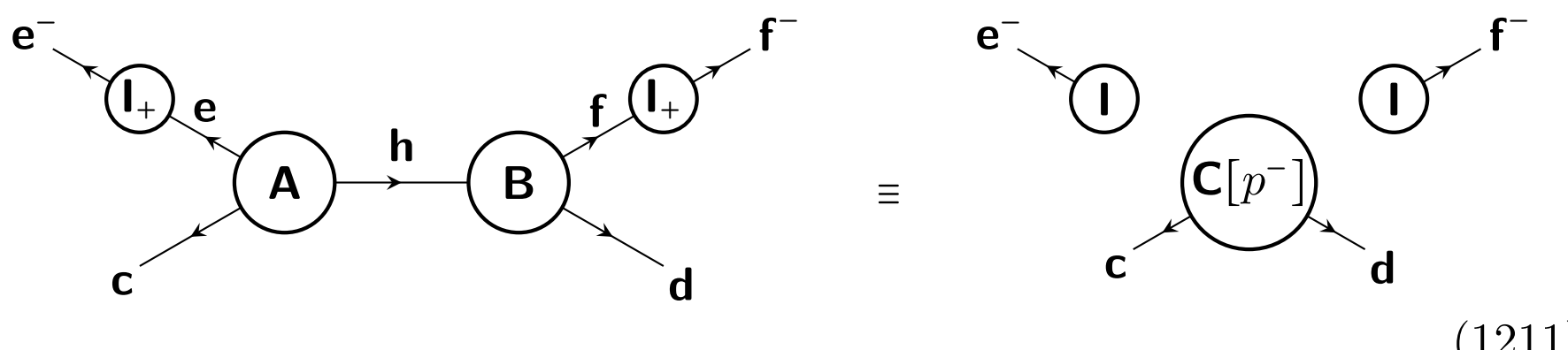

(1211)

Now we will consider taking a "bite" out of the $\mathbf{A}$ or $\mathbf{B}$ part of the causal diagram. We will illustrate this by taking a bite out of the $\mathbf{B}$ part of the causal diagram in the example in (1210) using the synchronous partition $q$. To facilitate this taking a bite, we write the expression on the left of (1211) in the interconvertible form

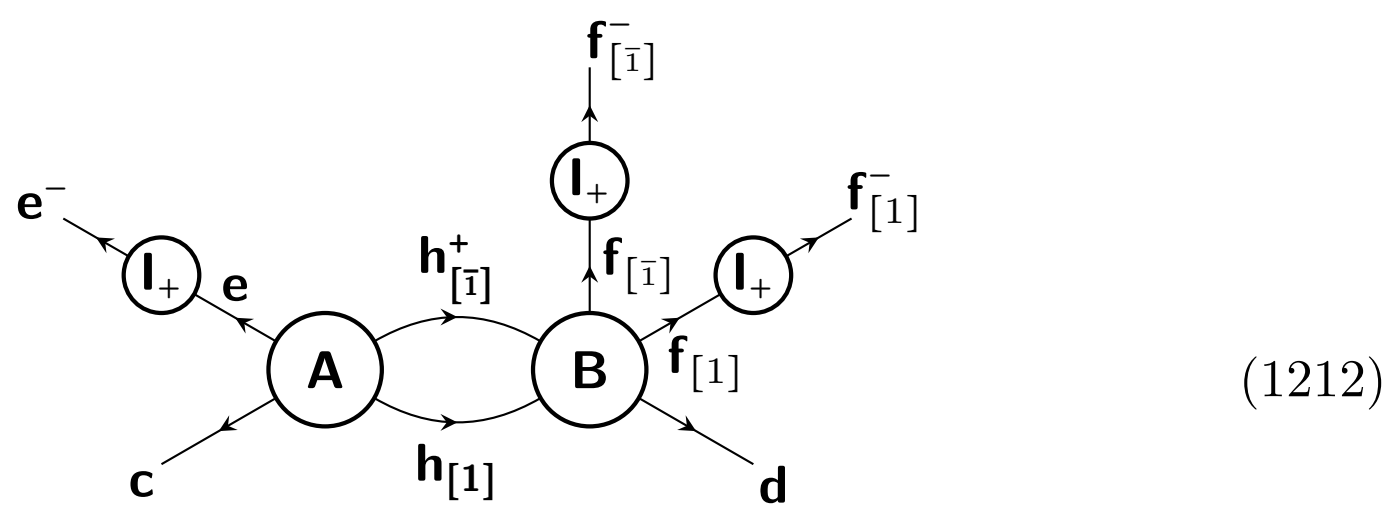

(1212)

where $\mathbf{h}_{[1]}\mathbf{h}^{+}_{[\bar{1}]} = \mathbf{h}$ and we suppose the causal diagram for $\mathbf{B}$ can be written

$$\begin{array}{c} \delta \\ \mathbf{h}^{+}_{[\bar{1}]}\mathbf{f}_{[\bar{1}]} \\ \uparrow Q \\ \text{- - - - - - } q \\ \gamma \\ \mathbf{h}_{[1]}\mathbf{f}_{[1]}\mathbf{d} \end{array} \tag{1213}$$

where $Q$ is a list of labels for the causal arrows. We can always find a synchronous partition, $q$, that takes a bite like this out of either $\mathbf{A}$ or (as in our example) $\mathbf{B}$. Crucially this bite has a system being inputted from the other side (this is the $\mathbf{h}^{+}_{[1]}$ system in our example). We have achieved this by imposing a total order on the fusion nodes (as clear in the example (1210)). Now $\mathbf{B}$ satisfies forward causality (according to our theorem statement). By applying forward causality on $q$ we see that the expression in (1212) is equivalent to

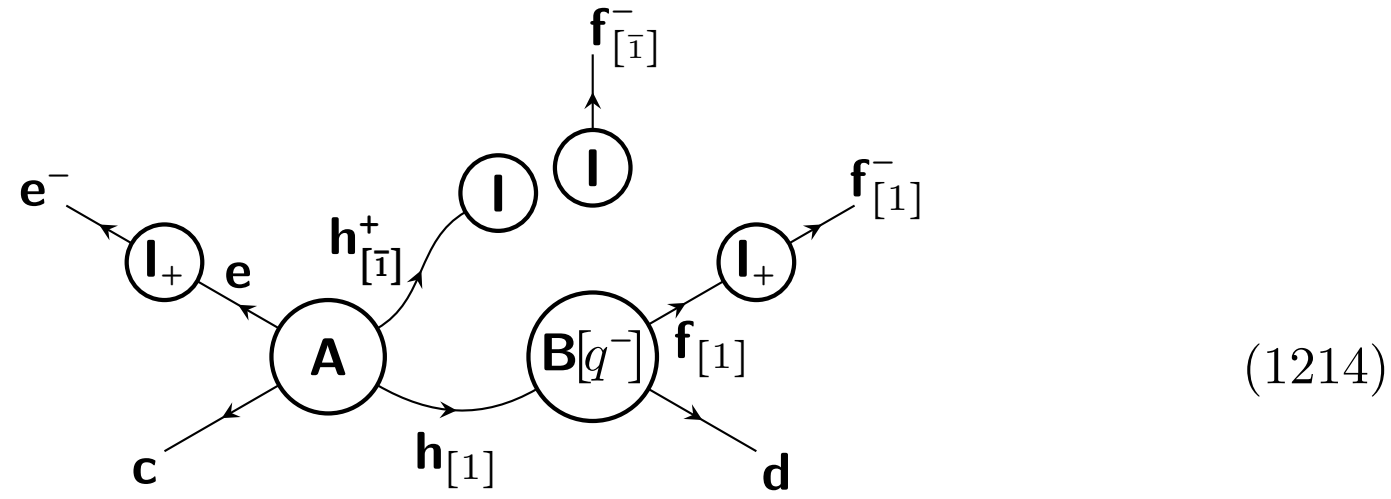

$$(1214)$$

which we write as

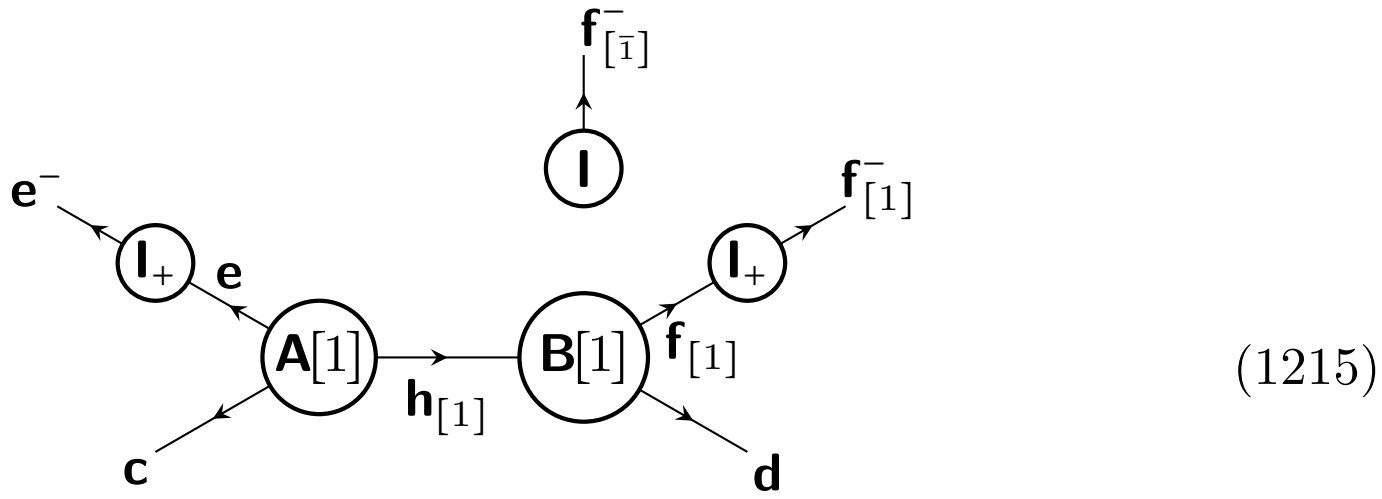

$$(1215)$$

where $\mathsf{A}[1]$ is equal to $\mathsf{A}$ with the $\mathsf{I}$ attached, and $\mathsf{B}[1] = \mathsf{B}[q^-]$. It is easy to show that $\mathsf{A}[1]$ satisfies forward causality (given that $\mathsf{A}$ does according to the theorem statement). Further, it follows from the residua causality theorem in Sec. 49.4.7, that $\mathsf{B}[1]$ satisfies forward causality. Hence, we are back where we started (compare (1215) with the expression on the left of (1211)) except that we have some [1] labels and we have the "crumb" $\mathsf{I}^{f^-_{[\bar{1}]}}$. Next we can take a bite out of the other side (i.e. out of $\mathsf{A}$). We can keep taking bites until we reach the scynchronous partition, $p$. We will accumulate crumbs such that, when we reach $p$, we will satisfy (1211) (where we use the property that ignore boxes factorise (1100)). In our example, it is clear we can reach $p$ after a total of three such bites and that the crumbs will give rise to $\mathsf{I}^{\mathsf{e}^-}$ and $\mathsf{I}^{\mathsf{f}^-}$. This proof technique works for any synchronous partition, $p$. We can prove the backwards causality property in the same way. If we have pointer systems then the diagram becomes more complicated (every physical system wire needs to be accompanied by a pointer system wire and every $\mathsf{I}$ should be accompanied by an $\mathsf{R}$) but the graphical manipulations are the same. If we join together more than two operations we can proceed by placing any order on them, use above technique to prove that the first two satisfy double causality, then join the third (treating the first two as a single operation) and prove that this satisfies double causality, and so on.

We can, similarly, prove a composition theorem for joining operations some or all of which may be nondeterministic.

> **Causality composition theorem (general case).** Consider joining two or more causally complex operations. Then the following statements are true.

**Forward.** If each of these operations satisfies general forward causality (1204) then the resulting network will also satisfy general forward causality.

**Backward.** If each of these operations satisfies general backward causality (1205) then the resulting network will also satisfy general backward causality.

The proof of this is similar to the proof above for the deterministic case except where, now, we have non-bold A and B, and after each use of the forward (or backward) causality for nondeterministic operations, we have an inequality ($\leqq_T$) rather than equivalence. Thus, the steps in the above proof starting at (1207) and going to (1212) are the same except that, now, A and B are non-bold and (1211) (which is the thing we are trying to prove) should have a $\leqq_T$ rather than an $\equiv$. Now consider the step from (1212) to (1214). Now this step involves use of general forward causality on A with respect to $q$ (shown in (1209)) where A is not necessarily deterministic. Consequently, this step becomes

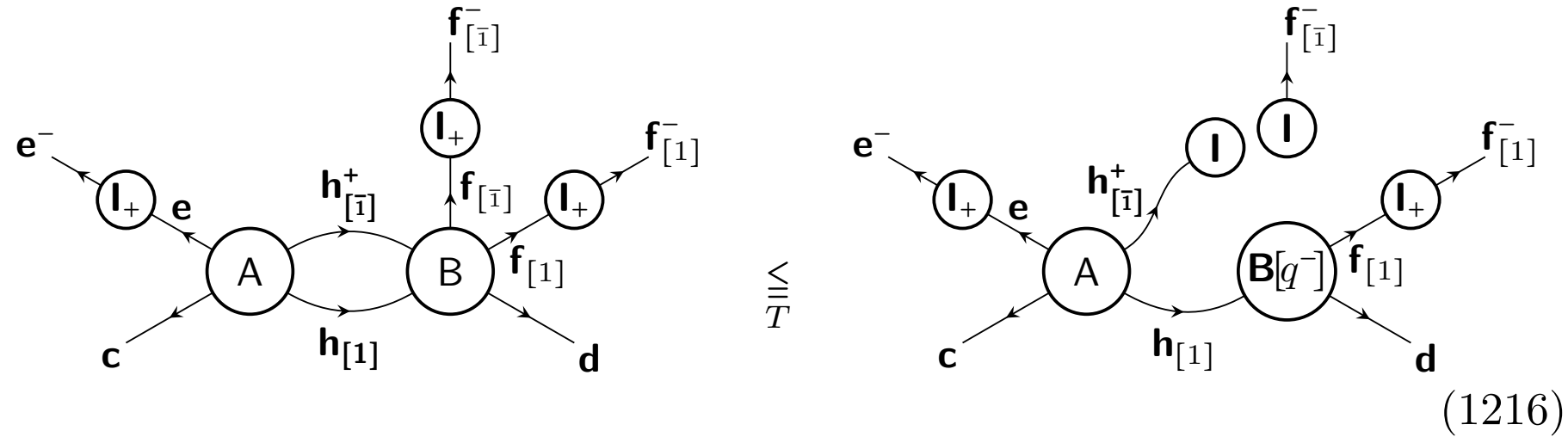

(1216)

Note, in particular, that the $\mathbf{B}[q^-]$ is, in fact, deterministic. We can write the right hand side of this inequality as

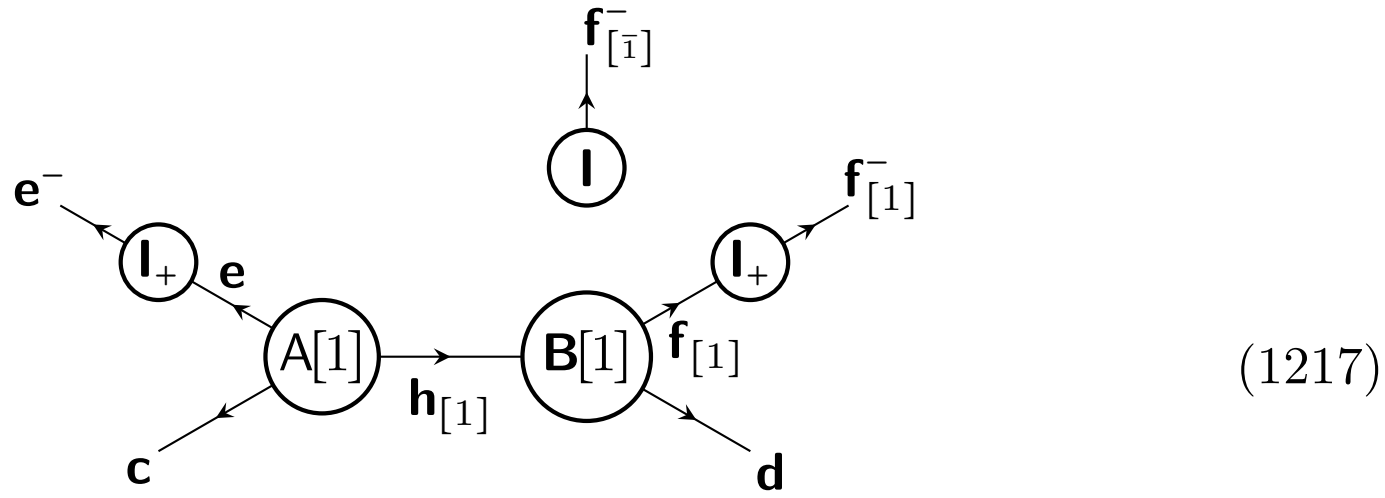

(1217)

where $\mathbf{B}[1]$ is set equal to $\mathbf{B}[q^-]$. This provides the first iterative step. Subsequent steps invoke use of either the general forward causality condition (1204) for the non-deterministic case or, the forward causality condition (1162) for the deterministic case (for example, if applied to $\mathbf{B}[1]$ here). Of course, once we introduce $\leqq_T$, we will have inequality even if some steps invoke $\equiv$. With these revisions to the proof of the composition theorem for the deterministic case we prove the composition theorem for this general case.

### 49.4.12 Residuum of a network

While we are here it is useful to prove a theorem that will be important later when we discuss Sorkin's impossible measurements in Sec. 56. This concerns the residuum of a network composed of two deterministic operations with respect to a synchronous partition that only intersects the causal diagram associated with one of the two operations.

> **Network residuum: special forward case**. Consider deterministic operations **A** and **B**, each satisfying forward causality, and subject to causal diagrams
>
> 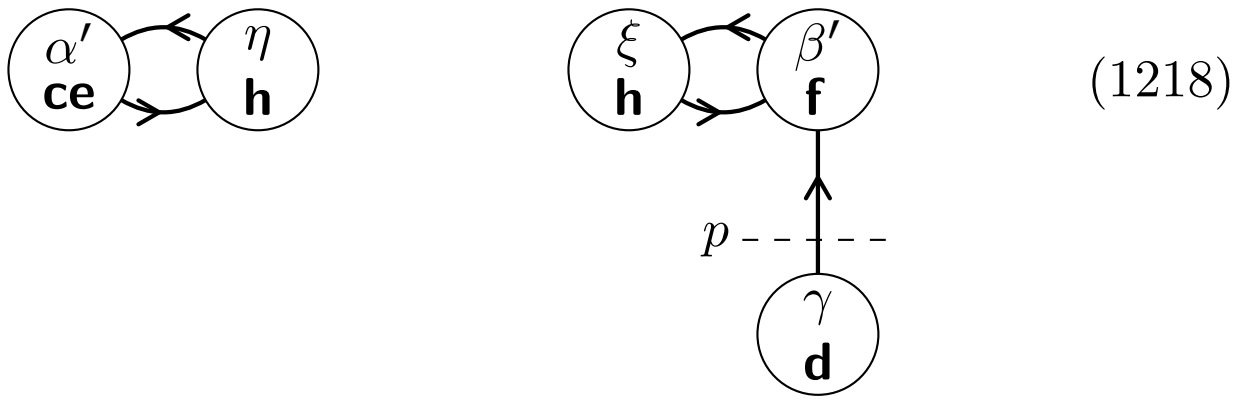
>
>
> $$(1218)$$
>
> respectively. Further, consider the network
>
> 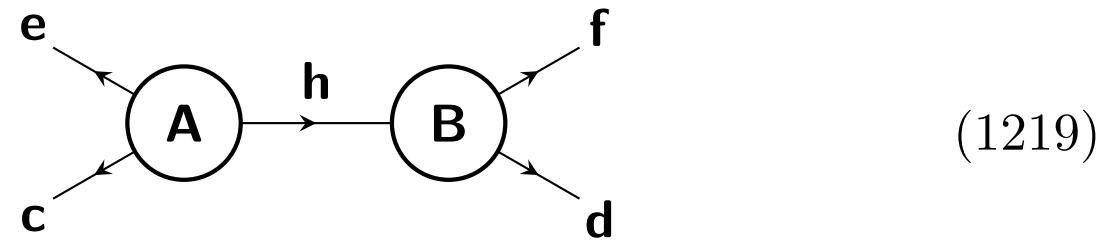
>
>
> $$(1219)$$
>
> which has causal diagram
>
> 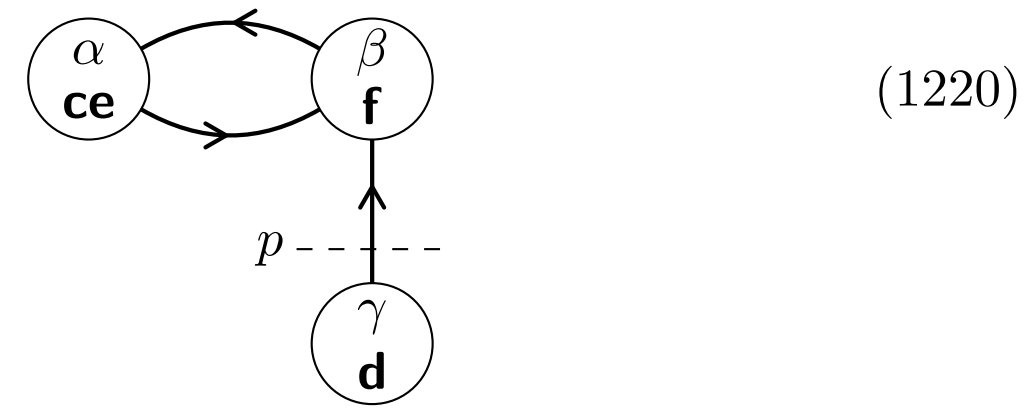
>
>
> $$(1220)$$
>
> Then the past residuum of the network in (1219) with respect to $p$ (shown in (1220)) is equal to $\mathbf{B}[p-]$ where this is the past residuum of **B** with respect to $p$ (shown in (1218)). In particular, note that the past residuum of the network, in these circumstances, does not depend on **A**.

Note that when we join the causal diagrams in (1218) we get a central node $\eta \circ \xi$ which can be absorbed to the left into $\alpha'$ or to the right into $\beta'$ (or split up and partially absorbed into both). The proof of this theorem follows by the same technique as we used to prove the causality composition theorem (deterministic case) in Sec. 49.4.11 by taking bites out of the causal diagrams until we reach $p$. Since $p$ is below any causal links connecting the two halves of the causal diagram (associated with **h**) the part of the causal diagram associated with **A** will be fully eaten up by these bites leaving just crumbs (as we follow the iterative process outlined in Sec. 49.4.11. In the final step we will end up with

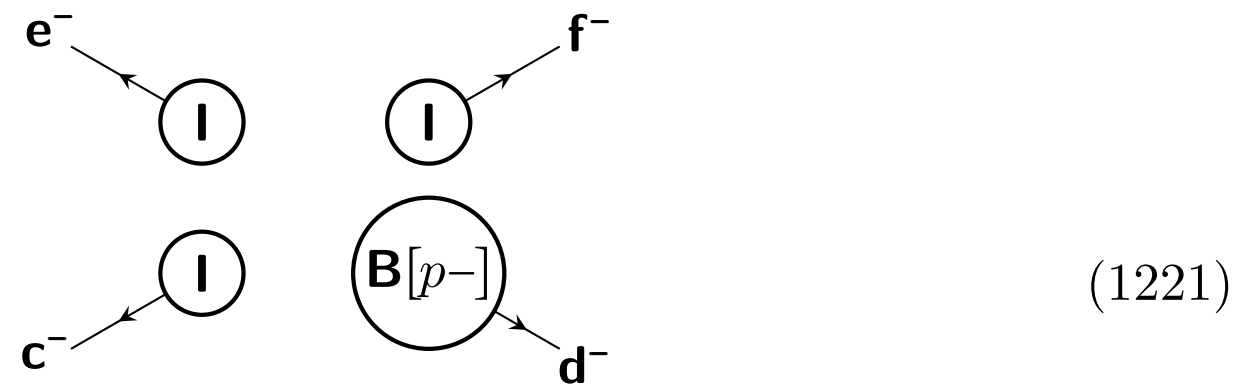

(1221)

which does not depend on $\mathbf{A}$. Hence the residuum is independent of $\mathbf{A}$ and, in fact, equal to $\mathbf{B}[p-]$ as stated in the theorem. This proves the theorem. It is, perhaps, useful to consider how this works with respect to the example causal diagram in (1210) given in Sec. 49.4.11 but where now the $p$ synchronous partition is as follows

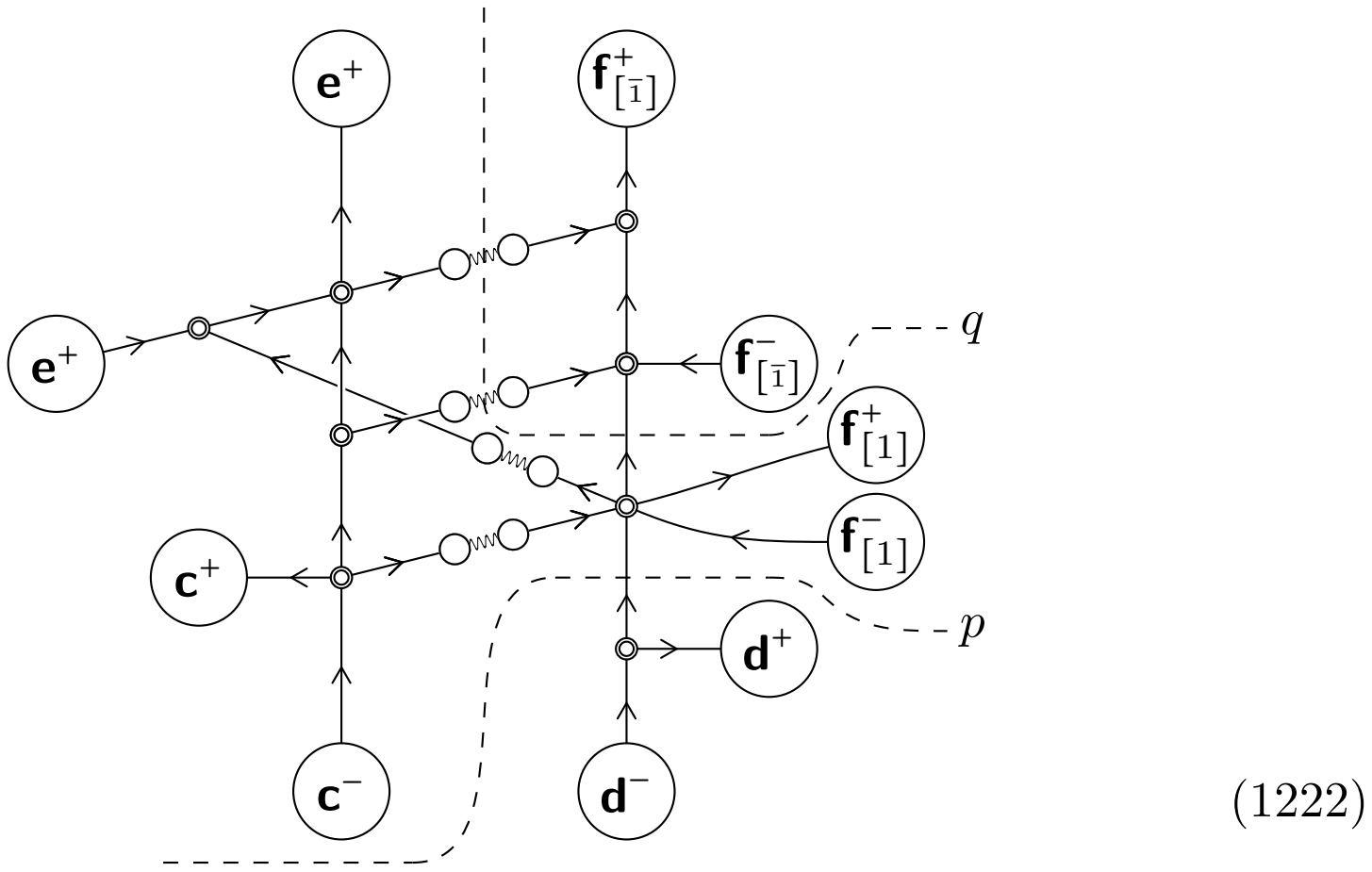

(1222)

We can see, in this example, that after taking bites out of the causal diagram to get to $p$ we end up with only crumbs on the $\mathbf{A}$ side of the causal diagram (as in (1221)).

For completeness, we also state the backward version of this theorem

> **Network residuum theorem: special backward case**. Consider deterministic operations $\mathbf{A}$ and $\mathbf{B}$, each satisfying forward causality, and subject to causal diagrams

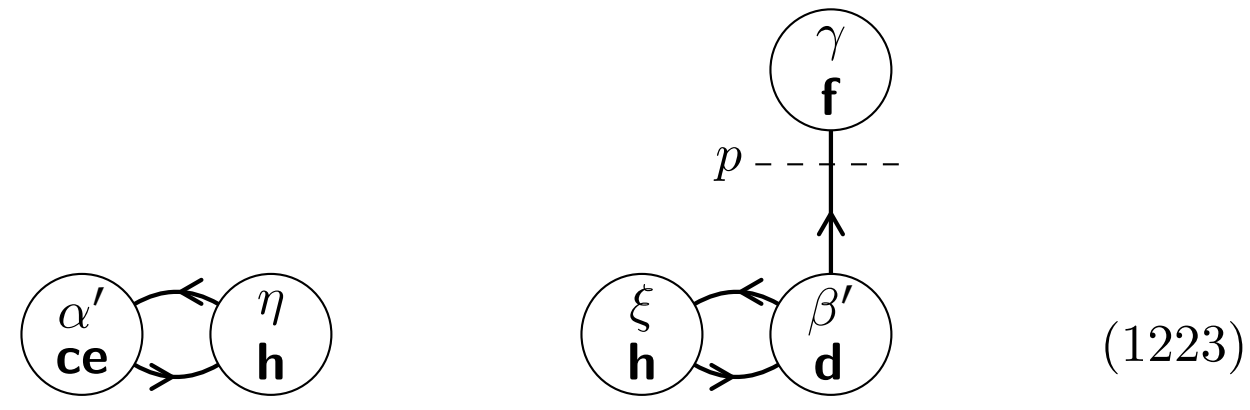

(1223)

respectively. Further, consider the network

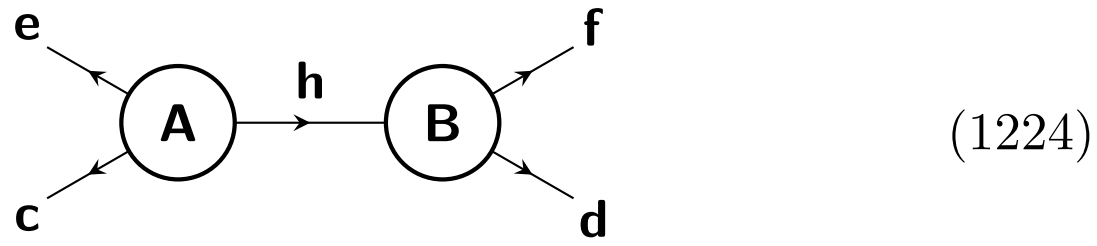

$$(1224)$$

which has causal diagram

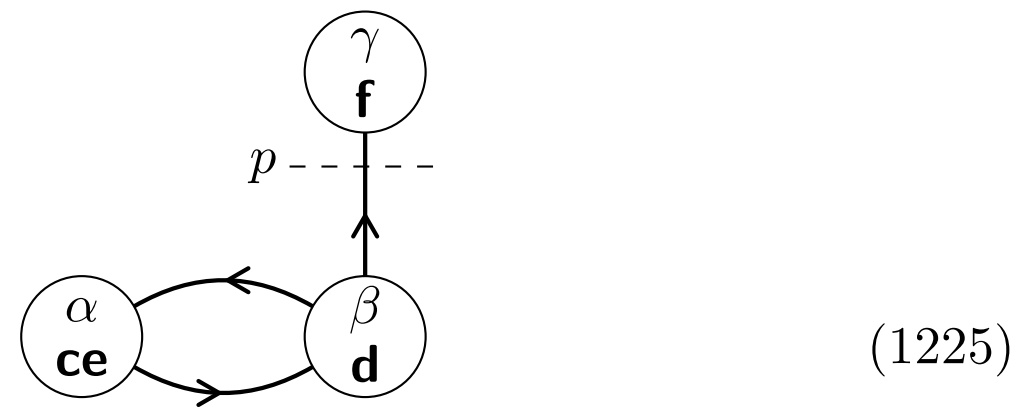

$$(1225)$$

> Then the future residuum of the network in (1219) with respect to $p$ (shown in (1220)) is equal to $\mathsf{B}[p+]$ where this is the future residuum of $\mathsf{B}$ with respect to $p$ (shown in (1218)). In particular, note that the future residuum of the network, in these circumstances, does not depend on $\mathsf{A}$.

The proof of this follows by a similar technique.

### 49.4.13 Subunity of circuits

The following follows immediately from the causality composition theorem

> **Subunity theorem.** Any circuit built from complex operations satisfying general double causality has probability less than or equal to one.

If we build a circuit, $C$, out of operations satisfying double causality then the causality composition theorem tells us that $C$ will also satisfy general double causality. Application of general double causality to a circuit, $\mathsf{C}$, simply yields $\mathsf{C} \leqq_T \mathbf{C}$ where $\mathbf{C}$ is a deterministic circuit (since there are no open wires to provide $\mathbf{I}$ and $\mathbf{R}$ crumbs). Applying the $p(\cdot)$ function to both sides gives the subunity theorem.

It is easy to see that it follows from similar reasoning to the theorem in Sec. 7.16 that, if all operations satisfy general double causality, then any of the following are both necessary and sufficient conditions for $\mathsf{B}$ to be deterministic: (i) that $\mathsf{B}$ passes the determinism test (see (1105)), (ii) that $\mathsf{B}$ satisfies deterministic forward causality (see (1162)), or (iii) that $\mathsf{B}$ satisfies deterministic backward causality (see (1163)).

### 49.4.14 Physicality of complex networks

We can combine the positivity composition theorem (from Sec. 49.2.6) and causality composition theorems (from Sec. 49.4.11) into one theorem as follows

**Physicality composition theorem.** Under a certain assumption (that pure preparations and pure results satisfy tester positivity), any network built from physical complex operations is, itself, physical.

This follows from the above mentioned composition theorems. However, it is logically course-grained (for example, to prove the network satisfies general forward causality we only need to assume that the components satisfy general forward causality).

# 50 Double maximality

## 50.1 The double maximality property

We introduced the idea of double maximality in Sec. 8 in the simple case. Double maximality is a property which holds in Quantum Theory and Classical Probability Theory in the time symmetric cases. (In the time forward temporal frame we just have forward maximality as discussed in Sec. 11.11 and in the backward temporal frame we just have backward maximality). Now we discuss double maximality in the complex case. We can translate the double maximality property from Sec. 8.1 into the language of the complex case.

**Double maximality** is the property that, for every pointer type, $\mathbf{x}^+$, we have maximal operations

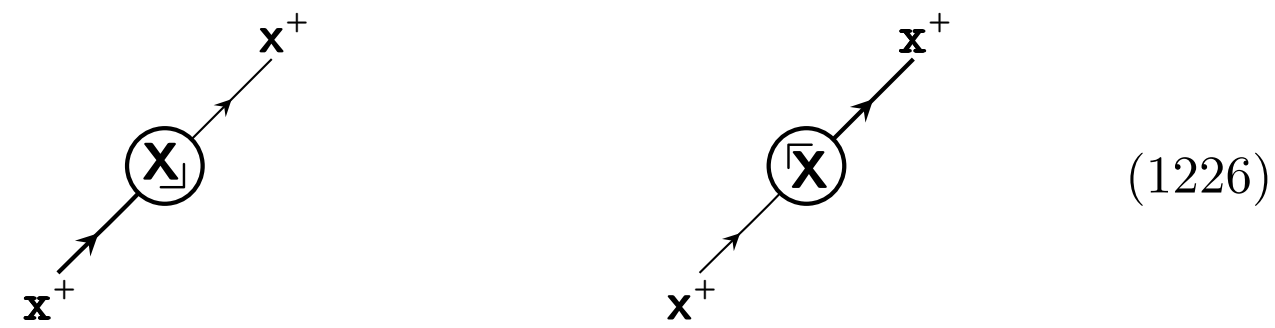

(1226)

having the double maximality property

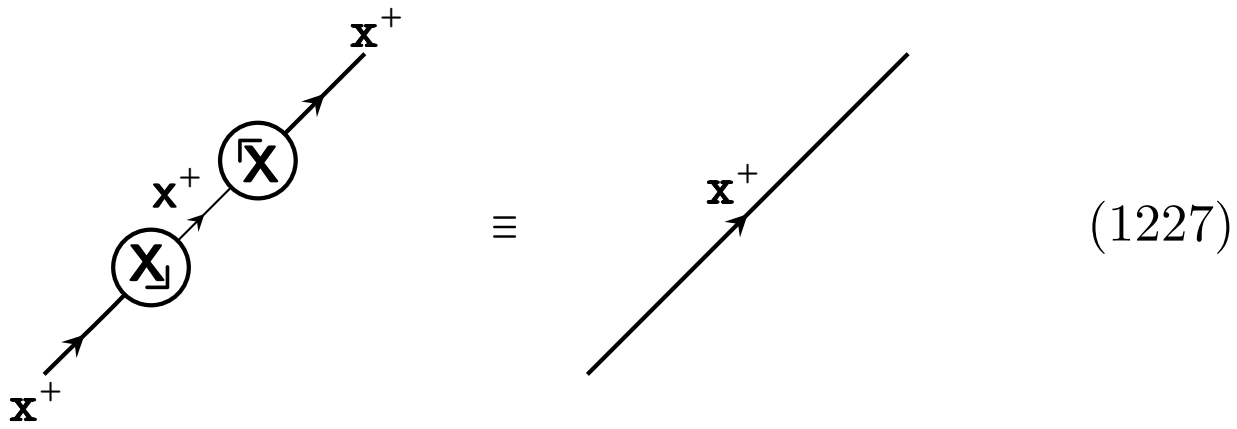

(1227)

where, for the system type $\mathbf{x}^+$ there does not exist another pointer type, $\mathbf{z}^+$, having the same property but with $N_{\mathbf{z}^+} > N_{\mathbf{x}^+}$.

A few comments on the notation are warranted here. First, it is important to note that we have two different fonts in use for the wire labels. For the system type we have $\mathbf{x}^+$ (where the wire is thin), and for the pointer type we have $\mathbf{x}^+$ (where the wire is thick). Second, we use $\mathbf{X}_\lrcorner$ when we have an in pointing pointer type arrow (represented by a thick wire) and an out pointing system type (represented by a thin wire). We use $^\ulcorner\mathbf{X}$ for the other case.

We can write this for $\mathbf{x}$ instead of just $\mathbf{x}^+$. Then the double maximality property is that, for each pointer type, $\mathbf{x}$, there exists maximal operations

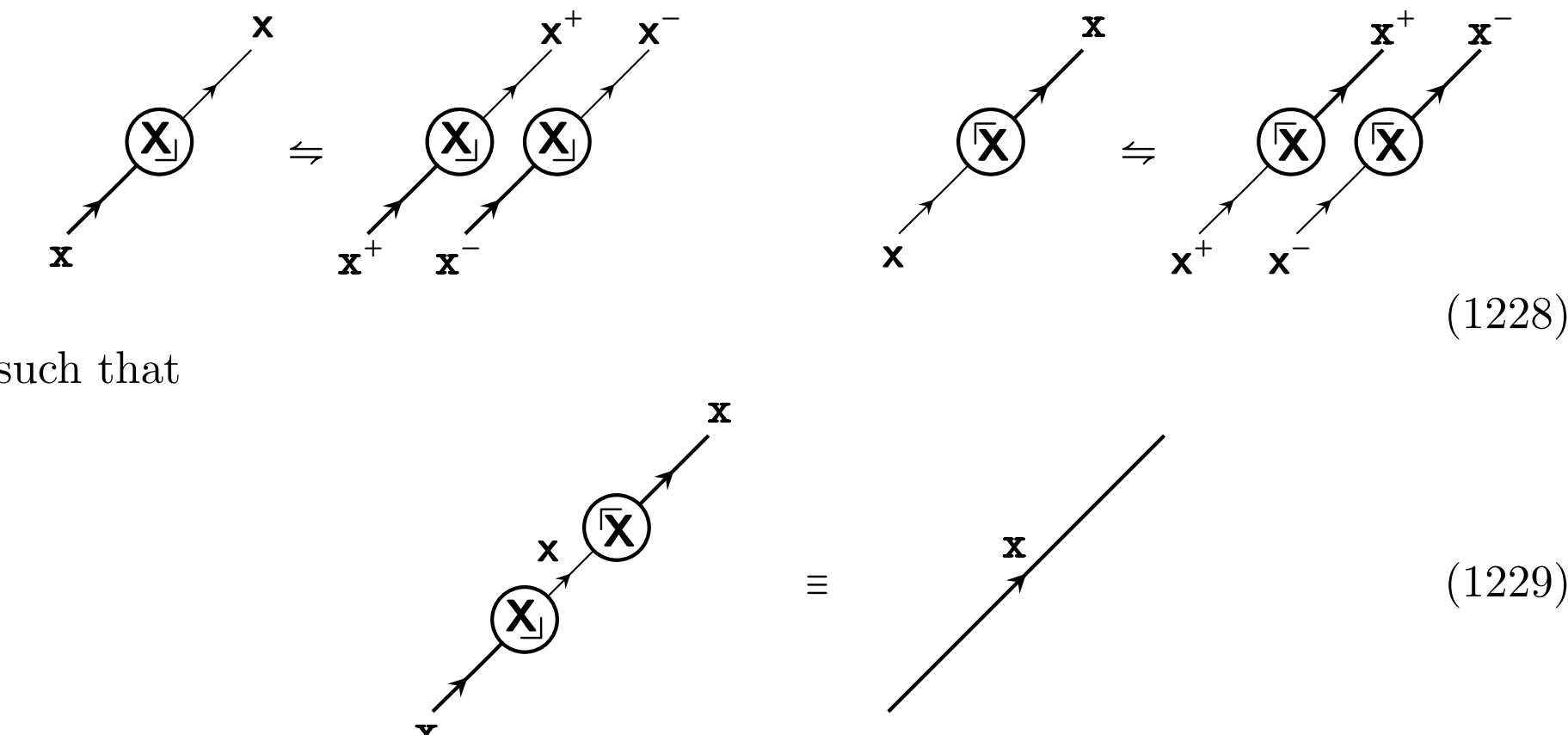

(1228)

such that

(1229)

where, for the physical type $\mathbf{x}$ there does not exist another pointer type, $\mathbf{z}$, having the same property but with $N_\mathbf{z} > N_\mathbf{x}$. Here we are, in fact, writing down the double maximality condition twice, once for $\mathbf{x}^+$ and once for $\mathbf{x}^-$. Note that, whether we use $\mathbf{X}_\lrcorner$ or $^\ulcorner\mathbf{X}$ depends on the directions of the arrows of the attached wires.

Maximal elements are necessarily deterministic. This is clear since, if we apply $\mathbf{I}$ operations to both ends of (1229), we obtain a circuit that is equivalent to 1.

## 50.2 Causal diagrams for maximal elements

We will assume maximal elements have simple causal structure. Thus,

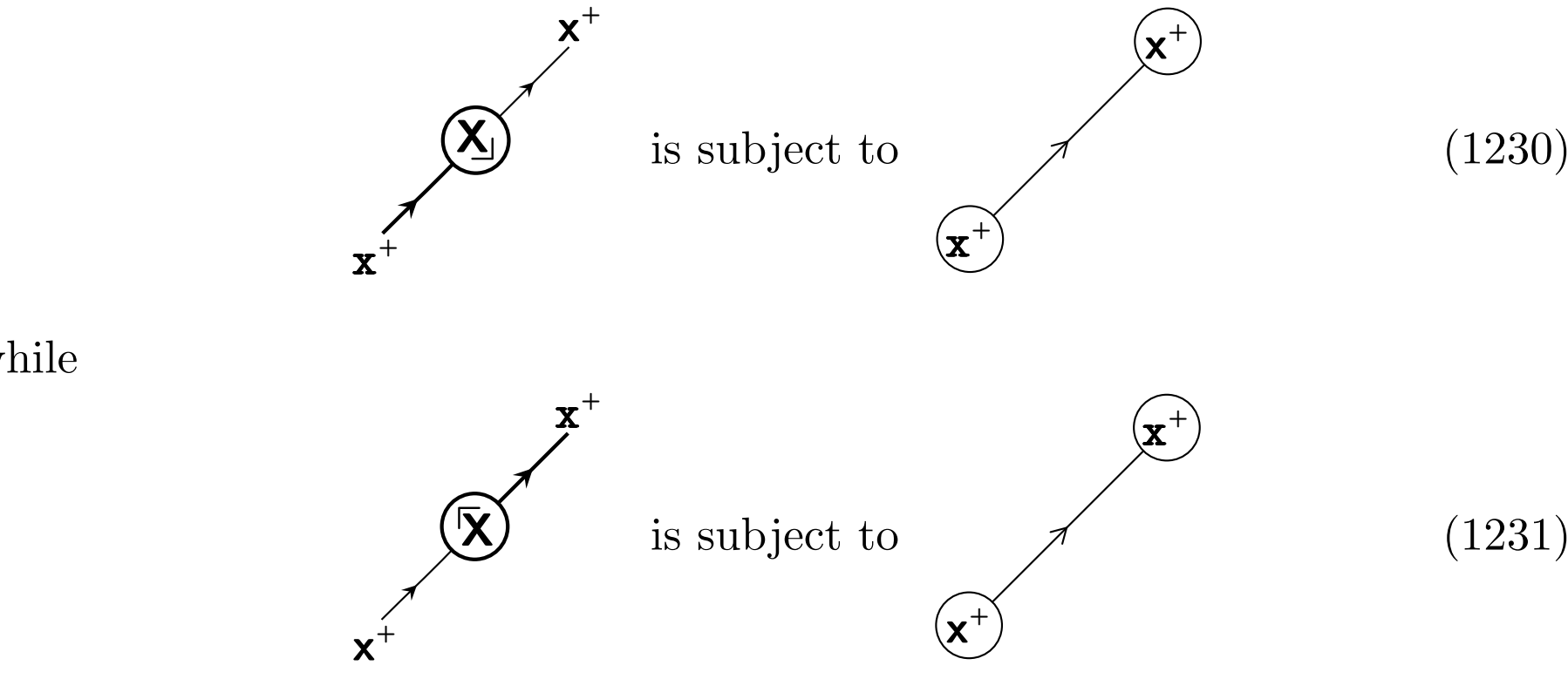

(1230)

while

(1231)

In the compact notation we have

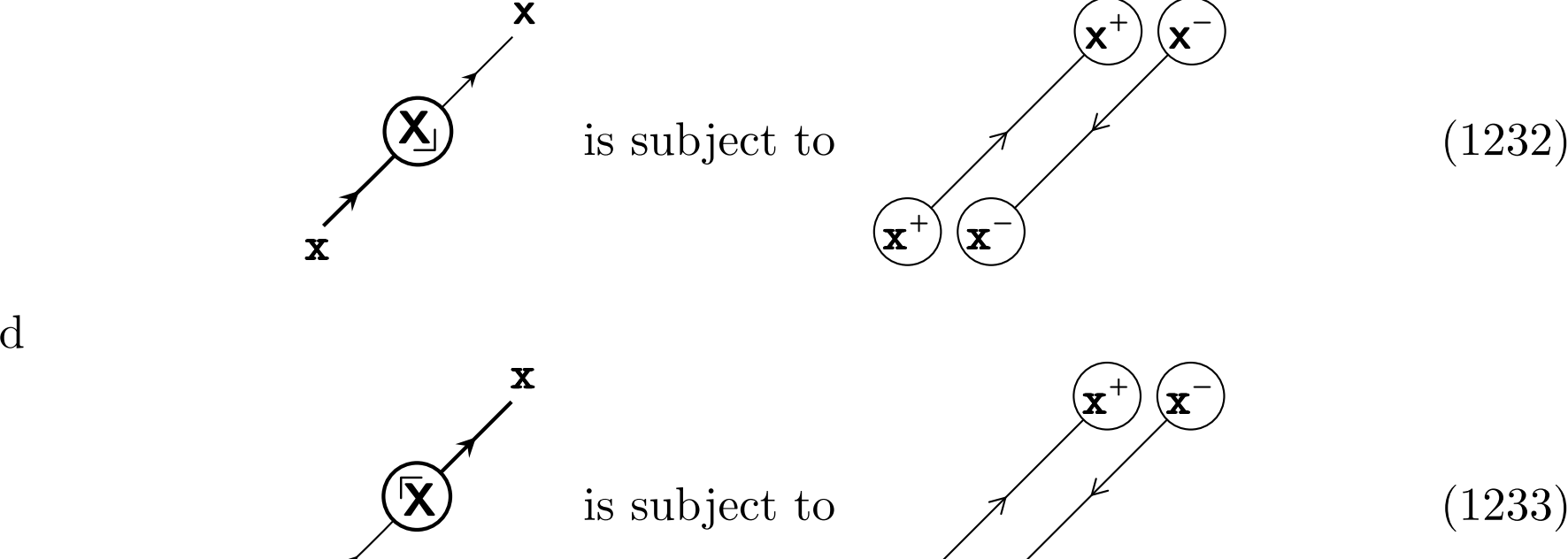

and

We could construct maximal elements associated with operations having more complex causal structure. However, the above simple choice is sufficient for our purposes.

## 50.3 Physicality of maximal elements

We require that maximal elements are physical. This means they must satisfy tester positivity and double causality. We discussed this already in the simple case in Sec. 8.2. Now we provide a similar discussion adapted to the complex case.

The tester positivity constraints are

$$\mathsf{D} \xrightarrow{\mathbf{h}^+} \mathsf{C} \xleftarrow{\mathbf{x}^+} \mathbf{X}_\lrcorner \xleftarrow{\mathbf{x}^+} x^+ \xleftarrow{\mathbf{x}^+} \mathbf{R} \quad \geqq \quad 0 \qquad (1234)$$

and

$$\mathsf{E} \xleftarrow{\mathbf{h}^+} \mathsf{A} \xleftarrow{\mathbf{x}^-} \mathbf{X}_\lrcorner \xleftarrow{\mathbf{x}^-} x^- \xleftarrow{\mathbf{x}^-} \mathbf{R} \quad \geqq \quad 0 \qquad (1235)$$

for all pure $\mathsf{A}$, $\mathsf{C}$, $\mathsf{D}$, $\mathsf{E}$, and $\mathbf{h}^+$ (these conditions can be read off of (1132)). From this we immediately obtain the following theorem

**Maximal on pure theorem.** We have

$$\xrightarrow{\mathbf{h}^+} \mathsf{C} \xleftarrow{\mathbf{x}^+} \mathbf{X}_\lrcorner \xleftarrow{\mathbf{x}^+} x^+ \xleftarrow{\mathbf{x}^+} \mathbf{R} \quad \underset{T}{\geqq} \quad 0 \qquad (1236)$$

and

$$\xleftarrow{\mathbf{h}^+} \mathsf{A} \xleftarrow{\mathbf{x}^-} \mathbf{X}_\lrcorner \xleftarrow{\mathbf{x}^-} x^- \xleftarrow{\mathbf{x}^-} \mathbf{R} \quad \underset{T}{\geqq} \quad 0 \qquad (1237)$$

for all pure preparations, A, and pure results, C.

This follows from the definition of tester positivity and the tester positivity conditions in (1234, 1235).

Since we assume that maximal elements have simple causal structure (see Sec. 50.2, we need only impose simple double causality. The simple double causality constraints are

$$\xleftarrow{\mathbf{x}^\pm} \textcircled{\mathbf{X}_\lrcorner} \xleftarrow{\mathbf{x}^\pm} \textcircled{\mathbf{R}} \quad \equiv \quad \xleftarrow{\mathbf{x}^\pm} \textcircled{\mathbf{I}} \tag{1238}$$

and

$$\textcircled{\mathbf{I}} \xleftarrow{\mathbf{x}^\pm} \textcircled{\mathbf{X}_\lrcorner} \xleftarrow{\mathbf{x}^\pm} \quad \equiv \quad \textcircled{\mathbf{R}} \xleftarrow{\mathbf{x}^\pm} \tag{1239}$$

We can write these in compact form as follows

$$\xleftarrow{\mathbf{x}^\mp} \textcircled{\mathbf{I}_\pm} \xleftarrow{\mathbf{x}} \textcircled{\mathbf{X}_\lrcorner} \xleftarrow{\mathbf{x}} \textcircled{\mathbf{R}^\mp} \xleftarrow{\mathbf{x}^\pm} \quad \equiv \quad \xleftarrow{\mathbf{x}^\mp} \textcircled{\mathbf{I}} \;\; \textcircled{\mathbf{R}} \xleftarrow{\mathbf{x}^\pm} \tag{1240}$$

Note we have used $\mathbf{R}^\pm$ defined in (1092) since the $\mathbf{x}$ arrow is pointing into $\mathbf{X}$.

## 50.4 Maximal representation theorem

As in the simple case (see Sec. 8.3) we can state a maximal representation theorem.

> **Maximal representation theorem.** If we have the double maximality property (1229) then any operation, B, can be written as
>
> $$\xleftarrow{\mathbf{z}} \textcircled{\mathsf{B}} \xrightarrow{\mathbf{c}} \quad \equiv \quad \xleftarrow{\mathbf{z}} \textcircled{{}^\ulcorner\mathbf{Z}} \xleftarrow{\mathsf{z}} \textcircled{{}_\llcorner\mathsf{B}^\urcorner} \xrightarrow{\mathbf{c}} \tag{1241}$$
>
> where
>
> $$\xleftarrow{\mathsf{z}} \textcircled{{}_\llcorner\mathsf{B}^\urcorner} \xrightarrow{\mathbf{c}} \quad := \quad \xleftarrow{\mathsf{z}} \textcircled{\mathbf{Z}_\lrcorner} \xleftarrow{\mathbf{z}} \textcircled{\mathsf{B}} \xrightarrow{\mathbf{c}} \tag{1242}$$
>
> Furthermore
>
> (i) ${}_\llcorner\mathsf{B}^\urcorner$ has the same causal diagram as B (where $\mathbf{z}$ is replaced by $\mathsf{z}$).

(ii) ${}_{|}\mathsf{B}^{|}$ satisfies tester positive if and only if $\mathsf{B}$ satisfies tester positivity.

(iii) ${}_{|}\mathsf{B}^{|}$ satisfies general forward causality at any synchronous partition if and only $\mathsf{B}$ satisfies it at the same partition. And, similarly, ${}_{|}\mathsf{B}^{|}$ satisfies general backward causality at any synchronous partition if and only $\mathsf{B}$ satisfies it at the same partition.

(iv) ${}_{|}\mathsf{B}^{|}$ has the same physical norm as $\mathsf{B}$.

A consequence of the above properties is that ${}_{|}\mathsf{B}^{|}$ is physical if and only $\mathsf{B}$ is physical.

First note that (1241) is easily proved by putting (1242) into the right hand side of (1241) and applying (1227). Before proving that $\mathsf{B}$ and ${}_{|}\mathsf{B}^{|}$ have the same causal diagrams let us make a note. Both $\mathbf{z}$ and $\mathbf{c}$ may be composite. The causal diagram of $\mathsf{B}$ may have separate nodes for each of these components. Thus, when we replace $\mathbf{z}$ in $\mathsf{B}$'s causal diagram by $\mathbf{z}$ to get ${}_{|}\mathsf{B}^{|}$'s causal diagram, it is the corresponding components that are replaced. The proof of property (i) is then straightforward. The causal diagram for $\ulcorner\bar{\mathbf{Z}}$ can be separated out into separate causal diagrams (each of which is simple - see Sec. 50.2) for each of the components. Then, when we join these factorised simple causal diagrams for the maximal elements to the causal diagram for $\mathsf{B}$ then we simply get back the same causal diagram but with the components of $\mathbf{z}$ replaced by the corresponding components of $\mathbf{z}$. Proof of (ii) follows along exactly the same lines as the proof in the simple case (in Sec. 8.3). Rather than repeating this proof, we simply note that we can write $\mathsf{B}$ in (1241) as

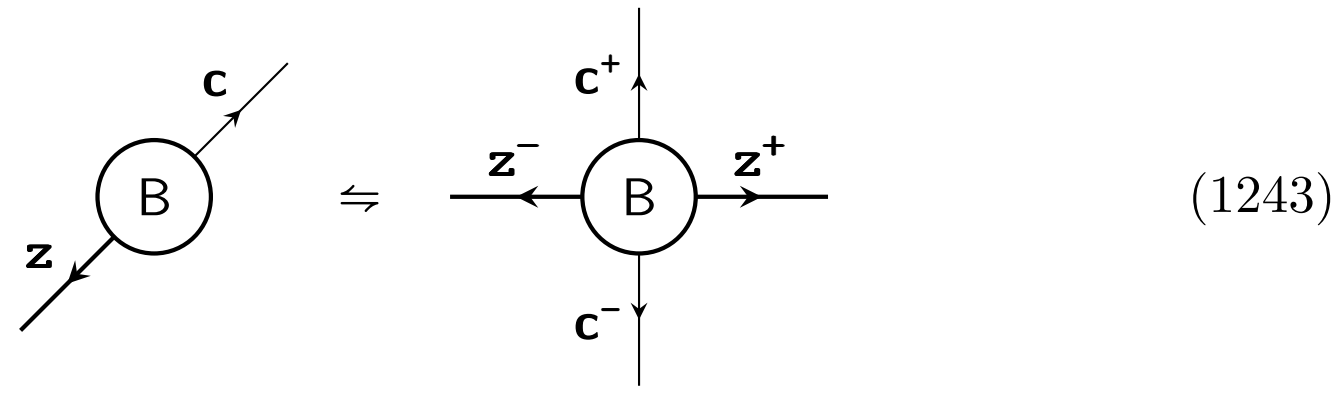

(1243)

which has the same form as

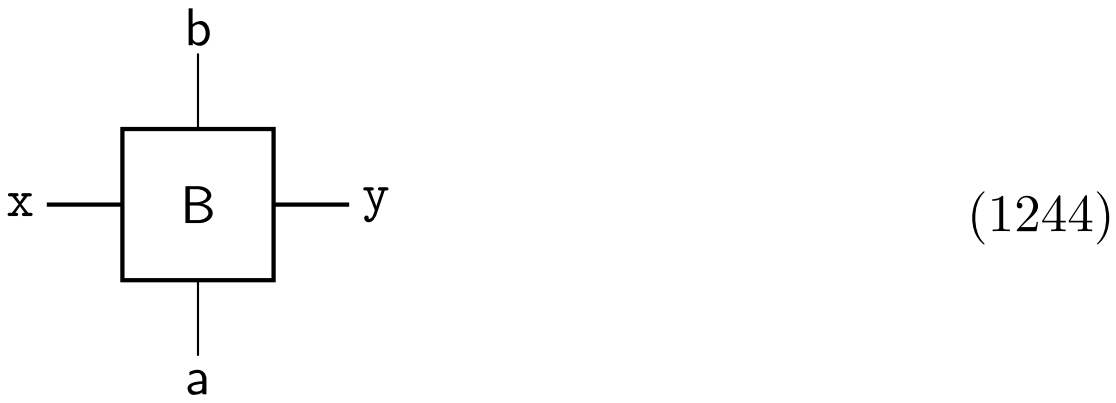

(1244)

in (176) used in the corresponding proof in the simple case from Sec. 8.3. The condition for tester positivity in (1133) has the same form as the condition for tester positivity in (149). Consequently (ii) follows. Now consider statement

(iii) in the theorem. To prove this we first write

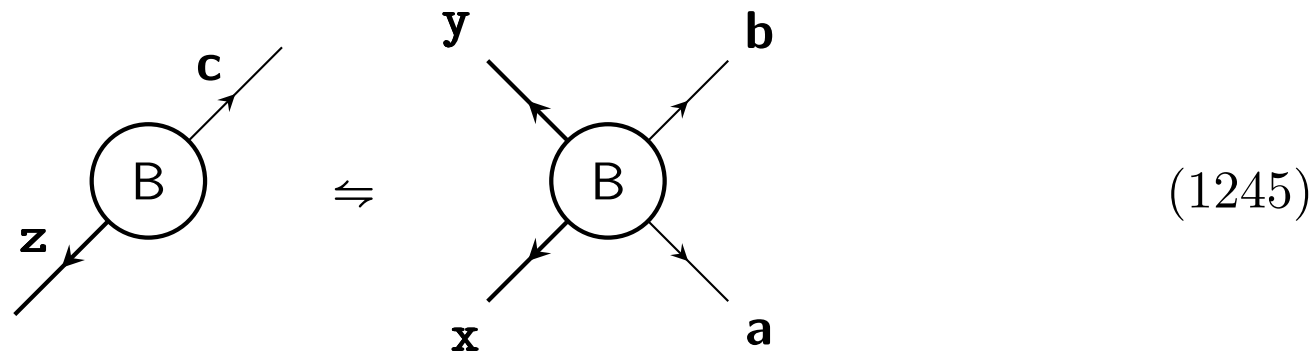

(1245)

(where we can assume that yb is to the future of **xa** in the causal diagram). Then we have to prove that this satisfies the general double causality in (1204, 1205) if and only if

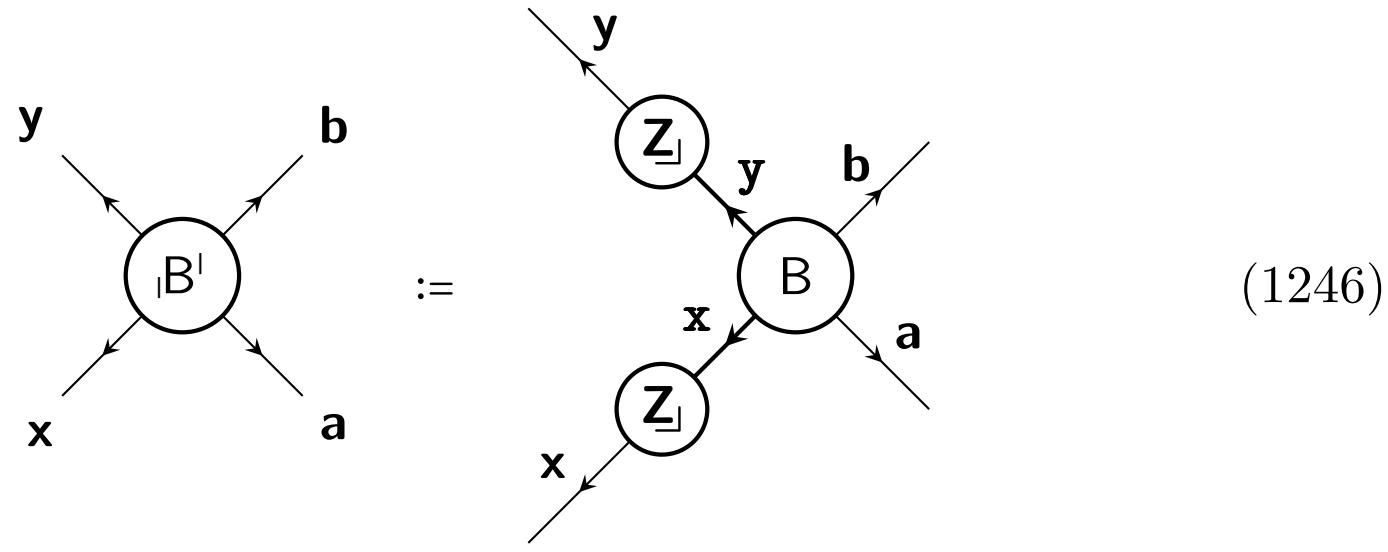

(1246)

satisfies general double causality. This is easily proven to be true using

(1247)

(1248)

and (1238, 1239). Statement (iv) about the physical norm follows since

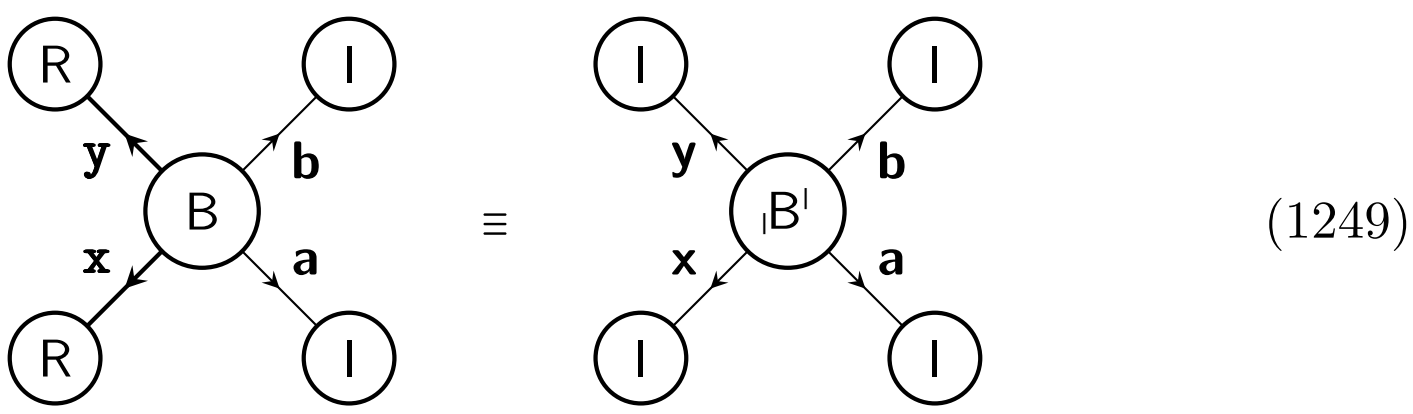

(1249)

using (1238, 1239). Now consider the final statement about physicality. For physicality to hold we require that the operation satisfies tester positivity and general double causality for every synchronous partition. Since $\mathsf{B}$ and ${}_{|}\mathsf{B}^{|}$ have the same causal diagram, it follows from the results proven above that $\mathsf{B}$ is physical if and only if ${}_{|}\mathsf{B}^{|}$ is physical.

# 51 Fiducial operation expansion framework

## 51.1 Fiducial elements

We can expand causally complex operators in terms of fiducial operators in the same way as we did for the simple case (though adapted to our new notation). We start by defining fiducial elements for physical systems and for pointer systems.

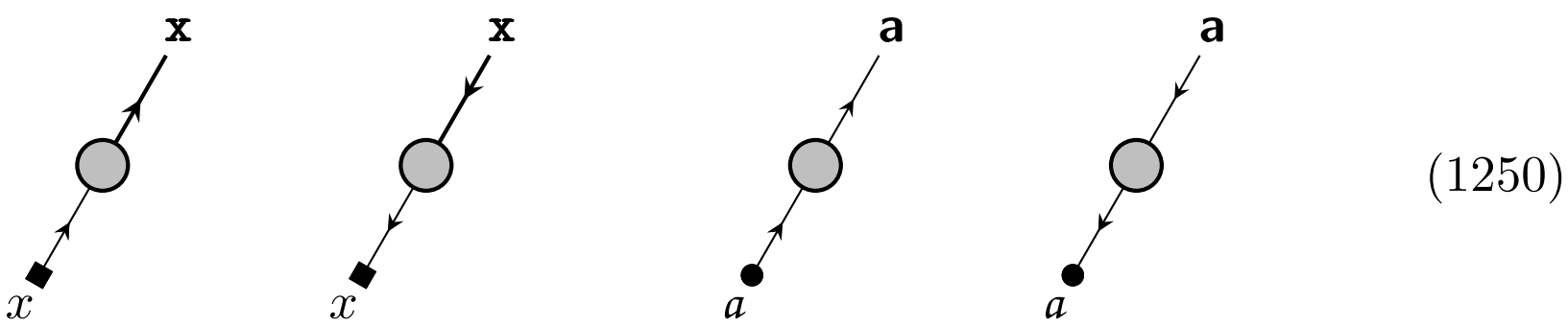

(1250)

where $x$ labels the pointer fiducial elements and $\mathsf{a}$ labels the system fiducial elements.

In the simple case we assumed tomographic locality (this was the name we used to refer colloquially to a host of equivalent assumptions - see Sec. 9.7). If we have this property we can put $x = x^+x^-$ with $x^\pm = 1, 2, \ldots, N_{\mathbf{x}^\pm}$ and $\mathsf{a} = \mathsf{a}^+\mathsf{a}^-$ with $\mathsf{a}^\pm = 1, 2, \ldots, K_{\mathbf{a}^\pm}$ with $N_{\mathbf{x}} = N_{\mathbf{x}^+}N_{\mathbf{x}^-}$ and $K_{\mathbf{a}} = K_{\mathbf{a}^+}K_{\mathbf{a}^-}$. Further, under tomographic locality we can assume that + and − parts of the fiducial elements factorise so we have

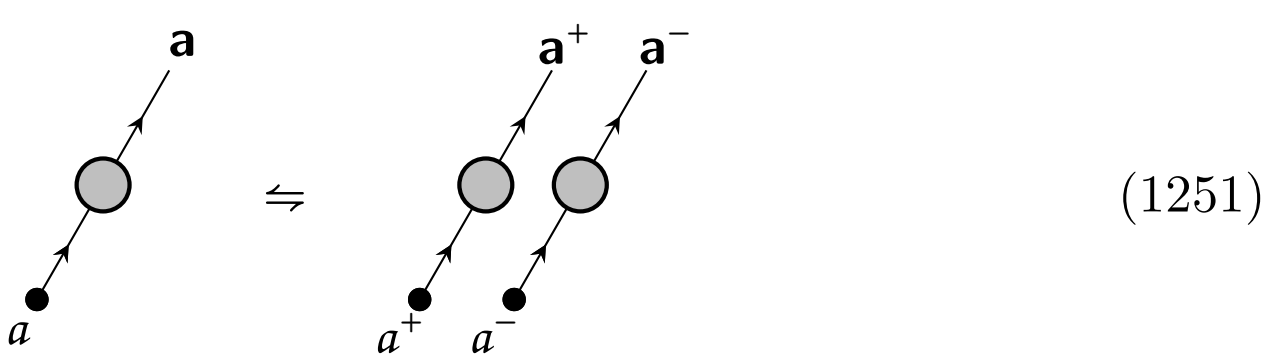

(1251)

and similarly for the other three cases in (1250). We will discuss tomographic locality further in Sec. 51.3.

## 51.2 Fiducial matrices

We can form fiducial matrices as follows

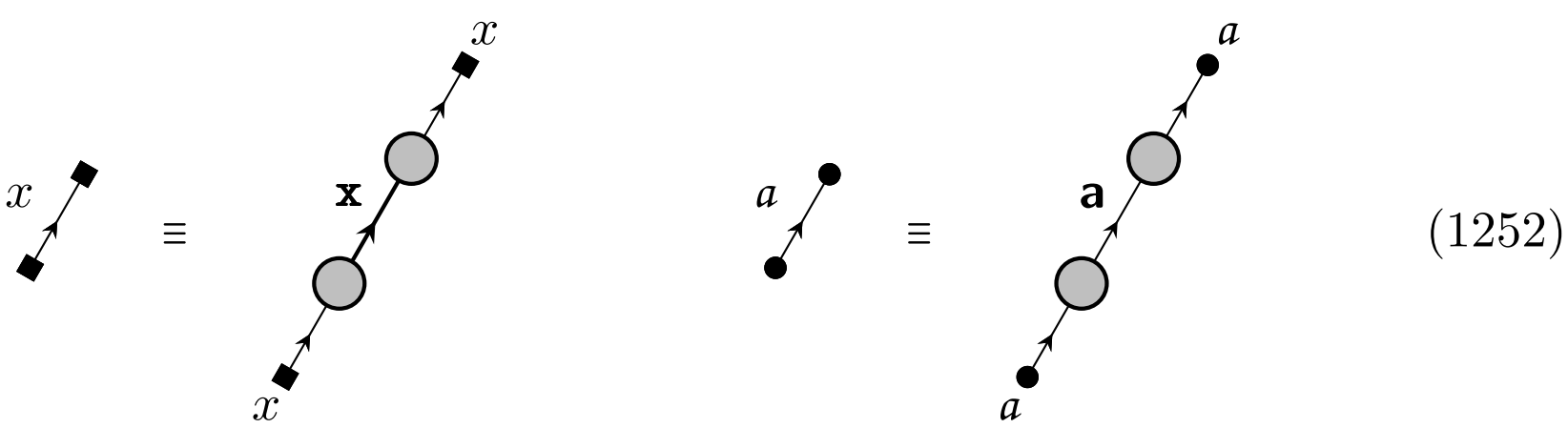

(1252)

The entries of a fiducial matrices are all probabilities. If we assume tomographic locality and choose the fiducial elements so that they factorise between $\mathsf{a}^+$ and $\mathsf{a}^-$ parts (see (1251)), then the probabilities do also so we have

$$x \quad \leftrightharpoons \quad x^+ \; x^- \qquad\qquad a \quad \leftrightharpoons \quad a^+ \; a^- \qquad (1253)$$

We represent the inverses of the fiducial matrices by

$$x \qquad\qquad a \qquad (1254)$$

Then, following the same reasoning as in Sec. 9.3 and Sec. 9.5, we have the following identities

$$= \quad = \qquad\qquad = \quad = \qquad (1255)$$

These means we can introduce and delete pairs of black and white dots in either order.

## 51.3 Tomographic locality

In Sec. 9.7 we discussed a number of equivalent assumptions in the simple case collected under the umbrella term "tomographic locality". Those assumptions can be taken over to the complex case. Here we will just discuss two of them.

The *decomposition locality* assumption (see Sec. 9.7.2) says we can expand an arbitrary operation in terms of fiducials. For example

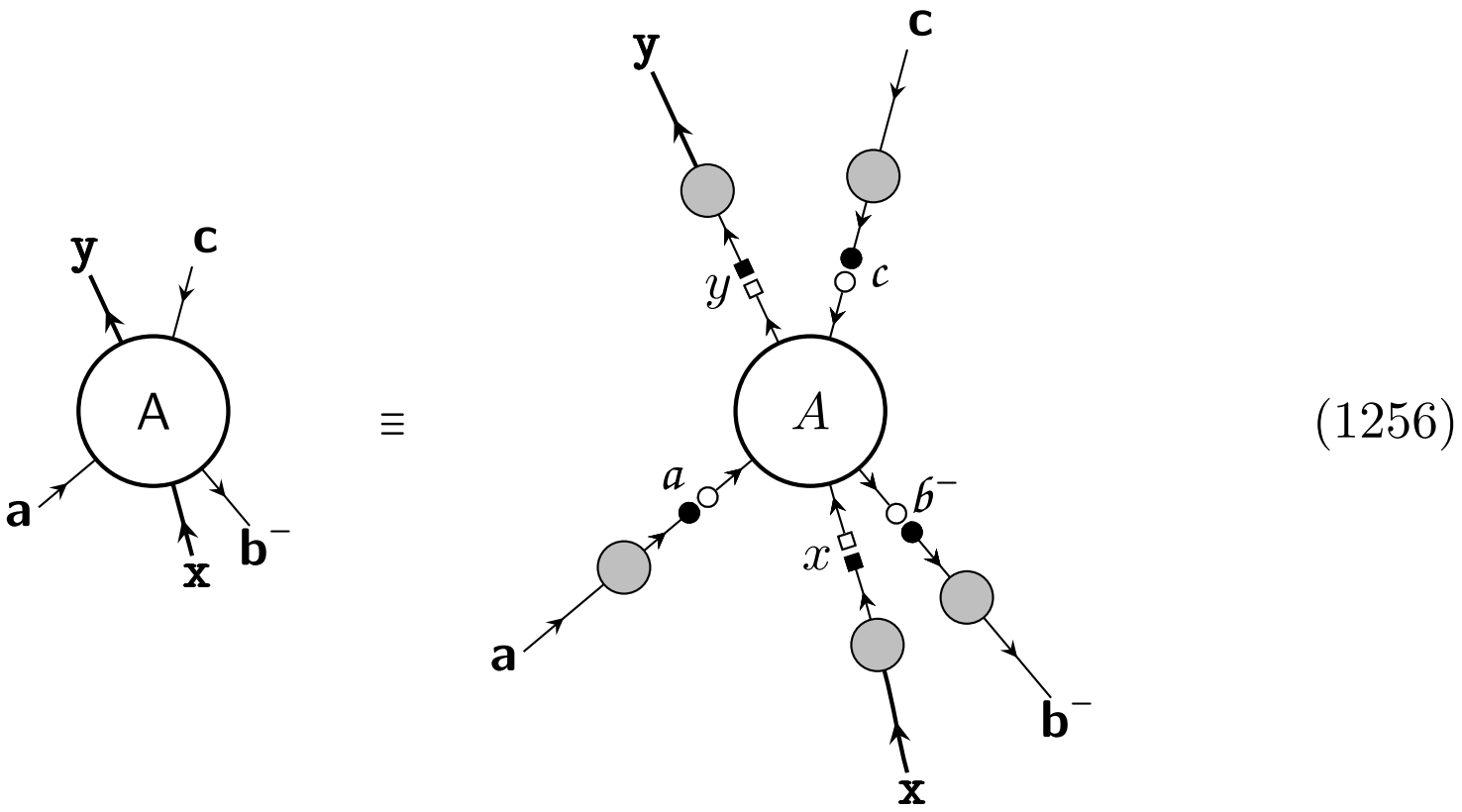

(1256)

The duotensor providing the weights has all white dots.

If we add fiducial elements to the ends of (1256) we obtain

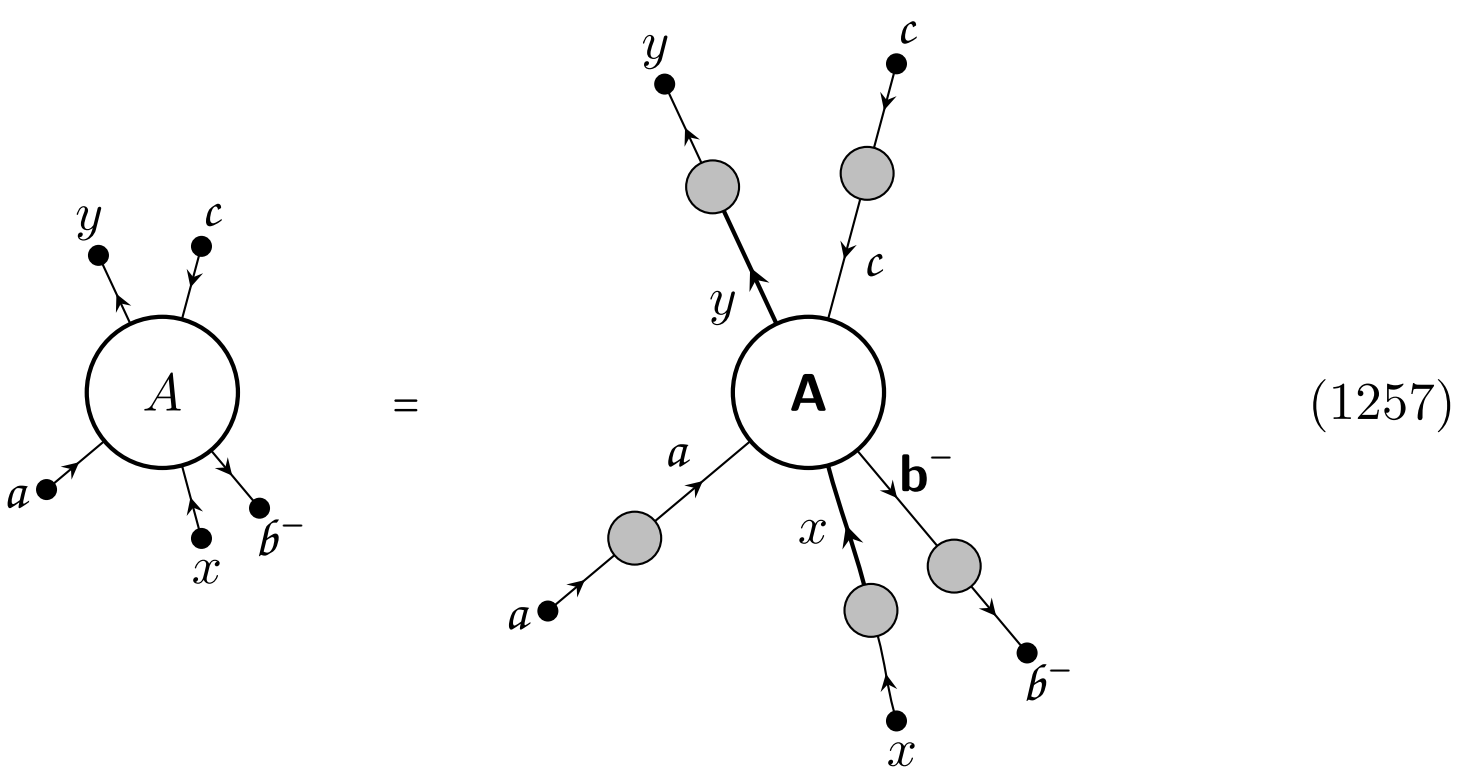

(1257)

This is a duotensor with all black dots. Here we have done local process tomography on an operation to obtain a duotensor. The assumption of *process tomographic locality* is that the probabilities obtained by doing local process tomography are sufficient to fully characterise the operation (up to its equivalence class). If the decomposition locality assumption above is true then we can use the duotensor in (1257) to obtain an expression equivalent to the operation by first changing the colour of the dots from black to white then plugging it into (1256). Thus, this duotensor characterises the operation and so decomposition locality implies process tomographic locality. It is easy to prove the converse (by similar reasoning to that in Sec. 9.7.3).

We can also write down the assumption of *wire decomposition locality* in the complex notation. This is also equivalent to decomposition locality. In complex notation, wire decomposition locality can be written as follows.

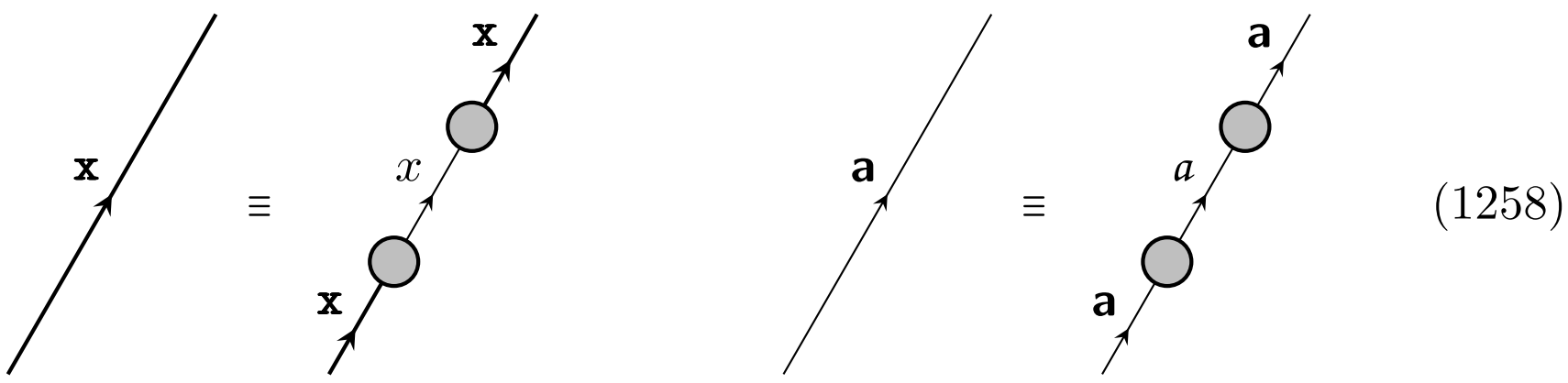

(1258)

Compare this with (218) in the simple case. Note that, in fact, we can assume a little less and allow some arbitrary expansion matrices (compare with in (219) in the simple case). Then we can show that these expansion matrices are such that (1258) holds. The proofs are analogous to the simple case and so we omit them here.

## 51.4 Duotensors for special cases

In Sec. 9.9 we presented an explicit form for some special duotensors in the simple case. We can do the same for the complex case. We start by setting the pointer fiducial elements

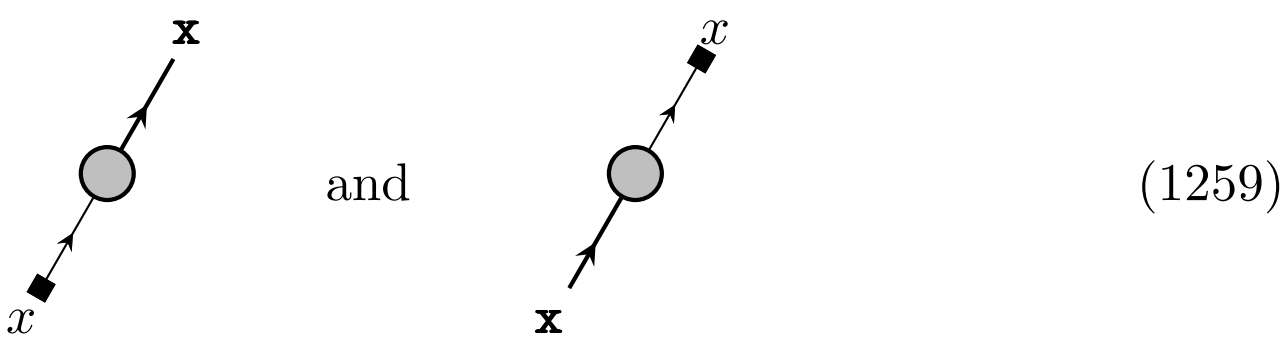

(1259)

to be equal to

x x $x$ $R$ and $R$ x $x$ x (1260)

respectively for each $x$ (c.f. (193) in simple case). With this choice the pointer fiducial matrix and its inverse take the form

$$x \;=\; \frac{1}{N_{\mathbf{x}}}\begin{pmatrix} 1 & & & \\ & 1 & & \\ & & \ddots & \\ & & & 1 \end{pmatrix} \qquad\qquad x \;=\; N_{\mathbf{x}}\begin{pmatrix} 1 & & & \\ & 1 & & \\ & & \ddots & \\ & & & 1 \end{pmatrix} \qquad (1261)$$

using (1252). Note, to write down this matrix, we impose some order on the $x$ (which are represented by integer pairs $x = x^+x^-$). The choice in (1260) for pointer fiducials immediately gives

$$R \;x \;=\; \frac{1}{N_{\mathbf{x}}}\begin{pmatrix} 1 \\ 1 \\ \vdots \\ 1 \end{pmatrix} \qquad\qquad R \;x \;=\; \frac{1}{N_{\mathbf{x}}}\begin{pmatrix} 1 \\ 1 \\ \vdots \\ 1 \end{pmatrix} \qquad (1262)$$

Using the inverse fiducial matrix above we obtain

$$= \begin{pmatrix} 1 \\ 1 \\ \vdots \\ 1 \end{pmatrix} \qquad\qquad = \begin{pmatrix} 1 \\ 1 \\ \vdots \\ 1 \end{pmatrix} \tag{1263}$$

Finally, expanding a readout box in terms of a duotensor and following the same reasoning as in Sec. 9.9 we obtain

$$= \begin{pmatrix} 0 & & & & \\ & 0 & & & \\ & & \ddots & & \\ & & & 1 & \\ & & & & \ddots \end{pmatrix} \tag{1264}$$

for the readout box duotensor where the 1 is in position $x$ along the diagonal.

# 52 Duotensor calculations

Consider the circuit

$$\tag{1265}$$

We can calculate the probability for this circuit by turning it into an equivalent duotensor calculation. We can do this by substituting the local decomposition for each operation then substituting in the fiducial matrices (as we did in Sec. 10.1). This gives

$$\tag{1266}$$

We can, using (1255), replace black and white dot pairs by a line giving

$$\tag{1267}$$

To actually do this duotensor calculation we need to insert black and white dot pairs. We can do this in any way we like. We can also get from (1265) to (1267) by using the decomposition of wires as given in (1258).

# 53 Time forward complex operational probabilistic theories

So far we have looked at the causally complex case in the time symmetric temporal frame of reference. In Sec. 11 we presented time forward causally simple operational probability theories. We can adapt those techniques to the time forward complex case. We will be able to prove that time symmetric complex operational probabilistic theories can be modelled by time forward complex operational probabilistic theories using the same technique as for the simple case. However, proof of the converse statement in the simple case does not obviously go over to the complex case. Consequently, we do not know whether time forward complex operational probabilistic theories can be modelled by time symmetric ones. This leaves open the exciting possibility that time symmetric complex operational probabilistic theories are more constrained than time forward ones.

## 53.1 Complex operations in the time forward case

A time forward operation has no incomes so is represented as a, $\overline{\mathsf{B}}$, $\mathbf{x}^+$ $\leftrightharpoons$ a, $\overline{\mathsf{B}}$, $\mathbf{y}^-$ (1268) where $\mathbf{x} = \mathbf{y}^R$. We can model time forward complex operations in the time symmetric framework just as we did in the simple case.

First it is useful to define

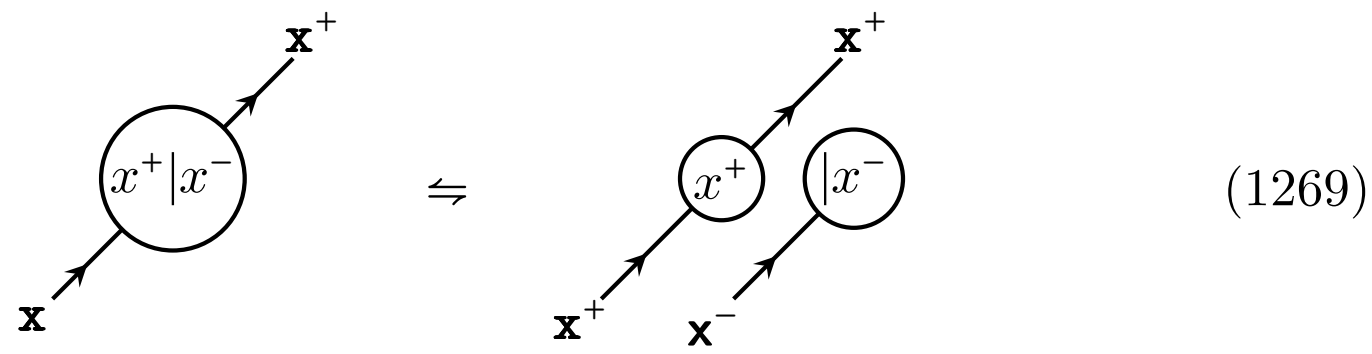

(1269)

where

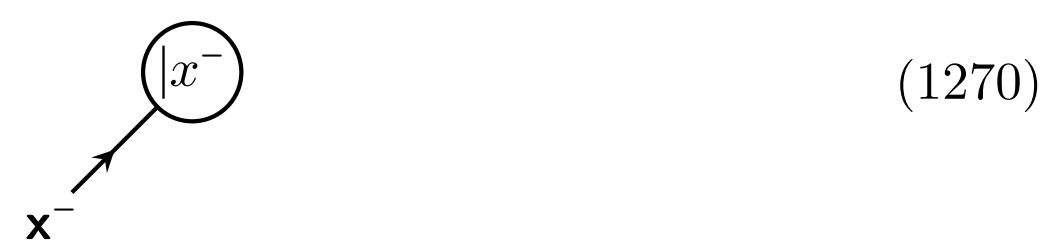

(1270)

means that we condition on $x^-$ (paying attention to the – sign superscripts we see this is actually preselection). We can also define

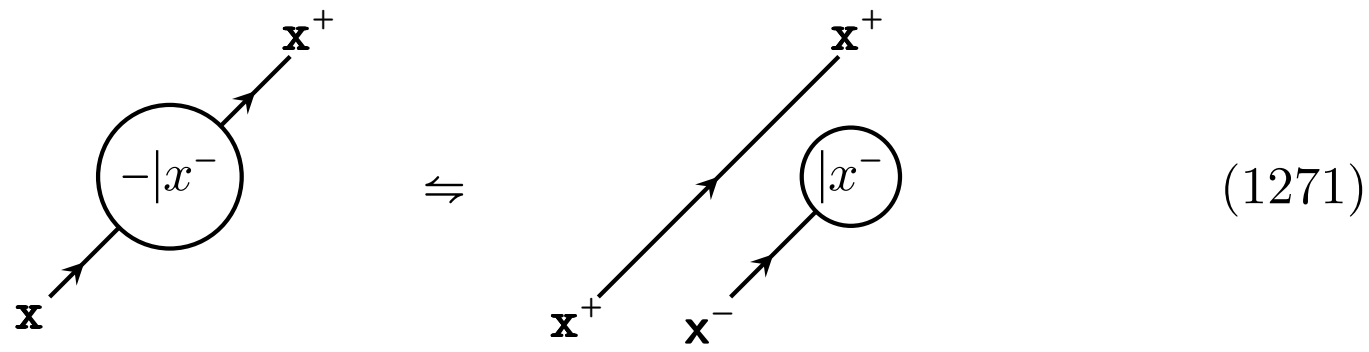

(1271)

where the $\mathbf{x}^+$ has no readout box on it.

We can model a time forward complex operation, $\overline{\mathbf{B}}$, in terms of a time symmetric complex operation, $\mathbf{C}$, as follows

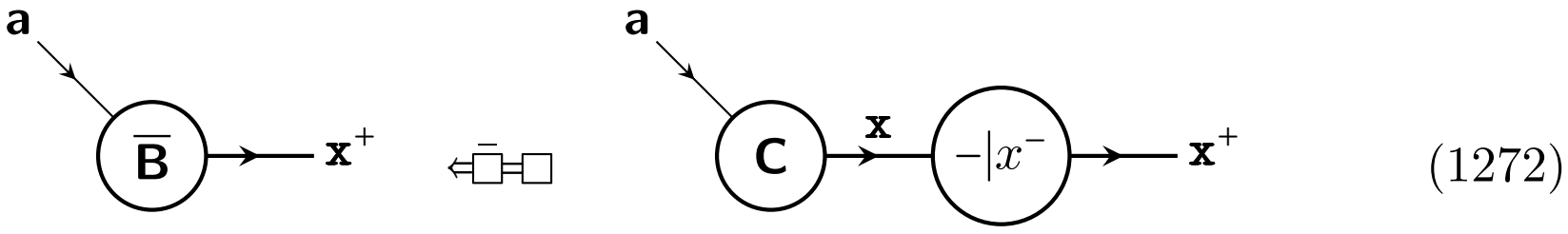

(1272)

(where $\Leftarrow\overline{\square}\!\!\Rightarrow\!\square$ means *modelled by*) or, for the other interconvertible form in (1268),

(1273)

We can make the outcome implicit. For example,

(1274)

where $\overline{\mathsf{D}}$ is not in bold since, now, the operation is non-deterministic.

## 53.2 Circuits in forward frame

Consider a circuit in the time forward frame such as

(1275)

We can replace each time forward operation by a time symmetric model for it

(1276)

Here we have conditioned on the incomes $x^-$ and $y^+$. Using the fact that

$$\text{prob}(x^+y^-|x^-y^+z^+) = \frac{\text{prob}(xyz^+)}{\text{prob}(x^-y^+z^+)} \tag{1277}$$

we can calculate the probability for the circuit in (1275) as follows

(1278)

where

(1279)

The expression in the denominator can be rewritten as

(1280)

where $\mathbf{D}$ is defined in terms of $\mathbf{A}$, $\mathbf{B}$, and $\mathbf{C}$ as well as $\mathbf{R}$ operations associated with $\mathbf{x}^+$ and $\mathbf{y}^-$. Since $\mathbf{D}$ is composed of only deterministic operations it is itself deterministic. Furthermore, it has only incomes. Hence, $\mathbf{D} = \mathbf{R}$ (pointer results are unique). Thus, this $\mathbf{R}$ factorises into three $\mathbf{R}$ operations associated with each of $\mathbf{x}^-$, $\mathbf{y}^+$, and $\mathbf{z}^+$. It follows that the denominator in (1278) is equal to $\frac{1}{N_{\mathbf{x}^-}N_{\mathbf{y}^+}N_{\mathbf{z}^+}}$. We can associate each of the $N$ factors with one of the incomes. For example, the $\mathbf{x}$ case can be written

(1281)

whilst the $\mathbf{z}^+$ case can be written

(1282)

The $N_{\mathbf{x}^-}$ and $N_{\mathbf{z}^+}$ in the circles are overall coefficients. These equation represent *temporal reference frame transformations*. Substituting (1281) for the $\mathbf{x}$ part and a similar expression for the $\mathbf{y}$ part into (1276) gives us the probability for the time forward circuit (1275) we started with

$$N_{\mathbf{x}^-} N_{\mathbf{y}^+} N_{\mathbf{z}^+} \text{prob}\left( \text{R} \xleftarrow{\mathbf{x}} x \xleftarrow{\mathbf{x}} \text{A},\ \text{A} \xrightarrow{\mathbf{b}} \text{C},\ \text{C} \xleftarrow{\mathbf{y}} y \xleftarrow{\mathbf{y}} \text{R},\ \text{C} \xrightarrow{\mathbf{a}} \text{B},\ \text{B} \xrightarrow{\mathbf{c}} \text{A},\ \text{B} \xleftarrow{\mathbf{z}^+} z^+ \xleftarrow{\mathbf{z}^+} \text{R} \right) \qquad (1283)$$

This shows us that we can obtain probabilities in the time forward case from those in the time symmetric case. We can also, of course, obtain the time symmetric probabilities from the time forward ones.

Note that the reference frame transformations in (1281) and (1282) work as long as we do not condition on any outcomes. They are, thus, subject to the replacement rule warning discussed in Sec. 11.5.

## 53.3 Positivity in the forward frame

The positivity condition is given by adapting the results of Sec. 49.2.2 to the case where there are no incomes as follows

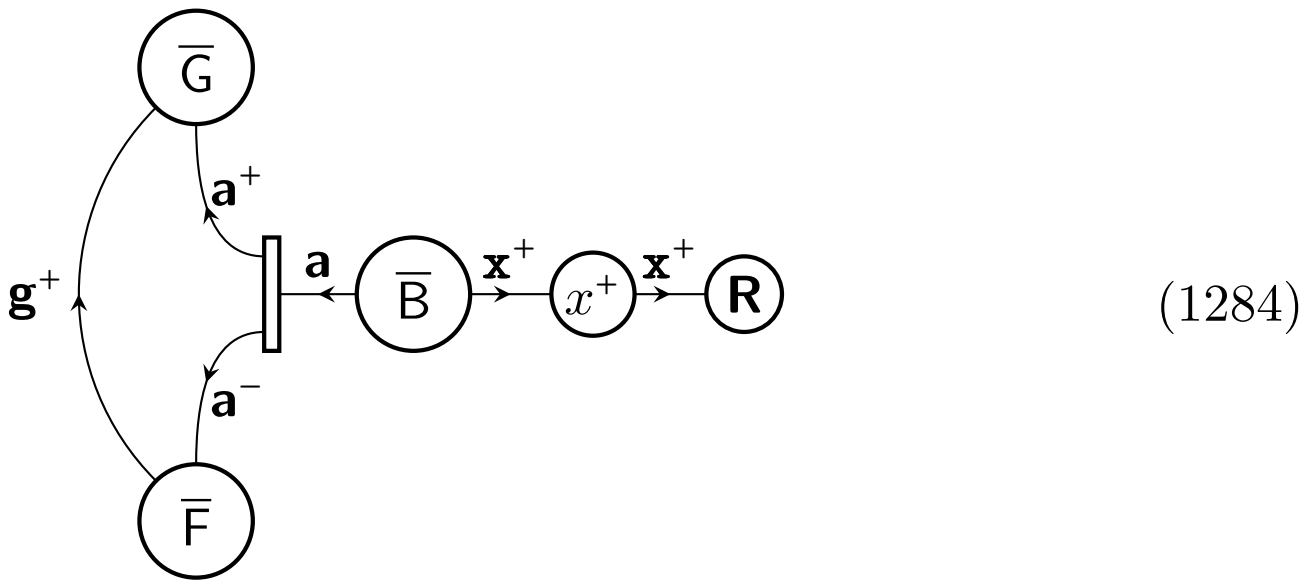

(1284)

for all pure preparations, $\overline{\mathsf{F}}$ and pure results, $\overline{\mathsf{G}}$. If we model the time forward operations $\overline{\mathsf{B}}$, $\overline{\mathsf{F}}$, and $\overline{\mathsf{G}}$ by time symmetric operations by conditioning on the incomes (as in (1274)), then by applying (1281), we get an overall factor equal to the product of the corresponding $N$'s. Thus, the positivity properties are the same as in the time symmetric case. We can write this tester positivity condition as

$$0 \underset{T}{\leq} \quad \xleftarrow{\mathbf{a}} \overline{\mathsf{B}} \xrightarrow{\mathbf{x}^+} \qquad (1285)$$

where $\underset{T}{\leq}$ means we have positivity with respect to the tester

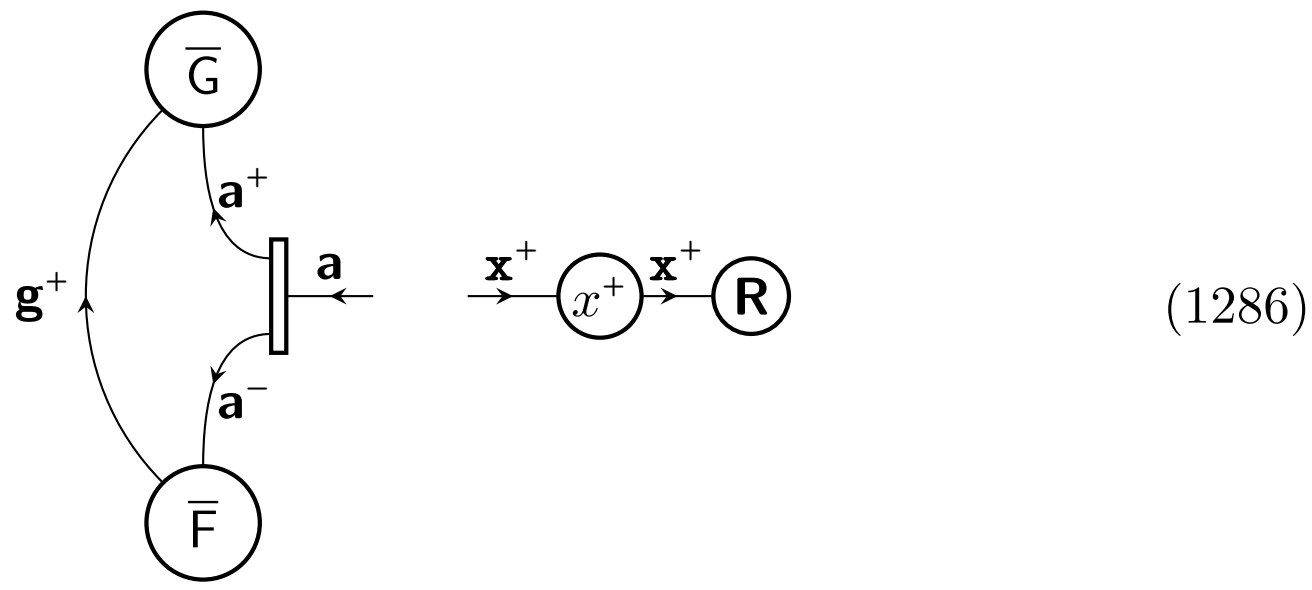

(1286)

for all pure preparations $\overline{\mathsf{F}}$ and pure results $\overline{\mathsf{G}}$.

Since we can model time forward operations in terms of time symmetric operations, the positivity under composition theorem in Sec. 7.12.3 goes through for the time forward case as well.

## 53.4 Causality in the forward frame

We can obtain causality conditions in the forward frame from those in the time symmetric frame by conditioning on incomes.

Consider the forward causality condition (1162). We can use the $\mathbf{B}$ appearing in this condition to model a forward time operation by conditioning on outcomes

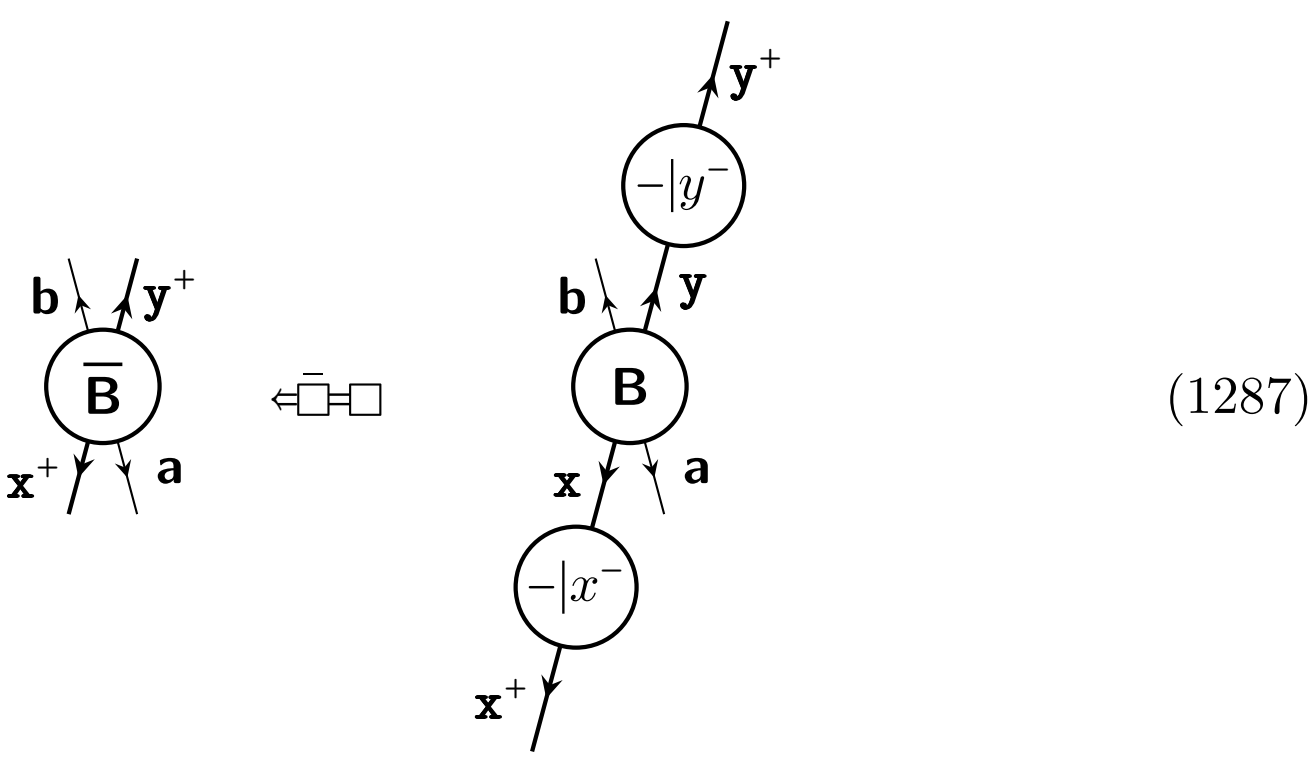

(1287)

(where, for ease of notation, the $x^-$ and $y^-$ that have been conditioned on are taken to be implicit in the notation $\overline{\mathbf{B}}$). Now condition on the incomes of the forward causality condition (1162)

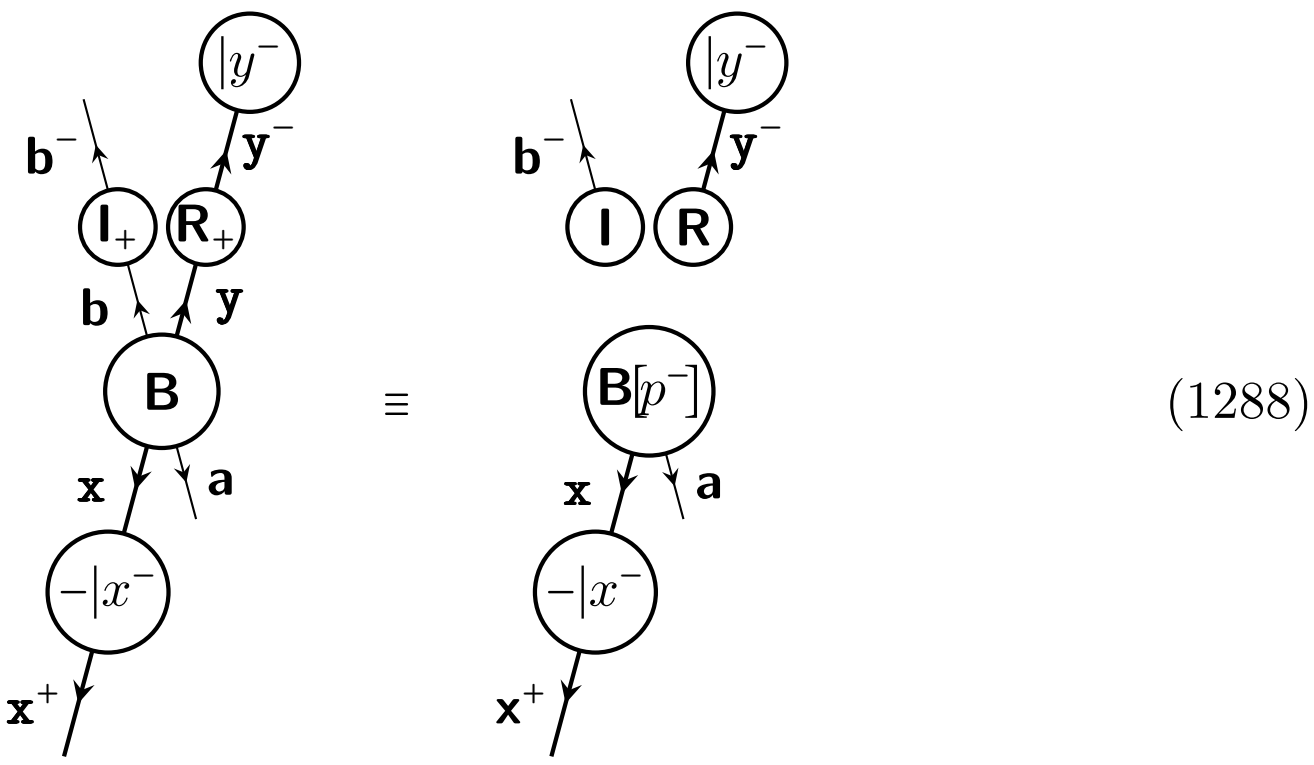

(1288)

This gives

(1289)

(Here, $\overline{\mathbf{B}}[p^-]$ has been conditioned on $x^-$.) This is the forward causality condition in the time forward temporal frame.

Now consider applying the backward causality condition (1163) in the time forward case. We are blocked from conditioning on the income $\mathbf{x}^-$ because of the $\mathbf{R}$ operation. Thus, we can only apply backward causality if $\mathbf{x} = \mathbf{x}^+$ (so that $\mathbf{x}^-$ is null). In other words, we can only apply the backward causality condition in the forward frame in the special case when $\overline{\mathbf{B}}$ can be modelled as follows

(1290)

Importantly, there is no implicit income conditioning before the synchronous surface, $p$ (this is denoted by the subscript $*p$). In this case, conditioning on the incomes of the backward causality condition (1163) just means conditioning

on $y^-$ which gives

$$\text{[diagram: } \mathbf{B} \text{ with } -|y^-] \equiv \text{[diagram: } \mathbf{B}[p^+], \mathbf{R}, \mathbf{I}] \tag{1291}$$

This gives

$$\text{[diagram: } \overline{\mathbf{B}}_{*p}, \mathbf{I}_-] \equiv \text{[diagram: } \overline{\mathbf{B}}[p^+], \mathbf{R}, \mathbf{I}] \tag{1292}$$

If we want to apply backward causality again, now to $\overline{\mathbf{B}}[p^+]$ and with respect to a later hypersurface, $q > p$, then we require that $\overline{\mathbf{B}}[p^+]$ has no implicit income-conditioning before $q$.

Since time forward operations can be modelled in terms of time symmetric operations the causality composition theorem (in Sec. 49.4.11) goes through in the forward case as well.

## 53.5 The forward seatbelt identity for complex case

In the simple case we used the forward seatbelt identity (277) (and the summed version in (278)) to analyse the connection between time symmetric and time forward cases. We can write down a similar identity in the complex case which (though it does not look like a seatbelt) we will also call the forward seatbelt identity.

First, we define the flag as

$$\xrightarrow{\mathbf{x}^+} [x^+ \quad := \quad \xrightarrow{\mathbf{x}^+} x^+ \xrightarrow{\mathbf{x}^+} \mathbf{R} \tag{1293}$$

(compare with (257) in the simple case). Then, using (1282) and (1117), we can write

$$\xrightarrow{\mathbf{x}^+} [x^+ \quad |x^+ \xrightarrow{\mathbf{x}^+} \quad = \quad \xrightarrow{\mathbf{x}^+} x^+ \xrightarrow{\mathbf{x}^+} \tag{1294}$$

This is the forward seatbelt identity (compare with (277)). Given our convention that systems labeled with a minus sign travel against the direction of the arrow, we can also write the forward seatbelt identity as

$$\xleftarrow{\mathbf{x}^-}\bigcirc\!\!\!\!\!\!\;[x^-\;\;\bigcirc\!\!\!\!\!\!\;|x^-\xleftarrow{\mathbf{x}^-} \;=\; \xleftarrow{\mathbf{x}^-}\bigcirc\!\!\!\!\!\!\;x^-\xleftarrow{\mathbf{x}^-} \tag{1295}$$

Since we use the replacement rule in (1282) to derive the forward seatbelt inequality, we can only use the latter with confidence when the replacement rule warning is heeded (see Sec. 53.2 and Sec. 11.5). This warning is heeded in the time forward frame since then we do not postselect.

We can sum the forward seatbelt identity (1294) over $x^+$ giving

$$\sum_{x^+} \xrightarrow{\mathbf{x}^+}\bigcirc\!\!\!\!\!\!\;[x^+\;\;\bigcirc\!\!\!\!\!\!\;|x^+\xrightarrow{\mathbf{x}^+} \;=\; \xrightarrow{\quad\mathbf{x}^+\quad} \tag{1296}$$

Similarly, we can sum the forward seatbelt identity (1295) over $x^-$ giving

$$\sum_{x^-} \xleftarrow{\mathbf{x}^-}\bigcirc\!\!\!\!\!\!\;[x^-\;\;\bigcirc\!\!\!\!\!\!\;|x^-\xleftarrow{\mathbf{x}^-} \;=\; \xleftarrow{\quad\mathbf{x}^-\quad} \tag{1297}$$

We will use this below to prove physical TS operations can be modelled by physical TF operations.

## 53.6 Physical TS modellable by physical TF

In Sec. 11.10 we proved equivalence between the TS and TF approaches - namely that physical simple operations in TS can be modeled by physical simple operations in TF and vice versa. This proof consisted of an easy proof (that physical TS is modellable by physical TF) and a hard proof (that physical TF is modellable by physical TS). In the complex case the simple proof goes through in the same way. However, the hard proof does not go through in the same way and we leave it as an open question as to whether the TS and TF approaches are equivalent. An exciting possibility is that TSCOPT is more constrained than TFCOPT. If this is the case, and if the TS approach is more fundamental, then this provides an interesting clue as to how to develop Quantum Theory going forward.

We will prove the following theorem

> **Modelling TS by TF.** Any physical complex operation in the time symmetric theory can be modelled by a set of physical complex operations in the time forward theory.

The proof of this is along the same lines as in the simple case in Sec. 11.10. Consider a general physical complex operation

$$\overset{\mathsf{c}\uparrow}{\textsf{B}}\xrightarrow{\mathsf{z}} \;\leftrightharpoons\; \xleftarrow{\mathsf{z}^-}\overset{\mathsf{c}\uparrow}{\textsf{B}}\xrightarrow{\mathsf{z}^+} \tag{1298}$$

in the time symmetric theory. Using the summed forward seatbelt indentity (1297) we have

$$\overset{\mathbf{c}}{\mathbf{z}^- \bigcirc\!\!\!\!\!\mathsf{B}\;\, \mathbf{z}^+} \equiv \sum_{x^-} \mathbf{x}^- \left[x^-\right. \quad \left|x^-\right. \mathbf{x}^- \overset{\mathbf{c}}{\bigcirc\!\!\!\!\!\mathsf{B}\;\, \mathbf{z}^+} \tag{1299}$$

which we can write as

$$\overset{\mathbf{c}}{\mathbf{z}^- \bigcirc\!\!\!\!\!\mathsf{B}\;\, \mathbf{z}^+} \equiv \sum_{x^-} \mathbf{x}^- \left[x^-\right. \quad \overset{\mathbf{c}}{\overline{\mathsf{B}}(x)\;\, \mathbf{z}^+} \tag{1300}$$

Thus, we can write any complex operation in the time symmetric theory as a sum over operations, $\overline{\mathsf{B}}(x)$, in the time forward theory. Further, if $\mathsf{B}$ is physical in the time symmetric, then each $\overline{\mathsf{B}}(x)$ must be physical in the time forward theory as discussed in Sec. 53.3 for the positivity condition and Sec. 53.4 for causality. This proves the theorem.

Now consider the question of whether we can model any physical TF operation by a TS operation in the complex case. In the simple case we were able to do this by writing down a particular form for the time symmetric operation, $\mathsf{C}$, that could model the given time forward operation (see (289)) with an extra term involving $\overline{\mathsf{D}}$ (which is given by (291)). The form of this extra term is chosen so that the time symmetric operation causality will satisfy simple backward causality. The problem in applying this approach to the complex case is that there is one backward causality condition for each synchronous partition so we would need to write down an expression for the time symmetric operation that simultaneously satisfied all these backward causality conditions. It is possible this can be done, but it would be a much harder proof than the simple case.

In the simple case we showed that the time symmetric and time forward approaches are equivalent (and therefore the time backward case is also equivalent). We can write this schematically as

$$\text{TBSSOPT} \equiv \text{TSSOPT} \equiv \text{TFSOPT} \tag{1301}$$

We have not proven this in the complex case. Rather we have proven

$$\text{TBSCOPT} \supseteq \text{TSCOPT} \subseteq \text{TFCOPT} \tag{1302}$$

We might conjecture that

$$\text{TBSCOPT} \supset \text{TSCOPT} \subset \text{TFCOPT} \qquad \text{conjecture} \tag{1303}$$

If this conjecture is true then, for complex theories, we have to choose between time symmetric and time forward (or backwards) approaches with the time symmetric theory being more constrained. If we choose the time symmetric case then there would be certain operations in the time forward theory that cannot be modelled in the time symmetric theory. In a similar vein, we can say

$$\text{TSSOPT} \subseteq \text{TSCOPT} \tag{1304}$$

$$\text{TSSOPT} \subseteq \text{TSCOPT} \tag{1305}$$

$$\text{TSSOPT} \subseteq \text{TSCOPT} \tag{1306}$$

because simple networks can be modelled as complex operations but we do not know that all complex operations can be modelled as simple networks. Further, we could have strict inclusions here whereby, for example, TSSOPT cannot model TSCOPT. It is possible that these equivalences and strict inclusions depend on further details of the theory being modelled. It is possible that, in the classical probability, case we have always have equivalence whilst, in the Quantum case, we may have $TSSOQT \subset TSCOQT \subset TFCOQT$ for example. The causal dilation theorems to be discussed in Sec. 78 shed some light on these questions.

# 54 Time backward temporal frame

## 54.1 Time backward complex operations

The time backward case is just the reverse of the time forward case where, now, we do not have outcomes. A time backward operation can be modelled by conditioning on the outcomes of a time symmetric operation.

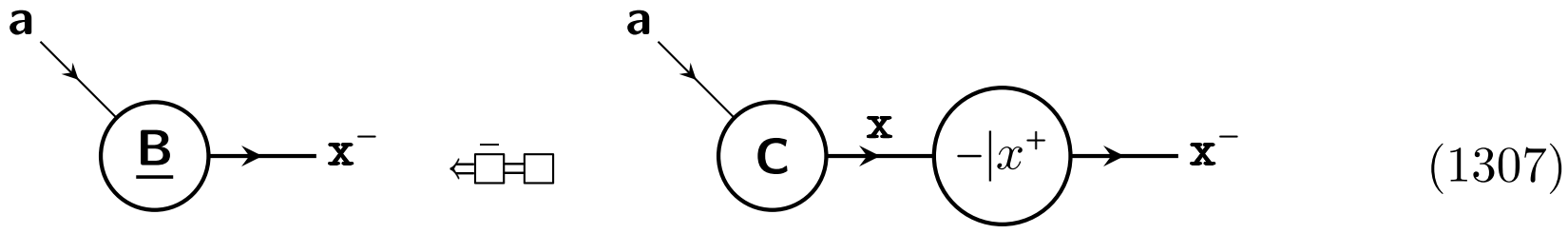

(1307)

The postselection on $x^-$ is implicit in the notation $\overline{\mathbf{B}}$.

## 54.2 Time backward complex circuits

We can wire such time forward operations together to form a circuit. For example

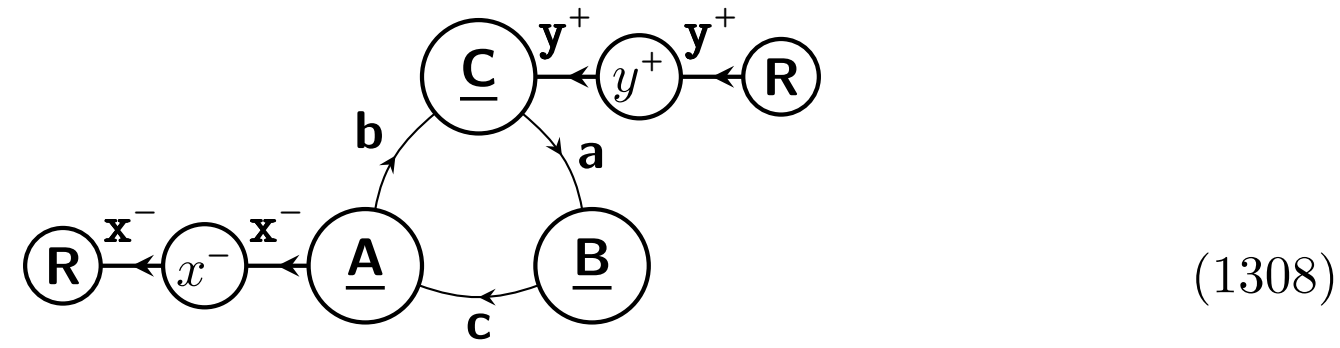

(1308)

This can be modelled by conditioning on outcomes of a time symmetric circuit

(1309)

Following similar reasoning as in Sec. 53.2 for the time forward case, we can calculate the probability for this by making the following replacement

(1310)

This represents a temporal frame transformation.

## 54.3 Positivity in the time backward frame

A time backward complex operation, $B$, satisfies $\underline{T}$-positivity if

$$0 \underset{\underline{T}}{\leq} \qquad (1311)$$

with respect to the tester

(1312)

for all pure preparations $\underline{\mathsf{F}}$ and pure results $\underline{\mathsf{G}}$.

Since we can model time backward operations in terms of time symmetric operations, the positivity under composition theorem in Sec. 7.12.3 goes through for the time backward case as well.

## 54.4 Causality in the time backward frame

We can apply the time forward and time backward causality conditions to the time backward frame by conditioning on outcomes.

First, we will consider the time backward causality condition (1162). We can use the **B** appearing in this condition to model a backward time operation by conditioning on outcomes

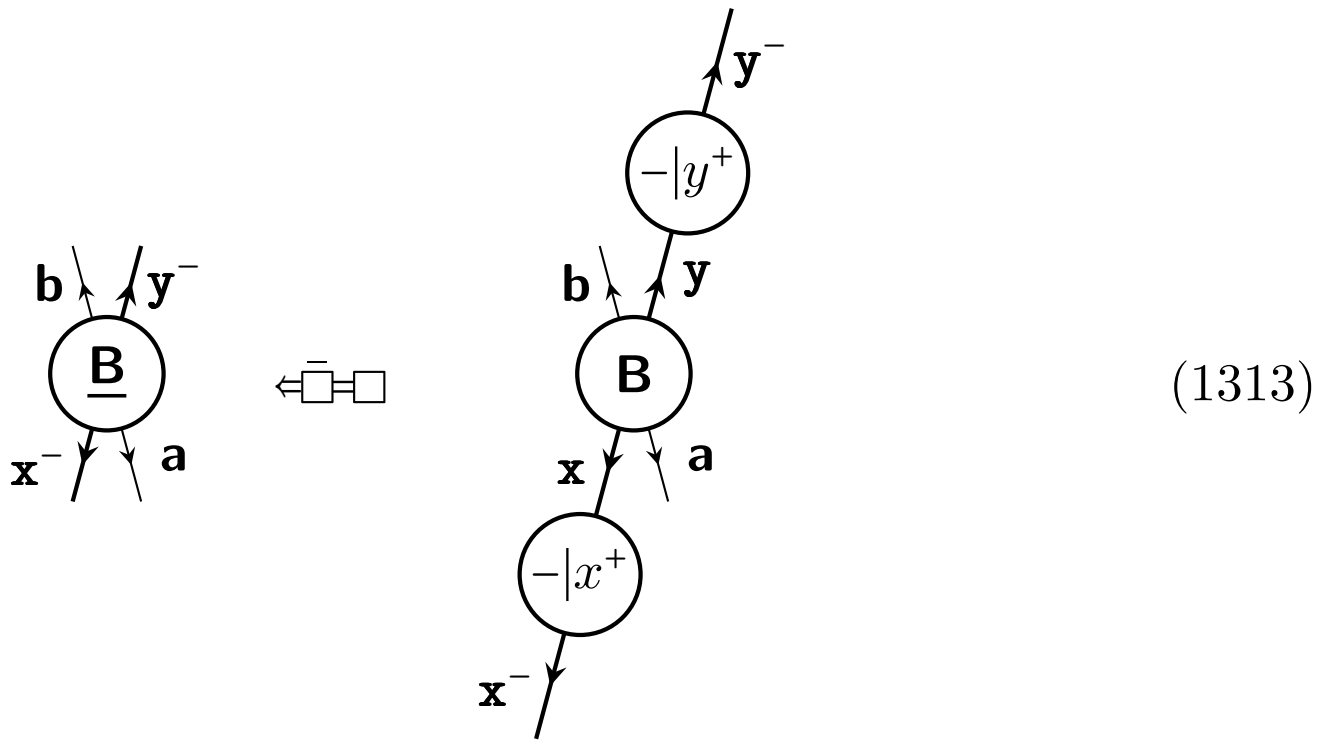

(1313)

(where postselection on $x^+$ and $y^+$ is implicit in the $\overline{\mathsf{B}}$ notation). By conditioning on outcomes for the time backward causality condition (1163) we obtain

(1314)

The technique here is similar to how we obtained (1288) in the time forward case.

The time forward causality condition (1162) can only be conditioned on outcomes if $\mathbf{x} = \mathbf{x}^-$ (so then $\mathbf{x}^+$ is null). In this case, the forward causality condition becomes

(1315)

This only applies in the special case that there are is no implicit outcome conditioning after $p$ (as denoted by the $*p$ superscript).

Since time backward operations can be modelled in terms of time symmetric operations the causality composition theorem (in Sec. 49.4.11) goes through in the backward case as well.

# 55 Modelling Spacetime - Field Theories

We can model spacetime in $1+1$ dimension by considering a number of evenly spaced similar physical systems moving from left to right, whilst an equal number of similarly evenly spaced similar physical systems move from right to left such that systems can interact when they intersect. We can represent this situation as follows

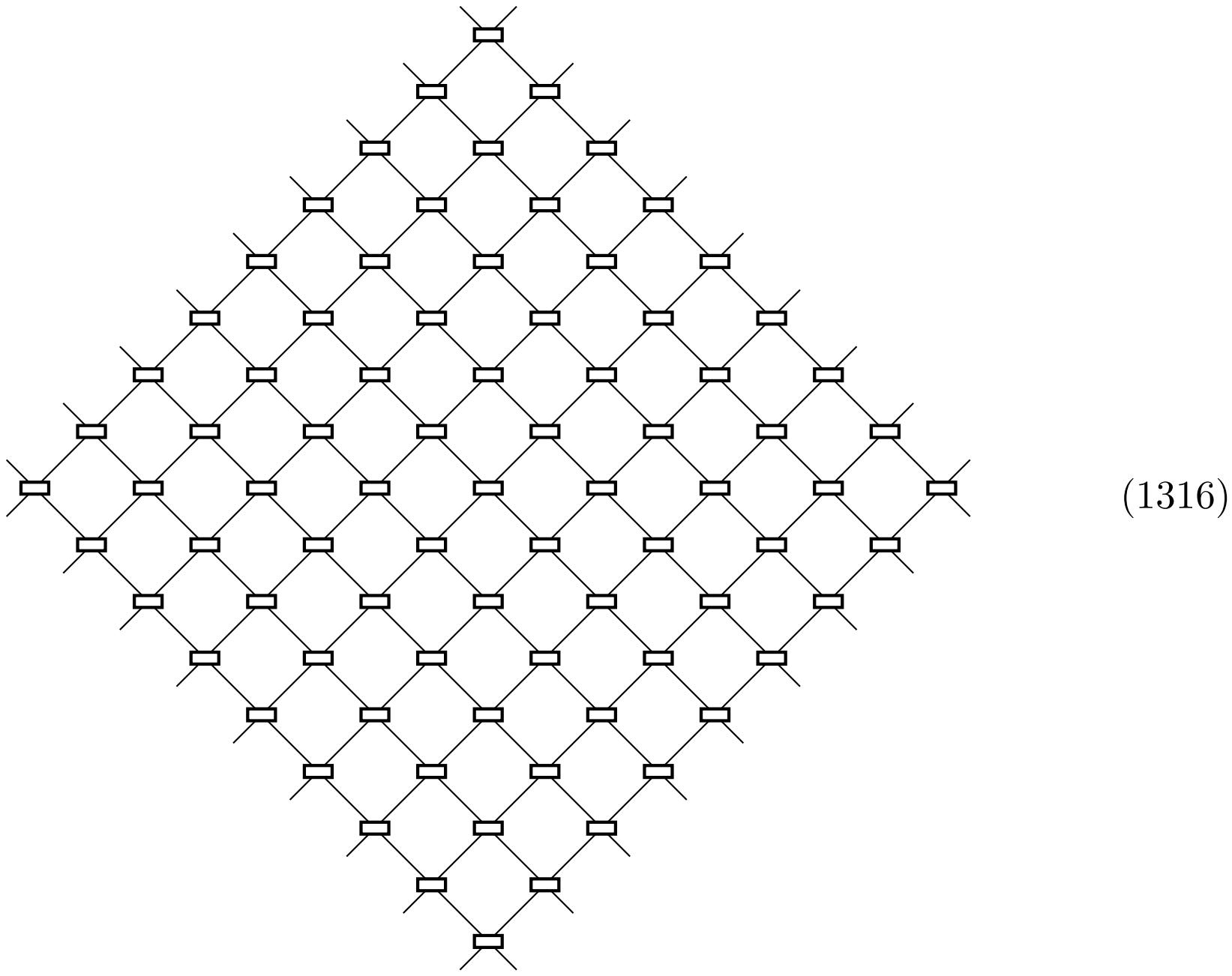

$$(1316)$$

where time goes up. We can imagine this network extending arbitrarily in space and time (so we have just shown a small part of it). There are simple operations at the places were the systems intersect as shown. Each of these operations has two systems going in and two systems coming out. Further, they each have a setting, an income, and an outcome. We can, indeed, think of these as providing a setting field, an income field, and an outcome field over spacetime (these being defined at the vertices only). We can easily extend this picture to model higher dimensional spacetimes by suitable latices of interacting qubits.

We can draw a region, $A$, which contains some of these operations as follows

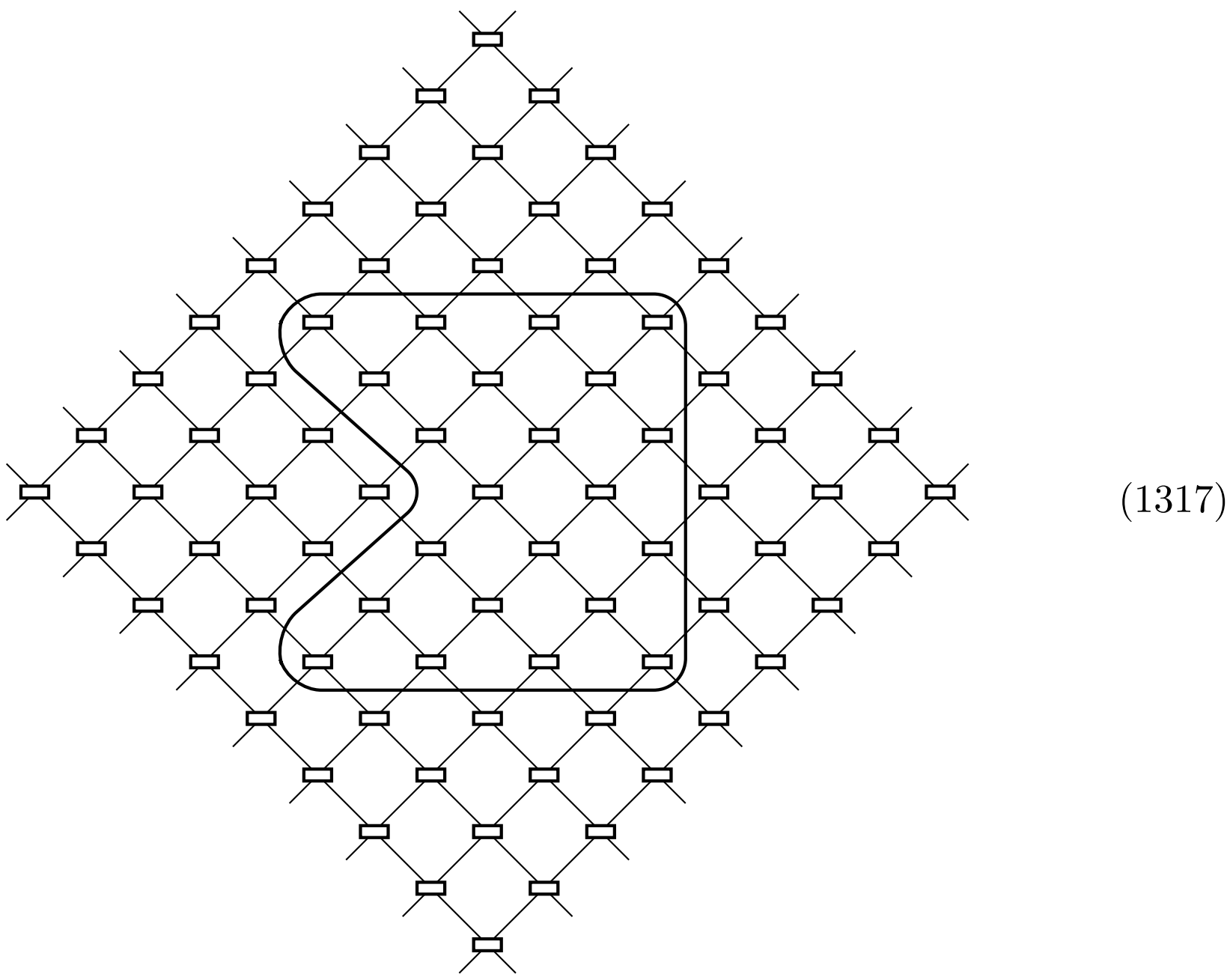

(1317)

We can imagine shrinking the size of the lattice (whilst keeping the area, $A$, as is) so that the wires and operations become more and more densely packed. If it is possible to take this limit in an appropriate way, this would offer a way to study continuous spacetime. We will not pursue that goal in this book. However, we can treat the case where the grid size is as small as we like (as long as it is non-zero).

Region $A$ above encloses the following simple network

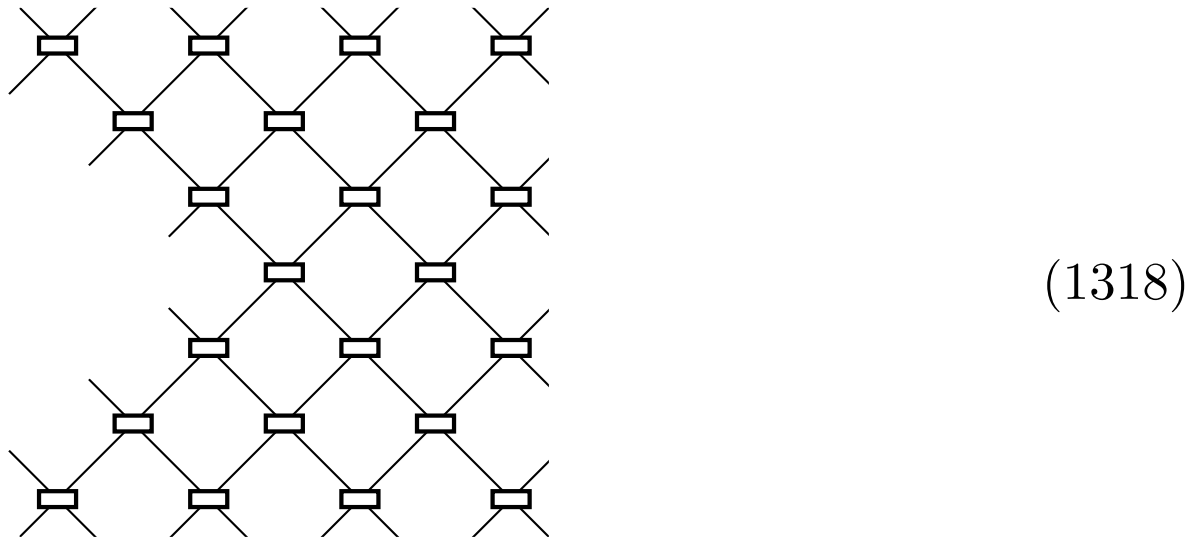

(1318)

We can treat this network as a complex operation. At the top all the systems are outputted from $A$ both to the left and to the right. At the bottom they are all inputted from the left and from the right. Along the right side we notice that there are both inputted (from the right) and outputted (to the right) systems. On the left side, where there is an indent, the lower part of this indent has outputted systems moving to the left and the upper part has inputted systems coming from the left. In the complex operational framework, our systems can be constituted of both inputted and outputted subsystems so we can, for example, treat write the (composite) system associated with the right side as a single system. As we shrink the grid size, these systems passing through the boundary will become more densely packed.

The boundary of $A$ intersects through system wires. We can divide this boundary up into different segments corresponding to different disjoint sets of wires. For example

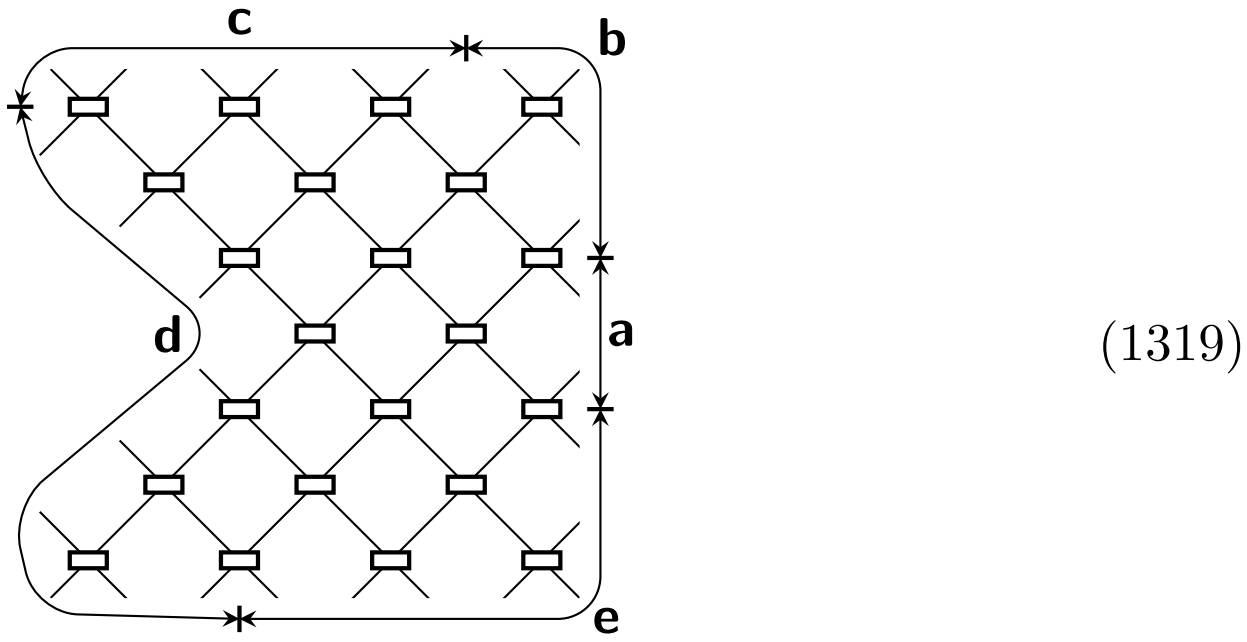

(1319)

Here **a** corresponds to the wires that intersect the indicated segment of boundary. We can write the complex operation associated with $A$ as

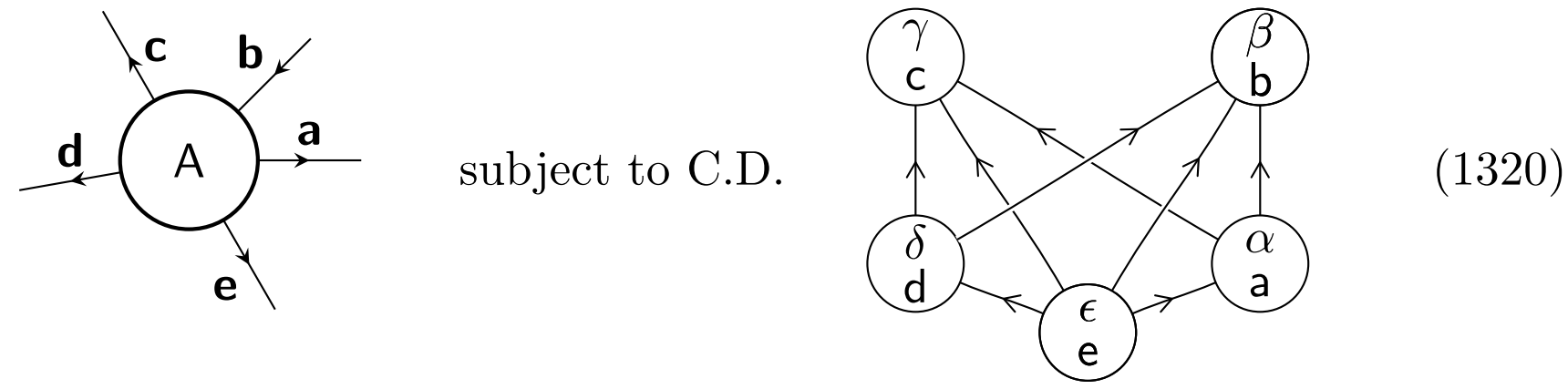

(1320)

where **abcde** consists all the systems passing through the boundary. It is instructive to look at different parts the boundary where we see different types of behaviour. Each of the systems can have positive and negative parts defined with respect to the direction of the arrow. Note that **c** only has a positive part (so we can write **c** = **c**$^+$). Further, note that we have defined **b** to be pointing in. This means that the systems in the associated boundary segment belong to **b**$^+$ if they are inputs (in the example, there is one such system) and **b**$^-$ if they are outputs (there are two such systems in the example).

The causal diagram in (1320) can be obtained from the network in (1319). To obtain this we can use the techniques in Sec. 47.9 which involves starting with a causal diagram obtained by fusing the causal diagrams associated with the simple operations then simplifying using the causal spider identities. In this spacetime context, however, it is also instructive to notice how the spacetime modelled by the lattice can be used to read-off the causal relations. For example we see that **e** < **b** from (1319) by which we mean that there are forward paths through the operations starting from some subsystems associated with **e** to subsystems associated with **b**. This means there has to be an arrow from **e** to **b** in the causal diagram. On the other hand, we see that **d** ~ **a** as there are no forward paths from **d** to **a** or from **a** to **d**. One causal relationship, not represented in the causal diagram above can be written $\mathbf{f} \gtrless \mathbf{g}$. This is where we can find a forward path from some subsystems associated with **f** to some subsystems associated with **g** and we can also find a forward path from some (other) subsystems associated with **g** to **f**. This is a fairly generic situation and is represented in the causal diagram by having an arrow going from the **f** node to the **g** node and another arrow going from the **g** node to the **f** node. The various causal relationships we have just discussed will maintain as we shrink the size of the grid since they depend on the underlying spacetime we are modelling. Thus, in the continuous limit (if this limit can be shown to behave well), we can use these techniques to write down the causal diagram. Indeed, in our example, the causal diagram will continue to be that in (1320) as we take this limit.

If we have two or more regions that share some portion of their boundaries then we can represent them by joining the corresponding wires of the complex operations. For example, the situation

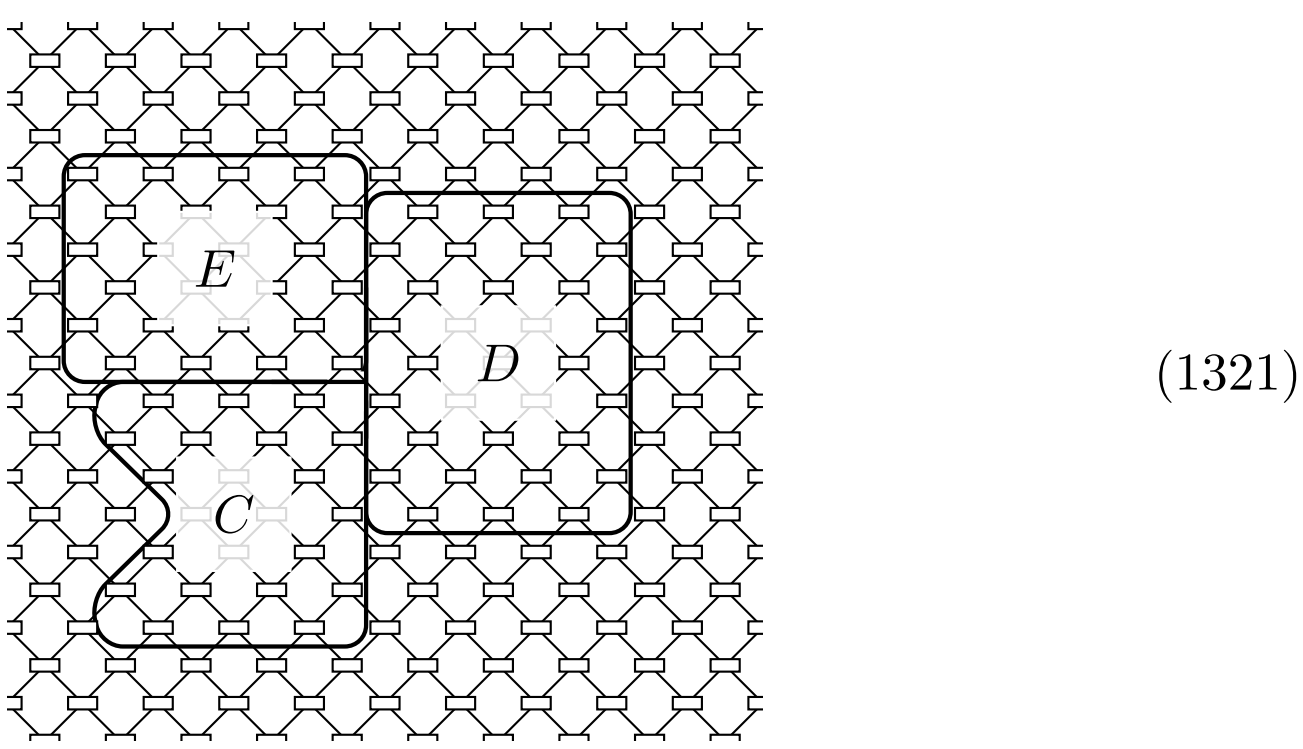

(1321)

can be written as

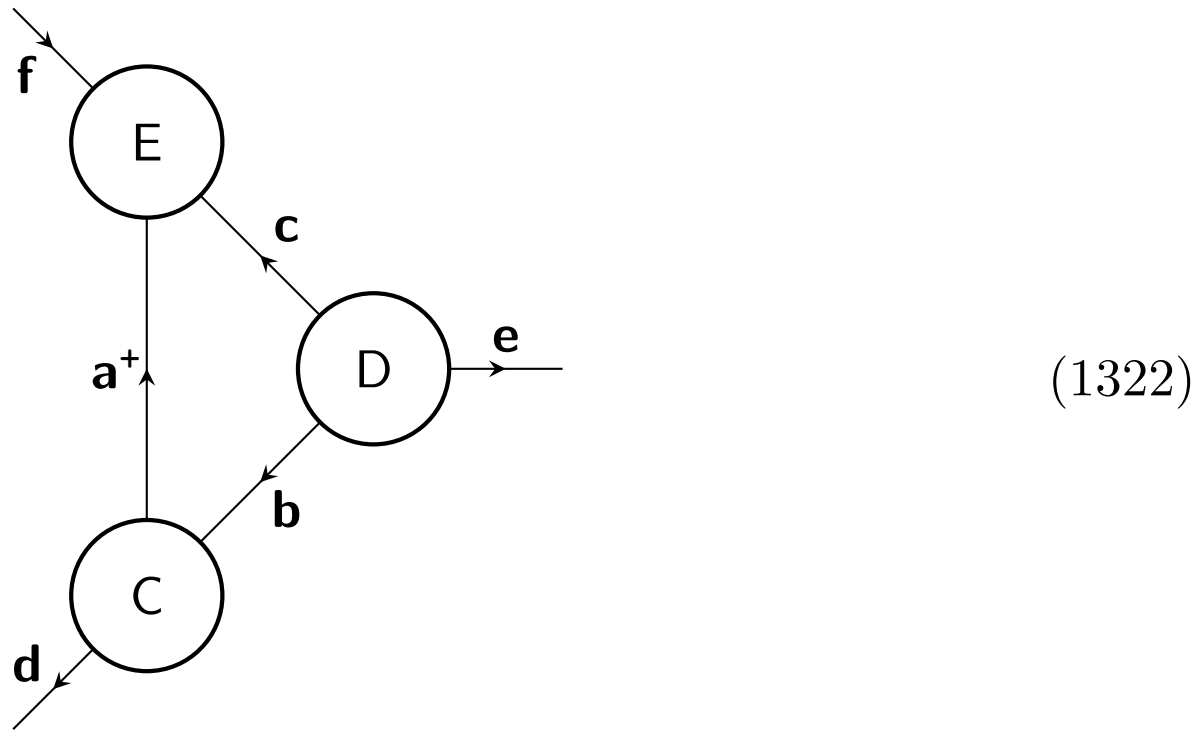

(1322)

where the systems that join the operations correspond to the shared portions of the boundaries. Note that the boundary between $C$ and $E$ has only systems travelling from $C$ to $E$ so we can write the associated system as $\mathbf{a}^+$.

The approach described here is very natural and it is anticipated that it can form a basis to model Quantum Field Theory. However, that project will not be attempted in this book. There remain many challenges for such a program. In particular note that the continuous limit has not been shown to be mathematically well behaved. The continuum is always understood by taking a limit in mathematics. If we are to be true to the spirit of the current work then we would need to understand such a limit in operational terms (perhaps corresponding to a limit of operational descriptions with ever increasing accuracy). Further, we have not shown how to model even very simple Quantum Field Theories such as the Klein Gordon equation. We might better call this a proto-Field Theory approach.

This way of thinking about field theories in spacetime is, in part, inspired by Robert Oeckl's general boundary formulation Oeckl [2003] from 2003 which was partly motivated by the problem of Quantum Gravity. There he associated objects with general boundaries (as we do here). In his earlier work, these objects are linear in Quantum amplitudes. He was later influenced by modern approaches to Operational Quantum Theory and developed an approach where the objects are is linear in probabilities which he called the positive formalism Oeckl [2013, 2014, 2016]. There is a similarity in spirit between his approach and the field approach outlined here.

# 56 Sorkin's impossible measurements

## 56.1 Introduction

Rafael Sorkin showed that, in Quantum Field Theory, it is possible to signal faster than light if we allow arbitrary ideal measurements (see Sorkin [1993]. The scenario he considers consists of operations (in our language) in three non-overlapping regions, $A$, $B$, and $C$ (which may meet at their boundaries) such that region $C$ is spacelike separated from $A$ whilst $B$ overlaps with the future light-cone of $A$ and the past light-cone of $C$ as depicted below

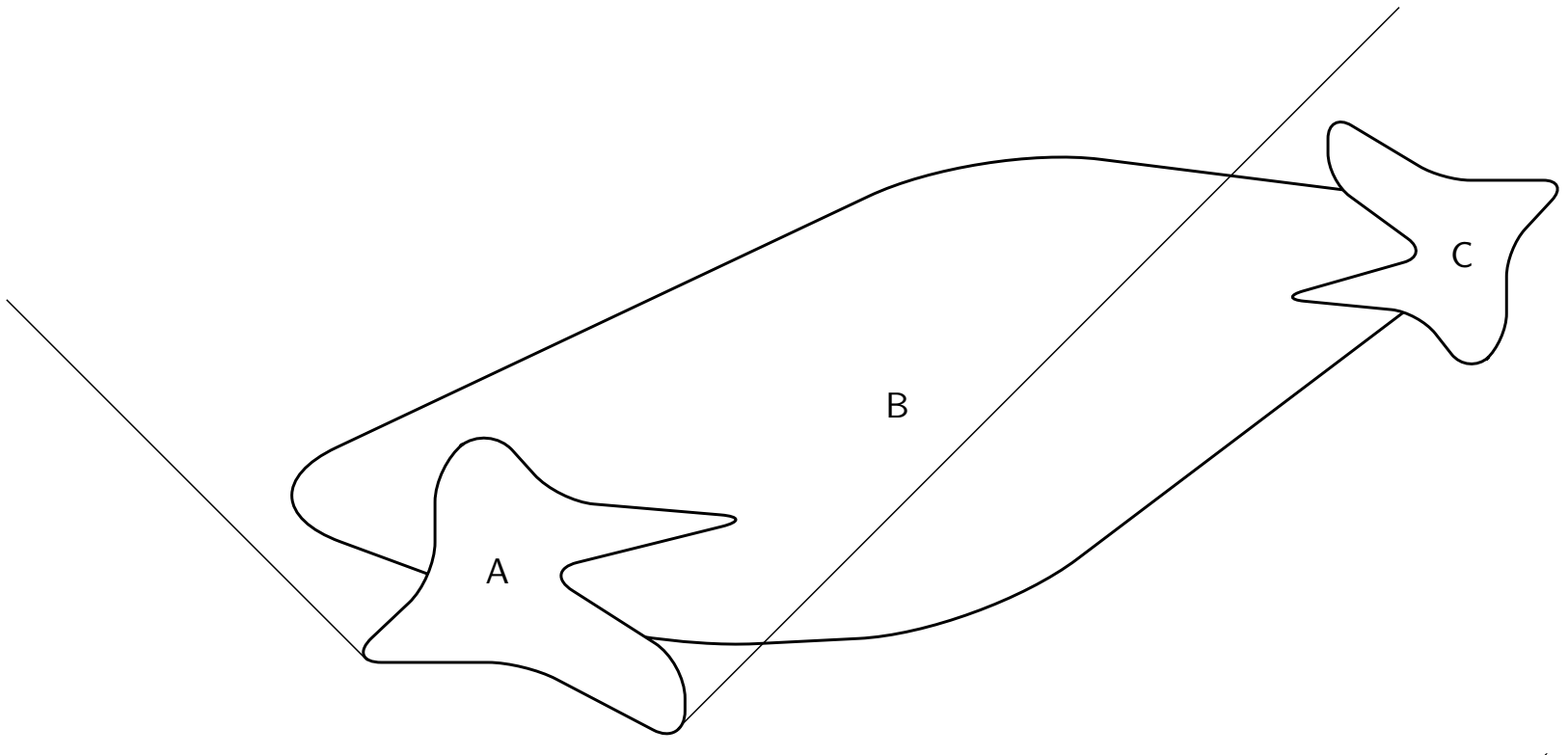

(1323)

(we have shown the forward light-cone from $A$). Sorkin, in particular, considers region $B$ to stretch out infinitely in space whilst having some short but finite duration (this is an example of the sort depicted in the above picture but where $B$ is unbounded in its spatial extent). Then region $B$ is sufficient to perform an idea projection measurement which projects onto the associated eigenstate or the complement of this. The outcome of this measurement is ignored. We can, Sorkin shows, find scenarios where probability of a certain outcome in $C$ depends on a choice of setting in $A$ (even though $C$ is outside the forward light-cone of $A$). Given that projection measurements are very strong this is, perhaps, not surprising. Further examples of this type, but where region $B$ is bounded, are supplied by Borsten, Jubb, and Kells [2021]. These authors also point out that the standard microcausality conditions of Quantum Field Theory are not sufficient to rule out this kind of signalling.

If we are unconstrained in what measurements we can make in region $B$ then it is, perhaps, not surprising that we can signal from $A$ to $C$. For example, if different choices of preparation in $A$ effects preparation of one state or another impinging on the boundary with $B$ then an arbitrary measurement in $B$ can shunt this information to $C$ enabling signalling. What is required of any theory is physically motivated constraints so that these impossible measurements are prohibited. Sorkin's impossible measurements is an important challenge for any approach to formulating physics in spacetime.

We will see how the framework developed in this book resolves this problem very naturally at this general operational level (even before we use correspondence to implement the Quantum version of this theory) if we prohibit nonphysical operations. The physicality constraint builds in the necessary causality conditions that exactly forbid signalling in such a scenario. Although we have not shown how this approach can actually model a Quantum Field Theory (even a simple one such as afforded by the Klein-Gordon equation) this approach, nevertheless, is promising because it is rooted in a very simple understanding of causality.

Many authors have addressed Sorkin's impossible measurements in various frameworks. In the context of the Algerbraic Quantum Field Theory framework (AQFT) which goes back to Haag and Kastler [1964]. A modern perspective has been taken by Fewster and Verlach Fewster and Verch [2020] who, in particular, model how measurements can be implemented by introducing probe fields. This makes their approach suitable for investigating Sorkin's problem. Impossible measurements have been addressed in papers by Bostelmann et al. [2021], Jubb [2022], Fewster et al. [2023]. The dissertation of van der Lugt [2021] is an accessible introduction to this work. A more recent article by Papageorgiou and Fraser [2024] provides a very useful survey of this work.

Another approach has been taken by Gisin and Del Santo Gisin and Del Santo [2024]. They set up a finite dimensional Hilbert space version of Sorkin's problem where, in the first place, region $B$ is modelled by a bipartite quantum system. They show that nonlocal measurements on this bipartite system can lead to signalling. Then, leveraging results on such nonlocal measurements, they study nonlocal measurements that can implemented with entanglement resources and classical signalling.

The approach of Gisin and Del Santo is more readily comparable to the approach in this book than the AQFT since, like Gisin and Del Santo, here we take a Quantum Information based perspective. The approach here is, perhaps, complementary to the approach of Gisin and Del Santo in that, here, no-faster-than-light signalling is ensured by the forward causality part of the physicality constraints. This does not require modeling how such measurements might be implemented. Additionally, we are able to model arbitrarily shaped regions of spacetime through the complex operations framework. Further, we are able to resolve Sorkin's problem at the level of the operational theory even before we introduce Hilbert spaces and Quantum Theory. All these results will then pass over to the quantum case because of the correspondence principle.

There has been even more recent work which combines Quantum Information and AQFT intuitions in Simmons et al. [2025]. They consider factorisation conditions on the local $S$-matrix that guarantee measurement operations in quantum field theory are compatible with relativistic causality (i.e., they cannot produce superluminal signalling). Using probe-based measurement models, they show that physically realizable measurements satisfy these conditions.

Huhtala and Vilja [2025] analyzes the impossible measurement setup in a non-relativistic model and calculates the amount of superluminal signalling it would produce. It then identifies phenomenological constraints—motivated by relativistic causality—that must restrict measurement operations so that such signalling cannot occur.

Other important work on relativistic constraints in Quantum Theory is due to Aharonov and Albert [1981] who showed how it is possible to measure non-localised observales and Beckman et al. [2001] who showed that, in Quantum Theory, there are causal operations which are not locally implementable.

## 56.2 Modeling Sorkin's problem as a complex network

We can model Sorkin's problem as a complex network by associating operations **A**, **B**, and C with regions $A$, $B$, and $C$ so we have

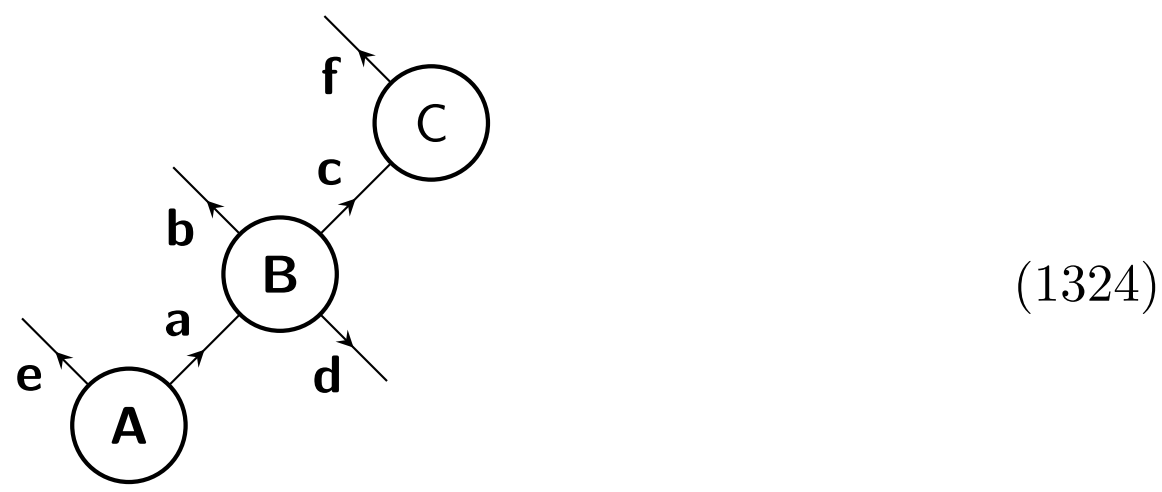

(1324)

where, in particular, we impose that where **B** is subject to the causal diagram

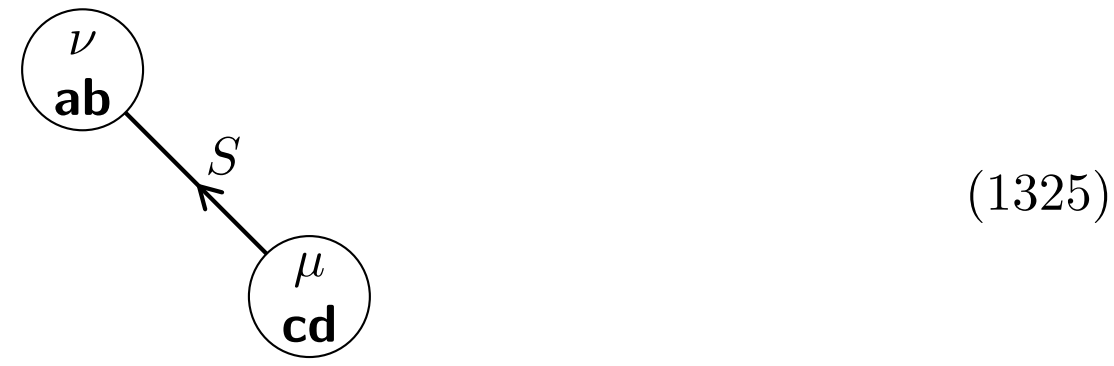

(1325)

We see there is no forward arrow in this causal diagram from system **a** which is between **A** and **B** and system **c** which is between **B** and C. This models, in an appropriate way, the causal situation shown in the space time diagram in (1323). We can regard these complex operations originate from an network lattice as discussed in Sec. 55. Then the spacetime diagram in (1323) can be interpreted as an example of the sort shown in (1321). However, the theorem below pertains very generally and is not rooted in that model of spacetime.

Given that **B** has the causal diagram shown in (1325), the causal diagram for the network shown in (1324) must take the form

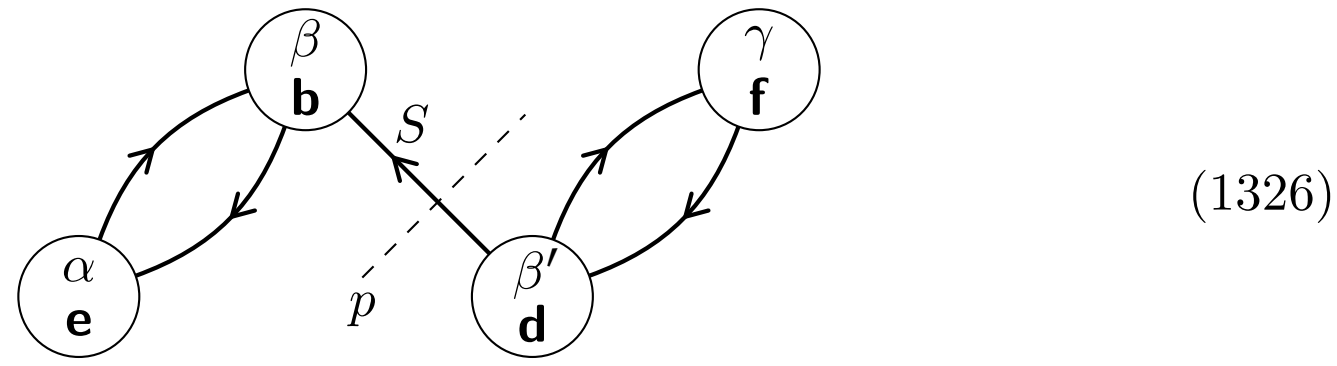

(1326)

Here $\alpha$ comes from A, $\beta$ and $\beta'$ come from **B**, and $\gamma$ comes from **C**. Since **a** may be highly composite and have both positive and negative parts, we have drawn thick arrows in both directions between the $\alpha$ and $\beta$ nodes, and similarly for **c**. We have omitted the symbols for the sets of wire labels on these thick wires to keep the figure tidy. The synchronous partition, $p$, can be thought of as corresponding to the part of the light cone that intersects region $B$ as depicted in (1323). This causal diagram suitably models the causal structure shown in (1323). Note that **A** and **B** are deterministic whilst C is not. This is because we do not look at any readouts for **A** or **B** but we are interested in the readout at C (which is, therefore, not deterministic). Now, according to the basic forward causality principle (see Sec. 7.5), if we are to send information forward in time we need to preselect on some income into **A**. Thus we allow preselection in region $A$. For generality we also allow preselection in regions $B$, and $C$. In fact we can consider an even more general situation, namely

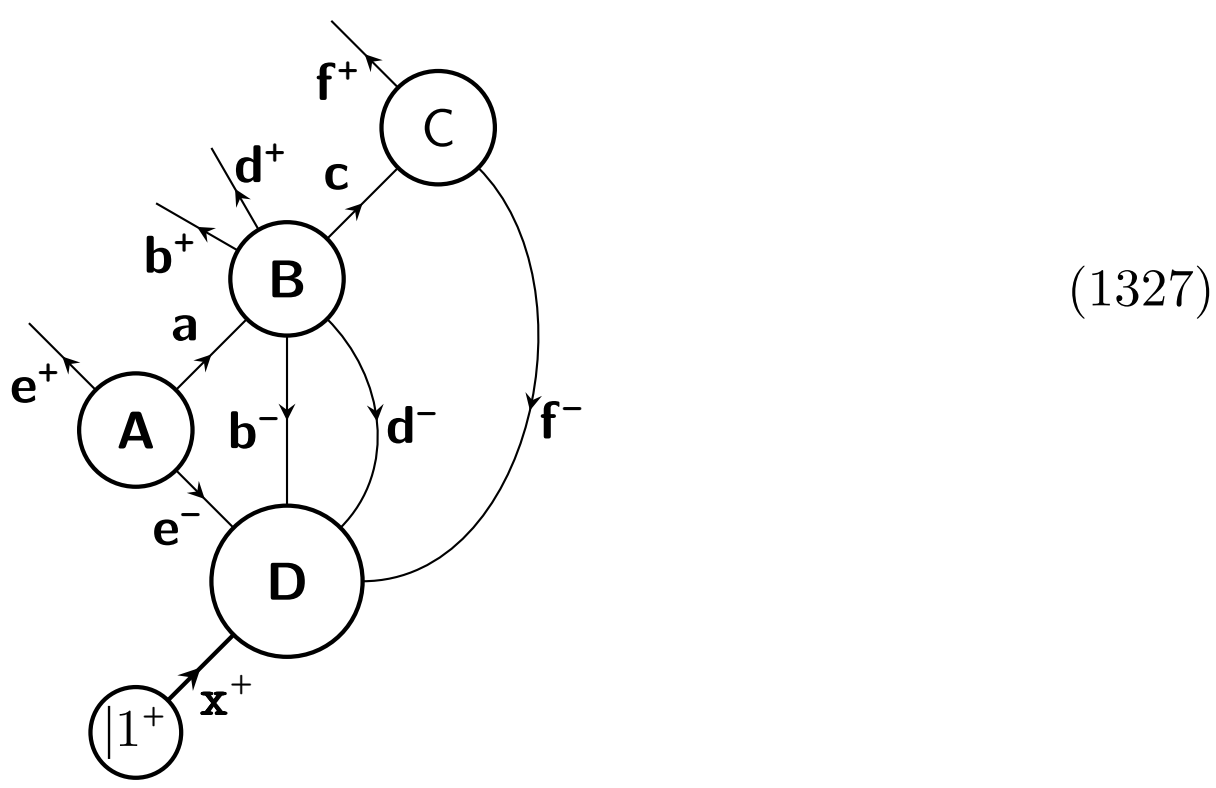

(1327)

Note that **D** is joined to **A**, **B**, and C by forward pointing wires. The circle with the $|1^+$ in it preselects on $x^+ = 1^+$ (these preselection boxes were introduced in (1270)). Note further that, with an appropriate choice for the new operation **D**, we could "channel" preselection separately to each of **A**, **B**, and C should we wish. However, the theorem below will work for this very general situation.

## 56.3 Sorkin's impossible measurements are unphysical

We can now prove the following theorem

> **Sorkin's impossible measurements are unphysical.** If **A** and **B** satisfy forward causality, and **B** has causal diagram
>
> $\nu$ **ab** — $S$ — $\mu$ **cd** (1328)
>
> then
>
> 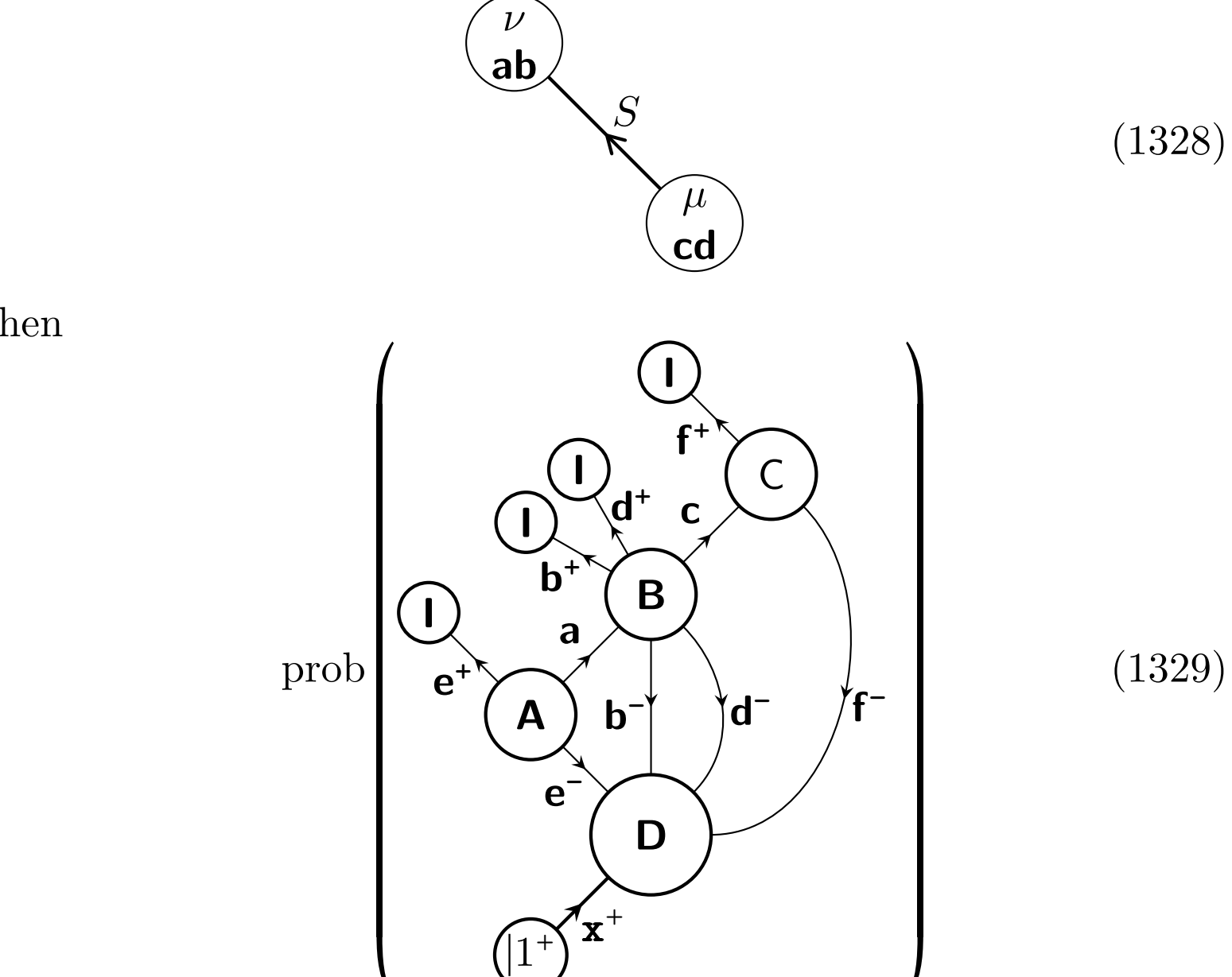
>
>
> (1329)
>
> is independent of the choice of **A**. Here **A**, **B**, and **D** are deterministic. [Note this theorem also holds if we include a pointer wire following a similar path to each system wire. Then we include some **R** results to absorb future going pointer wires where the corresponding system wires are absorbed by **I** boxes. We have not shown these pointer wires to keep the diagram tidy.]

This means we cannot signal from region $A$ to region $C$ by varying the choice of operation in **A** as long as **A** and **B** satisfy forward causality (which they must if they are physical). The proof of this result is very simple given the network residuum theorem (special forward case) from Sec. 49.4.12 (which, itself, was nontrivial to prove). First note that, given that **B** has the causal diagram shown in (1328), we know that

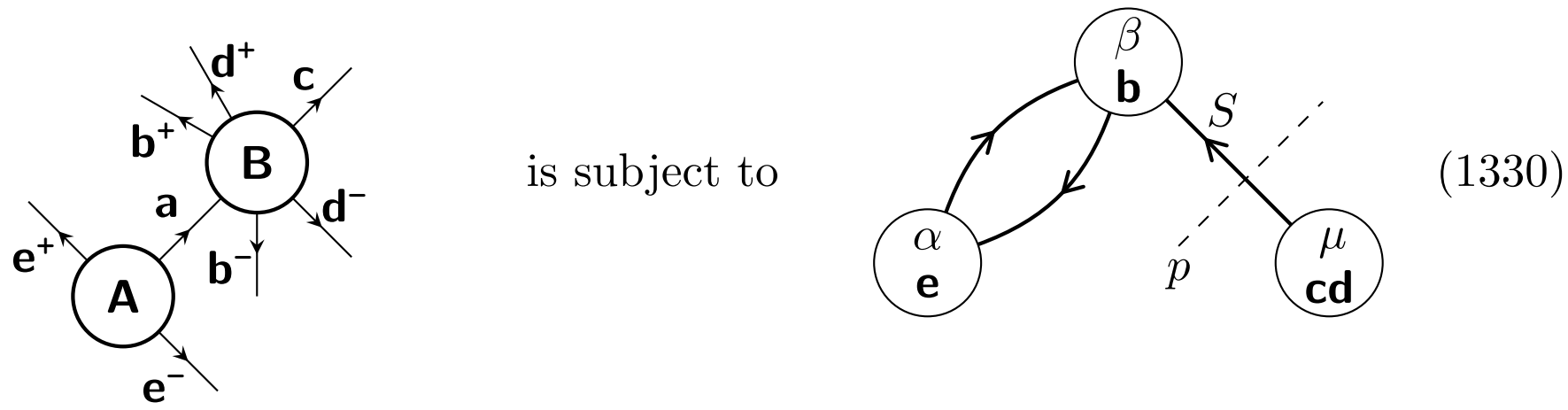

(1330)

Consider applying forward causality with respect to $p$. Given the form for the causal diagram in (1330) and given that both **A** and **B** are taken to satisfy forward causality, we can apply the network residuum theorem just mentioned. Application of this theorem gives

(1331)

If we substitute this into (1329) we immediately see that dependance on **A** drops out. This proves the theorem.

This theorem can be applied to the forward frame. In that case the incomes on every operation are preselected. As discussed below (1327), we can channel preselection to each of **A**, **B**, and C by appropriate choice of **D**. Thus we the above theorem means we cannot use the network

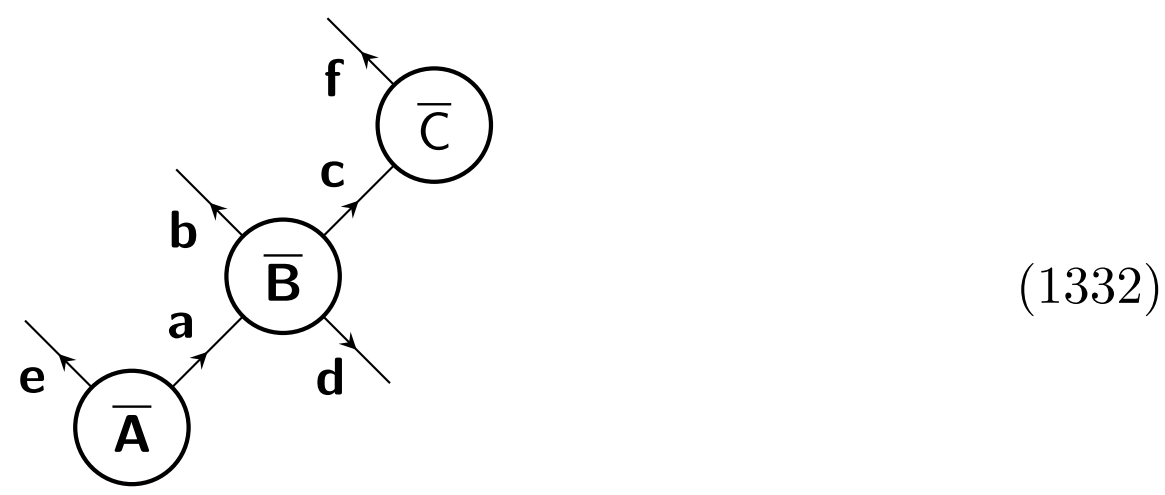

(1332)

to signal from $A$ to $C$ by varying the operation $\overline{\textbf{A}}$.

## 56.4 Backward Sorkin scenario

Interestingly, there is also a backward scenario. For completeness we will state this also.

> **Sorkin's impossible measurements are unphysical (backward version).** If **B** and **C** satisfy backward causality, and **B** has causal diagram

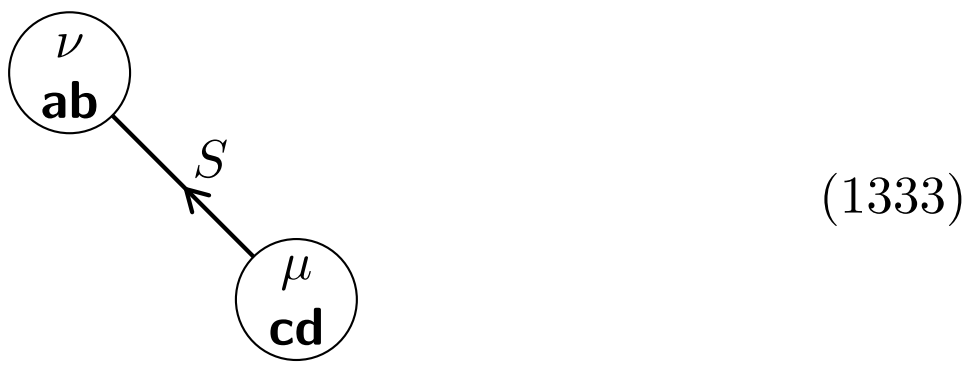

(1333)

then

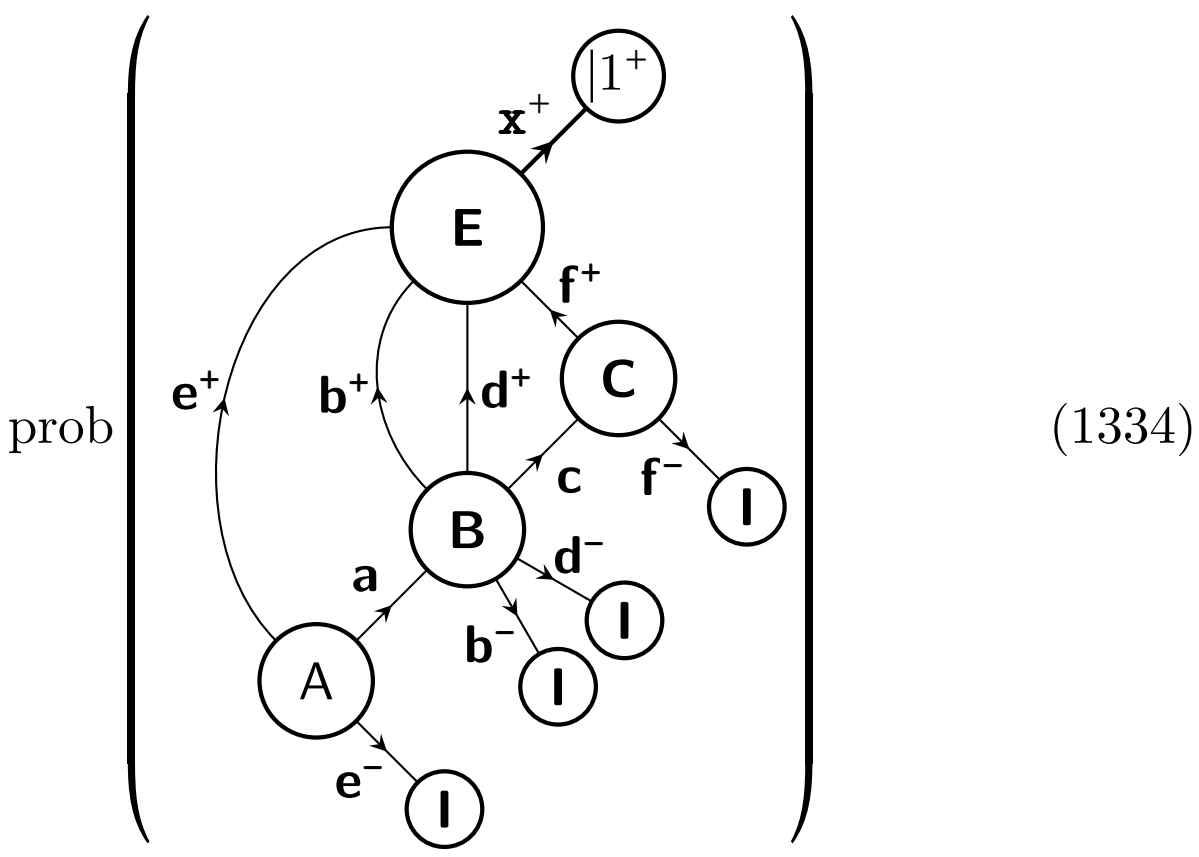

is independent of the choice of **C**. Here **B**, **C**, and **E** are deterministic. [Note this theorem also holds if we include a pointer wire following a similar path to each system wire. Then we include some **R** results to absorb past going pointer wires where the corresponding system wires are absorbed by **I** boxes. We have not shown these pointer wires to keep the diagram tidy.]

The proof of this is along the same lines as the previous theorem. This theorem fits naturally with the backward temporal frame where we postselect on outcomes. We can channel the postselection in (1334) back through **E** so we have separate postselection on each of A, **B**, and **C**. Then we have

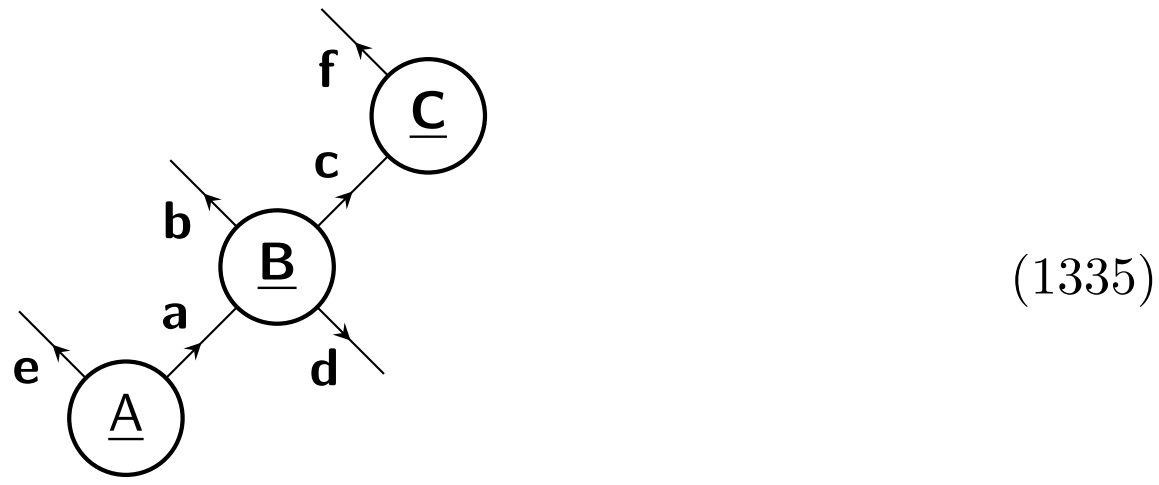

In this case we see that, in the backward temporal frame, we cannot signal from region $C$ to region $A$ by varying $\underline{\mathbf{C}}$.

## 56.5 Discussion

This framework resolves Sorkin's impossible measurement problem in a very natural way - impossible measurements are not physical. The proof is a little non-trivial since it uses the theorem about residua from Sec. 49.4.12 which required taking bites out of the causal diagram of a network.

One point worth mentioning is that the regions $A$, $B$, and $C$ do not need to be convex. For example, regions $A$ and $B$ are depicted as non-convex in (1323). This means that the system **a** and which connects **A** and **B** (see (1324)) can have future pointing (+ve) and past pointing (−ve) parts. Similarly, the system **c** connecting **B** and **C** can have future and past pointing parts.

We have proven these results at the operational level. They go over to Quantum Theory under correspondence (which will be introduced in Sec. 61.1). Showing that we can model full Quantum Field Theory (even for a simple case) is beyond the scope of this book. Consequently we cannot claim to have fully addressed Sorkin's problem. Nevertheless, this approach is very natural and flexible. Note, for example, that we are able to model causally non-convex regions of spacetime and we can envisage some limiting process to get to arbitrarily dense causal diagrams that may model continuous field theories like Quantum Field Theory in the limit (this is discussed in Hardy [2018]).

# Part V
# Causally Complex Operational Quantum Theory

## 57 Introduction

Now we have set up the framework for causally complex operational probability theories ($t$COPT) we can consider specializing to causally complex operational Quantum Theory ($t$COQT). This works in a similar way to the simple case ($t$SOPT → $t$SOQT). Here is a flowchart showing how we obtain $t$COQT by appending structure to $t$COPT.

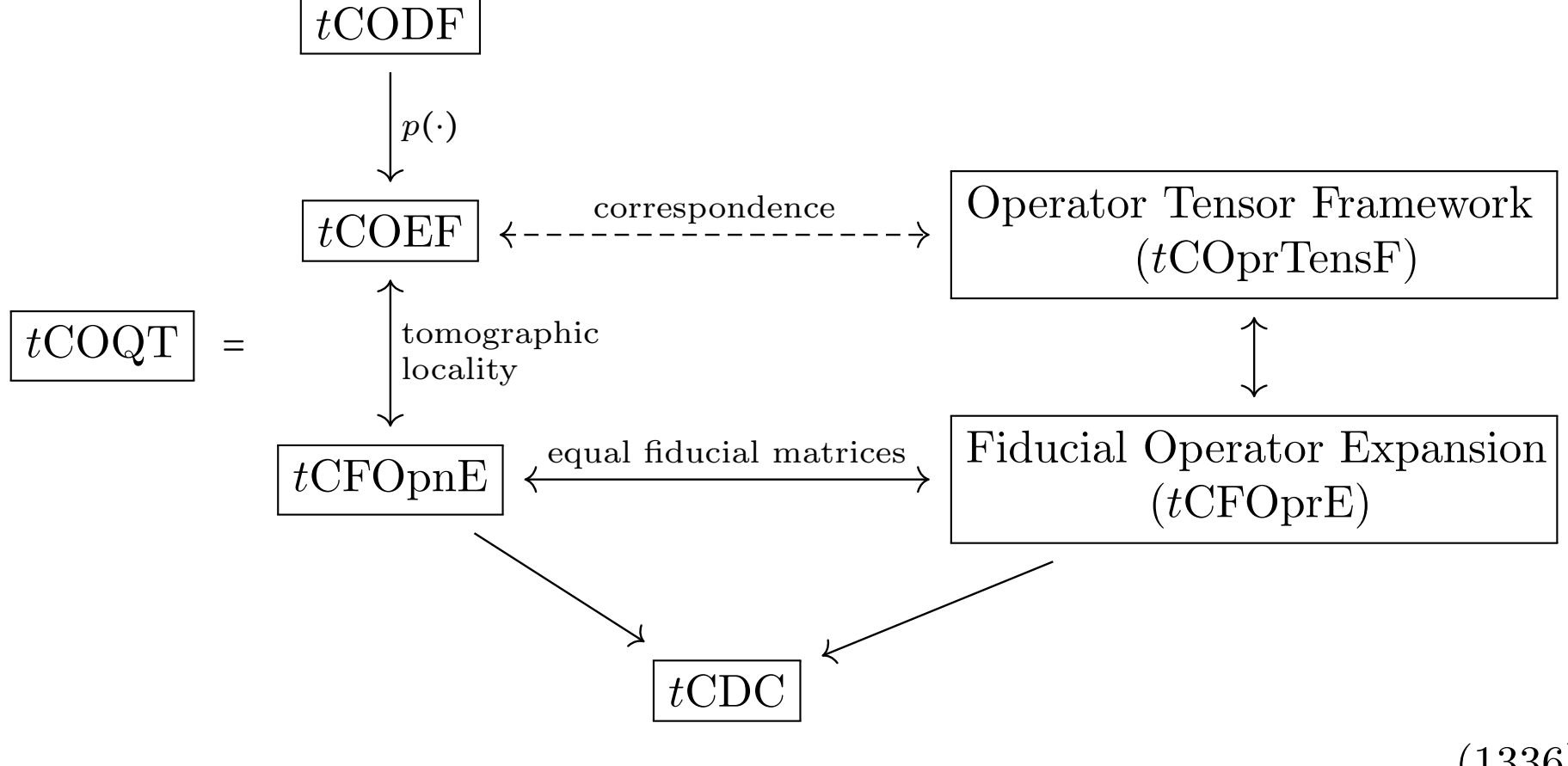

(1336)

This is the flowchart in (37) with $x$ =C (a discussion of the general strategy was given there). We will, in the first place, work in a time symmetric temporal frame ($t$ =TS).

# 58 Complex operator

## 58.1 Basic idea

In Sec. 46.2 we introduced a complex operation as a pair of objects

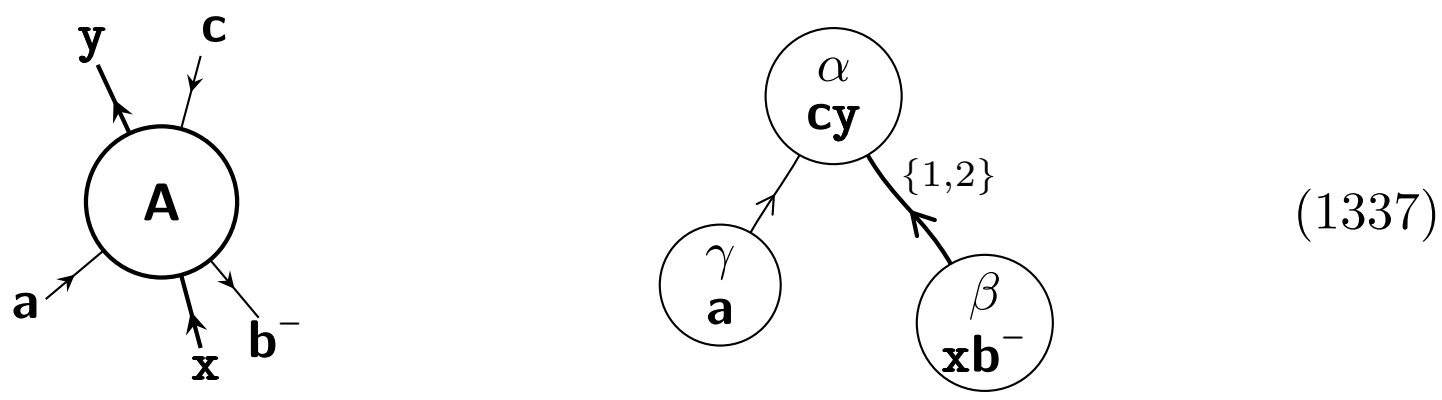

$$(1337)$$

The first object in the pair is the "operation itself" and the second object is the causal diagram. We wish to set up a correspondence between complex operations and complex operators. Thus, we want complex operators to also be represented by a pair of objects

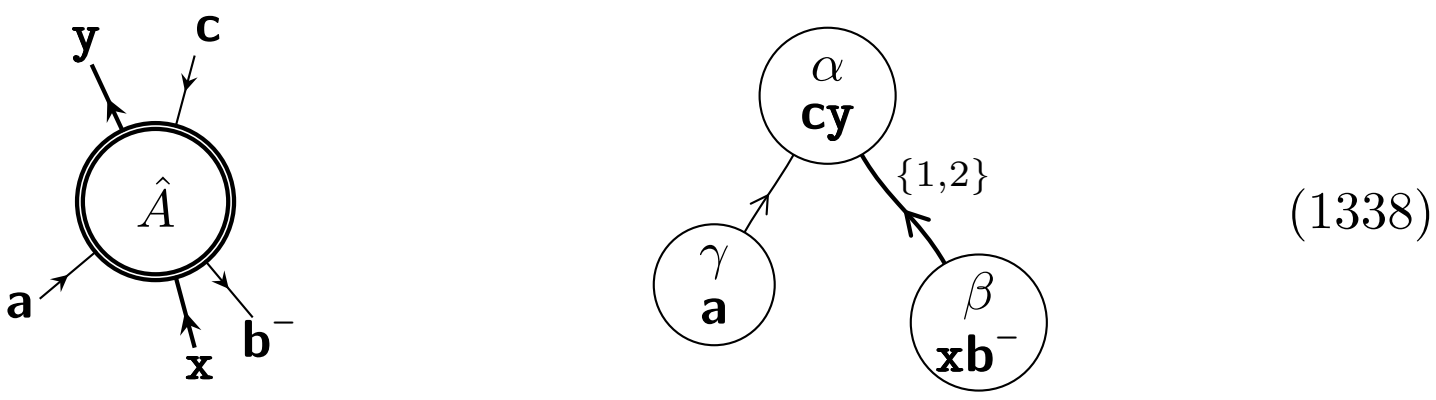

$$(1338)$$

We will see in Sec. 61.1 how to construct the first object in the pair (the operator itself) using fiducial expansions. The second object in the pair is, under correspondence, the *same* causal diagram. Thus complex operators have the same causal diagram as the complex operations they are associated with.

## 58.2 The space complex operator tensors live in

A causally complex operator tensor is represented as follows

$$\hat{B}^{\mathbf{b}_4\mathbf{v}_5}_{\mathbf{a}_1\mathbf{c}^-_2\mathbf{u}_3} \tag{1339}$$

with the diagrammatic representation on the left and the symbolic representation on the right. This operator tensor is an element of the space

$$\mathcal{L}_{\mathbf{a}_1}\otimes\mathcal{L}_{\mathbf{c}^-_2}\otimes\mathcal{P}_{\mathbf{u}_3}\otimes\mathcal{L}^{\mathbf{b}_4}\otimes\mathcal{P}^{\mathbf{v}_5} \tag{1340}$$

Later, in Part VI, we will provide a diagrammatic way of understanding how the $\mathcal{L}$ spaces can be built out of underlying Hilbert spaces. For the moment we will provide definitions that are sufficient for our present purposes. First,

$$\mathcal{L}^{\mathbf{b}_4} = \mathcal{H}^{\mathbf{b}_4} \otimes \overline{\mathcal{H}}^{\mathbf{b}_4} \tag{1341}$$

where $\mathcal{H}^{\mathbf{b}_4} = \mathcal{H}^{\mathbf{b}_4^+} \otimes \mathcal{H}^{\mathbf{b}_4^-}$ is a Hilbert space of dimension $N_{\mathbf{b}} = N_{\mathbf{b}^+} N_{\mathbf{b}^-}$. The Hilbert space $\mathcal{H}^{\mathbf{b}_4^+}$ is interpreted as pointing forward in time since the $\mathbf{b}_4^+$ is a superscript *and* it has a + sign. The Hilbert space $\mathcal{H}^{\mathbf{b}_4^-}$ is interpreted as pointing backward in time since it is a superscript and the $\mathbf{b}_4^-$ has a minus sign. Second,

$$\mathcal{L}_{\mathbf{a}_1} = \mathcal{H}_{\mathbf{a}_1} \otimes \overline{\mathcal{H}}_{\mathbf{a}_1} \tag{1342}$$

where $\mathcal{H}_{\mathbf{a}_1} = \mathcal{H}_{\mathbf{a}_1^+} \otimes \mathcal{H}_{\mathbf{a}_1^-}$. The Hilbert space $\mathcal{H}_{\mathbf{a}_1^+}$ is interpreted as pointing backward in time since the $\mathbf{a}_1^+$ is a subscript *and* it has a + sign. The Hilbert space $\mathcal{H}_{\mathbf{a}_1^-}$ is interpreted as pointing *forward* in time since it is a subscript and the $\mathbf{a}_1^-$ has a minus sign. Further, $\mathcal{P}_{\mathbf{x}_3}$ is a real vector space of dimension $N_{\mathbf{x}}$ and $\mathcal{P}^{\mathbf{y}_4}$ is a real vector space of dimension $N_{\mathbf{y}}$. In the case that there are no inputs or outputs we will omit the hat over the $A$ symbol.

| $(R)\!\rightarrow\!\!-\ \mathbf{x}$ and $(R)\!-\!\!\leftarrow\ \mathbf{x}$ | $\boldsymbol{R}^{\mathbf{x}_1}$ and $\boldsymbol{R}_{\mathbf{x}_1}$ | Flat distribution operators |
|---|---|---|
| $\mathbf{x}\rightarrow(x)\rightarrow\mathbf{x}$ | $B^{\mathbf{x}_2}_{\mathbf{x}_1}[x]$ | Readout box operator |
| $\mathbf{x}\nearrow(\hat{X})\nearrow\mathbf{x}$ and $\mathbf{x}\swarrow(\hat{X})\swarrow\mathbf{x}$ | $\hat{\boldsymbol{X}}^{\mathbf{x}_2}_{\mathbf{x}_1}$ and $\hat{\boldsymbol{X}}^{\mathbf{x}_1}_{\mathbf{x}_2}$ | Maximal operators |
| $(\hat{I})\!\rightarrow\!\!-\ \mathbf{a}$ and $(\hat{I})\!-\!\!\leftarrow\ \mathbf{a}$ | $\hat{\boldsymbol{I}}^{\mathbf{a}_1}$ and $\hat{\boldsymbol{I}}_{\mathbf{a}_1}$ | Ignore operators |

Table 7: Causally complex operators that play a special role. We provide diagrammatic and symbolic notation for these operators.

As in the simple case, we will see that circuit reality requires that operator tensors are Hermitian so the above object will actually live in the space

$$\mathcal{V}_{\mathbf{a}_1} \otimes \mathcal{V}_{\mathbf{c}_2^-} \otimes \mathcal{P}_{\mathbf{u}_3} \otimes \mathcal{V}^{\mathbf{b}_4} \otimes \mathcal{P}^{\mathbf{v}_5} \tag{1343}$$

where

$$\mathcal{V}^{\mathbf{b}_4} \subset \mathcal{L}^{\mathbf{b}_4} \qquad \text{and} \qquad \mathcal{V}_{\mathbf{a}_1} \subset \mathcal{L}_{\mathbf{a}_1} \tag{1344}$$

are the Hermitian subsets. This is proven in Sec. 59.3.

Special complex operators are shown in Table 7.

### 58.3 Complex operator networks and circuits

We can wire together complex operators to form complex operator networks and circuits. For example

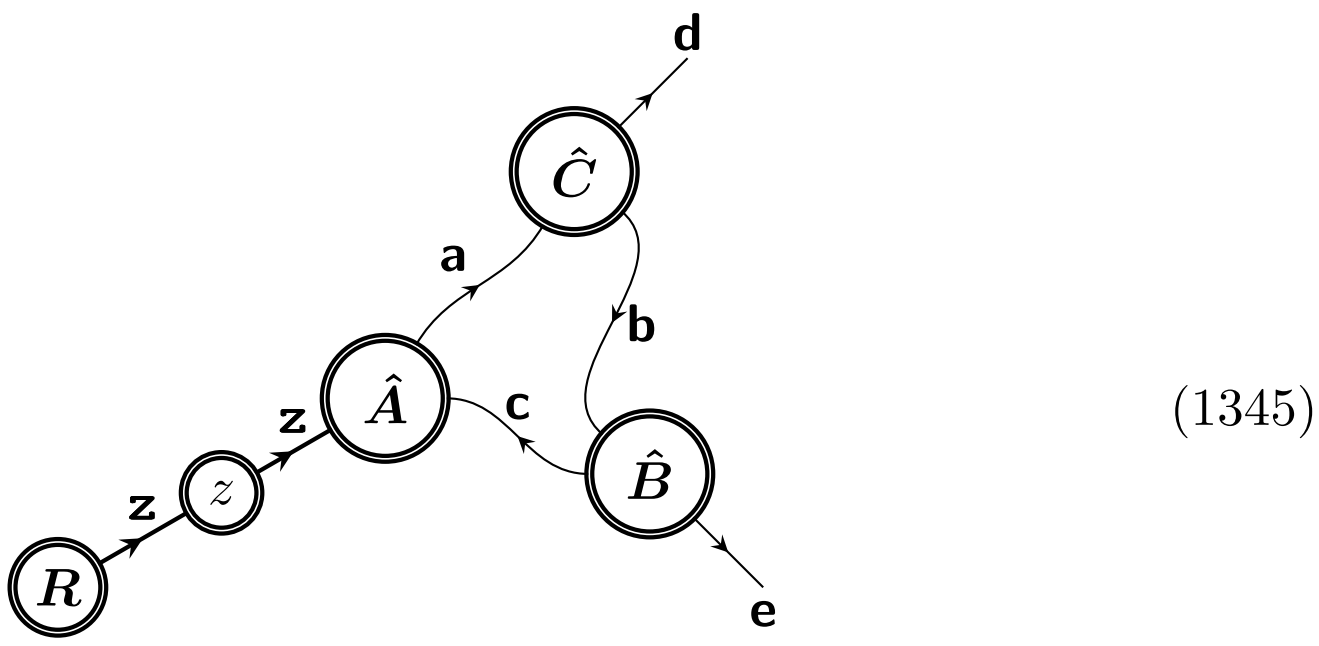

(1345)

and

(1346)

We simplify these expressions by taking the partial trace over the physical wires and the "partial dot product" (i.e. standard tensor summation) over the pointer wires as discussed in the simple case in Sec. 17. An operator network is equal to an operator in the space associated with the open wires.

One way to evaluate operator networks and circuits is to use fiducial expansions and duotensors as we will explain in Sec. 60.

## 59 Correspondence, circuit reality, and Hermiticity

### 59.1 Correspondence from operations to operators

We intend to use complex operator circuits to calculate the probability associated with circuits from the operational description. To this end we define a notion of correspondence

> **Correspondence.** We have established correspondence between operations and operators if, under this correspondence, any circuit composed of operations has probability equal to the corresponding operator circuit comprised of operators.

Recall that an operation consists of the operation itself and a causal diagram (as illustrated in (1036)) and it corresponds to a operator which consists of the operator itself plus the same causal diagram as for the operation (as illustrated in (1338)). If we have correspondence then, for example, we have

$$\text{prob}\left( \begin{array}{c} \mathsf{C} \xleftarrow{\mathbf{y}} (y) \xleftarrow{\mathbf{y}} \boldsymbol{R} \\ \mathsf{A} \xrightarrow{\mathbf{b}} \mathsf{C} \xrightarrow{\mathbf{a}} \mathsf{B} \xrightarrow{\mathbf{c}} \mathsf{A} \end{array} \right) = \begin{array}{c} \hat{C} \xleftarrow{\mathbf{y}} (\hat{y}) \xleftarrow{\mathbf{y}} \boldsymbol{R} \\ \hat{A} \xrightarrow{\mathbf{b}} \hat{C} \xrightarrow{\mathbf{a}} \hat{\boldsymbol{B}} \xrightarrow{\mathbf{c}} \hat{A} \end{array} \tag{1347}$$

where we use the convention that corresponding operators are represented by the same letter (e.g. $\mathsf{A}$ corresponds to $\hat{A}$, $\mathsf{B}$ corresponds to $\hat{\boldsymbol{B}}$, etc.). We will show how to explicitly set up such a correspondence in Sec. 61.1 using fiducial expansions.

## 59.2 Purity correspondence assumption

As in Sec. 16.1 we make the following assumption.

**Purity correspondence assumption.** We have the following

*Pure preparations.* Pure system preparations, $\mathsf{A}^{\mathsf{a}_1^+}$, correspond to rank one operators, $\hat{A}^{\mathsf{a}_1^+} = |A\rangle^{\mathsf{a}_1^+}\langle A|$ (where $|A\rangle^{\mathsf{a}_1^+}$ is not necessarily normalised to 1). Furthermore, all appropriately normalised rank one operators, $\hat{A}^{\mathsf{a}_1^+} = |A\rangle^{\mathsf{a}_1^+}\langle A|$, have a pure preparation, $\mathsf{A}^{\mathsf{a}_1^+}$, that they correspond to.

*Pure results.* Similarly, pure system results, $\mathsf{C}_{\mathsf{a}_1^+}$, correspond to rank one operators, $\hat{C}_{\mathsf{a}_1^+} = |C\rangle_{\mathsf{a}_1^+}\langle C|$ (where, again, $|C\rangle_{\mathsf{a}_1^+}$ is not necessarily normalised to 1). Furthermore, all appropriately normalised rank one operators, $\hat{C}_{\mathsf{a}_1^+} = |C\rangle_{\mathsf{a}_1^+}\langle C|$, have a pure result, $\mathsf{C}_{\mathsf{a}_1^+}$, that they correspond to.

This assumption helps us establish that circuit reality implies Hermiticity as we will see in Sec. 59.3 and is important in setting up the tester positivity condition.

## 59.3 Circuit reality implies Hermiticity

Consider a general operator

$$\mathbf{a} \leftarrow \hat{B} \xrightarrow{\mathbf{x}} \tag{1348}$$

This is general since $\mathbf{a}$ and $\mathbf{x}$ can be composite. We will write

$$\mathbf{a} \leftarrow \hat{B}(x) \quad := \quad \mathbf{a} \leftarrow \hat{B} \xrightarrow{\mathbf{x}} (\hat{x}) \xrightarrow{\mathbf{x}} \boldsymbol{R} \tag{1349}$$

Consider the operator circuit

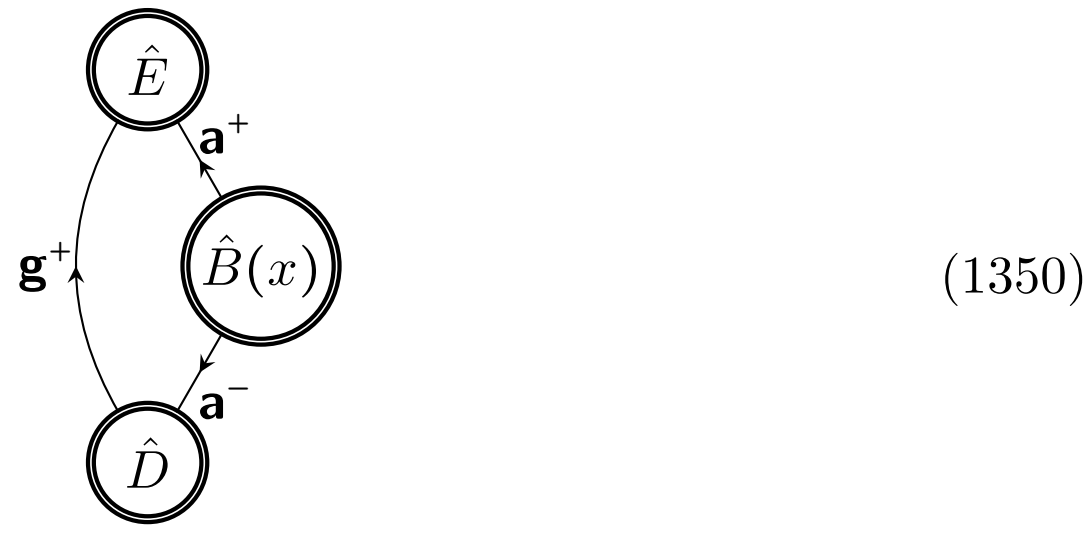

(1350)

where $\hat{D}$ and $\hat{E}$ are pure. We can write $\hat{B}(x)$ as the sum of a Hermitian and an anti-Hermitian part.

$$\hat{B} = \hat{B}_H + \hat{B}_A \quad \text{where} \quad \begin{array}{ll} B_H = \frac{1}{2}(B + B^\dagger) & \text{is Hermitian} \\ B_A = \frac{1}{2}(B - B^\dagger) & \text{is anti-Hermitian} \end{array} \tag{1351}$$

Then the operator circuit becomes

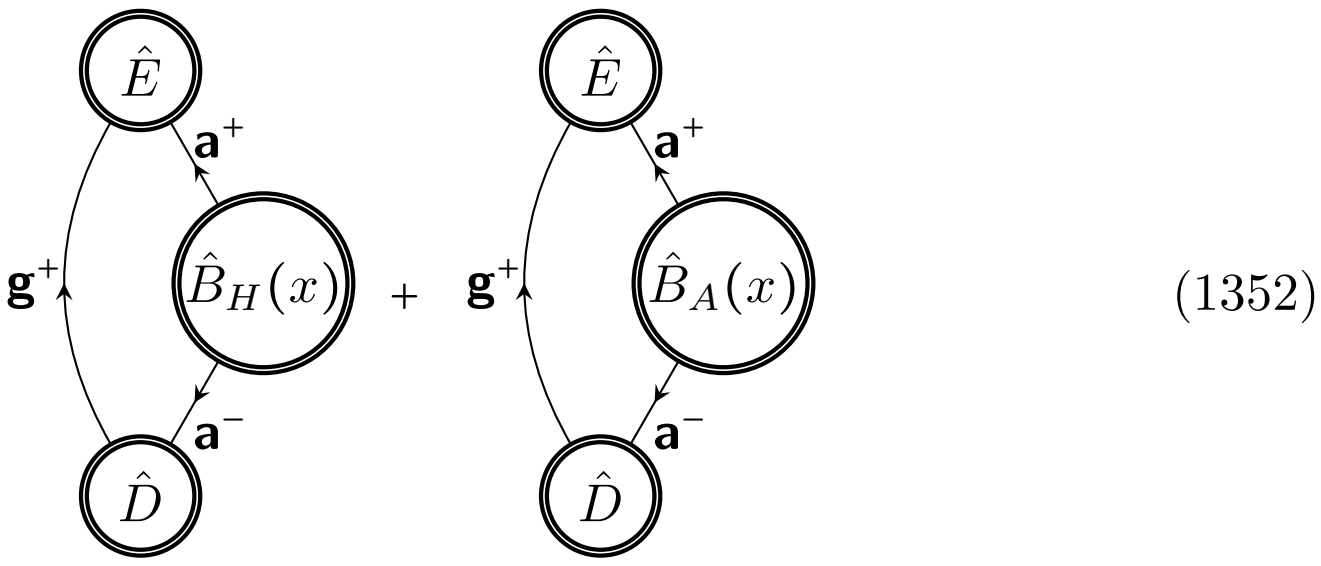

(1352)

Since $\hat{D}$ and $\hat{E}$ correspond to a preparation and a result respectively, the purity correspondence assumption tells us they are rank one Hermitian operators. We will show in Sec. 75.1 that the term on the left (with the Hermitian operator, $\hat{B}_H$) must be real. We show in Sec. 75.3 that the term on the right (with the anti-Hermitian operator, $\hat{B}_A$) will be pure imaginary for some choices of $\hat{D}$ and $\hat{E}$ unless $\hat{B}_A = 0$. Since circuit reality implies, under correspondence, that operator circuits are real. Consequently, the anti-Hermitian part of $\hat{B}$ must be equal to zero. Hence operators, $\hat{B}$, must be Hermitian.

# 60 Fiducial operators

We can define fiducial operators for the complex case as follows

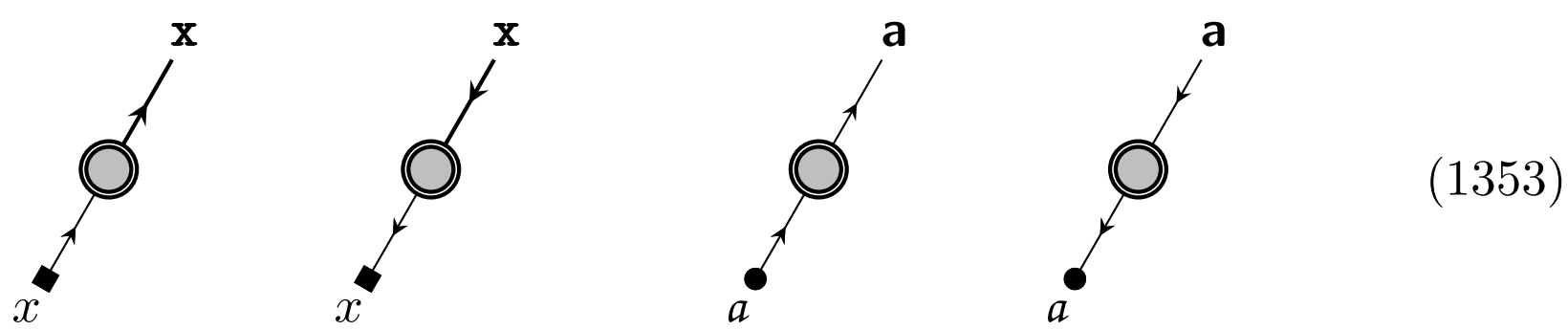

(1353)

The two pointer fiducial operators on the left each represent a spanning set of vectors in $\mathcal{P}^{\mathbf{x}_1}$ and $\mathcal{P}_{\mathbf{x}_1}$ respectively. The two physical fiducial operators on the right each represent a spanning set of operators in $\mathcal{V}^{\mathbf{a}_1}$ and $\mathcal{V}_{\mathbf{a}_1}$.

## 60.1 Fiducial matrices

We can form fiducial matrices as follows

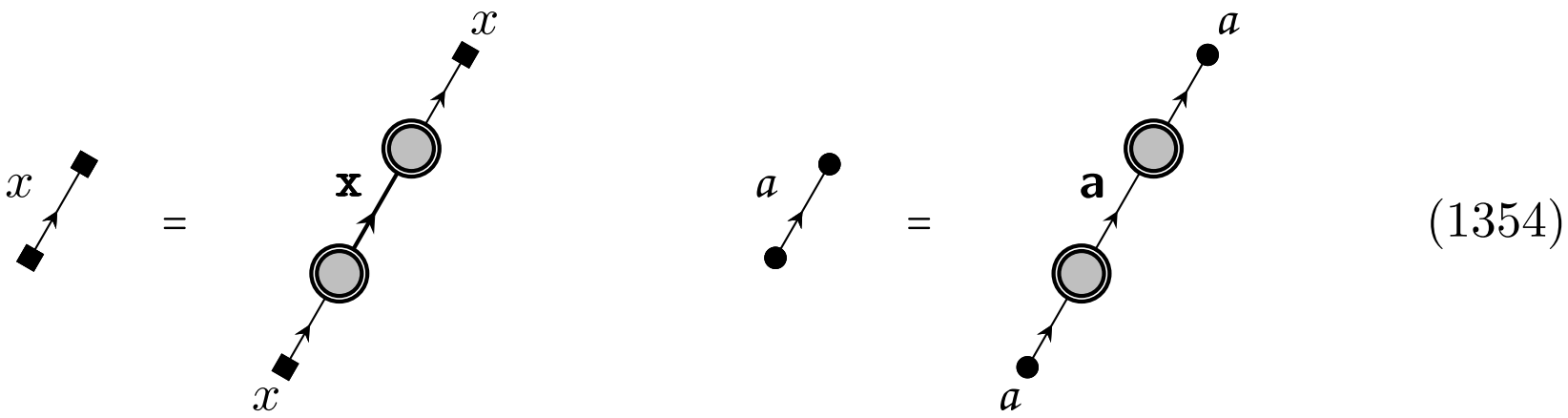

(1354)

Following the same reasoning as in Sec. 9.3 and Sec. 9.5, we have the following identities

(1355)

These mean we can introduce and delete pairs of black and white dots in either order.

## 60.2 Fiducial expansion of operator

We can expand any operator in terms of the fiducial operators as follows

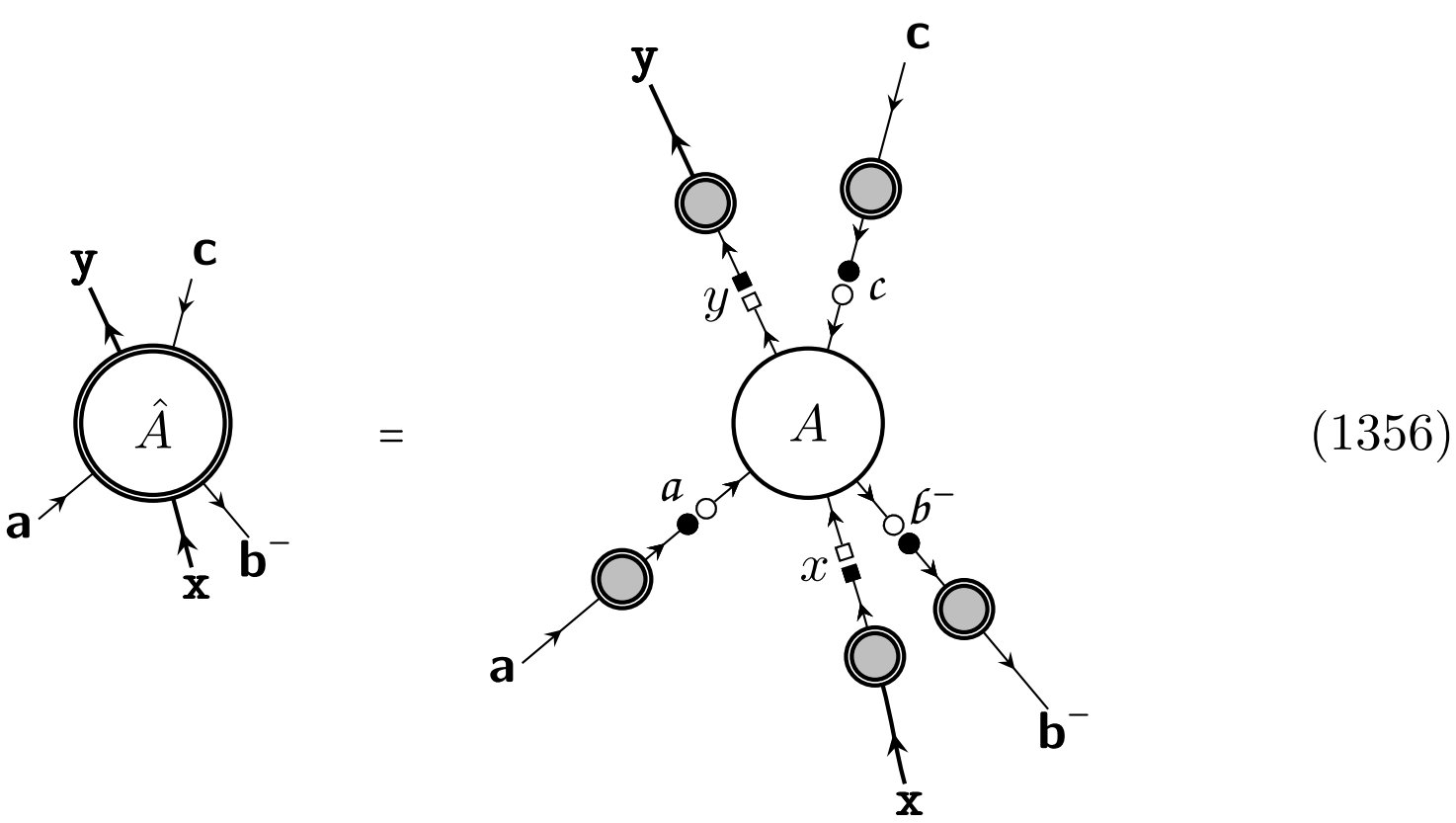

(1356)

(compare with (1256)).

## 60.3 Wire decomposition

We can decompose wires as follows

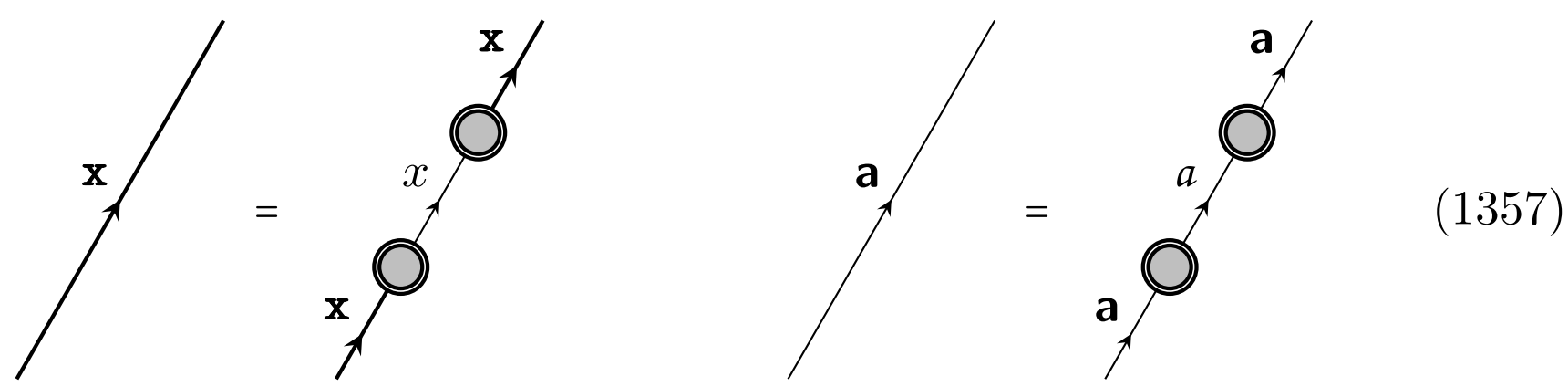

(1357)

The proof of this is analogous to the proof in Sec. 9.7.4.

## 60.4 Duotensor Calculations

We can do duotensor calculations to evaluate operator networks and operator circuits. For example, consider the operator circuit

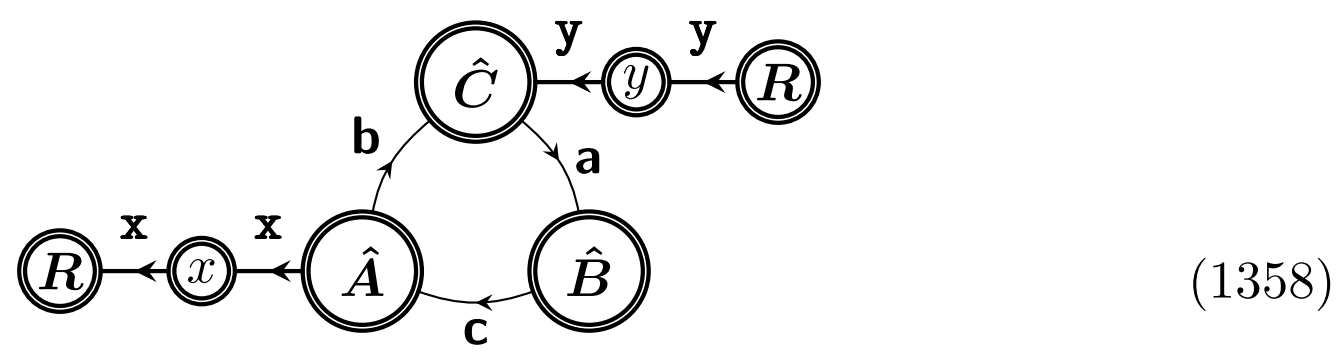

(1358)

We can calculate the probability for this circuit by turning it into an equivalent duotensor calculation. We can do this by substituting the local decomposition (1356) for each operator then substituting in the fiducial matrices from (1354). This is the same procedure as we followed in Sec. 22. This gives

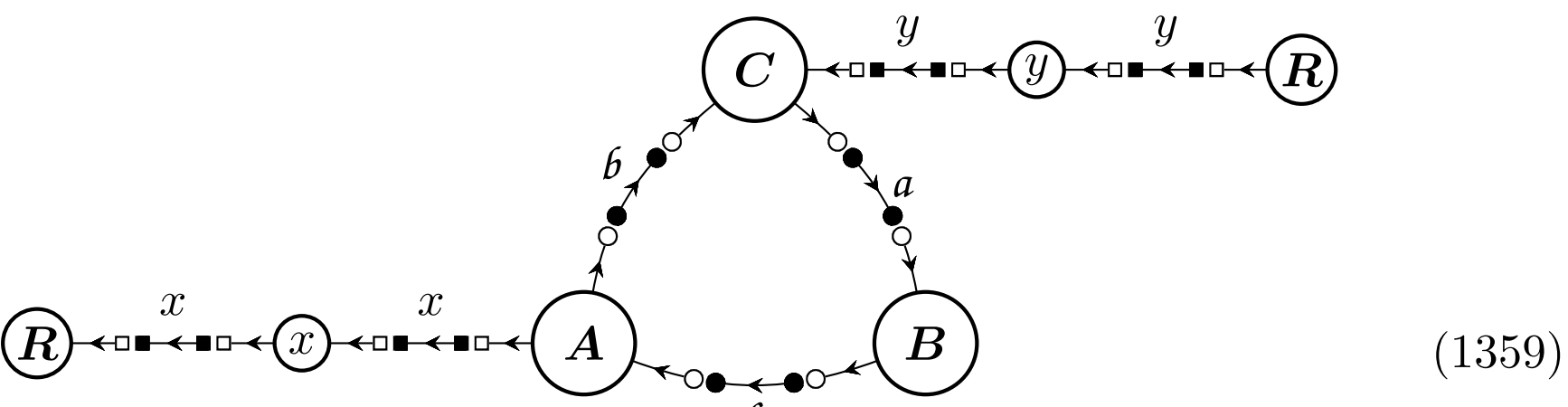

(1359)

We can, using (1355), replace black and white dot pairs by a line giving

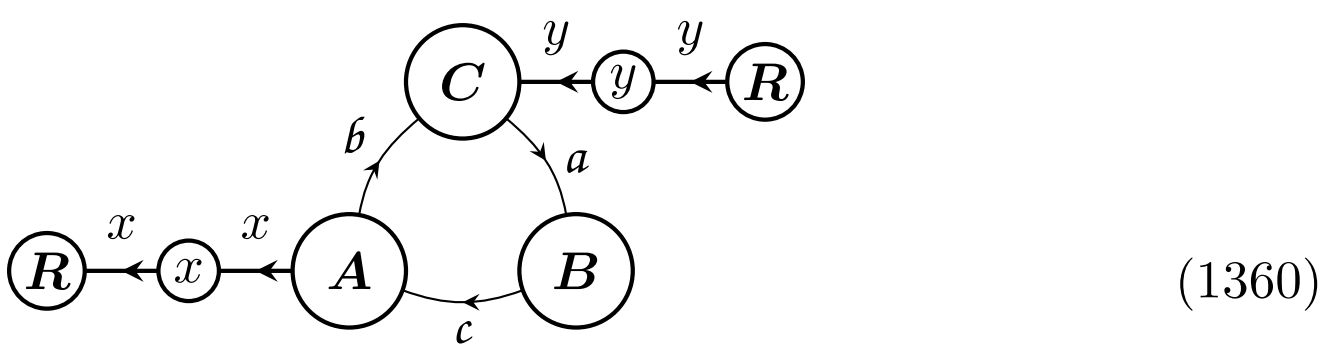

(1360)

To actually do this duotensor calculation we need to insert black and white dot pairs. We can do this in any way we like. We can get from (1358) to (1360) by using the decomposition of wires as given in (1357).

# 61 Correspondence

## 61.1 Correspondence theorem

We discussed correspondence between simple operations and simple operators in Sec. 19 and proved a correspondence theorem in Sec. 23.1. We can proceed in the same way in the complex case. Recall we have correspondence between operations and operators if, under this correspondence, we get the probability for a circuit when we replace all operations in the circuit with operators. We have the following correspondence theorem

> **Correspondence theorem.** If the fiducial matrices coming from operations are the same as those coming from operators, i.e.
>
> $$\equiv \qquad := \tag{1361}$$
>
> and
>
> $$\equiv \qquad := \tag{1362}$$
>
> then the operations correspond to operators when they have the same duotensor in their expansions.

This means that, when the fiducial matrices are equal as described above, the operation

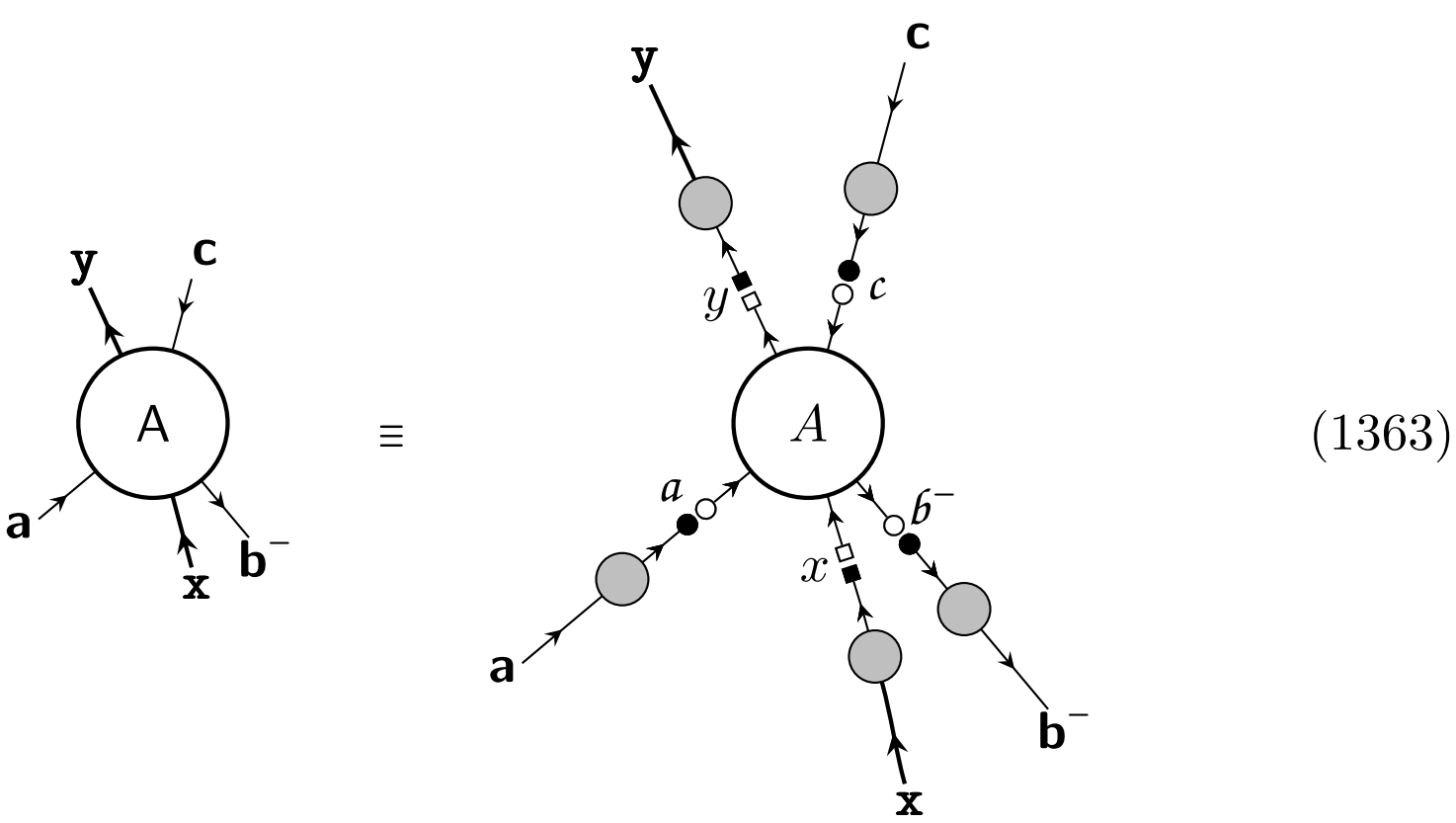

(1363)

corresponds to the operator

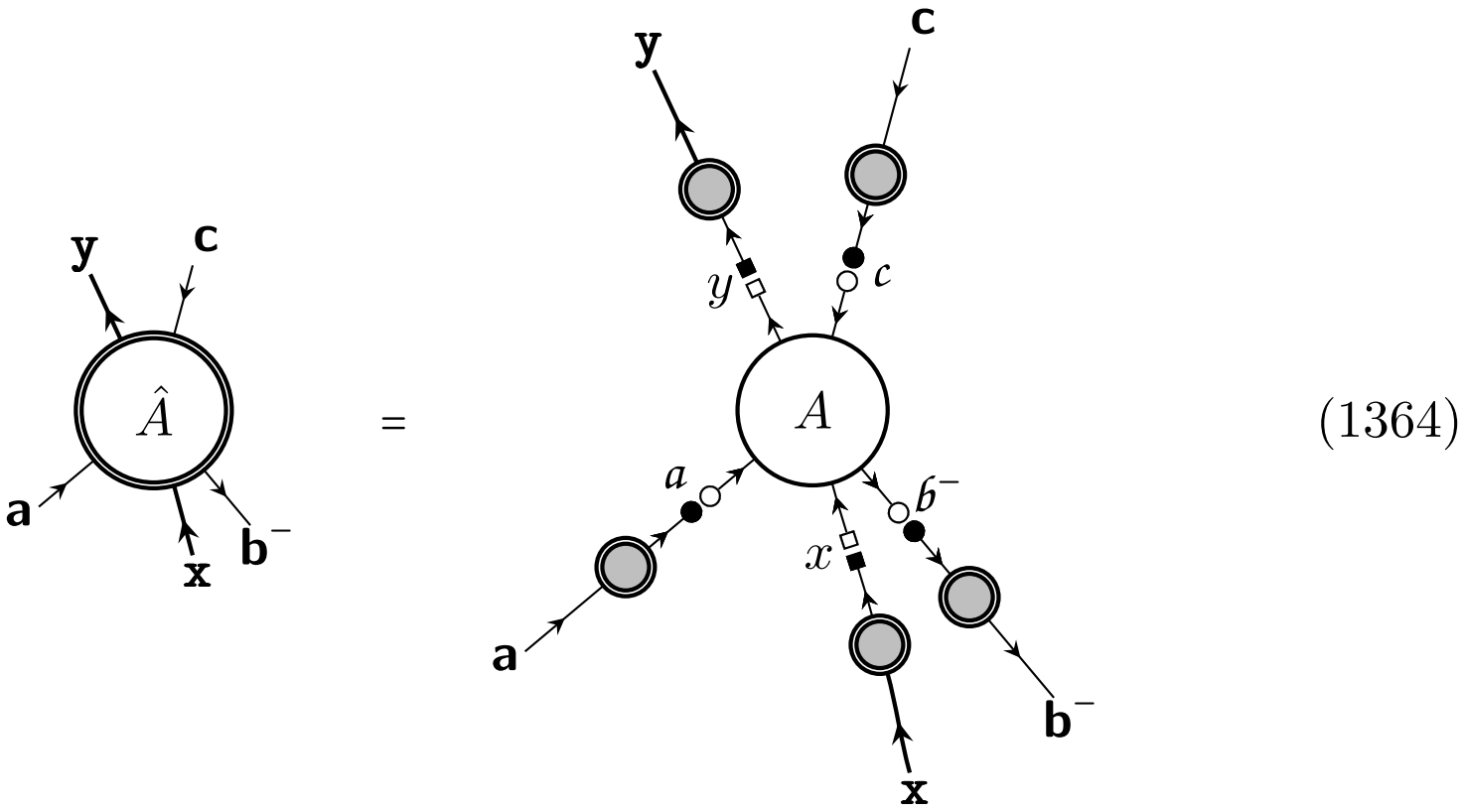

(1364)

(note that the duotensors are the same). The accompanying causal diagrams are the same. The proof of this is analogous to that in Sec. 23.1 (though it is obvious since the intermediate step in (1266) for the duotensor calculation with operations also appears as an intermediate step in (1359) in the calculation with operators).

## 61.2 Correspondence rule

In Sec. 23.3 we proved a correspondence rule between operation expressions and operator expressions for the causally simple theory. We have a correspondence rule of exactly the same kind for the causally complex theory. We repeat the correspondence rule here for completeness

> **Correspondence rule.** Equivalences and inequalities, which are respectively of the form
>
> $$\text{exprn}_1 \equiv \text{exprn}_2 \qquad\qquad \text{exprn}_3 \underset{T}{\leqq} \text{exprn}_4 \tag{1365}$$
>
> between operation expressions, go over to equations and inequalities of the form
>
> $$\widehat{exprn}_1 = \widehat{exprn}_2 \qquad\qquad \widehat{exprn}_3 \underset{T}{\leq} \widehat{exprn}_4 \tag{1366}$$
>
> between operator expressions under the correspondence
>
> $$\text{exprn}_i \Longmapsto \widehat{exprn}_i$$
>
> for $i = 1, 2, 3, 4$.

The proof of this correspondence rule in the complex case proceeds in the same way as in the simple case. This correspondence rule is important because we can take over equivalences/inequalties in the operational theory and use them to obtain equalities/inequalities in the operator theory.

## 62 Special operator tensors

Following the logic of Sec. 24 (for the simple case) we can use correspondence to write down explicit forms for the various special operator tensors. To do this we first choose a form for the pointer fiducials. A natural choice is

$$\mathbf{x} \;\; x \;=\; \beta^{\mathbf{x}} \begin{pmatrix} 1 & & \\ & 1 & \\ & & \ddots \end{pmatrix} \qquad\qquad x \;\; \mathbf{x} \;=\; \beta_{\mathbf{x}} \begin{pmatrix} 1 & & \\ & 1 & \\ & & \ddots \end{pmatrix} \tag{1367}$$

These follow by the same reasoning as (425) in the simple case. Here $\beta^{\mathbf{x}}$ and $\beta_{\mathbf{x}}$ are normalisation gauge parameters which are taken to be real with the normalisation condition

$$\beta^{\mathbf{x}}\beta_{\mathbf{x}} = \frac{1}{N_{\mathbf{x}}} \tag{1368}$$

With this normalisation condition, we get the correct pointer fiducial matrix.

We can use correspondence to obtain the operator tensor for the special operator tensors. We can write

$$\mathbf{x} \;\; R \;=\; \mathbf{x} \;\; x \;\; R \tag{1369}$$

Using (1367) and (1263) we obtain the equation on the left below

$$\mathbf{x} \;\; R \;=\; \beta^{\mathbf{x}} \begin{pmatrix} 1 \\ 1 \\ \vdots \\ 1 \end{pmatrix} \qquad \text{and} \qquad R \;\; \mathbf{x} \;=\; \beta_{\mathbf{x}} \begin{pmatrix} 1 \\ 1 \\ \vdots \\ 1 \end{pmatrix}^{\top} \tag{1370}$$

The equation on the right is obtained by similar techniques (we indicate the transpose with $\top$ as the arrow is pointing in). We have used correspondence since the duotensors used in these operator expansions are taken from the operation expansions.

The readout box can be obtained using the duotensor expansion

$$= \begin{pmatrix} 0 & & & & \\ & 0 & & & \\ & & \ddots & & \\ & & & 1 & \\ & & & & \ddots \end{pmatrix} \qquad (1371)$$

where we use the duotensor (1264) from the operation expansion and the normalisation condition (1368).

We expect maximal operators to satisfy

$$(1372)$$

by analogy with (1227). We can sandwich this between pointer fiducials giving

$$(1373)$$

We can obtain this if we associate maximal operators with projection operators onto some basis, $|x\rangle$ as follows

$$\xrightarrow[x]{\text{given}} \alpha^{\mathsf{x}}|x\rangle^{\mathsf{x}}\langle x| \qquad \xrightarrow[x]{\text{given}} \alpha_{\mathsf{x}}|x\rangle_{\mathsf{x}}\langle x| \qquad (1374)$$

where the normalisation gauge parameters, $\alpha^{\mathsf{x}}$ and $\alpha_{\mathsf{x}}$, are real (so that $X$ is Hermitian) and where, further, we impose the normalisation condition,

$$\alpha^{\mathsf{x}}\alpha_{\mathsf{x}} = \frac{1}{N_{\mathsf{x}}} \qquad (1375)$$

This normalisation condition guarantees we obtain (1373). We can introduce the normalisation gauge parameters, $\alpha_{\mathsf{x}^{\pm}}$, because they cancel when we insert these expressions into (1373).

Finally, ignore operators can be written as

$$\hat{\boldsymbol{I}}^{\mathsf{x}} = \hat{\boldsymbol{X}}^{\mathsf{x}} \text{ with } \boldsymbol{R} \xrightarrow{x} \qquad\qquad \hat{\boldsymbol{I}}_{\mathsf{x}} = \hat{\boldsymbol{X}}_{\mathsf{x}} \text{ with } \xrightarrow{x} \boldsymbol{R} \tag{1376}$$

Using (1263) and correspondence we obtain

$$\hat{\boldsymbol{I}}^{\mathsf{x}} = \boxed{\alpha^{\mathsf{x}}}\, \hat{\mathbb{1}}^{\mathsf{x}} \qquad\qquad \hat{\boldsymbol{I}}_{\mathsf{x}} = \boxed{\alpha_{\mathsf{x}}}\, \hat{\mathbb{1}}_{\mathsf{x}} \tag{1377}$$

where $\mathbb{1}^{\mathsf{x}_1} = \sum_n |x\rangle^{\mathsf{x}_1}\langle x|$ and $\mathbb{1}_{\mathsf{x}_1} = \sum_n |x\rangle_{\mathsf{x}_1}\langle x|$ are identity operators.

# 63 Physicality conditions

The physicality conditions for complex operational probability theory are that complex operations satisfy tester positivity and $t$-causality (so double causality $t$ =TS, forward causality for $t = TF$, and backward causality for $t$ =TB). We will now discuss how to use correspondence to obtain physicality conditions on complex operators.

From the tester positivity condition on operations in (1134) and the purity correspondence assumption in Sec. 59.2 we obtain the tester positivity condition on complex operators

$$0 \underset{T}{\leq} {}_{\mathsf{a}}\hat{B}^{\mathsf{x}} \tag{1378}$$

where $\underset{T}{\leq}$ means we have positivity with respect to any tester of the form

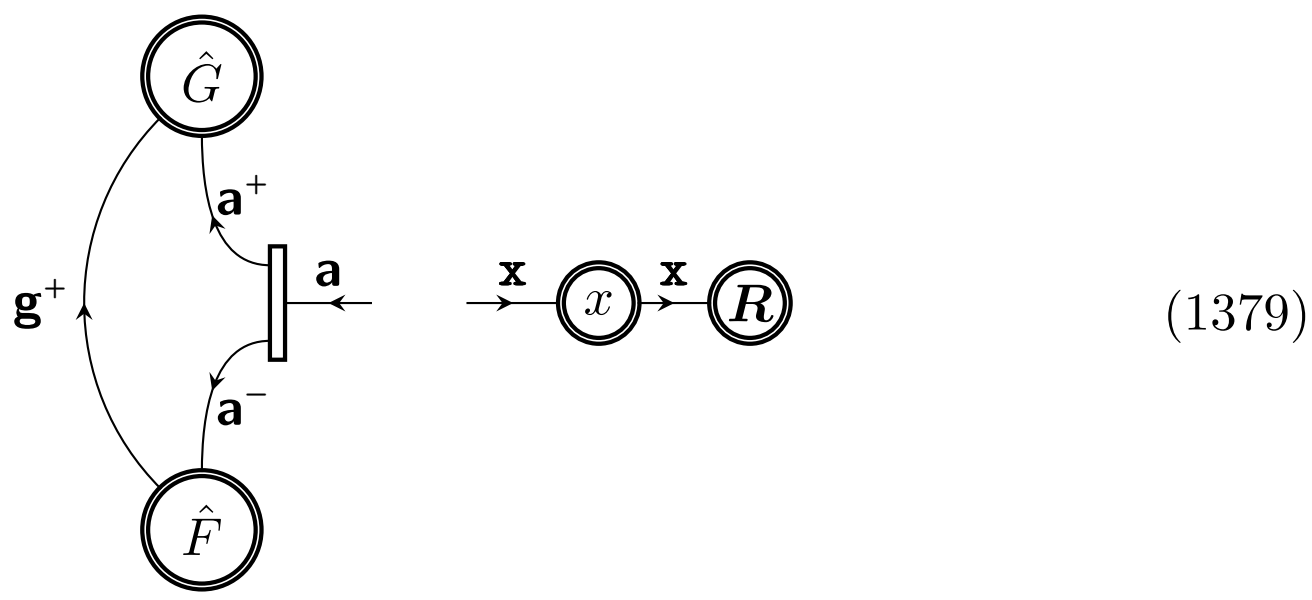

(1379)

where $\hat{G}$ and $\hat{F}$ are rank one projectors. This condition can be given equivalently in two further forms (this will be discussed in Sec. 76).

For the time symmetric case ($t = TS$) we have double causality conditions coming from correspondence with the double causality conditions (1162, 1163) discussed in Sec. 49.4.5. Consider a deterministic complex operator $\hat{\boldsymbol{B}}$ where

$$\text{with c.d.} \qquad (1380)$$

then we have the *forward causality* condition

$$= \qquad (1381)$$

and the *backward causality* condition

$$= \qquad (1382)$$

where $\hat{\boldsymbol{B}}[p^{\pm}]$ also satisfy these double causality conditions (1381, 1382) in place of **B** (this follows under correspondence from the residua causality theorem in Sec. 49.4.7).

There are also $t$-causality conditions on operators for time forward and time backward theories. These can be obtained by correspondence with the time forward case for operations (discussed in Sec. 53.4) and by correspondence with the time backward case for operations (discussed in Sec. 54.4). We will not study these cases in any depth in this book.

# 64 Axioms for Complex Operational Quantum Theory

In Sec. 27 we considered axioms for Simple Operational Quantum Theory. The axioms for COQT have the same form as the axioms for the SOQT case (though we state Axiom 0 in a different way). They are as follows

**Axioms for Complex Operational Quantum Theory.**

0 The causal diagram for any circuit is empty.

1 All realisable operations are physical.

2 Every realisable operation has a physical operator corresponding to it.

3 Every physical operator has a realisable operation corresponding to it.

Compare with the axioms for Simple Operational Quantum Theory in Sec. 27. Axiom 0 here is a way to impose that circuits do not have closed time like loops since any such structure will not simplify when we fuse the causal diagrams of the complex operations that comprise the circuit. Physicality is understood to be with respect to the temporal frame we are working in.

- *In the* $t = TS$ *temporal frame,* physicality for operations means we have tester positivity (1134) and the double causality conditions (1162, 1163). For operators we have the corresponding conditions.

- *In the* $t = TF$ *temporal frame,* physicality for operations means we have $T$-positivity and the double causality conditions as applicable to $\overline{B}$ as discussed in Sec. 53.3 and Sec. 53.4. For operators we have corresponding conditions.

- *In the* $t = TB$ *temporal frame,* physicality means we have $T$-positivity and the double causality conditions as applicable to $\underline{B}$ as discussed in Sec. 54.3 and Sec. 54.4 with corresponding conditions in the operator case.

In the simple case we were able to prove positivity theorems (that tester positivity is equivalent to twofold positivity and twist positivity). We were also able to prove a dilation theorem. To prove these theorems we set up the theory of simple Hilbert objects in Part III. We want to prove some similar theorems in the complex case and so next we will set up the theory of *complex Hilbert objects*.

# Part VI

# Complex Hilbert Objects

## 65 A new diagrammatic notation

### 65.1 Toggle and flip

In Part III we introduced a diagrammatic notation for elements of Hilbert spaces to be used in studying causally simple operator tensors. This was similar to the notation used by Coecke and Kissinger [2017] and others in that inputs go into the bottom of a box while outputs come out of the top of a box. In this notation we are able to implement the normal vertical adjoint by vertically flipping Hilbert objects (see Sec. 32.4). Further, we can implement the normal transpose by using cups and caps on a left or right Hilbert object. Then (following 574) we can think of "sliding" boxes along wires and through a cup or cap leading them to be inverted (see Sec. 32.3) then using the yanking equation to remove unnecessary cups and caps in the wires. One of the selling points of the pictorial approach of Coecke et al. is that these important mathematical transformations (like taking the transpose) have a tactile realisation.

In the causally complex case inputs and outputs are combined into single wires which have an arrow (to establish a convention). These wires attach to circles and, as we will see, the curved part of semicircles. There are no tops and bottoms of boxes and so we do not have the same kind of vertically oriented structure we have in the simple case (though we will maintain a left/right oriented structure). To interchange inputs and outputs we reverse arrows (since this reverses the convention) instead of flipping vertically. We will denote the reVerse transformation by $V$ so we can continue to use our previous notation for conjupositions. If we also conjugate then we implement the normal reverse adjoint ($\overline{V}$).

(1383)

This is the Hilbert square for complex left and right objects. We can imagine that the arrows and small square boxes are white on one side and black on the other. When we "flip" them over we reveal the colour on the other side.

Further, we can imagine that the small squares in the corners represent a switch that can be "toggled" up and down. When this is toggled the direction of the arrows is changed while maintaining the colour. In this way we have a tactile undertanding of the transformations wherein *flip* corresponds to $\overline{H}$ (normal horizontal adjoint) and *toggle* corresponds to the $\overline{V}$ (normal reVerse adjoint).

We will begin in the normal picture (associated with the normal conjuposition group). Later we will study the natural picture (associated with the natural conjuposition group).

## 65.2 Doubling up notation

We will employ the following doubling up notation

$$\mathsf{a} \uparrow \; = \; \mathsf{a} \blacktriangle \triangle \mathsf{a} \tag{1384}$$

This is similar to the simple case (456).

## 65.3 Basic objects

As in the simple case, we have four basic objects

$$\begin{array}{cc} B \;\; \in \mathcal{H}_{\mathsf{a}_1} & B \;\; \in \overline{\mathcal{H}}_{\mathsf{a}_1} \\ A \;\; \in \mathcal{H}^{\mathsf{a}_1} & A \;\; \in \overline{\mathcal{H}}^{\mathsf{a}_1} \end{array} \tag{1385}$$

with

$$\begin{array}{cc} \mathcal{H}_{\mathsf{a}_1} = \mathcal{H}_{\mathsf{a}_1^+} \otimes \mathcal{H}_{\mathsf{a}_1^-} & \overline{\mathcal{H}}_{\mathsf{a}_1} = \overline{\mathcal{H}}_{\mathsf{a}_1^+} \otimes \overline{\mathcal{H}}_{\mathsf{a}_1^-} \\ \mathcal{H}^{\mathsf{a}_1} = \mathcal{H}^{\mathsf{a}_1^+} \otimes \mathcal{H}^{\mathsf{a}_1^-} & \overline{\mathcal{H}}^{\mathsf{a}_1} = \overline{\mathcal{H}}^{\mathsf{a}_1^+} \otimes \overline{\mathcal{H}}^{\mathsf{a}_1^-} \end{array} \tag{1386}$$

We call the objects on the left of (1385) *left Hilbert space objects* whilst the objects on the right are *right Hilbert space objects*. The interpretation of the Hilbert spaces associated with $\mathsf{a}^{\pm}$ is given in Table 8. The main idea is that if $\mathsf{a}^+$ is a superscript then it is associated with forward in time while, if it is in a subscript, it is associated with backward in time (and this is reversed for $\mathsf{a}^-$). Also shown in the table is the old (antiquated) bra/ket notation for each of these spaces. The antiquated symbolic notation is useful to consider because of its familiarity. However, if we want to use symbolic notation for left and right

| Hilbert space | old notation | new notation | time direction | Hilbert space | old notation | new notation |
|---|---|---|---|---|---|---|
| $\mathcal{H}^{\mathsf{a}_1^+}$ | $\vert C\rangle^{\mathsf{a}_1^+}$ | ${}^{\mathsf{a}_1^+}\langle\!\langle C\vert$ | forward | $\bar{\mathcal{H}}^{\mathsf{a}_1^+}$ | ${}^{\mathsf{a}_1^+}\langle C\vert$ | $\vert C\rangle\!\rangle^{\mathsf{a}_1^+}$ |
| $\mathcal{H}^{\mathsf{a}_1^-}$ | ${}^{\mathsf{a}_1^-}\langle D\vert$ | ${}^{\mathsf{a}_1^-}\langle\!\langle D\vert$ | backward | $\bar{\mathcal{H}}^{\mathsf{a}_1^-}$ | $\vert D\rangle^{\mathsf{a}_1^-}$ | $\vert D\rangle\!\rangle^{\mathsf{a}_1^-}$ |
| $\mathcal{H}_{\mathsf{a}_1^+}$ | ${}_{\mathsf{a}_1^+}\langle E\vert$ | ${}_{\mathsf{a}_1^+}\langle\!\langle E\vert$ | backward | $\bar{\mathcal{H}}_{\mathsf{a}_1^+}$ | $\vert E\rangle_{\mathsf{a}_1^+}$ | $\vert E\rangle\!\rangle_{\mathsf{a}_1^+}$ |
| $\mathcal{H}_{\mathsf{a}_1^-}$ | $\vert F\rangle_{\mathsf{a}_1^-}$ | ${}_{\mathsf{a}_1^-}\langle\!\langle F\vert$ | forward | $\bar{\mathcal{H}}_{\mathsf{a}_1^-}$ | ${}_{\mathsf{a}_1^-}\langle F\vert$ | $\vert F\rangle\!\rangle_{\mathsf{a}_1^-}$ |

Table 8: This table shows the time direction (forward or backward) interpretation of the Hilbert spaces associated with $\mathsf{a}_1^\pm$. It also shows elements of these Hilbert spaces in the old (antiquated) bra/ket symbolic notation and the new symbolic notation

Hilbert space objects then we need a new symbolic notation. A possible notation which is in the same spirit as the diagrammatic notation is the following

$$
\begin{aligned}
&{}_{\mathsf{a}_1}\langle\!\langle B| \in \mathcal{H}_{\mathsf{a}_1} \qquad\qquad && |B\rangle\!\rangle_{\mathsf{a}_1} \in \bar{\mathcal{H}}_{\mathsf{a}_1} \\
&{}^{\mathsf{a}_1}\langle\!\langle A| \in \mathcal{H}^{\mathsf{a}_1} \qquad\qquad && |A\rangle\!\rangle^{\mathsf{a}_1} \in \bar{\mathcal{H}}^{\mathsf{a}_1}
\end{aligned}
\tag{1387}
$$

This notation models the diagrammatic notation. The objects in (1385) can be entangled between the $\mathsf{a}_1^+$ and $\mathsf{a}_1^-$ parts. For example, ${}_{\mathsf{a}_1}\langle\!\langle B|$ can be written (in the old notation) as

$$
{}_{\mathsf{a}_1}\langle\!\langle B| = \sum_{a^+a^-} B^{a^+a^-} \; {}_{\mathsf{a}_1^+}\langle a^+| \otimes |a^-\rangle_{\mathsf{a}_1^-}
\tag{1388}
$$

where $|a^\pm\rangle_{\mathsf{a}_1^\pm}$ are orthonormal basis elements. This demonstrates the need for new notation. Similarly, ${}^{\mathsf{a}_1}\langle\!\langle A|$ can be written as

$$
{}^{\mathsf{a}_1}\langle\!\langle A| = \sum_{a^+a^-} A_{a^+a^-} |a^+\rangle^{\mathsf{a}_1^+} \otimes {}^{\mathsf{a}_1^-}\langle a^-|
\tag{1389}
$$

We will, for the most part, use diagrammatic notation for left and right Hilbert objects since it is more versatile and the interpretation is more immediate.

## 65.4 Scalar products

We can take the scalar product of two left Hilbert space objects (or two right Hilbert space objects) when the arrows allow. For the objects in (1385) we can

form the scalar products

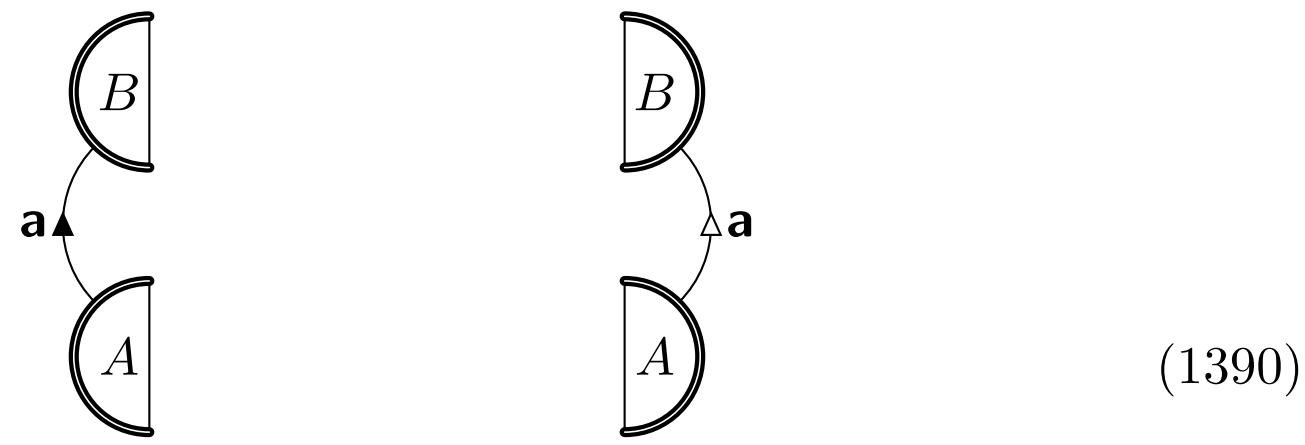

(1390)

These each evaluate to an element of $\mathbb{C}$. We can do this calculation, in symbolic notation, by using (1388, 1389) above. We obtain

$$\sum_{a^+,a^-} A_{a^+a^-} B^{a^+A^-} \tag{1391}$$

where we have used standard bra-ket products

$$_{\mathsf{a}_1^+}\langle a^+|a'^+\rangle^{\mathsf{a}_1^+} = \delta_{a^+a'^+} \qquad\qquad ^{\mathsf{a}_1^-}\langle a^-|a'^-\rangle^{\mathsf{a}_1^-} = \delta_{a^-a'^-} \tag{1392}$$

We will not usually use this antiquated notation - we do so here to establish the connection with the standard presentations. We will have more to say about taking scalar products in Sec. 67.2 below.

## 65.5 Sums over Hilbert space elements

We can take sums over Hilbert space elements. For example

(1393)

(which is an element of the left Hilbert space $\mathcal{H}_{\mathsf{a}_1} \otimes \mathcal{H}^{\mathsf{c}_2} \otimes \mathcal{H}^{\mathsf{a}_3}$ with a label $r$). This has horizontal adjoint

(1394)

(this is an element of the right Hilbert space $\overline{\mathcal{H}}_{\mathsf{a}_1} \otimes \overline{\mathcal{H}}^{\mathsf{c}_2} \otimes \overline{\mathcal{H}}^{\mathsf{a}_3}$ with label $r$). The full name for the horizontal adjoint is the *normal horizontal adjoint* as it is an element of the normal conjuposition group as we will see in Sec. 69. Note that the label wires ($r$, $k$, $l$, $n$ in the above diagram) have an arrow on them. This will be essential in what follows. The idea is that each label has a + part (which we can think of as being associated with the direction of the arrow) and a − part (which we can think of as being associated with the opposite direction to the arrow). We will write, for example, $r = r^+r^-$ (where this is shorthand for $r = (r^+, r^-)$).

## 65.6 Left and right Hilbert circuits

A left (right) Hilbert circuit is formed by wiring together left (right) Hilbert elements so there are no system wires left open (though there can be open label wires). For example

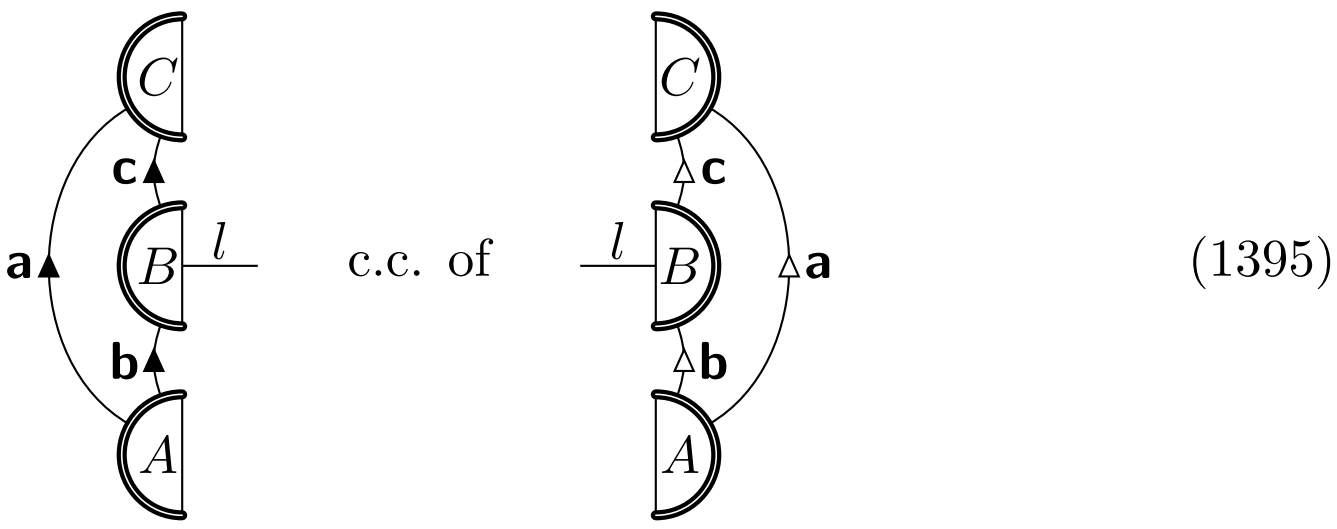

are left and right Hilbert circuits.

## 65.7 General Hilbert objects

In the simple case we defined general Hilbert objects as objects that are composed from left and right Hilbert objects (see Sec. 28.7). We can do the same thing for the causally complex case. An example of a general Hilbert object is the following

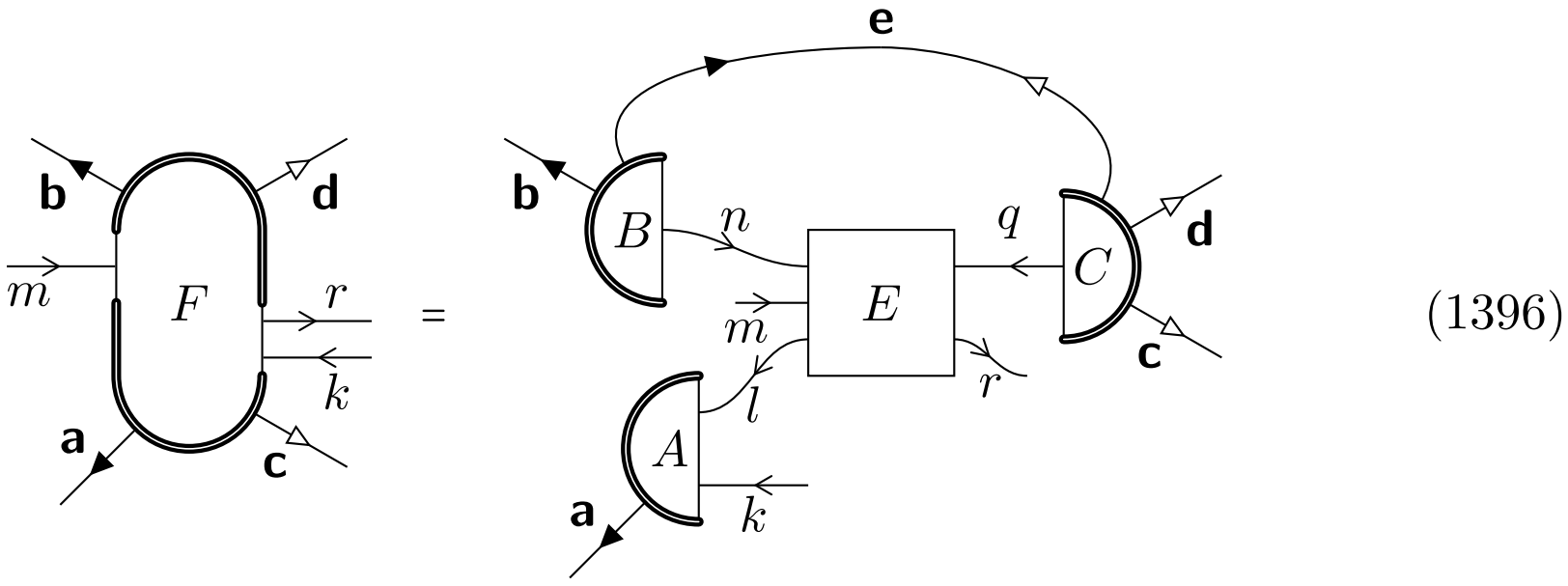

This example has a black and white arrow meeting on wire $\mathsf{e}$. This is allowed and will be discussed shortly. Left Hilbert objects are special cases of general Hilbert objects (as are right Hilbert objects). If a general Hilbert object has both left and right wires coming out of it (like the above example) then we will call it a *hybrid Hilbert object*. Operator tensors are a special case of hybrid Hilbert objects where left and right wires are matched.

# 66 Conjuposition of hypermatrices

In (1393) we had a hypermatrix

$$k \quad n \quad \beta \quad r \quad l \tag{1397}$$

This is different from the hypermatrices we studied in the simple case in Sec. 29.1 because we have arrows on the wires and the labels are tuples with + and − parts. Here we will set up the conjuposition group for these objects. The important difference is that we do not vertically flip these hypermatrices. Rather, we reverse the arrows. This has the effect of interchanging the + and − parts the labels. We will use a $V$ to denote reVersing the arrows to emphasize the similarity with the simple case.

On a point of nomenclature, it is sometimes clumsy to refer to these objects as hypermatrices and so we may refer to them simply as "matrices". This is particularly when the pertinent properties we are using are matrix like properties.

We need notation for reversing arrows. We will write

$$a \;:=\; a \;\boxed{\delta}\; a \qquad\qquad a \;:=\; a \;\boxed{\delta}\; a \tag{1398}$$

where these $\delta$ matrices are equal to 1 if both label wires are equal, and 0 otherwise (so, for example, if we have $a = (3, 2)$ on both sides of the box, then the matrix returns a 1). Consider the equation on the left of (1398). For this equation, if we have $a = (3, 2)$ then the "3" is associated with the rightward direction on the left side but with the leftward direction on the right side. We can think of the cross that appears in the equation on the left as resulting from pushing the two in pointing arrowheads together till they meet. We can think of the diamond that appears in the equation on the right in a similar fashion.

Now consider the hypermatrix

$$a \;\boxed{M}\; b \tag{1399}$$

We include a small black dot in the corner so we can track the conjupositions. We define the horizontal transpose, reVerse transpose, and transpose as follows

| $I$ | identity | $\overline{I}$ | conjugate |
|---|---|---|---|
| $H$ | horizontal transpose | $\overline{H}$ | horizontal adjoint |
| $V$ | reVerse transpose | $\overline{V}$ | reVerse adjoint |
| $T$ | transpose | $\overline{T}$ | adjoint |

Table 9: The conjuposition transformations on a hypermatrix in the causally complex case.

$:=$ horizontal transpose $H$ (1400)

$:=$ reVerse transpose $V$ (1401)

$:=$ transpose $T$ (1402)

If we include the identity transformation, $I$, then we have the group of non-conjugating transformations $\{I, H, V, T\}$. This is a subset of the full set, $\mathcal{C}$, of conjupositions. To get to this full set we need to allow conjugations. This is where we take the complex conjugate of all the elements of the matrix, $M$. We denote the resulting matrix by $\overline{M}$. This gives us the full set of eight hypermatrices

(1403)

(1404)

(1405)

(1406)

The conjugate transformation, $\overline{I}$, transforms between the left and right hand sides here. This full set of eight hypermatrices are related by the full conjuposition group

$$\mathcal{C} = \{I, \overline{I}, H, \overline{H}, V, \overline{V}, T, \overline{T}\} \tag{1407}$$

We can define the same subgroups, $\mathcal{S}_{BI}$, $\mathcal{S}_{side}$, and $\mathcal{S}_{\text{nonconjugating}}$ as discussed in the simple case in Sec. 29.1.

It is important to note that, unlike in the simple case, we do not vertically flip matrices. When we perform the reVerse transformation of

(1408)

we obtain

(1409)

We can think of the small black dot "toggling" the arrow reversal when it is moved vertically. We do, however, flip horizontally. The horizontal transpose of (1408) is

(1410)

Here we do simply flip the diagram over (horizontally).

If we wire together a bunch of hypermatrices forming a hypermatrix network then, just as in the simple case studied in Sec. 29.2, a given conjuposition of the whole network is given by taking the same conjuposition of each component hypermatrix in the network and wiring these together in the appropriate way. The proof of this is along the same lines as in the simple case but now, when we have a reVerse component we simply reverse the arrows rather than performing a vertical flip (so this part of the proof is actually graphically simpler).

# 67 Orthonormal bases and associated objects

## 67.1 Orthonormal bases

Orthonormal basis sets are represented as follows

(1411)

where $a = a^+a^-$ (by which we mean $a = (a^+, a^-)$) and $a^\pm = 1$ to $N_{\mathsf{a}^\pm}$ (so $a$ takes $N_{\mathsf{a}} = N_{\mathsf{a}^+}N_{\mathsf{a}^-}$ values). We have included small black and white squares. These will play an important role when we come to consider conjupositions of Hilbert objects (as already discussed a little in Sec. 65.1).

We have the following relations

$$(1412)$$

which enforce orthonormality.

These orthonormal basis elements can be chosen to be any set of orthogonal elements of the corresponding Hilbert space. For example, the basis elements on the bottom left of (1411) are elements of $\mathcal{H}^{\mathsf{a}_1} = \mathcal{H}^{\mathsf{a}_1} \otimes \mathcal{H}_{\mathsf{a}_1}$. A natural choice is that they have product form $|a^+\rangle^{\mathsf{a}_1}{}_{\mathsf{a}_1} \otimes \langle a_-|$ (so then $a = a^+a^-$ with $a^\pm = 1$ to $N_{\mathsf{a}^\pm}$). They could, however, be chosen so that some or all of them are entangled. Then we would have entanglement between forward and backward going systems. We can go between the case where the basis elements have product form and the case where they have entangled form by means of a basis transformation. We will discuss how to implement basis transformations in Sec. 67.3

## 67.2 Identity decompositions

There are four decompositions of the identity. These are

$$(1413)$$

(which are analogous to (506)) and

$$(1414)$$

(which are analogous to the cap and cup equations in (508, 509)). Interestingly, (1414) allow black and white arrows to meet in opposing directions on a line. We will later see how to use flip and toggle to remove such opposing arrows (or, indeed, to insert opposing arrows).

In Sec. 65.4 scalar products we saw how to take the scalar product between two left Hilbert space objects or two right Hilbert space objects. We can use (1414) to take the scalar product between a left and a right Hilbert space object.

For example

(1415)

This is analogous to using the a cap or cup in the causally simple case to take the scalar product between a left and a right Hilbert space object.

## 67.3 Basis transformations

We can implement a change of basis using unitary matrices as follows.

(1416)

where the unitary matrices satisfies

(1417)

(1418)

These unitarity properties may appear a little strange. To gain insight into them let us separate out the + and − wires and (just for the moment) use the notation we used for hypermatrices in the simple case. Then the first unitarity property (in (1417)) becomes

(1419)

Now, let us consider the special case where our basis is a product state between forward and backward moving systems (as discussed at the end of Sec. 67.1). Further, let us consider the case where the unitary factorises between forward and backward parts such that

(1420)

In this special case the unitarity defining property in (1419) becomes

(1421)

This is what we expect in this special case - that we get a unitary basis change on the + and − parts separately. Similar remarks apply, of course, to the second unitarity property (in (1418)).

With the unitarity properties in (1417, 1418) we can see immediately that the decomposition of the identity properties in Sec. 67.2 are basis independent. To see this note that, using (1414) and the unitarity property in (1417), we obtain

(1422)

The left equation follows immediately. To get the right equation we use

(1423)

which follows from (1417) on reversing the direction of the arrows (taking $V$ on each component). Similarly, we can show that

(1424)

follow from (1413) and (after some suitable manipulations) the first unitarity property (1417). We can go backwards from the shaded to the unshaded decompositions of the identity using the second unitarity property (1418).

It is also easy to show that the new (shaded) basis satisfy orthonormality conditions

(1425)

follow from orthonormality in the unshaded basis (the properties in (1412)) and the second unitarity property (1418). To see this note that (1418) can be written in either of the following forms

$$(1426)$$

(we leave the proof of this exercise in hypermatrix manipulation to the reader). Furthermore, the other unitarity property (in (1417)) can be used to show that orthonormality in the unshaded basis follows from orthonormality in the shaded basis.

## 67.4 Tugging equations

In Sec. 30.3 we considered the yanking equations. The analogous equations here take a different form. We will call them *tugging* equations. There are two tugging equations

$$(1427)$$

We could imagine building a physical model from string and cardboard arrowheads painted black on one side and white on the other. We can join three arrow heads with the string all pointing the same way and all showing the same colour. Then we could turn the arrowhead in the centre over about an axis perpendicular to the string so it is now pointing the opposite direction and showing the opposite colour. If we tug gently on the ends this central arrowhead will turn back over revealing its original colour and pointing, again, in the original direction. Three arrow heads all of the same colour pointing in the same direction has the same meaning as just one arrow head of that colour pointing in that direction.

The proof of these tugging equations is similar to the proof of the yanking equations for the simple case. For the tugging equation on the left, the proof is as follows

$$(1428)$$

In the first step we simply insert an extra white arrow (two white arrows pointing the same way has the same meaning as one). In the second step we use both equations in (1414). In the third step we use orthonormality (the right equation in (1411) but with the arrows pointing in the opposite direction). This gives the first tugging equation in (1427) (it is oriented vertically rather than horizontally but these equations have the same meaning no matter how they are oriented). The second tugging equation is obtained similarly.

## 67.5 Twist objects

We have four identity decompositions given in Sec. 67.2. There are a further six ways to combine bases together at their label wires giving us the *twist objects*:

(1429)

(1430)

(1431)

These twist operators are not basis independent.

Just as in the simple case, we have the $w$-annihilation property. Any pair of $w$ circles will annihilate one another because of the orthonormality condition for the bases. For example

(1432)

These are tugging equations with twists (see the tugging equations without twists in (1427)). The first example is the direct analogue of the yanking equation with twists (519) in the simple case.

## 67.6 Ignore operators

In (1377) we saw that ignore operators can be defined in terms of the identity operator. Using (1414) we obtain the following ways to represent ignore operators: We can write the ignore operators in this Hilbert space notation as follows

$$\text{[diagram]} \tag{1433}$$

and

$$\text{[diagram]} \tag{1434}$$

where we have introduced the $\gamma$ normalisation gauge parameters satisfying

$$\gamma^{\mathsf{a}}\overline{\gamma}^{\mathsf{a}} = \alpha^{\mathsf{a}} \qquad\qquad \gamma_{\mathsf{a}}\overline{\gamma}_{\mathsf{a}} = \alpha_{\mathsf{a}} \tag{1435}$$

such that

$$|\gamma^{\mathsf{a}}\gamma_{\mathsf{a}}|^2 = \frac{1}{N_{\mathsf{a}}} \tag{1436}$$

using (1375).

# 68 Complex operator tensors

## 68.1 Expansion in terms of left and right Hilbert objects

We introduced complex operator tensors in Sec. 58. We can write these operator tensors in terms of left and right Hilbert objects. An example of a complex operator tensor is

$$\text{[diagram]} \tag{1437}$$

On the right there is a black and white triangle arrow wire associated with each wire on the left. We use the convention that each wire on the right hand side of this equation attaches either at the same angle as on the left hand side or, if that is not possible, at the point given by reflecting about the vertical axis.

We can expand an operator tensor out in terms of orthonormal basis elements. For example,

$$\text{[diagram]} \tag{1438}$$

Note, we will follow the convention in Sec. 16.3 and use the notation $\lceil C \rfloor$ to indicate that we have an expansion matrix in terms of orthonormal basis elements for an operator tensor.

## 68.2 Normal operator tensors

We can follow the same reasoning as in the simple case (see Sec. 31.2). If the orthonormal basis expansion matrix is normal then it is unitarily diagonalisable and we can write

(1439)

where the diamond shaped box represents a diagonal matrix with eigenvalues $\lambda_l$ (which, in general, can be complex). The $\underset{\leftarrow}{B}$ and $\underset{\rightarrow}{B}$ objects represent eigenvectors. They can be expanded as

(1440)

where the $\underset{\leftarrow}{B}$ and $\underset{\rightarrow}{B}$ matrices are horizontal adjoints of one another and obey the unitarity properties

(1441)

(compare with (534)). These unitary properties on $\underset{\leftarrow}{B}$ and $\underset{\rightarrow}{B}$ are equivalent to

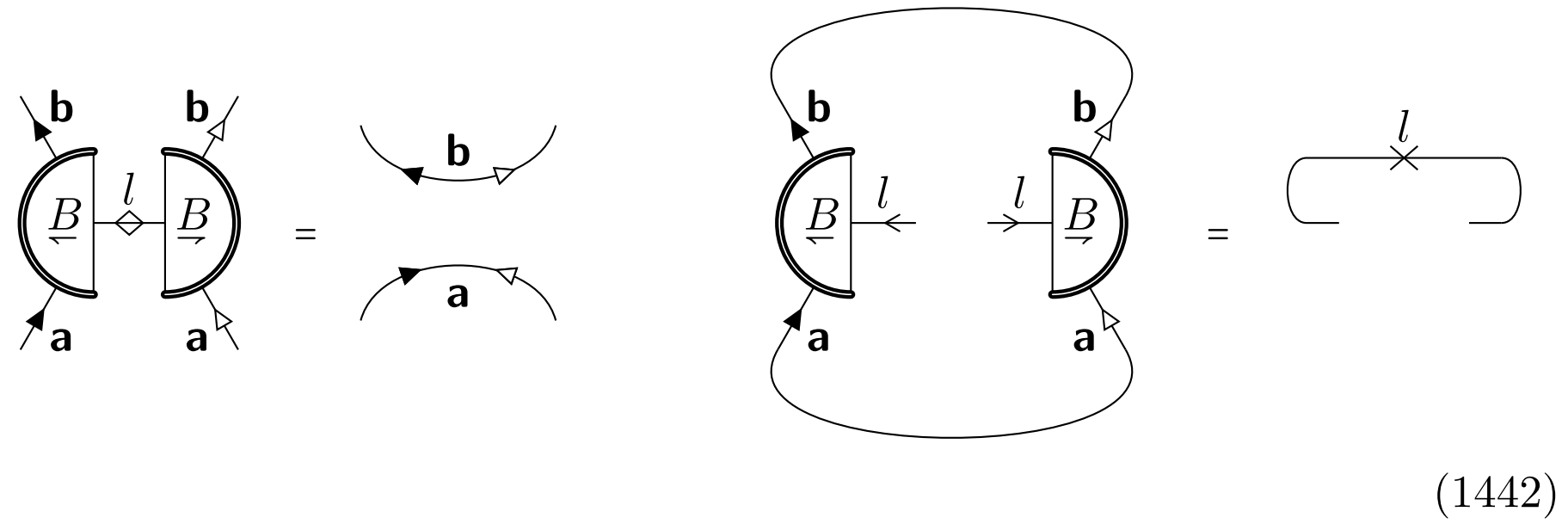

(1442)

for the associated Hilbert objects.

## 68.3 Hermitian operator tensors

We are only interested in normal operators which are Hermitian. In the complex case, Hermitian operator tensors can be expanded as

(1443)

where the eigenvalues, $\lambda_l$, are real.

## 68.4 Twofold positive operator tensors

We say that an operator tensor satisfies *twofold positivity* if it can be written in the following *twofold form*

(1444)

It is easy to see that such operators are Hermitian and that eigenvalues of the expansion matrix, $\ulcorner\! B \!\urcorner$, are nonnegative. Further, we observe that we can immediately write a Hermitian operator tensor in this form if all the $\lambda_l$ in (1443) are positive since they can then be absorbed (as $\sqrt{\lambda_l}$) into the $B$'s and we can write

(1445)

where $\ulcorner\! B$ and $B\!\urcorner$ are horizontal adjoints of one another. If we obtain our twofold form in this way then the label $l$ runs over $N_{\mathsf{a}}N_{\mathsf{b}}$ values (since $\underleftarrow{B}$ and $\underrightarrow{B}$ are unitary and $\ulcorner\! B$ and $B\!\urcorner$ have been formed by absorbing the $\sqrt{\lambda_l}$'s). Thus, we can write

(1446)

Of course, it is possible that some of the $\sqrt{\lambda_l}$'s are zero. In this case we obtain the 0 element of the Hilbert space for some values of $ab$.

The comments here mirror the comments in the simple case discussed in Sec. 31.5. There we stated two theorems and the same theorems apply here too.

First, we have the theorem that there is unitary freedom in the representation of a twofold operator as a sum of left and right Hilbert objects. In Sec. 31.5 we stated, but did not prove, this theorem. Having developed notation for the complex case, we are in a good position to prove this theorem. The proof is given in Sec. 68.7.

Second, we proved a theorem concerning the rank of operators in twofold representation. In Sec. 68.5 we will state that theorem for the complex case (the proof is essentially identical) and we will also prove a additional theorem concerning the rank of operator networks.

## 68.5 Rank considerations

First, we define the rank

> **The rank of twofold operator.** The rank of a twofold operator tensor
>
> $$\hat{B}^{\mathsf{b}}_{\mathsf{a}} = B^{\mathsf{b}}_{\mathsf{a}} \overset{l}{\diamond} B^{\mathsf{b}}_{\mathsf{a}} \tag{1447}$$
>
> is the minimum size of the set $\{l\}$ over all such twofold decompositions. We will write this as $\text{rank}(\hat{B})$.

This definition can be extended to operators having incomes and outcomes using the maximal representation, $\llcorner\hat{B}\urcorner$ as discussed at the end of Sec. 78.2.

We have then have the following minmax theorem

> **Minmax theorem for twofold representation.** The rank of a twofold operator, $\hat{B}$, is equal to the number of nonzero eigenvalues of the expansion matrix, $\ulcorner B \lrcorner$, and satisfies $\text{rank}(\hat{B}) \leq N_{\mathsf{a}} N_{\mathsf{b}}$. Further, there exist twofold operators which necessarily saturate this upper bound.

The proof of this theorem is along exactly the same lines at the proof in the simple case, as given in Sec. 31.5 and we will not repeat it here.

Next we state the following theorem which places an upper bound on the rank of operator networks.

> **Operator networks rank theorem.** The rank of an operator network is less than or equal to the product of the ranks of its components. Further, this bound is saturated when the components of the network are not connected by wires.

To prove this consider the case of a network with two components. Any such network can be written as

(1448)

where we put $K := |\{k\}| = \text{rank}(\hat{A})$ and $L := |\{l\}| = \text{rank}(\hat{B})$. Thus, there always exists an twofold decomposition of this network as a sum over $KL$ cases. Given the definition of rank, this establishes an upper bound on the rank of this network. In the case where there is no **c** wire we need all $KL$ terms because, in that case, we have the tensor product of the $A_k$ and $C_l$ left Hilbert objects on the left side (and similarly on the right side) so the bound is saturated. This reasoning clearly generalises to networks with more than two components.

## 68.6 Homogeneous operator tensors

By analogy with the definitions in Sec. 31.6 for the causally simple case, we call any Hermitian (causally complex) operator tensor that can be written in the form

(1449)

*homogeneous*. Homogeneous operators are twofold operators. Furthermore, we call twofold operator tensors that cannot be written in this form *heterogeneous*. The name heterogeneous is chosen because any such twofold operator tensor must be written as a sum of homogeneous operator tensors that are not proportional to one another (as in (447) where $l$ runs over at least two values). Homogeneous operators have rank equal to 1. Heterogeneous operators have rank greater than 1.

## 68.7 Unitary freedom in the twofold representation

The twofold representation of a given operator is not unique. However, there is a simple relationship between the left (or right) Hilbert objects in different such representations. We will prove a few theorems here to elucidate this before proving the unitary freedom theorem at the end of this subsection.

First we prove the following

**Twofold operators orthogonal representation: Part A.** For any twofold representation of an operator

$$(1450)$$

we can write

$$(1451)$$

where $\lambda_c \geq 0$ (these are the diagonal entries of the diagonal matrix $\lambda$) and where

$$(1452)$$

Further, we have the property

$$(1453)$$

where $s(\lambda)$ is a diagonal matrix having entries along the diagonal equal to 0 if the corresponding $\lambda_c = 0$ and 1 if the corresponding $\lambda_c > 0$. Note that $\lambda^{\frac{1}{2}}$ is a diagonal matrix with entries $\lambda_c^{\frac{1}{2}}$ along the diagonal.

Recall that the shaded Hilbert objects in (1451) and (1452) above are an orthonormal basis as represented in (1416). We can read (1453) as an isometry property when we restrict to $c$ having $\lambda > 0$ on the open wires. We prove this as follows. As $\hat{B}$ is twofold, it must be normal and so we can write it in the form

$$(1454)$$

To see this compare with (1439). Note that we have replaced the sum over $l$ there with a sum over $c$. We can do this because the expansion matrix, $\underleftarrow{B}$, is unitary (and so must be square). Using (1440) applied to the present case, we can write

$$(1455)$$

where the last step follows since the expansion matrix, $\underleftarrow{B}$, is unitary (see (1441)) and so can be used to effect a change of basis (see (1416)). This proves we can write $\hat{B}$ in the form in (1451). Next, we can partition the Hilbert space, $\mathcal{H}_{\mathbf{c}}$, as

$$\mathcal{H}_{\mathbf{c}} = \mathcal{H}_{\mathbf{c}}^{>0} \oplus \mathcal{H}_{\mathbf{c}}^{=0} \tag{1456}$$

where $\mathcal{H}_{\mathbf{c}}^{>0}$ is spanned by the reVerse adjoint of the basis vectors defined in (1455) for which $\lambda_c > 0$ and $\mathcal{H}_{\mathbf{c}}^{=0}$ is spanned by those for which $\lambda_c = 0$. Consider

$$E \;\; \in \mathcal{H}_{\mathbf{c}}^{=0} \tag{1457}$$

Then, using (1451), we obtain

$$= \quad = \quad 0 \tag{1458}$$

where the equality in the last step follows because $\lambda_c = 0$ for $E \in \mathcal{H}_{\mathbf{c}}^{=0}$. Thus,

$$B \;\; \in \;\; \mathcal{H}_{\mathbf{c}}^{>0} \tag{1459}$$

This means we can expand these $B$'s in terms the shaded basis vectors for which $\lambda_c > 0$. In other words, we can expand each of the $B$'s exactly as in (1452) for some $V'$. If we substitute the expansion in (1452) into the following equation

$$= \tag{1460}$$

(which we obtain from (1451)) then, using (1425), we obtain

$$\lambda^{\frac{1}{2}} \;\; V' \;\; \overline{V}' \;\; \lambda^{\frac{1}{2}} \;\; = \;\; \lambda \tag{1461}$$

(where we have also used the fact that we can bend the label wires straight). Equation (1453) follows immediately from this equation.

Before stating the next theorem, we will define a useful object. Consider labels, $k \in \Gamma_K$ and $m \in \Gamma_M$ where $\Gamma_K \subseteq \Gamma_M$ such that $K = |\Gamma_K| \leq M = |\Gamma_M|$. Then we define the *padding matrix*

$$k \quad m \quad = \begin{cases} 1 & \text{if } k = m \\ 0 & \text{else} \end{cases} \tag{1462}$$

(we defined this in the simple case, without arrows, in (493)). We are mostly interested in the case where $K < M$ (rather than when $K = M$) and, in such cases, we choose the flat side of the "cut diamond" shape to be associated with the label whose set has more elements ($m$ in the above case). We can apply the non-conjugating conjupositions in $(I, H, V, T)$, as defined in Sec. 66, giving the objects

$$k \quad m \qquad\qquad m \quad k \qquad\qquad k \quad m \qquad\qquad m \quad k \tag{1463}$$

respectively. These all have the same property that elements having $k = m$ are equal to 1 and all other elements are equal to 0. Note that we get the same matrices if we apply the conjugating conjupositions $(\overline{I}, \overline{H}, \overline{V}, \overline{T})$ instead since the entries are real. As an example, consider the case where $K = 3$ and $M = 5$. Then we have

$$k \quad m \quad = \begin{pmatrix} 1 & 0 & 0 & 0 & 0 \\ 0 & 1 & 0 & 0 & 0 \\ 0 & 0 & 1 & 0 & 0 \end{pmatrix} \tag{1464}$$

Note that we have the following property

$$k \quad m \quad k \quad = \quad k \tag{1465}$$

(so it is, in fact, an example of an isometry). The reason for calling the matrix introduced in (1462) a "padding" matrix is that we can write

$$\mathbf{c} \quad B' \quad m \quad := \quad \mathbf{c} \quad B \quad k \quad m \tag{1466}$$

This is a set of $M$ left Hilbert objects which are equal to 0 for $m \in \Gamma_M - \Gamma_K$, and equal to the original left Hilbert objects, $B$, when $m \in \Gamma_K$.

We will also need a variant of the padding matrix. We define the $f$-padding matrix as

$$m \quad f \quad k \quad = \begin{cases} 1 & \text{if } m = f(k) \\ 0 & \text{else} \end{cases} \tag{1467}$$

where $f$ is a one-to-one function from the elements of $\Gamma_K$ to the elements of $\Gamma_M$ (where $|\Gamma_K| \leq |\Gamma_M|$). This is useful if $\Gamma_K$ is not a subset of $\Gamma_M$. Note the

larger set, $\Gamma_M$, is associated with the flat side of the cut diamond. We can take conjupositions of this object as before yielding

(1468)

where the flat side of the diamond always represents the bigger set (note that the arrows establish a convention - they are not there to indicate the direction in which the function, $f$, is acting). We also have the property

(1469)

since $f$ is a one-to-one function.

Now we are in a position to prove the following theorem

> **Twofold operators orthogonal representation: Part B.** Consider an operator tensor, $\hat{B}$, that can be written in twofold form as in (1450) and, therefore, according to Part A of this theorem, we can write the left object $B$ as in (1452). Further, let $\Gamma_{\mathsf{c}}^{>0}$ be the set of labels, $c$, for which $\lambda_c > 0$. We will write $c^{>0} \in \Gamma_{\mathsf{c}}^{>0}$ on wires where we sum only over the elements of $\Gamma_{\mathsf{c}}^{>0}$. Let $K$ be the number of elements of the set of $k$ labels. Then we have the following three properties.
>
> **Isometry.** We can write
>
> (1470)
>
> where
>
> (1471)
>
> (so $V$ is an isometry). Note also that $\text{rank}(\hat{B}) = |\Gamma_{\mathsf{c}}^{>0}| \leq K$. Thus, the elements of any decomposition of $\hat{B}$ can be written by acting on a set of $\text{rank}(\hat{B})$ orthogonal elements with an isometry.
>
> **Unitary.** We can write
>
> (1472)
>
> where $U$ satisfies
>
> (1473)

$$(1474)$$

(i.e. it is unitary).

**Bigger unitary.** We can write

$$(1475)$$

where

$$(1476)$$

(for any choice of one-to-one function $h$) such that

$$(1477)$$

where $U$ satisfies

$$(1478)$$

$$(1479)$$

(i.e. it is unitary).

The isometry property follows immediately from part A of the theorem since the matrix $s(\lambda)$ "projects down" to the set $\Gamma_{\mathbf{c}}^{>0}$. We will write $N_{\mathbf{c}}^{>0} = |\Gamma_{\mathbf{c}}^{>0}|$. It follows from this isometry property that $N_{\mathbf{c}}^{>0} \leq K$ (this is to be expected since not all the $B_k^{\mathbf{c}}$'s need be linearly independent). This proves that $\text{rank}(\hat{B}) = N_{\mathbf{c}}^{>0}$ as stated in the theorem. To prove the unitary property note that (1472) yields (1470) if

$$(1480)$$

holds. We can easily arrange for this to hold as follows. $V$ is a rectangular matrix with $N_{\mathbf{c}}^{>0}$ rows and $K$ columns. Since $V$ is an isometry, the rows are orthonormal vectors. We can add extra rows to $V$ to form a square matrix choosing these extra rows to complete the orthonormal set. This gives us a unitary matrix which has the property in (1480). Note that this works for any choice of the one-to-one function, $f$. To obtain the bigger unitary property note that (1475) yields (1470) if

$$c^{>0} \quad f \quad m \quad U' \quad m \quad h \quad k \quad = \quad c^{>0} \quad V \quad k \tag{1481}$$

where we have applied (1469) to (1476) as the first step in proving this. The unitary matrix, $U'$, is easily obtained by applying columns and rows to $V$ as follows. We add columns of 0's to $V$ to turn it into a $N_{\mathsf{c}}^{>0}$ by $M$ matrix. The rows of our new matrix are all orthonormal by virtue of the fact that $V$ is isometric. Now we append extra rows to complete the orthonormal set giving a square matrix. This is then a unitary matrix. The matrix so constructed satisfies (1481).

The following corollary immediately follows from the isometry case of the above theorem.

**Isometries and orthogonal twofold representations.** Any operator can be written in the form

$$\mathsf{c} \quad \hat{B} \quad = \quad \mathsf{c} \quad B_\perp \quad n \quad B_\perp \quad \mathsf{c} \tag{1482}$$

where the Hilbert objects

$$\mathsf{c} \quad B_\perp \quad n \tag{1483}$$

are orthogonal and $|\{n\}| = \mathrm{rank}(\hat{B})$. Further, for any twofold representation of $\hat{B}$ such that

$$\mathsf{c} \quad B \quad k \quad B \quad \mathsf{c} \quad = \quad \mathsf{c} \quad B_\perp \quad n \quad B_\perp \quad \mathsf{c} \tag{1484}$$

we can write

$$\mathsf{c} \quad B \quad k \quad := \quad \mathsf{c} \quad B_\perp \quad n \quad V \quad k \tag{1485}$$

where

$$n \quad V \quad k \quad \overline{V} \quad n \tag{1486}$$

so $V$ is an isometry and $|\{k\}| > |\{n\}|$.

This seen by equating $n$ with $c^{>}$ from the proof above, and then writing

$$\mathbf{c}\; B_{\perp}\; c^{>} \;=\; \mathbf{c}\; c\; \lambda^{\frac{1}{2}}\; c\; c^{>0} \tag{1487}$$

These Hilbert objects are orthogonal since they are proportional to an orthonormal set and $|\{c^{>}\}| = \text{rank}(\hat{B})$. This corollary, then, follows immediately from the isometry part of the above theorem.

Finally we prove

**Unitary freedom in twofold representation.** We have

$$\mathbf{c}\; A\; m\; A\; \mathbf{c} \;=\; \mathbf{c}\; B\; k\; B\; \mathbf{c} \tag{1488}$$

(where $m \in \Gamma_M$, $k \in \Gamma_K$ and we take $|\Gamma_M| \geq |\Gamma_K|$) if and only if there exists $U$ such that

$$\mathbf{c}\; A\; m \;=\; \mathbf{c}\; B\; k\; f\; m\; U\; m \tag{1489}$$

where $U$ is unitary (for any one-to-one choice of $f$ from $\Gamma_K$ to $\Gamma_M$).

It follows immediately from unitarity and the padding matrix property in (1465) that, if we have (1489), (1488) follows. This proves the "if" direction. We obtain "only if" direction from the Part B theorem above as follows. We can construct $B'$ as in (1476) using some one-to-one function, $h$. It now follows from the Part B theorem that $A$ and $B$ can each be obtained by a unitary acting on

$$\mathbf{c}\; c\; \lambda^{\frac{1}{2}}\; c\; c^{>0}\; g\; m \tag{1490}$$

for any choice of one-to-one function $g$. Let these unitaries be $U_A$ and $U_B$. Then it follows that we go from $B$ to $A$ using the unitary $U = U_A U_B^{-1}$. This proves the theorem.

In Sec. 31.5 we stated the equivalent theorem for the simple case. The theorem stated there follows from the theorem we have just proven since the simple case is a special case of the complex case (where we have simple causal structure). It is worth commenting here that the notation for complex operators is more powerful since we are able to build a composite system, $\mathbf{c}$, from inputs and outputs. This allows a more direct proof of this unitary freedom.

# 69 Normal Conjuposions of Hilbert objects

## 69.1 Eight normal conjupositions

As in the simple case, we can define *normal conjupositions* by expanding Hilbert objects in terms of an orthonormal basis then performing conjupositions on the expansion hypermatrices. A general left Hilbert object can be written as a sum over orthonormal basis elements as follows

(1491)

This is general because, if there are multiple systems, they can be combined into one composite system which is then represented by **a**. We have included a square dot so we can track the normal conjupositions. We can represent the eight normal conjupositions of this object by basis expansions

(1492)

(1493)

(1494)

(1495)

Recall that the notation ${}^{\ulcorner}_{\llcorner}B$ and $B^{\urcorner}_{\lrcorner}$ notation plays the same role as the $M$ and $\overline{M}$ notation in Sec. 66. Thus, the eight expansion hypermatrices in (1492-1495) cover all possible conjupositions of the original expansion hypermatrix in (1491). We have had to introduce some new notation concerning the small black and white squares. In the expressions in (1492, 1493), the small squares are their natural colours (black on the left and white on the right). However, in the expressions in (1494, 1495) the left objects have white squares and the right objects have black squares (i.e. not their natural colours). This is to deal with taking nonconjugating tranformations and will be discussed in Sec. 69.3.

In the simple case we saw that taking the normal conjugate, $\overline{I}$, of a basis element leaves it unchanged (see (591)). The same is true here. We have

$$(1496)$$

This is clear from (1494, 1495) above since, when $B$ is a basis element, its expansion matrix [$B$ in (1491) is just the identity.

## 69.2 The Hilbert square

The four objects in (1492) and (1493) form a Hilbert square as follows.

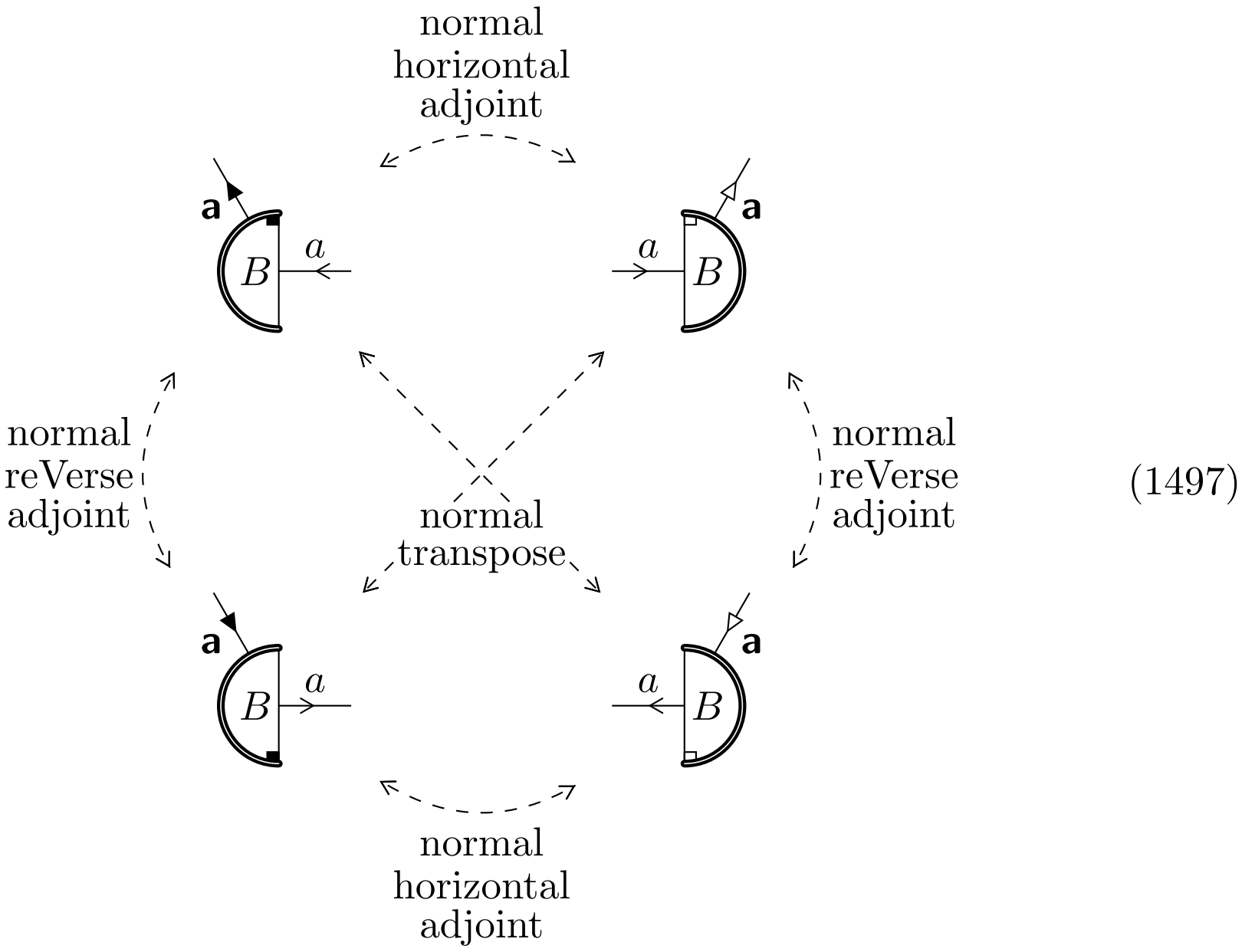

(1497)

These are mapped into one another by the subgroup of the normal conjupositions in $\{I, \overline{H}, \overline{V}, T\}$. These normal conjupositions are basis independent as will now see. It is sufficient, in fact, to show that $\overline{H}$ and $T$ are basis independent since $\overline{V} = T\overline{H}$.

To see the normal horizontal adjoint is basis independent note that we can

write

(1498)

We obtain the first expression below by taking the normal horizontal adjoint of the last expression above.

(1499)

we obtain the second expression above by noting that $\overline{U}$ acting on the shaded right basis element gives us the unshaded right basis element. Thus, the normal horizontal adjoint is independent of the orthonormal basis used. This reasoning is analogous to that in Sec. 32.2 in the simple case.

The normal transpose operation, $T$, will take us from the left object in (1492) to the right object in (1495). To see that this is basis independent note that we can take the normal transpose as follows

(1500)

Here we are using the object, consisting of a black and white arrow pointing in opposite directions, that we introduced in (1414). This object is basis independent and consequently the transpose operation is basis independent. To check that (1500) works we can substitute in the expansion on the left as follows

(1501)

To make the last step we note that, in the penultimate expression, the label wires attached to the $\ulcorner\!\!\llcorner B$ matrix are bent back and the arrow directions are reversed. This takes the transpose (see (1402)).

In the simple case the normal transpose is implemented by appending cups and caps, sliding, then straightening out using the yanking equation. In the complex case we can implement the normal transpose by adding an opposing arrow of the opposite colour to the system wire and bending the label wire with a diamond arrow or an X arrow as appropriate. The net effect of taking the normal transpose can be thought of in a tactile sense as implementing a *toggle and flip* action. We toggle the switch (moving the square to the other location and reversing the arrows) then we flip the object over horizontally revealing the opposite colours for the arrows and square.

## 69.3 The nonconjugating transformations

The nonconjugating transformations are $\{I, H, V, T\}$. These can be implemented diagrammatically by adding wires and using twist objects. We have already seen how this can be done for $T$. To implement $H$ we use a twist object as follows

$$\mathbf{a}\ B\ a \quad \xrightarrow[\text{transpose, } H]{\text{normal horizontal}} \quad a\ B\ \mathbf{a} \quad = \quad \mathbf{a}\ (w)\ \mathbf{a}\ B\ a \tag{1502}$$

This takes us from the left object in (1492) to the right object in (1494). This transformation is basis dependent since it used the twist object (see Sec. 67.5) which is basis dependent. Note also, that under $H$, we use notation such that the colour of the small black square does not change.

We can obtain the normal reVerse transpose, $V$, by applying the normal horizontal transpose, $H$, then the normal transpose, $T$. Thus, all four elements of the nonconjugating subgroup of normal conjupositions can be implemented by attaching wires (and using twist objects). Since this used the twist object, $V$ is also a basis dependent operation.

## 69.4 The Hilbert cube

Just as in the simple case, it is instructive to arrange the left and right objects obtained from one another under conjupositions in the form of a Hilbert cube as follows.

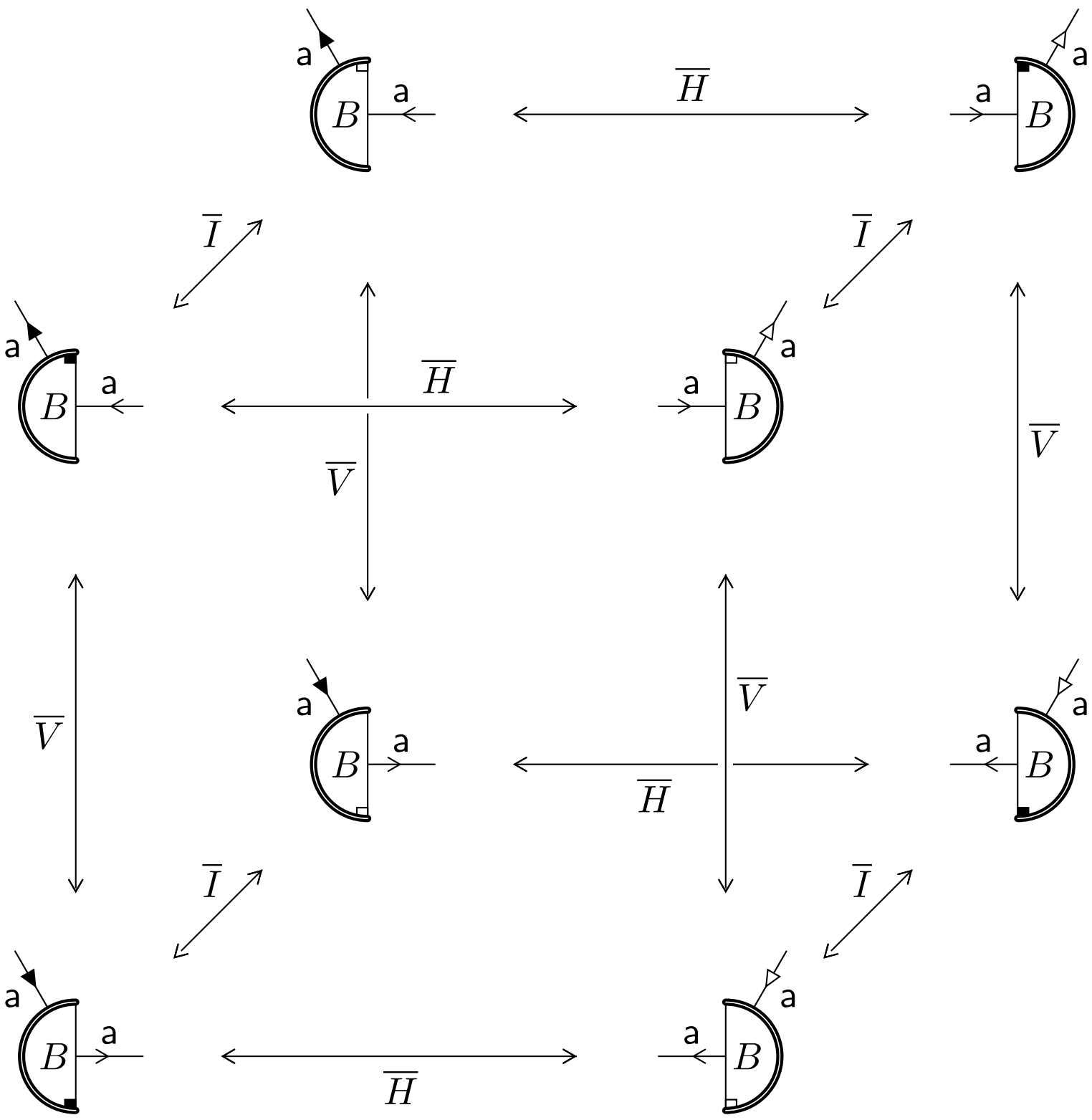

(1503)

The edges show some of the transformations. There are also transformations associated with the diagonals (which are not shown to prevent overcrowding). The diagonals on the front and back squares are associated the transpose, $T$. The diagonals on the side squares are associated with the reverse transpose, $V$. The diagonals on the top and bottom squares are associated with the horizontal transpose, $H$. The four interior diagonals are associated with the adjoint, $\overline{T}$.

## 69.5 Normal conjuposions of composite objects

We can implement a normal conjuposition of an object composed of multiple left and right objects. To do this we can expand everything out in terms of orthonormal basis elements. This will result in a bunch of bases connected to a composite hypermatrix network. We can then implement the given conjuposition on this hypermatrix network and change the basis elements accordingly. Since conjuposition respects the compositional structure of hypermatrix networks it is clear that, by this procedure, it will respect the compositional structure of left and right objects.

These remarks are best illustrated by an example. Consider a general Hilbert object

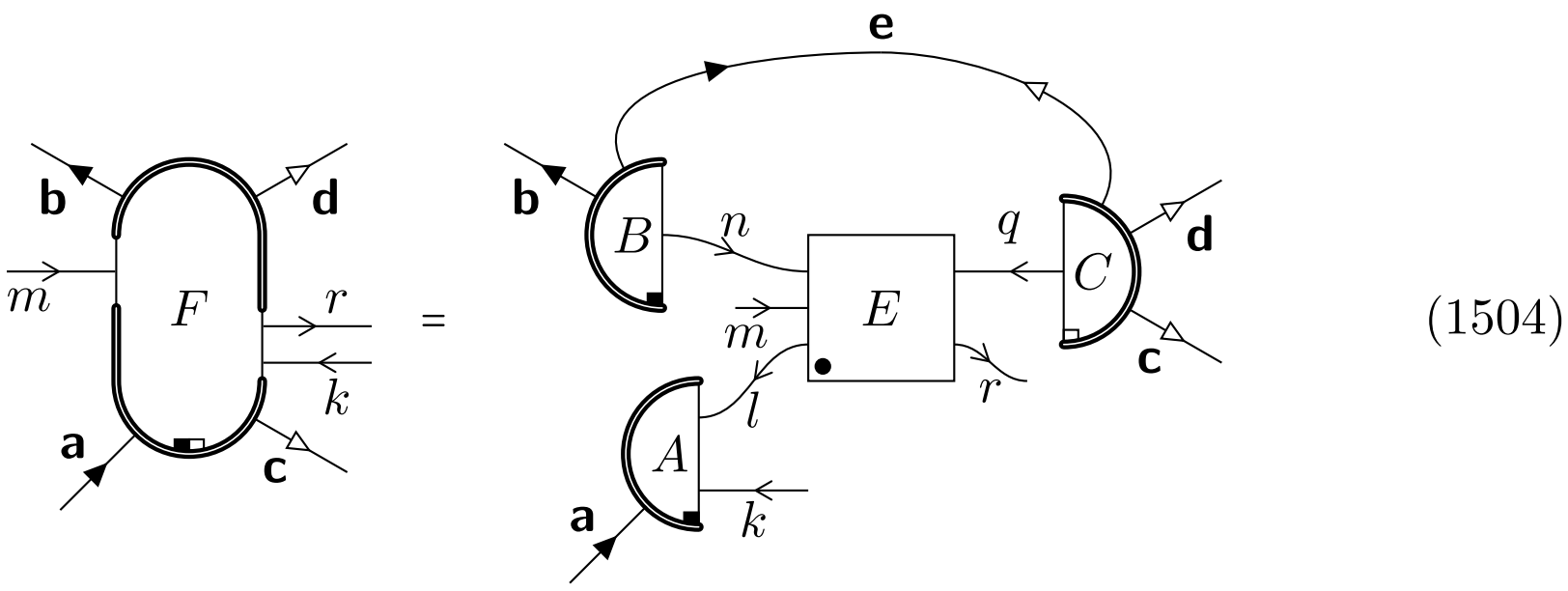

(1504)

Note, on the left, we have introduced a pair of black and white dots to help us track the conjuposition. We can take the normal transpose of this by taking the normal transpose of every component object. The resulting normally transposed general Hilbert object is

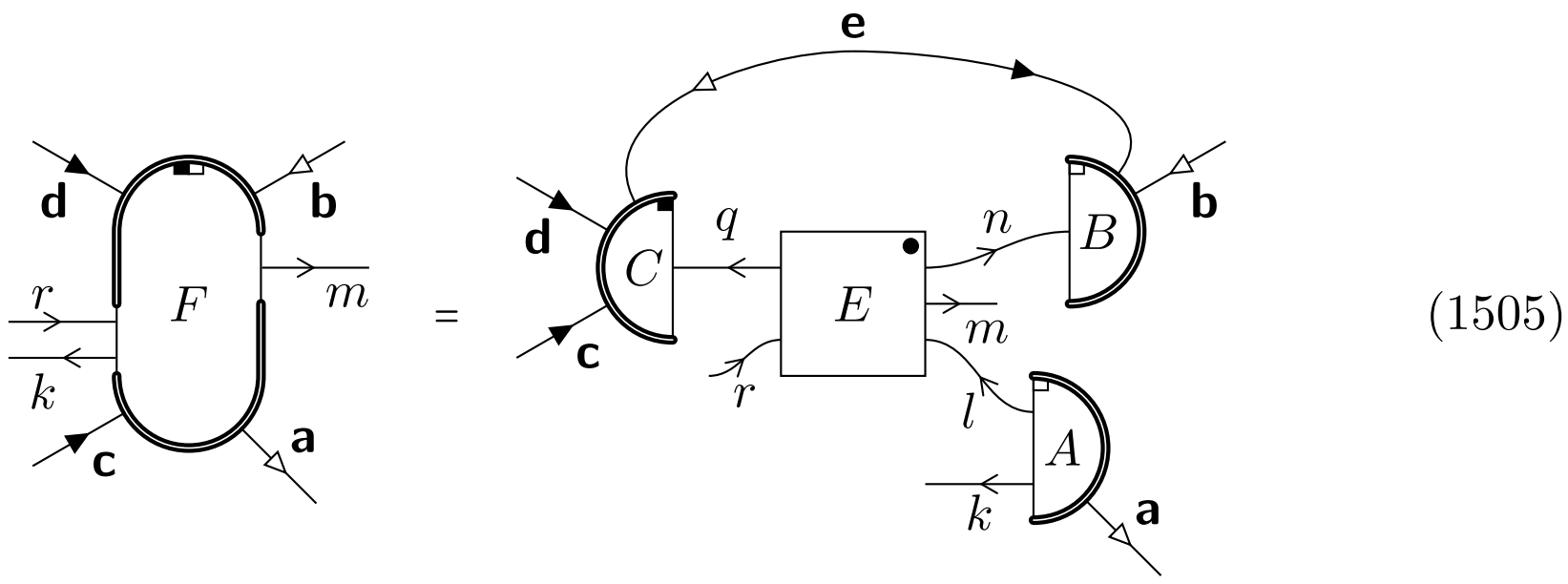

(1505)

Note that every switch has been toggled and every object has been flipped (so black arrows and black squares become white arrows and white squares and vice versa). The black-white square pair appearing at the bottom has been toggled (moved from the bottom to the top) then flipped. Since the black square has white on the back and the white has black on the back, when we flip horizontally the black-white pair appears unchanged. Also note that we have taken the transpose of the hypermatrix, $E$.

By starting with a given general Hilbert object and applying normal conjupositions we can obtain eight general Hilbert objects (including the original one). These can be arranged in a Hilbert cube.

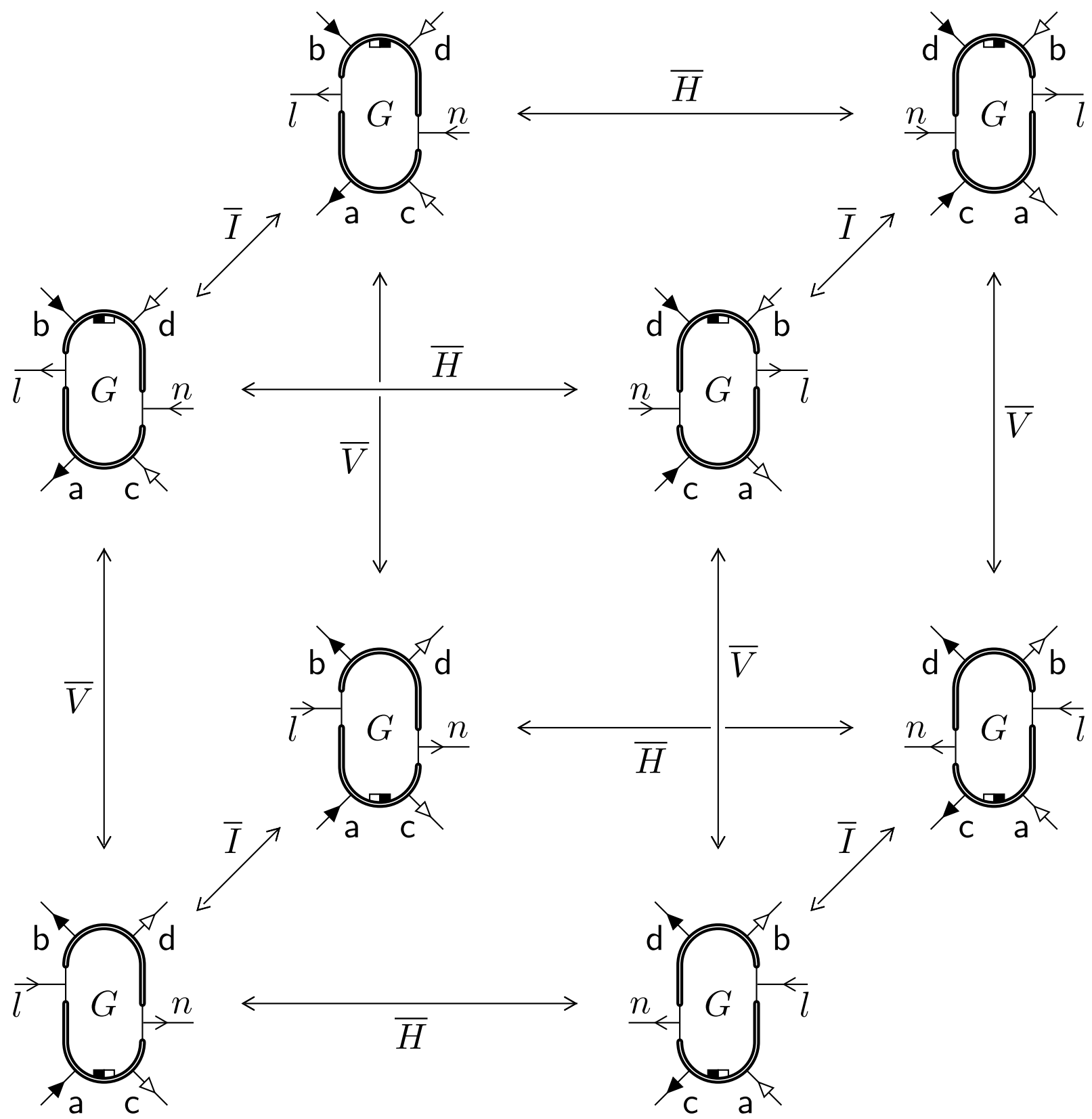

(1506)

Note that the objects at the front have black-white square markings while the objects at the back have white-black square markings. Note that, when we flip horizontally the shape of the object is flipped also (observe the position of the double versus single line parts of the border).

## 69.6 Normal Conjuposions of operator tensors

Operator tensors are, from a mathematical point of view, special cases of general Hilbert objects. They are of particular physical interest since they correspond to operations. Since they are special cases of general objects we can, clearly, act on them with normal conjupositions. We will restrict our attention to the normal conjupositions in the basis independent subgroup $\mathcal{S}_{BI} = \{I, \overline{H}, \overline{V}, T\}$.

Consider an operator tensor

(1507)

where we have expanded in terms of a basis. Since the circle is left right symmetric this removes an expressive element in our notation. If we flip the circle horizontally it looks the same (given that there is white on the back of the black square and vice versa). We will, in any case, be interested in operator tensors that are Hermitian and so invariant under this horizontal flipping so this notational deficiency might actually be regarded as a positive feature. In any case, for the moment, we will indicate taking the normal horizontal adjoint by a subscript, $\overline{H}$. Thus we have

(1508)

(1509)

We are interested in operator tensors that satisfy twofold positivity since this is one of the physicality conditions. This means they are Hermitian. We can see immediately from the above equations that, if $\hat{B}$ is Hermitian, then taking the horizontal transpose leaves it unchanged (this is the same as discussed in the simple case in Sec. 32.14). Thus, $\overline{V}$ and $T$ have the same effect. So, effectively we only have two transformations - the identity $I$ which leaves the expression unchanged and $\overline{V}$ which reverses all the arrows.

If we have a network built out of Hermitian operator tensors such as on the left below

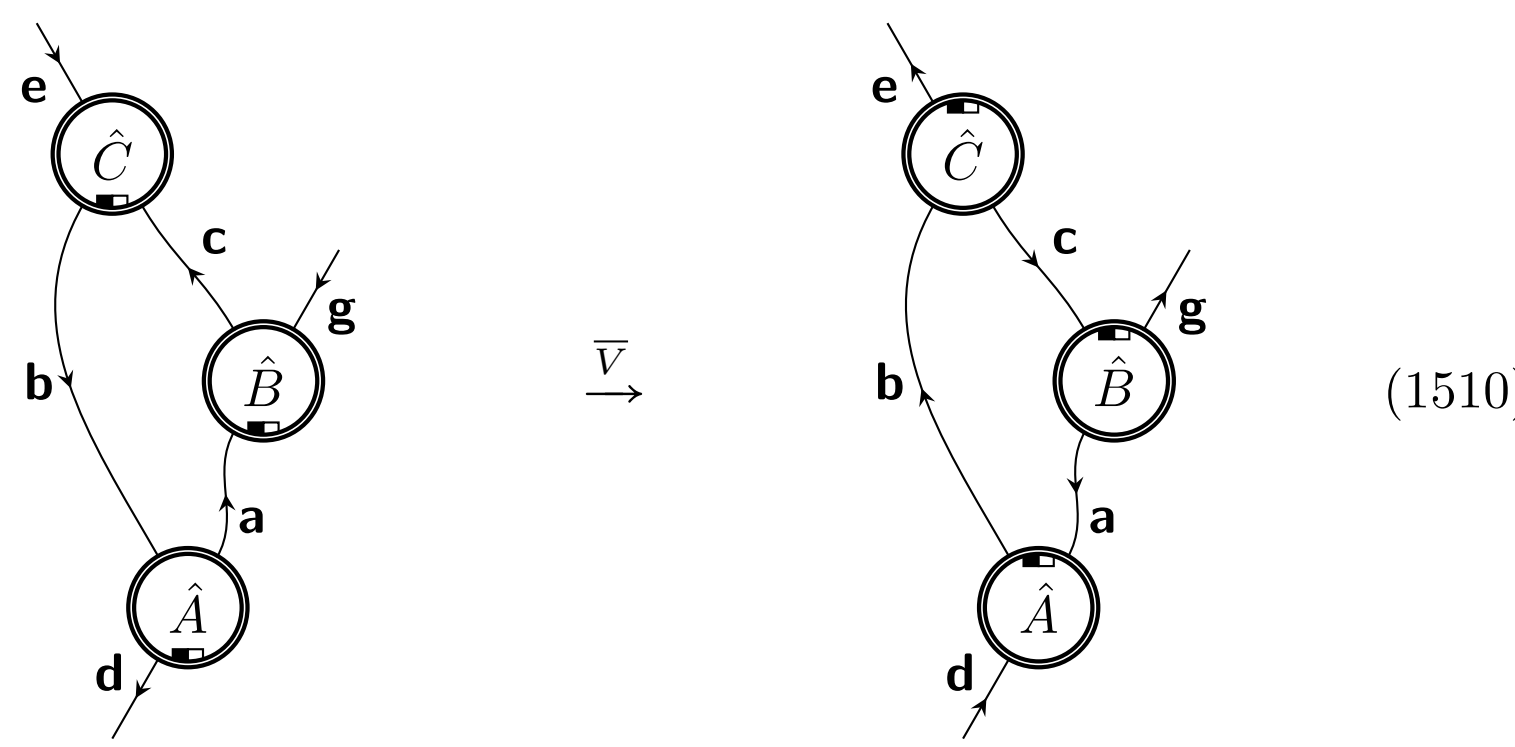

(1510)

we can implement $\overline{V}$ (or equivalently $T$) on the whole network simply by reversing the arrows (giving the network on the right above). This strongly suggests that $\overline{V}$ implements time reversal (as discussed in Sec. 48.4). However, as in the simple case, there is something not quite right about this. Namely that the ignore operators (introduced in Sec. 67.6) do not transform into one another under $\overline{V}$ (or $T$). Indeed,

$$\hat{I}^{\mathsf{a}} \xrightarrow{\overline{V}} \boxed{\frac{\gamma^{\mathsf{a}}}{\gamma_{\mathsf{a}}}}\,\hat{I}^{\mathsf{a}}\,\boxed{\frac{\overline{\gamma}^{\mathsf{a}}}{\overline{\gamma}_{\mathsf{a}}}} \tag{1511}$$

as is easily verified from (1433,1434).

As in the simple case, we will introduce the idea of a ortho-physical basis along with the complementary ortho-deterministic basis. Then we can define natural conjupositions with respect to this ortho-physical basis (this can, equivalently, be done with respect to the ortho-deterministic basis). Under the natural reverse adjoint ignore operators will transform into one another. Then we can choose the natural reverse adjoint (or equivalently, the natural transpose) to implement time reversal.

# 70 Natural conjupositions of Hilbert objects

## 70.1 Ortho-physical bases

We define the following *ortho-physical* basis elements for the complex case

$$\mathsf{a}\ \ a = \boxed{\gamma_{\mathsf{a}}}\ \mathsf{a}\ \ a \qquad\qquad a\ \ \mathsf{a} = a\ \ \mathsf{a}\ \boxed{\overline{\gamma}_{\mathsf{a}}}$$

$$\mathsf{a}\ \ a = \boxed{\gamma^{\mathsf{a}}}\ \mathsf{a}\ \ a \qquad\qquad a\ \ \mathsf{a} = a\ \ \mathsf{a}\ \boxed{\overline{\gamma}^{\mathsf{a}}} \tag{1512}$$

These are analogous to the ortho-physical basis elements defined in (624) in Sec. 33.1 for the simple case. Our discussion here will mirror the simple case discussed in that section.

The gauge normalisation condition is

$$|\gamma^{\mathsf{a}}\gamma_{\mathsf{a}}|^2 = \frac{1}{N_{\mathsf{a}}} \tag{1513}$$

It is, however, convenient to impose the following *ortho-physical gauge normalisation condition*

$$\gamma^{\mathsf{a}}\gamma_{\mathsf{a}} = \frac{1}{N_{\mathsf{a}}^{\frac{1}{2}}} = \overline{\gamma}^{\mathsf{a}}\overline{\gamma}_{\mathsf{a}} \tag{1514}$$

so that physical conjuposition behaves appropriately on composite objects.

Using the ortho-physical gauge normalisation condition (1514) we get the following orthogonality relations

$$\boxed{\frac{1}{\sqrt{N_{\mathsf{a}}}}} \qquad\qquad \boxed{\frac{1}{\sqrt{N_{\mathsf{a}}}}} \tag{1515}$$

(so these ortho-physical basis elements are orthogonal and subnormalised). Furthermore we obtain the following scaled decompositions of the identity

$$:= \qquad = \boxed{\frac{1}{\sqrt{N_{\mathsf{a}}}}} \qquad\qquad := \qquad = \boxed{\frac{1}{\sqrt{N_{\mathsf{a}}}}} \tag{1516}$$

$$:= \qquad = \qquad \boxed{\gamma_{\mathsf{a}}\overline{\gamma}_{\mathsf{a}}} \tag{1517}$$

and

$$:= \qquad = \qquad \boxed{\gamma^{\mathsf{a}}\overline{\gamma}^{\mathsf{a}}} \tag{1518}$$

We note that this allows us to write the ignore operators defined in (1433) and (1434) as

$$\hat{I} \qquad = \qquad = \tag{1519}$$

$$\hat{I} \qquad = \qquad = \tag{1520}$$

(recall, as commented in Sec. 68.1, that when doubling up the system wire with a black and white arrow, we allow them to attach at angles reflected about the vertical axis). Thus, crucially, the ignore operators are naturally expressed in terms of the ortho-physical basis.

We can also define the physical twist objects as follows

$$\widetilde{w} \; := \qquad\qquad \widetilde{w} \; := \tag{1521}$$

(1522)

(1523)

Compare with the twist objects in Sec. 67.5.

## 70.2 Natural conjupositions of Hilbert objects

We can take the natural conjuposition of Hilbert objects by expanding in terms of the ortho-physical basis then applying the conjuposition to the expansion matrix. Thus, if we start with

(1524)

then we have the following eight natural conjupositions

(1525)

(1526)

(1527)

(1528)

(compare these with the normal conjupositions in (1492-1495). As in the simple case, we represent the elements of the natural conjuposition group by placing a tilde under the symbol

$$\mathcal{C}_{\rm phys} = \{I, \underset{\sim}{\overline{I}}, \underset{\sim}{H}, \underset{\sim}{\overline{H}}, \underset{\sim}{V}, \underset{\sim}{\overline{V}}, \underset{\sim}{T}, \underset{\sim}{\overline{T}}\} \tag{1529}$$

We can represent the natural conjupositions on a Hilbert Cube as follows

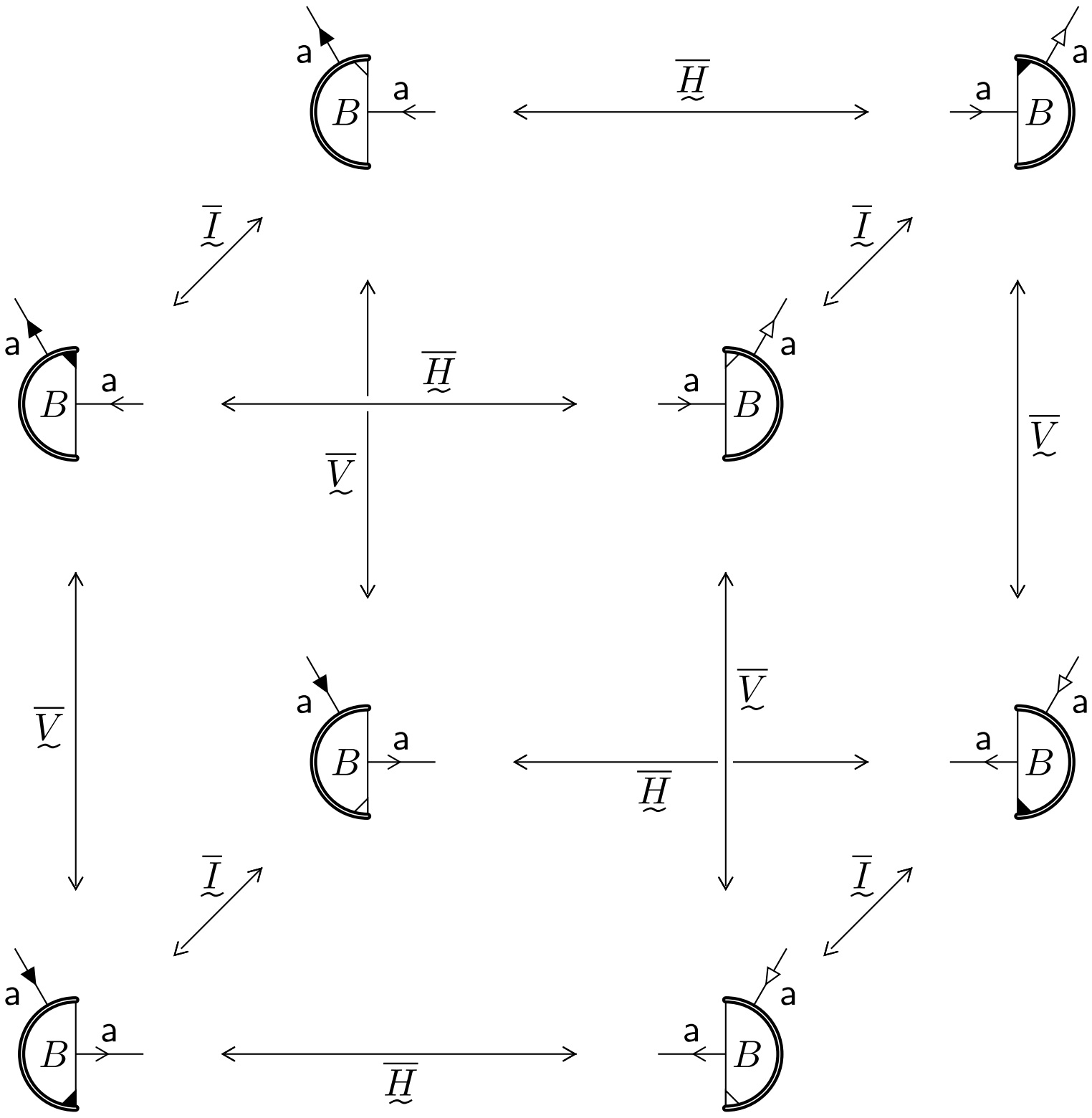

(1530)

where, to prevent overcrowding, we have not explicitly labeled the natural conjupositions associated with the diagonals.

As in the simple case (see Sec. 33.3) we have that $\underset{\sim}{\overline{H}} = \overline{H}$ (this can be seen by comparing (1525) with (1492)). Furthermore, we will see in Sec. 70.5 that the natural transpose, $\underset{\sim}{T}$, can be taken in a basis independent way (by applying a special object). Since $\underset{\sim}{\overline{V}} = \underset{\sim}{T}\underset{\sim}{\overline{H}}$ this means that natural conjupositions in $\{I, \underset{\sim}{\overline{H}}, \underset{\sim}{\overline{V}}, \underset{\sim}{T}\}$ are basis independent.

The basis elements are unchanged under the natural conjugation transfor-mation, $\overline{\underset{\sim}{I}}$

(1531)

This is clear because, if we consider expanding the basis element in the form (1524) then the expansion matrix is the identity.

Furthermore, in the special case where the operation is of the form

(1532)

(so the input system is the same as the output system) then any natural conjuposition has the same effect as the corresponding normal conjuposition.

We could adopt the $\gamma(N)$ gauge (discussed in Sec. 33.4 for the simple case) whereby

$$\gamma^{\mathsf{a}} = \gamma(N_{\mathsf{a}}) \quad \text{and} \quad \gamma_{\mathsf{a}} = \frac{1}{\sqrt{N_{\mathsf{a}}}}\gamma^{-1}(N_{\mathsf{a}}) \tag{1533}$$

where $\gamma(N)$ is any (non-zero) complex-valued function of the integer, $N$, satisfying the completely multiplicative property

$$\gamma(MN) = \gamma(M)\gamma(N) \tag{1534}$$

for all $M$ and $N$. This partial gauge fixing has the added bonus that, if we impose it, then natural conjuposition commutes with interconversion (as denoted by $\leftrightharpoons$ in Sec. 47.2). That is to say, if we have some Hilbert object, $B$, and perform some natural conjuposition on it then some interconversion, we will get the same Hilbert object as if we first perform the interconversion then do the natural conjuposition. We will not actually use the $\gamma N$ gauge as it would obscure some of the interesting mathematical structures under study.

## 70.3 Ortho-deterministic bases

We can introduce an ortho-deterministic basis in the complex case as follows

$$\eta_{\mathsf{a}} \qquad\qquad \overline{\eta}_{\mathsf{a}} \qquad\qquad \eta^{\mathsf{a}} \qquad\qquad \overline{\eta}^{\mathsf{a}} \tag{1535}$$

This is analogous to (664) introduced in the simple case. We define $\eta_{\mathsf{a}}$ and $\eta^{\mathsf{a}}$ such that

$$\tag{1536}$$

and

$$\tag{1537}$$

which means

$$\gamma^{\mathsf{a}}\eta_{\mathsf{a}} = 1 \qquad\qquad \eta^{\mathsf{a}}\gamma_{\mathsf{a}} = 1 \tag{1538}$$

If we use the ortho-physical gauge normalisation condition in (1514) then we obtain

$$\eta_{\mathsf{a}} = \sqrt{N_{\mathsf{a}}}\gamma_{\mathsf{a}} \qquad\qquad \eta^{\mathsf{a}} = \sqrt{N_{\mathsf{a}}}\gamma^{\mathsf{a}} \tag{1539}$$

(compare with the simple case (667)). We can see immediately from this that, if we define a conjuposition group acting on Hilbert objects using the ortho-deterministic bases, we obtain the same group as using the ortho-physical bases. Thus, we obtain the natural conjuposition group using the ortho-deterministic basis as well.

We can decompose the identity using ortho-physical and ortho-deterministic

bases as follows

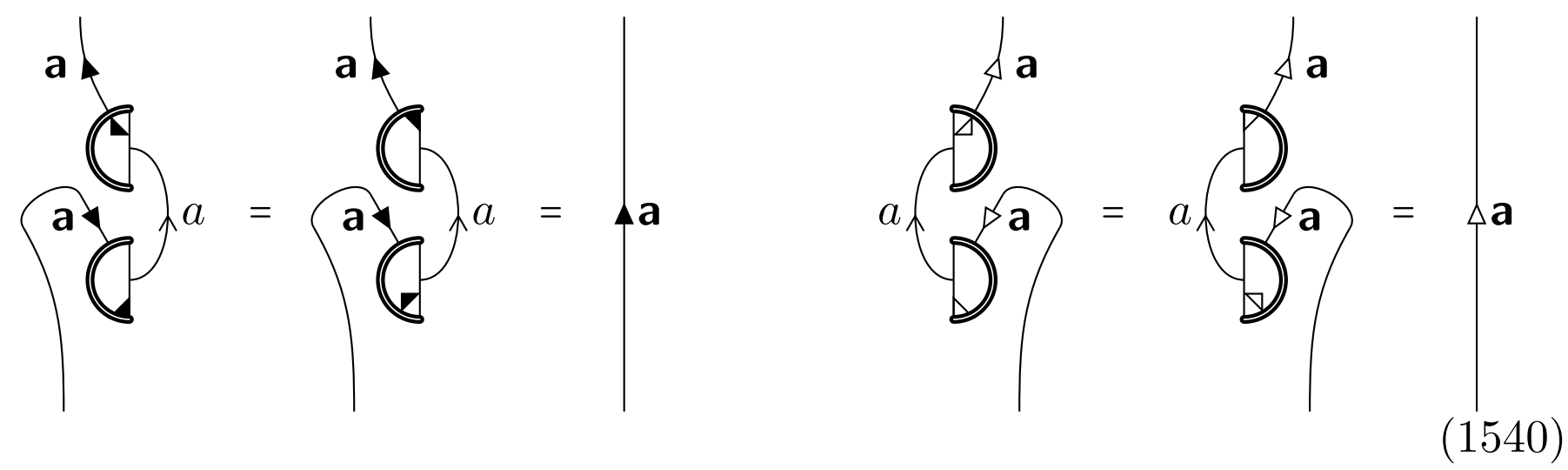

(1540)

(compare with (668)).

We note here the following useful result

(1541)

which follows from (1519, 1520) and (1540). This will be used in proving the simple dilation theorem in Sec. 78.3.

## 70.4 The specials

We saw, in the simple case, that there are ten distinct ways of joining an ortho-physical and an ortho-deterministic basis element at their label wires (as discussed in Sec. 33.10 and displayed in Table 5). It is, similarly, true that we have ten distinct ways of combining ortho-physical and ortho-deterministic basis elements in the complex case. We call these *the specials*. Two of the specials have been displayed already above in (1540) - they correspond to the decomposition of the identity. We can also define

(1542)

Note that, if we use the ortho-physical gauge condition, then we can swap the positions of the ortho-physical and the ortho-deterministic basis elements above (this follows from (1539)). This means that, if we join these specials end to end then we just get a wire with a black or a white arrow accordingly). These are the tugging equations for the specials (analogous to the yanking equations (705)). Further we have the *special twists*

(1543)

(1544)

(1545)

This gives us all the ten specials. Note that the six special twists are basis dependent objects whilst the other four specials are basis independent (this follows from similar reasoning to that used in Sec. 67.2).

We can use the special objects in (1542) to transform between ortho-physical bases. For example, we have

(1546)

and

(1547)

## 70.5 Using special objects to take the natural transpose

One important application of the specials in (1542) is that they can be used to take the natural transpose. Thus, for example, we have

(1548)

This is easily verified by substituting in the expansion of $B$ given in (1524). Since the special object used here is basis independent the natural transpose is basis independent.

## 70.6 Natural conjuposition of general Hilbert objects

We can take the natural conjuposition of a general Hilbert object as long as we assume the ortho-physical gauge (1514). This works for the same reasons as explained in the simple case (see Sec. 33.14). Consider the general Hilbert object

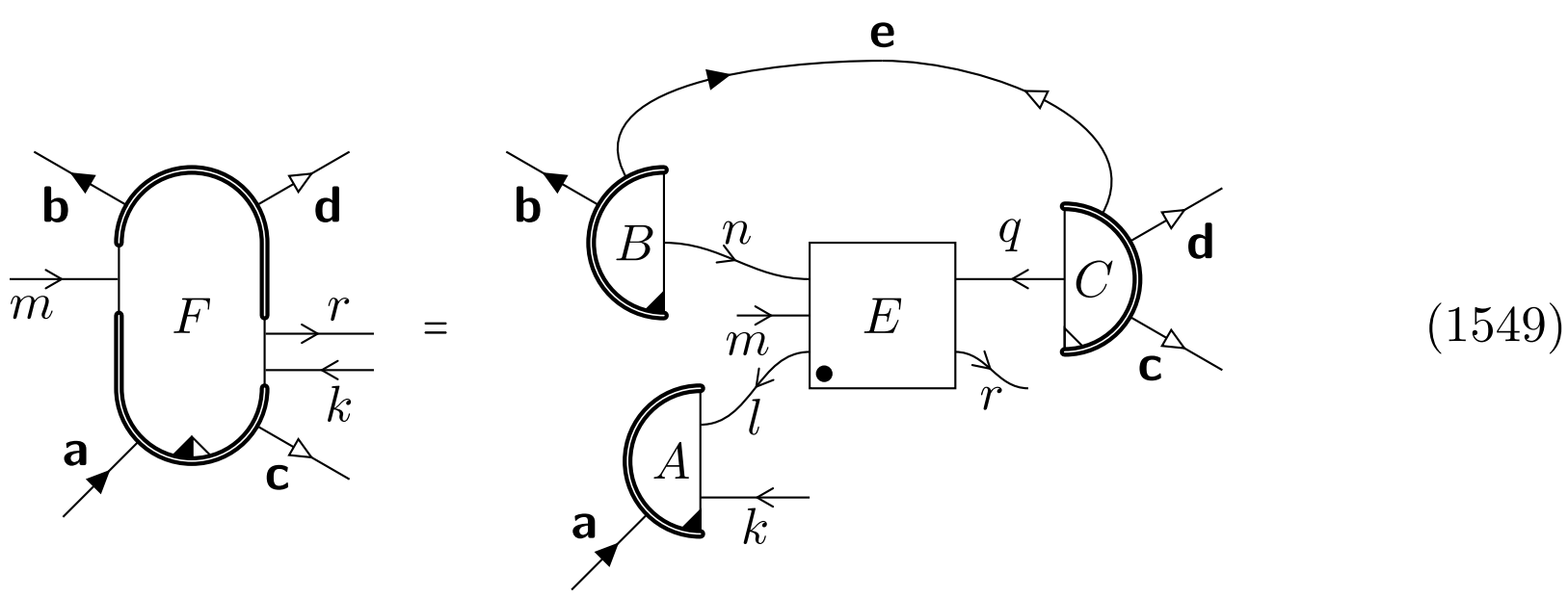

(1549)

(note we have included black/white triangle pair markings to track conjupositions). The natural transpose of this is

(1550)

(as long as we use the ortho-physical gauge). The Hilbert cube for general Hilbert objects under natural conjupositions is as follows

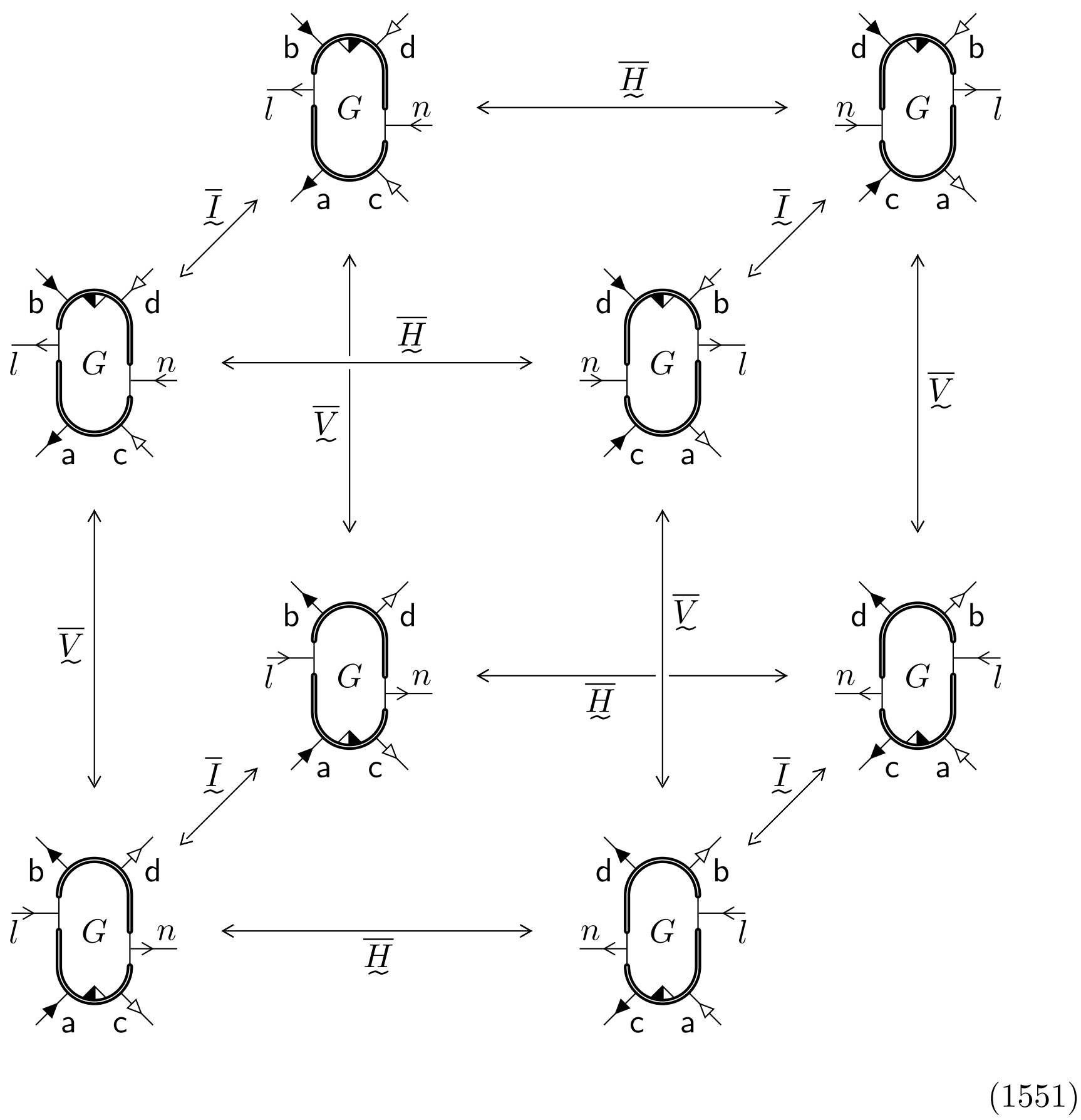

(1551)

Compare this with (1506).

## 70.7 Natural transpose of operator tensors

In Sec. 33.15 we saw how to take the natural transpose of simple operator tensor by attaching special system cups and caps and special pointer cups and caps. This worked even when the operator has pointer wires. We can take natural transpose of a complex operator in a similar way.

We define *special system objects* (analogous to the cups and caps defined in

(703)) by doubling up special objects as follows

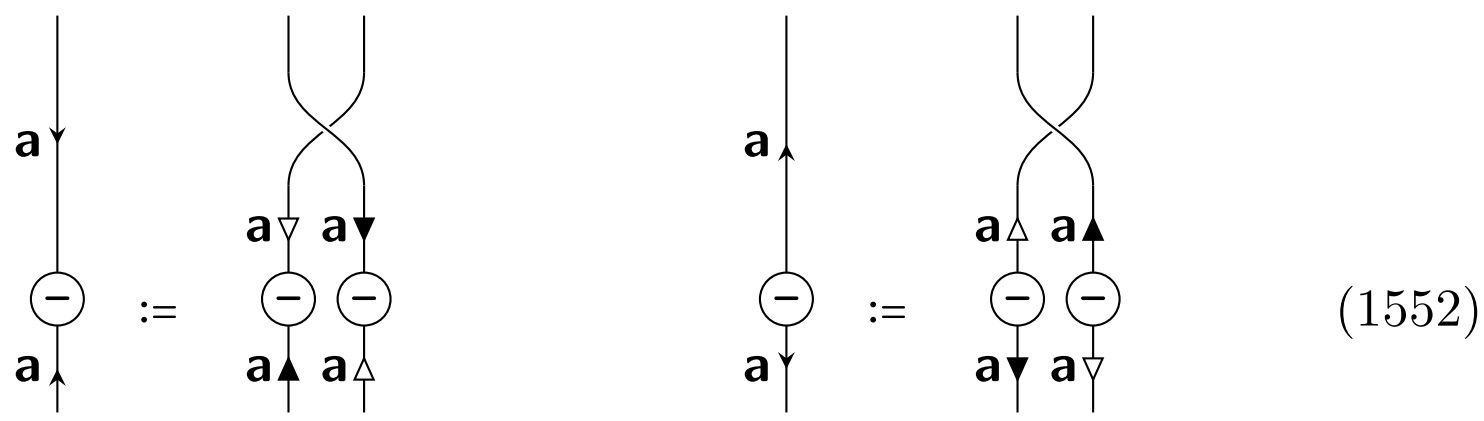

(1552)

We will define

(1553)

so we can compare the natural objects in (1552) with the corresponding normal objects. If we assume the ortho-physical gauge condition (1514) then we can show that

(1554)

(compare these equations with (704) from the simple case). The following tugging equation

(1555)

is satisfied (this analogous to the yanking equation in (705)).

We define *special pointer objects* (analogous to the cups and caps defined in (709)) as follows

(1556)

where

$$:= \begin{pmatrix} 1 & 0 & \cdots & 0 \\ 0 & 1 & \cdots & 0 \\ \vdots & \vdots & \ddots & \vdots \\ 0 & 0 & \cdots & 1 \end{pmatrix} \qquad \in \mathcal{P}_{\mathbf{x}_1} \otimes \mathcal{P}_{\mathbf{x}_2} \tag{1557}$$

and

$$:= \begin{pmatrix} 1 & 0 & \cdots & 0 \\ 0 & 1 & \cdots & 0 \\ \vdots & \vdots & \ddots & \vdots \\ 0 & 0 & \cdots & 1 \end{pmatrix} \qquad \in \mathcal{P}^{\mathbf{x}_1} \otimes \mathcal{P}^{\mathbf{x}_2} \tag{1558}$$

The coefficients in (1556) ensure that, when we can use these to convert between an $\boldsymbol{R}$ boxes with the arrow pointing out and an $\boldsymbol{R}$ box with the arrow pointing in (see (1370)). The special pointer objects satisfy a tugging equation

$$= \tag{1559}$$

as the coefficients in (1556) are reciprocals. This is analogous to the pointer yanking equation in (712).

To take the natural transpose we simply attach special objects as illustrated in the following example

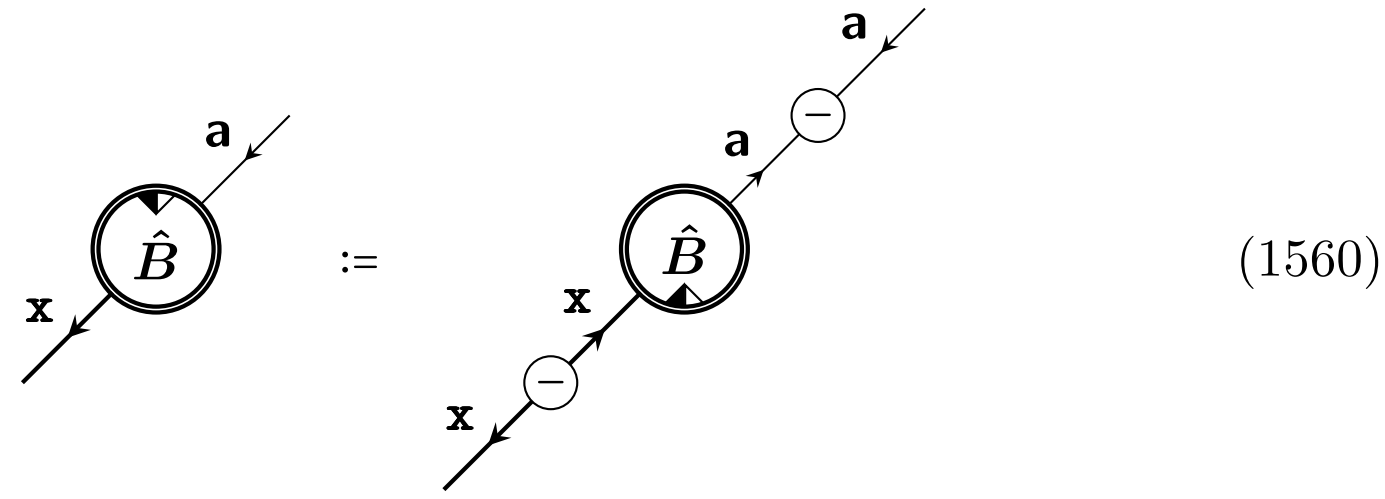

(1560)

(compare with the simple case in (713)). This reverses the direction of the wires. We have included black white triangular markings to indicate that the natural transpose has been taken.

The natural transpose of an operator network respects the compositional structure. In fact, we simply reverse the direction of all the arrows and change the position of the black/white triangular markings. For example

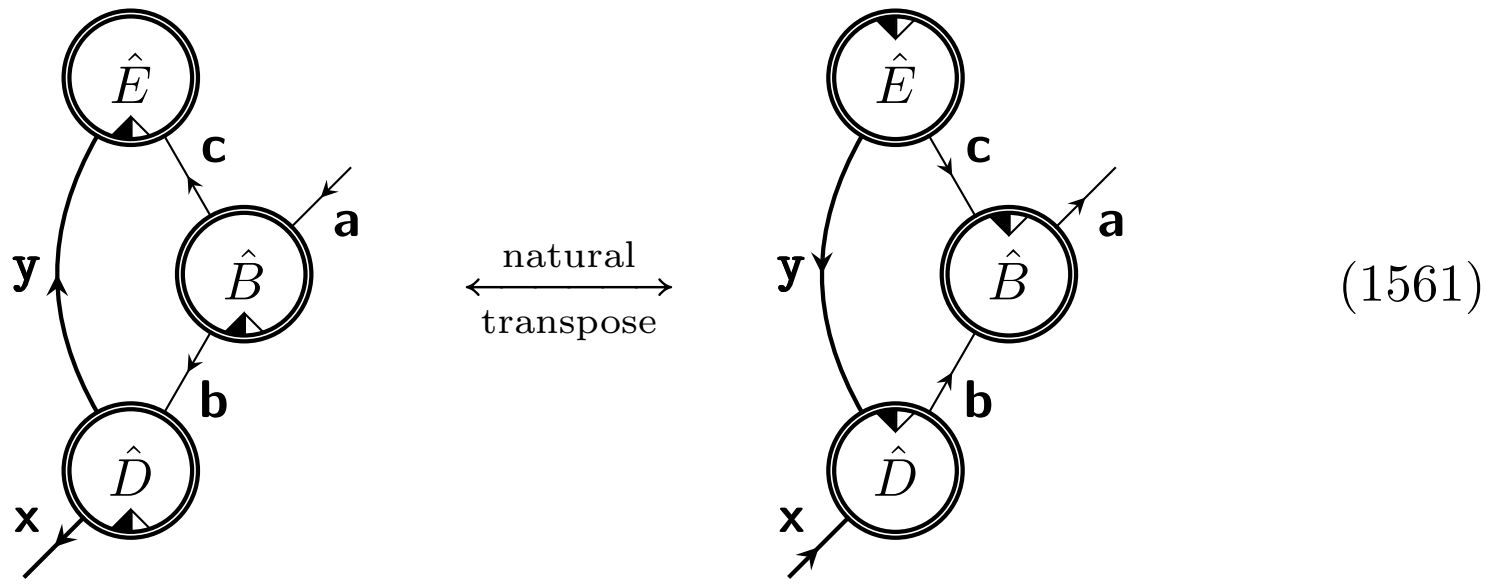

(1561)

This is obtained by applying the special system and special pointer objects to the open wires (as in (1560)). Then we can substitute in left hand side of the tugging equations (1555, 1559) for each closed wire. Now each operation is surrounded by $\ominus$'s and so we can substitute in the natural transpose.

It is worth noting that the notation here lacks expressiveness - we do not have a way to notate the effect of all of the basis independent natural conjupositions. Both $I$ and $\overline{\underset{\sim}{H}}$ leave the markings unchanged whilst $\overline{\underset{\sim}{V}}$ and $\underset{\sim}{T}$ vertically flip the markings. Since we are only interested in Hermitian operators this conflation does not cause us any problems. It is worth contrasting this with the notation we used for a general Hilbert object illustrated in the Hilbert cube in (1551) where the notation was more expressive since we had the flat sides where the label wires joined which enabled us to distinguish the effect of elements of the basis independent natural conjupositions (the front face) and, indeed, all natural conjupositions.

## 70.8 Natural transpose as time reverse

In Sec. 33.16, in the simple case, we saw that the natural conjupositions on operators are symmetries of the physics. Further, the natural transpose can be interpreted as implementing the time reverse of an operator. Similar remarks hold in the complex case. We will focus here on showing that the natural transpose can be treated as the time reverse.

Consider the following operator

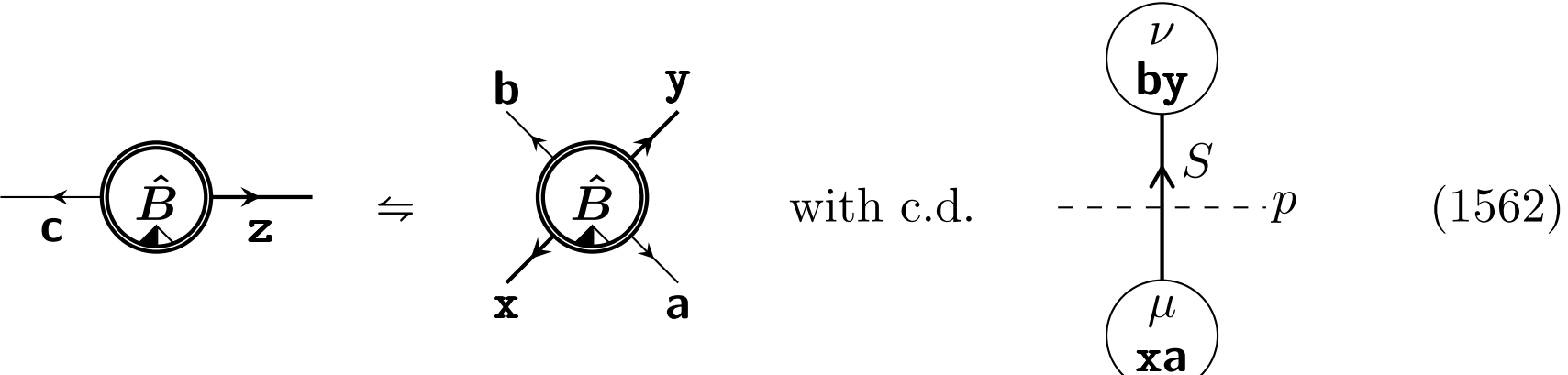

(1562)

We have included black-white triangle markings to track conjupositions. However, we have a notational deficiency here (that will not trouble us too much).

The natural transpose of this can be written

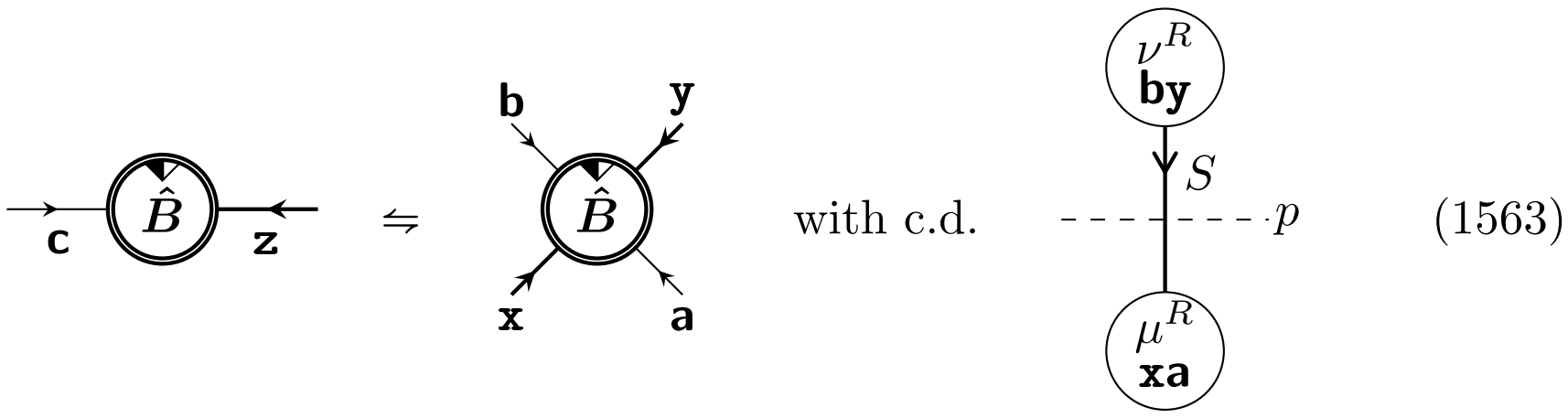

(1563)

where we have also reversed the arrows on the causal diagram since we intend to interpret this natural transpose as the time reverse (here $\mu^R$ and $\nu^R$ indicate reversing the arrows in these implicit causal structures).

The conditions for $\hat{B}$ to be physical that it satisfies tester positivity and double causality. Consider tester positivity first. The condition for tester positivity is given in (1378) (with respect to the tester in (1379)). Now consider double causality. The general forward causality condition is

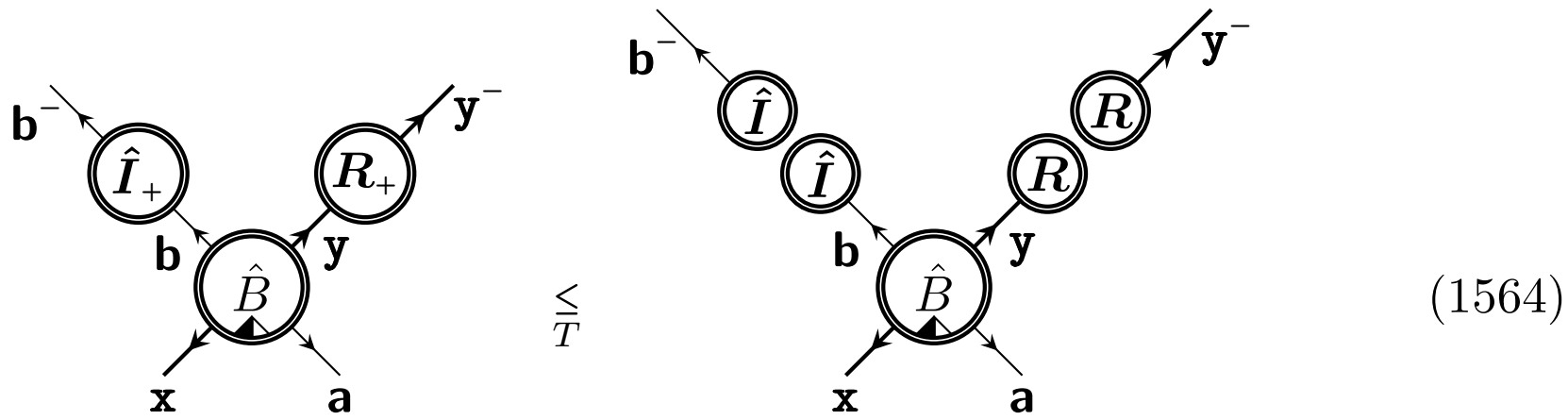

(1564)

The general backward causality condition is

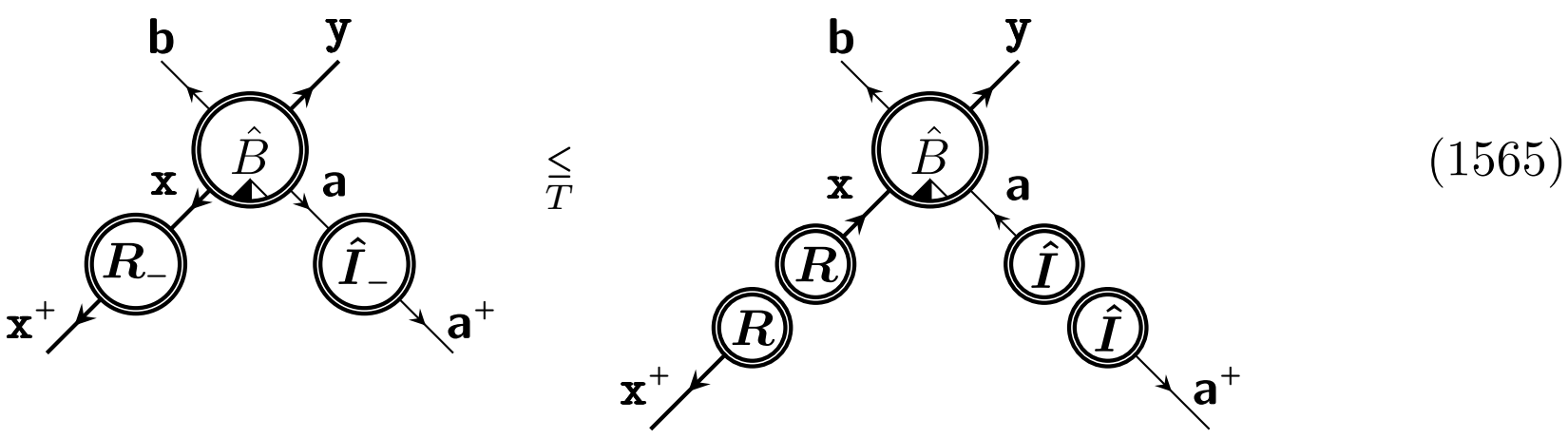

(1565)

These are the general double causality conditions (1204, 1205) under correspondence. The inequalities are saturated for deterministic operations. We have substituted in the expressions (1182) (under correspondence) for the residua, to avoid some notational confusion that might otherwise arise when we take the time reverse.

We laid out the conditions for a transformation on operations to constitute a time reverse in Sec. 48.4. Under correspondence to operators, these conditions require that the time reverse transformation (1) must take physical operators to physical operators , (2) must take readout boxes to readout boxes, and (3) the probability of any circuit is unchanged. We will see that both these conditions hold if we interpret the natural transpose as the time reverse transformation.

First, we can look at physicality. It is easy to verify that the tester positivity condition for the operator in (1562) is equivalent to the tester positivity condition for its natural transpose in (1563). Consider next the double causality conditions. These conditions use $\boldsymbol{R}_{\pm}$ and $\hat{\boldsymbol{I}}_{\pm}$. It is easy to verify that

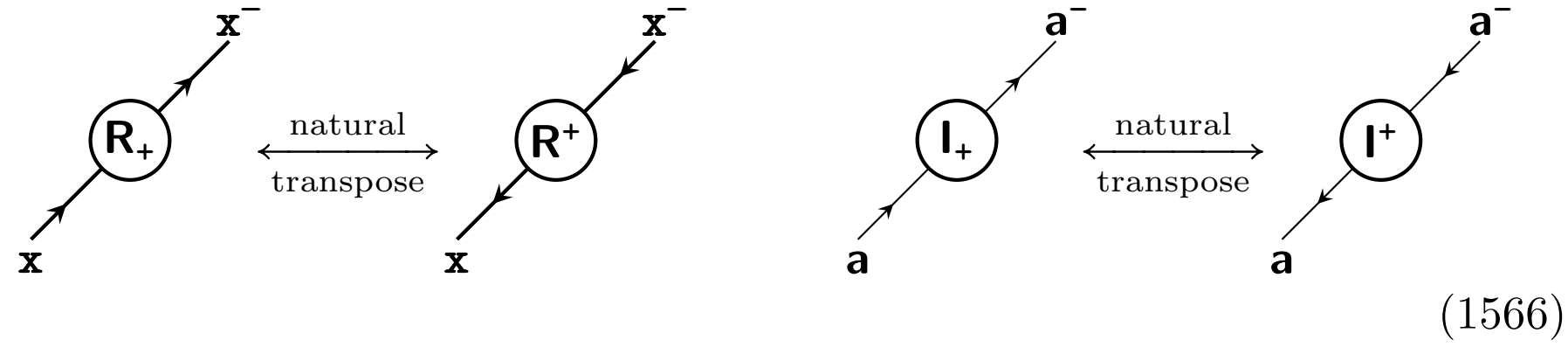

(1566)

using the definitions in (1091, 1092, 1103, 1104). If we take the natural transpose of forward causality condition above we obtain

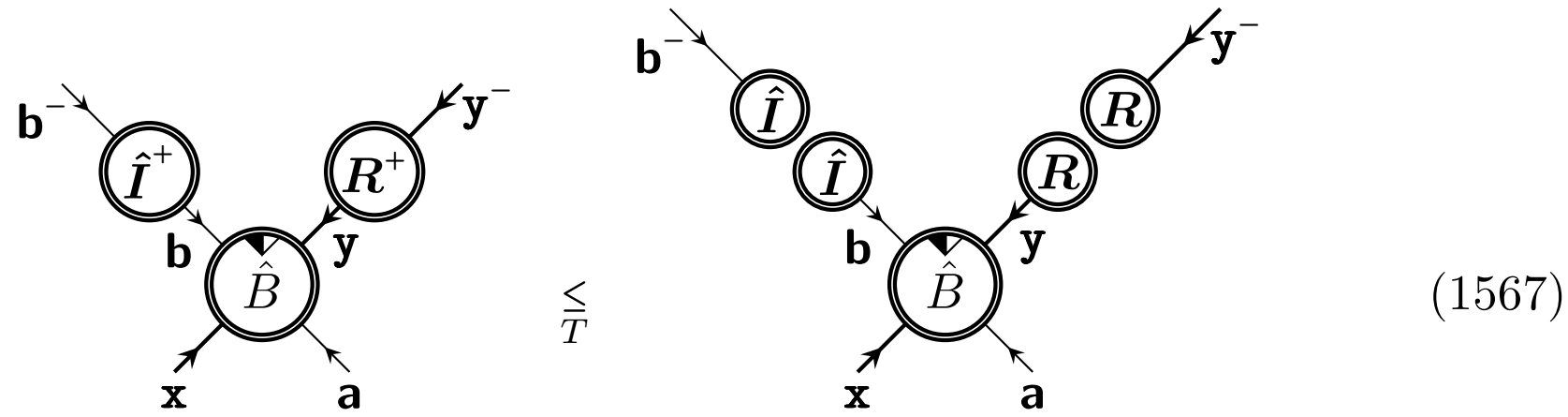

(1567)

which is the backward causality condition for the operator shown in (1563) (i.e. for the natural transpose of the operator we started with). If we take the natural transpose of the backward causality condition above we obtain

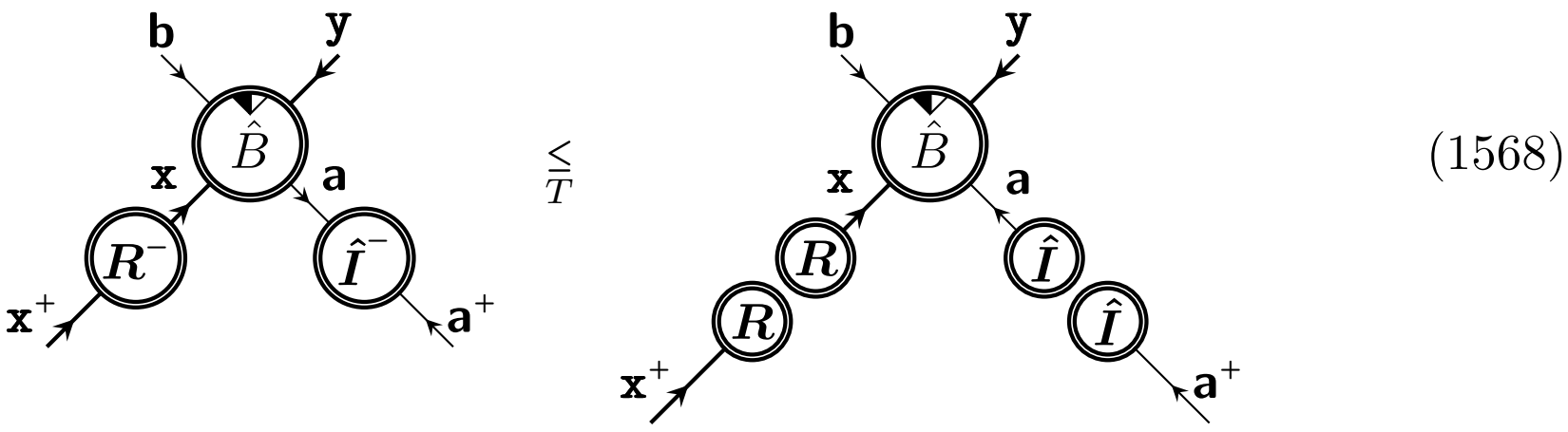

(1568)

which is the forward causality condition (1563). Thus, we satisfy the first requirement for the natural transpose to constitute a time reverse.

It is easily verified that the readout box transforms to a readout box under the natural transpose using (1371). So we satisfy the second requirement.

Furthermore, it is easy to see that an operator circuit is unchanged in value if we take the natural transpose of every element. For example,

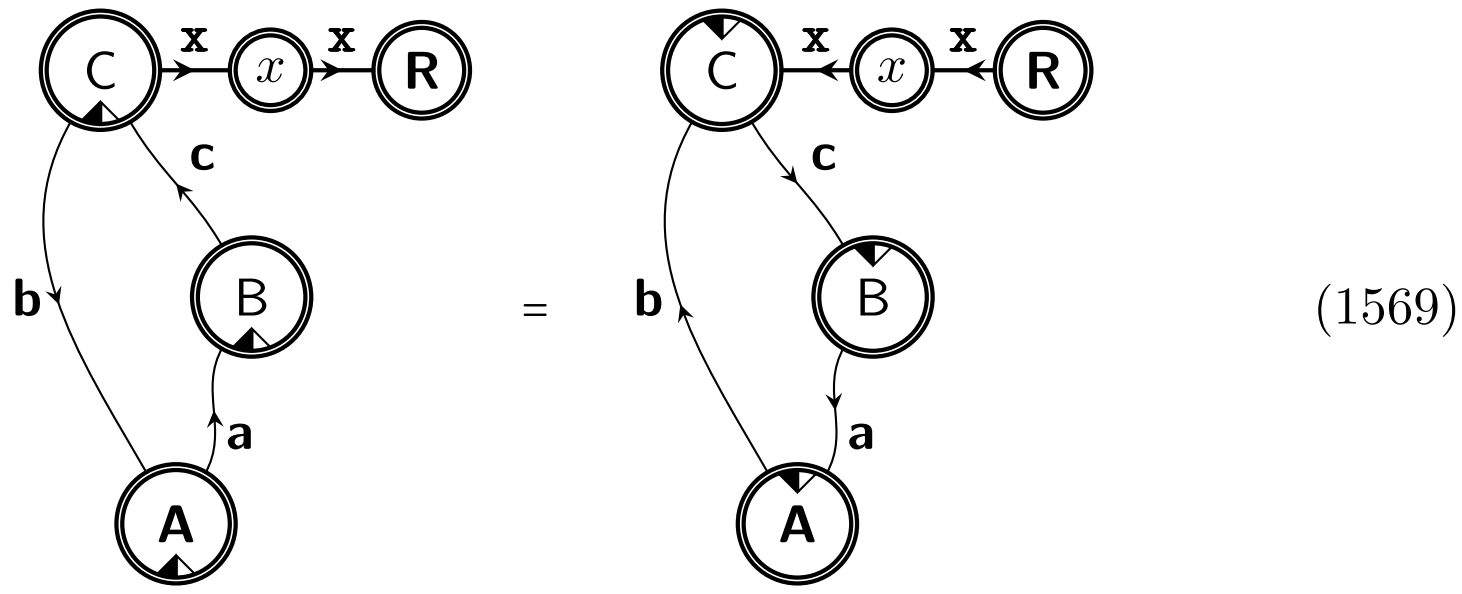

(1569)

To verify this we can substitute the expression for the transpose for each operator on one side. The specials cancel by virtue of the tugging equations. Thus, we satisfy the third requirement and so we have proven that the natural transpose can consistently be treated as the time reverse transformation.

Here we just discussed the natural transpose. It is easy to consider the effect of the other natural conjupositions (at least in the case where there are no pointers). Since the compositional structure of equations is preserved under natural conjupositions (as long as we assume the ortho-physical gauge condition) it is easy to show that the other natural conjupositions are also symmetries of the physics (arguing along the same lines as in Sec. 33.16).

# 71 Normal Mirrors

In this section we will introduce normal mirrors for complex Hilbert objects. In the next section we will discuss physical mirrors associated with natural conjupositions along with ortho-physical bases.

The underlying ideas are the same as for mirrors in the simple case. Mirrors are associated with conjugating conjupositions. However, since the notational conventions are a little different from the simple case, some of the diagrammatics will look different too. In particular, $\overline{V}$ stands for reVerse adjoint and so we reverse the arrows rather than flipping the diagram vertically (as we did in the simple case). Further, we now have arrow heads and square markings that can be black or white.

We can associate mirrors with the elements of the conjugating subset of the standard conjupositions. As in the simple case we can associated mirrors with the conjugating conjupositions. We associate regular normal mirrors with $\{\overline{H}, \overline{V}\}$ and we associate twist normal mirrors with $\{\overline{I}, \overline{T}\}$. We will provide definitions for all these mirrors. We will do this for general Hilbert objects. The Hilbert cube illustrated in (1506) will provide a useful reference to follow these definitions.

## 71.1 Regular normal mirrors

We have two types of mirror here. Associated with $\overline{H}$ is a vertical mirror (that reflects horizontally) such that

(1570)

The image in the mirror is the horizontal transpose. Here we have shown a mirror that is reflective on the left. We could also have a mirror that is reflective on the right. Here are some special applications of this equation involving bases that can be used to extend the notation

(1571)

(1572)

(1573)

and

(1574)

The last expression in the above special applications allows us to extend the notation such that wires with black and white arrowheads can attach to a vertical mirror.

Associated with $\overline{V}$ is a horizontal mirror (that acts vertically)

(1575)

Note that, unlike in the simple case, the image is not flipped vertically. Rather, we reverse all the arrows (as required by the reVerse adjoint, $\overline{V}$). The resulting image is, therefore, not strictly a "reflection" as such. We have shown a mirror that is "reflective" on the bottom. We could also have one that is reflective on the top. Here are some special applications of this equation involving bases

(1576)

(1577)

(1578)

and

(1579)

This extended notation means we can attach wires with black and white arrow heads to horizontal mirrors.

## 71.2 Twist normal mirrors

We can also define mirrors that have images generated by $\overline{I}$ and by $\overline{T}$. (In the simple case we thought of these mirrors as being filled with a special kind of smoke and we can do this here also.)

The mirror with an image generated by $\overline{I}$ is defined as follows

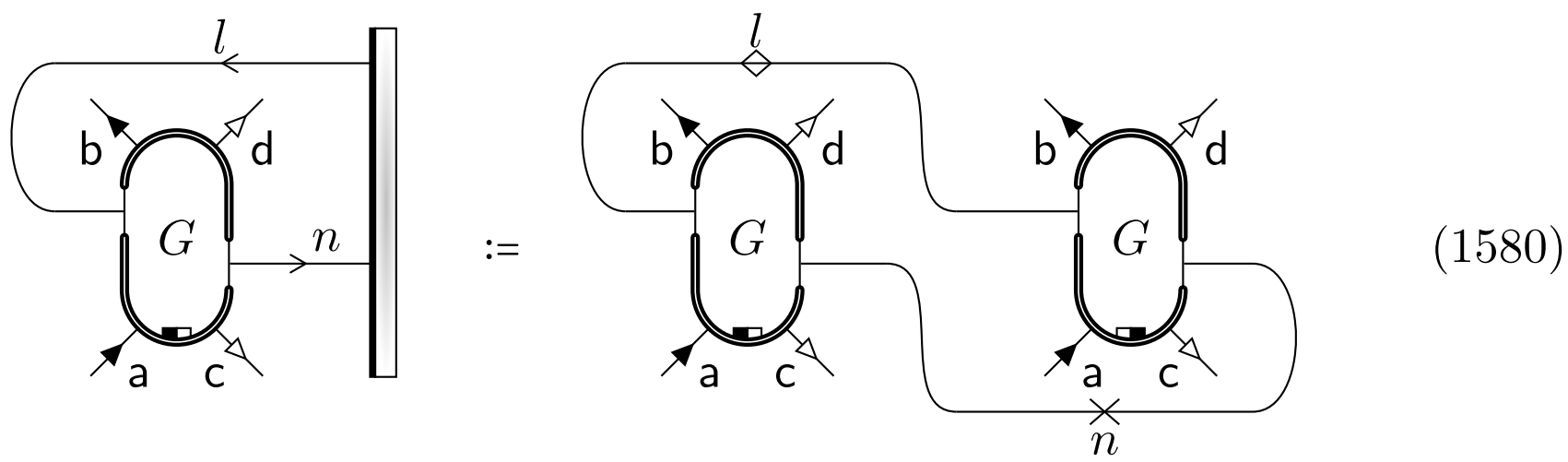

(1580)

Special applications involving bases are as follows

(1581)

(1582)

(1583)

(1584)

These allow us to extent the notation as indicated in the last expression in each case. Note that system wires that attach to the mirror acquire a twist, $w$, when reflected out.

The defining equation of the twist normal mirror associated with $\overline{T}$ is

(1585)

Special applications involving bases are as follows

(1586)

(1587)

(1588)

(1589)

These allow us to extend the notation as indicated by the last expression in each case. Again, note that wires that attach to the twist normal mirror acquire a twist, $w$, when reflected out.

## 71.3 Mirror theorem

The mirrors in the complex case are, formally, the same as the mirrors in the simple case albeit with some changes in notational convention. Thus we have a mirror theorem in the complex case as we did in the simple case (see Sec. 34.5). Here we will simply state the mirror theorem (the proof being along the same lines as in the simple case).

> **Normal mirror theorem.** Consider an equation, expressed such that the left hand side has a general Hilbert object, $C$, reflected in a conjugating normal mirror and the right hand side has another general Hilbert object, $D$, reflected in the same kind of mirror. This equation is equivalent to any equation formed such that the left hand side is expressed in terms of any conjuposition acting on $C$ reflected in any conjugating normal mirror and the right hand side has the same conjuposition acting on $D$ reflected in the same kind of mirror. Further, we require that the attached wires in the original equation are "dragged" so they attach to the mirror on each side of the equation in the corresponding way.

This is the same theorem statement as in the simple case. However, now we are referring to the mirrors for complex Hilbert objects. It is worth noting that the simple mirror theorem is "powered" by the properties illustrated in (761) concerning non-conjugating conjupositions on the image of a mirror expression. Similar properties pertain in the complex case and this can be used to prove the above theorem.

Let us illustrate the mirror theorem above with a simple example. If we have the equation

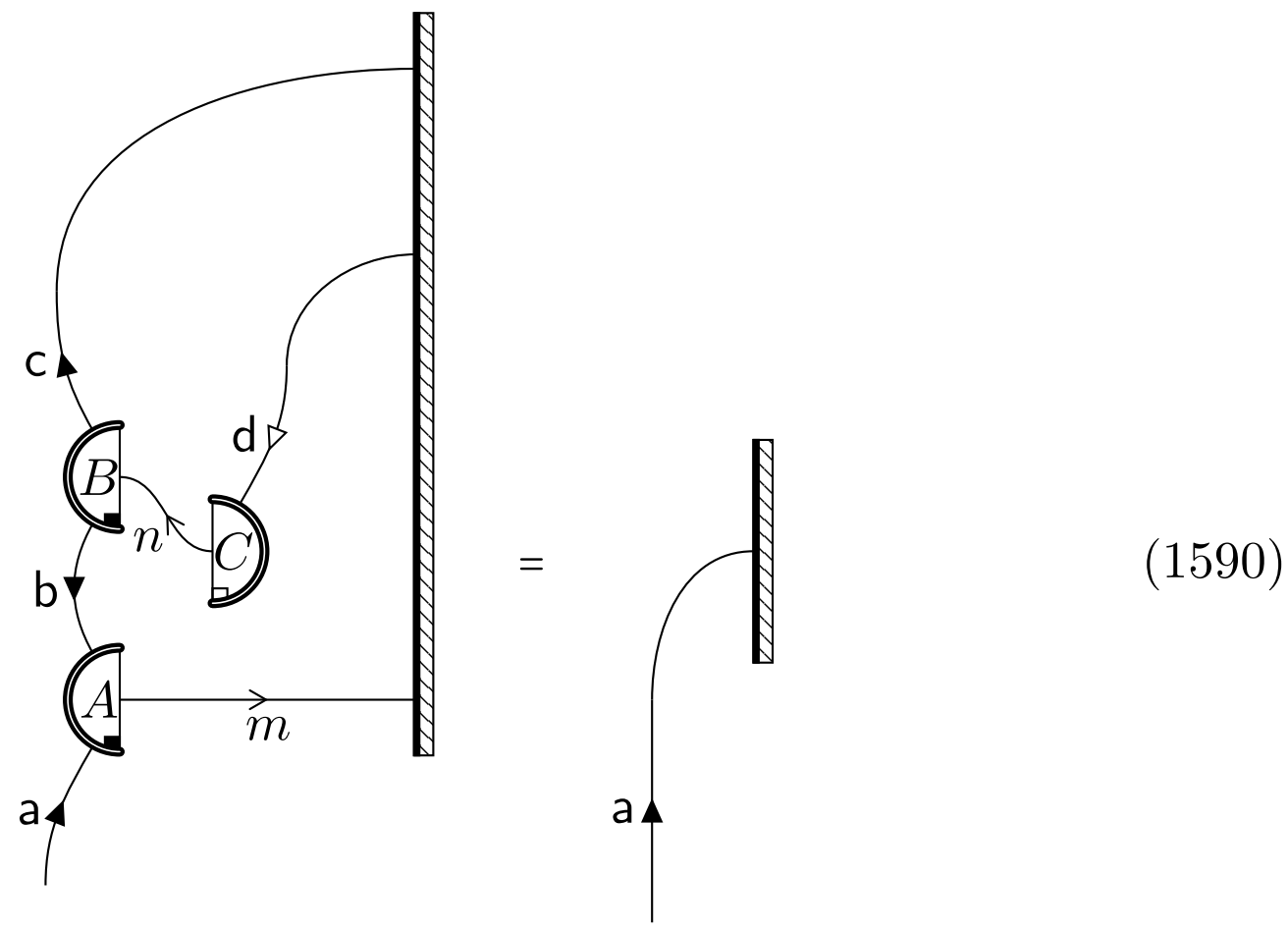

(1590)

then it follows from the normal mirror theorem that

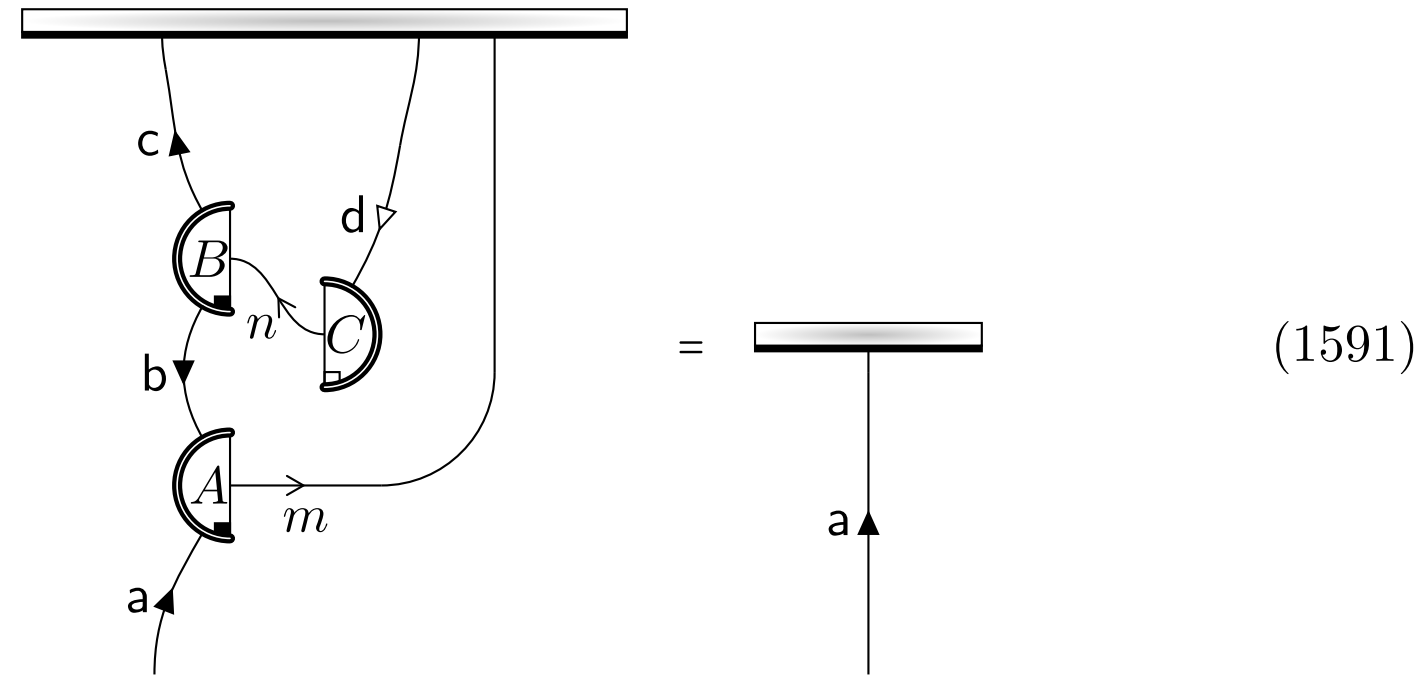

(1591)

is an equivalent equation.

### 71.4 Pulling out a wire

In Sec. 34.7 we discussed pulling out a wire in front of a regular normal mirror in the simple case. We can do the same for the complex case. We obtain, by similar reasoning,

$$\text{a} \quad = \quad \boxed{N_{\mathsf{a}}^{\frac{1}{2}}} \qquad (1592)$$

This follows from (1411) and the definition of a regular normal mirror given in (15711594, 1572). We can use the normal mirror theorem to find equivalent forms of the equation above (as we did in Sec. 34.7) in the simple case).

## 72 Physical mirrors

Physical mirrors are based on the conjugating members of the natural conjuposition group, namely $\{\overline{\underset{\sim}{H}}, \overline{\underset{\sim}{V}}, \overline{\underset{\sim}{I}}, \overline{\underset{\sim}{T}}\}$ and ortho-physical bases.

### 72.1 Regular physical mirrors

Regular physical mirrors form images with respect to $\overline{\underset{\sim}{H}}$ and $\overline{\underset{\sim}{V}}$.

First consider the regular physical mirror forming an image with respect to

$\overline{\underset{\sim}{H}}$. This is defined as follows

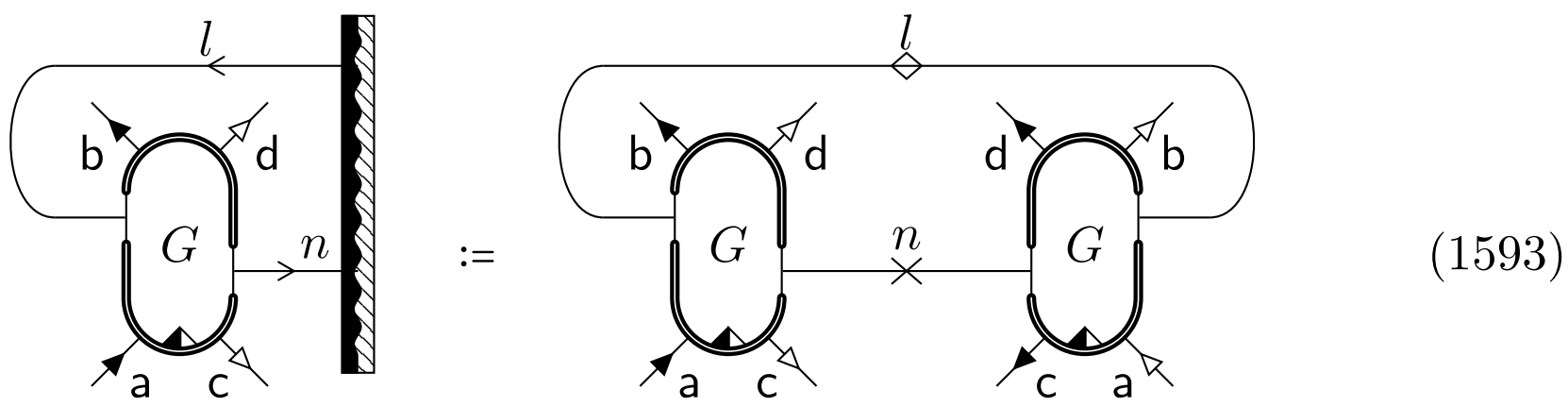

(1593)

and it has special applications

(1594)

(1595)

(1596)

and

(1597)

The last expression in each case extends the notation allowing system wires to connect to mirrors directly. Note that the properties of ortho-physical bases come into play here.

The regular physical mirror associated with $\overline{\underset{\sim}{V}}$ is defined as follows

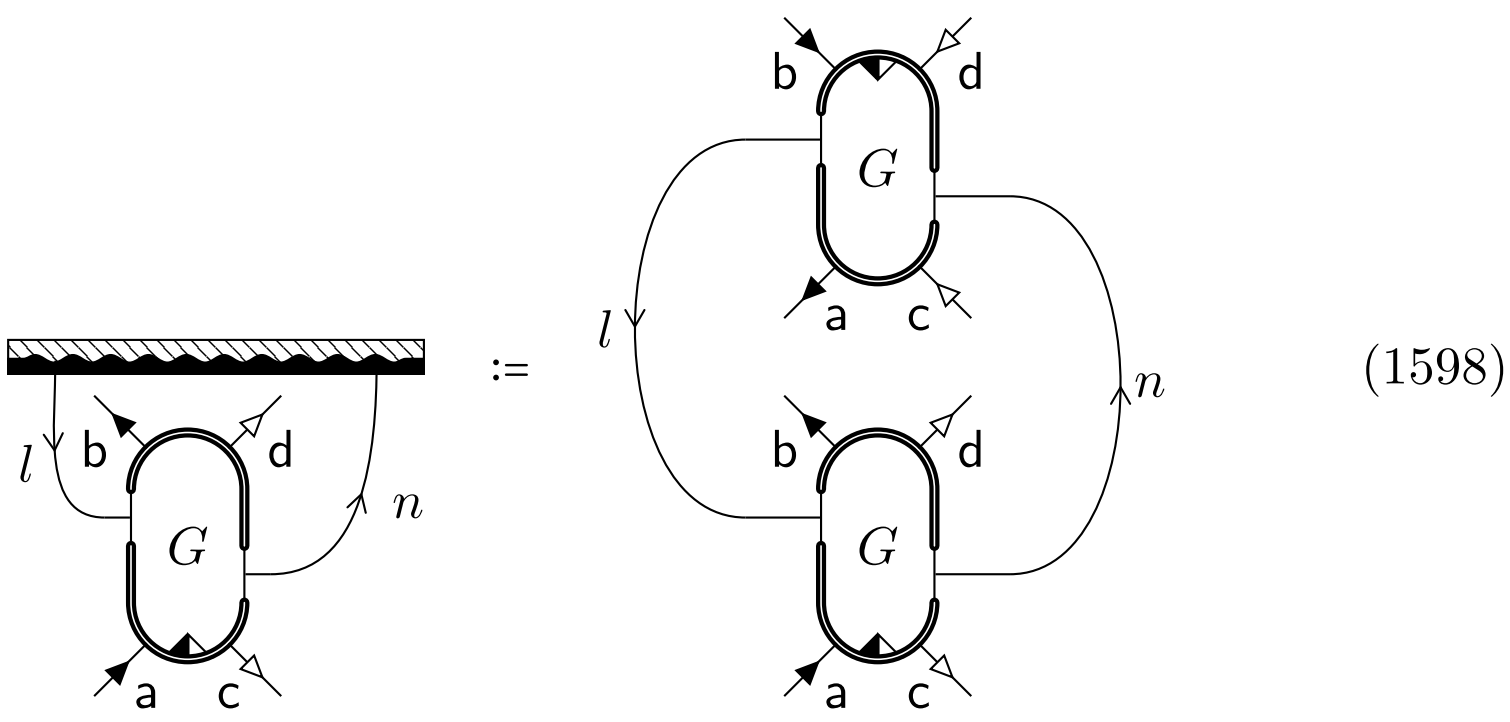

(1598)

Applying this to bases we have

(1599)

(1600)

(1601)

and

(1602)

In the final step in each case we see how to extend the notation (again, using the properties of ortho-physical bases).

It is interesting to compare (1599-1602) in the complex physical case above with (796-799) in the simple case where we needed to use special cups. In the simple case we only had two ways of connecting a system wire to the mirror

(1603)

However, in the complex case, we have four ways of connecting a system wire to the mirror

(1604)

For this reason we do not need to use special objects. This simplifies the physical mirror theorem in the complex case (see Sec. 72.3).

## 72.2 Twist physical mirrors

Now we define the twist physical mirrors. These are associated with $\overline{\underset{\sim}{I}}$ and $\overline{\underset{\sim}{T}}$.

The twist physical mirror associated with $\overline{\underset{\sim}{I}}$ is defined as follows

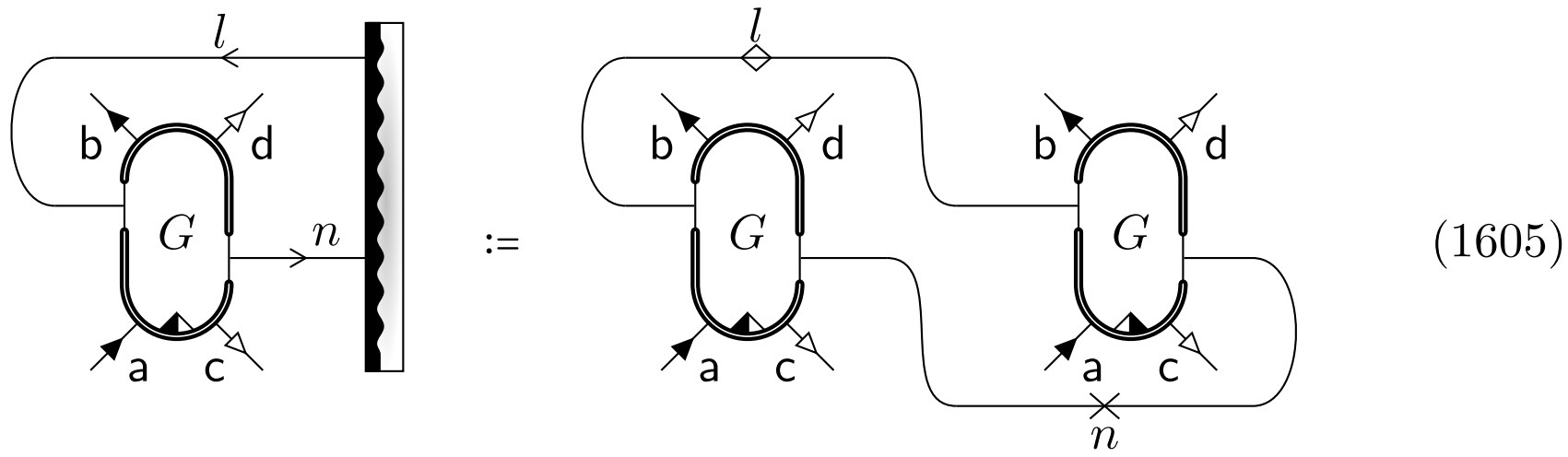

(1605)

Applying this to bases we obtain the following

(1606)

(1607)

(1608)

(1609)

The last expression in each case extends the notation. In particular, note the appearance of the special objects. This is analogous to what happened in the simple case.

The twist physical mirror associated with $\overline{\underset{\sim}{T}}$ is defined as follows

$$ (1610) $$

If we apply this to bases, we can obtain the following

$$ (1611) $$

$$ (1612) $$

$$ (1613) $$

$$ (1614) $$

The last expression in each case extends the notation.

## 72.3 Physical mirror theorem

We can obtain a physical mirror theorem in the complex case. Since the steps are formally similar to the physical mirror theorem in the simple case (as discussed in Sec. 35.3) we will simply state this theorem. It is worth noting, however, that the simple physical mirror theorem was "powered by the properties illustrated in (808) concerning taking non-conjugating physical conjupositions on the image of mirror expressions. Similar properties follow for the complex case. There is, however, an important difference which we will discuss below. The theorem is as follows.

> **Physical mirror theorem.** Consider an equation, expressed such that the left hand side has a general Hilbert object, $C$, reflected in a conjugating physical mirror and the right hand side has another general Hilbert object, $D$, reflected in the same kind of mirror. This equation is equivalent to any equation formed such that the left hand side is expressed in terms of any physical conjuposition acting on $C$ reflected in any conjugating physical mirror and the right hand side has the same physical conjuposition acting on $D$ reflected in the same kind of mirror. Further, we require that the attached wires in the original equation are "dragged" so they attach to the mirror on each side of the equation in the corresponding way.

The difference between this and the physical mirror theorem statement in the simple case is that we do not need to use special cups (or caps). The reason for this is because, in the complex case, we have black and white arrows and so we have four ways of connecting to the mirrors (as shown in (1604)) where as in the simple case we have no arrows and so only have two ways of connecting (see (1603)) and so need special cups or caps to deal with the other two cases that arise. See end of Sec. 72.1 for some discussion of this.

We will illustrate this with an example. The physical mirror equation

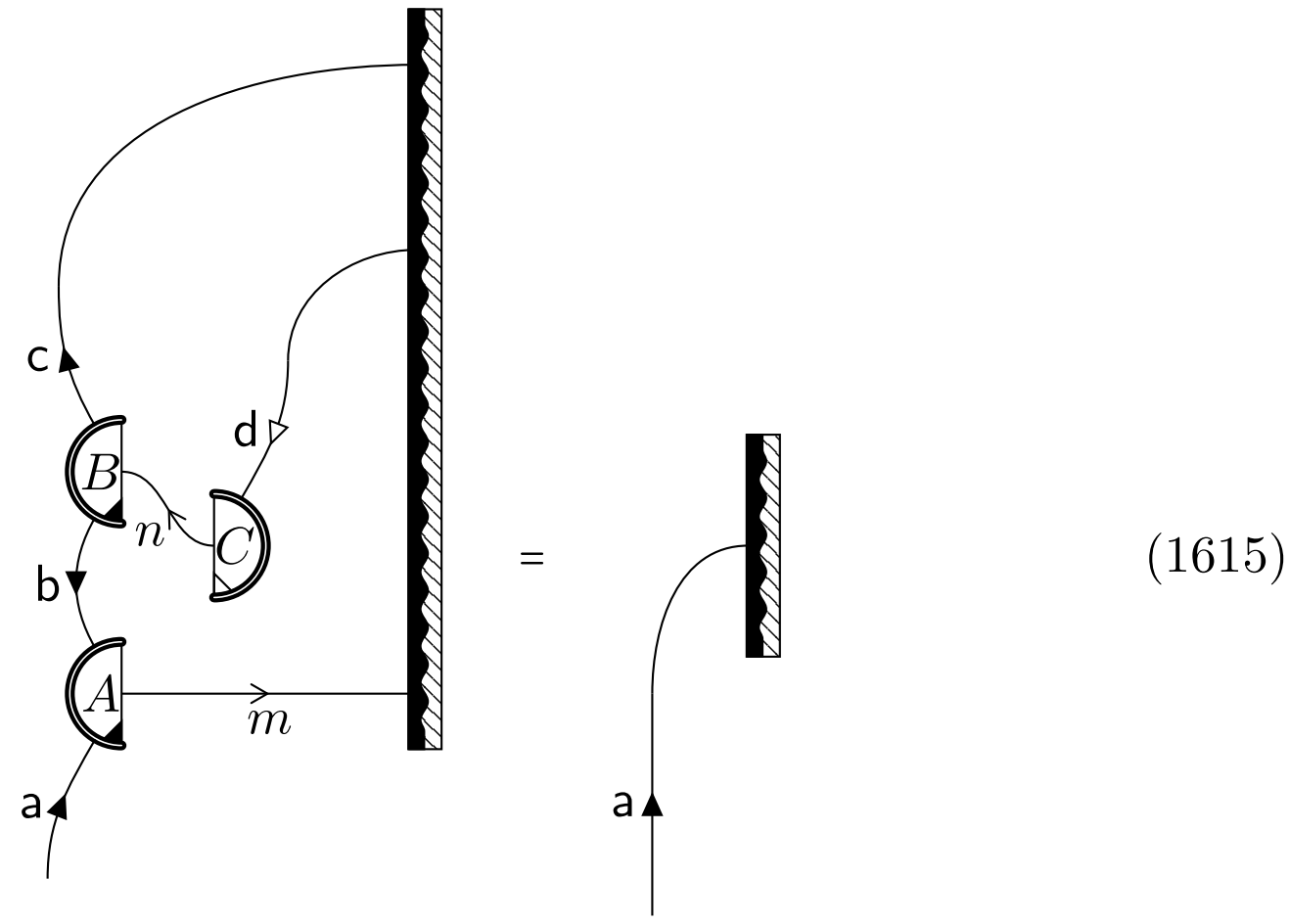

(1615)

is equivalent to the equation

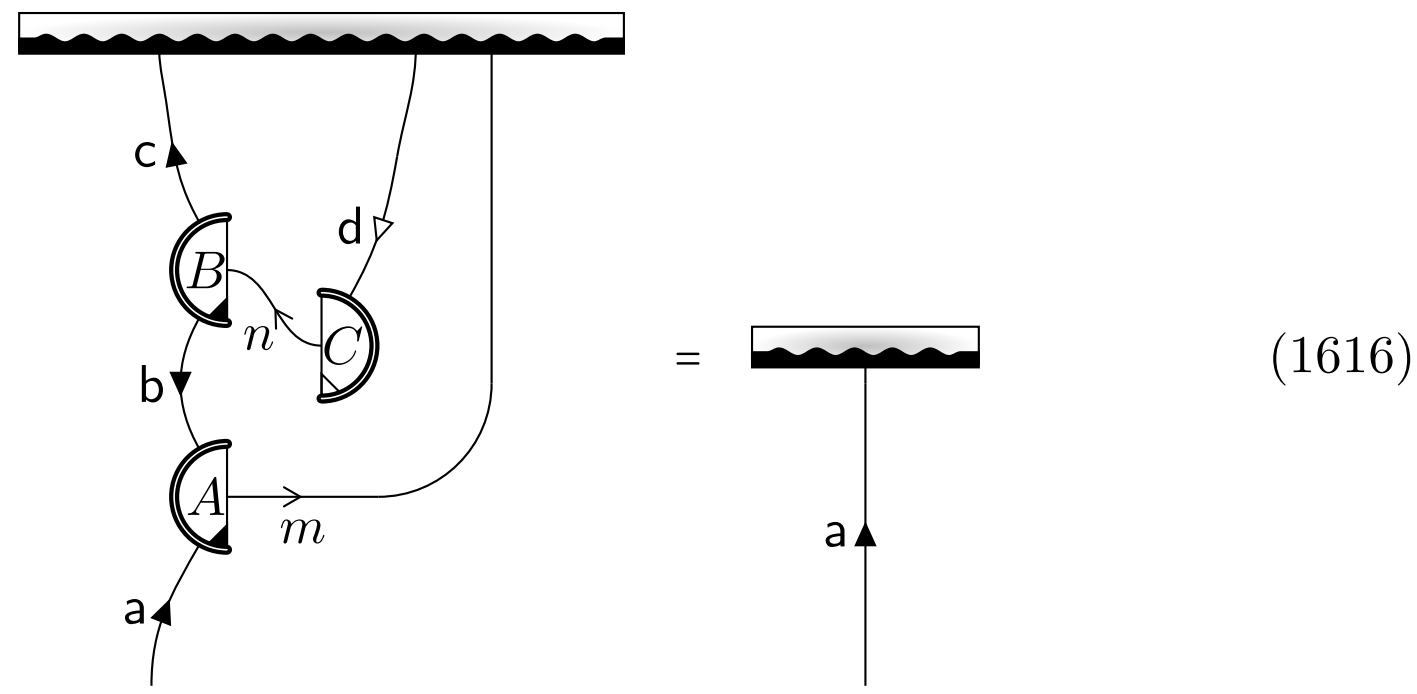

(1616)

Note that, unlike in the case of applications of the physical mirror theorem in the simple case, there are no special caps or cups (to see this compare the above two equivalent equations with the equivalent equations (809) and (814) where the latter equation does have special cups).

### 72.4 Pulling out a wire

In Sec. 35.5 we discussed pulling out a wire in front of a physical mirror in the simple case. We can do the same for the complex case. We obtain, by similar reasoning,

a = 1 (1617)

This follows from (1515) and the definition of a physical mirror given in (1594, 1595). We can use the physical mirror theorem to find equivalent forms of the equation above (as we did in Sec. 35.5) in the simple case).

## 73 Isometries, Coisometries, and Unitaries

In the simple case, we defined both normal and natural isometries, coisometries, and unitarities. We can do the same in the complex case also. There are three key differences however.

The first key difference is that, in the simple case, the system wires are either inputs or outputs whereas, in the complex case, the system wires have both positive and negative parts. It is natural to define normal and natural unitaries that act on these systems. These unitaries, in the causally complex case, can act on a basis set whose elements are a product between past and future pointing parts leading to a transformed basis set whose elements are entangled between these past and future pointing parts. This is fine if we think of this unitary as simply implementing a basis change. However, if we are interested in unitaries that correspond to actual evolution (that satisfy the physicality conditions) then we need to restrict ourselves to the special case of *temporal unitaries* where we separate out the positive and negative parts of the wires in the definition.

The second key difference relates to the first. In the simple case the unitary is always defined with respect to the given input output structure. However, in the complex case, we have to specify which *bipartition* we are using to define the unitary. For example, we might use the bipartition $(\mathbf{a}, \mathbf{b})$, or we might use some other bipartition associated with an interconvertible form for the given Hilbert object. This point will be clear shortly when we define the notion of a unitary for the complex case.

The third key difference is that in the simple case our operations have simple causal structure and so physicality entails satisfying $T$-positivity and simple (general) double causality. A consequence of this was that, in the simple case, all natural unitary operators are, in fact, physical. In the complex case we have the full set of double causality conditions applying to each synchronous partition through the causal diagram. We will see that a certain condition, natural temporal unitarity, is a necessary condition for physicality (it corresponds to deterministic simple double causality). However, it is not a sufficient condition because of the complex causal structure. Thus, we cannot say that natural temporal unitaries are necessarily physical (unless we have simple causal structure).

## 73.1 Normal case

Now we will provide definitions with respect to an orthonormal basis. Consider a left object

(1618)

We say this is a *normal unitary* with respect to the *bipartition* $(\mathbf{a}, \mathbf{b})$ if it satisfies the conditions

(1619)

The condition on the left is *normal isometry* and the condition on the right is *normal coisometry* . It is worth noting that normal isometry with respect to the bipartion $(\mathbf{a},\mathbf{b})$ is equivalent to normal coisometry with respect to the bipartition $(\mathbf{b},\mathbf{a})$ so we really have only one concept here.

In fact, we can show that the normal unitarity conditions (1619) can only be satisfied for a bipartition $(\mathbf{a},\mathbf{b})$ if $N_{\mathbf{a}} = N_{\mathbf{b}}$. The proof of this is best seen by writing the conditions in (1619) in the following equivalent form

(1620)

(using the normal mirror theorem). The condition on the left is *normal isometry* and the condition on the right is *normal coisometry*. Intuitively we can see that if $N_{\mathbf{b}} < N_{\mathbf{a}}$ then we have a channel capacity problem. In more detail: if we send in a basis set of cardinality $N_{\mathbf{a}}$ at the bottom of the equation on the left then, after passing through the first $U$, these must be projected down to a set which spans no more than $N_{\mathbf{b}}$ dimensions at the intermediate stage. There is no way to recover the original basis set of $N_{\mathbf{a}}$ elements after passing through the second $U$. A similar argument goes through for $N_{\mathbf{a}} < N_{\mathbf{b}}$ case using the equation on the right. Hence, we must have $N_{\mathbf{b}} = N_{\mathbf{a}}$ for any bipartition satisfying normal unitarity.

The normal unitary conditions above can also be expressed using horizontal mirrors

(1621)

We will use this form to motivate the definition of *natural unitaries* in Sec. (73.5).

It may turn out that $U$ satisfies normal unitarity with respect to one bipartition but not another. For example, if we have

(1622)

it might be that the normal unitarity conditions are satisfied for the bipartition $(\mathbf{a},\mathbf{b})$ but not the bipartition $(\mathbf{c},\mathbf{d})$. Indeed, this would have to be the case if $U$ was normal unitary with respect to the $(\mathbf{a},\mathbf{b})$ bipartition and we have $N_{\mathbf{c}} \neq N_{\mathbf{d}}$.

## 73.2 Transforming basis using a normal unitary

We can use a normal unitary

(1623)

with respect to the bipartition $(\mathsf{a}, \mathsf{a})$ to transform an orthonormal basis as follows

(1624)

We can see, using the definition of normal unitarity in (1620), that the orthogonality conditions in (1412) are satisfied for this new basis. We can also see, using these normal unitarity conditions, that the decompositions of the identity in (1413) and (1414) are invariant under a change of basis.

We can, in fact, use normal unitaries to implement a more general notion of basis change, where we go from a basis for system of type $\mathsf{a}$ to a basis for a system of type $\mathsf{b}$ (where $N_{\mathsf{a}} = N_{\mathsf{b}}$). We can use a normal unitary

(1625)

with respect to bipartition, $(\mathsf{a}, \mathsf{b})$, to implement the basis transformation

(1626)

where $\pi$ is a permutation matrix (having a single 1 in each column and each row, and 0's elsewhere). We can verify that this gives an orthonormal basis for the system type $\mathbf{b}$ (by verifying the orthonormality relations in (1416) hold for $\mathbf{b}$ for this basis). What is particularly interesting here is that, if $N_{\mathbf{a}}$ is a composite number, $\mathbf{a}$ and $\mathbf{b}$ may factorise into +ve and −ve parts differently (so that $N_{\mathbf{a}^+} \neq N_{\mathbf{b}^+}$ and $N_{\mathbf{a}^-} \neq N_{\mathbf{b}^-}$. For example, $\mathbf{a}$ could have both future and past moving parts, while $\mathbf{b}$ could have only a +ve part.

We can use (1626) to obtain the following

(1627)

where we have used (1413) and the fact that the horizontal transpose of a permutation matrix is its inverse. Thus, any normal unitary can be written in terms of basis elements for the systems and a permutation matrix.

## 73.3 Normal unitary operator tensors

Given any normal unitary left object

(1628)

(with respect to bipartition $(\mathbf{a}, \mathbf{b})$). we can define a *normal unitary operator tensor*

(1629)

(with respect to the same bipartition). This is homogeneous. The normal unitarity properties in (1619) can be written as

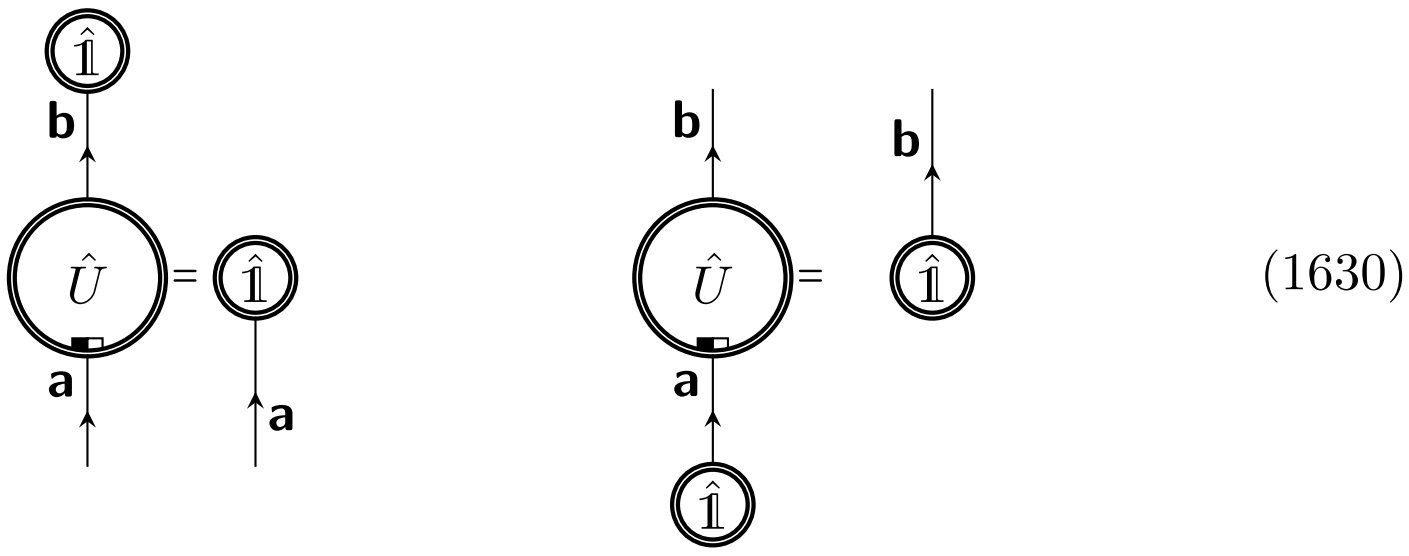

(1630)

under the constraint that $\hat{U}$ is homogeneous (the definition of $\hat{\mathbb{1}}$ is implicit in (1433, 1434)). We can also write the conditions for unitarity as

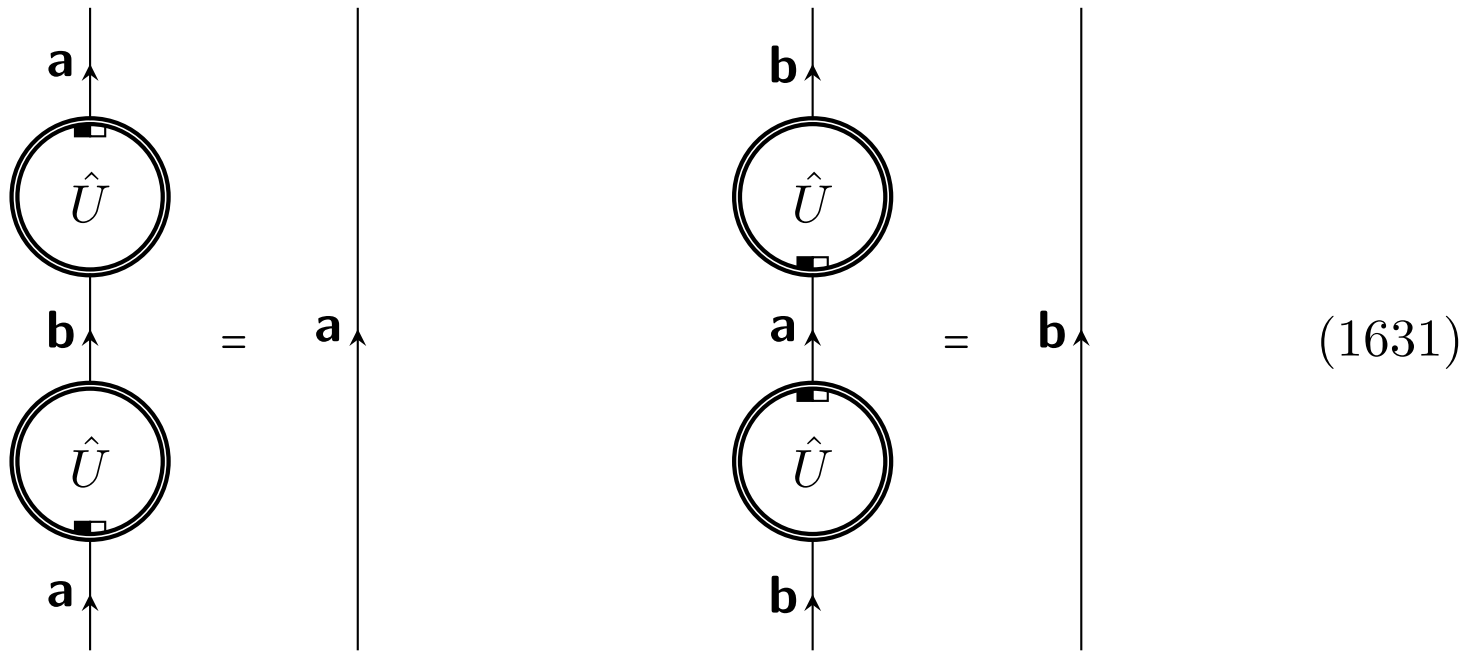

(1631)

under the assumption, again, that $\hat{U}$ is homogeneous. This follows by doubling up (1620).

## 73.4 Normal temporal case

Normal unitary operators are not necessarily physical. One requirement for physicality is that simple double causality is satisfied with respect to what we will call a temporal bipartition where we separate out the input and output systems. A second requirement is that we use ortho-physical bases rather than orthonormal bases and this will take us to the case of natural temporal unitary operators (see Sec. 73.6 and Sec. 73.7). For the time being, however, it is interesting to study normal temporal unitary operators.

We say that we have *normal temporal unitarity* if we have unitarity with respect to the bipartition of the form $(\mathbf{a}^+, \mathbf{b}^+)$ (where $\mathbf{a}^+$ represents the input and $\mathbf{b}^+$ represents the output). The bipartition into input and output systems is unique so we do not need to specify the bipartition. The temporal normal unitarity conditions can be written in the form

(1632)

We call the condition on the left *forward temporal normal unitarity* and the condition on the right *backward temporal normal unitarity*. These essentially the same as the normal unitarity conditions (857) in the simple case (where a temporal bipartition is built into the framework). For a temporal normal unitary, the bipartition is always with respect to inputs (i.e. positive on arrows pointing in and negative on arrows pointing out) and outputs (negative on arrows pointing in and positive on arrows pointing out). As such, the bipartition is unique and so does not need to be specified. Using the normal mirror theorem, the temporal normal unitarity conditions in (1632) can be put in the following equivalent form

(1633)

We call the condition on the left *temporal normal isometry* (it is equivalent to forward temporal normal unitarity) and the condition on the right *temporal normal coisometry* (it is equivalent to backward temporal normal unitarity). It is clear that the conditions for normal unitarity can only be satisfied if $N_{\mathsf{a}^+} = N_{\mathsf{b}^+}$. For completeness, here is the condition for temporal normal unitarity in mirror form

(1634)

We could get an equivalent form by rotating the mirrors clockwise by 90°.

Consider a homogeneous operator

(1635)

The bipartition into forward and backward moving systems is unique and so we can write down conditions for this to be a temporal normal unitary operator. These conditions are

(1636)

Note that these conditions can only be satisfied if $N_{\mathbf{b}^+} = N_{\mathbf{b}^-}$.

## 73.5 Natural case

The defining properties of a normal unitary, with respect to bipartition $(\mathbf{a}, \mathbf{b})$, given in (1620), can be written in mirror form as

(1637)

To define *natural unitaries* we replace the normal mirrors in this definition by physical mirrors. This gives the defining properties of natural unitaries, with respect to bipartition $(\mathbf{a}, \mathbf{b})$, as

(1638)

The condition on the left is the *natural isometry* condition (from which it can be proven that $N_{\mathbf{b}} \geq N_{\mathbf{a}}$ using the channel capacity argument of Sec. 36). The condition on the right is the *natural coisometry* condition (from which it can be proven that $N_{\mathbf{b}} \leq N_{\mathbf{a}}$). As in the normal case, natural isometry with respect to the bipartion $(\mathbf{a}, \mathbf{b})$ is equivalent to natural coisometry with respect to the bipartition $(\mathbf{b}, \mathbf{a})$ (so, really, we have only one concept here).

Therefore natural unitaries (which satisfy both conditions) have $N_{\mathbf{a}} = N_{\mathbf{b}}$. By arguing as we did in Sec. 38 for the simple case, we can prove that the notions of normal unitaries and natural unitaries are equivalent if either (i) the two systems of the bipartition are actually of the same type, or (ii) we use the $\gamma(N)$ gauge in (1533). For unitaries we can reflect out the mirrors in (1638) and use the fact that $N_{\mathbf{a}} = N_{\mathbf{b}}$ to cancel overall constants (see the analogous discussion in Sec. 38.2 in the simple case concerning equations 870 and (871)) to obtain

(1639)

Note that these are different conditions from the conditions in (1620) for normal unitaries because we take the reverse natural adjoint, $\underset{\sim}{\overline{V}}$, rather than the normal reverse adjoint, $\overline{V}$, to convert the lower $U$ to the upper $U$ in each equation.

We can use natural unitaries with respect to a bipartition, $(\mathbf{a}, \mathbf{b})$, to transform between deterministic bases of $\mathbf{a}$ and $\mathbf{b}$ as follows

(1640)

The argument for this is the same as for (878) (in the simple case). The $\pi$ boxes are permutation matrices. We can prove that the orthogonality conditions (1515) are satisfied for the shaded basis given that they are satisfied for the unshaded basis (we need to use the above established fact that $N_{\mathbf{a}} = N_{\mathbf{b}}$ to prove this). Further, it follows from this that we can write any natural unitary

as

(1641)

(compare with (1627)). This is proven in the same way as (879) was proven in the simple case (i.e. by appending ortho-physical basis elements at the label wire to both sides of (1640) and using (1540) to simplify the right hand side).

We can define a *natural unitary operator tensor* as

(1642)

This is homogeneous. The natural unitarity conditions in (1638) can be written as

(1643)

under the condition that $\hat{U}$ is homogeneous (to see this, it is best to rotate the physical mirrors in (1638) to obtain equivalent equations using the physical mirror theorem). Alternatively, the natural unitarity conditions can be written as

(1644)

under the condition that $\hat{U}$ is homogeneous. This follows by doubling up the equation in (1639).

It is worth noting that, if we impose the $\gamma(N)$-gauge fixing condition (1533) then (1643) and (1630), from the case of normal unitary operators, become equivalent. Similarly, under the same gauge fixing condition, (1631) and (1644) become equivalent.

## 73.6 Natural temporal case

The natural temporal case is where we use the (unique) bipartition into input and output systems. A natural temporal unitary satisfies

(1645)

The condition for natural temporal isometry is on the left and for natural temporal coisometry is on the right. We require $N_{\mathbf{a}^+} = N_{\mathbf{b}^+}$ for a natural temporal unitary (when both of these conditions are satisfied).

Since the bipartition is unique for the temporal case, we can provide natural temporal unitarity conditions for any homogeneous operator. Thus,

(1646)

is a natural temporal unitary operator iff

(1647)

These conditions can only be satisfied if $N_{\mathbf{b}^+} = N_{\mathbf{b}^-}$. Since $\mathbf{b}$ can be composite, this definition works for any homogeneous operator.

## 73.7 Natural temporal unitary operators

Here we will discuss *natural temporal unitary operators*. If the causal structure is simple, then these operators are physical (as well as being deterministic).

They are, then, an important object for us since they go beyond being merely mathematical objects.

First, let us consider an operator built out of natural temporal unitaries as follows

(1648)

where $\hat{U}$ satisfies the natural temporal unitarity conditions in (1645). We will call it a *natural temporal unitary operator*. Using the physical mirror theorem (see Sec. 72.3) we can write these conditions as

(1649)

If we reflect these equations out and use (1648) and (1520) then we obtain

(1650)

These are the simple double causality conditions for $\hat{U}$ (see by correspondence with the conditions in Sec. 49.4.1). These are necessarily satisfied for any deterministic physical operator. If we have simple causal structure of the form

(1651)

(see Sec. 47.16 for more discussion) then the simple double causality conditions are the full set of double causality conditions. Since $\hat{U}$ is homogeneous it is necessarily $T$-positive (as it is in twofold form). Consequently, if an operator built from natural temporal unitaries as in (1648) has simple causal structure then it is physical.

More generally, if the causal structure is not simple, then there are more double causality conditions to be satisfied.

## 73.8 Physical deterministic homogeneous operators

We can use the ideas above to prove the following theorem

> **Deterministic homogeneous operator tensors.** Deterministic homogeneous operators satisfying simple double causality are natural temporal unitary operators and vice versa.

The "vice versa" part of this was already proven in Sec. 73.7 (note that determinism follows from the fact that the forward (or backward) simple causality condition is satisfied as we saw in (1650)). To complete the proof of this theorem, note that a homogeneous operator tensor without pointer wires has the form

(1652)

If it is physical and deterministic then it satisfies the two simple double causality conditions

(1653)

We can write

(1654)

where $\mathsf{a}^+ = (\mathsf{b}^-)^R$ Then the simple double causality conditions (1653) become

(1655)

Given that $\hat{B}$ is homogenous it is easy to see that it satisfies which are the natural temporal unitarity conditions in (1649) with $\hat{B}$ in place of $U$. This completes the proof of the theorem.

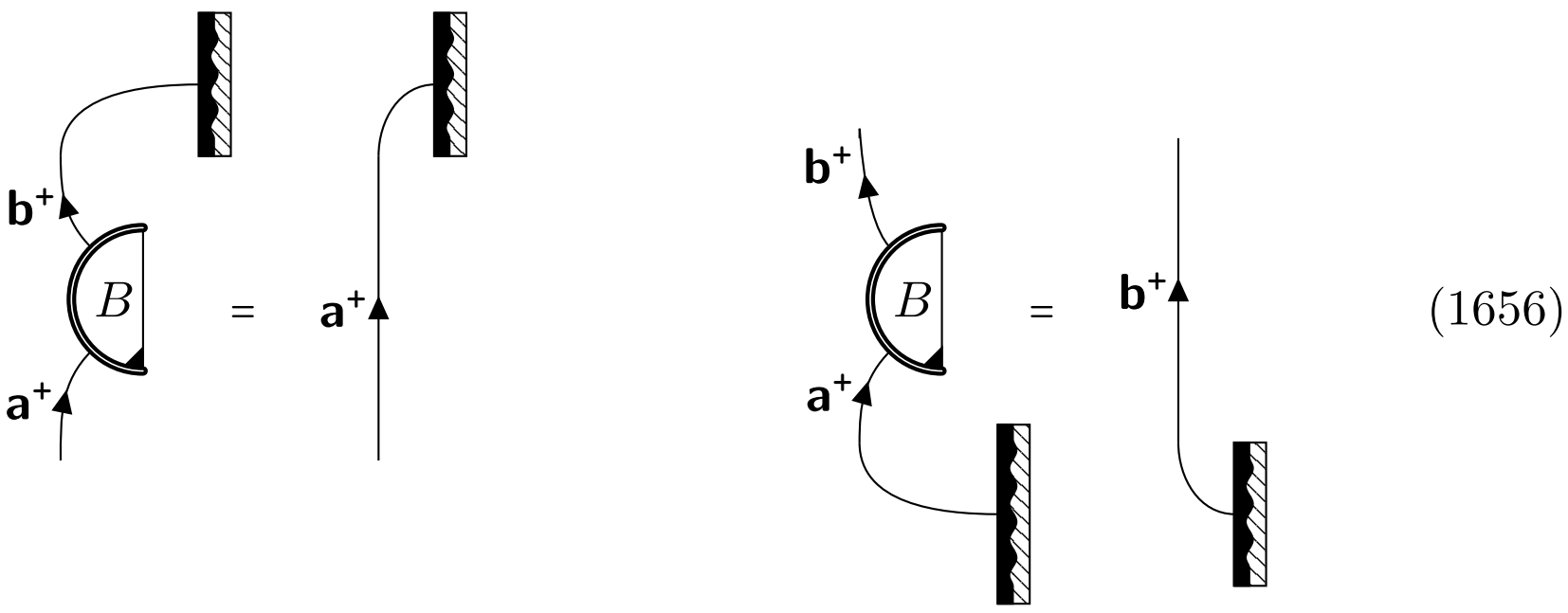

We can use the ideas above to prove the following important theorem

> **Physical deterministic homogeneous operator tensors.** All physical deterministic homogeneous operator tensors without pointer wires are natural temporal unitary operator tensors.

This follows from the above theorem since physicality implies double causality which implies simple double causality.

## 73.9 Factored natural temporal unitary operators

The above discussion goes through if we have a bipartition $(\mathbf{a}^-, \mathbf{b}^-)$ instead of $(\mathbf{a}^+, \mathbf{b}^+)$. We can, indeed, consider a $\hat{U}$ which factors as follows

(1657)

(where we are using $\mathbf{c}$ and $\mathbf{d}$ for pedagogical purposes as will become clear). If both $\hat{U}[+]$ and $\hat{U}[-]$ are natural temporal unitary operators then so is $\hat{U}$. Indeed, this factored form of $\hat{U}$ is a special case. This can be seen by putting

$$\mathbf{a}^+ = \mathbf{c}^+(\mathbf{d}^-)^R \qquad \mathbf{b}^+ = (\mathbf{c}^-)^R\mathbf{d}^+ \tag{1658}$$

Then the simple double causality conditions (1650) are satisfied (and these are, for homogeneous operator tensors, equivalent to the conditions for natural temporal unitary operators).

## 73.10 Natural reversing unitaries

An interesting natural unitary is

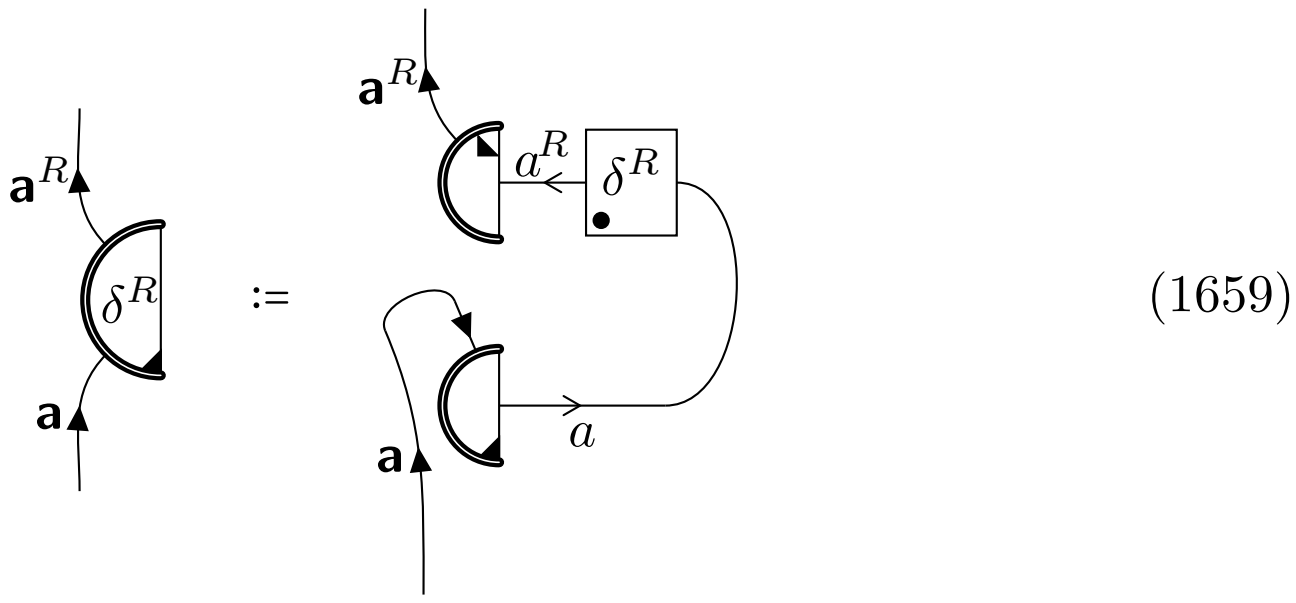

(1659)

where $\delta^R$ is the permutation matrix that simply interchanges $a^+$ and $a^-$. We will call this the *natural reversing unitary*. This is clearly a natural unitary since it is of the form in (1641). This unitary simply inverts the direction of the system from $\mathsf{a}$ to $\mathsf{a}^R$ (if we read forward along the arrows). It is clearly related to the special twist in (1545).

We can form the natural reversing unitary operator,

(1660)

by doubling up. Since this is a natural unitary operator, we note that

(1661)

from (1644) and

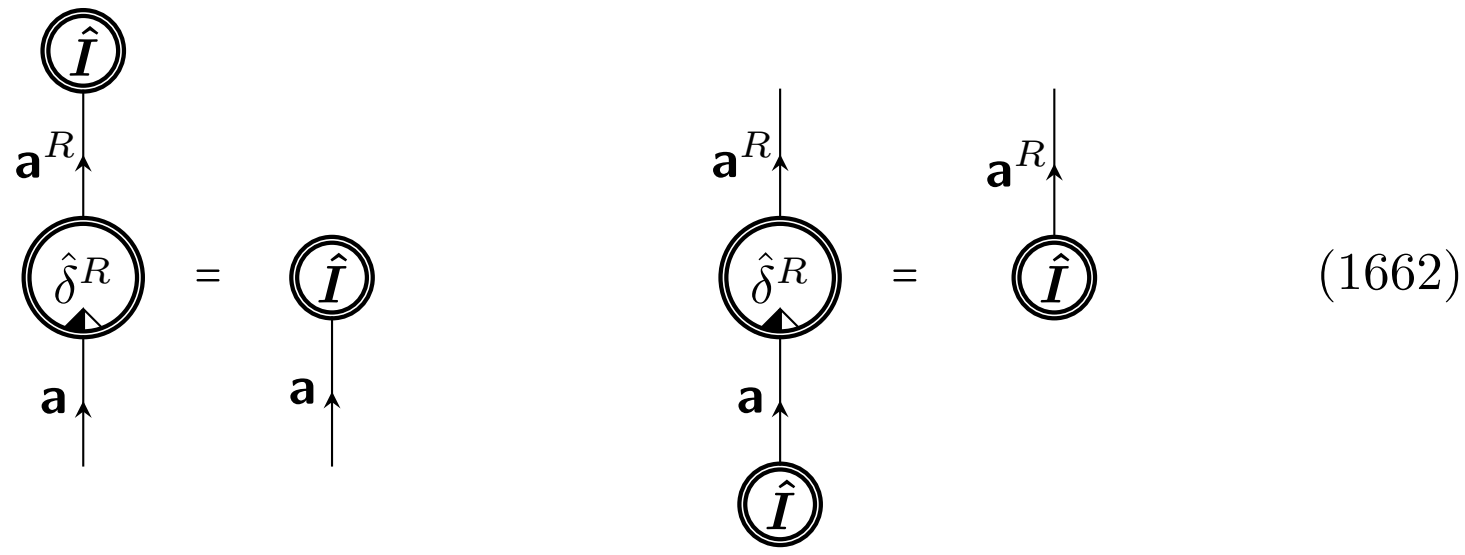

(1662)

from (1643).

Special cases of interest are the *natural reversing temporal unitary operators*

(1663)

The operator on the right has only outputs. Looking at (1659) we see that it does, in fact, correspond to a maximally entangled state. The natural reversing operator on the left has only inputs and corresponds to a maximally entangled effect. This teaches us that maximally entangled states and effects are, in fact unitaries. However, if they are normalised to be deterministic (as are natural unitaries) then they are not physical. We will use these natural reversing temporal unitary operators in Sec. 78.20 where we study causal snakes.

# 74 Maxometries

In the simple case we defined natural and normal maxometries. We can do the same here. Further, we will define temporal maxometries (which are with respect to a bipartition separating out the inputs and outputs) for the natural case. Natural temporal maxometries will play an important role in the dilation theorems we will prove in Sec. 78.

## 74.1 Natural maxometries

We define natural maxometries as follows

**Natural maxometries.** If the left object

(1664)

has the properties

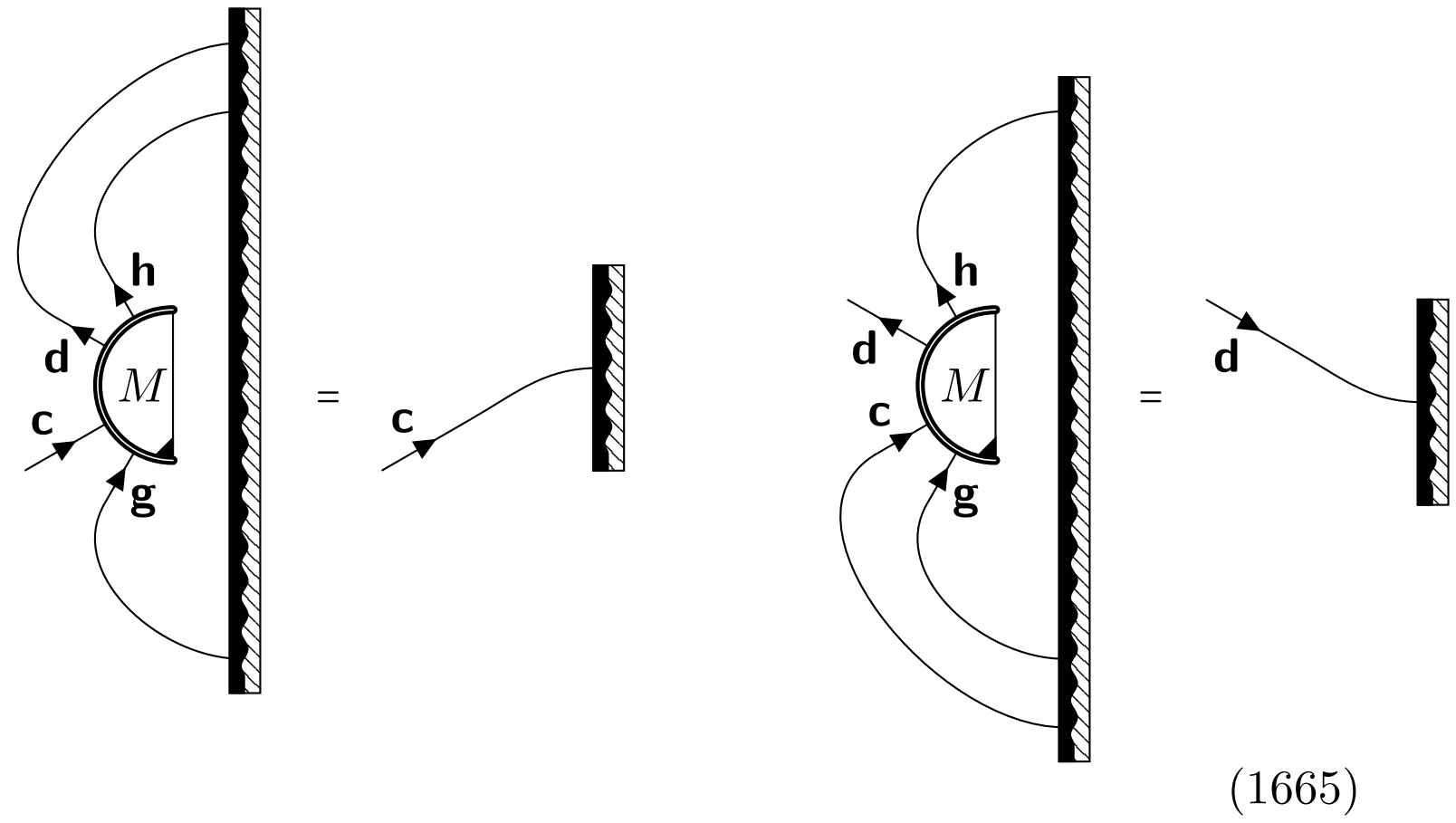

(1665)

then we say it is a natural maxometry with respect to ancillae **g** and **h** and bipartition $(\mathbf{cg}, \mathbf{dh})$. Note, further, that $|\underset{\sim}{M}| = 1$

The last statement is easily proven by attaching the open wire in either condition in (1665) to the mirror and using (1617). Note that the physical norm, $|\underset{\sim}{M}|$, is defined through

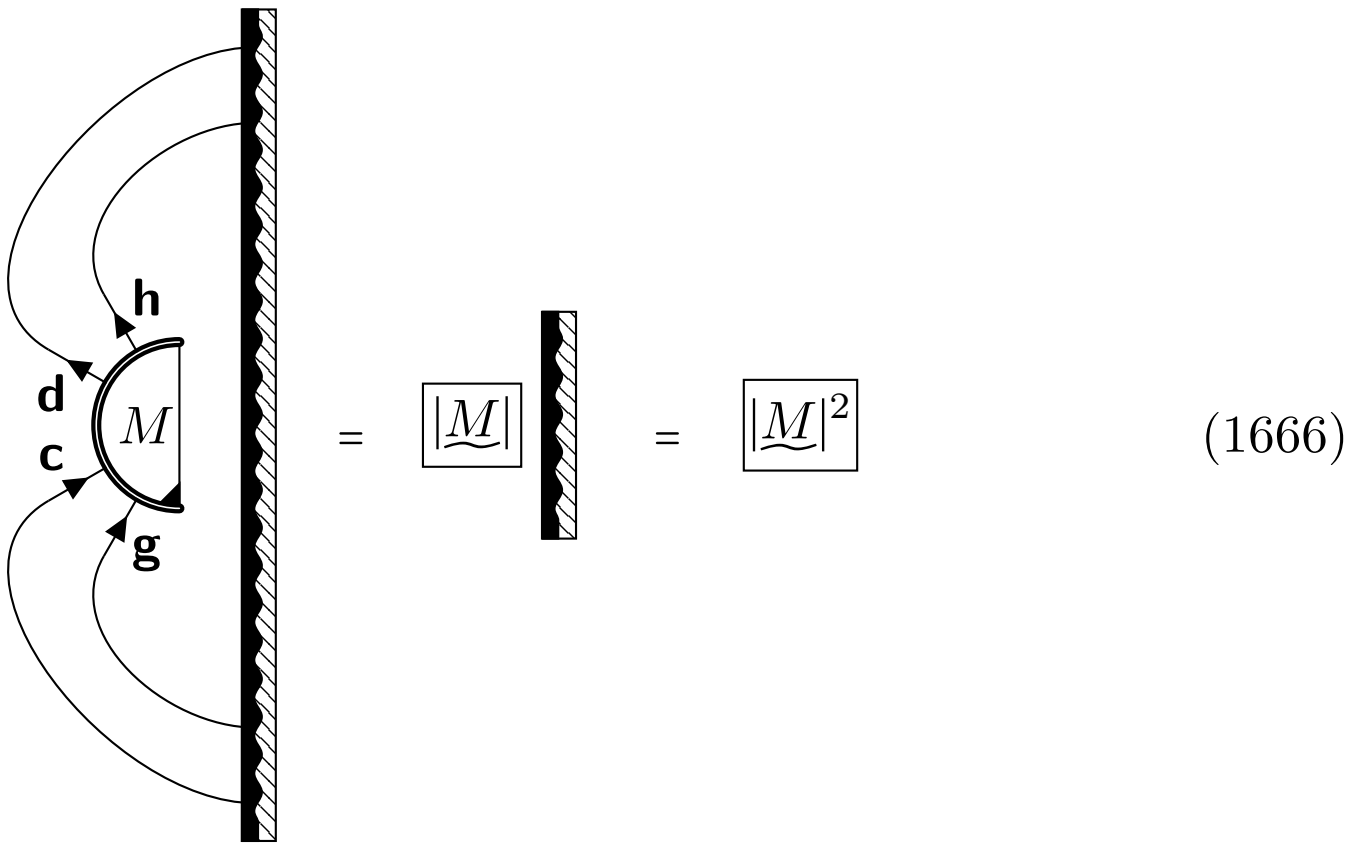

(1666)

It is worth noting that the notation in the complex case allows us to combine the two ancilla systems into a single system, $\mathbf{hg}^R$. However, it is useful to maintain two ancilla in the definition for the proof, in Sec. 74.6, that natural unitaries are a special case of natural maxometries.

As in the simple case, we can hit a natural maxometry with a maximally entangled Hilbert object and produce a natural isometry or coisometry. In the complex case we need to be clear how the bipartition is reassigned after such a hit. Consider, then, the natural maxometry $M$. It is easily verified that the

object

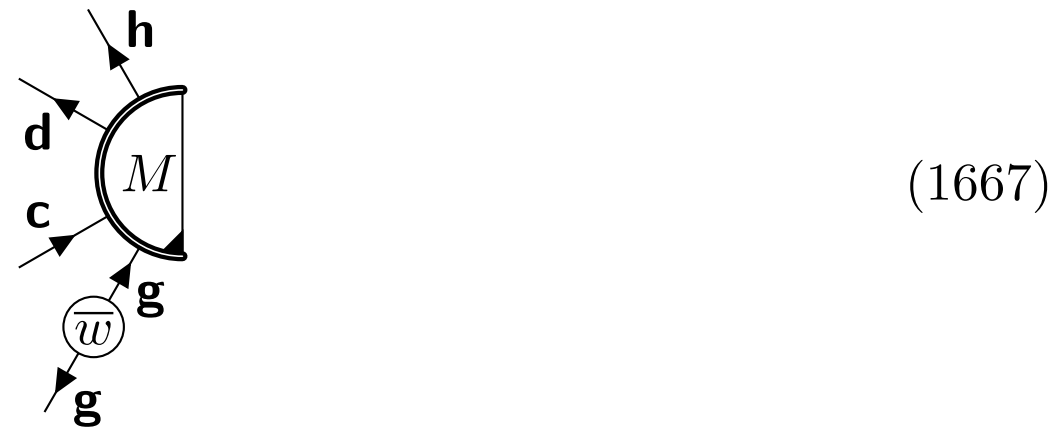

is a natural isometry with respect to the bipartition $(\mathbf{c}, \mathbf{dgh})$. Similarly, it can be shown that the object

(1668)

is a natural coisometry with respect to the bipartition $(\mathbf{cgh}, \mathbf{d})$. The $\overline{w}$ here are the requisite maximally entangled objects (defined in (1544) and (1545).

## 74.2 Normal maxometries

We define a normal maxometry as follows

**Normal maxometries.** If the left object

(1669)

has the properties

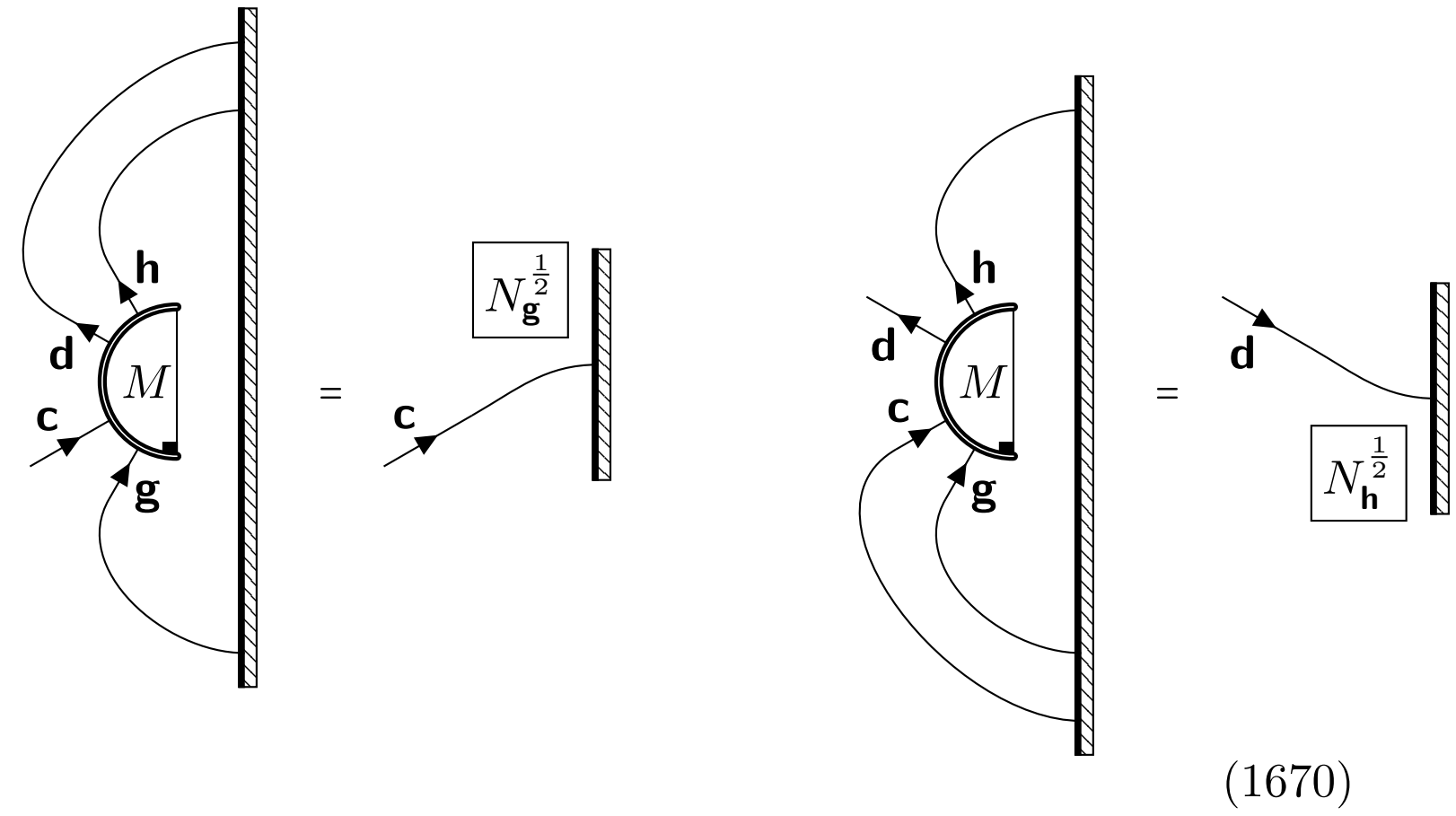

(1670)

then we say it is a normal maxometry with respect to ancillae **g** and **h** and bipartition $(\mathbf{cg}, \mathbf{dh})$. Note further that

$$|M|^2 = N_{\mathbf{c}} N_{\mathbf{g}} = N_{\mathbf{d}} N_{\mathbf{h}} \tag{1671}$$

where $|M|$ is the norm of $M$.

Note that the norm $|M|$ is defined as

$$= \boxed{|M|} \quad = \boxed{|M|^2} \tag{1672}$$

We will see in Sec. 74.6 that normal unitaries (defined in Sec. 73.1) are a special case of normal maxometries.

## 74.3 Natural temporal maxometries

Now we consider restricting this definition to the *temporal* case. This is where the bipartition is such that the first entry corresponds to inputs and the second to outputs. If we have already nominated our ancilla then there is one unique such partition.

We define

**Natural temporal maxometries.** If the left object

(1673)

has the properties

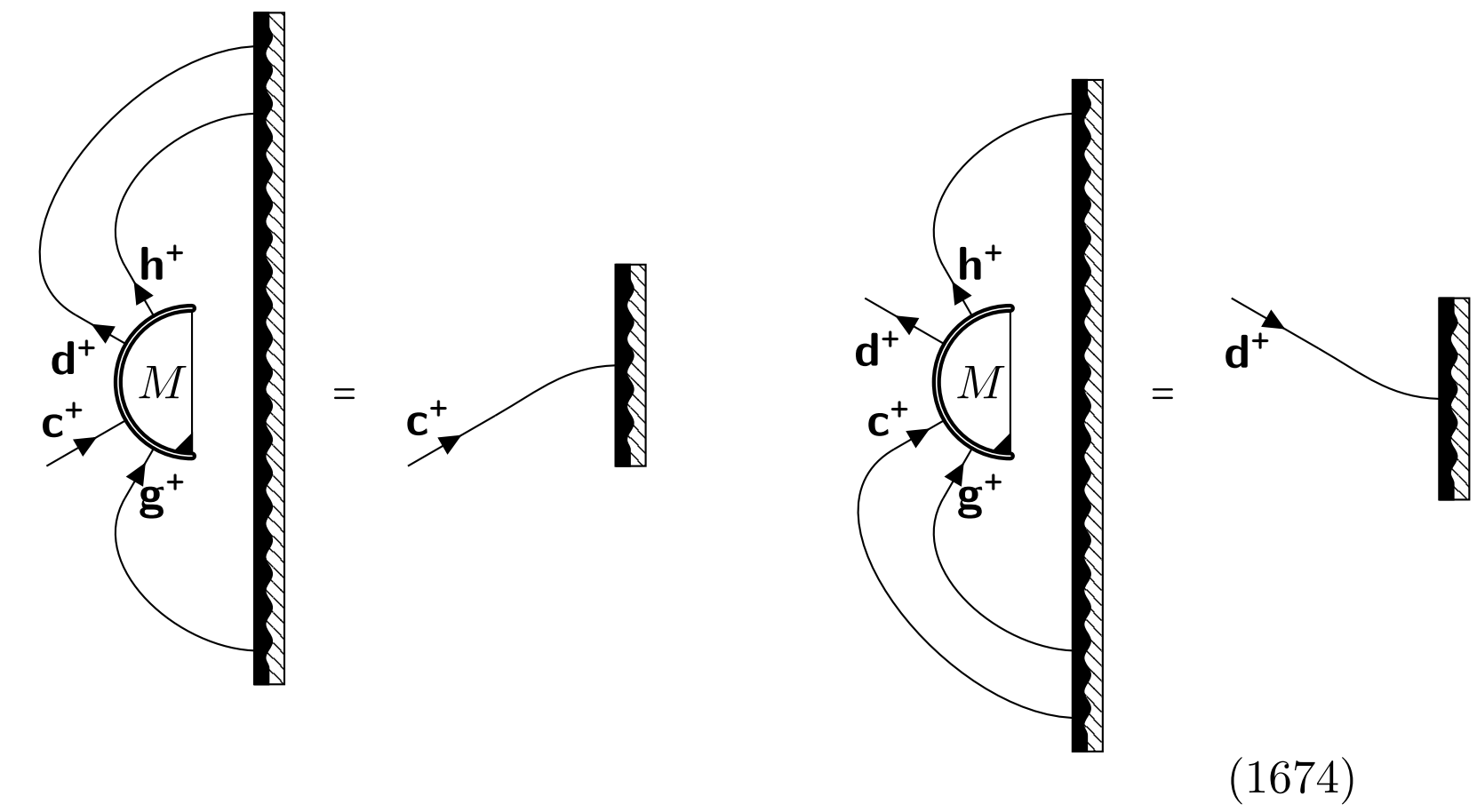

(1674)

then we say it is a natural temporal maxometry with respect to ancillae $\mathbf{g}^+$ and $\mathbf{h}^+$.

In the temporal case the bipartition we use to define the maxometry is unique. This means the conditions for natural temporal maxometry for a general left object

(1675)

with respect to ancilla **f** are

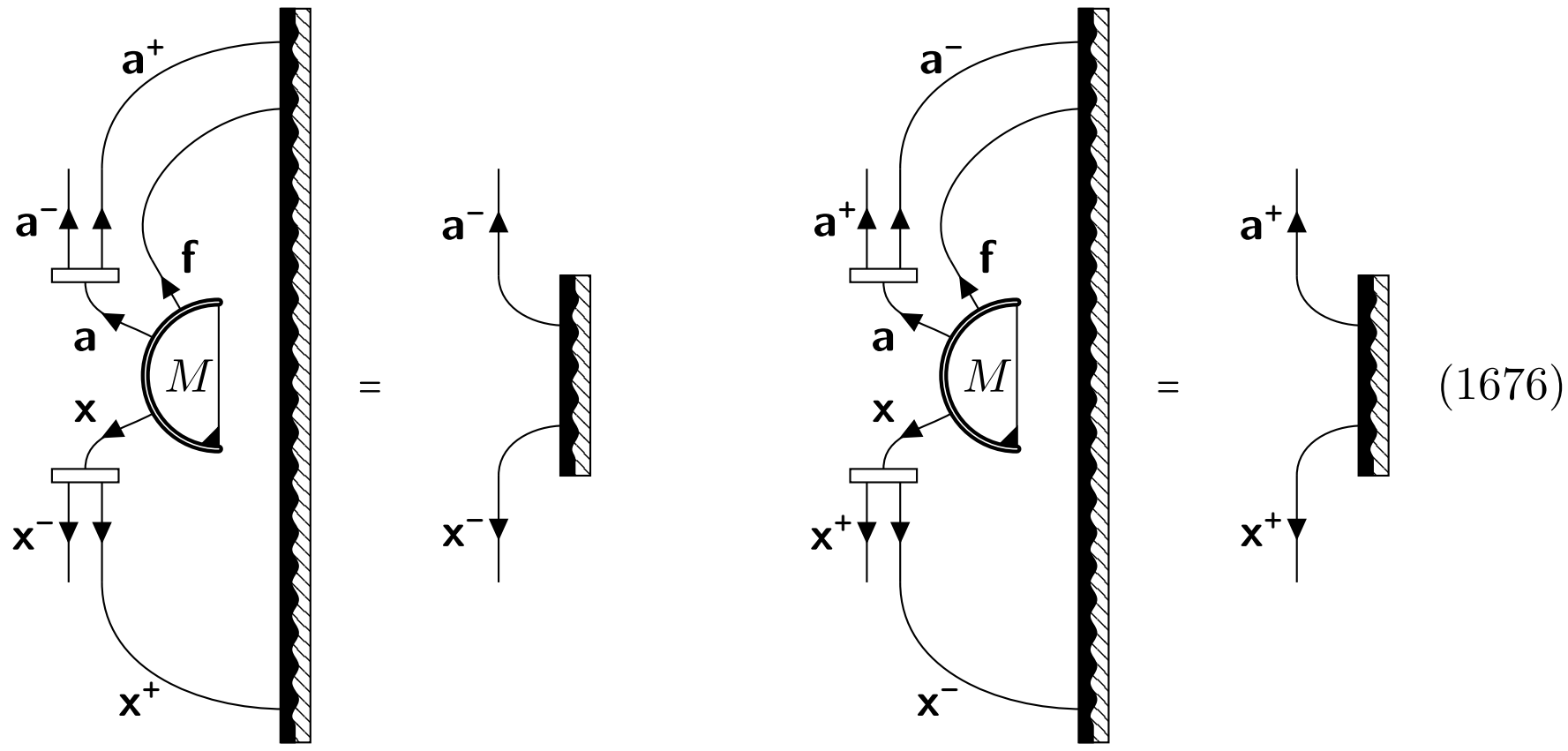

(1676)

Natural temporal maxometries have $|\underset{\sim}{M}| = 1$ since they are special cases of natural maxometries.

## 74.4 Pseudo-physical operators and deterministic temporal maxometries

We say that the operator

(1677)

is *pseudo-physical* with respect to ancilla **f** if

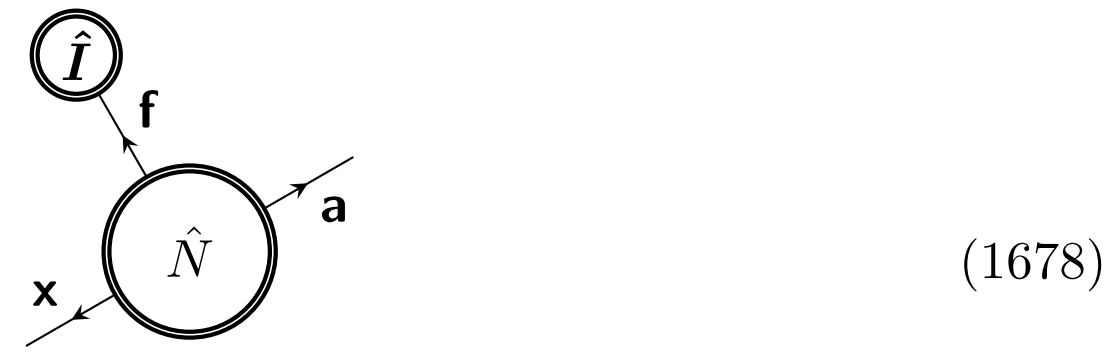

(1678)

is physical. For any pseudo-physical operator, $\hat{N}$, we have the *associated physical operator* (as given in (1678)). Further, we say $\hat{N}$ is a *pseudo-physical deterministic operator* if the associated physical operator in (1678) is both physical and deterministic.

In the special case that the associated physical operator in (1678) has simple causal structure with causal diagram

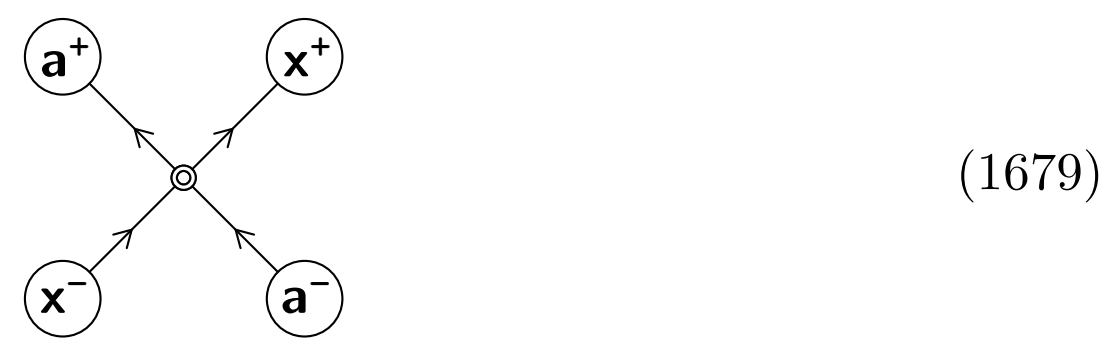

(1679)

then the conditions for determinism and physicality of this operator (and, thereby, pseudo-physical determinism of $\hat{N}$) are that the operator in (1678) is $T$-positive and that $\hat{N}$ satisfies the simple double causality conditions for deterministic operators

(1680)

(since the causal structure is simple, these conditions are sufficient for double causality and determinism - see Sec. 49.4.9).

Now we prove the following theorem

> **Natural temporal maxometries theorem with simple causal structure.** The condition that a homogeneous operator tensor of the form
>
> (1681)
>
> is a pseudo-physical deterministic operator (with respect to ancilla $\mathsf{f}$) given that the associated physical operator
>
> 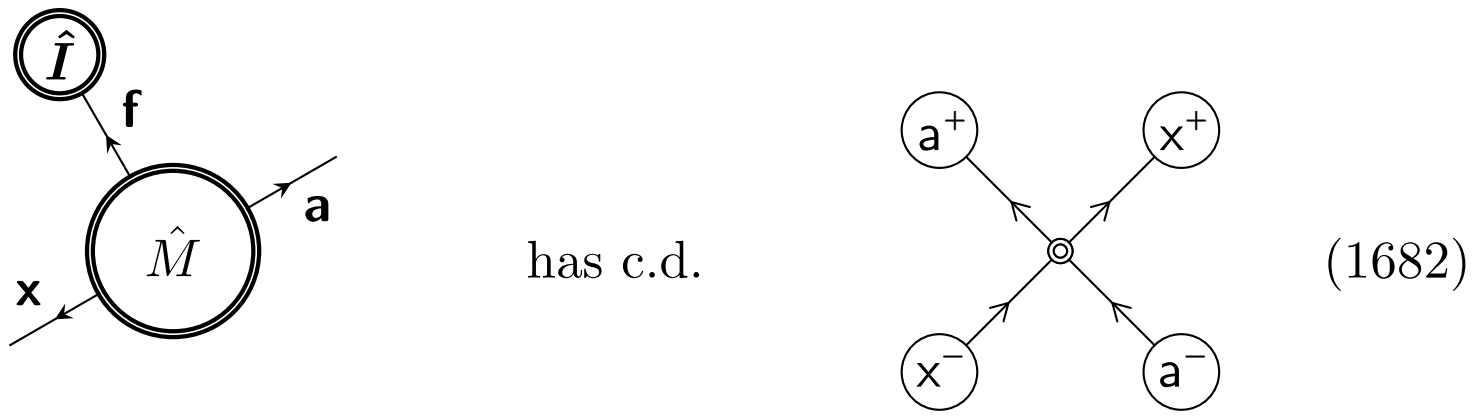
>
>
> is equivalent to the condition that $M$ is a natural temporal maxometry (with respect to ancillae $\mathsf{g}$ and $\mathsf{h}$).

Note we are assuming simple causal structure in (1682). The proof of this is straightforward. First note that $\hat{M}$ is twofold positive and hence satisfies $T$-positivity (and consequently the associated physical operator does too). Second, it is easy to show that the conditions for a natural temporal maxometry (these are (1676) are equivalent to the simple double causality conditions in (1680) (with $\hat{M}$ in place of $\hat{N}$) when we substitute in (1681).

For the case where we have general causal structure we can prove the following theorem

**Natural temporal maxometries theorem.** Consider a homogeneous operator tensor of the form

(1683)

The condition that the operator

(1684)

satisfies the deterministic simple double causality constraints

(1685)

is equivalent to the condition that $M$ is a natural temporal maxometry (with respect to ancillae g and h).

This is true for the same reason as the previous theorem was true. The simple double causality conditions in (1685) are, when applied to (1683), are equivalent to the conditions for natural temporal maxometry given in (1676).

## 74.5 Transforming pseudo-physical operators

We can perform a natural unitary transformation on the ancillae of a pseudo-physical operator (or, as a special case, a maxometric operator). This, in turn, transforms to a new operator having the same associated physical operator. Let us state this property as a theorem.

**Transforming pseudo-physical operators.** If

(1686)

is a pseudo-physical operator with ancilla **f** then

$$\hat{N}' \;:=\; \hat{U}\,\hat{N} \qquad (1687)$$

(where $\hat{U}$ is a natural unitary operator with respect to bipartition, $(\mathbf{f}, \mathbf{g})$) is a pseudo-physical operator with ancilla **g** where

$$\hat{I}\,\hat{N}' \;=\; \hat{I}\,\hat{N} \qquad (1688)$$

i.e. $\hat{N}'$ and $\hat{N}$ have the same associated physical operator. Furthermore, $\hat{N}$ is a natural temporal maxometry (with respect to ancilla **f**) if and only if $\hat{N}'$ is a natural temporal maxometry (with respect to ancilla **g**).

That (1688) holds follows immediately from (1643). The statement about maxometries is obvious since both maxometries and unitaries are homogeneous, and also the unitaries can be inverted.

One immediate application we will use later is the following.

**Transforming a natural temporal maxometric operator.** If

$$\hat{M} \qquad (1689)$$

is a natural temporal maxometric operator then

$$\hat{M}' \;:=\; \hat{U}\,\hat{M} \qquad (1690)$$

(where $\hat{U}$ is a natural unitary with respect to bipartition $(\mathbf{f}, \mathbf{g}^{\pm})$) is also a natural temporal maxometric operator with the same associated physical operator as $\hat{M}$.

This is an immediate consequence of the transforming pseudo-physical operators theorem above where $N_{\mathbf{g}^\pm} = N_{\mathbf{h}}$ because $U$ is a natural unitary with respect to bipartition $(\mathbf{f}, \mathbf{g}^\pm)$.

## 74.6 Unitaries are maxometries

Here we will prove that natural unitaries are natural maxometries and that normal unitaries are normal maxometries.

First we prove

> **Natural unitaries are natural maxometries** with respect to the same bipartition (for any choice of ancillae). Thus, if
>
> (1691)
>
> is a natural unitary with respect to the bipartition $(\mathbf{cf}, \mathbf{dh})$ then it is also a natural maxometry with respect to the same bipartition for ancillae $\mathbf{f}$ and $\mathbf{h}$.

To prove this note that it follows from natural unitarity that

(1692)

where we have used (1617). A similar property follows if we attach the $\mathbf{c}$ wire (instead of the $\mathbf{d}$ wire) to the mirror on the left.

Similarly, we have

> **Normal unitaries are normal maxometries** with respect to the same bipartition for any choice of ancillae. Thus, if
>
> (1693)

is a normal unitary with respect to the bipartition $(\mathbf{cf}, \mathbf{dh})$ then it is also a normal maxometry with respect to the same bipartition for ancillae **f** and **h**.

To prove this we note

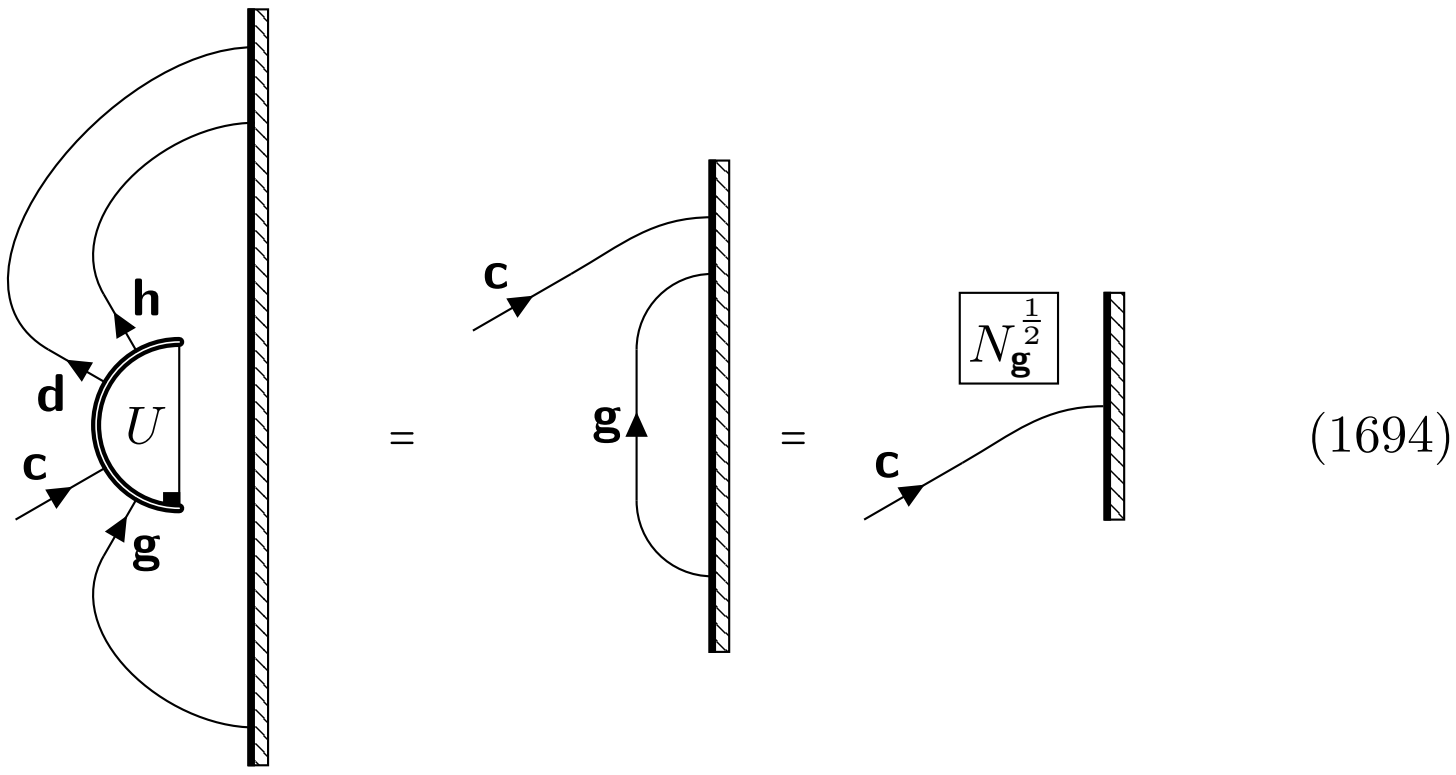

(1694)

where we have used (1592). A similar property follows if we attach the **c** wire (instead of the **d** wire) to the mirror on the left.

The above theorems apply to any partition. In particular, they apply to the temporal case. Thus, we have the theorem

> **Temporal unitaries.** Natural temporal unitaries are natural temporal maxometries. Temporal normal unitarities are temporal normal maxometries.

The first case is of particular interest to us since natural temporal unitary operators satisfy the double causality conditions (and, therefore, these too are deterministic temporal maxometries).

Not all natural maxometries are natural unitaries, and not all normal maxometries are normal unitaries. This is easily proven by adapting the counterexample employed in the simple case in Sec. 41.2.

## 74.7 Natural (co)isometries from natural unitaries

In Sec. 74.7 we proved a little theorem which was used to show that any natural isometry operator, $\hat{V}$, can be written as

d
$\hat{V}$
c
=
d
$\hat{\boldsymbol{I}}$
c
$\hat{\boldsymbol{U}}$
d
c
$\hat{A}_{\text{det}}$

(1695)

(see (899)) where $\hat{\boldsymbol{U}}$ is a natural unitary and $\hat{A}_{\rm det}$ is homogeneous and deterministic (though not physical). Similarly, we saw that any natural coisometry operator, $\hat{V}$, can be written as

$$\hat{V} \;=\; \hat{\boldsymbol{U}} \quad (\text{with } \hat{C}_{\rm det},\ \hat{\boldsymbol{I}};\ \mathsf{c},\ \mathsf{d}) \tag{1696}$$

(see (901)) where $\hat{\boldsymbol{U}}$ is a natural unitary and $\hat{C}_{\rm det}$ is homogeneous and deterministic (but not physical).

The formal properties of (co)isometries, unitaries, ignore operators ($\hat{\boldsymbol{I}}$), $\hat{\boldsymbol{A}}_{\rm det}$, and $\hat{\boldsymbol{C}}_{\rm det}$ that enable the proofs of these results also hold in the complex case. Thus, we can go through these proofs replacing squares with semicircles/circles as appropriate. In so doing we prove that we can write a natural isometry operator, $\hat{V}$, in the complex case as follows

$$\hat{V} \;=\; \hat{U} \quad (\text{with } \hat{\boldsymbol{I}},\ \hat{A}_{\rm det};\ \mathbf{c},\ \mathbf{d}) \tag{1697}$$

where $\hat{U}$ is a natural unitary and $\hat{A}_{\rm det}$ is a homogeneous deterministic operator. Similarly we can write any natural coisometry operator, $\hat{V}$, in the complex case as follows

$$\hat{V} \;=\; \hat{U} \quad (\text{with } \hat{C}_{\rm det},\ \hat{\boldsymbol{I}};\ \mathbf{c},\ \mathbf{d}) \tag{1698}$$

where $\hat{U}$ is a natural unitary and $\hat{C}_{\rm det}$ is a homogeneous deterministic operator.

Any isometry can be written in the form on the right hand side of (1697). However, it is not the case that any operator taking the form on the right hand side of (1697) is necessarily an isometry. Similarly, any coisometry can be written in the form on the right hand side of (1698) but not any operator taking this form is necessarily a coisometry.

The natural unitaries in (1697) and (1698) may not be physical - unlike in the simple cases (1695) and (1696) where the unitaries are physical.

Consider a natural maxometric operator, $\hat{M}$, having ancilla $\mathsf{h}$ and bipartition $(\mathsf{x}, \mathsf{ah})$. This is a natural isometry with respect to the same bipartition (see discussion at the end of Sec. 74.1). Consequently, we can write

$$(1699)$$

where $\hat{U}$ is a natural unitary with respect to the bipartition (ingoing arrows, out going arrows). Similarly, we have a natural maxometric operator, $\hat{M}$, with ancilla $\mathsf{g}$ and bipartition $(\mathsf{xg}, \mathsf{a})$ then this is a natural coisometry with respect to the same bipartition and we can write

$$(1700)$$

where $\hat{U}$ is a natural unitary with respect to the bipartition (ingoing arrows, out going arrows).

For the causal dilation theorems in Sec. 78 we will be interested in the case where we have a temporal maxometry with an ancilla that is either pure output or pure input. In these cases we have

$$(1701)$$

where $\mathsf{u}^+ = \mathsf{a}^+\mathsf{x}^+\mathsf{h}^+$ and $\mathsf{v}^+ = (\mathsf{a}^-\mathsf{x}^-)^R$ for the left expression, whilst $\mathsf{q}^+ = \mathsf{a}^+\mathsf{x}^+$ and $\mathsf{s}^+ = \mathsf{g}^+(\mathsf{a}^-\mathsf{x}^-)^R$. Since we have temporality here, the bipartition for the natural maxometries and unitaries are unique.

# 75 Evaluating operator circuits using Hilbert objects

If we have correspondence, then corresponding to any circuit is an operator circuit equal to the probability for that circuit. In this section we are interested in whether operator circuits are real, and whether they are non-negative. We can write any circuit as a positive weighted sum of fully regularized circuits using the control and midcome identities (see Sec. 49.1.2). Hence, if we are interested in reality and non-negativity, it is sufficient to consider fully regularised circuits. For example we have

prob = (1702)

We will show that (i) operator circuits with Hermitian operators are real, (ii) operator circuits with twofold positive operators are non-negative, and (iii) operator circuits with an anti-Hermitian operator can be pure imaginary. The first and third of these results completes the proof that circuit reality implies Hermiticity in Sec. 59.3.

## 75.1 Operator circuits with Hermitian operators are real

If every operator in the operator circuit on the right of (1702) is Hermitian then we can write

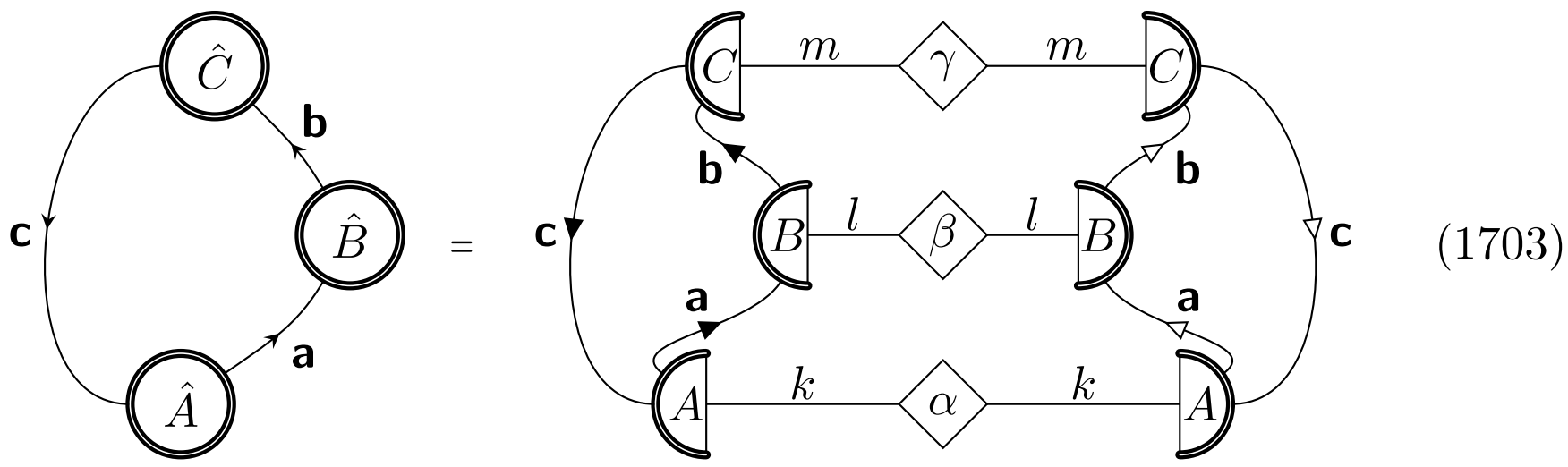

(1703)

where the eigenvalues $\alpha_k$, $\beta_l$, and $\gamma_m$ are real. This is real since it is of the form $\sum_{klm} D_{klm}\overline{D}_{klm}\alpha_k\beta_l\gamma_m$ where $D_{klm}$ is a complex number equal to the left Hilbert circuit in (1703). This clearly works for any circuit.

## 75.2 Operator circuits with twofold positive operators are non-negative

If the operators are all twofold positive in the circuit in (1702) then we can write

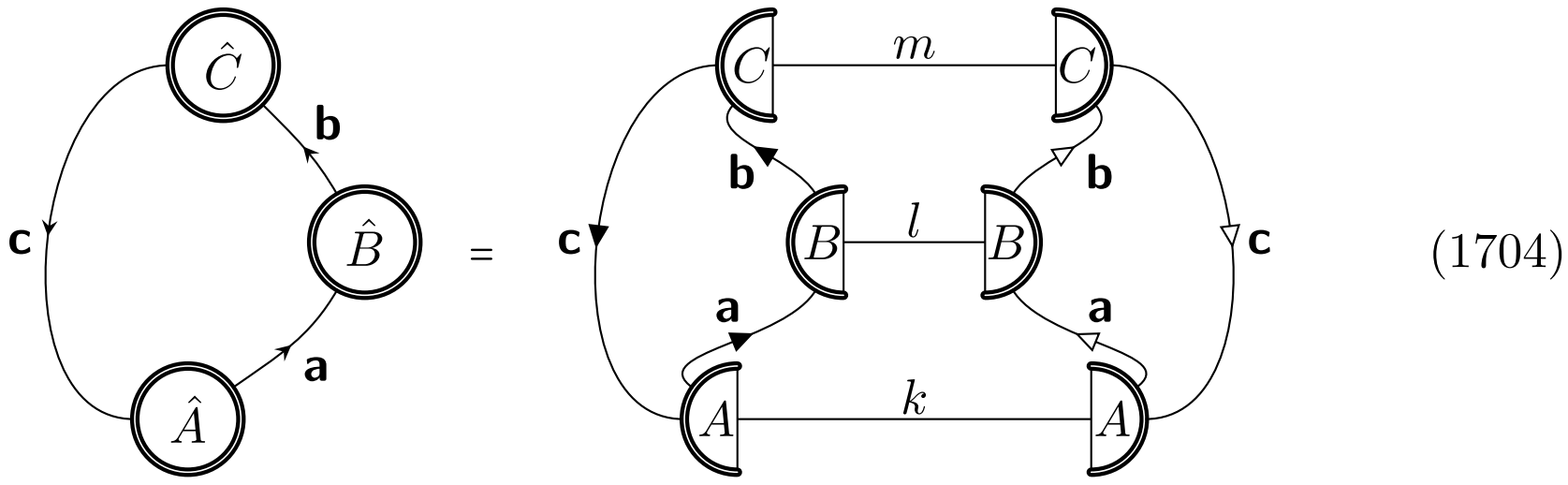

(1704)

This is of the form $\sum_{klm} D_{klm}\overline{D}_{klm}$ which is non-negative. This clearly works for any circuit.

## 75.3 Operator circuits with an anti-Hermitian operator

Consider a circuit containing an anti-Hermitian operator, $\hat{B}_A(x)$. We can write

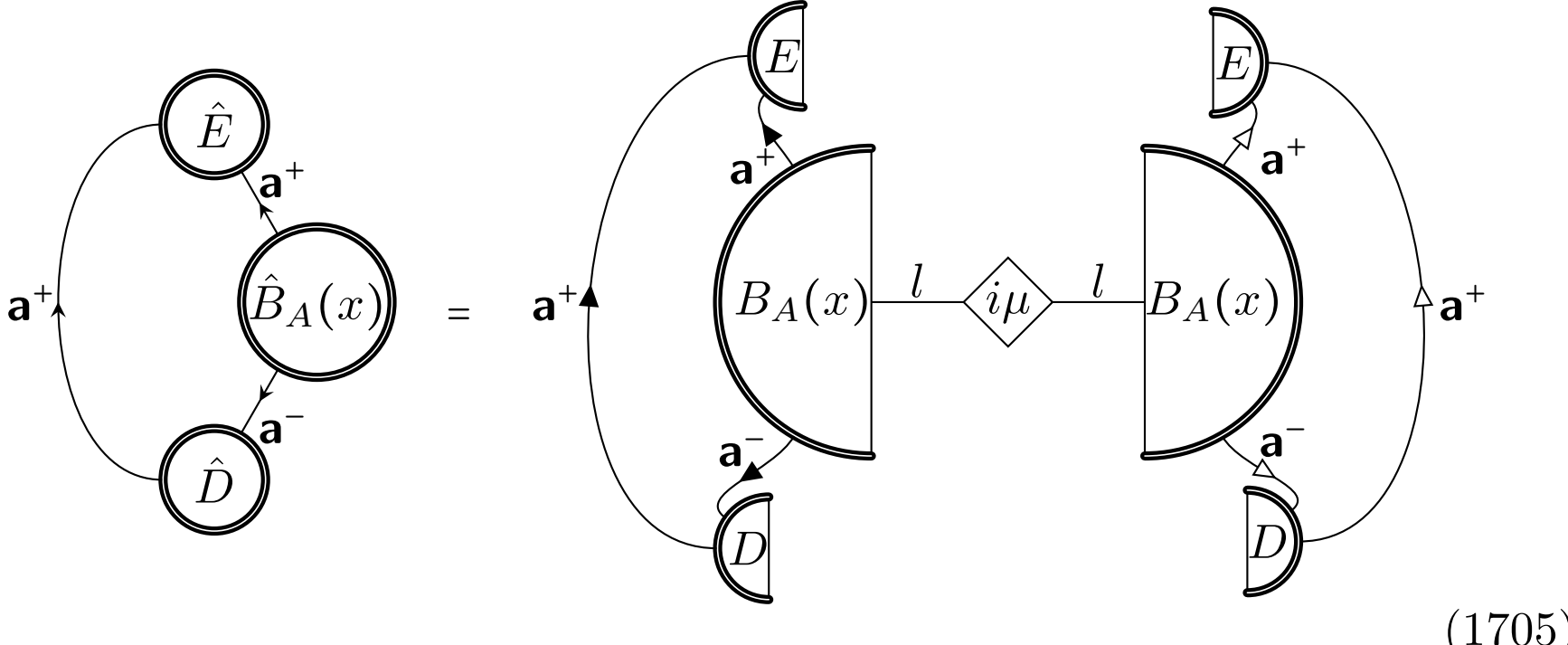

(1705)

where $\mu$ is real. We have chosen $\hat{D}$ and $\hat{E}$ to be homogeneous. It is easy to show, following the same techniques as in the simple case in Sec. 42.3, that the only way to guarantee this is real is if $\mu_l = 0$ for all $l$. Given that Hermitian operator circuits must be real, circuit reality implies, by the argument in Sec. 59.3, that operator tensors are necessarily Hermitian.

# 76 Positivity conditions

## 76.1 Input and output twist

We define the *input twist* of an operator tensor by taking the twist over only the inputs. In the simple case it was sufficient to attach twists to the input wires to take the input twist. In the complex case, however, wires have both input and output parts and so we have to address these separately. To this end, we define $w_\pm$ operators as follows

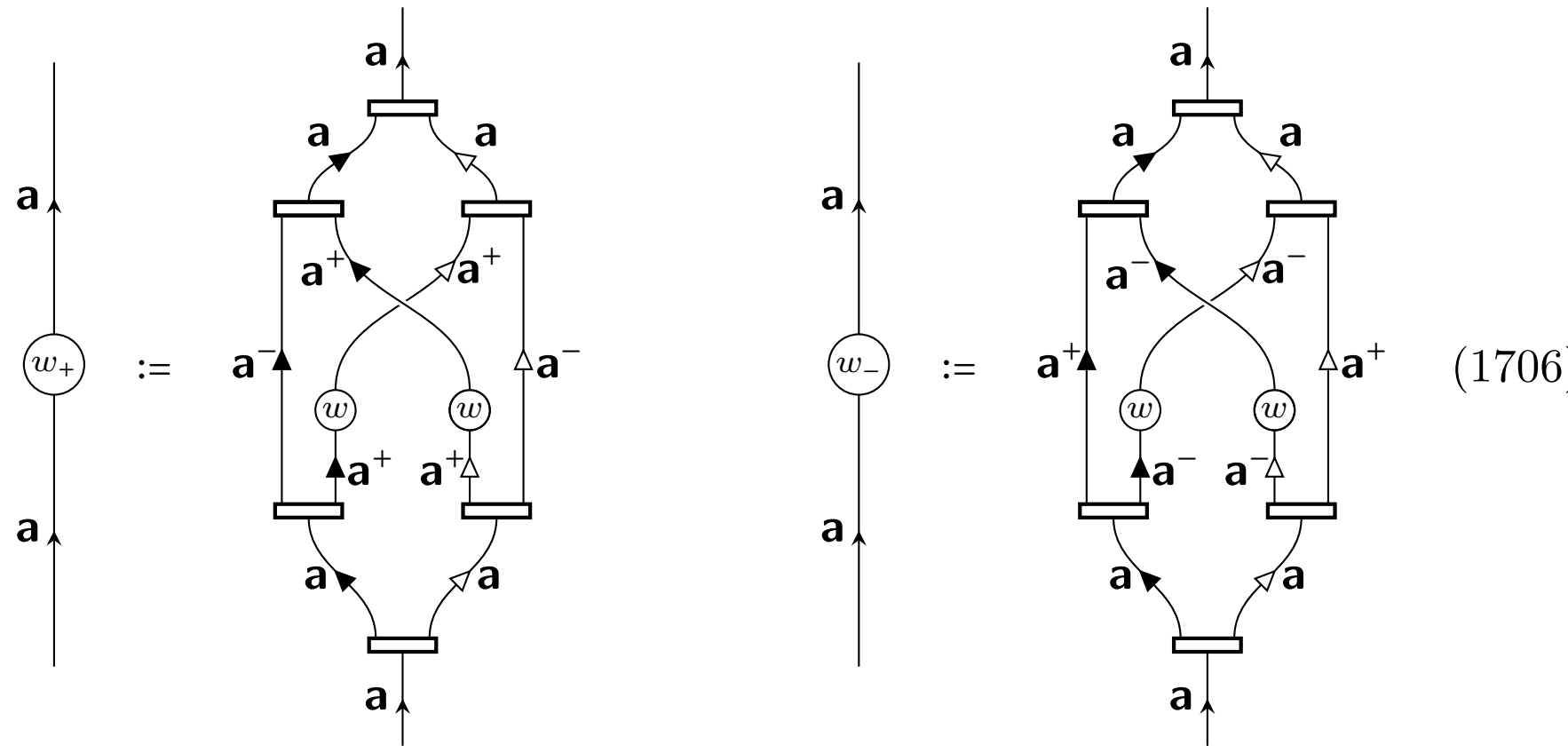

(1706)

The operator $w_\pm$ takes the twist only on the $\pm$ part (using the twist defined in (1429)). We can use this to take the input twist of an operator as follows

$$\hat{B}_{H_{\text{in}}} \;=\; \hat{B} \text{ with } w_+ \text{ on input wire } \mathbf{a} \text{ and } w_- \text{ on output wire } \mathbf{b} \qquad (1707)$$

Note that we have $w_+$ on the wire with the arrow pointing in since, there, the input is in the direction of the arrow.

The output twist of $\hat{B}$ is denoted $\hat{B}_{H_{\text{out}}}$ and is obtained by taking the transpose of the output wire as follows.

$$\hat{B}_{H_{\text{out}}} \;=\; \hat{B} \text{ with } w_- \text{ on input wire } \mathbf{a} \text{ and } w_+ \text{ on output wire } \mathbf{b} \qquad (1708)$$

If we take both the input twist and the output twist this amounts to taking the twist and it implements the horizontal transpose (as discussed in Sec. 32.6). Thus, the input/output twist might also be called the input/output horizontal transpose. This explains the notation, $H_{\text{in}}$, $H_{\text{out}}$ above.

## 76.2 Positivity theorem

We will write

$$\mathsf{a}\ \hat{B}(x) \quad := \quad \mathsf{a}\ \hat{B}\ \mathsf{x}\ x\ \mathsf{x}\ \hat{R} \tag{1709}$$

for our subsequent convenience.

We can state four positivity conditions on $\hat{B}(x)$.

**Strong tester positivity** is the condition

$$0 \le \quad \hat{E}\ \mathsf{a}\ \hat{B}(x) \tag{1710}$$

for all pure $\hat{E}$ (this comes from correspondence with (1129)).

**Tester positivity** is the condition

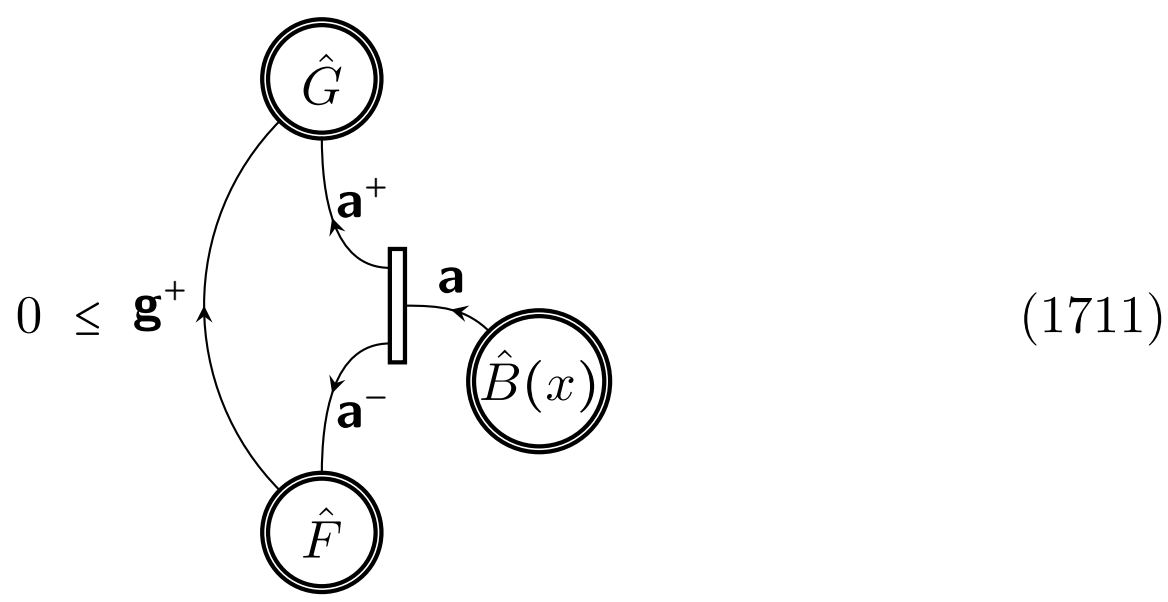

(1711)

for all pure states, $\hat{F}$, and effects, $\hat{G}$. This comes from correspondence with the TSCOPT (see Sec. 63).

**Twofold positivity** is the condition

$$\mathsf{a}\ \hat{B}(x) \quad = \quad \mathsf{a}\ B(x)\ l\ B(x)\ \mathsf{a} \tag{1712}$$

which comes from the Hilbert space theory (see Sec. 68.4).

**Input Twist positivity** is the condition

$$0 \leq \hat{B}_{H_{\mathrm{in}}(x)} \quad (\mathsf{a}) \tag{1713}$$

This utilizes the definitions in Sec. 76.1. Input twist positivity is equivalent to output twist positivity.

These conditions are each of interest for different reasons. The strong tester and tester positivity conditions connect positivity with operational notions. The twofold condition connects positivity with the structure of Hilbert space and the input transpose positivity allows us to test for positivity by looking at eigenvalues.

We can prove the following theorem

> **Operator positivity theorem.** For complex operator tensors, strong tester positivity, tester positivity, twofold positivity, and input twist positivity conditions are all equivalent.

The proof of this theorem for tester positivity, twofold positivity, and input twist positivity is along exactly the same lines here in the complex case as in the simple case as discussed in Sec. 43. To prove that the strong tester positivity condition is equivalent to the other three conditions it suffices to prove it is equivalent to twofold positivity. First we prove that twofold positivity implies strong tester positivity holds. Since $\hat{E}$ is pure (and therefore homogeneous), we see that twofold positivity implies that

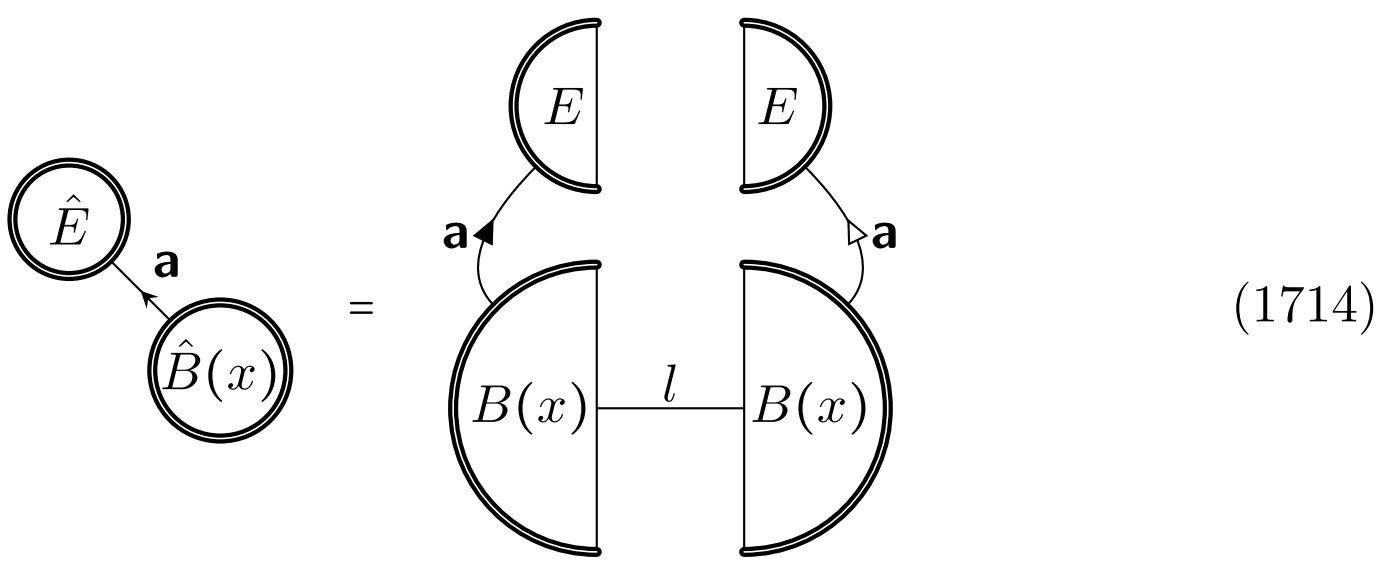

(1714)

Clearly the right hand side is non-negative (as it is of the form $\sum_l F_l \overline{F}_l$ where $F_l$ is the left Hilbert circuit). Hence twofold positivity implies strong tester positivity. Now we will show that strong tester positivity implies twofold positivity. First note that strong tester positivity implies non-negativity which implies reality. This in turn implies Hermiticity (see Sec. 59.3). Strong tester positivity

implies

$$0 \;\leq\; \hat{E}[l] \xleftarrow{\mathsf{a}} \hat{B}(x) \;=\; \begin{matrix} E[l] & & E[l] \\ \mathsf{a}\uparrow & & \triangle\,\mathsf{a} \\ \underleftarrow{B}(x) \overset{l}{\text{---}} & \lambda & \overset{l}{\text{---}} \underrightarrow{B}(x) \end{matrix} \tag{1715}$$

where we have used the form in (1443) to expand the Hermitian operator in terms of a diagonal matrix with real eigenvalues, $\lambda_l$, and left and right eigenvectors labeled with $\underleftarrow{B}$ and $\underrightarrow{B}$. Further, we have introduced the $l$ label on $E$ because we wish to consider the following set of left Hilbert objects, $E[l]$

$$\underset{\mathsf{a}}{E[l]} \;=\; \begin{matrix} & \mathsf{a}\overset{a}{\text{---}}\mathsf{a} \\ l \overset{l}{\text{---}} & \underrightarrow{B}(x) \\ \nu & \end{matrix} \tag{1716}$$

where $\nu$ is a constant (that allows $E[l]$ to be physical). Inserting this and its conjugate (obtained by flipping) into (1715) we obtain

$$0 \;\leq\; \begin{matrix} \nu & & \bar{\nu} \\ l \overset{l}{\text{---}} & \lambda & \overset{l}{\text{---}} l \end{matrix} \qquad \forall l \tag{1717}$$

where we have used the second of the unitarity properties for eigenvectors in (1442). This implies that the eigenvalues, $\lambda_l$, are all non-negative and hence twofold positivity follows.

# 77 Inequalities

In Sec. 49.2.3 we saw that

$$\text{exprn}_1 \leqq \text{exprn}_2 \quad \Rightarrow \quad \text{exprn}_1 \underset{T}{\leqq} \text{exprn}_2 \tag{1718}$$

Further, we saw that the reverse implication follows for expressions that are real weighted sums of operations with simple causal structure.

Under correspondence, the reverse implication in Quantum Theory is

$$\overparen{exprn}_1 \leqq \overparen{\text{exprn}}_2 \quad \Leftarrow \quad \overparen{exprn}_1 \underset{T}{\leqq} \overparen{exprn}_2 \tag{1719}$$

Given the positivity theorem, it is easy to show that this reverse implication is true in Quantum Theory. We can rewrite (1719)

$$0 \leqq \widehat{\mathrm{exprn}}_2 - \widehat{exprn}_1 \quad \Leftarrow \quad 0 \underset{T}{\leqq} \widehat{exprn}_2 - \widehat{exprn}_1 \tag{1720}$$

The operator positivity theorem tells us that positivity with respect to testers is equivalent to positivity with respect to strong testers (which is, as pointed out in Sec. 49.2.3, equivalent to positivity with respect to arbitrary complement networks). Hence we have proven the reverse implication.

It is a curious feature of Quantum Theory that it allows us to make this reverse implication.

# 78 Causal dilation theorems

## 78.1 Introduction

In the simple case we saw that a simple operator can be dilated in terms of a natural maxometric operator, appropriate maximal operators, and ignore operators (see Sec. 44). In Sec. 78.3 we will prove a similar theorem here for complex operators with simple causal structure. It is natural to conjecture that a complex operation with any causal structure can be dilated in a way that reflects this causal structure. This would be a *causal dilation*. Explicit formulation and proof of such a general causal dilation theorem remains open. In this book we will, however, be able to prove more limited causal dilation theorems for special cases. First, by adapting a proof technique Gutoski and Watrous [2007] on quantum strategies and a similar proof from the work of Chribella, D'Arioano and Perinotti (CDP) on quantum combs Chiribella, D'Ariano, and Perinotti [2009a], we will prove a *basic causal dilation theorem*. This is used to prove some more causal dilation theorems. We prove a dilation theorem for *causal ladders* (see Sec. 78.14). This provides a time symmetric version of a representation theorem provided in the aforementioned works on quantum strategies and quantum combs. Then, using these ideas, we will prove a dilation theorem for *causal snakes* (see Sec. 78.19). Finally, by application of the no correlation without causation assumption (from Sec. 48.9 we prove a *many ancestors theorem* and *many descendants theorem* in Sec. 78.21.

## 78.2 Maximal representation of operators

By correspondence with Sec. 50.4 we obtain the following theorem for the maximal representation of complex operators

**Maximal representation theorem.** Any operator, $\hat{B}$, can be

written as

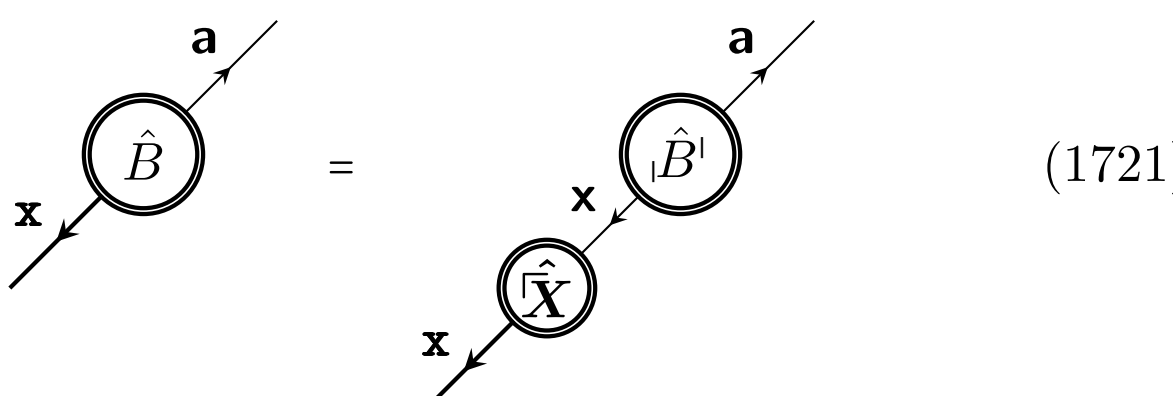

(1721)

where

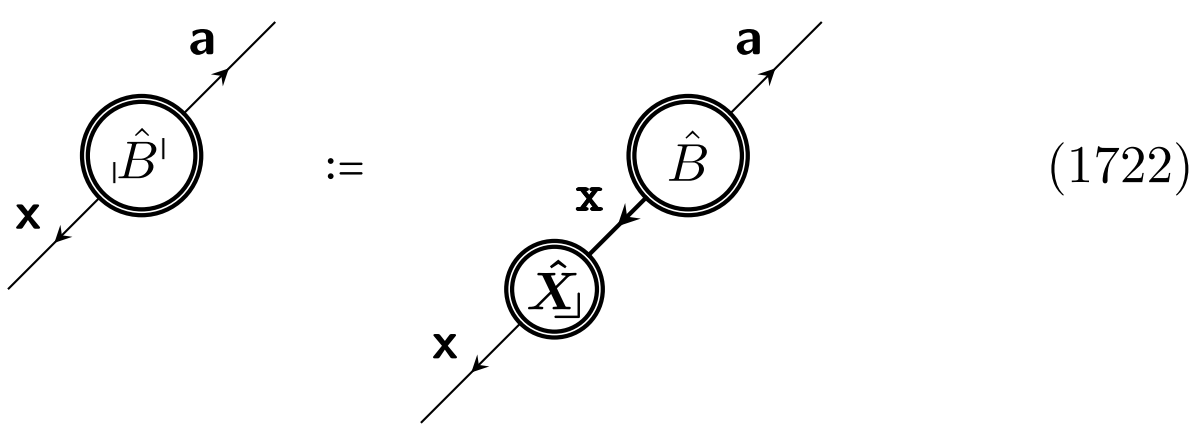

(1722)

Furthermore

(i) ${}_{|}\hat{B}^{|}$ has the same causal diagram as $\hat{B}$ (where **z** is replaced by **z**).

(ii) ${}_{|}\hat{B}^{|}$ is $T$-positive if and only if $\hat{B}$ is $T$-positive.

(iii) ${}_{|}\hat{B}^{|}$ satisfies general forward causality at any synchronous partition if and only $\hat{B}$ satisfies it at the same partition, and ${}_{|}\hat{B}^{|}$ satisfies general backward causality at any synchronous partition if and only $\hat{B}$ satisfies it at the same partition

(iv) ${}_{|}\hat{B}^{|}$ and $\hat{B}$ have the same physical norm.

A consequence of the above properties is that ${}_{|}\hat{B}^{|}$ is physical if and only $\hat{B}$ is physical, and they have the same physical norm.

We prove this using correspondence with the operation case discussed in Sec. 50.4.

The maximal representation theorem is useful because it allows us to prove causal dilation theorems for operators without pointer wires and then convert back to operators with pointer wires using maximal elements. This is useful because we have the twofold representation for positive operators having no wires.

We defined the rank of twofold positive operator tensors at the end of Sec. 68.4 for the case of operator tensors having only inputs and outputs. We can use the maximal representation to extend the definition of rank to operator tensors having incomes and outcomes. We say that $\text{rank}(\hat{\boldsymbol{B}}) = \text{rank}({}_{|}\hat{\boldsymbol{B}}^{|})$.

## 78.3 Simple dilation theorem

We will now provide a simple dilation theorem - essentially this is the same as the dilation theorem in Sec. 44. The theorem is as follows.

**Simple dilation theorem.** An operator tensor,

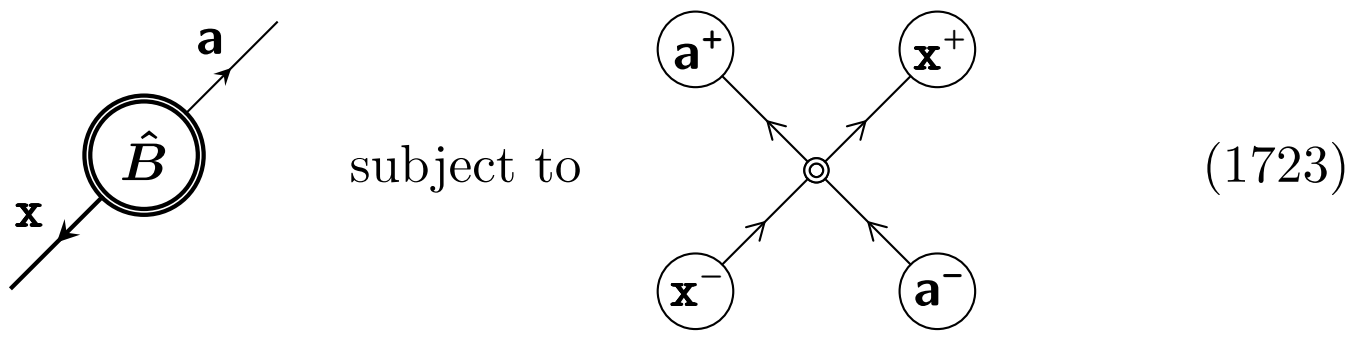

(1723)

(i.e. subject to simple causal structure) is physical and deterministic if and only if it can be written in dilated form

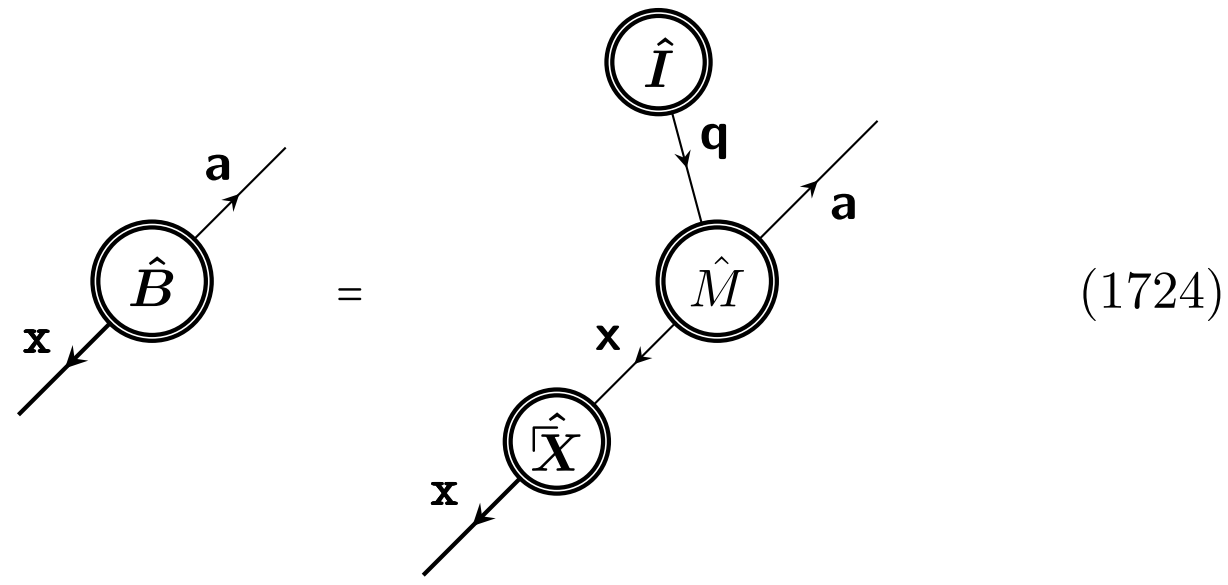

(1724)

where $\hat{M}$ is a natural temporal maxometric operator with respect to ancilla $\mathsf{q}$. The ancilla $\mathsf{q}$ can be any system having $N_{\mathsf{q}} \geq \text{rank}(\hat{\boldsymbol{B}})$.

The proof is along the same lines as the proof in the simple case in Sec. 44.3. First we note that, since we have simple causal structure, the double causality conditions are just the simple double causality conditions (see Sec. 49.4.9). Now, it follows from the maximal representation theorem that $\hat{\boldsymbol{B}}$ is deterministic and physical if and only if $\lfloor\hat{\boldsymbol{B}}\rceil$ is deterministic and physical because they must have the same norm. Also, note that from this theorem, it follows that $\hat{\boldsymbol{B}}$ and $\lfloor\hat{\boldsymbol{B}}\rceil$ have the same causal diagrams (though with the pointer-system replacements mentioned in the maximal representation theorem. Since we have simple causal structure, the statement that $\lfloor\hat{\boldsymbol{B}}\rceil$ is deterministic and physical is equivalent to the conjunction of $T$-positivity and simple double causality being satisfied. We can write $T$-positivity as the condition that $\lfloor\hat{\boldsymbol{B}}\rceil$ can be written in twofold form

(1725)

Note that we can always put $\mathsf{q} = \mathsf{ax}$ since the rank of this operator cannot be greater than $N_{\mathsf{ax}}$. As long as $N_{\mathsf{q}} \geq \text{rank}({}_\llcorner\hat{\boldsymbol{B}}^\urcorner)$, we can use (1541) giving

$$(1726)$$

where

$$(1727)$$

with

$$(1728)$$

The statement that ${}_\llcorner\hat{\boldsymbol{B}}^\urcorner$ is deterministic and physical is equivalent to the statement that $\hat{M}$ is deterministic and pseudo-physical (with respect to ancilla $\mathsf{q}$). Further, we note that $\hat{M}$ is homogeneous. It follows from the theorem in Sec. 74.4 that, for homogeneous $\hat{M}$ with simple causal structure, the statement $\hat{M}$ is deterministic and pseudo-physical is equivalent to the statement that $\hat{M}$ is a natural temporal maxometry with respect to ancilla $\mathsf{q}$. This proves the simple dilation theorem (and that it works for any ancilla, $\mathsf{q}$ with $N_{\mathsf{q}} \geq \text{rank}({}_\llcorner\hat{\boldsymbol{B}}^\urcorner)$).

It is worth noting the following

> **Simple dilation corollary 1.** The natural temporal maxometric operator in the dilation of $\hat{\boldsymbol{B}}$ is given by
>
> $$(1729)$$

with respect to ancilla $\mathsf{q}$.

The form for $M$ in this corollary appears as a step (see (1728)) in the proof of the simple dilation theorem.

The special case where $\hat{\boldsymbol{B}}$ is homogeneous deserves its own corollary.

**Simple dilation corollary 2.** If $\hat{\boldsymbol{B}}$ is of rank one (so $\llcorner\hat{\boldsymbol{B}}\urcorner$ is homogeneous) then a necessary and sufficient condition for $\hat{\boldsymbol{B}}$ to be physical and deterministic is that we can write

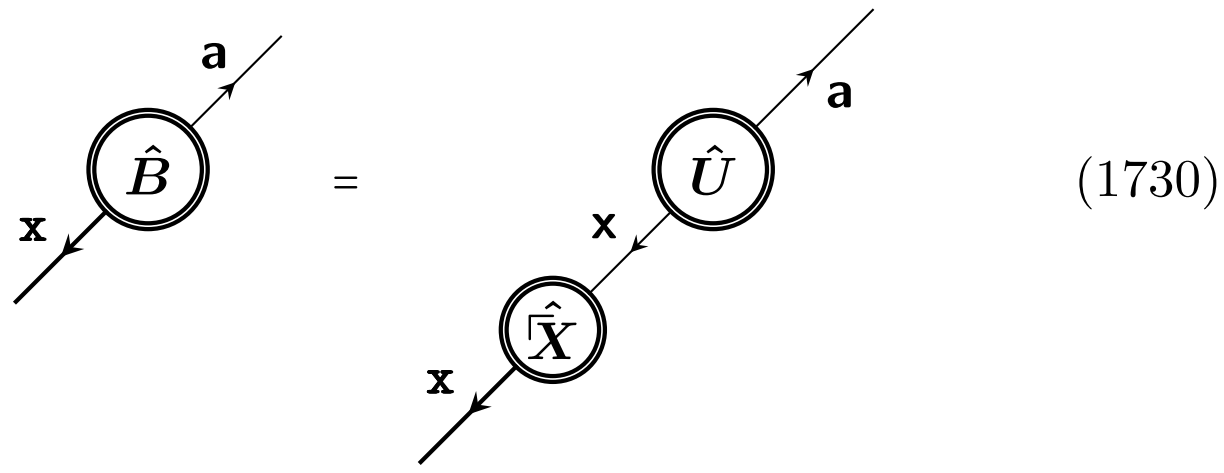

(1730)

where $\hat{\boldsymbol{U}}$ is a natural temporal unitary.

This follows almost immediately from the simple dilation theorem. If $\text{rank}(\hat{\boldsymbol{B}}) = 1$ we can, according to the simple dilation theorem, choose $N_{\mathsf{q}} = 1$. This means that $\mathsf{q}$ is null and can be omitted. In this case there is no $\hat{\boldsymbol{I}}$ and $\hat{M}$ has no ancillae. A natural temporal maxometry without ancillae is a natural temporal unitary (this follows from the basic definitions). Hence, we have (1730).

The simple dilation theorem above was for deterministic physical operations having simple causal structure. For such operations the simple double causality conditions (for deterministic operations) are the full set of double causality conditions. This led to the simple dilation theorem being a necessary and sufficient condition ("if and only if"). If we consider deterministic physical operations not necessarily having simple causal structdure then the simple double causality conditions are still necessary conditions (see Sec. 49.4.8). This means we can proof that the dilation is a necessary condition (but, in general, not a sufficient condition). Thus we have the following theorem:

**Dilation theorem - necessary condition** Any deterministic physical operator tensor, $\hat{\boldsymbol{B}}$ can be written in dilated form

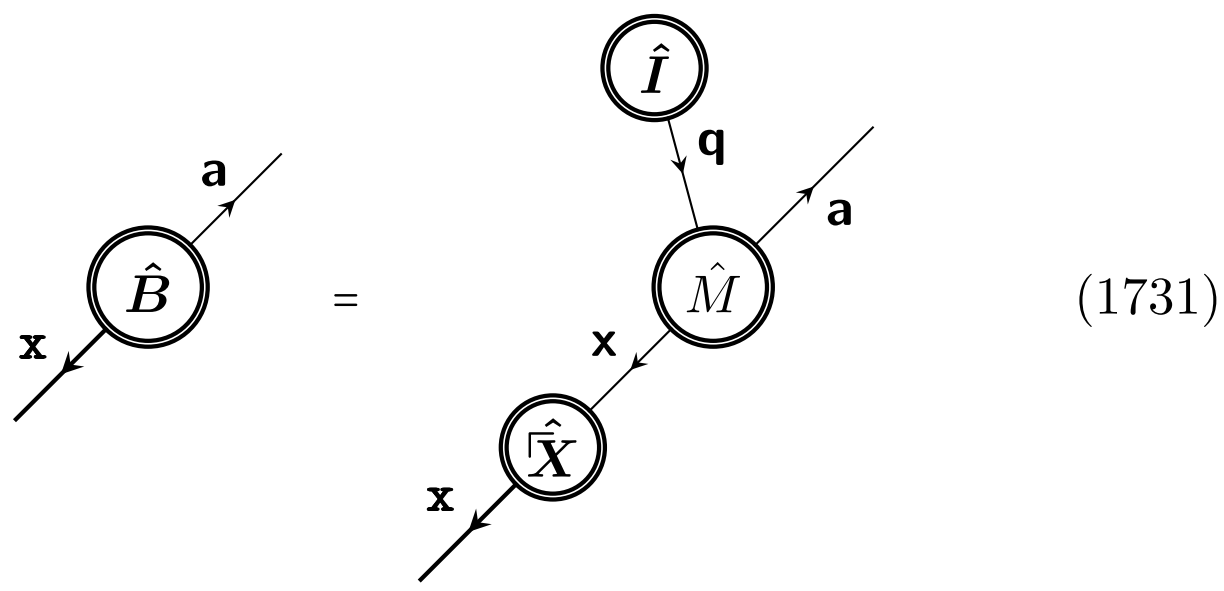

(1731)

where $\hat{M}$ is a natural temporal maxometric operator with respect to ancilla $\mathsf{q}$ (where the ancilla can be any system satisfying $N_{\mathsf{q}} \geq \text{rank}(\llcorner\hat{\boldsymbol{B}}\urcorner)$). However, except when $\hat{\boldsymbol{B}}$ has simple causal structure, this is not a sufficient condition for $\hat{\boldsymbol{B}}$ to be deterministic and physical.

This theorem follows by just looking at the "only if" part of the proof of the simple dilation theorem above.

In Sec. 45.1 we discussed constructible and unconstructible operators in the context of the simple framework. Constructible operators are deterministic physical operators that can be modelled by a dilation where the natural temporal maxometry, $\hat{M}$, is a natural unitary, and unconstrucible operators are ones that cannot. In Sec. 45.2 we made some conjectures concerning whether unconstructible operators actually exist. These conjectures apply equally to the discussion of the simple dilation theorem here. The conjectures become even more interesting for the causal dilation theorems we will discuss next when we may have non-simple causal structure.

## 78.4 Simple dilation theorems with sufficient pairs

In Sec. 44.4 we saw how to state the dilation theorem for simple operators with sufficient pairs in two different ways - using (co)isometries and using unitaries. It follows we can do the same here for simple causal structure.

First we note that, according to the above dilation theorems, we can choose any $\mathsf{q}$ with $N_{\mathsf{q}}$ greater than or equal to the rank of $\llcorner\hat{\boldsymbol{B}}\urcorner$. This fact is consistent with the theorem in Sec. 74.5. In particular, this means we can choose $\mathsf{q}$ to be all input, or all output as follows.

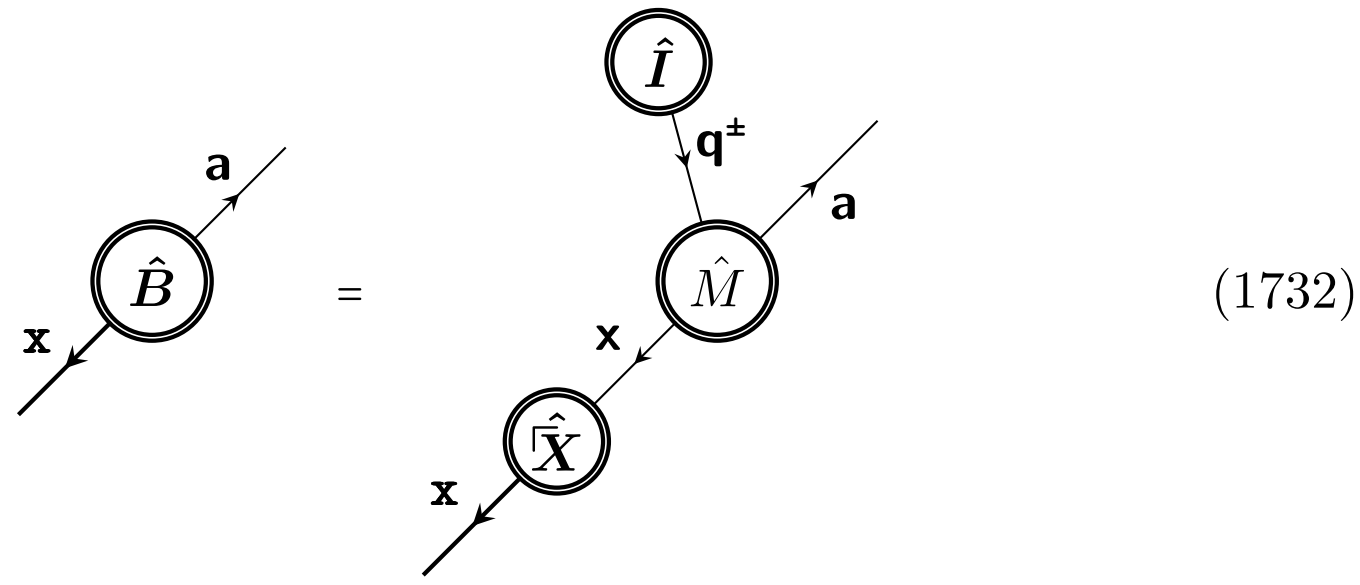

We will now use this to prove the two versions of a sufficient pair theorem which apply to an operator

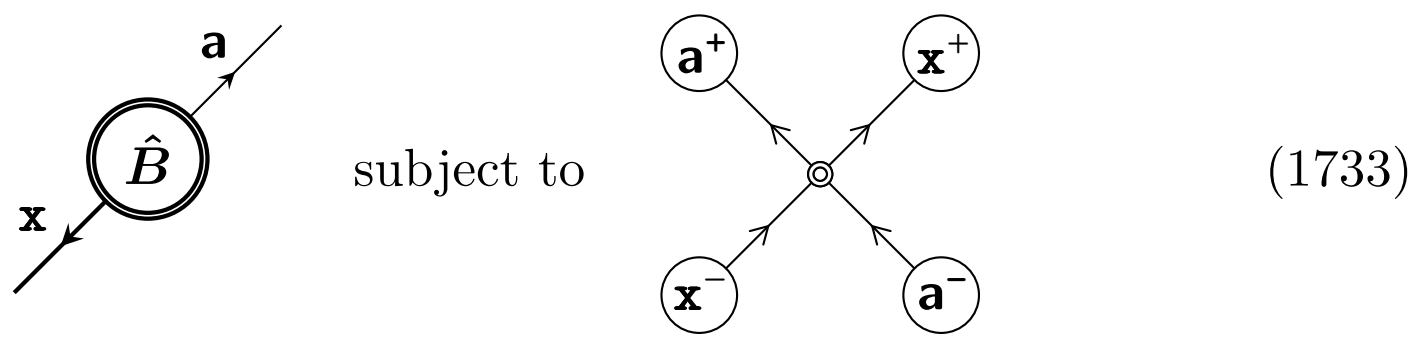

- this has simple causal structure.

First consider the (co)isometry case.

> **Simple dilation theorem with (co)isometric sufficient pairs.** An operator tensor with simple causal structure as shown in (1733) is deterministic and physical if and only if it can be written as being equal to *both*

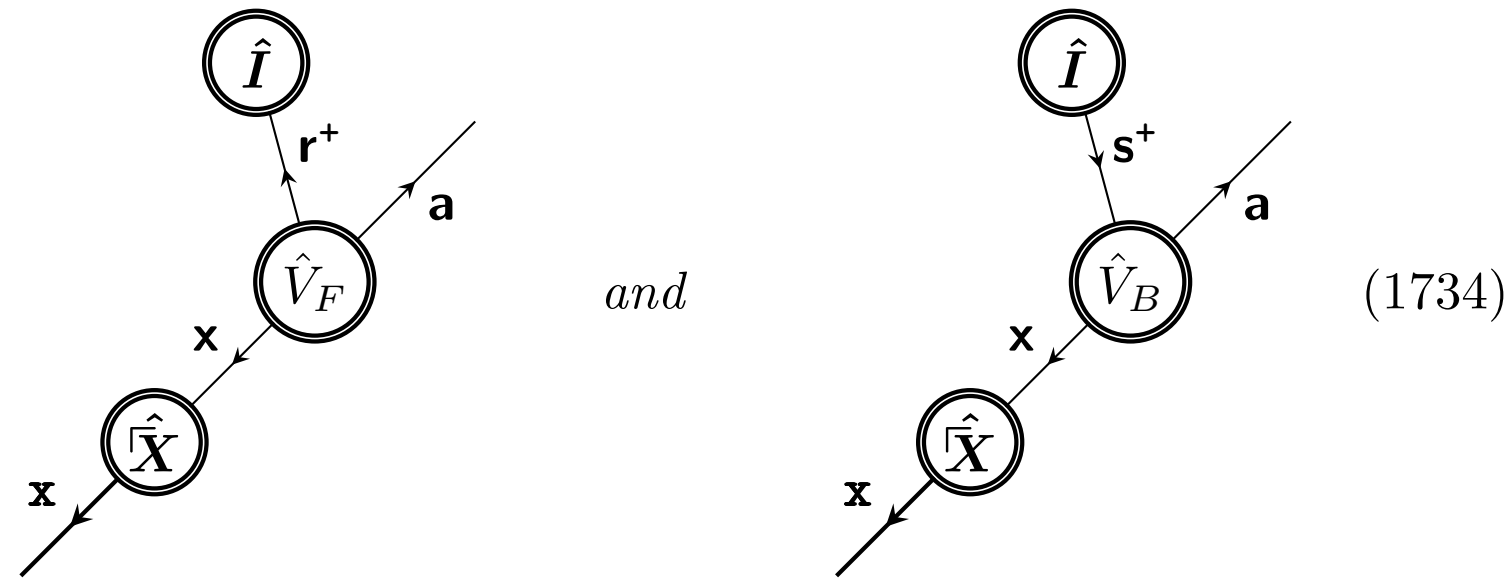

> where $\hat{\boldsymbol{V}}_F$ is a natural temporal isometric operator and $\hat{\boldsymbol{V}}_B$ is a natural temporal coisometric operator. Further, $\mathsf{r}^+$ and $\mathsf{s}^+$ can be any systems satisfying $N_{\mathsf{r}^+} \geq \mathrm{rank}(\hat{\boldsymbol{B}})$ and $N_{\mathsf{s}^+} \geq \mathrm{rank}(\hat{\boldsymbol{B}})$.

The proof of this is straightforward. To obtain the expression on the left we can choose the ancilla of $\hat{M}$ to be all output then use the fact that a natural temporal maxometry with only output ancilla is a natural temporal isometry. To obtain the expression on the right we take the ancilla to be all input and use the fact that a natural temporal maxometry with only input ancilla is a natural temporal coisometry. Then it is easy to verify that both expressions satisfy $T$-positivity and, further, the left expression satisfies forward causality and the right expression satisfies backward causality.

We can use this to prove the following statement of the dilation theorem

> **Simple dilation theorem with unitary sufficient pairs.** An operator tensor with simple causal structure as shown in (1733) is deterministic and physical if and only if it can be written as being equal to *both*

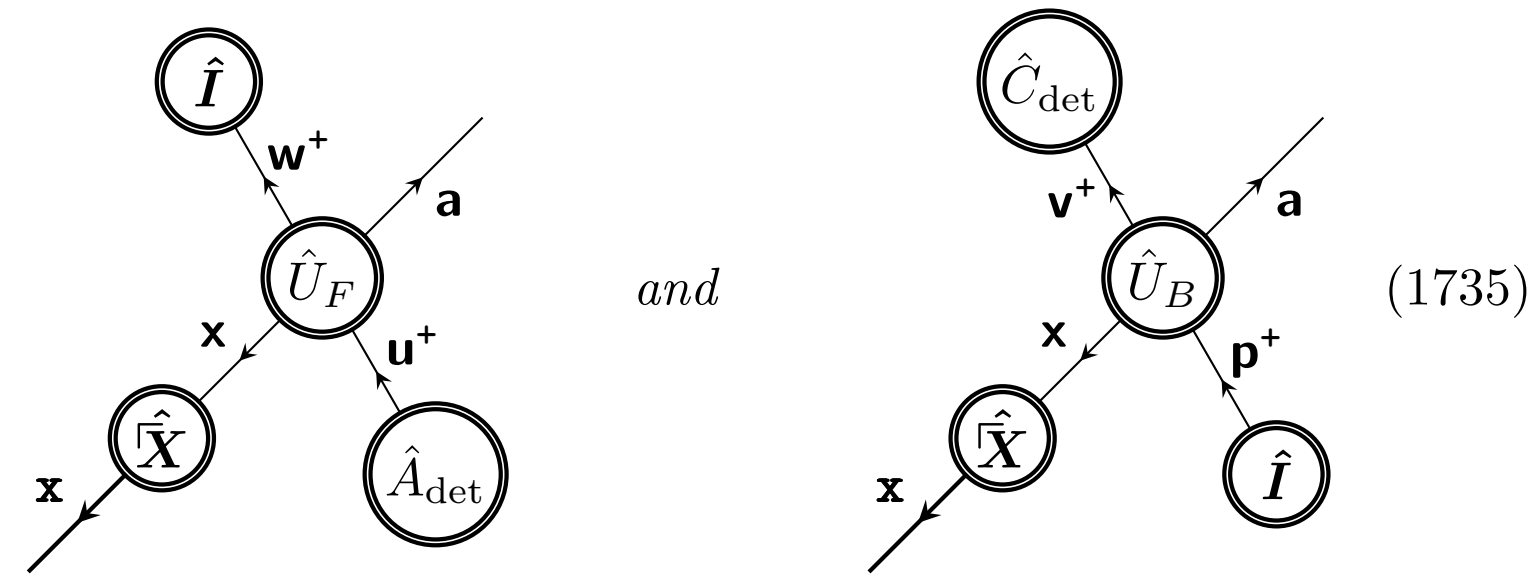

> where $\hat{A}_{\mathrm{det}}$ and $\hat{C}_{\mathrm{det}}$ are homogeneous and deterministic and $\hat{\boldsymbol{U}}_F$ and $\hat{\boldsymbol{U}}_B$ are natural temporal unitary operators.

Note that $\hat{A}_{\rm det}$ and $\hat{C}_{\rm det}$ are, necessarily, not physical. On the other hand, $\hat{\boldsymbol{U}}_F$ and $\hat{\boldsymbol{U}}_B$ are necessarily deterministic and physical because we have simple causal structure. The proof of this theorem is also straightforward. First we note that, using (1701) in (1732) we immediately obtain (1735) as necessary conditions (where, in each case, we combine two $\hat{\boldsymbol{I}}$'s into a single $\hat{\boldsymbol{I}}$). Next we note that if $\hat{\boldsymbol{B}}$ takes the left form in (1735) it satisfies forward causality, and if it takes the right form in (1735) it satisfies backward causality. Since these forms also guarantee that $T$-positivity is satisfied we have sufficient conditions for $\hat{\boldsymbol{B}}$ to be deterministic and physical.

It is worth noting that the (co)isometric sufficient pair has simple inequality constraints relating the dimension of the ancillary systems and the rank of the operator. In particular, if the operator is homogeneous then we can set these dimensions equal to one such that the systems are null. Then the ignore operators, $\hat{\boldsymbol{I}}$, do not appear. The same is not true for the unitary sufficient pair. The selling point for the unitary sufficient pair is that all the nonphysicality appears in $\hat{A}_{\rm det}$ and $\hat{C}_{\rm det}$. Further, the forward dilation (on the left), is familiar as the Stinespring dilation from standard time forward Quantum Theory (and the backward dilation is then the time reverse of the Stinespring dilation).

## 78.5 The basic unitary causal dilation theorem

The two theorems in the previous section leverage only the simple double causality conditions (which stem from synchronous partitions that are either "before" or "after" the whole causal diagram - see Sec. 49.4.8). In this subsection and the next we will provide dilation theorems that pertain to a general synchronous partition, $p$, through a causal diagram for an operator

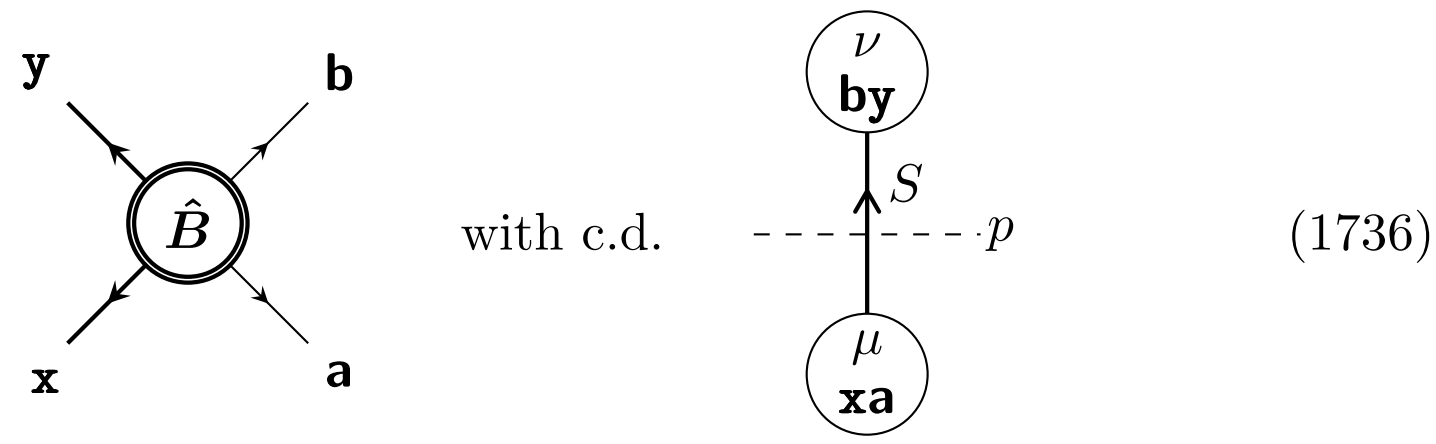

(1736)

subject to the forward causality condition at $p$

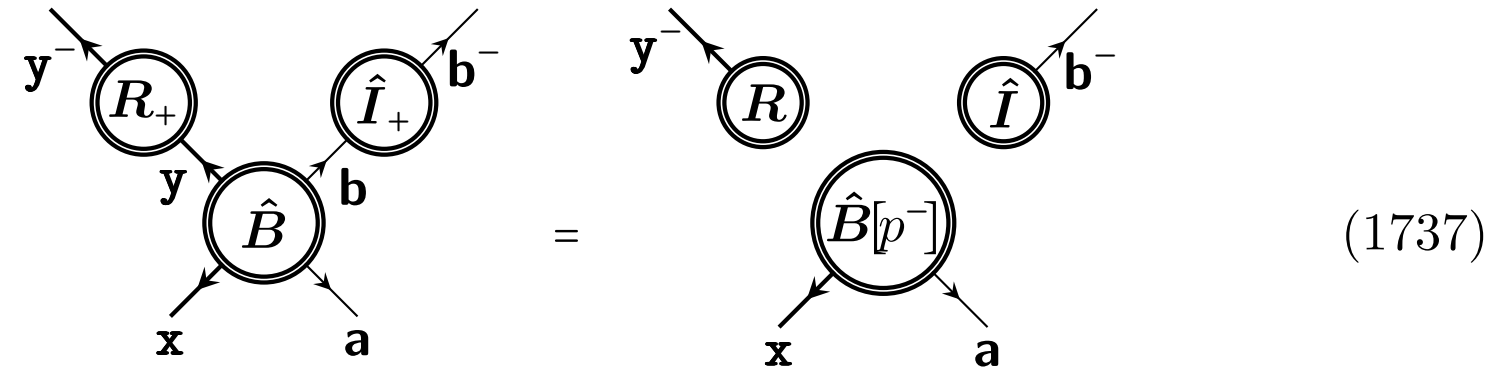

(1737)

and the backward causality condition at $p$

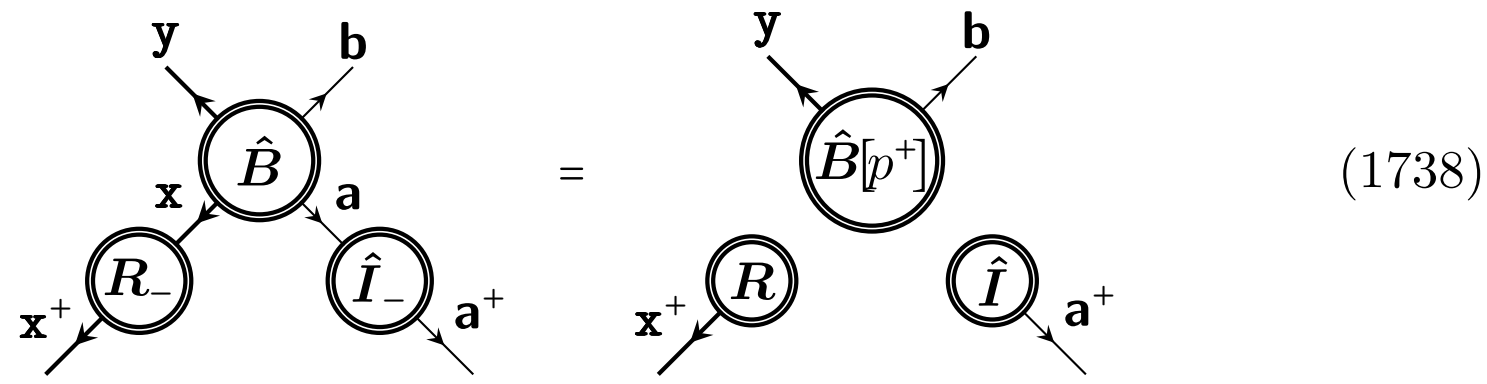

(1738)

Here $\hat{\boldsymbol{B}}[p\pm]$ are the future and past residua respectively. In Sec. 49.4.7 we proved theorems for the residua of operations which carry over to operators under correspondence. In particular, it follows (under correspondence) from the double causality residua theorem in Sec. 49.4.7 that, if $\hat{\boldsymbol{B}}$ satisfies double causality for every synchronous partition, then do the residua. This means that the residua should satisfy the simple double causality conditions (since the latter are a special case of double causality conditions). Thus, simple double causality of residua is a very minimal consequence of physicality.

In the theorem below we provide a dilation that is equivalent to the forward causality condition and another dilation that is equivalent to the backward causality condition. These dilations each employ a natural temporal unitary dilation such that the only nonphysical element is the maxometry that also appears in each of these dilations. In Sec. 78.6 we will prove a theorem where we have a natural temporal isometric operator in the forward dilation and a natural temporal coisometric operator in the backward dilation.

These theorems are useful for proving subsequent theorems on some specific causal diagrams. To prove this theorem we adapt the ingenious technique at the heart the proof in Theorem 6 of Gutoski and Watrous [2007] and Theorem 3 of the work in Chiribella, D'Ariano, and Perinotti [2009a] so it is suitable for the time symmetric setting - meaning we have a forward and a backward condition and also we employ maxometries. A key step in these proofs uses the fact that there is a isometry between two equal decompositions of an operator. This step appears as (1764) in Sec. 78.6 and as (1749) in the current section (where we make this isometry into an unitary). To prove the unitary case we also have to extend the technique so that the matrix, $U$, below is a unitary (actually a natural temporal unitary) rather than just an isometry (as in the aforementioned works). Further, we present a pictorial proof (rather than a symbolic one as used in the aforementioned work). Whilst this takes more page space, it elucidates the mathematical steps required. The steps in the proof below are modeled on specifically on the steps in the proof of Theorem 3 by CDP.

We will prove the following theorem.

**Basic unitary causal dilation theorem**. Consider a twofold pos-itive operator, $\hat{\boldsymbol{B}}$, satisfying

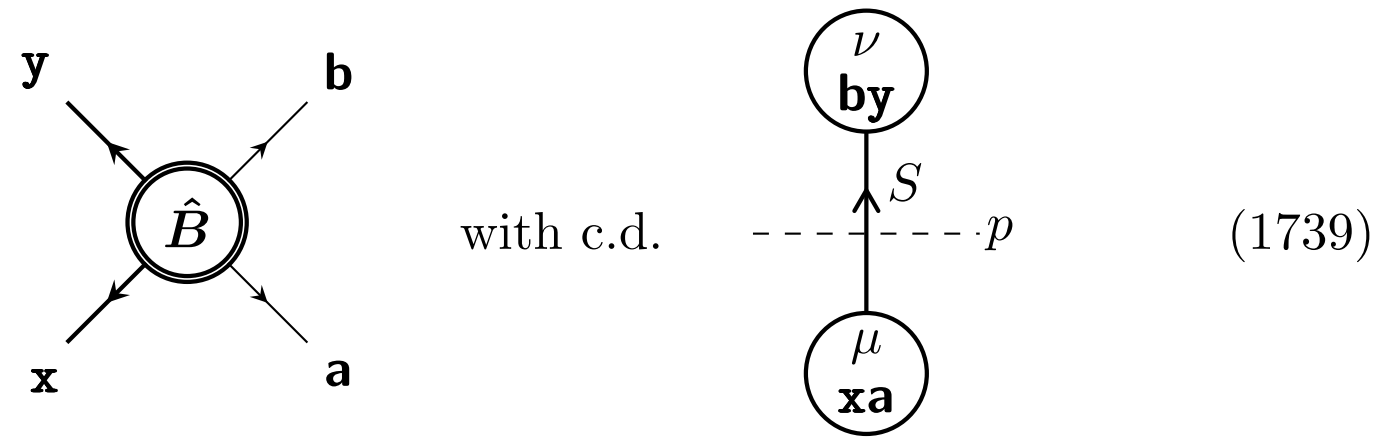

(1739)

Then the following statements are true.

**Forward case.** The forward causality condition (1737) at $p$ is equivalent to

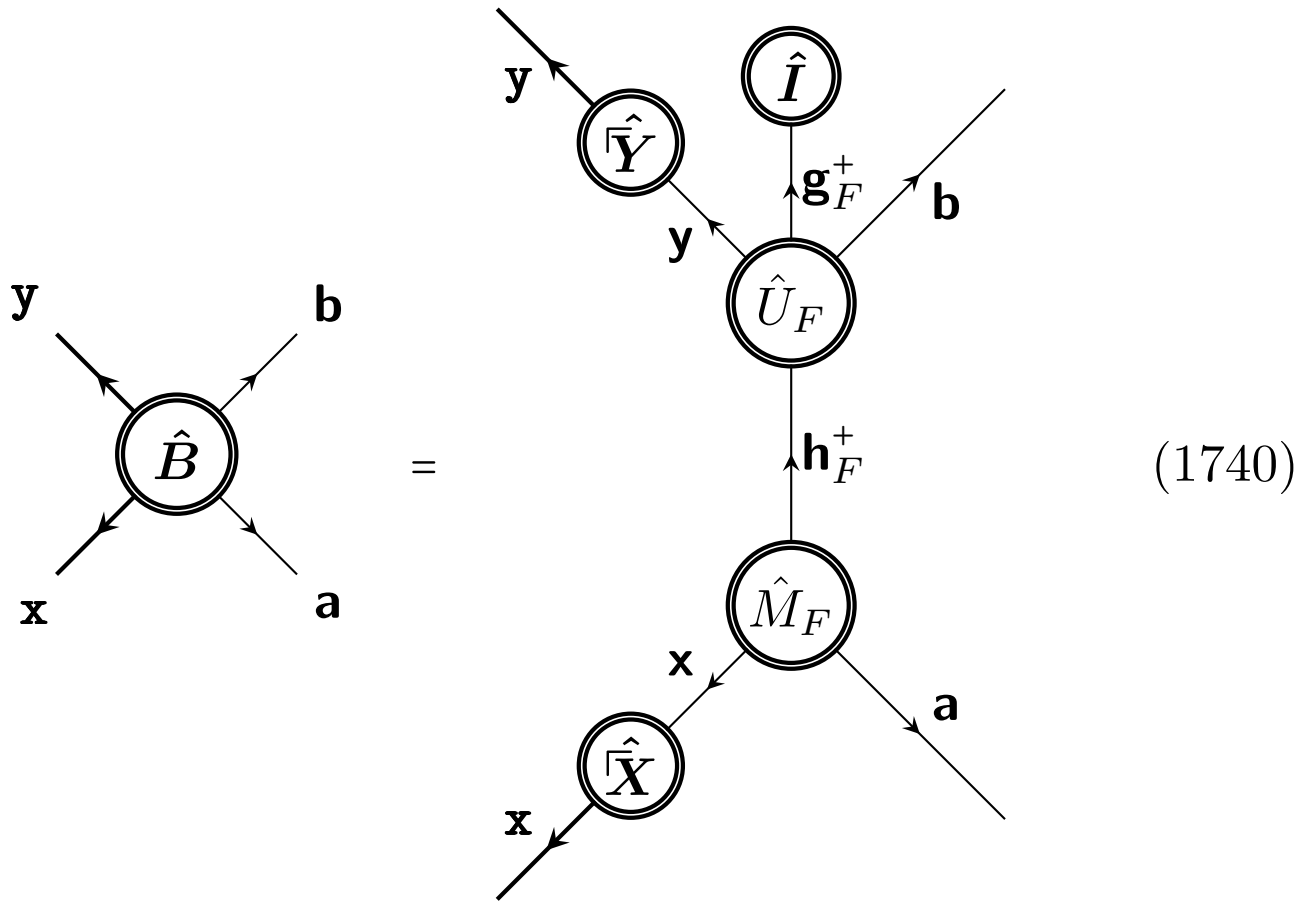

(1740)

where $\hat{U}_F$ is a natural temporal unitary, and where $\hat{M}_F$ is homogeneous and satisfies

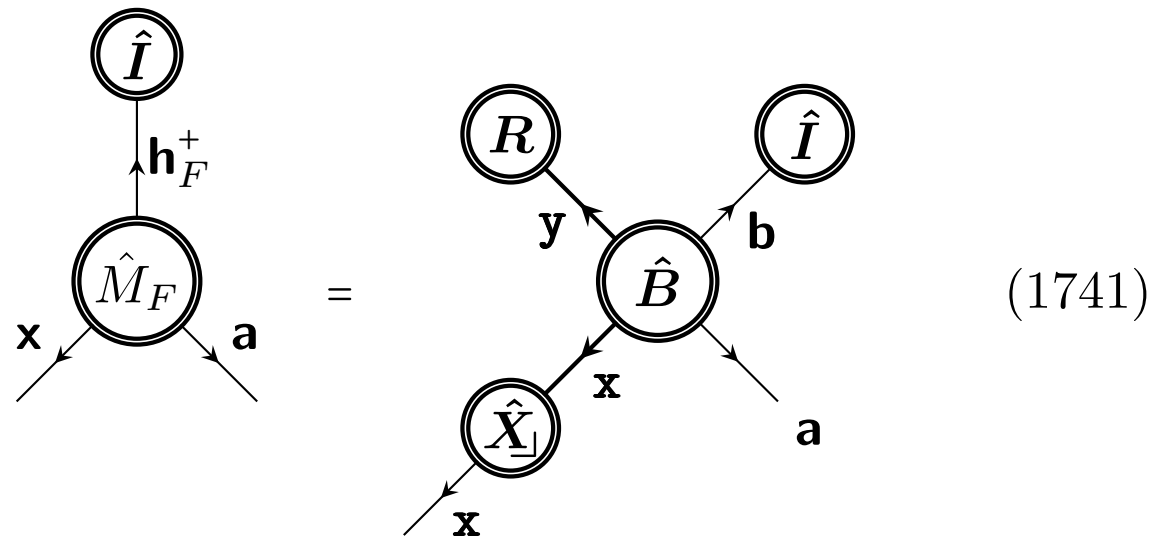

(1741)

Further, $\hat{\boldsymbol{B}}[p^-]$ (the past residuum of $\hat{\boldsymbol{B}}$ at $p$) satisfies the simple double causality conditions if and only if $\hat{M}_F$ is a natural temporal maxometry (with respect to ancilla $\mathsf{h}_F^+$).

**Backward case.** The backward causality condition (1738) at $p$ is equivalent to the condition

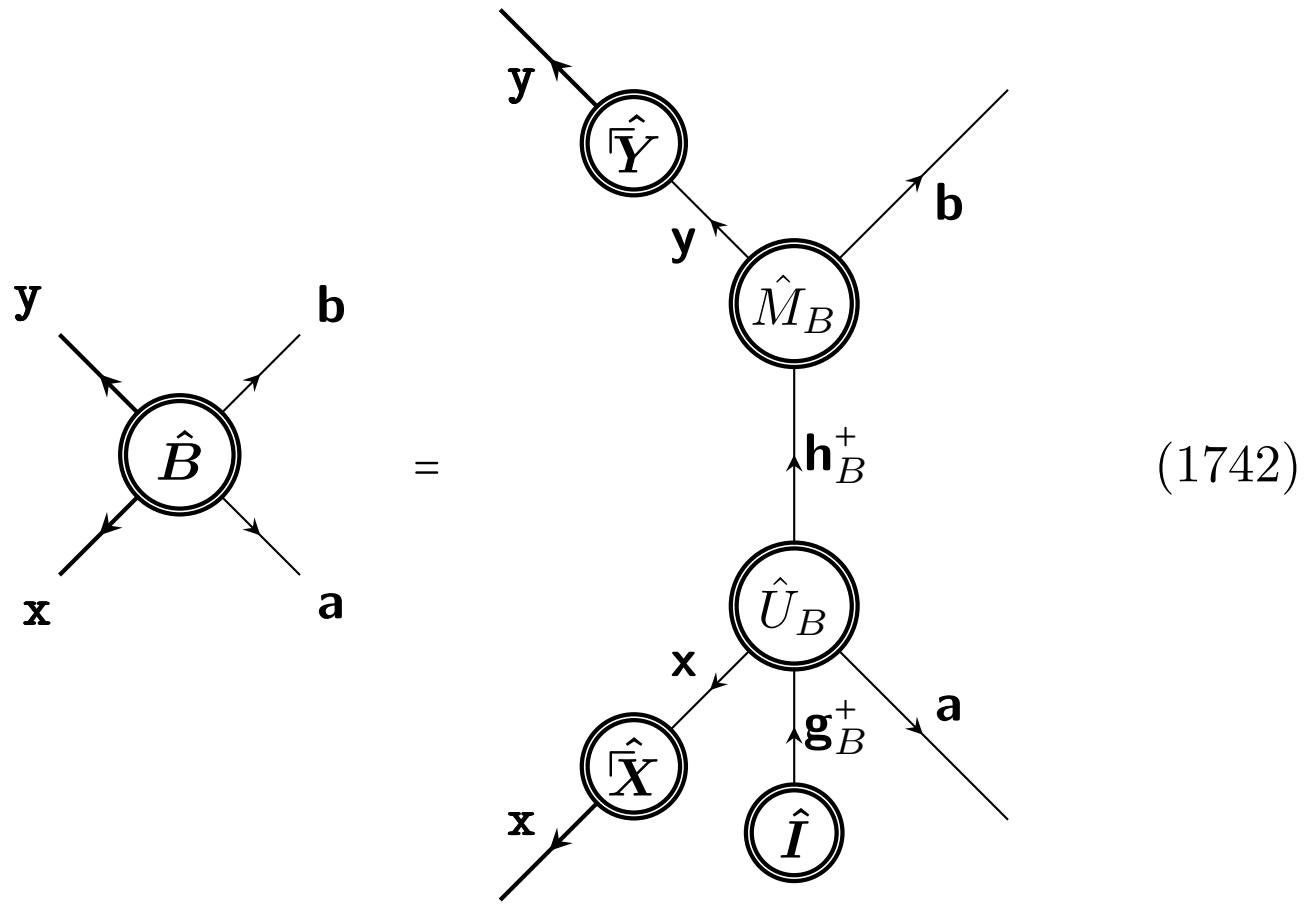

(1742)

where $\hat{U}_B$ is a natural temporal unitary, and where $\hat{\boldsymbol{M}}_B$ is homogeneous and satisfies

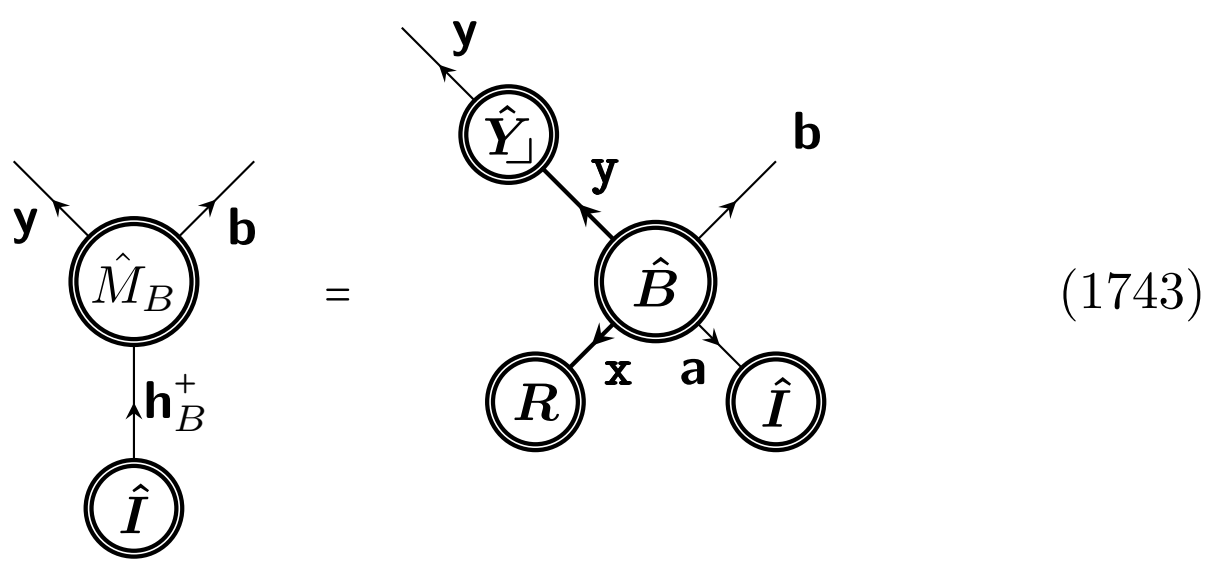

(1743)

Further, $\hat{\boldsymbol{B}}[p^+]$ (the future residuum of $\hat{\boldsymbol{B}}$ at $p$) satisfies the simple double causality conditions if and only if $\hat{M}_B$ is a natural temporal maxometry (with respect to ancilla $\mathsf{h}_B^+$).

Although we will always apply this to situations where $\hat{\boldsymbol{B}}$ is physical and deterministic (meaning that it satisfies double causality conditions at all synchronous partitions), note that the main parts of the theorem statement only pertain, effectively, to double causality conditions being imposed at a single synchronous partition. We will prove the forward case. The backward case follows by similar reasoning. First we will prove that (1740) follows from (1737). To simplify steps in the proof, we will use $\llcorner\hat{\boldsymbol{B}}\urcorner$ (as defined in (1721)). Afterwards, we can convert back to $\hat{\boldsymbol{B}}$ by applying maximal elements. The maximal representation theorem in Sec. 78.2 guarantees that $\llcorner\hat{\boldsymbol{B}}\urcorner$ has same pertinent properties as $\hat{\boldsymbol{B}}$. We will

work with

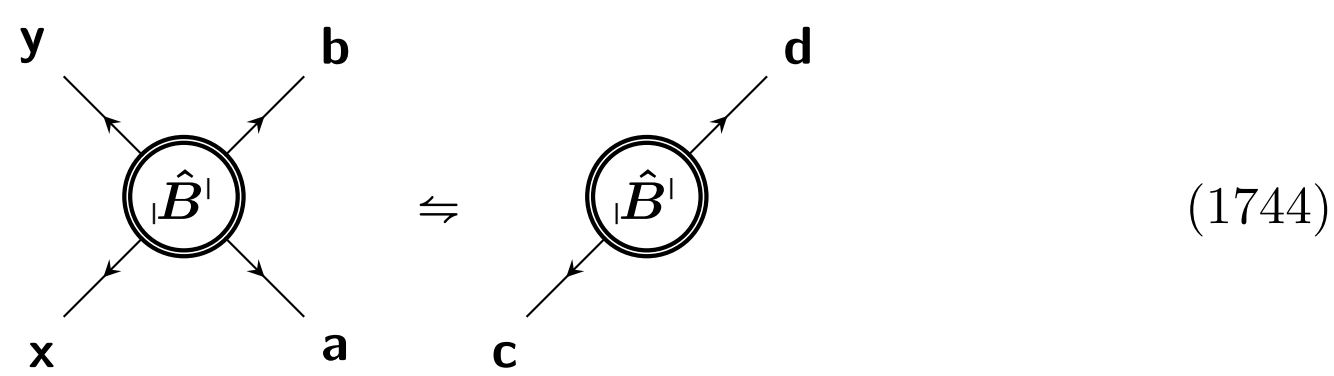

$$(1744)$$

where $\mathsf{c} = \mathsf{ax}$ and $\mathsf{d} = \mathsf{by}$. Under this move, we have to prove that

$$(1745)$$

(this is the forward causality condition (1737) under the conversion) is equivalent to

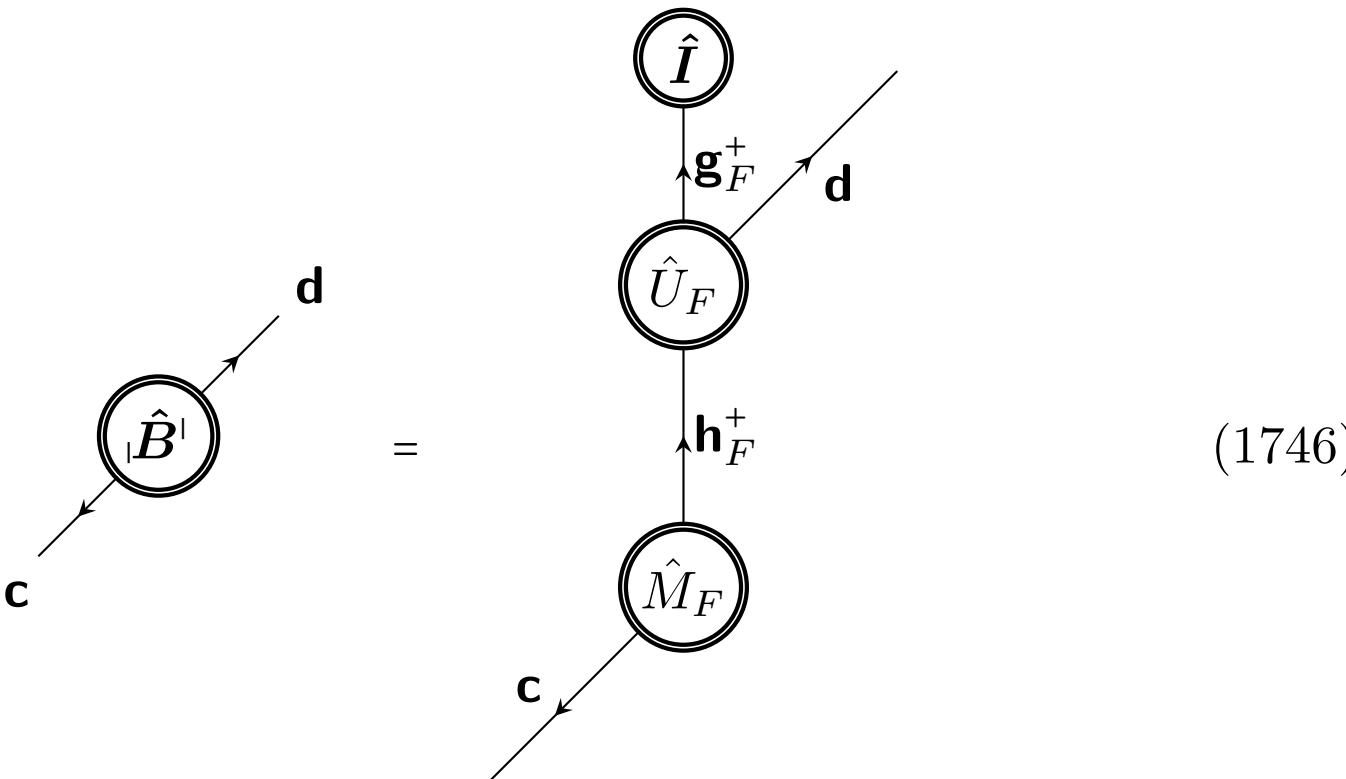

$$(1746)$$

(this is (1740) under the conversion). If we can prove this then, by reapplying the maximal elements, we prove that (1737) and (1740) are equivalent. Since, according the theorem statement, $\hat{\boldsymbol{B}}$ is twofold positive, ${}_{|}\hat{\boldsymbol{B}}^{|}$ is also twofold positive (this was proven in the maximal representation theorem). Further, it follows that ${}_{|}\hat{\boldsymbol{B}}^{|}[p^-]$ is also twofold positive (using (1745), the fact that twofold positivity is equivalent to tester positivity as proven in the positivity theorem in Sec. 76.2, and the **RI** partial contraction positivity theorem in Sec. 49.2.4 under correspondence). Consequently, the forward causality condition in (1746) can

be written as

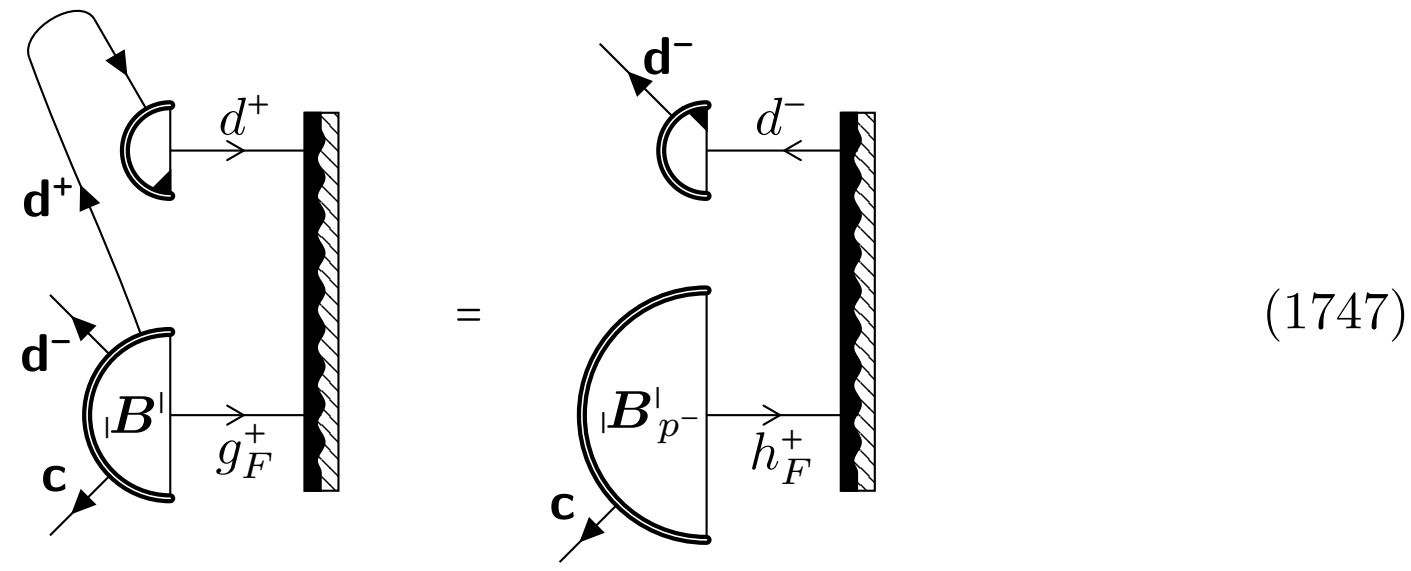

(1747)

where we have written ${}_\shortmid\boldsymbol{B}^\shortmid[p^-] = {}_\shortmid\boldsymbol{B}^\shortmid{}_{p^-}$ for space management reasons (this is the left object associated with the past residuum, ${}_\shortmid\hat{\boldsymbol{B}}^\shortmid[p^-]$). Here we have used the definition of $\hat{\boldsymbol{I}}_+$ (which is through correspondence with the definition of $\mathsf{I}_+$ in (1103)) and then we have written the equation in twofold form and used (1519, 1520) to replace the $\hat{\boldsymbol{I}}$'s. We must have

$$|\{g_F^+\}| \geq \text{rank}({}_\shortmid\hat{\boldsymbol{B}}^\shortmid) \qquad \text{and} \qquad |\{h_F^+\}| \geq \text{rank}({}_\shortmid\hat{\boldsymbol{B}}^\shortmid[p^-]) \tag{1748}$$

(note, we are free to choose the label wires $g_F^+$ and $h_F^+$ to have + superscripts - this turns out to be important in the proof below). Applying (1489) from the unitary freedom theorem for twofold operators from Sec. 68.7 to (1747) gives

(1749)

where we choose $g_F^+$ and $h_F^+$ such that

$$|\{g_F^+\}||\{d^+\}| = |\{h_F^+\}||\{d^-\}| \tag{1750}$$

so that we can use the unitary freedom theorem. It is clearly always possible to choose integers $|\{g_F^+\}|, |\{h_F^+\}| \geq 1$ such that both (1748) and (1750) are satisfied. [Aside: In Sec. 78.6 we will state the theorem where we do not impose (1750) and use an isometry rather than a unitary at this point - this is what Gutoski and Watrous [2007] do in their Theorem 6 and Chiribella et al. [2009a] do in

their Theorem 3.] From (1749) we obtain

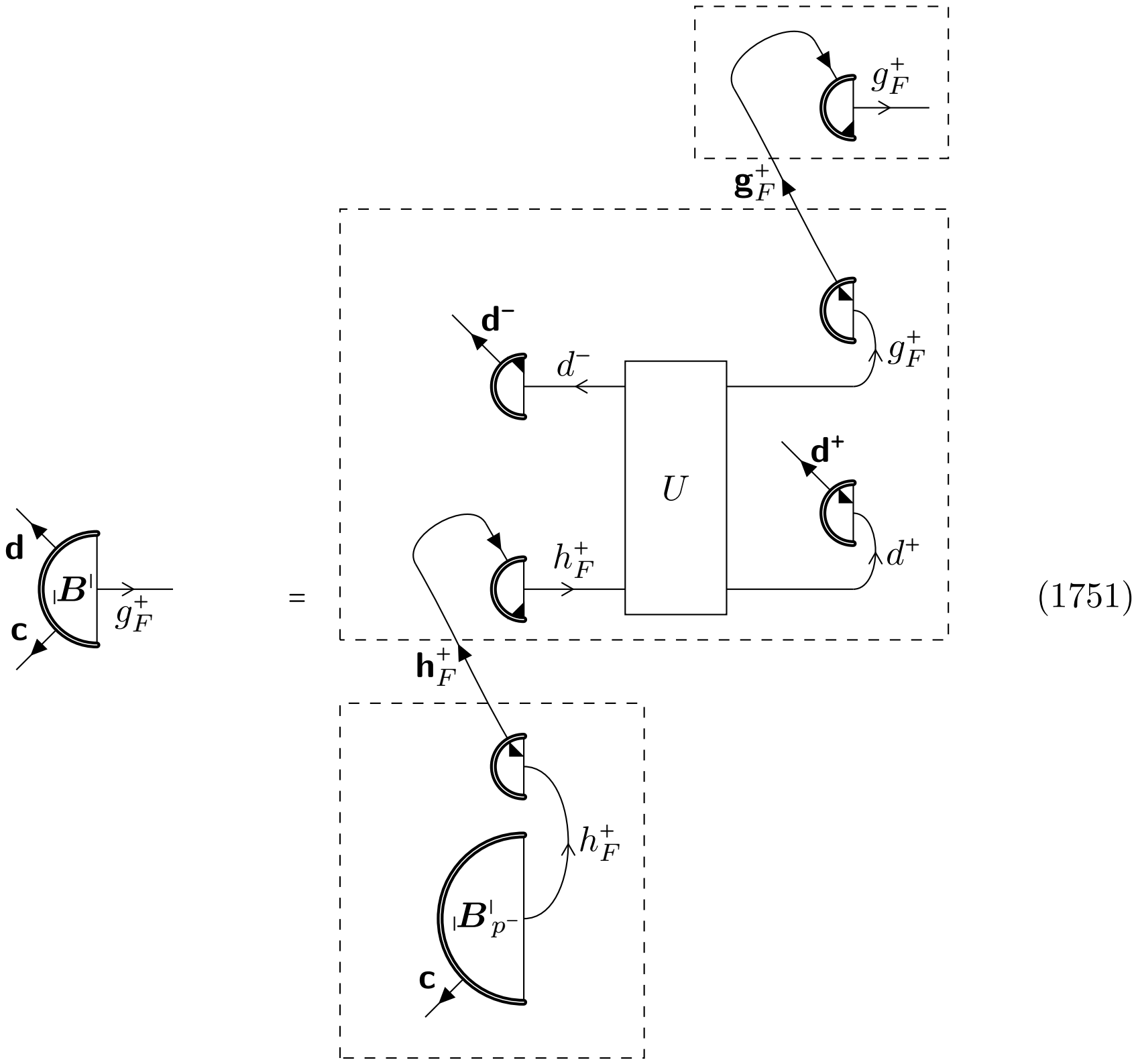

(1751)

To get this from (1749) involves three steps. First, we have hit the open $d^+$ label on both sides of (1749) with

(1752)

On the left hand side, we have used (1540) to simplify. Then we have recombined the $\mathbf{d}^+$ and $\mathbf{d}^-$ wires into a $\mathbf{d}$ wire. Second, we have used (1536) to replace the $h_F^+$ wire with an object containing the scalar product of $\mathbf{h}_F^+$ basis elements. Third, we have used (1536) on the $g_F^+$ label wire to introduce the scalar product of $\mathbf{g}_F^+$ deterministic and physical basis elements (also note that we have interchanged the positions of the $d^+$ and $g_F^+$ wires on the right of the $U$ box for space management reasons). We can write (1751) as follows

$$(1753)$$

The left object, $M_F$, corresponds to the left object in the lower dashed box in (1751). The left object, $U_F$, corresponds to the left object in the middle dashed box. It is easy to prove that $U_F$ is a natural temporal unitary. Looking at the expression for this in the middle dashed box in (1751), we see that the unitary matrix, $U$ has + wires going in on the left (the $d^-$ label wire is pointing out, but it has a minus sign, so it actually corresponds to a system moving in). Further, the wires on the left are all moving out. Thus, the left object satisfies the properties for temporality. We can prove that it is a natural temporal unitary (by substituting the expression for $U_F$ in the middle dashed box into (1649)). We can now reflect the expression in (1753) in a regular physical mirror. This gives (1746) as required. This proves that (1746) is a *necessary* consequence of (1745). To prove they are equivalent we need to prove that (1746) is *sufficient* for (1745) to be true. This is, in fact, easily proven by substituting the right hand of (1746) into the right hand side of (1745). In so doing, we obtain

$$(1754)$$

This proves equivalence of (1745) and (1746). Invoking the maximal representation theorem we thereby prove equivalence of the forward causality condition (1737) and (1740). To prove (1741) we hit the left and right hand side of (1740) with an $\boldsymbol{R}$ on $\mathbf{y}$ and an $\hat{\boldsymbol{I}}$ on $\mathbf{b}$ and use the fact that $\hat{U}_F$ is a natural temporal unitary. It follows from the natural temporal maxometries theorem in Sec. 74.4 that ${}_{\lfloor}\hat{\boldsymbol{B}}^{\rceil}[p^-]$ satisfies simple double causality if and only if $\hat{M}_F$ is a natural temporal maxometry. We can convert this back to the statement for $\hat{\boldsymbol{B}}[p^-]$ in the theorem statement by applying maximal elements as necessary. This completes the proof of the forward part of the theorem. The backward part of the theorem follows similarly.

It is useful, as in the case of the simple dilation theorem, to add a corollary about the form of the maxometries used

**Basic unitary causal dilation corollary.** We have a forward and a backward case for this corollary.

**Forward case.** The natural temporal maxometric operator, $\hat{M}_F$, in the dilation of $\hat{\boldsymbol{B}}$ in (1740) is formed from the natural temporal maxometry

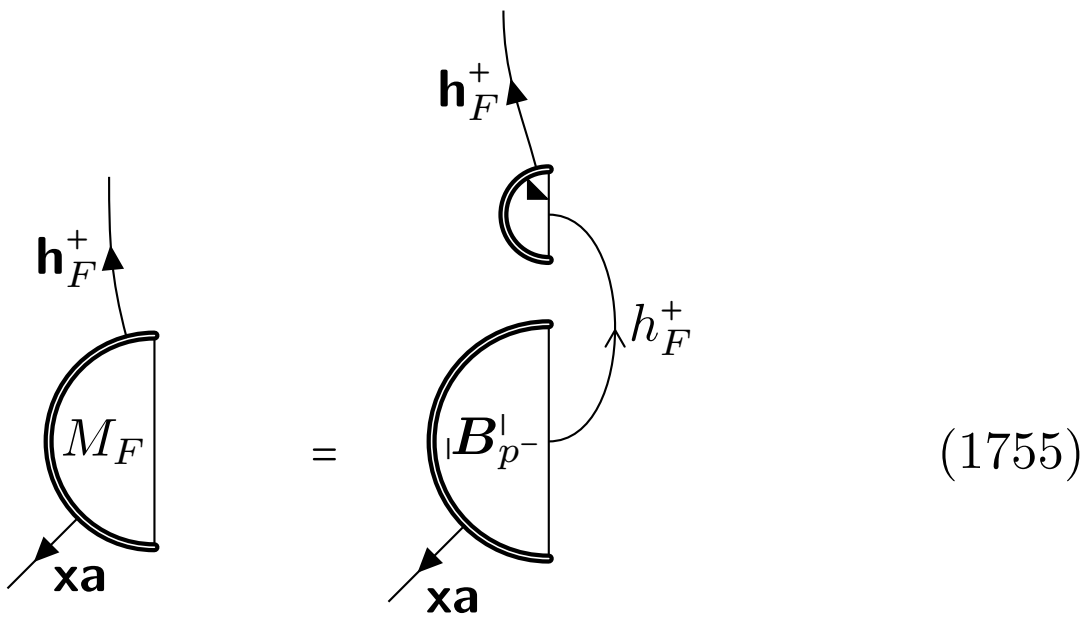

(1755)

having ancilla $\mathsf{h}_F^+$. Note that ${}_{|}\boldsymbol{B}^{|}_{p-}$ is shorthand for ${}_{|}\boldsymbol{B}^{|}[p^-]$ (this is a left object associated with the past residuum, ${}_{|}\hat{\boldsymbol{B}}^{|}[p^-]$ at $p$).

**Backward case.** The natural temporal maxometric operator, $\hat{M}_B$, in the dilation of $\hat{\boldsymbol{B}}$ in (1742) is formed from the natural temporal maxometry

(1756)

having ancilla $\mathsf{h}_B^+$. Note that ${}_{|}\boldsymbol{B}^{|}_{p+}$ is shorthand for ${}_{|}\boldsymbol{B}^{|}[p^+]$ (this is a left object associated with the future residuum, ${}_{|}\hat{\boldsymbol{B}}^{|}[p^+]$ at $p$).

The proof of the forward case follows from noting that $M_F$ is defined to be the object inside the lower dashed box in (1751). The backward case follows similarly.

## 78.6 Basic (co)isometric causal dilation theorem

In proving the basic unitary causal dilation theorem we did a bit of extra work to have unitaries. In particular, we noted that we can always find $|\{g_F^+\}|$ and $|\{h_F^+\}|$ that satisfy (1748) and (1750) so that $\hat{\boldsymbol{U}}_F$ in the forward case is a natural temporal unitary (and we proceed similarly in the backward case). The motivation for having natural temporal unitaries is that, then, all the nonphysicality is in the maxometry $\hat{M}_F$ (and $\hat{M}_B$ in the backward case).

There are, however, also good reasons to arrange to have an natural temporal isometry, $\hat{V}_F$, in the forward case (and a natural temporal coisometry, $\hat{V}_B$, in the backwards case). In particular, we will see that the rank of the ignore operator, $\hat{\boldsymbol{I}}$, appearing in these dilations can be set equal to the rank of $\hat{\boldsymbol{B}}$. This is particularly useful when we want to study homogeneous $\hat{\boldsymbol{B}}$ (since then we do not need an ignore operator). Further, the dimension of the system $\mathsf{h}_F^+$ ($\mathsf{h}_B^+$ in the backward case) connecting the past and future parts can be set equal to the rank of the past (future) residuum at $p$.

In fact, Gutoski and Watrous [2007] in their Theorem 6 and Chiribella, D'Ariano, and Perinotti [2009a] in their Theorem 3 give the isometric case (since they are looking at time forward Quantum Theory, they do not study the backwards case which has a coisometry here). The key difference with the basic unitary causal dilation theorem (in Sec. 78.5 is that we do not need to find $|\{g_F^+\}|$ and $|\{h_F^+\}|$ that satisfy the equality (1750). First, let us state the theorem.

**Basic (co)isometric causal dilation theorem**. Consider a twofold positive operator, $\hat{\boldsymbol{B}}$, satisfying

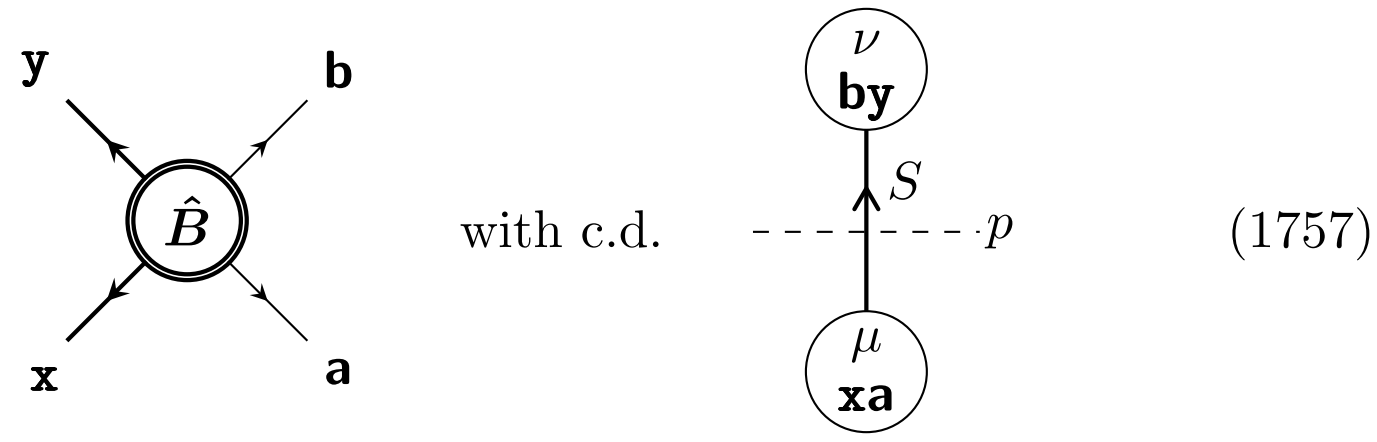

(1757)

Then the following statements are true.

**Forward case.** The forward causality condition (1737) at $p$ is equiv-

alent to

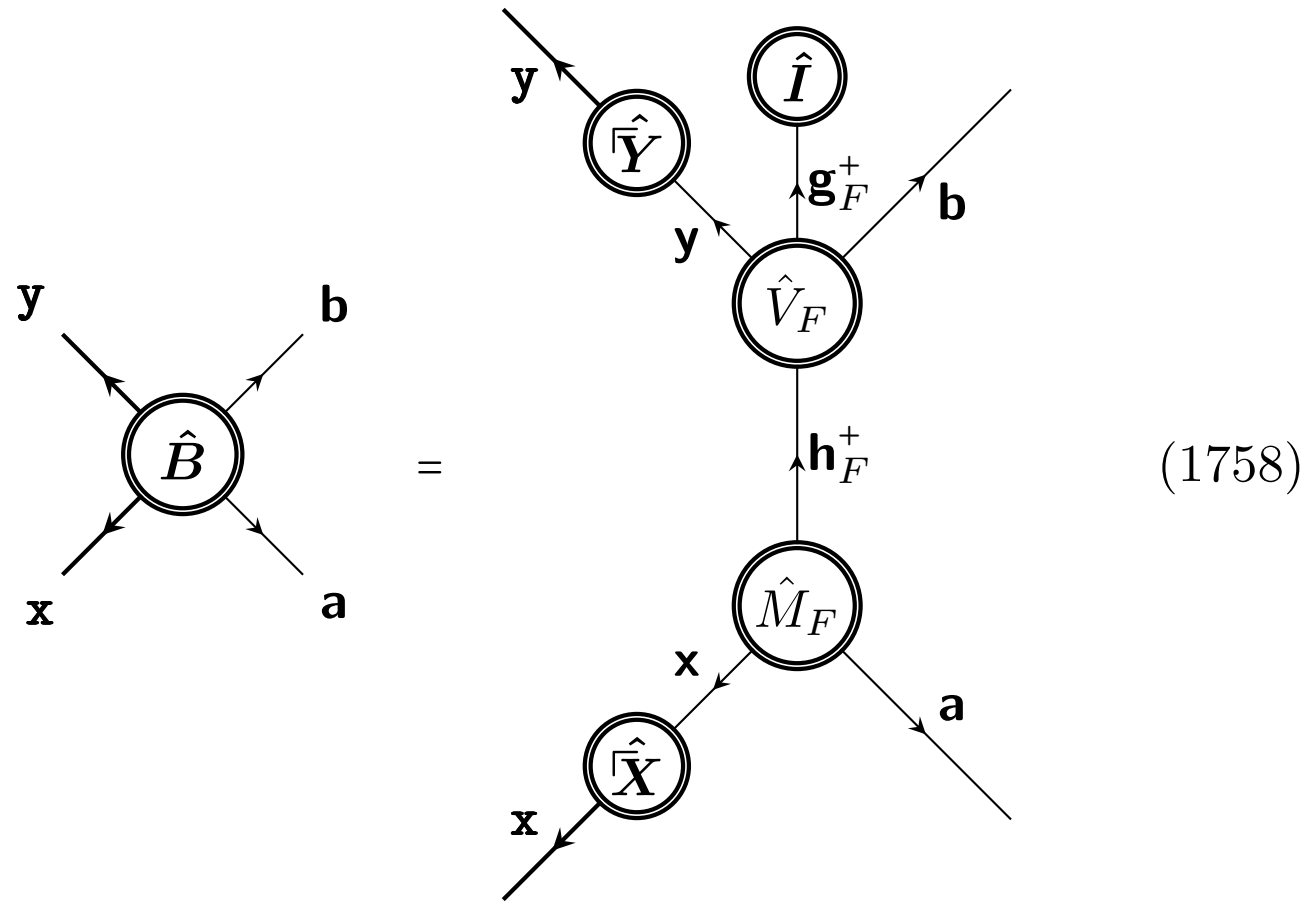

$$(1758)$$

where (i) $\hat{V}_F$ is a natural temporal isometry, (ii) we can choose $N_{\mathsf{g}_F^+} = \mathrm{rank}(\hat{\boldsymbol{B}})$, (iii) $\hat{M}_F$ is homogeneous and satisfies

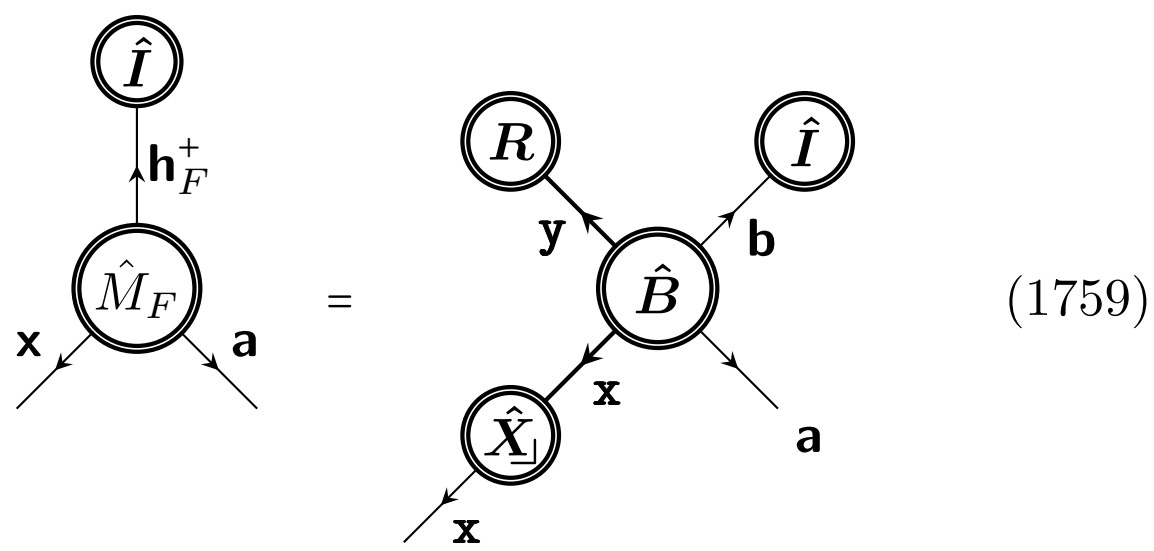

$$(1759)$$

(iiii) $\hat{\boldsymbol{B}}[p^-]$ (the past residuum of $\hat{\boldsymbol{B}}$ at $p$) satisfies the simple double causality conditions if and only if $M_F$ is a natural temporal maxometry (with respect to ancilla $\mathsf{h}_F^+$), and (v) we can choose $N_{\mathsf{h}_F^+} = \mathrm{rank}(\hat{\boldsymbol{B}}[p^-])$.

**Backward case.** The backward causality condition (1738) at $p$ is equivalent to the condition

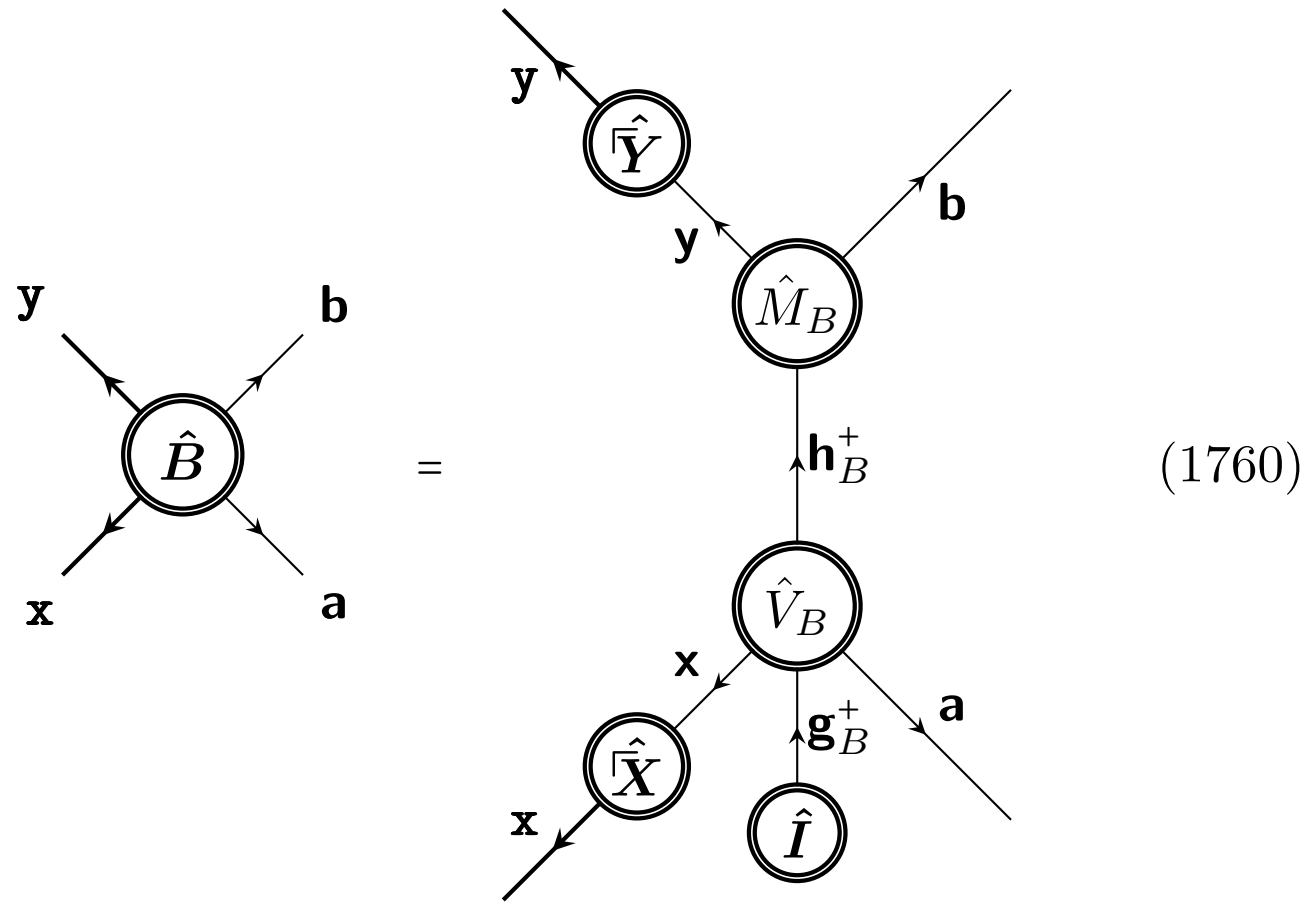

(1760)

where (i) $\hat{V}_B$ is a natural temporal isometry, (ii) we can choose $N_{\mathsf{g}_B^+} = \text{rank}(\hat{\boldsymbol{B}})$, (iii) $\hat{\boldsymbol{M}}_B$ is homogeneous and satisfies

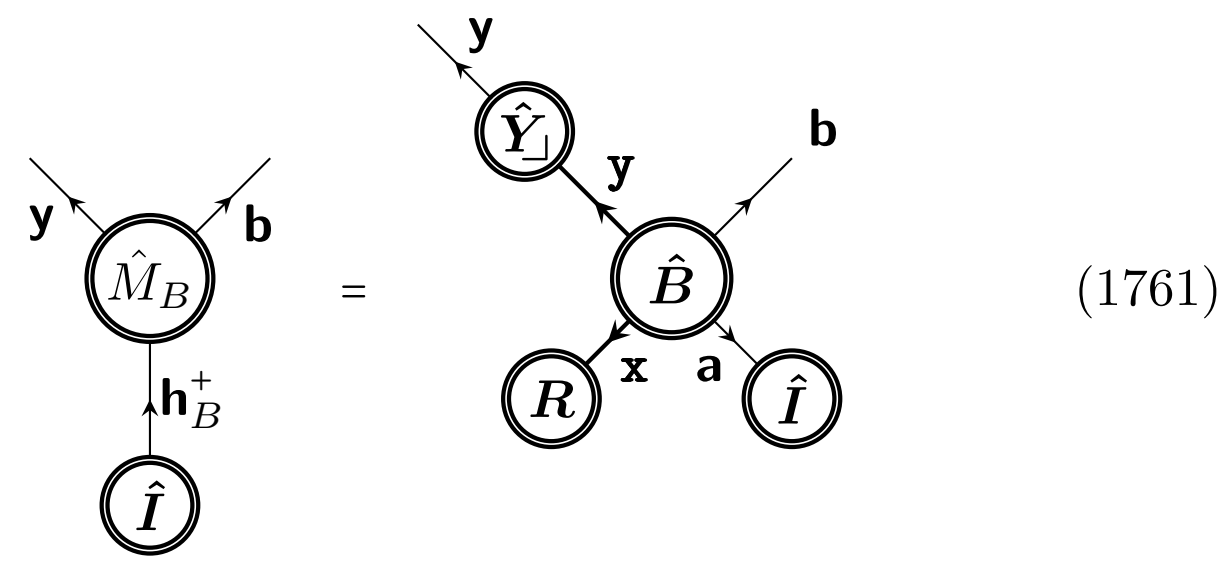

(1761)

(iiii) $\hat{\boldsymbol{B}}[p^-]$ (the past residuum of $\hat{\boldsymbol{B}}$ at $p$) satisfies the simple double causality conditions if and only if $M_B$ is a natural temporal maxometry (with respect to ancilla $\mathsf{h}_B^+$), and (v) we can choose $N_{\mathsf{h}_B^+} = \text{rank}(\hat{\boldsymbol{B}}[p^+])$.

It is easy to prove this by adapting the proof of the basic causal dilation theorem in Sec. 78.5. First note from the operator network rank theorem in Sec. 68.5 that

$$|\{d^+\}|\text{rank}(\hat{\boldsymbol{B}}) \geq |\{d^-\}|\text{rank}(\hat{\boldsymbol{B}}[p^-]) \tag{1762}$$

follows from the forward causality condition at $p$ for $\hat{\boldsymbol{B}}$ (given in (1745)). Following Chiribella et al. [2009a] we choose an orthogonal decomposition of $\llcorner\hat{\boldsymbol{B}}^{\urcorner}[p^-]$ (this is always possible according to the corollary in Sec. 68.7 - see (1487)).

Thus, we can write (1747) that appears in that proof as

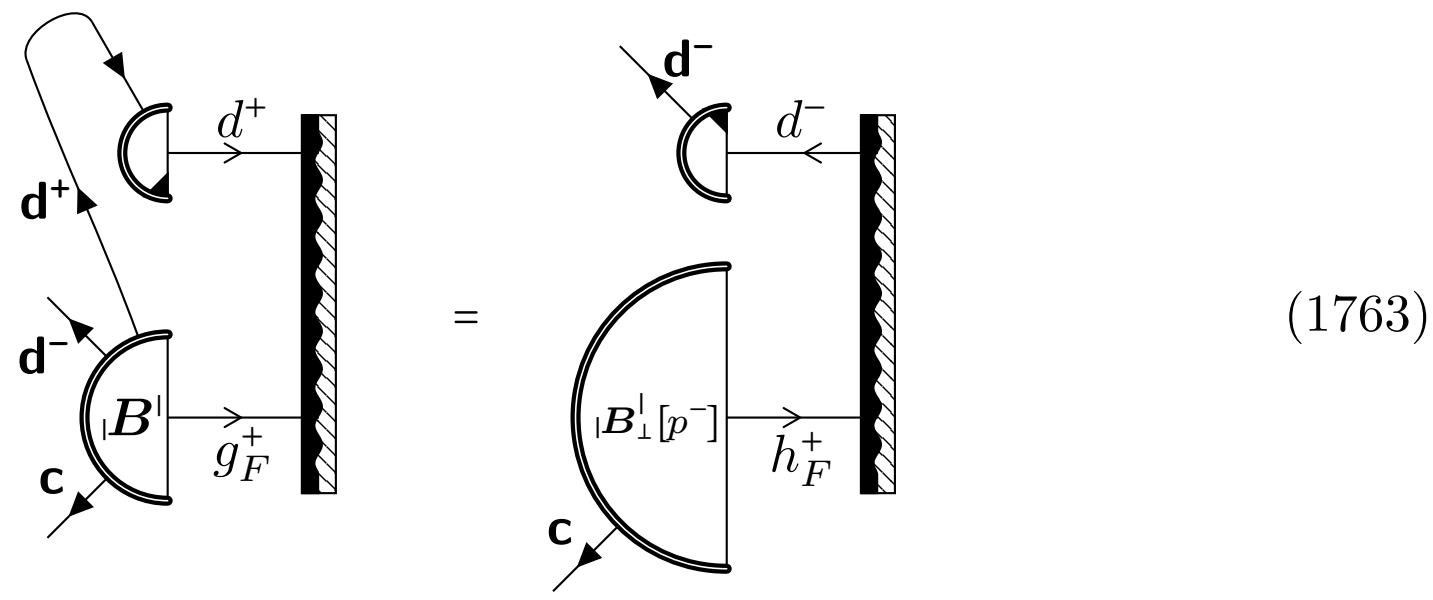

$$(1763)$$

where $\perp$ denotes that these Hilbert objects are orthogonal. On the right hand side of (1763), to the left of the mirror, we have the tensor product of ${}_{|}\boldsymbol{B}^{|}{}_{\perp}[p^-]$ with the orthogonal $\mathsf{d}^-$ basis elements. Thus, these tensor product objects do, themselves, constitute an orthogonal set. Therefore, it is possible to use the fact that the orthogonal elements in such a representation are related by an isometry to the representation on the left hand side (see the corollary in Sec. 68.7). Now, instead of a unitary, $U$, as in (1749) we would have an isometry, $V$ as follows

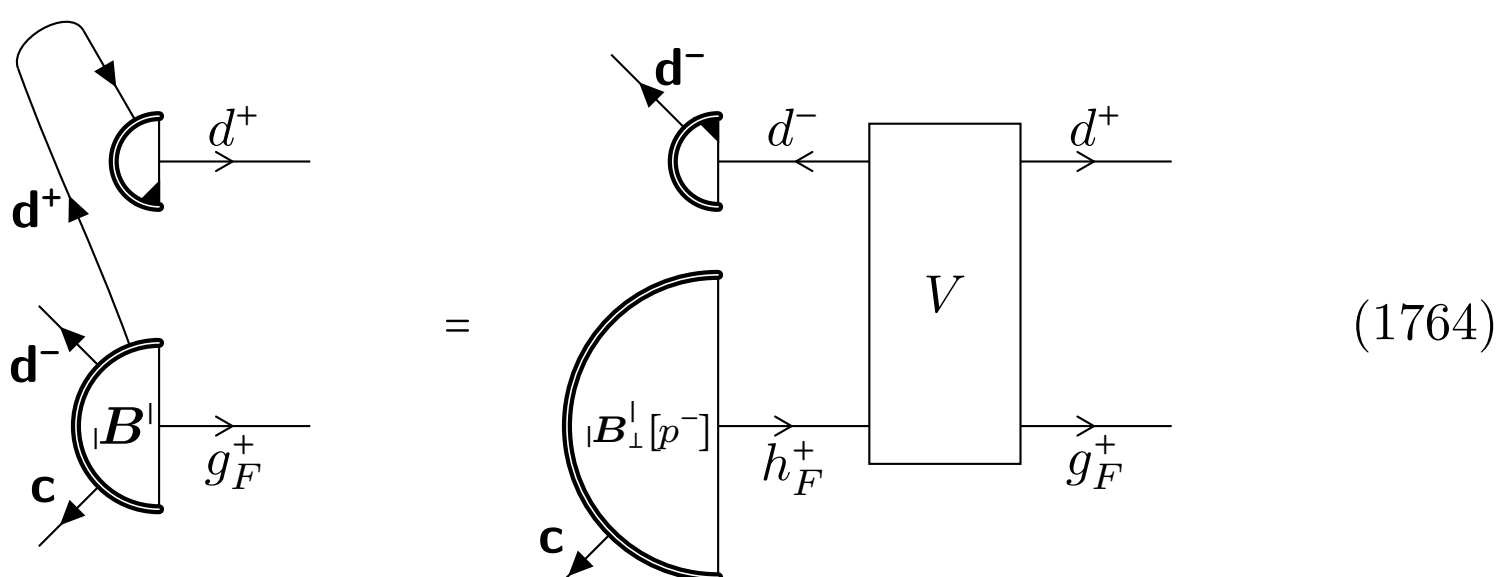

$$(1764)$$

where we require

$$|\{g_F^+\}||\{d^+\}| \geq |\{h_F^+\}||\{d^-\}| \tag{1765}$$

so that $V$ can be an isometry. We can, according to (1762), satisfy the above inequality by choosing $|\{g_F^+\}| = \text{rank}(\hat{\boldsymbol{B}})$ and $|\{h_F^+\}| = \text{rank}(\hat{\boldsymbol{B}}[p^-])$. If we now track through the remaining steps of the proof of the basic causal dilation theorem (but where $V_F$ replaces $U_F$) we obtain (1758) where $\hat{V}_F$ is a natural temporal isometry. Further, we have $N_{\mathsf{g}_F^+} = \text{rank}(\hat{\boldsymbol{B}})$ and $N_{\mathsf{h}_F^+} = \text{rank}(\hat{\boldsymbol{B}}[p^-])$. Paying attention to the time direction of the wires we can show similarly that $\hat{V}_B$ is a natural temporal coisometry in (1760) and that we can have $N_{\mathsf{g}_B^+} = \text{rank}(\hat{\boldsymbol{B}})$ and $N_{\mathsf{h}_B^+} = \text{rank}(\hat{\boldsymbol{B}}[p^+])$. This completes the necessary part of this proof. The sufficient part of the proof here is exactly as in the basic causal dilation theorem in Sec. 78.5.

## 78.7 Constructibility and unconstructibility

Having provided the basic causal dilation theorems above we will reconnect with two topics that were raised earlier in this book. These are the constructibility and unconstructibility conjectures of Sec. 45.2 and the synchronous partition assumptions discussed in Sec. 49.2.5, Sec. 49.4.3 and Sec. 49.4.4.

First, let us discuss the synchronous partition assumptions. We used these to prove the positivity composition theorem in Sec. 49.2.6 and the double causality theorem in Sec. 49.4.5 which provided the double causality conditions. A natural question is whether we can "close the circle" and prove these synchronous partition assumptions within the quantum framework we have developed.

To prove the positivity composition theorem we used the *synchronous partition with positivity assumption.* This assumption does, indeed, follow from any of the basic unitary dilation theorems (proved in Sec. 78.5 or Sec. 78.6.

To prove the double causality conditions it was sufficient to use the pruned down weak synchronous partition assumption (in Sec. 49.4.4). It is clear that this assumption follows also from any of the basic dilation theorems.

Thus, we do close the circle in the above sense. In fact, these basic dilation theorems go a bit further than the synchronous partition with positivity assumption or the weak synchronous partition assumption. However, neither of these basic dilation theorems recovers the full (not weak) synchronous partition assumption (in Sec. 49.4.3).

To see that the basic (co)isometric causal dilation theorem recovers the weak synchronous partition assumption note that the top part of (1758) consisting of $\hat{V}_F$, $\hat{\boldsymbol{I}}$, and $\ulcorner\hat{\boldsymbol{Y}}$ can be identified with $\mathbf{B}_F[+]$ (from (1168) and satisfies simple forward causality. This, along with a similar result for $\mathbf{B}_B[-]$, recovers the weak synchronous partition assumption. We do, however, have a bit more. The bottom part of (1758) consisting of $\hat{M}_F$ and $\ulcorner\hat{\boldsymbol{X}}$ can be identified with $B_F[-]$ in (1168) and also satisfies simple forward causality. Similar remarks apply for the backwards dilation in (1760).

The basic unitary causal dilation theorem goes a bit further since, in this case, $\mathbf{B}[+]$ is identified with the top of (1740) consisting of $\hat{\boldsymbol{U}}_F$, $\hat{\boldsymbol{I}}$, and $\ulcorner\hat{\boldsymbol{Y}}$. The latter operator is composed of deterministic physical operators and so is, itself, deterministic and physical (so it satisfies simple double causality).

The bottom part of (1740) (consisting of $\hat{M}_F$ and $\ulcorner\hat{\boldsymbol{X}}$) only satisfies simple forward causality and so we do not recover the (not weak) synchronous partition assumption.

Now let us consider the issue of constructibility and unconstructibility. In Sec. 45.1 we considered simple dilations. We saw that some deterministic physical operators that used a natural maxometry that was not a natural unitary in their dilation (so, on the face of things, the looked unconstructible) actually had a constructible dilation. However, we do not have a proof that a constructible dilation is always possible. In response to this we provided a number of conjectures in Sec. 45.2. Those conjectures apply to simple dilations of the form considered in Sec. 44 and Sec. 78.3. Following on from the conjectures in Sec.

45.2, we can form constructibility and unconstructibility conjectures concerning the basic unitary causal dilations appearing in (1740) and (1742).

> **Basic causal constructibility conjecture.** Any basic unitary causal dilation of the form in (1740) (or (1742)) can be rewritten in such a way that $\hat{M}_F$ (or $\hat{M}_B$) is a natural unitary.

or

> **Basic causal unconstructibility conjecture.** There exist deterministic physical operators, $\hat{\boldsymbol{B}}$, for which there exists no basic unitary causal dilation of the form in (1740) (or (1742)) where $M_F$ (or $M_B$) is a natural unitary.

We could also write down conjectures concerning whether constructible dilations (should they exist) can have ancillae, $\mathsf{h}^+_{F,B}$ and $\mathsf{g}^+_{F,B}$ of finite dimension, or whether we must allow this dimension to be arbitrarily large (using similar wording to the conjectures in Sec. 45.2).

## 78.8 Causal ladders

The simple causal structure in (1723) can be written as

$$\mathsf{a}^-\mathsf{x}^- \rightarrow \circledcirc \rightarrow \mathsf{a}^+\mathsf{x}^+ \qquad (1766)$$

(where we are using the notation in (1052)). In general, we may be interested in an operator having a causal diagram which is what we will call an *L-causal ladder* as follows

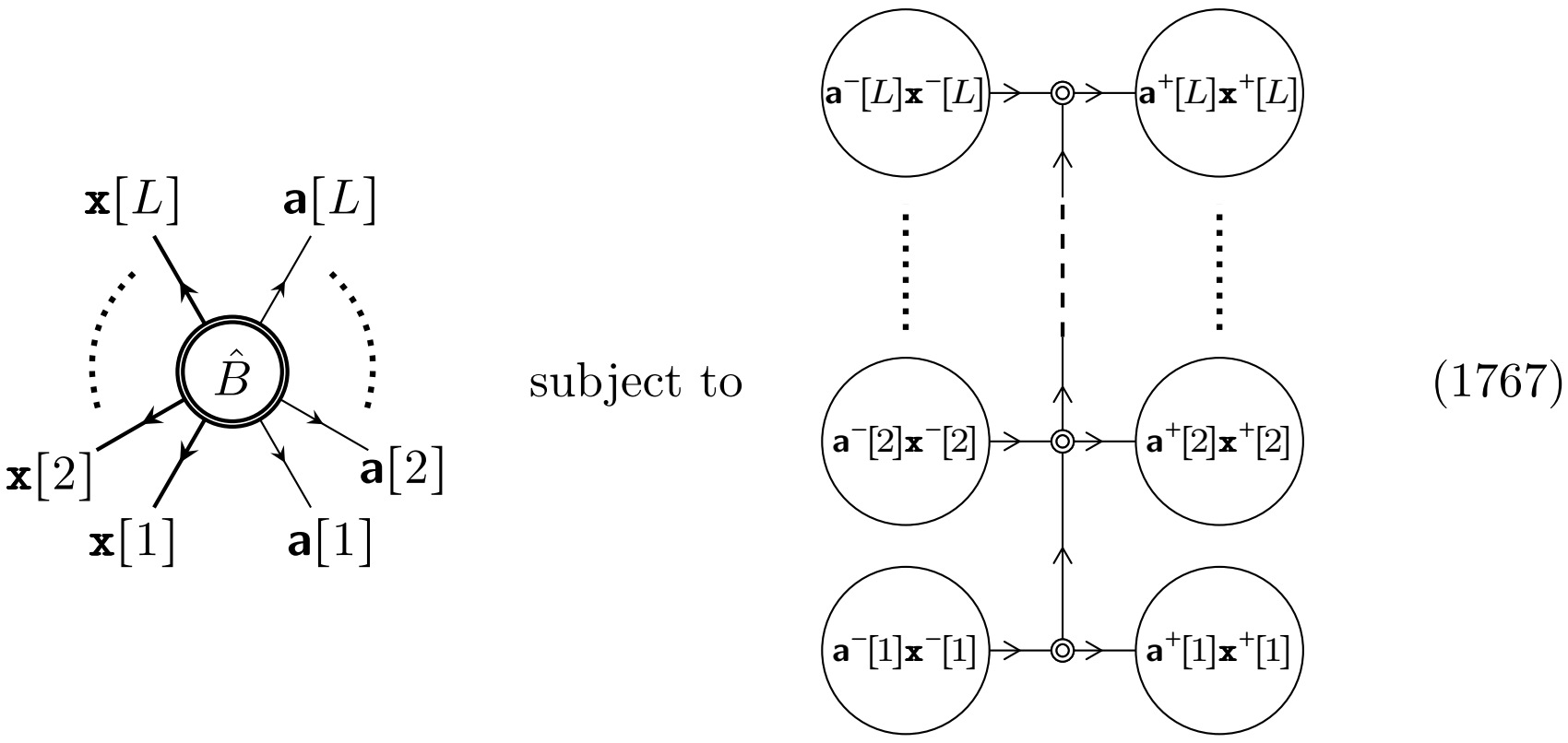

Note that the case of simple causal structure shown in (1766) is the case when $L = 1$. So a 1-causal ladder is the same as a simple causal structure.

The following simple operator network has an $L$-causal ladder as its causal diagram.

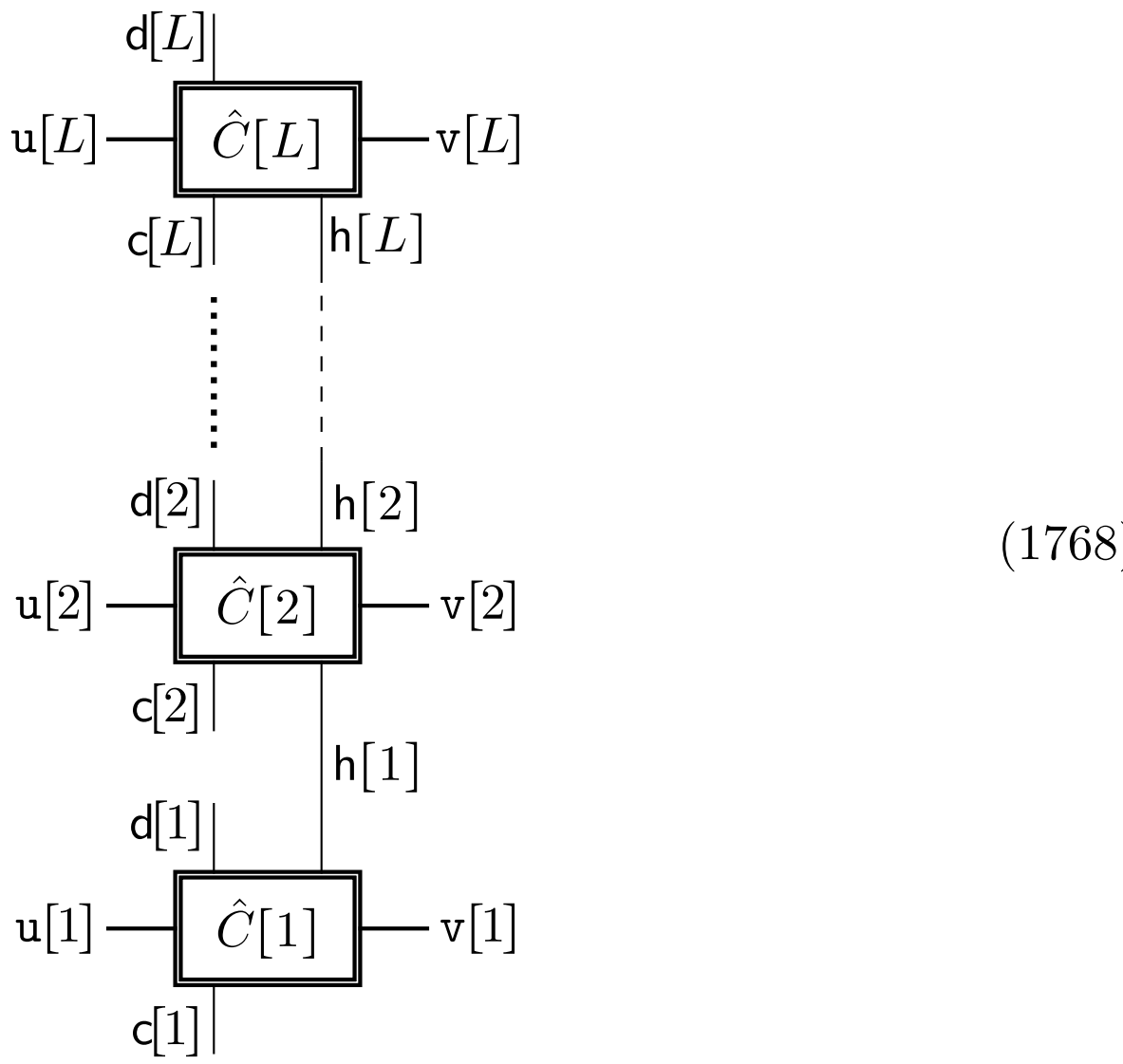

(1768)

where we identify $\mathsf{c}[l]$ with $\mathbf{a}^-[l]$, $\mathsf{d}[l]$ with $\mathbf{a}^+[l]$, $\mathtt{u}[l]$ with $\mathbf{x}^-[l]$, and $\mathtt{v}[l]$ with $\mathbf{x}^+[l]$. A natural question is whether any (complex) operator subject to the physicality conditions is, necessarily of this form. We will address this question below.

The quantum strategies of Gutoski and Watrous [2007] (their Theorem 6) and quantum combs of Chiribella et al. [2009a] (their Theorem 3) in prove that any such quantum strategy/comb can be represented by a network consisting of $L$ isometric operations in sequence if certain positivity and causality conditions are satisfied. The positivity condition in these papers is essentially equivalent to the positivity condition used here. The recursive normalisation conditions they use are, essentially, equivalent to the forward causality conditions here. Since they work in the standard (time forward) point of view, they do not have the time backward causality conditions. We have already borrowed the essential proof idea from these works to prove the basic causal dilation theorem in Sec. 78.5. Borrowing techniques from these papers, we will now use the basic causal dilation theorem to prove some results concerning operators whose causal diagrams are $L$-causal ladders.

## 78.9 Dilation for 2-causal ladders

Since the proofs for the $L$-causal ladder theorems are little involved, we will warm up by first considering the case of 2-causal ladders. We have the following theorem

**Dilation theorem for 2-causal ladders.** An operator tensor,

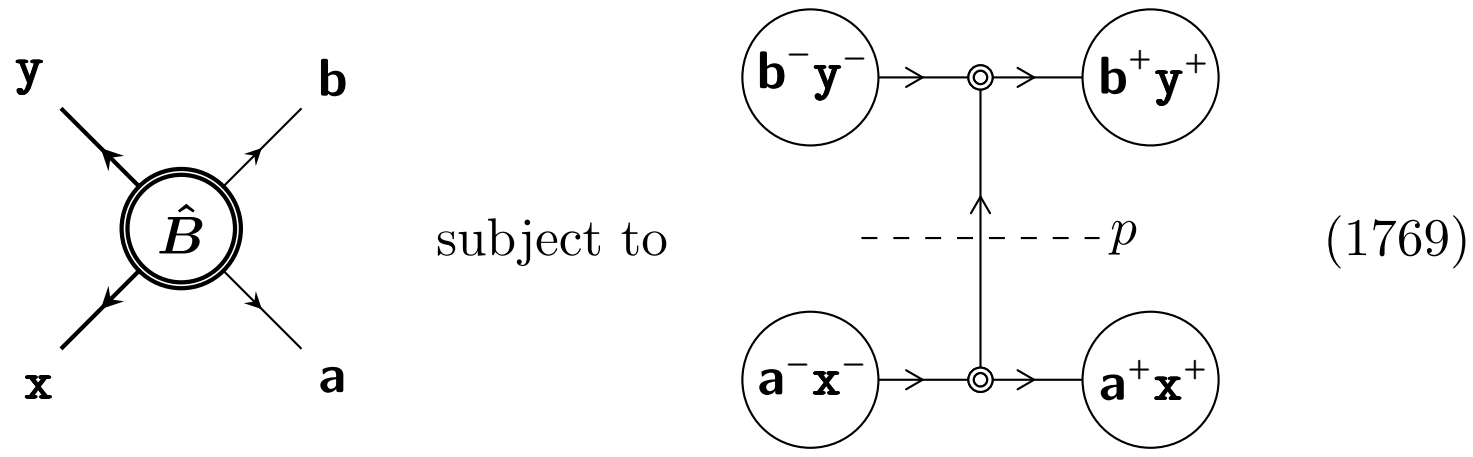

is deterministic and physical if and only if we can write $\hat{\boldsymbol{B}}$ as being equal to *both*

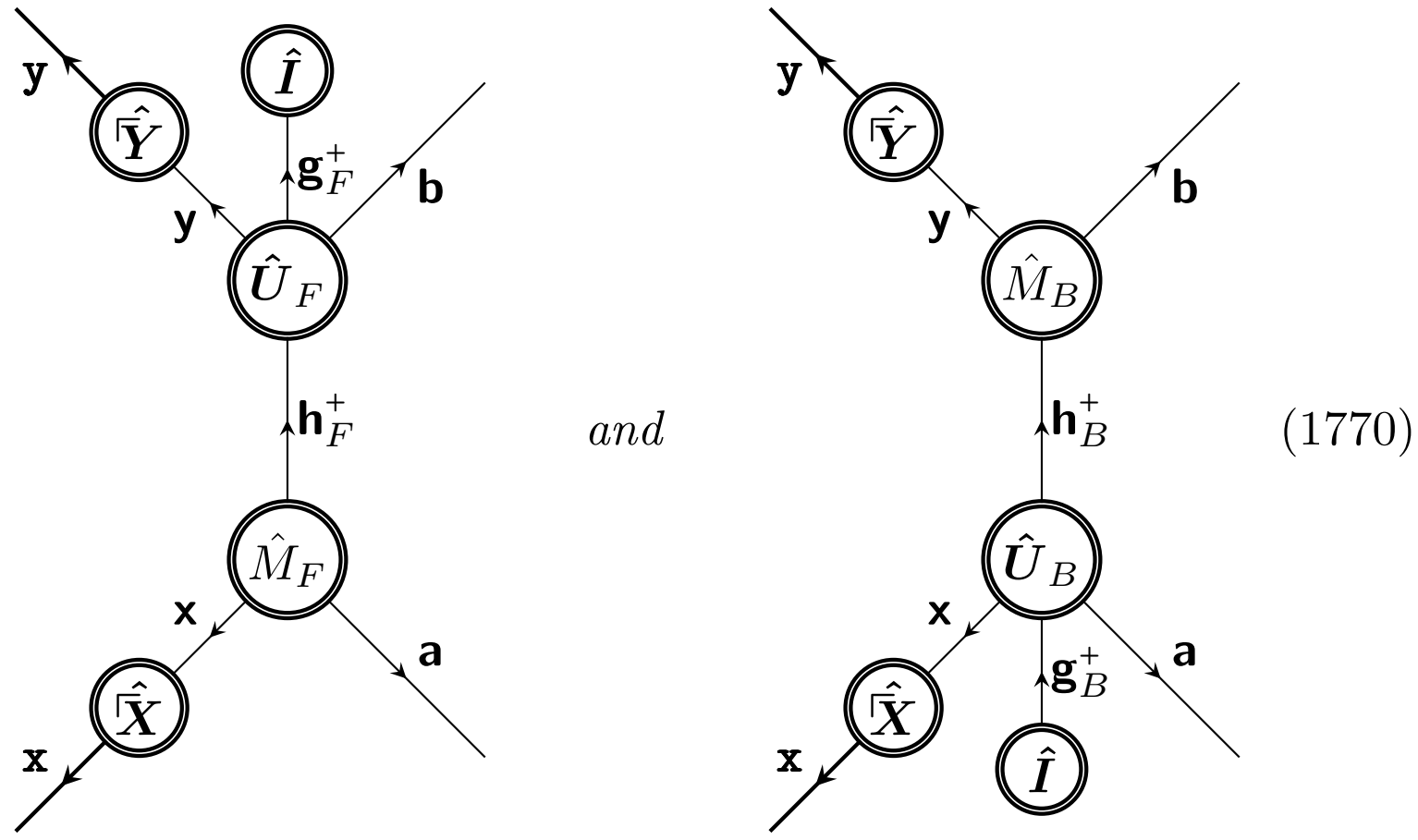

where $\hat{\boldsymbol{U}}_F$ and $\hat{\boldsymbol{U}}_B$ are natural temporal unitary operators, and $\hat{M}_F$ and $\hat{M}_B$ are natural temporal maxometric operators with, respectively, ancillae $\mathbf{h}_F^+$ and $\mathbf{h}_B^+$.

We have to prove both necessity and sufficiency. Importantly, both conditions in (1770) are necessary, and both conditions are required for sufficiency. First we prove **necessity**. In fact, the hard work was already done when we proved necessity in the basic unitary causal dilation theorem. We can apply that theorem at $p$ and immediately obtain the two necessary forms for $\hat{\boldsymbol{B}}$ in (1770). Now we turn to **sufficiency**. The hard work for this has also, to a large extent, already been done through the positivity and causality composition theorems in Sec. 49.2.6 and Sec. 49.4.11 respectively. To prove sufficiency we need to prove that if $\hat{\boldsymbol{B}}$ can be written in both of the forms (1770), then this is sufficient to guarantee it is a deterministic physical operator. To prove physicality and determinism we need to prove $T$-positivity and the double causality conditions. $T$-positivity follows simply since each of the component operators in either form in (1770) are $T$-positive and so, through correspondence with the positivity composition theorem in Sec. 49.2.6 these forms are $T$-positive. Now turn to proving that double causality is satisfied. We require that, if $\hat{\boldsymbol{B}}$ can be represented as in both forms in (1770), then it must satisfy the double causality conditions for all synchronous partitions. We will prove that, if it can be written as the left expression in (1770) then it satisfies forward causality for all synchronous partitions (and, it follows similarly, if it can be written in as the right expression, it satisfies backward causality for all synchronous partitions). We note that we can write the left expression in (1770) as

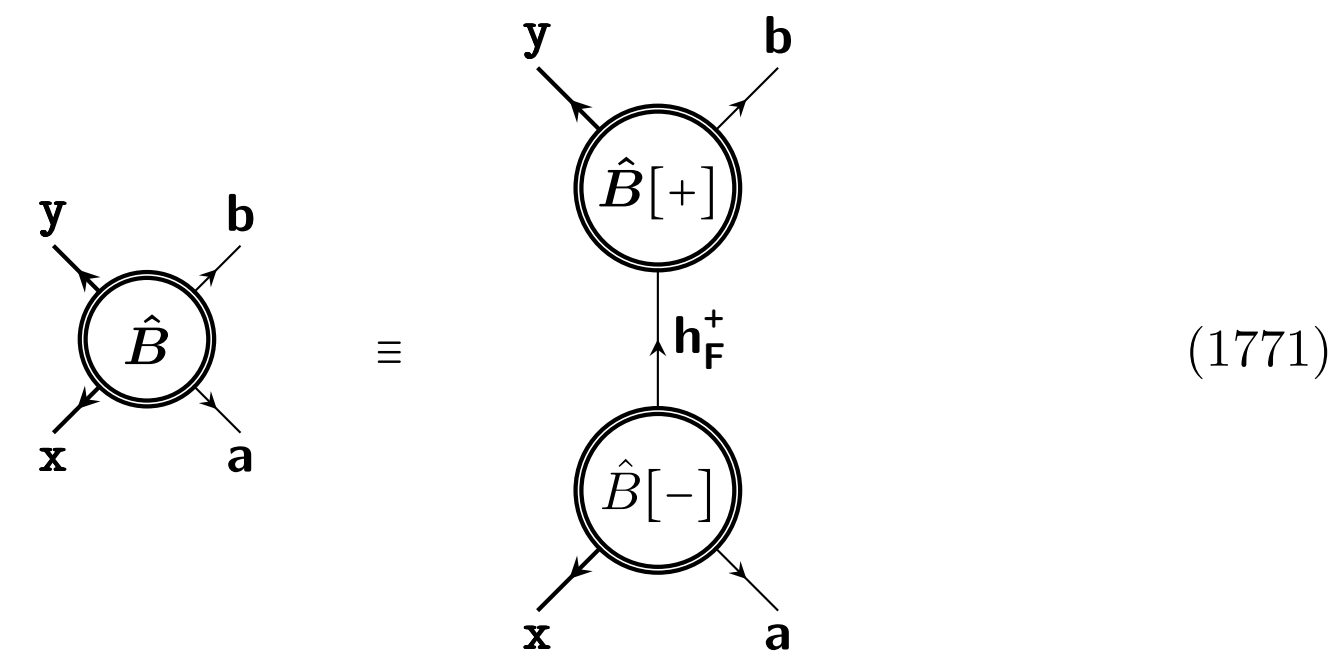

(1771)

where

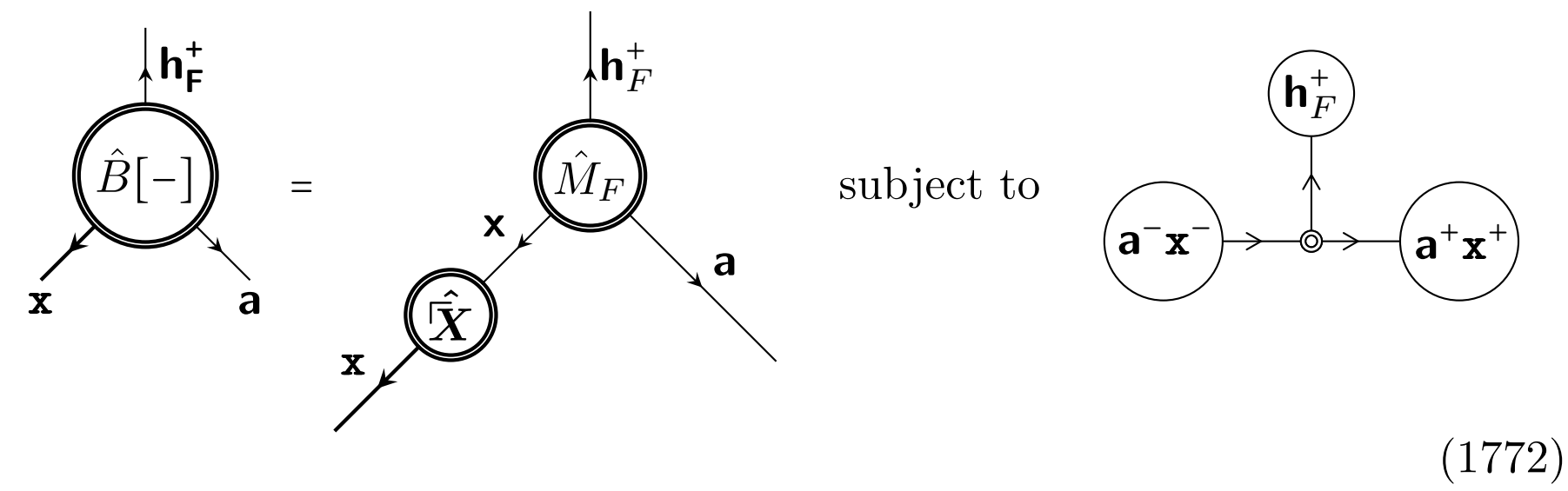

(1772)

and

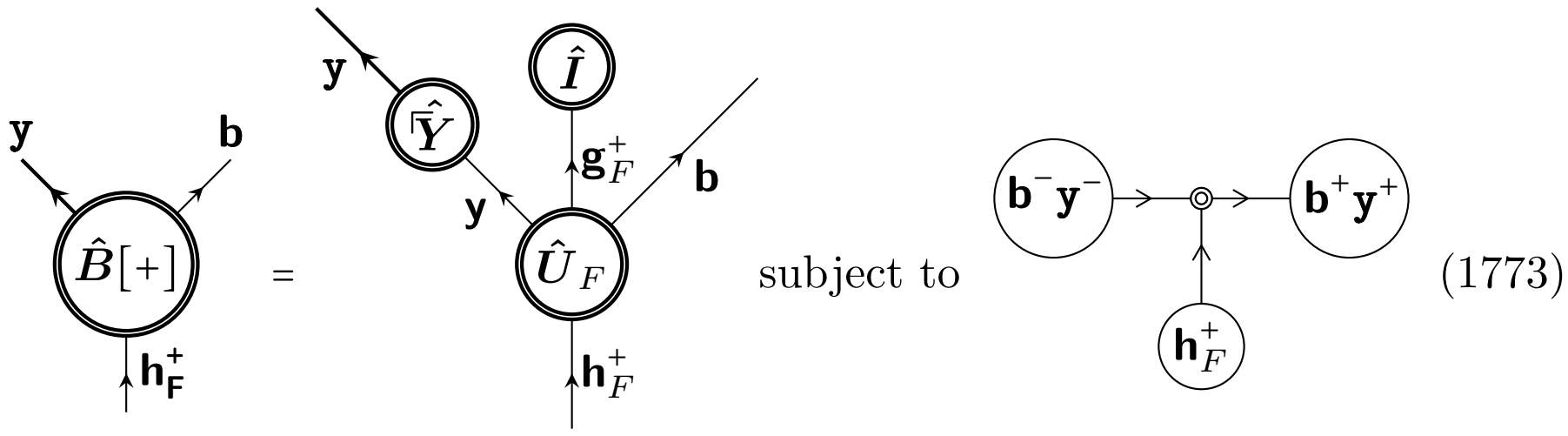

(1773)

We can regard $\hat{\boldsymbol{B}}[+]$ as being composed of $\hat{\boldsymbol{U}}_F$ (having simple causal structure) and $\hat{\boldsymbol{I}}$, both of which are physical (in the case of $\hat{\boldsymbol{U}}_F$, see the comments in Sec. 73.7). Thus, it follows from the causality composition theorem (under correspondence) that $\hat{\boldsymbol{B}}[+]$ satisfies double causality for all synchronous partitions. In particular, this means it satisfies forward causality for all synchronous partitions. The expression for $\hat{B}[-]$ (in (1773)) satisfies forward causality for all synchronous partitions. This is clear because its causal structure is simple and

$\hat{M}_F$ is a natural temporal maxometry with ancilla $\mathsf{h}_F^+$. The latter means that

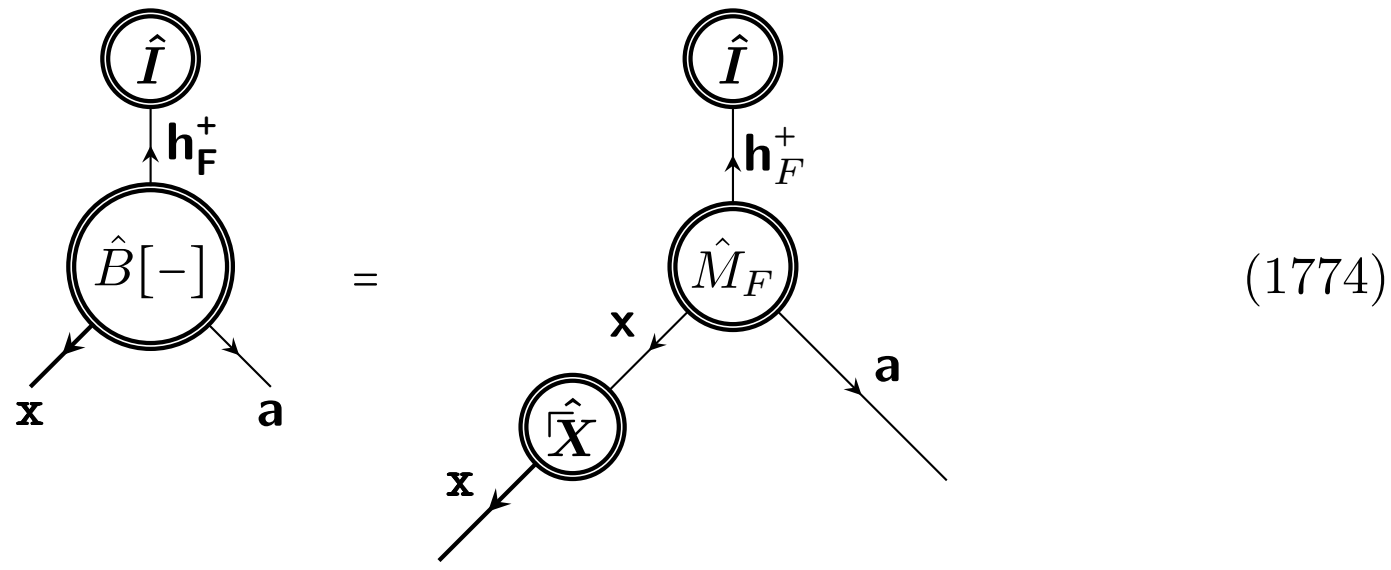

(1774)

satisfies simple double causality. Therefore, the object in (1772) satisfies simple forward causality. It then follows from the reasoning in Sec. 49.4.9 that $\hat{B}[-]$ satisfies forward causality for all synchronous partitions. It now follows from the causal composition theorem $\hat{\boldsymbol{B}}$ satisfies forward causality for all synchronous partitions. We can prove that $\hat{\boldsymbol{B}}$ satisfies backward causality for all synchronous partitions by similar reasoning. Since we have already proven $T$-positivity, we have proven that $\hat{\boldsymbol{B}}$ the conditions in (1770) are sufficient for deterministic and physical. This completes the proof of the theorem.

It is worth commenting that we can view the *simple dilation theorem* in Sec. 78.3 as a special case of the 2-causal ladder dilation theorem using the form of the simple dilation given in (1732) (whereby $\mathsf{q}$ can be pure input or pure output as $\mathsf{q}^\pm$). If we put that $\mathsf{b}$ and $\mathsf{y}$ are null, then the left forms for $\hat{\boldsymbol{B}}$ in (1770) becomes equivalent to the simple dilation in (1732) where the ancilla is pure output (and also, the causal diagram in (1769) becomes simple). Similarly we put $\mathsf{a}$ and $\mathsf{x}$ are null, then the left forms for $\hat{\boldsymbol{B}}$ in (1770) takes the same form as the simple dilation in (1732) where the ancilla is pure input. We can use the theorem in Sec. 73.2 to give the ancilla mixed input and output parts to get the more general form in (1724).

Perhaps even more interesting is that we can view the *simple dilation theorem with sufficient pairs* in Sec. 78.4 as a special case of the 2-causal ladder dilation theorem. To see this note that if we make $\mathsf{x}$ and $\mathsf{a}$ null in the expression on the left of (1770) then $\hat{M}_F$ becomes a homogeneous deterministic preparation and can be written as $\hat{A}_{\text{det}}$. Thus we obtain the an expression having the same form as the left expression in (1735). We obtain the right expression in (1735) in a similar way from the right expression by putting $\mathsf{y}$ and $\mathsf{b}$ null in (1770).

## 78.10 Dilation for 2-ladders with minimal sufficient pairs

The statement of the 2-causal ladder dilation theorem involves a sufficient pair as shown in (1770). Interestingly it is possible to weaken each member of the pair whilst maintaining sufficiency. Thus, we can state the theorem as follows

**Dilation theorem for 2-causal ladders with minimal suffi-**

**cient pairs.** An operator tensor,

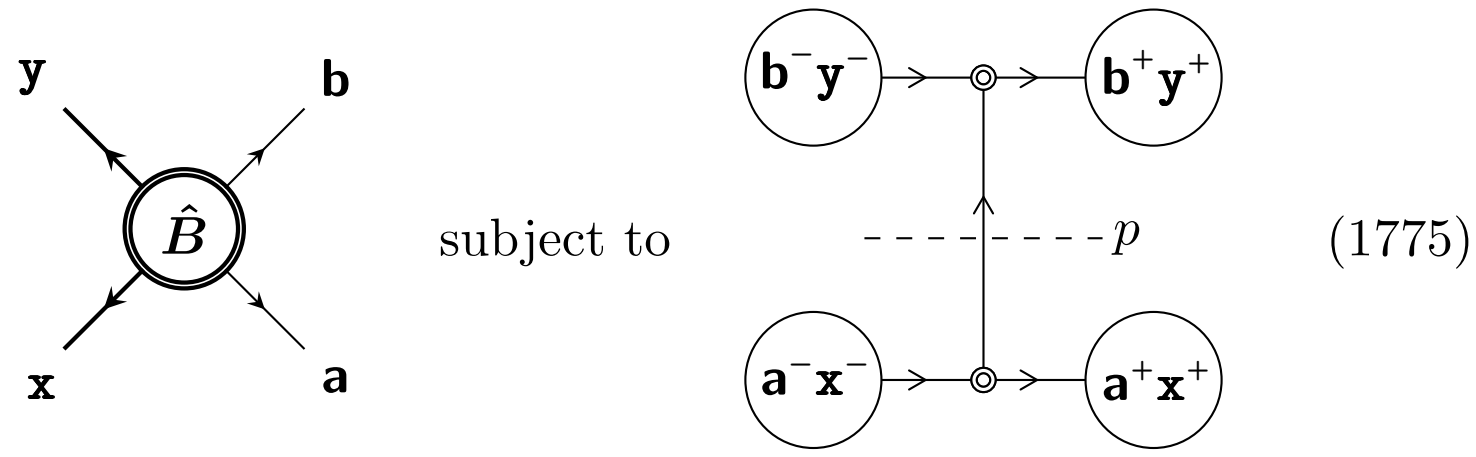

is deterministic and physical if and only if we can write $\hat{\boldsymbol{B}}$ as being equal to *both*

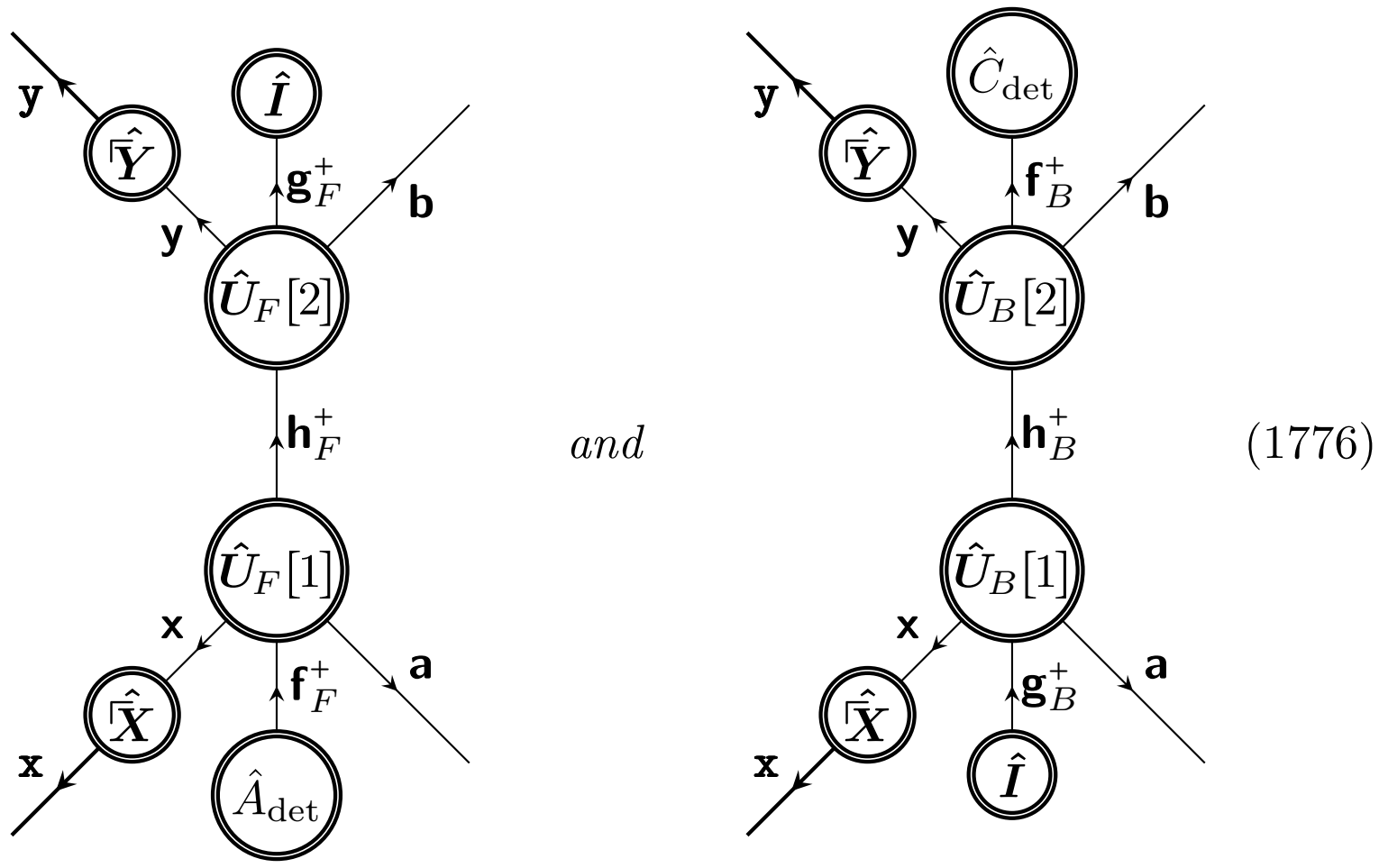

where $\hat{\boldsymbol{U}}_F[l]$ and $\hat{\boldsymbol{U}}_B[l]$ (for $l = 1, 2$) are natural temporal unitary operators and $\hat{A}_{\rm det}$ and $\hat{C}_{\rm det}$ are homogeneous and deterministic.

The proof of this is easy. First prove necessity. Since $\hat{M}_B$ in (1770) has only output ancilla, we can replace it by (1701) whilst maintaining necessity. This

gives

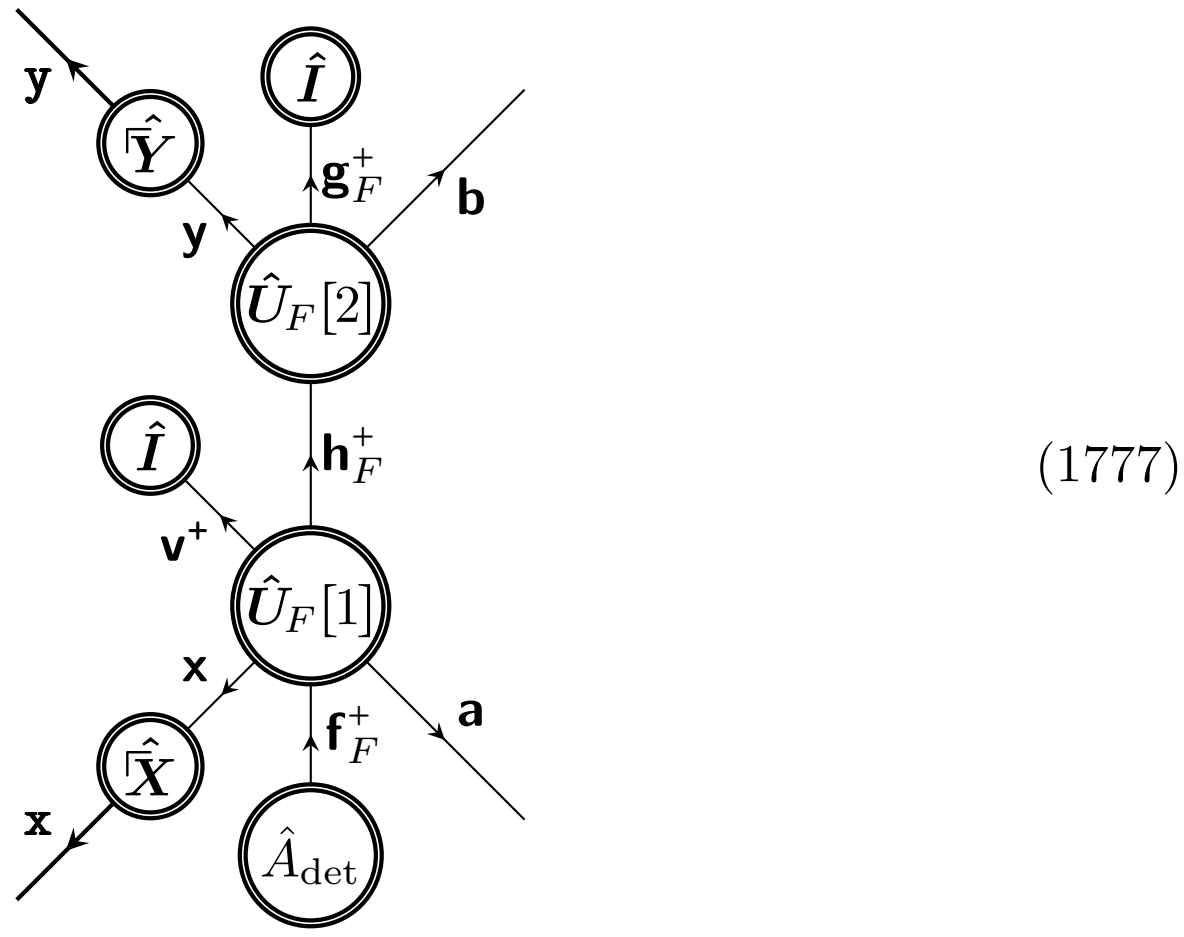

(1777)

(where we have suitably renamed the natural temporal unitaries) This is not quite the same as the left expression in (1776) since it has the extra $\hat{\boldsymbol{I}}$. However, this can be absorbed into the $\hat{\boldsymbol{I}}$ at the top by suitably redefining the unitary, $\hat{\boldsymbol{U}}_F[2]$, to incorporate the $\mathsf{v}^+$ and then suitably redefining the $\mathsf{h}_F^+$ and $\mathsf{g}_F^+$ wires to include the $\mathsf{v}^+$ wire. This gives the left expression in (1776). The right expression can be obtained similarly. This proves that both expressions are necessary. We can prove sufficiency by the same technique we used in Sec. 78.9. The left expression in (1776) satisfies forward causality because, when this is partitioned into two parts at wire $\mathsf{h}_F^+$ then both the top part and the bottom part satisfy forward causality. Hence, by the causality composition theorem, this left expression satisfies forward causality. The right expression in (1776) satisfies backward causality by similar reasoning. Since both expressions are clearly $T$-positive, these expressions are sufficient (taken together) for determinism and physicality.

Since each member of the sufficient pair in (1776) is more minimal than the corresponding member of the sufficient pair in (1770) we call (1776) a *minimal sufficient pair* for the sake of having a name. However, it is possible that an even more minimal sufficient pair exists.

## 78.11 Dilation for 2-causal ladders using isometries and coisometries

The theorem for 2-causal ladders can be redone using isometries and coisometries rather than unitaries. The advantage of doing this is that, then, the ancillary systems have dimensions that can be set equal to the pertinent ranks. We have

**Dilation theorem for 2-causal ladders using isometries and**

**coisometries.** An operator tensor,

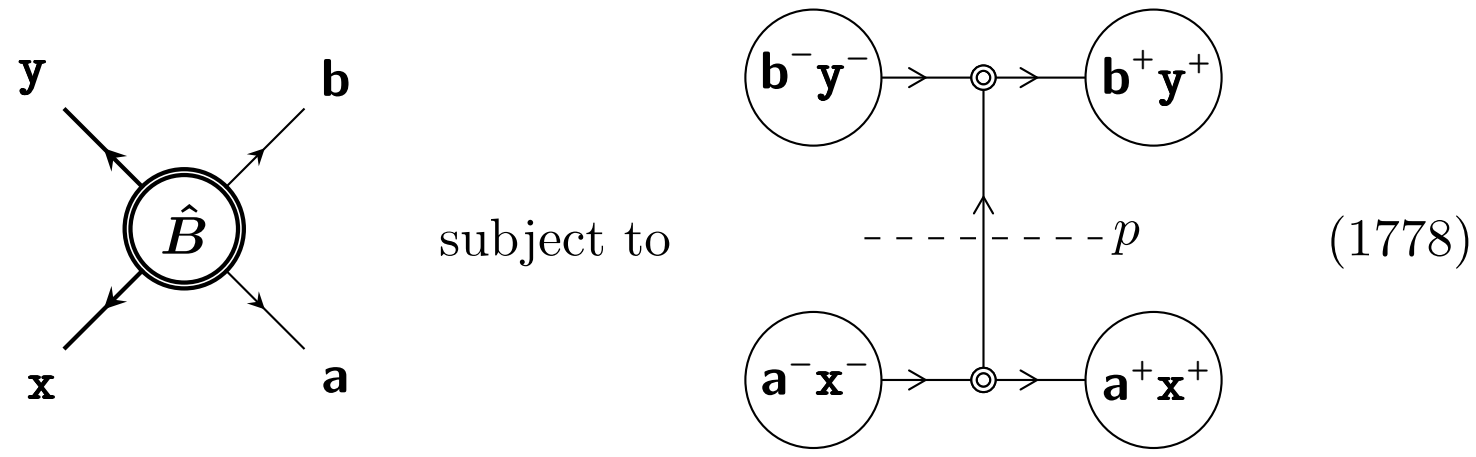

(1778)

is deterministic and physical if and only if we can write $\hat{\boldsymbol{B}}$ as being equal to *both*

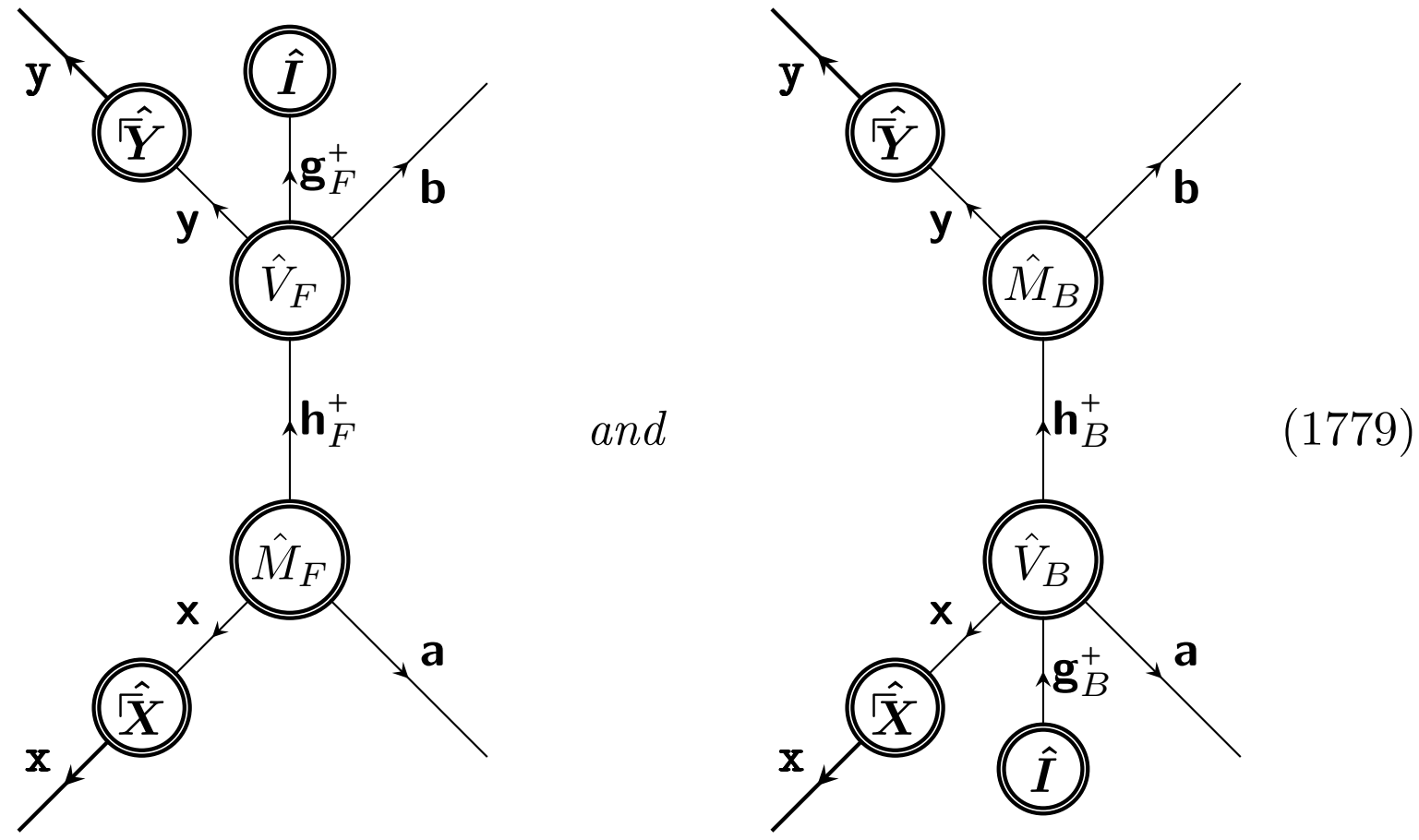

(1779)

where $\hat{V}_F$ is a natural temporal isometric operator, $\hat{V}_B$ is a natural temporal coisometric operator, and $\hat{M}_F$ and $\hat{M}_B$ are natural temporal maxometric operators with, respectively, ancillae $\mathbf{h}_F^+$ and $\mathbf{h}_B^+$. Furthermore we can choose $N_{\mathbf{g}_F^+} = N_{\mathbf{g}_B^+} = \text{rank}(\hat{\boldsymbol{B}})$, $N_{\mathbf{h}_F^+} = \text{rank}(\hat{\boldsymbol{B}}[p^-])$, and $N_{\mathbf{h}_B^+} = \text{rank}(\hat{\boldsymbol{B}}[p^+])$.

The proof for this theorem is essentially the same as the proof of the 2-ladder theorem with unitaries in (78.9). To prove **necessity** we use the basic (co)isometric causal dilation theorem. The extra detail is that, using this theorem, we obtain the facts about ranks at the end of the theorem statement. The proof of **sufficiency** is exactly the same as in the 2-ladder theorem.

## 78.12 Dilation for 2-causal ladders with minimal (co)isometric sufficient pairs

We can weaken the elements of the sufficient pair in the previous theorem slightly to obtain a theorem which is slightly stronger from a sufficiency point of view (but slightly weaker from a necessity point of view).

**Dilation theorem for 2-causal ladders with minimal sufficient (co)isometries pairs.** An operator tensor,

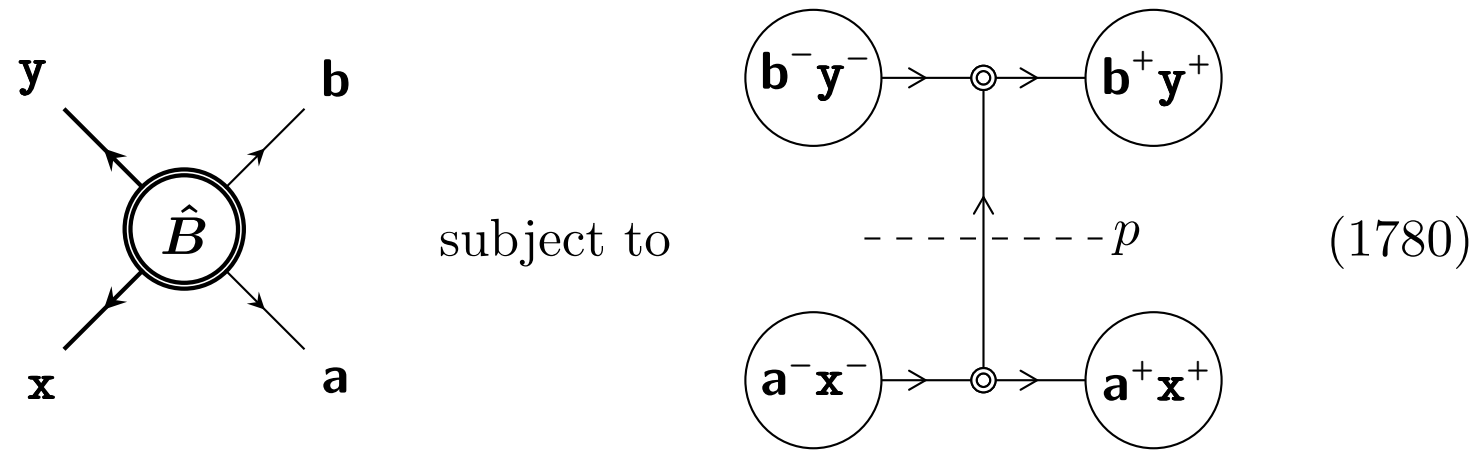

(1780)

is deterministic and physical if and only if we can write $\hat{\boldsymbol{B}}$ as being equal to *both*

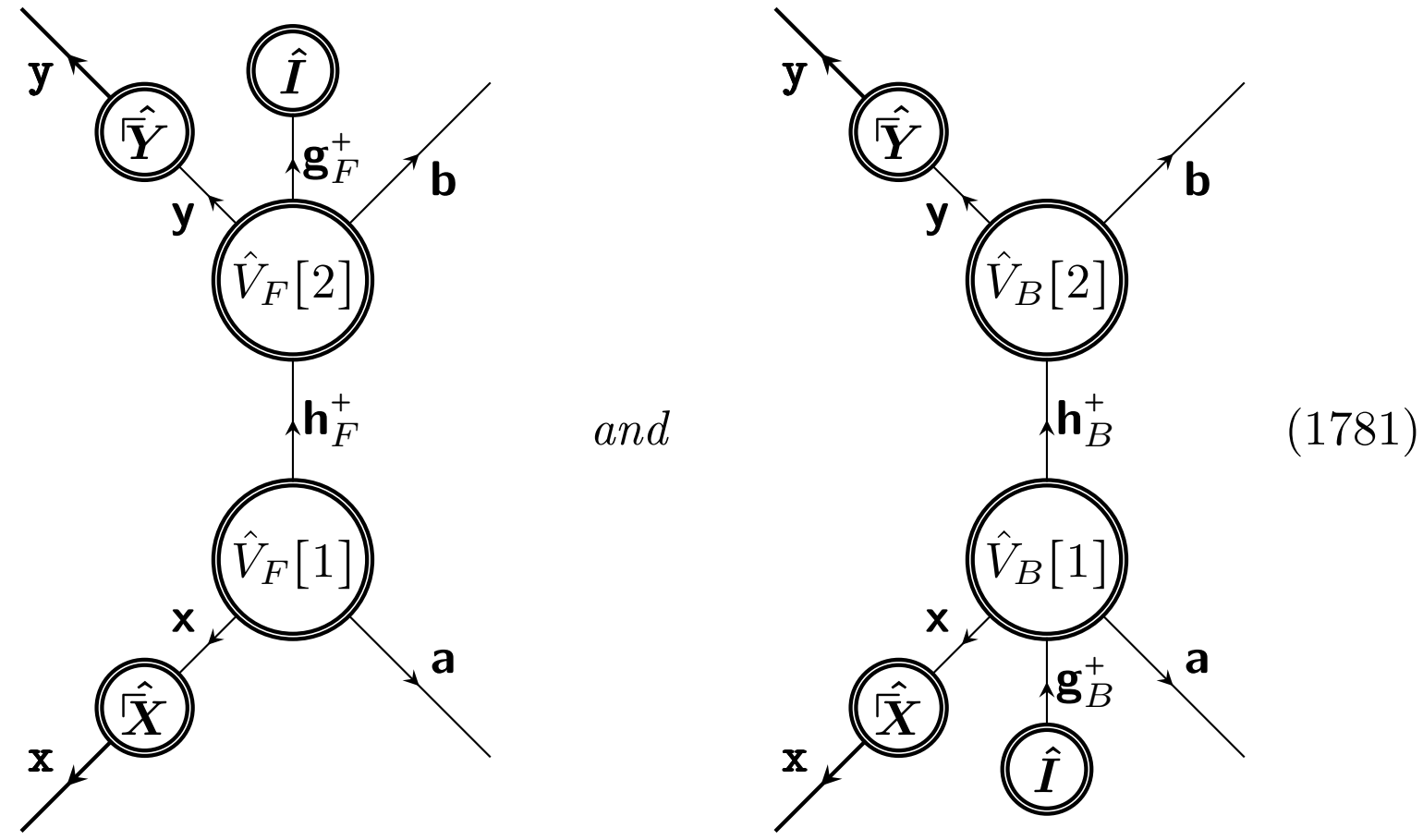

(1781)

where $\hat{V}_F[l]$ (for $l = 1, 2$) are natural temporal isometric operators, and $\hat{V}_B[l]$ (for $l = 1, 2$) are natural temporal coisometric operators. Further more we can choose $N_{\mathsf{g}_F^+} = N_{\mathsf{g}_B^+} = \text{rank}(\hat{\boldsymbol{B}})$, $N_{\mathsf{h}_F^+} = \text{rank}(\hat{\boldsymbol{B}}[p^-])$, and $N_{\mathsf{h}_B^+} = \text{rank}(\hat{\boldsymbol{B}}[p^+])$.

This follows from the previous theorem. Necessity follows immediately since the $\hat{M}_F$ maxometry has only output ancilla and so is an isometry whilst the $\hat{M}_B$ maxometry has only an input ancilla so is a coisometry. Sufficiency follows by the same arguments as before.

Note that the conditions for sufficiency are now weaker than in the previous theorem (which makes the theorem stronger from a sufficiency point of view). However, the necessary conditions are weaker (which makes the theorem weaker from a necessity point of view).

## 78.13 Homogeneous 2-causal ladders dilation

The fact that the ancillary system, $\mathsf{g}_{F,B}^+$, in the 2-causal ladder theorem with (co)isometries can be set equal to the rank of the $\hat{\boldsymbol{B}}$ is very useful when we come to consider homogenous operator tensors. We have the following theorem

**Homogeneous 2-causal ladder theorem.** An operator tensor

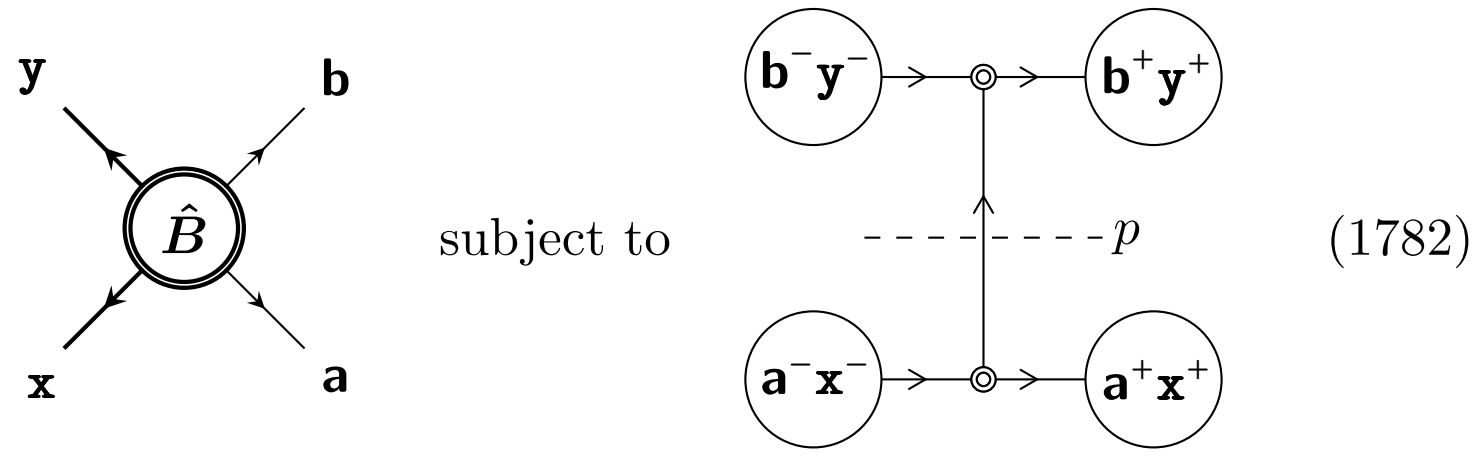

(1782)

is homogeneous, deterministic, and physical if and only if we can write it as

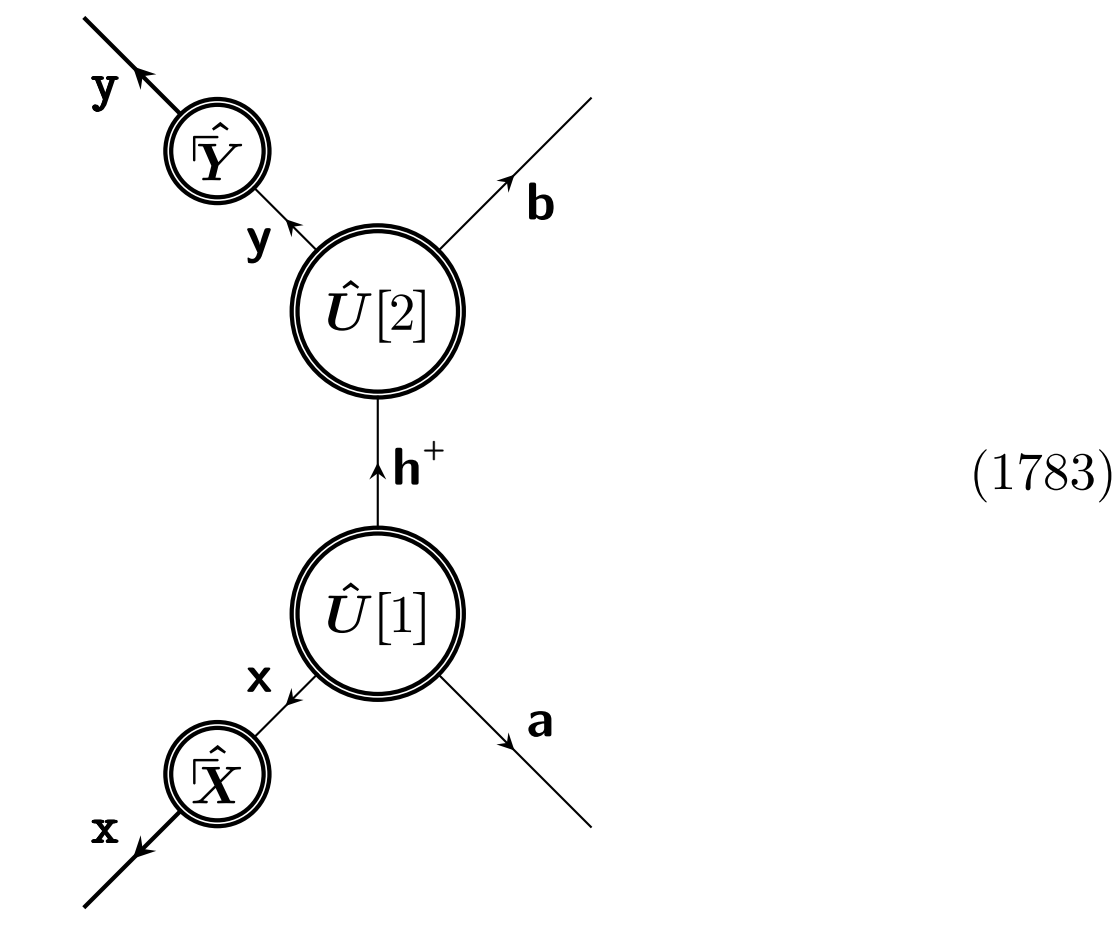

(1783)

with

$$\frac{N_{\mathbf{a}^-}N_{\mathbf{x}^-}}{N_{\mathbf{a}^+}N_{\mathbf{x}^+}} = \frac{N_{\mathbf{b}^+}N_{\mathbf{y}^+}}{N_{\mathbf{b}^-}N_{\mathbf{y}^-}} = N_{\mathbf{h}^+} \tag{1784}$$

where $\hat{\boldsymbol{U}}[l]$ (for $l = 1, 2$) are natural temporal unitary operators.

Note that the constraints (1784) are simply the necessary constraints on the dimensions of systems going into and going out of a natural temporal unitary. To prove this theorem we note that, according to the theorem in Sec. 78.12 we can set $N_{\mathbf{g}_F^+} = N_{\mathbf{g}_B^+} = 1$ since $\hat{\boldsymbol{B}}$ is homogeneous. Using the fact that $\hat{V}_F[1]$ and $\hat{V}_F[2]$ in (1781) are isometric, we must have

$$N_{\mathbf{b}^+}N_{\mathbf{y}^+} \geq N_{\mathbf{b}^-}N_{\mathbf{y}^-}N_{\mathbf{h}_F^+} \qquad \text{and} \qquad N_{\mathbf{a}^+}N_{\mathbf{x}^+}N_{\mathbf{h}_F^+} \geq N_{\mathbf{a}^-}N_{\mathbf{x}^-} \tag{1785}$$

Further, $\hat{V}_F[1]$ and $\hat{V}_F[2]$ are natural temporal unitary operators if these inequalities are saturated. Similarly, from the fact that $\hat{V}_B[1]$ and $\hat{V}_B[2]$ are coisometric, we must have

$$N_{\mathbf{a}^-}N_{\mathbf{x}^-} \geq N_{\mathbf{a}^+}N_{\mathbf{x}^+}N_{\mathbf{h}_B^+} \qquad \text{and} \qquad N_{\mathbf{b}^-}N_{\mathbf{y}^-}N_{\mathbf{h}_B^+} \geq N_{\mathbf{b}^+}N_{\mathbf{y}^+} \tag{1786}$$

where $\hat{V}_B[1]$ and $\hat{V}_B[2]$ are natural temporal unitary operators if these inequalities are saturated. Rearranging these equations we see that we require both

$$\frac{N_{\mathbf{b}^+}N_{\mathbf{y}^+}}{N_{\mathbf{b}^-}N_{\mathbf{y}^-}} \geq N_{\mathbf{h}_F^+} \geq \frac{N_{\mathbf{a}^-}N_{\mathbf{x}^-}}{N_{\mathbf{a}^+}N_{\mathbf{x}^+}} \qquad \text{and} \qquad \frac{N_{\mathbf{a}^-}N_{\mathbf{x}^-}}{N_{\mathbf{a}^+}N_{\mathbf{x}^+}} \geq N_{\mathbf{h}_B^+} \geq \frac{N_{\mathbf{b}^+}N_{\mathbf{y}^+}}{N_{\mathbf{b}^-}N_{\mathbf{y}^-}} \tag{1787}$$

which can only be satisfied when (1784) hold with $N_{\mathsf{h}^+} = N_{\mathsf{h}^+_F} = N_{\mathsf{h}^+_B}$. Further, we then must then saturate (1785) and (1786) and so we have natural temporal unitary operators as in the theorem (with $\mathsf{h}^+ = \mathsf{h}^+_F = \mathsf{h}^+_B$). This proves that the form (1783) is necessary. It is immediately clear that the expression in (1783) is sufficient for $\hat{\boldsymbol{B}}$ to be physical and deterministic since all the components are physical and deterministic (and so we can use the composition theorem in Sec. 49.4.11 under correspondence).

This theorem is interesting in that (i) homogeneity forces unitarity in the 2-ladder causal situation along with the necessary relationship between the dimensions of the systems. Further (ii) in the homogeneous case a single dilation is sufficient - we do not need a sufficient pair. It is also worth noting that the proof of this theorem needs both the forward and backward dilations in (1781) which come from the forward and backward causality conditions. Thus, this proof uses elements which are explicit in the time symmetric formulation of Quantum Theory but not in the time forward formulation (where there is no backward causality condition). It is not clear that a correspondingly simple proof of the same fact could go through in the standard time forward formulation.

This theorem generalises to the $L$-ladder as we will see in Sec. 78.18.

## 78.14 Dilation theorem for $L$-causal ladders

Now we will come to the general $L$-causal ladder case. Theorem 6 of Gutoski and Watrous [2007] and Theorem 3 of Chiribella, D'Ariano, and Perinotti [2009a] pertained to this case though, as mentioned already, in a time forward rather than time symmetric temporal frame. These authors employed induction to prove that such a case can be modelled by network composed by a sequence of isometries. The induction technique we use here is based specifically on that of CDP (though GW use a similar technique). However, we need to work a little bit harder because we want to prove we have a sequence comprised of a natural temporal maxometric operator followed (or preceded) by natural temporal unitary operators as befitting the time symmetric case. It is also possible to prove that we have a maxometric operator followed (or preceded) by a sequence of isometric operators. We will treat that case in Sec. 78.16. Further, in Sec. 78.17 we prove that we can have a sufficient pair consisting of one element having only a sequence isometries and the other having only sequence of coisometries (the forward case of this theorem contains the above mentioned Theorem 6 of GW and Theorem 3 of CDP).

We will prove the following theorem.

**Dilation theorem for $L$-causal ladders.** An operator tensor

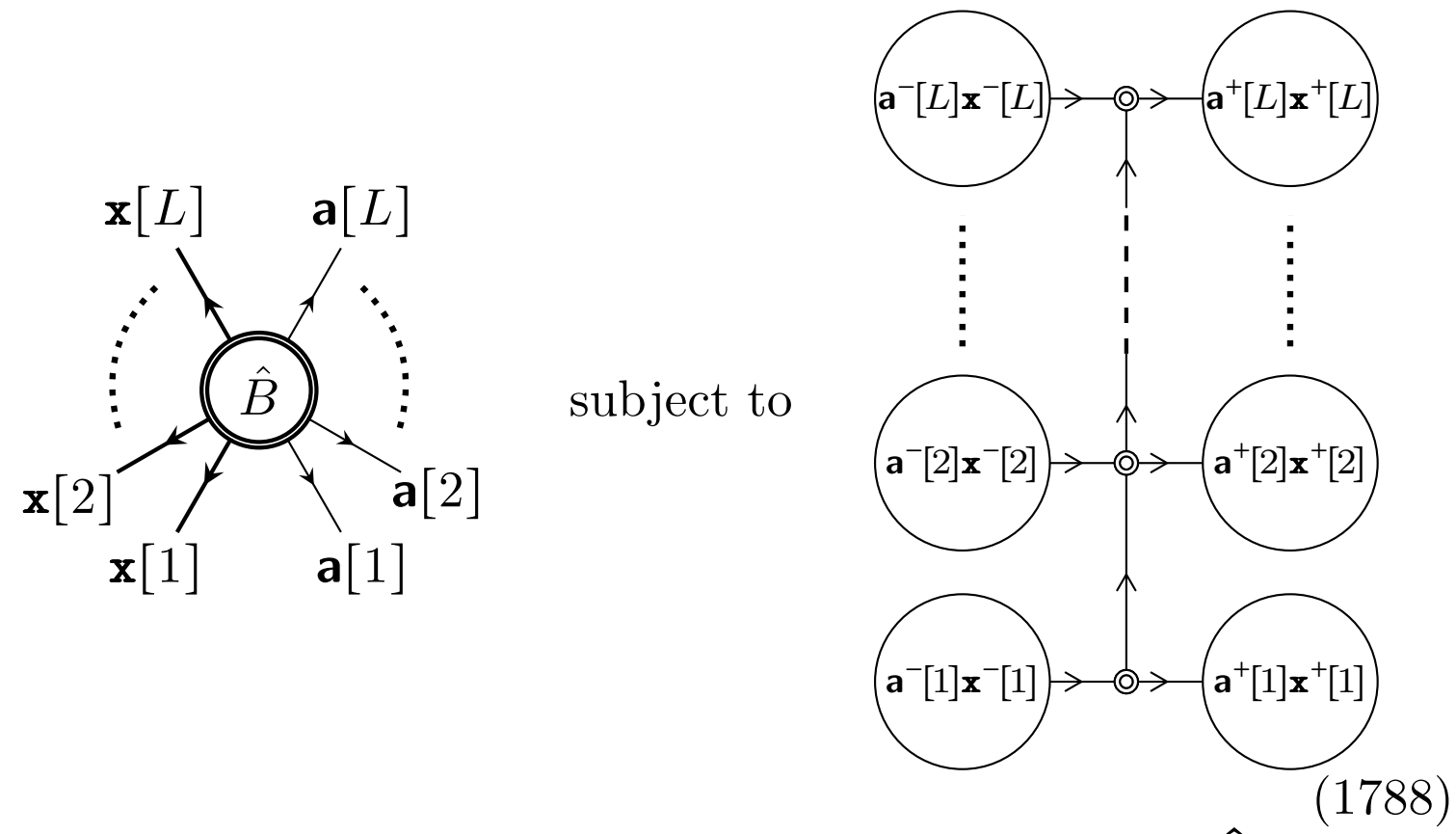

(1788)

is deterministic and physical if and only if we can write $\hat{\boldsymbol{B}}$ as being equal to *both*

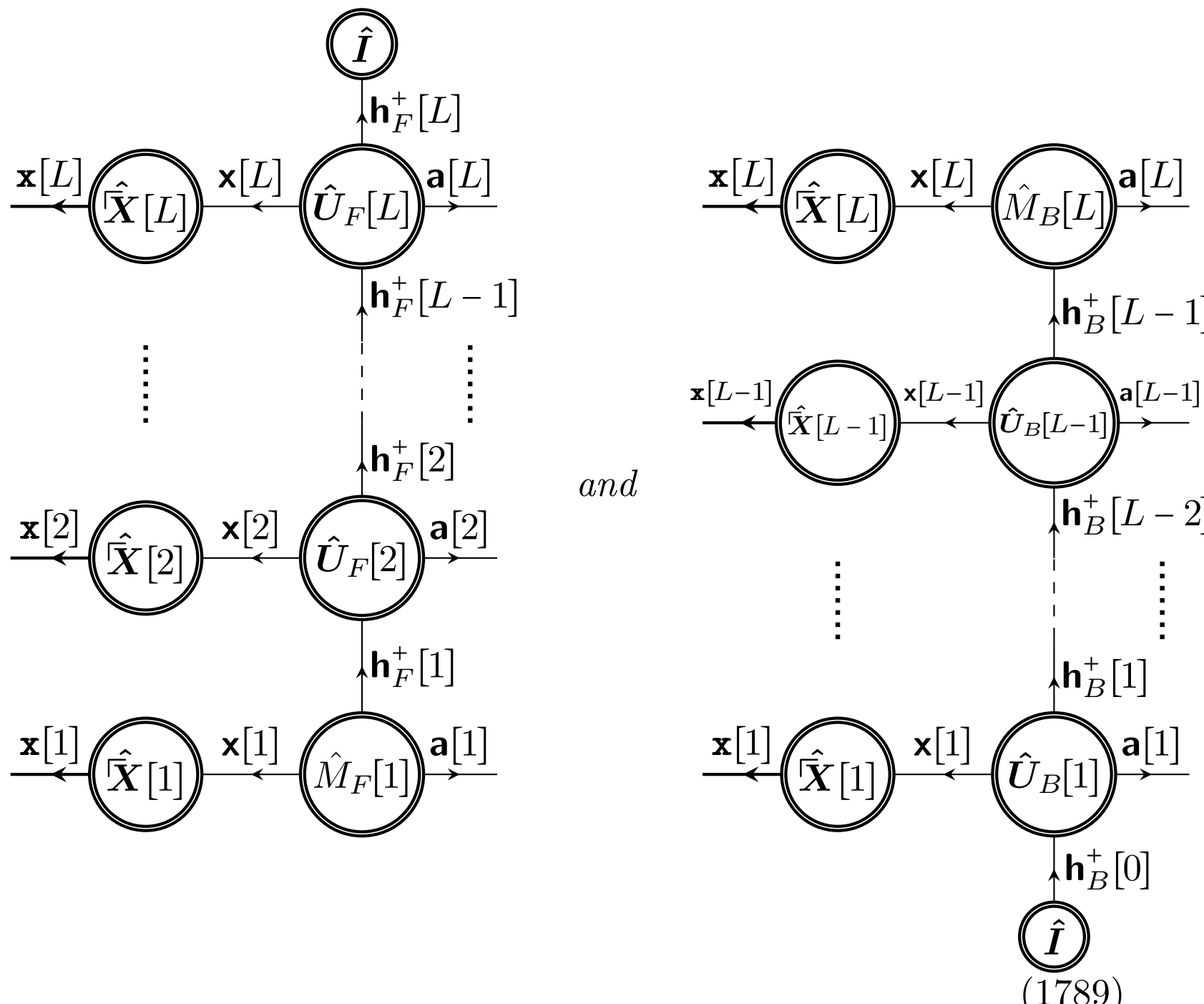

(1789)

where $\hat{M}_F[1]$ and $\hat{M}_B[L]$ are natural temporal maxometries with ancillas $\mathsf{h}^+_F[1]$ and $\mathsf{h}^+_B[L-1]$ respectively, and $\hat{\boldsymbol{U}}_F[l]$ (for $l = 2$ to $L$) and $\hat{\boldsymbol{U}}_B[l]$ (for $l = 1$ to $L-1$) are natural temporal unitaries.

As in previous dilation proofs, we will prove the theorem for $\llcorner\hat{\boldsymbol{B}}\urcorner$ which is deterministic and physical if and only if $\hat{\boldsymbol{B}}$ is (by the maximal representation theorem in Sec. 78.2). We will write

$$\mathbf{c}[l] = \mathbf{x}[l]\mathbf{a}[l] \tag{1790}$$

We start by defining forward and backward sequential networks

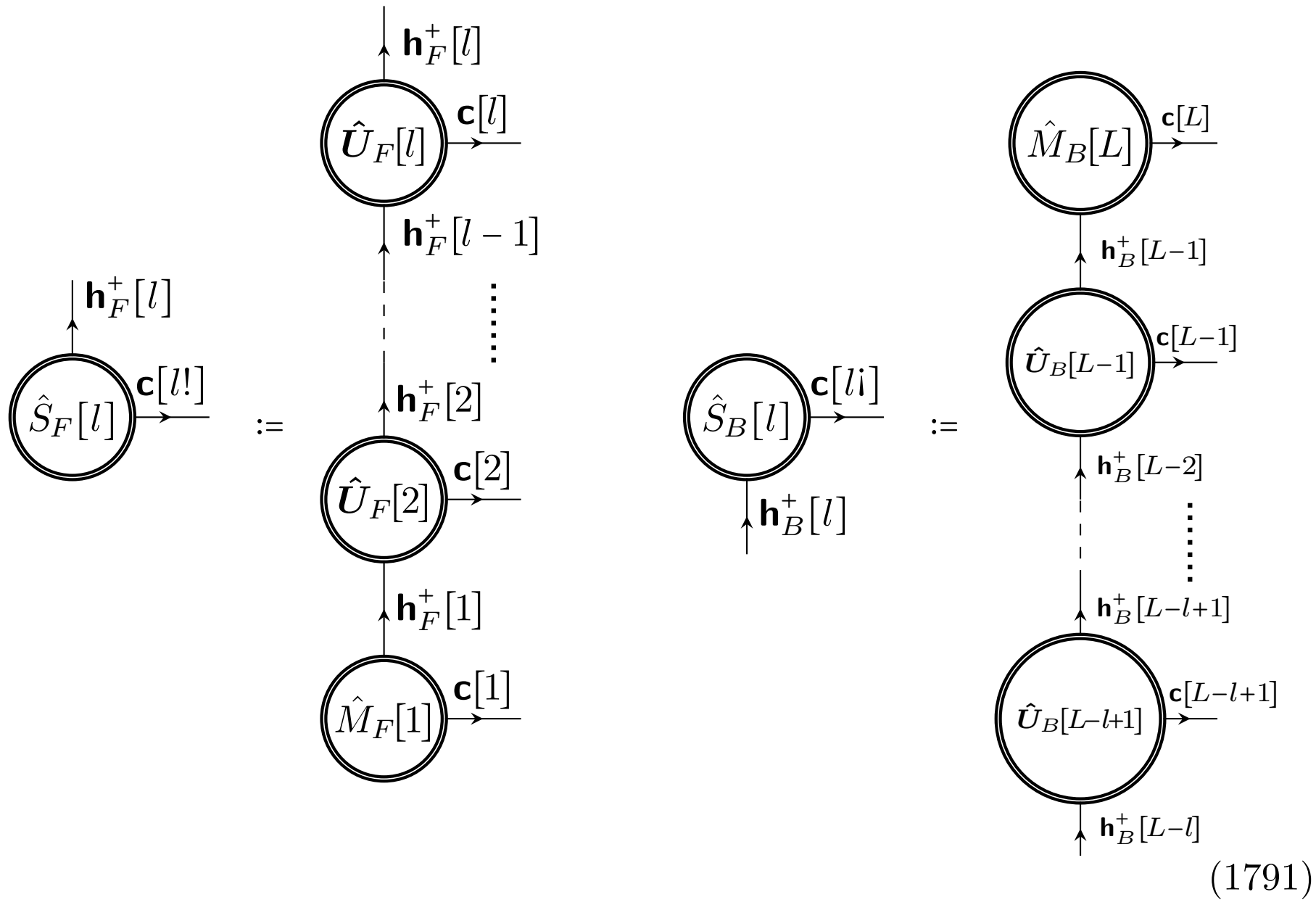

(1791)

where we are using the notation

$$\mathbf{c}[l!] = \mathbf{c}[1]\mathbf{c}[2]\ldots\mathbf{c}[l] \qquad \text{and} \qquad \mathbf{c}[l¡] = \mathbf{c}[L-l+1]\ldots\mathbf{c}[L-1]\mathbf{c}[L] \tag{1792}$$

For the forward sequential network (on the left), $\hat{\boldsymbol{U}}_F[k]$ are natural unitaries and $\hat{M}_F[1]$ and are natural temporal maxometries with ancillae $\mathbf{h}_F^+[1]$. For the backward sequential network (on the right), $\hat{\boldsymbol{U}}_B[k]$ are natural temporal unitaries and $\hat{M}_B[L]$ are natural temporal maxometries with ancillae $\mathbf{h}_B^+[L-1]$. Note that the networks in (1791) are homogeneous (since unitaries and maxometries are homogeneous). Given the maximal representation theorem, our task is equivalent to proving that

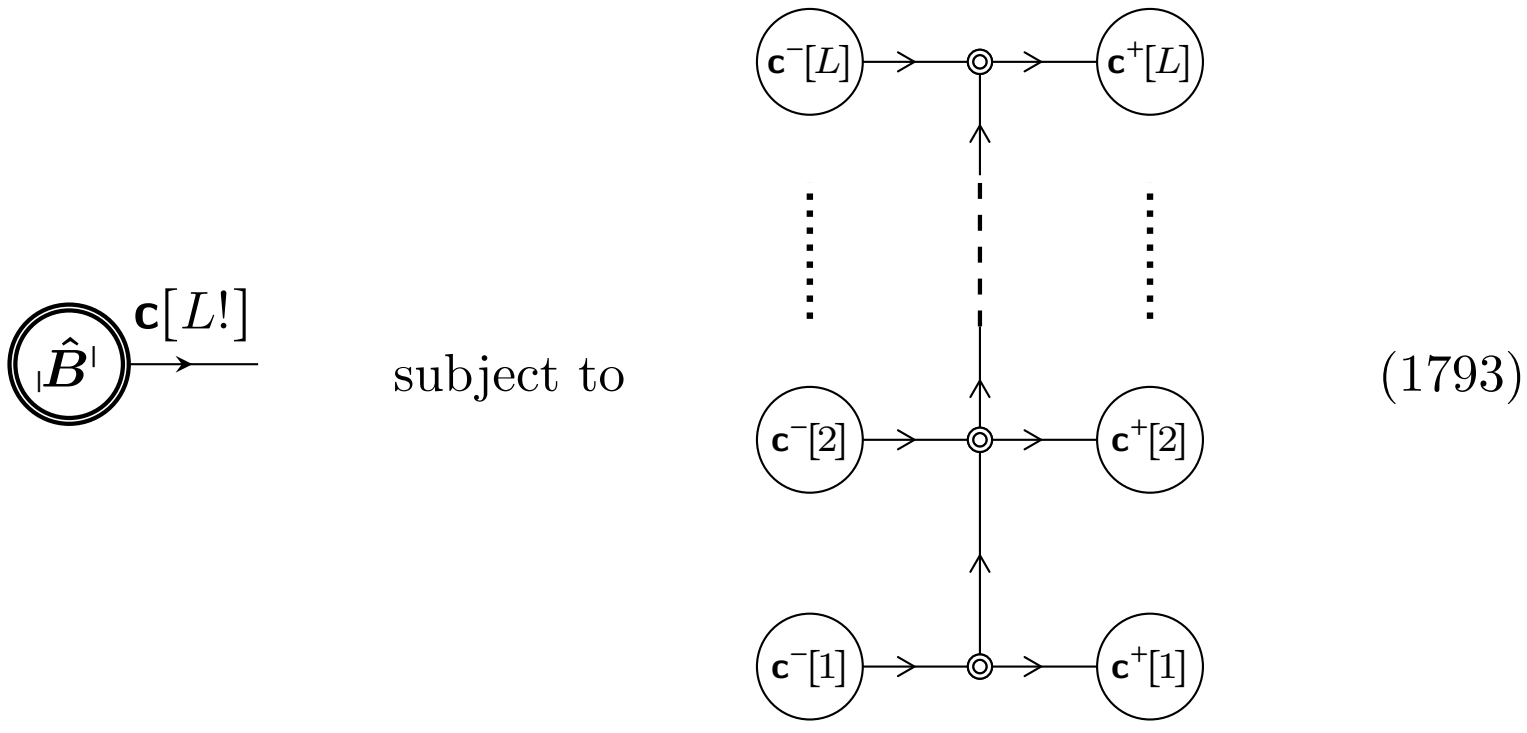

(1793)

(compare with (1788)) is deterministic and physical if and only if we can write

$$
{}_{|}\hat{\boldsymbol{B}}^{|}\ \xrightarrow{\ \mathbf{c}[L!]\ } \quad = \quad \hat{I} \xleftarrow{\ \mathbf{h}^{+}_{F}[L]\ } \hat{S}_F[L] \xrightarrow{\ \mathbf{c}[L!]\ } \quad = \quad \hat{I} \xrightarrow{\ \mathbf{h}^{+}_{B}[0]\ } \hat{S}_B[L] \xrightarrow{\ \mathbf{c}[L!]\ } \tag{1794}
$$

(compare with (1789)). To prove this theorem we need to prove both necessity and sufficiency. Necessity means that we can always write ${}_{|}\hat{\boldsymbol{B}}^{|}$ in both of the given forms. Sufficiency means that, if we can write it in both of these forms, then it is deterministic and physical. To prove necessity and sufficiency we will set up an induction hypothesis with base case given by the simple dilation (following CDP). In fact we have two induction hypotheses. A forward one (which we will state explicitly) and a backward one which is the time reverse of the forward one. First, we define

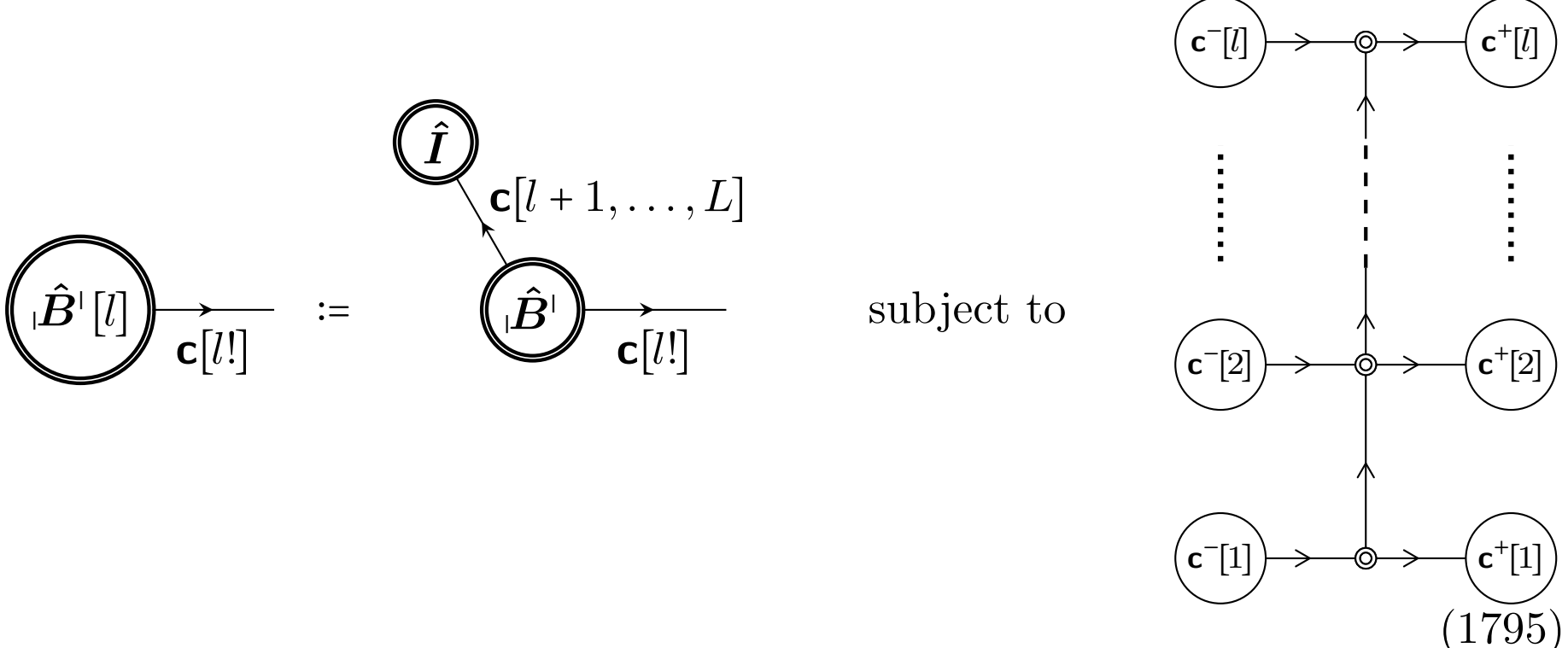

(1795)

(where we have converted ${}_{|}\hat{\boldsymbol{B}}^{|}$, given in (1793), to a convenient form so we can hit some of the systems in $\mathbf{c}[L!]$ with $\hat{\boldsymbol{I}}$). By application of the first theorem in Sec. 49.4.7 (under correspondence) we see that ${}_{|}\hat{\boldsymbol{B}}^{|}[l]$ is the past residuum of ${}_{|}\hat{\boldsymbol{B}}^{|}$ at $p[l]$ (where $p[l]$ is the synchronous partition going through the $l$th of the vertical wires of the causal diagram above). By application of the second theorem in the same section, we see that the causal diagram of ${}_{|}\hat{\boldsymbol{B}}^{|}[l]$ is, indeed, the one given in (1795). Note that ${}_{|}\hat{\boldsymbol{B}}^{|} = {}_{|}\hat{\boldsymbol{B}}^{|}[L]$. The *forward inductive hypothesis* is the following. *If the following are true*

(a) we have

$$
{}_{|}\hat{\boldsymbol{B}}^{|}[l] \;\xrightarrow{\mathbf{c}[l!]}\quad = \quad \hat{\boldsymbol{I}} \xleftarrow{\mathbf{h}_F^{+}[l]} \hat{S}_F[l] \;\xrightarrow{\mathbf{c}[l!]} \tag{1796}
$$

where

$$
|\{h_F^{+}[l]\}| \geq \operatorname{rank}({}_{|}\hat{\boldsymbol{B}}^{|}[l]) \tag{1797}
$$

and

(b) ${}_{|}\hat{\boldsymbol{B}}^{|}[l+1]$ satisfies forward causality at $p[l]$, i.e.

$$
\xleftarrow{\mathbf{c}^{-}[l+1]} \hat{\boldsymbol{I}}_{+} \xleftarrow{\mathbf{c}[l+1]} {}_{|}\hat{\boldsymbol{B}}^{|}[l+1] \;\xrightarrow{\mathbf{c}[l!]}\quad = \quad \xleftarrow{\mathbf{c}^{-}[l+1]} \hat{\boldsymbol{I}} \qquad {}_{|}\hat{\boldsymbol{B}}^{|}[l] \;\xrightarrow{\mathbf{c}[l!]} \tag{1798}
$$

(note that ${}_{|}\hat{\boldsymbol{B}}^{|}[l]$ is necessarily the past residuum here),

*then the following are true*

(A) we have that

$$
{}_{|}\hat{\boldsymbol{B}}^{|}[l+1] \;\xrightarrow{\mathbf{c}[(l+1)!]}\quad = \quad \hat{\boldsymbol{I}} \xleftarrow{\mathbf{h}_F^{+}[l+1]} \hat{\boldsymbol{U}}_F[l+1] \left(\xrightarrow{\mathbf{c}[l+1]}\right) \xleftarrow{\mathbf{h}_F^{+}[l]} \hat{S}_F[l] \;\xrightarrow{\mathbf{c}[l!]}\quad = \quad \hat{\boldsymbol{I}} \xleftarrow{\mathbf{h}_F^{+}[l+1]} \hat{S}_F[l+1] \;\xrightarrow{\mathbf{c}[(l+1)!]} \tag{1799}
$$

where

$$
|\{h_F[l+1]\}||\{c^{+}[l+1]\}| = |\{h_F[l]\}||\{c^{-}[l+1]\}| \tag{1800}
$$

(this condition is necessary so that $\hat{\boldsymbol{U}}_F[l+1]$ can be a unitary - compare with (1750)) with

$$
|\{h_F^{+}[l+1]\}| \geq \operatorname{rank}({}_{|}\hat{\boldsymbol{B}}^{|}[l+1]) \tag{1801}
$$

(which is necessary since $\hat{S}_F[l+1]$ is homogeneous), and

(B) ${}_{|}\hat{\boldsymbol{B}}^{|}[l+1]$ satisfies forward causality for all synchronous partitions.

The base case for the induction process is the simple dilation

$$
\left({}_{|}\hat{\boldsymbol{B}}^{|}[1]\right) \xrightarrow[\mathsf{c}[1]]{} \quad = \quad \left(\hat{M}_F[1]\right) \xrightarrow[\mathsf{c}[1]]{}, \qquad \left(\hat{M}_F[1]\right) \xrightarrow{\mathsf{h}_F^+[1]} \left(\hat{\boldsymbol{I}}\right) \tag{1802}
$$

(which must have $h_F[1] \geq \text{rank}({}_{|}\hat{\boldsymbol{B}}^{|}[1])$) for the case of simple causal structure (compare this with (1732) - also note that we can have the ancilla system as an input instead of as an output from the maxometry which is needed for the backward induction proof). According to the simple dilation theorem (Sec. 78.3) this is deterministic and physical and therefore satisfies forward causality for all synchronous partitions. Thus, we satisfy (a) in the forward induction hypothesis as required for the base case. Note that (b) must be true since, according to the residua causality theorem in Sec. 49.4.7, residua satisfy double causality for all synchronous partitions. We will prove the forward induction hypothesis below (we can prove an induction hypothesis in the backward direction by similar techniques). Before providing this proof, let us see how these induction hypotheses can be used to prove the theorem.

1. First we prove that the system of equations appearing in the forward induction hypothesis

$$
|\{h_F[l+1]\}| \{c^+[l+1]\}| = |\{h_F[l]\}| \{c^-[l+1]\}| \qquad \text{for} \quad l = 1 \text{ to } L-1 \tag{1803}
$$

with

$$
|\{h_F^+[l]\}| \geq \text{rank}({}_{|}\hat{\boldsymbol{B}}^{|}[l]) \qquad \text{for} \quad l = 1 \text{ to } L \tag{1804}
$$

have solutions for $|\{h_F^+[l]\}|$ for $l = 1$ to $L$.

2. To prove the necessity part of the theorem note that we can use the forward induction step to prove that ${}_{|}\hat{\boldsymbol{B}}^{|}$ can be written as in the middle expression in (1794) (involving $\hat{S}_F[L]$). We start with the base case - this is the simple dilation in (1802). Then, by induction, we can prove that ${}_{|}\hat{\boldsymbol{B}}^{|} = {}_{|}\hat{\boldsymbol{B}}^{|}[L]$ has the form in the middle expression in (1794) as required. By employing the corresponding backward induction hypothesis we can prove that ${}_{|}\hat{\boldsymbol{B}}^{|} = {}_{|}\hat{\boldsymbol{B}}^{|}[L]$ is equal to the right expression in (1794) (involving $\hat{S}_B[L]$).

3. Now consider the sufficiency part of the theorem. Consider the forward induction hypothesis. The base case is the simple dilation and we have already proved (in the simple dilation theorem) that this satisfies forward causality for all synchronous partitions. It then follows by induction from (B) that ${}_{|}\hat{\boldsymbol{B}}^{|} = {}_{|}\hat{\boldsymbol{B}}^{|}[L]$ satisfies forward causality for all synchronous partitions. That backward causality is satisfied for all synchronous partitions

follows from the backward induction hypothesis by similar arguments. This means that double causality is satisfied for all synchronous partitions. To complete the proof of sufficiency, we note that the expressions in (1794) are necessarily twofold positive since the components are. Thus we have proven that, if ${}_{\shortmid}\hat{\boldsymbol{B}}^{\shortmid}$ can be written as being equal to both of the expressions in (1794), then this is sufficient to guarantee that it is deterministic and physical.

To complete the proof of the theorem, then, we need to show that the $h_F^+[l]$ exist according to point 1 above and to prove the forward induction hypothesis (the proof of the backward induction hypothesis follows by similar reasoning). We start by proving that the $h_F^+[l]$ satisfying (1803,1804) exist. We will do this by induction. We introduce $h_{F_i}^+[l]$ which have an induction parameter $i$ such that $h_{F_L}^+[l] = h_F^+[l]$. Consider the system of equations above, but now labeled by our induction parameter as follows.

$$|\{h_{F_i}[l+1]\}||\{c^+[l+1]\}| = |\{h_{F_i}[l]\}||\{c^-[l+1]\}| \qquad \text{for} \quad l = 1 \text{ to } i \tag{1805}$$

with

$$|\{h_{F_i}^+[l]\}| \geq \text{rank}({}_{\shortmid}\hat{\boldsymbol{B}}^{\shortmid}[l]) \qquad \text{for} \quad l = 1 \text{ to } i+1 \tag{1806}$$

We will show inductively how to construct solutions for the system (1805,1806) of equations. Starting with $i = 1$, as a base case, the system of equation is just

$$\begin{aligned} |\{h_{F_1}[2]\}||\{c^+[2]\}| =& |\{h_{F_1}[1]\}||\{c^-[2]\}| \\ |\{h_{F_1}^+[1]\}| \geq& \text{rank}({}_{\shortmid}\hat{\boldsymbol{B}}^{\shortmid}[1]) \\ |\{h_{F_1}^+[2]\}| \geq& \text{rank}({}_{\shortmid}\hat{\boldsymbol{B}}^{\shortmid}[2]) \end{aligned} \tag{1807}$$

We can always find

$$\left(|\{h_{F_1}^+[1]\}|, |\{h_{F_1}^+[2]\}|\right) \tag{1808}$$

satisfying these equations. Next, consider $i = 2$. Now we have

$$\begin{aligned} |\{h_{F_2}[2]\}||\{c^+[2]\}| =& |\{h_{F_2}[1]\}||\{c^-[2]\}| \\ |\{h_{F_2}[3]\}||\{c^+[3]\}| =& |\{h_{F_2}[2]\}||\{c^-[3]\}| \\ |\{h_{F_2}^+[1]\}| \geq& \text{rank}({}_{\shortmid}\hat{\boldsymbol{B}}^{\shortmid}[1]) \\ |\{h_{F_2}^+[2]\}| \geq& \text{rank}({}_{\shortmid}\hat{\boldsymbol{B}}^{\shortmid}[2]) \\ |\{h_{F_2}^+[3]\}| \geq& \text{rank}({}_{\shortmid}\hat{\boldsymbol{B}}^{\shortmid}[3]) \end{aligned} \tag{1809}$$

We can satisfy these equations by noting that we can always find an integer, $n_2 \geq 1$, such that

$$\left(|\{h_{F_2}^+[1]\}|, |\{h_{F_2}^+[2]\}|, |\{h_{F_2}^+[3]\}|\right) = \left(n_2|\{h_{F_1}^+[1]\}|, n_2|\{h_{F_1}^+[2]\}|, |\{h_{F_2}^+[3]\}|\right) \tag{1810}$$

since multiplying (1808) by $n_2 \geq 1$ gives new solutions that still satisfy (1807) and can serve as our solutions for the first equation in (1809). We can always choose $n_2$ and $|\{h^+_{F_2}[3]\}|$ such that the $i = 2$ system of equations in (1809) are satisfied by (1810). We can iterate this process until we have a set of solutions for the $i = L$ case. These satisfy the original system of equations in (1803,1804) as required. Now we will prove the forward induction hypothesis above. From (b) we have

$\hat{\boldsymbol{I}}$

$\mathbf{h}^+_F[l+1]$

$\hat{\boldsymbol{B}}[l+1]$ $\quad \mathbf{c}[(l+1)!]$ $\quad = \quad$ $\hat{\boldsymbol{U}}_F[l+1]$ $\quad \mathbf{c}[l+1]$ (1811)

$\mathbf{h}^+_F[l]$

$\hat{M}_F[l]$ $\quad \mathbf{c}[l!]$

by the basic unitary causal dilation theorem (this is an application of (1740) where we have no pointer systems) where $\hat{M}_F[l]$ is a natural temporal maxometry to be used in the induction process (such that, for $l = 1$, we connect with the base case) and $\hat{\boldsymbol{U}}_F[l+1]$ is a natural temporal unitary. Since $\hat{S}_F[l]$, defined in (1791), is homogeneous we can write

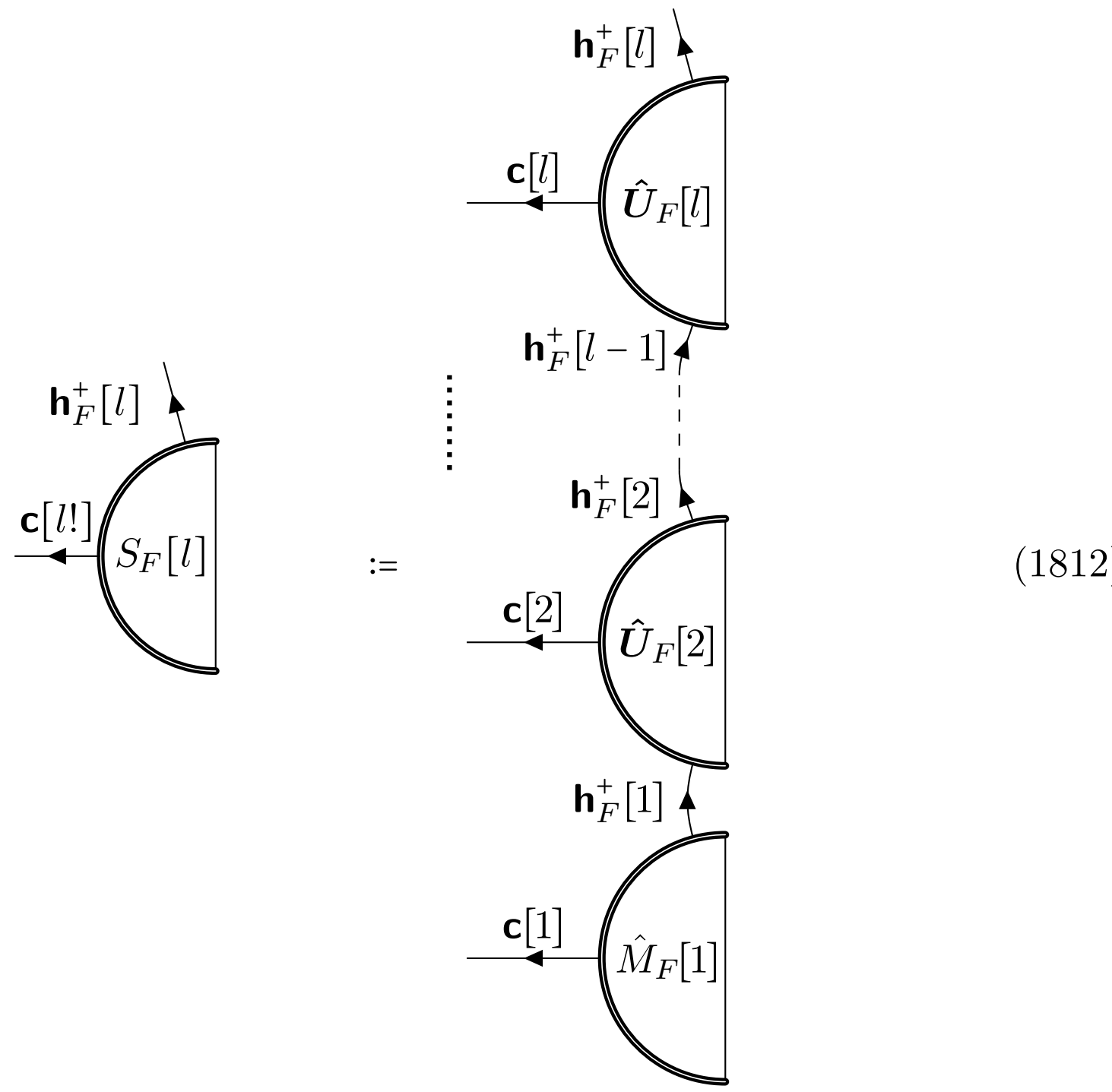

for the associated left object. From (a) we have

$$\mathsf{c}[l!] \;\; {}_\llcorner\boldsymbol{B}^\urcorner[l] \;\; h_F^+[l] \quad = \quad \mathsf{c}[l!] \;\; S_F[l] \;\; h_F^+[l] \tag{1813}$$

since, if this is reflected in a regular physical mirror we get back (1796) (where $|\{h_F^+[l]\}| \geq \text{rank}({}_\llcorner\boldsymbol{B}^\urcorner[l])$). Since ${}_\llcorner\hat{\boldsymbol{B}}^\urcorner[l]$ is the past residuum of ${}_\llcorner\hat{\boldsymbol{B}}^\urcorner[l+1]$ (see (1798)) we can, with an appropriate assignment of notation, substitute (1813) into (1755) from the basic unitary causal dilation corollary giving

$$\mathbf{h}_F^+[l] \;\; \mathsf{c}[l!] \;\; M_F[l] \quad = \quad \mathbf{h}_F^+[l] \;\; \mathsf{c}[l!] \;\; S_F[l] \tag{1814}$$

where we have used the decomposition of the identity property (1413) on system $\mathbf{h}_F^+[l]$ to obtain this from (1755). From (1814), we obtain

$$\mathbf{h}_F^+[l] \;\; M_F[l] \;\; \mathsf{c}[l!] \quad = \quad \mathbf{h}_F^+[l] \;\; S_F[l] \;\; \mathsf{c}[l!] \tag{1815}$$

Plugging this into (1811) gives (1799) which proves (A) as required. Now we will prove (B). To do this note that ${}_\llcorner\hat{\boldsymbol{B}}^\urcorner[l+1]$ can be written as in (1811). The top part of this (consisting of $\hat{\boldsymbol{U}}_F[l+1]$ and $\hat{\boldsymbol{I}}$) has simple causal structure and is physical (and so, in particular, it satisfies forward causality for all synchronous partitions through its associated causal diagram). If the lower part, $\hat{M}_F[l]$, also satisfies forward causality for all partitions of its associated causal diagram then, using the causality composition theorem (deterministic case) under correspondence (this theorem is in Sec. 49.4.11), we can say that ${}_\llcorner\hat{\boldsymbol{B}}^\urcorner[l+1]$ satisfies forward causality for all synchronous partitions. Now we note that $\hat{M}_F[1]$ appearing in the base case in (1802) associated with the base case satisfies forward causality for all synchronous partitions since it (i) it has simple causal structure and (ii) it is a natural temporal maxometry whose ancilla is an output. Consequently, by the causality composition theorem (under correspondence), $\hat{M}_F[l]$ satisfies forward causality (for all synchronous partitions). Thus, by (1811) and the causality composition theorem, $\llcorner\hat{\boldsymbol{B}}\urcorner[l+1]$ must satisfy forward causality for all synchronous partitions. This proves (B). This proves the forward induction hypothesis. To prove the corresponding backward induction hypothesis we employ similar techniques. Thus, the theorem is proven.

## 78.15 $L$-causal ladder dilation theorem with minimal sufficient pairs

As in the case of the 2-ladder causal dilation theorem, we can weaken each member of the sufficient pair to obtain a minimal sufficient pair version of the $L$-causal ladder dilation theorem.

> **$L$-causal ladder dilation theorem with minimal sufficient pairs.** An operator tensor
>
> 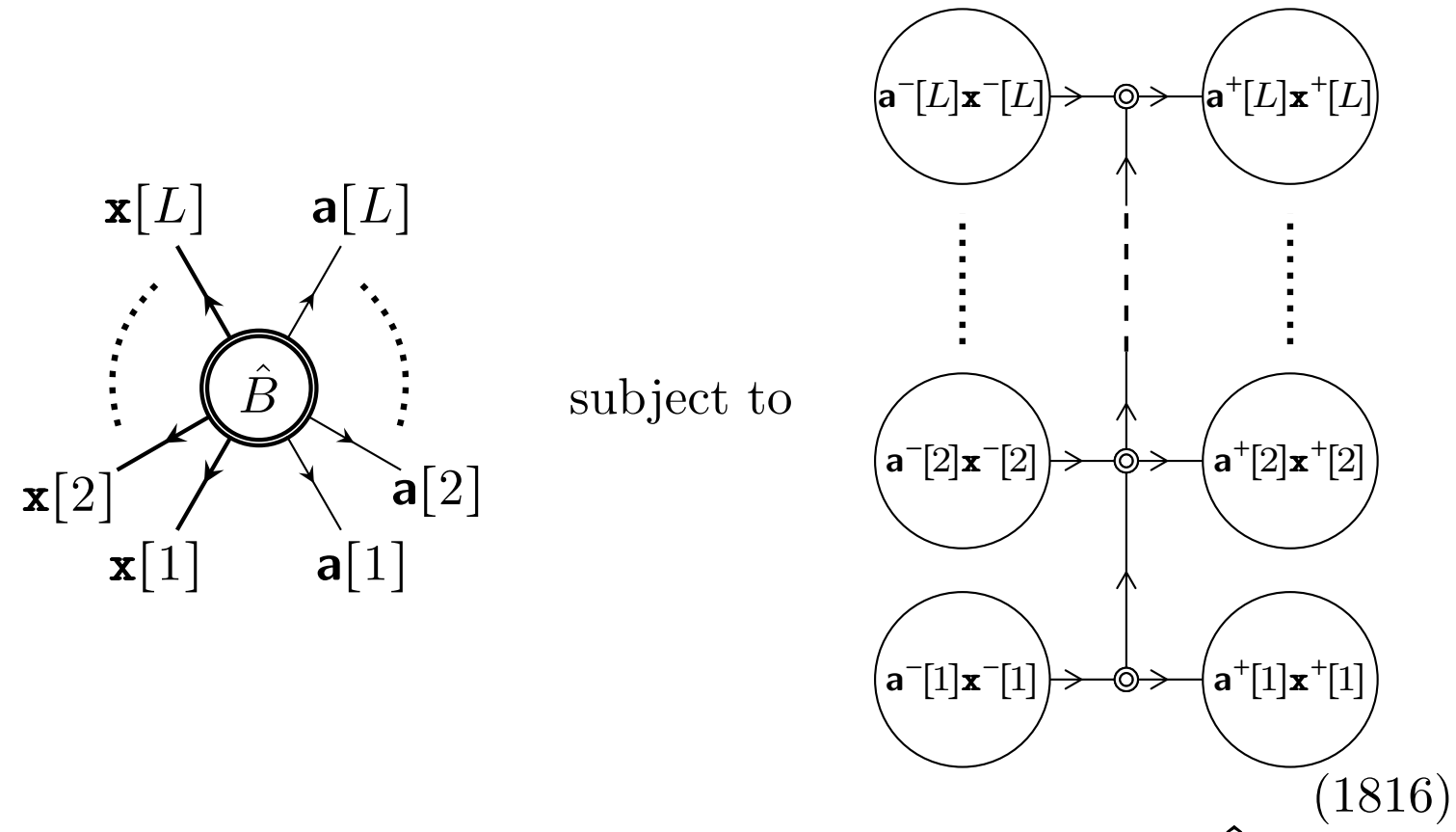
>
>
> (1816)
>
> is deterministic and physical if and only if we can write $\hat{\boldsymbol{B}}$ as being

equal to *both*

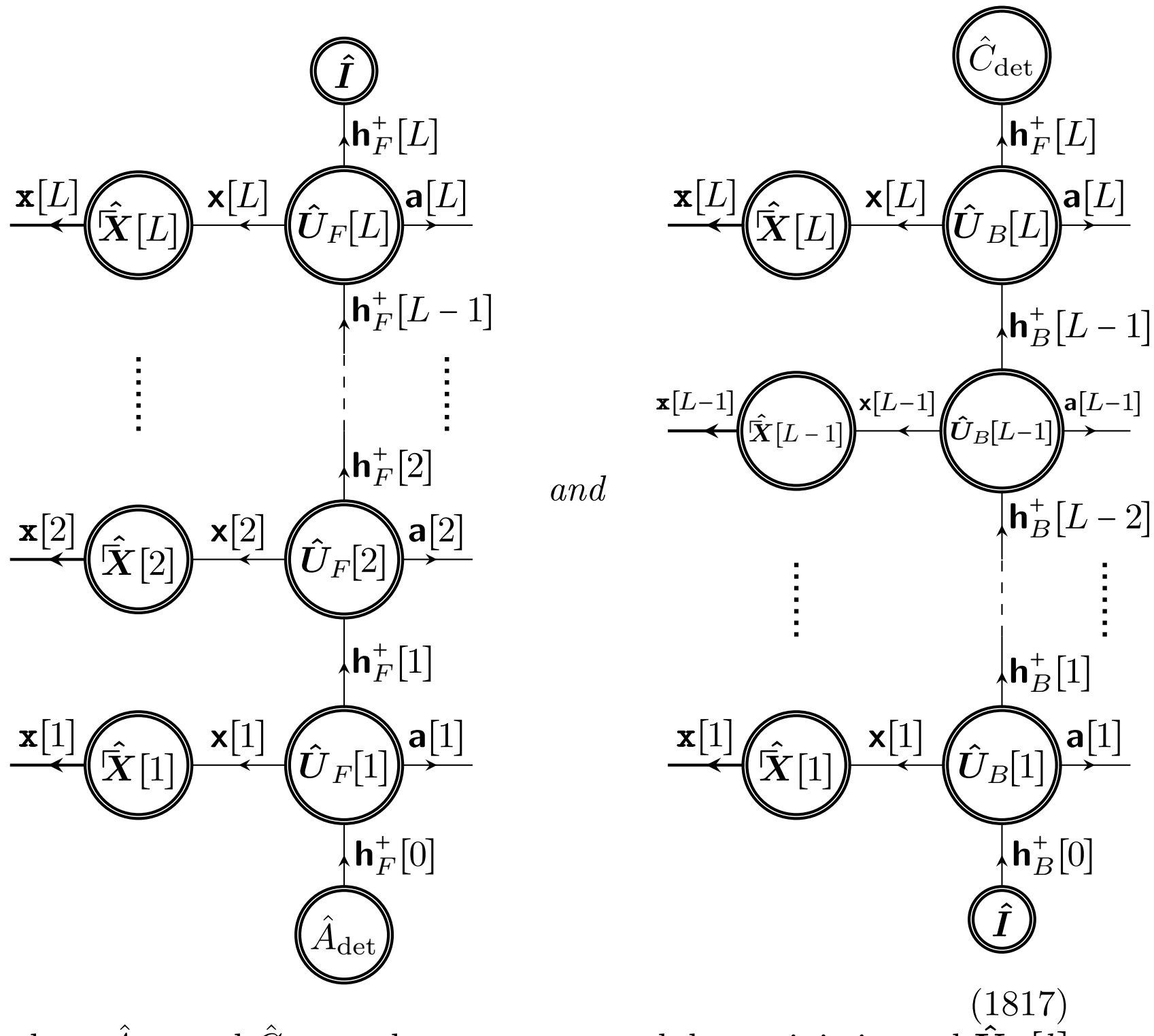

(1817)

where $\hat{A}_{\mathrm{det}}$ and $\hat{C}_{\mathrm{det}}$ are homogeneous and deterministic, and $\hat{\boldsymbol{U}}_F[l]$ and $\hat{\boldsymbol{U}}_B[l]$ (for $l = 1$ to $L$) are natural temporal unitaries.

The proof of this is the obvious generalisation of the proof of the 2-ladder dilation theorem with minimal sufficient pairs in Sec. 78.10. First we replace the natural temporal maxometries with the expression in (1701). Then we can pull the extra $\hat{\boldsymbol{I}}$ up the middle (or down the middle) so it joins the $\hat{\boldsymbol{I}}$ at the top (or at the bottom). This proves necessity. Sufficiency is proven by the same technique as for the $L$-causal ladder dilation theorem.

## 78.16 Dilation theorem for $L$-causal ladders with (co)isometries

Here we prove a dilation theorem for $L$-causal ladders that uses isometries and coisometries.

**Dilation theorem for $L$-causal ladders with isometries and**

**coisometries.** An operator tensor

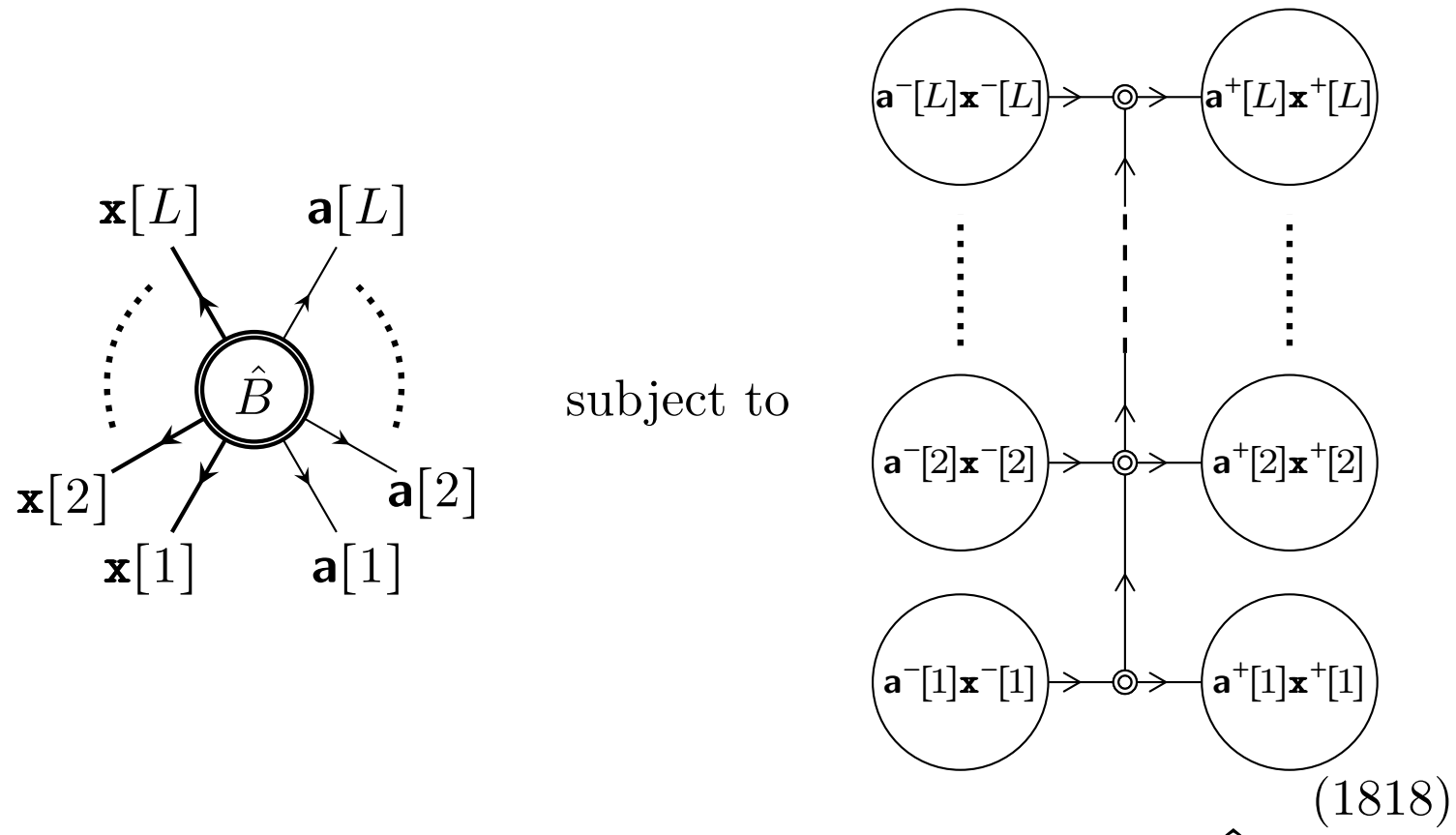

(1818)

is deterministic and physical if and only if we can write $\hat{\boldsymbol{B}}$ as being equal to *both*

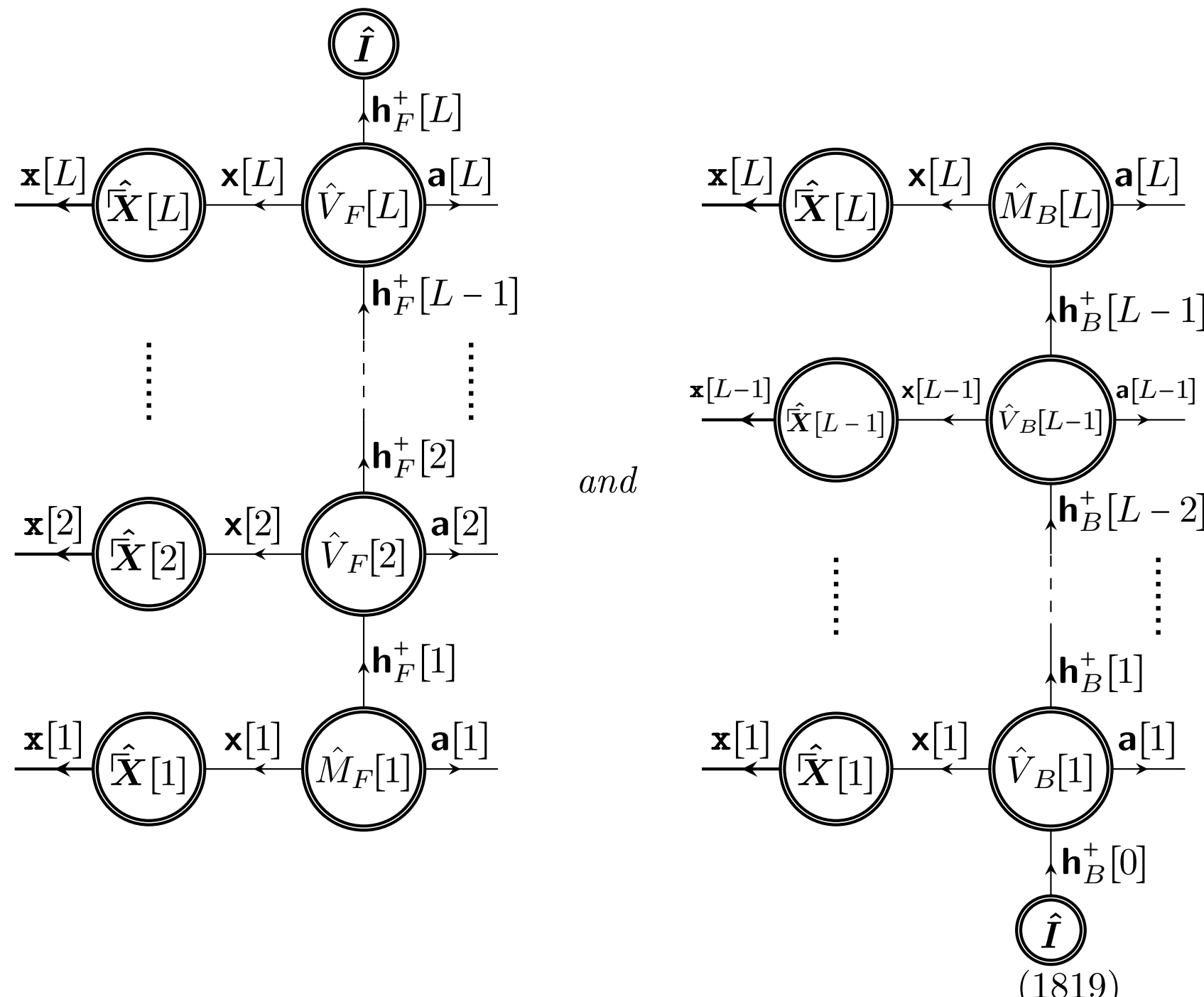

(1819)

where $\hat{M}_F[1]$ and $\hat{M}_B[L]$ are natural temporal maxometries with ancillas $\mathsf{h}_F^+[1]$ and $\mathsf{h}_B^+[L-1]$ respectively, and $\hat{\boldsymbol{V}}_F[l]$ (for $l = 2$ to $L$) are natural temporal isometries and $\hat{\boldsymbol{V}}_B[l]$ (for $l = 1$ to $L - 1$) are natural temporal coisometries. Furthermore, we can choose $N_{\mathsf{h}_F^+}[L] = \text{rank}(\hat{\boldsymbol{B}})$ and $N_{\mathsf{h}_B^+}[0] = \text{rank}(\hat{\boldsymbol{B}})$.

The proof of this theorem is essentially same as the proof of the dilation theorem for $L$-ladders with unitaries in Sec. 78.14. We can write down a forward induction hypothesis as in that theorem with some small differences. In (a) we have

$$|\{h_F^+[l]\}| = \text{rank}({}_{\shortmid}\hat{\boldsymbol{B}}^{\shortmid}[l]) \tag{1820}$$

instead of (1797). In (A) we have a natural temporal isometric operator, $V_F[l+1]$ instead of the natural temporal unitary operator, $U_F[l+1]$, that appears in (1799), we have

$$|\{h_F[l+1]\}||\{c^+[l+1]\}| \geq |\{h_F[l]\}||\{c^-[l+1]\}| \tag{1821}$$

instead of (1800), and we have

$$|\{h_F^+[l+1]\}| = \text{rank}({}_{\shortmid}\hat{\boldsymbol{B}}^{\shortmid}[l+1]) \tag{1822}$$

instead of (1801). This forward induction hypothesis is proven using the basic (co)isometric causal dilation theorem. Using this modified induction hypothesis we obtain the left expression in (1819) using a similar technique. Notably, we do not need to go through the exercise of solving the system of equations in (1803) and (1804) since we are not requiring the $V_F[l]$ to be unitary. The right expression in (1819) is obtained similarly. This proves necessity. Sufficiency follows from the causality composition theorem (see Sec. 49.4.11) in the same way as for the proof in Sec. 78.14. The remarks concerning rank follow from the fact that ${}_{\shortmid}\hat{\boldsymbol{B}}^{\shortmid} = {}_{\shortmid}\hat{\boldsymbol{B}}^{\shortmid}[L]$ for the forward case and from the corresponding construction in the backwards case.

## 78.17 Dilation for $L$-causal ladders with (co)isometric minimal sufficient pairs

Given the previous causal dilation theorem, it is trivial to see that we have the following theorem with minimal sufficient pairs.

**Dilation theorem for $L$-causal ladders with (co)isometric minimal sufficient pairs.** An operator tensor

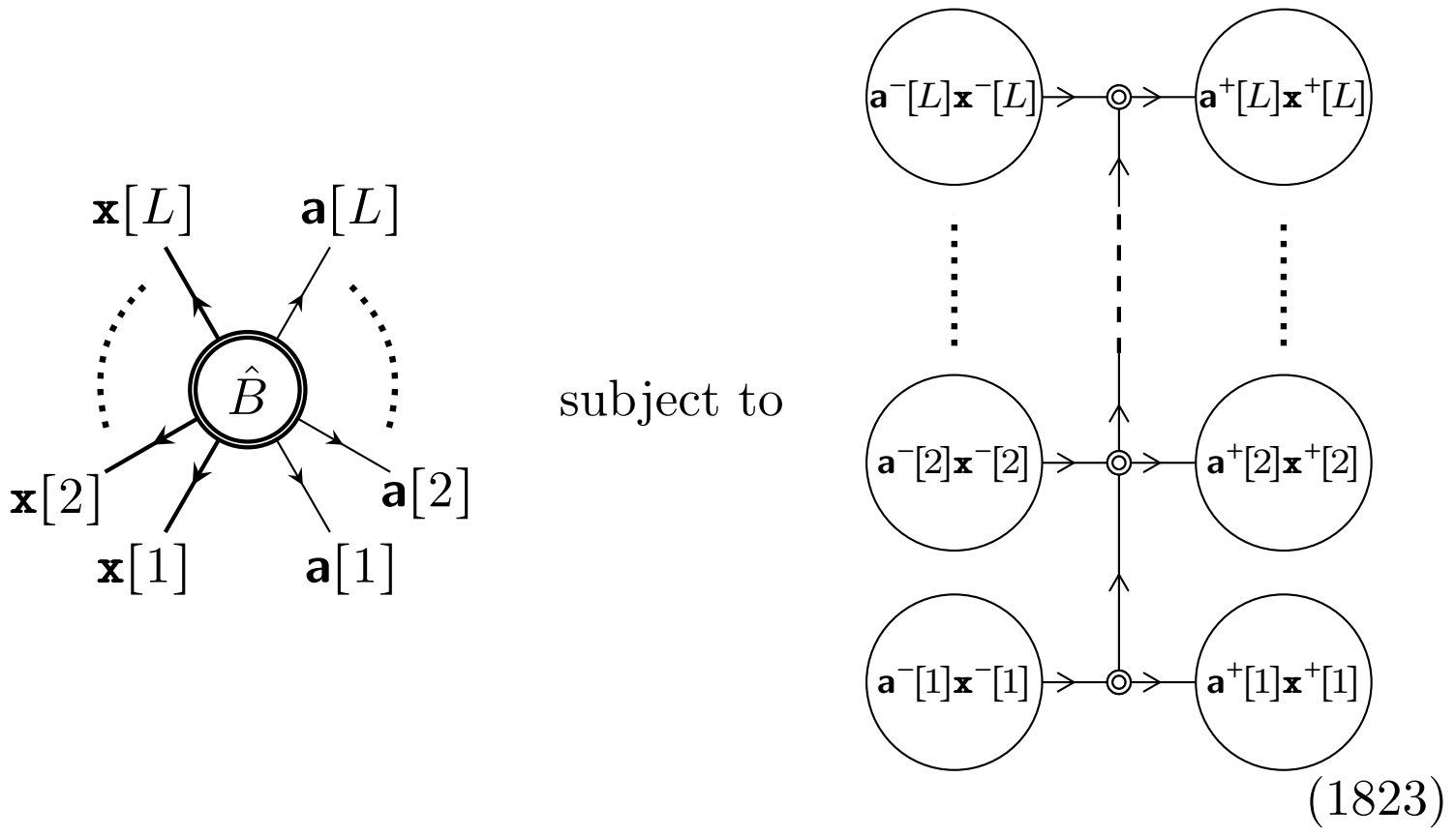

(1823)

is deterministic and physical if and only if we can write $\hat{\boldsymbol{B}}$ as being equal to *both*

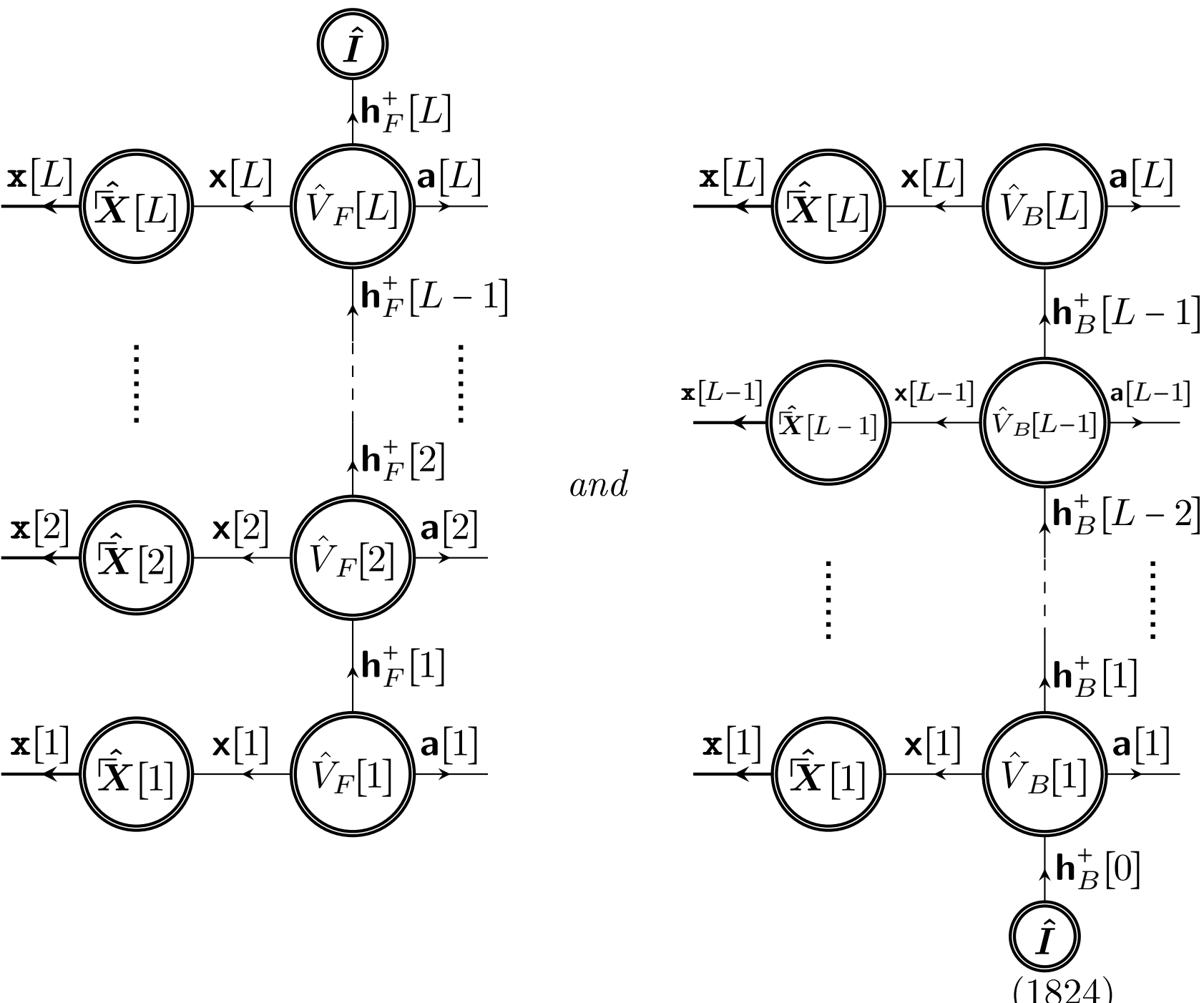

(1824)

where $\hat{V}_F[l]$ (for $l = 1$ to $L$) are natural temporal isometries and $\hat{V}_B[l]$ (for $l = 1$ to $L$) are natural temporal coisometries. Furthermore, we can choose $N_{\mathsf{h}_F^+}[L] = \text{rank}(\hat{\boldsymbol{B}})$ and $N_{\mathsf{h}_B^+}[0] = \text{rank}(\hat{\boldsymbol{B}})$.

We obtain this from the previous theorem by noting that the natural temporal maxometric operator, $\hat{M}_F[1]$, in the left expression of (1819) only has only output ancilla and is, hence, a natural temporal isometric operator. Similarly $\hat{M}_B[L]$ in the right expression of (1819) has only input ancilla and is, hence, a natural temporal coisometric operator. This establishes that (1824) are necessary conditions. That they are sufficient follows from application of the causality composition theorem as in the proof of the previous theorem.

This theorem is both weaker and stronger than the previous theorem. It is stronger in that we require weaker conditions for sufficiency. It is weaker in that the conditions imposed by necessity are weaker.

The forward version of this theorem is the same dilation that is found in Theorem 6 of Gutoski and Watrous [2007] and Theorem 3 of Chiribella et al. [2009a].

## 78.18 Homogeneous $L$-causal ladder dilation

In Sec. 78.13 we proved that homogeneous 2-causal ladders are necessarily equal to a sequence of natural temporal unitary operators. Here we prove a similar theorem for the more general case of $L$-ladders.

> **Homogeneous $L$-causal ladder dilation theorem.** An operator tensor
>
> 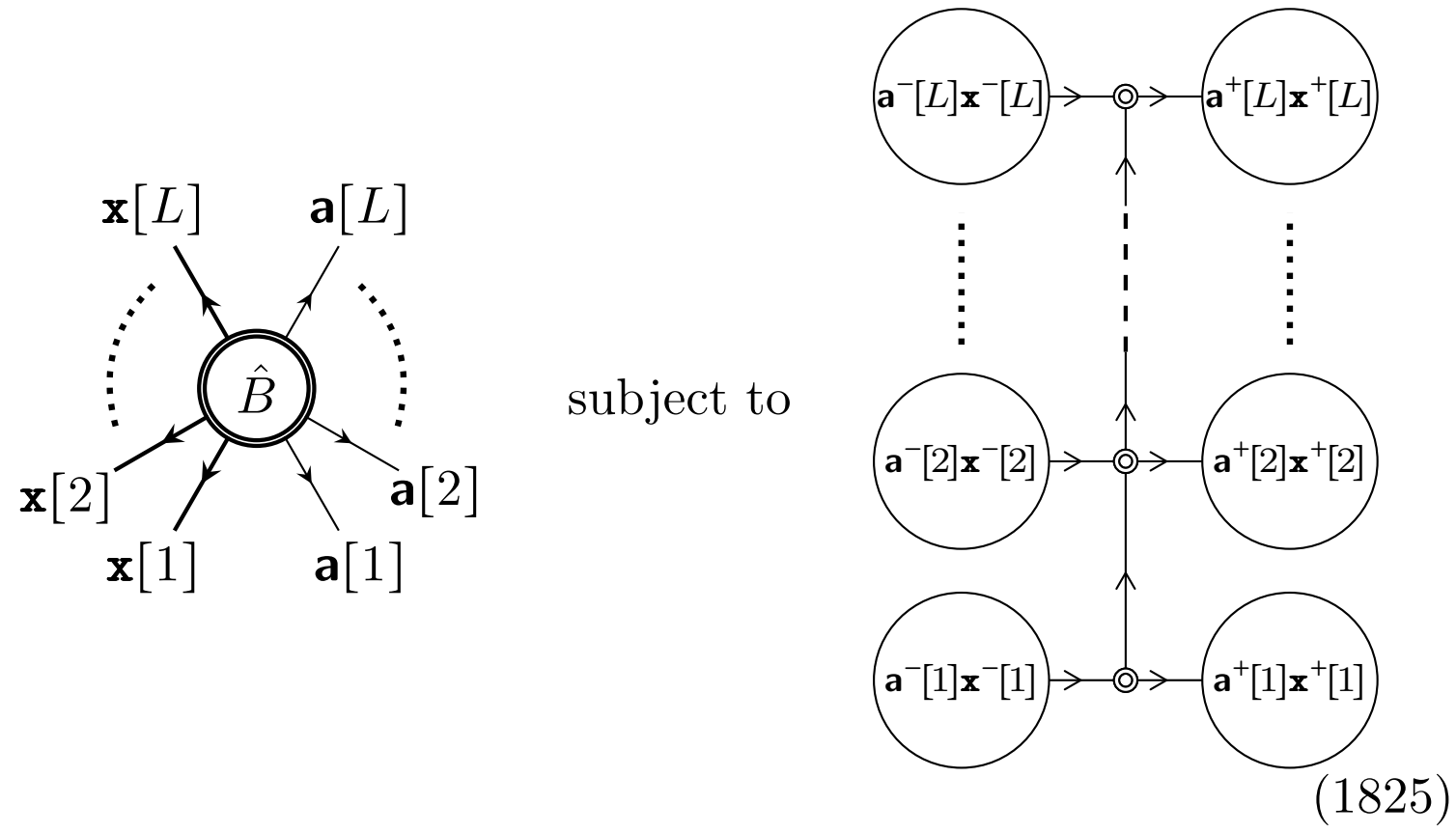
>
>
> (1825)
>
> is homogeneous, deterministic and physical if and only if we can write $\hat{\boldsymbol{B}}$ as being equal to
>
> 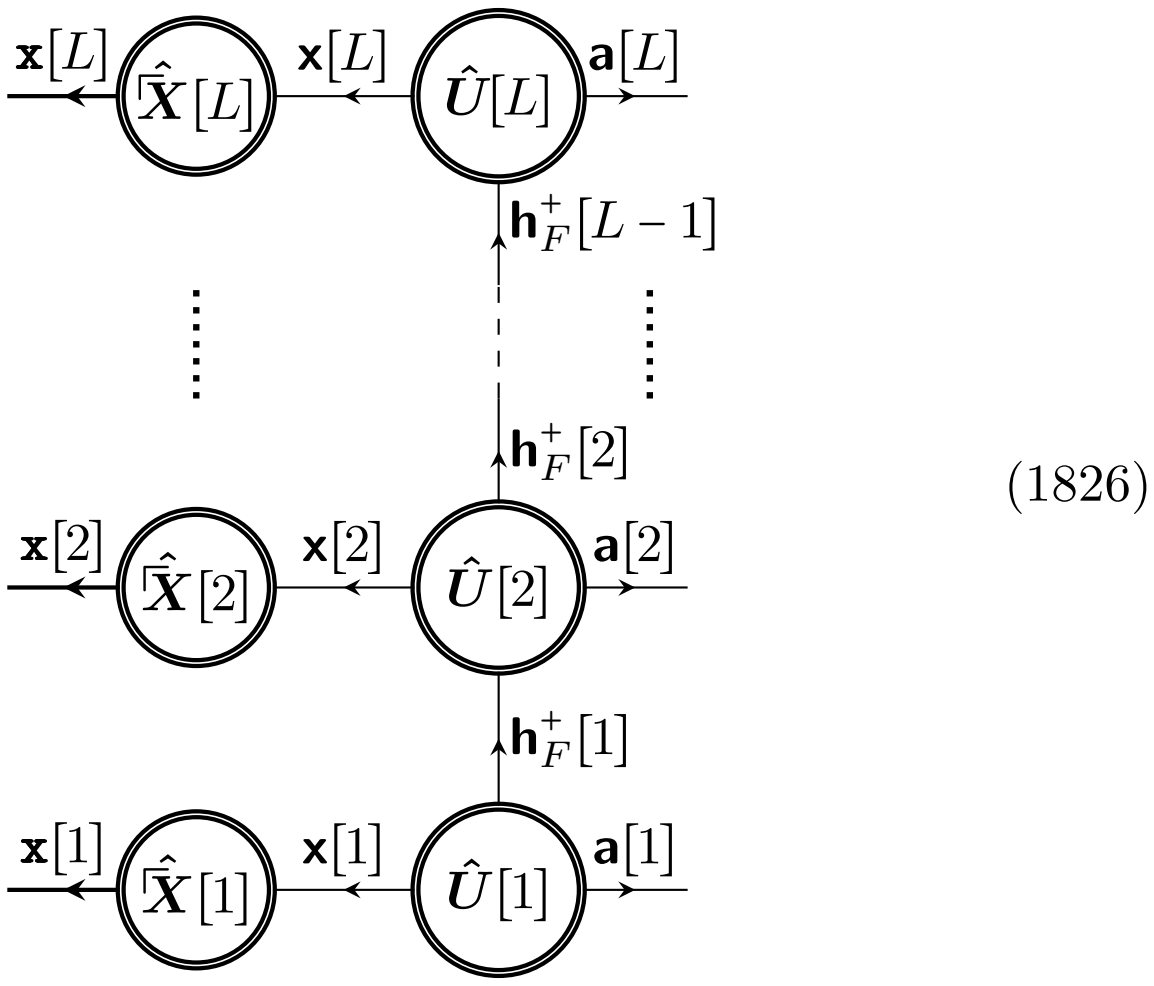
>
>
> (1826)
>
> where $\hat{\boldsymbol{U}}[l]$ (for $l = 1$ to $L$) are natural temporal unitaries.

We prove this using the dilation theorem for $L$-causal ladders with (co)isometric minimal pairs in Sec. 78.17. Since $\hat{\boldsymbol{B}}$ is homogeneous (i.e. of rank 1) we can put $N_{\mathsf{h}^+_F[L]} = N_{\mathsf{h}^+_B[0]} = 1$ for the dilations in (1824). We prove this theorem by induction starting with $l = L$ and then we work back to $l = 1$. To simplify the proof, we will use the maximal representation, $\llcorner\hat{\boldsymbol{B}}^\urcorner$ (see Sec. 78.2). Then we can put $\mathbf{a}[l]\mathbf{x}[l] = \mathbf{c}[l]$. We can, afterwards, convert back to $\hat{\boldsymbol{B}}$. We define

$$\hat{V}_F[l!] \;:=\; \hat{V}_F[l]\,\hat{V}_F[l-1]\cdots\hat{V}_F[2]\,\hat{V}_F[1] \qquad\qquad \hat{V}_B[l!] \;:=\; \hat{V}_B[l]\,\hat{V}_B[l-1]\cdots\hat{V}_B[2]\,\hat{V}_B[1] \tag{1827}$$

where $\hat{V}_F[l]$ are natural temporal isometries and $\hat{V}_B[l]$ are natural temporal coisometries. Consequently, $\hat{V}_F[l!]$ are natural temporal isometries and $\hat{V}_B[l!]$ are natural temporal coisometries. It is useful to proceed as if there are trivial systems, $\mathbf{h}^+_{F,B}[0]$ and $\mathbf{h}^+_{F,B}[L]$, having $N$ equal to 1. We have

$$\hat{\boldsymbol{B}} \;\; (\mathbf{c}[l!]) \;=\; \hat{V}_F[L!] \;\; (\mathbf{c}[l!]) \;=\; \hat{V}_B[L!] \;\; (\mathbf{c}[l!]) \tag{1828}$$

according to (1824). The general induction step uses the simple facts that, by definition,

$$\hat{V}_F[l!] \;=\; \hat{V}_F[l]\;\hat{V}_F[(l-1)!] \qquad\qquad \hat{V}_B[l!] \;=\; \hat{V}_B[l]\;\hat{V}_B[(l-1)!] \tag{1829}$$

Since $\hat{V}_F[l]$ and $\hat{V}_F[(l-1)!]$ are natural temporal isometric operators we have

$$N_{\mathbf{h}^+_F[l-1]} N_{\mathbf{c}^-[l]} \leq N_{\mathbf{h}^+_F[l]} N_{\mathbf{c}^+[l]} \qquad\qquad N_{\mathbf{c}^-[(l-1)!]} \leq N_{\mathbf{h}^+_F[l-1]} N_{\mathbf{c}^+[(l-1)!]} \tag{1830}$$

which give

$$\frac{N_{\mathsf{c}^-[(l-1)!]}}{N_{\mathsf{c}^+[(l-1)!]}} \leq N_{\mathsf{h}^+_F[l-1]} \leq \frac{N_{\mathsf{h}^+_F[l]} N_{\mathsf{c}^+[l]}}{N_{\mathsf{c}^-[l]}} \tag{1831}$$

Using the fact that $\hat{V}_B[l]$ and $\hat{V}_B[(l-1)!]$ are natural temporal coisometric operators, we can similarly obtain

$$\frac{N_{\mathsf{c}^-[(l-1)!]}}{N_{\mathsf{c}^+[(l-1)!]}} \geq N_{\mathsf{h}^+_B[l-1]} \geq \frac{N_{\mathsf{h}^+_B[l]} N_{\mathsf{c}^+[l]}}{N_{\mathsf{c}^-[l]}} \tag{1832}$$

(note the direction of the inequalities is reversed). Now, starting with $l = L$ we have $N_{\mathsf{h}^+_F[L]} = N_{\mathsf{h}^+_B[L]} = 1$. Therefore, for $l = L$, the inequalities (1831) and (1832) must be saturated with

$$N_{\mathsf{h}^+_F[L-1]} = N_{\mathsf{h}^+_B[L-1]} := N_{\mathsf{h}^+[L-1]} \tag{1833}$$

We continue this induction with $l = L - 1$ in the inequalities (1831) and (1832), and so on down to $l = 1$. At each step we obtain

$$N_{\mathsf{h}^+_F[l-1]} = N_{\mathsf{h}^+_B[l-1]} := N_{\mathsf{h}^+[l-1]} \tag{1834}$$

which keeps the induction working (since then the rightmost expression in (1831) will be equal to the rightmost expression (1832) in the next induction step). Since the inequalities are saturated, we obtain

$$N_{\mathsf{h}^+[l-1]} = \frac{N_{\mathsf{c}^-[(l-1)!]}}{N_{\mathsf{c}^+[(l-1)!]}} \tag{1835}$$

and

$$N_{\mathsf{h}^+[l-1]} = \frac{N_{\mathsf{h}^+[l]} N_{\mathsf{c}^+[l]}}{N_{\mathsf{c}^-[l]}} \tag{1836}$$

The latter equation is, simply,

$$N_{\mathsf{h}^+[l-1]} N_{\mathsf{c}^-[l]} = N_{\mathsf{h}^+[l]} N_{\mathsf{c}^+[l]} \tag{1837}$$

To obtain the form in (1826) we note that a natural temporal isometric operator, $\hat{V}_F[l]$ satisfying (1837) is, in fact, a natural temporal unitary operator. Thus, the left dilation in (1824) gives us (1826). We can argue in a similar way that the $\hat{V}_B[l]$ are, in fact, natural temporal unitary operators. Thus, the sufficient pair in (1824) both give rise to the form in (1826) (after reinserting the maximal elements to convert back from ${}_\llcorner\hat{\boldsymbol{B}}^\lrcorner$ to $\hat{\boldsymbol{B}}$). Hence the condition in (1826) is sufficient. This proves both necessity and sufficiency. Sufficiency also follows simply by application of the causality composition theorem from Sec. 49.4.11 noting that each of the components of (1826) is physical and deterministic.

There are some constraints on the dimensions of the systems in the above which we will collect into a corollary.

> **Homogeneous $L$-causal ladder dilation corollary.** The dimensions of the systems in the homogeneous dilation for the $L$-causal ladder dilation in (1826) must satisfy the following relations:

1. The relationship

$$N_{\mathbf{a}^-[l]} N_{\mathbf{x}^-[l]} N_{\mathbf{h}^+[l-1]} = N_{\mathbf{a}^+[l]} N_{\mathbf{x}^+[l]} N_{\mathbf{h}^+[l]} \tag{1838}$$

   (for $l = 1$ to $L$) where we put $N_{\mathbf{h}^+[0]} = N_{\mathbf{h}^+[L]} = 1$.

2. The dimension of the ancillary systems $\mathbf{h}^+[l]$ is given by

$$N_{\mathbf{h}^+[l]} = \frac{N_{\mathbf{a}^-[l!]} N_{\mathbf{x}^-[l!]}}{N_{\mathbf{a}^+[l!]} N_{\mathbf{x}^+[l!]}} \tag{1839}$$

   for $l = 1$ to $L$.

3. The following

$$N_{\mathbf{a}^-[L!]} N_{\mathbf{x}^-[L!]} = N_{\mathbf{a}^+[L!]} N_{\mathbf{x}^+[L!]} \tag{1840}$$

   is a necessary condition for the operator $\hat{\boldsymbol{B}}$ to be homogeneous, deterministic, and physical.

Equation (1838) is necessary for $U[l]$ to be unitary. Equation (1839) follows from (1835) and (1840) follows from (1839) and the fact that $N_{\mathbf{h}^+[L]} = 1$.

## 78.19 Causal snakes

Here we will define *causal snakes* which have a zigzag like causal structure (resembling a snake). To introduce this idea it is convenient, for the moment, to make our notation a little more compact. We write

$l$ := $\mathbf{a}^-[l]\mathbf{x}^-[l]$ $\mathbf{a}^+[l]\mathbf{x}^+[l]$ (1841)

An example of a causal snake diagram is the following

2 1 3 5 4 6 (1842)

Of course, we can make this snake as long as we want with multiple zigs and zags of varied lengths. In fact, it is useful to introduce a bit more new notation. We will write

$+$ := $-$ := (1843)

We can straighten out the causal snake above writing it as

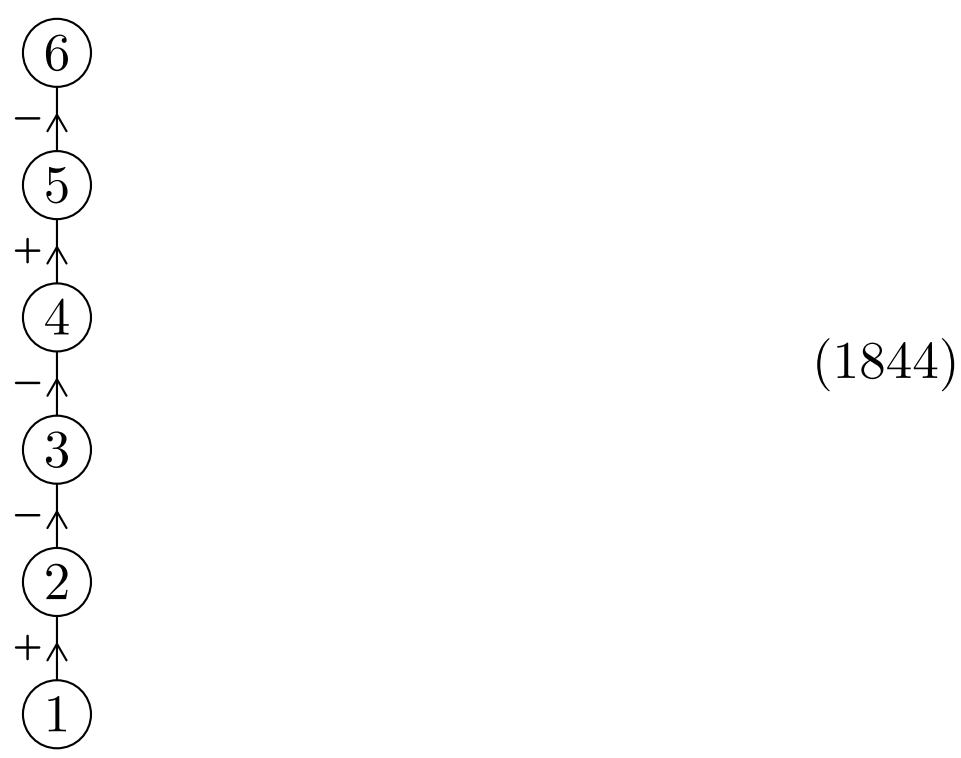

(1844)

A general causal snake can be written as

$L$ $s_{L-1}$ $s_2$ 2 $s_1$ 1 (1845)

where $s_l$ (for $l = 1$ to $L - 1$) takes values $+$ or $-$ indicating the direction of the causal arrow. Any sequence of $+$'s and $-$'s will give rise to a causal snake. Subsequences consisting all of $+$'s, or all of $-$'s will give rise to segments of corresponding lengths.

Reinserting the full notation from the right hand side of (1841), we see we can write a causal snake as

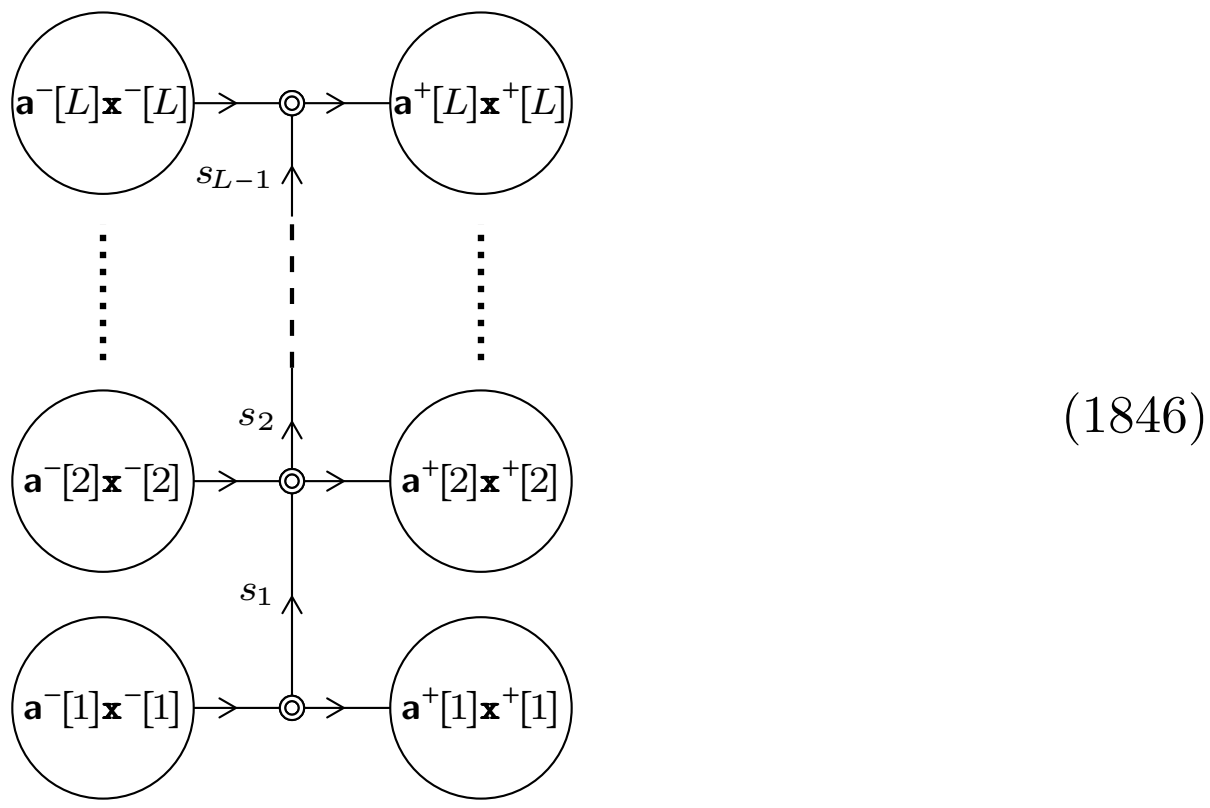

(1846)

Thus, a causal snake is a natural generalisation of a causal ladder as in (1767). We will see how to adapt the proof of the $L$-causal ladder dilation theorem with (co)isometries (from Sec. 78.16) to prove a similar theorem for causal snakes (see Sec. 78.20 below). By running along the snake from top to bottom or bottom to top we obtain a necessary pair of dilations. However, we are not able to prove sufficiency for this pair because some the natural (co)isometric operators in the dilations associated with the turning points in the snake are not temporal (so do not satisfy forward or backward causality which, in turn, blocks use of the causality composition theorem).

There is a different method for getting necessary dilations for causal snakes that have only one zig or zag in them - i.e. those that are V-shaped or upside down V-shaped. These are, in fact, examples of the many ancestors and many descendants causal diagrams which we will discuss in Sec. 78.21. The method for analysing these cases uses the fact that operators factorize when their causal diagrams are disjoint as discussed in Sec. 48.9. For such causal diagrams, some of the residua will have this disjointness property.

## 78.20 Dilation theorem for causal snakes

Here we will provide some necessary (but not sufficient) dilations for operators subject to causal structure given by a causal snake. We do this by adapting the technique from Sec. 78.16 where the dilation theorem for $L$-causal ladder with (co)isometries was proved. We could, using the same basic techniques, prove a dilation with unitaries along the same lines as in Sec. 78.14. Here is the theorem we will prove.

> **Dilation theorem for causal snakes.** A deterministic physical operator tensor

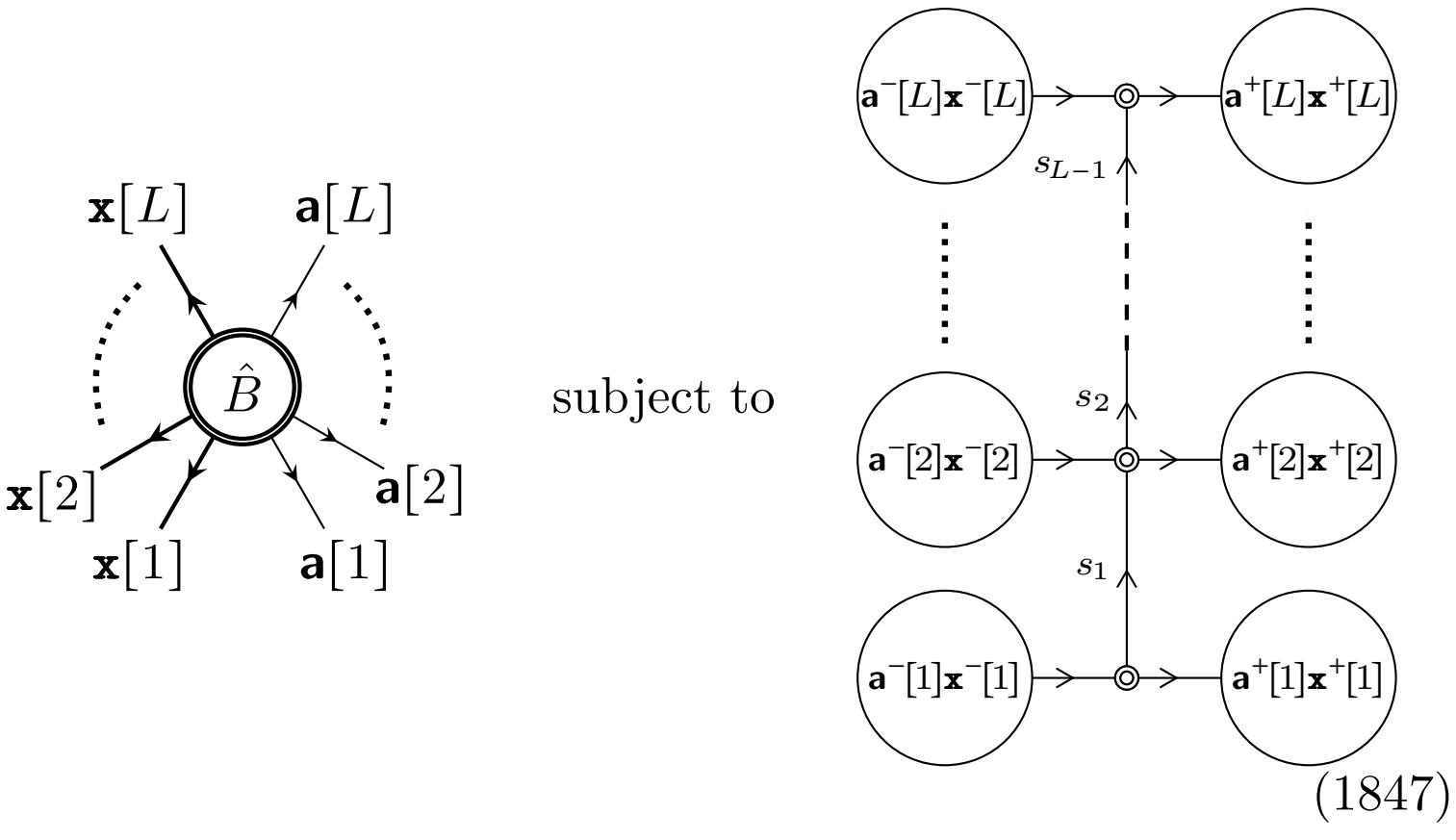

can be written as

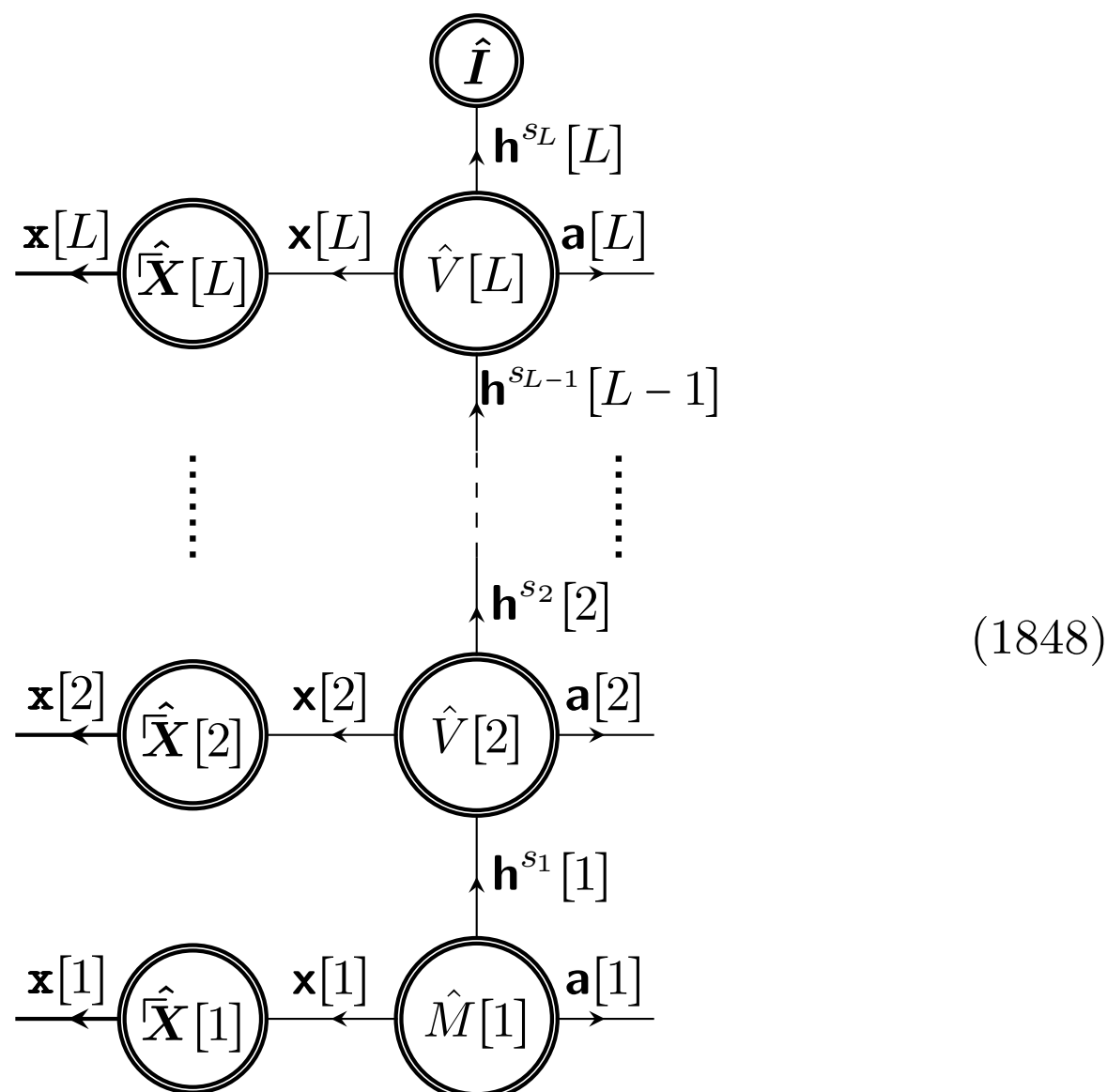

(1848)

Here $\hat{M}[1]$ is a natural temporal maxometry having ancilla $\mathsf{h}^{s_1}[1]$, and $\hat{V}[l]$ (for $l = 2$ to $L$) are natural isometries with respect to the bipartition

$$(\mathsf{a}^{\bar{s}_{l-1}}[l]\mathsf{x}^{\bar{s}_{l-1}}[l]\mathsf{h}^{s_{l-1}}, \mathsf{a}^{s_{l-1}}[l]\mathsf{x}^{s_{l-1}}[l]\mathsf{h}^{s_l}[l]) \tag{1849}$$

where $\bar{s}_l$ is defined to have the opposite sign to $s_l$. We are free to choose $s_L = +$ or $s_L = -$. Furthermore, we can choose $N_{\mathsf{h}_F^{s_L}[L]} = \text{rank}(\hat{\boldsymbol{B}})$.

Note that the natural isometries, $\hat{V}[l]$, are temporal only if $s_{l-1} = s_l$. Also note that we can have the maxometry at position $L$ by applying the theorem "backwards" (it does not matter which end of the snake we label with a 1 and which we label with an $L$). Finally, note that we are not able to prove that the dilation in (1848) is sufficient for physicality. The proof of this theorem is an adaptation of the proof of the (co)isometric $L$-causal ladder theorem from Sec. 78.16 (which was, itself, presented as an adaptation of the proof of the unitary $L$-causal ladder dilation theorem from Sec. 78.14). The main idea driving this proof is that we can keep progressing along the snake from one end to the other using the basic (co)isometric causal dilation theorem where, sometimes we use induction based on forward causality and sometimes we use induction based on backward causality. We can do this because we have a time symmetric formulation of QT. At the turning points in the snake we need to apply an (unphysical) natural unitary operator that reverses the time direction. We will work with the maximal representation, $\llcorner\hat{\boldsymbol{B}}\urcorner$, as in previous proofs. We can convert back to $\hat{\boldsymbol{B}}$ using the maximal representation theorem as before. We start by defining

$$
\begin{array}{c} \mathbf{h}^{s_l}[l] \\ \uparrow \\ \hat{S}[l] \xrightarrow{\mathbf{c}[l!]} \end{array}
\quad := \quad
\begin{array}{c} \mathbf{h}^{s_l}[l] \\ \uparrow \\ \hat{V}[l] \xrightarrow{\mathbf{c}[l]} \\ \uparrow \mathbf{h}^{s_{l-1}}[l-1] \\ \vdots \\ \uparrow \mathbf{h}^{s_2}[2] \\ \hat{V}[2] \xrightarrow{\mathbf{c}[2]} \\ \uparrow \mathbf{h}^{s_1}[1] \\ \hat{M}[1] \xrightarrow{\mathbf{c}[1]} \end{array}
\tag{1850}
$$

where $\hat{M}[1]$ is a natural temporal maxometry with respect to ancilla $h^{s_1}[1]$ Then our task is equivalent to proving that

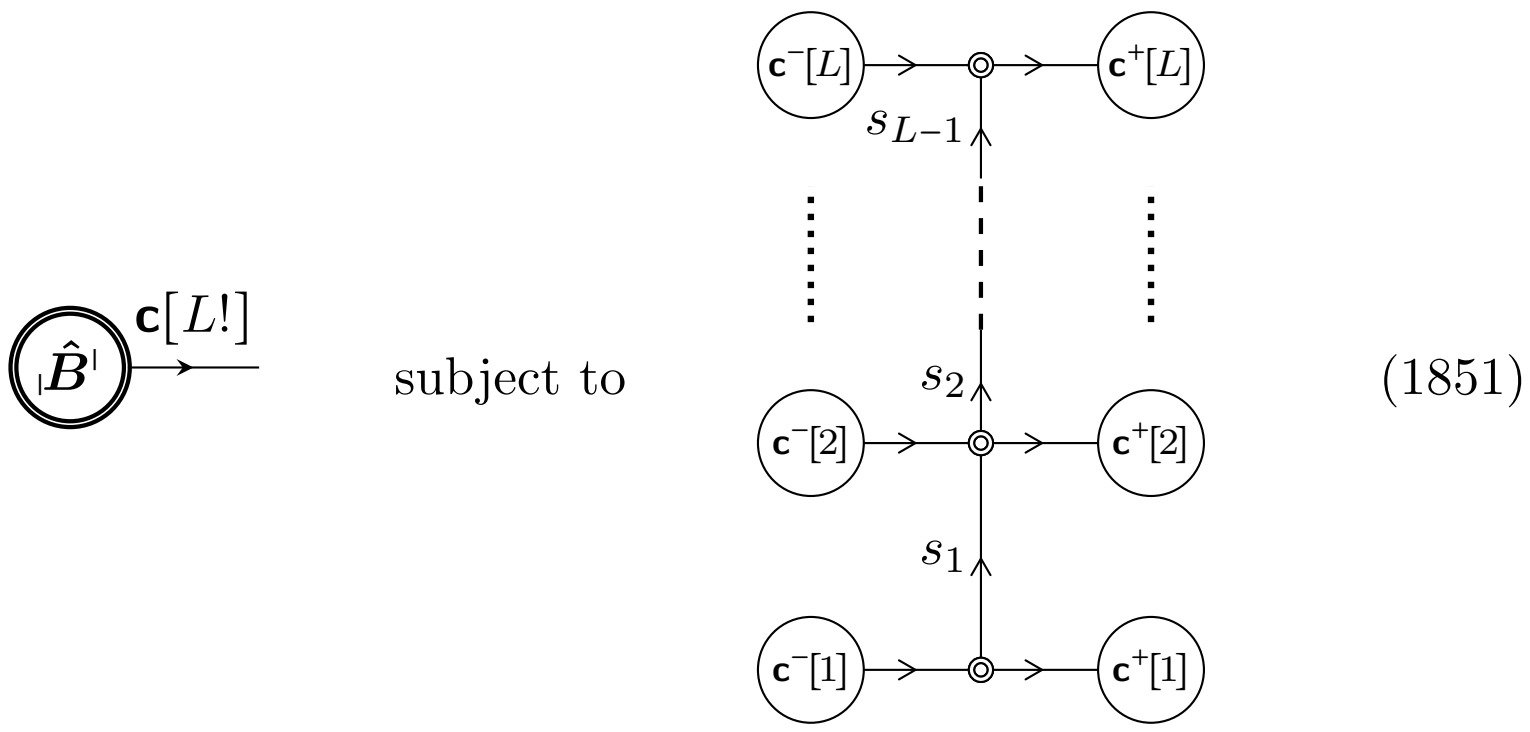

(1851)

can necessarily be written as

$$
\hat{\boldsymbol{B}} \xrightarrow{\mathbf{c}[L!]} \quad = \quad
\begin{array}{c} \hat{\boldsymbol{I}} \\ \uparrow \mathbf{h}^{s_L}[L] \\ \hat{S}_F[L] \xrightarrow{\mathbf{c}[L!]} \end{array}
\tag{1852}
$$

We define

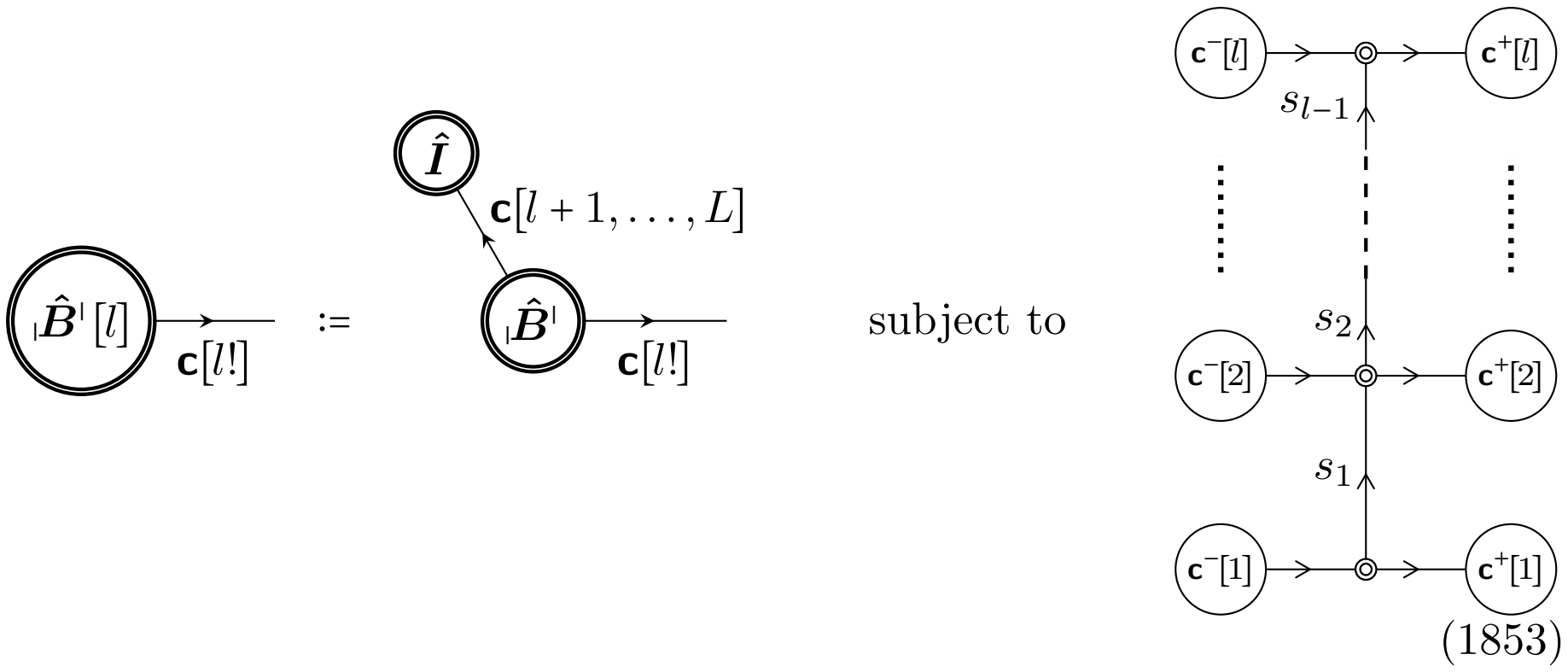

(1853)

It is useful below to substitute + for "forward" or "future" and − for "backward" or "past". Then, for example, $s_l$-causality means forward causality if $s_l = +$ and $s_l$-residuum means past residuum if $s_l = -$. Note that a synchronous bipartition is formed by intersecting any one of the vertical wires in the above causal diagram (this is, perhaps, clearer when we write the causal diagram in zig-zag form as ilustrated in the example (1842)). We label the $l$th such synchronous bipartition as $p[l]$. It follows by application of the first theorem in Sec. 49.4.7 that ${}_{|}\hat{\boldsymbol{B}}^{|}[l]$ above is the $\bar{s}_l$-residuum of ${}_{|}\hat{\boldsymbol{B}}^{|}$. Further, it follows by application of the second theorem in the same section that the causal diagram associated with this residuum is the one shown in (1853). We can now state our *induction hypothesis. If the following are true*

(a) we have

$$ {}_{|}\hat{\boldsymbol{B}}^{|}[l] \; \mathbf{c}[l!] \quad = \quad \hat{\boldsymbol{I}} \; \mathbf{h}^{s_l}[l] \; \hat{S}[l] \; \mathbf{c}[l!] \tag{1854} $$

where

$$ N_{\mathbf{h}^{s_l}[l]} = \mathrm{rank}({}_{|}\hat{\boldsymbol{B}}^{|}[l]) \tag{1855} $$

(b) ${}_{|}\hat{\boldsymbol{B}}^{|}[l+1]$ satisfies $\bar{s}_l$-causality at $p[l]$, i.e.

$$ \mathbf{c}^{\bar{s}_l}[l+1] \; \hat{\boldsymbol{I}}_{s_l} \; \mathbf{c}[l+1] \; {}_{|}\hat{\boldsymbol{B}}^{|}[l+1] \; \mathbf{c}[l!] \quad = \quad \mathbf{c}^{\bar{s}_l}[l+1] \; \hat{\boldsymbol{I}} \; {}_{|}\hat{\boldsymbol{B}}^{|}[l] \; \mathbf{c}[l!] \tag{1856} $$

(where ${}_{|}\hat{\boldsymbol{B}}^{|}[l]$ is necessarily the $\bar{s}_l$-residuum here as explained above),

then the following is true

(A) we have that

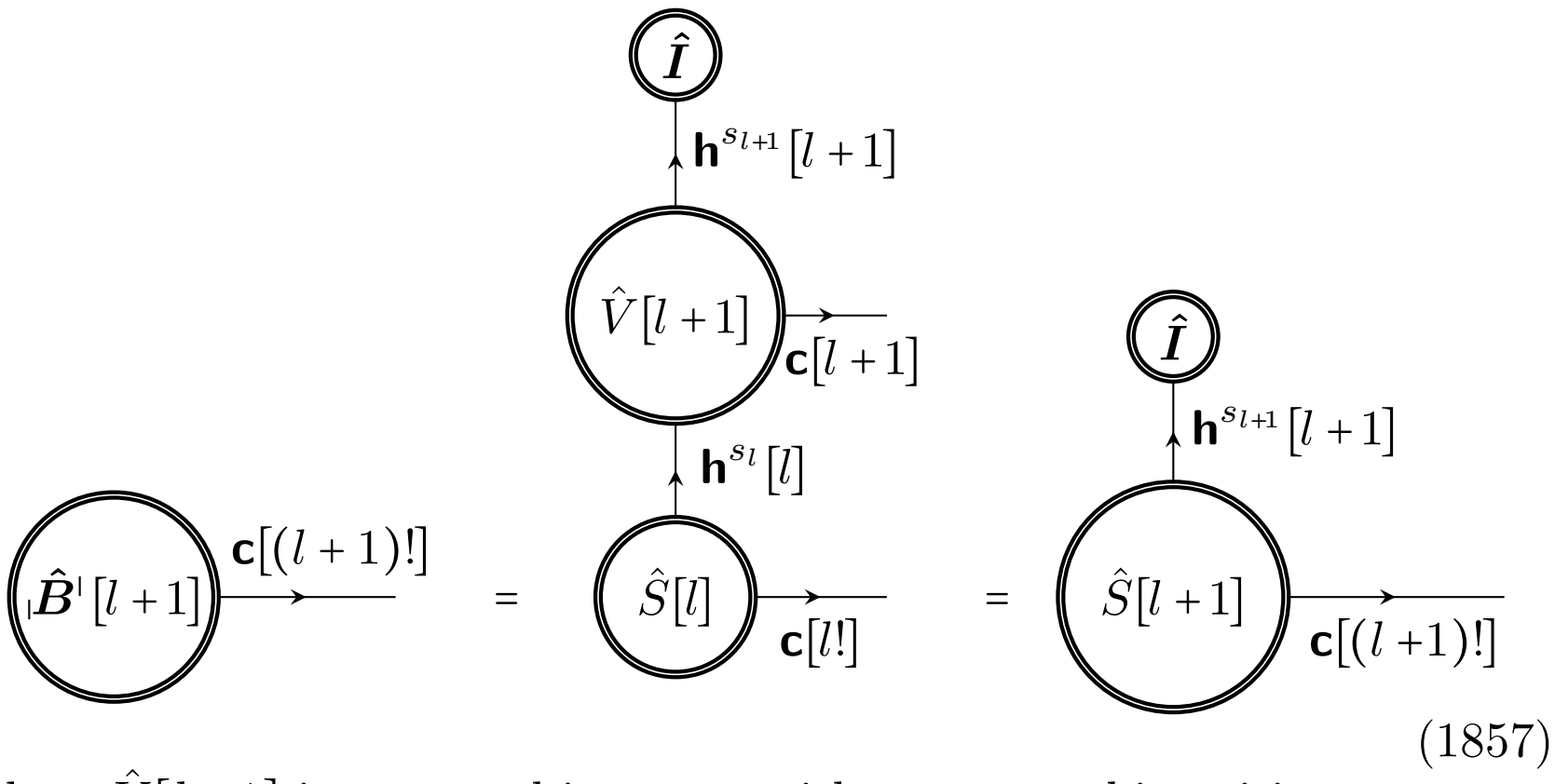

(1857)

where $\hat{V}[l+1]$ is a natural isometry with respect to bipartition

$$(\mathbf{c}^{\bar{s}_l}[l+1]\mathbf{h}^{s_l}[l], \mathbf{c}^{s_l}[l+1]\mathbf{h}^{s_{l+1}}[l+1]) \tag{1858}$$

and ancillae $\mathbf{h}^{s_l}[l]$ and $\mathbf{h}^{s_{l+1}}[l+1]$ and

$$N_{\mathbf{h}^{s_{l+1}}[l+1]} = \text{rank}({}_{|}\hat{\boldsymbol{B}}^{|}[l+1]) \tag{1859}$$

Note the right most expression in (1857) simply follows by the definition in (1850).

The base case for the induction is

$$\hat{\boldsymbol{I}} \quad \mathbf{h}^{s_1}[1] \qquad {}_{|}\hat{\boldsymbol{B}}^{|}[1] \;\; \mathbf{c}[1] \quad = \quad \hat{M}[1] \;\; \mathbf{c}[1] \tag{1860}$$

This follows because ${}_{|}\hat{\boldsymbol{B}}^{|}[1]$ has simple causal structure so we can use the simple dilation discussed in Sec. 78.3 where we choose the ancilla to be all output (the $s_1 = +$ case) or all input (the $s_1 = -$ case) as previously illustrated in (1732). Further, we can choose

$$N_{\mathbf{h}^{s_1}[1]} = \text{rank}({}_{|}\hat{\boldsymbol{B}}^{|}[1]) \tag{1861}$$

(by the simple dilation theorem). If the above induction hypothesis is true, then starting with the base case with $l = 1$ and proceeding by induction, we easily obtain the dilation in (1852) above along with the fact that we can have $N_{\mathbf{h}^{s_L}[L]} = \text{rank}({}_{|}\hat{\boldsymbol{B}}^{|})[L] = \text{rank}({}_{|}\hat{\boldsymbol{B}}^{|})$ which gives us the requirement at the end of the theorem above. Thus, to complete the proof of this theorem we need only to

prove the induction hypothesis. It follows from the $s_l$-causality condition (given in (1856) above) by the basic dilation theorem with (co)isometries that

$$\hat{\boldsymbol{B}}^{|}[l+1] \xrightarrow{\mathbf{c}[(l+1)!]} \quad = \quad \begin{array}{c} \hat{\boldsymbol{I}} \\ \uparrow \mathbf{h}^{s_l}[l+1] \\ \hat{V}'[l+1] \xrightarrow{\mathbf{c}[l+1]} \\ \uparrow \mathbf{h}^{s_l}[l] \\ \hat{M}_F[l] \xrightarrow{\mathbf{c}[l!]} \end{array} \tag{1862}$$

follows from the basic (co)isometric causal dilation theorem in Sec. 78.6(compare this with (1811) in the proof of the $L$-causal ladder (co)isometric dilation theorem). Here $\hat{V}'[l+1]$ is a natural temporal isometry (as proven in the aforementioned theorem). However, note that the output ancilla is $\mathbf{h}^{s_l}[l+1]$ but we want it to be $\mathbf{h}^{s_{l+1}}[l+1]$ to prove the theorem. To rectify we consider two cases (i) where $s_{l+1} = s_l$ and (ii) where $s_{l+1} = \bar{s}_l$ separately. Consider case (i). We set

$$\mathbf{h}^{s_{l+1}}[l] = \mathbf{h}^{s_l}[l] \quad \text{ if } \; s_{l+1} = s_l \tag{1863}$$

and then we can define

$$\begin{array}{c} \uparrow \mathbf{h}^{s_{l+1}}[l+1] \\ \hat{V}[l+1] \xrightarrow{\mathbf{c}[l+1]} \\ \uparrow \mathbf{h}^{s_l}[l] \end{array} \quad := \quad \begin{array}{c} \uparrow \mathbf{h}^{s_l}[l+1] \\ \hat{V}'[l+1] \xrightarrow{\mathbf{c}[l+1]} \\ \uparrow \mathbf{h}^{s_l}[l] \end{array} \qquad \text{if } \; s_{l+1} = s_l \tag{1864}$$

In this case $V[l+1]$ is a natural temporal isometry. If we substitute (1864) into (1862) then we obtain (1857) as required where $V[l+1]$ is a natural isometry with respect to the bipartition in (1858). Now consider case (ii). We put

$$\mathbf{h}^{s_{l+1}}[l] = (\mathbf{h}^{s_l}[l])^R \quad \text{ if } \; s_{l+1} = \bar{s}_l \tag{1865}$$

and then we can define

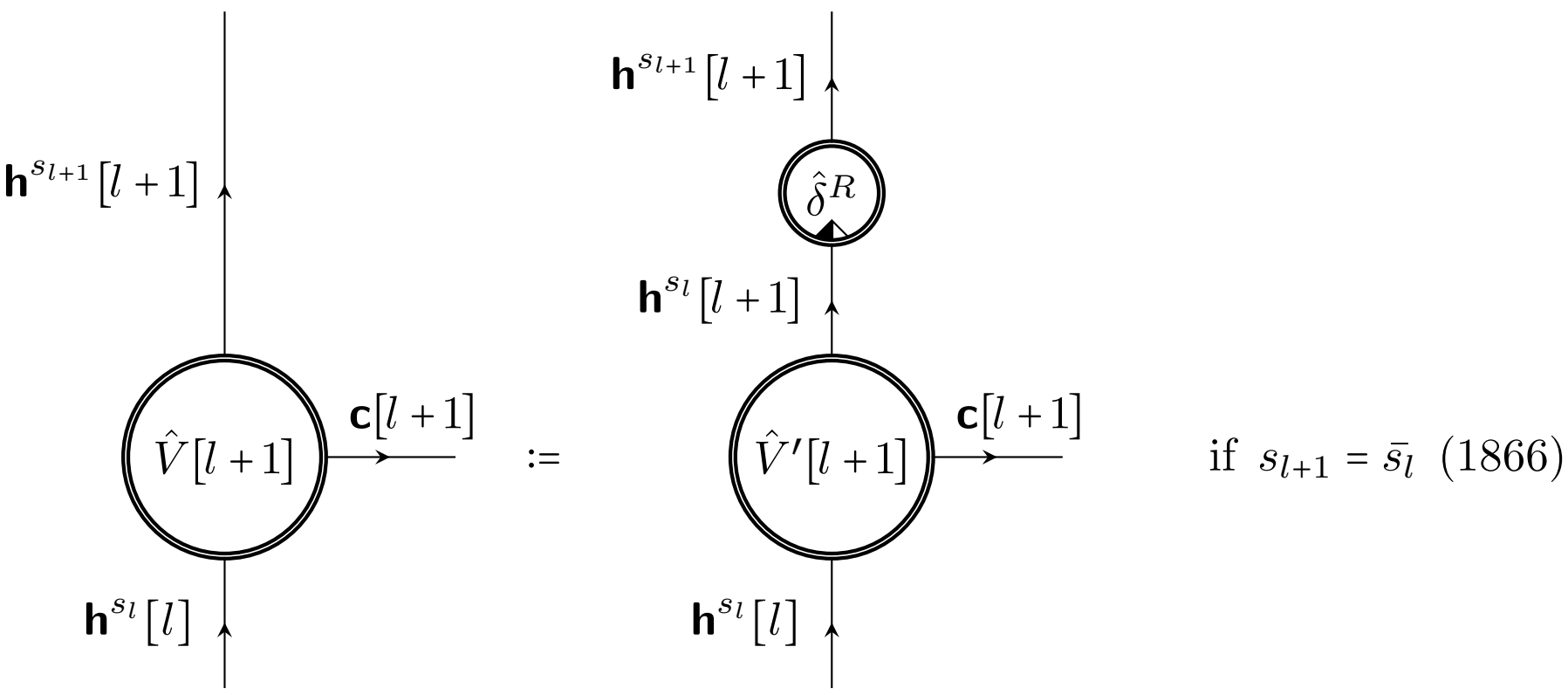

where $\delta^R$ is a natural reversing unitary (as defined in Sec. 73.10). Here $\hat{V}[l+1]$ is a natural isometry with respect to the bipartition in (1858). It is not temporal because we have employed the natural reversing unitary. Employing (1661) we obtain

$\hat{\boldsymbol{I}}$

$\mathbf{h}^{s_l}[l+1]$

$\hat{\delta}^R$

$\mathbf{h}^{s_{l+1}}[l+1]$

$\hat{\boldsymbol{B}}[l+1]$ $\mathbf{c}[(l+1)!]$ = $\hat{V}[l+1]$ $\mathbf{c}[l+1]$ (1867)

$\mathbf{h}^{s_l}[l]$

$\hat{M}_F[l]$ $\mathbf{c}[l!]$

from (1862) and (1866). Finally, we can absorb the $\delta^R$ into the $\hat{\boldsymbol{I}}$ using (1662). This gives (1857) for case (ii). Thus, we have proven the induction hypothesis for both cases which completes the proof of this theorem. Now let us prove that we obtain (1859) if we have (1855). This follows from the application of the basic (co)isometric causal dilation theorem we used to obtain (1862): Equation (1855) follows from point (v) of that theorem and equation (1859) follows from point (ii). This completes the proof of the causal snake dilation theorem.

We proved this theorem by analogue with the $L$-causal ladder theorem with (co)isometries. We could equally have proven a theorem by analogue with the $L$-causal ladder theorem with unitaries from Sec. 78.14 obtaining a necessary (but not sufficient) dilation in terms of a natural temporal maxometry at the begin-ning of the snake followed by natural unitary operators as we go along the snake. This would entail solving the analogue system of equations to (1805,1806).

## 78.21 Many ancestors/descendants causal dilation theorems

In Sec. 78.5 we proved the basic unitary causal dilation theorem, and in Sec. 78.6 we proved the basic isometric causal dilation theorem. These theorems (which were key to proving the subsequent dilation theorems) can be regarded as special cases of the many ancestors/descendents causal dilation theorems we will now discuss.

An operation subject to a *many ancestors* causal diagram looks like this

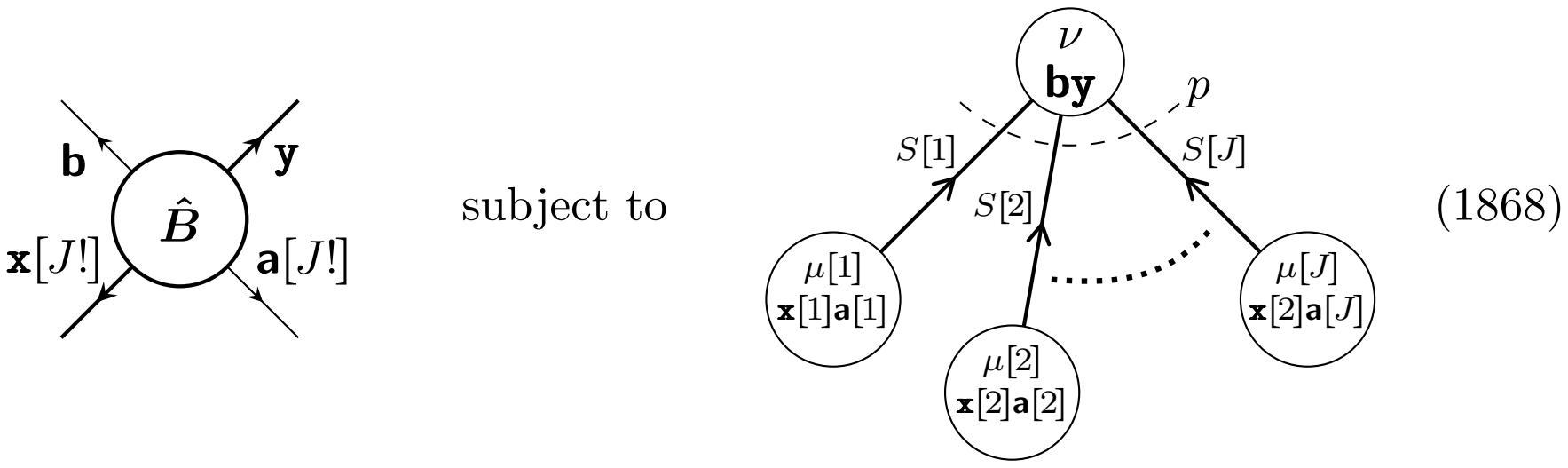

An operation subject to the *many descendants* causal diagram looks like this

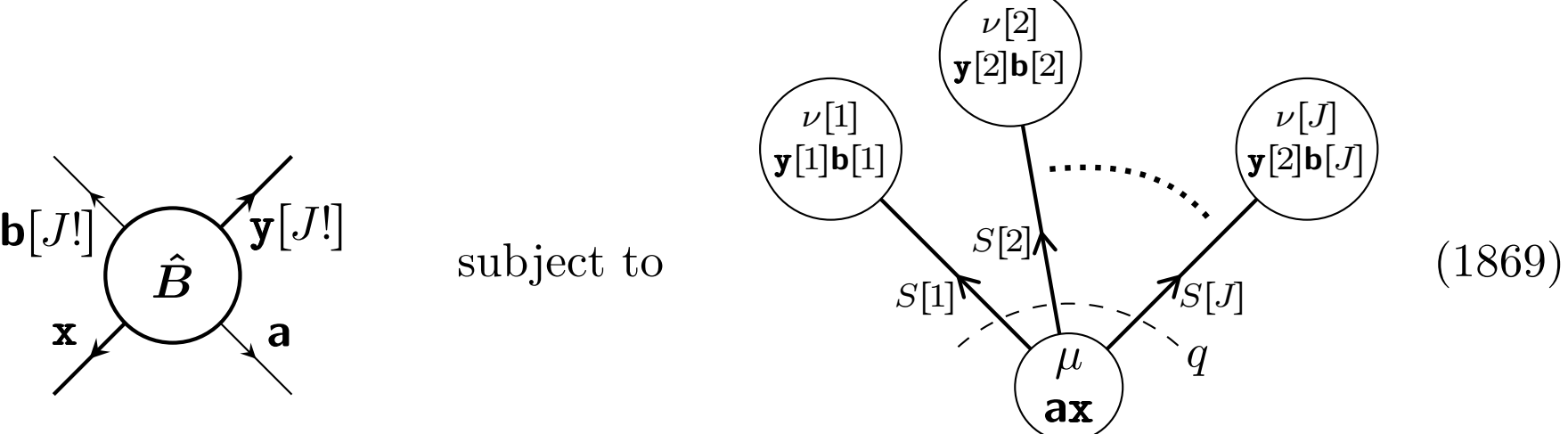

Clearly the causal diagram in (1736) is a special case of these causal diagrams.

We will prove some dilation theorems concerning these which are relatively simple applications of the ideas used to prove the basic causal dilation theorems where, crucially, we use the no correlation without causation assumption and the subsequent theorem about operations factorising for disjoint causal diagrams when we assume weak tomographic locality (from Sec. 48.9). Consider the many ancestors diagram in (1868). If we close $\mathbf{b}^+$ and $\mathbf{y}^+$ (by appending ignore operators to them) the resulting causal diagram is

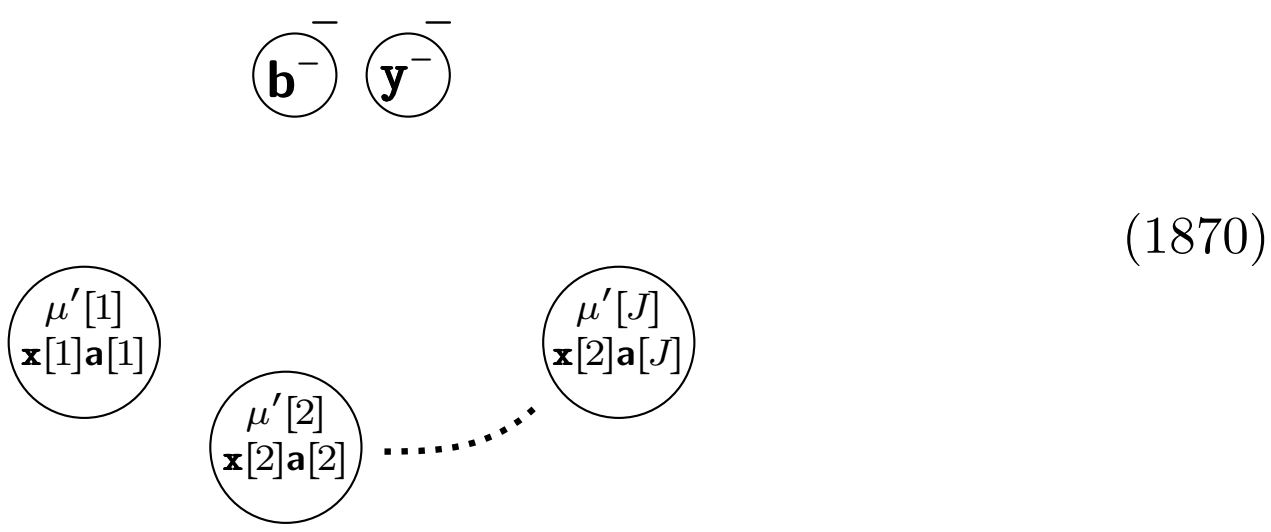

(1870)

The reason for this is that $\mathbf{b}^-$ is an input and $\mathbf{y}^-$ is an income and so these can only have causal arrows pointing out of them. However, since we have closed $\mathbf{b}^+$ and $\mathbf{y}^+$, there is nowhere for such causal arrows to go. This means that, when we apply forward causality at $p$, the past residuum factorises (now employing the theorem in Sec. 48.9 and the fact that Quantum Theory satisfies the weak tomographic locality assumption). Similar properties hold for the many descendants case.

Before stating the theorems we introduce the following *dot notation* for compactness

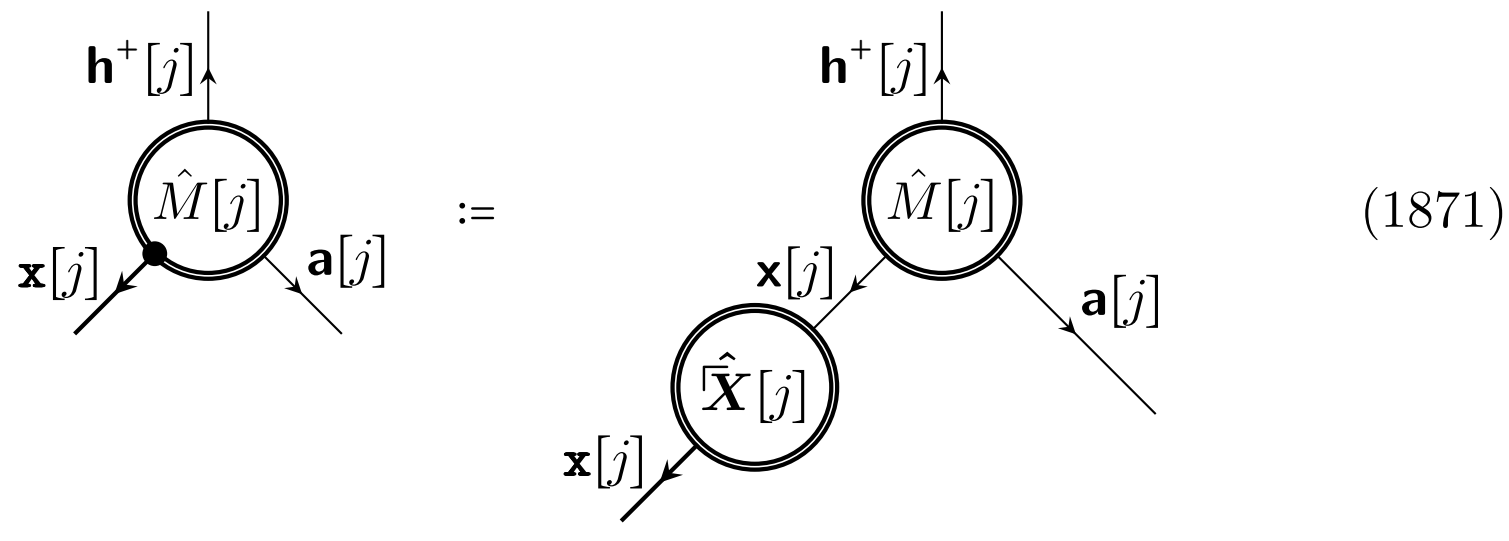

(1871)

and

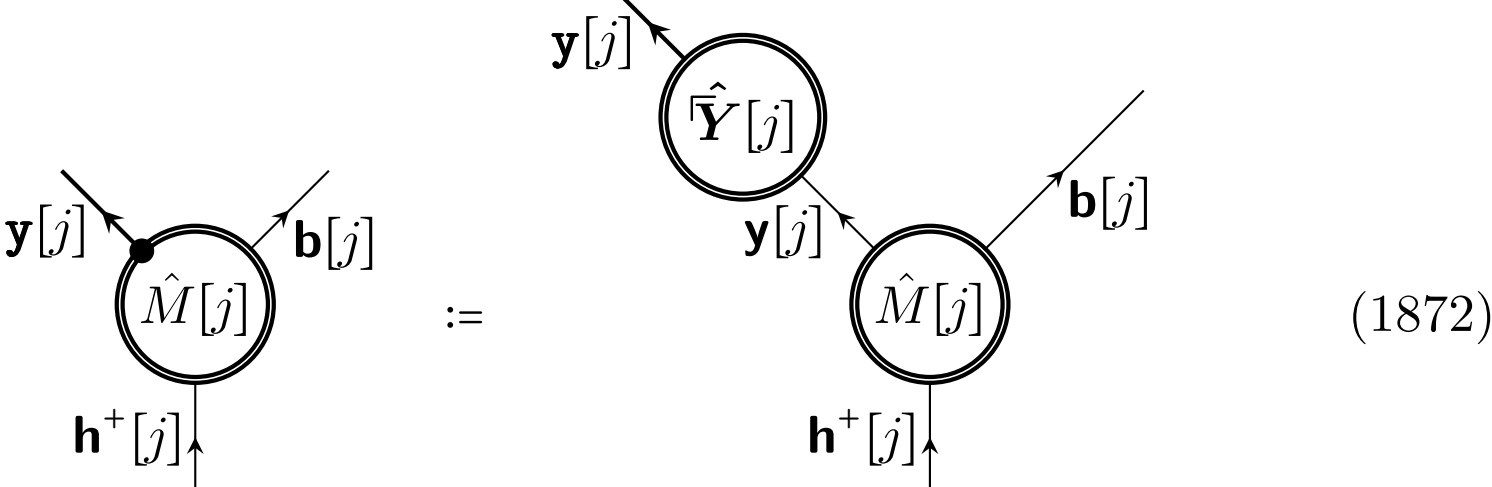

(1872)

which we will use in the statement of the theorems below.

Let us start with the following dilation theorem.

**Many ancestors unitary causal dilation theorem**. Consider a twofold positive operator

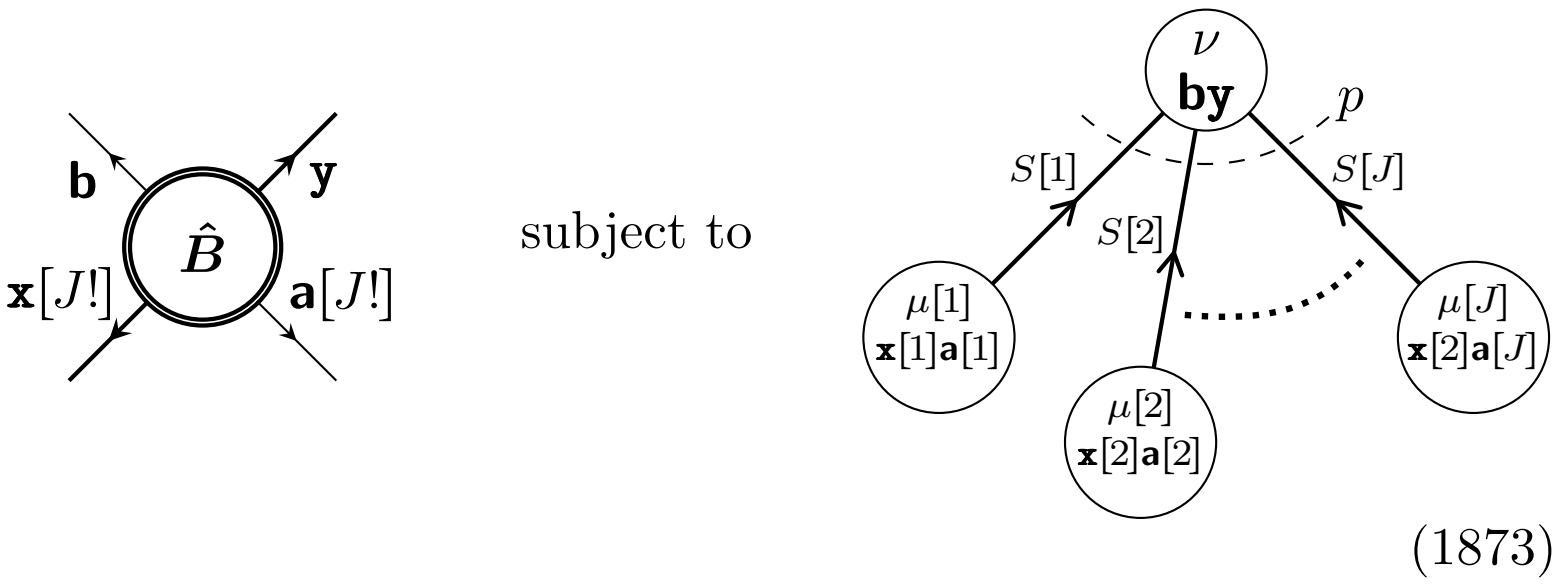

(1873)

If we assume the no correlation without causation assumption is true then the forward causality condition at $p$ is equivalent to the condition that we can write

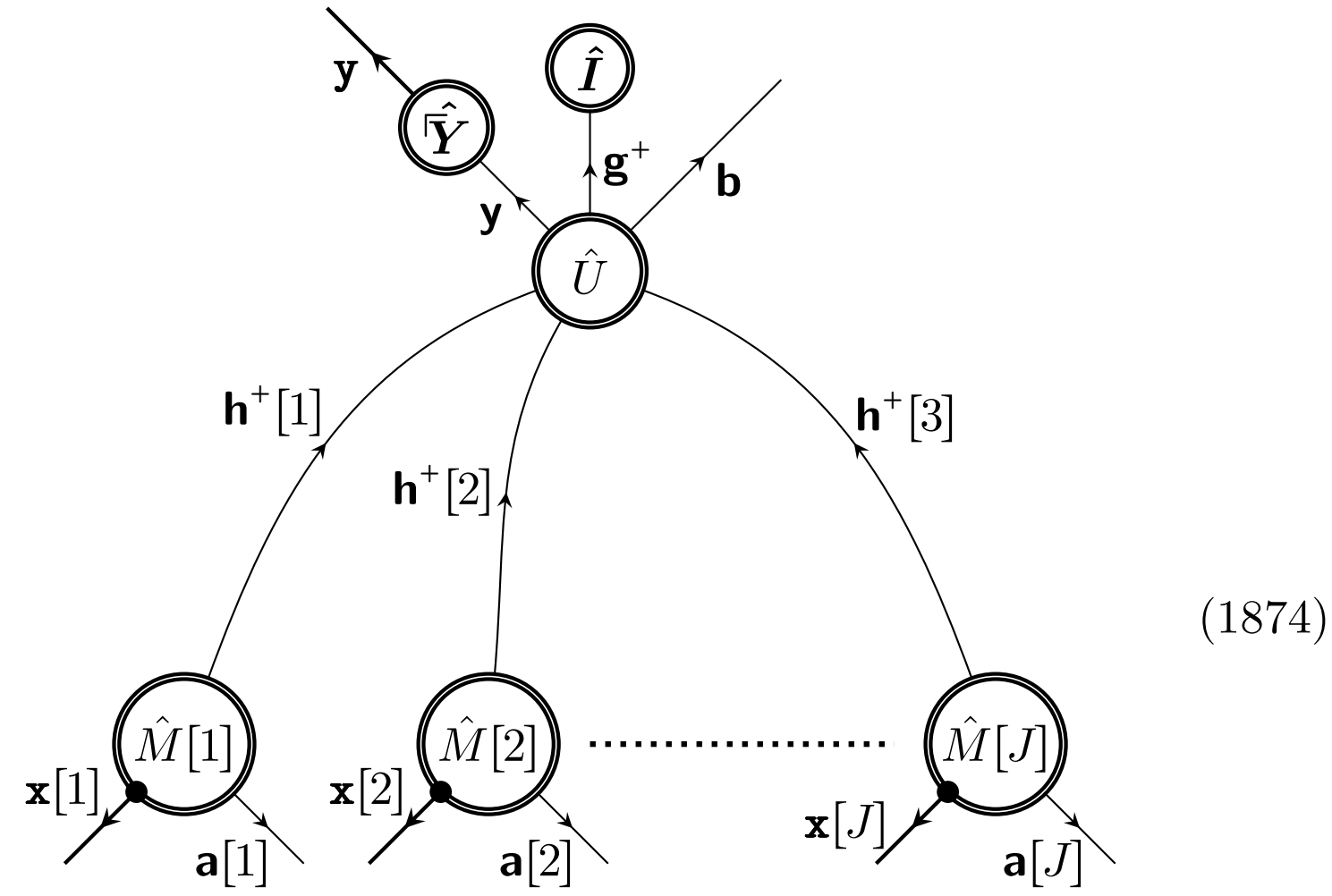

(1874)

where $\hat{U}$ is a natural temporal unitary and $\hat{M}[j]$ are natural temporal maxometries for $j = 1$ to $J$.

This theorem follows from the basic unitary causal dilation theorem along with corollary to that theorem (from Sec. 78.5). Since the past residuum factorises (for the reasons given above) the maxometry in the basic unitary causal dilation theorem given in (1755) also factorises. Thus, we obtain the form (1874) above.

Similarly, we have the following theorem.

**Many descendants unitary causal dilation theorem.** Consider a twofold positive operator

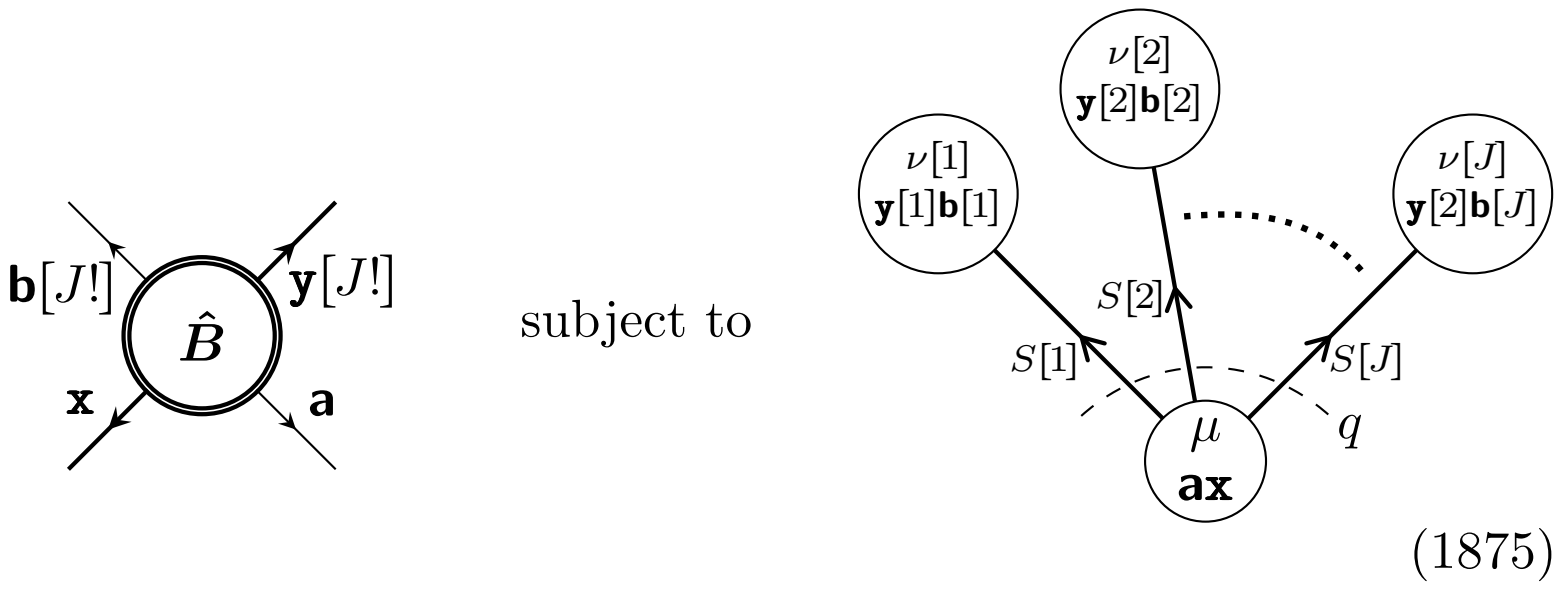

(1875)

If we assume the no correlation without causation assumption is true then the backward causality condition at $q$ is equivalent to the condition that we can write

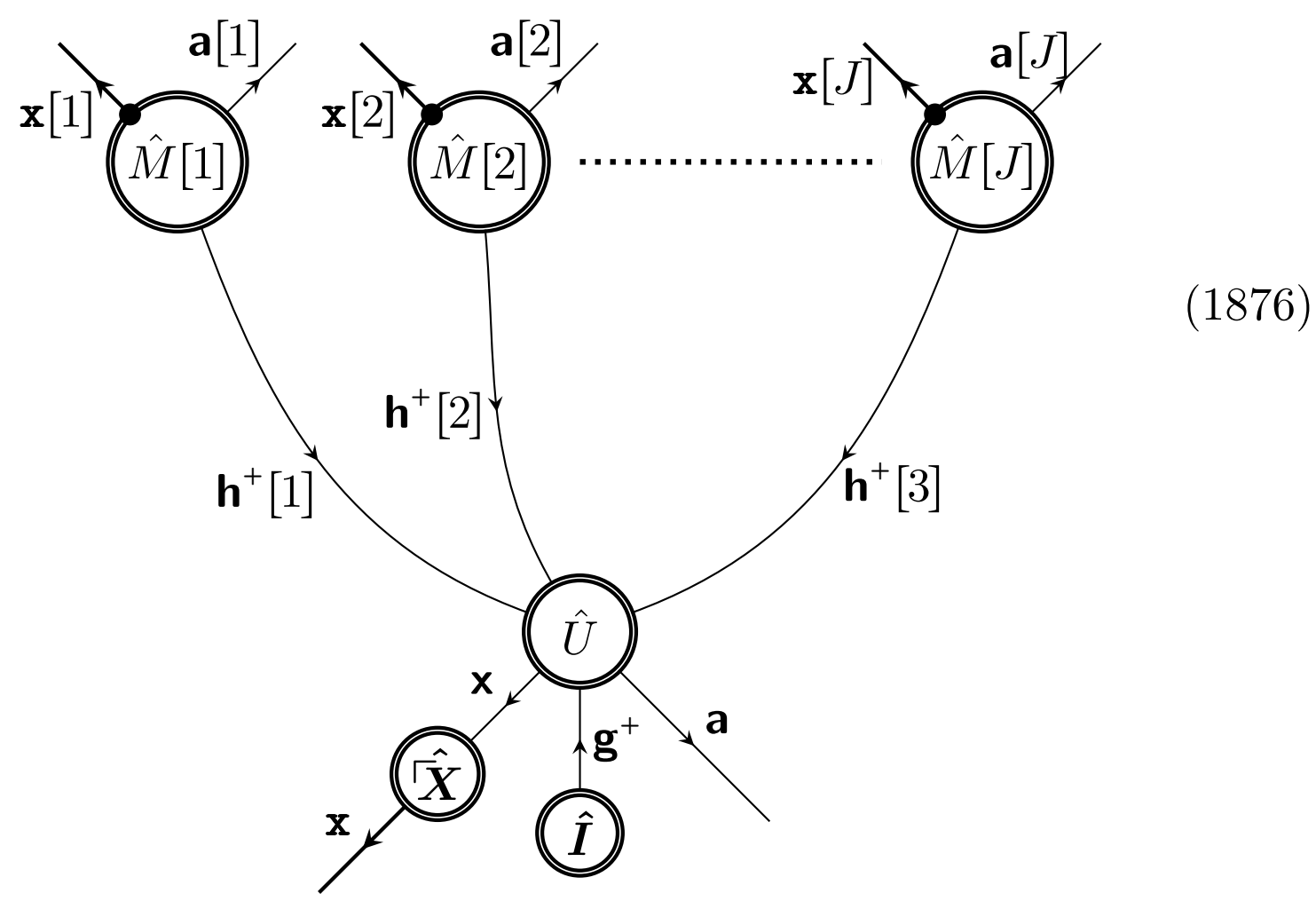

(1876)

where $\hat{U}$ is a natural temporal unitary and $\hat{M}[j]$ are natural temporal maxometries for $j = 1$ to $J$.

This is proven in a similar way to the previous theorem - though using backward causality. This theorem would be difficult to prove in the time forward temporal frame of reference and so can be regarded as an application of the time symmetric theory.

We can also state a *many ancestors isometry causal dilation theorem* where $\hat{U}$ in (1874) is replaced with a natural temporal isometry, $\hat{V}$. The theorem statement is as follows

**Many ancestors isometric causal dilation theorem**. Consider a twofold positive operator

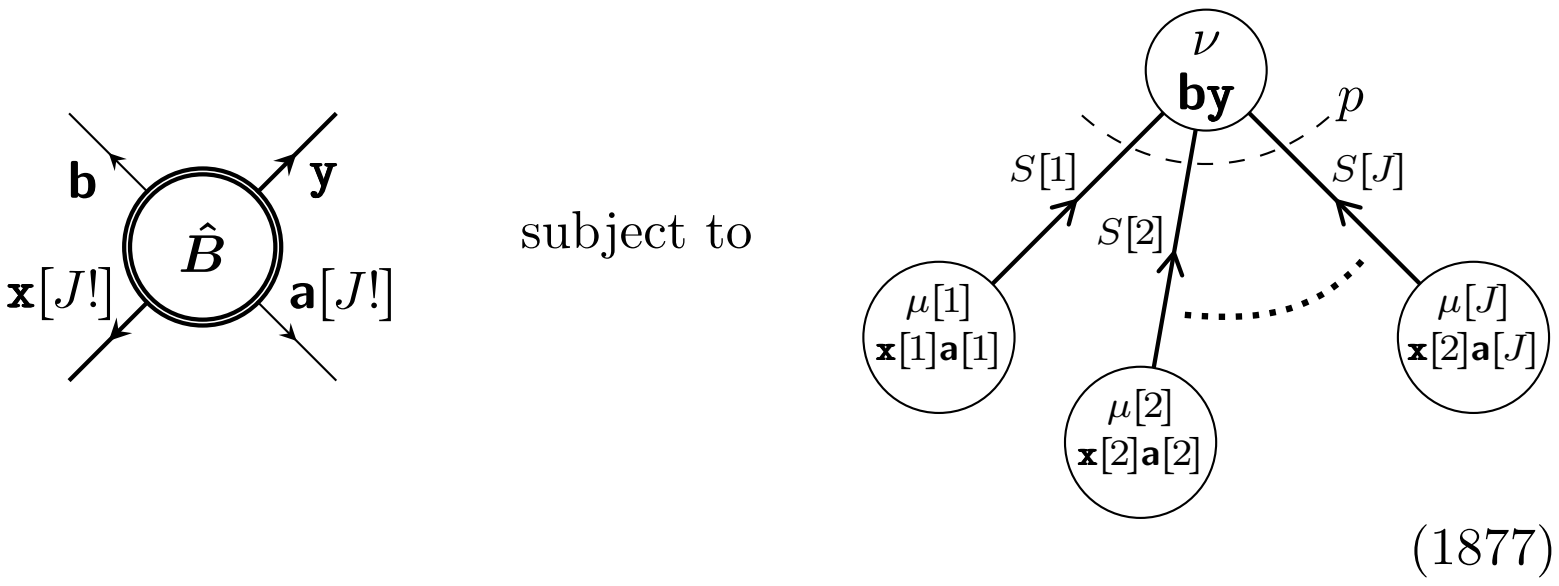

(1877)

If we assume the no correlation without causation assumption is true then the forward causality condition at $p$ is equivalent to the condition that we can write

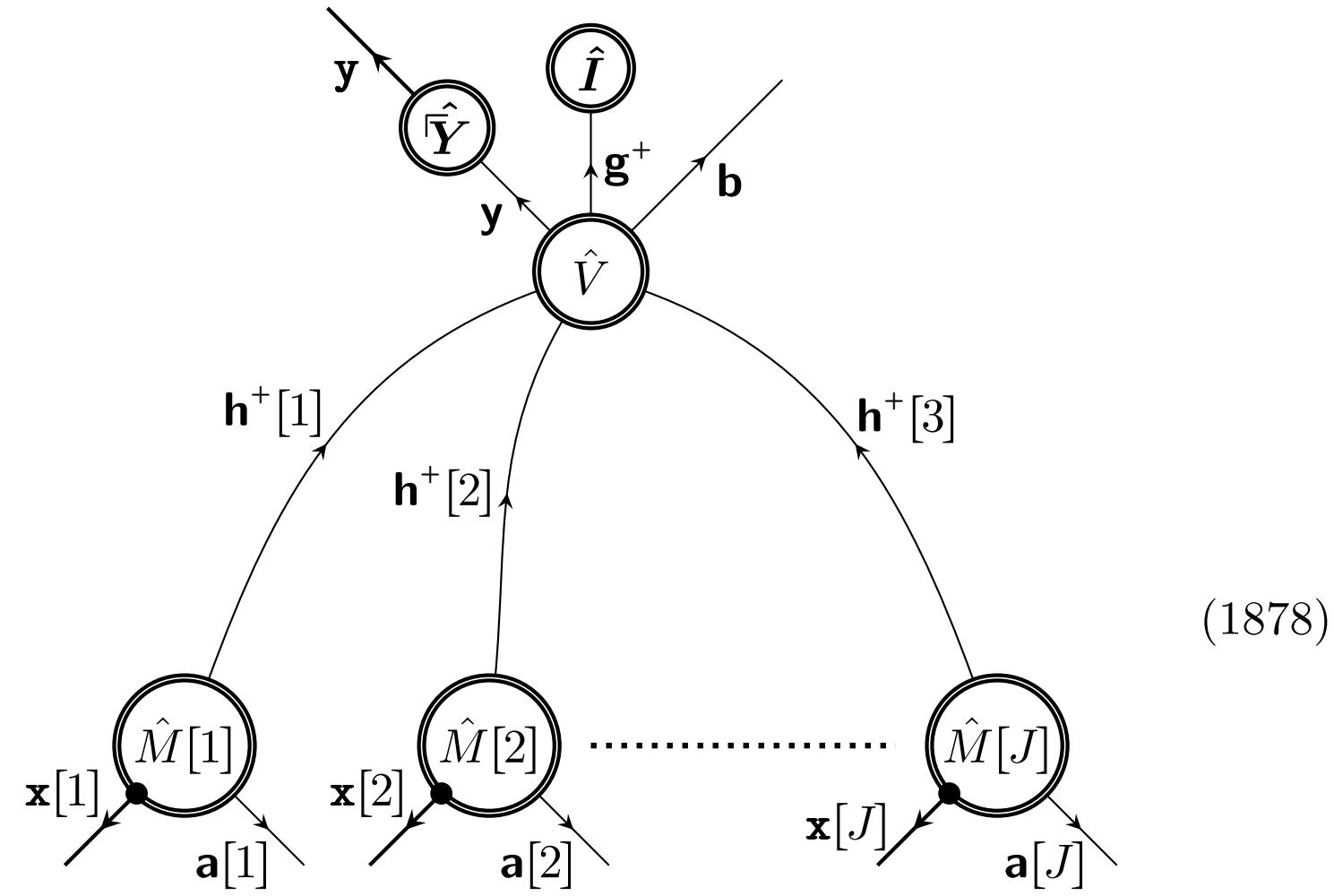

(1878)

where $\hat{V}$ is a natural temporal unitary and $\hat{M}[j]$ are natural temporal maxometries for $j = 1$ to $J$. Furthermore, we can chose $\mathbf{g}^+$ such that $N_{\mathbf{g}^+} = \text{rank}({}_\shortmid\hat{\boldsymbol{B}}^\shortmid)$ and $\mathbf{h}^+[j]$ such that $N_{\mathbf{h}^+[j]} = \text{rank}({}_\shortmid\hat{\boldsymbol{B}}^\shortmid[p_j^-])$ where ${}_\shortmid\hat{\boldsymbol{B}}^\shortmid[p_j^-]$ is the past residuum at $p_j$ (this being the synchronous partition passing only through the wires labelled with the set $S[j]$).

To prove this theorem we use the isometric part of the basic (co)isometric causal dilation theorem from Sec. 78.6 in a similar way to the way we proved the many ancestors unitary causal dilation theorem above. The comments concerning rank follow in a straightforward way.

Finally, we can also state a *many descendants coisometry causal dilation theorem* where $\hat{U}$ in (1876) is replaced with a natural temporal coisometry, $\hat{V}$

**Many descendants coisometric causal dilation theorem.** Con-sider a twofold positive operator

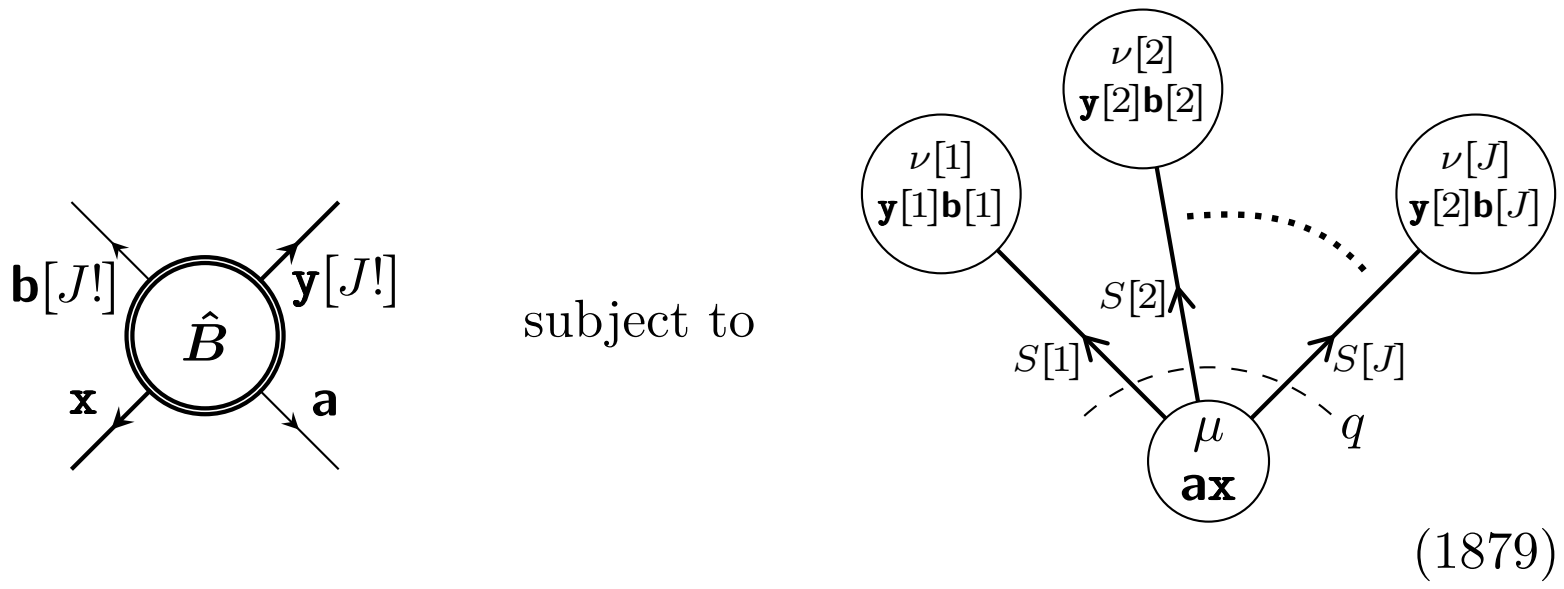

(1879)

If we assume the no correlation without causation assumption is true then the backward causality condition at $q$ is equivalent to the condition that we can write

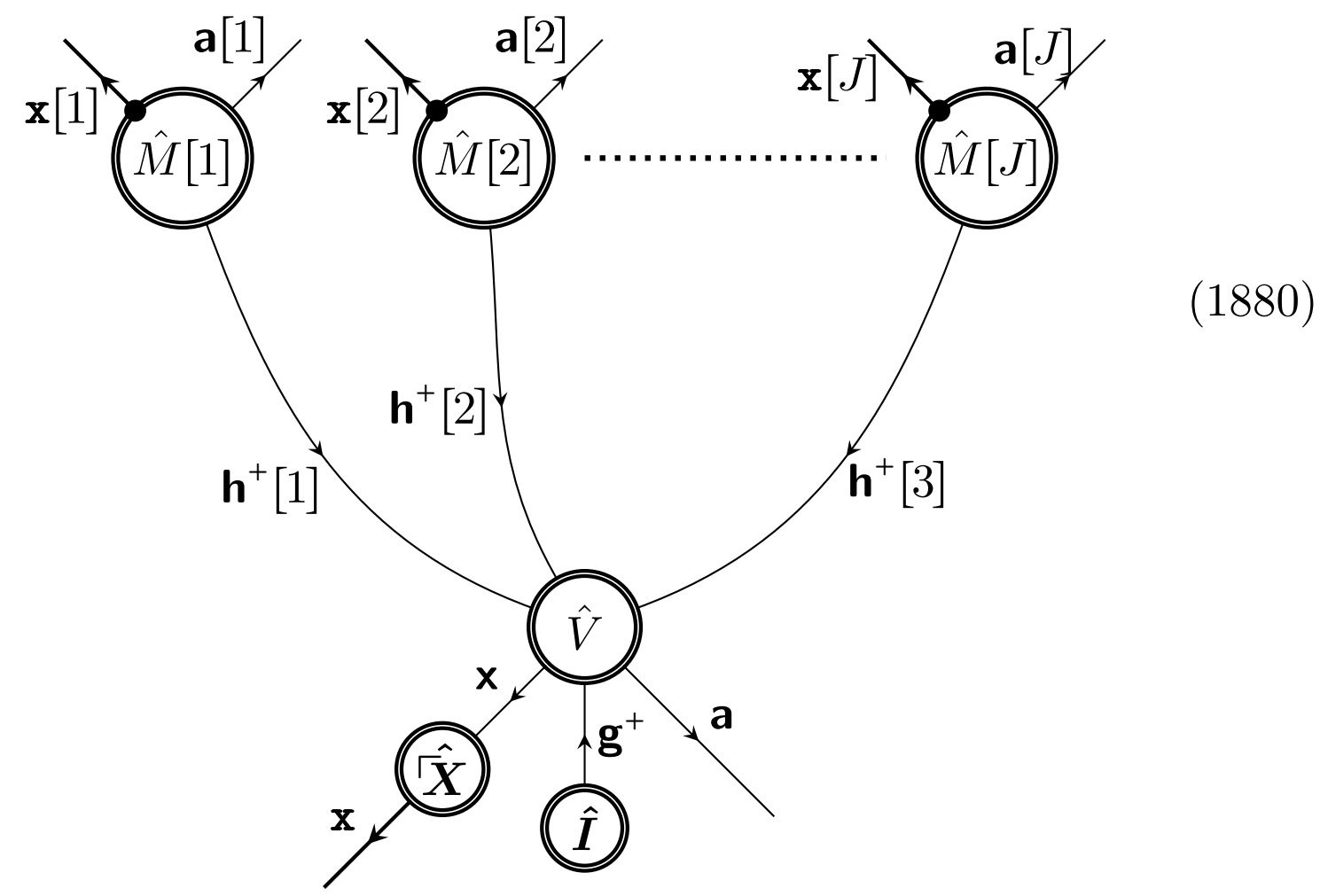

(1880)

where $\hat{V}$ is a natural temporal coisometry and $\hat{M}[j]$ are natural temporal maxometries for $j = 1$ to $J$. Furthermore, we can chose $\mathbf{g}^+$ such that $N_{\mathbf{g}^+} = \text{rank}({}_{|}\hat{\boldsymbol{B}}^{|})$ and $\mathbf{h}^+[j]$ such that $N_{\mathbf{h}^+[j]} = \text{rank}({}_{|}\hat{\boldsymbol{B}}^{|}[p_j^+])$ where ${}_{|}\hat{\boldsymbol{B}}^{|}[q_j^+]$ is the future residuum at $q_j$ (this being the synchronous partition passing only through the wires labelled with the set $S[j]$).

This follows by application of the coisometric part of the basic (co)isometric causal dilation theorem from Sec. 78.6.

We can apply either of the descendants dilation theorems above to the case where each of these descendants is a ladder such as in the following example

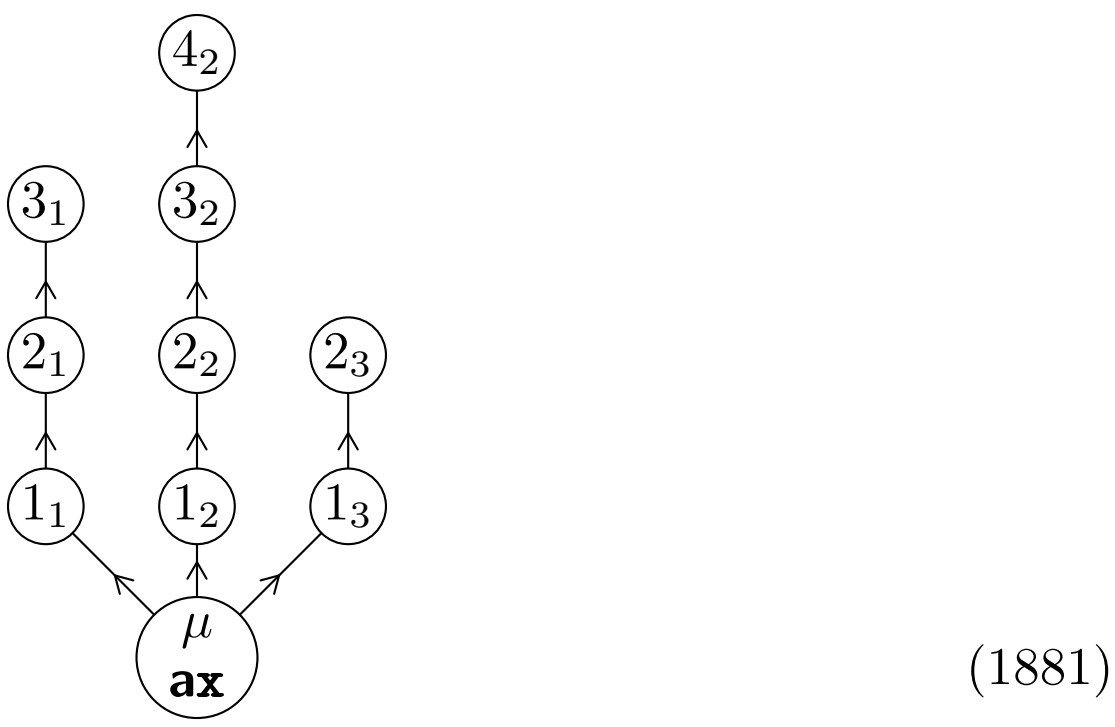

(1881)

where we are using the compact notation introduced in (1841). We call this the many ladder ancestors scenario. Each of the ladders could be dilated using one of the ladder dilation theorems. Thus, the above example can be dilated as

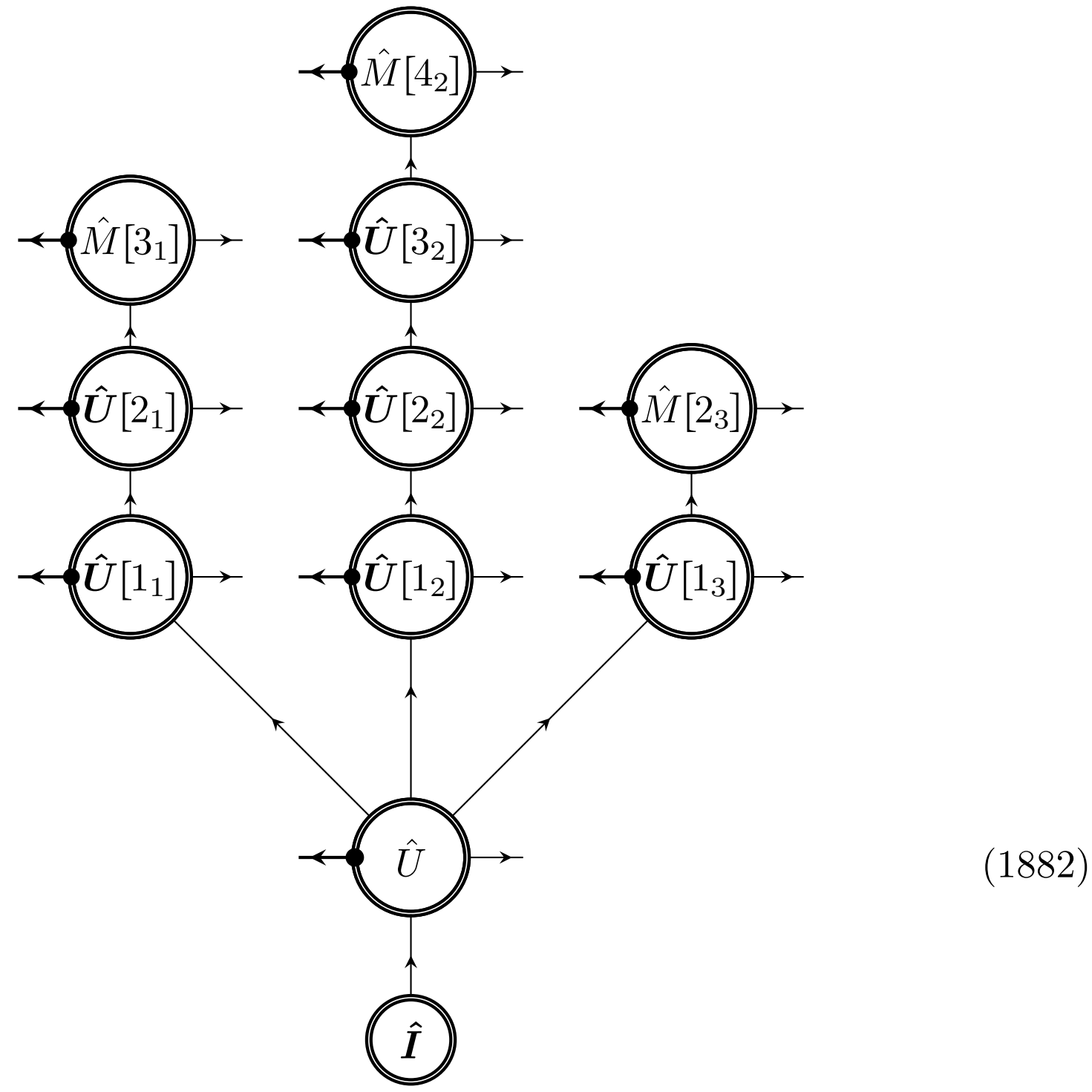

(1882)

(where we have omitted wire labels to prevent overcrowding and we have used the dot notation from earlier) To do this, we dilate the future residua, $\llcorner\hat{\boldsymbol{B}}^{\urcorner}[p_j^+]$ (which are deterministic and physical and subject to a ladder causal diagram), then use the corollary in Sec. 78.5 to obtain $\hat{M}[j]$ from these residua. This technique clearly generalises to arbitrary numbers of ladders, to the many ancestors scenario, and to (co)isometric dilations as well.

One point of comparison is the theorem of Renner and Wolf [2023] who prove a representation theorem for the situation where, if two operations acting on some shared system commute (meaning that the order in which they are applied does not matter) then this situation can be represented such that wherein the shared system splits into two systems with each operation acting separately on one of these. The causal diagram for this example is similar to the simplest version of the many descendants theorem. Furthermore, the authors of this work make a certain unitality assumption which, from our point of view, relates strongly to the assumption of double causality.

## 78.22 What can we say about general causal diagrams?

We have presented a zoo (really more of a jungle) of dilation theorems for different causal diagrams. In some cases we have given a sufficient pair of dilations that, taken together, are both necessary and sufficient for the given operation to be deterministic and physical. In some cases we have a necessary dilation, but have not proven sufficiency. We would, ultimately, like to have a single constructible dilation (rather than a sufficient pair) in terms of physical operations that is both necessary and sufficient dilations for operations with any given causal structure to be deterministic and physical. If we had this, we could assert that the given dilation is, in some sense, the true dilation that nature uses. However, this may not be possible - indeed, we have already seen some evidence that constructible dilations may not be possible even for simple causal diagrams.

In the absence of such dilations, it may be interesting to attempt to find causal dilations for operations having causal diagrams not discussed already. Indeed, here are a few diagrams that might be of interest

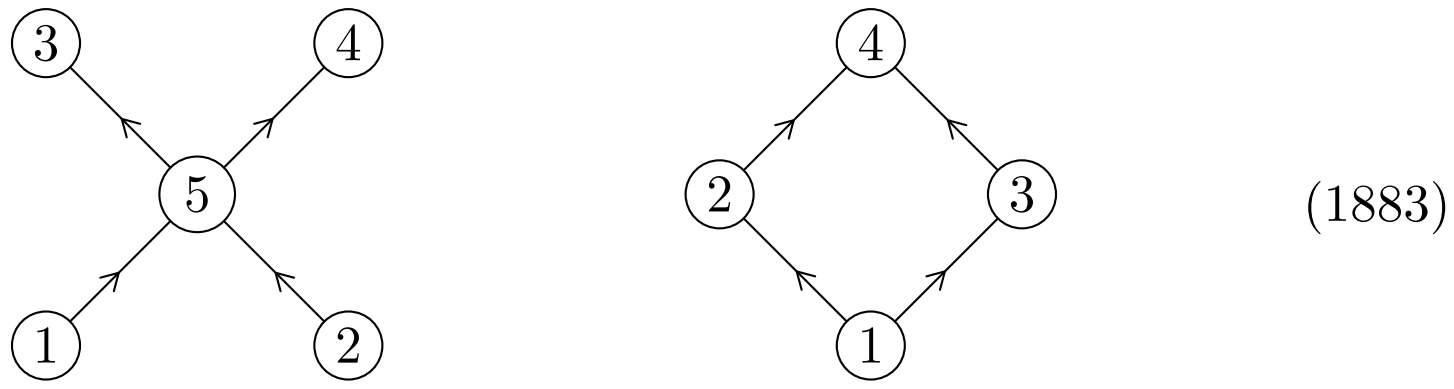

(1883)

It would also be interesting to look for causal dilations for lattice causal diagrams such as

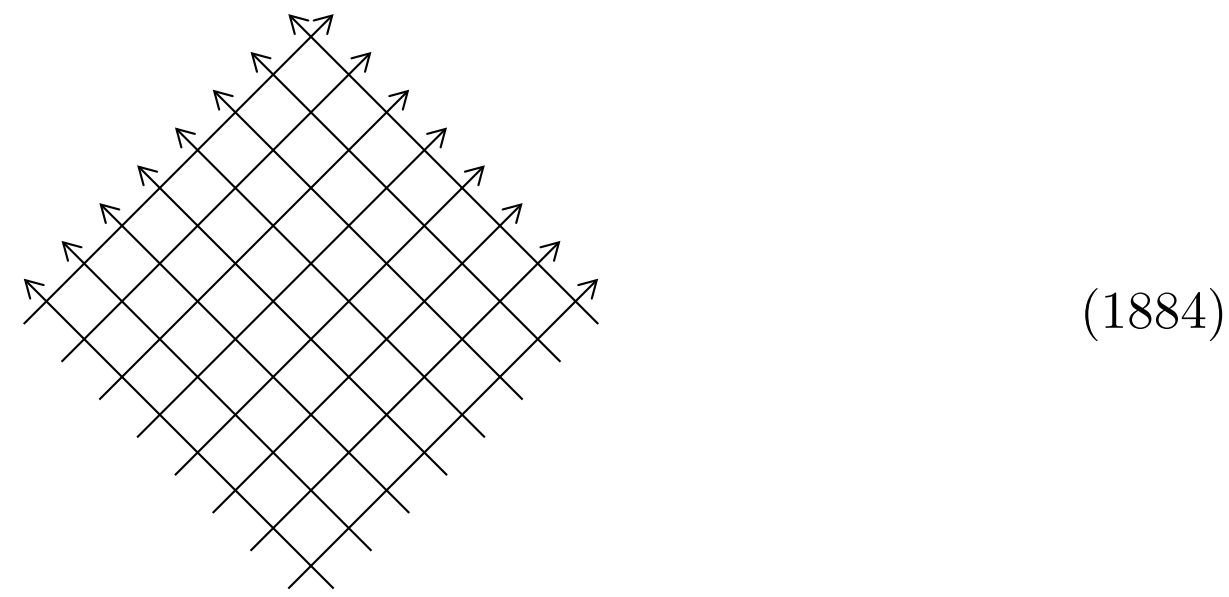

(1884)

where each vertex is understood to have a simple causal structure module as in (1841) (and we have not drawn all the arrow heads to keep the diagram tidy). since these are good models for spacetime.

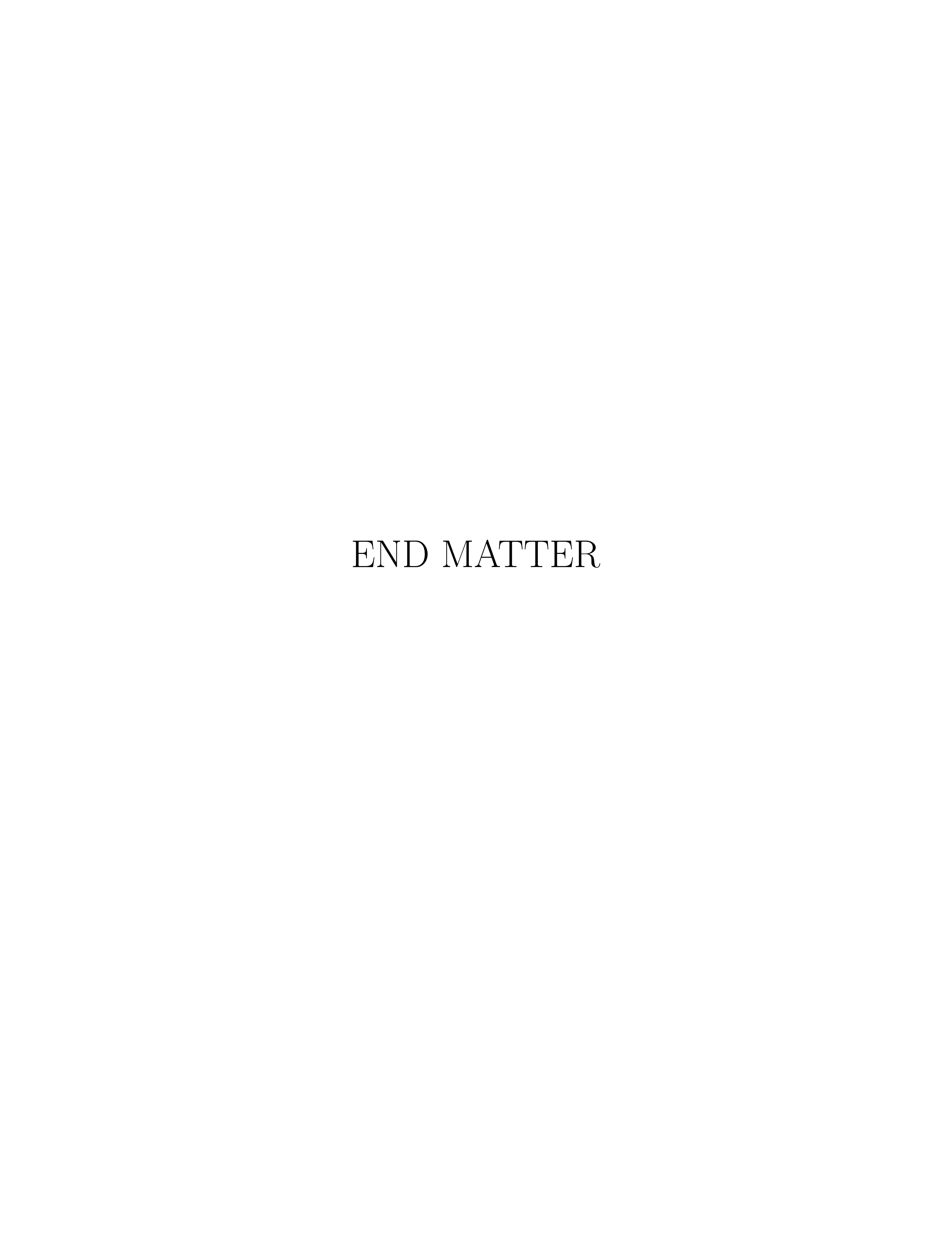

# END MATTER

# 79 Concluding remarks

This book was divided into a simple book and a complex book corresponding to dealing with operations having simple and complicated causal structure respectively. We saw, in each case, how to formulate Operational Probabilistic Theories (OPTs). We mostly take a time symmetric perspective, but also consider the time forward perspective (which is the standard point of view) and the time backward perspective. Networks are built out of operations in these theories. Circuits are networks with no open wires. In the complex case, operations come with equipped with causal diagrams which play an important role. Operations are taken to be physical meaning that they satisfied a certain positivity condition and certain causality conditions. Physicality guarantees that circuits have probabilities between 0 and 1 as well as guaranteeing certain causality conditions. In both the simple and the complex cases we were able to prove various composition theorems proving that, if each operation in a network is physical then the network itself is physical. These composition theorems become more interesting and more intricate in the complex case.

In the simple case were able to prove that the time symmetric perspective is equivalent to both the time forward and the time backward perspectives. In the complex case we were able to prove that the time forward (or the time backward) theory can be used to model the time symmetric theory. Whether the time symmetric theory can model the time forward (or time backward) theory remains an open question. In the complex case we invoke a Reichenbach-style principle that there is no correlation without causation. This principle, although very reasonable, does not play much of a role in the subsequent developments (except for proving the many ancestors and many descendants theorems at the very end of the book).

The complex OPT enables us to model physics in a discrete spacetime in terms of a set of left moving systems interacting with a set of right moving systems (this looks like a diamond lattice when viewed in $1+1$ dimensions). We view this set up as a discrete field theory (a precursor to Quantum Field Theory). Within this framework we are able to study Sorkin's impossible measurements and show that such measurements are ruled out by physicality.

Operational Quantum Theory (OQT) can be modelled within these operational frameworks through the machinery of operator tensors (with duotensors mediating). In particular, there is a correspondence theorem which allows us to cleanly take results from the OPT over to the corresponding OQT. This includes the physicality conditions. This allows us to state axioms for QPT simple and complex cases. The axioms in the simple case are equivalent to the usual axioms of Operational Quantum Theory found in textbooks. The axioms in the complex case have not been well "road tested" and we might ask whether they are sufficient. In fact, there are two questions we can ask. First, if it turns out, as mentioned above, that the time symmetric case cannot model the time forward (or time backward) case, then which theory should we adopt as the fundamental one? The inclination of this author is to take the time symmetric theory as fundamental. The second question is whether the principle of no correlation without causation, just mentioned, should be added as an axiom or whether it is actually derivable from the existing axioms.

We then set up the diagrammatic theory of Hilbert objects (based on the process theory work of Coecke and others). We adopted and adapted this for the present work introducing new notation. In the complex case we introduce new elements into this notation to enable "toggle" and "flip". We introduced conjupositions of hypermatices. These were used to set up normal and natural conjuposition groups acting on the Hilbert objects (resulting in Hilbert cubes). We also introduced mirror notation associated with each of these groups. Further, we stated mirror theorems which provide a way to manipulate between equivalent equations. Within the framework of Hilbert objects we introduced normal and natural isometries, normal and natural coisometries, normal and natural unitaries, and normal and natural maxometries. The natural objects, in each case, are more physically motivated. Importantly, the time reverse operation is the natural transpose (rather than the normal transpose). Natural maxometries become necessary in the time symmetric formalism and play a role in dilation theorems.

The last part of both the simple and the complex book considered dilation theorems. In the simple case we provide a time symmetric version of the well known Stinespring dilation. Interestingly, this involves a maxometric operator instead of a unitary operator as in the standard Stinespring dilation. Maxometric operators are not, in general, physical. This raises the question of whether such dilations are, in general, constructible or unconstructible. This remains an open question. If the answer is that they are unconstructible (meaning they do not, in general, provide a prescription to build the given operation out of basic elements in the laboratory) then we break the peaceful coexistence between the Church of the Larger Hilbert Space and the Church of the Smaller Hilbert Space in the time symmetric setting. Resolution of such theological questions might be important for setting the correct attitude for developing new fundamental theories. In the complex book we proved a number of causal dilation theorems. However, we were unable to prove a general dilation theorem. What we would really like is a theorem proving a necessary and sufficient condition for dilations that works for any causal diagram. Furthermore, we would like to resolve whether such dilations are constructible or not.

There have been many reconstructions of Quantum Theory in terms of operationally motivated axioms or principles (here is a small selection – Hardy [2001], Dakic and Brukner [2009], Masanes and Müller [2011], Chiribella et al. [2011], and Barnum et al. [2020]). These reconstructions have taken a time asymmetric point of view (a notable exception is that of Jia et al. [2018]. A central contention of this book is that the time symmetric perspective is more fundamental. If that is the case, then a reconstruction should mirror this. Indeed, we already have a number of principles that narrow us down to a time symmetric OPT. The question is, are there some reasonable axioms we can add to this to get Quantum Theory? We might also ask whether it is more natural to reconstruct the full complex Operational Quantum Theory which has a closer connection to Quantum Field Theory.

In this book we introduce tomographic locality early on so we can set up the theory of duotensors which, in turn, allows us to set up correspondence between OPTs and OQTs. However, it is not clear whether tomographic locality is ultimately a property nature will respect. In particular, D'Ariano et al. [2014] showed that fermionic theories do not respect this principle. To this end, a natural and broader framework for tomographically nonlocal OPTs and OQTs is required.

An obvious question, in a book on time symmetry, is where does the time asymmetry in our experience of the world come from? We refined this into the question of where does strong preselection come from. We provided some ideas on how to resolve this. One of those ideas is that the coalescence of humans and experimental apparatusses from basic constituents is, contingently, necessary for us to set up our operational approach. Thus, the probabilities we consider are necessarily conditions on this coalescence introducing an apparent time asymmetry. It remains as an open problem to describe the detail of such a coalescence mechanism which actually provides strong preselection without throwing the time symmetric "baby" out with the bath water. This coalescence picture suggests that the world might, in fact, always be in a maximally mixed state but with a view from the inside allowing an operational description that accounts for our experience of the world. If such a view is tenable, it would be interesting to attempt to account for the second law of thermodynamics. Fundamentally, the entropy would always be equal to its maximum. However, we could attempt to account for the second law in a contingent way using this view from the inside.

As mentioned above, we applied the tools in this book to model systems in spacetime. This provides a way to think about Quantum Field Theory, though only within a kind of toy model (where the spacetime is effectively discrete and all the Hilbert spaces are finite). An important goal is to find ways to take the limit of continuous spacetime and infinite dimensional Hilbert spaces and use these to model actual Quantum Field Theories. One glimmer of hope in this direction is that the ignore preparation and result, central to the time symmetric perspective, have inner product equal to 1 even as we take the limit to an infinite dimensional Hilbert space (rather than being unbounded if we take the inner product of identity operators).

This project was motivated by the search for a theory of Quantum Gravity. As the framework developed here stands, there are a number of hurdles in the way of such an application. Perhaps the most immediate is that causal diagrams, as treated in this book, provide for a fixed background. To treat the problem of Quantum Gravity we would want to relax this. This is an interesting direction for further research.

# Appendices

## A The Hilbert hypercube

The Hilbert cube plays a role in this work. We can go further and define a *Hilbert hypercube* for arbitrary dimensions. Such hypercubes appear as a mathematical curiosity - it is not clear what role they might play in formulating Quantum Theory or doing calculations in the theory. Correspondingly, this appendix is an aside.

To set up a $4D$-hypercube we choose another set of bases for a and b (let these bases be shaded as in (498)). We can use these to define new twist objects (we will use a prime to denote objects formed under this new twist object below). We can use these new twist objects find a new shadow of the original Hilbert square (call this $s'HS$) and a shadow of the old shadow Hilbert square (call this $s'sHS$). These two new squares can form their own cube. This new cube along with the original cube can be used to form a $4D$-hypercube as follows

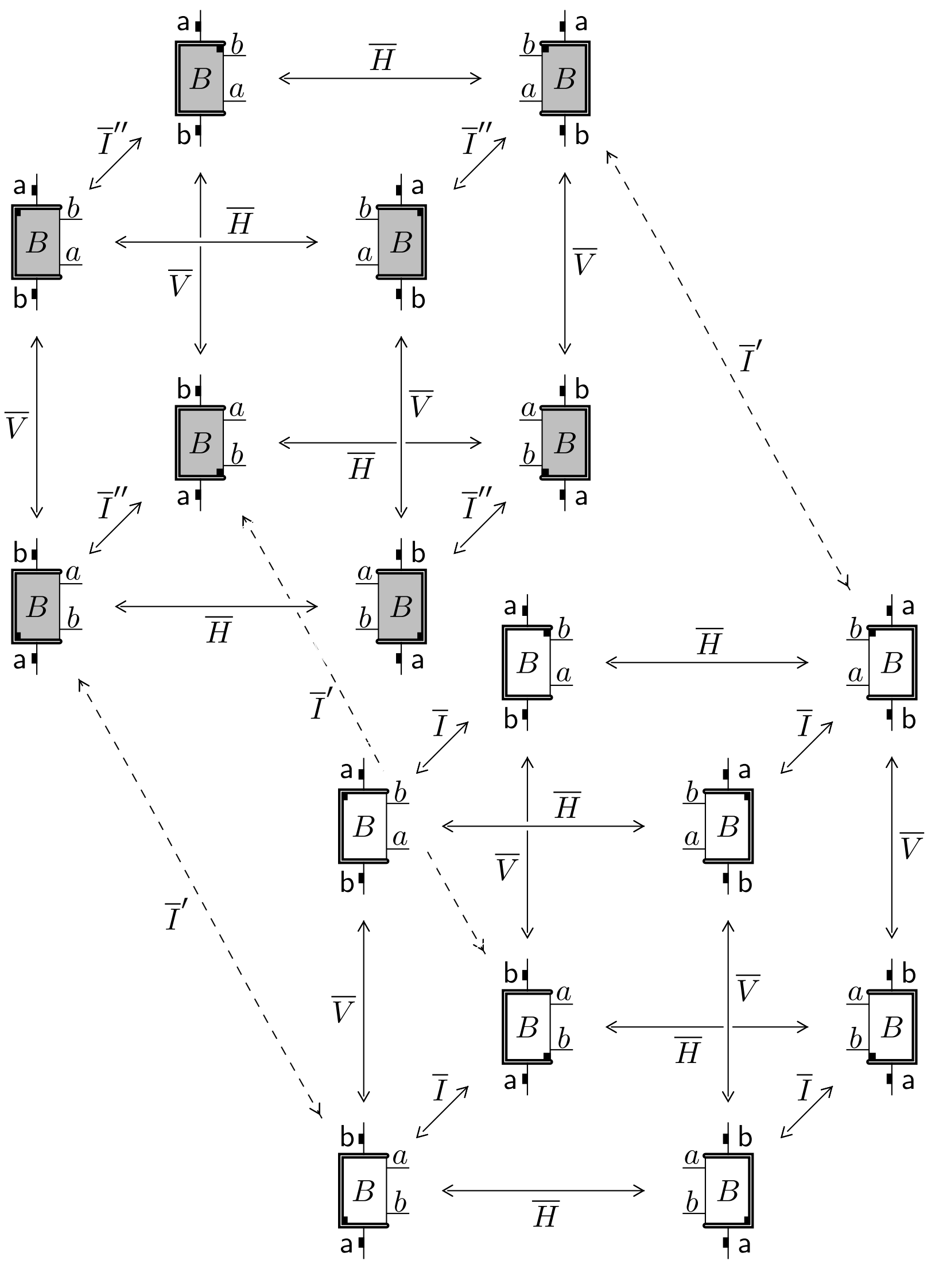

$$(1885)$$

The elements of the new cube are shaded. We have shown some of the $\overline{I}'$ transformations connecting the old and new cubes (these corresponding to conjugation in the new basis) but have omitted most transformations to prevent the diagram becoming overcrowded. At this stage we can see that the elements of this hypercube are not necessarily transformed into one another by a group of transformations. This is evident as each vertex in the old cube has $\{I, \overline{I}, \overline{I}'\}$ (plus transformations involving $H$, $V$, and $T$) whilst each vertex in the new cube has $\{I, \overline{I}'', \overline{I}'\}$ (plus transformations involving $H$, $V$, and $T$). To force that the set of transformations is a group we must have the same transformations at each vertex. This means we need $\overline{I}'' = \overline{I}$. Noting that $\overline{I}'' = \overline{I}'\overline{I}\overline{I}'$ (we can read this off the diagram) we must have $\overline{I} == \overline{I}'\overline{I}\overline{I}'$. Staring at the hypercube in (1885) above we can see that there is a similar requirement that $H = H'HH'$. The $H$ operation is directly implemented by the twist object (for example (579)). Thus, we require

(1886)

with similar constraints for b. We have, of course, used shading to indicate the twist objects with respect to the shaded basis. These are equivalent to the constraints

(1887)

So the original and new twist objects must commute. These constraints guarantee that $H'' = H'HH' = H$. This fixes the other conditions - that $\overline{I}'' = \overline{I}$, $V'' = V$, and $\overline{T}'' = \overline{T}$. To see this note $\overline{I}'' = H''\overline{H}$. Therefore $H'' = H$ implies $\overline{I}'' = \overline{I}$. Thus, if the twist objects commute, we have a group of transformations associated with this hypercube.

Is it possible to construct nontrivial Hilbert hypercubes? By nontrivial we require that the twist objects are distinct. To answer this note that

(1888)

using (501). For the original and new twists to be distinct we require

(1889)

Further, if we insert (1888) into (1886) we obtain the condition

$$\overset{a}{—}\boxed{\bullet U}\overset{a}{—}\boxed{U\bullet}\overset{a}{—}\boxed{\bullet U}\overset{a}{—}\boxed{U\bullet}\overset{a}{—} \quad = \quad \overset{a}{———} \tag{1890}$$

If our systems are qubits then one way to satisfy (1889) and (1890) is to set

$$\overset{a}{—}\boxed{\bullet U}\overset{a}{—} \quad = \quad \begin{pmatrix} 0 & i \\ i & 0 \end{pmatrix} \tag{1891}$$

This unitary performs a $\frac{\pi}{2}$ rotation in the $xy$ plane (where the $z$ axis is aligned with the basis). It is equal to its own transpose. If we apply it twice we get a $\pi$ rotation which is not equal to the identity transformation (so (1889) is satisfied). If we apply it four times then we get a $2\pi$ rotation which is the identity (so (1890) is satisfied).

We see it is straightforward to construct a 4D Hilbert hypercube. We could attempt to form higher dimensional Hilbert hypercubes by choosing yet more sets of bases which provide additional twist objects which we can use to generate additional cubes. We require pairwise commutation between the twist objects introduced so we have the requisite group structure. If we enlarge the dimension of the Hilbert spaces for our systems then we can do this by using block diagonal unitaries where each new twist rotates through $\frac{\pi}{2}$ in a different two-dimensional subspace. This requires Hilbert space dimensions equal to $2(d-3)$ where $d$ is the dimension of the Hilbert hypercube (to provide two new dimensions of Hilbert space for each new dimension of the hypercube). It is not clear whether there exist more interesting examples. For example, is it possible to build a nontrivial $d$-dimensional Hilbert hypercube using Hilbert spaces of dimension less than $2(d-3)$?

# B Condition for projected normal unitary to be a (co)isometry

In Sec. 39 we showed how normal isometries and normal coisometries can be written as projected normal unitaries. However, not all projected normal unitaries are isometries or coisometries. Here we prove a theorem indicating the required form for a projected normal unitary to be an normal isometry and/or normal coisometry. It is straightforward to adapt this proof for the natural (rather than normal) case.

**Condition for projected normal unitary to be a normal**

**(co)isometry.** The projected normal unitary

(1892)

(where $U$ is a normal unitary and $K$ and $K$ are normalised), is a normal isometry iff

(1893)

and a normal coisometry iff

(1894)

It is easy to prove sufficiency. Consider sufficiency of the condition in (1893) for a normal isometry (the argument for a normal coisometry is similar). We have

(1895)

We can verify the first equality by substituting in the expression in (1893) (and its vertically flipped version). The second equality follows from the normal unitarity of $U$. Equality of the first expression with the last is the normal isometry condition for the expression in (1893). To prove necessity takes a little more work. We will prove the normal isometric case. The normal coisometric case follows by similar arguments. First note that we can always write

(1896)

by expanding in terms of a basis for the s system. Further, we are free to choose this basis such that

(1897)

Substituting (1896) into (1892) and using (1897) we obtain

(1898)

Unitarity of $U$ entails

(1899)

(using the orthonormality property of the shaded bases). Isometry of $V$ entails

(1900)

Hence

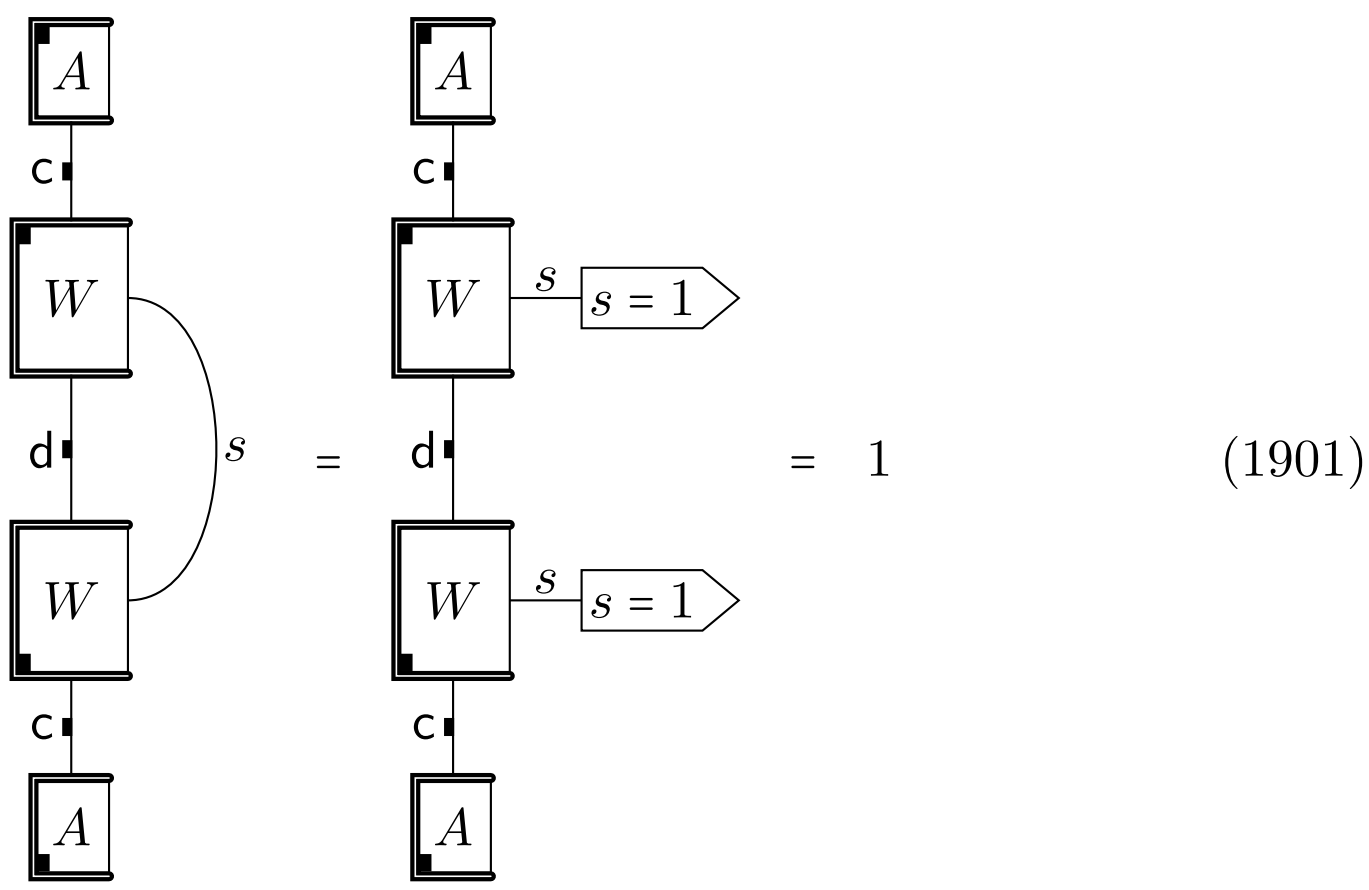

for all normalised $A$. The left most expression is a sum of over $s$ of non-negative terms (that the terms are positive follows from the (783)). The expression in the middle is equal to just one of these terms (for $s = 1$). Hence, all the terms for which $s \neq 1$ must actually be equal to zero. Since this is true for all $A$, this implies that

d
W
s
$s \neq 1$
=
d
0
c
c

(1902)

where the 0 box on the right is the zero element of the $\mathcal{H}_{\mathsf{c}_1} \otimes \mathcal{H}^{\mathsf{c}_2}$ Hilbert space. Hence, the expression in (1896) is necessarily equal to (1893).

It is worth noticing that the form for the normal isometric case in (1893) appeared as the step in (882) in the proof of the little theorem in Sec. 39.1.

# C Natural maxometries as projected natural unitaries

Now we prove that natural maxometries can be written as projected natural unitaries using the results of Sec. 39.

**Natural maxometries as projected natural unitaries.** We can write any natural maxometry as

(1903)

where $\mathsf{g}$ and $\mathsf{h}$ are the ancillae and $U$ is a natural unitary.

To prove the theorem we use the first of the conditions in (913) which says that if we send half a maximally entangled state (represented by a twisted special cup) into the input ancilla of $M$ we obtain a natural isometry. We can then apply the theorem in Sec. 40.2 which says that natural isometries can always be written as projected natural unitaries. We obtain

(1904)

Hence

(1905)

where we have used the special cup and cap yanking equation (688). The object in the dashed box is a natural unitary. Hence we have shown that $M$ can be written as in (1903). This proves the theorem. Note that we have only used the natural isometry property required of natural maxometric objects (the first condition in (913)). We could go through a similar argument using the natural coisometry property instead (the pictures would be vertically flipped). This would also show that the natural maxometry can be written in projected form.

While all natural maxometries are projected natural unitaries, the converse is not true (even when normalisation is taken into account).

**Not all projected natural unitaries are proportional to natural maxometries.** The projected natural unitary

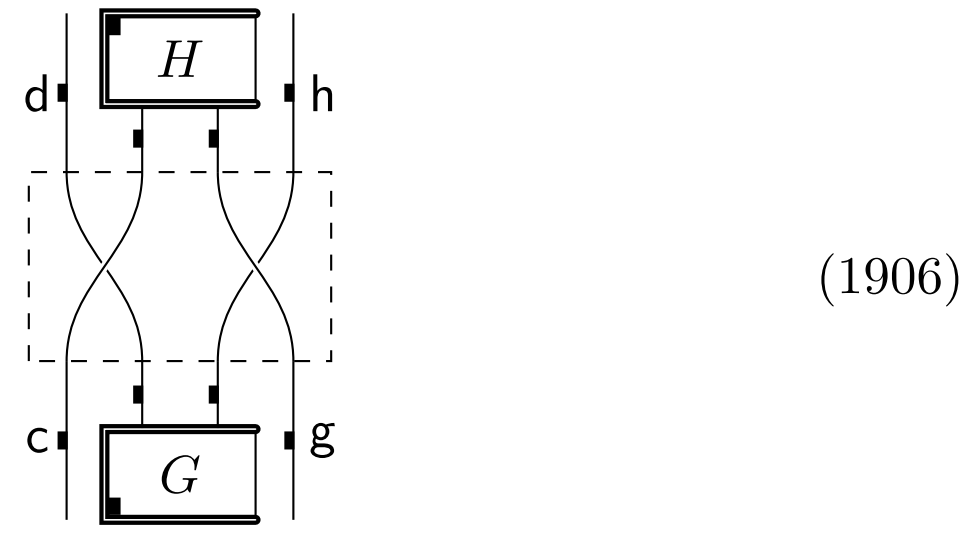

(1906)

is not a natural maxometry with respect to the ancillae g and h except for special choices of $G$ and $H$.

First note that it is clear the expression in (1906) is a projected natural unitary since the object inside the dashed box is a natural unitary. Now, in fact, this example is just example (ii) in (968). This example is only a natural maxometry if the conditions (969, 970) are satisfied. In general, if $G$ and $H$ are not maximally entangled in an appropriate sense then these conditions will not be satisfied (even if we rescale $G$ and $H$). This proves the above theorem.

# Acknowledgements

I gave a talk titled "Causality constraints in Quantum Field Theory from an operational perspective" at the conference *Mathematical and Conceptual Aspects of Quantum Theory* organised by Robert Oeckl in June of 2022. This book started out life as an attempt to write a short (~ 35 pages) paper based on that talk. That, clearly, turned out to be impossible. At a certain point I had to admit that this was a book rather than a paper. Anyway, I am very grateful to Robert Oeckl for inviting me to that conference and for many discussions over the years that have clearly influenced this work. I am grateful to Andrea Di Biagio for a talk he gave on time symmetry over Zoom during the early days of the Covid pandemic and to others involved in that discussion – Carlo Rovelli, Flaminia Giacomini, Doreen Fraser, Francesca Vidotto, and Laurent Freidel. I have benefitted from conversations with Mauro D'Ariano, Paulo Perinotti, and Giulio Chiribella whose work has played a big role in motivating ideas in the present work. I am grateful to Bob Coecke for numerous discussions and for writing the book Picturing Quantum Processes with Aleks Kissinger which helped me gain some understanding of the picturalism program (which is fully embraced in this book). Having written (though not yet published) the present book, I am beginning to gain some insight into what kind of endeavour this must have been.

I am, further, grateful to Ognyan Oreshkov, Samson Abramsky, Tein van der Lugt, Elie Wolfe, Huw Price, Adrian Kent, Nicholas Ormrod, Roberto Baldijao, Neil Turok, David Schmid, Robert Spekkens, Chris Fuchs, Nathan Argaman, Jonathan Barrett, Časlav Brukner, Sandu Popescu, Ding Jia, Nitica Sakharwade, Lee Smolin, Rafael Sorkin, Ian Jubb, Matthew Leifer, and very many others for discussions related to time (and other topics) at one time or another. I am grateful for Luca Mrini for pushing me to provide a full proof of equivalence between the time symmetric and time forward cases. I am grateful to the people at OpenAI for ChatGPT which I used to help with compiling citations to work in the literature and even more grateful to Yìlè Yīng for pointing out errors in these references.

A book like this is written in space and time (and with caffeine). The spaces included many cafes at which I spent a lot of time and drank very many cappuccinni. I am grateful to the people at Covenant Cafe, Seven Shores, Cafe O, Market Cafe Elmira, the Cafe Nero at Terminal 2 arrivals lounge at Heathrow airport, and many Tim Horton's dotted around the Kitchener-Waterloo region. I also have not forgotten the Aroma Cafe on Erb Street West, or Casa Mia Deli and Cafe on Father David Bauer Drive, in which places the foundations of much of this work was laid. I am also grateful to Gabe and Ruth at WholeTone for maintaining a table I could work at whilst my kids had music lessons. A favourite place of mine to work is the Upper Bistro at Perimeter (where I am right now). I am grateful to Dan Lynch for showing me how to make my own cappuccino using the machine here at Perimeter. Thank you to all the Bistro staff, without whom Perimeter would not be the magical place it is.

Talking of Perimeter, I am deeply grateful to Mike Lazaridis for the vision and drive to set up this institute in the first place (and donating a substantial part of his own personal fortune to make it happen). Interestingly, the original seed for this institute was a lecture in 1982 by Lynn Watt at the University of Waterloo about the experiment of Alain Aspect et al. confirming the theory of John Stewart Bell that nature is nonlocal. At a recent event celebrating 25 years of Perimeter, Mike said "We were astounded. At first we had a hard time believing what Professor Watt was saying about what it implied." Anyway, here we are some 44 years after Lynn's lecture still trying to make sense of it all. I am also grateful to Howard Burton who, as founding director, pulled off the miracle of conjuring this institute into existence.

I am very grateful to Zivy for her support throughout this project, and for pointing out, with simple everyday examples, that the world is not time symmetric. I am trying to deal with this problem (see the bit on coalescence in this book). My worry is that this might lead to a radical many worlds ontology - something she dislikes even more viscerally than I do (don't worry, I don't believe it for a moment). I am grateful to Vivienne and Helen for whom my writing of this book has constituted a large part of their childhood. I am looking forward to their mathematical and philosophical feedback on this work.

Research at Perimeter Institute is supported in part by the Government of Canada through the Department of Innovation, Science and Industry Canada and by the Province of Ontario through the Ministry of Colleges and Universi-ties This work was made possible through the support of grants from the John TempletonFoundation for the “The Quantum Information Structure of Spacetime (QISS)” project and the “Without SpaceTime (WOST)” project.